\documentstyle[color,amsfonts,amsmath,amssymb,mathrsfs,verbatim,11pt,epsf]{report}
\def\inn{\in}
\def\innf{\in}

\def\ccdot{ \!\! \mbox{{\raisebox{0.5ex}{\makebox[1em]{$\centerdot$}}}} \!\!}
\def\sigsum{\mbox{{\Large $\sum$}$ \!\!\!\!\!\!  {\raisebox{-2.4ex}{\makebox[1em]{\scriptsize $r\neq 
1,q$}}} \; $}}
\def\pqg{\mbox{$ \!\! \mbox{\Large \_} \! $}}
\def\pqo{\mbox{$ \!\!\!\!\; \mbox{\textbf {\Large \_}} \! $}}
\def\sp{\; \!}
\def\sb{\;\;}
\def\kl{k \pqo l}
\def\jl{j \pqo l}
\def\coha{\cos\frac{\alpha}{2}}
\def\siha{\sin\frac{\alpha}{2}}
\def\il{i \mbox{$ \!\!\!\!\; \:\! \mbox{\textbf{\Large \_}} \! \!\!\; $} l}
\def\scirc{\mbox{\raisebox{0.2ex}{\scriptsize {$\circ$}}}}
\def\sscirc{\mbox{\raisebox{0.12ex}{\tiny {$\circ$}}}}
\def\TM{\mbox{\it TM}}
\def\FM{\mbox{\it FM}}

\def\cont{\ulcorner \!\! \mbox{\raisebox{0.48ex}{\textbf{--$\!$---$\!$--}}} \!\!\!\urcorner}
\def\contd{\!\!\ulcorner \!\! \mbox{\raisebox{0.48ex}{\textbf{--$\!$---$\!$--}}} \!\!\!\urcorner}
\def\conte{\ulcorner \!\! \mbox{\raisebox{0.48ex}{\textbf{---$\!$--$\!$---}}} \!\!\!\urcorner}
\def\Gbar{G\!\!\!\!\!\!\!\; \mbox{{\raisebox{0.3ex}{\small{--}\,}} }}
\def\Sbar{S\mbox{$\!\!\!\!\!\: \mbox{{\raisebox{0.3ex}{\small{--}\!\!\small{--}\!\!}} }$}}
\def\Sbard{\dot{S}\mbox{$\!\!\!\!\!\: \mbox{{\raisebox{0.3ex}{\small{--}\!\!\small{--}\!\!}} }$}}
\def\Sbart{\frac{{\dot{S}\!\!\!\!\!\!\;\:\! \mbox{{\raisebox{0.0ex}{\small{--}\!\!}} }}^2_l}{2}}
\def\bGbar{\mbox{\boldmath $G$}\!\!\!\!\!\!\!\!\; \mbox{{\raisebox{0.3ex}{\textbf{\small{--}}\,}} }}
\def\rrr{{\mathbb{R}}}
\def\ccc{{\mathbb{C}}}
\def\hhh{{\mathbb{H}}}
\def\ooo{{\mathbb{O}}}
\def\kkk{{\mathbb{K}}}

\def\zzz{{\mathbb{Z}}}
\def\kkk{{\mathbb{K}}}

\def\b1{\mbox{\boldmath $1$}}
\def\v{\mbox{\boldmath $v$}}

\def\ba{\mbox{\boldmath $a$}}

\def\bh{\mbox{\boldmath $h$}}
\def\bk{\mbox{\boldmath $k$}}
\def\bc{\mbox{\boldmath $c$}}
\def\bp{\mbox{\boldmath $p$}}
\def\bsl{\boldsymbol}
\def\omp{\omega_{\bsl{p}}}

\def\bq{\mbox{\boldmath $q$}}
\def\br{\mbox{\boldmath $r$}}
\def\bv{\mbox{\boldmath $v$}}
\def\bvh{\hat{\bv}}

\def\bx{\mbox{\boldmath $x$}}
\def\by{\mbox{\boldmath $y$}}
\def\bu{\mbox{\boldmath $u$}}
\def\bU{\mbox{\boldmath $U$}}

\def\chc{\check c}
\def\che{\check e}
\def\chg{\check g}

\def\bB{\mbox{\boldmath $B$}}

\def\bE{\mbox{\boldmath $E$}}
\def\bR{\mbox{\boldmath $R$}}
\def\bT{\mbox{\boldmath $T$}}

\def\bWW{\mbox{\boldmath $W$}}

\def\bsig{\boldsymbol {\sigma}}
\def\bvare{\boldsymbol {\varepsilon}}
\def\lv{L(\v)=1}
\def\lvh{L(\hat{\v})=1}

\def\dmlvh{D_{\mu}L({\hat{\v}}) = 0}
\def\lvth{L(\v_3)=1}
\def\lvf{L(\v_4)=1}
\def\lvfh{L(\v_4) =h^2}

\def\lvfi{L(\v_5)=1}
\def\lvte{L(\v_{10})=1}
\def\lvt{L(\v_{27})=1}
\def\lvfs{L(\v_{56})=1}
\def\lvtfe{L(\v_{248})=1}
\def\lvn{L(\v_n)=1}

\def\htwc{\mbox{h}_2\ccc}
\def\hthc{\mbox{h}_3\ccc}
\def\htwh{\mbox{h}_2\hhh}

\def\htwo{\mbox{h}_2\ooo}
\def\htho{\mbox{h}_3\ooo}
\def\htwk{\mbox{h}_2\kkk}

\def\glnrp{\mbox{GL}^+(n,\rrr)}
\def\glnra{\mbox{gl}(n,\rrr)}
\def\glmr{\mbox{GL}(m,\rrr)}
\def\glmra{\mbox{gl}(m,\rrr)}
\def\glfr{\mbox{GL}(4,\rrr)}
\def\glfra{\mbox{gl}(4,\rrr)}
\def\glsr{\mbox{GL}(7,\rrr)}
\def\glfrp{\mbox{GL}^+(4,\rrr)}

\def\gltr{\mbox{GL}(27,\rrr)}

\def\VCKM{V_{\mbox{{\scriptsize CKM}}}}
\def\trec{t_{\mbox{{\scriptsize rec}}}}
\def\tinf{t_{\mbox{{\scriptsize inf}}}}
\def\marrow{\mbox{
{\raisebox{-1.0ex}{ 
$\stackrel{\mbox{\small{$\nearrow$}}}{\mbox{\small{$\searrow$}}}$
}}
}\!\!\!\!\!\!\!\!\!\!\!\to}

\def\btzt{\dot{B}_{t \pqg z}^{1} + 2\dot{B}_{t \pqg z}^{2}}

\def\sltc{\mbox{SL}(2,\ccc)}
\def\sltca{\mbox{sl}(2,\ccc)}
\def\sltwoh{\mbox{SL}(2,\hhh)}
\def\slthc{\mbox{SL}(3,\ccc)}
\def\slthca{\mbox{sl}(3,\ccc)}
\def\sltho{\mbox{SL}(3,\ooo)}

\def\sltwoo{\mbox{SL}(2,\ooo)}
\def\sutwoo{\mbox{SU}(2,\ooo)}
\def\suthoo{\mbox{SU}(3,\ooo)}
\def\slthoa{\mbox{sl}(3,\ooo)}
\def\sofi{\mbox{SO}(5)}
\def\sofia{\mbox{so}(5)}
\def\sox{\mbox{SO}(6)}
\def\soth{\mbox{SO}(3)}
\def\sotha{\mbox{so}(3)}
\def\sotw{\mbox{SO}(2)}

\def\uo{\mbox{U}(1)}
\def\uoa{\mbox{u}(1)}
\def\doo{\mbox{D}(1)}
\def\sutw{\mbox{SU}(2)}
\def\sutwa{\mbox{su}(2)}
\def\suth{\mbox{SU}(3)}
\def\sutha{\mbox{su}(3)}
\def\spot{\mbox{Spin}^+(1,3)}
\def\spotn{\mbox{Spin}^+(1,9)}

\def\soot{\mbox{SO}^+(1,3)}
\def\soota{\mbox{so}^+(1,3)}
\def\sootf{\mbox{SO}^+(1,5)}
\def\sootn{\mbox{SO}^+(1,9)}

\def\olg{\mbox{L}_{+}^{\uparrow}}
\def\gt{\mbox{G}_2}
\def\ff{\mbox{F}_4}
\def\ee{\mbox{E}_8}
\def\eeg{\mbox{E}_{8(-24)}}
\def\ese{\mbox{E}_7}
\def\eseg{\mbox{E}_{7(-25)}}
\def\esi{\mbox{E}_6}
\def\esig{\mbox{E}_{6(-26)}}

\def\SM{\mbox{SU}(3) \times \mbox{SU}(2) \times \mbox{U}(1)}
\def\SML{\mbox{SU}(3)_c \times \mbox{SU}(2)_L \times \mbox{U}(1)_Y}
\def\stab{\mbox{Stab}(\TM_4)}
\def\stabse{\mbox{Stab}_7(\TM_4)}
\def\stabee{\mbox{Stab}_8(\TM_4)}
\def\stabto{\mbox{Stab}_2(\TM_4)}
\def\staba{\mbox{stab}(\TM_4)}

\def\thetmt{\theta_{\! M^2}}
\def\thetmth{\theta_{\! M^3}}
\def\thX{{\theta^{1}_{\! \mbox{\tiny{$X$}}}}}
\def\thY{{\theta^{1}_{\! \mbox{\tiny{$Y$}}}}}
\def\phX{{\phi^{1}_{\! \mbox{\tiny{$X$}}}}}
\def\psX{{\psi^{1}_{\! \mbox{\tiny{$X$}}}}}

\def\teta{e^{a}_{\phantom{i}\mu}(x)}
\def\tetu{e^{\mu}_{\ph{\mu}a}(x)}

\def\md{\mbox{d}}
\def\mD{\mbox{D}}
\def\past{{}^{\,\ast\!}}
\def\ph{\phantom}
\def\pho{\phantom{\mu}}
\def\pht{\phantom{\mu\nu}}
\def\tem{T^{\mu\nu}_{\mathrm{em}}}
\def\tef{T^{\mu\nu}_{\!\epsilon}}
\def\gmo{G^{\mu\nu}_{\pht \! ;\mu}=0}
\def\tmo{T^{\mu\nu}_{\pht \! ;\mu}=0}
\def\gmyv{G_{\mu\nu} = f(Y,\bvh)}
\def\dmo{D_{\mu}L(\bv_{27})=0}
\def\dmofs{D_{\mu}L(\bv_{56})=0}
\def\dmoh{D_{\mu}L(\bvh)=0}
\def\lag{{\mathcal L}}

\def\mcM{{\mathcal M}}
\def\mcT{{\mathcal T}}
\def\mcX{{\mathcal X}}
\def\mcY{{\mathcal Y}}
\def\mcW{{\mathcal W}}
\def\mcZ{{\mathcal Z}}
\def\mcx{{x}}
\def\mcy{{y}}
\def\Mfi{{\mathcal M}_{fi}}
\def\pal{\partial}
\def\fh{\frac{1}{2}}
\def\fhs{\mbox{\small{$\fh$}}}

\def\fots{\mbox{\small{$\frac{1}{3}$}}}
\def\ftts{\mbox{\small{$\frac{2}{3}$}}}

\def\fhb{\mbox{\small{$\frac{\beta}{2}$}}}
\def\fhy{\mbox{\small{$\frac{Y}{2}$}}}

\def\tra{\mbox{\tiny{$T$}}}
\def\splus{\mbox{\tiny{+}}}

\def\ssl{S \!\!\! \mbox{{\raisebox{+0.3ex}{\footnotesize{$\backslash$}}}}}
\def\ssls{S \!\!\! \mbox{{\raisebox{+0.25ex}{\tiny{$\backslash$}}}}}
\def\dssl{\dot{S} \!\!\! \mbox{{\raisebox{+0.3ex}{\footnotesize{$\backslash$}}}}}
\def\ol{\overline}
\def\ul{\underline}
\def\GN{G_{\! N}}

\def\Asl{\mbox{$/\!\!\!\!\,\! A$}}
\def\pasl{\mbox{$/\!\!\!\!\!\;\pal$}}

\def\cstr{c^{\alpha}_{\ph{\alpha}\beta\gamma}}
\def\Gamabc{\Gamma^a_{\ph{a}bc}}
\def\Gamalp{\Gamma^{\alpha}_{\ph{\alpha}\beta\gamma}}
\def\uP{\underline{P}}

\def\uF{\underline{F}}

\def\und{\underline}

\def\ralpha{\mathring{\alpha}}
\def\rbeta{\mathring{\beta}}

\def\gpath{ }

\def\maxwidth{14.4cm}

\def\setb{\setlength{\baselineskip}{0.625\baselineskip}}

\begin{document} 

{\setlength{\baselineskip}{0.625\baselineskip}

\begin{titlepage}
\bigskip

\begin{center}
\bigskip
 
 {\Huge{\bf  Unification in One Dimension }} \\

\bigskip
\bigskip
\vspace{40pt}

\mbox {{\Large David J. Jackson} }  \\

  \vspace{20pt}

 { \large }  
  
  \vspace{10pt}
 
 { \large February 15, 2016 }

 \vspace{60pt}

{\bf  Abstract}

\end{center}
 
   A physical theory of the world is presented under the unifying principle that all of nature is laid out 
before us and experienced through the passage of time. The one-dimensional progression in time is opened 
out into a multi-dimensional mathematically consistent flow, with the simplicity of the former giving rise 
to symmetries of the latter. The act of perception identifies an extended spacetime arena of intermediate 
dimension, incorporating the symmetry of geometric spatial rotations, against which physical objects are 
formed and observed. The spacetime symmetry is contained as a subgroup of, and provides a natural breaking 
mechanism for, the higher general symmetry of time. It will be described how the world of gravitation and 
cosmology, as well as quantum theory and particle physics, arises from these considerations.

\pagebreak

{ 
\parskip -0.05pt plus 1pt minus 1pt

\tableofcontents
}

\end{titlepage}


\pagebreak

\chapter{Introduction}
\label{intro}

  In establishing a conceptual framework for a physical theory of the world one of the 
most fundamental questions concerns the nature of the ultimate entity out of which the 
world is made. From the earth, water, air and fire of the ancient Greeks, through 
various manifestations of elementary extended or point-like particle theories, to the 
quantum fields of $20^{\mathrm{th}}$ century high energy physics, the notion of a 
fundamental form of matter behaving according to laws of nature, to be identified 
empirically or through powers of reason, has strongly influenced the development of 
scientific theories of the world. The general trend has been to dig deeper into the 
layers of matter such that macroscopic objects are taken to be composed of discrete 
particle-like entities which in turn are composed of more basic particles which have 
become themselves to be considered as merely the states detected in physics 
experiments as a manifestation of yet deeper underlying entities, such as fields or 
strings. It is a trend which ever poses the question of what may be uncovered at the 
next layer down, or whether we may reach the ultimate bedrock of the world.

         The view taken in the present investigations is that the world can be built 
out of the one entity within which all our experiments, experiences, perceptions and 
indeed our thoughts in general are conducted, that is through progression in time. 
This universal nature of time applies both to inner thought experiences in the mind, 
as well as outer thoughts of the physical world; for example, perception of a cloud 
passing by or of a book on the table. With the basic structure of time being 
identified with, or isomorphic to,  that of the real numbers $\rrr$ this gives an 
immediate connection to a purely mathematical world. The mathematical possibility to 
express an \textit{inner} one-dimensional sense of time in the form of an 
\textit{outer} multi-dimensional space as an intrinsic and elementary property of the 
real numbers provides a significant motivation for this study. The aim will be to 
demonstrate how the external world of experience can result directly from the 
mathematical structure of temporal flow itself without the need to interpose or 
postulate the notion of an underlying `material' substratum of any form.
  
      It may be helpful to begin with an analogy using a familiar example in which 
multi-dimensional structures are enfolded within a lower-dimensional entity, namely a 
child's `pop-up' book of cartoon zoo animals, although, of course, it should not be 
taken too literally to represent the theory to be presented here. We can consider such 
a book, when closed, to be an essentially 2-dimensional object in space. When opened 
fully on a given page a figure will `pop-up', perhaps an elephant, extended in 
3-dimensional space; on another page a 3-dimensional crocodile may appear, and so on. 
It is down to the creativity and origami skills of the bookmakers to form such 
3-dimensional structures that can be perfectly folded away into the 2-dimensional 
plane (when the book closes) within the fixed constraints of the possibilities allowed 
by the laws of Euclidean geometry.

    It is the contention of these investigations that the 1-dimensional flow of time 
itself naturally opens out, according to necessary mathematical and geometrical forms, 
into a higher-dimensional space. The mechanism will be somewhat different to that in 
the above metaphor since time, unlike a book in space, is not experienced `all at 
once' and indeed space itself will need to be unfolded out of the temporal flow. 
However, within the 1-dimensional flow of time we shall find implicitly contained not 
only the appropriate mathematical structures for 3-dimensional space and 4-dimensional 
spacetime but also still higher-dimensional possibilities. The intermediate 3 and 
4-dimensional cases can be interpreted as subspaces of the higher-dimensional forms, 
with the properties of physical objects perceived in spacetime being largely 
determined by the nature of the general higher-dimensional structures. It is claimed 
that the opening out of the progression of time in this way into a mathematically 
determined multi-dimensional flow is responsible for not only our perceptions of 
objects, from books to real elephants themselves, but of the entire physical universe 
around us. 

    In a similar way that the laws of geometry constrain the design of pop-up books, 
so mathematics will constrain the way in which the physical world can open out from 
the flow of time and hence determine the laws of physics. It is the main aim of this 
paper to show how far the consequences of this idea resemble the observed laws of 
nature of the actual world. In traditional theories properties are assigned to 
underlying particles or fields, out of an enormous range of conceivable choices of 
such properties, largely for pragmatic reasons to match the empirically observed 
world. Here, on the contrary, we expect the present theory to make a much more 
thorough and direct contact with the structure of the physical world.

  In aiming for  an inclusive and complete theory, as well as accounting for the basic 
observed scientific phenomena from particle physics to cosmology, the theory might 
also address the everyday direct manner through which we actually encounter and 
experience the world. We shall touch upon all these areas, all of which would benefit 
from further study, in an attempt to gain an overall consistent worldview.

 Although this paper is lengthy all of the contents relate to a single unified theory, 
rather than to a collection of independent ideas, as will be clear from the 
progression of sections and the mutual cross-referencing within the text. Here we 
review the contents of the paper to guide the reader towards the sections which may be 
of most interest.
While the overall order of the text has been designed to introduce the various facets 
of the theory in a reasonably logical sequence there are four main areas in which 
progress on the theory has been made essentially in parallel. Each area addresses a 
particular question and related set of issues which might be asked of any candidate 
for a unified physical theory. The four areas correspond generally to subsets of the 
subsequent chapters of this paper:

\begin{itemize}
 \item Chapters 2--5: 
    The main goal here is to describe how an extended spacetime arena may be 
identified together with an external and internal curvature and the manner in which 
they are mutually constrained. First, beginning with a one-dimensional temporal 
interval in chapter~\ref{sym},  we  make precise the notion of the multi-dimensional 
flow of time by deriving in elementary terms what is considered to be its general 
mathematical form and give several relevant examples. We also emphasise the fact that 
we are dealing here with a general symmetry of time, in contrast to the symmetry of a 
higher-dimensional spacetime found in a different class of theories. 
The notion of perception as identified with the interpretation of certain mathematical 
forms, implicit in the multi-dimensional flow of time, in the shape of a geometrical 
spacetime. That is, we describe how an extended external spacetime arena for the world 
can `pop-out' from the temporal flow. This structure motivates the employment of more 
sophisticated mathematical tools, and in chapter 3 standard textbook material on 
differential geometry and general relativity is reviewed. Papers in the literature 
from around the mid-1970s to mid-1980s regarding non-Abelian Kaluza-Klein theories, 
which also describe a unified approach relating external and internal curvature within 
a similar mathematical framework, are then reviewed in chapter 4. In chapter 5 we then 
pick up the thread from end of chapter 2, in light of the mediating chapters, using 
the constraints of the present theory to study the relationship
 between the external and internal curvature in 4-dimensional spacetime and consider  
further implications such as  constraints on the equations of motion.     
 
 \item Chapters 6--9:
  Here we consider higher-dimensional forms of temporal flow with the main aim of 
establishing a connection with the structures of the Standard Model of particle 
physics in the breaking of the full symmetry of time over the 4-dimensional base 
manifold. Crucial to this investigation are the references  concerning the structure 
of the Lie group $\esi$ acting on the space $\htho$ of $3 \times 3$ Hermitian octonion 
matrices which have been published within the past ten years. The relevant details 
from this literature are followed closely and reviewed in chapter 6 in the context of 
the present theory, in particular with the determinant preserving action of $\esi$ on 
the elements of $\htho$ 
 interpreted as a particularly rich symmetry of a 27-dimensional cubic form of 
temporal flow. 
  In chapter 7 the principle features of the Standard Model, and their relation to 
mathematical models based on unification groups, are reviewed in order describe the 
physical structures to be accounted for by the present theory and the kind of 
theoretical structures which may be relevant. In chapter 8 we investigate the extent 
to which the $\esi$ symmetry action on $\htho$ in the context of the present theory 
can account for the properties of the Standard Model. Several successes are noted in 
terms of a correlation between the transformation properties of components of $\htho$ 
under the external and identified internal symmetries and corresponding properties of 
Standard Model particle states. The need to incorporate further particle properties 
and the natural extension to the larger structure of an $\ese$ action preserving a 
quartic from on the space $F(\htho)$, interpreted as  a symmetry of a 56-dimensional 
form of temporal flow,  leads to some further success in chapter 9  and also the 
suggestion of
 investigating  yet higher-dimensional forms.

 \item Chapters 10--11:
   In addition to uncovering Standard Model features here the ambition is to 
understand how 
 the present theory might accommodate the empirical observations of HEP experiments, 
in terms of cross-sections and decay rates for example, and  incorporate quantum 
phenomena in general.
 The essential textbook aspects of quantum field theory are reviewed in chapter 10, 
with in particular the structure of cross-section calculations analysed into its basic 
elements in order to establish a correspondence with the present theory. This 
correspondence is described in chapter 11 in which the conceptual origins of quantum 
phenomena within the context of the present theory are established. In one sense this 
involves generalising the relation between the external geometry and a particular 
internal gauge field, as assessed in relation to Kaluza-Klein theories in the earlier 
chapters, for an external geometry expressed in terms of a degeneracy of underlying 
field solutions. These underlying fields include both gauge fields deriving from the 
symmetry of time and fields deriving from  components of the multi-dimensional form of 
temporal flow itself, mutually related by a set of implicit constraint equations 
rather than via a dedicated Lagrangian function. Within the scope of these 
investigations two main and related points concern the conceptual nature of physical 
particle states, as analysed in laboratory experiments, and the manner in which the 
phenomena of quantum theory and general relativity 
 coexist as aspects of the same unified theory.

 \item Chapters 12--14:
  The theory has been developed with not only laboratory phenomena in mind but also 
the large scale structure of the universe with the goal of understanding the extent to 
which observations in cosmology might also be accounted for. To this end in chapter 12 
textbook material on both the standard cosmological model and inflationary theory is 
reviewed. A new feature of the present theory is described in chapter 13 concerning 
the possibility of non-trivial intrinsic curvature for the spacetime manifold arising 
from the elementary properties of its projection out of the multi-dimensional form of 
time. The extent to which this, combined with additional features of the theory 
identified earlier, might account for both the phenomena of the dark sector in 
cosmology and the structure of the very early universe is then considered, with the 
question of uniqueness for the theory in general also discussed. The study of the Big 
Bang epoch raises broader questions, in addition to the need to describe physical 
properties, relating to the reason why the universe should exist at all. In the 
context of the present theory, with everything constructed through a multi-dimensional 
form of progression in time, this inevitably leads to the question concerning the 
origin of time itself. In addressing this issue the speculations of chapter 14 include 
areas which are not  necessarily within the traditional bounds of physics but touch 
upon other scientific fields of study. This detour is however of value in providing an 
opportunity to elaborate upon the possibility of identifying a firm foundation for the 
full physical theory. 

\end{itemize}

   Following the discussion of the foundations of the theory in chapter 14 in the 
concluding chapter 15 we look outwards to the prospects for the further development of 
the theory in the  four main areas outline above, which are also  depicted as the four 
fronts in figure~\ref{fronts4}. In section~\ref{secsafd} the mutual relations between 
all aspects of this unified theory will also be described. 
    
   Broadly, the four subsets of chapters for the four main branches of the theory 
listed above are each presented with a structure to some extent analogous to a PhD 
thesis, in terms of combination of the presentation of new ideas and reviews of 
established material. In particular summaries of standard textbook material and other 
cited literature are presented mainly in the subsections, sections and chapters: 
2.2.2, 3, 4, 6, 7, 9.2 (up to equation~\ref{psidacta}), 10 and 12, although always in 
the context of the present theory. On the other hand the main novel theoretical 
developments follow a trail through the sections and chapters: 2, 5, 8, 9, 10.1, 11, 
12.1, 13, 14 and 15, with reference to the standard material of the above intervening 
chapters and section and further citations discussed in the course of the 
presentation. 

  Generally speaking the central chapters 6--11 deal more with the microscopic and 
laboratory scale  while the outer sections through to chapter 13 pursue a thread more 
closely associated with the macroscopic and large scale features typically studied 
under general relativity. However all aspects of the present theory are relevant for 
all scales, as further discussed in the concluding chapter. The current point of 
closest approach between the present theory and empirically established features of 
the physical world is in terms of a relationship with the Standard Model of particle 
physics, regarding in particular transformation properties under the external Lorentz 
symmetry group and the internal $\SML$ gauge group, as alluded to in the synopsis of 
Chapters 6--9 listed above. The shortest path from introducing the basic ideas of the 
theory to an elaboration of this connection with the Standard Model is to follow 
sections: 2.1, the opening of 2.2, 8.1, 8.2, 9.2 (after equation~\ref{psidacta}) and 
with further discussion in section 9.3.

The purpose of these investigations can be described as an enquiry into the extent to 
which the form of the physical world can be determined purely from the fact that it is 
perceived in time; that is, the extent to which the world can be constructed out of 
the pure mathematical nature of the progression in time itself. 
To this end we establish in the following section an expression for the potential 
\textit{multi-dimensional} flow of time in the appropriate general mathematical form. 
This will later provide the means to incorporate 4-dimensional spacetime together with 
the structures of `extra dimensions' in a naturally unified way.
 We begin with a particular example of a multi-dimensional expression implicit in a 
finite interval of time which exhibits an apparent geometric symmetry.


\pagebreak

\chapter{The Symmetry of Time}
\label{sym}

\section{General Form of Temporal Flow}
\label{gfotf}

    A finite interval of time represented by the real number $s\inn \rrr$ can be 
algebraically expressed in terms of other real numbers $x^a$ $(a=1,2,3\ldots)$ in an 
endless variety of ways. For example $s=x^1+x^2$ composes time intervals in series, 
while $s=x^1x^2$ might represent a rescaling of the temporal unit, or more generally 
we can have $s=x^1(x^2x^3+x^4)$ and so on simply by employing the basic arithmetic 
structure of the real line. More specifically, writing the square of the interval in 
the form, familiar since Pythagoras, $s^2=(x^1)^2+(x^2)^2+(x^3)^2$, the interval $s$ 
is then invariant under transformations of the set of real numbers 
$\{x^1,x^2,x^3\}$~$\inn\rrr^3$ by the orthogonal rotation group O(3), as depicted in 
figure~\ref{pythag}. In this case the set of possible numbers \{$x^a$\} consistent 
with this form for $s$ exhibits the mathematical symmetry of a vector in 3-dimensional 
Euclidean space that maintains a fixed length under rotations. This is exactly the 
mathematical structure within which the physical objects of our perceptions appear to 
us spatially; thus providing a simple example of how the geometric properties of space 
can be algebraically embedded within structures implicit in the arithmetic properties 
of time as represented by the real line.
\vspace{5pt}

\begin{figure}[htbp]  
\centering
\epsfxsize=10cm
\leavevmode
\epsffile[0 0 1081 800]{\gpath aPfig21e}
\caption{\setb With a fixed finite 1-dimensional interval of time $s$ expressed in the 
form $\mbox{$s=\sqrt{(x^1)^2+(x^2)^2+(x^3)^2}$}$ \hspace{0.5mm} the \textit{numerical} 
morphisms of the three numbers \{$x^1,x^2,x^3$\} under which $s$ is invariant can be 
interpreted as mapping out a spherical shell in a 3-dimensional \textit{geometrical} 
space.}
\label{pythag}
\end{figure}

    The broad range of possible expressions for a finite interval $s$ in terms of an 
arbitrary number of variables \{$x^a$\}, $a = 1\ldots n$, will be constrained to a 
more  
 restrictive structure  in the limit of infinitesimally small temporal intervals. We 
first consider this limit for the trivial case with the flow of time $s$ expressed in 
terms of a single real variable $x^1$ only for which we have simply $s=x^1$. This can 
symbolically be written as $\delta s=\delta x^1$ as we approach the limit of 
infinitesimal intervals. We then express the rate of change of $x$ with respect to $s$ 
in this limit as:   
\begin{equation}
   \label{vonedo}
   v^1= \frac{dx^1}{ds} \equiv \frac{\delta x^1}{\delta s}\bigg{\vert}_{\delta s \to 
0}  = 1 	   
\end{equation}
   For the case with multiple real numbers $\{x^a\}\inn\rrr^n$ representing the flow 
of time $s$ each will be associated with a corresponding rate of change $v^a=dx^a/ds$ 
with respect to pure time. For example, we may consider the propagation of time 
expressed for an infinitesimal interval as:
\begin{eqnarray}
 (\delta s)^2 & = & (\delta x^1)^2 + (\delta x^2)^2 + (\delta x^3)^2  \label{flow3del} 
\\
         & = & \eta_{ab}  \delta x^a\delta x^b  \qquad \mbox{with}  \quad
		         \eta_{ab} = \mbox{diag}(+1,+1,+1)  \label{etamet}
\end{eqnarray}
  where $\{a,b\} = \{1,2,3\}$ (and with the conventional summation over repeated 
indices implied throughout this paper). Dividing by $(\delta s)^2$ and taking the 
limit $\delta s \to 0$ this can be written as $\eta_{ab}v^a v^b=1$ or 
$(v^1)^2+(v^2)^2+(v^3)^2=1$, which is invariant under the group, O(3), of orthogonal 
transformations in three dimensions applied to $\bv_3 = (v^1,v^2,v^3) \inn \rrr^3$.
     This is simply the infinitesimal case of the situation depicted in 
figure~\ref{pythag}, with $s\to \delta s/ \delta s = 1$ and $x^a \to \delta x^a / 
\delta s = v^a$ (for $\delta s \to 0$), and is again open to a similar Euclidean 
spatial interpretation, here for the components $\{v^a\} \inn \rrr^3$. The question is 
then how to express the general case for the composition and symmetries of a 
multi-dimensional set of velocities $\{v^a\} \inn \rrr^n$.

     The infinitesimal elements of time can be written most generally, taking care to 
balance the order of the vanishing elements in each term, as:
\begin{equation}
\label{sbits}
  \delta s = \alpha_a\delta x^a + \sqrt{\alpha_{bc}\delta x^b \delta x^c} +
  \;\,  {}^{\substack{ 3 \vspace{0.6mm} \\ {} }} \!\!\!\!\sqrt{\alpha_{def}\delta x^d 
\delta x^e \delta x^f} + \ldots\ldots
\end{equation}
                   Here the coefficients $\alpha_{abc\ldots}$ are each equal to $\pm 
1$ or $0$ since we wish to express the $\delta s$ purely in terms of simple arithmetic 
relations of the $\delta x^a$. In equation~\ref{sbits} each term divides $\delta s$ 
into a separate portion of time:
\begin{equation}
  \delta s \quad =\quad \delta s_1 \qquad +\qquad \delta s_2
   \qquad +\qquad \delta s_3 \qquad+\qquad \ldots
\end{equation}
                   where each term $\delta s_p$ is the $p^{\mathrm{th}}$-root of a 
homogeneous polynomial of order $p$ in the \{$\delta x^a$\}. Taking each term in turn, 
dividing by the interval $\delta s_p$ in each case and taking the limit $\{ \delta s_p 
, \delta x^a \} \to 0$ we find:
\begin{eqnarray}
         \delta s_p & = & \;\,  {}^{\substack{ p \vspace{0.6mm} \\ {} }} \!\!\!\! 
\sqrt{\alpha_{abc\ldots}\delta x^a \delta x^b \delta x^c\ldots} \\   
    \mbox{divide by $\delta s_p$:$\;\;$} \qquad \quad \qquad \qquad   1 & = & \;\,  
{}^{\substack{ p \vspace{0.6mm} \\ {} }} \!\!\!\! \sqrt{\alpha_{abc\ldots}v^a v^b 
v^c\ldots} \label{pone} \\ 
             \mbox{that is:}   \qquad    \alpha_{abc\ldots}v^a v^b v^c\ldots  & = & 1  
\label{ptwo} \\ 
         \mbox{which we write:} \qquad  \qquad \qquad    L(\v) & = & 1   \label{lv}
\end{eqnarray}      
                   where $L$ is a homogeneous polynomial of order $p$ in the 
components $v^a$; it can be considered as a map from the elements of a real 
$n$-dimensional vector space $\v \inn \rrr^n$ onto the unit $1\inn\rrr$.

   The $p^{\mathrm th}$-root is dropped in stepping from equation~\ref{pone} to 
equation~\ref{ptwo} since, trivially, $1^p = 1$.
 If the equality in equation~\ref{pone} involved a variable quantity on the left-hand 
side rather than  unity, such as in the case of
 finding a `path of extremal length' on an extended manifold for a quadratic form, or 
metric, using a variational method then the root would be needed, as will be described 
later for equation~\ref{extgeo}.  
  Further, the components of a local `metric' $\eta_{ab} = \alpha_{ab} \in \{\pm 1,0\} 
$ may be mapped onto a general metric involving components $g_{\mu\nu}(x) \notin \{\pm 
1,0\}$  under a transformation from `local coordinate' variables \{$x^a$\} to a 
`general coordinate system' on such an extended manifold, as we shall describe leading 
up equation~\ref{metten}.
 (In principle this observation could also apply to the other coefficients 
$\alpha_{abc\ldots}$ of equation~\ref{sbits} considered as generalised `metrics' for 
the corresponding extended dimensions).

Equation~\ref{lv} is taken to express the general mathematical form of 
multi-dimensional temporal flow and it is the central equation of this paper.   
  The symmetries of $L(\v)= 1$ will be represented by groups acting on the vector 
space $\rrr^n$ such that for all elements $g$ of the group $G$ and all vectors $\v\inn 
\rrr^n$ satisfying $\lv$ we have $L(\sigma_g(\v)) = L(\v') = 1$ where $\sigma_g(\v)$ 
represents the action of the group element $g \inn G$ on the vector $\v\inn \rrr^n$. 
As $G$ acts on $\rrr^n$ over a continuous range of elements beginning at the identity 
$g=e\inn G$ we can think of this as a continuous morphism of the real numbers $v^a$.
 (This is equivalent to the symmetry over the 2-sphere in the example with finite 
intervals $\{x^a\}$ in figure~\ref{pythag}).
 This morphism is always consistent with the dissolving of the fundamental temporal 
flow $s$  into the possible rates of change $v^a$ of the multi-dimensional real 
quantities $x^a$  conforming to the requirement $\lv$ and hence may be termed an 
\textit{isochronal} symmetry, of which we next describe several examples. 

     Quadratic forms in general, including the 4-dimensional example of the expression 
$L(\bv) = \eta_{ab}v^a v^b$, with $\bv \inn \rrr^4$, Minkowski metric 
$\eta_{ab}=\mbox{diag}(+1,-1,-1,-1)$ and $\{a,b\}=\{0,1,2,3\}$, and the norm of an 
element of a division algebra ($\rrr,\ccc,\hhh$ or $\ooo$ as introduced below), 
together with their symmetry groups, are expected to be particularly significant forms 
of $\lv$. This is due to their close relation to Clifford algebras and Euclidean 
spatial geometry, describing for example the space within which we perceive objects.  
Other possible forms of $\lv$ include the determinants of matrices, which are 
homogeneous polynomials in the matrix elements.

The complex numbers $\ccc$ had been studied by Hamilton in the 1830s in a manner 
consistent with his  view of algebra as the science of pure time. This program in part 
led to his discovery of the quaternions in the 1840s, which also however led him to 
essentially abandon the notion of a close relation between algebra and time owing to 
the non-commutative property of the quaternion algebra.  Subsequently an 8-dimensional 
algebra, the octonions $\ooo$, was discovered independently by Graves and Cayley in 
the mid-1840s and completed the unique series, $\rrr$, $\ccc$, $\hhh$ and $\ooo$, of 
normed division algebras~\cite{Baez1}, as will be reviewed in section~\ref{oaags}.
 In fact division algebras only exist over vector spaces of dimension $1,2,4$ or 8.
 An algebra $A$ is a division algebra if $ab=0$ implies $a=0$ or $b=0$, with $a,b\inn 
A$; it is a normed division algebra if $A$ is also a normed vector space with $\vert 
ab \vert = \vert a \vert \vert b \vert$. This latter property naturally provides a 
source of structures of form of equation~\ref{lv} together with the corresponding 
symmetries.

	For example, the quaternion algebra $\hhh$ may be used to compose a possible 
multi-dimensional form of progression in time. On the space of unit norm elements 
$\v\inn\hhh$, with $L(\v)=\vert\v\vert = 1$, the symmetry group $G$ composed of 
quaternions of unit norm operating on $\v$ under left and right algebra multiplication 
forms the two-to-one cover of $\mbox{SO}(4)$. The 1-dimensional character of temporal 
flow is represented by the `norm' function $L$ applied collectively to the components 
of $\bv \in \hhh \equiv \rrr^4$; with the non-commutative behaviour of the $\sigma_g$ 
symmetry operations within  $L(\ldots \sigma_{g''}\sigma_{g'}\sigma_g (\v)) = 1$ 
describing the properties of the multiple apparently `internal' temporal dimensions. 
For the case in which $G$ is homomorphic to an orthonormal rotation group (as is the 
case for $\hhh$ representing three or four dimensional space, with for example the 
three imaginary units of the quaternions associated with 3-dimensional Euclidean 
space) the non-commutative algebraic properties correlate directly with the 
non-commutative property of \textit{spatial} rotations for $n>2$.

    The fourth division algebra, the octonions $\ooo$, being non-associative, do not 
themselves form a group in such a direct way as for the complex numbers or the 
quaternions; they will however play a significant role in the symmetry of time and  
hence in physics as will be explained in this paper. Here the division algebras will 
be combined with matrix algebras in considering the 27-dimensional real vector space 
of $3\times 3$ Hermitian matrices $\htho$ over the octonions with the determinant 
required to be unity: $L(\v_{27})=\mbox{det}(\mcX)=1$, with $\mcX \inn \htho$. The 
group $G$ of  determinant preserving symmetry transformations on $\htho$ is the 
exceptional Lie group E$_6$. This group is well known to be of interest for 
unification models and will be discussed in detail in the context of the present 
investigations in chapters~\ref{esihtho}--\ref{chapesb}.

   With various different forms of progression in time to be considered, in general 
the subscript $n$ in the notation $L(\bv_n)=1$ indicates collectively the vector space 
$\rrr^n$, the implied form $L$ and the corresponding symmetry group $G$ (respectively 
$\bv_{27} \equiv \mcX \inn \htho \equiv \rrr^{27}$, $L(\bv_{27}) = \det(\mcX) = 1$ and 
$G=\esi$ in the above example for $n=27$), where any case of ambiguity will be 
clarified in the text.

   Given a possible $n$-dimensional form of progression in time, $L(\bv_n)=1$, the 
vector $\bv_n \inn \rrr^n$ may be written as the ordered set of velocities: 
\begin{eqnarray}
      \bv_n & = & \, \{ \: v^1, \quad  v^2,\ldots \quad v^n \, \}  \label{vnset}  \\   
            & = &  \bigg\{ \frac{dx^1}{ds},  \frac{dx^2}{ds},\ldots  \frac{dx^n}{ds} 
\bigg\}  
\end{eqnarray}  
    the values of which are unchanged by a numerical translation of the real 
variables,
\begin{equation}
    x^a \to x^a + r^a
\end{equation}
    for any constant set $\{ r^a \} = \br_{\! n} \inn \rrr^n$, or for a subset of 
$\rrr^n$. 
	Above we described a possible symmetry of $\lv$ with the action of a group $G$ 
mixing the numerical components $v^a$, which represent elements of the temporal flow 
$dx^a/ds$. Here we have a further symmetry implicit in $\lv$ with respect to 
translations of the numerical variables as $x^a \to x^a + r^a$.
	That is, we also have trivially:
\begin{eqnarray}
  \label{rspill}
      \bv_n & = &  \bigg\{\frac{d(x^1+r^1)}{ds},\frac{d(x^2+r^2)}{ds},\ldots  
\frac{d(x^n+r^n)}{ds}\bigg\}.  
\end{eqnarray}  
  satisfying $L(\bv_n) = 1$. For the 1-dimensional case of equation~\ref{vonedo} the 
symmetry $v^1= \frac{d(x^1+r^1)}{ds}$ can be readily visualised as a flow $v^1$ 
present everywhere on the real line parametrised by $r^1\inn\rrr$ (rather than at a 
single arbitrary point for example).  In the general case
since equation~\ref{rspill} is equally valid for all possible $\br_{\! n} \inn \rrr^n$ 
the temporal flow, under the condition $L(\bv_n)=1$, effectively occupies the entire 
$\rrr^n$ manifold as depicted in figure~\ref{spillout}.
\begin{figure}[htbp]  
\centering
\epsfxsize=9cm
\leavevmode
\epsffile[0 0 878 733]{\gpath aPfig22e}
\caption{\setb Since the real variables $\{ x^a \} \innf  \rrr^n$ are arbitrary, the 
flow described within $L(\bv_n)=1$ applies equally for the particular value $\bx_0 
\innf \rrr^n$ as  for $\bx^{\prime} = \bx_0 + \br_{\! n}$ and over the 
range $-\infty < r^a > \infty$, for $a = 1\ldots n$. This `translation' symmetry is 
implied within the form $\lvn$ and is depicted here for $n=3$.}
\label{spillout}
\end{figure}

   This $n$-dimensional freedom in $\rrr^n$ forms a continuous $n$-dimensional 
parameter space, which may be considered to form an implicit `base manifold' $M_n$, 
upon which the vector $\bv_n$ naturally resides in the tangent space $T_xM_n$ at every 
point $x\inn M_n$.
Hence the internal structure of the form $\lvn$ and its symmetries  contain the 
skeletal form of a mathematical framework for the description of an apparently 
external and extended spatial structure.
  
  In other theories and models a higher-dimensional \textit{symmetry of spacetime} is 
considered, extending beyond our familiar 4-dimensional spacetime arena to one with a 
total of, for example, five or ten spacetime dimensions.
 Such models, initiated by Kaluza and Klein,  will be described in more detail in 
chapter~\ref{kktheory}. 
 In these theories 
 it is necessary to explain how our 4-dimensional spacetime world is embedded in the 
larger arena, and the means by which the `extra dimensions' are \textit{compactified} 
or otherwise evade direct observation.

   As described in the introductory chapter we are familiar with the idea that not 
only all of our scientific experiments but also everything we experience in the world 
takes place in time. Relative to 4-dimensional spacetime the flow of pure time is an 
apparently `lower-dimensional' structure which pervades all observations and events in 
the universe. This is in contrast to hypothetical extra dimensions, above the four of 
space and time, which are beyond our domain of experience. Here we begin on a firm 
footing by treating one-dimensional temporal flow as the fundamental entity of the 
world.

   Hence, in contrast with Kaluza-Klein theory, for the theory presented in this paper 
we deal instead with a general higher-dimensional \textit{symmetry of time}, and it is 
here necessary to explain how the large scale  \textit{extended}   4-dimensional 
spacetime geometry and physical structures of the universe can arise from a 
fundamentally 1-dimensional temporal flow. This phenomenon, and the internal 
mathematical identifications that give rise to it, will intimately involve the nature 
of \textit{perception}. It is the means through which time experienced as a purely 
1-dimensional progression can be experienced simultaneously as a multi-dimensional 
flow of physical objects in an extended spacetime. The mathematical basis for 
obtaining such an extended base manifold will be found in the application of the 
symmetry described in figure~\ref{spillout} to a 4-dimensional spacetime subset of the 
translational degrees of freedom of a higher-dimensional temporal form. 
   
   For the case considered for the real world, in addition to the 27-dimensional space 
$\htho$ described above another important example of a form of time involving both a 
matrix and a division algebra is identified in the determinant of elements of the 
4-dimensional real vector space $\htwc \equiv \rrr^4$, that is the $2\times 2$ 
Hermitian matrices over the complex numbers, together with the action of the 
determinant preserving group SL$(2,\ccc)$. This group is the double cover of the 
Lorentz group and will also be significant in this paper since $\htwc$ is naturally 
embedded as a subspace of $\htho$, with the symmetry group $\sltc$ being a subgroup of 
E$_6$.

    Applying the translation symmetry of equation~\ref{rspill} in four dimensions 
only, corresponding to the $\htwc$ components, provides a natural mechanism for 
breaking the symmetry of the larger group through the necessary identification of a 
4-dimensional background manifold $M_4$ upon which the Lorentz group acts locally, and 
to a good approximation globally over extended regions of spacetime. Under the overall 
normalisation $\lvt$ the 4-dimensional form will take more general values $L(\bv_4) = 
h^2 \inn \rrr$ for the subcomponent $\bv_4 \subset \bv_{27}$ local tangent vectors on 
$M_4$ (in this paper the relation $\bv' \subset \bv$ between two vectors will denote 
the \textit{projection} of $\bv'$ out of $\bv$). Further consequences of the  symmetry 
breaking are associated with the necessary choice of a particular direction for the 
vector field $\v_4(x)\inn\mbox{h}_2\ccc$, locally a 1-dimensional flow embedded within 
a 4-dimensional manifold. Comparisons between these symmetry breaking structures and 
the Standard Model of particle physics will be made in chapter~\ref{chapesb}.

The relation between the `translation symmetry' of $\lv$ and the `rotation symmetry', 
more generally denoted by the action $\sigma_g(\bv)$ for $g\inn G$, is key to the 
development of the geometrical structure of the theory and motivates the review of 
elements of textbook geometry in chapter~\ref{rogaeom}.
 We begin in the following two sections by describing a simple model universe, based 
on a small number of dimensions in order to elaborate upon the nature of the geometric 
structures involved, in particular concerning the identification of the 
 base manifold. The geometric properties of this manifold, which are significant in 
general relativity, are intrinsically related to the geometry and symmetries of the 
residual dimensions, which are significant for gauge theories,  resulting from the 
projection of a higher-dimensional form of $\lv$ over the base manifold and 
corresponding symmetry breaking pattern, as will be described in section~\ref{hdasb}. 
This development of the theory will be continued in chapter~\ref{chaputtf} where 
   the relation between the external gravitational field and internal gauge fields 
over a 4-dimensional spacetime manifold in the context of the present theory will be 
described.

  
\section{Perception in Space and Time}
\label{perc}

   The fact that all of our experiences in the world are encompassed within the 
passage of time motivated the formulation of the general expression for temporal 
progression, $\lv$, presented as equation~\ref{lv} of the previous section.
However it is also necessary to  account for the fact that all of our experiences of 
such a physical world appear to be distributed through an extended \textit{manifold}, 
with the immediate and necessary location of observed physical objects in 
\textit{space}, as well as in \textit{time}. 
While the general mathematical form for the flow of time may be exemplified by a wide 
range of  mathematical structures and  symmetry groups it is the identification of 
relatively simple structures, those which may be most readily suited to the 
organisation and  understanding of experiences in the world with respect to a 
background arena of space as well as time, that will be designated by the term 
\textit{perception}.

     The apparent physical form of the world is shaped out of the interplay between 
these two basic notions: that of the mathematical form of temporal flow and that of a 
necessary form of perception. It is the act of interpreting \textit{algebraic} 
structures within the temporal flow $\bv$ in terms of an extended coherent 
\textit{geometrical} structure that breaks the symmetry of the general flow of time 
described by $\lv$. 
 
   In this and the following section the discussion will be maintained largely at a 
general level with a simplified model universe, a world with two spatial dimensions 
only, being used to make the presentation more concrete for a case which is 
mathematically simpler than our own world and, in particular, one which may be more 
readily visualised. 
  The notion of a base manifold may be introduced by considering how it would be 
possible for physical objects in a spatially 2-dimensional world to be perceived 
propagating in time.
 This situation brings to mind the image pictured in figure~\ref{twodworld}. (Such 
illustrations clearly also serve by analogy to represent our own world, with one 
spatial dimension being suppressed. Indeed, throughout this chapter the model universe 
described should be considered both as a metaphor for the general case and for our own 
world in particular).

\begin{figure}[htbp]  
\centering
\epsfxsize=7.9cm
\leavevmode
\epsffile[0 0 696 712]{\gpath aPfig23e}
\caption{\setb A representation of a model universe with 2-dimensional physical 
objects propagating through a 3-dimensional base manifold.}
\label{twodworld}
\end{figure}

  Objects in such a world are here depicted by figures in a 2-dimensional plane which 
are animated, presumably according to certain laws of physics in the form of equations 
of motion, as they propagate through the third dimension on a 3-dimensional base 
manifold $M_3$. The geometrical structure of the 2-dimensional plane may be considered 
to be compatible with the notion of \textit{spatial perception} of objects by beings 
in this model world if it possesses, at least to a good approximation, an SO(2) 
rotational symmetry about any point as well as translational symmetry in this plane. 
Hence the local symmetry group $G$ of the  manifold $M_3$ must: 
\begin{itemize}
\item[i)] contain as a subgroup the symmetry of the purely spatial structure of the 
world; here the group SO(2),
\item[ii)] act on a space of one dimension higher than that of the spatial geometry; 
in this case 3-dimensional, and
\item[iii)] be a possible symmetry group or subgroup of a form $\lv$ in order to 
conform with the present conceptual ideas.
\end{itemize}

  For our model universe we begin with the 3-dimensional form of temporal flow:
\begin{equation}
L(\bv_3)  =  (v^1)^2+(v^2)^2+(v^3)^2 = 1  \label{flow3d}    
\end{equation}  
  that is with $L(\bv_3)  = \eta_{ab}v^av^b = 1$ and the 3-dimensional metric 
$\eta_{ab}$ of equation~\ref{etamet} as introduced in the previous section.
 The full 3-dimensional translational symmetry of this form depicted in 
figure~\ref{spillout}
 provides the framework for an extended 2-dimensional `spatial' environment, in 
addition to the temporal one, constituting the background manifold $M_3$. Ultimately a 
metric with a `spacetime' signature will be required in order to incorporate causal 
structure on the base manifold, however this feature is neglected for the simple model 
presented in this chapter. 
 For the case of the model world an unbroken external symmetry SO(3) will be described 
in this section, before extending to a larger symmetry SO(5) over the same 
3-dimensional base manifold in the following section.

 
\subsection{The Base Manifold}
\label{base}

    The metric $\eta_{ab}$ implies the existence of an orthonormal basis $\{e_a\}$ 
with respect to which the pure temporal flow $\bv_3$ can be expressed in terms of the 
components $\{v^a\} \inn \rrr^3$ as:
\begin{equation}
\label{spillrx}
 \bv_3 = v^a e_a = \frac{dx^a}{ds} e_a = \frac{d(x^a + r^a)}{ds} e_a
\end{equation}
    With a 3-dimensional translation symmetry $-\infty < r^a > \infty$ as depicted in 
figure~\ref{spillout} the orthonormal basis projects over the base manifold as an 
orthonormal frame field on $M_3$. 
This smooth differentiable manifold naturally possesses a tangent space $T_xM_3$ at 
each point $x\inn M_3$, that is the space $M_3$ has the properties of a 
$3$-dimensional base manifold of a tangent bundle space, as we shall discuss further 
in section~\ref{riegeo}  for the general and 4-dimensional spacetime cases.

The assignment $\bv_3(x) = v^a e_a$ is valid for a local orthonormal coordinate basis 
or a frame field (with  index $a$ for such  an orthonormal frame, here $a = 
\{1,2,3\}$). General coordinates on the manifold naturally give rise to a coordinate 
basis for the tangent space $\{\partial_{\mu}\}$, with $\partial_{\mu} \equiv 
\partial/\partial x^{\mu}$, (with  index $\mu$ for general coordinates, here $\mu = 
\{1,2,3\}$).
 Relabelling the parameters $\{r^a \} \inn \rrr^3$ in equation~\ref{spillrx}  as a 
particular set of `general coordinates' $x^{\mu} = \delta^{\mu}_{\ph{\mu}a}r^a$  on 
$M_3$,  there is an implied coordinate frame on the base manifold $\{\partial_{\mu} 
\}$ such that $\bv_3 = v^a e_a$ of equation~\ref{spillrx} can be expressed as $\bv_3 = 
v^{\mu} \partial_{\mu} = v^a \tetu \partial_{\mu}$, with the `triad' components $\tetu 
= \delta^{\mu}_{\phantom{\mu}a}$. 

More generally under a passive reparametrisation to any general coordinates 
$\{x^{\mu}\} \inn \rrr^3$ in a region of $M_3$
a frame field consists of a triad of vector fields $e_a = \tetu\partial_{\mu}$ (for 
$a=1,2,3$) with components with respect to the general coordinate frame given by the 
matrix function $\tetu$ which points to the local Euclidean metric structure at any 
$x\inn M_3$.
 The set of components $\tetu$ contains the same information as its matrix inverse 
$\teta$, and either of these matrices are sometimes referred to as the `triad' itself. 
These matrices transform both under general coordinate transformations  and local, or 
\textit{gauge}, SO(3) transformations.

  The kernel symbol $\bv$ will usually denote a vector or vector field corresponding 
to the fundamental flow of time in the form $\lv$, while the kernel symbol $\bu$ will 
denote
 arbitrary tangent vector fields, such as $\bu(x) = u^{\mu}(x)\partial_{\mu}$, as 
indicated in figure~\ref{spill}. Either type of vector field may be expressed either 
in a local orthonormal frame or in a general coordinate frame.
 The components a vector field $\bu(x)$ belong to the space $\rrr^3$ whether presented 
in a local or a general coordinate basis; these two possibilities are related by the 
matrix $\teta \inn \mbox{GL}(3,\rrr)$ such that $u^a(x) = \teta u^{\mu}(x)$.

\begin{figure}[htb]  
\centering
\epsfxsize=11cm
\leavevmode
\epsffile[0 0 1341 1056]{\gpath aPfig24e}
\vspace{-17pt}
\caption{\setb A local orthonormal basis $e_a \equiv \partial/\partial x^{a}$ for the 
vector field $\bv_3(x) = v^a(x) e_a(x)$ is related to any other such basis of the same 
orientation at each point $x\in M_3$ by the action of the symmetry group $G= \soth$, 
which can vary arbitrarily over $M_3$. In general a tangent vector field $\bu(x)$ on 
the manifold may be expressed in terms of an arbitrary frame field, or a particular 
orthonormal or general coordinate frame.}
\label{spill}
\end{figure}

Through a frame field $e_a(x)$ on $M_3$ the flow of time described numerically by 
$\bv_3(x)$ is isomorphic to an \textit{external} tangent vector field which may be 
described in terms of general coordinates on $M_3$, and may be considered to be a flow 
of time on this manifold space itself, even for the case in which the global geometry 
is not Euclidean. This latter situation will arise when the local tangent space on 
$M_3$ is embedded within a higher-dimensional form of temporal flow, as described in 
the following section. In this case $M_3$ will necessarily be treated as a 
differentiable manifold with  finite curvature in general for which  only the local 
geometry at any point $x\inn M_3$ will be isomorphic to the Euclidean geometry of 
$\rrr^3$.

  Via the triad field $\teta$ the internal space constant metric $\eta_{ab}$ of 
equation~\ref{etamet} implied in equation~\ref{flow3d}, similarly as for the vector 
components $v^a$, may be expressed on the tangent space for a general coordinate 
basis. This determines the metric tensor:
 \begin{equation}
 \label{metten}
  g_{\mu\nu}(x) = \teta e^{b}_{\phantom{i}\nu}(x) \eta_{ab}.
 \end{equation}  
   In the theory of general relativity it is the freedom of the metric field 
$g_{\mu\nu}(x)$, or equivalently the tetrad field $\teta$, on a $4$-dimensional 
spacetime base manifold as will be described for equation~\ref{geeeta}, with respect 
to an arbitrary coordinate system that describes gravitation in the world, as we shall 
review in section~\ref{gcatep}. While in general the components of $g_{\mu\nu}(x)$ 
differ from those of $\eta_{ab}$ in a general coordinate system, even for a flat 
spacetime, in general relativity it is the absence of \textit{any} global coordinate 
basis such that $g_{\mu\nu}(x) = \delta^{a}_{\ph{i}\mu} \delta^{b}_{\ph{i}\nu} 
\eta_{ab}$ everywhere that is responsible for gravitational effects.

 In order to consider the \textit{curvature} of the base manifold it is necessary to 
formalise the notion of \textit{parallelism}. 
 The question concerns the way in which the base manifold $M_3$ originates out of the 
flow of time as depicted in figure~\ref{spillout}, specifically with $n=3$ for the 
model case here, such that the symmetry $G=\mbox{SO(3)}$, acting upon individual 
vectors $\bv_3(x) \inn \rrr^3$ in the equation $L(\bv_3) = 1$ can act as an 
approximately \textit{global} symmetry over scales that are large compared with the 
objects being perceived. In section~\ref{gfotf} we began with finite 
\textit{intervals} of multi-dimensional time, as depicted in figure~\ref{pythag}, and 
then went on to the infinitesimal case $s \to \delta s$ in order to derive the 
relation $\lv$ of equation~\ref{lv}. We here need to understand how the symmetry of 
such infinitesimal intervals can apply coherently over finite \textit{distances} on 
the manifold $M_3$. This is required on the manifold in order to frame stable 
perceptions of 2-dimensional spatial objects propagating through such a world as 
depicted in figure~\ref{twodworld}.

     In terms of the model world the point is that since we are locally free to choose 
an orthonormal frame within which to specify the numerical values $v^a(x)$ for the 
components of $\bv_3(x)\inn TM_3$, the values themselves have no absolute meaning. In 
particular, no conclusion concerning the equality, or parallelism, of two sets of 
vector components $v^a(x_1)$ and $v^a(x_2)$ at two different points $x_1 , x_2\inn 
M_3$ in figure~\ref{spill} may be drawn since the bases of local frames at $x_1$ and 
$x_2$ may be chosen independently, within the local SO(3) freedom of the relation 
$L(\sigma_{g(x)}(\v_3(x))) = 1$. A triad frame (such as  $e_a$ described in 
equation~\ref{spillrx} for the coordinate frame $x^{\mu} = 
\delta^{\mu}_{\phantom{\mu}a} r^a$) could be declared to specify a parallelism on the 
manifold (that is, $\bv_3(x_1)$ is parallel to $\bv_3(x_2)$ if each of the components 
agree, $v^a(x_1) = v^a(x_2)$, in the specified triad frame $e_a(x)$). However for the 
case of a curved space or spacetime no global frame field $e_a(x)$ exists in a manner 
compatible with the parallelism, since the latter now depends on the path taken 
between $x_1$ and $x_2$.

   The underlying notion of parallelism is more generally defined in terms of a 
\textit{connection} 1-form on the manifold, which readily extends to the case of 
non-global parallelism.  The connection is a mathematical object, a Lie algebra valued 
1-form, that mutually relates the bases on the manifold, with respect to a given path 
connecting the points $x_1,x_2 \inn M_3$,  and hence determines whether any two 
vectors at these locations  are parallel with respect to the path. With such a 
structure the relative values of the components of two vectors at differing locations 
does acquire meaning. Hence we wish to identify an SO(3) \textit{connection form} 
$A(x)$ on the base manifold $M_3$. We describe how a \textit{flat} connection arises 
\textit{canonically} on $M_3$ through it's relation to the symmetry group $G$ in 
subsection~\ref{mgjoin}  (and further in section~\ref{cacfc} in the context of the 
principle bundle structure). In the following subsection we first review the standard 
geometry on a group manifold itself.


\subsection{The Group Manifold}
\label{tgman}

   While the $\{r^a \}\inn \rrr^3$ translational symmetry of $\lvth$ gives rise to the 
base manifold $M_3$, the algebraic structure of the rotational symmetry constitutes a 
second differentiable manifold which is identified with the Lie Group $G=\mbox{SO(3)}$ 
itself. This manifold is also intimately related to the temporal flow $\bv_3$ through 
the expression $L(\bv_3) = L(\sigma_{g}({\bv_3})) = 1$, with the action $g \inn G$ 
realised on the subspace of unit norm vectors in $\rrr^3$.

   Elements of a general Lie group $g \inn G$ also act as diffeomorphisms on the 
manifold $G$ itself \cite{kob,amp,fecko}. An example is the diffeomorphism $L_g:G\to 
G$, mapping the point $h \to gh$ with $g,h\inn G$, called `left translation' on the 
manifold. Due to the nature of the algebraic properties of a symmetry group a Lie 
group manifold $G$ exhibits distinctive canonical geometrical structures. The 
significance of `canonical' (in the sense of intrinsic or naturally existing) 
structures, where relevant, is that they carry the mathematical development of a 
theory forward in a necessary and non-arbitrary way. As for the base manifold 
described above, the group $G$ as a manifold also has a tangent space $T_gG$ at each 
point $g \inn G$, through which a tangent vector field $V(g)$ may be described on $G$. 
Smooth vector fields $X(g)$ belonging to the subset which satisfy the relation:
  \begin{equation}
 L_{g\ast}X(h) = X(gh) \label{linvact}
  \end{equation}
  for all $g,h \inn G$, where $L_{g\ast}$ is the `tangent mapping', or differential, 
of the left translation $L_g$ (acting upon objects defined on the tangent space of 
$G$), are said to be left-invariant. The set of left-invariant vector fields together 
with their multiplication in terms of the commutator $[X,Y]$ (considering the vector 
fields $X,Y$ as mappings in the space of scalar functions $f(g)$ on $G$), which itself 
describes a left-invariant vector field, defines the Lie algebra $L(G)$ of the Lie 
group $G$. As a vector space $L(G)$ is isomorphic to the set of tangent vectors at any 
location on $G$, and in particular to the space $T_eG$, where $e\inn G$ is the 
identity element of the group.
Given any  point $h \inn G$ the orbit of left translation for all $g \inn G$ covers 
the entire group manifold, as a consequence of the transitive property of 
multiplication within a Lie group, and hence the corresponding tangent mapping 
$L_{g\ast}$ carries any vector $V(h)$ into a left-invariant vector field on $G$.

   In general a 1-form $\omega (x)$, or covector field, on a differentiable manifold 
$M$ maps a vector field $V(x)$ into the space of real functions on $M$; this map may 
be denoted by:
\begin{equation}
   \langle \omega(x),V(x) \rangle = f(x)
     \label{formap}
\end{equation} 
  at any point $x\inn M$.   Over the manifold $G$ a linearly independent set of 
left-invariant vector fields, $\{X_{\alpha} \}$ with $\alpha = 1\ldots n_G = 
\mbox{dim}(G)$, forms a global frame field on $G$.
 A dual basis of 1-forms $\{\theta^{\alpha}\}$ with $\alpha = 1\ldots n_G$ such that 
$\langle \theta^{\alpha},X_{\beta} \rangle = \delta^{\alpha}_{\ph{\alpha}\beta}$, 
constitutes a coframe field on $G$. These covector fields are also left-invariant with 
$L^{\ast}_g\, \theta^{\alpha}(gh)=\theta^{\alpha}(h)$ for the `pull-back' $L^{\ast}_g$ 
of the left translation by $g\inn G$. 

    The `exterior algebra' of differential forms includes the exterior product 
`$\wedge$' and exterior derivative `$\mbox{d}$' which act on  1-forms such as 
$\omega(x) = \omega_{\mu} \mbox{d}x^{\mu}$ and $\sigma(x) = \sigma_{\mu} 
\mbox{d}x^{\mu}$ to produce 2-forms
such as:
\begin{eqnarray}
  \omega \wedge \sigma & = & \omega \otimes \sigma - \sigma \otimes \omega  
\label{wedgep} \\
   \mbox{and} \qquad \mbox{d}\omega & = & 
\frac{\pal{\omega_{\mu}}}{\pal{x^{\nu}}}\;\mbox{d}x^{\nu}\wedge\mbox{d}x^{\mu} 
\label{ddef}
\end{eqnarray}
   For any diffeomorphism between manifolds, $f:M \to N$ (where it may be that $M=N$), 
the pull-back map $f^{\ast}$ is a structure preserving homomorphism of the exterior 
algebra.
 Hence, as for the 1-form basis covectors $\theta^{\alpha}$ ($\alpha = 1\ldots n_G$),  
the 2-forms $\mbox{d}\theta^{\alpha}$ and $\theta^{\beta} \wedge \theta^{\gamma}$  are 
also left-invariant on $G$ and are therefore related via left-invariant, that is 
constant, scalar coefficients $c^{\alpha}_{\ph{\alpha}\beta\gamma}$ as defined in:
\begin{equation}
\label{maca1}
   \mbox{d}\theta^{\alpha} + \frac{1}{2} c^{\alpha}_{\ph{\alpha}\beta\gamma} 
\theta^{\beta} \wedge \theta^{\gamma} = 0.
\end{equation}

   This is the Maurer-Cartan equation which also serves to define the Lie algebra 
structure constants $c^{\alpha}_{\ph{\alpha}\beta\gamma}$, with respect to the basis 
$\{ \theta^{\alpha}\}$. It is equivalent to the definition of the structure constants 
in terms of the dual basis of vector fields \{$X_{\alpha}$\}, which represents the Lie 
algebra itself, in the relation: 
\begin{equation} 
[X_{\beta}, X_{\gamma}] = c^{\alpha}_{\ph{\alpha}\beta\gamma} X_{\alpha}. 
\label{xxcomcx}
\end{equation}
 
    The Maurer-Cartan 1-form $\theta$ is a single, basis independent, canonical object 
on the manifold $G$ that expresses the properties of the collective set of $n_G$ 
1-forms $\{\theta^{\alpha}\}$. It is a Lie algebra-valued 1-form defined by its action 
on a \textit{general} tangent vector field $V = V^{\alpha}(g)X_{\alpha}$ on $G$ with 
$\langle \theta, V(g)\rangle := V^{\alpha}X_{\alpha} \inn L(G)$, where $V^{\alpha}$ 
are the component values of $V$ at $g$ and
  $V^{\alpha}X_{\alpha}$ 
 is hence the Lie algebra element corresponding to the left-invariant vector field on 
$G$ with the tangent vector $V(g)$ at the given point $g\inn G$. 

    The Maurer-Cartan form can be written as $\theta =  \theta^{\alpha}X_{\alpha}$ in 
terms of the dual bases. The canonical form $\theta$ encapsulates the parallelisable 
nature of any Lie group manifold by defining a consistent global parallelism on $G$. 
That is, $\theta$ represents a single reference frame for each of the tangent spaces 
which resolves each vector $V(g)$ at any $g\inn G$ into its components with respect to 
a left-invariant frame field $\{X_{\alpha}\}$.
For a matrix group such as SO(3), with a matrix basis $\{E_{\alpha}\}$ for $L(G)$,  
the Maurer-Cartan form can also be expressed as the left-invariant matrix of 1-forms 
$\theta = g^{-1} \mbox{d}g$. In this case $\langle \theta, V(g)\rangle$ is the Lie 
algebra element $V^{\alpha}E_{\alpha}$ represented in matrix form in terms of the 
components of the corresponding left-invariant vector field at $T_eG$. 

  In terms of  $\theta$ the Maurer-Cartan equation~\ref{maca1} can be written as: 
\begin{equation}
  \label{maca2}
  \mbox{d}\theta + \frac{1}{2} [\theta,\theta]    =  0
\end{equation}
   where the bracket denotes the `exterior product' for Lie algebra valued 1-forms. 
Such a product may be defined on vector-valued $p$-forms in general provided there is 
a product defined on the vector space of the values. This is the case for Lie algebra 
valued forms where, with $\theta = \theta^{\alpha}X_{\alpha}$ and $\phi = 
\phi^{\beta}X_{\beta}$ and with $\{X_{\alpha}\}$ a basis for $L(G)$, the product is 
defined as:
\begin{equation}
\label{thphiw}
   [\theta,\phi]  :=  \theta^{\alpha} \wedge \phi^{\beta} \; [X_{\alpha}, X_{\beta}]. 
\end{equation}   
   For a matrix basis $\{E_{\alpha}\}$ and the case of a single 1-form as for 
equation~\ref{maca2} the product $\fh[\theta,\theta] = \theta \wedge \theta$ which 
implicitly incorporates the multiplication of the $E_{\alpha}$ matrices.

  On the group manifold $G$ each left-invariant field $X$ \textit{generates} a 
one-parameter subgroup described by the flow $\phi_t = \exp(tX_e)$, where $X_e \inn 
T_e G$ denotes the tangent vector belonging to the field $X$ at the identity $e \inn 
G$ and `$\exp$' is the `exponential map' from $L(G)$ into the manifold $G$. The action 
of this one-parameter group on any point $h \inn G$ is by right translation, as 
indicated in figure~\ref{grightran}.
\begin{figure}[htbp]  
\centering
\epsfxsize=8cm
\leavevmode
\epsffile[0 0 782 746]{\gpath aPfig25e}
\vspace{-20pt}
\caption{\setb The integral curves of a left-invariant field on $G$ generate right 
translations.}
\label{grightran}
\end{figure}
  
  Alternatively, a left-invariant field $X^A$ on $G$ can be \textit{induced} by the 
right action $R_g: h \to hg$ of elements of the one-parameter group $g(t)= \exp(tA)$, 
with $A \inn T_e G$ such that:
\begin{equation}
   \label{atoxa}
    X_h^A(f) = \frac{d}{dt} \: f(h \, \exp(tA)) \, \vert_{t=0}
\end{equation}
	where $f$ is a real-valued function on the manifold $G$. 
 Left-invariant fields are sometimes denoted by a label `$R$' since they are generated 
by right translations; hence $X^R$ denotes a left-invariant field.

  Since a right-invariant field $Y^L$ (which can be generated by left translation) is 
by definition invariant under right translations the Lie derivative of $Y^L$ with 
respect to the vector field $X^R$ vanishes:
\begin{eqnarray}
    {\mathcal L}_{X^{\! R}} Y^L = \frac{d}{dt} \:
	                 \phi_t^{\ast}\, Y^L \: \vert_{t=0} &  =  &  0  \\
	\mbox{that is:} \qquad         [X^R,Y^L]          &  =  &  0              
\label{xyrbra}
\end{eqnarray} 
    
	For each $g\inn G$ a further diffeomorphism on the group manifold called the 
adjoint map can be defined by $\mbox{Ad}_g(h) = ghg^{-1}$ for all $h \inn G$, that is 
$\mbox{Ad}_g = L_g \,\scirc\, R_{g^{-1}} \equiv R_{g^{-1}} \,\scirc\, L_g$ by the 
associative property of group composition. The adjoint map is an \textit{automorphism} 
of the group composition. Since $\mbox{Ad}_{g_2}\,\scirc\,\mbox{Ad}_{g_1} = 
\mbox{Ad}_{g_2\:\!g_1}$ this is a left action of $G$ on itself.

  The adjoint map applied to elements near $e\inn G$ gives rise to the group 
representation
    $\mbox{Ad}_g = L_{g\ast} \,\scirc\, R_{g^{-1}\ast}\vert_{e} \equiv R_{g^{-1}\ast} 
\,\scirc\, L_{g\ast}\vert_{e}$ acting upon the Lie algebra of the group. For a group 
represented by matrices  this takes the form $\mbox{Ad}_g(Y) = gYg^{-1}$ for any $Y 
\inn L(G)$. The adjoint representation $g \to \mbox{Ad}_g$ is a group homomorphism of 
$G$ into $\mbox{GL}(L(G))$. The `derived homomorphism' of this representation induces 
the corresponding adjoint representation for the Lie algebra elements with 
$\mbox{ad}_{X}Y = \lbrack X, Y \rbrack$ for $X,Y\inn L(G)$, as an automorphism of the 
Lie bracket algebra, which naturally involves the structure constants of the group 
through equation~\ref{xxcomcx}.
	
   Finally we note that a left-invariant field $X^R$, which also generates right 
translations, itself transforms under right translation as $R_{g\ast} X^R = 
\mbox{Ad}_{g^{-1}}X^R$  by the definition of left-invariance and the adjoint 
representation (while $L_{g\ast} X^L = \mbox{Ad}_{g}X^L$ for a right-invariant field 
$X^L$). These group properties will be important for the structure of principle 
bundles described in section~\ref{fibre}.

   
\subsection{Relating the Base and Group Manifolds}
\label{mgjoin} 

   As for the basis $\{X_{\alpha}\}$ on the manifold $G$, a frame field $\{e_a \}$, 
with $a = 1 \ldots n$, may be introduced on any $n$-dimensional differentiable 
manifold $M_n$, forming a linearly independent set of tangent vectors at each point of 
the manifold. The real quantities $c^{a}_{\ph{a}bc}(x)$ in the relation:
\begin{eqnarray}
      [e_{b}, e_{c}] &   =  &  c^{a}_{\ph{a}bc}(x)e_{a} \label{cabconm} \\
  \mbox{or equivalently in:} \qquad  \quad
 \mbox{d} e^{a} & = & - \frac{1}{2}c^{a}_{\ph{a}bc}(x) e^{b} \wedge e^{c} 
     \qquad \qquad  \label{mccoeff}
\end{eqnarray}
  in terms of the dual coframe field $\{e^a \}$, are here variables called structure 
\textit{coefficients} (or `coefficients of anholonomy') rather than \textit{constants} 
as for equations~\ref{maca1} and \ref{xxcomcx}. Given a general coordinate chart $\{ 
x^{\mu} \}$ on $M_n$ and a holonomic frame $e_{\mu} = \partial_{\mu}$ the 
corresponding coefficients $c^{\rho}_{\ph{\rho}\mu\nu}(x)$ are all zero, while 
$c^{a}_{\ph{a}bc}(x) \neq 0$ implies that a non-coordinate frame is being employed.

   On the manifold $G$ frames composed of \textit{left-invariant} vector fields $\{ 
X_{\alpha}(g) \}$ were identified as being particularly important owing to the group 
structure. On the base manifold $M_3$, possessing the metric of equation~\ref{metten}, 
basis vectors forming  \textit{orthonormal} frames $\{e_a(x)\}$ are particularly 
significant. As described in subsection~\ref{base} such a triad frames the components 
of $\bv_3$ subject to the pure numerical relation $\lvth$ of equation~\ref{flow3d}, 
which implicitly contains the local $3\times 3$ Euclidean metric $\eta$ as expressed 
in equation~\ref{etamet}.
 (In this paper indices $a,b,c\ldots$ for a basis $\{e_a\}$ will denote an arbitrary 
smooth frame field, as for $M_n$ above, or an orthonormal frame field, as for $M_3$ 
here, or even a coordinate basis depending on the context; while indices 
$\mu,\nu,\rho\ldots$ for a basis $\{e_{\mu}\} \equiv \{\pal_{\mu} \}$ will always 
denote a coordinate frame).

 The two manifolds $M_3$ and $G = \soth$, representing the translational and 
rotational symmetries of the form $\lvth$, as described in section~\ref{gfotf} and in 
the two subsections above, are linked through the mapping $g(x): M_3 \to G$. An 
initial orthonormal frame field $\{e_a(x)\}$ can be transformed to any other 
orthonormal frame field $\{e'_b(x)\}$ by the matrix action $e'_b = e_a g^a_{\ph{a}b}$  
via the group element $g(x)\inn\soth$ at every $x\inn M$. The map $g(x): M_3 \to G$, 
as depicted in figure~\ref{mtogmap}, expresses the local choice of an orthonormal 
frame field $\{e_a(x)\}$, essentially the choice of local $\{x^1,x^2,x^3\}$ axes of 
figure~\ref{pythag} at each point $x\inn M_3$, as a basis for tangent vectors $\bv_3 
\inn TM_3$. It is this `gauge' freedom $g(x) \inn \soth$ in the choice of local 
orthonormal frames that prevents a particular frame $\{e_a(x)\}$ from directly 
representing parallelism on the base manifold, as described towards the end of 
subsection~\ref{base}. 
\vspace{-3pt}
\begin{figure}[htbp]  
\centering
\epsfxsize=10.5cm
\leavevmode
\epsffile[0 0 1587 1405]{\gpath aPfig26e}
\vspace{-6pt}
\caption{\setb The gauge choice of a frame at each $x\in M_3$ described as a map into 
elements $g\in G$ between the two manifolds. }
\label{mtogmap}
\end{figure}

  Since the operations of the exterior algebra of $p$-forms are preserved under the 
pull-back of forms through diffeomorphism maps on manifolds the Lie algebra-valued 
1-form:
\begin{equation}
  \label{aegasth}
 A = g^{\ast}\theta
\end{equation}
  on $M_3$ captures the structural properties of the Maurer-Cartan 1-form $\theta$ on 
$G$ relative to the map $g$. While on $G$ we have the linear map $\langle \theta, 
V(g)\rangle \inn L(G)$ from $V(g)\inn T_gG$ into the Lie algebra of $G$, on $M_3$ we 
have the linear map $\langle A(x), \bu(x) \rangle \inn L(G)$ from $\bu(x) \inn T_xM_3$ 
into the same Lie algebra. The Lie algebra-valued 1-form $A(x)$  may be written as 
$A(x) = A^{\alpha}_{\ph{\alpha}\mu}(x) \, X_{\alpha} \,\mbox{d}x^{\mu} $ where $\{ 
\mbox{d}x^{\mu} \}$ is a coordinate basis of 1-forms on $M_3$ and $\{X_{\alpha}\}$ is 
a basis for $L(G)$. 
 In the appropriate $3 \times 3$ matrix representation the $\soth$ generators can be 
denoted by $\{E_{\alpha}\} \equiv \{ {L}_{p \pqg q}\}$, labelled by a single index 
mnemonic double letter symbol $p \pqg  q = \{1\pqg 2,1\pqg 3,2\pqg 3\}$, with: 
  \begin{equation}
   \label{somat}
     ({L}_{p \pqg q})_{ab} = \delta_{pa} \delta_{qb} - \delta_{pb} \delta_{qa}  
  \end{equation}
  where $a$ and $b$ label the matrix rows and columns respectively, that is:
\begin{equation}
 \label{m3}
  {L}_{1 \pqg 2} = \left( \begin{array}{ccc}
        0  &  1  &  0  \\
       -1  &  0  &  0  \\
        0  &  0  &  0  
          \end{array}  \right),       \qquad 
  {L}_{1 \pqg 3} = \left( \begin{array}{ccc}
        0  &  0  &  1  \\
        0  &  0  &  0  \\
       -1  &  0  &  0  
          \end{array}  \right),       \qquad 
  {L}_{2 \pqg 3} = \left( \begin{array}{ccc}
        0  &  0  &  0  \\
        0  &  0  &  1  \\
        0  & -1  &  0  
          \end{array}  \right)       \qquad 
\end{equation}

   Unlike the canonical 1-form $\theta =  \theta^{\alpha}X_{\alpha}$ on $G$, the 
1-form $A$ on $M_3$ has variable real coefficients $A^{\alpha}_{\ph{\alpha}\mu}(x)$ 
which, however, are not arbitrary but depend upon the choice of gauge function $g(x)$ 
as well as upon the choice of coordinates $\{x^{\mu}\}$ on $M_3$. Explicitly, for a 
matrix group $G$, the 1-form $A = g^{\ast}\theta$ on $M_3$ can be expressed as:
\begin{equation}
  \label{puga}
     A(x) = g^{-1} \mbox{d}g = g^{-1} \frac {\pal g}{\pal x^{\mu}} \mbox{d}x^{\mu}
\end{equation}
   It is this canonical mathematical object that serves as a \textit{connection 
1-form} on the base manifold $M_3$, formalising the notion of parallelism in manner 
which will naturally generalise for the case of finite curvature. Here it is possible 
to choose a gauge with $A(x) = 0$ everywhere on $M_3$, simply by taking $g(x)$ to be 
constant in equation~\ref{puga}, and hence we have a flat connection. Indeed, this 
connection can always be written in terms of `pure gauge', as it is in 
equation~\ref{puga}, which is one way of defining a flat connection (to be described 
in more detail in section~\ref{cacfc}). Given a connection $A(x) \neq 0$ a gauge 
transformation via any $g(x)$ transforms the connection in the standard way as: 
\begin{equation}
   \label{atrana}
 A \to A' = g^{-1} A g + g^{-1} \mbox{d}g.
\end{equation} 
which can be expressed as pure gauge $A' = g'^{-1}\mbox{d}g'$, that is in the form of 
equation~\ref{puga}, in terms of an appropriate gauge function $g'(x)$.  
 
 By the homomorphism of exterior algebra relations across the pull-back map the Lie 
algebra-valued 1-form $A = g^{\ast}\theta$ is also subject to a structure equation 
corresponding to equation~\ref{maca2}, that is:
\begin{equation}
 \label{maca3}
   \mbox{d}A + \frac{1}{2} [A,A]    =  0  
\end{equation}
In general the curvature 2-form $F$ on the base manifold can be expressed as:
\begin{equation}  
   F =  \mbox{d}A + \frac{1}{2} [A,A]    \label{fdefaaa}
\end{equation}
    which transforms under a gauge change $g(x)$ as $F\to F' = g^{-1} F g$. 
Equations~\ref{maca3} and \ref{fdefaaa} then immediately show that the curvature is 
equal to zero, with $F=0$ in any gauge, and further expresses the global parallelism 
implied by the canonical flat connection of equation~\ref{aegasth}.

  While the connection 1-form can be written as  $A(x) = 
A^{\alpha}_{\ph{\alpha}\mu}(x)  E_{\alpha} \mbox{d}x^{\mu}$ and curvature 2-form can 
be written as $F(x) = \fh F^{\alpha}_{\ph{\alpha}\mu\nu}(x) E_{\alpha} \mbox{d}x^{\mu} 
\wedge \mbox{d}x^{\nu} $ (where the factor of $\fh$ arises from the convention of 
equation~\ref{wedgep} and the double counting implicit on the right-hand side since 
the set of asymmetric $\mbox{d}x^{\mu} \wedge \mbox{d}x^{\nu}$ 2-forms, with nine 
values of $\{\mu,\nu\} = \{1,2,3\}$, does not describe a linearly independent basis).

	Interest in the group $G=\soth$ arose as a symmetry action on the form $\lvth$ and 
hence the Lie algebra values of $A(x)$ and $F(x)$ are composed of elements
 $\{E_{\alpha}\} \equiv \{ {L}_{p \pqg q}\}$ of equation~\ref{m3}
 in a representation of $L(G)$ acting naturally upon the vectors $\bu \inn TM_3$, that 
is on the tangent space of the base manifold, and in particular on the vector $\bv_3 
\inn TM_3$ originating in the form $\lvth$. The mathematical objects involved are 
hence intimately associated with each other, with the base space $M_3$ and the flat 
connection $A(x) = g(x)^{\ast} \theta$ of equation~\ref{aegasth} upon it arising out 
of the translation and rotation symmetry properties of $\lvth$, with the vectors 
$\bv_3$ themselves being tangent to $M_3$.

  As an example of this association the constancy of the scalar function $\lvth$ on 
$M_3$ can be expressed as $\partial_{\mu} L(\bv_3) = 0$, or, consistent with the gauge 
transformations of $\bv'_3 = g^{-1}\bv_3$ and equation~\ref{atrana}, covariantly as:
\begin{eqnarray}
  D_{\mu} L(\bv_3) & = & D_{\mu}(\bv_3 \cdot \bv_3) = 2\bv_3 \cdot D_{\mu}\bv_3 = 0   
\label{dlvth1}   \\
  \mbox{with} \;\;\;\;\;  D_{\mu} \bv_3 & = & \partial_{\mu} \bv_3 + A_{\mu} \bv_3    
\label{dlvth2}
\end{eqnarray}
  The `covariant derivative' $D_{\mu}$ relating to a connection $A_{\mu}$ will be 
defined more precisely in the following chapter, leading to equation~\ref{covv}. These 
above two equations show how the connection field $A_{\mu}(x)$ explicitly acts on the 
vector field $\bv_3(x)$ in a constraining relation, and hence there is a `coupling' 
between these fields over the base manifold $M_3$.

   A vector field $\bu(x)$ which satisfies $D_{\mu} \bu = 0$ everywhere represents a 
parallel vector field on the manifold.
   A frame field $\{e_{a} \}$ that satisfies $D_{\mu} e_{a} = 0$ for each value of $a$ 
defines a parallel frame field -- in which case the frame field itself may be used to 
define the parallelism on the manifold, which is only possible for a flat connection. 
With respect to the original global coordinates defined in terms of a parametrisation 
of the translation symmetry of equation~\ref{spillrx} on $M_3$ the triad field with 
components $\tetu = \delta^{\mu}_{\phantom{\mu}a}$ was identified. The covariant 
derivative of the corresponding orthonormal basis vectors $e_a$ is constant with 
respect to the connection form $A(x) = 0$, with constant $g(x)$ in 
equation~\ref{puga}, and hence $\{e_a\}$ defines the parallelism in this case. The 
geometric objects $\tetu = \delta^{\mu}_{\phantom{\mu}a}$ and $A(x) = 0$ may both be 
associated with the constant gauge function $g(x)$ on $M_3$ taken as the identity 
element $e\inn G = \soth$ of the group. A gauge change by a constant $g(x) \neq e \inn 
G$ changes the frame $\{e_{a} \}$ but not the connection 
$A(x) = 0$. Under a general gauge change $g(x)$ the connection, with $A(x) \neq 0$ in 
general, can be written as an explicit function of the triad field.

   In summary the canonical connection $A = g^{\ast}\theta$, constructed as depicted 
in figure~\ref{mtogmap}, defines a global parallelism on $M_3$ (as $\theta$ does on 
the manifold $G$), such that the parallel transport of a vector $\bu(x_1)$ from $x_1$ 
to another point $x_2$ on the base manifold, see for example figure~\ref{spill}, 
results in a definite vector $\bu(x_2)$ independent of the path taken.
 This formalises the notion of parallelism on the base manifold in a manner which can 
be generalised for the case of non-global parallelism.

    Since $A(x)$ is SO(3)-valued in acting on the tangent space $TM_3$ it also 
describes a `metric compatible' connection. In being completely determined by the 
triad field the connection $A(x)$ is also torsion-free, where torsion will be defined 
in section~\ref{riegeo}. In fact since $A(x)$ is a particular case of a linear 
connection acting on the tangent space  the curvature 2-form $F$ of 
equation~\ref{fdefaaa} may be identified with the Riemann tensor and  hence denoted  
$\bR$.  
   The so(3)-valued curvature tensor $\bR = \fh R^{\alpha}_{\ph{\alpha}\mu\:\!\nu} 
E_{\alpha} \mbox{d}x^{\mu} \wedge \mbox{d}x^{\nu} $ on $M_3$, with $\alpha = 1\ldots 
n_G$, then has components $R_{a\:\!b\:\!\mu\:\!\nu} = R^{p \pqg 
q}_{\ph{pq}\mu\:\!\nu}({L}_{p \pqg q})_{a\:\!b}$. Via the triad field the same tensor 
can be expressed either fully in a local orthonormal frame or fully in a general 
coordinate frame -- in the latter case with four general coordinate indices as:
\begin{equation}
    R_{\rho\:\!\sigma\:\!\mu\:\!\nu} = e^a_{\ph{a}\rho}  e^b_{\ph{a}\sigma} 
R_{a\:\!b\:\!\mu\:\!\nu} 
\end{equation}
   Hence we have constructed a zero Riemann curvature tensor with all components 
$R_{\rho\sigma\mu\nu} = 0$ as implicit in the identification of a canonical flat 
connection $A = g^{\ast}\theta$.

   In section~\ref{fibre} we shall review the geometry of a principle fibre bundle, 
hence incorporating  $M_3$ as the base manifold and $G$ as the structure group 
together in a single manifold, before reviewing Riemannian geometry itself. A non-zero 
Riemannian curvature will ultimately be obtained on the original base manifold by 
expanding the form $\lv$ into a higher-dimensional temporal flow with a larger 
symmetry group, as we provisionally describe in the following section.


\section{Higher Dimensions and Symmetry Breaking} 
\label{hdasb}

   In the previous section the construction of a model world required that we drew 
attention to the particular form of temporal flow $\lvth$, as expressed in 
equation~\ref{flow3d}, as an example of the general $n$-dimensional case. It was shown 
how $\bv_3$ could be interpreted as a tangent vector field over a base manifold $M_3$, 
represented in figure~\ref{spill}, which in turn may be parametrised by a set of real 
number coordinates ${x^{\mu}} \inn \rrr^3$, and with a choice of a local orthonormal 
reference frame $\{e_a(x) \}$ determined within the freedom of the local $\soth$ 
symmetry.

 However, in general there are many higher symmetry groups acting upon vector spaces 
of a larger dimension, with elements conforming to $\lv$, which we have no 
mathematical reason to neglect. Indeed, the reasoning of section~\ref{gfotf} is 
consistent with the flow of time being channelled into a space of arbitrarily large 
dimension.
 Hence,  mathematically, there is nothing to prevent the 3-dimensional space of 
parameters $\bv_3 \inn \rrr^3$, representing a 3-dimensional flow of time, from 
further dividing into a larger multi-dimensional space of parameters described by the 
vector $\bv_n \inn \rrr^n$ ($n>3$) subject to a new form $L(\bv_n) = 1$ with a higher 
symmetry group $G$.
  (In later chapters the expression $\lvh$ will denote the full form of temporal flow 
being considered, while the full symmetry group, excluding translations, may be 
denoted $\hat{G}$ for clarity, as for the remainder of this section).

  The original $\soth$ geometric symmetry group may now be identified as a subgroup  
$\ol{H}\subset \hat{G}$, with the `overline' denoting an \textit{external} symmetry,  
acting on the 3-dimensional flow $\bv_3$ which is  projected onto the tangent space of 
the base manifold $M_3$ out of the higher-dimensional temporal flow. 
  In this section we begin to consider the conceptual implications and mathematical 
possibilities of this generalisation for the necessary existence of such a higher 
symmetry group $\hat{G}$ acting upon a higher-dimensional form of temporal flow 
$L(\bv_n) = 1$. 

Since it will ultimately be required to mathematically support the kind of situation 
depicted in figure~\ref{twodworld}, in which the smaller symmetry
 $\ol{H}\subset \hat{G}$ is treated in a distinctive way in giving rise to the global 
geometrical nature of a perceived universe of physical objects, we shall expect to be 
dealing with a natural mechanism for breaking the higher symmetry. The full `rotation' 
symmetry of the form $\lvn$, as a generalisation of that depicted in 
figure~\ref{pythag}, is now broken since only the degrees of freedom of a  
\textit{subset} of the possible dimensions of translation symmetry, depicted in 
figure~\ref{spillout}, is employed to locally construct the base manifold

 In subsequent chapters, for the real world, we shall motivate the choice of $\hat{G}$ 
as the Lie group $\esi$, acting on a 27-dimensional form $\lvt$, with a Lorentz 
subgroup acting on the local tangent space of the 4-dimensional spacetime base 
manifold $M_4$. In the meantime here, for the model world, we shall take $\hat{G}$ to 
be a symmetry group of a form of $\lvn$ large enough to contain $\soth$, the 
orthonormal frame symmetry group, as a subgroup of $\hat{G}$, while retaining the 
3-dimensional base space $M_3$.

    For the case of the model universe $\ol{H}=\soth$, acting upon $\bv_3 \inn 
\rrr^3$, could be taken to be embedded within various kinds of larger groups, for 
example $\hat{G}=\suth$ acting upon the 6 real components of $\bv_6$ corresponding to 
a 3-dimensional complex vector $\bc_3 \inn \ccc^3$ with $L(\bv_6) = \bc_3^{\dag}\bc_3$ 
= 1. However, here we consider the vectors $\bv_3 \inn \rrr^3$ of section~\ref{perc} 
to be vectors in a subspace of $\rrr^n$ with $n>3$ upon which the group SO(3) is a 
straightforward subgroup of SO($n$), the latter being a perfectly acceptable symmetry 
of $L(\bv_n) = 1$, acting upon the vectors $\bv_n \inn \rrr^n$. In particular we 
choose $n=5$ and consider the Lie group $\hat{G}=\sofi$ acting on the form $\lvfi$:
 \begin{eqnarray}
 L(\bv_5) =  \bv_5\cdot\bv_5 & = & (v^1)^2+(v^2)^2+(v^3)^2+(v^4)^2+(v^5)^2 = 1   
\label{vimp1}  \\
                                        & = & \eta_{ab}v^av^b +(v^4)^2+(v^5)^2 = 1 
\label{vimp} 
 \end{eqnarray}
  where $a,b = 1\ldots 3$ and $\eta_{ab} = \mbox{diag}(+1,+1,+1)$ represents the 
3-dimensional Euclidean metric, which was introduced in equation~\ref{etamet} of 
section~\ref{gfotf}.  

 The 5-dimensional vector $\bv_5$ has components $v^{{a}} = dx^{{a}}/ds$, with 
${a}=1\ldots 5$.  Hence the vectors of $\lvfi$ implicitly contain the 5-dimensional 
translational freedom $x^{{a}} \to x^{{a}} + r^{{a}}$, $\{r^a\}\inn \rrr^5$, as a 
particular example of equation~\ref{rspill} and figure~\ref{spillout} which 
generalises the 3-dimensional case of equation~\ref{spillrx}. However, we consider 
only the 3-dimensional freedom of this parameter space and continue to take $M_3$ to 
be the base space as we did in section~\ref{perc} and as depicted for the present case 
in figure~\ref{mtogmaph}. This choice will ultimately be justified by the 
identification of geometrical structures on $M_3$ which may then be interpreted as the 
base space for perception of physical events as sketched in figure~\ref{twodworld}.  
The model described in figure~\ref{mtogmaph} provides a convenient picture for a 
provisional discussion of the symmetry breaking structure which  will be picked up 
again in section~\ref{reaic} for the more realistic case over a 4-dimensional 
spacetime base manifold $M_4$.

\vspace{10pt}
\begin{figure}[htbp]  
\centering
\epsfxsize=\maxwidth
\leavevmode
\epsffile[0 0 1882 964]{\gpath aPfig27e}
\vspace{-10pt}
\caption{\setb (a) The gauge choice at each $x\in M_3$ depicted in 
figure~\ref{mtogmap} is extended for $\hat{G}=\sofi$ with (b) only the 
\textit{subgroup}  $\soth \subset \sofi$ now acting on $TM_3$. }
\label{mtogmaph}
\end{figure}

   A basis for the 10-dimensional Lie algebra  $\sofia$ (with the lower case `so' 
denoting the Lie algebra corresponding to the SO(5) Lie group), as represented on a 
5-dimensional vector space as the generators of a symmetry of $\lvfi$, is provided by 
the set of ten $5\times 5$ matrices $({L}_{p \pqg q})_{{a}{b}}$ of the type described 
in equation~\ref{somat}, now with ten distinct labels composed out of $p,q = 1\ldots 
5, p < q$. The three so(5) Lie algebra elements:   
\begin{equation}
 \def\vgap{-7pt}
  {L}_{1 \pqg 2} = \left( \begin{array}{@{}ccc|cc@{}}
                                        \vspace{\vgap}
        0  &  1  &  0  &  0  &  0  \\   \vspace{\vgap}
  \! -1 \, &  0  &  0  &  0  &  0  \\   \vspace{-2pt}
        0  &  0  &  0  &  0  &  0  \\   \hline  \vspace{-27pt}
		   &     &     &     &     \\   \vspace{\vgap}  
		0  &  0  &  0  &  0  &  0  \\  
		0  &  0  &  0  &  0  &  0  
          \end{array}  \right),       \quad \!
  {L}_{1 \pqg 3} = \left( \begin{array}{@{}ccc|cc@{}}
                                        \vspace{\vgap}
        0  &  0  &  1  &  0  &  0  \\   \vspace{\vgap}
        0  &  0  &  0  &  0  &  0  \\   \vspace{-2pt}
  \! -1 \, &  0  &  0  &  0  &  0  \\   \hline  \vspace{-27pt}
           &     &     &     &     \\   \vspace{\vgap}    
		0  &  0  &  0  &  0  &  0  \\  
		0  &  0  &  0  &  0  &  0  
          \end{array}  \right),       \quad \!
  {L}_{2 \pqg 3} = \left( \begin{array}{@{}ccc|cc@{}}
                                        \vspace{\vgap}
        0  &  0  &  0  &  0  &  0  \\   \vspace{\vgap}
        0  &  0  &  1  &  0  &  0  \\   \vspace{-2pt}
   0  & \! -1 \, &  0  &  0  &  0  \\   \hline  \vspace{-27pt} 
           &     &     &     &     \\   \vspace{\vgap} 
		0  &  0  &  0  &  0  &  0  \\  
		0  &  0  &  0  &  0  &  0  
          \end{array}  \right)       \label{m5ext}
\end{equation}
     generate an $\soth \subset \sofi$ subgroup, as can be see directly as guided by 
the horizontal and vertical lines drawn into the matrices in equation~\ref{m5ext} and 
by comparison with equation~\ref{m3}. This $\soth$ subgroup can be taken to act on the 
tangent space of $M_3$ and hence upon the subspace of vectors $\ol{\bv}_3 \subset 
\bv_5$ projected onto the base space. Of the other seven $\sofi$ generators one acts 
purely on the complementary 2-dimensional subspace $\ul{\bv}_2  \subset \bv_5$, 
namely:
\begin{equation}
\def\vgap{-7pt}
    {L}_{4 \pqg 5} = \left( \begin{array}{@{}ccc|cc@{}}
                                        \vspace{\vgap}
        0  &  0  &  0  &  0  &  0  \\   \vspace{\vgap}
        0  &  0  &  0  &  0  &  0  \\   \vspace{-2pt}
        0  &  0  &  0  &  0  &  0  \\   \hline  \vspace{-27pt} 
           &     &     &     &     \\   \vspace{\vgap} 
		0  &  0  &  0  &  0  &  1  \\  
		0  &  0  &  0  &  \! -1 \,  &  0  
          \end{array}  \right)       \label{m5int}     
\end{equation}
   With the subspace of vectors $\ol{\bv}_3 \subset \bv_5$ with $\ol{\bv}_3 \inn TM_3$ 
in the \textit{external} space the SO(2) generator ${L}_{4 \pqg 5}$ can be said to act 
upon the \textit{internal} 2-dimensional space of vectors $\ul{\bv}_2 \inn \rrr^2$, 
which can be considered to represent \textit{extra dimensions} over those required to 
describe the extended base space $M_3$.
 In general while an `overline' denotes an external object an `underline' will denote 
an object defined in the internal space.
 In this representation basis the remaining six $\sofia$ Lie algebra elements are:
\begin{equation}
\def\vgap{-7pt}
  {L}_{1 \pqg 4} = \left( \begin{array}{@{}ccc|cc@{}}
                                        \vspace{\vgap}
        0  &  0  &  0  &  1  &  0  \\   \vspace{\vgap}
        0  &  0  &  0  &  0  &  0  \\   \vspace{-2pt}
        0  &  0  &  0  &  0  &  0  \\   \hline  \vspace{-27pt}
		   &     &     &     &     \\   \vspace{\vgap}  
  \! -1 \, &  0  &  0  &  0  &  0  \\  
		0  &  0  &  0  &  0  &  0  
          \end{array}  \right),       \quad 
  {L}_{1 \pqg 5}, \:  {L}_{2 \pqg 4}, \:{L}_{2 \pqg 5}, \:{L}_{3 \pqg 4}, \quad
  {L}_{3 \pqg 5} = \left( \begin{array}{@{}ccc|cc@{}}
                                        \vspace{\vgap}
        0  &  0  &  0  &  0  &  0  \\   \vspace{\vgap}
        0  &  0  &  0  &  0  &  0  \\   \vspace{-2pt}
        0  &  0  &  0  &  0  &  1  \\   \hline  \vspace{-27pt} 
           &     &     &     &     \\   \vspace{\vgap} 
		0  &  0  &  0  &  0  &  0  \\  
		0  &  0  &  \! -1 \,  &  0  &  0  
          \end{array}  \right)       \label{m5mix}
\end{equation}
       This final set of six matrices generate $\sofi$ group elements that 
\textit{mix} the external $\ol{\bv}_3$ and internal $\ul{\bv}_2$ parts of the full 
5-dimensional temporal flow $\bv_5 = \binom{\ol{\bv}_3}{\ul{\bv}_2}$. The vectors 
$\ol{\bv}_3$ and $\ul{\bv}_2$ are physically distinct with respect to the $M_3$ base 
space. The full SO(5) symmetry is hence broken  down to:
\begin{equation}
   \soth \times \sotw \subset \sofi
\end{equation}
  as the external SO(3) symmetry, represented by the generators of 
equation~\ref{m5ext} `locks on' to the tangent space $TM_3$ leaving the residual 
internal symmetry SO(2), represented by the generator in equation~\ref{m5int}, as 
depicted in figure~\ref{mtogmaph}(b). Hence only four of the original ten generators 
of SO(5) survive the symmetry breaking.

 For the original unbroken full symmetry the constancy of $L(\bv_5)=1$ can be 
expressed, in comparison with equations~\ref{dlvth1} and \ref{dlvth2}, as the 
vanishing of the covariant derivative of $\lvfi$ on the base manifold:
\begin{equation}
   \hat{D}_{\mu}  L(\bv_5) = 0 \quad \Rightarrow \quad 
   \bv_5 \cdot \partial_{\mu} \bv_5 \;\; + \;\;
                                                   \bv_5 \cdot \hat{A}_{\mu} \bv_5 = 0
												   \label{dlvfi}
\end{equation}
   where the `hat' on $\hat{D}_{\mu}$ and $\hat{A}_{\mu}(x)$ signifies that the 
unbroken 10-component so(5)-valued connection 1-form on $M_3$ is being considered. 
With the identification of a Riemannian curvature tensor on $M_3$ six of the gauge 
field generator degrees of freedom are lost and the broken, physical, form of 
equation~\ref{dlvfi} can be written as:
\begin{equation}
   D_{\mu}  L(\bv_5) = 0 \quad \Rightarrow \quad
    \bv_5 \cdot \partial_{\mu} \bv_5 \;\; + \;\;
        \ol{\bv}_3 \cdot A_{\mu} \ol{\bv}_3   \;\; + \;\;
	    \ul{\bv}_2 \cdot Y_{\mu} \ul{\bv}_2  = 0
												   \label{dlvfib}
\end{equation}
   where $A_{\mu}(x)$ represents the external  so(3)-valued connection 1-form and the 
gauge field $Y_{\mu}(x)$ describes an internal so(2)-valued  connection 1-form. The 
final term in equation~\ref{dlvfib} expresses an `interaction' between the gauge field 
$Y_{\mu}(x)$ and the internal temporal components $\ul{\bv}_2(x)$ which follows 
directly from the `minimal coupling' between them implicit in the covariant derivative 
(as a generalisation from the purely external field coupling described for 
equations~\ref{dlvth1} and \ref{dlvth2}). The structure of the pattern of interactions 
for the breaking of the $\esi$ symmetry of $\lvt$ over the base manifold $M_4$ for the 
real world will be described in chapter~\ref{chapesb}.

  Here for the model, a 3-dimensional projection of the full temporal flow $\ol{\bv}_3 
\subset \bv_5$ will form a  tangent vector field on the base manifold $M_3$.
While $\vert \bv_5 \vert = \sqrt{L(\bv_5)}$ is fixed in equation~\ref{vimp1} the 
quantity $\vert \ol{\bv}_3 \vert$ in principle has a variable magnitude, however here 
we mainly focus on the breaking of the group symmetry action and in particular the 
relation between the geometry of the resulting external and internal curvature.

 Considering first the case of figure~\ref{mtogmaph}(a) with an unbroken set of
 ten SO(5) Lie group generators the algebra product is given by the following 
expression, which is valid in general for the orthogonal SO($n$)  groups:
\begin{equation}
 [{L}_{p \pqg q}, {L}_{r \pqg s}] = \delta_{qr}{L}_{p \pqg s} -  \delta_{qs}{L}_{p 
\pqg r}
                              - \delta_{pr}{L}_{q \pqg s} +  \delta_{ps}{L}_{q \pqg r}
							     \label{malg}
\end{equation}
 From this the so(5) algebra structure constants $c^{p \pqg q}_{\ph{pq}r \pqg s \: t 
\pqg u}$ in this basis can be read off, for example  $c^{2 \pqg 3}_{\ph{23}1 \pqg 2 \: 
1 \pqg 3} = -1$ since $[{L}_{1 \pqg 2}, {L}_{1 \pqg 3}] = - {L}_{2 \pqg 3}$.
   Since the group $\sofi$ is connected and compact any element $g \inn \sofi$ in this 
matrix representation can be expressed as 
\begin{equation}
  \label{gexpam}
g=\exp(\alpha_{p \pqg q} {L}_{p \pqg q})
\end{equation}
  with ten real coefficients $\alpha_{p \pqg q}$ (summation is implied over repeated 
index combinations, for the ten labels $\{p,q = 1\ldots 5, p < q\}$, however the 
`upper' or `lower' location of these indices is of no significance). The `exponential 
map' was described in the discussion around figure~\ref{grightran} in the context of 
left-invariant vector fields on the group manifold. Here with the Lie algebra 
represented by real matrices $L =\alpha_{p \pqg q} {L}_{p \pqg q} \inn L(\hat{G})$ the 
exponential map may be explicitly written as $\exp (L) = \sum^{\infty}_{k=0} 
\frac{1}{k!}L^k$ which converges to a map from $L(\hat{G}) \to \hat{G}$.

 The elements of equation~\ref{gexpam} satisfy the relation $gg^T = \b1_5$ (where 
$\b1_5\inn \hat{G}$ is the identity element of the group as represented by the $5 
\times 5$ unit matrix) and $\det(g) = 1$, as required for the special orthogonal group 
$\sofi$. As an example, with the notation $c_{p \pqg q} = \cos\alpha_{p \pqg q}$ and 
$s_{p \pqg q} = \sin\alpha_{p \pqg q}$ the element $g_{1 \pqg 4}=\exp(\alpha_{1 \pqg 
4} {L}_{1 \pqg 4})$ has the form:
\begin{equation}
\def\vgap{-7pt}
    g_{1 \pqg 4} = \left( \begin{array}{@{}ccc|cc@{}}
                                        \vspace{\vgap}
      c_{1 \pqg 4}  &  0  &  0  &  s_{1 \pqg 4}  &  0  \\   \vspace{\vgap}
        0  &  1  &  0  &  0  &  0  \\   \vspace{-2pt}
        0  &  0  &  1  &  0  &  0  \\   \hline  \vspace{-27pt} 
           &     &     &     &     \\   \vspace{\vgap} 
     -s_{1 \pqg 4}  &  0  &  0  &  c_{1 \pqg 4}  &  0  \\  
		0  &  0  &  0  &  0  &  1  
          \end{array}  \right)       \label{g14}     
\end{equation}

The Maurer-Cartan 1-form $\theta$, now defined on the manifold of the \textit{full} 
gauge symmetry group $\hat{G}=\sofi$, can be pulled-back onto the base manifold $M_3$ 
as described in subsection~\ref{mgjoin}. In this way we canonically identify a flat 
so(5)-valued connection 1-form $\hat{A}(x) = g^{\ast} \theta$ and  an so(5)-valued 
curvature 2-form $\hat{F}(x) = 
 \mbox{d}\hat{A} + \frac{1}{2} [\hat{A},\hat{A}]  =0$, following equations~\ref{maca3} 
and \ref{fdefaaa}, where the `hat' on $A$ and $F$ here again denote quantities 
involving the \textit{full} $\sofi$ symmetry group.

  Beginning with a choice of constant gauge $g_c(x): M_3 \to \sofi$, with fixed $g_c 
\inn \sofi$, we have $\hat{A}(x) = g_c^{\ast} \theta = 0$ from equation~\ref{puga}.
Under a more general gauge transformation of the form of equation~\ref{gexpam} but 
with in particular $g_{1 \pqg 4} = \exp (\alpha_{1 \pqg 4} {L}_{1 \pqg 4})$ and with 
small values of the function $\alpha_{1 \pqg 4}(x)$, we have from 
equation~\ref{atrana}:
\begin{eqnarray}
        \hat{A}'(x) & = & g^{-1}_{1 \pqg 4} \hat{A}(x) \, g_{1 \pqg 4}  + 
	                       g^{-1}_{1 \pqg 4} \mbox{d} g_{1 \pqg 4}   \\
                    & \simeq &   \mbox{d} \alpha_{1 \pqg 4}(x)  {L}_{1 \pqg 4} 
					\qquad \; \mbox{for} \; \hat{A}(x)=0
\end{eqnarray}
   This can be written $\hat{A}'(x) = {A}^{1 \pqg 4}(x) {L}_{1 \pqg 4}$ with ${A}^{1 
\pqg 4}(x) =  \mbox{d} \alpha_{1 \pqg 4}(x) = \partial_{\mu}  \alpha_{1 \pqg 4}(x) 
\mbox{d}x^{\mu}$ as the 1-form coefficient of the Lie algebra element ${L}_{1 \pqg 
4}$.
 Applying a full sequence of all six `mixing' actions in the order $g_6 = g_{3 \pqg 5} 
\, g_{3 \pqg 4}\, g_{2 \pqg 5}\, g_{2 \pqg 4}\, g_{1 \pqg 5}\, g_{1 \pqg 4}$ (that is 
with $g_{1 \pqg 4}$ first and $g_{3 \pqg 5}$ last, 
 each generated by an element of equation~\ref{m5mix} and
all being functions of $x\inn M_3$) a 1-form coefficient for each of the ten Lie 
algebra basis elements ${L}_{p \pqg q}$ may be found.  This is equivalent to taking 
$\hat{A}(x) = g_6^{\ast} \theta = g^{-1}_{6} \mbox{d} g_{6}$.  To order $O(\alpha_{p 
\pqg q} \mbox {d} \alpha_{r \pqg s})$ for small transformations the ten coefficients 
of this connection $\hat{A}(x) = {A}^{p \pqg q}(x) {L}_{p \pqg q}$ are found as:
\begin{equation}
    {A}^{1 \pqg 2}  =  -\alpha_{2 \pqg 4} \, \mbox{d}\alpha_{1 \pqg 4} 
	                                -\alpha_{2 \pqg 5} \, \mbox{d}\alpha_{1 \pqg 5}, 
\;\:
	{A}^{1 \pqg 3}  =  -\alpha_{3 \pqg 4} \, \mbox{d}\alpha_{1 \pqg 4} 
	                                -\alpha_{3 \pqg 5} \, \mbox{d}\alpha_{1 \pqg 5}, 
\;\:
	{A}^{2 \pqg 3}  =  -\alpha_{3 \pqg 4} \, \mbox{d}\alpha_{2 \pqg 4} 
	                                -\alpha_{3 \pqg 5} \, \mbox{d}\alpha_{2 \pqg 5}, 
		\nonumber 
\end{equation}
\begin{equation}							
	{A}^{1 \pqg 4}  =   \mbox{d} \alpha_{1 \pqg 4},  \quad 
	{A}^{1 \pqg 5}  =   \mbox{d} \alpha_{1 \pqg 5},  \quad   								
	{A}^{2 \pqg 4}  =   \mbox{d} \alpha_{2 \pqg 4},  \quad    
	{A}^{2 \pqg 5}  =   \mbox{d} \alpha_{2 \pqg 5},  \quad  
	{A}^{3 \pqg 4}  =   \mbox{d} \alpha_{3 \pqg 4},  \quad   								
	{A}^{3 \pqg 5}  =   \mbox{d} \alpha_{3 \pqg 5},      \nonumber
\end{equation}
\begin{equation}
	{A}^{4 \pqg 5}  =  -\alpha_{1 \pqg 5} \, \mbox{d}\alpha_{1 \pqg 4} 
	                            -\alpha_{2 \pqg 5} \, \mbox{d}\alpha_{2 \pqg 4}
							    -\alpha_{3 \pqg 5} \, \mbox{d}\alpha_{3 \pqg 4}
		   \label{a10}
\end{equation}
   The  full so(5)-valued connection 1-form $\hat{A}(x) = {A}^{p \pqg q}(x) {L}_{p 
\pqg q}$, summing over the ten 1-form coefficients in equation~\ref{a10}, is 
`unphysical' in the sense that it is `pure gauge' with respect to the full $\sofi$ 
symmetry, and merely presents the same original flat connection, for which we had all 
ten ${A}^{p \pqg q}(x)=0$, in a different choice of gauge, namely $g_6(x) \inn \sofi$. 

   Of more significance is the structure and interpretation of the ten components of 
the curvature 2-form $\hat{F}$ in the new gauge. These can be written using the 
general expression for the curvature 2-from coefficients in terms of the connection 
1-form coefficients (consistent with equation~\ref{fdefaaa}):
  \begin{equation}
    \hat{F}^{\alpha} = \mbox{d}{A}^{\alpha} + \frac{1}{2} 
\hat{c}^{\alpha}_{\ph{a}\beta\gamma} {A}^{\beta} \wedge {A}^{\gamma} \label{fdahcaa}
  \end{equation}
    The Lie algebra basis indices for $L(\sofi)$ are here denoted by $\{\alpha , 
\beta, \gamma\} = 1\ldots 10$ (with $\alpha = 1,2\ldots$ corresponding to $p \pqg q = 
1 \pqg 2, 1 \pqg 3  \ldots$) and the structure constants may be read off from 
equation~\ref{malg}. For the internal curvature 2-form coefficient $\hat{F}^{4 \pqg 
5}(x)$ in the gauge $g_6(x)$, using the asymmetry in the $\beta\gamma$ indices of the 
structure constants and of the exterior product of 1-forms, we find:
\begin{eqnarray}
  & \!\!\!\!\!  \hat{F}^{4 \pqg 5} \!\!\! & = \;  \mbox{d}{A}^{4 \pqg 5} + \frac{1}{2} 
\hat{c}^{4 \pqg 5}_{\ph{a}\beta\,\gamma} \, {A}^{\beta} \wedge {A}^{\gamma}  \nonumber  
\\
       & = \!\!\! & \mbox{d}{A}^{4 \pqg 5} 
	+ \hat{c}^{4 \pqg 5}_{\ph{a}1 \pqg 4\,1 \pqg 5} \, {A}^{1 \pqg 4} \wedge {A}^{1 
\pqg 5}
	+ \hat{c}^{4 \pqg 5}_{\ph{a}2 \pqg 4\,2 \pqg 5} \, {A}^{2 \pqg 4} \wedge {A}^{2 
\pqg 5}
	+ \hat{c}^{4 \pqg 5}_{\ph{a}3 \pqg 4\,3 \pqg 5} \, {A}^{3 \pqg 4} \wedge {A}^{3 
\pqg 5}  \nonumber \\
	 & = \!\!\! & \mbox{d}{A}^{4 \pqg 5} -  {A}^{1 \pqg 4} \wedge {A}^{1 \pqg 5}
	   - {A}^{2 \pqg 4} \wedge {A}^{2 \pqg 5} - {A}^{3 \pqg 4} \wedge {A}^{3 \pqg 5}
	                                        \nonumber \\
    & = \!\!\! &   \mbox{d}\alpha_{1 \pqg 4} \! \wedge \! \mbox{d}\alpha_{1 \pqg 5} 
         +  \mbox{d}\alpha_{2 \pqg 4} \! \wedge \! \mbox{d}\alpha_{2 \pqg 5}
		 +  \mbox{d}\alpha_{3 \pqg 4} \! \wedge \! \mbox{d}\alpha_{3 \pqg 5}   
		  - \mbox{d}\alpha_{1 \pqg 4} \! \wedge \! \mbox{d}\alpha_{1 \pqg 5} 
         -  \mbox{d}\alpha_{2 \pqg 4} \! \wedge \! \mbox{d}\alpha_{2 \pqg 5}
		 -  \mbox{d}\alpha_{3 \pqg 4} \! \wedge \! \mbox{d}\alpha_{3 \pqg 5}   
\nonumber \\ 
    & = \!\!\! &  0		 
		     \label{ffull}
\end{eqnarray}
   where in the penultimate line the connection coefficients from equation~\ref{a10} 
have been substituted into this expression. This result is as expected and indeed zero 
curvature, $\hat{F}^{p \pqg q} = 0$ for all ten 2-form coefficients, is associated 
with the connection 1-form ${A}^{p \pqg q}$ coefficients, in any gauge choice such as 
that for equation~\ref{a10}, so long as  
 the full 10-dimensional Lie algebra of $\sofi$ is retained.
 
 At each point on $M_3$ the Lie algebra so(5) acts partly on the external space 
$T_xM_3$ of the base manifold in figure~\ref{mtogmaph}(a), via an $\soth \subset 
\sofi$ subgroup, and partly on the internal space through the complementary $\sotw 
\subset \sofi$, while the remaining generators of equation~\ref{m5mix} straddle the 
external and internal parts of $\bv_5 \inn \rrr^5$. By choosing an SO(5) gauge in 
figure~\ref{mtogmaph}(a) and then breaking the symmetry through a projection onto the 
structure in figure~\ref{mtogmaph}(b)
the aim here is to demonstrate how  non-zero curvature may be associated with the 
SO(3) and SO(2) subgroups.

  While an so(5)-valued connection $\hat{A}(x)$ provides a means for the parallel 
transport of a 5-dimensional vector $\bv_5(x)$ over $M_3$ for 
figure~\ref{mtogmaph}(a), a single component such as ${A}^{4 \pqg 5}(x){L}_{4 \pqg 5} 
$ can be interpreted in the \textit{restricted} sense of a representation in the 
subgroup $\sotw \subset \sofi$, that is $\hat{A} \to \ul{A} = {A}^{4 \pqg 5} {L}_{4 
\pqg 5}$ with ${L}_{4 \pqg 5} \inn L(\sotw)$,  acting on the subspace of 2-dimensional 
vectors $\ul{\bv}_2 \subset \bv_5$. In this case $\ul{A}(x)$  is a 1-form connection 
describing the parallel transport of vectors $\ul{\bv}_2(x)$ in the internal vector 
space over $M_3$ for figure~\ref{mtogmaph}(b).

Hence treating the generator ${L}_{4 \pqg 5}$ in isolation from the other nine 
generators as a purely $\sotw$ action on the internal space of vectors $\ul{\bv}_2 
\inn \rrr^2$ a curvature 2-form $\ul{F} = \ul{F}^{4 \pqg 5} {L}_{4 \pqg 5}$ may be 
identified for this \textit{restricted} internal $\sotw$ symmetry.  In comparison with 
equation~\ref{ffull}, for the restricted SO(2) subgroup a finite curvature may be 
obtained:
\begin{eqnarray}
    \ul{F}^{4 \pqg 5} & = & \mbox{d}{A}^{4 \pqg 5}
	                    \nonumber \\  
                 & = &   \mbox{d}\alpha_{1 \pqg 4} \wedge \mbox{d}\alpha_{1 \pqg 5} 
         +  \mbox{d}\alpha_{2 \pqg 4} \wedge \mbox{d}\alpha_{2 \pqg 5} 
		 +  \mbox{d}\alpha_{3 \pqg 4} \wedge \mbox{d}\alpha_{3 \pqg 5} \nonumber \\
		   & \neq & 0   \qquad \mbox{in general.} \label{fint}
\end{eqnarray}
  That is, under a suitable choice of gauge parameters $\alpha_{p \pqg q}(x)$ in the 
full $\sofi$ symmetry it is possible to identify \textit{non-zero} curvature 
components with respect to subgroups such as $\sotw$, which may be interpreted as a 
structure with finite internal physical curvature over the base space manifold.

 To obtain  equation~\ref{fint} we effectively took  a restricted set of structure 
constants, which is trivial for the Abelian subgroup $\sotw \subset \sofi$ with a 
single generator and hence there are no $\ul{c}^{\alpha}_{\ph{a}\beta\gamma}$ terms in 
place of the $\hat{c}^{\alpha}_{\ph{a}\beta\gamma}$ terms of equation~\ref{fdahcaa}. 
This results in a non-zero internal curvature $\ul{F}^{4 \pqg 5} \neq 0$ for the 
subgroup $\ul{H}$ under what was purely a change of gauge from the point of view of 
the full group $\hat{G}$. In this way a finite SO(2) curvature is essentially carved 
out of the degrees of freedom implicit within the structure of the unbroken flat SO(5) 
connection.

Complementary to the subgroup $\sotw \subset \sofi$ the $\soth$ subgroup is generated 
by the Lie algebra elements of equation~\ref{m5ext}.
  Acting on the external vector components $\ol{\bv}_3(x) \inn T_xM_3$ the connection 
1-forms  $A^{1 \pqg 2}{L}_{1 \pqg 2}$, $A^{1 \pqg 3}{L}_{1 \pqg 3}$ and $A^{2 \pqg 
3}{L}_{2 \pqg 3}$, associated with the three generators of $\soth$, define parallelism  
on the external tangent space of the manifold $M_3$. However here we temporarily 
follow the same approach applied to the $\sotw$ case in leading to 
equation~\ref{ffull}, hence treating this SO(3) as an `internal' symmetry in order to 
examine any new features that arise for a non-Abelian gauge group such as $\soth$. 
Again, initially substituting the \textit{full} set of $\sofi$ connection coefficients 
from equation~\ref{a10} into equation~\ref{fdahcaa}, we find $\hat{F}^{1 \pqg 2} = 
\hat{F}^{1 \pqg 3} = \hat{F}^{2 \pqg 3} = 0$  as for all ten so(5)-valued curvature 
components as discussed above.

   A purely SO(3) Lie algebra-valued curvature 2-form can be obtained, similarly as 
for the SO(2) case, by using only the restricted Lie algebra values of the connection 
acting purely on the subspace of  vectors $\ol{\bv}_3 \subset \bv_5$ to identify the 
curvature tensor $\ol{F} = \ol{F}^{p \pqg q}{L}_{p \pqg q}$, with $p \pqg q = 1 \pqg 
2, \, 1 \pqg 3, \, 2 \pqg 3$, over $M_3$. The three components  are obtained from 
equation~\ref{fdahcaa}  by curtailing the summations to include only the so(3)-valued 
parts with a restricted set of structure constants 
  $\{ \ol{c}^{\alpha}_{\ph{a}\beta\gamma} \} \subset
   \{ \hat{c}^{\alpha}_{\ph{a}\beta\gamma} \}$
 describing the external SO(3) symmetry only. Following a similar procedure that led 
to equation~\ref{fint} a set of three so(3)-valued curvature coefficients is found to 
lowest non-trivial order:
\begin{eqnarray}
    \ol{F}^{1 \pqg 2} & = &   \mbox{d}\alpha_{1 \pqg 4} \wedge \mbox{d}\alpha_{2 \pqg 
4} 
         +  \mbox{d}\alpha_{1 \pqg 5} \wedge \mbox{d}\alpha_{2 \pqg 5}     \nonumber 
\\
    \ol{F}^{1 \pqg 3} & = &   \mbox{d}\alpha_{1 \pqg 4} \wedge \mbox{d}\alpha_{3 \pqg 
4} 
         +  \mbox{d}\alpha_{1 \pqg 5} \wedge \mbox{d}\alpha_{3 \pqg 5}     \nonumber 
\\
    \ol{F}^{2 \pqg 3} & = &   \mbox{d}\alpha_{2 \pqg 4} \wedge \mbox{d}\alpha_{3 \pqg 
4} 
         +  \mbox{d}\alpha_{2 \pqg 5} \wedge \mbox{d}\alpha_{3 \pqg 5}      
\label{rext}
\end{eqnarray}

 Hence the SO(3) curvature is also non-zero in general and clearly correlated with the 
$\sotw$ curvature of equation~\ref{fint}, with the correlation mediated through the 
six mixing gauge functions $\alpha_{p \pqg q}(x)$, with $p \pqg q = 1 \pqg 4$, $1 \pqg 
5$, $2 \pqg 4$, $2 \pqg 5$, $ 3 \pqg 4$, $3 \pqg 5$ under the full $\sofi$ symmetry. 
Either the SO(3) or the SO(2) curvature may be non-zero while the other remains zero 
for a suitable choice of the $\alpha_{p \pqg q}(x)$, while both $\ol{F}=0$ and 
$\ul{F}=0$ are simultaneously attained under any choice of SO(5) gauge with constant 
$\alpha_{p \pqg q}(x)$ for example.

 The six full $\sofi$ curvature components $\hat{F}^{p \pqg q} = 0$ for $p \pqg q = 1 
\pqg 4$, $1 \pqg 5$, $2 \pqg 4$, $2 \pqg 5$, $ 3 \pqg 4$, $3 \pqg 5$ are not  directly 
associated with subgroup restrictions giving rise to further finite curvature  
components. Rather  these six mixing degrees of gauge symmetry are lost or 
\textit{broken} in the projection of the full $\sofi$ symmetry over $M_3$  through 
which finite physical curvature for the four components in equations~\ref{fint} and 
\ref{rext} is identified.
  The four  functions $\alpha_{p \pqg q}(x)$, with $p \pqg q = \{1 \pqg 2$, $1 \pqg 
3$, $2 \pqg 3$\} and $4 \pqg 5$, corresponding to four  actions $g_{p \pqg q} = 
\exp(\alpha_{p \pqg q}{L}_{p \pqg q})$ of the gauge symmetry,  \textit{survive} the 
symmetry breaking and are retained as the gauge symmetries associated with the 
non-Abelian SO(3) and Abelian SO(2) subgroups respectively.

  However the mechanism of symmetry breaking over $M_3$ itself implies that there is 
   a more fundamental difference between the $\soth$ and $\sotw$ subgroups in the 
context of the model world we are considering. The former acts \textit{externally} on 
the tangent space of the base manifold, that is on $\ol{\bv}_3 \inn TM_3$ as depicted 
in figure~\ref{mtogmaph}(b), and is therefore closely related to the geometry of the 
background space itself and to the existence of a \textit{linear connection} on the 
base manifold. Hence the correlation observed above between the SO(3) and SO(2) 
curvature, with both effectively treated as \textit{internal} symmetries, merely 
provides a provisional motivation for seeking a unified framework in which SO(2) 
remains as an internal gauge symmetry while SO(3) is considered as an external 
symmetry on $TM_3$.   
  A non-zero curvature for the internal symmetry $\ul{F}^{4 \pqg 5} \neq 0$ in 
equation~\ref{fint} was 
  obtained by considering the generator of the subgroup $\ul{H}=\sotw$ within the full 
gauge symmetry group $\hat{G}=\sofi$ over a fixed base space $M_3$. A non-zero 
Riemannian curvature with components $R^\rho_{\ph{\rho}\sigma\mu\nu}(x)\neq 0$ on 
$M_3$ will transform  locally under the complementary external subgroup $\soth \subset 
\sofi$ acting on the tangent space $TM_3$. 

In this chapter the base manifold $M_3$ and group manifold $G$
 (where here $G$ may be the full symmetry $\hat{G}$ of the full form $\lvn$ or either 
the internal $\ul{H}$ or external $\ol{H}$ subgroups)
 have been treated as largely independent geometric objects, however their mutual 
relationship  is more precisely defined in terms of a single manifold in the form of a 
principle bundle with base space $M_3$ and structure group $G$. Hence the standard 
properties of these geometric objects, together with a review of Riemannian geometry 
and the Lagrangian approach to obtaining equations of motion for the corresponding 
field entities, with be presented in the following chapter.

  A relationship between the internal gauge curvature and external Riemannian 
curvature might be determined through such a principle fibre bundle $P = (M_3, 
\hat{G})$ with base space $M_3$ and structure group $\hat{G}$, based on the picture in 
figure~\ref{mtogmaph}(a), with the canonical zero full curvature $\hat{F}=0$ providing 
the constraint that relates the internal and external geometry.
 However the physical situation is represented by figure~\ref{mtogmaph}(b) which leads 
to a consideration of two detached bundle spaces, $\ol{P} = (M_3, \soth)$ and $\ul{P} 
= (M_3, \sotw)$. While $\ol{P}$ directly only contains information about the external 
symmetry and curvature, the bundle space $\ul{P}$ explicitly contains both the 
structure of the external geometry on $M_3$ and that of the internal $\ul{H}=\sotw$ 
curvature in the bundle space. 
As a unifying framework for combining  external and internal symmetries this latter 
structure is very similar to that employed in non-Abelian Kaluza-Klein theories, which 
are hence reviewed in chapter~\ref{kktheory}.

  In chapter~\ref{chaputtf}  we consider how the above structures of principle bundles 
and Kaluza-Klein theory might be adapted for the present theory. There we shall 
upgrade the model presented in this chapter by considering the real world situation 
with a 4-dimensional base space $M_4$ with local Lorentz symmetry. This will be 
embedded in the `full' symmetry group taken as $\hat{G}=\sootn$, acting on a 
10-dimensional form of temporal flow $\lvte$,  which will be broken to $\soot \times 
\sox$ in the projection onto the spacetime base $M_4$. This symmetry breaking 
structure naturally embeds in the further higher-dimensional extensions considered 
from chapter~\ref{esihtho}, which will provide a more realistic framework for the 
details of the internal structures also.


\pagebreak

\chapter{Review of Geometry and Equations of Motion}
\label{rogaeom}

\section{Principle Bundle Structure}
\label{fibre}

  In section~\ref{perc} we introduced \textit{two} independent differentiable 
manifolds, the base space $M_3$ and Lie group $G=\soth$ with points labelled by $x 
\inn M_3$ and $g \inn G$
 respectively, both of which are associated with the 3-dimensional form of temporal 
flow $\lvth$ through the respective `translational' $\{r^a\} \inn \rrr^3$ and 
`rotational' $\sigma_g : g \inn G$  symmetries:
  \begin{equation}
     \label{lv2syms}
	   L\left(\sigma_g \left\{ \frac{d(x^a + r^a)}{ds} \right\} \right) = 1
  \end{equation}
    The map between the manifolds $g:M_3 \to G$ described in figure~\ref{mtogmap}, 
mapping $x \to g(x)$ (where \{$x^{\mu}$\} may be taken as general coordinates on 
$M_3$) represents a local choice of gauge, or orthonormal frame, in which to express 
the tangent vector $\bv_3(x)$ on $M_3$.  This association between $M_3$ and $G$ may be 
examined more precisely through the construction of a \textit{single} differentiable 
manifold, namely a principle fibre bundle $P$, which combines the geometric properties 
of the base space $M_3$ and Lie group $G$ together with their mutual relation.

   In the general case the structure group $G$ of a principle fibre bundle $P$ does 
not need to be related to a symmetry on the tangent space to the base manifold $M$, as 
it is for the SO(3) model as implied in equation~\ref{lv2syms}. Indeed in the case of 
figure~\ref{mtogmaph}(a) the symmetry group $G=\sofi$  acts only partially  on the 
tangent space of $M_3$, while for figure~\ref{mtogmaph}(b) the group $G=\sotw$ does 
not act on the external tangent space at all.
 Hence it is the generalisation in which $G$ and $M$ are initially  introduced 
independently  that we shall review here for the benefit of the subsequent application 
to the case of a higher symmetry group such as presented in section~\ref{hdasb}.

    A principle bundle $P$ is a $G$-manifold, that is a differentiable manifold upon 
which the transformation group $G$ acts, with a particular structure as described, 
with reference to figure~\ref{pbundle}, by the following properties (see for example 
\cite{kob}, \cite{amp} chapter Vbis, \cite{fecko}):
\begin{figure}[htbp]  
\centering
\epsfxsize=\maxwidth
\leavevmode
\epsffile[0 0 1583 709]{\gpath aPfig31e}
\caption{\setb The relations between three differentiable manifolds: a principle fibre 
bundle $P$, the base space $M$ and the structure group $G$.}
\label{pbundle}
\end{figure} 
\begin{itemize}
\item[1:] There is a surjective map $\pi : P \to M$ projecting from the bundle space 
onto the base manifold. Given a section $\sigma(x):M \to P$ then 
  $\pi \,\scirc\, \sigma(x) = x$ is the identity map on points $x \inn M$.
\item[2:] For each $x \inn M$ the submanifolds $\pi^{-1}(x) \subset P$, called the 
fibres of $P$, are diffeomorphic to each other and to the Lie group $G$.
\item[3:] The right action of $g \inn G$ on points $p \inn P$, that is $R_g \! : P \to 
P$ with $R_{gh} = R_h \,\scirc\, R_g$, preserves the fibres of $P$, that is $\pi 
\,\scirc\, R_g = \pi$, and is free and transitive on each fibre.
\item[4:] There exist local trivialisations over each open subset $U_{r} \subset M$, 
consisting of maps $\psi_{r}: \pi^{-1}(U_{r}) \to U_{r} \times G$ with $\psi_{r}: p 
\to (x,h)$, such that   $\psi_{r}: pg \to (x,hg)$ -- that is the right action on $P$ 
is compatible with the right action on $G$. 
\end{itemize}

    While the right action of $G$ on itself induces left-invariant fields as described 
in equation~\ref{atoxa} of subsection~\ref{tgman}, the right action of $G$ on the 
manifold $P$ induces `vertical' vector fields in the tangent space $\mbox{\it TP}$ as 
(where $f$ is now a real-valued function on the bundle space):  
\begin{equation}
   \label{atova}
    V_p^A(f) = \frac{d}{dt} \: f(p \, \exp(tA)) \, \vert_{t=0}
\end{equation}
    where $A\inn L(G)$ and $V_p^A$ is a tangent vector to the fibre of $P$ at the 
point $p$. The map $A \to V_p^A$ described in equation~\ref{atova} represents an 
\textit{isomorphism} of the Lie algebra $L(G)$ into the space of vector fields 
residing in the vertical tangent space $\mbox{\it VP}\subset \mbox{\it TP}$. That is, 
the Lie algebra bracket structure $[X^A,X^B] = X^{[A,B]}$ of equation~\ref{xxcomcx} 
for the corresponding left-invariant fields $\{X^A\}$ on the manifold $G$ is respected 
by the Lie bracket on the $P$ bundle with:    
 \begin{equation}
    \label{vavbvab}
    [V^A,V^B] = V^{[A,B]}
\end{equation}

   This structure relates to $\mbox{\it VP}$, the space of vectors tangent to the 
individual fibres of $P$. Different fibres may be related by an additional structure 
on $P$ called a \textit{connection} which, conceptually, is smooth assignment of a 
`horizontal' subspace $H_pP$ of the full tangent space $T_pP$ at each point $p\inn P$ 
such that:
\begin{eqnarray}
              T_pP &  =  &  V_pP \oplus H_pP   \label{vplush}  \\
   R_{g\ast} H_pP  &  =  &  H_{pg}P         \label{racthp}     
\end{eqnarray}
  where compatibility of the horizontal subspaces on $P$ with the right action of $G$ 
is assured by the latter requirement.

   At every point $p \inn P$ a basis for the tangent space of the principle bundle can 
be expressed in terms of these complementary subspaces. Such a basis $\{ {\acute e}_i 
\} = \{{\acute e}_{\alpha}, \, {\acute e}_a\}$ consists of the subset $\{{\acute 
e}_{\alpha}\} \inn \mbox{\it VP}$ (that is vectors of the form $V_p^A$ in 
equation~\ref{atova}, tangent to the fibres $G_x$ over each point $x\inn M$) and the  
subset
$\{{\acute e}_a \}\inn \mbox{\it HP}$, where ${\acute e}_a $ is the `horizontal lift' 
of the basis vector $e_a \inn T_xM$ to the point $p \inn P$ such that $\pi_{\ast} 
{\acute e}_a = e_a$.
 The `acute' mark above the kernel symbol, such as for ${\acute e}$, denotes an object 
defined on a principle bundle space in the horizontal lift basis. In all cases 
 the indices $\{i,j,k\ldots \}$ correspond to basis elements for $\mbox{\it TP}$ in 
the total space; $\{ \alpha,\beta,\gamma\ldots \}$ in the fibre space on $P$ or on the 
manifold $G$; and $\{ a,b,c\ldots \}$ in a complementary subspace on $P$ or on the 
base space $M$.

  It should be noted that ${\acute e}_a$ and $e_a$ are not only different vector 
fields but are also defined on two different manifolds, $P$ and $M$ respectively, 
although there is a one-to-one correspondence between them. Similarly, there is a 
one-to-one correspondence between a vector field ${\acute e}_{\alpha}$ on $P$ and a 
vector field $X_{\alpha}$ on $G$, for example as generated by the same element $A\inn 
T_eG$ in equations~\ref{atova} and \ref{atoxa} respectively. The relations between 
these vector fields are indicated in figure~\ref{pbunbasis}. 
\begin{figure}[htbp]  
\centering
\epsfxsize=12cm
\leavevmode
\epsffile[0 0 1246 728]{\gpath aPfig32e}
\vspace{-10pt}
\caption{\setb Vertical and horizontal basis vectors in a local trivialisation $U 
\times G$ of a principle bundle  $P$, together with their associated basis vectors on 
the group space $G$ and the base manifold $M$ respectively.}
\label{pbunbasis}
\end{figure}

   The specification of a connection on the principle bundle allows `parallel 
transport' \textit{between} the fibres to be defined by a path in $P$ for which the 
tangent vector at any $p\inn P$ always lies within the horizontal subspace $H_pP$. 
This notion of parallelism over $M$ is used in turn to define a \textit{covariant 
derivative} for associated fields $\phi(x)$ on the base space that transform under a 
representation of the structure group $G$, by tracking a parallel basis for the field 
$\phi(x)$ over any curve $C$ on the base manifold (technically, $\phi(x)$ is a section 
in a fibre bundle associated with $P$).

  As depicted in figure~\ref{paratran} given a point $p_1\inn P$ with $\pi(p_1)=x_1$ 
and a curve $C$ on $M$ from $x_1$ to $x_2$ a connection on $P$ specifies a unique 
horizontal lift of the curve $C$ to the curve $C'$ on $P$, by advancing locally within 
the horizontal subspace $\mbox{\it HP} \subset \mbox{\it TP}$. The path $C'$ then 
represents the `parallel transport' of $p_1$ mapped to the unique point $p_2 \inn P$, 
with $\pi(p_2)=x_2$. 

\begin{figure}[htbp]  
\centering
\epsfxsize=12cm
\leavevmode
\epsffile[0 0 1496 1097]{\gpath aPfig33e}
\caption{\setb Parallel transport $C'$ between fibres on a principle fibre bundle $P$ 
above the curve $C$ on the base manifold from the point $x_1\innf M$ to $x_2 \innf 
M$.}
\label{paratran}
\end{figure}

 The geometric structure developed in section~\ref{perc} corresponds to a particular 
kind of principle bundle, namely a frame bundle with structure group $G=\soth$ over 
the base space $M_3$, which may be denoted $P = (M_3, \soth)$. In this case the 
mapping from $p_1$ to $p_2$ in figure~\ref{paratran} provides a unique, path $C$ 
dependent, transport of an orthonormal basis frame from $x_1$ to $x_2$ on the base 
manifold (such basis frames $\{e_a\}$ are shown in figure~\ref{spill} for the model on 
$M_3$). With respect to such a parallel frame the difference between the values of the 
vector field $\bv_3(x)$ (belonging to the vector representation of $\soth$)  at the 
two base points of the associated vector bundle can be determined. 
 In particular vectors $\bv_3(x_1)$ and $\bv_3(x_2)$, as originally depicted in 
figure~\ref{spill}, are defined to be parallel with respect to a given path $C$ and 
connection $\mbox{\it HP}$ if each of their components coincide in an orthonormal 
reference frame transported from $x_1$ to $x_2$ along $C$ via the horizontal lift 
$C'$. This definition of parallelism is independent of the choice $p_1 \inn 
\pi^{-1}(x_1)$ of initial frame. In addition, any vector $\bv_3(x_1)$ may be `parallel 
transported' to any point of the curve $C$ by maintaining constant vector components 
in the corresponding transported frame of the principle bundle at each point along 
$C$. 

       This construction generalises for an arbitrary structure group $G$ acting via a 
group representation on the field $\phi(x)$ over the base manifold $M$. The covariant 
derivative of the field $\phi(x)$ is defined in a such a way as to quantitatively 
indicate deviations of the value of the field function from that of the parallel 
transported field for infinitesimal displacements on the base manifold -- i.e. the 
extent to which the field is not self-parallel along a path on the base manifold. That 
is, if $\bc$ is a vector at $x_0$ tangent to the curve $C$ parametrised by $\lambda 
\to C(\lambda)$ on $M$ with $C(0)=x_0$, see figure~\ref{paratran}, then the covariant 
derivative of the field $\phi(x)$ along $C$ at $x_0$ is defined as:
\begin{equation}
   \label{covv}
    D_{\mathbf{c}}\phi \, \vert_{x_0}=\lim_{\lambda\to 0}(T_{\lambda,0}\,\phi 
(x_{\lambda})-\phi (x_0))/\lambda
\end{equation}
      where $T_{\lambda,0}\,\phi(x_{\lambda})$ is the field value $\phi 
(x_{\lambda})\equiv\phi(C(\lambda))$ parallel transported along $C$ from $x_{\lambda}$ 
to $x_0$.
The covariant derivative for the SO(3) connection applied to vectors $\bv_3(x)$ in 
equation~\ref{dlvth1} and \ref{dlvth2}  was denoted $D_{\mu}$ corresponding to 
derivatives with respect to general coordinate parameters $\{x^{\mu}\}$ on $M_3$.

     If the connection is such that, for all $p\inn P$ and all $X,Y \inn H_pP$, the 
bracket composition on $H_pP$ is closed, that is: 
\begin{equation}
 [X,Y] \inn H_pP
 \label{frobcon}
\end{equation}
 then Frobenius criterion is satisfied and $P$ is `foliated' into a family of 
integrable `leaves'. Each leaf is `horizontal section' of $P$, with tangent space 
$\mbox{\it HP}$, that is a smooth $n$-dimensional submanifold of $P$, where $n$ is the 
dimension of the base space $M$. In this case all horizontal lift curves $C'$ of 
figure~\ref{paratran} effectively follow the contours of a single horizontal section 
submanifold, \textit{globally} defined over $M$, and parallelism is independent of the 
path $C$ taken between any two points on the base manifold. A `flat' connection is 
defined by this property, as will be described in more detail in the following 
section.

  Generally a horizontal subpace can be specified by a connection 1-form $\omega$ on 
$P$. 
   The defining structure for $\mbox{\it HP}$ of equation~\ref{vplush} and 
\ref{racthp} can be attained via a smooth Lie algebra-valued 1-form $\omega \inn L(G) 
\otimes T^{\ast}\!P$, mapping vectors $X\inn \mbox{\it TP}$ into elements of $L(G)$,  
with the properties (the first of which is essentially the reverse of 
equation~\ref{atova}):
\begin{eqnarray}
  & \mbox{(i)}  &   \omega (V^A) = A \qquad \mbox{with} \; A\inn L(G)  \label{wvaalg} 
\\
   & \mbox{(ii)}  &  R^{\ast}_g \omega = \mbox{Ad}(g^{-1}) \omega \quad \mbox{i.e.} \;
       R^{\ast}_g \omega_{pg}(X) = g^{-1} \omega_p(X)g   \quad \mbox{with} \; X\inn 
T_pP  \qquad \label{rwgog}  \\
  & \mbox{where} & H_pP \equiv \{ X \inn T_pP \mid \omega(X) = 0 \} 
\end{eqnarray}

is the horizontal subspace.
  A set of trivialisations $\{U_r,\psi_r \}$ consists of an atlas \{$U_r$\} covering 
the base manifold $M$ together with a mapping $\psi_r$ of each $\pi^{-1}(U_{r}) 
\subset P$ onto $U_{r} \times G$ such as depicted figure~\ref{pbundle} and described 
in the subsequent `item 4:'. Each trivialisation $\psi_{r}$ is canonically associated 
with a section in $\pi^{-1}(U_{r})$ which can be written as $\sigma_{r}(x) = 
\psi_{r,x}^{-1} \cdot \iota(x)$, where the map $\iota:U_r \to U_r \times G$ sends $x 
\to (x,e)_r$, with $e\inn G$ being the identity element of the group and $\psi_{r,x}$ 
is the map $\psi_{r}$ restricted to the space $\pi^{-1}(x)$. That is, $\sigma_{r}(x) 
\inn P$ corresponds to the identity element $e\inn G$ under the local trivialisation 
map $\psi_{r}$, as depicted in figure~\ref{uusection}. More generally we have the map 
$\psi_{r,x}: P \to (U_{r}\times G)  $  mapping between the points $\sigma_{r}(x)h \to 
(x,h)_r$, with $h\inn G$.

\begin{figure}[htbp]  
\centering
\epsfxsize=\maxwidth
\leavevmode
\epsffile[0 0 1972 919]{\gpath aPfig34e}
\caption{\setb Two trivialisations, denoted $\{U_{r},\psi_{r}\}$ and 
$\{U_{s},\psi_{s}\}$, with an overlap region, on a principle bundle $P$ over a base 
space $M$.}
\label{uusection}
\end{figure}

   In the overlap regions on $M$, for $x\inn U_{r} \cap U_{s}$,  transition mappings 
$g_{rs}: \: U_{r} \cap U_{s} \to G$ between such trivialisations are defined as the 
functions on $M$:
\begin{equation}
   x \to g_{rs}(x) = \psi_{r,x} \,\scirc\,  \psi_{s,x}^{-1}  \inn G
\end{equation}
   The transition functions act on the left on a fibre such that 
$\psi_{r,x}^{-1}(x,g_{rs}h)_r = \psi_{s,x}^{-1}(x,h)_s$.
   These relate the corresponding canonical sections $\sigma_{r}(x)$ via the right 
action of the structure group on $P$, which commutes with the left action, in a way 
that is consistent with both $\psi_{s}(\sigma_{s}(x)h) = (x,h)_{s}$ and 
$\psi_{r}(\sigma_{r}(x)h') = (x,h')_{r}$ in the respective trivialisations, as:
 \begin{equation}
 \label{sitosj}
   \sigma_{s}(x) = \sigma_{r}(x) g_{rs}(x)
\end{equation}

   Given a general connection 1-form $\omega$ on $P$ and  a set of trivialisations 
$\{U_{r},\psi_{r} \}$  a unique family $A_{r}$ of connection 1-forms may be defined on 
$M$.
 Under a particular section $\sigma_r$ the connection 
$A_r(x)=\sigma_r^{\ast}\omega(p)$ on $M$ can be expressed as $A_r(x)= A^{\alpha}(x) 
X_{\alpha} = A^{\alpha}_{\ph{\alpha}\mu}(x) X_{\alpha} \mbox{d}x^{\mu}$ where 
$\{X_{\alpha} \}$ is a basis for $L(G)$. 
 The field of connection coefficients $A^{\alpha}_{\ph{\alpha}\mu}(x)$ link the basis 
$\{ X_{\alpha} \}$ for the Lie algebra, typically expressed in the appropriate 
representation (such as the set of matrices $\{E_{\alpha}\}$  of equation~\ref{m3} for 
the case of the vector representation of $G=\soth$), with indices $\alpha = 1\ldots 
\mbox{dim}(G)$, to a coordinate basis of 1-forms $\{ \mbox{d}x^{\mu} \}$  with  
indices $\mu = 0,1,2,3$ in the case of a 4-dimensional spacetime base manifold $M_4$. 
Further,
given an $L(G)$-valued 1-form $A(x)$ on $M$ and any section $\sigma(x)$ then there 
exists a unique connection 1-form $\omega(p)$ on $P$ such that $A = \sigma^{\ast} 
\omega$.

 In a  gauge theory, that is a theory which is invariant under transformations of  the 
gauge group $G$ which describes a local internal symmetry,
 the local Lie algebra-valued 1-form $A(x)$ on the base manifold $M$ is also known as 
a Yang-Mills field or `gauge potential'. Such fields will be generically denoted 
$Y(x)$ in this paper, as for example in equation~\ref{dlvfib}. The notation $A(x)$ may 
refer to a general connection 1-form, as described above, the gauge field associated 
with an internal $\uo$ gauge symmetry, as for electromagnetism, or a connection 
associated with an orthonormal frame in the external space, as was the case in 
subsection~\ref{mgjoin} and as will be the case in relation to general relativity, 
depending on the context. A gauge theory bases upon an internal symmetry, through the  
notion of a connection 1-form, involves similar mathematical structures as found in 
general relativity based upon an external symmetry.

  For the present theory it will be assumed that the structure of principle bundles 
with a trivial global topology
  will be sufficient. In this case the bundle $P=(M,G)$ can be expressed as
   $P \equiv U \times G$ where a single `subset' $U$ of figure~\ref{uusection} may be 
identified with the entire base manifold $M$. This triviality is implied in deriving 
the bundle structure through the symmetries of $\lv$ as described for example in 
figures~\ref{spillout}, \ref{mtogmap} and \ref{mtogmaph}.
 In this case the `overlap region' for a change of trivialisation, or gauge 
transformation, may consist of the entire volume of the base space $M$, rather than a 
limited patch as depicted in figure~\ref{uusection}.

In order to study the dynamics of the gauge fields it is helpful to introduce the 
\textit{exterior covariant derivative} on the bundle space, 
which will be important for equations in physics.
 This derivative essentially combines the properties of 
 the exterior derivative, introduced for equation~\ref{ddef}, with the structure of 
the partial derivative  $\pal_{\mu}$ as augmented to the covariant derivative 
$D_{\mu}$, as described for equation~\ref{covv}.

  More explicitly, on a principle bundle $P$ the exterior covariant derivative D maps 
a $V$-valued $r$-form $\phi$, which acts upon $r$ vector fields $\{X_1\ldots X_r\}$ on 
$P$, to a $V$-valued $(r+1)$-form D$\phi$, where $V$ is the representation space 
associated with $G$. The action of D is defined as:
\begin{equation}
 \mbox{D}\phi = (\mbox{d}\phi) \,\scirc\, \mbox{hor}    \label{extcovd}
\end{equation}
  where `hor' first maps vectors $X$ on the tangent space of $P$ to their horizontal 
components (that is, $\mbox{hor}: \, X \to X_h$, such that $X_h \subset  H_pP$ of 
equation~\ref{vplush} and $\omega(X_h)=0$, with $\omega$ the connection 1-form on 
$P$), and \mbox{d} is the exterior derivative map acting on the $r$-form $\phi$.

    In general, for a $V$-valued $r$-form $\phi$ on $P$ which is \textit{horizontal} 
(that is if any of the $r$ vectors $X$ on which $\phi$ acts is purely vertical then 
the map is zero, $\phi (X, \ldots) = 0$) and \textit{equivariant} of type $\rho$ (that 
is $\phi$ in the associated bundle transforms as $R^{\ast}_g \phi = \rho(g^{-1})\phi$ 
under the right action by $g(x)$ in the $\rho$ representation) then the exterior 
covariant derivative of $\phi$, equation~\ref{extcovd}, takes the simplified form:
  \begin{equation}
     \mbox{D}\phi  =  \mbox{d}\phi \, + \, \rho'(\omega) \wedge \phi. \label{dphiecd}
  \end{equation} 
  where $\rho'(\omega)$ denotes the appropriate representation of the Lie algebra 
acting on $V$.

\section{Curvature and Flat Connection}  
\label{cacfc}

The curvature 2-form $\Omega$ can be \textit{defined} as the exterior covariant 
derivative of the connection 1-form, that is $\Omega = \mbox{D}\omega$, on the 
principle bundle. The connection 1-form $\omega$ is equivariant, of type $\mbox{Ad}$ 
as seen in equation~\ref{rwgog}, but it is clearly not a horizontal form, as seen in 
equation~\ref{wvaalg}. However, for this particular case the exterior covariant 
derivative of the connection 1-form  can also be expressed in a simplified form 
directly in terms  of $\omega = \omega^{\alpha}X_{\alpha}$ itself (with 
$\{X_{\alpha}\}$ a basis for $L(G)$) through the Cartan structure equation for the 
curvature 2-form $\Omega = \Omega^{\alpha}X_{\alpha}$ on $P$:
\begin{eqnarray}
   \Omega(X,Y) = \mbox{D}\omega (X,Y)
    & = & \mbox{d}\omega (X,Y) + [\omega(X),\omega(Y)]  \label{omegado1}  \\
    & = &    \mbox{d}\omega (X,Y) + \fh[\omega,\omega](X,Y)    \label{omegadoh}       
\\
 \mbox{that is} \qquad  \Omega^{\alpha}(X,Y) & = &  \mbox{d}\omega^{\alpha}(X,Y) + 
\frac{1}{2}
             \cstr \omega^{\beta} \wedge \omega^{\gamma}(X,Y)                  
\label{omegado2}
\end{eqnarray} 
   acting upon any pair of tangent vectors $X,Y \inn \mbox{\it TP}$ (the meaning of  
$[\omega,\omega]$ is explained in the discussion around equation~\ref{thphiw}).
   The Lie algebra-valued curvature 2-form $\Omega$ on $P$ is defined in such a way as 
to be quantitatively sensitive to deviations of the connection 1-form $\omega$, and 
hence horizontal subspace $\mbox{\it HP}$, from the condition of flatness. This can be 
seen by substituting any $X,Y \inn \mbox{\it HP}$ as arguments for the curvature 
2-form $\Omega$ in equation~\ref{omegado1}, or equivalently for any $X,Y \inn 
\mbox{\it TP}$  
 in terms of  $\Omega = \mbox{D}\omega$ we have:
\begin{eqnarray}
  \Omega(X,Y) & = & \mbox{D}\omega(X,Y)   \\
               & = & \mbox{d}\omega(X_h,Y_h)  \label{doxhyh} \\
			   & = & X_h \langle \omega, Y_h \rangle - Y_h \langle \omega, X_h \rangle
			             - \langle \omega, [X_h, Y_h] \rangle  \label{doform}   \\
			   & = &   - \langle \omega, [X_h, Y_h] \rangle 
\end{eqnarray}
   where equation~\ref{doxhyh} follows directly from equation~\ref{extcovd}, 
equation~\ref{doform} follows from the standard definition of the exterior derivative 
of a 1-form and the map $\langle\;,\;\rangle$ was defined in equation~\ref{formap}. 
Hence it follows that $\Omega$ is non-zero only if the local horizontal subspaces on 
$P$ defined by $\omega$ are non-integrable, that is the Frobenius criterion of 
equation~\ref{frobcon} is not satisfied, and hence a non-zero curvature $\Omega$ 
indeed indicates a non-flat connection.

   The Lie algebra valued curvature 2-form $\Omega$ on $P$ is equivariant of type 
$\mbox{Ad}$, that is it transforms under the adjoint representation of $G$ as 
$R^{\ast}_g \Omega = \mbox{Ad}(g^{-1})\Omega$. However, unlike the connection 
$\omega$, the curvature $\Omega$ is also a horizontal form on $P$. Hence $\Omega$, 
unlike $\omega$, is a \textit{tensorial} form meaning that, for a given choice of 
gauge or cross-section $\sigma$ over a region of the base manifold, it can be mapped 
via the pull-back $\sigma^{\ast}$ to a geometrical object on the base manifold that 
transforms homogeneously as a representation of the gauge group.
 Such quantities may be more naturally equated in the expressions of physics. For the 
model universe of the previous chapter the curvature form of the $\soth$ connection on 
$P$ is tensorial of type (Ad, $\sotha$), that is it takes values in the $\soth$ Lie 
algebra and transforms under the adjoint representation, as indicated after 
equation~\ref{fdefaaa} for the curvature form $F$ on the base manifold.

   In a trivialisation $P \equiv U\times G$ on the principle bundle a 
  direct product basis  $\{ {\ddot e}_i \} = \{{\ddot e}_{\alpha}, \, {\ddot e}_a\}$ 
for the tangent space 
   consists of the subset $\{{\ddot e}_{\alpha}\} \inn \mbox{\it VP}$, tangent to the 
fibres $G_x$ over each point $x\inn M$, and the  subset
$\{{\ddot e}_a \}$  with ${\ddot e}_a = \sigma_{\ast} e_a $ for each basis vector $e_a 
\inn T_xM$ such that $\pi_{\ast} {\ddot e}_a = e_a$.
 Each vector ${\ddot e}_a$ defined on the section $\sigma(x)$ is Lie transported via 
the right action of $G$ on $P$ such that the basis covers the entire principle bundle.
 The `double dot' mark above the kernel symbol, such as for ${\ddot e}$, denotes an 
object defined on a principle bundle space in the direct product basis.

   Since $P$ itself is a differentiable manifold  equation~\ref{cabconm} applies for 
any frame field $\{e_i\}$ on $P$ and is here expressed as:
\begin{equation}
   [{e}_j, {e}_k] = {c}^i_{\ph{i}jk}(p) {e}_i    \label{eecefull}
\end{equation}
    with real-valued structure coefficients ${c}^i_{\ph{i}jk}(p)$.
  In the direct product basis the bracket relations
  $[\ddot{e}_j, \ddot{e}_k] = \ddot{c}^i_{\ph{i}jk}(p) \ddot{e}_i$ are simply:  
\begin{eqnarray}
  \lbrack \ddot{e}_{\alpha}, \ddot{e}_{\beta} \rbrack & = &
                        c^{\gamma}_{\ph{\gamma}\alpha\beta}\ddot{e}_{\gamma} 
\label{barb1}  \\
  \lbrack \ddot{e}_{\alpha}, \ddot{e}_{b} \rbrack & = &  0   \label{barb2} \\
  \lbrack \ddot{e}_{a}, \ddot{e}_{b} \rbrack & = &
                        0   \label{barb3}
\end{eqnarray}
   where $c^{\gamma}_{\ph{\gamma}\alpha\beta}$ are the structure constants of the 
group $G$. The zero coefficients for the second equation follow as the vector fields 
$\ddot{e}_{\alpha}$ generate the right translations which Lie transport the vectors 
$\ddot{e}_{b}$ over $P$, and those in the final equation correspond to the choice of a 
coordinate basis on $M$.

    By contrast the horizontal lift basis ${\acute e}_i = ({\acute e}_{\alpha}, 
{\acute e}_a)$
for the tangent space $\mbox{\it TP}$, introduced after equation~\ref{vplush}, is 
adapted to a given connection $\omega$ such that ${\acute e}_{\alpha} \inn V_p$ and  
${\acute e}_a \inn H_p$, as was depicted in figure~\ref{pbunbasis}, with 
$\omega(\acute{e}_{\alpha}) = X_{\alpha}$ and $\omega(\acute{e}_{a}) = 0$, by the 
definition of the horizontal lift basis. Given a trivialisation the horizontal lift 
basis $\{ \acute{e}_i \}$ can be expressed in terms the direct product basis $\{ 
\ddot{e}_i\}$ on $P$ 
 via the coefficients $\omega^{\alpha}_{\ph{\alpha}a}(x,g)$ with:
\begin{eqnarray}
   \acute{e}_{\alpha}  =  \ddot{e}_{\alpha}, & \qquad &
   \acute{e}_a  = \ddot{e}_a \, - \, \omega^{\alpha}_{\ph{\alpha}a} \ddot{e}_{\alpha}  
    \label{ehtoeb}   \\
   \acute{e}^{\alpha} = \ddot{e}^{\alpha} \, + \,
       	\omega^{\alpha}_{\ph{\alpha}a} \ddot{e}^{a},
		 & \qquad &  \acute{e}^a  = \ddot{e}^a  
		   \label{dualetoe}
\end{eqnarray}
  where $\{ {\acute e}^i \} = \{{\acute e}^{\alpha}, {\acute e}^{a}\}$ is the dual 
basis  defined as usual such that $\langle {\acute e}^i,{\acute e}_j \rangle = 
\delta^i_{\ph{i}j}$.    The relation between the horizontal lift basis 
$\{\acute{e}_i\}$ on $P$ and a direct product basis $\{\ddot{e}_i\}$ on $U\times G$ is 
indicated in figure~\ref{trivconn}.

\begin{figure}[htb]  
\centering
\epsfxsize=12cm
\leavevmode
\vspace{-10pt}
\epsffile[0 0 1291 760]{\gpath aPfig35e}
\vspace{-15pt}
\caption{\setb The adapted tangent space basis $\{\acute{e}_i\}$ on $P$ with respect 
to a particular local trivialisation $\psi : P \to U \times G$ and the corresponding
 direct product basis $\{\ddot{e}_i\}$.}
\label{trivconn}
\end{figure}

   Acting on both sides of the second expression in equation~\ref{ehtoeb} with the
1-form  coefficients $\omega^{\beta}$ of the Lie algebra-valued connection 1-form 
   $\omega = \omega^{\beta}X_{\beta}$  on $P$ determines the connection coefficients  
$\omega^{\beta}_{\ph{o}a}(x,g) = \omega^{\beta}(\ddot{e}_a)$ on $U \times G$, as 
depicted in figure~\ref{trivconn}.
From the transformation property of the connection 1-form $\omega$ on $P$ under 
$R^{\ast}_g$ in equation~\ref{rwgog}  and with the vector field $\acute{e}_{\alpha}$ 
generating right actions on $P$, it follows that:
\begin{equation}
\acute{e}_{\alpha} \omega^{\beta}_{\ph{o}a} =
\ddot{e}_{\alpha} \omega^{\beta}_{\ph{o}a}  = 
 - c^{\beta}_{\ph{o}\alpha\gamma}\omega^{\gamma}_{\ph{o}a}  \label{omeginf}
\end{equation}
  as the infinitesimal form of the adjoint transformation under the right action of 
the group.

 Covariant differentiation on the base space is intimately related to the directional 
derivative $\acute{e}_a$ on the principle bundle. Using equation~\ref{ehtoeb} 
 the bracket $[\acute{e}_a, \acute{e}_b]$ may be expressed in a direct product basis 
as:
\begin{eqnarray}
  & \!\!\!\!\! \lbrack \acute{e}_{a}, \acute{e}_{b} \rbrack \!\! & = \;  
        \lbrack  (\ddot{e}_{a} - \omega^{\alpha}_{\ph{\beta}a} \ddot{e}_{\alpha}),
		 (\ddot{e}_{b} - \omega^{\beta}_{\ph{\beta}b} \ddot{e}_{\beta}) \rbrack 
\nonumber \\
   & = \!\! & \!\! \lbrack \ddot{e}_{a}, \ddot{e}_{b} \rbrack  -
     \lbrack \omega^{\alpha}_{\ph{\alpha}a}\ddot{e}_{\alpha}, \ddot{e}_{b} \rbrack  -
	 \lbrack \ddot{e}_{a}, \omega^{\beta}_{\ph{\beta}b}\ddot{e}_{\beta} \rbrack  +
	 \lbrack \omega^{\alpha}_{\ph{\alpha}a}\ddot{e}_{\alpha},
	      \omega^{\beta}_{\ph{\beta}b}\ddot{e}_{\beta}  \rbrack  \nonumber \\
  & = \!\! & \!\! 0 \, + \, \ddot{e}_{b}(\omega^{\alpha}_{\ph{\alpha}a}) 
\ddot{e}_{\alpha}
                              \, - \,
                  \ddot{e}_{a}(\omega^{\beta}_{\ph{\beta}b}) \ddot{e}_{\beta} \, + \,
	\omega^{\alpha}_{\ph{\alpha}a}\omega^{\beta}_{\ph{\beta}b}
	                    c^{\gamma}_{\ph{\gamma}\alpha\beta}\ddot{e}_{\gamma} \, + \,
 \omega^{\alpha}_{\ph{\alpha}a}(\ddot{e}_{\alpha}\omega^{\beta}_{\ph{\beta}b})\ddot{e}
_{\beta} \, - \,
 \omega^{\beta}_{\ph{\beta}b}(\ddot{e}_{\beta}\omega^{\alpha}_{\ph{\alpha}a})\ddot{e}_
{\alpha}
           \nonumber \\
  & = \!\! & \!\! \ddot{e}_{b}(\omega^{\gamma}_{\ph{\gamma}a}) \ddot{e}_{\gamma} \, - 
\,
                  \ddot{e}_{a}(\omega^{\gamma}_{\ph{\gamma}b}) \ddot{e}_{\gamma} \, + 
\,
	(\omega^{\alpha}_{\ph{\alpha}a}\omega^{\beta}_{\ph{\beta}b}
	                    c^{\gamma}_{\ph{\gamma}\alpha\beta}\ddot{e}_{\gamma} -
	 \omega^{\alpha}_{\ph{\alpha}a}c^{\gamma}_{\ph{\gamma}\alpha\beta}
                 \omega^{\beta}_{\ph{\beta}b}
	                    \ddot{e}_{\gamma}  +
\omega^{\beta}_{\ph{\beta}b}c^{\gamma}_{\ph{\gamma}\beta\alpha}\omega^{\alpha}_{\ph{\a
lpha}a}
	                    \ddot{e}_{\gamma} )   \nonumber \\
   & = \!\! & \!\! \big( \ddot{e}_{b}\omega^{\gamma}_{\ph{\gamma}a} 
                              \, -  \, \ddot{e}_{a}\omega^{\gamma}_{\ph{\gamma}b} \, - 
\,
    \omega^{\alpha}_{\ph{\alpha}a}\omega^{\beta}_{\ph{\beta}b}
	                    c^{\gamma}_{\ph{\gamma}\alpha\beta} \big)\ddot{e}_{\gamma}   
\nonumber \\
  & = \!\! & \!\! -\Omega^{\gamma}_{\ph{\gamma}ab}\acute{e}_{\gamma}    \label{eehatf} 
\end{eqnarray}
 using the first of equations~\ref{ehtoeb} and where 
\begin{equation} 
  \Omega^{\gamma}_{\ph{\gamma}ab}(x,g)  =  \ddot{e}_{a}\omega^{\gamma}_{\ph{\gamma}b} 
                   \, -  \, \ddot{e}_{b}\omega^{\gamma}_{\ph{\gamma}a} \, + \,
 c^{\gamma}_{\ph{\gamma}\alpha\beta}\omega^{\alpha}_{\ph{\alpha}a}
              \omega^{\beta}_{\ph{\beta}b}	          
              \label{fewewww}
\end{equation}
    are the curvature components on the principle bundle expressed in a particular 
trivialisation, as can be shown explicitly by substituting $(\ddot{e}_a,\ddot{e}_b)$ 
for $(X,Y)$ in equation~\ref{omegado2}.
 At any point $p\inn P$ the components of $\Omega(p)$ are numerically the same in the 
horizontal lift basis as for a direct product basis, that is 
 $\Omega^{\gamma}_{\ph{\gamma}ab} = 
 \Omega^{\gamma}(\acute{e}_a,\acute{e}_b) = \Omega^{\gamma}(\ddot{e}_a,\ddot{e}_b)$, 
since $\Omega$ is a horizontal form and $\acute{e}_a$  and $\ddot{e}_a$ differ only by 
a vertical vector, as seen in the second of equations~\ref{ehtoeb} and 
figure~\ref{trivconn}. 
 From equation~\ref{fewewww}, using equation~\ref{omeginf}, it can be shown that:   
\begin{equation}
\acute{e}_{\alpha} \Omega^{\beta}_{\ph{\beta}ab}(x,g) =
\ddot{e}_{\alpha} \Omega^{\beta}_{\ph{\beta}ab}(x,g)  = 
 - c^{\beta}_{\ph{\beta}\alpha\gamma}\Omega^{\gamma}_{\ph{\gamma}ab}(x,g)  
\label{fabinf}
\end{equation}
   again transforming infinitesimally under the adjoint representation, as for the 
gauge field $\omega^{\alpha}_{\ph{\alpha}a}(x,g)$, on the principle bundle.

  In summary in the horizontal lift basis the full set of structure coefficients on 
$P$ are considered with: 
\begin{eqnarray}
  \lbrack \acute{e}_{\alpha}, \acute{e}_{\beta} \rbrack & = &
                   c^{\gamma}_{\ph{\gamma}\alpha\beta}\acute{e}_{\gamma} \label{hatb1}  
\\
  \lbrack \acute{e}_{\alpha}, \acute{e}_{b} \rbrack & = &  0   \label{hatb2} \\
  \lbrack \acute{e}_{a}, \acute{e}_{b} \rbrack & = &
           \acute{c}^{\alpha}_{\ph{i}ab}\acute{e}_{\alpha}
		   =-\Omega^{\alpha}_{\ph{i}ab}\acute{e}_{\alpha}  \label{hatb3}
\end{eqnarray}
 Equation~\ref{hatb2} follows directly from equations~\ref{barb2} and \ref{ehtoeb}.
  Since right translations induce the basis vectors of the subspace $\mbox{\it VP}$, 
via equation~\ref{atova}, 
 equation~\ref{hatb2} expresses the right-invariance of the fields $\acute{e}_b \inn 
\mbox{\it HP}$, consistent with equation~\ref{racthp}, and may be compared with 
equation~\ref{xyrbra} in which $Y^L$ is right-invariant.
    For the third equation the structure coefficients $\acute{c}^d_{\ph{d}ab}$ are set 
to zero since here a coordinate basis is taken for $\{e_a\}$ on the base manifold $M$ 
in order to simplify the expressions.
 The fibre dependence of the structure coefficients $\acute{c}^{\alpha}_{\ph{i}ab}$ 
may be deduced by application of the Jacobi identity with:
\begin{eqnarray}
 & &  \lbrack \acute{e}_{\alpha}, \lbrack \acute{e}_{a}, \acute{e}_{b} \rbrack \rbrack  
+
 \lbrack \acute{e}_{a}, \lbrack \acute{e}_{b}, \acute{e}_{\alpha} \rbrack \rbrack  +
 \lbrack \acute{e}_{b}, \lbrack \acute{e}_{\alpha}, \acute{e}_{a} \rbrack \rbrack  = 0   
   \nonumber  \\
 & = & \lbrack \acute{e}_{\alpha},\,
  \acute{c}^{\beta}_{\ph{o}ab}\acute{e}_{\beta} \rbrack +
     \qquad\, 0 \qquad + \qquad\, 0 \qquad = 0   \nonumber \\
 & \Rightarrow & (\acute{e}_{\alpha} \, 
    \acute{c}^{\beta}_{\ph{o}ab})\acute{e}_{\beta} +
       \acute{c}^{\gamma}_{\ph{o}ab}
	      c^{\beta}_{\ph{o}\alpha\gamma}  \acute{e}_{\beta} = 0  
	                                          \nonumber \\
 & \Rightarrow &   \acute{e}_{\alpha} \, \acute{c}^{\beta}_{\ph{o}ab} = 
         - c^{\beta}_{\ph{o}\alpha\gamma} \,    
		 \acute{c}^{\gamma}_{\ph{o}ab}  \label{ecderv}
\end{eqnarray}
  The final expression describes the directional derivative of the coefficients 
$\acute{c}^{\beta}_{\ph{o}ab}$ with respect to the vector field  $\acute{e}_{\alpha}$, 
and hence expresses the transformation of  $\acute{c}^{\beta}_{\ph{o}ab}$ under the 
action of right translation, that is the gauge transformation generated by 
$\acute{e}_{\alpha}$. This is consistent with the transformation property in 
equation~\ref{fabinf}, for the components the curvature 2-form under infinitesimal 
gauge transformations, as expected since by equations~\ref{eehatf} and \ref{hatb3} we 
have simply:
\begin{equation}
   \acute{c}^{\alpha}_{\ph{\alpha}ab} = - \Omega^{\alpha}_{\ph{\alpha}ab}
     \label{fabecab}
\end{equation}

    Given a curvature 2-form $\Omega(p)$ on a principle bundle $P$ and a local section 
$\sigma(x)$ on $P$, for $x\inn U \subset M$, the local representative of $\Omega$ on 
the base space is defined by the pull-back map as the 2-form 
 $F(x)=\sigma^{\ast}\Omega(p)$, which also takes values in the Lie algebra, that is 
$F(x) = F^{\alpha}(x) X_{\alpha}$.

   Another significant property of the curvature on the principle bundle $P$ is that 
the exterior covariant derivative of $\Omega$ itself vanishes as a consequence of the 
definitions used to construct it, that is $\mbox{D}\Omega =0$, which is called the 
\textit{Bianchi} identity. The object $\mbox{D}\Omega = 0$ is also a tensorial form on 
$P$, like $\Omega$ itself, and since the exterior algebra structure pulls back through 
a section map $\sigma(x)$ we have a similar property for the corresponding object on 
$M$, that is on the base space we have $\mbox{D} F = 0$, which is also referred to as 
the Bianchi identity.

 Through the section map $\sigma$ the structure equation for the curvature 2-form
 $\Omega$ on $P$, for example in equation~\ref{omegadoh}, pulls back to the base space 
$M$ as:
\begin{equation}
  \label{fdapaa}
   F  =  \mbox{d}A + \frac{1}{2} [A,A]  
\end{equation}
  which was introduced in equation~\ref{fdefaaa}. In a particular trivialisation the 
components of the  `Yang-Mills field strength' on the base manifold $M$ are 
$F^{\alpha}_{\ph{\alpha}ab}(x) = \Omega^{\alpha}_{\ph{\gamma}ab}(x,e)$, while the 
`gauge potentials' are   $A^{\alpha}_{\ph{\alpha}a}(x) = 
\omega^{\alpha}_{\ph{\alpha}a}(x,e)$.
  Consistent with equation~\ref{fewewww} the above expression for $F$ can be written 
in components, in a coordinate basis on $M$, as:
\begin{equation}
 \label{falfbcp}
  F^{\alpha}_{\ph{\alpha}\mu\nu}(x) = \partial_{\mu}A^{\alpha}_{\ph{a}\nu} - 
	         \partial_{\nu}A^{\alpha}_{\ph{a}\mu} + \cstr A^{\beta}_{\ph{a}\mu} 
A^{\gamma}_{\ph{a}\nu}
\end{equation}  
  while the 2-forms $F^{\alpha}$ are related to the 1-forms $A^{\alpha}$ according to 
equation~\ref{fdahcaa}.

 For a connection on a principle bundle 
  for which the structure group $G$ as a subgroup of $\mbox{GL}(m,\rrr)$ exhibits a 
matrix representation acting upon objects $\bv(x) \inn V$ of an $m$-dimensional  
vector space (where $m$ is not necessarily equal to the dimension $n$ of the base 
manifold)
   the vector and curvature fields transform under a change of gauge $g(x)\inn G$ on 
the base space $M$ as:
\begin{eqnarray}
     \bv \to \bv' & = & g^{-1} \, \bv   \label{gaugev}  \\
	    F \to F' & = & g^{-1}\, F \, g      \label{ftogfg}
\end{eqnarray}  
   This form of transformation follows from the choice of a \textit{right} action of 
$G$ on $P$, as featuring for example in equation~\ref{sitosj}, and in turn ultimately 
on the choice for $L(G)$ to be represented by \textit{left}-invariant vector fields on 
$G$ as described in subsection~\ref{tgman}.

  Connection 1-forms $A_{r}(x) = \sigma_{r}(x)^{\ast}\omega$ on the base manifold with 
respect to different trivialisations are related under the local gauge transformations 
by $g_{rs}(x)$ between the sections of equation~\ref{sitosj} as:
 \begin{equation}
  A_{s}(x) = \mbox{Ad}(g_{rs}^{-1}(x)) A_{r}(x) + (g_{rs}^{\ast}\theta)_x
  \nonumber
\end{equation} 
  where Ad is the transformation of the adjoint representation on the Lie algebra 
values of $A_{r}(x)$ and $\theta$ is the Maurer-Cartan 1-form on the group manifold 
$G$, which here is pulled back onto $M$ via the transition function map $g_{rs}(x):M 
\to G$.
 For a matrix representation, dropping the subscript labels, this transformation 
can be written as:
  \begin{equation}
  \label{omeginhomo}
  A \to A'  = g^{-1} A g
                 + g^{-1} \mbox{d}g 
\end{equation} 
where the second term is needed to take into account general gauge changes $g(x)$ 
between sections over $M$ since $\omega$ is not a horizontal form on $P$. 
  Under a change of section $\sigma'(x) = \sigma(x)g(x)$ via the local gauge function 
$g(x)$,
  the transformations of equations~\ref{gaugev}--\ref{omeginhomo} 
    are considered a \textit{passive} symmetry from a physical point of view.

    The connection 1-form $\omega$ on the principle bundle, which is a Lie algebra 
valued map on the tangent space $T_pP$ of equation~\ref{vplush}, may be restricted to 
a mapping on elements of $V_pP$ tangent to the fibres of the bundle space, as it is in 
equation~\ref{wvaalg} for example. Under this restriction the properties of $\omega$ 
are equivalent to the Maurer-Cartan 1-form $\theta$, described in 
subsection~\ref{tgman}, which maps left-invariant vector fields on the manifold $G$ as 
$\theta (X^A) = A$ and which transforms under right translation as $R^{\ast}_g \, 
\theta = \mbox{Ad}(g^{-1})\theta$, to be compared with equations~\ref{wvaalg} and 
\ref{rwgog}.

   Indeed, for a trivial bundle we have  $P = M \times G $ and through the natural 
projection $\pi_2: M \times G \to G$, the canonical Maurer-Cartan 1-form $\theta$ on 
$G$ can be pulled back to $\omega =  \pi_2^{\ast}\, \theta$ on $P$. Since the 
pull-back map captures the structure of the exterior algebra as seen through the map 
itself the Maurer-Cartan equation, that is equation~\ref{maca2}, pulls back to:
\begin{equation}
  \mbox{d}\omega + \frac{1}{2} [\omega,\omega]    =  0
\end{equation}
   By comparison with equation~\ref{omegadoh} it can be seen that for this connection 
the curvature vanishes, $\Omega =0$, that is $\omega$ is the canonical flat connection 
on $P$.

   In general for a continuous map between two differentiable manifolds $f:M\to N$, 
with a vector field $\bu$ on $M$ and a 1-form $\xi$ on $N$, the pull-back of the 
1-form $\xi$ onto $M$ can be \textit{defined} as $\langle f^{\ast}\xi, \bu \rangle_x = 
\langle \xi, f_{\ast} \bu \rangle_{f(x)} $. For the present case the canonical flat 
connection on the base manifold $M$, expressed as $A(x) =  \sigma^{\ast} \omega = 
A^{\alpha}_{\ph{\alpha}\mu}(x)X_{\alpha}\mbox{d}x^{\mu}$  is a Lie algebra-valued map 
on tangent vectors $\bu \inn T_xM$ and we have:
\begin{equation}
 \langle A, \bu \rangle_x = 
     \langle \sigma^{\ast} \scirc\,  \pi_2^{\ast} \, \theta, \bu \rangle_x   =
   \langle \theta , \pi_{2\ast}  \scirc\, \sigma_{\ast} \bu
                    \rangle_{g \,=\, \pi_2  \,\sscirc\, \sigma (x)}
\end{equation}
  where in the latter expression the vector $\bu \inn T_xM$ has been `pushed forward' 
through the two maps to a vector in the tangent space of the group manifold. In 
general $\langle A, \bu \rangle \neq 0$, even for a flat connection, since an 
arbitrary trivialisation can be used to define the section map $\sigma_{r}(x) \equiv 
\psi^{-1}_{r}(x,e)_r$.
 However, for the canonical flat connection on $P$ the horizontal subspace is 
everywhere tangent to a submanifold $M \times \{g\}$ for some $g\inn G$ and the 
Frobenius criterion of equation~\ref{frobcon} is satisfied. Hence in this case the 
 section map from $M$ to $P$ may be chosen to coincide with the horizontal section of 
the canonical flat connection and we have:
 \begin{equation}
 \langle A, \bu \rangle_x = 
     \langle \sigma^{\ast} \omega, \bu \rangle_x   =
   \langle \omega,  \sigma_{\ast} \bu \rangle_{p\, =\, \sigma (x)}  = 0
\end{equation} 
  since for all $\bu \inn T_xM$ we have $\sigma_{\ast} \bu \inn H_pP$ in this case, 
and hence we have $A(x) = 0$ in this choice of gauge section.
In general the cross-section $\sigma$ and horizontal subspace $H_pP$ are distinct 
objects on $P$, as indicated for example in figure~\ref{trivconn}, relating to the 
gauge choice $g(x)$ and connection $\omega$ respectively. As can be seen from 
equations~\ref{ehtoeb} and \ref{dualetoe} if it is possible to choose a direct product 
basis to coincide with the horizontal lift basis on $P$ then 
$\omega^{\alpha}_{\ph{\alpha}a}(x,g) = 0$, that is all connection coefficients vanish 
for this choice of section.

 Here we have described  the flat connection that was introduced in 
equations~\ref{aegasth} and \ref{maca3} directly on the base manifold without 
constructing the principle fibre bundle. The use of the principle bundle will be more 
significant for the case of an enlarged  symmetry group of $\lv$ as introduced in 
section~\ref{hdasb} and studied further in section~\ref{reaic}.


\section{Riemannian Geometry}
\label{riegeo}

  Any $n$-dimensional differentiable manifold $M$ is canonically associated with the 
principle fibre bundle of frames \textit{FM}, with structure group $\glnrp$, which 
preserves the orientation of the frames, over $M$ as the base manifold. 
  A linear connection $\widetilde{\omega}$ can be defined on a frame bundle as a 
$\glnra$-valued 1-form on \textit{FM}
which may be written $\widetilde{\omega} = 
\widetilde{\omega}^a_{\ph{a}b}E^b_{\ph{b}a}$. The
 quantities $\widetilde{\omega}^a_{\ph{a}b} = \widetilde{\omega}^a_{\ph{a}bi}e^i$ 
(with $\{e^i\}$ a basis of 1-forms on the frame bundle) are a set of $n^2$ 1-forms on 
\textit{FM}. Each 1-form $\widetilde{\omega}^a_{\ph{a}b}$ is associated with a  basis 
element of $\glnra$ represented by the $n\times n$ matrix $E^b_{\ph{b}a}$ for which 
the only non-zero entry is a `1' in the $a^{\mathrm th}$-row and $b^{\mathrm 
th}$-column, that is $(E^b_{\ph{b}a})^d_{\ph{d}c} = \delta^b_c\,\delta^d_a$ (where 
$\{a,b\}$ label the matrices and $\{ c,d \}$ label the matrix elements. By comparison 
the generators of SO$(n)$, as described in equation~\ref{somat}, form a subalgebra of  
$\glnra$ with matrices of the form ${L}_{p \pqg q} = E^q_{\ph{q}p} - E^p_{\ph{p}q}$).
 
 The frame field $\{e_a\}$ on the base space $M$ is a general basis which in some 
situations may be taken to be an orthonormal or coordinate basis. A section $\sigma$ 
on \textit{FM} corresponds to a choice of frame, that is a basis $\{{e}_a\}$, at each 
point of the base space $M$, with the pull-back ${\Gamma} = 
\sigma^{\ast}\widetilde{\omega}$ being the representative of $\widetilde{\omega}$ 
under this section. This linear connection 1-form $\Gamma$ on $M$ has 
 components $\Gamma^{a}_{\ph{a}b} = {\Gamma}^{a}_{\ph{a}bc}{e}^c$, where $\{e^a\}$ is 
a coframe basis for $T^{\ast}M$.

  In general for a gauge symmetry group with generators represented by $m \times m$ 
matrices $E_{\alpha} \inn L(G)$  the
 connection components, for an arbitrary coframe \{$e^a$\} on the base manifold,  may 
be written $A^{r}_{\ph{r}s} = A^{\alpha}_{\ph{\alpha}a}(E_{\alpha})^{r}_{\ph{r}s}e^a = 
A^{r}_{\ph{r}sa}e^a$, with $\{r,s\} = 1\ldots m$, composing a matrix of 1-forms. In 
the case of a linear connection  on $M$, with $\bu(x)$ as any tangent vector field, 
$\Gamma^{a}_{\ph{a}b}(\bu)=\Gamma^{a}_{\ph{a}bc}u^c$
 is a  matrix element with $\Gamma^{a}_{\ph{a}bc}(x)$ being the components of the 
linear connection. 

  The covariant derivative $D_{a}$ for the case of a linear connection on the external 
tangent space will be denoted by the kernel symbol $\nabla$.
 With respect to a general frame field $\{e_a\}$, the components of the corresponding 
linear connection $\Gamma^a_{\ph{a}bc}$ satisfy the relation $\nabla e_b = 
\Gamma^a_{\ph{a}bc} e^c \otimes e_a$, that is:
\begin{equation}
  \begin{array}{rcl}
     \nabla_{\! c \,} e_b & = & \Gamma^a_{\ph{a}bc} e_a    \\
	 \mbox{and hence,} \qquad  \Gamma^a_{\ph{a}bc} & = &
	    \langle e^a, \nabla_{\! c \,} e_b \rangle
		\end{array}  \label{gamene} 
\end{equation}  
   where in the final term the angular brackets, defined in equation~\ref{formap}, 
denote the 1-form $e^a$ mapping the vector field $\nabla_{\! c \,} e_b$ into the space 
of real numbers, that is the coefficients $\Gamma^a_{\ph{a}bc}$.

 The linear connection coefficients  ${\Gamma}^a_{\ph{a}bc}$  transform under a 
general change of basis  to $e_{b'} = e_a \, e^a_{\ph{a}b'}(x)$, with primed indices 
denoting the new frame and the matrix $e^a_{\ph{a}b'}(x) \inn \glnrp$, as: 
 \begin{equation}
   \Gamma^{a'}_{\ph{'a}b'c'} = (e^{-1})^{a'}_{\ph{a}d}\, e^e_{\ph{e}b'} \, 
e^f_{\ph{f}c'} \, \Gamma^d_{\ph{d}ef}
                       + (e^{-1})^{a'}_{\ph{a}d} \, e_{c'} \, e^d_{\ph{d}b'}   
\label{gammap}
 \end{equation}
   Compared with the gauge transformation of equation~\ref{omeginhomo} an extra $e^{ 
f}_{\ph{f}{ c'}}$ factor appears here for the 3-index affine connection to reflect the 
tensor-like transformation law of the 1-form part of the connection under  a local 
change of frame on the  manifold $M$.

 A  subset of frames is provided by a general coordinate chart on the patch $U\subset 
M$ for which a section of the general frame bundle $\sigma(x): \: U \to \mbox{\it FM}$ 
is given by the coordinate basis $x \to \{\partial_{\mu} \}_x$. This defines a 
holonomic frame $\{\partial_{\mu} \}$, with $[\partial_{\mu}, \partial_{\nu}] = 0$, 
through which a 
local representative of the linear connection $\Gamma = \sigma^{\ast} 
\widetilde{\omega}$ may be obtained.
   A second general coordinate chart with coordinate frame section $\{\partial_{\mu'} 
\}$  defines a further representative of the linear connection $\Gamma' = 
\sigma'^{\ast} \widetilde{\omega}$. The transition function $j(x): \: M \to \glfrp$ 
for all $x\inn M$ relates coordinate frames as:  
\begin{equation}
  \label{jacobj}
   \partial_{\mu'} (x) = \partial_{\nu} (x) \, j^{\nu}_{\ph{\nu}\mu'}(x)
\end{equation}  
   where $j^{\nu}_{\ph{\nu}\mu'}(x) = \partial x^{\nu} / \partial x^{\mu'}$  is the 
Jacobian matrix of the general coordinate transformation. These transformations form a 
special case for equation~\ref{gammap} corresponding to a change of coordinate system 
$\{x^{\mu}\} \to \{x^{\mu'}\}$ on $M$.

    If $M$ is an $n$-dimensional Riemannian or pseudo-Riemannian manifold $(M,g)$, 
that is given a metric field with components $g_{\mu\nu}(x)$ on the manifold,  a 
subset of distinguished frames may be identified which are \textit{orthonormal} with 
respect to the metric. This subset of frames over $M$ reduces the total space of 
{\it FM} to a submanifold $\mbox{\it OM} \subset \mbox{\it FM}$ which is
 itself a principle fibre bundle with structure group SO$^+(p,q)$ (or more generally 
O$(p,q)$) with $p+q = n$.
 There is a one-to-one correspondence between metric fields $g_{\mu\nu}(x)$ on $M$ and 
reductions of the structure group $\mbox{GL}^+(n,\rrr)$ to SO$^+(p,q)$ on \textit{FM}, 
with each choice of field $g_{\mu\nu}(x)$ isolating one out of the many possible 
isomorphic copies of principle SO$^+(p,q)$-bundles.

  From the above general case we next consider specifically the spacetime symmetry of 
a 4-dimensional manifold  $M_4$.
 Matrices $l^a_{\ph{a}b}(x) \inn \soot$  of the Lorentz group  describe spacetime 
orientation preserving gauge transformations between sections of the principle bundle 
of orthonormal frames. 
With the set of vector fields $\{e_a(x)\}$ for each $x\inn M$ now representing such an  
orthonormal frame, any other orthonormal frame can be expressed as: 
\begin{equation}
  e_{b'}(x) =  e_a(x) \; l^a_{\ph{a}b'}(x)
 \label{orthtran}
\end{equation} 
  while the dual coframe 
   transforms as $e^{b'}(x) = (l^{-1})^{b'}_{\ph{b}a}(x) \: e^a(x) $. 
 Equation~\ref{orthtran} expresses the right action of elements of the Lorentz group 
on the frame field. Since the set of orthonormal frames on the tangent space at any 
one point $x\inn M_4$ is isomorphic to the Lorentz group, through 
equation~\ref{orthtran},  a principle fibre bundle over $M_4$ is obtained, with both 
the fibre space and structure group being $\soot$ itself. It is a reduction of the 
principle bundle of general linear frames $\mbox{\it FM}_4$, the latter having fibres 
isomorphic to the larger group $\glfrp$.

 We can consider a tetrad field $\teta$ as describing an element of a restricted set 
of the gauge group $\glfrp$ of all possible orientation-preserving frame 
transformations over $M_4$ or,
in bridging local orthonormal frames with general coordinate frames,
 as a mapping between the principle bundle of Lorentz frames and the principle bundle 
of coordinate frames. That is, $\teta$ relates a section of orthonormal frames 
$\{e_a\}_x$ with a coordinate frame basis $\{\partial_{\mu} \}_x$ via the right 
action:
\begin{equation}
\label{eatopmu}
  \partial_{\mu}(x) = e_a(x) \, \teta
\end{equation}
   with $\teta \inn \glfrp$,
 which can be directly compared to equation~\ref{orthtran}  with the transformation 
$l^{a}_{\ph{a}b}(x) \inn \soot$.

  For the spacetime metric $g(x)$ on $M_4$ any  local orthonormal frame $\{e_a\}$ is 
associated with the
  Minkowski metric $\eta_{ab} = g(e_a,e_b) =  \mbox{diag}(+1,-1,-1,-1)$, while in a 
general coordinate system the components of the metric are determined by the tetrad 
field $\teta$ (similarly as we had in equation~\ref{metten} for the
 3-dimensional model):
\begin{equation}
 \label{geeeta}
   g_{\mu\nu}(x) = e^a_{\ph{a}\mu}(x)e^b_{\ph{b}\nu}(x) \eta_{ab}
\end{equation}

   The $\soot$ bundle $\mbox{\it OM}_4$ may be \textit{extended} to the frame bundle 
$\mbox{\it FM}_4$ with an $\soot$-valued Lorentz connection $A(x)$ uniquely inducing a 
linear connection $\Gamma(x)$ for the extended bundle space. Such a $\glfrp$-valued 
linear connection $\Gamma$ is compatible with the metric, that is $\nabla g = 0$, 
 while $\Gamma$ and $g$ need not be related in the general case.
  The principle bundle of orthonormal frames $\mbox{\it OM}_4$, equipped with a 
Lorentz connection, as a subbundle of the principle bundle of general linear frames 
$\mbox{\it FM}_4$ over the base manifold hence induces a unique metric connection on 
the latter space.

   Expressing the Lorentz connection in a coordinate basis on $M_4$ as
    $A(x) = A_{\mu}(x) \mbox{d}x^{\mu}$ 
  the tetrad components may be considered as a local gauge transformation -- that is 
as a change from a choice of local orthonormal Lorentz frames to the general 
coordinate frames over the base manifold, within the 
 $\glfrp$ freedom of the principle bundle $\mbox{\it FM}_4$. In this way, and by 
comparison with equation~\ref{omeginhomo} for example,
   the metric preserving linear connection $\Gamma$ for a general coordinate system 
may be defined by:
\begin{equation}
\label{AtoG}
        \Gamma^{\lambda}_{\ph{\rho}\mu\nu}=
       e^{\lambda}_{\ph{\lambda}a}  \: A^{a}_{\ph{a}b\nu} \:   e^b_{\ph{a}\mu}  +
		e^{\lambda}_{\ph{\lambda}a}   \partial_{\nu}e^{a}_{\ph{a}\mu} 
\end{equation}

  The identification of the linear connection $\Gamma$ in this form implies that the 
covariant derivative of the tetrad field vanishes identically:
\begin{equation}
   \nabla_{\mu} e^{a}_{\ph{a}\nu} = \partial_{\mu} e^{a}_{\ph{a}\nu} + 
A^{a}_{\ph{a}b\mu} e^{b}_{\ph{a}\nu} - \Gamma^{\lambda}_{\ph{\lambda}\nu\mu} 
e^{a}_{\ph{a}\lambda} = 0
\end{equation}
This condition itself implies that $A$ and $\Gamma$ are \textit{compatible} 
connections, regardless of  the value of the torsion (defined below).
   In this case the tetrad field $\teta$ `commutes' with the operation  $\nabla$ of 
covariant differentiation. This means that the operation of interchanging between 
local field components, such as $u^a(x)$, and general
 coordinate tangent space field components, such as $u^{\mu}(x)$, via the tetrad field 
$\teta$, applies in a straightforward manner even for equations involving covariant 
derivatives.

  In particular, since $g_{\mu\nu}(x)$ has the form of equation~\ref{geeeta} and the 
Minkowski metric is a constant, the metric field $g(x)$ is preserved by covariant 
differentiation defined in terms of the linear connection $\Gamma(x)$, which in turn 
is defined in terms of the Lorentz connection through equation~\ref{AtoG}, that is 
$\nabla g = 0$ as cited above. If $A^{a}_{\ph{a}b\mu}(x)$ is chosen to be the unique 
torsion-free Lorentz connection for a given tetrad field $\teta$, then the 
corresponding linear connection $\Gamma$ is the unique torsion-free metric connection 
expressed in a general coordinate system. This is the Levi-Civita connection, 
significant for  general relativity, which  can be written uniquely as a function of 
the metric tensor components $g_{\mu\nu}(x)$ as:  
\begin{equation}
  \label{gtoGam}
    \Gamma^{\sigma}_{\ph{\sigma}\mu\nu} = \frac{1}{2} g^{\sigma\rho} 
(\partial_{\mu}g_{\rho\nu} + \partial_{\nu}g_{\mu\rho} - \partial_{\rho}g_{\mu\nu}) 
\end{equation}

   On the space of the frame bundle over any $n$-dimensional differentiable manifold 
$M$, even without a metric, a canonical $\rrr^n$-valued 1-form $\theta_{\mathrm{C}} = 
\theta^a E_a$ can be identified, with each $\theta^a$ being a 1-form on $\mbox{\it 
FM}$ and $\{E_a\}$ a basis for $\rrr^n$,  such that at any point $f\inn \mbox{\it FM}$ 
and for any vector $X \inn T_f\mbox{\it FM}$ we have:
 \begin{equation}
    \langle \theta^a , X \rangle := \langle e^a , \pi_{\ast} X \rangle = 
(\pi_{\ast}X)^a
 \end{equation}
  which is just the components of the projection of $X$ onto the base space $M$ in the 
frame $f = \{e_a\}$ itself. Given a section $\sigma(x) = f$ on $F\!\!\: M$ the 
pull-back $e^a = \sigma^{\ast} \theta^a$ describes the dual basis vectors of the 
general $\mbox{GL}(n,\rrr)$ frame $f$.

  The canonical 1-form $\theta_{\mathrm{C}}$ is therefore horizontal and equivariant 
and hence a tensorial form  on $\mbox{\it FM}$. Given a linear connection 
$\widetilde{\omega}$ on $\mbox{\it FM}$  the exterior covariant derivative $\Theta = 
\mbox{D}\theta_{\mathrm{C}}$ is called the \textit{torsion} 2-form on $\mbox{\it FM}$. 
With $\Theta = \Theta^a E_a$, and following equation~\ref{dphiecd}, the torsion can be 
expressed as:
\begin{equation}
    \Theta^a = \mbox{d}\theta^a  +  \widetilde{\omega}^a_{\ph{a}b} \wedge \theta^b
\end{equation}
  This object in turn pulls back to the torsion 2-form $\bT = \sigma^{\ast} \Theta$ on 
the base manifold $M$ with coefficients $T^a_{\ph{a}bc}$ defined in $T^a = 
\frac{1}{2}T^a_{\ph{a}bc} e^b \wedge e^c$, with: 
\begin{eqnarray}
   T^a & = & \mbox{d} e^a  +  \Gamma^a_{\ph{a}b} \wedge e^b  \\
       & = & -\frac{1}{2}c^a_{\ph{a}bc} e^b \wedge e^c + \Gamma^a_{\ph{a}bc} e^c 
\wedge e^b  \\
	   & = & (-\frac{1}{2}c^a_{\ph{a}bc} - \frac{1}{2}(\Gamma^a_{\ph{a}bc} - 
\Gamma^a_{\ph{a}cb} )) 
	            e^b \wedge e^c       
\end{eqnarray}
   where each term above is a 2-form.
   Hence for a general linear connection on the manifold $M$ the torsion components 
can be written as:
\begin{equation}
     T^a_{\ph{a}bc} = - 2\Gamma^a_{\ph{a}[bc]} - c^a_{\ph{a}bc} \label{t2gc}
\end{equation}   
  with  $\lbrack \ldots \rbrack$ denoting $\frac{1}{n!}$ times the antisymmetrised sum 
of the $n!$ terms obtained through permuting the $n$ enclosed indices.  
 Via the vielbein field $\teta$ this may be written in a general coordinate frame as:
 \begin{equation}
    T^{\rho}_{\ph{\rho}\mu\nu} = - \Gamma^{\rho}_{\ph{\rho}\mu\nu}  +  
\Gamma^{\rho}_{\ph{\rho}\nu\mu}
	 \label{trmngg}
 \end{equation} 
    since $[\partial_{\mu},\partial_{\nu}]=0$ for such a holonomic frame.

    The curvature of the linear connection may also  be defined on the frame bundle 
$\mbox{\it FM}$ as $\widetilde{\Omega} = \mbox{D}\widetilde{\omega}$, that is as the 
exterior covariant derivative of the connection in the usual way, to obtain the 
tensorial form $\widetilde{\Omega}$ of type $(\mbox{Ad}, \glnra)$. However, here we 
deal directly with objects on the base manifold $M$ for an arbitrary frame field 
$\{e_a\}$ and study the Riemannian curvature $\bR = \sigma^{\ast}\widetilde{\Omega} = 
R^a_{\ph{a}b}E^b_{\ph{b}a}$, where the matrices $E^b_{\ph{b}a}$ were defined in the 
opening of this section. From the definition of the curvature 2-form in 
equations~\ref{omegado1}--\ref{omegado2} and the $\glnra$ commutators $[E^b_{\ph{b}a}, 
E^d_{\ph{d}c}] = \delta^b_c E^d_{\ph{d}a} - \delta^d_a E^b_{\ph{b}c}$
(which can be compared with the commutators for the $L_{p \pqg q}$ matrices describing 
the so($n$) subalgebra in equation~\ref{malg}) 
 the components of curvature $R^a_{\ph{a}b}$ may be written for any linear connection 
$\Gamma$ in any choice of  frame field as:
\begin{eqnarray}
   R^a_{\ph{a}b} & = & \mbox{d}\Gamma^a_{\ph{a}b} \, + \, \Gamma^a_{\ph{a}d}
                      \wedge  \Gamma^d_{\ph{d}b}    \nonumber   \\
   & = & (\mbox{d}\Gamma^a_{\ph{a}bc})e^c \, + \, \Gamma^a_{\ph{a}bd}\mbox{d}e^d
       \, + \,  \Gamma^a_{\ph{a}dc} e^c \wedge  \Gamma^d_{\ph{d}be}e^e   \nonumber \\
   & = & (e_e \Gamma^a_{\ph{a}bc}) e^e \wedge e^c \, - \,
    \fhs  \Gamma^a_{\ph{a}bd}c^d_{\ph{d}ce} e^c \wedge e^e \, + \,
	  \Gamma^a_{\ph{a}dc}   \Gamma^d_{\ph{d}be}  e^c \wedge e^e   \nonumber \\
  &= & \fhs (e_c {\Gamma}^a_{\ph{a}be} - e_e {\Gamma}^a_{\ph{a}bc}
		      + {\Gamma}^a_{\ph{a}dc} {\Gamma}^d_{\ph{d}be} - 
			    {\Gamma}^a_{\ph{a}de} {\Gamma}^d_{\ph{d}bc} 
    -  {\Gamma}^a_{\ph{a}bd}c^d_{\ph{d}ce}) e^c \wedge e^e   \label{omdggg}
\end{eqnarray}
   In terms of the components of the rank-4 Riemann tensor the curvature can be 
expressed as $\bR = \fh R^a_{\ph{a}bcd} e^c \wedge e^d E^b_{\ph{b}a}$. Hence the 
curvature components on the base manifold $M$ can be written in terms of the linear 
connection and structure coefficients as: 
\begin{equation} 
{R}^a_{\ph{a}bcd} 
		  =   e_c {\Gamma}^a_{\ph{a}bd} - e_d {\Gamma}^a_{\ph{a}bc}
		      + {\Gamma}^a_{\ph{a}ec} {\Gamma}^e_{\ph{e}bd} - 
			    {\Gamma}^a_{\ph{a}ed} {\Gamma}^e_{\ph{e}bc} 
			              - c^e_{\ph{e}cd} {\Gamma}^a_{\ph{a}be}  \label{rabcd}
\end{equation}

If a metric $g$ is also defined on $M$ then \{$e_a$\} may represent a local 
orthonormal frame field. 
  In the dual covector basis $\{e^a\}$ the Riemann tensor may be written as:
\begin{eqnarray}
    \bR & = & \frac{1}{2} R^{p \pqg q}_{\ph{p \pqg q}cd}
	\, L_{p \pqg q} \, e^c \wedge e^d    \nonumber \\
	  & = &       R^{p \pqg q}_{\ph{p \pqg q}cd}
	\, L_{p \pqg q}  \, e^c \otimes e^d 
	  \label{boldr}
\end{eqnarray}
 where the latter follows due to the asymmetric arrangement of the $\{c,e\}$ indices 
for the coefficients in the final line of equation~\ref{omdggg}.
    Under the group SO$^+(p,q)$ this object transforms as a rank-4 tensor which can be 
expressed in components in several equivalent ways, including:
\begin{eqnarray}
    R^a_{\ph{a}bcd} & = &  R^{p \pqg q}_{\ph{p \pqg q}cd}\,  (L_{p \pqg 
q})^a_{\ph{a}b} \nonumber \\
   \mbox{and} \qquad	R_{abcd}   & = &   \eta_{ae}    R^e_{\ph{e}bcd}        
\label{rabcdsym}
\end{eqnarray}
   This latter object is asymmetric in the indices $\{a,b\}$ as well as in $\{c,d \}$.
   The Riemann tensor in a general coordinate system, as described towards the end of 
section~\ref{perc} in the context of the SO(3) model on $M_3$,
 may be obtained through the vielbein field $\teta$, with the resulting components:
\begin{equation}
   R_{\rho\sigma\mu\nu} = e^{a}_{\ph{a}\rho} e^{b}_{\ph{b}\sigma} e^{c}_{\ph{c}\mu} 
              e^{d}_{\ph{d}\nu}   R_{abcd}  
			  \label{rcoordset}
\end{equation}

Both the curvature and torsion  may be considered properties of a linear connection 
$\Gamma$ in general. Although they are related through the Ricci and Bianchi 
identities, respectively:  
\begin{eqnarray}
R^{\rho}_{\ph{\rho}[\sigma\mu\nu]} &=& -T^{\rho}_{\ph{\rho}[\sigma\mu ;\nu]}
                                  - T^{\rho}_{\ph{\rho}\kappa[\sigma} 
T^{\kappa}_{\ph{\kappa}\mu\nu]}
                              \label{riccirt}  \\
R^{\rho}_{\ph{\rho}\sigma[\mu\nu ; \tau]} &=& 
                          - R^{\rho}_{\ph{\rho}\sigma\kappa[\tau} 
T^{\kappa}_{\ph{\kappa}\mu\nu]}   
                              \label{bianchirt}
\end{eqnarray}
  (where $;\tau$ denotes the covariant derivative $\nabla_{\tau}$ with respect to the 
$x^{\tau}$ coordinate)  
   the curvature and torsion are independent geometric concepts where either one may 
be non-zero while the other is zero. 	
  For example for the complete parallelism exhibited on a Lie group manifold $G$  in 
terms of the  self-parallel frame composed  of left-invariant vector fields 
$X_{\alpha}$ on  $G$, with each $\Gamma^{\alpha}_{\ph{a}\beta\gamma} = 0$, the 
curvature vanishes, as can be seen trivially from equation~\ref{rabcd}, while the 
torsion is finite, with $ T^\alpha_{\ph{\alpha}\beta\gamma} =  - \cstr $, as 
determined directly by equation~\ref{t2gc}. On the other hand for the linear 
connection $\Gamma^{\alpha}_{\ph{a}\beta\gamma} = -\fh 
c^{\alpha}_{\ph{a}\beta\gamma}$, in the same basis on $G$, the curvature is finite 
while the torsion vanishes, as can also be seen from equations~\ref{rabcd} and 
\ref{t2gc}. This latter case is the unique Levi-Civita connection  on a group manifold 
defined in terms of the Killing metric on $G$. In general the identities of 
equations~\ref{riccirt} and \ref{bianchirt} clearly simplify for the torsion-free 
case.

  Returning to the case of 4-dimensional spacetime $M_4$
   the quantities $R_{\rho\sigma\mu\nu}$ of equation~\ref{rcoordset} are the 
components of a general coordinate frame rank-4 tensor with transformations 
$j^{\mu}_{\ph{\mu}\nu'}\inn \glfrp$, introduced in equation~\ref{jacobj},  acting on 
all indices under a change of coordinates.
 The most general rank-4 tensor on a 4-dimensional manifold has $4^4= 256$ independent 
components. However the geometric origin and structure of the Riemann tensor results 
in considerably less freedom. In components $R_{\rho\sigma \;\! \mu\nu}$ is asymmetric 
in the first two indices $\{\rho,\sigma\}$ since it derives from a Lorentz-valued 
metric connection and also asymmetric in the final two indices $\{\mu,\nu\}$ since the 
curvature originates as a 2-form object. This reduces the number of free components 
down to $(6 \times 6) = 36$.   For the torsion-free case considered here the Ricci 
identity in a general coordinate system of equation~\ref{riccirt} reduces to simply:
\begin{eqnarray}
    R_{\rho[\sigma\mu\nu]} &=& 0    \label{ricciz}
                               \\
   \mbox{or} \qquad R_{\rho\sigma\mu\nu} \, + \, R_{\rho\nu\sigma\mu} \, + \, 
                       R_{\rho\mu\nu\sigma}   &=& 0    \nonumber
\end{eqnarray}
  where the second equation follows from the asymmetry of $R_{\rho\sigma\mu\nu}$ in 
the final two indices. This further constraint results in a final total of 20 
independent components for the Riemann curvature tensor for the metric and 
torsion-free case.

 The Ricci tensor may be defined as the `trace' of the Riemann tensor $R_{\sigma\mu} = 
R^{\rho}_{\ph{\rho}\sigma\mu\rho}$. This is also termed a `contraction' of upper and 
lower indices in $R^{\rho}_{\phantom{\mu}\sigma\mu\nu}$, which transform in a dual 
manner to each other under the action of $\glfrp$. Also for the Lorentz curvature 
tensor components $R^a_{\phantom{a}b\mu\nu}$ transformations in the $\{a,b\}$ indices 
via the group $\soot$ are closely related to those in the $\{ \mu,\nu \}$ indices via 
the holonomic  subgroup of  $\glfrp$  through the components of the tetrad field 
$\teta$, and it is through the latter field that tensor contractions are again 
possible. In both cases the Lie algebra valued part  of the  curvature form possesses 
a transformation symmetry closely related to that of the $r$-form part in the tangent 
space of the base manifold. This, of course, is not the case for curvature forms 
derived for general principle bundles with the symmetry group composing the fibres 
unrelated to the local symmetry of the base space manifold, and hence an equivalent 
contraction does not exist for a gauge theory based on such an internal symmetry.

 The Ricci tensor is symmetric and hence possesses 10 independent degrees of freedom, 
including the scalar curvature $R=g^{\mu\nu}R_{\mu\nu}$ (as distinct from the Riemann 
tensor denoted by a bold $\bR$, as on the left-hand side equation~\ref{boldr}). 
 The utility of such expressions follows from the fact that the operation of 
contraction maps a tensor object onto another tensor, that is the contracted tensor 
also transforms as a representation of $\glfrp$. This tensor preserving property is 
shared by the operations of the covariant derivative and exterior algebra as we 
described earlier, and hence all of these operations are useful for identifying the 
equations of physics.

  The remaining 10  components of the Riemann tensor, the non-Ricci part, are 
described by the Weyl tensor $C_{\rho\sigma\mu\nu}$, it is the trace-free part of 
$R_{\rho\sigma\mu\nu}$ (all contractions are zero) with which it shares the same 
symmetries.
 The trace-free property implies ten relations $C_{\sigma\mu} = C^{\rho}_{\pho 
\sigma\mu\rho} = 0$ between the components of the Weyl tensor  $C_{\rho\sigma\mu\nu}$ 
and hence only ten of them  are independent.
  The Weyl tensor is also the conformally invariant part of the Riemann tensor, that 
is it is unchanged under a conformal transformation of the metric $g_{\mu\nu}(x) \to 
f(x)g_{\mu\nu}(x)$ where $f(x)$ is any smooth real function on $M_4$.  
  The twenty components of the Riemann tensor can be decomposed explicitly in terms of 
those of the Weyl tensor and Ricci tensor  as:
\begin{equation}
   \label{rdecom}
  R_{\rho\sigma\mu\nu} = C_{\rho\sigma\mu\nu} + \frac{1}{2}(g_{\rho\mu}R_{\sigma\nu}
      - g_{\rho\nu}R_{\sigma\mu} - g_{\sigma\mu}R_{\rho\nu} + 
g_{\sigma\nu}R_{\rho\mu})
	   +\frac{1}{6}(g_{\rho\nu}g_{\sigma\mu} - g_{\rho\mu}g_{\sigma\nu})R \;\;
\end{equation}
\begin{equation}
   \mbox{that is:} \qquad
   R^{\rho\sigma}_{\ph{\rho\sigma}\mu\nu} = C^{\rho\sigma}_{\ph{\rho\sigma}\mu\nu} +
   2R^{[\rho}_{\ph{[\rho}_{[\mu}}g^{\sigma ]}_{\ph{\sigma ]}_{\nu ]}} -
   \frac{1}{3}Rg^{[\rho}_{\ph{[\rho}\mu}g^{\sigma ]}_{\ph{\sigma ]}\nu}
   \qquad \qquad \qquad \qquad \qquad
   \nonumber 
\end{equation}

  The Bianchi identity of equation~\ref{bianchirt} for the curvature tensor in the 
torsion-free case is simply:
\begin{eqnarray}
      R^{\rho}_{\ph{\rho}\sigma[\mu\nu ; \tau]}  & = & 0  \label{rbianc}  \\
	\Rightarrow \qquad
	(R^{\mu\nu} - \frac{1}{2} R g^{\mu\nu})_{;\mu} & = & 0   \qquad\quad  
\label{rrbianc}
\end{eqnarray}
  where the latter expression follows from the double contraction of the former. 
  The Einstein tensor is defined as $G^{\mu\nu} := R^{\mu\nu} - \frac{1}{2} R 
g^{\mu\nu}$.
   Hence the Einstein tensor $G^{\mu\nu}$, unlike its `dual' geometric object the 
Ricci tensor $R^{\mu\nu}$, represents an identitically conserved quantity, that is 
$G^{\mu\nu}_{\ph{\mu\nu};\mu} = 0$, which is the origin of its central importance in 
the field equation of general relativity.

 For general relativity in regions of `empty space' with $T^{\mu\nu}=0$ by the 
Einstein equations~\ref{Eins} we also have $G^{\mu\nu}=0$ and hence $R^{\mu\nu}=0$ and 
the manifold is said to be `Ricci flat'.
  In this Ricci vacuum the Riemann tensor is simply $R^{\rho}_{\pho \sigma\mu\nu} = 
C^{\rho}_{\pho \sigma\mu\nu}$, as can be seen explicitly from equation~\ref{rdecom}.  
The spacetime curvature is then described in terms of the Weyl tensor 
$C_{\rho\sigma\mu\nu}$, yet in a way dependent upon the matter content in other 
spacetime regions as will be reviewed alongside equation~\ref{cbian} in 
section~\ref{subwal}.

  We note here that the various possible sign conventions for the expressions of 
general relativity can be distilled down to the $\pm$ sign used for the right-hand 
side of just three expressions in the Riemannian geometry: 
\begin{itemize}
 \item[1)] The metric tensor:  
   \begin{equation}
     \label{metcon}
      \eta_{ab} = \mbox{diag}(+1,-1,-1,-1)
   \end{equation}  
	   With `$+1$' for the time component this is a natural convention for the present 
theory based on forms of temporal flow.
 \item[2)] The Riemann tensor:
  \begin{equation}
     \label{ritencon}
   R^{\rho}_{\ph{\rho}\sigma\mu\nu} = \partial_{\mu} 
\Gamma^{\rho}_{\ph{\rho}\sigma\nu} - \partial_{\nu} \Gamma^{\rho}_{\ph{\rho}\sigma\mu} 
+  \Gamma^{\rho}_{\ph{\rho}\lambda\mu} \Gamma^{\lambda}_{\ph{\lambda}\sigma\nu}  - 
\Gamma^{\rho}_{\ph{\rho}\lambda\nu} \Gamma^{\lambda}_{\ph{\lambda}\sigma\mu}
  \end{equation}
 Where the final term of equation~\ref{rabcd} is zero when expressed in a coordinate 
frame as is the case here.
 \item[3)]  The Ricci tensor: 
 \begin{equation}
   \label{riccicon}
    R_{\mu\nu} = R^{\rho}_{\ph{\rho}\mu\nu\rho} \quad 
(=-R^{\rho}_{\ph{\rho}\mu\rho\nu}) 
  \end{equation}  
   This is equivalent to choosing the sign convention for 
the Einstein field equation as $G^{\mu\nu}=-\kappa T^{\mu\nu}$ with positive 
normalisation constant $\kappa$ (as will be justified after equation~\ref{gtruu}).
\end{itemize}

  The convention for these three signs chosen here is the same as used for example 
in~(\cite{Pea} p.24) that is with signs `$(-+-)$'  relative to the original discussion 
of these conventions in~\cite{MTW}. The Einstein equation, and general relativity 
itself, will be reviewed in the following section.



\section{General Relativity}
\label{gcatep}

  In his 1854 work `On the Hypotheses which lie at the Foundation of Geometry' 
Riemann, building upon the study of the intrinsic curvature of 2-dimensional surfaces 
by Gauss,
considered more generally spaces of $n$-dimensions and
 introduced tensor analysis, in particular incorporating the  metric tensor and the 
Riemann curvature tensor. At the same time Riemann also speculated on the possible 
curvature for the space of our own world, both on small and large scales, and its 
possible physical implications.    

  At around the same time (1861,1865) Maxwell, building upon the `field' concept 
introduced earlier by Faraday based on empirical observations, formulated the
   equations of motion for the electromagnetic field,  providing a unified description 
of electric fields, magnetic fields and also the properties of light.

  The mathematical structure of general relativity was developed leading up to 1915 as 
an application of Riemann's work in geometry, with the dimension of \textit{time} now 
included along with \textit{space} in a 4-dimensional spacetime manifold. 
 Influenced by the work of Maxwell on electromagnetism
objects such as the metric and Riemann curvature tensor, as mathematical functions 
describing the phenomena of gravitation, were now considered as \textit{fields} in 
spacetime.

  In search of a relativistic gravitational field equation consistent with the 
`equivalence principle', defined below, and under the empirical guidance that the 
Newton-Poisson equation  $\nabla^2 \Phi = 4 \pi G_{\! N} \rho$ (a second order 
differential equation, with Laplacian operator $\nabla^2 = \pal_x^2 + \pal_y^2 + 
\pal_z^2$, relating the gravitational scalar potential $\Phi$, via Newton's constant 
$G_{\! N}$, to the scalar mass density distribution $\rho$)
   should emerge in the non-relativistic limiting case for small distortions from a 
flat spacetime, Einstein converged in 1915 upon the field equation:
\begin{equation}
\label{Eins}
       G^{\mu\nu}= -\kappa T^{\mu\nu}
\end{equation}  
    with $\kappa$ a constant  and $T^{\mu\nu}$ the energy-momentum tensor for the 
distribution of matter in 4-dimensional spacetime. From the limit of Newtonian gravity 
   the normalisation constant is found to be $\kappa = \frac{8\pi G_{\! N}}{c^4}$.

  In general relativity, it is considered always possible to have a local 
\textit{inertial} coordinate system on $M_4$ that is valid within a sufficiently small 
region of curved 4-dimensional spacetime -- strictly an infinitesimal neighbourhood 
about any point $x\inn M_4$, with local metric $\eta = \mbox{diag}(+1,-1,-1,-1)$.

   The \textit{strong equivalence principle} states that within such a local 
coordinate system, within a sufficiently small region about the point $x\inn M_4$, all 
laws of physics, other than gravity, take the same form that applies for special 
relativity in an unaccelerated Cartesian coordinate system in the absence of gravity.
  These assumptions augment the \textit{weak} form of the equivalence principle for 
which the `laws of physics' are limited to `the laws of motion of freely falling 
particles' corresponding to the equivalence of gravitational and inertial mass, and 
the observation of the apparent lack of gravitational effects within a freely falling 
lift.

   The motion of a freely falling particle in such a local inertial coordinate system 
$\{x^a\}$ satisfies the equation $d^2x^a/d\tau^2 = 0$, in choosing the proper time 
$\tau$ to parametrise the trajectory. Transforming to a general coordinate system 
$\{x^{\mu}\}$ this becomes: 
\begin{equation}
 \label{geotra}
  \frac{d^2 x^{\lambda}}{d\tau^2} + 
\Gamma^{\lambda}_{\ph{\lambda}\mu\nu}\frac{dx^{\mu}}{d\tau} \frac{dx^{\nu}}{d\tau}  = 
0
\end{equation}
 which is called the \textit{geodesic} equation of motion and which is valid also in 
an extended curved spacetime. The quantities $\Gamma^{\lambda}_{\ph{\lambda}\mu\nu}$ 
are the coefficients of the linear connection and the proper time $\tau$ itself can be 
defined in terms of an integral of the invariant intervals $d\tau = (g_{\mu\nu} 
dx^{\mu}dx^{\nu})^{1/2}$ along the trajectory. In terms of the 4-velocity $u^{\mu} = 
dx^{\mu}/d\tau$ the above geodesic equation can be written as simply:
 \begin{equation}
 \label{geotrau}
   u^{\mu}\nabla_{\mu}u^{\nu} = 0
\end{equation}

 The equivalence principle states that all gravitational effects can be locally 
transformed away and can be interpreted to mean that we may always choose a local 
inertial  coordinate frame at any $x\inn M_4$ such that all the coefficients 
$\Gamma^{\lambda}_{\ph{\lambda}\mu\nu} = 0$. Hence, although the coefficients of the 
non-tensor object $\Gamma$ will be frame dependent the torsion tensor $\bT$ vanishes 
in all reference frames, by equation~\ref{trmngg}.
  This torsion-free assumption
 for Einstein's theory of general relativity  has the benefit of simplifying some of 
the mathematics of the theory, as for example in equations~\ref{ricciz} and 
\ref{rbianc} of the previous section.

   Given a metric  $g_{\mu\nu}(x)$  on $M_4$  the Levi-Civita connection is the unique 
metric ($\nabla g = 0$), torsion-free ($\bT=0$) linear connection. The corresponding 
connection coefficients may be written in a general coordinate frame uniquely in terms 
of those of the metric tensor as described in equation~\ref{gtoGam}.
  For such a connection equations~\ref{geotra} and \ref{geotrau} describe the 
trajectory which extremises the path length between any given end points:
\begin{equation}
 \label{extgeo}
 L = \int (g_{\mu\nu} u^{\mu} u^{\nu})^{1/2} d\tau
\end{equation}
   and hence 
earns the name `geodesic'. Further, for this connection with $\Gamma(x)$  determined 
uniquely by $g(x)$, as implied by the equivalence principle, the metric alone 
determines all gravitational effects and hence can be considered to \textit{be} the 
gravitational field for Einstein's general relativity. 
  Since the tetrad field $\teta$ may be considered to be the `square-root' of the 
metric, with $g_{\mu\nu} = e^a_{\ph{a}\mu}e^b_{\ph{b}\nu}\eta_{ab}$ in 
equation~\ref{geeeta}, the tetrad field itself, which everywhere exhibits the presence 
of the local inertial frames, may also be considered to represent the gravitational 
field.

  As well as being able to express the metric as $g_{\mu\nu} = 
\mbox{diag}(+1,-1,-1,-1)$ at any spacetime location
  there is sufficient freedom under coordinate transformations such that at any $x\inn 
M_4$ all 40 components of the metric derivatives can be set to zero, that is 
$g_{\mu\nu ,\rho}(x) = 0$, corresponding to coordinate frames with $\Gamma=0$ as can 
be seen from equation~\ref{gtoGam}. However there is insufficient freedom under 
general coordinate transformations to set all 100 second derivative quantities 
$g_{\mu\nu ,\rho\sigma}(x)$ to zero and there remain 20 irreducible degrees of freedom 
which are described by the Riemann curvature tensor, as deduced earlier after 
equation~\ref{ricciz}.

The components of the  metric tensor field $g_{\mu\nu}(x)$ may be determined by 
solving the  second order differential field equation $G^{\mu\nu} = -\kappa 
T^{\mu\nu}$ for a  distribution of matter described by the energy-momentum tensor 
$T^{\mu\nu}$, in practice by  introducing `boundary conditions' as described in the 
following section. For a
 particular physical state for the geometry of the world there will be a range of 
possible solutions for $g_{\mu\nu}(x)$ and $e^a_{\ph{a}\mu}(x)$ in spacetime (over and 
above the local Lorentz freedom for the latter field) all with equivalent physical 
content.

  Essentially there is only \textit{one} `coordinate system' $\rrr^4$ through which 
any  region of spacetime may be described, as depicted in figure~\ref{onecoord}(a), as 
a simple space of 4 independent real parameters upon which a solution for the field 
$g_{\mu\nu}(x)$ may be inscribed. 
  
\begin{figure}[htbp]  
\centering
\epsfxsize=\maxwidth
\leavevmode
\epsffile[0 0 1731 572]{\gpath aPfig36e}
\caption{\setb (a) Alternative metric solutions on $\rrr^4$ for the same physical 
state and (b) as apparently represented through an `alternative' coordinate system 
overlaid upon the `original' coordinates.}
\label{onecoord}
\end{figure} 
  
   An alternative expression of the \textit{same} physical solution then corresponds 
to a different metric function $g'_{\mu\nu}(x)$ inscribed upon the \textit{same} 
$\rrr^4$ space. For example in the Schwarzschild solution for the metric field 
associated with a single massive body located at one point in space, to be presented 
in equation~\ref{ttrtp}, the physical point where the curvature scalar $R$ is largest, 
and perhaps even singular, will in general have different coordinate values $x\inn 
\rrr^4$ under a `coordinate transformation', as indicated by the two small circles in 
figure~\ref{onecoord}(a). However the transformed solution could be conceived of as a 
new set of `curvilinear' coordinates overlaying the \textit{same} physical 
configuration (explicitly represented by the \textit{same} metric field) as shown in 
figure~\ref{onecoord}(b).     
  
   In general it is less useful to think of any coordinates as curvilinear, indeed it 
is always the case that $[\partial_{\mu}, \partial_{\nu}] = 0$ with all structure 
coefficients $c^{\rho}_{\ph{\rho}\mu\nu}=0$. In this sense all coordinate systems can 
be pictured as a `flat' purely \textit{mathematical} parameter space, which for the 
case of $\rrr^4$ can be visualised as the set of `Euclidean' real number parameters as 
represented in figure~\ref{onecoord}(a). \textit{Physical} curvature is a property of 
the \textit{fields} on $M_4$ itself  with the Riemann curvature tensor describing the 
geometrical structure and warping of the corresponding physical spacetime. The set of 
components $R^{\rho}_{\ph{\rho}\sigma\mu\nu}(x)$ are given at points on the manifold 
labelled  $x \inn M_4$ under the coordinate chart map $\phi : M \to \rrr^4$, or on a 
$U \subset M$ subset. A general coordinate transformation is then a mapping between 
solutions represented  on different choices of the \textit{map} $M \to \rrr^4$ onto  a 
unique $\rrr^4$  (assuming here a non-degenerate Jacobian matrix 
$j^{\mu}_{\ph{\mu}\nu}(x)$, that is neglecting the artificial difference of a 
`coordinate singularity' for example for polar coordinates at the corresponding 
Cartesian coordinate origin).
  
  In general relativity a general coordinate system $\{ x^{\mu} \}$ is of no physical 
significance; all the physics is in the `fields' on the manifold~(see for example 
\cite{Rov} chapter~2), with the gravitational field $\teta$ giving rise to the 
spacetime geometry of the manifold. It is the possibility of relating field quantities 
on $M_4$, such as the coincidence of physical events or the equating of the Einstein 
tensor with the energy-momentum tensor, that determines the physical content of the 
theory.

  While the coordinate system plays a passive unphysical role, in particular 
circumstances it may be associated with physical structure. This is true in the case 
of the Schwarzschild solution in which the origin of a polar coordinate system is 
associated with the central massive object. This is an example with non-zero Riemann 
curvature in which the exact spherical symmetry of the physical state is assumed to be 
exhibited by the metric for which a solution may be found in a greatly simplified form 
in a naturally preferred system of spherical polar coordinates. For similar reasons, 
but with finite 4-dimensional curvature considered on a much larger scale, 
cosmological models also employ a preferred system of coordinates to study solutions 
of Einstein's field equation, as we shall describe in section~\ref{sectsmoc}.

  In general, however, there will be no preferred solutions and hence no privileged 
coordinate systems on the base manifold. In this sense all coordinate systems are 
`equally bad', or at least on a equal footing, and this expresses the relevance of 
\textit{general covariance} for general relativity. Other theories may also be 
`generally covariant', but if there is \textit{always} a particular kind of 
distinguished coordinate reference frame then the general covariance may be of no 
relevance. This is the case for special relativity and also for Newtonian mechanics 
formulated against a flat absolute  background of an independent space and time. 
  
   Even for general relativity, if the curvature is very small, as it is in practice 
in a laboratory on the surface of the Earth or even locally within the solar system 
with respect to the `fixed stars' of the galaxy, then there will be `preferred' 
solutions with everywhere $\teta \simeq \delta^a_{\ph{a}\mu}$ and $g_{\mu\nu}(x) 
\simeq \mbox{diag}(1,-1,-1,-1)$ found for a coordinate system which is then implicitly 
pseudo-Euclidean to a very good approximation. In the limit of flat Minkowski 
spacetime there is a preferred coordinate systems with $g_{\mu\nu}(x) = 
\mbox{diag}(1,-1,-1,-1)$ exactly. The corresponding tetrad field is 
$\teta = \delta^a_{\ph{a}\mu}$, within a global Lorentz transformation (which leaves 
the metric invariant).
 In this case a  coordinate transformation such that in general $g'_{\mu\nu}(x) \neq 
\mbox{diag}(1,-1,-1,-1)$, while the Riemann tensor necessarily remains zero, may be 
considered as an introduction of a new `curvilinear' coordinate system, as pictured 
for example by the transformation in figure~\ref{onecoord}(b).

 Newtonian mechanics in Euclidean spacetime takes its simplest form when expressed 
using Cartesian coordinates; however even for a flat space the description of parallel 
transport and the form of the covariant derivative is non-trivial when expressed in a 
curvilinear coordinate system. It is only for the choice of a Cartesian coordinate 
system
that a trivial linear connection may be adopted.
In general it is the fact that we can not assume a `flat' geometry over macroscopic 
distances that necessitates the introduction of the more general notion of parallelism 
as described by a connection form, as is also the case for a gauge theory based on an 
internal symmetry as described in section~\ref{fibre}. For an externally curved 
geometry  the lack of a preferred coordinate system, with a preferred description of 
parallelism, highlights the significance of general covariance for the theory of 
general relativity.

 As described above 
 a choice of coordinates may be useful in order to express some metric solutions in a 
simple mathematical form but they are a non-physical, and in this sense a `gauge', 
artifact that drop out of all expressions for observable quantities. 
Working with a general coordinate system and  the corresponding use of holonomic 
reference frames $\{ \partial_{\mu}\}$ does not allow for arbitrary frame 
transformations as elements of $\glfrp$. Rather the transition functions 
$j^{\mu}_{\ph{\mu}\nu}(x)$ of equation~\ref{jacobj} are restricted to a `holonomic 
subgroup' of all possible $\glfrp$ transformations over the manifold, sometimes called 
the `Einstein gauge', and this to some extent disguises underlying gauge  structure of 
general relativity.

 Although  the `coordinate invariance' symmetry of the kind implied by general 
covariance is mathematically rather different from the usual concept of a `gauge 
invariance' symmetry, there is a close analogy between them.
 In both cases there is a \textit{loosening} of a global symmetry or absolute 
structure that would otherwise be arbitrarily imposed. In both cases also the 
equations of motion, together with their solutions, are mapped on to equally valid 
equations and solutions under the coordinate or gauge transformations. Further, while 
a particular choice of coordinates greatly assists with calculations for some 
solutions in general relativity a particular choice of gauge is frequently employed to 
assist with calculations in a gauge theory.

 For general relativity to be considered in terms of a $\glfrp$ gauge theory of 
gravity, within the framework of general covariance, the equivalence principle is 
needed to distinguish the local Minkowski metric as being physically significant in 
that it marks the transition to special relativity in local inertial coordinate 
frames. That is, the metric or tetrad field needs to be introduced everywhere on $M_4$ 
(there is no equivalent of such fields for an internal symmetry gauge theory). This 
implies the possibility to contract the $\glfrp$ structure group down to the Lorentz 
subgroup (which is then the \textit{holonomy} group of the general frame bundle). 
The local Lorentz symmetry itself has mathematical properties very closely related to 
those of the local symmetry of a gauge theory.

 Indeed, while gravitation in Einstein's original theory of 1915 is described through 
the freedom of the metric field $g_{\mu\nu}(x)$ field, together with its relation to 
the Levi-Civita connection $\Gamma(x)$,
 an equivalent formulation of general relativity can be given in terms of the tetrad 
field $\teta$ together with a Lorentz connection $A(x)$.  This latter approach was 
introduced in 1956 by Utiyama~\cite{Uti} in which general relativity is considered as 
a type of gauge theory invariant under local Lorentz transformations. Such local 
transformations are displayed in equation~\ref{orthtran} and map one local inertial 
coordinate frame onto another. 
  As well as tensor representations the Lorentz group also has \textit{spinor} 
representations. Hence spinor fields can be introduced on a spacetime manifold with an 
arbitrary metric $g_{\mu\nu}(x)$ via the  tetrad field $\teta$. This also permits 
gravitation to be considered in terms of an $\sltc$ gauge theory, where $\sltc$ is the 
double cover of the Lorentz group, as will be described in section~\ref{dynkin}.

 The fundamental structures on the base manifold are the local Minkowski spaces, 
together with their mutual relations through the Lorentz connection on $M_4$.
    With respect to a given coordinate system either the tetrad $\teta$ or metric 
$g_{\mu\nu}(x)$ field identifies the local inertial frames.
In 1920 Einstein postulated that the metric field should be considered to be the 
fundamental entity of general relativity, referring to it as the `new ether'.
 However, whichever fields are considered as fundamental, field equations are still 
required in order to determine the nature of the field dynamics.
 At the same time that Einstein arrived at equation~\ref{Eins} via the heuristic 
arguments outlined in the opening of this section Hilbert was in the process of 
deriving the same equation via a Lagrangian approach. This latter argument, and the 
employment of Lagrangian methods more generally, will be reviewed in the following 
section.


\section{Lagrangian Formalism}
  \label{subfal}

    In this section we review the standard use of the Lagrangian formalism to derive 
physical equations of motion, including those for general relativity and gauge 
theories. 
 In the 4-dimensional spacetime of general relativity the scalar curvature $R$ is 
adopted as the principle geometric contribution to the total scalar Lagrangian 
function,  with 
 the field equations  determined from the Einstein-Hilbert action 
integral~(\cite{HawkEl} p.75):
\begin{equation}
   I = \int (\alpha(R - 2\Lambda) + \lag)   \sqrt{\vert g \vert}\; d^4 x  
\label{einhil}   
\end{equation}
  Here   $\Lambda$ is the cosmological constant, $\lag$ is the Lagrangian function for 
matter fields and $\alpha$ is a normalisation constant.
 The magnitude of the metric determinant 
$\vert g \vert$ is employed in the 4-dimensional invariant volume element $\sqrt{\vert 
g \vert}\; d^4 x$.
  The vacuum equations for general relativity, that is with $\lag=0$ and $\Lambda = 
0$, are obtained by requiring that $\delta I = 0$ in equation~\ref{einhil} under 
variation of the metric $\delta g_{\mu\nu}$. With $\delta g^{\mu\nu}=-\delta 
g_{\mu\nu}$ to first order, $\delta \sqrt{\vert g \vert } = \frac{1}{2}\sqrt{\vert g 
\vert } \: g^{\mu\nu}\delta g_{\mu\nu}$ and with $R=R_{\mu\nu}g^{\mu\nu}$ we have:
\begin{eqnarray}
  \delta I & = & \int \alpha (R \, \delta \sqrt{\vert g \vert } 
                  \;  + \; R_{\mu\nu} \,\delta g^{\mu\nu} \,\sqrt{\vert g \vert } \; +  
\; \delta R_{\mu\nu}\, g^{\mu\nu} \sqrt{\vert g \vert })
                                                             \;  d^4 x    
\label{delI1}   \\
           & = & \int \alpha (\frac{1}{2}R\, g^{\mu\nu} - R^{\mu\nu})\,\delta 
g_{\mu\nu}\, \sqrt{\vert g \vert } \; d^4 x   \label{delI2} 
\end{eqnarray}
  where the final term in equation~\ref{delI1} contributes zero to the integral since
 $g^{\mu\nu} \delta R_{\mu\nu} = (g^{\mu\nu}\delta \Gamma^{\rho}_{\ph{\rho}\mu\nu} - 
g^{\mu\rho}\delta \Gamma^{\nu}_{\ph{\nu}\mu\nu})_{;\rho}$ and $\delta 
\Gamma^{\rho}_{\ph{\rho}\mu\nu}$ vanishes on the boundary of 
integration~(\cite{HawkEl} p.75). Requiring the action to be stationary, $\delta I = 
0$, for any variation of the metric, $\delta g_{\mu\nu}$, leads directly from 
equation~\ref{delI2} to the Einstein vacuum equation:
\begin{equation}
    \label{lagtoein}
    G^{\mu\nu} := R^{\mu\nu} - \frac{1}{2}R \, g^{\mu\nu} = 0
\end{equation}

  For the non-vacuum case 
   the energy momentum tensor $T^{\mu\nu}$ for a general matter Lagrangian $\lag \neq 
0$ can be defined under variations of the metric $\delta g_{\mu\nu}$ through:
\begin{equation}
   \delta I \quad = \quad \delta \!\! \int \lag \, \sqrt{\vert g \vert } \, d^4x \quad 
= \quad
	   \int \fhs \, T^{\mu\nu} \, \delta g_{\mu\nu} \, \sqrt{\vert g \vert } \, d^4x 
\end{equation}
 Hence for the full action integral of equation~\ref{einhil} stationarity $\delta I = 
0$ under the metric variation gives Einstein's field equation for the general case, 
with $\kappa \equiv \frac{-1}{2\alpha}$  adopted as the normalisation constant:
\begin{equation}
   \label{einlamt}
    G^{\mu\nu} + \Lambda g^{\mu\nu} = - \kappa T^{\mu\nu}
\end{equation}
  Assuming that the matter Lagrangian may be  a function of $g_{\mu\nu}(x)$, but not 
of the metric derivatives,
   the energy-momentum tensor itself, consistent with these equations, can be 
expressed directly in terms of the matter Lagrangian as:
\begin{equation}
   \label{tengr}
    T^{\mu\nu}\, = \,	\frac{2}{\sqrt{\vert g \vert}}
	 \frac{\pal ( \lag \sqrt{\vert g \vert})}{\pal g_{\mu\nu}}
	\, = \, 2 \frac{\pal \lag}{\pal g_{\mu\nu}} \, - \, \lag g^{\mu\nu}
\end{equation}

  A  simpler application of the principle of least action in the context of general 
relativity was described earlier for equation~\ref{extgeo} regarding the derivation of 
the geodesic equation of motion for a body moving in a gravitational field.
 Generalising from equation~\ref{extgeo} for a body with mass $m$ and charge $q$ 
moving in a curved spacetime through an electromagnetic field with 4-potential 
$A_{\mu}(x)$  an action $S$ may be constructed including both the kinematic and an 
interaction Lagrangian term respectively in:
\begin{equation}
  \label{slorlag}
   S= \int \left(m \, (g_{\mu\nu}u^{\mu} u^{\nu})^{1/2} 
      \; + \; q u^{\mu} A_{\mu} \right) d\tau
\end{equation}
  Requiring $\delta S = 0$ under variation of the trajectory of the charged body leads 
to the equation of motion:
 \begin{equation}
  \label{grellor}
   m \left(\frac{du^{\lambda}}{d\tau} + \Gamma^{\lambda}_{\ph{\lambda}\mu\nu} 
        u^{\mu}u^{\nu}   \right)   =  + F^{\lambda}_{\ph{\lambda}\sigma}J^{\sigma} 
 \end{equation}
 where $F^{\lambda}_{\ph{\lambda}\sigma}$ are components of the electromagnetic field 
tensor and  $J^{\mu} = q u^{\mu}$ is the 4-current of the charged body having 
4-velocity $u^{\mu}$ with respect to the proper time $\tau$. The above equation hence 
describes a correction to the purely geodesic trajectory of equation~\ref{geotra}. In 
the limit of a flat Minkowski spacetime, and will respect to a Cartesian coordinate 
frame, equation~\ref{grellor} simplifies to: 
  \begin{equation}
    \frac{\partial p^b}{\partial \tau} = + F^b_{\ph{b}c}J^c \label{rellor}
  \end{equation} 
  which is the relativistic Lorentz force law, for the charged body with 4-momentum 
$p^b = m u^b$.
  Further, in the non-relativistic limit equation~\ref{rellor} becomes $m\ba = q(\bE + 
\bv \times \bB)$, the original form of the Lorentz force law, where $\bv$ and $\ba$ 
are the 3-velocity and 3-acceleration of the body respectively.

 These examples, for the trajectory of a body in a gravitational and/or 
electromagnetic field,   demonstrate the flexibility and generality of the Lagrangian 
approach.
As well as applying to macroscopic physical bodies the use of Lagrangian functions is 
a standard tool in classical field theory.
  In general the form of the Lagrangian ${\mathcal L}$, a function of the fields such 
as $\phi(x)$, guided by considerations of symmetry, is constructed in order that the 
requirement for the action integral $S = \int {\mathcal L}(\phi, 
\partial_{\mu}\phi)\omega$ (where $\omega$ is the volume 4-form) to be stationary, 
$\delta S = 0$, under variations of the fields, such as $\delta \phi$, yields the 
required equations of motion for the fields via the Euler-Lagrange equation:
\begin{equation}
 \label{eula}
   \partial_{\mu} \frac{\partial \lag}{\partial(\partial_{\mu} \phi)} 
                           - \frac{\partial \lag}{\partial \phi} = 0.
\end{equation}  

   In a flat spacetime, 
    in terms of the electromagnetic curvature tensor $F_{\mu\nu}$, Maxwell's  
equations  are:
  \begin{eqnarray}
    F_{[\mu\nu , \rho]} & = & 0  \label{maxo} \\
	 F^{\mu\nu}_{\ph{\mu\nu},\mu} & = & +J^{\nu}
	  \label{maxj} 
  \end{eqnarray}
  which can also be written as $\md F  =  0$ and $\md \past F = \past J$ respectively
  (where `${}^{\ast}$' denotes the `Hodge dual' as employed in equation~\ref{hodged}).
 These equations are equally valid in a curved spacetime on replacing the partial 
derivatives `$,\rho$'  by the covariant derivatives `$;\rho$', as an application of 
the strong principle of equivalence.
  The first of these equations is simply the Bianchi identity, introduced in 
section~\ref{cacfc}, for the curvature tensor of a $\uo$ gauge theory.
	Here working in the Lorenz gauge with $\pal_{\mu} A^{\mu} = 0$  the inhomogeneous 
Maxwell equation~\ref{maxj} can be written as:
 \begin{equation}
   \label{maxaj}
     \square A^{\mu} = +J^{\mu}
 \end{equation} 
 
  The Maxwell Lagrangian for the electromagnetic field is constructed as:
\begin{equation}
  \label{lagem}
   \lag_{\mathrm{em}} \, = \, - \frac{1}{4}F_{\mu\nu}F^{\mu\nu}
\end{equation}
 Under variation of the electromagnetic gauge field $A_{\mu}(x)$ the Euler-Lagrange 
equation for $\lag_{\mathrm{em}}$ yields Maxwell's equation for the source-free 
$J^{\nu} = 0$ case, that is $F^{\mu\nu}_{\ph{\mu\nu},\mu} = 0$.
 In combining the Lagrangian of equation~\ref{lagem} with the final term of that in 
equation~\ref{slorlag}, hence including a term coupling the electromagnetic field to a 
classical charged body, the corresponding  
Euler-Lagrange equation for $\delta A_{\mu}(x)$  yields equation~\ref{maxj} with the 
source term on the right-hand side.

  The form of the Lagrangian for non-Abelian gauge theory is  guided by the Abelian 
case of electromagnetism, motivating the   Lorentz and gauge invariant  Yang-Mills  
Lagrangian:
\begin{equation}
 \label{lagym}
 \lag_{\mathrm{YM}} \, = \, - \frac{1}{4} F_{\alpha \, \mu\nu} F^{\alpha \, \mu\nu}
\end{equation}
  as a direct generalisation of equation~\ref{lagem}.
   For the non-Abelian case there is a further contraction over the index $\alpha = 
1\ldots n_G$, for the group generators, between the adjoint and coadjoint 
representations, which are related by the Killing metric $K_{\alpha \beta}$ (which in 
a suitable basis  is simply $-\delta_{\alpha\beta}$ for the compact simple Lie groups 
relevant for the internal gauge symmetries in particle physics).  In this case the 
Euler-Lagrange equation~\ref{eula} for $\lag_{\mathrm{YM}}$ under variation of the 
gauge field components $Y^{\alpha}_{\ph{\alpha}\mu}(x)$ yields the non-linear second 
order differential equation:
\begin{equation}
  \label{ymol}
  D_{\mu}F^{\alpha \,\mu\nu} = \partial_{\mu} F^{\alpha \,\mu\nu}
     + \cstr Y^{\beta}_{\ph{\beta}\mu} F^{\gamma \,\mu\nu}  = 0 
\end{equation}
 where $D_{\mu}$ is the gauge covariant derivative, which also appears in the Bianchi 
identity 
 $D_{[ \rho}^{ } F_{\ph{\alpha}\mu\nu ]}^{\alpha} = 0$ as the non-Abelian 
generalisation of equation~\ref{maxo}. 
The immediate distinctive feature of  equation~\ref{ymol}, in comparison with the 
Maxwell equation~\ref{maxj}, is the additional non-linear term of the form $[Y,F]$ 
appearing for the non-Abelian case. Such terms are interpreted as self-interactions of 
the gauge fields $Y_{\mu}(x)$, which do not occur for the Maxwell theory. 
 This self-interaction is intrinsically geometric in origin and is implied in the 
Lagrangian of equation~\ref{lagym} itself given that the curvature for a non-Abelian 
gauge field has the form of equation~\ref{falfbcp} with non-trivial structure 
constants.

 Additional terms in the Lagrangian, either for the Maxwell or Yang-Mills case, may 
lead to further sources of interactions.
     In the Standard Model of particle physics interactions between fermion and gauge 
fields in the corresponding equations of motion are introduced through the `minimal 
coupling' in the covariant derivative terms included in a Lagrangian.
   For example by including $\lag_{\mathrm{YM}}$ alongside the Dirac Lagrangian for a 
massless spinor field $\psi(x)$, which transforms as a multiplet under the internal 
symmetry, together with the conjugate field $\ol{\psi} = \psi^{\dag}\gamma^0$ (where 
the $\gamma$-matrices will be defined in section~\ref{lsspin}), we have combined:
\begin{eqnarray}
  \lag_{\mathrm{YMD}} & = & \; - \; \frac{1}{4} F_{\alpha \, \mu\nu} F^{\alpha \, 
\mu\nu} \; - \; \overline{\psi} \gamma^{\mu} D_{\mu} \psi \;  \label{lagdym} \\
   \mbox{where} \qquad \qquad \quad \;
    D_{\mu} & = & \pal_{\mu} + Y^{\alpha}_{\ph{\alpha}\mu} E_{\alpha}
                             \nonumber \\ 
   \mbox{and} \qquad \quad \; \overline{\psi} \gamma^{\mu} D_{\mu} \psi & = &
        \overline{\psi} \gamma^{\mu} \partial_{\mu} \psi  \; + \;      
Y^{\alpha}_{\ph{\alpha}\mu} j^{\mu}_{\ph{\mu}\alpha} \nonumber  \\
    \mbox{with} \qquad \qquad \quad \;\; j^{\mu}_{\ph{\mu}\alpha} & = & 
\overline{\psi} \gamma^{\mu} E_{\alpha} \psi \label{jglocur}
\end{eqnarray}
 where in the appropriate representation the $E_{\alpha}$ are $n \times n$ matrices 
acting on the $n$-dimensional field $\psi$ in the internal space.
 Here a lower case `$j$' will generally denote a current such as the Lorentz vector in 
equation~\ref{jglocur} composed of elementary fields, as opposed to the upper case 
analogue  
 $J^{\mu} = qu^{\mu}$ for the macroscopic current featuring in equation~\ref{grellor} 
for example.
Under variation of the gauge field $Y^{\alpha}_{\ph{\alpha}\mu}(x)$ the extra term 
$Y^{\alpha}_{\ph{\alpha}\mu} j^{\mu}_{\ph{\mu}\alpha}$ in this Lagrangian leads to a 
modification of equation~\ref{ymol} with the source $j^{\mu}_{\ph{\mu}\alpha}(x)$ now 
appearing in the right-hand side to give:
\begin{equation}
  \label{ymols}
  D_{\mu}F^{\alpha \,\mu\nu} = \partial_{\mu} F^{\alpha \,\mu\nu}
     + \cstr Y^{\beta}_{\ph{\beta}\mu} F^{\gamma \,\mu\nu}  = \, j^{\nu\,\alpha}
\end{equation}

  In practice factors of $i=\sqrt{-1}$ and differing $\pm$ signs in the above 
equations will depend upon the conventions adopted, with coupling constants such as 
$g$ also appearing in expressions for specific applications in the Standard Model as 
will be reviewed 
in section~\ref{ewtatsm}.
 In addition to the requirements of symmetry the form of the scalar Lagrangian 
function is typically heavily guided by the need to obtain the desired equations of 
motion. As a further example the above Lagrangian of equation~\ref{lagdym}, augmented 
with a fermion mass term $+m\overline{\psi}\psi$, under variation of the field 
$\overline{\psi}(x)$ yields the Euler-Lagrange equation:
\begin{equation}
  \label{diracl}
   (\gamma^{\mu}D_{\mu} \, - \, m)\psi  = 0
\end{equation}
   which is the Dirac equation for the spinor field $\psi$ (within conventional 
factors of $i$). The interaction between the fermion field $\psi(x)$ and the gauge 
field $Y_{\mu}(x)=Y^{\alpha}_{\ph{\alpha}\mu}(x) E_{\alpha}$ is here found in the 
`minimal coupling' in the action of the covariant derivative $D_{\mu}\psi = 
\partial_{\mu}\psi + Y_{\mu}\psi$ in the kinetic term of the Lagrangian in 
equation~\ref{lagdym}. 

   As for the case of the charged macroscopic body in equation~\ref{slorlag} here also 
the mass $m$ for the field $\psi$ in equation~\ref{diracl} has been  introduced  
through a Lagrangian mass term, in this case with $+m\overline{\psi}\psi$ appended to 
equation~\ref{lagdym}.
  Mass terms are generally added to the Lagrangian by hand in this way, 
 although this may not be  straightforward to achieve. For example, a corresponding 
Lagrangian term such as $m^2 Y_{\mu}Y^{\mu}$ for a gauge field mass is prohibited by 
the requirement of gauge invariance, and even the fermion mass term $m \overline{\psi} 
\psi$   is prohibited in the Standard Model Lagrangian due to the left-right asymmetry 
of the $\sutw_L$-valued gauge field relating to electroweak interactions. In both 
cases mass terms are incorporated into the Lagrangian  through interactions with the 
Higgs field and spontaneous symmetry breaking, 
 involving the addition of further, apparently ad hoc, terms to the Lagrangian, as 
will be described in section~\ref{ewtatsm}.

 As described above interactions may be introduced into the Lagrangian by the 
requirement of invariance under a \textit{local} gauge symmetry. Such a local symmetry 
incorporates a corresponding global symmetry of the equations of motion and hence 
 Noether's theorem applies. The theorem states that each \textit{global} continuous 
symmetry  is associated with a conserved current, written in terms of the field 
$\phi(x)$ as:
\begin{equation}
  \label{noether}
  j^{\nu}_{\ph{\nu}\alpha} := \left( \frac{\partial \lag}{\partial( \pal_{\nu} 
\phi^a)} \right) (E_{\alpha})^a_{\pho b} \phi^b
\end{equation}
   for each generator $E_{\alpha}$ of the global symmetry.
 For the Dirac Lagrangian with a $\uo$ gauge symmetry, that is the final term of 
equation~\ref{lagdym} for the Abelian case, the global $\uo$ symmetry with a single 
generator  is associated with the Dirac current:
 \begin{equation}
   \label{jmupgp}
    j^{\mu} = \overline{\psi} \gamma^{\mu} \psi
 \end{equation}
  and the 
 conservation law  is simply $\pal_{\mu}j^{\mu} = 0$.

   In contrast to the case of an \textit{internal} global symmetry of the Lagrangian 
applying Noether's theorem for the \textit{external} symmetry of global translational 
invariance of $\lag (\phi)$ in a flat Minkowski spacetime  leads to the quantity 
(\cite{Kaku} p.27):
\begin{equation}
  \label{tmnneo}
   t^{\mu\nu} =  \left( \frac{\partial \lag}{\partial (\partial_{\mu}\phi)}\right)
       \partial^{\nu} \phi - \lag \eta^{\mu\nu}
\end{equation}
 which satisfies the conservation law $\partial_{\mu}t^{\mu\nu} = 0$, again owing to 
the Euler-Lagrange field equation. In field theory equation~\ref{tmnneo} can be taken 
as a definition of the energy-momentum tensor. There are four `conserved charges' 
associated with $t^{\mu\nu}$, namely the 4-momentum $P^{\mu} = \int d^3 \bx \; t^{\mu 
0}$. 
 These include the Hamiltonian $H=P^0$ and the 3-vector {\boldmath$P$} which is 
interpreted as the physical 3-momentum carried by the field.

 However in general the form of $t^{\mu\nu}$ defined in equation~\ref{tmnneo} is 
neither symmetric nor gauge invariant. For example, with the Lagrangian for the 
electromagnetic field $\lag_{\mathrm{em}} = -\frac{1}{4} F_{\mu\nu} F^{\mu\nu}$ of 
equation~\ref{lagem}, as a function of $A_{\mu}(x)$,  equation~\ref{tmnneo} yields:
\begin{equation}
  \label{temlaga}
  t^{\mu\nu} = - F^{\mu\rho} \partial^{\nu}A_{\rho} + 
\frac{1}{4}\eta^{\mu\nu}F_{\rho\sigma}F^{\rho\sigma}
\end{equation}
   for which the lack of symmetry is clear in the $\mu\nu$ indices in the first term 
and the lack of gauge invariance is clear from the form of the explicit $A_{\rho}$ in 
this term. The standard interpretation of this observation is that in Lagrangian field 
theory the energy-momentum $t^{\mu\nu}$ is \textit{not} a directly measurable quantity  
and the corresponding ambiguity   allows for the addition of a extra terms, leading 
for example to the quantity (\cite{Kaku} p.101):
\begin{equation}
 \label{temlag}
   T^{\mu\nu} = t^{\mu\nu} + \partial_{\rho}(F^{\mu\rho}A^{\nu})
\end{equation}
  For the source-free case considered here with $\pal_{\mu}F^{\mu\nu} = 0$ 
  this produces a symmetric gauge invariant form of the Maxwell energy-momentum 
tensor, in fact in the form of equation~\ref{temkk} below with $\eta^{\mu\nu}$ in 
place of $g^{\mu\nu}$. However, in addition to the ad hoc nature this procedure is 
clearly flawed in that it is incompatible with general relativity. That is, for any 
$T^{\mu\nu} \neq 0$ the spacetime geometry, described by the Einstein 
equation~\ref{einlamt}, is not flat and hence the assumption of spacetime translation 
symmetry which led to equation~\ref{tmnneo} itself is invalid.

   On the other hand, the electromagnetic energy-momentum tensor $T^{\mu\nu}$ can be 
derived directly by a different standard procedure, in general relativity, from the 
stationarity of the matter Lagrangian $\delta \int \lag = 0$ with respect to variation 
in the metric tensor $g_{\mu\nu}(x)$, as described towards the opening of this 
section.  Substituting the Maxwell Lagrangian of equation~\ref{lagem} into 
equation~\ref{tengr} gives directly:
\begin{equation}
  \label{temkk}
      T^{\mu\nu} = + F^{\mu\rho}F_{\rho}^{\ph{\rho}\nu} \, + \, \frac{1}{4}
	      g^{\mu\nu} F^{\rho\sigma}F_{\rho\sigma}
\end{equation}
   This general relativistic method yields an energy-momentum tensor $T^{\mu\nu}$ 
which is symmetric, gauge invariant and complies necessarily with the Einstein 
equation~\ref{einlamt} since it derives from the Einstein-Hilbert action of 
equation~\ref{einhil}.

  For general relativity the four relations $\gmo$ of the contracted Bianchi identity 
of equation~\ref{rrbianc}, together with the identity $(\Lambda g^{\mu\nu})_{;\mu} 
=0$, places four constraints $\tmo$ on the energy-momentum tensor for the general case 
via the Einstein equation~\ref{einlamt},  which in turn implies that only  
 six of the ten field equations are independent. Hence the metric $g_{\mu\nu}(x)$ is 
not determined uniquely by $G^{\mu\nu} + \Lambda g^{\mu\nu}  = -\kappa T^{\mu\nu}$, 
but rather four degrees of freedom remain for arbitrary coordinate transformations.
  Indeed, the field equation is only required to define $g_{\mu\nu}(x)$ up to an 
equivalence class $(M_4,g)$ of geometries on the manifold $M_4$ related by coordinate 
transformations $\theta$ such that $(M_4,g)$ and $(M_4, \theta^{\ast}g)$ are 
physically equivalent, as described in the discussion of figure~\ref{onecoord}
 in the previous section.

   Within the Lagrangian framework it is also possible to derive the contracted 
Bianchi identity $\gmo$ itself. Taking $\lag$ to be the Ricci scalar $R$ the 
Einstein-Hilbert action $I = \int R \sqrt{\vert g \vert }\, d^4 x$  
(equation~\ref{einhil} for the vacuum case and setting $\alpha=1$) is a scalar 
quantity and hence invariant under coordinate transformations.
 Indeed while the variational method can be employed, via the Einstein equation, to 
determine the 4-dimensional spacetime geometry it is unable to deduce a specific 
choice of metric function and coordinates, by the principle of general covariance.  
However,
 the fact that $\delta I = 0$ for  coordinate transformations can be shown (\cite{MTW} 
p.503) to imply the  identity $\gmo$  of equation~\ref{rrbianc}.

   The examples of this section have shown some of the great variety of circumstances 
in which the Lagrangian method may be employed. These include cases in Newtonian 
mechanics, special relativity and general relativity as well as for electromagnetism 
and non-Abelian gauge theories. However all of these examples also rest on the 
assumption of the validity of the Lagrangian approach. One of the aims of the present 
theory is to derive all equations of motion \textit{without} employing a Lagrangian

    Already it has been described for equation~\ref{rrbianc} how the relation $\gmo$ 
is a geometric \textit{identity} which stands alone as a `conserved' geometric 
quantity \textit{without} the need for a Lagrangian formulation. In the present theory 
it stands at the head as central to the derivation of physical equations of motion, as 
we shall investigate in section~\ref{subwal}. This is universally true both for 
equations of motion at the effective macroscopic level, relating to classical 
phenomena such as the Lorentz force law, and also at the microscopic level of the 
fundamental underlying fields, relating to quantum phenomena, where the constraint of 
the full form of temporal flow $\lvh$ will also prove central to the physics.

  In contrast to the Lagrangian approaches in general relativity via 
equation~\ref{tengr} and in field theory via equation~\ref{tmnneo}
in the present theory the Einstein equation will essentially be interpreted  as the 
definition of energy-momentum, that is $T^{\mu\nu} := -\frac{1}{\kappa}G^{\mu\nu}$, 
where a possible
 $\frac{\Lambda}{\kappa}g^{\mu\nu}$ term may be implicitly included in the left-hand 
side.
 (In subsequent chapters this relation may be written simply as $T^{\mu\nu} := 
G^{\mu\nu}$ to emphasise the equivalence of the two tensors, with the implied 
normalisation factor of $-\kappa$ explicitly introduced for practical applications). 
   Since the geometric content of $G^{\mu\nu}$ \textit{is} measurable in general 
relativity, in principle at least as the gravitational influence on test bodies, 
defining the energy-momentum tensor this way does have an unambiguous meaning.
 In principle the structure of the energy-momentum tensor in such a theory may be 
uniquely specified,   distinguishing  between equations~\ref{temlaga} and \ref{temkk} 
in the example of the electromagnetic field.

  This is the case for Kaluza-Klein theory in which equation~\ref{temkk}, generalised 
for non-Abelian internal symmetry, is \textit{derived}  from the structure of a
  higher-dimensional geometry  as will be reviewed in the following chapter leading to  
equation~\ref{einym}. 
  With the Yang-Mills equation~\ref{ymol} also being derived in equation~\ref{yangmk} 
within this framework the Kaluza-Klein approach achieves a degree of unification with 
less dependence upon the introduction of Lagrangian terms, such as 
equation~\ref{lagym}.
 In section~\ref{reaic} we describe how the techniques of Kaluza-Klein theory might be 
adopted within the present theory before continuing in section~\ref{subwal}  
  to explore some of the consequences of these structures in terms of avoiding the 
need to postulate  Lagrangian functions.


\pagebreak
\chapter{Kaluza-Klein Theory}

 \label{kktheory}
  
\section{General Relativity with Extra Dimensions}  
\label{lccop}
  
  Theories with an extra spatial dimension were initially proposed \cite{Kaluza,Klein} 
within a few years of the publication of the general theory of relativity, with the 
aim of accounting for non-gravitational forces of nature through the
 higher-dimensional geometry, at a time when only two fundamental forces were known, 
namely gravitation and electromagnetism.
 A generalisation of the original Kaluza-Klein theory  for the case of a non-Abelian 
internal symmetry, incorporating further dimensions, was elaborated in detail around 
half a century later (\cite{Cho}, see also \cite{Kerner}, \cite{ChMaMa} sections I--V 
and \cite{Orzalesi}).

This unifying framework for gravitation and gauge theories, reviewed here, is 
constructed in the mathematical setting of a principle fibre bundle. Keeping within 
the spirit of Einstein's original 4-dimensional spacetime theory of gravitation and 
the extension to a 5-dimensional arena by Kaluza and Klein, the geometric unification 
with non-Abelian gauge theory is founded upon a metric tensor $\check{g}$, now defined 
upon the manifold of the principle bundle $P=(M_4,G)$ itself (with the `check' on  
$\check{g}$ denoting an object on the bundle space).

  We note that conventions vary in the literature -- in particular with respect to the 
assignment of index labels such as $\{a,b,\ldots\}$, $\{\alpha,\beta,\ldots\}$, and 
$\{i,j,\ldots\}$ which in this paper are associated with objects on the  manifolds 
$M_4$, $G$ and $P$ respectively, in the manner described shortly before and in  
figure~\ref{pbunbasis}. The conventional order of the indices for the linear 
connection coefficients $\Gamma^a_{\ph{a}bc}$ also varies, with the convention of 
equation~\ref{gamene} adopted here,
while the sign of the Ricci tensor $R_{\mu\nu} = R^{\rho}_{\ph{\rho}\mu\nu\rho}$  of 
equation~\ref{riccicon} also differs in some of the references. Hence in turn a number 
of derived expressions here will have signs differing to those in  the literature.

   In addition to the metric $g_{ab}$ on the base manifold $M_4$ a natural metric for 
the group manifold $G$  is provided by the Killing form $K$, which as a matrix of 
components is invertible provided $G$ is a semi-simple Lie group and negative definite 
if $G$ is compact. In the latter case   a basis for the Lie algebra can be chosen such 
that the Killing form has components $K_{\alpha\beta} = -\delta_{\alpha\beta}$, is 
described after equation~\ref{lagym}.
 Here we choose metric components 
 $g_{\alpha\beta} = +K_{\alpha\beta}$ in order to match the signature convention of 
equation~\ref{metcon}, with spacelike components having a negative norm.

  The $\mbox{Ad}(G)$-invariant Killing form 
 defines a bi-invariant metric on the manifold $G$; that is with both the left $L_a$ 
and right $R_a$ group actions, for any $a\inn G$, being isometries on $G$, with for 
example $(R^{\ast}_a \, g)_b(X,Y) = g_b(X,Y)$ for all $X,Y\inn T_bG$  for the Killing 
metric $g$ at any point $b\inn G$ (subsequently the Killing metric will often be 
denoted by $g_{\alpha\beta}$, rather than simply the kernel letter $g$, as the 
notation used for the indices helps identify the space to which the object belongs).
 In particular, in terms of the group structure constants $\cstr$ in a left-invariant 
basis $\{X_{\alpha}\}$ on the group manifold, the components of the Killing metric 
are:  
\begin{equation}
  \label{killmet}
g_{\alpha\beta} = K_{\alpha\beta} =  
c^{\rho}_{\ph{\rho}\alpha\sigma}c^{\sigma}_{\ph{\sigma}\beta\rho} 
\end{equation}

 A gauge connection 1-form $\omega$ on a principle bundle $P$ specifies a 
right-invariant horizontal subspace $H_pP$ for all points $p\inn P$, as described in 
section~\ref{fibre}. A unique metric $\check{g}$ may be defined on such a principle 
bundle space, aligned with the gauge connection structure with:
 \begin{equation}   
  \check{g} (X,Y) = g (\pi_{\ast} X, \pi_{\ast} Y)  +  K (\omega(X), \omega(Y))  
\label{ggxyk}
\end{equation}
  where $X,Y \inn \mbox{\it TP}$, while here $g$ and $K$ are the metrics on the base 
space $M_4$ and group space $G$ respectively. This construction yields an intuitively 
natural metric on the bundle space in the sense that the vertical $\mbox{\it VP}$ and 
horizontal $\mbox{\it HP}$ subspaces of the tangent space of $P$, as depicted in 
figure~\ref{pbunbasis}, are then \textit{orthogonal} with respect to $\check{g}$, with 
$\check{g}(X,Y)=0$ if $X \inn \mbox{\it VP}$ and $Y\inn \mbox{\it HP}$ for example.

   Alternatively, and perhaps more in the spirit of the original Kaluza-Klein theory, 
the metric $\check {g}$ rather than the connection $\omega$ can be taken as the 
fundamental entity on $P$. That is, the bundle is initially endowed with a 
pseudo-Riemannian metric $\check{g}$ with \textit{certain restrictions} -- namely 
compatibility with a metric $g_{ab}$ on $M_4$ and metric $g_{\alpha\beta}$ on the 
fibres $G_x$ and the requirement of invariance under the right action of $G$ on $P$:
\begin{equation}
 \label{rtrang}
  R^{\ast}_a \, \check{g}_{pa}(X,Y) =   \check{g}_p(X,Y) = \check{g}_{pa}(R_{\ast}X, 
R_{\ast}Y)
\end{equation} 
 for any $p\inn P$, $a\inn G$ and $X,Y \inn \mbox{\it TP}$.
  This latter property then implies the existence of a subspace {\it HP}, orthogonal 
to {\it VP}, which is right-invariant and hence is equivalent to the existence of a 
connection 1-form $\omega$ on the bundle $P$, which is related to $\check{g}$ as 
described in equation~\ref{ggxyk}.

  From either perspective from the relation of $\check{g}$ to $\omega$ in 
equation~\ref{ggxyk} in the horizontal lift basis $\acute{e}_i = \{\acute{e}_{\alpha}, 
\acute{e}_a\}$,  with  $\acute{e}_{\alpha} \inn \mbox{\it VP}$ and $\acute{e}_a \inn 
\mbox{\it HP}$, for the tangent space on $P$ the metric $\acute{g}$, and its inverse, 
take respectively the simple forms:
\begin{equation}
  \acute{g}_{ij} =  \left( \begin{array}{c|c}
                 g_{ab} &    \\
				          \hline
				        & g_{\alpha\beta} 
          \end{array}  \right)    
		       \qquad \mbox{and} \qquad 
  \acute{g}^{ij} = \left( \begin{array}{c|c}
                 g^{ab} &    \\
	            		  \hline
				        & g^{\alpha\beta}   
          \end{array}  \right)     \label{ggghlb}
\end{equation}
   That is with the components of the metric on the base space $M_4$ being $g_{ab} = 
\acute{g}(\acute{e}_a, \acute{e}_b)$ and those of the Killing metric on the group 
space being $g_{\alpha\beta} = \acute{g}(\acute{e}_{\alpha}, \acute{e}_{\beta})$. The 
off-diagonal components in equation~\ref{ggghlb} are all zero, with for example 
$\acute{g}(\acute{e}_a, \acute{e}_{\beta}) = 0$ describing the orthogonality of any 
$X\inn H_pP$ to any $Y\inn V_pP$  with respect to this right-invariant metric 
$\acute{g}$.

   Under a change of frame to a direct product basis $\{\acute{e}_i\} \to 
\{\ddot{e}_i\}$, that is the reverse of equation~\ref{ehtoeb} with $\ddot{e}_{\alpha} 
= \acute{e}_{\alpha}$ and $\ddot{e}_a = \acute{e}_a + 
\omega^{\alpha}_{\ph{\alpha}a}\acute{e}_{\alpha}$  for a choice of trivialisation 
$\psi: P \to U \times G$, see figure~\ref{trivconn}, we have: 
\begin{equation}
  \ddot{g}_{ij} =  \left( \begin{array}{c|c}
                 g_{ab} + 
g_{\alpha\beta}\omega^{\alpha}_{\ph{\alpha}a}\omega^{\beta}_{\ph{\beta}b}  & 
				                   \omega^{\alpha}_{\ph{\alpha}a} g_{\alpha\beta}    
\\
				          \hline
	 g_{\alpha\beta}\omega^{\beta}_{\ph{\beta}b}  & g_{\alpha\beta} 
          \end{array}  \right)    
		       \quad \mbox{and} \quad 
  \ddot{g}^{ij} = \left( \begin{array}{c|c}
                 g^{ab} &  -g^{ab}\omega^{\beta}_{\ph{\beta}b}  \\
	            		  \hline
				 -\omega^{\alpha}_{\ph{\alpha}a}g^{ab}      & g^{\alpha\beta}  +
				      g^{ab}\omega^{\alpha}_{\ph{\alpha}a}\omega^{\beta}_{\ph{\beta}b} 
          \end{array}  \right)      \label{gmetug}
\end{equation}

   In this latter basis the non-Abelian gauge fields 
$\omega^{\alpha}_{\ph{\alpha}a}(p)$ on $P$  for the internal symmetry are found 
alongside the external spacetime metric elements $g_{ab}$ framed within the components 
of the full metric $\ddot{g}_{ij}$ on the bundle space. This is a generalisation of 
the original 5-dimensional Kaluza-Klein theory in which the electromagnetic 4-vector 
potential $A_{a}$ appears alongside the components of the spacetime metric $g_{ab}$ 
within the extended $5 \times 5$ metric tensor.

   Any differentiable manifold $M_n$ is canonically associated with a principle bundle 
of linear frames $\mbox{\it FM}_n$ with structure group $\mbox{GL}(n,\rrr)$, where $n$ 
is the dimension of the base manifold $M_n$, as described in the opening of 
section~\ref{riegeo}. This includes the case in which the base manifold is actually 
the space of a given principle fibre bundle $P$ itself.

  While the metrics $g$ and $K$ on the manifolds $M_4$ and $G$ can be naturally 
extended to the metric $\check{g}$ of equation~\ref{ggxyk} on the principle bundle $P 
= (M_4,G)$ with a connection $\omega$,   
  linear connections on the manifolds $M_4$ and $G$ may also be generalised to the 
domain of the larger manifold $P$. As described for equation~\ref{gamene}
such a linear connection $\check{\Gamma}$ will define covariant differentiation with 
$\check{\nabla} \check{e}_i = \check{\Gamma}^j_{\ph{j}i}\check{e}_j
   \equiv \check{\Gamma}^j_{\ph{j}ik} \check{e}^k \otimes  \check{e}_j$ in a general 
tangent space basis $\{\check{e}_i\}$ for {\it TP} with dual basis $\{\check{e}^i\}$ 
for $T^\ast \! P$. 
The identification of the smooth symmetric gauge covariant rank-2 tensor $\check{g}$ 
everywhere on $P$  endows the principle bundle itself with the structure of a 
pseudo-Riemannian manifold. In  
 turn a connection $\check{\Gamma}$ compatible with the metric $\check{g}$, and hence 
with the geometric structure of the underlying manifold $P$, may be extended from the 
notion of a metric connection on $M_4$.

 Indeed, and further guided by Einstein's general theory of relativity in 
4-dimensional spacetime, the unique linear connection  which is torsion-free, 
$\check{T}^i_{\ph{i}jk}=0$, and compatible 
 with the metric, $\check{\nabla}_k \check{g}_{ij} = 0$, that is the Levi-Civita 
connection, may be defined on the bundle space $P$. The corresponding  connection 
coefficients can be expressed, with $\Gamma_{ijk} = g_{il}\Gamma^l_{\ph{i}jk}$ and  
$c_{ijk} = g_{il}c^l_{\ph{i}jk}$, as:
\begin{equation}
    \check{\Gamma}_{ijk} = \frac{1}{2} (\che_j(\chg_{ik}) + \che_k(\chg_{ij}) - 
\che_i(\chg_{jk}))
	                    -\frac{1}{2} (\chc_{ijk} + \chc_{kji} + \chc_{jki})   
\label{hgamijk}
\end{equation}
  which expresses equation~\ref{gtoGam} in a general frame. These coefficients take a 
relatively simple form in the  horizontal lift basis, as employed for the metric in 
equation~\ref{ggghlb}, while a coordinate basis will also be adopted on the base space 
$M_4$. 
 In this basis the connection coefficients $\Gamma^a_{\ph{a}bc}$ on the base space 
$M_4$ contribute to the set in equation~\ref{hgamijk} with:
\begin{equation}
  \acute{\Gamma}^a_{\ph{a}bc} = \Gamma^a_{\ph{a}bc} = \frac{1}{2} g^{ad}(e_b(g_{cd}) + 
e_c(g_{bd}) - e_d(g_{bc}))   \label{hgamext}
\end{equation}
which is simply equation~\ref{gtoGam}, since the structure coefficients on $P$  are 
related to the structure coefficients on the base manifold with $\acute{c}_{abc} = 
c_{abc} = 0$
in this basis (and with the corresponding term hence absent in equation~\ref{hatb3}).
 The connection coefficients $\check{\Gamma}_{ijk}$ are also related to the internal 
curvature through equation~\ref{hgamijk} since in the horizontal lift basis, by 
equation~\ref{fabecab},  we have  ${\acute c}^{\alpha}_{\ph{\alpha}ab} = 
-\Omega^{\alpha}_{\ph{\alpha}ab}$.
 Here we adopt the convention of denoting the components of curvature 
$\Omega^{\alpha}_{\ph{\alpha}ab}$ on the principle bundle by 
$F^{\alpha}_{\ph{\alpha}ab}$, which may then represent the curvature components on $P$ 
or $M_4$ depending on the context, in order to match the notation in many of the 
references. Ultimately the curvature $F^{\alpha}_{\ph{\alpha}ab}$ will feature in 
gauge invariant expressions on the base manifold.
 From equation~\ref{hgamijk} we find in the horizontal lift basis on the bundle $P$ 
 terms such as (see \cite{Cho} equation~(22)): 
\begin{equation}
 \acute{\Gamma}^{\alpha}_{\ph{\alpha}ab} = +\frac{1}{2} {F}^{\alpha}_{\ph{\alpha}ab}
   \qquad \mbox{and} \qquad
 \acute{\Gamma}^a_{\ph{a}b\alpha} = \acute{\Gamma}^a_{\ph{a}\alpha b}  = 
      +\frac{1}{2}g^{ac} g_{\alpha\beta} {F}^{\beta}_{\ph{\beta}bc}   \label{hgamint}
\end{equation} 
  The complete set of coefficients for the Levi-Civita connection on $P$ are listed 
  under `Cho~\cite{Cho}' as the first case in table~\ref{Gamsetsr} in the following 
section.

 Hence  the Levi-Civita connection of equation~\ref{hgamijk} on the total bundle space 
$P$ is intimately related to  the external curvature on the base space as well as the 
internal curvature of the gauge group.
 In turn the components of the Riemann curvature tensor $\check{R}^i_{\ph{i}jkl}$ 
calculated for this Levi-Civita connection on $P$ according to equation~\ref{rabcd} is 
intimately related to \textit{both} the \textit{external} curvature on $M_4$ via 
equation~\ref{hgamext} and the \textit{internal} curvature, associated with gauge 
group $G$, which is drawn into the Riemannian geometry through equation~\ref{hgamint}.

  It is important to clarify the relation between the linear connection 
$\check{\Gamma}(p)$ and gauge connection $\omega(p)$ on the manifold $P$. In fact from 
the point of view of frame bundles and principle fibre bundles in general a 
\textit{linear} connection $\widetilde{\omega}$ (see the opening of 
section~\ref{riegeo}) on {\it FP}  would be the same kind of object as the 
\textit{gauge} connection $\omega$ on $P$.
 Here we are dealing with Riemannian geometry of the manifold $P$ itself, which is 
hence the \textit{base space} upon which the $\glmra$-valued 1-form $\check{\Gamma}(p) 
= \Sigma^{\ast} \widetilde{\omega}$ is defined, where $m = \mbox{dim}(P) = 
\mbox{dim}(M_4) + \mbox{dim}(G)$ and $\Sigma(p)$ is a section map  
$P \to \mbox{\it FP}$ over $P$. The same manifold $P$ is also the \textit{principle 
bundle} upon which the gauge connection $\omega$ is defined, with $A(x) = 
\sigma^{\ast}\omega(p)$ being the gauge field on $M_4$, for a section map $\sigma(x): 
M_4 \to P$ over the space $M_4$.

 Having the metric $\chg_{ij}$ on $P$ the Ricci tensor $\check{R}_{jk} = \chg^{il} 
\check{R}_{ijkl}$ (equation~\ref{riccicon}) and scalar curvature $\check{R} = 
\chg^{ij} \check{R}_{ij}$ may also be computed, where the latter is found to be (with 
differing sign convention to \cite{Cho}):   
\begin{equation}
  \check{R} = R_M + R_G + \frac{1}{4}F^2    \label{hrscal}
\end{equation}
 Here $R_M$ is the usual scalar curvature on the base manifold (which varies with the 
point $x = \pi(p) \inn M_4$ under $p\inn P$) and $R_G$ is the scalar curvature on the 
group manifold $G$ (which can be interpreted as a, problematically very large, 
cosmological constant in this version of Kaluza-Klein theory).
 The term $F^2 = {F}^{\alpha}_{\ph{\alpha}ab}{F}_{\alpha}^{\ph{\alpha}ab}$, 
constructed from the non-Abelian gauge fields, is also gauge invariant and  
the curvature components ${F}^{\alpha}_{\ph{\alpha}ab}(p)$  on $P$ can be interpreted 
as the corresponding gauge covariant curvature components 
$F^{\alpha}_{\ph{\alpha}ab}(x)$ on the base space $M_4$, for example in 
table~\ref{Gamsetsr}.

  As a scalar $\check{R}$ in equation~\ref{hrscal} is a quantity which is independent 
of the basis $\{\che_i\}$ in which it is determined (for example in the direct product 
or horizontal lift basis respectively  for equations (17) and (24) of reference 
\cite{Cho}). The equations of motion for the theory are then derived by adopting the 
Lagrangian function $\check{R}\sqrt{\vert \chg \vert}$, where $\vert \chg \vert$ is 
the magnitude of the determinant of the metric $\chg_{ij}$ on $P$, in the 
Einstein-Hilbert action integral:
\begin{equation}
A_{m} = \int \check{R}\sqrt{\vert \chg \vert} \; d^4 x \; d^{n_G} G  \label{einhilf}
\end{equation}
 with $m=4+n_G$.
 The integration over the group manifold $G$, with volume $V_G$, is trivial and the 
above expression reduces to the 4-dimensional action integral:
\begin{equation}
   A_4 = V_G \int  \check{R}\sqrt{\vert g \vert} \; d^4 x    \label{einhilk}
\end{equation}
  where $\vert g \vert$ is here the determinant of the metric $g_{ac}$ on $M_4$. The 
variational principle is then applied under the constraint $\delta A_{m} = 0$, and 
hence $\delta A_{4} = 0$, with respect to restricted variations of the metric 
$\delta\chg$ on the bundle space, consistent with equation~\ref{rtrang}, as explained 
before equation~\ref{einfiek} in the following section. Within this restriction 
 this again follows the prescription for the original theory of general relativity on 
a 4-dimensional spacetime manifold $M_4$ with scalar curvature $R$ for which the field 
equations can be determined from the Einstein-Hilbert action integral of 
equation~\ref{einhil}.

 By comparison of equations~\ref{hrscal} and \ref{einhilk} with \ref{einhil} the 
constant $R_G$ of the Kaluza-Klein theory indeed appears as a cosmological constant 
term (albeit too large by a factor of $\sim 10^{120}$ if a natural normalisation  is 
used with the length scale of the group space $G$ taken to be of order the Planck 
length~\cite{Cho}), while
  the term $F^2$  effectively contributes the content for the matter Lagrangian 
$\lag$. Hence, as a particularly elegant feature of Kaluza-Klein theory, the 
\textit{geometry} of the 4-dimensional spacetime manifold along with a \textit{matter} 
contribution is identified within a single \textit{geometrical} object in the form of 
$\check{R}$ on the principle bundle space.

\section{Theories with Torsion on the Bundle}
\label{olcop}

 One way to remove the problematic cosmological term $R_G$ in equation~\ref{hrscal} 
would be to redefine the Lagrangian for the Kaluza-Klein theory by simply adding by 
hand a \textit{counter}-cosmological constant term $\Lambda_c$ to $\check{R}$ in 
equation~\ref{einhilf} to cancel $R_G$. However this would be an ad hoc measure, 
similar in spirit to the inclusion of the original cosmological constant term 
$\Lambda$ in equation~\ref{einhil}, contrived largely to match empirical observation.
 
    However there is flexibility within the Kaluza-Klein approach on a principle fibre 
bundle if the metric $\chg_{ij}$ is \textit{not} treated as the fundamental object of 
the theory (see for example  \cite{Kopcz,OrzPau,Kalin,Katan}).
 While the same natural metric $\chg_{ij}$ of equation~\ref{ggghlb} is employed
 the linear connection $\check{\Gamma}^i_{\ph{i}jk}$ on $P$ may be defined with  some 
independence from $\chg_{ij}$, unlike for the Levi-Civita connection of 
equation~\ref{hgamijk}.  In this case it is possible to derive a curvature scalar 
$\check{R}$ on $P$ such that the cosmological term $R_G$ vanishes and 
equation~\ref{hrscal} reduces to simply:
\begin{equation}
  \label{hrsnorg}
   \check{R} = R_M + \frac{1}{4}F^2
\end{equation}

   One way to achieve this is to require the linear connection 
$\check{\Gamma}^i_{\ph{i}jk}$ to incorporate a description of absolute parallelism on 
the bundle fibres $G_x$.
 As reviewed in subsection~\ref{tgman}
 on the manifold $G$ itself the list of canonical geometric objects include a basis of 
left-invariant vector fields $\{X_{\alpha}\}$ and the Maurer-Cartan 1-form $\theta = 
\theta^{\alpha}X_{\alpha}$ as well as the structure constants $\cstr$ and the Killing 
form metric $g_{\alpha\beta}$ of equation~\ref{killmet}.
As described below equation~\ref{bianchirt}
 in the basis $\{X_{\alpha}\}$ 
the choice of linear connection coefficients $\Gamalp = 0$ is equivalent to inducing 
parallel transport on the group manifold via the left action $L_a$ of $G$ on itself, 
for any $a\inn G$, that is with parallelism  defined by the left-invariant vector 
fields $\{X_{\alpha}\}$ on $G$,  while $\Gamalp = -\cstr$ corresponds to the 
parallelism described by a right-invariant frame field under $R_a$. In either case  
 the resulting Riemann curvature is zero with 
$R^{\alpha}_{\ph{\alpha}\beta\gamma\delta} = 0$, as can be shown using 
equation~\ref{rabcd} together with the Jacobi identity expressed in terms of the 
structure constants.

  More generally, employing the derivative action of the left-invariant basis vectors 
$\{X_{\alpha}\}$, the right-invariance of the Killing metric implies that the 
covariant derivative of the metric on $G$ vanishes:
\begin{eqnarray}
    \nabla_{\alpha} g_{\beta\gamma}  & = & 
	       X_{\alpha} g_{\beta\gamma} - 
\Gamma^{\delta}_{\ph{\delta}\beta\alpha}g_{\gamma\delta}
	                               - 
\Gamma^{\delta}_{\ph{\delta}\gamma\alpha}g_{\beta\delta}
								    \;  = \; 0   \\
  \mbox{provided} \quad	\Gamalp & = & -\rho\, \cstr 
 \quad \mbox{for any} \quad \rho \inn \rrr		\label{gamlamc}					 
\end{eqnarray}
 since $X_{\alpha} g_{\beta\gamma} = 0$ and by the antisymmetry of the 
$c_{\alpha\beta\gamma}$
  indices. 
  Hence for any value of $\rho$ this linear connection is metric compatible, with 
  $\nabla g = 0$ on $G$. The torsion is zero only for $\rho = \fh$ which hence 
represents the unique Levi-Civita connection on $G$.
  On the other hand 
    the Riemannian curvature is zero on $G$ only for the cases of $\rho = 0$ and 
$\rho=1$, which with finite torsion are not Levi-Civita connections. However these 
latter two cases in describing  an absolute parallelism on $G$ can be considered as 
geometrically natural metric connections on $G$.

  For the linear connection $\Gamalp = 0$ or $\Gamalp = -\cstr$   employed on the 
bundle fibres $G_x$ a subset of the torsion components on $P$ are also necessarily 
non-zero,  with  $\check{T}^{\alpha}_{\ph{\alpha}\beta\gamma} \neq 0$. Hence with the 
torsion allowed to be non-zero on the bundle space $P$ this version of Kaluza-Klein 
theory resembles the Einstein-Cartan theory on 4-dimensional spacetime for which 
$\Gamma$ and $g$ are treated as independent geometric objects. Here we review four 
such approaches in the literature.

  In Kopczy\'{n}ski \cite{Kopcz} a $G$-invariant linear connection $\check{\Gamma}$ is 
constructed in terms of the structure on the principle bundle
with a gauge connection $\omega$ without reference to a metric and with non-zero 
torsion. This generalises from the Levi-Civita connection described in the previous 
section (as employed by $\cite{Cho}$ and others) with the `gravitational field' on $P$ 
now being described by the combination of both $\chg$ of equation~\ref{ggxyk} and the 
components of $\check{\Gamma}$ as listed in the corresponding column under `Kop 
\cite{Kopcz}' in table~\ref{Gamsetsr}. With these components the scalar curvature on 
$P$ is found to be $\check{R} = R_M + \alpha (\alpha - 1)K^2$, with $K^2 = 
K^{\alpha\beta}K_{\alpha\beta}$. For the case $\beta = 0$ the connection is metric 
compatible, resembling Einstein-Cartan general relativity in 4-dimensional spacetime. 
 While this reference shows that the connection coefficients can be greatly simplified 
compared with the Levi-Civita case, listed under `Cho \cite{Cho}' in the first column 
of table~\ref{Gamsetsr}, in order to achieve the correct dynamics a more complicated 
Lagrangian function is postulated with 
 $\lag = \check{R} + \frac{\mu}{2} \, \check{T}^i_{\ph{i}jk}\check{T}_i^{\ph{i}jk}$ 
including a quadratic torsion term.
The cosmological constant $\Lambda$ obtained in this approach is arbitrary, and may be 
set to be zero or very small by a suitable choice of the parameters $\alpha$ and 
$\mu$.

 In Orzalesi and Pauri~\cite{OrzPau} the main motivation is to describe a linear 
connection $\check{\Gamma}$ on the principle bundle which is gauge covariant.
 In particular requiring the Ricci curvature on the fibre space to be gauge invariant 
implies the adoption of zero curvature on the group manifold, that is the case $\rho = 
0$ or $\rho = 1$ as described above for equation~\ref{gamlamc}. This form differs in a 
relatively  
 minimal way from the Levi-Civita connection, as can be seen by comparing the entries 
of column~\cite{OrzPau} with column~\cite{Cho} in table~\ref{Gamsetsr}. Here the 
simple scalar Lagrangian $\lag = \check{R}$ on the bundle space is again adopted, 
 with the resulting vanishing of  $\Lambda \equiv R_G $ on the base space $M_4$  
interpreted as a consequence of the underlying gauge $G$-symmetry of the Riemannian 
geometry on $P$. Without  an $R_G$ term the vacuum solution corresponds to zero 
external curvature $R_M = 0$ together with zero internal curvature $F=0$.

  In Kalinowski \cite{Kalin} the linear connection 1-forms 
   $\check{\Gamma}^{i}_{\ph{i}j} = \check{\Gamma}^{i}_{\ph{i}jk} \che^k$ on $P$ are 
defined as
     the horizontal part of the  Levi-Civita connection 1-forms 
${\Gamma}^{i}_{\ph{i}j}$
 of equation~\ref{hgamijk}, that is
  $\check{\Gamma}^{i}_{\ph{i}j} = \mbox{hor}({\Gamma}^{i}_{\ph{i}j})$
   (with `hor' introduced in equation~\ref{extcovd}) which maps the vertical component 
of tangent vectors on {\it TP} to zero.
 The components of this linear connection $\check{\Gamma}^{i}_{\ph{i}jk}$ in the 
horizontal lift basis are listed in column~\cite{Kalin} of table~\ref{Gamsetsr}. The 
factors of $\lambda$ arise as here the metric on $G$ is taken to be $g_{\alpha\beta} = 
\lambda^2 K_{\alpha\beta}$.
 This linear connection $\check{\Gamma}^{i}_{\ph{i}jk}$ is metrical, invariant under 
the $G$-action, again with non-zero torsion and, while motivated in the context of 
gauge derivatives of spinor fields, again leads to a vanishing cosmological constant.

 In Katanaev \cite{Katan} an initially completely general 
$\check{\Gamma}^i_{\ph{i}jk}$ on the principle bundle is considered. Four conditions 
are postulated for $\check{\Gamma}$ in 
a geometrically meaningful way related to
 the structure group $G$ over $P$  and, as for the previous reference, with emphasis 
on  horizontal propagation. In particular for column \cite{Katan} of 
table~\ref{Gamsetsr}
 on taking $c=1$ for entry `5)' $\acute{\Gamma}^{\alpha}_{\ph{\alpha}ab}= c 
F^{\alpha}_{\ph{\alpha}ab}$   the change in a tangent vector to $P$ under parallel 
transport using these linear connection coefficients equals the change in the vector 
due to the basis transformation under parallel transport of the fibres using the gauge 
connection, with the latter depicted in figure~\ref{paratran}. The entry `4)' in this 
column is included for metric compatibility.
The coefficients listed represent the case presented in  \cite{Katan} with finite 
torsion and the absence of a cosmological constant term, although a different choice 
of $\check{\Gamma}$ consistent with the postulates is possible. A further possibility 
within this framework would be to set the first two entries, `1)' and `2)', equal to 
zero in column \cite{Katan}. This reference is of significance for the present paper 
in that it highlights the possibility of a geometric origin of $\check{\Gamma}$ on $P$ 
without any appeal to the Levi-Civita connection.

   The complete set of linear connection coefficients for reference \cite{Cho}, 
augmenting equation~\ref{hgamint}, are collected in the first column of 
table~\ref{Gamsetsr}. These are listed alongside the linear connection coefficients  
$\check{\Gamma}^{i}_{\ph{i}jk}$ on the bundle space $P$ for the above four cases with 
non-zero torsion. Where necessary signs have been aligned to the conventions used 
here, with for example linear connection 1-forms     
   $\check{\Gamma}^{i}_{\ph{i}j} = \check{\Gamma}^{i}_{\ph{i}jk} \che^k$. The 
motivation for the final column headed `minimal' will be explained in 
section~\ref{reaic}.

\begin{table}[htbp]
\centering
\hspace*{-16pt}
\begin{tabular}{|l||c|c|c|c|c|c|}
 \hline
    $\quad \acute{\Gamma}^i_{\ph{i}jk}$ & Cho~\cite{Cho} 
  &	Kop~\cite{Kopcz}  &   O$+$P~\cite{OrzPau} 
    & Kal~\cite{Kalin}  & Kat~\cite{Katan}  
	 & minimal \\  
 \hline				
1)  $\acute{\Gamma}^{\alpha}_{\ph{\alpha}\beta\gamma}$  &
 $-\fhs \cstr$ & $\,\:\;-\alpha \cstr\,\:\;$ & $-\cstr$ or $0$ & 0 & $-\cstr$ 
   & 0\\ 
 \hline
2)  $\acute{\Gamma}^{\alpha}_{\ph{\alpha}\gamma a}$  &
 0 & 0 & 0 & 0 & $-\omega^{\beta}_{\ph{\beta}a}
                      \acute{\Gamma}^{\alpha}_{\ph{\alpha}\beta\gamma}  $ 
  & 0 \\ 
  \hline
3)  $\acute{\Gamma}^{a}_{\ph{a}b\gamma}$  & 
   $\!\fhs g^{ac}g_{\gamma\beta}F^{\beta}_{\ph{\beta}bc}\!$ & 0 &
      $\!\fhs g^{ac}g_{\gamma\beta}F^{\beta}_{\ph{\beta}bc}\!$ & 0 & 0 
	  & $\! \gamma g^{ac}g_{\gamma\beta}F^{\beta}_{\ph{\beta}bc}\!$ \\
 \hline
4)   $\acute{\Gamma}^{a}_{\ph{a}\gamma b}$  & 
   $\!\fhs g^{ac}g_{\gamma\beta}F^{\beta}_{\ph{\beta}bc}\!$ & 0 &
   $\!\fhs  g^{ac}g_{\gamma\beta}F^{\beta}_{\ph{\beta}bc}\!$ &      
    $\!\frac{\lambda}{2}g^{ac}g_{\gamma\beta}F^{\beta}_{\ph{\beta}bc}\!$
	   &  $\! c g^{ac}g_{\gamma\beta}F^{\beta}_{\ph{\beta}bc}\!$
	    & 0 \\
\hline
5)   $\acute{\Gamma}^{\alpha}_{\ph{\alpha}ab}$  & 
   $ \fhs F^{\alpha}_{\ph{\alpha}ab}$ & $\beta F^{\alpha}_{\ph{\alpha}ab}$  &
   $\fhs F^{\alpha}_{\ph{\alpha}ab}$ & $\frac{\lambda}{2}F^{\alpha}_{\ph{\alpha}ab}$ &
   $c F^{\alpha}_{\ph{\alpha}ab}$ 
    & 0   \\
 \hline
6)   $\acute{\Gamma}^a_{\ph{a}bc}$   & $\Gamabc$ & $\Gamabc$ & $\Gamabc$ 
       & $\Gamabc$ & $\Gamabc$  
	    & $\Gamabc$ \\
	   \hline 
  \end{tabular}
  \caption{\setb Linear connection components $\acute{\Gamma}^i_{\; jk}$ on a 
principle bundle extracted from \protect\cite{Cho} equation~(22), \protect\cite{Kopcz} 
p.367, \protect\cite{OrzPau} equation~(19), \protect\cite{Kalin} equation~(29), the 
case in \protect\cite{Katan} with non-zero torsion on $G$ and for a `minimal' model.
 All components are expressed in the horizontal lift basis and 
$\acute{\Gamma}^{a}_{\;\beta\gamma}= \acute{\Gamma}^{\alpha}_{\; a\gamma}= 0$ in all 
six cases. Each of
 $\lambda > 0$, $\alpha$, $\beta$, $c$ and $\gamma$, where used as coefficients, are 
real constant parameters.}
\label{Gamsetsr}
\end{table} 

   Only the first case in table~\ref{Gamsetsr} describes a torsion-free linear 
connection, yet each of the six cases is a Kaluza-Klein theory providing a unifying 
framework for general relativity together with gauge field theory. The purpose of 
collecting together the range of linear connection coefficients is to demonstrate that 
a significant degree of flexibility is possible within Kaluza-Klein theory while still 
maintaining this unified framework.

  The derivation of Einstein's equations on 4-dimensional spacetime from the 
Einstein-Hilbert action of equation~\ref{einhil} was described in the opening of 
section~\ref{subfal}. In the vacuum case with $\lag=0$ and $\Lambda = 0$ variation of 
the metric $\delta g_{\mu\nu}$ on $M_4$ leads to the equation of motion $G^{\mu\nu} 
=0$ of equation~\ref{lagtoein}.
  For the Kaluza-Klein extension to the scalar curvature $\check{R}$ for the 
Levi-Civita connection on a principle bundle space the same steps lead
 to the requirement of the stationarity of 
 the action integral over the full bundle space in equation~\ref{einhilf}, that is 
$\delta A_{m}=0$, under variation of the extended metric $\chg_{ij}$ on $P$, which 
results in the expression:
\begin{equation}
    \check{G}^{ij} = \check{R}^{ij} - \frac{1}{2}\check{R} \, \chg^{ij} = 0   
\label{einvace}
\end{equation}
  In some versions of Kaluza-Klein theory, in particular for the 5-dimensional case, 
equation~\ref{einvace}, which implies $\check{R}^{ij} = 0$, is quoted as an ansatz at 
the outset in order to derive equations of motion for the 4-dimensional world by 
imposing this higher-dimensional `vacuum equation' (see for example~\cite{Wess}, in 
which the 5-dimensional metric $\chg_{ij}(x)$ may depend on the  $5^{\mathrm th}$ 
coordinate).

   However for the extended Kaluza-Klein theories described here the variations in the 
metric $\chg_{ij}$ are  \textit{not} arbitrary since the structure of the symmetries 
of $\chg_{ij}$ on the bundle space $P$ need to be preserved under the variations 
$\delta \chg_{ij}$. That is, the right-invariance of $\chg_{ij}$ of 
equation~\ref{rtrang} and the general form of the metric in equation~\ref{gmetug} 
should be preserved. This limits the metric variations to the components $\delta 
g_{ac}$ and $\delta \omega^{\alpha}_{\ph{\alpha}a}$ on $P$ and leads to two equations 
of motion on the base manifold $M_4$. Applying the variation $\delta g_{ac}$ under 
$\delta A_{m} = 0$ for the action in equation~\ref{einhilf}, with the curvature 
$\check{R}$ of equation~\ref{hrsnorg}, in a general coordinate basis on $M_4$ leads to 
(\cite{Kalin} equation~38):
\begin{eqnarray}
  G^{\mu\nu} \: = \:  R^{\mu\nu} - \frac{1}{2}R \, g^{\mu\nu}
      & = & -\kappa T^{\mu\nu}
      \label{einfiek}    \\
	\mbox{with} \qquad  \qquad -2\kappa T^{\mu\nu}  & = & - F^{\alpha \mu}_{\ph{\alpha 
\mu}\rho}F_{\alpha}^{\ph{\alpha}\rho\nu}
	                -\frac{1}{4} g^{\mu\nu} \, F^{\alpha}_{\ph{\alpha}\rho 
\sigma}F_{\alpha}^{\ph{\alpha}\rho\sigma}    \label{einym}       
\end{eqnarray}
     The left-hand side of the top line would read $G^{\mu\nu} + R_Gg^{\mu\nu}$ if the 
scalar curvature of equation~\ref{hrscal} based on a Levi-Civita connection is used 
instead.  On the other hand the variation $\delta \omega^{\alpha}_{\ph{\alpha}a}$ 
leads to
\begin{equation}
	D_{\mu}F^{\alpha \,\mu\nu} \: = \:  0    \label{yangmk}
\end{equation}
   Equation~\ref{einfiek} is the Einstein field equation with the energy-momentum 
tensor $T^{\mu\nu}$ composed as equation~\ref{einym} purely from the gauge fields, 
with the latter being subject to equation~\ref{yangmk} which is the Yang-Mills field 
equation (or Maxwell's equation $F^{\mu\nu}_{\ph{\mu\nu};\mu}=0$ in the case of the 
Abelian internal symmetry group $G=\uo$, see also the discussion after 
equation~\ref{maxj}).  Hence the source-free Yang-Mills field equation~\ref{ymol} has 
been derived \textit{without} the explicit introduction of the Yang-Mills Lagrangian 
of equation~\ref{lagym}. Rather 
such a `Lagrangian term' $F^2 = F_{\alpha \, \mu\nu} F^{\alpha \, \mu\nu}$ has been 
incorporated  within the Einstein-Hilbert action based purely on the geometry of the 
bundle space.
 In this way the non-Abelian Kaluza-Klein theory provides a unified framework for the 
combined Einstein-Yang-Mills field equations.

\section{Theories with Homogeneous Fibres}
\label{thwhf}

  A further generalisation of Kaluza-Klein theory is also of relevance for the 
framework presented in this paper. In the present theory the symmetry group $G$ rather 
than being motivated independently is introduced in terms of the set of symmetry 
actions on a form $\lv$ of multi-dimensional temporal flow.  
   This structure is reminiscent of Kaluza-Klein theories with homogeneous fibres in 
which $G$ acts on a $k$-dimensional manifold $S_k$. A bundle space $E$ is constructed 
with each fibre being a copy of $S_k$  over the base space $M_4$. Based on the  
references 
\cite{Perc,Lucietal,CLLM}
 this approach will be collectively summarised in this section.

  In these models  the bundle $E=(M_4, S_k)$ is constructed over the base space $M_4$ 
while the fibres $S_k$ may be considered to represent $k$ `extra dimensions'. 
 For our purposes it is sufficient to consider the trivial bundle with $E \equiv M_4 
\times S_k$. While either a left or right action may be considered here
 we take the gauge group $G$ to act on the space
 $S_k$ on the left (as for reference \cite{Perc} for example, and as will be the case 
for the $\esi$ action on the space $\htho$ as discussed alongside 
equation~\ref{rrbrac}) such that each Lie algebra element $X_{\alpha} \inn L(G)$ 
generates a vector field $K_{\alpha}$ on $S_k$ with the bracket composition exhibiting 
the negative of the structure constants $\cstr$ of $L(G)$, that is:
\begin{equation}
 \label{negcstr}
 \lbrack K_{\beta}, K_{\gamma} \rbrack = -\cstr K_{\alpha}
\end{equation}

  The group actions may also be considered to be one-to-one with the isometry 
transformations for an inner product defined on the tangent space $\mbox{\it TS}_k$. 
That is, a $G$-invariant metric may be defined on $S_k$ with Killing vector fields:
 \begin{equation}
   K_{\alpha} = K^{\ralpha}_{\ph{\alpha}\alpha}e_{\mathring{\alpha}}
     \label{killcomp}
 \end{equation}
   where $K^{\mathring{\alpha}}_{\ph{\alpha}\alpha}$ are the components of 
$K_{\alpha}$ in a linearly independent tangent space basis $\{e_{\mathring{\alpha}}\}$  
on $S_k$, with indices $\mathring{\alpha} = 1\ldots k$ and $\alpha = 1 \ldots 
\mbox{dim}(G)$.
Such a $G$-invariant metric $g_{\ralpha\rbeta}$ on $S_k$ may be induced from the 
Killing metric $K_{\alpha\beta}$ on $G$ itself.

If $G$ acts upon $S_k$ \textit{transitively} then $S_k$ is a homogeneous space. 
  Given any point $y_0 \inn S_k$ the elements $h\inn G$ for which $h \cdot y_0 = y_0$ 
under the left action  of the group form the isotropy subgroup $H$, with $h \inn H 
\subset G$.
 The homogeneous space $S_k$ is then diffeomorphic  to the space of left cosets $gH$  
   as identified for varying $g \inn G$, 
that is $S_k \equiv G/H$ where $H$ is the isotropy subgroup of the isometry group $G$. 
As a vector space the Lie algebra of $G$ may be decomposed as:
\begin{equation}
  \label{lglhb}
 L(G) = L(H) + B
\end{equation}
  with $\lbrack L(H), L(H) \rbrack \subset L(H)$ and $\lbrack L(H), B \rbrack \subset 
B$,  where $B \equiv T(G/H)$ forms a basis for the tangent space at $y_0 \inn S_k$.

  Such a linearly independent basis $\{e_{\mathring{\alpha}}\}$ for $\mbox{\it TS}_k$ 
forms a basis for the vertical subspace of the tangent space on the fibre bundle $E$. 
A complete `horizontal basis' on $E$, written
 $\acute{e}_{i}(x,y) = \{\acute{e}_{\ralpha}, \acute{e}_a\}$, in place of the 
horizontal lift basis  for the principle bundle $P$ of figure~\ref{pbunbasis}, can be 
expressed as:
\begin{equation}
 \acute{e}_{\ralpha}  =  \ddot{e}_{\ralpha},  \qquad 
  \acute{e}_a  = \ddot{e}_a \, - \, A^{\alpha}_{\ph{\alpha}a}(x) 
K^{\ralpha}_{\ph{\alpha}\alpha}(y)
			    \ddot{e}_{\ralpha}(x,y)     
	\;\; \equiv \;\; \ddot{e}_a  \, - \,
              A^{\alpha}_{\ph{\alpha}a}(x) 
			    \ddot{e}_{\alpha}    \label{depae}
\end{equation}
   in terms of a direct product basis $\ddot{e}_{i}(x,y) = \{\ddot{e}_{\ralpha}, 
\ddot{e}_a\}$ on $E$, by comparison with equation~\ref{ehtoeb} and 
figure~\ref{trivconn},
 using the Killing vector components $K^{\ralpha}_{\ph{\alpha}\alpha}$ defined in 
equation~\ref{killcomp}. As implied in equation~\ref{depae} the construction of such a 
horizontal basis on $E$  corresponds to the introduction of a connection form $\omega$ 
on the associated principle bundle $P \equiv M_4 \times G$. This connection form is 
written in terms of the coefficients $A^{\alpha}_{\ph{\alpha}a}(x)$ rather than 
$\omega^{\alpha}_{\ph{\alpha}a}(x,g)$ since the vertical basis is here defined through 
the \textit{left} action of $G$ (see the discussion in~\cite{Perc} after equation 
(7.2) for example).

  Consistent with the horizontal basis of equation~\ref{depae} a natural metric on the 
bundle space $E$ may be defined, for which horizontal and vertical vectors are 
mutually orthogonal, and expressed in a direct product basis as:  
\begin{equation}
  \ddot{g}_{ij} =  \left( \begin{array}{c|c}
       g_{ab}(x) - g_{\ralpha\rbeta}(y)
	   K^{\ralpha}_{\ph{\alpha}\alpha}(y) A^{\alpha}_{\ph{\alpha}a}(x)
	   K^{\rbeta}_{\ph{\beta}\beta}(y) A^{\beta}_{\ph{\beta}b}(x)      & 
	   K^{\ralpha}_{\ph{\alpha}\alpha}(y) A^{\alpha}_{\ph{\alpha}a}(x) 
g_{\ralpha\rbeta} \\
				          \hline
	  K^{\rbeta}_{\ph{\beta}\alpha}(y) A^{\alpha}_{\ph{\alpha}b}(x) 					  
	 g_{\ralpha\rbeta}      &          g_{\ralpha\rbeta}(y) 
          \end{array}  \right)      \label{gmetone}   
\end{equation}
  which may be compared with equation~\ref{gmetug} for the case of a principle fibre 
bundle. Changes in the vertical coordinates $\{y\}$ on $E$ described by the 
infinitesimal isometries $\varepsilon^{\alpha}(x)$ induce changes in the metric 
components with respectively:
\begin{equation}
  \begin{array}{rcl}
  y^{\ralpha} & \to & y^{\ralpha}
             \; + \; \varepsilon^{\alpha}(x)K^{\ralpha}_{\ph{\alpha}\alpha}(y)   \\
  A^{\alpha}_{\ph{\alpha}a} & \to & A^{\alpha}_{\ph{\alpha}a}
   \; + \; \pal_a  \varepsilon^{\alpha}(x)  \, + \,
    \cstr \varepsilon^{\beta}(x) A^{\gamma}_{\ph{\gamma}a}
	\end{array}
	\label{coogutr} 
\end{equation}
   Hence such isometries effectively simulate
    non-Abelian gauge transformations  with $A^{\alpha}_{\ph{\alpha}a}(x)$ identified 
as the Yang-Mills gauge field on the base space.

 Following the Kaluza-Klein prescription described  in section~\ref{lccop} the 
Levi-Civita connection, that is the unique torsion-free linear connection compatible 
with the metric, and curvature can be constructed on the manifold $E$ based on the 
metric $\ddot{g}_{ij}(x,y)$ of equation~\ref{gmetone}. In turn an action principle may 
be employed on this $(4+k)$-dimensional space with action $A_{4+k} = \int 
\check{R}\sqrt{\vert \chg \vert} \; d^4 x \; d^k y$  in comparison to 
equation~\ref{einhilf} where now the curvature scalar $\check{R}$ on
 the bundle $E \equiv M_4 \times S_k$ is found to take the form:
\begin{equation}
  \check{R} = R_{M} + R_{S_k} + \frac{1}{4}g_{\ralpha\rbeta}
      K^{\ralpha}_{\ph{\alpha}\alpha}K^{\rbeta}_{\ph{\beta}\beta}
	  F^{\alpha}_{\ph{\alpha}ab}F^{\beta ab}     \label{rscalone}
\end{equation} 
  where $R_{S_k}$ is the scalar curvature of the homogeneous space $S_k$.
With $F^{\alpha}_{\ph{\alpha}ab}$ being the gauge curvature components for the 
internal symmetry group $G$ the above equation again demonstrates a relation between 
the external Riemann curvature with scalar $R_M$ and a quadratic term  in the internal 
curvature. This relationship derived for $G$ acting on homogeneous fibres is hence in 
turn similar to that obtained in equations~\ref{hrscal} and \ref{hrsnorg} with $G$ 
itself composing the fibres of a principle bundle. 
A linear connection on $E$ differing from the Levi-Civita connection may be employed 
to remove the cosmological term $S_k$ by analogy with the examples cited in the 
previous section. 
	The Einstein-Yang-Mills equations also follow from a prescription analogous to 
that described for equations~\ref{einfiek}--\ref{yangmk}.

	 For models with homogeneous fibres in which the metric $g_{\ralpha\rbeta}(y)$ is 
replaced by the more general field components  
  $g^{\ralpha\rbeta}(x,y) = K^{\ralpha}_{\ph{\alpha}\alpha}(y)
  K^{\rbeta}_{\ph{\beta}\beta}(y)\Phi^{\alpha\beta}(x)$, which depend on $x\inn M_4$ 
and where $\Phi$ is a non-Killing metric on $G$, a set of scalar fields are introduced 
into the theory with a number of further terms featuring the derivatives $\pal_a 
g^{\ralpha\rbeta}(x,y)$ appearing in the corresponding generalisation of 
equation~\ref{rscalone} (see for example~\cite{Perc} equation (8.6)).

  On the other hand on constructing $\ddot{g}_{ij}$ in equation~\ref{gmetone} for the 
case of trivial isotropy group $H = \{e\}$, where $e\inn G$ is the identity element, 
then by equation~\ref{lglhb} we have $B = L(G)$ and the $\ralpha, \ldots$ indices can 
be replaced by $\alpha, \ldots$ indices, with $K^{\ralpha}_{\ph{\alpha}\alpha} = 
\delta^{\ralpha}_{\ph{\alpha}\alpha}$. In this case the theory simplifies to that 
described  in section~\ref{lccop} based on the metric of equation~\ref{gmetug}  with 
the set of vector fields $\{\acute{e}_{\alpha}\}$  spanning the vertical subspace of 
the tangent space on $P \equiv M_4\times G$ (with care for the convention choice of a 
left or right group action, see for example~\cite{Cho} equations ($8'$) and (12)).

  Even for the case with $H\neq \{e\}$ the full $G$-symmetry Yang-Mills dynamics is 
obtained so long as $G$ acts \textit{effectively} on the fibres $S_k$. This is also 
the case for $G$ acting on $\lv$ for the present theory in this paper, and in 
particular for the $\esi$ action to be described in chapter~\ref{esihtho}, and for  
the broken internal subgroups of $G$.

The action of $G$ on the set of elements $\bv$ underlying the form $\lv$ is also 
transitive, and hence  this set forms a homogeneous space, motivating the review of 
this section. However with the observation that the Kaluza-Klein unification achieved 
with homogeneous fibres, given an effective group action, is closely related to that 
achieved on the associated principle bundle, in the following section we apply some of 
the observations of the previous section to the present theory.
 This in particular picks up from the development of section~\ref{hdasb} with the goal 
of relating the external curvature to the internal curvature in the context of the new 
theory.


\pagebreak
\chapter{Geometry Unified through Temporal Flow}
\label{chaputtf}

\section{Relating External and Internal Curvature}
\label{reaic}

   In this section, ultimately guided by the framework of Kaluza-Klein theories 
described in the previous chapter, the aim is to determine a relation between the 
external and internal geometry over the base manifold arising out of the symmetries of 
a form of temporal flow $\lv$, building upon the structures described in 
chapter~\ref{sym}.
  In place of the base space $M_3$ with local symmetry SO(3), introduced for the model 
universe in section~\ref{perc} with the 3-dimensional form $\lvth$ of 
equation~\ref{flow3d}, here we consider the form:
\begin{equation}
L(\bv_4)  = (v^0)^2- (v^1)^2-(v^2)^2-(v^3)^2 = 1  \label{flow4d}    
\end{equation}    
   that is $L(\bv_4) = \eta_{ab}v^av^b = 1$ with Minkowski metric
    $\eta = \mbox{diag}(+1,-1,-1,-1)$, with Lorentz $\soot$ symmetry, projected into 
the 4-dimensional base space $M_4$.
  Over the spacetime manifold $M_4$ a globally defined orthonormal basis arises in the 
manner of equation~\ref{spillrx} with the natural parallelism on $M_4$ described by a 
linear connection with components $\Gamabc(x) = 0$ in this  basis.
 With the local symmetry group $\soot$ over the base manifold $M_4$ the principle 
bundle  $P = (M_4, \soot)$ is implicitly identified in this structure, which in fact 
can be expressed as the manifold product
$P \equiv M_4 \times \soot$ owing to the triviality of the bundle as described towards 
the end of section~\ref{fibre}. 

  However, following section~\ref{hdasb}, here we study initially  the geometry of the 
principle fibre bundle $P \equiv M_4\times \hat{G}$, where  $\hat{G}=\sootn$ is 
provisionally taken as the full symmetry group for the form $L(\bvh)=\lvte$, which in 
turn is the 10-dimensional extension of equation~\ref{flow4d}.  The extended base 
manifold $M_4$ now arises out of four of the ten translational degrees of freedom of 
$\lvte$, in the manner described in equation~\ref{rspill}.
 In place of figure~\ref{mtogmaph} for the SO(5) model over $M_3$ described earlier, 
for this more realistic model we now have the structures described in 
figure~\ref{mtogmaphr}.

\vspace{5pt}
\begin{figure}[htbp]  
\centering
\epsfxsize=\maxwidth
\leavevmode
\epsffile[0 0 1882 964]{\gpath aPfig51e}
\vspace{-30pt}
\caption{\setb (a) The full symmetry group $\hat{G}=\sootn$ over the base space $M_4$
(b) broken to the internal symmetry SO(6) with external subgroup $\soot \subset 
\sootn$
 acting on the tangent space $\TM_4$. }
\label{mtogmaphr}
\end{figure} 

  The structure of figure~\ref{mtogmaphr} is associated with a canonical flat 
connection on $M_4$, as described by $A(x) = g^{\ast}\theta$ of equation~\ref{aegasth} 
where here $\theta$ is the Maurer-Cartan 1-form on the group manifold 
$\hat{G}=\sootn$. This canonical flat connection defines a horizontal lift basis 
$\{\acute{e}_i\}$ on the corresponding principle bundle structure $P \equiv M_4 \times 
\sootn$, as a particular case of figure~\ref{pbunbasis}. In turn  a collection of 
affine connection coefficients $\acute{\Gamma}^i_{\ph{i}jk}$ can be defined in this  
basis on $P$. 

 While $\Gamma^{a}_{\ph{a}bc} = 0$ represents the initial parallelism on $M_4$
   the set $\Gamma^{\alpha}_{\ph{\alpha}\beta\gamma}  =  0 $
  describes an absolute parallelism on the manifold $\hat{G}$, as described in 
section~\ref{olcop}.
 Extending to the full bundle space $P$ here we provisionally consider the  
$\acute{\Gamma}^i_{\ph{i}jk}$ set of reference~\cite{OrzPau} listed in the third 
column of table~\ref{Gamsetsr}. This set of linear connection coefficients are gauge 
covariant on $P$ and compatible with the metric of equation~\ref{ggxyk} deriving from 
the gauge connection $\omega$ on $P$, that is with $\nabla \check{g} = 0$. On adopting 
such a linear connection on $P$, based on compatibility with the structures of the 
form $\lvte$ here, we then consider the implications of incorporating this element of 
Kaluza-Klein theory into the present framework.

   The components of the Riemann curvature on the manifold $P\equiv M_4 \times \sootn$ 
can be written in terms of the linear connection and structure coefficients, such as 
the set $\acute{\Gamma}^i_{\ph{i}jk}$ described above, directly from 
equation~\ref{rabcd} as: 
\begin{equation} 
{\acute R}^i_{\ph{i}jkl} 
		  =   \acute{e}_k \acute{\Gamma}^i_{\ph{i}jl} - 
		  \acute{e}_l \acute{\Gamma}^i_{\ph{i}jk}
		      + \acute{\Gamma}^i_{\ph{i}mk} \acute{\Gamma}^m_{\ph{m}jl} - 
			    \acute{\Gamma}^i_{\ph{i}ml} \acute{\Gamma}^m_{\ph{m}jk} 
			              - \acute{c}^m_{\ph{m}kl} \acute{\Gamma}^i_{\ph{i}jm}  
\label{rijkl}
\end{equation}
   In the present theory we begin with the translational symmetry of the form $\lvte$ 
over the manifold $M_4$ with a flat Minkowski metric $g_{ac}(x) = \eta_{ac}$ and the 
canonical flat $\sootn$-valued connection $\omega$ on $P$. 
 As described in chapter~\ref{sym}, initially in equations~\ref{maca3} and 
\ref{fdefaaa},
 this 
 latter property means that the full curvature is zero $\hat{F}=0$, or in  components
  $\hat{F}^{\alpha}_{\ph{\alpha}ab}=0$. 
Hence, given that  $\acute{\Gamma}^{\alpha}_{\ph{\alpha}\beta\gamma} =
 \Gamma^{\alpha}_{\ph{\alpha}\beta\gamma}  =  0 $ and
  $\acute{\Gamma}^{a}_{\ph{a}bc} = \Gamma^{a}_{\ph{a}bc} = 0$, 
 \textit{all} the linear connection coefficients  in column `O+P \cite{OrzPau}' of 
table~\ref{Gamsetsr}
  are  zero, $\acute{\Gamma}^i_{\ph{i}jk} = 0$, and in turn all components of the 
Riemann curvature tensor of equation~\ref{rijkl}  vanish on the principle bundle 
manifold $P$.

 Here the natural absolute parallelism on $M_4$ and $G$ has been generalised to a 
natural parallelism on $P \equiv M_4 \times G$ with $\acute{\Gamma}^i_{\ph{i}jk} = 0$ 
for \textit{all} coefficients of the linear connection in the horizontal lift basis. 
 In fact for the canonical zero full curvature
 $\hat{F}^{\alpha}_{\ph{\alpha}ab} =0$ on the principle bundle all five 
non-Levi-Civita choices for $\acute{\Gamma}^i_{\ph{i}jk}$ in table~\ref{Gamsetsr} lead 
via equation~\ref{rijkl} to $\acute{R}^i_{\ph{i}jkl} = 0$, which as the components of 
a tensor vanish for any frame field on $P$, expressed generally as:
\begin{equation}
 \check{R}^{i}_{\ph{i}jkl} = 0
  \label{rvani} 
\end{equation}
  On the other hand there are torsion components with $\check{T}^i_{\ph{i}jk} \neq 0$ 
and hence the torsion is finite,
 as it is for the case of a self-parallel frame composed out of left-invariant vector 
fields on a Lie group manifold $G$, a copy of which here forms part of the total 
parallel frame field on $P$, as described in section~\ref{olcop}.

   In this way the full zero gauge curvature $\hat{F} =0$ for $\hat{G}$ over $M_4$ has 
been translated into zero Riemannian curvature $\check{R}^{i}_{\ph{i}jkl} = 0$ on the 
bundle space $P$.  The question then remains regarding how this structure might 
provide the link
 through which the external gravitational field will relate to the internal gauge 
fields over the base space $M_4$ when the full symmetry is broken.

   On the principle bundle $P \equiv M_4 \times \sootn$ a trivialisation may be chosen 
such that the corresponding direct product basis for the tangent space {\it TP} is 
identical to the horizontal lift basis associated with the canonical flat connection 
$\omega$, which in turn is derived from the full symmetry group. 
 Such a trivialisation represents a gauge choice for which the section $\sigma_0$, 
depicted in figure~\ref{omgamonp}, on the principle bundle $P$ coincides with the 
submanifolds of the integrable horizontal subspaces of $\omega$ on $P$, and hence with
  gauge connection components  $\omega^{\alpha}_{{\ph{\alpha}}a}(x,g) = 0$ on the 
bundle space. The linear connection components are identical 
$\ddot{\Gamma}^i_{\ph{i}jk} = \acute{\Gamma}^i_{\ph{i}jk} = 0$ in the respective 
direct product and horizontal lift bases for this choice of gauge, describing the 
 absolute parallelism defined in the frame field adapted to this section on $P$.

\begin{figure}[htbp]  
\centering
\epsfxsize=13.5cm
\leavevmode
\epsffile[0 0 1357 676]{\gpath aPfig52e}
\caption{\setb Geometric objects $\check{\Gamma}$ and $\omega$ on the principle bundle 
$P$ in relation to the linear connection $\Gamma$ on the base space $M_4$ and  
Maurer-Cartan 1-form $\theta$ on the group manifold $\hat{G}$. }
\label{omgamonp}
\end{figure} 

  While the canonical flat connection $\omega = \pi^{\ast}_2 \theta$ on $P$ describes 
a unique horizontal subspace and the corresponding horizontal lift basis, a direct 
product basis may be defined in terms of \textit{any} section on the bundle. Indeed,
  more generally geometric objects over the base space $M_4$ may be described with 
respect to a choice of gauge on the bundle $P$, as for example determined by the 
section
$\sigma' = \sigma_0 g$, with $g(x) \inn \hat{G}=\sootn$, as also represented in 
figure~\ref{omgamonp}.
 The gauge connection components $\omega^{\alpha}_{\ph{\alpha}a}(x,g) \neq 0$ in the 
new trivialisation are such that the vectors of the horizontal lift basis $\{ 
\acute{e}_i \}$ are expressed as for equation~\ref{ehtoeb} with:
\begin{eqnarray}
   \acute{e}_a & = & \ddot{e}_a \, - \, \omega^{\alpha}_{\ph{\alpha}a} 
\ddot{e}_{\alpha}
      \qquad \;\; \mbox{with} \qquad 
	   \omega^{\alpha}_{\ph{\alpha}a} \neq 0  \label{eewep} \\
	 \mbox{while} \qquad   \lbrack \acute{e}_{a}, \acute{e}_{b} \rbrack & = &
	 -\hat{F}^{\alpha}_{\ph{\alpha}ab}\acute{e}_{\alpha}
	     \qquad\qquad \mbox{with} \qquad \!\!  \hat{F}^{\alpha}_{\ph{\alpha}ab} = 0 
\label{eefp}
\end{eqnarray} 
   since the full $\sootn$ zero curvature is a gauge independent structure.
   However the full group $\sootn$ does not act purely as an internal symmetry but is 
broken by the action of the subgroup $\soot \subset \sootn$ on the external tangent 
space $\TM_4$.
 While the choice of gauge $g(x) \inn \sootn$ 
remains arbitrary with respect to the full unbroken symmetry  it
 will affect the physics of the broken symmetry over $M_4$. 
 For the restricted set of internal SO(6) generators  
  the horizontal lift vectors extracted from equations~\ref{eewep} and \ref{eefp} have 
the properties:
\begin{eqnarray}
   \acute{\ul{e}}_a & = & \ddot{\ul{e}}_a \, - \, \ul{\omega}^{\alpha}_{\ph{\alpha}a} 
\ddot{\ul{e}}_{\alpha}
      \qquad \;\; \mbox{with} \qquad 
	   \ul{\omega}^{\alpha}_{\ph{\alpha}a} \neq 0  \label{eeweps} \\
	 \mbox{while} \qquad   \lbrack \acute{\ul{e}}_{a}, \acute{\ul{e}}_{b} \rbrack & = 
&
	 -\ul{F}^{\alpha}_{\ph{\alpha}ab}\acute{\ul{e}}_{\alpha}
	     \qquad\qquad\! \mbox{with} \qquad\!  \ul{F}^{\alpha}_{\ph{\alpha}ab} \neq 0 
\label{eefps}
\end{eqnarray} 
  Here the $\ul{\omega}^{\alpha}_{\ph{\alpha}a}$ are the components of an so(6)-valued 
connection 1-form, with the
    sums over $\alpha$  restricted to the SO(6) generators, resulting in a generally 
non-zero internal curvature $\ul{F}^{\alpha}_{\ph{\alpha}ab}$, as was
   demonstrated  in equation~\ref{fint} for the finite internal SO(2) curvature 
achieved for small $\sofi$ gauge transformations over $M_3$ for the model world of 
section~\ref{hdasb}.
 Here we are reproducing the symmetry breaking approach of section~\ref{hdasb} in the 
light of the principle bundle structure and Kaluza-Klein theories described in the 
previous two chapters.

  As well as the transformation of the gauge connection $\omega$ for a different 
choice of basis on $P$ the linear connection $\check{\Gamma}$ also transforms.
 For any change of frame $e_{i'} = e_i \, e^i_{\ph{i}i'}$ with $e^i_{\ph{i}i'} \inn 
\glmr$ 
  the transformation of a linear connection, displayed in equation~\ref{gammap}, can 
be written as:
\begin{equation}
   \Gamma^{i'}_{\ph{i}j'k'} = e^{i'}_{\ph{i}i}\, e^j_{\ph{j}j'} \, e^k_{\ph{k}k'} \, 
   \Gamma^i_{\ph{i}jk}
        \, + \, e^{i'}_{\ph{i}l} \, e_{k'} \, e^l_{\ph{l}j'}   \label{gammatran}
\end{equation}
  The gauge choice associated with the section $\sigma' = \sigma_0 g(x)$ on $P$ 
corresponds to 
  a transformation from the horizontal lift basis to an arbitrary direct product basis 
$\ddot{e}_{i'} = \acute{e}_i \, e^i_{\ph{i}i'}$ on a principle bundle, that is the 
reverse of equation~\ref{ehtoeb} or \ref{eewep}, and we have:
\begin{equation}
  \left( \begin{array}{cc}  \ddot{e}_{a'} & \ddot{e}_{\alpha'} \end{array} \right) \; 
= \;
  \left( \begin{array}{cc}  \acute{e}_{a} & \acute{e}_{\alpha} \end{array} \right) \,
  \left( \begin{array}{cc}  \delta^a_{\ph{a}a'} & 0  \\
                  \omega^{\alpha}_{\ph{\alpha}a'} & 
\delta^{\alpha}_{\ph{\alpha}\alpha'}
       \end{array} \right)  
\end{equation}
\begin{equation}
\mbox{and hence} \quad e^i_{\ph{i}i'} \, = \,
    \left( \begin{array}{cc}  \delta^a_{\ph{a}a'} & 0  \\
                  \omega^{\alpha}_{\ph{\alpha}a'} & 
\delta^{\alpha}_{\ph{\alpha}\alpha'}
       \end{array} \right)   \quad \mbox{with inverse} \quad
	     (e^{-1})^{i'}_{\ph{i}i} \, = \,
    \left( \begin{array}{cc}  \delta^{a'}_{\ph{a}a} & 0  \\
                  -\omega^{\alpha'}_{\ph{\alpha}a} & 
\delta^{\alpha'}_{\ph{\alpha}\alpha}
       \end{array} \right)
	   \label{eeinv}
\end{equation}
 
 As a consistency check the same transformation is applied to the full set of 
Levi-Civita connection coefficients in the horizontal lift basis 
$\acute{\Gamma}^{i}_{\ph{i}jk}$ as listed in the `Cho \cite{Cho}'  column of 
table~\ref{Gamsetsr} (as extracted from \cite{Cho} equation~(22)). The expressions for 
the $\ddot{\Gamma}^{i'}_{\ph{i}j'k'}$ components obtained in the direct product basis 
using equations~\ref{gammatran} and \ref{eeinv} is found to agree with the original 
reference (the $\bar{\Gamma}$ components in the notation of \cite{Cho} equation~(15)).

  The general aim of this approach is to use equations~\ref{rijkl} and \ref{rvani}, 
with $\check{R}^{i}_{\ph{i}jkl} = 0$ deriving from the full zero curvature $\hat{F} = 
0$, on the principle bundle $P \equiv M_4 \times \sootn$ as a mutual constraint on the 
form of the external and internal curvature that results from the symmetry breaking. 
Once the full symmetry is broken non-zero internal gauge curvature components 
$\ul{F}^{\alpha}_{\ph{\alpha}ab} \neq 0$ from equation~\ref{eefps} will be introduced 
quadratically into the  terms of equation~\ref{rijkl} via
 relations to the linear connection $\check{\Gamma}$ of the kind listed in  
table~\ref{Gamsetsr} (by adopting for example the coefficients of column \cite{OrzPau} 
as provisionally suggested above)
 and hence into correlation with the external curvature $R^a_{\ph{a}bcd} \neq 0$ on 
the base manifold as identified within the appropriate components of ${\check 
R}^i_{\ph{i}jkl} = 0$ in a suitable basis. 

  These structures emerge in the symmetry breaking as represented by the transition 
from figure~\ref{mtogmaphr}(a) to (b). The structure of figure~\ref{mtogmaphr}(a) 
implies that the total symmetry $\sootn$ of $\lvte$ is associated with a canonical 
flat gauge field $A = g^{\ast} \theta$ with full curvature $\hat{F}=0$, under which a 
correlation between the external curvature $\ol{\bR}$ and internal curvature $\ul{F}$ 
is implied in the symmetry breaking to the structure of figure~\ref{mtogmaphr}(b), in 
particular with the case of both 
$\ol{\bR}=0$ and $\ul{F}=0$ simultaneously possible.

 In this picture a non-zero external curvature $\ol{\bR} \neq 0$ on $M_4$ is absorbed 
under $\check{R}^i_{\ph{i}jkl}=0$ on the extended bundle space as the `buckling' of 
the  geometry of the base manifold is countered by a corresponding finite internal 
curvature $\und{F} \neq 0$. The external and internal curvature is hence generated in 
a necessarily mutually consistent way under the symmetry of $\lvte$ in a choice of 
$\sootn$ gauge  over the base space $M_4$.
The invariance of the zero Riemann tensor of equation~\ref{rvani} under a change of 
frame adapted to choice of section is analogous to the invariance of the action 
integral of equation~\ref{einhilf}, defined in terms of a scalar curvature 
$\check{R}$, under variations of the metric $\chg^{ij}$ of the kind described in 
section~\ref{olcop}.
 This motivates the \textit{conjecture} that this framework leads to a similar 
unification of the Einstein-Yang-Mills equations of motion, that is 
equations~\ref{einfiek}--\ref{yangmk}, as found for non-Abelian Kaluza-Klein theory 
but ultimately \textit{without} the need to \textit{postulate} a Lagrangian function, 
coupled with the variational principle for the corresponding action integral, to 
obtain these equations. 

 Resulting from the projection of the structure of figure~\ref{mtogmaphr}(a) over that 
of figure~\ref{mtogmaphr}(b), with a choice of an $\sootn$ gauge section over $M_4$  
for the former,
 two further bundle structures, associated with the latter figure, 
 may be identified and considered separately.
 The subgroup $\soot$ is  distinguished in that it acts on  tangent space vectors 
$\ol{\bv}_4 \inn \TM_4$ of the base manifold, as depicted in 
figure~\ref{mtogmaphr}(b),  and therefore is designated as an \textit{external} 
symmetry, with the residual $\sox$ acting
 on the remaining components of $\ul{\bv}_6 \subset \bv_{10}$ of the form $\lvte$ and 
constituting an \textit{internal} symmetry.
 This results in consideration of the complementary \textit{subbundles}
  $\ol{P} \equiv M_4 \times \soot$ 
  and  $\ul{P}\equiv M_4\times\sox$ which  effectively \textit{decouple} from each 
other as mathematical structures, although related through the correlated geometrical 
structures they support, as they are mutually \textit{carved out} of the initial 
unbroken bundle $P=M_4 \times \sootn$.

  Indeed it is the extraction of the subgroup  $\soot \subset \sootn$, with the action 
of $\soot$ identified as the external symmetry and absorbed into the local tangent 
space geometry on $M_4$, that \textit{breaks} the full $\sootn$ symmetry.
 The base space $M_4$ is naturally associated with the frame bundle $\FM_4$, which is 
itself a particular type of principle fibre bundle as described in 
section~\ref{riegeo}.
 The bundle space $\ol{P} \equiv M_4 \times \soot$, obtained as a \textit{restriction} 
of the $P \equiv M_4 \times \sootn$ bundle, can also be interpreted as a 
\textit{reduction} of the $\FM_4$ frame bundle. In turn 
  an $\soota$-connection on $\ol{P}$ may be \textit{extended} to a 
$\mbox{gl}(4,\rrr)$-connection on the frame bundle, together with the associated  
  tetrad $e^{\mu}_{\ph{\mu}a}(x)$ and metric $g_{\mu\nu}(x)$ fields on $M_4$, as 
familiar in the theory of general relativity and also described in 
section~\ref{riegeo}.

  As described towards the end of section~\ref{gcatep} 
 the $\soot$ symmetry can be treated by analogy with  an `internal' Yang-Mills gauge 
structure. Indeed, as described above for the full $\sootn$ symmetry,  quadratic terms 
in the external $\soot$  `gauge curvature' $\ol{F}^{\alpha}_{\ph{\alpha}ab}$  will 
appear in the third and fourth terms of equation~\ref{rijkl}
 (essentially  as described in $\cite{MaCh}$, which adopts the Levi-Civita connection 
on the bundle space, leading to equations (3.14) and (3.15) there). 
 However this same external geometry, from the action of $\soot$ on $\TM_4$,
  is  represented by the Riemannian curvature $R^{a}_{\ph{a}bcd}(x)$, which is 
\textit{also}  contained within the corresponding  $\check{R}^{i}_{\ph{i}jkl}$ 
components in a suitable basis (as also described in \cite{MaCh}). Hence the bundle 
$\ol{P}$  appears to incorporate a redundant description of the external geometry 
while lacking an explicit reference to the internal curvature.

 On the other hand the
  subbundle $\uP \equiv M_4 \times \sox$ is closely  related to both  the frame bundle 
$\FM_4$, upon  which the external $\soot$ geometry is expressed in terms of fields 
such as
 $g_{\mu\nu}(x)$,  as well as the structures of the internal $\sox$ geometry with the 
associated gauge field $Y^{\alpha}_{\ph{\alpha}\mu}(x)$ and curvature 
$\uF^{\alpha}_{\ph{\alpha}\mu\nu}(x)$ components constructed on $M_4$.
  Hence in principle all the necessary geometric structures
 for relating the external and internal curvature  can be identified on the bundle 
$\ul{P}$.

 Rather than dealing with a connection form $\omega$ over $M_4$ for the full $\hat{G} 
= \sootn$ symmetry it is precisely through the symmetry breaking action, with the 
degrees of freedom of the $\soot \subset \sootn$ subgroup part of the \textit{gauge} 
connection being converted into the freedom of a \textit{linear} connection on $M_4$, 
that the bundle space 
$\uP \equiv M_4 \times \sox$ emerges. This in turn implies that the structure of the 
zero curvature $\hat{F} = 0$ for the full canonical flat connection does 
 \textit{not} explicitly survive the symmetry breaking transition from 
figure~\ref{mtogmaphr}(a) to (b).

 This motivates the study of a unified framework on the space $\uP \equiv M_4 \times 
\sox$ considered from now as a principle bundle standing independently by itself, and 
not as subbundle `carved out' of a larger bundle space such as $P \equiv M_4 \times 
\sootn$.
  It remains then to 
 explicitly define the mathematical nature of the constraint between the internal 
$\sox$ curvature and external $\soot$ geometry in terms of the bundle $\uP$.

  Earlier in this section an absolute parallelism on the bundle $P \equiv M_4 \times 
\sootn$ was constructed in the horizontal lift basis with all 
$\acute{\Gamma}^i_{\ph{i}jk} = 0$, taken from the set of reference~\cite{OrzPau} 
listed in the third column of table~\ref{Gamsetsr} for the canonical zero full 
curvature $\hat{F} = 0$, implying the identity of equation~\ref{rvani}. Now, beginning 
directly on the bundle $\ul{P} = M_4 \times \sox$ in itself, the question arises 
concerning the possible definition of a linear connection on this space. Since there 
is a gauge connection (which now derives from the internal SO(6) symmetry and in 
general is \textit{not} flat) on $\ul{P}$ the horizontal lift basis may be employed, 
and in turn the natural metric structure
with components $\acute{g}_{ij}$  of equation~\ref{ggghlb} introduced.
  
  Hence it is \textit{possible} to define the unique Levi-Civita connection on this 
bundle, as described in section~\ref{lccop}, with the components of 
equation~\ref{hgamijk} as listed for the horizontal lift basis in the first column  of 
table~\ref{Gamsetsr} under `Cho \cite{Cho}'. However in the present theory at no stage 
is $\ul{P}$ considered to be a \textit{physical} space or spacetime structure, hence 
neither the metric $\acute{g}_{ij}$ nor a linear connection 
$\acute{\Gamma}^i_{\ph{i}jk}$ on $\ul{P}$ have a physical geometric meaning, as they 
do on the base space $M_4$. Hence the unique metric-compatible torsion-free 
Levi-Civita connection is not here considered to be a natural structure on the bundle 
space as it is for the base manifold, and an alternative argument for the form of 
$\check{\Gamma}$ on $\ul{P}$ is sought.

  In particular 
  the linear connection on $\ul{P}$ is expected to be closely associated with the  
linear connection $\Gamma$ on the base space $M_4$, which does describe a physical 
geometry. Since this is a gl$(4,\rrr)$-valued 1-form $\Gamma(x) = \Gamma^{a}_{\ph{a} 
bc} E^{b}_{\ph{b} a} e^c$ on $M_4$, with respect to the distinguished horizontal lift 
basis on $\ul{P}$
 the components   $\acute{\Gamma}^{a}_{\ph{a} bc}$ and  $\acute{\Gamma}^{a}_{\ph{a} 
b\gamma}$ alone may be favoured for a linear connection $\Gamma$ on $M_4$ in some 
sense lifted onto $\uP$, and hence  
the only non-trivial coefficients of  $\acute{\Gamma}^i_{\ph{i}jk}$  on $\uP$ might be 
taken to be:
\begin{equation}
 \label{gamggfp}
   \acute{\Gamma}^{a}_{\ph{a} bc} \; =  \; {\Gamma}^{a}_{\ph{a} bc}
   \qquad \mbox{and} \qquad
   \acute{\Gamma}^{a}_{\ph{a}b\gamma} \; =  \;
         \gamma \,  g^{ac}g_{\beta\gamma}F^{\beta}_{\ph{\beta}bc}
\end{equation}   
 as listed as the `minimal' set in the final column of table~\ref{Gamsetsr}. The form 
of $\acute{\Gamma}^{a}_{\ph{a}b\gamma}$ in the equation above and the third row of the 
table as adopted from the other models in the table, consistent with the  requirement 
that $\acute{\Gamma}$ should transform in a  gauge covariant manner on $\ul{P}$ as 
appropriate for any object relating to a physical entity on the base manifold $M_4$.
 As will be described below this proposal will amount to a minimal structure on $\uP$ 
linking the present theory with Kaluza-Klein theory with a manifest correlation 
between the external Riemannian geometry and internal gauge curvature.

 While the Levi-Civita connection, $\Gamma = f(g)$ of equation~\ref{gtoGam},  on $M_4$ 
provides a unique description of the geometry on the base manifold in terms of the 
metric $g_{\mu\nu}(x)$, the linear connection $\acute{\Gamma}$ of 
equation~\ref{gamggfp} represents an attempt to extend this structure onto $\uP$ while 
maintaining the character of the connection $\Gamma$ on $M_4$, concerning in 
particular the $\glfra$-valued property. 
 However any linear connection on $\ul{P}$ is intrinsically a 
$\mbox{gl}(m,\rrr)$-valued 1-form (where $m = 4+15$ for the internal SO(6) gauge 
group). For example under the transformation to a direct product basis, as described 
in equations~\ref{gammatran}--\ref{eeinv}, the components of equation~\ref{gamggfp} in 
general give rise to linear connection coefficients 
$\ddot{\Gamma}^{\alpha}_{\ph{\alpha} b c} \neq 0$ and 
$\ddot{\Gamma}^{\alpha}_{\ph{\alpha}b\gamma} \neq 0$  in addition to 
$\ddot{\Gamma}^{a}_{\ph{a} bc} \neq 0$ and $\ddot{\Gamma}^{a}_{\ph{a}b\gamma} \neq 0$.
  Since the character of being $\mbox{gl}(4,\rrr)$-valued cannot be upheld for a 
linear connection on $\uP$ an alternative proposal, and one for which  parallel 
transport in the  horizontal and vertical directions on $\uP$ more directly reflects 
the geometry of the base manifold $M_4$, will be considered.

  A direct way to obtain a linear connection $\check{\Gamma}(p)$ on $\ul{P}$ closely 
related to $\Gamma(x)$ on $M_4$ would be to define $\check{\Gamma} = \pi^{\ast}\Gamma$ 
as the pull-back of the $\glfra$-valued 1-form $\Gamma$ through the bundle projection 
$\pi:\uP \to M_4$, by analogy with the identification of the canonical Lie 
algebra-valued 1-form $\omega = \pi^{\ast}_2 \theta$ as the pull-back of the 
Maurer-Cartan 1-form $\theta$ through the projection map 
$\pi_2: P \equiv M_4 \times \hat{G} \to \hat{G}$ for the full bundle $P$ as described 
in figure~\ref{omgamonp}. Indeed, the $\glfra$-valued linear connection $\Gamma$ on  
$M_4$, associated with the external symmetry, and $L(\hat{G})$-valued 1-form $\theta$ 
on $\hat{G}$, associated with the full symmetry, each describe the parallelism on 
their respective manifolds.

 The canonical flat connection $\omega = \pi^{\ast}_2 \theta$ on $P \equiv M_4 \times 
\sootn$ itself is an unambiguous geometric object, completely independent of any 
particular choice of section or gauge over the base manifold. It derives purely from 
the properties of $\theta$ on the gauge group $\hat{G}=\sootn$. Similarly, a linear 
connection $\check{\Gamma} = \pi^{\ast}\Gamma $ on $P$ or $\uP$ has no physical 
significance in itself other than that derived from its relation to a linear 
connection $\Gamma$, and the related Riemannian geometry, on the base manifold $M_4$.
 For the case $\check{\Gamma}(p) = \pi^{\ast}\Gamma(x)$ on $\uP$ and for any vector 
field $X(p) \inn \mbox{\it T}\uP$
 we have:
 \begin{equation}
   \langle \check{\Gamma},X \rangle_p \; = \; 
   \langle \pi^{\ast}\Gamma,X \rangle_p \; = \;
   \langle  \Gamma,\pi_{\ast} X \rangle_x
 \end{equation}
  with $X(p)$ projected in the final line onto the vector $\pi_{\ast} X \inn \TM_4$ 
 and with $\pi: p\inn \uP \to x \inn M_4$. Hence for any vector in the vertical 
subspace $X(p) \inn \mbox{\it V}\uP$ we have $\langle \check{\Gamma},X \rangle_p = 0$ 
since $\pi_{\ast} X = 0$.
This structure is related to the linear connection on the bundle described for the 
case of Kalinowski \cite{Kalin}  in section~\ref{olcop} for which all tangent vectors 
are mapped onto their horizontal parts, again with the property $\langle 
\check{\Gamma},X \rangle_p = 0$ for any vertical vector $X$, and hence again with 
emphasis on the horizontal structure, which in turn is closely associated with the 
geometry of the base space $M_4$.
In fact consideration of all cases collected in table~\ref{Gamsetsr}  leads to the 
following proposal for the properties of  $\check{\Gamma}$ on $\uP$ appropriate for 
the present theory:

\begin{itemize}
 
 \item[a)]  parallel propagation via $\check{\Gamma}$ in the vertical directions is 
taken to be trivial in the manner of  \cite{Kalin} in the fourth column of 
table~\ref{Gamsetsr}.

 \item[b)] parallel propagation via $\check{\Gamma}$ in the horizontal directions on 
$\uP$ is taken to relate to the contours of the gauge curvature over the base space 
$M_4$, following \cite{Katan} in the fifth column of table~\ref{Gamsetsr} for the case 
$c=1$.

 \item[c)]  with a view to deriving physical equations on the base space $M_4$
  compatibility with gauge covariance should be observed, as emphasised in 
\cite{OrzPau}.
 
 \item[d)]  consistent with c) the linear connection may be compatible with the 
natural, but non-physical, metric $\check{g}_{ij}$ of equation~\ref{ggghlb}, although 
the torsion may be arbitrary as initially emphasised in~\cite{Kopcz}.
 
 \item[e)] the bundle $\uP$ serves as an arena to relate the external and internal 
symmetry structures compatible with the simultaneous possibility of $\ol{\bR} = 0$ and 
$\ul{F} = 0$, as derived from consideration of figure~\ref{mtogmaphr} for the present 
theory.

\end{itemize} 
 
  Based on these observations and the broader discussion of Kaluza-Klein theory in 
chapter~\ref{kktheory} the conjectured linear connection components on $\uP$, as 
extracted from table~\ref{Gamsetsr}, can be summarised as:
\begin{equation}
 \label{gamsetkk}
   4)\; \acute{\Gamma}^{a}_{\ph{a}\gamma b} = 
g^{ac}g_{\gamma\beta}F^{\beta}_{\ph{\beta}bc},
   \qquad
   5)\; \acute{\Gamma}^{\alpha}_{\ph{\alpha}ab} = F^{\alpha}_{\ph{\alpha}ab},
   \qquad
   6)\; \acute{\Gamma}^a_{\ph{a}bc} = \Gamabc
\end{equation}
 with all other $\acute{\Gamma}^i_{\ph{i}jk} = 0$. Hence these are essentially the set 
of 
\cite{Kalin} in the fourth column of table~\ref{Gamsetsr} with $\lambda = 2$, with the 
motivation for employing this latter value derived from the geometrical argument in 
\cite{Katan}. This latter argument also has the benefit of fixing the geometry of 
$\check{\Gamma}$ without any reference to the Levi-Civita connection on $\uP$.

  The whole purpose of constructing a linear connection $\check{\Gamma}$ on $\uP$, as 
described above, is to provide a means through which a correlation between the 
external and internal curvature may be explicitly described. On the spacetime manifold 
$M_4$ any relationship between the external geometry, expressed in terms of the 
Einstein tensor with components $G_{\mu\nu}(x)$, and the internal geometry, expressed 
in terms of the gauge curvature with components 
$\ul{F}^{\alpha}_{\ph{\alpha}\mu\nu}(x)$, must transform covariantly both under 
general coordinate transformations and under gauge transformations, as described in 
particular in section~\ref{gcatep}. One technique for obtaining such a relation is to 
first identify a scalar `Lagrangian' function which has these invariance properties, 
as described in section~\ref{subfal}. This approach, again following the Kaluza-Klein 
theories, will be adopted provisionally here, although a more direct geometric 
argument leading to equation~\ref{einfiek}--\ref{einym},  which itself has the desired 
symmetry properties, would ultimately be preferred. 
(Since in the following components such as $F^{\alpha}_{\ph{\alpha}\mu\nu}(x)$ will 
always refer to the purely internal gauge curvature we now omit the underscore  for 
these objects).

While earlier in this section the Riemannian curvature ${\acute R}^i_{\ph{i}jkl}$ was 
constructed on the full bundle $P \equiv M_4 \times \sootn$ we are now focusing on the 
bundle $\uP \equiv M_4 \times \sox$, upon which the gauge curvature is generally 
finite.
  For any linear connection on the bundle space $\uP$, such as defined by any of the 
six sets of connection coefficients $\acute{\Gamma}^i_{\ph{i}jk}$ listed in 
table~\ref{Gamsetsr}, the Riemann curvature tensor can be determined according to 
equation~\ref{rijkl}, which is specified  in the horizontal lift basis. The 
corresponding
 Ricci  curvature components $\acute{R}_{\alpha\beta}$ and $\acute{R}_{ac}$ are listed 
here in the first and fourth rows of table~\ref{Rsetsr} for the six familiar examples.
In all cases the entries in this table calculated here agree with the corresponding 
equations of the given references -- within the sign conventions such as that of 
equation~\ref{riccicon} and as alluded to near the opening of chapter~\ref{kktheory}.

\begin{table}[htbp]
\centering
 \hspace*{-19pt}
\begin{tabular}{|r||c|c|c|c|c|}
 \hline
     & Cho~\cite{Cho} 
  &	Kop~\cite{Kopcz}  &   O$+$P~\cite{OrzPau} 
    & Kal~\cite{Kalin}  & $\!\!$Kat \cite{Katan}/min$\!\!$ 
	  \\  
 \hline				
 $\acute{R}_{\alpha\beta} = R_{(G)\alpha\beta} \, + \!$  &
 $-\frac{1}{4}F_{\alpha bd}F_{\beta}^{\ph{\beta}bd}$      &
  0 & $-\frac{1}{4}F_{\alpha bd}F_{\beta}^{\ph{\beta}bd}$  & 0 & 0 \\ 
 \hline
  $ R_{(G)\alpha\beta} = \!$  &
   $\frac{1}{4}K_{\alpha\beta}$ &  $\!(\alpha - \alpha^2) K_{\alpha\beta}\!$
    & 0 & 0 & 0 \\ 
  \hline
  $K^{\alpha\beta}\acute{R}_{\alpha\beta} = \!$  & 
   $R_G - \frac{1}{4}F^2$  &  $R_G  $ &
      $- \frac{1}{4}F^2$ & 0 & 0 \\
 \hline
  $\acute{R}_{ac} = R_{(M)ac} \, + \!$  & 
   $\fh F^{\beta}_{\ph{\beta}ad}F_{\beta c}^{\ph{\beta c}d}$ & 0 &
      $\fh F^{\beta}_{\ph{\beta}ad}F_{\beta c}^{\ph{\beta c}d}$  &
	  $\frac{\lambda^2}{4} F^{\beta}_{\ph{\beta}ad}F_{\beta c}^{\ph{\beta c}d}$
	  & $c^2 F^{\beta}_{\ph{\beta}ad}F_{\beta c}^{\ph{\beta c}d}$
     \\
\hline
   $g^{ac}\acute{R}_{ac} = R_M\, + \!$   & 
   $\fh F^2$ & 0  &
   $\fh F^2$ & $\frac{\lambda^2}{4}F^2$ &
   $c^2 F^2$  \\
 \hline
  $\acute{R} = \acute{g}^{ij}\acute{R}_{ij} = \!$   &
     $\! R_{\! M} \! + \! R_{\! G} \! + \! \frac{1}{4}F^2 \!$  & $R_M + R_G$   &
	  $R_M+\frac{1}{4}F^2$ &  $R_M+\frac{\lambda^2}{4}F^2$  &
	    $R_M+c^2F^2$   
  \\
	   \hline 
  \end{tabular}
  \caption{\setb Composition of the scalar curvature $\acute{R}$ on the bundle space 
for the six cases of table~\ref{Gamsetsr}. Contributions to the components of the 
Ricci curvature on the bundle include 
 $R_{(G)\alpha\beta}$ and $R_{(M)ac}$ from the group manifold and base space 
respectively,
 with  
 $R_G = K^{\alpha\beta}R_{(G)\alpha\beta}$ and $R_M = g^{ac}R_{(M)ac}$ being the 
respective scalar curvatures.
 The results for the sixth, `minimal', case in table~\ref{Gamsetsr} are identical to 
those listed for  \protect\cite{Katan} in the final column above with $c^2 = \gamma$.}
\label{Rsetsr}
\end{table} 

   The scalar curvature constructed in the horizontal lift basis on the principle  
bundle space can be written as:
\begin{equation}
   \acute{R}\, = \, \acute{g}^{ij}\acute{R}_{ij} \,  = \,
   g^{ac}\acute{R}_{ac} \, + \, K^{\alpha\beta}\acute{R}_{\alpha\beta} 
    \label{rtwobits}
\end{equation}
  owing to the simple form of the metric $\acute{g}$ in this basis as expressed in 
equation~\ref{ggghlb}. Hence the Ricci curvature components $\acute{R}_{a\alpha}$ and 
$\acute{R}_{\alpha a}$ are not required in order to determine the scalar curvature on 
the bundle.
  
  If each of the four factors of $\fh$ in the `Cho~\cite{Cho}' column in 
table~\ref{Gamsetsr}, 
   for the case of Levi-Civita connection coefficients $\acute{\Gamma}^i_{\ph{i}jk}$ 
on the bundle, listed in rows 1), 3), 4) and 5) are replaced by the real factors 
$f_1$, $f_3$, $f_4$ and $f_5$ respectively then the scalar curvature in the horizontal 
lift basis is found to be:
\begin{eqnarray}
  \label{rrrffff}
   \acute{R} & = & R_M \, + \, R_G \, + \, \chi F^2  \\
   \mbox{with} \quad \chi & = &  f_3 - f_3f_4 + f_4f_5 - f_3f_5   \label{rrrffff2}
\end{eqnarray}
  This expression agrees with the scalar curvature for the Levi-Civita case, with each 
$f_i = \fh$, as quoted originally in equation~\ref{hrscal}, and with each subsequent 
case of table~\ref{Gamsetsr} as quoted in the final row of table~\ref{Rsetsr}. 
Equations~\ref{rrrffff} 
 and \ref{rrrffff2} show that $f_3$ is the only coefficient which is sufficient in 
itself to introduce a non-trivial $F^2$ term, alongside $R_M$, into the scalar 
curvature $\acute{R}$, and this observation in part motivated the consideration of 
this simplest set of   $\acute{\Gamma}^i_{\ph{i}jk}$ coefficients, as listed in the 
`minimal' column of table~\ref{Gamsetsr}
and described in equation~\ref{gamggfp} above. While perhaps not developed as a 
serious physical proposal this minimal model further demonstrates the flexibility 
within the Kaluza-Klein framework, obtaining the appropriate link between the external 
geometry and internal curvature with a seemingly much simpler linear connection on the 
bundle compared with the Levi-Civita case.  More generally, equations~\ref{rrrffff} 
and \ref{rrrffff2} display the mutual consequences of the non-zero 
$\acute{\Gamma}^i_{\ph{i}jk}$ terms for the models listed in table~\ref{Gamsetsr}.

   Since $\acute{R}(p)$ is a scalar field  on the bundle at any given point $p$ it 
takes the same value in any local frame. Hence for example in a direct product basis, 
corresponding to a section $\sigma$ on $\uP$, the scalar value is simply $\ddot{R}(p) 
= \acute{R}(p)$. Further, since each of the scalar terms in the bottom line of 
table~\ref{Rsetsr} is gauge invariant, a corresponding scalar function on the base 
space $M_4$ may be deduced as:
\begin{equation}
  \label{ronxbase}
   \tilde{R}(x) \, = \, \sigma^{\ast}\ddot{R}(p) \, = \, R_M + R_G + \chi F^2
\end{equation}
   which is equivalent to $\acute{R}(p)$ for any $p\inn \uP$ such that $\pi(p)=x \inn 
M_4$.  
 Hence $\tilde{R}(x)$ is a real scalar function on $M_4$ which contains information 
about both the external and internal geometry, is invariant both under coordinate and 
gauge transformations on the base space, and therefore makes a suitable `Lagrangian' 
candidate on $M_4$.
Whether or not $R_G$ vanishes  and the real value $\chi$ in equation~\ref{ronxbase} 
depend upon the particular model, as can be seen for the examples of 
table~\ref{Rsetsr} and via equation~\ref{rrrffff2} respectively.
 For the case of most interest for the present theory, with non-zero linear connection 
coefficients listed in equation~\ref{gamsetkk}, corresponding to setting $\lambda =2$ 
in the `Kal~\cite{Kalin}' columns of tables~\ref{Gamsetsr} and \ref{Rsetsr},
  we have simply $\tilde{R}(x)  = R_M +  F^2$.

    The starting point for the Kaluza-Klein theories reviewed in sections~\ref{lccop} 
and \ref{olcop} is the mathematical structure of a principle fibre bundle $P = (M_4, 
G)$, such as described in section~\ref{fibre} and pictured in figure~\ref{pbundle}. 
This structure features an extended base space $M_4$ over which a gauge connection may 
be introduced on the bundle space $P$ transforming under the internal symmetry gauge 
group $G$. In these theories the bundle space is typically interpreted as a 
higher-dimensional \textit{physical} spacetime. For example in reference 
(\cite{OrzPau} p.190) the authors write: `Our general attitude is to regard the $n_G$ 
vertical dimensions as physically real, and hence the vertical Einstein equations as 
true dynamical equations of the $(n+n_G)$-theory.'

  A similar perspective is generally adopted for the theories with homogeneous fibres, 
described in section~\ref{thwhf}, in this case for the bundle space $E=(M_4,S_k)$. In 
the introduction of reference~\cite{CLLM} the authors write: `Kaluza-Klein theories 
are theories in which the gravitational potential together with the gauge potentials 
of various interactions are interpreted as manifestations of (pseudo-) Riemannian 
structure of the Universe which is $4+k$ dimensional.' The analogy between  coordinate 
transformations in general relativity and  gauge transformations in gauge theory, 
discussed in section~\ref{gcatep}, is more explicitly realised in these theories as 
demonstrated for example in equations~\ref{coogutr}.

 In Kaluza-Klein theories restrictions on the form of the metric $\check{g}_{ij}$ on 
the  higher-dimensional space, in particular a necessary conformity with 
equation~\ref{rtrang}, induce a  `dimensional reduction' or `spontaneous 
compactification'  of the larger space. The latter is then interpreted as a bundle 
structure with fibres, corresponding to the $n_G$-dimensional gauge group $G$ or an 
associated $k$-dimensional homogeneous space $S_k$, over the smaller $n=4$-dimensional 
spacetime $M_4$.

 The origin of the bundle structure in Kaluza-Klein theories hence contrasts sharply 
with that for the present theory. Here the geometric structure $M \times G$ arises out 
of the symmetries of a general form of temporal flow $\lv$ as described in 
chapter~\ref{sym}. In particular for the 10-dimensional form $\lvte$, considered in 
this section and employed in figure~\ref{mtogmaphr}, the base space $M_4$ arises out 
of a parametrisation of a 4-dimensional subset of the `translational' degrees of 
freedom of the components $\bv_{10}$ under $\lvte$, with gauge fields  drawn over the 
base space out of the `rotational' degrees of freedom of the same temporal form.

Here the only \textit{physical} space is the manifold $M_4$, providing the arena for 
general relativity in a 4-dimensional spacetime, with no `compactification' from a 
higher-dimensional extended spacetime required. The spacetime geometry on $M_4$ 
derives from the local Minkowski  metric $\eta_{ab}$ implicit in the 4-dimensional 
temporal form of equation~\ref{flow4d},
 now written $\lvfh$ in the projection out of the higher-dimensional form $\lvte$. 
 On the other hand the Killing metric $g_{\alpha\beta} = K_{\alpha\beta}$ does not 
describe the geometry of a physical space, either on the group manifold $G$ or bundle 
space $\uP$. It 
  relates the Lie algebra adjoint and coadjoint representations as usual, with for 
example $F_{\alpha ab} = g_{\alpha \beta} F^{\beta}_{\ph{\beta}ab}$, and it may be 
employed as a mathematical structure on $\uP$ in the derivation of scalar quantities,
 as for example in equation~\ref{rtwobits}.

  The  roles of the metric $g_{\mu\nu}(x)$ and gauge field 
$Y^{\alpha}_{\ph{\alpha}\mu}(x)$ in the laws of physics on $M_4$ are well defined. 
When lifted to the principle bundle $\ul{P}$ these objects can be augmented by the 
Killing metric $g_{\alpha\beta}$ on $G$ to define a metric $\chg_{ij}$ in the form of 
equation~\ref{gmetug} on the bundle space.
 This latter metric could be employed on $\ul{P}$, for example to construct a
 curvature scalar $\check{R}$ from the Riemann tensor $\check{R}^i_{\ph{i}jkl}$ based 
on a 
 Levi-Civita connection $\check{\Gamma}^i_{\ph{i}jk}$, but no \textit{physical} 
significance should be attached to the geometric connotations of the metric 
$\chg^{ij}$ introduced in this way. 

  Indeed $\chg^{ij}$, as described in equation~\ref{ggghlb}, consists of an unnatural 
marriage with the external local metric $\eta_{ab}$ originating \textit{within} the 
form of $\lvte$ \textit{upon} which the group $G$, with Killing metric 
$g_{\alpha\beta}$, acts.
 This is the case whether the group describes full symmetry $\hat{G}=\sootn$, as 
considered earlier in this section, or the internal symmetry $G=\sox$ as considered 
here.
Such  a hybrid metric $\chg_{ij}$, composed of parts of quite different character, 
hence seems an unnatural object to endow with a physical geometric meaning.
 Hence here  the construction of a Levi-Civita connection on the bundle space as 
described in subsection~\ref{lccop} is not well motivated, with the bundle not 
considered as representing an extension of general relativity to a higher-dimensional 
space.
On the other hand with this unifying framework taking the shape of a principle fibre 
bundle over the base space the present theory is naturally related to  Kaluza-Klein 
theories, in particular those of the kind reviewed in section~\ref{olcop}.

 While the structure of these Kaluza-Klein theories rests on a deliberate 
\textit{extension} of the formalism of general relativity into a space with extra 
dimensions, in the present theory the construction of a linear connection on the 
bundle space is motivated rather as a mathematical means to relate the physical 
Riemannian curvature on the base space $M_4$ to that of the internal gauge fields. 
 Indeed it is still possible to define a linear connection $\check{\Gamma}$ on the 
bundle
  which is 
  closely associated with both the
  linear connection $\Gamabc$ on the base space $M_4$ and the internal gauge curvature 
$F^{\alpha}_{\ph{\alpha}ab}$, however only the
   linear connection $\Gamma$ on $M_4$ has a significance in terms of describing  a 
physical space. 

  As for other branches of this theory, including its connections with the Standard 
Model, quantum theory and cosmology to be presented subsequently in this paper, the 
aim is to develop the theory naturally out of the basic conceptual ideas. Here it is 
the basic geometric structures relating to the symmetries of $\lv$, in particular in 
the symmetry breaking  over the $M_4$ base manifold
 pictured in figure~\ref{mtogmaphr}, that provides the unified framework for the 
external and internal curvature. 
 The resulting geometric structure, exemplified here by the principle bundle $\uP 
\equiv M_4 \times \sox$, while not forming a physical spacetime itself, provides the 
mathematical arena for a unification of the external and internal geometry arising out 
of the breaking of the full $\lvte$ symmetry over the base space $M_4$.

 The general form of the relation between the external Riemannian geometry $\bR$ and 
internal gauge curvature $F$ is conjectured to arise naturally in this framework, in a 
generally and gauge covariant manner, essentially taking the form of 
equation~\ref{einfiek}--\ref{einym}. This relation is provisionally derived here via 
the scalar function $\tilde{R}(x)$ of equation~\ref{ronxbase}, interpreted as a 
geometric perturbation to the Einstein-Hilbert action on the base space $M_4$ arising 
from the higher-dimensional form of temporal flow $\lvte$. In particular, from the 
range of models studied, with linear connection coefficients 
$\acute{\Gamma}^i_{\ph{i}jk}$ on the bundle listed in table~\ref{Gamsetsr} and the 
corresponding scalar curvature $\acute{R}$ determined in table~\ref{Rsetsr}, the 
argument outlined in points `a) -- e)' earlier in this section leads to the proposed 
set of equation~\ref{gamsetkk}. This argument  focuses on the horizontal transport in 
$\uP$ skirting over the base manifold $M_4$, and in appealing in particular to 
references~\cite{Kalin} and \cite{Katan} meets half-way with Kaluza-Klein theory.
 Further progress might be made for example by placing more complete emphasis on point 
`b)' with a full set of $\acute{\Gamma}^i_{\ph{i}jk}$ coefficients defined in terms of 
the parallel transport associated with the internal gauge curvature as described for 
figure~\ref{paratran}.

  In standard Kaluza-Klein theory the action $A_m$ for the scalar curvature 
$\check{R}$ defined on the bundle space $P$ in equation~\ref{einhilf} reduces to the 
4-dimensional action integral $A_4$ of equation~\ref{einhilk} owing to the trivial 
integration over the fibre degrees of freedom.  The point of view adopted here is that 
the scalar field
  $\tilde{R}(x) = R_M + \chi F^2$ of equation~\ref{ronxbase} (with $R_G=0$ and $\chi = 
1$ for the model of equation~\ref{gamsetkk} constructed here) is defined 
\textit{directly} on the base space $M_4$ itself. In turn the action integral is 
defined directly on the base space as:
\begin{equation}
   \tilde{I} = \int  (R_M + \chi F^2)    \sqrt{\vert g \vert}\; d^4 x  \label{teinhil}   
\end{equation} 
  as a coordinate and gauge invariant expression with all fields defined on $M_4$. As 
denoted by the `tilde' on $\tilde{I}$ this function is considered as a perturbation of 
the Einstein-Hilbert action for the vacuum case,  equation~\ref{einhil} with $\alpha = 
1$, $\Lambda = 0$ and $\lag = 0$, which was described in the opening of 
section~\ref{subfal}. That is, equation~\ref{teinhil} incorporates the perturbation  
to the scalar curvature $R_M(x) \to \tilde{R}(x)$ on the base space $M_4$. The full 
Einstein-Hilbert action of equation~\ref{einhil} can be written:
\begin{equation}
   I = \int (\alpha R_M  + \lag)   \sqrt{\vert g \vert}\; d^4 x  \label{einhilnol}   
\end{equation}
  where the cosmological constant $\Lambda$ has been dropped in correspondence with 
the lack of a finite $R_G$ term in equation~\ref{teinhil}.  Further comparison between 
the above two equations shows that equation~\ref{teinhil} describes a perturbation to 
general relativity equivalent to the introduction of a Lagrangian term $\lag = +\alpha 
\chi F^2$ in the original Einstein-Hilbert action.
  While the mathematical conclusion is identical to Kaluza-Klein theory, here the 
interpretation involves a more minimal impact on the arena of general relativity in 
4-dimensional spacetime, namely \textit{without} a physical augmentation
  into a higher-dimensional extended spacetime.

  The choice of $\chi=\frac{1}{4}$ and $\alpha=\frac{-1}{16\pi G_{\! N}}$ respectively 
in the two equations above represents the standard normalisation for the incorporation 
of gauge fields into the Einstein-Hilbert action, as described in 
section~\ref{subfal}. This standard action if also discussed in
  (\cite{Pen} section 20.6) where the shortcomings of the Lagrangian approach are 
highlighted. The intention of the present theory is ultimately to avoid any direct 
reference to the Lagrangian formalism entirely. For the present case the form of 
$\tilde{R}=R_M + \chi F^2$ in equation~\ref{teinhil}, in deriving from 
equation~\ref{ronxbase}, arises from the geometry on the bundle $\uP = M_4 \times 
\sox$ in a physically meaningful way in terms of entities on the base space $M_4$. 
This structure can be considered as a perturbation to general relativity deriving from 
the need to take into account the internal space of the form $\lvte$ and the geometric 
structures entailed.

  If the 4-dimensional form $\lvf$ of equation~\ref{flow4d} alone is considered no 
symmetry breaking is involved in the identification of the bundle $P \equiv M_4 \times 
\soot$ out of the symmetries of this form. As described in section~\ref{perc}, in the 
context of the SO(3) model, this structure incorporates a \textit{canonical} flat 
connection with zero curvature, that is $R^{\rho}_{\ph{\rho}\sigma \mu\nu}(x)=0$, 
\textit{without} any reference to a Lagrangian. This result is however identical to 
that achieved in equation~\ref{lagtoein}  for the vacuum case using the stationarity 
of the Einstein-Hilbert action under variation of the metric field $g_{\mu\nu}(x)$ on 
$M_4$; since $R^{\rho}_{\ph{\rho}\sigma \mu\nu}(x)=0$
 if $G^{\mu\nu}(x)$ vanishes everywhere in spacetime.
 Hence the conjecture here is that a perturbation to this Einstein-Hilbert action, in 
the form of equation~\ref{teinhil},    carries with it the consequences for the 
Riemannian geometry on $M_4$ that follow from an embedding in the structures of a 
larger form of temporal flow such as $\lvte$.

 Here the provisional adoption of a `Lagrangian function' has a direct conceptual 
motivation. This is unlike for example the case of the Standard Model Lagrangian for 
particle physics, elements of which will be reviewed in section~\ref{ewtatsm}, for 
which both the fields and Lagrangian terms are generally introduced and contrived by 
hand with the aim of achieving the desired equations of motion and particle 
interactions for the known phenomena of high energy physics. The means of bypassing 
the Standard Model Lagrangian for the present theory will then be described in 
subsequent chapters, while the avoidance of a necessary Lagrangian to derive classical 
equations of motion will be considered further here in the following section.

  Within this caveat for the employment of a Lagrangian approach, the equation of 
motion obtained by requiring 
  $\delta \tilde{I}=0$ for equation~\ref{teinhil}, under variations $\delta 
g_{\mu\nu}(x)$ of the metric on $M_4$, follows the derivation of 
equation~\ref{einfiek}--\ref{einym} and can be written here as:  
\begin{equation}
 \label{gchift}
  G^{\mu\nu} \: = \: 
  2\chi(  - F^{\alpha \mu}_{\ph{\alpha\mu} \rho}F_{\alpha}^{\ph{\alpha}\rho\nu}
	                -\frac{1}{4} g^{\mu\nu} \, F^{\alpha}_{\ph{\alpha}\rho 
\sigma}F_{\alpha}^{\ph{\alpha}\rho\sigma})   \: =: \:  -\kappa T^{\mu\nu}   
\end{equation}
   At the purely theoretical level the factor of $\chi$ in this equation arises 
directly  in equations~\ref{rrrffff} and \ref{rrrffff2}, which in turn derive from  
the relation of linear connection $\acute{\Gamma}^i_{\ph{i}jk}$ on the bundle 
 to the gauge curvature $F^{\alpha}_{\ph{\alpha}ab}$ as
listed in the columns of table~\ref{Gamsetsr}. For the present theory
 the correlation between  the external and internal geometry in the breaking of the 
full form $\lvh$
 over the base space $M_4$ has been considered provisionally in terms of the set of 
linear connection coefficients of equation~\ref{gamsetkk}, and hence with $\chi = 1$.

 With gravitational and gauge field phenomena historically studied independently in 
practice the normalisation factor connecting the left-hand side and central 
expressions of equation~\ref{gchift} is a matter for empirical convention, as for the 
factor of $\kappa= \frac{8\pi G_{\! N}}{c^4}$ on the right-hand side of this equation. 
 Here for normalisation in practice we shall set $\chi = \frac{\kappa}{2}$ implying a 
choice of physical units such that the energy-momentum tensor can be expressed 
directly in terms of the  gauge curvature, as will be the case for the electromagnetic 
field tensor $F_{\mu\nu}$ in the following section (see for example 
equation~\ref{tffgff}).

  Equation~\ref{gchift} reduces to the vacuum solution $G^{\mu\nu}(x)=0$ for the case 
in which curvature of the internal gauge field vanishes $F^{\alpha }_{\ph{\alpha} 
\mu\nu}=0$. More generally, with the Einstein tensor $G^{\mu\nu} = R^{\mu\nu} - 
\frac{1}{2}R_Mg^{\mu\nu}$, contracting the equation~\ref{gchift} with $g_{\mu\nu}$ 
leads to the conclusion  $R_M = 0$,
 the standard vanishing of the scalar curvature associated with a classical gauge 
field, while the Ricci curvature is generally finite with $R^{\mu\nu} = G^{\mu\nu} 
\neq 0$.

 Hence while for a general solution we have $\tilde{R} = \chi F^2 \neq 0$, the full 
expression $\tilde{R} =R_M +  \chi F^2 \neq 0$ is needed in equation~\ref{teinhil} in 
order to derive the field equation~\ref{gchift} through the method of variation. A 
similar observation applies for the vacuum equations of general relativity, namely the 
derivation of equation~\ref{lagtoein}, and further suggests that the Lagrangian 
approach may not be entirely satisfactory. Ideally the aim here would be to derive 
equation~\ref{gchift} purely by geometrical means and without reference to a 
Lagrangian. In the meantime, by further considering $\delta \tilde{I}=0$ for the 
action in equation~\ref{teinhil}, now with respect to variation in the gauge fields
 $\delta Y^{\alpha}_{\ph{\alpha}\mu}(x)$, leads, as described earlier for 
equation~\ref{yangmk}, to the Yang-Mills vacuum equation:
\begin{equation}
	D_{\mu}F^{\alpha \,\mu\nu} \: = \:  0    \label{delyym}
\end{equation}  
  For the case of an Abelian internal $\uo$ symmetry this relation expresses Maxwell's 
equation for a source-free electromagnetic field.

  While the unification has been described here in terms of the principle bundle space 
$\uP \equiv M_4 \times \sox$, for the broken group symmetry action, a bundle of 
homogeneous fibres
 $E \equiv M_4 \times S_k$ might also be constructed, with fibres composed of the 
purely internal  $\ul{\bv}_6$ components of $\lvte$, complementary to the projection 
onto the external spacetime  with
 $\ol{\bv}_4 \inn \TM_4$ as pictured in figure~\ref{mtogmaphr}(b). 
  A  \textit{transitive} action of $\sootn$ on the space underlying $\lvte$
  can be identified, as for the action of $\sox$ on the internal space which hence 
forms the
  homogeneous space $S_k$ employed for the fibres. Since these actions are also 
\textit{effective} the complete internal gauge symmetry dynamics will be represented 
for the theory formulated in terms of a bundle with homogeneous fibres, rather than 
the principle fibre bundle, as was reviewed in section~\ref{thwhf}.

 In the models of section~\ref{thwhf} the internal group $G$ can be considered as a 
global  \textit{isometry}, that is a symmetry preserving a metric $g_{\ralpha\rbeta}$ 
on $S_k$, with $H \subset G$ as the \textit{isotropy} subgroup leaving any point $y_0 
\inn S_k$ fixed. By contrast for the present theory $\hat{G}=\sootn$ can be considered 
as an \textit{isochronal} symmetry preserving the temporal form $\lvte$  with $\ol{H} 
= \soot \subset \hat{G}$ as the local \textit{isometry} subgroup preserving the metric 
on $\TM_4$, while the complementary $\ul{H} = \sox \subset \sootn$ leaves any vector 
$\ol{\bv}_4 \inn \TM_4$ fixed.  
The bundle structures on $E \equiv M_4 \times S_k$ may ultimately shed further light 
on the derivation of equation~\ref{gchift} together with the theoretical value of 
$\chi$.

 While a consistent and rigorous \textit{mathematical} framework needs to be 
established a full understanding of the appropriate \textit{conceptual} picture for 
the extraction of the geometry on the base manifold derived from, and breaking, the 
symmetries of the full form $\lvh$ is also required. It is out of the marriage of 
these mathematical and conceptual ideas that an ultimate form for the relationship 
between the external Riemannian  curvature $\bR$ and internal gauge curvature $F$ on 
the base space $M_4$ might be arrived at.
 This section has described the evolution of ideas arising out of the symmetries of 
$\lv$ described in chapter~\ref{sym}, steered by the structures of differential 
geometry and Kaluza-Klein theory as described in chapters~\ref{rogaeom} and 
\ref{kktheory}, aiming towards such a unification. Attempting to justify all the steps 
along the way, via the linear connection on the bundle of equation~\ref{gamsetkk}, 
scalar function on the base space of equation~\ref{ronxbase} (with $R_G = 0$) and 
action integral of equation~\ref{teinhil}, the aim has been to arrive provisionally at 
the relation of equation~\ref{gchift} with minimal assumptions.
 This equation shows how  a relation between the external and internal curvature might 
be achieved in the present theory with non-zero values 
for $\bR\neq 0$ and $F \neq 0$ closely correlated. The possibility of deriving 
equation~\ref{gchift} via purely geometric means without any reference to a Lagrangian 
formulation remains as a conjecture of the theory.

  It should be further noted that only classical fields have been considered so far 
and it may be that, given the symmetry of the classical picture described originally 
in figure~\ref{spillout}, a quantum field description of the theory will be required 
to provide the mechanism through which non-flat structures ultimately arises on the 
base manifold in general. This in turn relates to the concept of `many solutions' for 
the geometry $G^{\mu\nu}(x)$ on the base space as will be described in 
chapter~\ref{newapp}. 
In the meantime, given the Kaluza-Klein relation of equation~\ref{gchift} itself, 
 a number of further  equations of motion
 may be deduced without the need for a Lagrangian formalism. Hence  these consequences 
are conjectured also to apply in the present theory, as we  review in the following 
section.



\section{Equations of Motion for Fields and Matter}
  \label{subwal}

   In standard field theory the Lagrangian, being a scalar, provides a means to 
introduce arbitrary, although generally empirically motivated, symmetries into the 
theory with such symmetries generally preserved in the resulting equations of motion, 
as reviewed in section~\ref{subfal}.
 In the Lagrangian approach the compatibility of the equations of motion with 
 energy-momentum conservation $\pal_{\mu}T^{\mu\nu} = 0$ is ensured through the 
Euler-Lagrange equation if the energy-momentum tensor is defined according to 
equation~\ref{tmnneo},
 as an application of Noether's theorem.
 
  In the present theory  equation~\ref{gchift} 
   emerges out of the constraint of the simple form $\lvh$ projected over the base 
space $M_4$, in principle without the need for a Lagrangian formalism, as described in 
the previous section for a model based on the form $\lvte$.
 The new theory avoids the ambiguity inherent in the choice of a scalar Lagrangian 
function and  replaces the need to impose the principle of extremal action with a firm 
conceptual grounding in the physical manifestation of the full  form of temporal flow 
$\lvh$ and its symmetries. 
 Hence in contrast to the Lagrangian approach
   here  we \textit{begin} with  $T^{\mu\nu}_{\pht ;\mu} = 0$ as a direct consequence 
of the definition of energy-momentum as $T^{\mu\nu} := G^{\mu\nu}$, within a 
conventional normalisation factor in relations such as equation~\ref{gchift}, together 
with the contracted Bianchi identity $\gmo$. In the limit of vanishingly small 
spacetime curvature, with a linear connection
 $\Gamma \to 0$ in a suitable choice of coordinates,
 this constraint can be written as 
  $\nabla_{\mu}T^{\mu\nu}=0 \to \pal_{\mu}T^{\mu\nu} = 0$
  and interpreted as energy-momentum conservation.
 The question then regards the extent to which this constraint determines the 
equations of motion, both in a curved spacetime and in the limit of flat Minkowski 
spacetime, for the entities which apparently compose $T^{\mu\nu}$, without appealing 
to a Lagrangian structure.

  This  also contrasts with a more standard approach to general relativity, reviewed 
in section~\ref{gcatep}, in which the Einstein tensor $G^{\mu\nu}$ is first equated 
with a generic energy-momentum tensor, $G^{\mu\nu} = -\kappa T^{\mu\nu}$ in 
equation~\ref{Eins}, via a normalisation constant $\kappa$. In the meantime various 
examples  of possible forms $T^{\mu\nu}$ may be postulated, or deduced from a 
Lagrangian method, for example for the energy-momentum of a perfect fluid or an 
electromagnetic field, again with appropriate normalisation factors. Only 
\textit{then} are the Einstein tensor and the chosen form for $T^{\mu\nu}$ linked 
together via equation~\ref{Eins}. This standard  approach distances the relation 
between the external curvature $G^{\mu\nu}$ and internal curvature 
$F^{\alpha\,\mu\nu}$ by the insertion of the apparently \textit{mediating} object 
$T^{\mu\nu}$, which may be considered to act as a `source' for the gravitational 
field. It is this structure that motivates the form of equations~\ref{einfiek} and 
\ref{einym}. One of the main reasons for considering $T^{\mu\nu}$ to \textit{be} the 
source term in the Einstein equation is that material phenomena (such as the 
properties of  everyday tables and chairs) are generally more readily observable than 
their counterparts in the warping of the spacetime geometry, particularly within the 
local laboratory environment.

 In the present theory the more intimate relation of equation~\ref{gchift} arises 
\textit{directly} from the basic conceptual ideas of the theory, as described in the 
previous section, with the symmetry groups of both the external and internal geometry 
mutually related through the unifying symmetry of the full form $\lvh$. The motivation 
for the right-hand side of equation~\ref{gchift} to subsequently be interpreted as an 
energy-momentum tensor corresponding to $G^{\mu\nu}$ will be found in the empirical 
usefulness of such a concept. This will be more apparent when `quantum effects' are 
introduced and augment the possible forms of $T^{\mu\nu}$ beyond that of continuous 
classical fields, as we alluded to at the end of the previous section.

   Here, beginning from the unified point of view for classical fields, the external 
and internal curvatures appear on a similar footing in equation~\ref{gchift}, with the 
contracted Riemann curvature on the left-hand side \textit{equated identically} with 
terms quadratic in the internal curvature in the central expression. The great 
difference in the relative strengths of the respective physical forces encountered 
empirically in nature will later need to be accounted for through the respective 
interactions and couplings of the fields to be identified in the theory. These will 
give rise to a variety of laboratory phenomena and will lead to normalisation factors 
replacing $\chi$ in relations such as equation~\ref{gchift} once practical units are 
employed for measured quantities.
 While the bare mathematical relations are needed to understand the theoretical basis 
of the unification, for a discussion of the empirical consequences here we set $\chi = 
\frac{\kappa}{2}$ as suggested following equation~\ref{gchift} in the previous 
section.

   The tensor $T^{\mu\nu}$ is composed of effective macroscopic quantities or as a 
function of fundamental fields, to be determined in the theory, which in turn mutually 
constrains the form of $G^{\mu\nu}$. Here the initial aim will be to demonstrate the 
extent to which the equations of motion for both  external gravitational and internal 
gauge fields are implied within the unifying form of equation~\ref{gchift}.

  First  we consider the classical  field for the  
	 particular case of $\uo$ as the internal symmetry,
	 that is the case of electromagnetism.
	 In terms of the components $F_{\mu\nu} = \partial_{\mu}A_{\nu} - \partial_{\nu} 
A_{\mu}$  of the electromagnetic field tensor
  the components of the Einstein tensor $G^{\mu\nu}$ of equation~\ref{gchift}, with a 
single generator for the internal group, can be written as:
\begin{equation}
   -\frac{1}{\kappa} G^{\mu\nu}   =  F^{\mu}_{\ph{\mu}\rho}F^{\rho\nu}
	        + \frac{1}{4} g^{\mu\nu} \, F_{\rho\sigma}F^{\rho\sigma}   \label{geinm}    
\end{equation}	
 Hence through this equation direct contact is made between gravitation in the form of 
the geometric curvature of spacetime and the familiar laboratory phenomena of the 
electromagnetic field. The fact that powerful electromagnetic effects may be observed 
for which the associated gravitational field is immeasurably small is an indication of 
the need to explain the origin of laboratory normalisation units, as mentioned above.

    Given the tetrad field components $\tetu$ of a local orthonormal frame field 
$\{e_a(x)\}$ the components of the electromagnetic curvature tensor in a local Lorentz 
frame
$F_{ab} = e^{\mu}_{\ph{\mu}a}e^{\nu}_{\ph{\nu}b} F_{\mu\nu}$ may be written out as  
the $4 \times 4$ asymmetric  matrix:
\begin{equation}
   \label{febmat}
   [F]_{ab} = \left( \begin{array}{cccc}
              0 & E_1  &  E_2 & E_3  \\
		   -E_1 & 0    & -B_3 & B_2  \\
		   -E_2 & B_3  &  0   & -B_1 \\
		   -E_3 & -B_2 &  B_1 &  0    \end{array} \right).
\end{equation}
  This is also the conventional form for the electromagnetic field tensor defined 
globally for the flat Minkowski spacetime of special relativity. The special symbols 
$E_i$ and $B_i$ ($i = 1,2,3$ with $E_i = F_{0i} = F(e_0, e_i)$ and 
$-\varepsilon_{ijk}B_k = F_{ij} = F(e_i, e_j)$) for the six independent components of 
the electromagnetic curvature 2-form $F$ in a particular Lorentz frame $\{e_a\}$ 
represent the electric and magnetic fields respectively. These six components 
transform non-trivially under external Lorentz transformations but are trivially 
unchanged under an internal $g(x) \inn \uo$ gauge transformation, 
equation~\ref{ftogfg}, since $g^{-1}F g = F$ for an Abelian group.

  Historically it was realised that Maxwell's equations~\ref{maxo} and \ref{maxj} 
exhibit a $\uo$ symmetry  before an understanding of gauge theories had been 
developed, although it was not considered to be a fundamental physical symmetry of 
nature since it is not a spacetime symmetry. However in the present theory fundamental 
symmetries are not of spacetime (in any dimension) but of multi-dimensional forms of 
temporal flow expressed as $\lv$. These include \textit{both} the familiar 
4-dimensional spacetime symmetry associated with perception on an extended manifold 
$M_4$ and equally the gauge symmetry groups, including the $\uo$ of electromagnetism 
that arises here as will be described in section~\ref{intsym}. 
Here both external and internal symmetries, together with their respective physical 
phenomena, originate naturally from the fundamental concepts of the theory.

	 In  an approximately Minkowskian spacetime  the electromagnetic field $F_{ab}$ 
may  be defined and measured operationally by observing the motion of a body of mass 
$m$ and charge $q$ in the field and using the Lorentz force law of 
equation~\ref{rellor}.
   In that equation $F^b_{\ph{b}c} = \eta^{ba}F_{ac}$ is a mixed index form of the 
electromagnetic curvature tensor. The metric is needed to define this tensor, as it is 
for 
 $F^{cd} = \eta^{ca}\eta^{db}F_{ab}$ and hence in turn to define the `Hodge dual' of 
the electromagnetic curvature tensor:
\begin{equation} 
   \label{hodged}
        \past F_{ab}  =  \frac{1}{2}\varepsilon_{abcd}F^{cd}  
		\quad\; \mbox{with} \quad\;
        [\past F]_{ab}  =  \left( \begin{array}{cccc}
                 0 & -B_1 & -B_2 & -B_3  \\
	    	   B_1 & 0    & -E_3 & E_2  \\
		       B_2 & E_3  &  0   & -E_1 \\
		       B_3 & -E_2 &  E_1 &  0    \end{array} \right).
\end{equation}
   In Minkowski spacetime $\varepsilon_{abcd} = \varepsilon_{[abcd]}$ are the 
components of the completely antisymmetric rank-4 tensor $\varepsilon \equiv e^0 
\wedge e^1 \wedge e^2 \wedge e^3$, with $\varepsilon_{0123} = \varepsilon (e_0, e_1, 
e_2, e_3) = +1$ implying the choice of right-handed orientation for the orthonormal 
basis $\{e_a\}$, while the cotensor components are simply 
 $\varepsilon^{abcd} = - \varepsilon_{abcd}$.
 In a general coordinate system, including the case of a curved spacetime, the metric 
volume form $\omega$  with components:
 \begin{eqnarray}
   \omega_{abcd} & = & \sqrt{\vert g \vert} \; \varepsilon_{abcd} \\
   \omega^{abcd} & = &  (1 / \sqrt{\vert g \vert}) \; \varepsilon^{abcd}
  \end{eqnarray}
  where $g(x)$ is the determinant of the metric $g_{\mu\nu}(x)$, is employed for the 
Hodge dual operator of equation~\ref{hodged} since $\varepsilon$, unlike $\omega$, 
does not transform as a tensor under general coordinate transformations.   
   The Levi-Civita symbol $\varepsilon_{abcd}$ is equivalent to the components of the 
volume form $\omega$ in Minkowski spacetime with global coordinates employed such that 
the metric $g_{\mu\nu}(x) = \delta^{a}_{\ph{a}{\mu}}\delta^{b}_{\ph{b}{\nu}}\eta_{ab}$ 
everywhere.

  In general on an $n$-dimensional manifold the space of $p$-forms has the same number 
of degrees of freedom as the space of $(n-p)$-forms with a canonical isomorphism 
between the two sets given by the metric volume form $\omega$. The isomorphism map is 
the Hodge dual of a form which contains precisely the same information reorganised 
into the components of the dual form. For example the map from $F$ in 
equation~\ref{febmat} to $\past F$ in equation~\ref{hodged} corresponds to a 
rearrangement of matrix components with $(E_i, B_j) \to (-B_i, E_j)$.

   The Einstein tensor $G^{\mu\nu} = R^{\mu\nu} - \frac{1}{2} R g^{\mu\nu}$
 is the `trace-reversed' Ricci tensor, it   
    can also be defined as the contraction (\cite{MTW} p.325):
\begin{eqnarray}
    G^{\beta}_{\ph{\beta}\gamma} & := & \Gbar^{\tau\beta}_{\ph{\tau\beta}\gamma\tau}
   \nonumber \\
  \mbox{with} \qquad  \Gbar^{\alpha\beta}_{\ph{\alpha\beta}\gamma\delta} & := &
    \frac{1}{2} \omega^{\alpha\beta\rho\sigma} \, 
R_{\rho\sigma}^{\ph{\rho\sigma}\mu\nu} \,
	\frac{1}{2} \omega_{\mu\nu\gamma\delta}  \nonumber
\end{eqnarray}
  and in this sense is `dual' to the Ricci tensor $R^{\mu\nu}$.  The tensor $\bGbar$ 
carries exactly the same information, and possesses the same rank-4 tensor symmetries, 
as the Riemann tensor $\bR$ and hence also has 20 independent components. It is 
analogous to the dual tensor ${}^{\ast}\! F$ for the electromagnetic curvature tensor 
$F$.

    The electromagnetic energy-momentum tensor identified with
	$T^{\mu\nu} := -\frac{1}{\kappa}G^{\mu\nu}$ for equation~\ref{geinm}, as guided by 
the Kaluza-Klein framework, is identical to that obtained in equation~\ref{temkk} in 
the Lagrangian formalism since effectively the same matter Lagrangian $\lag \sim F^2$ 
is introduced in both cases, via equations~\ref{teinhil} and \ref{lagem} respectively.
  This expression can also be written in an equivalent but more symmetric form 
(\cite{Pen}~p.456):
\begin{eqnarray}
   \label{geinms}
    T^{\mu\nu} & = & \frac{1}{2}(F^{\mu}_{\ph{\mu}\rho}F^{\rho\nu} \: + \:
	              \past F^{\mu}_{\ph{\mu}\rho}\past F^{\rho\nu})  \\ 
			   & = &	  
			    F^{\mu}_{\ph{\mu}\rho}F^{\rho\nu}
	        \; + \; \frac{1}{4} g^{\mu\nu} \, F_{\rho\sigma}F^{\rho\sigma} 
\label{tffgff}
\end{eqnarray}

   From either of these equations the energy density of the electromagnetic field is 
found to be $T^{00} = \frac{1}{2}(\bE^2 + \bB^2)$, as originally expressed by Maxwell. 
There are two Lorentz invariants of the electromagnetic field, the scalar norm 
$\frac{1}{2}F_{\mu\nu}F^{\mu\nu}= - (\bE^2 - \bB^2)$ and the pseudo-scalar 
$\frac{1}{4}F_{\mu\nu}\past F^{\mu\nu}= \bE \ccdot \bB$, although expressions of the 
latter kind (composing $F_{\mu\nu}$ with its dual) do not feature in $T^{\mu\nu}$. 
Both of these quantities are functions on the spacetime manifold which locally take 
the same value in any Lorentz frame and are also invariant under (orientation 
preserving) general coordinate transformations.

  The energy-momentum tensor for the electromagnetic field is also traceless, 
$T^{\mu}_{\ph{\mu}\mu}=0$, from which the trace of the Einstein equation implies that 
the scalar curvature vanishes, $R = 0$, and hence in this case the Einstein equation 
can be written $G^{\mu\nu} = R^{\mu\nu} = -\kappa T^{\mu\nu}$, as described shortly 
after equation~\ref{gchift} in the previous section.  Hence in the Einstein-Maxwell 
theory while the Maxwell tensor $F_{\mu\nu}$ and its dual $\past F_{\mu\nu}$ appear in 
a \textit{symmetric} way in equation~\ref{geinms} the Einstein tensor $G^{\mu\nu}$ is 
\textit{identical} to its `dual' $R^{\mu\nu}$.

From this underlying theoretical point of view electromagnetism arises as a $\uo$ 
gauge theory with the electromagnetic field tensor being the \textit{exact} 2-form $F 
= \md A$ as defined in terms of the $\uo$ connection 1-form $A(x)$. Hence by the 
exterior algebra property $\md^2 = 0$ the curvature 2-form is in turn necessarily 
\textit{closed} $\md F = 0$ as an identity that gives immediately the homogeneous 
Maxwell equations summarised in equation~\ref{maxo}.

  With the electric current 1-form  \textit{defined} as $J := \past \md \past F$ (that 
is $\past J := \md \past F$ consistent with  the inhomogeneous Maxwell 
equation~\ref{maxj}) from the property $\md^2 = 0$ it also follows immediately that 
 $\md \md \past  F = 0$ and we also find the identity $\md \past J = 0$. In Minkowski 
spacetime this in turn  implies that $\partial_a J^{a} = 0$ corresponding to the 
conservation of electric charge expressed in terms of the components of the conserved 
current $J$ associated with the internal $\uo$ symmetry.
 This is very closely analogous to the fact that \textit{defining} the energy-momentum 
tensor to be $T^{\mu\nu} := G^{\mu\nu}$ leads immediately to the local conservation of 
energy-momentum $T^{\mu\nu}_{\ph{\mu\nu};\mu}=0$ via the contracted Bianchi identity 
for the Einstein tensor $G^{\mu\nu}$. Hence  Noether's theorem, based on a Lagrangian 
approach as described in section~\ref{subfal}, is not needed to identify either of 
these conserved quantities, which are both purely
 geometric in origin.

  It can be shown (\cite{MTW} p.472) that for the case $J=0$ the Einstein equation, in 
the form of equation~\ref{geinm}, mutually constrains the evolution of both the 
gravitational and electromagnetic field, with the latter usually expressed by the 
source-free Maxwell equation $\md \past F = 0$, that is equation~\ref{maxj} for $J=0$, 
as we review here. Applying the identity $\gmo$ to both sides of equation~\ref{geinm} 
gives:
\begin{eqnarray}
   0 & = & F^{\mu\tau}_{\pht ;\mu}F^{\pho \nu}_{\tau}
   + F^{\mu\tau} F^{\pho \nu}_{\tau \ph1 ;\mu} + \frac{1}{2}g^{\mu\nu}F_{\rho\sigma 
;\mu}F^{\rho\sigma} 
          \nonumber \\
   & = & F^{\pho \nu}_{\tau}F^{\mu\tau}_{\pht ;\mu} 
       + g^{\mu\nu} F^{\rho\sigma} F_{\sigma\mu ;\rho}
	    + \frac{1}{2}g^{\mu\nu}F_{\rho\sigma ;\mu}F^{\rho\sigma}  \nonumber \\
   & = & F^{\pho \nu}_{\tau}F^{\mu\tau}_{\pht ;\mu} +
   \frac{1}{2}g^{\mu\nu}F^{\rho\sigma} (F_{\sigma\mu ;\rho}+F_{\mu\rho 
;\sigma}+F_{\rho\sigma ;\mu})  \nonumber   \\ 
   & \Rightarrow &  F^{\pho \nu}_{\tau}F^{\mu\tau}_{\pht ;\mu} = 0 
     \label{trmaxw}
\end{eqnarray}
      The final term in the penultimate equation vanishes by the identity $\md F = 0$, 
that is the homogeneous Maxwell equation~\ref{maxo}, or $F_{[\sigma\mu;\rho]} = 0$ in 
components (again here `$;\mu$' is the covariant derivative with respect to the linear 
connection $\Gamma$ in a general curved spacetime). The remaining expression in the 
bottom line involves a linear combination of the four quantities $F^{\mu\tau}_{\pht 
;\mu}$. The determinant of the coefficients $F^{\pho\nu}_{\tau}$ is the Lorentz 
pseudo-scalar $\vert F^{\pho\nu}_{\tau} \vert = - (\bE \ccdot \bB)^2$ (\cite{MTW}  
p.472). For a general electromagnetic field this quantity is non-zero, except that it 
may vanish on hypersurfaces, and hence in general the source-free form of the Maxwell 
equation~\ref{maxj} does not need to be imposed, rather it may instead be deduced from 
the Einstein equation for the electromagnetic field that:
\begin{equation}
  \label{maxfein}
   F^{\mu\tau}_{\pht ;\mu} = 0
\end{equation}  
   
   On defining $J^{\tau} = F^{\mu\tau}_{\pht ;\mu}$ this result shows that vanishing 
current $J=0$ is implied for the relation of equation~\ref{geinm} under the Bianchi 
identity $\gmo$.
 For this vacuum case $J=0$ both the curvature $F$ and its dual $\past F$ satisfy a 
similar equation, $\md F = 0$ and $\md \past F =0$ respectively,  while for the 
external curvature there is a greater symmetry with $G^{\mu\nu}$ \textit{equal} to its 
`dual' $R^{\mu\nu}$, as described above.

   A similar argument may be followed for the non-Abelian case, beginning with 
equation~\ref{gchift} and following the sequence of expressions leading to 
equation~\ref{trmaxw} except with $F_{\mu\nu} \to F^{\alpha}_{\ph{\alpha}\mu\nu}$  and 
an extra contraction over the index $\alpha$, representing the group generators, for 
each quadratic term in the internal curvature.
  Sandwiched between the two complementary constraining identities for the external 
and internal curvature, that is the Bianchi identities $\gmo$ and $\mD F =0$ 
respectively,
this leads to the Yang-Mills equation $ D_{\mu}F^{\alpha \,\mu\nu} =  0 $, which was 
derived from a Lagrangian in equation~\ref{ymol},
 and includes self-interaction terms for the non-Abelian gauge field 
$Y^{\alpha}_{\ph{\alpha}\mu}(x)$.
 The same equation was 
   also derived as a consequence of Kaluza-Klein theory in equation~\ref{yangmk} from 
the stationarity of the action integral of equation~\ref{einhilf} on a principle 
bundle. 
  Generally for the non-Abelian case,
 as for the Abelian case of Maxwell's equations, a conserved current can be obtained 
in terms of a geometric identity.

 For the present theory the Maxwell and Yang-Mills equations are also proposed to 
arise
 through a purely geometric argument, similar to that described for 
equation~\ref{trmaxw},
  directly from the identity $\gmo$ as applied to equation~\ref{gchift}.
  This relation itself arose in equation~\ref{einfiek}-\ref{einym}  under the 
stationarity of an action integral in Kaluza-Klein theory, although in the previous  
section we described how equation~\ref{gchift} might be obtained ultimately in the 
present theory without any appeal to the Lagrangian formalism. Here 
equation~\ref{gchift} is considered to arise as a perturbation to the Einstein vacuum 
equations, derived for equation~\ref{lagtoein} in terms of the stationarity of the 
Einstein-Hilbert action under variations of the metric $\delta g_{\mu\nu}(x)$. 
Consistent with this approach the above discussion suggests that the variation of the 
gauge field $\delta Y^{\alpha}_{\ph{\alpha}\mu}(x)$ is \textit{not} needed in order to 
derive the vacuum Yang-Mills equation~\ref{delyym}; rather, as for general relativity, 
only the $\delta g_{\mu\nu}(x)$ variation is needed in order to derive 
equation~\ref{gchift}, which in turn itself implies the relation of 
equation~\ref{delyym} as a consequence of the geometric structure. With 
equation~\ref{gchift} itself conjectured to arise inevitably out of the geometric 
constraints implied in the breaking of the full $\lvh$ symmetry over $M_4$ any 
explicit reference to the Lagrangian formalism might be avoided entirely.

In the present framework  non-Abelian symmetries arise, as for the case of $\uo$ 
above, within the internal symmetry action on the full form $\lvh$.
 The symmetry breaking is pictured in figure~\ref{mtogmaphr} for the $\lvte$ model, 
for which 
 the internal symmetry is identified simply as $\sox$. Internal symmetries deriving 
from yet higher-dimensional forms of $\lvh$ will be considered in 
chapters~\ref{chapesb} and \ref{secfd}.

 Returning to the Abelian case of electromagnetism, 
  more generally for $J\neq 0$, in applying to the Maxwell tensor $F$ and not to the 
dual tensor $\past F$ the Bianchi identity $\md F = 0$ introduces a clear break in the 
mathematical symmetry between these two tensors. This in turn is directly associated 
with the empirical asymmetry between the observed roles of the electric and magnetic 
fields. The field components $(E_i, B_j)$ are oriented within the Maxwell tensor in 
equation~\ref{febmat} such that they are distinguished by the particular properties 
that $\nabla \ccdot \bB = 0$ while $\nabla \ccdot \bE = \sigma$,  where $\sigma$ is 
the charge density for the case of static fields .
  (From the historical empirical point of view the asymmetry between the expressions 
for $\md F$ in equation~\ref{maxo} and $\md \past F$ in equation~\ref{maxj} is a 
physical observation in the sense it `might have been' observed that $\md F = \past 
J_M$  with a `magnetic monopole current' $J_M$, however empirically such a current has 
never been seen.)

   Here we next consider how equations of motion describing the broad macroscopic 
properties of matter arise. The microscopic details of fields and quantum physics 
which underlie these properties need not be considered in any detail here. Rather the 
general freedom inherent in the Einstein equation, beyond a specific form such as 
equation~\ref{geinm}, will be opened up to a more general structure $G^{\mu\nu} = 
-\kappa T^{\mu\nu}_{\!\epsilon}$, where $\epsilon$ here denotes an effective  
energy-momentum tensor describing coarse macroscopic phenomena. This macroscopic form 
of $T^{\mu\nu}_{\!\epsilon}$
 will include terms for the effective flow of physical matter, either charged or 
uncharged, as well as for the original electromagnetic field, all combinations of 
which will be collectively subject to   $T^{\mu\nu}_{\!\epsilon\, \pho ;\mu} = 0$ 
through the Einstein equation.

 Under the symmetry transformations of a higher-dimensional form of temporal flow 
$\lvh$ the projection over the base manifold $M_4$, as described in the previous 
section, leads to a relation between classical external and internal fields 
culminating in a relation of the form of equation~\ref{gchift}, which may be written:
\begin{equation} 
  G^{\mu\nu} = f(Y)
  \label{gety}
\end{equation}

    The identity  $G^{\mu\nu}_{\ph{\mu\nu};\mu} = 0$ then leads to constraints on the 
equations of motion for the internal gauge fields $Y(x)$, that is the 
Yang-Mills-Maxwell equations, as described above.
A particular form for the energy-momentum tensor is identified as 
$T^{\mu\nu}:= -\frac{1}{\kappa}G^{\mu\nu}$,  that is via the Einstein equation.

   So far we have considered only the case in which $G^{\mu\nu}(x)$ is equated with a 
function of the curvature $F^{\alpha}_{\ph{\alpha}\mu\nu}(x)$, in turn derived from a 
classical continuous  gauge field $Y^{\alpha}_{\ph{\alpha}\mu}(x)$, in the form of 
equation~\ref{gety},
 which  exhibits a relatively even significance for the external gravitational field 
on the left-hand side and the internal gauge field on the right-hand side.
 This structure was motivated to obtain $G^{\mu\nu}$ on the left-hand side of 
equation~\ref{gchift} corresponding to a  global continuous \textit{external} linear 
connection field $\Gamma(x)$ as required to define a geometric perceptual arena on the 
base manifold as described in section~\ref{perc}.

   More generally a continuous \textit{internal} gauge field $Y(x)$ is only a local 
requirement so long as the central expression of equation~\ref{gchift} can be modified 
in a manner compatible with the identity $G^{\mu\nu}_{\pht ;\mu} = 0$. 
With the components of the internal symmetry gauge fields 
$Y^{\alpha}_{\ph{\alpha}\mu}(x)$ coupled with the internal temporal components,  
through a relation of the form of  equation~\ref{dlvfib}, only the \textit{combined} 
effect is required to be compatible with the necessary smooth geometric structure on 
the left-hand side of equation~\ref{gety} and we can write:
\begin{equation}
  \label{getypsi}
 G^{\mu\nu} = f(Y,\hat{\bv})
\end{equation} 
  implying in turn a more flexible expression for the energy-momentum tensor
  $T^{\mu\nu}:= -\frac{1}{\kappa}G^{\mu\nu}$.
  This extra freedom, not tied to the constraint of a continuous internal gauge field 
on $M_4$, allows for field exchanges between the internal gauge connection $Y(x)$ and 
components of temporal flow $\hat{\bv}(x)$, which will be of the kind described in 
chapters~\ref{chapesb} and \ref{secfd}  for more realistic forms $\lvh$ in comparison 
with the observations of high energy physics experiments.
   The possibility of multiple solutions for $G^{\mu\nu}$ involving exchanges between 
the field values of $Y$ and $\bvh$ will be interpreted as quantum and particle 
phenomena  via the local indistinguishability of the field components, 
  as will be described chapter~\ref{newapp}.

   While equation~\ref{gety} might be expressed as $G^{\mu\nu} = -\kappa 
T^{\mu\nu}(Y)$
   the more general non-classical extension to equation~\ref{getypsi} can also be 
written as
   $G^{\mu\nu} = -\kappa T^{\mu\nu}(Y,\hat{\bv})$ with 
   the identification of the rank-2 tensor fields on either side of this expression  
remaining valid since both sides transform the same way and the contracted Bianchi 
identity will still apply to both. While $G^{\mu\nu}$ and $T^{\mu\nu}$ are identical 
in form they denote and possess a differing internal  compositions; while the 
right-hand side can be interpreted as a \textit{source} in terms of the fragmented 
temporal flow composed of apparent `\textit{matter} fields', $Y(x)$ and 
$\hat{\bv}(x)$,  the left-hand side represents the same mathematical object 
interpreted as the Einstein tensor for  a linear connection  describing the external 
\textit{geometry}, as required for perception.

 Equation~\ref{getypsi}  expresses the relation between the gravitational field 
described by the metric $g_{\mu\nu}(x)$ underlying $G^{\mu\nu}$ and the matter fields 
$Y_{\mu}(x)$ and $\hat{\bv}(x)$, together with the implicit interaction between these 
latter `microscopic' fields themselves. Alternatively the term `matter field' can 
refer to an effective macroscopic form for the energy-momentum tensor such as 
$T^{\mu\nu}_{\epsilon}$ averaging over the microscopic field interaction effects.
We begin by looking more generally at properties of the symmetric Einstein tensor 
$G^{\mu\nu}$ in terms of $T^{\mu\nu}_{\!\epsilon} := -\frac{1}{\kappa}G^{\mu\nu}$.
   A timelike eigenvector $\bu$ may be defined for the energy-momentum tensor such 
that (\cite{Synge} p.174):
\begin{equation}
  \label{ueigen}
    T_{\!\epsilon}^{\mu\nu}u_{\nu} = \rho u^{\mu}
\end{equation}
    with the vector field $\bu(x)$ normalised as $\vert \bu \vert = 
g_{\mu\nu}u^{\mu}u^{\nu}=u^{\mu}u_{\mu} = +1$ such that $\rho = 
T_{\!\epsilon}^{\mu\nu}u_{\mu}u_{\nu} \; (=\rho u^{\mu}u_{\mu})$ which will be 
identified as the effective `proper energy density' or mass density, effectively 
averaging over underlying microscopic field interactions. In the general case:
\begin{equation}
   \label{gruus}
    T_{\!\epsilon}^{\mu\nu} = \rho u^{\mu} u^{\nu} - S^{\mu\nu}
\end{equation} 
   defines the stress tensor $S^{\mu\nu}$ (\cite{Synge} p.175). This is a symmetric 
tensor with four constraints $S^{\mu\nu}u_{\nu} = 0$ (as can be seen by contracting 
equation~\ref{gruus} with $u_{\nu}$) and hence with six degrees of freedom.    The 
simplest example is that in which the effective energy-momentum tensor represents a 
pressureless perfect fluid (such as a dust cloud) with:
\begin{equation}
 \label{gtruu}
     T^{\mu\nu}_{\!\epsilon} = \rho u^{\mu} u^{\nu}. 
\end{equation} 
  In this case $G^{\mu\nu} = -\kappa\rho u^{\mu} u^{\nu}$ and 
   we have $g_{\mu\nu}G^{\mu\nu} = -\kappa g_{\mu\nu}\rho u^{\mu} u^{\nu} = 
-\kappa\rho$.
   With $G^{\mu\nu} = R^{\mu\nu}-\frac{1}{2}Rg^{\mu\nu}$ 
    this in turn implies  $R = +\kappa\rho$ with the matter density $\rho$ therefore 
directly associated with the spacetime scalar curvature $R$ and hence with 
gravitational effects.
 The sign convention of equation~\ref{riccicon}, with $G^{\mu\nu}=-\kappa T^{\mu\nu}$ 
and positive constant $\kappa$ determined in the Newtonian limit, is motivated in part 
by the resulting sign in the relation $R = +\kappa\rho$, that is such that 
\textit{positive} scalar curvature is associated with \textit{positive} matter 
density.

   Applying the contracted Bianchi identity $G^{\mu\nu}_{\pht ;\mu} = 0$ to the 
right-hand side of equation~\ref{gtruu} we then have (\cite{Synge} p.175):
\begin{eqnarray}
 T^{\mu\nu}_{\!\epsilon\, \pho ;\mu} = 0 \quad & \Rightarrow & \quad
         (\rho u^{\mu})_{;\mu}u^{\nu} + \rho u^{\mu}(u^{\nu})_{;\mu} = 0  \nonumber \\
 \Sigma_{\nu} (\times \,  u_{\nu})  \quad & \Rightarrow & \quad  (\rho u^{\mu})_{;\mu} 
= 0 \quad 
                                          \mbox{since} \;\; u_{\nu}(u^{\nu})_{;\mu} = 
0 \nonumber \\
 \mbox{hence}         & &     \rho u^{\mu}(u^{\nu})_{;\mu} = 0. \label{trgeod}
\end{eqnarray}
  Here the continuity equation $ (\rho u^{\mu})_{;\mu} = 0$, describing the 
conservation of mass-energy, in the second line is substituted back into the first 
line to deduce the expression in the final line. 
 From this we see that the form of equation~\ref{gtruu}, with $\rho \neq 0$, implies 
that $u^{\mu}(u^{\nu})_{;\mu} = 0$, that is the flow lines of the fluid are 
\textit{geodesics}. 
   Such a result could be derived from the simple Lagrangian of equation~\ref{extgeo}, 
with the requirement  $\delta L = 0$ under variation of the path implying 
equation~\ref{geotrau}.
However here in the case of a perfect fluid the geodesic law for the motion of bodies 
in general relativity is an inescapable consequence of the Einstein field equation and 
the Bianchi identity, which is a well-known result.

  More generally the effective energy-momentum tensor $T^{\mu\nu}_{\!\epsilon}$ can 
describe a perfect fluid with non-zero effective pressure $p$ in the form:
\begin{equation}
 \label{gtruup}
    -\frac{1}{\kappa}G^{\mu\nu} =: T^{\mu\nu}_{\!\epsilon} = (\rho + p) u^{\mu} 
u^{\nu} - p \, g^{\mu\nu}. 
\end{equation} 
  with, by comparison with equation~\ref{gruus}, $S^{\mu\nu} = p(g^{\mu\nu} - u^{\mu} 
u^{\nu})$ which satisfies $S^{\mu\nu} u_{\nu} = 0$.
   The material flow $\bu$ is again subject to $\vert \bu \vert = 1$  with $\rho$ and 
also now $p$ as effective macroscopic terms irrespective of the classical or quantum 
fields underlying this structure. 
   Again here the structure of matter perceived in spacetime is constrained by the 
geometrical properties of $G^{\mu\nu}$. Applying the Bianchi identity 
$G^{\mu\nu}_{\pht ;\mu}=0$ to the right-hand side of equation~\ref{gtruup}, similarly 
as above for equation~\ref{gtruu} leading to equation~\ref{trgeod}, we now find that 
in general $u^{\mu}(u^{\nu})_{;\mu}$ is non-zero and proportional to the pressure 
gradient (\cite{Synge} p.176), as a deviation from pure geodesic flow of the fluid due 
to the pressure term.

  Alternatively we may consider a pressureless fluid carrying charge, that is a fluid 
with energy density $\rho$ and also a charge density $\sigma$. 
Here we are dealing with continuous classical fields and bodies corresponding to the 
motions of macroscopic entities, where $T^{\mu\nu}_{\epsilon}$  may represent charged 
metal plates, wires and so on and $T^{\mu\nu}_{\mathrm{em}}$  describes a classical 
electromagnetic field, for example in a laboratory setting.
 For the original case with the classical electromagnetic gauge field only and $\tem 
:= -\frac{1}{\kappa} G^{\mu\nu}$ from equation~\ref{geinm} consistency with 
$G^{\mu\nu}_{\pht ;\mu}=0$ required that $J^{\nu} := F^{\mu\nu}_{\pht ;\mu} = 0$, as 
described for equation~\ref{maxfein}. It is then through the introduction of effective 
matter terms that the equations for the electromagnetic field allow for a charged 
current $J \neq 0$ in combination with energy-momentum in the form $\tef = \rho 
u^{\mu}u^{\nu}$, both of which are composed in terms of the effective matter content.

We have \textit{defined} $T^{\mu\nu} := G^{\mu\nu}$ and argued, following the previous 
section, that for an internal $\uo$ symmetry identified within the full symmetry of 
$\lvh$ this naturally leads to $\tem$ in the form of equation~\ref{geinm}. Similarly 
here with $J^{\nu} :=  F^{\mu\nu}_{\pht ;\mu}$ we would like to understand the form of 
$J$  that results as microscopic field transitions over $M_4$ are considered such that 
equation~\ref{geinm} breaks down giving:
\begin{equation}
 \label{gneqjneq}
  -\frac{1}{\kappa}G^{\mu\nu} = T^{\mu\nu}(Y,\bvh) \neq  
F^{\mu}_{\ph{\mu}\rho}F^{\rho\nu}
	        + \frac{1}{4} g^{\mu\nu} \, F_{\rho\sigma}F^{\rho\sigma}
    \quad \mbox{\underline{and}} \quad  J^{\nu} =  F^{\mu\nu}_{\pht ;\mu} \neq 0
\end{equation} 
   with a specific form for the first equation relating to a  specific form for the 
latter. In the phenomenological macroscopic limit the effective energy-momentum tensor 
$\tef = \rho u^{\mu}u^{\nu}$ arose as a possible form for a non-trivial  $G^{\mu\nu}$ 
field for the external spacetime geometry. With charge density defined by
 $\sigma := \nabla \ccdot \bE$ in the electrostatic limit, under a Lorentz 
transformation we may associate the 4-vector $J^{\nu} =\sigma u^{\nu}$ with a charged 
body, such that $\sigma = u_{\nu}J^{\nu}$ is closely analogous to $\rho = 
u_{\mu}u_{\nu}\tef$ for the matter density of a pressureless fluid.
Hence in addition to the 4-momentum density $\rho u^{\mu}$  the fluid carries an 
effective charge 4-current $J^{\nu} = \sigma u^{\nu}$, which is
 identified as a possible form of $F^{\mu\nu}_{\pht ;\mu}$ and with the identity 
$J^{\nu}_{\ph{\nu};\nu} = 0$ implying the conservation of charge.
 That is we consider the flow of matter to be simultaneously associated with:
\begin{eqnarray}
  -\frac{1}{\kappa}G^{\mu\nu} \; =: \; \tef   & = & \rho u^{\mu}u^{\nu} \label{gtruu2} 
\\
  +F^{\mu\nu}_{\pht ;\mu}\;\; =:\;\; J^{\nu}  & = & \sigma u^{\nu} \label{jeqsigu}  
\end{eqnarray}
  as the respective definitions of matter density $\rho$ and charge density $\sigma$.
  Here the 4-velocity $\bu(x)$ with $\vert \bu \vert =1$ represents a fluid carrying 
both the mass and the charge. The fluid body is interpreted  to be immersed in and 
passing through the electromagnetic field $F_{\mu\nu}$ such that the Einstein equation 
reads:
\begin{equation}
 \label{gtruum}
    -\frac{1}{\kappa}G^{\mu\nu} = T^{\mu\nu}_{\!\epsilon} = \rho u^{\mu} u^{\nu} \; + 
\; F^{\mu\tau}F_{\tau}^{\ph{\tau}\nu}
	        + \frac{1}{4} g^{\mu\nu} \, F_{\rho\sigma}F^{\rho\sigma}
\end{equation}    
   That is the form of the energy-momentum tensor for the electromagnetic field from 
equation~\ref{geinm} has been combined with the pressureless perfect fluid term.
 Here the 4-velocity $\bu$ of the fluid 
 differs from the 4-velocity eigenvector $\bU$ defined in $\tef U_{\nu} = \rho' 
U^{\mu}$ by equation~\ref{ueigen}.
 With $T^{\mu\nu}_{\!\epsilon} = \rho' U^{\mu}U^{\nu} - S^{\mu\nu}$
  from equation~\ref{gruus}, the 4-velocity
 $\bU$
  represents a synthesis of the charged fluid and the electromagnetic field 
(\cite{Synge} p.357).

   Applying $T^{\mu\nu}_{\!\epsilon\, \pho ;\mu} = 0$ the effect on the terms on the 
right-hand side of equation~\ref{gtruum} has already been worked out separately in 
equations~\ref{trgeod} and \ref{trmaxw} respectively. Combined together we find that 
under the Bianchi identity equation~\ref{gtruum} becomes (based on \cite{Synge}  
p.358):
 \begin{eqnarray}
 T^{\mu\nu}_{\!\epsilon\, \pho ;\mu} = 0 \quad & \Rightarrow & \quad
         (\rho u^{\mu})_{;\mu}u^{\nu} + \rho u^{\mu}(u^{\nu})_{;\mu} 
		        \; + \; F^{\pho \nu}_{\tau}F^{\mu\tau}_{\pht ;\mu}   = 0  \nonumber \\
 \Sigma_{\nu} (\times \, u_{\nu})  \quad & \Rightarrow & \quad  (\rho u^{\mu})_{;\mu} 
   \quad + \quad 0 \quad + \quad F^{\pho \nu}_{\tau} F^{\mu\tau}_{\pht ;\mu} \; 
u_{\nu} = 0 \nonumber \\
         & \Rightarrow & \quad  (\rho u^{\mu})_{;\mu} 
     \qquad \quad \! + \qquad \quad \! g_{\lambda\tau}F^{\lambda\nu} \, J^{\tau} \, 
u_{\nu} = 0 \nonumber \\
	     & \Rightarrow & \quad  (\rho u^{\mu})_{;\mu} 
     \qquad \quad \! + \qquad \quad \!  F^{\lambda\nu} \; \sigma u_{\lambda}u_{\nu} = 
0 \nonumber
\end{eqnarray}
       The final term in the fourth line above is asymmetric in the indices of 
$F^{\lambda\nu}$ while symmetric in the indices of $u_{\lambda}u_{\nu}$ and is 
therefore equal to zero. The same line then implies that $(\rho u^{\mu})_{;\mu} = 0$ 
(as for the second line of equation~\ref{trgeod}) which can be substituted into the 
first line giving:
\begin{equation}
          \rho u^{\mu}(u^{\nu})_{;\mu} \; + \; F^{\pho \nu}_{\tau} J^{\tau}  = 0. 
\label{trlore}  
\end{equation}
    Each term in equation~\ref{trlore} was found to be zero for the individual cases 
of a perfect pressureless fluid alone or an electromagnetic field alone, giving 
equation~\ref{trgeod} for geodesic  motion and Maxwell's vacuum equation~\ref{maxfein} 
respectively. However for the combined case only the total vanishes  and hence 
$G^{\mu\nu}_{\pht ;\mu}=0$ implies that:
\begin{equation}
   \label{rellorg}
          \rho u^{\mu}(u^{\nu})_{;\mu} \; = \; + F^{\nu}_{\pho\tau} J^{\tau} 
\end{equation}
    This is the  relativistic Lorentz force law for a charged fluid in a curved 
spacetime, which is equivalent to  the corresponding law of equation~\ref{grellor}  
for discrete bodies in the appropriate limit (\cite{Synge} p.359)  as is similarly the 
case for the geodesic motion of equation~\ref{trgeod} considered above.
 Again Lagrangian terms, such as those in equation~\ref{slorlag}, are not required.

  As we described earlier for the effective energy-momentum tensor of 
equation~\ref{gtruup} the geodesic flow of an uncharged fluid is modified by the 
pressure gradient. Similarly for the  energy-momentum tensor of equation~\ref{gtruum} 
for charged matter the geodesic law is modified by the presence of an electromagnetic 
field to a form, equation~\ref{rellorg}, which precisely gives the Lorentz force law 
of equation~\ref{grellor}. This law, typically in the flat spacetime limit of
 equation~\ref{rellor} or the further non-relativistic limit, can be used to determine 
the strength of charges and electromagnetic fields in the laboratory and establish 
appropriate empirical normalisation factors.

 The possibility of incorporating electromagnetism and the Lorentz force law within a 
higher-dimensional approach to general relativity is well known and dates back to 
Kaluza in 1921 (\cite{Kaluza} equation~12). There it was shown that the 
\textit{five}-dimensional geodesic equation automatically incorporates the Lorentz 
force law in 4-dimensional spacetime, in the approximation of low 5-velocity. In the 
present theory the internal gauge fields, such as that for electromagnetism, arise as 
a higher-dimensional form $\lvh$ is projected onto the base space $M_4$, with charged 
matter arising through the interaction properties of the internal fields underlying 
the smooth spacetime geometry.

 For the case in which there is no electromagnetic field $F_{\mu\nu} = 0$  or in which 
the material flow is uncharged $J^{\mu} = 0$ the geodesic flow is recovered from 
equation~\ref{rellorg}. On the other hand $J^{\nu} := F^{\mu\nu}_{\pht ;\mu} \neq 0$ 
represents the case for which $T^{\mu\nu}$ as a function of $F^{\mu\nu}$ only, 
equation~\ref{tffgff}, itself  \textit{is not} conserved, as can be seen from the 
inconsistency with equation~\ref{trmaxw}, while the total $T^{\mu\nu}_{\!\epsilon}$ of 
equation~\ref{gtruum}, augmented to include the flow of macroscopic charged matter, 
\textit{is} conserved. The Lorentz force law results from the consistency of this 
total energy-momentum tensor bound together under the requirement of 
$T^{\mu\nu}_{\!\epsilon\, \pho ;\mu} = 0$, which itself is a direct consequence of the 
definition $T^{\mu\nu} := G^{\mu\nu}$ and the Bianchi identity.

   While the Bianchi identity implies $T^{\mu\nu}_{\pht ;\mu} = 0$ further 
conservation laws follow from further geometric identities, principally of the form 
$\md^2 =0$ which for example given $J^{\nu} :=  F^{\mu\nu}_{\pht ;\mu}$  implies that 
$J^{\mu}_{\ph{\mu};\mu}=0$,
 as described in the discussion following equation~\ref{tffgff}.
 This 
  leads to conserved charges associated with the internal symmetries both for Maxwell 
and Yang-Mills theories.
  However, while the Maxwell equations with source $J \neq 0$ imply the conservation 
of charge, this conservation law is limited to physical entities that carry charge. 
This marks a fundamental difference with the consequences of the Einstein equation, 
which can be interpreted as
  $T^{\mu\nu} :=  G^{\mu\nu}$, in that, assuming that all fields are associated with 
energy-momentum defined this way, \textit{all} fields are covered under the identity 
$G^{\mu\nu}_{\pht ;\mu} = 0$ and in principle `no physical entity escapes this 
surveillance' (\cite{MTW} p.475).

  In the above only the \textit{contracted} Bianchi identity for the Riemann curvature 
tensor has been employed. Further,
the Einstein equation $G^{\mu\nu} = -\kappa T^{\mu\nu}$ only directly yields certain 
linear combinations of the Riemann curvature tensor components.
  However,
although the Weyl tensor, introduced before equation~\ref{rdecom}, is that part of the 
Riemann tensor which is not directly equated with matter $T^{\mu\nu}$ in the Einstein 
equation it is not arbitrary. Applying the \textit{full} Bianchi identity
 $R_{\rho\sigma[\mu\nu;\tau]} = 0$ of equation~\ref{rbianc} to equation~\ref{rdecom}, 
rearranging the terms and contracting once leads to (\cite{HawkEl} p.85):
\begin{equation}
\label{cbian}
   C^{\rho\sigma\mu\nu}_{\pht\pht ;\nu} \, = \,
        R^{\mu[\rho;\sigma]} + \frac{1}{6} g^{\mu[\sigma}R^{;\rho]} \, =: \, 
K^{\rho\sigma\mu}
\end{equation} 
     Hence the full Bianchi identity, which contains more information than the 
contracted form, can be regarded as a field equation for the Weyl tensor in which the 
source $K^{\rho\sigma\mu}$  is defined as a function of the Ricci tensor. This is 
analogous to the Maxwell equation~\ref{maxj} for the electromagnetic field, which can 
be written in a curved spacetime as $F^{\mu\nu}_{\pht;\mu} = J^{\nu}$, with the 
electromagnetic current $J^{\nu}$ as the source. For equation~\ref{cbian} the source 
$K^{\rho\sigma\mu}$ depends on $R^{\mu\nu}$ which in turn is intimately  related to 
the matter content $T^{\mu\nu}$ through the Einstein equation, which can be written 
$R^{\mu\nu} = -\kappa (T^{\mu\nu} - \frac{1}{2}Tg^{\mu\nu})$ where $T = 
g_{\mu\nu}T^{\mu\nu}$. Hence, by substituting $T^{\mu\nu}$ into equation~\ref{cbian}, 
the Weyl curvature at any given location on $M_4$ depends on the matter content 
elsewhere in spacetime, in a similar way that the electric and magnetic fields depend 
on the charges elsewhere. The Weyl tensor represents the non-flat part of the Riemann 
tensor in the matter vacuum, this includes the phenomena of gravity waves (in analogy 
with electromagnetic waves) as well as gravitational tidal forces and lensing effects. 
Further, since gravitational waves carry energy even in regions of spacetime where 
$G^{\mu\nu}=0$ the association of $T^{\mu\nu} := G^{\mu\nu}$ with `energy-momentum' 
itself has a degree of ambiguity, while being of great value for many practical 
applications.

   In this section we have reviewed how a number of equations of motion arise out of 
the geometry of the Bianchi identities for the external and internal symmetries, given 
the relation of equation~\ref{gchift} obtained by comparison with Kaluza-Klein theory. 
However the equations of motion are derived we note that in order to empirically test 
a theory 
solutions of the field equations need to be determined and compared with actual 
observations in the world. This in turn requires the specification of initial 
conditions, or more general boundary conditions, in order to obtain such solutions. 
With care for the role of the implicit degrees of freedom of gauge and general 
coordinate transformations the `initial value problem' is well posed for both 
classical electromagnetism and general relativity respectively. 
   The evolution of the spacetime geometry is in principle fully obtainable from 
Einstein's equation and the equations of motion for the matter fields together with 
suitable boundary conditions.

   It is generally not possible to begin with a given source term on the right-hand 
side of  the Einstein field equation $G^{\mu\nu} = -\kappa T^{\mu\nu}$ since a 
coordinate system is required in order to specify the components of $T^{\mu\nu}(x)$, 
and further the distribution of matter itself is dynamically intertwined with the 
spacetime geometry through which it propagates. 
 One procedure would be to begin with arbitrary metric functions $g_{\mu\nu}(x)$ and 
catalogue $(g_{\mu\nu}(x), T^{\mu\nu}(x))$ pairs via equations~\ref{gtoGam}, 
\ref{ritencon} and the field equation with
 $-\kappa T^{\mu\nu} := G^{\mu\nu} = f(g_{\mu\nu})$, 
 in an attempt to converge upon a particular physical system.

 In practice exact solutions for the metric $g_{\mu\nu}(x)$ have been found for the 
cases in which $T^{\mu\nu}$ represents the vacuum ($T^{\mu\nu} = 0$), a perfect fluid  
or the electromagnetic field 
 (or a combination of the latter two, as described for a pressureless fluid in 
equation~\ref{gtruum}) and then only for spaces with a high degree of symmetry with a 
simple form of matter content. 
 All solutions in general relativity consist of a metric description for a complete 
spacetime geometry, which will be relevant for the study of cosmology, while only a 
limited region of the manifold may be of physical interest in other cases such as the 
study of planetary orbits using the Schwarzschild solution, described in the following 
section, for example.

   In summary, many of the equations of motion derived from a Lagrangian in 
section~\ref{subfal} have been shown to arise directly 
 as a consequence of the identity $\gmo$ given a solution for $G^{\mu\nu}(x)$ for 
example in the form of
 equation~\ref{gchift}. This latter relation itself arose as guided by Kaluza-Klein 
theory and  equation~\ref{einfiek}--\ref{einym} through the employment of a single 
`Lagrangian function' on a principle bundle space.  As described in the previous 
section in the present theory it is conjectured that the Lagrangian approach might be 
ultimately side-stepped entirely and that
 this one remaining pivotal Lagrangian, in the action of equation~\ref{teinhil}, may 
also be discarded. 
  In principle it may always be possible to work backwards from the present theory to 
obtain  apparent Lagrangian functions for the theory, but from the present point of 
view the Lagrangian method is ultimately effective due to its conformity with $\gmo$  
through the compatibility of the Euler-Lagrange equation with the requirement 
$T^{\mu\nu}_{\pht ;\mu} = 0$, as described in the opening of this section.

  In this section the form of the 4-current $J^{\nu} := + F^{\mu\nu}_{\pht ;\mu}$, in 
equations~\ref{maxj} and \ref{jeqsigu}, has been taken to emerge macroscopically and 
does not necessarily apply for `elementary particles'. 
 The origin and role of 4-currents for microscopic fields of the form 
   $j^{\mu}_{\ph{\mu}\alpha}  =  \overline{\psi} \gamma^{\mu} E_{\alpha} \psi$ in 
    equation~\ref{jglocur}, as well as the  Dirac equation~\ref{diracl},  in the 
present theory will be addressed in section~\ref{secdos}, in particular as exemplified 
by the Abelian $\uo$ case of electromagnetism.
  In order to consider the properties of microscopic elementary particles (electrons, 
photons etc.) it will first be necessary to address the more fundamental questions 
concerning the quantisation of the theory and  the concept of an elementary particle 
itself. 
   In addition to the identity $\gmo$ the full form of temporal flow $\lvh$, projected 
over the $M_4$ base space, will provide constraints on possible field interactions 
which are closely analogous to those provided by the Lagrangian for the Standard Model 
of particle physics, as will be described in chapters~\ref{chapesb} and \ref{secfd}.



\section{Spacetime Manifold and Time Dilation}
\label{fdandtd}

  Here we consider some of the geometric properties on the 4-dimensional spacetime 
manifold $M_4$ as arising in the present theory and in relation to general relativity. 
Here the 4-dimensional base manifold $M_4$ carries the four coordinate degrees of 
freedom of our spacetime experience of physical objects in the universe. The symmetry 
of the Lorentz group fits naturally on such a manifold since it acts on a 
4-dimensional vector space which corresponds to the tangent space of $M_4$. Selecting 
a 4-dimensional base space in this way is a provisional empirical input. It is 
empirical for the obvious reason and provisional since at this point the choice of 
four dimensions seems theoretically arbitrary and there remains the question of 
whether a base space of a different dimension could in principle be considered as a 
background for experience in another possible world. We shall return to this issue, 
and the question of the uniqueness of the theory in general, in section~\ref{secuni}.

  Hence the study of the Lorentz symmetry is motivated by the fact that it contains 
SO(3), the rotational symmetry of the background space within which we perceive 
physical objects, together with its respect for temporal causality, as well as its 
central importance in established physical theories of the world. In conformity with 
the present theory 
 $\soot$ is also the symmetry of a possible form of progression in time, denoted 
$\lvf$ and presented explicitly in equation~\ref{flow4d}, over a 4-dimensional vector 
space.

 Here we are considering the proper orthochronous Lorentz group $\soot$, sometimes 
denoted $\olg$, which is the part of the full Lorentz group that is continuously 
connected to the identity element. It is hence a continuous symmetry group acting on 
vectors in the 4-dimensional vector space $\rrr^{1,3}$, denoting the space $\rrr^4$ 
with Minkowski metric $\eta_{ab}= \mbox{diag}(+1,-1,-1,-1)$, as will be reviewed in 
more detail in section~\ref{lsspin}.
 Elements of the Lorentz group $l\inn\soot$  generate the symmetry  transformations
 $\sigma_l:\bv_4\to\bv'_4$ such that  $L(\bv_4)=L(\bv'_4)$ as an invariant form of 
temporal flow in four dimensions. At any $x\inn M_4$ on the spacetime manifold 
$\bv_4(x) \inn \TM_4$ is a vector in the local tangent space.

The base space $M_4$ itself originates out of the four dimensions of the  translation 
symmetry of the form $\lvf$ which is trivially invariant under $x^a \to x^a + r^a$ for 
the four components $v_4^a = dx^a/ds$ with $a=\{0,1,2,3\}$, as described more 
generally in equations~\ref{vnset}--\ref{rspill} of section~\ref{gfotf} and in section 
\ref{perc} for the model world. Here 
 the set of four numbers $r^a \inn \rrr^4$ can be identified with an initial set of 
four coordinates $x^{\mu} \inn \rrr^4$, with $x^{\mu} = \delta^{\mu}_{\ph{\mu}a}r^a$. 

 The Lorentzian structure of the vector space to which $\bv_4(x)$ belongs is 
transferred onto the tangent space of the parameter space $M_4$ and hence the latter 
acquires the properties of a 4-dimensional pseudo-Riemannian manifold. That is, since 
the flow $\bv_4(x)$ necessarily exists on the manifold, with components $v^a = 
dx^a/ds$, on $M_4$ the metric $\eta_{ab}$ derives locally  from the form $L(\bv_4) = 
\eta_{ab}v^a v^b$, and it is described by the metric $g_{\mu\nu}$ in a general 
coordinate system  via a tetrad field $\teta$ as:
 \begin{equation}
    g_{\mu\nu} = e^{a}_{\ph{a}\mu} e^{b}_{\ph{a}\nu} \eta_{ab}  \label{guneen}
  \end{equation}
  
  Hence the manifold $M_4$ inherits its pseudo-Riemannian structure \textit{from} the 
Lorentz symmetry of $L(\bv_4)$; with the SO(3) subgroup implying the possibility of a 
suitable 3-dimensional background space which \textit{appears} to us to be of a more 
fundamental \textit{a priori} existence than the objects we perceive moving through 
it.

  For such a manifold in which there exist global coordinates such that $g_{\mu\nu}(x) 
= \mbox{diag}(+1,-1,-1,-1)$ for all $x \inn M_4$, that is the constant Minkowski 
metric, we have the 4-dimensional spacetime of special relativity. In this case the 
local $\eta_{ab}$ metric has been drawn out and made global through the existence of 
large scale coordinates with respect to which the tetrad field can be simply be 
expressed as $\teta = \delta^a_{\ph{a}\mu}$.
For such a Minkowski spacetime manifold the $\soth$ subgroup of $\soot$, now acting 
globally, provides the symmetry of the 3-dimensional space through which a physical 
world of objects might be perceived.

  The point of view taken in this paper is that it is the nature of perception itself 
that implicitly requires an approximately flat background manifold, at least for the 
extended neighbourhood of the observer, and hence essentially inflates the local 
Minkowski metric into the extended spacetime arena and thus draws the Lorentz 
structure of the form $L(\bv_4)$ out onto an approximately uniform background 
spacetime within which objects are perceived. The mathematical expression for such a 
spacetime structure, to be utilised by perception, arises spontaneously out of the 
translational symmetry of the form $L(\bv_4)$.

   While the identification of the 4-dimensional spacetime manifold through the 
4-dimensional form of temporal flow $\lvf$ will result in a \textit{flat} spacetime 
geometry, as described for the model world in subsection~\ref{mgjoin},
     ultimately the base manifold $M_4$ will be obtained through a subset of four 
translational degrees of freedom  breaking the
 symmetry of a higher-dimensional form of temporal flow $\lvn$ with $n>4$, 
representing the full form $\lvh$. A specific expression of for $\lvh$ will be 
introduced in the following chapter, extending beyond the case of $\lvte$ described in 
section~\ref{reaic}. This results in general in a non-zero external Riemannian 
curvature, complemented by a non-zero internal gauge curvature, as described in 
sections~\ref{hdasb} and \ref{reaic}. 
On the $M_4$ manifold the Lorentz form of equation~\ref{flow4d}, now embedded within 
the full form $\lvh$, locally expresses the relation between the components 
$v^a=dx^a/ds$ of tangent vectors in an ordered orthonormal basis of the tangent space. 
Such a local basis, or frame field, $\{e_a\}$ satisfies $g(e_a,e_b)=\eta_{ab}$, and 
with $\bv_4(x) = v^a(x) e_a(x)$  the local relation of equation~\ref{flow4d} is 
replaced by the looser constraint on the four components $v^{a}(x)$ projected onto 
$\TM_4$ with: 
   \begin{equation}
 L(\v_4)=(v^0)^2- (v^1)^2-(v^2)^2-(v^3)^2= \eta_{ab} v^a v^b = \eta(\bv_4,\bv_4) = 
h^2.
 \label{lorform2}
\end{equation}
  with $h\inn \rrr$.
     While local coordinates $\{x^a\}$ necessarily exist to express the form $\lvfh$  
we may also introduce an arbitrary global coordinate system $\{x^{\mu}\}$ over $M_4$ 
which naturally gives rise to a coordinate frame basis, denoted $\{\partial_{\mu}\}$, 
for the tangent space at any $x\inn M_4$. The coordinate frame is related to the 
orthonormal frame
 with $\partial_{\mu} = e_a \, \teta$ as described in equation~\ref{eatopmu} and  
section~\ref{riegeo}.
   
    In addition to the observation that in general $h^2 \neq 1$ in 
equation~\ref{lorform2} the further consequence of the embedding in the larger form 
  $\lvh$ is the possibility of finite Riemannian curvature $\bR \neq 0$ as alluded to 
above. This implies a warping of the geometry such that global coordinates no longer 
exist such that $\teta = \delta^a_{\ph{a}\mu}$ in general. The tetrad field $\teta$ 
now describes the necessarily non-trivial 
relation between global and local coordinates. As described in section~\ref{gcatep} 
the unphysical nature of general coordinates is implied under general covariance, 
while a
 tetrad field with respect to a set of coordinates $\rrr^4$, as depicted in 
 figure~\ref{onecoord}(a), indicates physically distinguished local orthonormal frames
  as utilised by the
  equivalence principle.

In the present theory `general covariance' is significant since in general the Lorentz 
symmetry of the form $L(\bv_4) = h^2$ cannot  be expressed globally with respect to a 
single coordinate chart  on the manifold. Without such a preferred global reference 
frame all arbitrary coordinate systems are equally valid for the description of the 
equations of physics on the manifold.
   In the context of this theory the metric $g_{\mu\nu}(x)$ has particular physical 
significance for the nature of perception and describes the geometric form through 
which we literally \textit{see} the world,
 motivating its prominent role as the gravitational field; described as the `new 
ether' by Einstein as discussed at the end of section~\ref{gcatep}.
 
 The general \textit{global} coordinates do not correspond to an underlying Euclidean 
or any other geometric structure on the manifold. However, the manifold exists as a 
space for the flow $\bv_4(x)$ of $\lvfh$ itself  and we naturally have a frame field 
$\{e_a(x)\}$ of local orthonormal basis vectors and local coordinates $\{x^a\}$ with 
respect to which this flow can be written with the components $v^a=dx^a/ds$, 
corresponding to the tangent vector components $v^{\mu} = dx^{\mu}/ds$ in a general 
coordinate system, and hence we necessarily have a \textit{local} Lorentzian structure 
on $M_4$.
 While in principle the torsion on such a manifold may be finite the geometry 
described above is compatible with the
 `equivalence principle' which may hence be adopted, together with the implication of 
vanishing torsion, as a provisional simplifying assumption which will be discussed 
further in section~\ref{secuni}.

 With respect to a set of general coordinates $\{x^{\mu}\}$ on the $M_4$ manifold 
arbitrary vector fields $\bu(x)$, that is cross-sections of the tangent bundle 
$\TM_4$, can be expressed as $u^{\mu}(x)\partial_{\mu}$ with the numbers $u^{\mu} 
\inn\rrr^4$  regarded as the components of a tangent vector on the 4-dimensional 
manifold $M_4$. The situation is similar to that depicted in figure~\ref{spill}, 
except now for a 4-dimensional manifold.
For any vector field $\bu(x)$ on $M_4$  the quantity $g(\bu,\bu) = 
g_{\mu\nu}(x)u^{\mu}(x)u^{\nu}(x)$ may be determined at any point $x \inn M_4$ and  
the vector $\bu(x)$ described  as `timelike', `null' or `spacelike' according to 
whether this quantity is positive, zero or negative respectively. This range of 
possibilities is also the origin of the name `space-time' manifold. The `time' in 
`spacetime' refers to the existence of timelike vectors and coordinates rather than 
explicitly to the actual pure temporal flow $s$ which underlies the particular field 
$\bv_4(x)$ as constrained by the equation $\lvh$.

Since the Lorentzian manifold structure arises \textit{out of} the flow of time the 
light cone geometry of the tangent space is time-orientable over the 4-dimensional 
volume of the spacetime manifold $M_4$. That is, the time-orientation of the light 
cones is necessarily continuous on $M_4$ as determined by the directed line element 
field $\bv_4(x)$ of temporal flow itself as an extension of the original 1-dimensional 
progression in time. 
 This time-directed vector field is locally $\soth$ invariant and provides a local 
$(1+3)$-dimensional decomposition of spacetime for all $x\inn M_4$ with temporal and 
spatial parts identified in the local reference frames.

  Choosing the local coordinate $x^0$ to be aligned with $\bv_4(x)$, with components 
$v^a=(\frac{dx^0}{ds},0,0,0)$, then $x^0$ effectively acts as a parameter for the pure 
values of time, that is $ds=dx^0/h$ for $\lvfh$, which is a particular case of the 
more general local expression described in equation~\ref{vvvhids} below.  Three 
spacelike local coordinates $x^1,x^2$ and $x^3$ can also be constructed orthogonal to 
each other and to $x^0$ with respect to $\eta_{ab}$, with
local spatial frames related via the $\soth$ subgroup.

  Whereas  embedding the perceptual background of an effective 3-dimensional space and 
1-dimensional time within the symmetry structures of the mathematical form $\lvf$ led 
to their incorporation into the 4-dimensional Minkowski spacetime of special 
relativity, that is with zero Riemannian curvature, extracting the same base manifold 
out of a higher-dimensional form of temporal flow $\lvh$  results in a more flexible 
and dynamic 4-dimensional spacetime structure as employed in general relativity.
 With $M_4$ itself still originating  out of a 4-dimensional translational symmetry of 
$\lvh$,
 even for the generalisation in which the external geometry is expressed in terms of 
underlying interacting fields as implied equation~\ref{getypsi}, 
   the Minkowski metric $\eta_{ab}$ implicit in the form $\lvfh$ is sewn into the 
local tangent space structure everywhere on the base manifold. This defines a possible 
metric structure $g_{\mu\nu}(x)$ on $M_4$ associated in a one-to-one manner with the 
existence of an $\soot$ orthonormal frame bundle $OM_4$
 within the canonical $\glfr$ general frame bundle $\FM_4$ over the base manifold, as 
described in section~\ref{riegeo}.

 With the external geometry related to the internal geometry via 
equation~\ref{gchift}, or more generally with equation~\ref{gety} augmented to 
equation~\ref{getypsi},
 in principle the metric itself might be obtained by adopting the Levi-Civita linear 
connection on $M_4$. The connection is metric compatible, since it derives from the 
local $\soot$ symmetry of the form $\lvfh$, and assumed to be torsion-free as 
described above.
 Hence as for general relativity the metric itself may be extracted by solving the 
second order differential equation $G^{\mu\nu} = -\kappa T^{\mu\nu}$ given a form for 
the energy-momentum tensor $T^{\mu\nu}$ under appropriate boundary conditions, as 
described towards the end of the previous section.  An example is given in 
equation~\ref{ttrtp} below.

 The tetrad field $\teta$ with 16 independent components  carries two kinds of 
information. The 10 degrees of freedom of the symmetric metric field $g_{\mu\nu} = 
e^{a}_{\phantom{a}\mu} e^{b}_{\phantom{b}\nu} \eta_{ab}$ correspond to the 
gravitational field for the torsion-free metric connection in general relativity, and 
hence the tetrad field itself can be considered to represent the gravitational field. 
The remaining 6 degrees of freedom correspond to the local choice of Lorentz frames 
implicit in $\teta$. This local symmetry provides a link with the framework of local 
gauge theories as well as with the application of the spinor representations of the 
Lorentz group, as also alluded to towards the end of  section~\ref{gcatep}, which are 
important in particle physics as will be described in chapter~\ref{rotsm}.

  Here we consider the physical significance of a non-flat Riemannian geometry, 
described by the metric field $g_{\mu\nu}(x)$, in particular on the \textit{relative} 
passage of time itself.
 We also consider the relation of the original pure temporal flow $s$ with the proper 
time $\tau$ which may be recorded by physical objects such as clocks in the material 
flow of the world.

The underlying pure temporal flow $s$, subject to the full form $\lvh$, exists 
everywhere on the base manifold $M_4$.
 The $\bv_4 \subset \hat{\bv}$ projection onto the tangent space $\TM_4$ to the base 
manifold  is a timelike vector, as is the tangent to any world line on $M_4$, with 
components $v^a_4 = dx^a/ds$ restricted under $\lvh$ such that:
 \begin{equation}
  \label{vvvhids}
    \vert \bv_4 \vert^2 = \eta_{ab} v^av^b = h^2
	\quad \mbox{implying} \quad ds^2 = \frac{\eta_{ab}}{h^2}dx^{a}dx^{b}
 \end{equation}  
 However gravitational time dilation will not be directly observed from the 
perspective of the
  microscopic flow $\bv_4$. Indeed the underlying pure temporal flow $s$ is not  
\textit{measured directly} by physical instruments. Rather it is through the structure 
and  symmetries
  of the form $\lvh$ that the physical world emerges on $M_4$ through relations such 
as equation~\ref{gchift}, and with more general expressions for the apparent 
energy-momentum tensor as implied in equation~\ref{getypsi}.
 This more general apparent material world may be described empirically in part by the 
effective energy-momentum tensor   
   $T^{\mu\nu}_{\epsilon} = \rho u^{\mu}u^{\nu}$, as introduced in 
equation~\ref{gtruu} of the previous section and leading to the geodesic 
equation~\ref{trgeod}, where $\rho(x)$ is the matter density and the 4-velocity 
$u^{\mu}(x) = dx^{\mu}/d\tau$ is defined as the tangent vector at $x\inn M_4$ to the 
world line of the physical body, which may be an element of a pressureless fluid. It 
is through the motion of physical bodies, such as the hands of a mechanical clock, 
that time dilation effects may be observed.
  With the \textit{proper time} $\tau$ parametrising the motion of the body for a 
general coordinate system $\{x^{\mu}\}$ on $M_4$ we have: 
\begin{equation}
   \label{dtgxx}
         g_{\mu\nu}u^{\mu}u^{\nu} = 1
		 \quad \mbox{and with} \quad
		  d\tau^2 = g_{\mu\nu} \, dx^{\mu} dx^{\nu}
  \end{equation}
    identifying an \textit{interval} of proper time $d\tau$. These expressions are
   invariant under general coordinate transformations.  The normalisation for the 
components of the metric $g_{\mu\nu}(x)$  will depend on the choice of 
\textit{empirical} units adopted, for example seconds and metres for temporal and 
spatial dimensions, in recording the motions of the parts of a physical `clock'.

  The  local orthonormal coordinates $\{x^a\}$ constructed empirically for the 
macroscopic proper time interval with $d\tau^2 = \eta_{ab} dx^a dx^b$  will in general 
\textit{not} be identical to those of equation~\ref{vvvhids} arising directly out of 
the mathematical properties of the pure form of temporal flow $\lvh$.
 However with the physical world unfolding through the progression of the fundamental 
time parameter, and with $s$ and $\tau$ represented by the 4-vectors $\bv_4$ and $\bu$ 
in $\TM_4$ respectively,  both temporal parameters are subject to   time dilation 
effects in the same way. The proper time $\tau$, in 4-dimensional spacetime, is 
implicitly linearly proportional to the pure underlying temporal flow $s$, which may 
be expressed in any number of dimensions.
  This proportionality is expressed through the fixed parameter $\gamma$ in 
equation~\ref{taufroms} in section~\ref{secpotnt} where the relationship between 
$\tau$ and $s$ is further explored.

 Hence along a shared world line the fundamental time interval $ds$ is related to the 
proper time interval $d\tau$ by a constant scaling and the two temporal parameters are 
equivalent in this sense -- that is, within a fixed normalisation factor physical 
clocks \textit{do measure} the progression of pure time $s$.
 As described in the introductory chapter, and to be expanded in 
chapter~\ref{chaptoot}, the fundamental underlying mathematical time $s$ is ultimately 
identified with  
 `experienced' time, while proper time $\tau$ is associated with measurable empirical 
phenomena, which include for example  `physical brain processes'. Hence these 
subjective and objective temporal phenomena, which might be exemplified by an observer 
located within the same inertial frame as a physical clock, are intimately connected.
 We next consider a particular example of time dilation effects.

   The physical manifestation of the metric $g_{\mu\nu}(x)$ in a general coordinate 
system on $M_4$ resides in observable \textit{relative} temporal and spatial 
distortion effects at different locations on the manifold itself. For example the 
Schwarzschild solution for the metric of a spatially spherically symmetric geometry 
around a single massive body of mass $M$ is given by the line element:
\begin{equation}
 \label{ttrtp}
   d\tau^2 = \left( 1-\frac{2\GN M}{r} \right) dt^2 \: - \: \left( 1-\frac{2\GN M}{r} 
\right)^{-1}dr^2
           \: - \:  r^2 d\theta^2 \: - \:  r^2 \sin^2 \theta d \phi^2       
\end{equation} 
  in the 4-dimensional, spatially spherical polar, coordinates $\{t,r,\theta, \phi 
\}$, where $\GN$ is Newton's gravitational constant. In addition to the
 assumption of a spatially spherically symmetric metric
 this solution is obtained by imposing the boundary condition that $g_{\mu\nu}(x)$ 
approaches the flat Minkowski limit as $r \to \infty$ spatially. This limit can be 
seen explicitly on taking $r \to \infty$ in equation~\ref{ttrtp} (this example is 
closely analogous to the case of the Coulomb field for a central electric charge).

 The coordinate $r$ parametrises, but does not determine, radial distances. This is 
consistent with the arbitrary nature of coordinates and all coordinate systems in 
general, as described in section~\ref{gcatep} and figure~\ref{onecoord}. The actual 
radial distance, for given parameters $\{ t,  \theta, \phi \}$,  is measured by the 
integral of intervals $dR = (1 - 2\GN M/r)^{-1/2}dr$. Similarly a clock at a fixed 
coordinate location in space records the proper time $\tau$ elapsed along its world 
line through the intervals:
\begin{equation}
 \label{ttonly}
   d\tau \: = \: \left( 1-\frac{2\GN M}{r} \right)^{\frac{1}{2}} dt      
\end{equation} 
  relative to the time $\tau_{r \to \infty} = t$ measured by a clock in the flat 
spacetime limit at $r \to \infty$,
  and is a function of radial distance from the central mass, as parametrised by the 
coordinate $r$. While at any location $x \inn M_4$ it is possible to choose 
\textit{local} inertial coordinates $\{x^a\}$, for which $g_{\mu\nu}(x)
 = \mbox{diag}(1,-1,-1,-1)$, the absence of such a \textit{global}
  frame for non-zero mass $M > 0$ leads to a relative time dilation effect recorded by 
clocks at differing radial distances from the central massive object.

As described above this dilation effect applies for the fundamental temporal flow $s$ 
in exactly the same way as for the proper time $\tau$.
  Hence with the interval $ds \equiv d\tau$ the same metric $g_{\mu\nu}(x)$ represents 
the relative temporal dilation on $M_4$ for the fundamental flow of time $s$. An 
observer, named `twin $A$', accompanied by a clock measuring the physical temporal 
flow $\tau_{\! A}$ carries an equivalent universal time parameter $s_{\! A}$ through 
which the entire universe unfolds through the realisation and symmetry breaking of the 
full form of temporal flow $L(\hat{\bv}_{\! A}) = 1$, deriving from $s_{\! A}$ as 
described for equation~\ref{lv}. A second observer, `twin $B$', at a separate 
spacetime location carries a second personal temporal parameter $s_{\! B}$ through 
which $B$ perceives the same universe to unfold through the form
  $L(\hat{\bv}_{\! B}) = 1$ in a mutually consistent way. 
 This `dovetailing' of the `temporalisation' experienced by twins $A$ and $B$ as 
manifested in the same physical world will be described further in 
section~\ref{secauf}
 in the discussion of figure~\ref{alltog}.

  The same metric solution $g_{\mu\nu}(x)$ for the single consistent universe, 
expressed in a particular coordinate system (or equivalently a particular metric 
expression of a given geometry in terms of a unique set of coordinates $\rrr^4$, 
adopting the perspective of figure~\ref{onecoord}(a)), provides the relation between 
the intervals $ds_{\! A}$ and $ds_{\! B}$ and the equivalent gravitational temporal 
dilation effect observed between $d\tau_{\! A}$ and $d\tau_B$ measured by the clocks 
of twin $A$ and twin $B$ respectively. The time dilation effect is determined by the 
empirically constructed metric $g_{\mu\nu}(x)$ in the coordinate system $\{x^{\mu}\}$ 
since it implicitly determines local inertial coordinates $\{x^a\}$ which are related 
to those of equation~\ref{vvvhids} by a constant scale factor (again, as will be 
discussed further near the opening of section~\ref{secpotnt} and alongside 
equation~\ref{taufroms}). In turn the local coordinates of equation~\ref{vvvhids} 
directly parametrise the fundamental temporal flow $s$, within a factor of $h^2$, via 
the projection of the form $L(\bv_4)$ onto the tangent space of $M_4$.

 So far we have implicitly considered only the case of constant $h(x)$ in 
equation~\ref{vvvhids}. In this case all geometric time dilation effects can be 
considered as having a `source' in the right-hand side of the Einstein 
equation~\ref{Eins} in terms of the apparent energy-momentum $T^{\mu\nu}(x)$ of 
ordinary matter. This is the case for the Schwarzschild solution of 
equation~\ref{ttrtp} for a central massive body. On the other hand possible variations 
in the magnitude of $h(x)$ in equation~\ref{vvvhids} will act as conformal 
transformations of the geometry the possible consequences of which will be considered 
in section~\ref{secpotnt}, initially alongside figure~\ref{vtovary}.


\section{Beyond Kaluza-Klein Theory}

\label{secbkkt}

  For Kaluza-Klein theory, originating as a pure higher-dimensional spacetime 
extension of general relativity, to be interpreted as a unified theory of gravitation 
and gauge fields in a 4-dimensional spacetime the symmetry group of general coordinate 
transformations in the extended spacetime has to be \textit{broken} down to  
4-dimensional general covariance together with the local gauge symmetry. This is 
equivalent to placing restrictions on the metric of the extended space which then 
possesses a set of isometries described by Killing vector fields which have a 
one-to-one relationship with the left-invariant vector fields on the manifold of an 
apparent gauge group $G$. In this way a principle fibre bundle structure emerges on 
the extended space, exhibiting symmetries such that the freedom in variation of the 
metric $\ddot{g}_{ij}$, as expressed in a direct product basis in 
equation~\ref{gmetug}, is effectively reduced to the components $g_{ac}$ and 
$\omega^{\alpha}_{\ph{\alpha}a}$.
  The construction of an action integral on the bundle space then leads to 
corresponding  equations of motion such as those of 
equations~\ref{einfiek}--\ref{yangmk}.
 A dynamical mechanism for this process in which an extended 4-dimensional base 
manifold $M_4$ of general relativity survives while the \textit{extra} dimensions lose 
any sense of external spatial significance, sometimes called `spontaneous 
compactification', then remains to be specified, as alluded to in section~\ref{reaic}. 
That is, the origin of the above restrictions on the metric for the full space remains 
to be accounted for.

    The Kaluza-Klein models, reviewed in chapter~\ref{kktheory}, contrast with the 
idea presented in this paper since here the `extra dimensions', beyond four, are 
\textit{not} required to satisfy an explicitly geometric, or spacetime, symmetry.
 In turn for the present theory there is no need to explain such a `compactification', 
rather the base manifold $M_4$ is the \textit{only} physically extended manifold to 
consider as it emerges as a background arena for perception through the translational 
symmetry of the full form of temporal flow $\lvh$.
In section~\ref{reaic} we  presented these ideas as a \textit{mathematical} 
possibility taking as an  example the $\sootn$ symmetry of $\lvte$ projected over 
$M_4$, but the significant conceptual question concerning \textit{why} this situation 
should be found in nature also needs to be addressed. We review here the conceptual 
motivation that led to this framework in the context of this provisional $\sootn$ 
model world.

 Out of the purely \textit{algebraic} symmetries of $\lvte$ the possibility of a local 
$\soota$-valued connection 1-form gives \textit{geometric} meaning to $M_4$ as being 
not just a numerical parameter space for translational degress of freedom but rather 
implicitly possessing a Riemannian structure with local metric $g_{ac}(x)$ as an arena 
for the perception of physical objects in \textit{time} and \textit{space}.
 The identification of an extended base space is possible since there is a `spacetime' 
symmetry as a subgroup of the full symmetry of $\lvh$ which acts on the local tangent 
space of $M_4$.
This innate \textit{possibility} of such an interpretation is sufficient for such 
structures to `freeze out' from the full symmetry of $\lvh$ as a kind of `gestalt' 
through which by necessity the physical world is created and perceived.

Given this geometrical realisation of the perceptual `external' symmetry on the base 
manifold, out of the full symmetry there remain  `internal' residual gauge fields and 
surplus temporal components which will collectively contribute to the apparent 
`matter' content of the world through which the properties of physical entities will 
be perceived and identified on the base space.
 The symmetry of $\lvte$ is broken in the identification of the extended $M_4$ 
parameter space, with a $\ol{\bv}_4 \subset \bv_{10}$ component of the temporal flow 
projected onto the tangent space $\TM_4$ as depicted in figure~\ref{mtogmaphr}(b). 
Since the $\ol{\bv}_4$ components are distinguished in this way from the residual 
internal part $\ul{\bv}_6 \subset \bv_{10}$ the full symmetry of the original $\sootn$ 
action on $\bv_{10}$ is lost. The surviving symmetry, as gauge freedom over $M_4$, is 
resolved into two pieces with corresponding connection 1-forms identified for both the 
external and internal spaces.

   The combination of the general flow of time, expressed as $\lvte$, with the implied 
symmetry properties and canonical mathematical structures existing for these objects, 
together with the conceptual need for a perceptual base for observation in a world, 
all taken collectively, has resulted in the identification of a background manifold. 
The full symmetry of the temporal flow $\lvte$ has been `sacrificed' in the creation 
of the non-trivial external and internal geometrical entities, but remains as a 
`ghostly' presence through which these entities are related. This correlation between 
the external and internal curvature tensors $\ol{\bR} \neq 0$ and $\und{F} \neq 0$ 
(while both can be zero together) was described originally for the $\sofi$ model over 
$M_3$ in section~\ref{hdasb} 
  and for the $\sootn$ model over $M_4$, in the light of Kaluza-Klein theory, in 
section~\ref{reaic}. This latter structure will also apply to the full symmetry action 
considered for the real world from the following chapter.

The use of geometrical pictures, such as those of figure~\ref{mtogmaphr},  as a visual 
aid to understanding mathematical structures comes very naturally when the 
\textit{space} pictured represents the way we actually \textit{perceive} those 
structures in the world. However, the underlying properties of a purely 
\textit{mathematical} space, such as those demanded here by the concept of the 
symmetry of time, need to be worked out within the appropriate \textit{algebraic} 
rules, which are not necessarily visualisable even by analogy with lower-dimensional 
structures. Hence while possibly serving as a guide a reliance on such geometric 
pictures is ultimately likely to prove misleading. This in particular will be the case 
in the following chapter in which the internal dimensions will no longer have a 
spatial interpretation (\textit{unlike} the case for the 6-dimensional space of 
vectors $\ul{\bv}_6 \inn \rrr^6$ with an internal SO(6) rotational symmetry for the 
$\sootn$ model described above).

  On the other hand it can be asked what the \textit{perceived} part of the 
mathematics actually \textit{looks} like, and geometric pictures only really make  
sense in terms of a literal interpretation in this context. Perception is our window 
into the world of mathematical forms. It is a window which is both opened up and 
limited through the possibility of the internal mathematical relations which frame our 
experiences in a 4-dimensional spacetime. It is also part of the difficulty in 
theorising beyond the 4-dimensional world of general relativity, for which 
visualisation is a key tool.

   In conclusion then, here a spacetime geometric symmetry is only \textit{required} 
to exist on the base manifold, hence in four dimensions for our world. It is also 
required to be an approximately global symmetry,  such that the base manifold may be 
identified as a suitable arena for perception in the world, at least for extended 
regions on the scale of everyday observations although not necessarily on the larger 
scales considered in cosmology.

  In Kaluza-Klein theory, as described in chapter~\ref{kktheory}, while a unified 
framework is provided for gravity and gauge boson fields, 
equations~\ref{einfiek}--\ref{yangmk}, there is no energy-momentum tensor for fermion 
fields -- that is the matter fields for the leptons and quarks of our world are 
absent. These fields may be added by hand as sections of fibre bundles over $M_4$, 
associated to the principle bundle $P$, transforming as spinors under the external 
$\soot$ symmetry and in representation multiplets of the internal gauge symmetry 
group. Coupling between the gauge fields and fermions may then be introduced through 
interaction terms, also added by hand for example via `minimal coupling' involving 
covariant derivatives, in the Lagrangian constructed for the theory.

  A more mathematically self-contained approach is through a supersymmetric extension 
of the Kaluza-Klein framework (see for example \cite{ChMaMa} section VI, \cite{Manso}, 
\cite{Witten} and \cite{DuNiPo} sections 1 and 2). Fermions may be included for 
example through generalising the gauge group $G$ of the principle bundle to a 
`supergroup' by augmenting the Lie algebra $L(G)$ into a `graded' Lie algebra. Here 
the rule for multiplication in the Lie algebra by commutation of elements, as 
exemplified in equation~\ref{xxcomcx}, is extended algebraically to include 
anticommutation which can be used to accommodate the properties of fermion fields. The 
Einstein-Yang-Mills theory may be extracted as the purely bosonic sector of such 
extended \textit{supergravity} theories.

    Of the many formulations of supergravity the most attractive model involves a 
single supersymmetry generator, `$N=1$', so that each Standard Model particle has a 
single superparticle partner forming a supersymmetric doublet, and is constructed in 
an 11-dimensional spacetime, that is `$d=11$'.  The pairing of bosons with fermions 
through supersymmetry also tends to naturally lead to the attainment of finite 
calculations in the corresponding quantum field theory.
However, even for the most favourable version in 11-dimensional spacetime a fully 
renormalisable version of supergravity has not been realised~(\cite{Pen} p.880). 
Further generalisation of supergravity to a superstring theory, obtaining a finite 
theory of quantum gravity by modifying QFT at the Planck scale, addresses some of the 
technical difficulties.

  Through this geometrisation of matter in the spirit of Kaluza-Klein models based on 
higher dimensions of spacetime, extended to the supersymmetric theories of 
11-dimensional supergravity and 10-dimensional superstrings, the aim is to incorporate 
the degrees of freedom of the full set of Standard Model gauge interactions within the 
geometry of the 7 or 6 extra spatial dimensions. In some cases the extra dimensions 
are considered to be small and topologically compactified while in other models our 
own universe may be conceived as a 4-dimensional \textit{brane}-world embedded as a 
4-dimensional hypersurface within the higher-dimensional spacetime \textit{bulk} 
(see~\cite{Ponce} for a simpler case with a 5-dimensional bulk).

 Einstein's theory of gravitation based on a metric tensor in 4-dimensional spacetime 
hence stimulated a chain of extensions and generalisations that we have briefly 
reviewed above and summarise below in table~\ref{EtoS}.
\begin{table}[htbp]
\centering
\begin{tabular}{|l|l|}
 \hline
    Theoretical Framework    &    Physical Scope  \\  
 \hline					
  General Relativity  &   Gravitation    \\
  Kaluza-Klein in 5-dimensions $\qquad$  &  Electromagnetism       \\
  Non-Abelian Kaluza-Klein Theory  &  Non-Abelian Gauge Fields  \\			  
  (with non-Levi-Civita $\Gamma$ on $P$) & (avoid large Cosmological term) $\;\;$ \\	
  (with $G$ acting on homogeneous fibres) & (keep full $L(G)$-valued theory)  \\
  Supergravity                 &   Fermions as well as Bosons   \\
  Superstrings                 &   Finite Quantum Gravity    \\
  \hline
  \end{tabular}
  \caption{\setb A series of increasingly general frameworks is listed in the first 
column with their \textit{cumulative} extent of application listed in the second 
column for the non-parenthetical entries. The means of including fermions and quantum 
theory within the present framework will be described in section~\ref{extsym} and 
chapter~\ref{newapp} respectively.}
\label{EtoS}
\end{table}

 One of the attractions of using a symmetry of extra \textit{spatial} dimensions, as 
well as its intuitive appeal as an extension of 4-dimensional spacetime geometry, is 
that it limits the set of possible higher symmetries and 
mathematical structures
 to consider. 
  In this paper instead of considering arbitrary symmetries, general geometric 
symmetries or specifically the symmetry of a \textit{spacetime} in higher dimensions 
we consider general symmetries of pure \textit{time} alone, as expressed through the 
relation $\lv$ and described in chapter~\ref{sym}. This also greatly limits the choice 
of symmetry groups and their representations. As well as naturally extending to 
general higher-dimensional mathematical forms of the progression of time $\lv$, at the 
same time we retain the significance of the $(1+3)$-dimensional metrical manifold as a 
form of observation in the world as having a necessary and \textit{a priori} nature.

  The important point here is that the symmetry of the \textit{space} part of 
spacetime, such as that of the SO(3) subgroup of the Lorentz symmetry $\soot$ central 
to general relativity, can be \textit{experienced} in a different, geometrical, way 
compared with other higher symmetries of $\lv$. It may be that higher symmetries, such 
as $\sootn$, \textit{could} be interpreted in a geometrical way, but this feature is 
relatively incidental in comparison with the fundamental requirement that it must 
describe a symmetry of time.   

  However, through investigating possible symmetries of time a significant example is 
identified for the symmetry group $\sltwoo$ acting on the 10-dimensional space 
$\htwo$, constructed in terms of the octonion algebra 
  as described in the following chapter and in particular section~\ref{ltos}. 
  With $\sltwoo$ being the 
    covering group of the 10-dimensional Lorentzian symmetry $\sootn$ this structure 
will naturally correlate with some of the properties of models based on extra spatial 
dimensions for which 10-dimensional spacetime is significant. Further, the 
16-dimensional Majorana-Weyl spinor representation of the 10-dimensional Lorentz 
group, highlighted in table~\ref{LctoRso} of section~\ref{dynkin} and here represented 
by the $\theta$ components appearing in the extension to the space $\htho$ introduced 
in equation~\ref{xoct3} of section~\ref{esitran} and described near the opening of 
section~\ref{extsym}, is significant in various branches of string theory.

 In the present theory by exploring the physical interpretation of the 
higher-dimensional forms of $\lv$, together with the associated isochronal symmetry 
groups, expressed over a base space $M_4$,  contact is made with the series of 
generalisations  mid-way down table~\ref{EtoS}, with the items listed parenthetically, 
with a framework very similar to non-Abelian Kaluza-Klein theories. The use of a 
non-Levi-Civita $G$-invariant linear connection $\Gamma$ such as described for 
equation~\ref{gamsetkk} defined on a principle fibre bundle $P$, or on a bundle of 
homogeneous fibres $E$, makes a significant area of contact with the corresponding 
literature (including \cite{Cho,Kerner,ChMaMa,Orzalesi}, 
 \cite{Kopcz,OrzPau,Kalin,Katan}, 
\cite{Perc,Lucietal,CLLM}, \cite{MaCh}).
From this point we then immediately diverge
 away from the progression towards supersymmetry and string theory in table~\ref{EtoS}
 and in this context we shall need to explain how mathematical structures identified 
in the present theory correspond to the inclusion of fermion states as well as the 
physical concepts of quantum and particle phenomena in general.

  Here we describe how field interactions arise in the context of the $\sootn$ model.
  Returning to the bundle space  $\ul{P} \equiv M_4 \times \sox$  under the breaking 
of the full $\sootn$ symmetry of the model world, the local $\soot \subset \sootn$ 
symmetry acting on the tangent space $\TM_4$ is associated with the linear connection 
1-form $\Gamma(x)$ on $M_4$, which is central to the theory of general relativity and 
is subject to the Bianchi identity $\mbox{D} \ol{\bR} = 0$, while the so(6)-valued 
connection 1-form on $\ul{P}$ is interpreted as the gauge field $Y(x)$ on $M_4$,  
central to the gauge theory arising from the internal symmetry, and is subject to the 
Bianchi identity $\mbox{D} \und{F} = 0$.
 The structures of the external and internal geometry are correlated and the 
corresponding equations of motion constrained as described in sections~\ref{reaic} and 
\ref{subwal}, 
  with self-interactions arising for the gauge fields for the non-Abelian internal 
symmetry.

Further dynamical equations of motion will arise out of the full 10-dimensional 
temporal flow in the broken form of ${D}_{\mu}L(\bv_{10})=0$, by a direct 
generalisation of equations~\ref{dlvfi} and \ref{dlvfib} from the SO(5) model. For the 
$\sootn$ model the symmetry breaking leads to interactions between the gauge field 
$Y(x)$ and the internal degrees of freedom deriving from the components of $\ul{\bv}_6 
\subset \bv_{10}$.
That is, in comparison with equation~\ref{dlvfib}, we have:
\begin{equation}
   D_{\mu}  L(\bv_{10}) = 0 \quad \Rightarrow \quad
    \bv_{10} \cdot \partial_{\mu} \bv_{10} \;\; + \;\;
        \ol{\bv}_4 \cdot A_{\mu} \ol{\bv}_4   \;\; + \;\;
	    \ul{\bv}_6 \cdot Y_{\mu} \ul{\bv}_6  = 0
												   \label{dlvfibte}
\end{equation}
  where $A_{\mu}(x)$ is the external Lorentz connection on $M_4$.
 Through the interactions between the internal fields $Y_{\mu}(x)$ and $\ul{\bv}_6(x)$  
the apparent matter content of the world on the base manifold arises, together with 
its quantum properties, as outlined for equation~\ref{getypsi} and alluded to near the 
opening of this section.

 As described in section~\ref{reaic} the principle bundle $\ul{P} \equiv M_4 \times 
\sox$ is \textit{not} considered here to represent a physical space or spacetime, and 
neither is the associated bundle with homogeneous fibres. In the absence of a 
structure of extra \textit{spatial} dimensions in general the full form of purely 
\textit{temporal} flow $\lvh$  is not required to be associated with a metric 
geometry. The question then concerns the mathematical structure of the
  higher-dimensional forms of $\lv$ of relevance for the physical world. In the 
following chapter a 
  particular 27-dimensional  form  $\lvt$ together with its full symmetry group 
$\hat{G} = \esi$ will be introduced.

  Given the extra dimensions of the full  vector object $\bv_{27} \inn \htho$ the need 
to identify a  Riemannian curvature parametrised over  a locally approximately flat 
4-dimensional base manifold $M_4$ breaks the full $\esi$ symmetry. The geometry on 
$M_4$ drawn out of the underlying structures and symmetries implied in the form  
$\lvt$ can be described generically by the 4-dimensional  relation $- \kappa 
T^{\mu\nu}:=G^{\mu\nu} = f(Y,\bvh)$ of equation~\ref{getypsi}, with $\gmo$, shaping 
the perceptual background of our observable world. The external symmetry, acting on 
the extended manifold $M_4$ itself, is \textit{a priori} essential for perception in 
the world as geometrically described by the linear connection and Riemannian curvature 
which are smoothly dependent upon $x \inn M_4$.
 For the present theory this natural and necessary mechanism of symmetry breaking  
over the manifold $M_4$ forms a significant part of the conceptual framework through 
which the mathematical structures are realised in the physical world.

   For the full theory based on the action of $\esi$ on $\htho$ the internal coupling
   in the final term of equation~\ref{dlvfibte}
   will be replaced by an interaction between internal gauge fields and 
\textit{fermion} fields, where the latter are identified in the internal components of  
$\bv_{27} \inn \htho$ under the action of the external symmetry on $\lvt$ as will be 
explained in section~\ref{extsym}. Hence a particular form of extra dimensions can be 
identified for the present theory which ultimately provides the source for the 
interacting gauge $Y(x)$ and fermion $\psi(x)$ fields, each of which transforms in the 
appropriate way under the Lorentz symmetry on 4-dimensional spacetime, underlying the 
matter and particle effects observed in the real world. 
 A `supersymmetry' is not  required in order to introduce  gauge fields alongside  
fermions fields, together with their mutual interactions, in  the unified theory 
presented here.

 Having identified fermion states the question remains concerning the origin of more 
specific structures of the Standard Model of particle physics, as implemented through 
Lagrangian terms in the form of equation~\ref{lagdym} for example and as  reviewed 
more generally in chapter~\ref{rotsm}. The origin of a series of Standard Model 
properties in the context of the present theory through  the breaking of the full form 
$\lvh$ will be presented in chapters~\ref{chapesb} and \ref{secfd}. The constraints 
implied in the full form $\lvh$ augment the surveillance of the external geometry with 
$\gmo$, described at the end of section~\ref{subwal}, and the need to postulate any 
form of Lagrangian approach will recede further, implying ultimately  that it may be 
avoided entirely.  The full collection of constraints will also 
 be utilised in order to address the origin of quantum phenomena for the present 
theory in chapter~\ref{newapp}.

In the meantime, before considering the empirical implications for observed laboratory 
phenomena,
   in the following chapter we  leave the model worlds behind and motivate 
consideration of $\esi$ as the symmetry group
  acting upon   $\lvt$ as a natural higher-dimensional form of temporal flow.


\pagebreak
\chapter{$\esi$ Symmetry on $\htho$}

\label{esihtho}

\section{Early Formulations}
 \label{earlyo}

  In order to determine the physical effects, observable on the base manifold, of more 
general morphisms of the flow of time through a higher-dimensional form we shall need 
to consider a suitable larger symmetry group acting on an appropriate
  higher-dimensional vector space.  The motivation leading to the identity $\lv$ of 
equation~\ref{lv} as the general mathematical form acting on the $n$ real number 
components of temporal flow $\v$ in an $n$-dimensional vector space was described in 
chapter~\ref{sym}. 
We are particularly interested here in finding such an expression with $n$ somewhat 
larger than four (since the case of the external Lorentz symmetry of the form 
$L(\bv_4)$ in equation~\ref{lorform2}  corresponds to $n=4$) and with a significant 
degree of symmetry. The vector space $\htho$ is the set of $3\times 3$ Hermitian 
matrices over the octonions~\cite{Baez1} with elements:
\begin{equation}
 \label{htho}
   \mcX = 
   \left( \begin{array}{ccc}
       p & \bar{a} & c  \\
       a &   m     & \bar{b}        \\
 \bar{c} &   b     & n 
          \end{array}  \right)   \inn \htho
\end{equation}
  with $p,m,n\inn\rrr$ (here the component labels are chosen to conform with the 
notation in the relevant references, and $n$ here is of course not the dimension of 
any space), $a,b,c\inn\ooo$  and $\bar{a}$ denotes the octonion conjugate of $a$ 
reversing the sign of the 7-dimensional imaginary part (the octonion algebra is 
described in the following section). Hence the vector space $\htho$ is 27-dimensional 
over the real numbers. It is a space with particularly rich symmetry properties 
largely owing to the nature of the 8-dimensional octonion subspaces.

    The dimensions of the vector and spinor representations of the rotation group 
$\mbox{SO}(n)$ converge in the case of $n=8$. That is, as well as the 8-dimensional 
vector representation of SO(8) the 16-dimensional spinor representation reduces to two 
distinct 8-dimensional spinor spaces, dual to each other. The three 8-dimensional 
spaces undergo different SO(8) transformations, however mappings may be defined which 
interchange the transformation behaviour between the three spaces, with a two-to-one 
map from a spinor to the vector representation. The existence of such maps is due to a 
property known as the `principle of triality'~\cite{Gam,Man1} and it is unique to 
spaces of eight dimensions.

   Three such 8-dimensional spaces can be represented by three copies of the 
octonions, in particular under an appropriate SO(8) symmetry operation on the space 
$\htho$ in equation~\ref{htho}, as will be described later around 
equation~\ref{trialabc}. While the actions on the vector and two spinor 
representations differ for particular SO(8) transformations, collectively as three 
sets of transformation actions they are isomorphic by triality and it is a matter of 
convention which octonion space is assigned as the vector or spinor of either kind. 
Further, a 14-dimensional subgroup of the rotation group SO(8) (itself 28-dimensional) 
acts on the three octonion spaces in exactly the same way. This is $\gt$, the 
automorphism group of the octonion algebra. In fact $\gt \subset \mbox{SO}(7)$ as the 
automorphisms only act upon the seven imaginary units of the octonions. (In general an 
algebra automorphism $\Phi$ acts on any two elements $a,b$ of the algebra such that 
$\Phi(a+b) = \Phi(a) + \Phi(b)$ and $\Phi(ab) = \Phi(a)\Phi(b)$, with the order of the 
latter product being reversed in the case of an algebra anti-automorphism such as the 
map $a \to \bar{a}$ of equation~\ref{octabar} described in the following section).

   The elements of the vector space $\htho$ belong to a Jordan algebra for which the 
algebra product is given by:
\begin{equation}
 \mcX \circ \mcY = \frac{1}{2}(\mcX\mcY +\mcY\mcX)
  \label{joralg}
\end{equation}
  with $\mcX,\mcY \inn \htho$ and where $\mcX\mcY$ is the ordinary multiplication of 
the $3\times 3$ matrices, with the order of matrix components in products matching the 
order of the matrices in the product, since in general the components may not commute. 
The Jordan product can also be defined in terms of the triality mappings~(\cite{Baez1}   
p.30). This $\htho$ algebra is known as \textit{the} exceptional Jordan algebra since 
it cannot be expressed in terms of matrices with associative elements (such as real or 
complex numbers).  The algebra itself is commutative but non-associative (as is 
generally the case for all Jordan algebras) with the exceptional Lie group $\ff$ being 
the automorphism symmetry of the algebra.

    However, there is a larger symmetry group involving another structure which can be 
defined on the space  $\htho$ which is of particular interest here. This is a cubic 
norm, or determinant, $\mbox{det}(\mcX) \equiv (\mcX,\mcX,\mcX)$ for $\mcX\inn\htho$ 
which will be presented explicitly in section~\ref{esitran}. 
   A  subspace of the vectors $\bv \inn\rrr^{27}$ map onto elements $\mcX\inn\htho$ 
that satisfy the homogeneous cubic polynomial equation $\mbox{det}(\mcX) = 1$ which 
expresses a form  of the principle relation of equation~\ref{lv} denoted 
$L(\bv_{27})=1$.
  This subspace is locally 26-dimensional and hence may be denoted $S_{26}$, as a 
homogeneous space, following the convention in the opening of section~\ref{thwhf}, 
although here $S_{26}$ represents the full space of temporal flow rather than a purely 
internal fibre space.

 The symmetry of this 27-dimensional form  $\lvt$ corresponds to a group of morphisms 
of the elements of $\htho$ which preserve the unit cubic norm; that is the set of 
actions such as $A_{\lambda}(\mcX)$, parametrised by $\lambda \inn \rrr$, with 
$(A_{\lambda}(\mcX), A_{\lambda}(\mcX), A_{\lambda}(\mcX)) = (\mcX,\mcX,\mcX)$. With 
the identity transformation labelled by $\lambda = 0$ elements of the corresponding 
Lie algebra may be represented by the objects $D \equiv \partial A_{\lambda} / 
\partial \lambda \vert_{\lambda = 0}$. More generally the Lie algebra can be defined 
directly in terms of the set of operators $D$ that annihilate the norm, that is with:
\begin{equation}
  \label{Daaa}
   (D\mcX,\mcX,\mcX) + (\mcX,D\mcX,\mcX) + (\mcX,\mcX,D\mcX) = 0
\end{equation}
    These elements are found to comprise a 78-dimensional Lie algebra of rank 6 (the 
Cartan subalgebra consists of 6 mutually commuting generators); and these properties, 
together with the fact that it has a 27-dimensional representation, lead to the 
identification of the Lie algebra $\esi$ associated with the exceptional Lie group 
$\esi$.
 In fact it is one of four real non-compact forms of this Lie algebra denoted $\esig$ 
(since the Killing form signature is $-26$ $(=26-52)$. In this paper the same upper 
case kernel letter, such as `E',  is used to denote either the group or the algebra, 
depending on the context, for the exceptional Lie groups, although notation such as 
$L(\esi)$ may be used to emphasise the Lie algebra. Lower case kernel letters are used 
to denote a classical Lie algebra, such as so($n$), corresponding to a Lie group, such 
as SO($n$), although again whether a statement refers to the Lie group, its algebra or 
both should generally be clear from the context).

  The first construction of the $\esi$ Lie algebra in terms of action on the space 
$\htho$ dates from 1950 \cite{Chev} and combined the 52-dimensional algebra of 
derivations $D^R$
  of the Jordan algebra $\htho$
  (that is, the generators of the automorphism group $\ff$) 
  with the 26-dimensional set $D^B$ composed of operations of right action on $\htho$ 
by traceless elements of $\htho$ itself. The total set of elements in this 
$(52+26)=78$-dimensional space may be written:
\begin{eqnarray}
   D^{R,B} & = &  D^R  :  \{
     \mbox{$\ff$ automorphism group of $\htho$ Jordan algebra}  \} \; +
                                                          \nonumber \\
           &   &  D^B  :  \{ \mbox{maps }
		      \mcX\inn \htho \to \mcX \circ \mcx, \mbox{ with }      \mcx\inn \htho, 
\; \mbox{tr}(\mcx)=0 \}   \label{drbdecom}
\end{eqnarray}
 It can also be shown that the commutator $[D^B_{\mcx}, D^B_{\mcy}] = D^B_{\mcx} 
D^B_{\mcy} - D^B_{\mcy} D^B_{\mcx}\, \inn D^R$. (The $D^R$ and $D^B$ are analogous to 
the rotations and boosts, respectively, for the Lorentz group, as we shall see later 
in this chapter). All elements of the set $D^{R,B}$ have the property exhibited by $D$ 
in equation~\ref{Daaa} and are therefore associated with an $\esi$ group action that 
preserves det($\mcX$) for any $\mcX\inn \htho$ (\cite{Baez1} pp.44--46).

  Alternatively the Lie algebra $\esig$ can be expressed in terms of 
  mappings induced on $\htho$ by the 14 generators of $\gt$ acting on $\ooo$, 
supplementing 
 the actions of a basis of 64 independent tracefree $3 \times 3$ octonion matrices. 
This construction of a basis for the $\esi$ algebra, dating from the 1960s 
(\cite{Freud} pp.162--164), in terms of a $(14+64)=78$-dimensional decomposition  can 
be denoted by $D^{G,S}$ and is composed of two sets:
\begin{eqnarray}
   D^{G,S} & = &  D^G : \{ \mbox{$\gt$ automorphism group of $\ooo$ algebra} \}
   \;  +                                             \nonumber \\
           &   &   D^S : \{  \mbox{maps $\mcX\inn \htho \to \mcx_0 \mcX + \mcX 
\mcx_0^{\dagger}$, with $\mcx_0\inn \mbox{sl}_0(3,\ooo)$}  \} 
		    \label{dsgdecom}
\end{eqnarray}
  where sl$_0(3,\ooo)$ is the 64-dimensional set of traceless $3\times 3$ matrices 
over the octonions and $D^G$ is isomorphic to the 14-dimensional Lie algebra $\gt$. 
All elements of this combined set satisfy equation~\ref{Daaa} and hence 
$L(\bv_{27})=1$, in the form of det$(\mcX) = 1$, is preserved by the associated $\esi$ 
group action.

   More generally (\cite{Baez1} p.28, \cite{Sudb}) denoting by sl$_0(n,\kkk)$ the set 
of traceless $n \times n$ matrices with entries in the division algebra $\kkk = \rrr, 
\ccc, \hhh$ or $\ooo$ for $n>1$ the commutator of the set sl$_0(n,\kkk)$ is closed 
only if $\kkk$ is commutative and associative (i.e. for $\rrr$ and $\ccc$ only). 
However, in all cases sl$(n,\kkk)$ may be defined to be the Lie algebra of operators 
on $\kkk^n$ \textit{generated} by the elements of sl$_0(n,\kkk)$ under an appropriate 
matrix commutation rule (the Lie algebra $L(\esi)$ is identified with sl$(3,\ooo)$ in 
\cite{Sudb} p.950).  The Lie group SL$(n,\kkk)$ of operators on $\kkk^n$, with an 
associative multiplication even for $\kkk=\ooo$,  may be generated by the elements of 
this sl$(n,\kkk)$ Lie algebra. For the case $n=2$ the group SL$(2,\kkk)$ also has a 
representation on h$_2\kkk$ that preserves the determinant.

   However, this approach of first defining the Lie algebra purely in itself is 
\textit{not} followed here. Rather finite group transformations will be constructed 
first \cite{Man2,Wang,Man4,Man5,Wang2}. Here $\mbox{SL}(n,\kkk)$ will be defined 
principally in terms of a set of group transformations that preserve a particular norm 
on a vector space, essentially by generalisation from $\sltc$ as a determinant 
preserving action on h$_2\ccc$. The need for such a real-valued `norm' is here 
motivated by the form $\lv$. In obtaining the full set of symmetry actions on the form 
$\lv$ it is partly a matter of convention whether this group of transformations is 
given a name of the type $\mbox{SL}(n,\kkk)$. A representation of the corresponding 
Lie algebra $\mbox{sl}(n,\kkk)$ will be defined and derived subsequently through the 
group action on the representation space. In particular the $L(\esi)\equiv \slthoa$ 
Lie algebra will be described in terms of a basis of vector fields on the tangent 
space to the hypersurface $S_{26}$ embedded within the space $\htho$, which itself can 
be considered as a 27-dimensional manifold.

  Associativity is required of any \textit{group} of operations in general and it is 
the case for the elements of $\esi$ which will be described explicitly in 
section~\ref{esitran}. 
While in general multiplication between octonions is non-associative it is possible to 
use them in the construction of algebraic elements such that the multiplication 
defined between these latter elements is in fact associative. 
 Indeed it will  be possible to conceive of the elements of $\esi$ acting on $\mcX 
\inn \htho$ of equation~\ref{htho} in a manifestly associative way represented as a 
subgroup of $\gltr$ acting on $\v_{27} \inn \rrr^{27}$, with $\htho \equiv \rrr^{27}$ 
as vector spaces,  such that $L(\v_{27}) = \det(\mcX) = 1$ is invariant, similarly as 
for all forms of $\lv$ under symmetry operations.  
Since the construction of the Lie group $\esi \equiv \sltho$ here relies on the 
composition properties of the octonions in the following section we first turn to the 
octonion algebra itself.

\section{Octonion Algebra and Geometric Symmetries}
  \label{oaags}

     Having introduced the division algebras in section~\ref{gfotf}  in relation to 
possible multi-dimensional forms of temporal flow for equation~\ref{lv} here we focus 
on the largest such algebra. For the remainder of this chapter we follow 
references~\cite{Man2,Wang,Man4,Man5,Wang2}  in leading from the properties of 
octonions through their relation with Lorentz transformations to the construction of 
the symmetry group $\esi$ and the corresponding Lie algebra. The main reference for 
these latter structures in particular is (\cite{Wang} chapters 3 and 4). The above 
references are extensively reviewed in this chapter, 
 owing to their importance for the present work, and they also provide the source for 
much of the notation adopted here.

 We begin then with a general octonion which, as an element of an eight-dimensional 
vector space, has eight real parameters $\{a_1 \ldots a_8\}$ and can be written:
\begin{equation}
\label{octa}
  a \; = \; a_1 \; + \; a_2\,i \; + \; a_3\,j + \; a_4\,k \; + \; a_5\,{\kl}
        \; + \; a_6\,{\jl}   \; + \; a_7\,{\il} \; + \; a_8\,l
\end{equation}  
  The first term could be written as $a_1e$ with $e\equiv 1$ representing the real 
unit through which real numbers such as $a_1 \inn \rrr$ are embedded in the octonions 
as $a_1e \inn \ooo$. The seven imaginary units in this basis 
$\{i,j,k,\kl,\jl,\il,l\}$, with $i^2=j^2=\ldots=\il^2=l^2 = -1$, are mutually 
anticommuting, with $\il\sp j = - j \sp\il$ etc., with their full algebraic 
composition described in figure~\ref{octmult}.
\vspace{-8pt}
\begin{figure}[htb]  
\centering
\epsfxsize=12cm
\leavevmode
\epsffile[0 0 1223 1093]{\gpath aPfig61e}
\setlength{\unitlength}{25pt}
	    \begin{picture}(5.0,0.0)(0.0,0.0)
    \put(-2.93,1.6){\LARGE $i \!\!\!\;\!\!\: \mbox{\LARGE \_} \!\!\; l$}
	\put(2.35,1.6){\LARGE $k$}
	\put(7.7,1.6){\LARGE $j \!\!\!\;\!\!\: \mbox{\LARGE \_} \!\!\; l$}
	\put(-0.66,6.99){\LARGE $j$}
	\put(2.65,5.75){\LARGE $l$}
	\put(5.55,6.98){\LARGE $i$}
	\put(2.34,12.0){\LARGE $k \!\!\!\;\!\!\: \mbox{\LARGE \_} \!\!\; l$}
	 \end{picture}
\vspace{-30pt}	 
\vspace{-3pt}
\caption{\setb The multiplication of any two octonion units is given by $\pm$ the 
third octonion on the same directed line, with a $+$ (or $-$) sign for composition 
aligned with (or against) the arrow on the line, for example $\kl \sp \il = -j$.}
\label{octmult}
\end{figure}

   Hamilton's quaternions are contained as a subalgebra of the octonions with 
imaginary units $\{i,j,k\}$ composed as $i \sp j = k$ with cyclic permutations as 
represented by the   arrowed circle in figure~\ref{octmult}. The six other arrowed 
lines represent six further equivalent $\hhh$ subalgebras embedded in $\ooo$. As can 
be seen from examples such as $k\sp l = \kl$ the notation for the imaginary units is 
chosen as a mnemonic for these relations, where it should be understood that $\kl$ is 
a single imaginary base unit on equal footing with any of the other six, with $\kl \sp 
k = l$ and so on.      

   While multiplication within any of the seven quaternion subalgebras is associative, 
for example $(\kl \sp k) l = \kl (k \sp l) = -1$, multiplication between any three 
imaginary base units \textit{not} situated on the same line in figure~\ref{octmult} is 
\textit{anti}-associative, with for example $(i\sp j)l = - i(j \;\! l) = +\kl$. Care 
needs to be taken due to the possible ambiguity in expressions involving products of 
octonions due to this lack of general associativity. However, the algebra does satisfy 
the weaker condition that products involving only two distinct octonions $a,b \inn 
\ooo$ are associative, for example $(aa)b = a(ab)$, and hence the octonions form an 
\textit{alternative} algebra.     

   Octonion conjugation is defined as a real linear map $a \to \bar{a}$ on $\ooo$ such 
that for the real unit $a_1 \to a_1$ while for the seven imaginary units $a_h \to 
-a_h$ ($h = 2\ldots 8$). The octonion conjugate of $a \inn \ooo$ in 
equation~\ref{octa} is therefore:     
\begin{equation}
\label{octabar}
  \bar{a} \; = \; a_1 \; - \; a_2\,i \; - \; a_3\,j - \; a_4\,k \; - \; a_5\,{\kl}
        \; - \; a_6\,{\jl}   \; - \; a_7\,{\il} \; - \; a_8\,l
\end{equation} 
 which applies to any product as $\overline{ab} = \bar{b}\bar{a}$ and is hence an 
algebra anti-automorphism. For a given octonion $a$ the norm $\vert a \vert$ is a real 
number defined by:
\begin{equation}
\label{octanorm}
  \vert a \vert^2 \; = \; a\bar{a} \; = \; \sum_{h=1}^8 a_h^2
\end{equation} 
  which applies to any product as $\vert ab \vert = \vert a \vert \vert b \vert$ since 
the algebra is alternative. Hence the norm is compatible with octonion multiplication 
and it is these properties, which also imply the existence of a unique inverse:
\begin{equation} 
  \label{ainvoct}
 a^{-1} = \frac{\bar{a}}{\vert a \vert^2}
\end{equation}
 for any element $a\neq 0$, which make the octonions a `normed division algebra' 
(\cite{Baez1} p.9).  By a theorem of Hurwitz from 1898 only four such algebras, of 
real dimension 1, 2, 4 and 8, exist; these are $\rrr, \ccc, \hhh$ and $\ooo$ which 
hence form a unique set of algebras, as listed in section~\ref{gfotf}. With the 
octonions being the largest normed division algebra and possessing a rich symmetry 
structure they naturally find use in the present context for identifying possible 
forms $\lv$ for temporal flow together with the associated symmetries. In the simplest 
case for $a,b \inn \ooo$ with $\vert a \vert =1$ and $b \equiv \bv$ the composition 
$ab$ with $\vert ab \vert = \vert a \vert \vert b \vert$ provides a set of symmetry 
transformations leaving the form $L(\bv) := \vert b \vert^2 = 1$ invariant.

  Octonion conjugation can be used to extract the real and imaginary parts of $a \inn 
\ooo$ as $\mbox{Re}(a) = \frac{1}{2}(a+\bar{a})$ and $\mbox{Im}(a) = 
\frac{1}{2}(a-\bar{a})$ respectively. While for a complex number $z = x + yi \inn 
\ccc$ the imaginary part is usually defined such that $\mbox{Im}(z) = y$ is itself a 
\textit{real} number, as for example in equation~\ref{optith}, for the quaternion and 
octonion cases the imaginary part is defined as an \textit{imaginary} number, with for 
example $\mbox{Im}(a) = a_2 i + a_3 j + \ldots$ for equation~\ref{octa}, since such an 
object in general involves several distinct imaginary units.
 An inner product for any two octonions $a,b \inn \ooo$ may be defined by:
\begin{equation}
\label{octinner}
  \langle a,b \rangle \; = \; \frac{1}{2}(a\bar{b} + b\bar{a}) \; = \; 
\mbox{Re}(a\bar{b})
                     \; = \; \sum_{h=1}^8 a_h b_h.
\end{equation} 
  For a single octonion $\vert a \vert^2 = \langle a,a \rangle$, while for any two 
octonions $\langle a,b \rangle = \langle b,a \rangle$ and \textit{geometric} 
orthogonality can be defined by the \textit{algebraic} property $\langle a,b \rangle = 
0$. From equation~\ref{octinner} it can be seen that any real element of $\ooo$ is 
orthogonal to any imaginary element and also any pair of anticommuting octonions (such 
as $\{i,j\}$ with $i j = -j i$ etc.) are orthogonal to each other. In general a unit 
imaginary $s$ is an element $s\inn \mbox{Im}(\ooo)$ with unit norm $\vert s \vert =1$ 
which is not necessarily one of the basis units $\{i,j,k\ldots\}$. The real unit $1$ 
together with any two orthogonal imaginary units $s,s'$ define a quaternion subalgebra  
  with basis $\{1,s,s',ss'\} \inn \hhh$,
 which includes any of the seven $\hhh$ subalgebras as described by the seven lines in 
figure~\ref{octmult}. More generally any two non-parallel imaginary units in $\ooo$ 
generate a basis for a quaternion algebra.

   For any octonion $a$ with $\mbox{Im}(a)\neq 0$ the unit imaginary $s$ as the point 
on the 6-sphere of unit imaginary octonions in the direction of $\mbox{Im}(a)$ can be 
identified. Any such $s \inn S^6$, together with the real unit 1, generate a complex 
subalgebra of $\ooo$ with basis $\{1,s\}\inn \ccc$. In particular any $a\inn \ooo$ of 
equation~\ref{octa} may be written as $a = \vert a \vert e^{s\alpha}$, with $\alpha 
\inn \rrr$ and the Euler identity $e^{s\alpha} = \cos \alpha + s \sin \alpha$ applying 
in the complex subalgebra. Since any two octonions involve at most two complex 
subspaces, with bases $\{1,s\}$ and $\{1,s'\}$, it follows from the previous paragraph 
that any calculation involving only two octonions reduces to the case of the 
quaternion algebra, which being associate hence accounts for the alternative property 
of the octonion algebra. 

   As distinct from the `octonion conjugation' $a \to \bar{a}$, for each $q\inn \ooo$ 
with $q\neq 0$ a conjugation map on $a \inn \ooo$ is a linear transformation expressed 
by the following algebraic composition (which is well defined since $\ooo$ is an 
alternative algebra):
\begin{eqnarray}
     \phi_q: \; a & \to & qaq^{-1} \nonumber \\
 \mbox{that is} \qquad \phi_q: \; a & \to & qa\bar{q} \qquad \mbox{for } \vert q \vert 
= 1
   \label{phiconj}
\end{eqnarray}
 where the second expression follows using equation~\ref{ainvoct} and in fact 
describes the complete set of possible transformations since the first expression is 
insensitive to $\vert q \vert$. Selecting $\vert q \vert$ = 1 implies not only $q^{-1} 
= \bar{q}$ but also $q = e^{s\frac{\alpha}{2}}$ where $s\inn\ooo$ is a unit imaginary 
and $\alpha \inn \rrr$. Since
\begin{eqnarray}
  \vert \phi_q(a)\vert & = & \vert q \vert \vert a \vert \vert \bar{q}\vert = \vert a 
\vert \\
  \mbox{and} \qquad \langle a,b \rangle & = &
  \frac{1}{2}(\vert a+b \vert^2 - \vert a \vert^2 - \vert b \vert^2) \label{ababiso}
\end{eqnarray}
  (the latter by equations~\ref{octanorm} and \ref{octinner})  the map in 
equation~\ref{phiconj} represents an \textit{isometry} for the elements of $\ooo$, 
since geometric relations are preserved. This isometry, leaving $\mbox{Re}(a)$ 
invariant and being continuously connected to the identity transformation, represents 
an action of SO(7) upon the seven-dimensional space of imaginary octonions. 

  For the quaternion subalgebra, with $a,q \inn \hhh$ in the basis $\{1,i,j,k\}$, the 
map $\phi_q: a \to qa\bar{q}$, in $\hhh \to \hhh$, with $q = e^{i\frac{\alpha}{2}}$ 
rotates a vector $(a_1,a_2,a_3,a_4) \inn \rrr^4$ by the angle $\alpha$ radians in the 
$(j\mbox{-}k)$ plane (with $\alpha = 2\pi n$, $n\inn \zzz$, being the identity 
transformation). That is, applying equation~\ref{phiconj} and anticommutation for the 
imaginary units:
\begin{equation}
  \phi_{e^{i\frac{\alpha}{2}}}: \; a_1 + a_2 i + a_3 j + a_4 k \; 
    \to  \; a_1 + a_2 i + e^{i\alpha} (a_3 j + a_4 k)     \nonumber 
\end{equation}
\begin{equation} 
     \mbox{and hence} \qquad \left( \begin{array}{c} a_3 j \\ a_4 k \end{array} 
\right) \;  \to  \;  
	\left( \begin{array}{c}(a_3 \cos \alpha - a_4 \sin\alpha) j \\ (a_3 \sin \alpha + 
a_4 \cos\alpha) k
         \end{array} \right)	    \label{quatrot}   
\end{equation} 
   when considered as an \textit{active} transformation relative to a set of constant 
basis elements $\{1,i,j,k\}$, which is the point of view adopted for such 
transformations here
 (in contrast to \textit{passive} transformations such as exemplified in 
equations~\ref{gaugev}--\ref{omeginhomo} and described in sections~\ref{fibre} and 
\ref{cacfc} for gauge transformations on a principle bundle space).
On employing left and right multiplication by independent quaternions of unit norm
  on the full space $\hhh$, with 4 real dimensions,
  the two-to-one cover of SO(4) is obtained, as alluded to in section~\ref{gfotf}.

  Transformations in the space $\mbox{Im}(\hhh)$, in the form of 
equation~\ref{phiconj}, may be constructed about any unit imaginary element (taken as 
$i$ in the example of equation~\ref{quatrot}) as the axis of rotation. In this case 
for quaternions the map $\phi_q$ is a universal two-to-one covering map of $S^3$ into 
the group of rotations SO(3) (elements $q\inn\hhh$ with $\vert q \vert = 1$ describe 
the 3-sphere, with $q$ and $-q$ mapping to the same rotation).
  This is a group homomorphism from the algebraic composition of 
equation~\ref{phiconj} for quaternions with group structure $\phi_r \,\scirc\, \phi_s 
= \phi_{rs}$ into the geometric transformations SO(3) in a 3-dimensional space, which 
is also the automorphism group of the algebra $\hhh$. This group may also be generated 
by composing several actions of the form in equation~\ref{phiconj} requiring $\vert q 
\vert = 1$ \textit{and} $q \inn \mbox{Im}(\hhh)$, that is from the 2-sphere $S^2 
\subset S^3$ alone. Similar compositions of actions will actually be \textit{required} 
for the octonion case in order to construct the full symmetry, as described in the 
following.

 In the case of the octonions the map in equation~\ref{phiconj} for 
$\phi_{e^{i\frac{\alpha}{2}}}$ again fixes the $(1\mbox{-}i)$-plane but now rotates 
all \textit{three} mutually orthogonal planes, corresponding to the three quaternion 
subalgebras containing $i$ identified in figure~\ref{octmult}, simultaneously by 
$\alpha$ radians. Here the full set of maps $\phi_q$, with $q\inn\ooo$  and $\vert q 
\vert = 1$, does \textit{not} form a group homomorphism of the 7-sphere $S^7$ into 
SO(7) since in general there may be no value of $q\inn \ooo$ for which the map 
$\phi_q(a)$ is equivalent to the composition $\phi_r ( \phi_s (a))$, with $r,s \inn 
\ooo$, due to the non-associativity of the octonions. However it is precisely through 
this property of octonion composition that the set of maps $\phi_q$
 of equation~\ref{phiconj} can generate the full Lie group SO(7) by including ordered, 
or \textit{nested}, combinations such as $\phi_r(\phi_s(a))$ on $a\inn\ooo$ as 
elementary symmetry operations.

  In general a representation $R$ on a vector space $V$ is a structure preserving 
homomorphism from group  elements $\{g_1,g_2,g_3,e\} \inn G$ with $g_1g_2 = g_3$ into 
representation matrices with $R(g_1)R(g_2) = R(g_3)$ and $R(e) = \b1$, where the unit 
matrix $\b1$ represents the identity transformation on $V$. Since group structure is 
associative the above actions $\phi_r, \phi_s$, owing to the octonion 
non-associativity, do not technically \textit{represent} the Lie group SO(7).

  The group structure of these transformations can however be seen when these actions 
are instead represented by matrices $R(\phi) \inn \glsr$ acting on the vector space 
$\rrr^7$. Consider the example of $\phi_r(\phi_s(a)) = r(s(a)\bar{s})\bar{r}$ with $r 
= i \inn \ooo$ and $s = l \inn \ooo$. The map $\phi_l \! :  a \to la\bar{l}$ is a 
linear transformation of the components of $\mbox{Im}(a)$, that is $\{a_2,\ldots , 
a_8\}$ of equation~\ref{octa}, which can be represented on $\rrr^7$ by the action of 
the diagonal $7 \times 7$ matrix
 $R(\phi_l) = \mbox{diag}(-1,-1,-1,-1,-1,-1,+1) \inn \glsr$. Similarly the map $\phi_i 
\! :  a \to ia\bar{i}$ is represented by  $R(\phi_i) = 
\mbox{diag}(+1,-1,-1,-1,-1,-1,-1)$. These combine together  as the map 
$R(\phi_i)R(\phi_l) = \mbox{diag}(-1,+1,+1,+1,+1,+1,-1)$, 
 by matrix multiplication,
which does not correspond to any single conjugation action of equation~\ref{phiconj} 
but precisely represents the combined action $a \to i(l(a)\bar{l})\bar{i}$. Hence 
while $i(l(a)\bar{l})\bar{i} \neq
 (i \sp l) a (\bar{l} \sp \bar{i})$, due to the octonion non-associativity, the nested 
action can be representated in $\glsr$ with  
  $R(\phi_i)R(\phi_l) = R(\phi_i \,\scirc\, \phi_l)$,
    and with matrix compositions in general, which fully represents the SO(7) Lie 
group structure. (A similar situation is found for the spinor representation of SO(7) 
obtained from the one-sided composition action $a \to r(s(a))$, and also for the dual 
spinor using right actions alone).

    The octonion algebra provides a way to express these symmetry transformations in a 
compact algebraic form, which \textit{uses} the non-associativity in order to describe 
the full symmetry, and which may be unfolded into a more explicit group representation 
in terms of matrices in GL($n,\rrr$). In fact by using these octonion properties the 
full SO(7) rotation group can be generated with the elements  
$\vert r \vert = 1$ and $r \inn \mbox{Im}(\ooo)$, that is on the 6 sphere $S^6 \subset 
S^7$ alone, as described below.

   Firstly, setting $\alpha = \pm \pi$ for a single action the conjugation map 
$\phi_{e^{r\frac{\pm \pi}{2}}}$, with $r$ an imaginary octonion unit, corresponds to 
rotating the three planes orthogonal to the $(1\mbox{-}r)$-plane by $\pm 180^{\circ}$, 
hence reflecting, or `flipping' these three planes. This can be readily seen since 
$e^{r\frac{\pm \pi}{2}} = \cos \frac{\pi}{2} + r\sin\frac{\pm \pi}{2} = \pm r$ and 
hence equation~\ref{phiconj} is simply a conjugation map by a unit imaginary element 
$r \inn S^6$. For example with $r=i$ the map $\phi_{e^{i\frac{\pi}{2}}}(j) = ij\bar{i} 
= -j$ acts on $j$, as well as each of the other five imaginary units 
$\{k,\kl,\jl,\il,l\}$, as a sign flip. 

  Performing a second reflection based on the \textit{same} $(1\mbox{-}r)$-plane 
naturally cancels the first and leaves no total effect. However in performing the 
second flip with respect to a \textit{different} plane, namely the 
$(1\mbox{-}\small{(} r \cos \frac{\beta}{2} + s \sin \frac{\beta}{2} \small{)})$-plane 
with $s$ a unit imaginary orthogonal to $r$, while the combined reflections still 
cancel for most components a residual rotation by $\beta$ radians in the 
$(r\mbox{-}s)$-plane remains as the net effect on $\rrr^8$. That is the two 
reflections applied to any $a\inn \ooo$ as:
 \begin{equation}
    \phi_{r,s,\frac{\beta}{2}}(a) \; = \;
    ( r \cos \fhb + s \sin \fhb ) (-r \: a \: \mbox{\small{(}} \!\!\!\: - \!\!\!\: 
\bar{r} \mbox{\small{)}})
	(\overline{r \cos \fhb + s \sin \fhb})
   \label{twoflip}
 \end{equation}
 rotates the components of $a$ in the $(r\mbox{-}s)$-plane by $\beta$ radians, 
corresponding to two reflections in two mirror lines in this plane, while giving the 
identity map on the remaining components.  Although two \textit{discrete} flips are 
involved in this equation the total effect on vectors in $\ooo$ is of a rotation in 
the  $(r\mbox{-}s)$-plane  varying \textit{continuously} with the parameter $\beta 
\inn \rrr$, with the identity transformation for $\beta=0$. In the seven-dimensional 
space of $\mbox{Im}(\ooo)$ the 21 possible choices of rotation planes from the 21 sets 
of imaginary base unit pairs for $\{r,s\}$ describes the full Lie group SO(7). The 
14-parameter automorphism group of the octonions, that is the exceptional Lie group 
$\gt$, is contained as a subgroup of this SO(7) as will be discussed in 
section~\ref{esitran}.  

  The first rotation in equation~\ref{twoflip} is taken as $-\pi$, that is with 
$e^{r\frac{- \pi}{2}} = - r$,  followed by the second rotation by $+\pi$. The 
corresponding minus signs for the $-\pi$ rotation in the middle brackets on the 
right-hand side trivially cancel here but the minus sign is needed for the one-sided 
spinor actions $a \to (r \cos \fhb + s \sin \fhb ) (-r (a))$ in order for $\beta=0$ to 
correspond to the identity transformation in the spinor representation. This latter 
expression is also compatible with the identify transformation for the spinor case 
corresponding to $\beta =4\pi n$ with $n \inn \zzz$, rather than any multiple of 
$2\pi$ as for the vector representation of equation~\ref{twoflip}.

\section{Lorentz Transformations on Spacetime Forms}
\label{ltos}

  As well as rotations in spaces with a Euclidean metric, such as the case of SO(7) 
above, composition of division algebra elements can also be used to describe 
transformations in spaces with a Lorentzian metric, such as on the tangent space of a 
spacetime manifold. The content of this section is largely based on 
reference~\cite{Man2}. We begin here with an Hermitian $2 \times 2$ octonion matrix 
$X$ which may be written as:
\begin{equation}
\label{xoct}
  X  =  \left( \begin{array}{cc} t+z  & \bar{a}  \\
                                a   &  t-z   \end{array} \right)  \inn \htwo
\end{equation}
  with $a\inn \ooo$ in the general form of equation~\ref{octa} and $\{t,z\} \inn 
\rrr$, and hence $X$ is 10-dimensional over the real numbers. Since the components of 
$X$, involving only a single octonion $a$, can be taken to lie within a single complex 
subalgebra of $\ooo$ there are no problems with commutativity or associativity in 
unambiguously defining the determinant of the matrix $X$ in the usual way as:
\begin{equation}
  \det(X) \; = \; (t+z)(t-z) - a\bar{a} \; = \; t^2 - a_1^2 - a_2^2 \ldots - a_8^2 - 
z^2
\end{equation}
  This expression has the same form as the square of an invariant interval represented 
by a Lorentz 10-vector $\bx$ (or interval of `proper time' $\tau$), with 
10-dimensional spacetime metric $\eta = \mbox{diag}(+1, -1,\ldots, -1)$, which can be 
written as:
\begin{equation}
  \vert \bx \vert^2 \; = \; \bx^{\tra} \eta \, \bx \; = \; x_0^2 - x_1^2 - x_2^2 
\ldots - x_8^2 - x_9^2
\end{equation}
 Closely analogous structures are obtained for all four normed division algebras, 
$\kkk = \rrr, \ccc, \hhh$ or $\ooo$, with h$_2\kkk$ representing Lorentz vectors in 
$(k+2)$-dimensional spacetime where $k = \dim_{\rrr}(\kkk)$. For example in the 
familiar case of 4-dimensional spacetime a Lorentz 4-vector $(t,x,y,z)$ can be 
represented by:
\begin{equation}
    \label{xcomp}
  X = 
   \left( \begin{array}{cc} t+z   &  x-yi  \\
                               x+yi  &  t-z   \end{array} \right) 
	\; = \;\; t\sigma^0 + x\sigma^1 + y\sigma^2 + z\sigma^3  \;\; \inn \mbox{h}_2 \ccc
\end{equation}
  This case will be considered in more detail in the section~\ref{lsspin} where the 
$\sigma$-matrices are presented in equation~\ref{sigmas}. The above expression can be 
generalised by replacing $\sigma^2 = \binom{0\; -i}{i\;\;\;\; 0}$ in 
equation~\ref{sigmas} with $\sigma^q = \binom{0\; -q}{q\;\;\;\; 0}$ for $q=i,j$ and 
$k$ for the quaternion case or $q=i,j,k,\kl,\jl,\il$ and $l$ for the octonion case of 
equation~\ref{xoct}.

  A Weyl spinor can be expressed as the 2-component object $\theta = \binom{a}{b}\inn 
\kkk \oplus \kkk$, with the Hermitian conjugate  $\theta^{\dagger} = 
(\bar{a}\;\bar{b})$, and hence each spinor has 16 real components for the octonion 
case. As an element of h$_2\kkk$ the square of a spinor:
\begin{eqnarray}
 \theta\theta^{\dagger} & = &   \left(\!\!\! \begin{array}{cc}  \vert a \vert^2 & 
a\bar{b} \\
   b\bar{a} & \vert b \vert^2   \end{array}  \!\!\!\right) \label{ththvec} \\
 \mbox{has} \qquad  \det(\theta\theta^{\dagger}) & = & 0  \label{ththdet0}
\end{eqnarray}
  and hence corresponds to a null-vector in $(k+2)$-dimensional spacetime. 
   The `time' component of this null-vector can be expressed in a scalar spinor 
product $t = \frac{1}{2}\theta^{\dagger}\theta = \frac{1}{2}(\vert a \vert^2 + \vert b 
\vert^2)$, while the time component of a general element of $X \inn \mbox{h}_2 \kkk$ 
is given by $t = \frac{1}{2}\mbox{tr}(X)$, as can be seen in the examples of 
equations~\ref{xoct} and \ref{xcomp}.
  
  Lorentz transformations in $(k+2)$-dimensional spacetime are defined as actions 
$\Lambda$ which preserve proper time intervals, that is with $\vert \Lambda(\bx) \vert 
= \vert \bx \vert$. The subset of actions continuously connected to the identity 
transformation may be composed together to form the Lorentz group SO$^{+}(1,k+1)$. 
Since $\vert \bx \vert^2 \equiv \det(X)$ the rotations and boosts of these geometric 
spacetime symmetries can be associated with algebraic compositions in the relevant 
division algebra which preserve the determinant, and Hermitian property, of h$_2\kkk$. 
To represent a 10-dimensional Lorentz transformation the Hermitian requirement can be 
achieved by a conjugation map on $X\inn \htwo$ with the $2 \times 2$ matrix $M$ which 
is well defined if there is no associativity ambiguity:
\begin{equation}
 \label{mxmass}
     R\! : X \to  MXM^{\dagger} := (MX)M^{\dagger} = M(XM^{\dagger})
\end{equation}
   This in turn is achieved if the components of $M$ all belong to a single complex 
subspace of $\ooo$ (an alternative possibility is for the columns of $\mbox{Im}(M)$ to 
be real multiples of each other \cite{Man2} p.21). In this case $\det(M)$ is well 
defined and the further requirement that $\det(MM^{\dagger}) = 1$ is sufficient to 
ensure that the conjugation map $X \to MXM^{\dagger}$ leaves $\det(X)$ invariant. 
These two-sided transformations on the vector $X$ are required to be 
\textit{compatible} with the one-sided actions on the spinor $\theta$ and its 
Hermitian conjugate  $\theta^{\dagger}$, meaning that there should also be no 
associativity problems in relating these representations as:
\begin{equation}
  \label{compati}
   M(\theta \theta^{\dagger})M^{\dagger}  = (M\theta)(\theta^{\dagger}M^{\dagger}) = 
(M\theta)(M\theta)^{\dag}
\end{equation} 
   where on the left-hand side an octonionic vector is composed as $\theta 
\theta^{\dag}$, which is not a general element of $\htwo$ due to 
equation~\ref{ththdet0}. This compatibility, which will be needed in the following 
section for the $3 \times 3$ matrix case, is satisfied, along with 
equation~\ref{mxmass}, if the components of each individual $M$ all belong to the same 
complex subalgebra of $\ooo$ and also $\det(M) \inn \rrr$. Together with the 
requirement that the vector transformation preserves $\det(X)$ this implies that 
$\det(M) = \pm 1$. A complete set of such transformation matrices is listed in 
table~\ref{mtran45}. 
 
\begin{table}[htb]
\def\var{-10pt}
\centering
 \hspace*{-20pt}
\begin{tabular}{|l|}
 \hline  Category 1: Boosts $B_{t \pqg z}(\alpha),B_{t \pqg x}(\alpha)$ and $B_{t \pqg 
q}(\alpha)$ with: \\
     $ M_{t \pqg z}(\alpha)  = 
  \left(\!\!\! \begin{array}{cc} e^{+\frac{\alpha}{2}}  & 0 \\
                                        0 &  e^{-\frac{\alpha}{2}} \end{array}\!\!\! 
\right)\!, \;$
	 $M_{t \pqg x}(\alpha)  = 
  \left(\!\!\! \begin{array}{cc} \cosh\frac{\alpha}{2} & \sinh\frac{\alpha}{2} \\
                     \sinh\frac{\alpha}{2} & \cosh\frac{\alpha}{2}  \end{array}\!\!\! 
\right)\!, \;$
	 $M_{t \pqg q}(\alpha)  = 
  \left(\!\!\! \begin{array}{cc} \cosh\frac{\alpha}{2} & \! q\sinh\frac{\alpha}{2} \\
                  -q\sinh\frac{\alpha}{2} & \;\;\! \cosh\frac{\alpha}{2}  
\end{array}\!\!\!
					   \right)$	 				
       \\ \\  
 Category 2: Rotations $R_{x \pqg q}(\alpha),R_{x \pqg z}(\alpha)$ and $R_{z \pqg 
q}(\alpha)$ with:  
    \\ 
	$  M_{x \pqg q}(\alpha)  = 
  \left(\!\!\! \begin{array}{cc} e^{-q\frac{\alpha}{2}}  & 0 \\
                                        0 &  e^{+q\frac{\alpha}{2}} \end{array}\!\!\! 
\right)\!, \;$
	 $M_{x \pqg z}(\alpha)  = 
  \left(\!\!\! \begin{array}{cc} \cos\frac{\alpha}{2} & \sin\frac{\alpha}{2} \\
                     -\sin\frac{\alpha}{2} & \cos\frac{\alpha}{2}  \end{array}\!\!\! 
\right)\!, \;$
	 $M_{z \pqg q}(\alpha)  = 
  \left(\!\!\! \begin{array}{cc} \cos\frac{\alpha}{2} & -q\sin\frac{\alpha}{2} \\
                     -q\sin\frac{\alpha}{2} & \cos\frac{\alpha}{2}  \end{array}\!\!\!
					  \right) \!\!$		
      \\ \\ 
	  Category 3: Transverse Rotations $R_{r,s}(\alpha)$ with:  \\
	   $\;\,$
 $M_{r,s2}(\alpha) = \left(\!\!\! \begin{array}{cc} r\cos\frac{\alpha}{2} + 
s\sin\frac{\alpha}{2} & 0 \\
                  0 &    r\cos\frac{\alpha}{2} + s\sin\frac{\alpha}{2}  
\end{array}\!\!\! \right)$		
 \;\; \mbox{nested with} \;\; $M_{r,s1} = \left(\!\!\! \begin{array}{cc} -r & 0 \\
                                        0 & -r \end{array}\!\!\! \right)$
       \\ 
 \hline
  \end{tabular}
  \caption{\setb Three categories of matrices \protect\cite{Man2,Wang} for conjugation 
action on $X\innf \htwo$ preserving $\det(X)$; with $q,r,s \in 
\{i,j,k,\kl,\jl,\il,l\}$ there are $(1+1+7) = 9$ boosts, $(7+1+7)=15$ rotations and 21 
$R_{r,s}(\alpha)$ transverse rotations (where the subscript $r,s$
 denotes the ordered pair $\{r,s\}$ of imaginary units) representing the 
45-dimensional group of Lorentz transformations on a 10-dimensional spacetime.}

\label{mtran45}
\end{table} 
  
   In the first two categories $\det(M) = +1$ for each of the 24 actions. The third 
category is a simple $2 \times 2$ diagonal matrix form of equation~\ref{twoflip}, with 
the parameter $\beta$ replaced by $\alpha$. The action of the category 3 matrices is 
ordered by nesting the conjugation as:
\begin{equation}
 \label{rrsmat}
 R_{r , s}(\alpha)X = 
  M_{r , s2} \, (M_{r , s1} \, (X) \, M_{r , s1}^{\dag}) \, M_{r , s2}^{\dag}
\end{equation}
   with $\det(M_{r , s1}) = \det(M_{r , s2}) = -1$. Since the latter matrices are 
always combined in pairs, and hence are analogous to the action of a single matrix 
with a determinant of $+1$, the full group of transformations is denoted $\sltwoo$. It 
is composed of the 45 actions in table~\ref{mtran45}, each of which describes a 
one-parameter subgroup with $R(\alpha)R(\beta) = R(\alpha + \beta)$ and each of which 
represents transformations in a single 2-dimensional plane in 10-dimensional 
spacetime.

     For each of the 45 transformations with $\alpha = 0$ and $\alpha = 2\pi$ it can 
be seen that $M= \binom{1\;\;0}{0\;\;1}$ and $M = \binom{-1\;\;0}{\,0\!\;\;-1}$ 
respectively (this is effectively true for the category 3 case since these can be 
expressed by conjugation with a single such matrix $M$ for these $\alpha$ values). 
Hence $\sltwoo$ is the double cover of the 10-dimensional Lorentz group $\sootn$, that 
is $\sltwoo \to \sootn$ is a two-to-one homomorphism with kernel $\{M = \pm \b1_2\}$, 
where $\b1_2$ is the $2 \times 2$ identity matrix, since both cases for $M$ give the 
identity transformation on $X$ due to the two-sided action in equation~\ref{mxmass}.
  
   A number of subgroups may also be identified. The subgroup leaving $\mbox{tr}(X)$ 
invariant, composed of the 36 category 2 and 3 transformations, defines $\sutwoo$ 
which is the two-to-one cover of the purely rotational Lorentz subgroup SO(9), leaving 
the $t$-component in equation~\ref{xoct} invariant. In turn the 21 category 3 
transformations alone form the Spin(7) subgroup as the double cover of SO(7). The 
structure of these subgroups, including the SO(8) obtained by augmenting the SO(7) 
with an additional 7 $R_{x \pqg q}(\alpha)$ actions from category 2, will also be 
important for enlarging beyond $\sltwoo$ for the $3 \times 3$ matrix case in the 
following section.

  Finally in this section we note that for the quaternion case, obtained by 
restricting all transformations in table~\ref{mtran45} for $q,r,s \inn \{i,j,k\}$, 
there remain 15 transformations (5, 7 and 3 for category 1, 2 and 3 respectively) 
acting on h$_2\hhh$ forming $\sltwoh$ as the double cover of the Lorentz group 
$\sootf$ on 6-dimensional spacetime. Here, loosening the restriction $\det(M)= \pm 1$ 
for the transformation matrices, each of the three transverse rotations can  be 
achieved by a single unnested conjugation map such as:
\begin{equation}
    R_i(\alpha)(X) = M_i X M_i^{\dagger} \quad \mbox{with} \quad
	 M_i(\alpha)  = 
  \left(\!\!\! \begin{array}{cc} e^{i\frac{\alpha}{2}}  & 0 \\
                                        0 &  e^{i\frac{\alpha}{2}} \end{array}\!\!\! 
\right)
  \label{quatrot2}
\end{equation}
  which acts on the quaternion component $a\inn \hhh$ of $X=\binom{p \;\:\bar{a}}{a \; 
m} \inn \htwh$ by fixing the $(1\mbox{-}i)$-plane while performing a rotation in the 
$(j\mbox{-}k)$-plane of $\alpha$ radians as was described in equation~\ref{quatrot}. 
Taking a similar form to equation~\ref{quatrot2} the two actions $R_j(\alpha)$ and 
$R_k(\alpha)$ rotate the $(k\mbox{-}i)$-plane and $(i\mbox{-}j)$-plane respectively. 
This is possible for the quaternions since there is only \textit{one} imaginary plane 
orthogonal to each imaginary base unit. This is unlike the case for the octonions in 
which the nested transverse rotations are needed to describe all 21 such single plane 
rotations (as explained towards the end of the previous section) and hence account for 
the complete subgroup $\mbox{SO}(7) \subset \sootn$. In addition the restriction 
$\det(M)= \pm 1$ is imposed for the octonion case in order to meet the compatibility 
requirement of equation~\ref{compati}   as will be needed for extension to the $3 
\times 3$ case as noted after that equation.

   For the case of 4-dimensional spacetime the six Lorentz transformations are 
represented on h$_2\ccc$ by the six category 1 and 2 matrices $M$ in 
table~\ref{mtran45} with $q$ taking a single value such as $i$. This set of actions 
with $\det(M) = +1$ forms the group $\sltc$ as the double cover of the Lorentz group 
$\soot$, as will be studied in more detail in section~\ref{lsspin}. 
 In this case there are \textit{no} imaginary orthogonal planes for the transformation 
of equation~\ref{quatrot2} to rotate and the action $R_i(\alpha)$, which may be 
considered as a residue from the $2 \times 2$ matrix cases for $\hhh$ and $\ooo$, not 
only preserves $\det(X)$ but also leaves each \textit{component} of any $X \inn \htwc$ 
unchanged. In this sense the action in equation~\ref{quatrot2} may be interpreted as 
an \textit{internal} $\uo$ symmetry,  relative to the \textit{external} Lorentz 
symmetry of 4-dimensional spacetime, as will be relevant for the case of the embedding 
$\htwc \subset \htho$ in section~\ref{intsym}.


 
\section{$\esi$ Transformations on a Form of Time}
 \label{esitran}

  In this paper the  emphasis is on symmetries of forms of multi-dimensional temporal 
flow $\lv$, that is \textit{isochronal} symmetries as introduced in 
section~\ref{gfotf}, rather than on \textit{isometries} of a higher-dimensional space 
or spacetime, as described for equations~\ref{phiconj}--\ref{ababiso} and in the 
previous section for example.
 The  $\sltwoo$ action preserving $\det(X)$ with $X \inn \htwo$, described in the 
previous section, can be interpreted in either way, but there is no reason to restrict 
multi-dimensional forms of $L(\bv)$ to have such a spacetime interpretation, as it 
does in taking the quadratic form of $\det(X)$ for example. Further, with $\ooo$ being 
the largest division algebra, there is no clear extension of this construction based 
on $\htwk$ to a higher-dimensional \textit{spacetime} symmetry. This leads to the 
consideration of the extension of $\htwo$ to the 27-dimensional space of $3 \times 3$ 
Hermitian octonion matrices $\htho$ which has richer symmetry properties while still 
possessing an underlying structure appropriate for a form of  \textit{temporal} flow. 

   An element ${\mathcal X} \inn \htho$ of equation~\ref{htho} may be written as 
(again closely following \cite{Wang} chapters 3 and 4 together with 
\cite{Man4,Man5,Wang2} and generally adopting the notation therein):

\begin{equation}
  \label{xoct3}
  {\mathcal X} \; = \;     
	\left( \!\!\!\!\!\!\!\!\;\! \begin{array}{cc} 
               \begin{array}{cc}   p \,\, & \; \bar{a}  \\ a \,\, & \; m \end{array} 
\!\!\!\!\!\!   &
          \!     \begin{array}{c}    c  \\  \bar{b}   \end{array} \!\! 
				                         \\  
        \;\;\,\,\,\; \bar{c} \;\;\;\;\;\:\, b \!\!\!\!\!\!\!  \begin{array}{cc}        
&   \end{array}  &	  
		  \:  n      \end{array}  \right) \; = \;
	\left( \begin{array}{c|c} 
        \,\,\,\, X                \begin{array}{cc} &  \\  &  \end{array} \!\!\!   &
        \,  \theta  \begin{array}{cc} &  \\  &  \end{array} \!\!\!\!\!\!\!\!\!\! 
				                         \\  \hline
        \,\,\,\,\,\, \theta^{\dagger} \!\! \begin{array}{cc}        &   \end{array}   
&	  
		\,  n      \end{array}  \right)   
	\;	 \inn \htho  
\end{equation}
    with $p,m,n \inn \rrr$ and $a,b,c \inn \ooo$, while $X$ and $\theta$ have the 
structure of octonionic $2 \times 2$ vectors (equation~\ref{xoct}) and $1 \times 2$ 
spinors respectively, familiar from the previous section.
      
	  Under the Jordan product of equation~\ref{joralg} elements ${\mathcal X,Y} \inn 
\htho$ form the exceptional Jordan algebra. However it is the structure of a cubic 
norm, or determinant, which may be defined on $\htho$, without any ambiguity due to 
the non-associativity of the octonions, that is of interest here. The cubic norm is a 
homogeneous polynomial form in the components of $\htho$ as a mapping $\mcX \to 
\det({\mathcal X}) \inn \rrr$ into the real numbers, and hence has the correct 
structure for a form of $\lv$. This determinant may be expressed in several equivalent 
ways including:
\begin{eqnarray}
  \det({\mathcal X})  & = & \det(X)n + 2X\cdot (\theta\theta^{\dag})   \label{detx3}  
\\
                      & = & pmn - p\vert b \vert^2 - m\vert c \vert^2 - n\vert a 
\vert^2  
                            + 2 \mbox{Re}(\bar{a}\bar{b}\bar{c})     \label{detpmn}
\end{eqnarray}
   where the 10-dimensional Lorentz inner product $X\cdot Y = 
\frac{1}{2}(\mbox{tr}(X\circ Y) - \mbox{tr}(X)\mbox{tr}(Y))$, with $X,Y \inn \htwo$, 
in the first line together with equation~\ref{xoct3} can be used to derive the second 
line in which the cubic composition of components, consistent with the homogeneous 
form of equation~\ref{lv}, is explicitly seen.

   The $2 \times 2$ matrices $M$ of $\sltwoo$ actions listed in table~\ref{mtran45} 
can be embedded in the upper-left corner of $3 \times 3$ matrices ${\mathcal M}$ to 
obtain the conjugation action for the $3 \times 3$ case $R: \mcX \to \mcM \mcX 
\mcM^{\dag}$ with:
\begin{equation}
   \label{mxm3}
         \mcM \mcX \mcM^{\dag}     = 
		\left( \begin{array}{c|c} 
        \,\,\,\, M                \begin{array}{cc} &  \\  &  \end{array} \!\!\!   &
        \,     0  \begin{array}{cc} &  \\  &  \end{array} \!\!\!\!\!\!\!\!\!\! 
				                         \\  \hline
        \,\,\,\,\,\,  0  \!\! \begin{array}{cc}        &   \end{array}   &	  
		\,  1      \end{array}  \right)  \!\!
		\left( \begin{array}{c|c} 
        \,\,\,\, X                \begin{array}{cc} &  \\  &  \end{array} \!\!\!   &
        \,  \theta  \begin{array}{cc} &  \\  &  \end{array} \!\!\!\!\!\!\!\!\!\! 
				                         \\  \hline
        \,\,\,\,\,\, \theta^{\dagger} \!\! \begin{array}{cc}        &   \end{array}   
&	  
		\,  n      \end{array}  \right)  \!\!
		\left( \begin{array}{c|c} 
        \,\,\,\, M                \begin{array}{cc} &  \\  &  \end{array} \!\!\!   &
        \,  0  \begin{array}{cc} &  \\  &  \end{array} \!\!\!\!\!\!\!\!\!\! 
				                         \\  \hline
        \,\,\,\,\,\, 0 \!\! \begin{array}{cc}        &   \end{array}   &	  
		\,  1      \end{array}  \right)^{\!\!\mbox{\large $\dag$}}  =   
		\left( \begin{array}{c|c} 
       \!  MXM^{\dag}   \!\!\!\!\! 
	           \begin{array}{cc} &  \\  &  \end{array} \!\!\!   &
       \!   M\theta  \!\!\! \begin{array}{cc} &  \\  &  \end{array} 
\!\!\!\!\!\!\!\!\!\! 
				                         \\  \hline
        \,\, \theta^{\dagger}M^{\dag} \!\!\!\!\!\! \begin{array}{cc}        &   
\end{array}   &	  
		  \,  n      \end{array}  \right)  
\end{equation}
  This expression contains the vector $X \to R(X) = MXM^{\dag}$, spinor $\theta \to 
R(\theta) = M\theta$ and scalar $n \to 1 n$ representations of $\sltwoo$, each 
transforming in the appropriate way with the form of the action $R$ determined 
correspondingly. These transformations respect the $3 \times 3$ block structure, as do 
nested compositions in augmenting the $2 \times 2$ matrix actions such as 
equation~\ref{rrsmat} to expressions of the form:
\begin{equation}
 \label{nest3}
    \mcX \: \to \: R(\mcX) \, = \, 
\mcM_n(\ldots(\mcM_1(\mcX)\mcM_1^{\dag})\ldots)\mcM_n^{\dag}
\end{equation}
   which acts, for example, on the spinor as $\theta \to R(\theta) = 
M_n(\ldots(M_1(\theta)))$. As well as preserving $\det(X)$ with $X\inn\htwo$ the 45 
actions of $2\times 2$ matrices $M$ from table~\ref{mtran45} when embedded in the 
$3\times 3$ matrices $\mcM$ for equation~\ref{mxm3} also preserve $\det(\mcX)$ for 
$\mcX \inn \htho$ since, from equation~\ref{detx3}:
\begin{eqnarray}
   \det(R(\mcX)) & = & \det(R(X))n + 2R(X)\cdot (R(\theta)R(\theta^{\dag}))  \nonumber  
\\
                 & = & \det(R(X))n + 2R(X)\cdot R(\theta\theta^{\dag})    \nonumber  
\\
                 & = & \det(X)n + 2X \cdot \theta\theta^{\dag}  \nonumber \\
                 & = & \det(\mcX)      \label{hthoinv}
\end{eqnarray}
     where the second equality is a result of `compatibility', and motivates the 
introduction of this requirement in equation~\ref{compati}, and the third equality 
follows from the Lorentz symmetry of the $\sltwoo$ action. It is also by compatibility 
that the 45 $\sltwoo$ transformations act as one-parameter subgroups on the spinor 
$\theta$ (given the minus signs for the $M_1$ components for the transverse rotations, 
originating in equation~\ref{twoflip}, as for $M_{r , s1}$ in table~\ref{mtran45}) as 
well as on the vector $X$.

    These $\sltwoo$ actions, called Lorentz transformations when acting on $X\in 
\htwo$ representing 10-dimensional spacetime, also identify 45 one-parameter subgroups 
acting on $\mcX\inn \htho$, with $R(\alpha)R(\beta)\mcX = R(\alpha + \beta)\mcX$,  
preserving $\det(\mcX)$ (where $R(\alpha)\mcX$ denotes a particular action $R(\mcX)$,
 for example from table~\ref{mtran45}, for a particular transformation parameter 
$\alpha$). Hence these 45 actions are one-parameter subgroups of $\esi := \sltho$ 
which is defined as the group of symmetry transformations under which the determinant 
on $\htho$ is invariant. Again we emphasise that the key here is the structure of a 
higher-dimensional form of temporal flow which, while necessarily containing a 
4-dimensional form \textit{perceived} as spacetime, does not itself need to possess a 
higher-dimensional spacetime interpretation.

   The exceptional Lie group $\esi$ is 78-dimensional, as described in 
section~\ref{earlyo} and hence the 45 actions adopted from $\sltwoo$ represented on 
$\htho$ is only part of the full symmetry picture. However the scope of the $\sltwoo$ 
action can be enlarged by noting that there are \textit{three} similar and natural 
ways to embed the vector $X$, spinor $\theta$ and scalar $n$ representations of 
$\sltwoo$ in the $3 \times 3$ matrix $\mcX \inn \htho$. The original `type 1' action 
described in equation~\ref{mxm3} for the embedding depicted in equation~\ref{xoct3} 
can be written more explicitly in terms of the matrix components:
\begin{equation}
  \label{type1}           
	\mcM^{(1)} = 
	\left( \begin{array}{cc|c}  M_{11} & M_{12} & 0  \\
	                            M_{21} & M_{22} & 0   \\  \hline
					                0 &  0   &  1  \end{array}  \right) 	
		\quad \mbox{acting on} \quad							
	\left( \begin{array}{cc|c}  X_{11} & X_{12} & \theta_1  \\
	                            X_{21} & X_{22} & \theta_2   \\  \hline
					      \bar{\theta}_1 &  \bar{\theta}_2   &  n  \end{array}  
\right)	  
\end{equation}
  Maintaining the variables $p,m,n \inn \rrr$ and $a,b,c \inn \ooo$  in the same 
component locations of the $3\times 3$ matrix $\mcX \inn \htho$ in 
equation~\ref{xoct3} their placement within the $2 \times 2$ vector $X=\binom{X_{11} 
\; X_{12}}{X_{21} \; X_{22}}$
 and $1\times 2$ spinor $\theta=\binom{\theta_1}{\theta_2}$ under $\sltwoo$ may be 
reassigned by permuting the components of the matrices $\mcM$ as follows:      
\begin{equation}
  \label{tcycle3}
 \mcM^{(a)} =\mcT\mcM^{(b)}\mcT^{\dag} \quad \mbox{for } (a,b)=(2,1),(3,2),(1,3) \quad 
\mbox{with } 
     \mcT = 
	\left( \begin{array}{ccc} \;  0 \; & \; 0 \; & \; 1 \; \\
	                             1 & 0 & 0   \\  
					             0 & 1 & 0  \end{array}  \right)
\end{equation} 
  With $\mcT^{\dag} = \mcT^{-1} \: ( = \mcT^2)$ it can be seen that $\det(\mcT \mcX 
\mcT^{\dag}) = \det(\mcX)$ and since $\det(\mcT) = 1$ the action $\mcX \to \mcT \mcX 
\mcT^{\dag}$ can itself be considered as a transformation of the $\sltho$ symmetry. 
 The matrices $\mcM^{(2)}$  and $\mcM^{(3)}$ then correspond to `type 2' and `type 3' 
transformations respectively with:
\begin{equation}
  \label{type2}          
	\mcM^{(2)} = 
	\left( \begin{array}{c|cc}  1 & 0 & 0  \\  \hline
	                            0 & M_{11} & M_{12}    \\  
					            0 & M_{21} & M_{22}  \end{array}  \right) 
		\quad \mbox{acting on} \quad	
		\left( \begin{array}{c|cc} p & \bar{\theta}_1 & \bar{\theta}_2 \\   \hline
	                         \theta_1  & X_{11} & X_{12}    \\  
					         \theta_2  & X_{21} & X_{22}  \end{array}  \right)
\end{equation}
and
\begin{equation}
  \label{type3}          
	\mcM^{(3)} = 
	\left( \begin{array}{c|c|c} M_{22} & 0 & M_{21}  \\  \hline
	                              0 &    1 &   0    \\   \hline  
					            M_{12} & 0 & M_{11}  \end{array}  \right) 
		 \quad \mbox{acting on} \quad	
		 \left( \begin{array}{c|c|c} X_{22} & \theta_2 & X_{21} \\   \hline
	                         \bar{\theta}_2  & m &  \bar{\theta}_1   \\   \hline
					            X_{12} &  \theta_1  &  X_{11}  \end{array}  \right) 
\end{equation}

The three $\mcM^{(a)}$  represent three embeddings of the $2 \times 2$ matrix actions 
of table~\ref{mtran45} into a $3 \times 3$ matrix form acting on the same $\htho$ 
components of equation~\ref{htho}.
  Each type $1,2$ or 3 action, even for the nested case of equation~\ref{nest3} with 
$\mcM \to \mcM^{(a)}$ for $a=1,2$ or $3$, respects the corresponding block structure 
in equation~\ref{type1}, \ref{type2} or \ref{type3} respectively. Indeed the type 2 
and 3 cases are effectively obtained by a simple symmetric permutation of the three 
octonion and three real entries in $\htho$ under the original type 1 action of 
equation~\ref{mxm3}, and hence for all three types of transformation $\det(\mcX)$ is 
invariant, as was shown for the type 1 case in equation~\ref{hthoinv}.
(In addition to the discrete actions of equation~\ref{tcycle3} continuous type 
transformations may also be defined as described in \cite{Wang} section 4.4).

  With three possible embeddings of the 45-dimensional $\sltwoo$ transformations  
  there are now a total of $3 \times 45 =135$ $\mbox{det}(\mcX)$-preserving 
one-parameter subgroup actions for $\esi := \sltho$, which cannot be independent since 
$\esi$ is known to be a 78-dimensional group. A  basis for the $\esi$ actions on 
$\htho$ may be obtained by requiring linear independence at the Lie algebra level.
However a $G=\esi$ manifold is not well-defined in terms of the space of $3\times 3$ 
matrices $\mcM$ upon which to identify tangent vectors with Lie algebra elements.
 This is in contrast to a case such as $G=\soth$ represented by real $3 \times 3$ 
matrices $R \inn \mbox{SO}(3)$ acting on vectors $\bv_3 \inn \rrr^3$. In this case the 
space of matrices $R$, with $RR^T=\b1_3$ and $\mbox{det}(R)=1$ describing 
topologically the 3-sphere $S^3$ with antipodal points identified, defines the group 
space $G$ upon which tangent vector fields may represent the Lie algebra, as was 
depicted for the general case in figure~\ref{grightran}. The Lie algebra may  be 
described in terms of left-invariant vector fields on the group manifold or in terms 
of the tangent vectors at the identity $e\inn G$ through the isomorphism $L(G) \equiv  
T_eG$. (An example for the latter case was listed in the set of Lie algebra elements 
$\{ {L}_{p \pqg q}\}$ of equation~\ref{m3} for $G=\soth$).

  This situation can be understood by considering how a Lie group manifold, on the 
tangent space of which the Lie algebra may be defined, might also be identified for 
the Lorentz groups in $(k+2)$-dimensional spacetime, 
   with $k = \dim_{\rrr}(\kkk)$ and $\kkk = \ccc,\hhh,\ooo$,
    represented by the action of $\mbox{SL}(2,\kkk)$ on h$_2\kkk$ matrices. In the 
first case for $\kkk = \ccc$ since complex $2 \times 2$ matrices have $(4\times 2) = 
8$ real parameters and $\det(M) = 1 \inn \ccc$ represents two constraints on the 
matrices $M\inn \sltc$ these actions are described by $(8-2) = 6$ real parameters, 
which equals the dimension of the Lorentz group $\soot$. This set of $2\times 2$ 
matrices can take the form of the first six matrices in table~\ref{mtran45} for $q=i$ 
as described at the end of section~\ref{ltos}. Hence these six degrees of freedom of 
the matrices $\sltc$ fully describe the corresponding group manifold $G \equiv \sltc$ 
(as the double cover of $\soot$), upon which the Lie algebra of tangent vector fields 
may be constructed (having the same Lie algebra structure as $\soota$).
 This Lie algebra, in terms of tangent vectors at the identity $e \inn G$, will be 
explicitly listed as the set of $2\times2$ matrices $\{\dot{M}\}$ in 
equations~\ref{mrotgen} and \ref{mboogen} of section~\ref{extsym}.

  For $2 \times 2$ quaternion matrices under the constraint $\det(M) = 1 \inn \hhh$ 
there are $(4 \times 4)-4 =12$ free parameters for $M \inn \sltwoh$, insufficient 
alone to describe the 15-dimensional Lorentz group $\sootf$. However, including the 3 
transverse rotations via the $2\times 2$ matrix actions of equation~\ref{quatrot2} 
(one for each imaginary unit of $\hhh$) by loosening the constraint on the matrix 
determinant to $\vert \! \det(M) \vert = 1 \inn \rrr$ results in a total of $(4 \times 
4)-1 =15$ free parameters. Hence, as for the complex case, a subset of $2\times 2$ 
quaternion matrices can be identified with a group manifold structure as the double 
cover of the Lorentz group, here for a 6-dimensional spacetime, and the tangent space 
to this manifold hence used to describe the Lie algebra $\mbox{so}^+(1,5)$.

  However for the 45-dimensional Lorentz group in 10-dimensional spacetime the maximum 
of $(4\times 8) = 32$ parameters available in a $2\times 2$ octonion matrix are 
clearly insufficient to parametrise the full group, and hence the matrices of 
$\sltwoo$ in table~\ref{mtran45} cannot immediately be related to a Lie group manifold 
as they could for the complex ($\sltc$ with $\det(M) = 1$) and quaternion ($\sltwoh$ 
with $\vert \! \det (M) \vert = 1$) cases. Indeed this is why nested $\sltwoo$ actions 
are required in the octonion case to make up the extra transformations. Similarly for 
$\sltho$, with a maximum $(9\times 8) = 72$ real parameters available in the $3\times 
3$ octonion matrices, such objects are insufficient to represent the full 
78-dimensional group manifold for $\esi$.

   The nested action $\mcX \to \mcM_2 ( \mcM_1 \mcX \mcM_1^{\dag} ) \mcM_2^{\dag}$, in 
the form of equation~\ref{nest3}, for the case in which the elements of the matrices 
$\mcM_1$ and $\mcM_2$ belong to the same $\ccc \subset \ooo$ subspace
\textit{is} an associative composition, that is it is equal to $(\mcM_2 \mcM_1) \mcX 
(\mcM_1^{\dag} \mcM_2^{\dag})$. This is because each matrix  element of $\mcX$ 
involves at most only one further complex subspace, and hence each multiplicative 
action on these elements in the linear transformation on $\mcX$ takes place in an 
associative quaternion subalgebra. Hence these particular cases of nested 
transformations do behave like a group representation. More generally however, and as 
for the case of SO(7) generated by composition of the $\phi_q$ maps with $q\inn 
\mbox{Im}(\ooo)$ in equation~\ref{phiconj} as described in section~\ref{oaags}, here 
there does not exist a group homomorphism of the full set of $\esi$ transformations 
into the set of $3 \times 3$ octonionic matrices $\mcM^{(a)}$.

  However, associative group matrices could be constructed here by representing the 
linear transformations of the $\esi$ symmetry by $27 \times 27$ matrices in $\gltr$ 
acting on the space $\rrr^{27} \equiv \htho$, as was the case for SO(7) represented by 
matrices in $\glsr$ acting on $\rrr^7\equiv \mbox{Im}(\ooo)$ in section~\ref{oaags}. 
Indeed with such large matrices there is
plenty of freedom in which to express the full symmetry with elements $R(g) \inn 
\gltr$ which naturally form an associate algebra and with $R(g_1)R(g_2) = R(g_1 g_2)$ 
composing as a true representation of $\esi$.

 Given the 135 one-parameter subgroup actions on $\htho$,
 collectively implied in equations~\ref{type1}, \ref{type2} and \ref{type3},
  it would be straightforward, although laborious, to construct 135 matrices in 
$\gltr$ acting upon $\rrr^{27}$, with the latter containing the 27 parameters of an 
element of $\htho$ in equation~\ref{htho} drawn out into the real column vector 
$(p,m,n,a_1,\ldots,c_8)^T$. All such actions would preserve the cubic norm of 
equation~\ref{detpmn} considered as a map $\rrr^{27} \to \rrr$. 
The multiplication of such elements of $\esi$ represented as matrices in $\gltr$ is 
clearly associative, as only $\rrr$-valued matrices are involved. Together with the 
identity element given by the unit matrix $\b1_{27}$ and an inverse obtained for any 
matrix by reversing the transformation with real parameter $\alpha \to -\alpha$, the 
Lie group structure is evident. Combinations of the 135 one-parameter subgroup actions 
would carve out a $G \equiv \esi$ submanifold (for this non-compact real form of 
$\esi$) embedded within the $(27\times 27)$-dimensional space of $\gltr$.

    In principle left-invariant tangent vector fields, generated by \textit{right} 
translations on $G$ and associated with the one-parameter subgroups, could be 
constructed upon this $\esi$ group manifold, as depicted generically in 
figure~\ref{grightran} on a group manifold, and linear dependency used to reduce these 
to a basis set of 78 vector fields to describe the $\esi$ Lie algebra.
Hence in this representation the Lie algebra may also be identified in terms of the 
transformation matrices themselves, in the form of elements $D \equiv \partial 
A_{\lambda} / \partial \lambda \vert_{\lambda = 0}$ as described before 
equation~\ref{Daaa}, here with $A\inn \gltr$.
 Alternatively the \textit{left} translations of these symmetry transformations on 
$\rrr^{27}$ may be associated with vector fields in the tangent space $T\rrr^{27}$ 
which also represent the Lie algebra generators of the symmetry. This construction 
applies generally (see also the discussion in the opening of section~\ref{thwhf}) -- 
for example in the case of $\soth$ acting on $\rrr^3$ the Lie algebra $L(\soth)$ may 
be represented by vector fields in the space $T\rrr^3$ tangent to the 2-sphere $S^2$.

  This latter possibility of employing the left or right action of $G$ on the 
representation space itself to construct the Lie algebra can be employed for the 
representation of  $\esi$ of relevance here, that is on the space $\htho$.
Indeed the theoretical motivation for studying $\esi$ here is precisely owing to its 
representation on $\htho$, together with the subgroup representations on subspaces of 
$\htho$ obtained under symmetry breaking, rather than the pure $\esi$ group structure 
in itself. It is the fact that the space $\htho$ with unit determinant has the 
appropriate structure for a form of temporal flow $\lvt$ that provides the primary 
motivation, with $\esi$ identified in turn as the corresponding symmetry group.
 This symmetry is expressed in a very compact cubic form as the determinant preserving 
actions on $\htho$, such as described in equation~\ref{mxm3}, and indeed the origin of 
the \textit{very high} degree of symmetry, involving the triality relation for the 
largest division algebra $\ooo$, is evident explicitly in this form. These structures 
would be far from manifest in a $27 \times 27$ real matrix representation. The 
non-associativity of the octonion algebra is employed in folding the full set of 
$\esi$ actions into this highly compact $3 \times 3$ matrix form.

   For subgroups it will be possible to `straighten-out' or unfold this action into 
familiar group representation form. This will be the case for the  broken symmetry 
components, involving the external Lorentz group (in the form of the $\sltc$ subgroup 
already described above) and internal symmetry groups, as we shall study in 
chapter~\ref{chapesb} in comparison with the Standard Model of particle physics as 
reviewed in chapter~\ref{rotsm}. Since the representations of these subgroups are to 
be identified in the components of $\htho$, which is ultimately motivated as the space 
underlying $\lvt$, the tangent space $T\htho$ provides an apt arena for describing the 
$\esi$ Lie algebra. 

The homomorphism of the Lie algebra $L(\esi) \equiv \slthoa$ into the space of vector 
fields in $T\htho$, the tangent space to the 27-dimensional manifold of $\htho$, is in 
fact an isomorphism since the group action of $\sltho$ on $\htho$ is effective.
This isomorphism is used both to identify individual $\esi$ generators and also, as 
described in the following section, the Lie algebra structure itself in terms of the 
commutators of the algebra elements. Clearly the broken subgroups also act effectively 
on the components of $\htho$ and hence, following the discussion toward the end of 
section~\ref{thwhf}, the gauge field dynamics for the full internal symmetry group 
will be obtained.
 More generally the breaking of the isochronal symmetry of $\lv$ over the base space 
$M_4$ will ultimately need to be incorporated into the unification scheme described in 
section~\ref{reaic} with a structure in principle resembling Kaluza-Klein theory based 
on homogeneous  
fibres as reviewed in section~\ref{thwhf}.

  With the space $\htho$ considered as a manifold the map $R(\alpha)\mcX_0$, for any 
point $\mcX_0$ on $\htho$, is a left action on the manifold  with $R(0)\mcX_0 = 
\mcX_0$. 
  This action also describes a curve as a mapping from $\alpha \inn \rrr$ into $\htho$ 
which sends the real number $\alpha = 0$ to the point $\mcX_0$. Acting on all values 
of $\mcX \inn \htho$ this one-parameter group $R(\alpha)$ is associated with the 
tangent vector field:
\begin{equation}
\label{rdot}
   \dot{R} = \frac{\partial \: ( R(\alpha) \mcX  )}
      {\partial \alpha} \Big|_{\alpha = 0} \; \inn \; T\htho
\end{equation}
 where a `dot' over the kernel symbol such as for `$\dot{R}$' will generally denote a 
tangent vector field on the space $\htho$.
  The local tangent space on $\htho$ under the one constraint $\det(\mcX) = 1$ is 
26-dimensional, however the space of vector \textit{fields} over the 26-dimensional 
manifold $S_{26}$ is infinite. The task is then to identify the $\esi$ Lie algebra 
through a one-to-one isomorphic correspondence with a subset of 78 linearly 
independent vector fields in $T\htho$, of the form of equation~\ref{rdot}, within this 
$\infty$-dimensional space. The search is narrowed down by adopting a starting point 
based on the 135 one-parameter subgroup actions on $\htho$ obtained through the three 
types of $\sltwoo$ conjugation described in equations~\ref{mxm3} and 
\ref{type1}--\ref{type3}.

  The first stage, at the level of these one-parameter subgroups, is to find a 
convenient new basis for the 21 category 3 transverse rotations $R_{r , s}(\alpha)$ 
described in table~\ref{mtran45}. For each imaginary base unit $q$ in 
figure~\ref{octmult} the three pairs of imaginary units each describing a quaternion 
subalgebra with $q$ also form a right-handed 3-dimensional `coordinate frame' with 
$q$. That is, for example with $q=i$, we have $j k = +i$, $\kl \jl = +i$ and $l \il = 
+i$, matching the pairs listed in the top row of table~\ref{opairs}.
\begin{table}[htbp]
\centering
\begin{tabular}{|c|ccc|}
 \hline  
   $q \inn \mbox{Im}(\ooo)$ & 
  $\quad$ $1^{\mathrm{st}}$ pair $\quad$ & $2^{\mathrm{nd}}$ pair & $\quad$ 
$3^{\mathrm{rd}}$ pair $\quad$ \\
	\hline
	    $i$       &        $j,k$           &      $\kl,\jl$         &        $l,\il$            
\\ 	
		$j$       &        $k,i$           &      $\il,\kl$         &        $l,\jl$            
\\ 
		$k$       &        $i,j$           &      $\jl,\il$         &        $l,\kl$            
\\ 
		$\kl$     &        $\jl,i$         &      $j,\il$           &        $k,l$              
\\ 
		$\jl$     &        $i,\kl$         &      $\il,k$           &        $j,l$              
\\ 
		$\il$     &        $\kl,j$         &      $k,\jl$           &        $i,l$              
\\ 
		$l$       &        $\il,i$         &      $\jl,j$           &        $\kl,k$            
\\   
  \hline
  \end{tabular}
  \caption{\setb (Adopted directly from \protect\cite{Wang} p.107, table 4.2). The 3 
right-handed quaternion subalgebras for each imaginary octonion base unit $q$ ordered 
as $1^{\mathrm{st}}$, $2^{\mathrm{nd}}$ and $3^{\mathrm{rd}}$ (associated with the 
rotations $R_{q1}$, $R_{q2}$ and  $R_{q3}$ respectively) as appropriate for the new 
basis for transverse rotations listed in equations~\ref{arotn}--\ref{srotn}.}
\label{opairs}
\end{table} 
  
  For each choice of $q$ the associated $1^{\mathrm{st}}$, $2^{\mathrm{nd}}$ and 
$3^{\mathrm{rd}}$ planes, from the same row of the table, are mutually orthogonal and 
rotated independently by $R_{q1}(\alpha)$, $R_{q2}(\alpha)$ and  $R_{q3}(\alpha)$ 
respectively, where for example $R_{i1}(\alpha) = R_{j , k}(\alpha)$ by taking the 
appropriate pair, here $\{r,s\} = \{j,k\}$ from table~\ref{opairs}, to construct the 
corresponding category 3 transverse rotation $R_{r , s}(\alpha)$ from 
table~\ref{mtran45}. Adopting the point of view of \textit{active} transformations 
these individual plane rotations are in a clockwise sense about the $q$-axis for 
positive $\alpha$ and counterclockwise for negative $\alpha$.
They are then composed together in the following combinations:
\begin{eqnarray}
\label{arotn}
   A_q(\alpha) & = & R_{q1}(\alpha) \,\scirc\,
                                    R_{q2}(-\alpha) \\
\label{grotn}
   G_q(\alpha) & = & R_{q1}(\alpha) \,\scirc\, 
                   R_{q2}(\alpha) \,\scirc\, R_{q3}(-2\alpha)  \\
\label{srotn}
   S_q(\alpha) & = & R_{q1}(\alpha) \,\scirc\,
                   R_{q2}(\alpha) \,\scirc\, R_{q3}(\alpha)
\end{eqnarray}  
 Since in all cases each of the two or three plane rotations are independent of each 
other their order may be interchanged. (The three actions $A_q$, $G_q$ and $S_q$ may 
also be recombined to recover the original single plane rotations, for example 
$R_{q1}(\alpha) = A_q(\alpha/2) \,\scirc\, G_q(\alpha/6) \,\scirc\, S_q(\alpha/3)$).

  The $(3 \times 7)=21$ actions defined in equations~\ref{arotn}--\ref{srotn} hence 
provide a new basis for the Spin(7) transverse rotations applied in 
table~\ref{mtran45} on the space $\htwo$. Since each of these actions is represented 
by diagonal $2 \times 2$ matrices they also apply to the Spin(7) action on the space 
$\ooo$ itself, as the double cover of SO(7) acting on $\mbox{Im}(\ooo)$. However the 
mathematical motivation for introducing the new basis is seen when applied to the $3 
\times 3$ matrix case, implicitly due to the triality relation between the three 
octonion components of $\htho$. Indeed when embedded in the type 1, 2 and 3 actions of 
equations~\ref{type1}, \ref{type2} and \ref{type3} respectively and determining the 
tangent vectors of the new transverse rotations in $T\htho$ using equation~\ref{rdot} 
it can be shown by direct comparison that:
\begin{eqnarray} 
  \dot{A}_q^{1} \, = \, \dot{A}_q^{2}  \!\! & = & \!\!  \dot{A}_q^{3}  \label{agsdepa} 
\\ 
  \dot{G}_q^{1} \, = \, \dot{G}_q^{2}  \!\! & = & \!\!  \dot{G}_q^{3}  \label{agsdepg} 
\\ 
  \dot{S}_q^{1} \; + \; \dot{S}_q^{2}  \!\! & + & \!\!
       \dot{S}_q^{3} \, = \, 0  \label{agsdeps} 
\end{eqnarray}
 for each of the seven cases of $q \inn \{i,j,k,\kl,\jl,\il,l\}$. The superscript $a$ 
on tangent vectors, as for $\dot{A}^a$ here, will always denote the type and since 
raising such a vector to a power has no meaning the $a$ is not placed inside brackets 
(in cases of ambiguity  brackets will be used for the type index `$(a)$' as for $\mcM$ 
in equations~\ref{type1}--\ref{type3}). The new choice 
 of equations~\ref{arotn}--\ref{srotn} 
  for the category 3 actions on $\htho$ is hence justified by the manifest clarity of 
the linear dependencies seen in this basis.

    Each of the 14 independent generators $\{\dot{A}_q \equiv \dot{A}_q^{a}, 
\:\dot{G}_q \equiv \dot{G}_q^{a}\}$ (for any $a=1,2,3$) acts on the three octonion 
elements $a,b,c \inn \ooo$ in equation~\ref{htho} in exactly the same way (while 
vanishing on the $p,m,n \inn \rrr$ elements as for all transverse rotations). The 14 
corresponding group actions of equations~\ref{arotn} and \ref{grotn} preserve the 
multiplication table for $\ooo$ continuously as a function of the parameter $\alpha$, 
forming the proper continuous automorphism group of the octonions (this group is SO(3) 
for the quaternion case). Hence, taken together $A_q$ and  $G_q$ compose the 
exceptional group $\gt$, which justifies the notation `$G_q$' introduced in 
equation~\ref{grotn}.

    The notation `$A_q$' in equation~\ref{arotn} is introduced owing to the similarity 
of the kernel symbol to `$\lambda$' which denotes the Gell-Mann matrices, as listed in 
table~\ref{gellm}, which generate the Lie  group SU(3),
  a basis for the Lie algebra of which can also be composed of the 8 generators 
$\{\dot{A}_q , \dot{G}_l\}$, as will be described in section~\ref{intsym}.
 In fact the automorphism group of the octonions may be reduced to the subgroup $\suth 
\subset \gt$ by fixing an imaginary unit such as $l \inn \ooo$.
 The identification of this $\suth$ subgroup also provides a significant motivation 
for adopting the basis of equations~\ref{arotn}--\ref{srotn} from the potential 
physical perspective (see also the discussion following equation~\ref{extloract}).

  The notation `$S_q$' in equation~\ref{srotn} originates from the symmetric action of 
three synchronised rotations of $\alpha$ radians in three different planes. Applied to 
the $2 \times 2$ vector $X \inn \htwo$  this synchronised  action is identical to the 
original single action of equation~\ref{quatrot2} (with $i$ generalised to any $q\inn 
\{i,j,k,\kl,\jl,\il,l\}$) in rotating three planes of the $a\inn \ooo$ component of 
$X$, although due to the transformation of the spinor $\theta = \binom{c}{\bar{b}}$, 
as described below,  the action of ${S}_i^{(1)}$ for example, that is with $q=i$ in 
equation~\ref{srotn}, on $\htho$ in the $3 \times 3$ case is \textit{not} equivalent 
to the action of equation~\ref{quatrot2} embedded in equation~\ref{mxm3} or 
\ref{type1}.

    On the other hand equation~\ref{quatrot2} can be augmented to a single unnested $3 
\times 3$ matrix action $R(\alpha)\mcX = \mcM_{\ssls_q}^{(a)} \mcX 
\mcM^{(a)\dag}_{\ssls_q}$ with $\mcM_{\ssls_q}^{(a)}$, for the type $a=1$ case, 
expressed as:
\begin{equation}
  \mcM_{\ssls_q}^{(1)}(\alpha) \;= \;  \left( \begin{array}{ccc}
      e^{q\frac{\alpha}{2}}   &   0   &   0     \\
	    0   &   e^{q\frac{\alpha}{2}}   &   0   \\
	    0   &   0    &    e^{-q\alpha}       \end{array} \right)
  \label{sqdiag}
\end{equation}
   with a corresponding permutation of the diagonal entries for the type 2 and type 3 
cases. These actions are denoted by kernel symbol $\ssl$ with the 
`{\small{$\backslash$}}' as a mnemonic for the diagonal form of equation~\ref{sqdiag}. 
The action $\mcM_{\ssls_q}^{(1)}(\alpha) \, \mcX \, \mcM^{(1)\dag}_{\ssls_q}(\alpha)$  
may be considered as a `phase transformation' in rotating three orthogonal imaginary 
planes of the $a$ component of $\mcX$ in equation~\ref{xoct3} by the same angle 
$\alpha$. These actions are also related to a demonstration of triality in $\htho$ 
involving SO(8) transformations on the three octonion subspaces of $\htho$ (see the 
discussion below alongside equations~\ref{mqqbaro} and \ref{trialabc} and in 
\cite{Man4} around equation~43).

  Each of these three $3 \times 3$ matrices, including the type 1 case in 
equation~\ref{sqdiag} for a given $q$,  contains entries in a single complex 
subalgebra, satisfies $\det(\mcM_{\ssls_q}^{(a)}) = 1$ and preserves the form $\det 
(\mcX)$ of equation~\ref{detx3} or \ref{detpmn}, consistent with the requirements for 
an $\sltho$ action. However, these actions will not lead to elements of the preferred 
$\esi$ algebra basis under construction here since they are not of the form of 
equations~\ref{type1}--\ref{type3} with $\det (M) = \pm 1$ as required for the 
`compatible' $2 \times 2$ matrix actions described in the previous section. 
   Given the form of $M_{x \pqg q}$ in table~\ref{mtran45} and the type 1, 2 and 3 
embeddings of equations~\ref{type1}, \ref{type2} and \ref{type3} the diagonal matrix 
of equation~\ref{sqdiag} can be expressed by the matrix product:
\begin{eqnarray}
    \mcM_{\ssls_q}^{(1)}(\alpha) & = &  \mcM_{R_{x \pqg q}}^{(1)}(-\alpha)
	                        \times  \mcM_{R_{x \pqg q}}^{(2)}(-2\alpha)  \\
	\mbox{and hence} \qquad {\dssl}_q^{\, 1} & \;\; = \;\; &  - \dot{R}_{x \pqg q}^{1}
	      \quad\; - \quad\;  2\dot{R}_{x \pqg q}^{2}	\label{slinrr}
\end{eqnarray}

    The group action $S_q^{(1)}(\alpha)$ of equation~\ref{srotn} on the 10-dimensional 
subspace $\htwo \subset \htho$, consisting of three independent rotations of imaginary 
planes of the octonion $a$, is precisely the same as the $\ssl_q^{(1)}(\alpha)$ action 
using equation~\ref{sqdiag}. On the spinor components $\theta = \binom{c}{\bar{b}} 
\inn \ooo^2 \subset \htho$ these actions are only equivalent for small transformations 
to order $\alpha$ and diverge at O$(\alpha^2)$ and higher powers. However since all 27 
components of $\htho$ transform the same way to O$(\alpha)$ we have $\dot{S}_q^{1} = 
\dssl_q^1$ (and similarly for the type 2 and 3 cases) as vector fields in the space 
$T\htho$, and hence these two objects are interchangeable in expressions of linear 
dependence. 
   
   Since at the group level $S_q^{(a)}(\alpha)$ and $\ssl_q^{(a)}(\alpha)$ differ at 
O$(\alpha^2)$ the Lie bracket, to be described in the following section, of the 
corresponding generators with the same Lie algebra element $\dot{R}$ will also differ 
with $\lbrack \dot{S}, \dot{R} \rbrack \neq \lbrack \dssl, \dot{R} \rbrack$ in general 
even though $\dot{S} = \dssl$ as elements of a vector space. The Lie bracket in these 
two cases agrees for the $\htwo \subset \htho$ subspace but differs for the spinor 
components. 
The transformations on the components of the spinor $\theta$ are expected to be 
important for the internal symmetries in comparison with the Standard Model and hence 
care will need to be taken in choosing an appropriate Lie algebra basis.  
The $\esi$ Lie algebra table in \cite{Wang} uses the actions of $S_q^{(1)}(\alpha)$ 
from equation~\ref{srotn} rather than $\ssl_q^{(1)}(\alpha)$ based in 
equation~\ref{sqdiag}, which is hence significant for the Lie algebra structure, and 
in turn it is the former transformations which are also used in this paper.

   Adding the 7 actions $S_q^{(a)}$ for $a=1, 2$ or 3 to the set of 14 actions  
$\{A_q,G_q\}\equiv \gt$ completes a Spin(7) double cover of SO(7) for the type 1, 2 or 
3 transverse rotations respectively. These three SO(7)s are mutually related by  
equation~\ref{agsdeps}.
 In addition to equations~\ref{agsdepa}--\ref{agsdeps} further linear dependencies 
amongst the generators expressed on $T\htho$ are found (the second of which is 
equivalent to equation~\ref{slinrr}):
\begin{eqnarray}
 & & \dot{R}_{x \pqg q}^{1}  +  \dot{R}_{x \pqg q}^{2} 
                    + \dot{R}_{x \pqg q}^{3} = 0   \label{rsdeprrr}  \\
 & &  \dot{R}_{x \pqg q}^{2}  =  -\fhs\dot{R}_{x \pqg q}^{1} 
                    -\fhs\dot{S}_q^{1}   \label{rsdeprrs}  \\
 & &  \dot{S}_q^{2}  =  +\mbox{\small{$\frac{3}{2}$}}\dot{R}_{x \pqg q}^{1}
                   - \fhs\dot{S}_q^{1}    \label{rsdepsrs} 
\end{eqnarray}

    Appending the set of 7 actions $R_{x \pqg q}^{(a)}$ to the SO(7) of type $a$ (for 
$a=1,2$ or 3) completes a set of 28 actions forming the group SO(8). The three SO(8)s 
formed this way are actually the \textit{same} SO(8), that is they are composed of the 
same subset of $\esi$ transformations on $\htho$, due to the triality relation between 
the $\htho$ components. 
The triality symmetry is described explicitly in \cite{Wang,Man4,Man5}.
The transformations of the SO(8) subgroup of the type 1 $\sltwoo$ action in 
equation~\ref{type1} can be obtained by a nested composition with $3\times 3$ matrices 
of the form: 
\begin{equation}
 \label{mqqbaro}
  \mcM = \left( \begin{array}{ccc}
     q  & 0              & 0  \\
     0  &  \overline{q}  & 0        \\
     0  &   0            & 1  
          \end{array}  \right) 
\end{equation}
 with $q\inn\ooo$ and $\vert q \vert =1$. 
  The action of such type 1 transformations on an element $\mcX \inn \htho$ of 
equation~\ref{htho}  leaves the diagonal elements $\{p,m,n\}$ invariant while the 
three off-diagonal octonion elements transform non-trivially as (see \cite{Man5}  
equation~46 and discussion):
\begin{equation}
  \label{trialabc}
   \begin{array}{ccc}
   a & \to & \overline{q}a\overline{q} \\
   b & \to & bq  \\
   c & \to & qc  
          \end{array}  
\end{equation}  
  These generate and correspond to the three SO(8) 8-dimensional representations of 
vector, dual spinor and spinor exhibited via symmetric, right and left and octonion 
multiplication respectively, with an implicit triality mapping between the above three 
octonion actions identified by simply employing the same $q$ for each of the three 
actions, as alluded to in the opening paragraphs of section~\ref{earlyo}. 
Corresponding to the triality isomorphism the three actions of SO(8) are permuted into 
each other via the action of the matrices ${\mathcal T}$ in equation~\ref{tcycle3}, 
such that we effectively have the \textit{same} copy of SO(8) in common within each of 
the three types of $\sltwoo^a$ actions on $\htho$.

This subgroup SO$(8) \subset \esi$ is in fact precisely the subgroup of $\esi$ 
transformations on $\mcX \inn \htho$ that leaves invariant the diagonal entries, that 
is $\{p,m,n\}$ of equation~\ref{htho}. This unique SO(8) then contains three different 
SO(7)s, each built in turn on a unique $\gt$. 
 Only this subset of 14 $\gt$ transformations needs to be described in the form of 
nested actions while the remaining SO(8) transformations may be composed of seven 
unnested actions from $\ssl_q^{(a)}$ (as for example from equation~\ref{sqdiag}, and 
replacing the nested $S_q^{(a)}$ actions to obtain Spin(7) from $\gt$) together with 
seven $R_{x \pqg q}^{(a)}$ actions for type $a=1$, $2$ or $3$ \cite{Man4}.

       Here in this paper the initial importance of triality lies in the fact that it 
explains in part the rich symmetry of $\esi$ on $\htho$ as an expression of $\lv$. 
Indeed the triality symmetry is responsible for the large degree of redundancy in the 
set of $(3 \times 45) = 135$ generators for three types of $\sltwoo^a$ transformation 
described above. 
   The relations in equations~\ref{rsdeprrr}--\ref{rsdepsrs} show that given the type 
1 actions it is possible to exclude $\dot{S}_q^{2}$ (and hence, from 
equation~\ref{agsdeps}, also $\dot{S}_q^{3}$) as well as $\dot{R}_{x \pqg q}^{2}$ and  
$\dot{R}_{x \pqg q}^{3}$ from a linearly independent basis for the $\esi$ Lie algebra. 

  Building on the 28 generators of SO(8) (taking a type 1 basis) with any one of the 
three sets of 8 generators $\{\dot{R}_{x \pqg z}^{a}, \dot{R}_{z \pqg q}^{a}\}$, for 
type $a=1,2$ or $3$, leads to one of three copies of SO(9). Each of these $28+(3 
\times 8) = 52$ rotations preserves $\mbox{tr}(\mcX)$, with $\mcX \inn \htho$, and 
they  collectively define the group $\ff := \suthoo$. The trace on $\htho$ is 
analogous to the time component of the Lorentz vector represented by h$_2\kkk$ 
(described after equation~\ref{ththdet0} for equations~\ref{xoct} and \ref{xcomp}), 
but does not itself have a simple temporal interpretation here for the $3 \times 3$ 
case. 
 In the present theory the cubic norm $\mbox{det}(\mcX) \equiv \lvt$ itself expresses 
a multi-dimensional form of temporal flow, having the form of equation~\ref{lv} as 
introduced in section~\ref{gfotf}.

  Extending further to reproduce the type $a=1,2$ and $3$ Lorentz transformations 
$\sltwoo^a$ by including the 9 boost generators $\{\dot{B}_{t \pqg z}^{a}, \dot{B}_{t 
\pqg x}^{a}, \dot{B}_{t \pqg q}^{a}\}$ for each case, and taking into account the 
further linear dependence:
\begin{equation} 
 \label{bbbdep}
 \dot{B}_{t \pqg z}^{1} + \dot{B}_{t \pqg z}^{2} + \dot{B}_{t \pqg z}^{3} = 0
\end{equation} 
  a total of $52+(3\times 9) - 1=78$ actions are obtained, accounting for a complete 
basis of determinant preserving $\esi := \sltho$ transformations of $\htho$. The 
entire group is then described in terms of the actions of complex matrices $\mcM$ on 
the space $\htho$, with the preferred basis for the Lie algebra represented on 
$T\htho$ reproduced below in table~\ref{prefbas}.
\begin{table}[htbp]
\centering
\begin{tabular}{|ccc|r|}
 \hline  
 \multicolumn{3}{|l|}{Category 1: Boosts}    & \#  \\ \cline{4-4}
   $\qquad  \dot{B}_{t \pqg z}^{1} \qquad$
	 & $\qquad \dot{B}_{t \pqg x}^{1} \qquad$
	   &  $\qquad \dot{B}_{t \pqg q}^{1} \qquad$ &  $9$  \\
	$\dot{B}_{t \pqg z}^{2}$ & $\dot{B}_{t \pqg x}^{2}$ & $\dot{B}_{t \pqg q}^{2}$ &  
$9$  \\
                               & $\dot{B}_{t \pqg x}^{3}$ & $\dot{B}_{t \pqg q}^{3}$ &  
$8$  \\
 \hline
 \multicolumn{3}{|l|}{Category 2: Rotations} &   \\
    $\dot{R}_{x \pqg q}^{1}$ & $\dot{R}_{x \pqg z}^{1}$ & $\dot{R}_{z \pqg q}^{1}$ & 
$15$  \\
	                           & $\dot{R}_{x \pqg z}^{2}$ & $\dot{R}_{z \pqg q}^{2}$ &  
$8$  \\
	                           & $\dot{R}_{x \pqg z}^{3}$ & $\dot{R}_{z \pqg q}^{3}$ &  
$8$  \\
 \hline
 \multicolumn{3}{|l|}{Category 3: Transverse Rotations} &  \\ 
    $\dot{A}_q$                & $\dot{G}_q$                & $\dot{S}_{q}^{1}$        
& $21$  \\
 \hline
 \multicolumn{3}{|r|}{Total Generators}    &    78  \\
 \hline
  \end{tabular}
  \caption{\setb The complete basis for the Lie algebra of $\esi$, in terms of tangent 
vector fields on $T\htho$, reproduced from (\protect\cite{Wang} p.177, table A.1). The 
actual tangent vector fields are determined and listed in tables~\ref{lbrota} and 
\ref{ltrota} in the present paper at the end of the following section.}
\label{prefbas}
\end{table} 

  The generators, as described above, of the subalgebras corresponding to the various 
stages of the subgroup chain:
\begin{equation}
  \label{etossub}
\esi \supset \mbox{SO}^+(1,9)_{45} \supset \mbox{SO}(9)_{36} \supset \mbox{SO}(8)_{28} 
\supset \mbox{SO}(7)_{21} \supset (\gt)_{14} \supset \suth_{8}
\end{equation}
    (here, other than for the 78-dimensional $\esi$, the subscripts give the dimension 
of the algebra)  
 can be identified within the three type 1 lines of table~\ref{prefbas}. These can be 
built up from su$(3)_{8} \equiv \{\dot{A}_q, \dot{G}_l\}$ to $\mbox{so}^+(1,9)_{45}$ 
which includes $\{\dot{A}_q, \dot{G}_q\}$ together with all of the type 1 generators 
in table~\ref{prefbas}.
 
 The rotation subgroup of $\esi$, as the compact real form of $\ff := \suthoo$, is 
generated by the 52 category 2 and 3 transformations in table~\ref{prefbas}. The 
generator composition of a subalgebra chain leading down from $\esi \supset \ff$ is 
presented in (\cite{Wang} p.119, table 4.4). However, although both preserving  
$\mbox{tr}(\mcX)$ (for any $\mcX \inn \htho$)  and being the automorphism group of the 
exceptional Jordan algebra (equation~\ref{joralg}), the group $\ff$ is not of great 
significance in the present paper.

  At the group level in equation~\ref{etossub} each `SO' might more strictly be 
replaced by the corresponding double cover `Spin' group. 
  As the group $\sltho$ necessarily includes the one-sided spinor actions $\theta \to 
M \theta$ in equation~\ref{type1} (as well as in equations~\ref{type2} and 
\ref{type3}) the action for $M = -\b1_2$
(obtained for any of the category 2 rotations in table~\ref{mtran45} with $\alpha = 
2\pi$) on $\htho$ does \textit{not} give the identity transformation. However $\sltho$ 
is \textit{not} a double cover, rather it is a real simply connected form of $\esi := 
\sltho$ itself (\cite{Man5} section 2, with the same situation applying for $\ff := 
\suthoo$ acting on $\htho$). On the other hand  the action of $\sltwoo$
  is a double cover of the rotation group $\sootn \equiv \sltwoo / \zzz^2$, and 
similarly for the further rotation subgroups. With an awareness of these issues of 
group manifold topology groups such as $\sootn$ and $\soot$ can be considered to be 
embedded within the full group $\esi$.

\section{Lie Algebra of $\esi$}
\label{laofesi}

     At the group level the $\esi$ action on $\htho$ is composed of 52 rotations, that 
is the unitary $3 \times 3$ matrix actions with $\mcM \mcM^{\dag} = \b1_3$, and 26 
boosts, that is the Hermitian actions with $\mcM = \mcM^{\dag}$, as can be deduced 
from the embedded $2 \times 2$ matrices $M$ listed in table~\ref{mtran45} for the 
category 2 and 3 rotations and category 1 boosts respectively. At the Lie algebra 
level, in a normalised basis for which the Killing metric $K$ is diagonal with entries 
in $\{-1,+1\}$ 
(or more generally negative or positive entries for a diagonal but unnormalised 
Killing form, explicit values for which will be determined in 
subsection~\ref{strassy}), base vectors $X$ for which $K(X,X) = -1$ (or $< 0$) are 
called compact generators, corresponding to `rotations' of the Lie group, while those 
with  $K(X,X) = +1$ (or $> 0$) are called non-compact generators, corresponding to 
`boosts'.

 With Killing form signature of $-26$, also denoted $(52,26)$ for 52 rotations and 26 
boosts, the non-compact real form of $\esi$ constructed in the previous section may be 
denoted as $\esig$, and describes the generator composition $D^{R,B}$ introduced in  
section~\ref{earlyo} and displayed in equation~\ref{drbdecom}. 
The Killing form employed in equations~\ref{killmet} and \ref{ggxyk} of 
section~\ref{lccop} for Kaluza-Klein theory was chosen with components 
$K_{\alpha\beta} = -\delta_{\alpha\beta}$ corresponding to the choice of a compact 
gauge group. 
 In the symmetry breaking of $\esig$ over the base space $M_4$ such compact internal 
symmetry groups will be identified.

   An alternative description of $\esi$ in terms of 14 $\gt$ actions together with  
    64 non-$\gt$ transformations, composed from the actions of the 64 tracefree 
octonion $3\times 3$ matrices, was also introduced in section~\ref{earlyo} where 
$D^{G,S}$ denoted the generator composition as displayed in equation~\ref{dsgdecom}. 
In the previous section the $\gt$ subgroup was identified explicitly as the set of 14 
$\{A_q,G_q\}$ transverse rotations.
   As described shortly after equation~\ref{trialabc} the remaining 64 actions
  may be expressed with unnested compositions consisting for example of the 57 group 
actions corresponding to the category 1 and 2 generators of table~\ref{prefbas} 
together with seven 
   $\ssl_q^{1}$ actions from  equation~\ref{sqdiag} (in place of $S^1_q$). Hence both 
the (52+26) and (14+64) decompositions  can be clearly seen in table~\ref{prefbas} in 
terms of the respective subsets of generators. 

Here all 78 generators are explicitly presented in tables~\ref{lbrota} and 
\ref{ltrota}, for the category $\{1,2\}$ and 3 transformations respectively, as vector 
fields $\dot{R} \inn T\htho$ which, from equation~\ref{htho}, are of the form:
\begin{equation}
 \label{ththo}
  \dot{R} = 
   \left( \begin{array}{ccc}
       \dot{p} & \dot{\bar{a}} & \dot{c}  \\
       \dot{a} &   \dot{m}     & \dot{\bar{b}}        \\
 \dot{\bar{c}} &   \dot{b}     & \dot{n} 
          \end{array}  \right)   \inn T\htho 
\end{equation}

 These 78 matrices are themselves Hermitian and hence also belong to the space 
$\htho$. While there is no constraint on the determinant of any $\dot{R} \inn 
T{\htho}$ the matrices are tracefree for all of the category 2 rotations and category 
3 transverse rotations. The type 1 transformations act on the $\{p,m,n;a,b,c\}$ 
components on $\htho$ in the same way that the type 2 transformations act on the 
$\{m,n,p;b,c,a\}$ components and type 3 transformations act on the $\{n,p,m;c,a,b\}$ 
components as can be seen in equations~\ref{type1}, \ref{type2} and \ref{type3}, for 
example by following the explicit invariant components $n$, $p$ and $m$  respectively 
in these three equations. This same cyclic permutation, consistent with the action of 
$\mcT$ in equation~\ref{tcycle3}, is reflected in the tangent vectors in 
table~\ref{lbrota} and for $\dot{S}^a_q$ in table~\ref{ltrota}.

  These tables describe in intimate detail the anatomy of the $\esi$ action as 
expressed on the tangent space $T\htho$. With $p=t+z$ and $m=t-z$, embedding 
equation~\ref{xoct} into $\htho$, each type 1 tangent vector can be seen to `point' in 
the appropriate direction in the relevant $T\htwo$ components for the subspace $\htwo$ 
plane transformations resulting from the action of the matrices in  
table~\ref{mtran45}, with a similar correspondence identifiable for the type 2 and 3 
cases. For example the non-zero components of the category 1 boost and category 2 
rotation generators for the type 1 actions on the 10-dimensional subspace $\htwo$ are 
simply:
\begin{equation}
  \begin{array}{lll}
           \dot{B}_{t \pqg z}^{1}: \; \dot{t}=+z,\; \dot{z} = +t, \qquad 
        &  \dot{B}_{t \pqg x}^{1}: \; \dot{t}=+x,\; \dot{x} = +t,    \qquad
        &  \dot{B}_{t \pqg q}^{1}: \; \dot{t}=-a_q,\; \dot{a}_q = -t, 
		   \\
  	       \dot{R}_{x \pqg q}^{1}: \; \dot{x}=-a_q,\; \dot{a}_q = +x,
        &  \dot{R}_{x \pqg z}^{1}: \; \dot{x}=-z,\; \dot{z} = +x,
        &  \dot{R}_{z \pqg q}^{1}: \; \dot{z}=+a_q,\; \dot{a}_q = -z 
  \end{array}	 	  \label{sixlorbr}
\end{equation}
  where, here and in tables~\ref{lbrota} and \ref{ltrota}, $x \equiv a_x \equiv a_1$ 
and similarly $a_q$ refers to the real coefficient in equation~\ref{octa} 
corresponding to the imaginary unit $q$, (that is $a_l \equiv a_8$ etc.). The category 
3 transverse rotations of equations~\ref{arotn}, \ref{grotn} and \ref{srotn} each act 
on several planes in $\htwo$. The transformations of the spinor components of $\htho$ 
induced by the $3\times 3$ matrix action are also included in tables~\ref{lbrota} and 
\ref{ltrota}.

  In subsection~\ref{tgman} the Lie algebra of a group $G$ was defined in terms of the 
set of left-invariant vector fields on the group manifold $G$, as also recalled in the 
paragraphs leading to equation~\ref{rdot} in the previous section.
  Through any point $h\inn G$ each such vector field $X$ generates a one-parameter 
group of right translations 
$\phi_t(h) = h \exp (tA)$ where $A=X_e \inn T_eG$ is the vector of the field $X$ at 
the identity $e\inn G$, as depicted in figure~\ref{grightran} . If $G$ acts by 
\textit{right} translation on another manifold $M$ this realisation of $G$ induces 
vector fields $V^A \inn \TM$ such that $V^A_x(f) := \frac{d}{dt}f(x\exp 
(tA))\vert_{t=0}$, at $x\inn M$ where $f(x)$ is a real function on $M$, represents a 
homomorphism of the Lie algebra with $\lbrack V^A, V^B \rbrack = V^{\lbrack A,B 
\rbrack}$. If the action of $G$ on $M$ is effective  there is a one-to-one isomorphism 
between the Lie algebra $L(G)$ and the set of such vector fields $\{V^A\}$ in $\TM$
(as is the case for the action of $G$ on a principle fibre bundle $P$ as described in 
section~\ref{fibre}, see equations~\ref{atova} and \ref{vavbvab}).

   If $G$ acts on the manifold $M$ by \textit{left} translations then this 
relationship is an anti-homomorphism. This is the case for right-invariant vector 
fields on $G$ itself, which are generated by left translations. The structure 
constants $c^{\alpha}_{\ph{\alpha}\beta \gamma}$ for the Lie bracket of such 
right-invariant fields $\{Y^L_{\alpha}\}$ on $G$ are precisely the negative values of 
the Lie algebra structure constants defined in terms of the corresponding 
left-invariant fields $\{X^R_{\alpha}\}$ (which match the right-invariant fields as 
elements of the tangent vector space at the identity $e\inn G$, that is each 
$Y^L_\alpha(e) = X^R_{\alpha}(e)$). This anti-homomorphism was also noted for left 
translations applied to the space of homogeneous fibres for equation~\ref{negcstr} in 
the opening of section~\ref{thwhf}.

  In the present case the group manifold for $G=\esi$ is not constructed itself but 
rather the group 
 acts \textit{transitively} on $\htho$, which is hence a homogeneous space, such that 
$\det(\mcX) = 1$ is preserved for $\mcX \inn \htho$.
 The action of $\esi$ on the underlying space $\htho$ is also \textit{effective} and 
hence the Lie algebra $L(\esi)$ may be constructed in terms of vector fields on the 
tangent space $T\htho$. 
  The $\esi$ transformations composed as $\mcX \to \mcM \mcX \mcM^{\dag}$ are 
\textit{left} translations as opposed to \textit{right} translations, as has been 
described in the previous section, and as will be seen explicitly for subgroups such 
as $\sltc$ in  section~\ref{extsym}.
  The Lie algebra commutator, which determines the structure constants of the $\esi$ 
Lie algebra, for any two elements $\dot{R}_1, \dot{R}_2$ is defined through the action 
of the respective one-parameter subgroups $R_1(\alpha)$ and $R_2(\alpha)$ at any point 
$\mcX \inn \htho$:
\begin{equation}
 [\dot{R}_2,\dot{R}_1] \; = \; \frac{\partial}{\partial (\alpha^2)}
    [R_2(-\alpha)\,\scirc\, R_1(-\alpha)\,\scirc\,
	   R_2(\alpha)\,\scirc\, R_1(\alpha)\,\mcX] \Big|_{\alpha = 0}
\label{rrbrac}
\end{equation}
   Here the four $\pm \alpha$ signs inside the square brackets are chosen so that this 
Lie algebra structure deriving from \textit{left} translations is isomorphic to the 
standard definition of $L(G)$ of equation~\ref{xxcomcx} described in 
subsection~\ref{tgman}.   
  In the general case for a Lie group $G$ equation~\ref{rrbrac} holds with the 
\textit{opposite} signs for $\alpha$ in the square brackets for the \textit{right} 
translation mapping of one-parameter subgroup curves $\rrr \to G$ to the manifold of 
the Lie group space itself. These curves passing through the identity point $e\inn G$ 
allow a bracket to be constructed on the vector space $T_eG$ isomorphic to the Lie 
algebra of the group, leading for example to the basis of equation~\ref{m3} for the 
case $G=\soth$.

 In acting upon a representation space with a lower dimension than $G$, as is the case 
here for the group $\esi$ acting in the space $\htho$, the Lie bracket is constructed 
necessarily in terms of vector \textit{fields} on the representation space.  The 
choice of signs in equation~\ref{rrbrac} means that the various subalgebras will be 
defined in the usual way, equivalent to left-invariant fields on the broken subgroup 
manifolds.
 Indeed in principle the same $\esi$ Lie algebra could be constructed in terms of 
left-invariant fields on the submanifold of $\gltr$ identified as an $\esi$ group 
representation acting on $\rrr^{27}$ as described in the previous section.

  Here the term in square brackets on the right-hand side of equation~\ref{rrbrac} 
represents a curve that passes through any chosen point $\mcX_0 \inn \htho$ for 
$\alpha=0$. While the first derivative $\frac{\pal}{\pal \alpha}$ of this same term 
vanishes identically at $\alpha=0$ the second derivative $\big( \fh \frac{\pal^2}{\pal 
\alpha^2}$, or equivalently $\frac{\pal}{\pal (\alpha^2)} \big)$ is non-zero and 
yields a tangent vector field as $\mcX$ varies over $\htho$ corresponding to the Lie 
bracket of the two vector fields $\dot{R}_1$ and $\dot{R}_2$. For example, by direct 
calculation taking the type 1, 2 or 3 embeddings of the appropriate matrix actions 
from table~\ref{mtran45}, applying equation~\ref{rrbrac} and by comparison with 
tables~\ref{lbrota} and \ref{ltrota} the twelve brackets listed in table~\ref{ecomm8} 
are determined explicitly.

\begin{table}[htbp]
\centering
 \hspace*{-17pt}
\begin{tabular}{|l|l|l|}
 \hline 
  1)  $\lbrack \dot{R}^{1}_{x \pqg i}, \dot{R}^{1}_{x \pqg z}\rbrack \; = \; 
\dot{R}^{1}_{z \pqg i}      $
  & 5) $\lbrack \dot{R}^{1}_{x \pqg i}, \dot{B}^{1}_{t \pqg x}\rbrack \; = \; 
-\dot{B}^{1}_{t \pqg i}$	 				   
  & 9)  $\lbrack \dot{R}^{1}_{x \pqg i}, \dot{R}^{1}_{x \pqg l} \rbrack \; = \;
   -\frac{1}{3}\dot{G}_{\il} + \frac{1}{3}\dot{S}^{1}_{\il}$  
     \\
	  2)  $\lbrack \dot{R}^{1}_{x \pqg i}, \dot{R}^{1}_{z \pqg i}\rbrack  \; = \;
                                           -\dot{R}^{1}_{x \pqg z}$
  &  6)   $\lbrack \dot{R}^{1}_{x \pqg i}, \dot{R}^{2}_{x \pqg z}\rbrack  \; = \; 
                                          -\fh \dot{R}^{2}_{z \pqg i}$								   
  &  10) $\lbrack \dot{R}^{1}_{x \pqg i}, \dot{R}^{1}_{x \pqg j} \rbrack \; = \;
    \frac{1}{2}\dot{A}_{k}
   +\frac{1}{6}\dot{G}_{k} + \frac{1}{3}\dot{S}^{1}_{k}$    
       \\	   
	   3) $\lbrack \dot{B}^{1}_{t \pqg z}, \dot{B}^{1}_{t \pqg x}\rbrack \; = \; 
\dot{R}^{1}_{x \pqg z}$	   
  & 7)  $ \lbrack \dot{R}^{1}_{z \pqg i}, \dot{B}^{3}_{t \pqg i}\rbrack \; =  \; -\fh 
                             \dot{B}^{2}_{t \pqg x}$   
    & 11)  $\lbrack \dot{S}^{1}_{i}, \dot{R}^{1}_{x \pqg j}\rbrack  \; = \; 
\dot{R}^{1}_{x \pqg k}$
   \\
 4)  $\lbrack \dot{B}^{1}_{t \pqg z}, \dot{B}^{1}_{t \pqg i}\rbrack \; = \; 
-\dot{R}^{1}_{z \pqg i}$   
  & 8)  $\lbrack \dot{R}^{2}_{x \pqg z}, \dot{R}^{2}_{z \pqg l}\rbrack \; = \;
    \dot{R}^{2}_{x \pqg l}    $
   & 12) $\lbrack \dot{R}^{3}_{x \pqg z}, \dot{R}^{3}_{z \pqg i}\rbrack \; = \;
    \dot{R}^{3}_{x \pqg i}    $ 
    \\
	& $  \qquad\qquad\quad = -\fh \dot{R}^{1}_{x \pqg l} - \fh \dot{S}^{1}_{l}$
	& $  \qquad\qquad\qquad = -\fh \dot{R}^{1}_{x \pqg i} + \fh \dot{S}^{1}_{i}$  \\
  \hline
  \end{tabular}
  \caption{\setb The Lie algebra bracket composition determined for twelve cases by 
applying equation~\ref{rrbrac} to the left-hand sides and tracking terms to 
$O(\alpha^2)$ for the sequence of $R(\pm \alpha)\mcX$ compositions and matching the 
right-hand sides with elements in tables~\ref{lbrota} and \ref{ltrota}.}
\label{ecomm8}
\end{table} 

  All cases in table~\ref{ecomm8} were calculated in full with two exceptions: in 
`case 10)' the $b$ component only on the right-hand side was determined and for `case 
11)' the $a$ component only was determined and the action $\ssl_i^{(1)}(\alpha)$   was 
used in place of $S_i^{(1)}(\alpha)$ in the calculation since these actions are 
identical on the $\htwo \subset \htho$ subspace, as described in the discussion 
following equation~\ref{sqdiag}. The purpose of these calculations is to cross-check 
the notation and conventions used here. This is useful since there are several sign 
differences between quantities in this paper and the corresponding expressions in 
reference~\cite{Wang} as listed for example in table~\ref{werror}. (We also note that 
the conventions used in the present paper differ in the sign of $\pm \alpha$ for $R_{z 
\pqg q}(\alpha)$ and $B_{t \pqg q}(\alpha)$ with respect to (\cite{Wang2} table 1)).

 \begin{table}[htbp]
\centering
\begin{tabular}{|r|ccc|}
 \hline  
     Action: & $\quad R_{x \pqg q}(\alpha)$, $\quad$  & 
	            $\quad R_{x \pqg z}(\alpha)$, $\quad$ &  
				   $\quad R_{z \pqg q}(\alpha) \quad$  \\
				  \hline
	Sign:   &  $e^{\pm q\frac{\alpha}{2}}$, & 
	         $\pm\, \sin\frac{\alpha}{2}$, & $\pm\, q\sin\frac{\alpha}{2}$ \\
			    \hline
  \end{tabular}
  \caption{\setb Sign differences between table~\ref{mtran45} in the present paper and 
(\protect\cite{Wang} p.90, table~3.1).}
\label{werror}
\end{table}

  Of the 78 basis tangent vectors listed here in tables~\ref{lbrota} and \ref{ltrota} 
one is explicitly presented in reference~\cite{Wang}. The calculation of this tangent 
vector, namely  for $\dot{A}_l$ in table \ref{ltrota} here, differs by an overall 
$\pm$ sign from that presented in (\cite{Wang} p.112, equation~4.1), but agrees with 
the sign convention for the same components quoted on (\cite{Wang} p.121). There is 
also a factor of two difference between the expression for $[\dot{R}_2,\dot{R}_1]$ 
displayed here in equation~\ref{rrbrac} and that described in the equations of 
(\cite{Wang} p.109).

	Whether each of these discrepancies is due to a typographical error or the 
conventions used in \cite{Wang} the choice of signs and factors adopted in this paper 
is necessary in order that the calculations here in table~\ref{ecomm8} are both 
self-consistent and agree with the corresponding twelve entries in the full $\esi$ Lie 
algebra commutation table available in \cite{Wang} for which the full set of $(78 
\times 78 - 78)/2 = 3003$ independent entries were found by computer program, and 
which is used for this paper particularly in chapter~\ref{chapesb}.

  The references~\cite{Man2,Wang,Man4,Man5,Wang2}  are essential here for describing 
the anatomy of the $\esi$ action on $\htho$ in a tractable form which may be dissected 
for the analysis of symmetry breaking patterns. The 
 few inconsistencies in the notation as described above may be accounted for and will 
not affect the conclusions for physics. In this paper these conventions have been 
tuned for internal consistency and to be able to consistently read off entries from 
the full $L(\esi)$ table $\cite{Wang}$ as the principal point of reference. This in 
turn means that the correspondence between the generators of subgroups of $\esi$, such 
as an external Lorentz group or an internal $\suth$ gauge symmetry group, may not 
neatly match the conventions generally employed in physical theories, as will be seen 
in chapter~\ref{chapesb}, for example in equation~\ref{msigj}. Hence ultimately a new 
basis for $L(\esi)$ may be desired as tuned through a foreknowledge of the details of 
the physical application in the context of the present theory.

  In general applying equation~\ref{rrbrac} for any two basis vectors on $T\htho$ will 
itself result in a basis vector field as listed in table~\ref{prefbas} (or 
tables~\ref{lbrota} and \ref{ltrota}) or a linear combination of such elements as is 
the case for the brackets numbered 8), 9), 10) and 12) in table~\ref{ecomm8} above. 
While elements such as $\dot{R}^{1}_{z \pqg i}$ on the right-hand side of `case 1)' in 
this table may be `integrated up' to the group action ${R}^{(1)}_{z \pqg i}(\alpha)$ 
on $\mcX \inn \htho$ in general it is not straightforward to associate an element, or 
linear combination of elements, of the $\esi$ Lie algebra with a one-parameter action 
of the Lie group describing curves on $\htho$.  This is due to the non-associativity 
of the octonions and the necessary employment of a nested structure to describe the 
transverse rotations. This is hence unlike the case in general for Lie algebra 
elements defined on the tangent space of a group manifold such as $G = \soth$  which 
may be associated with Lie group elements by an `exponential map', as described 
alongside figure~\ref{grightran} and exemplified in equation~\ref{gexpam} for $\sofi$. 
 However, of interest here will be broken subgroups, such as the Lorentz group for 
4-dimensional spacetime and the $\suth$ colour symmetry which may be expressed without 
the above difficulties.
 (Again, alternatively, the full $\esi$ action could in principle be expressed in 
terms of $G \subset \gltr$ actions on $\rrr^{27}$ and the consequences of 
non-associativity and nested actions sidestepped completely).

 In the full Lie algebra table~\cite{Wang} with the basis vectors listed in 
table~\ref{prefbas} a total of six mutually commuting elements, that is with 
$[\dot{R}_2,\dot{R}_1] = 0$ for any pair of these six elements, can be identified as 
the set:
\begin{equation}
   \{\dot{B}_{t \pqg z}^{1}, \dot{B}_{t \pqg z}^{2}, \dot{R}_{x \pqg l}^{1},
     \dot{A}_l, \dot{G}_l, \dot{S}_l^{1} \}
	 \label{csaset}
\end{equation}
  which hence forms the Cartan subalgebra for the rank-6 Lie algebra $\esi$. There is 
some flexibility in this choice, due for example to equation~\ref{bbbdep}, with 
$(\dot{B}_{t \pqg z}^{2} - \dot{B}_{t \pqg z}^{3})$ replacing the second element 
$\dot{B}_{t \pqg z}^{2}$ in (\cite{Wang2} equation~3.8(15)).

  In \cite{Man4,Man5} `symmetry breaking' is considered in terms of  making a choice 
of a preferred $\htwo \subset \htho$ together with a preferred imaginary unit for the 
octonion element in $\htwo$.
 Here we  take a subspace $\htwc \subset \htho$, using the isomorphism of the 
4-dimensional space $\htwc$ to the space of the Lorentz vectors which in turn we have 
identified with tangent vectors on $M_4$, the base space for our perception of objects 
in the world; as described in the previous chapters. The Lorentz group, $\soot$, is 
seen correspondingly as a non-compact subgroup of $\esi$. With $\soot$ being an 
\textit{external} symmetry on $M_4$, which is also a global symmetry to a very good 
approximation in a laboratory setting, this will then provide the mechanism for the 
breaking of the $\esi$ symmetry down to  local gauge symmetries which may be compared 
with the $\SML$ gauge group and representations in the Standard Model of particle 
physics.

 The Lorentz subgroup for 
4-dimensional spacetime can be taken to be generated by the
 subset of
 Lie algebra elements in $L(\esi)$:
\begin{equation}
   \{\dot{B}_{t \pqg z}^{1}, \dot{R}_{x \pqg l}^{1}, \dot{B}_{t \pqg x}^{1},
     \dot{B}_{t \pqg l}^{1}, \dot{R}_{x \pqg z}^{1}, \dot{R}_{z \pqg l}^{1} \}
	 \label{extlor6}
\end{equation}
\begin{equation}
	\mbox{acting on} \quad
	\bh_2 = \left( \begin{array}{cc} t+z & x-yl \\ x+yl & t-z \end{array} \right)
	  \inn \htwc \subset \htho
	\label{extloract}
\end{equation}
  where here the first two generators for this rank-2 subgroup are taken from the 
Cartan subalgebra for $\esi$ in equation~\ref{csaset}. The octonion unit $l$ of the 
component $a$ in equation~\ref{htho} (rather than $i$ as for equation~\ref{xcomp} and 
as discussed at the end of section~\ref{ltos}) is chosen to represent an external 
spatial component since then the internal symmetry is more readily identified using 
the preferred basis of table~\ref{prefbas}, which in turn derived from the conventions 
of equations~\ref{arotn}-\ref{srotn} and table~\ref{opairs} in which $l$ only appears 
in the $3^{\mathrm{rd}}$ pair column. The use of the unit $l$, rather than $i$, in 
this way also serves as a reminder that the Lorentz transformations here are embedded 
within expressions based on the octonion algebra.
 This external Lorentz symmetry will be studied in detail in section~\ref{extsym}.

   In section~\ref{intsym} an internal symmetry will be provisionally defined here as 
any operation that fixes the external spacetime components $(t,x,y,z)$ of 
equation~\ref{extloract} for any Lorentz 4-vector. This will include in particular the 
subgroup $\suth\equiv \{{A}_q, {G}_l\}$, which from table~\ref{opairs} and 
equations~\ref{arotn} and \ref{grotn}
 leaves the $l$ component invariant, highlighting the significance of this basis 
choice for physics.
The full $\esi$ Lie algebra commutation table in \cite{Wang} can be used to identify 
further internal symmetry groups, 
 as we shall explore in chapter~\ref{chapesb}.

   In the following chapter we first review the Standard Model, and in particular the 
relationship  between the external and internal symmetries found there, before turning 
to the  group $\esi$ in general in section~\ref{dynkin} as a candidate for unification 
of these  symmetries as employed in particle physics. 
 Then in chapter~\ref{chapesb} the detailed structure of the action of $\esi$ on 
$\htho$, as reviewed in this chapter, will be applied to deduce the properties of the 
external Lorentz symmetry in relation to the complementary  internal symmetry for the 
present theory.

  Through the historical  development from the real numbers to the complex numbers, 
continuing on through the quaternions to the octonions, composed then in $2 \times 2$ 
and further in $3 \times 3$ matrix form, the construction of $\esi$ as a determinant 
preserving action on $\htho$ has been presented as an expression of the symmetry of 
temporal flow in the form of $\lvt$. 
 It is of course possible that there may be other, higher-dimensional, forms for 
$\lv$, with yet higher symmetry groups that will have consequences for the physics of 
the world.
 Further generalisation should be, however, a well defined mathematical problem.

 In chapter~\ref{secfd}  higher-dimensional forms of temporal flow and the possible 
role of the largest exceptional Lie groups E$_7$ and E$_8$ will be considered.
  For such cases the $\esi$ symmetry will be an intermediary on the way up to, or 
operate in some way parallel to, the larger symmetries for the 
 higher-dimensional forms of $\lv$. Even in this case, given that the richness of 
$\htho$ and its symmetries, as $3\times 3$ matrices expressing the triality relation 
between three elements of the largest normed division algebra, the octonions, is much 
greater than that of Lorentz 4-vectors and the symmetry of 4-dimensional spacetime, we 
might still hope to uncover elements of empirically established physical structure  in 
a study based on this $\esi$ symmetry, assuming that the overall conceptual framework 
that we are considering here broadly corresponds to the real physical world. This, in 
the very least, would provide a proof of principle for the conceptual scheme being 
developed in this paper.

\def\var{-12pt}
\begin{table}[htbp]
{\small
\centering
 \hspace*{-18pt}
\begin{tabular}{|ccc|}
 \hline  
   $\dot{B}_{t \pqg z}^{1}$ & $ \dot{B}_{t \pqg x}^{1}$ & $\dot{B}_{t \pqg q}^{1}$   
\\
  $\!\!\!\left(\!\! \begin{array}{ccc}
                    +p    &    0       &   +\fh c     \\
	                0    &   -m       &   -\fh \bar{b}      \\
                 +\fh \bar{c} &   -\fh b   &   0 
 \end{array} \!\!\right)\!\!\!$  & 
$\!\!\!\left(\!\! \begin{array}{ccc}
                  +a_x    &   \fh(p\!+\!m)       &    + \fh\bar{b}     \\
	         \fh(p\!+\!m)    &   +a_x        &   +\fh c             \\
                +\fh b   &   +\fh\bar{c} &   0
 \end{array} \!\!\right)\!\!\!$  &   
$\!\!\!\left(\!\!\!\! \begin{array}{ccc}
                  -a_q    &   \fh (p\!+\!m)q        &    +\fh q\bar{b}     \\
	        -\fh(p\!+\!m)q    &   -a_q        &  -\fh qc             \\
                -\fh bq    &   +\fh\bar{c}q  &    0
 \end{array} \!\!\!\right)\!\!\!$
    \\ \multicolumn{3}{|c|}{ }    \\  
	$\dot{B}_{t \pqg z}^{2}$ & $\dot{B}_{t \pqg x}^{2}$ & $\dot{B}_{t \pqg q}^{2}$   
\\
$\!\!\!\left(\!\! \begin{array}{ccc}
                    0   &  +\fh \bar{a}     &    -\fh c     \\
	                +\fh a    &    +m       &    0      \\
                    -\fh \bar{c} &    0   &    -n
 \end{array} \!\!\right)\!\!\!$ 	&   
$\!\!\!\left(\!\! \begin{array}{ccc}
                    0    &   +\fh c        &    +\fh\bar{a}     \\
	         +\fh\bar{c}  &   +b_x         &    \fh (m\!+\!n)      \\
                 +\fh a  &   \fh(m\!+\!n)    &    +b_x
 \end{array} \!\!\right)\!\!\!$  & 					
 $\!\!\!\left(\!\!\!\! \begin{array}{ccc}
                    0     &  -\fh cq       &   + \fh\bar{a}q     \\
	         +\fh q\bar{c} &  - b_q         & \fh(m\!+\!n)q           \\
                -\fh qa   &   -\fh(m\!+\!n)q    &   - b_q
 \end{array} \!\!\!\!\right)\!\!\!$  	
    \\    \multicolumn{3}{|c|}{ }   \\	 \cline{1-1} 
	 \multicolumn{1}{|c|}{ $\dot{B}_{t \pqg z}^{3}$ }               
  &  \multicolumn{1}{c}{  $\dot{B}_{t \pqg x}^{3}$ }
  &  \multicolumn{1}{c|}{ $\dot{B}_{t \pqg q}^{3}$ }   \\ 
     \multicolumn{1}{|c|}{  
           $\!\!\!\left(\!\! \begin{array}{ccc}
                   -p    &   -\fh \bar{a}        &    0     \\
	           -\fh a    &    0           &    +\fh \bar{b}      \\
                   0    &   +\fh b        &    +n
 \end{array} \!\!\right)\!\!\!$    }      &
      \multicolumn{1}{c}{  
$\!\!\!\left(\!\! \begin{array}{ccc}
                  +c_x    &   +\fh b       &    \fh(n\!+\!p)     \\
	         +\fh\bar{b}  &   0          &    +\fh a         \\
                  \fh(n\!+\!p)   &   +\fh\bar{a} &    +c_x
 \end{array} \!\!\right)\!\!\!$     } &
      \multicolumn{1}{c|}{
$\!\!\!\left(\!\! \begin{array}{ccc}
                  -c_q    &  -\fh qb          &    -\fh(n\!+\!p)q     \\
	       + \fh\bar{b}q  &   0             &    -\fh aq         \\
               \fh(n\!+\!p)q   &   +\fh q\bar{a} &    -c_q
 \end{array} \!\!\right)\!\!\!$  }
  \\
 \hline
   $\dot{R}_{x \pqg q}^{1}$ & $ \dot{R}_{x \pqg z}^{1}$ & $\dot{R}_{z \pqg q}^{1}$   
\\
$\!\!\!\left(\!\!\!\!\! \begin{array}{ccc}
                    0    &    -a_q\! - \! a_x q     &    -\fh qc     \\
	       -a_q\! + \! a_x q   &     0       &  +\fh q\bar{b}           \\
                  +\fh \bar{c}q  &   -\fh bq    &    0
 \end{array} \!\!\!\!\right)\!\!\!$  & 
$\!\!\!\left(\!\!\!\!\! \begin{array}{ccc}
                  +a_x    &  -\fh (p\!-\!m)       & + \fh\bar{b}     \\
	        -\fh(p\!-\!m)    &  -a_x        &    -\fh c            \\
                 +\fh b    &  -\fh\bar{c} &    0
 \end{array} \!\!\!\right)\!\!\!$  &
$\!\!\!\left(\!\!\!\!\! \begin{array}{ccc}
                 +a_q    &    \fh(p\!-\!m)q        &    -\fh q\bar{b}     \\
	        -\fh(p\!-\!m)q    &    -a_q         &    -\fh qc           \\
                 +\fh bq   &   +\fh\bar{c}q &    0
 \end{array} \!\!\!\right)\!\!\!$
    \\ \multicolumn{3}{|c|}{ }    \\   \cline{1-1} 
    \multicolumn{1}{|c|}{ $\dot{R}_{x \pqg q}^{2}$ }               
  & \multicolumn{1}{c}{  $\dot{R}_{x \pqg z}^{2}$ }
  & \multicolumn{1}{c|}{ $\dot{R}_{z \pqg q}^{2}$ }   \\
	\multicolumn{1}{|c|}{ 	
     $\!\!\!\left(\!\!\!\! \begin{array}{ccc}
                    0    &    +\fh\bar{a}q     &  -  \fh cq     \\
	           - \fh qa   &     0       & -b_q \!-\! b_xq            \\
                +\fh q\bar{c}  &  - b_q \!+\! b_xq    &    0
 \end{array} \!\!\!\!\!\right)\!\!\!$  } 	&
    \multicolumn{1}{c}{ 
$\!\!\!\left(\!\!\!\! \begin{array}{ccc}
                    0    &   +\fh c       &    -\fh\bar{a}     \\
	         +\fh\bar{c}  &   + b_x        &    -\fh(m\!-\!n)           \\
                  -\fh a  &   -\fh(m\!-\!n)    &    -b_x
 \end{array} \!\!\!\!\!\right)\!\!\!$ } &
    \multicolumn{1}{c|}{  					
 $\!\!\!\left(\!\!\!\! \begin{array}{ccc}
                    0     &   +\fh cq      &    +\fh\bar{a}q     \\
	        - \fh q\bar{c} &   +b_q         &   \fh(m\!-\!n)q       \\
                -\fh qa    &   -\fh(m\!-\!n)q    &    -b_q
 \end{array} \!\!\!\!\!\right)\!\!\!$ } 	
    \\ \multicolumn{1}{|c|}{ } & \multicolumn{2}{c|}{ }    \\	  
     \multicolumn{1}{|c|}{   $\dot{R}_{x \pqg q}^{3}$  }            
   & \multicolumn{1}{c}{  $\dot{R}_{x \pqg z}^{3}$ } 
   & \multicolumn{1}{c|}{  $\dot{R}_{z \pqg q}^{3}$  }  \\
   \multicolumn{1}{|c|}{  
  $\!\!\!\left(\!\!\!\! \begin{array}{ccc}
                    0    &  +\fh q\bar{a}     &  - c_q \!+\! c_xq     \\
	            -\fh aq   &     0       &  +\fh \bar{b}q          \\
               - c_q \!-\! c_xq     &   -\fh qb    &    0
 \end{array} \!\!\!\!\right)\!\!\!$ }  &
   \multicolumn{1}{c}{  
$\!\!\!\left(\!\!\!\! \begin{array}{ccc}
                 -c_x    &   -\fh b      &    -\fh(n\!-\!p)     \\
	        -\fh\bar{b}  &   0           &    +\fh a         \\
                 -\fh(n\!-\!p)   &   +\fh\bar{a} &   +c_x
 \end{array} \!\!\!\!\!\right)\!\!\!$ }  &
    \multicolumn{1}{c|}{ 
$\!\!\!\left(\!\!\! \begin{array}{ccc}
                  -c_q    &   -\fh qb          &    -\fh(n\!-\!p)q     \\
	     + \fh\bar{b}q   &   0             &   +\fh aq         \\
               \fh(n\!-\!p)q  &   -\fh q\bar{a}  &    +c_q
 \end{array} \!\!\!\!\right)\!\!\!$  } 
   \\
 \hline
  \end{tabular}
}
  \caption{\setb Vector fields on $T\htho$ generated by the 26 Category 1 Boosts and 
31 Category 2 Rotations from table~\ref{prefbas} in the form of equation~\ref{ththo} 
($\dot{B}_{t \pqg z}^{3},\dot{R}_{x \pqg q}^{2},\dot{R}_{x \pqg q}^{3} $ are non-basis 
elements).}
\label{lbrota}
\end{table}

\begin{table}[htbp]
\centering
\begin{tabular}{|l|}
 \hline   
$\begin{array}{lrrrrrrrr}
  \dot{A}_i:     & \; \dot{a} = &       & -a_4j & +a_3k & +a_6\kl & -a_5\jl &         
&     \\
  \dot{A}_j:     & \; \dot{a} = & +a_4i &       & -a_2k & -a_7\kl &         & +a_5\il 
&     \\ 
  \dot{A}_k:     & \; \dot{a} = & -a_3i & +a_2j &       &         & +a_7\jl & -a_6\il 
&     \\
  \dot{A}_{\kl}: & \; \dot{a} = & +a_6i & +a_7j &       &         & -a_2\jl & -a_3\il 
&     \\
  \dot{A}_{\jl}: & \; \dot{a} = & -a_5i &       & -a_7k & +a_2\kl &         & +a_4\il 
&     \\
  \dot{A}_{\il}: & \; \dot{a} = &       & +a_5j & +a_6k & -a_3\kl & -a_4\jl &         
&     \\
  \dot{A}_l:     & \; \dot{a} = & +a_7i & -a_6j &       &         & +a_3\jl & -a_2\il 
&     \\   
  \\  
  \dot{G}_i:     & \; \dot{a} = &       & -a_4j & +a_3k & -a_6\kl & +a_5\jl &-2a_8\il 
&+2a_7l  \\
  \dot{G}_j:     & \; \dot{a} = & +a_4i &       & -a_2k & +a_7\kl &-2a_8\jl & -a_5\il 
&+2a_6l  \\
  \dot{G}_k:     & \; \dot{a} = & -a_3i & +a_2j &       &-2a_8\kl & -a_7\jl & +a_6\il 
&+2a_5l  \\
  \dot{G}_{\kl}: & \; \dot{a} = & +a_6i & -a_7j &+2a_8k &         & -a_2\jl & +a_3\il 
&-2a_4l  \\
  \dot{G}_{\jl}: & \; \dot{a} = & -a_5i &+2a_8j & +a_7k & +a_2\kl &         & -a_4\il 
&-2a_3l  \\
  \dot{G}_{\il}: & \; \dot{a} = &+2a_8i & +a_5j & -a_6k & -a_3\kl & +a_4\jl &         
&-2a_2l  \\
  \dot{G}_l:     & \; \dot{a} = & +a_7i & +a_6j &-2a_5k &+2a_4\kl & -a_3\jl & -a_2\il 
&        \\
  \\  \end{array} $   \\  
 $\;\,\dot{S}_q^{1}:  
  \left\{\! \begin{array}{ll}
          \dot{a} = & q\sigsum a_r r  \\
		  \dot{b} = & +\frac{3}{2}b_q - \frac{3}{2}b_1q - \fh q \sigsum b_r r \\
		  \dot{c} = & -\frac{3}{2}c_q + \frac{3}{2}c_1q - \fh q \sigsum c_r r
                         \end{array} \right. $ 
  \\   \\
\hline 
 \vspace{-15pt} 
 \\
 {\footnotesize
 $\dot{S}_q^{2}:   \dot{a} = -\frac{3}{2}a_q + \frac{3}{2}a_1q - \fh q \sum a_r r, 
\quad\;\,
      \dot{b} =   q\sum b_r r, \quad\;\,
	     \dot{c} = +\frac{3}{2}c_q - \frac{3}{2}c_1q - \fh q \sum c_r r$   }
  \\
 {\footnotesize 
 $\dot{S}_q^{3}:  \dot{a} = +\frac{3}{2}a_q - \frac{3}{2}a_1q - \fh q \sum a_r r, 
\quad\;\,
     \dot{b} = -\frac{3}{2}b_q + \frac{3}{2}b_1q - \fh q \sum b_r r, \quad\;\,
	   \dot{c} = q\sum c_r r 
  $    }
 \vspace{-15pt} \\
 \\
\hline  
  \end{tabular}
  \caption{\setb Vector fields on $T\htho$ generated by the 21 Category 3 Transverse 
Rotations from the lower section of table~\ref{prefbas}. In the case of $\dot{A}_q$ 
and $\dot{G}_q$ the form of $\dot{b}=f(b)$ and $\dot{c}=f(c)$ is identical to 
$\dot{a}=f(a)$. With reference to equation~\ref{ththo}, in all cases $\dot{p} = 
\dot{m} = \dot{n} = 0$ with $\{\dot{\bar{a}}, \dot{\bar{b}}, \dot{\bar{c}}\}$  implied 
from $\{\dot{a}, \dot{b}, \dot{c}\}$.  ($\dot{S}_q^{2}$ and $\dot{S}_q^{3}$ are 
non-basis elements,
  with $\sum := \sum_{r \ne 1,q}$ here). }
\label{ltrota}
\end{table}


\pagebreak
\chapter{Review of the Standard Model}
\label{rotsm}

\section{Lorentz Symmetry and Spinors}
\label{lsspin}

 Having introduced the higher 27-dimensional form for the flow of time with symmetry 
group $\esi$ acting on $\htho$ in the previous chapter
 we shall address the embedding of the Lorentz symmetry $\soot$, acting on a 
4-dimensional spacetime associated with the components of the subspace $\htwc \subset 
\htho$, within the larger structure in the opening section of the following chapter.
 Here we first consider the properties of the group $\soot$ itself together with its 
representations.

   In general symmetries implicit in the form $\lv$ may include rotation groups,
   such as $\soth \subset \soot$,
   which are significant due to their geometrical interpretation as employed in the 
construction of the background manifold for perception.
 These rotation groups
 may also be embedded within a wider set of elements belonging to the Clifford algebra 
associated with the (pseudo-) Euclidean space to which the rotations apply. These 
algebras also have spinor representations, which are as mathematically natural as the 
vector representations.
  For example, as alluded to in the opening paragraphs of chapter~\ref{esihtho} and 
again explicitly in equation~\ref{trialabc}, the vector and spinor representations of 
SO(8) are equally significant  
 for the structure and symmetry of $\lvt$.
  
  Although most of the discussion below applies to Clifford algebras in general here 
we focus on the case of the 4-dimensional vector space $\rrr^{1,3}$, with real 
Clifford algebra $C(1,3)$ represented by $4 \times 4$ $\gamma$-matrices
 satisfying the relations:
\begin{equation}
  \label{gammalg}
   \gamma^a \gamma^b + \gamma^b \gamma^a = 2\eta^{ab} \, \b1_4
\end{equation}  
 with indices $\{a,b\}=0 \ldots 3$,
   Minkowski metric $\eta^{ab}$ and where $\b1_4$ denotes the $4 \times 4$ identity 
matrix.
  For any vectors $v,w \inn \rrr^{1,3}$ the associated algebra product with $v=v_a 
\gamma^a$ and $w = w_b\gamma^b$ as elements of $C(1,3)$ satisfies $vw+wv = 2\eta(v,w) 
\b1_4$, implying for example the relation $v^2 = \vert v \vert^2 \b1_4$ for all $v 
\inn \rrr^{1,3}$ which is also sufficient to generate the full algebra.
 A general element $u$ of the Clifford algebra $C(1,3)$ has the form:
\begin{equation}
 \label{uuuuu}
  \begin{array}{crlllll}
       &  u     &  = \quad u_0  &  + \quad u_a \gamma^a  &  + \quad u_{ab}\gamma^a 
\gamma^b & + \quad u_{abc} \gamma^a \gamma^b \gamma^c & + \quad \ldots \\
  \inn & C(1,3) &  = \quad C^0  & \oplus \quad C^1   & \oplus \quad C^2    & \oplus 
\quad C^3  & \oplus  \quad \ldots 
  \end{array}
\end{equation}
   with $u_0, u_a, u_{ab}\ldots \inn \rrr$, with index values ordered as $a < b < c 
\ldots $, and where $C^i$ denotes the subspace of $C(1,3)$ formed by the product of 
$i$ basis elements $\{\gamma^a\}$  in this representation.
Owing to equation~\ref{gammalg} the Clifford algebra itself has dimension $2^n$, with 
$n=4$ here, that is the elements of the $C(1,3)$ algebra describe a  vector space with 
16 linearly independent elements.
 
   The Clifford algebra itself does not form a group since in general an inverse 
element may not exist for any given $u \inn C(1,3)$. However, the elements belonging 
to the subset of $C(1,3)$ generated by elements $v \inn C^1$ with $\eta(v,v) = \pm 1$ 
do have an inverse and upon composition generate a group denoted Pin$(1,3)$. Further, 
given such elements $v\inn C^1 \cap \mbox{Pin}(1,3) $ the map $\phi_v$ from $w \inn 
C^1$ into $C^1$:
\begin{eqnarray}
  \label{phiconjc}
    \phi_v: \; w & \to & vwv^{-1}  \\
	   & = & 2\eta(v,w)v^{-1} - wvv^{-1}  \nonumber \\
	   & = & \frac{2\eta(v,w)v}{\eta(v,v)} - w    \label{phiconjr}
\end{eqnarray}
    is a reflection of $w$ through the line containing the origin and $v$ in the 
(psuedo-) Euclidean space $\rrr^{1,3}$. These reflections may be combined to describe 
a representation of Pin$(1,3)$ as orthogonal transformations on the space $\rrr^{1,3}$ 
(which is equivalent to $C^1(1,3)$ as a vector space). The application of Clifford 
algebra composition to induce representations of the rotation groups via 
equation~\ref{phiconjc} is similar to the use of the conjugation action for elements 
of a division algebra such as the octonions, described by equation~\ref{phiconj} in 
section~\ref{oaags}, also to represent rotations.

  In fact the Lie group Pin$(1,3)$ is the two-to-one cover of the full Lorentz group 
O$(1,3)$, which has four disconnected pieces. Restricting the elements of Pin$(1,3)$ 
to those in the \textit{even} subalgebra $C^{\mathrm{e}}(1,3) := \{C^i(1,3); i \; 
\mbox{even}\}$ of equation~\ref{uuuuu} identifies the subgroup Spin$(1,3)$, which has 
a representation on the space $\rrr^{1,3}$ as the group of \textit{special} orthogonal 
transformations SO$(1,3)$. In both cases these actions are two-to-one surjective 
homomorphisms $\pi$ with:
 \begin{eqnarray}
  & \pi : &  \; \mbox{Pin}(1,3) \to \mbox{O}(1,3) \\
  & \pi : &  \mbox{Spin}(1,3) \to \mbox{SO}(1,3) 
 \end{eqnarray}
  Hence the respective Lie algebras are isomorphic, for example $\mbox{spin}(1,3) = 
\mbox{so}(1,3)$.  
 The part of the group Spin$(1,3)$ as a manifold connected to the identity is in fact 
`simply connected' and is denoted $\spot$, the two-to-one covering group of 
$\mbox{SO}^+(1,3)$ -- which in turn is
 the part of the full Lorentz group (described above equation~\ref{mxmass} for the 
$k=2$ case) which preserves both the time and the space orientations, as well as the 
metric relations, of Lorentz 4-vectors.

 The set of matrices:
 \begin{equation}
   \label{sigmaab}
     \sigma^{a \pqg b} \, = \, \frac{1}{4}(\gamma^a \gamma^b - \gamma^b \gamma^a)
	   \, = \, \frac{1}{4}[\gamma^a,\gamma^b]
 \end{equation}
   with $a<b$ and $\gamma^a\gamma^b \inn C^2(1,3)$,
  is isomorphic to the Lie algebra $\mbox{spin}^+(1,3) = \mbox{so}^+(1,3)$ under 
matrix commutation of the six independent $\sigma^{a \pqg b}$ elements. 
   This algebra  generates group elements $R(\omega_{c \pqg d}) = \exp(\omega_{c \pqg 
d}\sigma^{c \pqg d})$ with $\omega_{c \pqg d} \inn \rrr$ (summing over the set of six 
index pairs with $c<d$, in a similar way to the group actions described in 
equation~\ref{gexpam}). These describe $\mbox{SO}^+(1,3)$ vector transformations on 
the $\gamma^a$ matrices themselves:
\begin{equation}
  \label{phiconjug}
    \phi_R: \; \gamma^a \to R(\omega_{c \pqg d}) \gamma^a R^{-1}(\omega_{c \pqg d}) 
\equiv
	 (A^{-1})^a_{\ph{a}b}\gamma^b
\end{equation} 
  with $A\inn \mbox{SO}^+(1,3)$, as well as the spinor   
 representation of $\soot$ on 4-component Dirac spinors $\psi \inn \ccc^4$:  
  \begin{equation}
  \label{psitrans1}
    \psi \; \to \; \psi'  \; = \; e^{\omega_{c \pqg d}\sigma^{c \pqg d}}
	   \psi \; = \; R(\omega_{c \pqg d}) \psi
  \end{equation}

  For a \textit{complex} Clifford algebra in any dimension $n$ this Dirac 
representation is irreducible. However for 
   the real forms of these algebras with even $n=p+q$  the space of the Dirac 
representation for the group $\mbox{Spin}^+(p,q)$ decomposes into two halves, known as 
chiral (\textit{left} and \textit{right}) spinors, upon which inequivalent 
representations act. This may be shown by defining the matrix $\Gamma^5 := \gamma^1 
\gamma^2 \ldots \gamma^n  \inn C^n(p,q)$ which anticommutes with each $\gamma^a$ and 
hence (by equation~\ref{sigmaab} for the general case) commutes with all elements of 
$\mbox{Spin}^+(p,q)$, and hence in turn by Schur's lemma the Dirac representation is 
reducible (unless $\Gamma^5$ is a proportional to the unit $2^{\frac{n}{2}} \times 
2^{\frac{n}{2}}$ matrix, which is generally not the case). 

   In the case of  $(1,3)$ spacetime, with $4 \times 4$ matrices $\gamma^a$ acting on 
the elements $\psi \inn \ccc^4$ of the spinor space, $\Gamma^5$ is denoted $\gamma^5$ 
and the usual convention is to take:
 \begin{equation}
   \gamma^5 = i \gamma^0 \gamma^1 \gamma^2 \gamma^3 \qquad \mbox{for which}
              \qquad (\gamma^5)^2 = +\b1_4
 \end{equation} 
    Due to the factor of $i$ this object does not belong to the \textit{real} Clifford 
algebra. However as a $4 \times 4$ matrix $\gamma^5$ does commute with each element of 
$\spot$ and can be used to extract the 
  chiral spinors $\psi_L$ and $\psi_R$ via the projection operators $P_L$ and $P_R$:
 \begin{eqnarray}
   \psi_L = P_L \psi  & \quad \mbox{with} \quad &  P_L = \fhs(1-\gamma^5)  
\label{plop}  \\
   \psi_R = P_R \psi  & \quad \mbox{with} \quad &  P_R = \fhs(1+\gamma^5)  
\label{prop}
 \end{eqnarray}
  By Schur's lemma this decomposition into left and right-handed spinors $\psi=\psi_L 
+ \psi_R \inn \ccc^4$ is maintained under the $4 \times 4$ matrix actions of the group 
$\spot$.
Hence the Dirac representation is reduced into two $\spot$ invariant and irreducible 
pieces called Weyl spinors.
 A suitable explicit representation for the $\gamma$-matrices
  is the Weyl basis with:
\begin{equation}
 \label{gamweyl}
 \gamma^0 =  \left( \begin{array}{cc} 0 & +\b1_2 \\ +\b1_2 & 0 \end{array}  \right), 
\qquad
 \gamma^a =  \left(
     \begin{array}{cc} 0 & +\sigma^a \\ -\sigma^a & 0 \end{array}  \right),  \qquad
  \gamma^5 =  \left( \begin{array}{cc} -\b1_2 & 0 \\ 0 & +\b1_2  \end{array}  \right)
\end{equation}
  where each entry is a $2 \times 2$ matrix and the three Pauli matrices $\sigma^a$ 
for $a=1,2,3$ are included in the following set:
\begin{equation}
 \label{sigmas}
  \sigma^0 =  \left( \begin{array}{cc} 1 & 0 \\ 0 & 1 \end{array}  \right), \quad
  \sigma^1 =  \left( \begin{array}{cc} 0 & 1 \\ 1 & 0 \end{array}  \right), \quad
  \sigma^2 =  \left( \begin{array}{cc} 0 &-i \\ i & 0 \end{array}  \right), \quad
  \sigma^3 =  \left( \begin{array}{cc} 1 & 0 \\ 0 &-1 \end{array}  \right) 
\end{equation}
In the $\gamma$-matrix basis of equation~\ref{gamweyl} the 
   $\spot$ action of equation~\ref{psitrans1} can be expressed on the Weyl spinors 
$\psi_L, \psi_R \inn \ccc^2$ simply as:
\begin{equation} 
\label{psitrans2}
\left(\!\! \begin{array}{c}  \vspace{-5mm} \\ \psi  \vspace{5mm} 
   \end{array}  \!\! \right)  =
      \left( \begin{array}{cc} \psi_L  \\ \psi_R  \end{array}  \right)
  \to  \left( \begin{array}{cc} R_L & 0 \\ 0 & R_R \end{array}  \right) 
      \left( \begin{array}{cc} \psi_L  \\ \psi_R  \end{array} \right)
\end{equation} 

  For particle states chirality itself is an observable only for massless fermions, 
that is $m_f = 0$, in which case it is equivalent to the particle helicity.

The `spin' group for the Clifford algebra of the real pseudo-Euclidean vector space 
$\rrr^{1,3}$ may also be approached directly via the group $\sltc$, which is closely 
related to the representations $R_L$ and $R_R$ in equation~\ref{psitrans2}. The 
6-dimensional Lorentz Lie algebra $\soota$ can be expressed in a conventional basis of  
anti-Hermitian rotation generators $\{J^1, J^2, J^3\}$ and Hermitian boost generators 
$\{K^1,K^2,K^3\}$ in terms of a $2 \times 2$ matrix basis for $\sltca$ in the form:
\begin{equation}
   J^a = -\frac{i}{2}\sigma^a \qquad \mbox{and} \qquad K^a = -\frac{1}{2}\sigma^a 
     \label{jsigksig}
\end{equation}
  for $a=1,2,3$. The signs are chosen such that the following algebra commutators 
hold:
\begin{eqnarray}
 \lbrack  J^a, J^b \rbrack & = & \varepsilon^{abc} J^c  \label{jjej} \\
 \lbrack  K^a, K^b \rbrack & = & -\varepsilon^{abc} J^c \label{kkej} \\
 \lbrack  J^a, K^b \rbrack & = & \varepsilon^{abc} K^c  \label{jkek}
\end{eqnarray} 
 with $\varepsilon^{123} = +1$.
  In other conventions the signs may vary, and factors of $i=\sqrt{-1}$ may appear if 
$J$ is defined to be Hermitian, as is the case in quantum mechanics in order to 
identify real observable quantities for angular momentum. In the standard treatment a 
general element of the group $\sltc$ is represented by the $2 \times 2$ matrix:
\begin{equation}
  \label{stjsk}
  S \: = \: e^{r_aJ^a + b_aK^a} \: = \: e^{\fh(-ir_a - b_a)\sigma^a}  
\end{equation} 
 with the rotations parametrised by $r_a \inn \rrr$, $a=1,2,3$, and the boosts 
parametrised by $b_a \inn \rrr$, $a=1,2,3$. For the complex linear combinations $A^a = 
\fh (J^a + iK^a)$ and $B^a = \fh (J^a - iK^a)$  the Lie bracket reads:
\begin{eqnarray}
 \lbrack  A^a, A^b \rbrack & = & \varepsilon^{abc} A^c \label{jpmeja} \\
 \lbrack  B^a, B^b \rbrack & = & \varepsilon^{abc} B^c \label{jpmejb} \\
 \lbrack  A^a, B^b \rbrack & = & 0     \label{jpmez}
\end{eqnarray} 
  demonstrating that the complexified Lie algebra of $\sltca$ is isomorphic to $\sutwa 
\oplus \sutwa$ (as will be represented in figure~\ref{dynkine}(d) and described in the 
accompanying text)
 which is used to label the representations of the Lorentz group by the half-integer 
values $(j_A, j_B)$. After the trivial (0,0) scalar case the two lowest-dimensional 
possibilities are the representations of $\sltc$ denoted $R_L(S)$ and $R_R(S)$ with: 
\begin{eqnarray}
 & &  (j_A,j_B)  =  (\fhs,0) \qquad \qquad \Rightarrow \quad 
                            R_L(S) = e^{\fh(-ir_a - b_a)\sigma^a} \label{leftrep}  \\
 & & (A^a  = -\fhs \sigma^a,\; B^a = 0; \qquad  J^a  =
     -\mbox{\small{$\frac{i}{2}$}} \sigma^a,\; K^a  = -\fhs \sigma^a)  \nonumber  \\ 
	       & &  \nonumber \\
\mbox{and} & &  (j_A,j_B)  =  (0,\fhs) \qquad \qquad \Rightarrow \quad 
                            R_R(S) = e^{\fh(-ir_a + b_a)\sigma^a}  \label{rightrep}  
\\
 & & (A^a  = 0,\; B^a = -\fhs \sigma^a; \qquad  J^a  =
     -\mbox{\small{$\frac{i}{2}$}} \sigma^a,\; K^a  = +\fhs \sigma^a)  \nonumber
\end{eqnarray}

 The first of these representations $R_L(S)$ can be identified with the original set 
of $2 \times 2$ matrices $S\inn \sltc$, that is $\{S \inn \ccc(2): \mbox{det}(S)=1\}$, 
as parametrised in the form of  equation~\ref{stjsk}. The representation  $R_R(S)$ in 
equation~\ref{rightrep} is a \textit{different} map from the \textit{same} complete 
set of $\sltc$ elements, considered as an abstract group, into $2 \times 2$ matrix 
transformations on a 2-dimensional complex vector space $\ccc^2$. The two 
representation spaces are given different subscript labels $L$ and $R$ to denote that 
they belong to different $\sltc$ representations with the left-handed Weyl spinor 
transforming as $\psi_L \to R_L(S) \psi_L$ and the right-handed Weyl spinor 
transforming as $\psi_R \to R_R(S) \psi_R$.

   Under a discrete parity transformation the sign of a Lorentz boost is reversed 
while the sign of a rotation is invariant. The naming convention of `left' and `right' 
representations originates since $R_L(S)$ and $R_R(S)$ are related by the sign of the 
boost generator contributions in equations~\ref{leftrep} and \ref{rightrep} and are 
hence interchanged under a parity transformation. Indeed in general the parity 
operation switches between the two Lorentz representations $(j_1,j_2)$ and 
$(j_2,j_1)$.

  Since there is no $2\times 2$ matrix $D$ such that $R_L(S) = DR_R(S)D^{-1}$ for all 
$S\inn \sltc$ the representations $R_L(S)$ and $R_R(S)$ are inequivalent. However the 
following relationships between equations~\ref{leftrep} and \ref{rightrep} hold  (with 
$\sigma^2$ defined in equation~\ref{sigmas}):
\begin{eqnarray}
  R_L^{\,\ast}(S) & = & \sigma^2 \; R_R(S) \; (\sigma^2)^{-1}  \label{rlssrrss}  \\
  R_L^{\dagger^{-1}}(S) & = & R_R(S) \label{rlsrrs}   \\
  {R_L^{}}^{\! \mbox{\tiny{$T$}}}(S) & = & \sigma^2 \; R_L^{\,-1}(S) \; 
(\sigma^2)^{-1}
\end{eqnarray}
   showing respectively that the complex conjugate of $R_L(S)$ is equivalent to 
$R_R(S)$, the contragredient of $R_L(S)$ is \textit{equal} to $R_R(S)$ and the 
transpose of $R_L(S)$ is equivalent to its inverse. 

The Dirac representation $R_D(S)$ of $\sltc$ has the reducible form $(\fh,0) \oplus 
(0,\fh)$, acting on spinors in the space $\ccc^4$, and via equation~\ref{rlsrrs} it 
can be written as:
\begin{equation}  
  \label{rdsrss}
   R_D(S) = \left( \begin{array}{cc} R_L(S) & 0 \\ 0 & R_R(S) \end{array}  \right)  =
            \left( \begin{array}{cc} S & 0 \\ 0 & S^{\dagger^{-1}} \end{array}  
\right) 
\end{equation}
  which is the same action as described in equation~\ref{psitrans2}, there derived 
from the Clifford algebra structure, with $R_L = R_L(S)$ and $R_R = R_R(S)$. 
  Hence $S\inn \sltc$ acts on the left-handed components of the $\ccc^4$ Dirac spinor 
and 
$S^{\dagger^{-1}}$  acts on the right-handed components as an inequivalent 
representation of $\sltc$.
 Equation~\ref{rdsrss} describes how the  $\spot$ Dirac representation can be 
constructed by combining left and right spinors as the $(\frac{1}{2}, 0) \oplus (0, 
\frac{1}{2})$ representation of $\sltc$.
  In fact $\spot$ is isomorphic to the group $\sltc$  (such isomorphisms for the spin 
groups only exist in low dimensions and for a handful of cases), each expressing the 
two-to-one cover of $\soot$.

 Equations~\ref{psitrans2} and \ref{rdsrss} and the comparison  of the Dirac 
representation 
  constructed as a reducible representation of $\spot$ via the Clifford algebra or as 
a combination of two representations of $\sltc$ expresses the relation between the 
4-component and 2-component spinor formalism. The 2-component Weyl spinors are more 
fundamental in the sense that $\psi_L$ and $\psi_R$  are treated differently in 
important features of the Standard Model, as we shall describe in the following 
section.

     The two-to-one relationship between $\sltc$ and $\soot$ may be exhibited by 
mapping a Lorentz vector $\bv_4 \inn \rrr^{1,3}$ into the space of $2\times 2$ complex 
Hermitian matrices as:
\begin{equation}
\label{vtoh}
  \v_4 = (v^0,v^1,v^2,v^3)\; \to\; \bh_2 = \bv_4 \! \cdot \! \bsig =
   \left( \begin{array}{cc} v^0+v^3 & v^1-v^2i \\ v^1+v^2i & v^0-v^3 \end{array}  
\right) \subset \htwc
\end{equation}
 where $\bsig$ denotes the $2\times 2$ identity matrix $\sigma^0$ together with the 
three Pauli matrices $\sigma^a$ of equation~\ref{sigmas}. This is the same object 
introduced in equation~\ref{xcomp} of section~\ref{ltos} and also in 
equation~\ref{extloract} of section~\ref{laofesi}, based on the imaginary unit $l$ in 
the latter case.
We see from this equation, and in comparison with section~\ref{ltos}, that 
$\mbox{det}(\bh_2) = (v^0)^2-(v^1)^2-(v^2)^2-(v^3)^2 =h^2$, with $h \inn \rrr$, which 
may be expressed as the form $L(\v_4) = h^2$
 (as employed in equation~\ref{lorform2}). While the fundamental representation of 
$\sltc$ acts on the space $\ccc^2$, the group action for elements $S\inn \sltc$ on the 
space $\htwc$ provides another representation given by: 
\begin{equation}
 \label{hshs}
  \bh_2 \to  \bh_2^{\prime} =  S \, \bh_2 \, S^{\dagger}
\end{equation} 
  This maps $\bh_2 \to \bh_2^{\prime}$ onto a new $2\times 2$ complex Hermitian matrix 
with the same determinant; hence mapping the components $v^a_4 \to v^{\prime a}_4$ 
according to a Lorentz transformation of the real 4-vector $\v_4\inn\rrr^{1,3}$.
 With $S\inn\{\b1_2,-\b1_2\}$ giving the identity transformation, $\bh_2' = \bh_2$, 
 the group $\sltc$, isomorphic to $\spot$ as described above, is the two-to-one 
covering spin group for $\soot$; that is $\soot = \sltc /\zzz^2$.

 In fact the components of $\bv_4$ transform under the 4-dimensional vector 
$(\frac{1}{2},\frac{1}{2})$  representation of $\sltc$. 
The matrix $\bh_2$, and hence the vector $\v_4$, can be considered to be constructed 
out of two 2-component left-handed Weyl spinors $\chi$ and $\phi$  such that: 
\begin{equation}
  \label{htwocp}
\bh_2 = \chi\chi^{\dag} + \phi\phi^{\dag}
\end{equation}
 as implied in equations~\ref{ththvec} and \ref{ththdet0} of section~\ref{ltos}, with 
the elements of the group $\sltc$ acting on the spinor components in the appropriate 
way. 

   This spinor substructure of vectors $\bv_4$
  has some similarity to the situation discussed in section~\ref{subwal} for the 
relation $-\frac{1}{\kappa}G^{\mu\nu} = \rho u^{\mu} u^{\nu} - S^{\mu\nu}$, as implied 
in equation~\ref{gruus} via the Einstein equation, which describes the possibility of 
composing the rank--2 Einstein tensor in terms of a substructure involving the 
apparent 4-dimensional \textit{macroscopic} vector flow $\bu(x)$ on the base manifold.
The  natural algebraic substructure of the field $\bv_4(x)$ in terms of the  spinor 
decomposition of equation~\ref{htwocp} may in turn  be intimately related to the 
possible field interactions implied within the higher-dimensional form of time $\lvt$ 
at the \textit{microscopic} level, underlying 
 the composition of the Einstein tensor  $G^{\mu\nu} = f(Y, \hat{\bv})$ as expressed 
in equation~\ref{getypsi}.

  Equation~\ref{hshs} describes the determinant preserving action $\bh_2 \to S \, 
\bh_2 \, S^{\dagger}$ of the elements $S\inn \sltc$ upon elements of the vector space 
of matrices $\bh_2 \inn \htwc$ that was extended in equations~\ref{xoct} and 
\ref{mxmass}, by augmenting the complex numbers to the octonions, to identify 
 an $\sltwoo$ action on $\htwo$ as
the covering group of the 10-dimensional Lorentz group, as an intermediary for the 
$\esi$ action on $\htho$. 
   For infinitesimal transformations we write $S=\exp(a) \simeq 1+a$, where $a\inn 
\sltca$  is an infinitesimal element of the Lie algebra of $\sltc$, and we have:    
\begin{eqnarray}
  \bh_2 \to (1+a)\bh_2(1+a^{\dagger})
   & \simeq &  \bh_2  + \delta \bh_2  \\
   \mbox{with} \qquad \delta \bh_2 & = & a\bh_2 + \bh_2 a^{\dagger}
\end{eqnarray}
   where $\delta \bh_2$ has the same form as the 64 $D^S$ actions on $\htho$ of 
equation~\ref{dsgdecom} in section~\ref{earlyo}, and may here be considered to 
represent the Lie algebra $\sltca$ on the tangent space $T\htwc$. This corresponds to 
a possible substructure embedding of $\htwc \subset \htwo \subset \htho$ with 
respective group actions $\sltc \subset \sltwoo \subset \sltho$. Before moving to the 
action of $\sltc$ on full space $\htho$ in the following chapter (see 
equation~\ref{hvvv}), we first here consider the action of $\sltc$ on the space 
$\hthc$. The $2 \times 2$ matrices $S\inn \sltc$ can be embedded in $3 \times 3$ 
matrices acting on $\mcX \inn \hthc$ as:
\begin{equation}
     \label{vinhinc}
          \mcX \;\, \to \;\,
		\left( \begin{array}{c|c} 
        \,\,\,\,\:\! S \;\!\!   \begin{array}{cc} &  \\  &  \end{array} \!\!\!   &
        \,     0  \begin{array}{cc} &  \\  &  \end{array} \!\!\!\!\!\!\!\!\!\! 
				                         \\  \hline
        \,\,\,\,\,\,  0  \!\! \begin{array}{cc}        &   \end{array}   &	  
		\,  1      \end{array}  \right) 
		\left( \begin{array}{c|c} 
        \,\,\,\, \bh_2     \!\!     \begin{array}{cc} &  \\  &  \end{array} \!\!\!   &
        \,  \psi_L  \begin{array}{cc} &  \\  &  \end{array} \!\!\!\!\!\!\!\!\!\! 
				                         \\  \hline
        \,\,\,\,\,\, \psi_L^{\dagger} \!\!\! \begin{array}{cc}    &   \end{array}   &	  
		\,  n      \end{array}  \right) 
		\left( \begin{array}{c|c} 
        \,\,\,\,\, S^{\dag} \!  \begin{array}{cc} &  \\  &  \end{array} \!\!\!   &
        \,  0  \begin{array}{cc} &  \\  &  \end{array} \!\!\!\!\!\!\!\!\!\! 
				                         \\  \hline
        \,\,\,\,\,\, 0 \!\! \begin{array}{cc}        &   \end{array}   &	  
		\,  1      \end{array}  \right) 
\end{equation}

   This combines the vector representation of $\sltc$ on $\bh_2 \inn \htwc$ and the 
spinor representation on $\psi_L \inn \ccc^2$, together with the scalar $n \inn \rrr$, 
in a single symmetry transformation which preserves $\det(\mcX) \inn \rrr$. In 
section~\ref{extsym} the spinor $\psi_L$ will be identified with $\theta_l \inn 
\ccc^2$ in a complex subspace of $\theta \inn \ooo^2$ under the full $\esi$ action on 
$\htho$, compatible with the embedding of the $\sltc$ action of equation~\ref{vinhinc} 
within the $\sltwoo \subset \esi$ action of equation~\ref{mxm3}.


\section{Internal Symmetry and Electroweak Theory}
\label{ewtatsm}

  Together with the external Lorentz symmetry internal gauge symmetries are key to the 
properties of particle states observed in the laboratory. In this section we review 
the internal symmetries of the Standard Model with a particular emphasis on 
electroweak  theory and the phenomenon of symmetry breaking (see for example 
\cite{Teub}).
  
  The quarks and leptons of one generation of Standard Model fermions transform under 
an internal symmetry described by the group product $\SML$ (with the subscripts `$c$', 
`$L$' and `$Y$' denoting colour, left-handed and hypercharge respectively).
 The corresponding representation of $\SML$ is composed as a sum of five irreducible 
pieces each labelled according to their transformation properties by 
$(n_3,n_2,n_1)_{L,R}$ with the subscript $L$ or $R$ denoting left or right chiral Weyl 
spinors, represented as four-component Dirac spinors, under the external Lorentz 
group. The five pieces are of dimension 6, 3, 3, 2 and 1 respectively  (without an 
extra piece $(1,1,0)_R$ for a right-handed neutrino $\nu_R$): 
\begin{equation}
 \begin{array}{lllll}
 \;\; (3,2,\mbox{\small{$\frac{1}{6}$}})_L \quad+& 
(3,1,\mbox{\small{$\frac{2}{3}$}})_R \quad+ & (3,1,-\mbox{\small{$\frac{1}{3}$}})_R 
\quad+& (1,2,-\mbox{\small{$\frac{1}{2}$}})_L \quad+& (1,1,-1)_R   \label{multsf}  \\
    q_L = \binom{u_L\,\,(\frac{2}{3})\,\,}{d_L(-\frac{1}{3})}   
    & \;\; u_R \; (\frac{2}{3}) & \;\; d_R \; ({-\frac{1}{3}}) &
	\!\! l_L = \binom{\nu_L\,\,(0)\,\,}{e_L(-1)} & \;\; e_R  \; (-1)   \\
\end{array}
\end{equation}
   with the corresponding set of 15 particle states named on the second line alongside 
their electromagnetic charges.   The components of particle multiplets transforming as 
triplets under $\suth_c$, ($n_3=3$), couple to the strong interaction and consist of 
$u$-type and $d$-type quarks, while the $\suth_c$ singlet components consist of the 
neutrino $\nu$ and electron $e$ leptonic states.

  In the Standard Model electroweak theory weak eigenstates, that is fields 
transforming according to definite $\sutw_L$ representations, are composed as 
left-handed doublets ($n_2=2$) and right-handed singlets ($n_2=1$), transforming for 
example in the case of leptons as $l_L \to l'_L = e^{-i\omega^{\alpha} \tau^{\alpha}} 
l_L$, with $\omega^{\alpha} \inn \rrr$ and
\begin{equation}
 \label{tauhsig}
  \tau^{\alpha} = \fh \sigma^{\alpha}
\end{equation}  
   for $\alpha=1,2,3$ (see equation~\ref{sigmas}, with Greek indices used here for the 
generators of a gauge group), and $e_R \to e'_R = e_R$. With left and right-handed 
fermions hence undergoing different interactions with the $\sutw_L$ gauge field this 
construction describes the empirical observation of parity violation in weak 
interactions. When additional generations of fermions are considered the weak 
eigenstates generally consist of a linear combination of physical mass eigenstates 
leading to the phenomena of mixing between the generations, as will be described 
towards the end of this section. 

  The electromagnetic charge of each particle in a multiplet is given by:
\begin{equation}
 \label{qethy}
    Q=T^3+\frac{Y}{2}
\end{equation}
  where $T^3$ is the eigenvalue under the third, diagonal, $\sutw_L$ generator and the 
hypercharge $Y$ labels the $\uo_Y$ representations, ($n_1=Y/2$) in 
equation~\ref{multsf}, which are all one-dimensional for this Abelian group. For the 
right-handed states  $T^3=0$ and the hypercharge is simply the electric charge of the 
fermion $Q(\psi_R) = \frac{Y}{2}(\psi_R)$. All fields transform as $\psi \to \psi' = 
e^{-i\omega \frac{Y}{2}(\psi)}\psi$, with $\omega \inn \rrr$, under the hypercharge 
gauge symmetry $\uo_Y$. The hypercharge $\frac{Y}{2}(\psi)$ itself is ultimately 
defined to give the correct electromagnetic charge $Q$, via the relation in 
equation~\ref{qethy}, which is the same for the $L$ and $R$ parts of each fermion type 
with $Q(e_L) = Q(e_R) = -1$ for example.  The charge $Q$ determines the coupling to 
the electromagnetic field corresponding to the $\uo_Q$ gauge symmetry that survives 
electroweak symmetry breaking. Equation~\ref{qethy} may  be considered as a relation 
either between the eigenvalues or the operators $Q$, $T^3$ and $\frac{Y}{2}$, 
depending on the context.

   The representations of equation~\ref{multsf} can be expressed purely in terms of 
left-handed fields by applying `charge conjugation' to the right-handed cases, under 
which $(3,1,\frac{2}{3})_R \to (\bar{3},1,-\frac{2}{3})_L$ for example. Having all 
fields expressed in terms of the same Lorentz representation in this way is useful for 
unification models, in which individual pieces of equation~\ref{multsf} are combined 
in a larger representation of a single unifying gauge group. Since gauge 
transformations commute with Lorentz transformations, without interchanging $L$ and 
$R$ states, such a unifying gauge group then respects Lorentz invariance in the 
theory. While the states in equation~\ref{multsf} are all considered as `particles' 
the action of charge conjugation also introduces `antiparticle' states. Hence both 
particle and antiparticle states may be combined in unified multiplets, as for the 
case of the SU(5) model \cite{GeoGla} cited regarding figure~\ref{dynkinb} in the 
following section.

  The dynamics of the Standard Model fields is heavily based on a Lagrangian 
formalism. The  Standard Model Lagrangian includes kinetic terms for the fermions in 
the form of the final term of equation~\ref{lagdym}, which for the lepton doublet 
$l_L$, with a conventional factor of $i$ and covariant derivative $D_{\mu}$, can be 
expressed as:
\begin{eqnarray}
 \lag_{\mbox{{\scriptsize kin}}} & = & i \bar{l}_L 
                 \gamma^{\mu} D_{\mu} l_L  \label{lagklep} \\
 \mbox{with} \quad D_{\mu} & = & \pal_{\mu}\, +\, ig \, W^{\alpha}_{\mu}(x)\, 
\tau^{\alpha}
                   \, + \,i g' \, B_{\mu}(x) \, \fhy(l_L)    \label{covdevlep}
\end{eqnarray}
 where $\tau^{\alpha}$ is defined in equation~\ref{tauhsig} (and with an additional 
$D_{\mu} = \ldots + ig_s  G^{\beta}_{\mu}(x) \lambda_{\beta}$ term, with $\beta = 
1\ldots 8$
 and the $\lambda_{\beta}$ matrices listed in table~\ref{gellm}, 
  for $\suth_c$ gauge interactions in the case of quarks). Hence the interaction 
between the gauge fields $W^{\alpha}_{\mu}(x)$, $B_{\mu}(x)$, with respective 
couplings $g$, $g'$, and left-handed leptons has the Lagrangian form: 
\begin{eqnarray}
  \lag_{\mbox{{\scriptsize int}}} \!\! & = & \!\!
   -\frac{g}{2}\big(\bar{\nu}_L \;\;\; \bar{e}_L\big) \gamma^{\mu}
    \left(\!\! \left(\!\! \begin{array}{cc}  W^{3}_{\mu}  & W^{1}_{\mu} -i W^{2}_{\mu}  
\\   
     W^{1}_{\mu} +i W^{2}_{\mu} & -W^{3}_{\mu} \end{array} \!\! \right)
  \! -\frac{g'}{g} \! \left(\!\!\! \begin{array}{cc} B_{\mu}  & 0 \\
                              0 & B_{\mu} \end{array}\!\! \right)\!\! \right) 
				\!\!\! \left(\!\! \begin{array}{c} \nu_L \\
                              e_L \end{array} \!\!\right)
							    \qquad \;\;  \label{lagwwwb} \\
   \lag_{\nu}  & \!\! = \!\! & \!  -\frac{g}{2} \, \bar{\nu}_L \gamma^{\mu} 
     (W^{3}_{\mu} -\frac{g'}{g} B_{\mu} ) \nu_L   \label{lagkneu}
\end{eqnarray} 
   where the part $\lag_{\nu}$  describes the gauge coupling to the neutrino alone, as 
implied in equation~\ref{lagklep}. Physical gauge boson fields $A_{\mu}(x)$ and 
$Z_{\mu}(x)$ are defined as a linear combination of  $B_{\mu}(x)$ and $W^3_{\mu}(x)$ 
via the orthogonal transformation:
\begin{equation}
 \begin{array}{lcr}
    A_{\mu}  & = &   \;\cos \theta_W \, B_{\mu} + \sin \theta_W \, W^{3}_{\mu} \\
	 Z_{\mu}  & = & \!\!\!-\sin \theta_W \, B_{\mu} + \cos \theta_W \, W^{3}_{\mu} \\
 \end{array} 
	 \label{azthwbw}
\end{equation}
  that is with:
\begin{equation}
 \begin{array}{lcr}
     B_{\mu}  & = &   \cos \theta_W \, A_{\mu} - \sin \theta_W \, Z_{\mu}   \\
	 W^{3}_{\mu}  & = & \sin \theta_W \, A_{\mu} + \cos \theta_W \, Z_{\mu}   \\
 \end{array}
     \label{bwthwaz}
\end{equation}
  where $\theta_W$ is the weak mixing angle. Hence from equation~\ref{lagkneu} the 
coupling of the neutrino to the physical gauge field $A_{\mu}$ is:
\begin{equation}
   \lag_{\nu A}  =    -\frac{g}{2} \, \bar{\nu}_L \gamma^{\mu} 
     ( \sin \theta_W \, A_{\mu} \, - \, \frac{g'}{g} \cos \theta_W \, A_{\mu})\nu_L 
	 \label{aonneut}
\end{equation}	
  which is zero for: 
\begin{equation}
        \tan \theta_W  =  \frac{g'}{g}   \label{tantw}
\end{equation}
    This value of the weak mixing angle $\theta_W$ hence describes the electric charge 
neutrality of the neutrino with $A_{\mu}(x)$ interpreted as the electromagnetic field, 
the quanta of which are photons. More generally the coupling terms for the photon can 
be extracted from the relevant part of the covariant derivative $D_{\mu}$, of the form 
in equation~\ref{covdevlep}, acting on any field $\psi(x)$ as (with $T^3$ representing 
the third component of su(2)$_L$ and the hypercharge $\frac{Y}{2}$ as operators acting 
on the field $\psi$):
\begin{eqnarray} 
 D_{\mu} \! & \! \sim \! & 
      ig \, W^3_{\mu}\, {T}^3(\psi)  + ig' \, B_{\mu} \, \fhy(\psi) 
         \qquad\qquad\quad  \mbox{retaining only $W^{3}_{\mu},B_{\mu}$ field parts} 
\nonumber \\  
       & = \! &   ig \sin \theta_W\, A_{\mu} \, {T}^3  
	         + ig'\cos \theta_W\, A_{\mu} \, \fhy
			   \qquad\;
			    \mbox{by equation~\ref{bwthwaz}, dropping $Z_{\mu}$ parts}  \nonumber 
\\ 
       & = \! &  ig \sin \theta_W\, A_{\mu} \, {T}^3 
	         + ig\sin \theta_W\, A_{\mu} \, \fhy
			   \qquad\;\;  \mbox{using equation~\ref{tantw}}   \nonumber  \\
	   & = \! & ig \sin \theta_W \, A_{\mu} \, ({T}^3 + \fhy)
	       \qquad \qquad \qquad \quad \! \equiv \; ie \, A_{\mu} Q   \label{amucoup}
\end{eqnarray}
    Hence the electromagnetic coupling of any particle state to the photon is always 
proportional to $eQ$ where the particle charge $Q$ is defined in equation~\ref{qethy} 
and the electromagnetic coupling $e$ is given by:
\begin{equation}
  \label{egsint}
     e = g \sin \theta_W
\end{equation} 
  As described after equation~\ref{qethy} the different values of $\frac{Y}{2}$  
compensate for the different $T^3$ values for the $L$ and $R$ states of a given 
particle such that the respective coupling of each chiral component to the gauge field 
$A_{\mu}(x)$ is the same, as can be seen for each particle type in 
equation~\ref{multsf}. Following the same lines of reasoning in equation~\ref{amucoup} 
except instead retaining the gauge field $Z_{\mu}(x)$ and dropping the $A_{\mu}(x)$ 
field parts in the second line leads to:
\begin{eqnarray} 
 D_{\mu} & \sim &    ig \cos \theta_W \, Z_{\mu} \,  {T}^3  
	         - ig'\sin \theta_W\, Z_{\mu} \, \fhy
			    \nonumber \\ 
       & = &  ig \cos \theta_W\, Z_{\mu} \, {T}^3
	         - ig  \frac{\sin^2 \theta_W}{\cos \theta_W}\, Z_{\mu} \,\fhy  \nonumber  
\\
	   & = & ig \, Z_{\mu} \, \left( \left( \cos \theta_W
	              + \frac{\sin^2 \theta_W}{\cos \theta_W}
	   \right) \, T^3 \, - \, \frac{\sin^2 \theta_W}{\cos \theta_W}\, 
	   \left( T^3 + \frac{Y}{2}  \right) \right)   \nonumber \\
	   & = & \frac{ig}{\cos \theta_W} \, Z_{\mu} \,
	    \left( T^3 \, - \, Q \sin^2 \theta_W \right)   \label{zmucoup}
\end{eqnarray}
   Hence there are two terms for the weak neutral interactions. The second term is 
proportional to the electromagnetic charge $Q$ and is hence the same for $L$ and $R$ 
particle states. However since the eigenvalues of $T^3$ are only non-zero for the 
left-handed states the first term only couples to the $\psi_L$ components.  The 
combination of the two terms in equation~\ref{zmucoup} implies that parity violation 
is only partial for neutral weak interactions. On the other hand for the charged weak 
interactions mediated via the $W^{\pm}_{\mu}(x)$ gauge fields, introduced in 
equation~\ref{wpmww} below and involving only $\sutw_L$ components, parity violation 
is maximal. 
In the Standard Model Lagrangian the left-handed chiral states $\psi_L$ are projected 
out of the Dirac spinor states for the fermions using the $P_L$ operator of 
equation~\ref{plop}, as seen for example in equations~\ref{lagqwint} and 
\ref{lagqwckm} at the end of this section.

 In addition to the spin-$\fh$ fermions and spin-$1$ gauge bosons the Standard Model 
also introduces a spin-$0$ Higgs field, which is massive itself and closely associated 
with the origin of  mass for the  $W^{\pm}$ and $Z^0$ gauge bosons as well as the 
fermion states.  
Indeed electroweak theory is inextricably linked to the Higgs sector with the breaking 
of the electroweak symmetry $\sutw_L \times \uo_Y$ to the $\uo_Q$ of electromagnetism 
mediated through the action of the gauge group on the Higgs field:
\begin{equation} 
 \phi= \left( \begin{array}{c}  \phi^+ \\ \phi^0 \end{array} \right) =
   \frac{1}{\sqrt{2}}  \left( \begin{array}{c} 
       \phi_1 + i\phi_2 \\ \phi_3 + i\phi_4 \end{array} \right)
	\label{higgsf}
\end{equation}
  Transforming as a scalar under the external Lorentz symmetry the Higgs field is also 
invariant under the internal $\suth_c$ symmetry. On the other hand the above complex 
doublet of scalar fields $\phi$ transforms as a doublet under $\sutw_L$ while also 
possessing hypercharge with $\fhy = +\fhs$, which also accounts for the notation 
$\phi^+$ and $\phi^0$ in equation~\ref{higgsf} by reference to equation~\ref{qethy}. 
This collection of properties may be denoted $(1,2,\fhs)_0$ by comparison with the 
list of Standard Model fermions in equation~\ref{multsf}. The Lagrangian for the Higgs 
sector is:
\begin{eqnarray}
   \lag_H & = & (D_{\mu}\phi)^{\dag}D^{\mu} \phi \; - \; V(\phi) \label{laghiggs} \\
   \mbox{where} \quad D_{\mu}\phi & = & \left( \pal_{\mu} \, + \, i\frac{g}{2} \,
    W^{\alpha}_{\mu} \,  \sigma^{\alpha} \,   
      + \, i\frac{g'}{2} \, B_{\mu} \, \sigma^0  \right) \phi   \label{covderphi}
\end{eqnarray}
 is the gauge covariant derivative
 which is similar in form to equation~\ref{covdevlep} except with $\fhy = +\fhs$ here, 
and  
  also $\sigma^0$ and $\sigma^{\alpha} = \{\sigma^1, \sigma^2, \sigma^3\}$ have been 
adopted directly from equation~\ref{sigmas}  rather than via equation~\ref{tauhsig}. 
The fields $W^{\alpha}_{\mu}(x)$ and $B_{\mu}(x)$ are the $\sutw_L$ and $\uo_Y$ gauge 
fields, with couplings $g$ and $g'$ respectively, as introduced in 
equation~\ref{covdevlep}.
 The breaking of the electroweak symmetry relies on the `Mexican hat' potential term 
in the Lagrangian of equation~\ref{laghiggs} with:
\begin{equation}
   \label{higgspot}
     V(\phi)\, = \, -\mu^2 \phi^{\dag}\phi\, + \, \lambda(\phi^{\dag}\phi)^2
\end{equation}
with real coefficients $\mu^2>0$ and $\lambda > 0$. From equation~\ref{higgsf} it can 
be seen that the potential $V(\phi)$ is a function of $\phi^{\dag}\phi = \fhs 
\sum_{i=1}^4 \phi_i^2$ only. The vacuum expectation value for this field $\langle \phi 
\rangle$, that is the minimum in the potential, can be taken without loss of 
generality (in the `unitarity gauge') to be:
\begin{equation}
   \label{higgsvac}   
    \langle \phi \rangle =  \frac{1}{\sqrt{2}}  \left( \begin{array}{c} 
       0 \\ v \end{array} \right)
	    \qquad \mbox{with} \qquad v =   \frac{\mu}{\sqrt{\lambda}}
\end{equation}
   This charge neutral component of the Higgs field $\phi^0 = \frac{v}{\sqrt{2}}$ is 
invariant under the action of the charge generator $Q = T^3 + \frac{Y}{2} = 
  \binom{\frac{1}{2} \; 0}{0 \, -\frac{1}{2}} + \binom{\frac{1}{2} \; 0}{0 
  \; \frac{1}{2}} = \binom{1 \; 0}{0 \; 0}$, from equation~\ref{qethy} applied for the 
Higgs field,  which remains unbroken. 
 Hence the gauge symmetry is broken from $\sutw_L \times \uo_Y$ down to $\uo_Q$, 
identified in a linear combination of the third component of $\sutwa_L$ and the 
hypercharge generator
   u$(1)_Y$, as the symmetry which leaves the vacuum value $\langle \phi \rangle$ in 
equation~\ref{higgsvac} invariant.

  Masses arise for the gauge fields corresponding to the broken  $\sutw_L \times 
\uo_Y$ generators from the kinetic term in the Higgs Lagrangian of 
equation~\ref{laghiggs}. Acting on the vacuum state the covariant derivative of 
equation~\ref{covderphi} can be written as:

\begin{equation}
  D_{\mu}\phi = \left( \begin{array}{cc}
    \pal_{\mu} +  \frac{i}{2}gW^3_{\mu} +  \frac{i}{2}g' B_{\mu} &
	                  \frac{i}{2} g(W^1_{\mu} - iW^2_{\mu})  \\
		  \frac{i}{2} g (W^1_{\mu} + iW^2_{\mu})	 & 
	\pal_{\mu} -  \frac{i}{2}gW^3_{\mu} +  \frac{i}{2} g' B_{\mu} \end{array} \right)
  \frac{1}{\sqrt{2}}	\left(\!\! \begin{array}{c} 0 \\ v \end{array} \!\!\right)
\end{equation}

  Hence $\lag_H$ in equation~\ref{laghiggs} contains the expression (for now 
neglecting fluctuations about the vacuum value $v$):
\begin{eqnarray}
   (D_{\mu}\phi)^{\dag}D^{\mu} \phi
 & = & \frac{g^2}{4}(W^1_{\mu} + iW^2_{\mu})(W^{1\mu} - iW^{2\mu})\frac{v^2}{2}
        \nonumber \\ & + &
   \frac{1}{4} \left(\!\!
    \begin{array}{cc} W^3_{\mu} & B_{\mu} \end{array} \!\! \right)
   \left(\!\!\begin{array}{cc} g^2 & -gg' \\ -gg' & g'^2  \end{array}\!\!\right)
    \left(\!\!\begin{array}{c} W^{3\mu} \\ B^{\mu}  
\end{array}\!\!\right)\frac{v^2}{2}
	\label{dmuphi2}
\end{eqnarray}

  With physical gauge fields $W^{\pm}_{\mu}(x)$ associated respectively with $2 \times 
2$ matrices $\sigma^{\pm}$ in the complexified $\sutw_L$ Lie algebra defined in turn 
as:
\begin{eqnarray}
  W^{\pm}_{\mu} & = & \frac{1}{\sqrt{2}}(W^1_{\mu} \mp iW^2_{\mu}) \label{wpmww} \\
  \sigma^{\pm} & = & \frac{1}{2} (\sigma^1 \pm i\sigma^2) \label{spmss}
\end{eqnarray}
  the relation:
\begin{equation}
   \frac{1}{\sqrt{2}}(W^+_{\mu}\sigma^+ \, + \, W^-_{\mu}\sigma^-)=
   \frac{1}{2}(W^1_{\mu}\sigma^1 \, + \, W^2_{\mu}\sigma^2)
\end{equation}
 may be substituted in for the $W^1_{\mu}, W^2_{\mu}$ piece of the covariant 
derivative in equation~\ref{covderphi}. In turn the first term in 
equation~\ref{dmuphi2} explicitly takes the form of a mass term for the 
$W^{\pm}_{\mu}(x)$ fields in the Lagrangian:
\begin{eqnarray}
  \lag_H & = & \frac{g^2v^2}{8}(W^+_{\mu}W^{+\mu} + W^-_{\mu}W^{-\mu}) \, + \, \ldots
      \nonumber \\
 \mbox{hence with} \quad M_W  & = & \frac{1}{2}gv  \label{masswgv}
\end{eqnarray}
   being the $W^{\pm}$ mass.
   
   The second term in equation~\ref{dmuphi2} contains a $2 \times 2$ mass matrix 
composed of quadratic terms in the couplings $g,g'$. Applying the same orthogonal 
transformation of equations~\ref{azthwbw} and \ref{bwthwaz} to the fields $W^3_{\mu}$ 
and $B_{\mu}$ with the weak mixing angle $\theta_W$ as specified in 
equation~\ref{tantw} diagonalises the mass matrix with respect to the fields $Z_{\mu}$ 
and $A_{\mu}$ such that:
\begin{equation}
   \lag_H = \ldots \, + \, \frac{1}{2}
    \left(\!\! \begin{array}{cc} Z_{\mu} & A_{\mu}  \end{array}  \!\!\right)
    \left(\!\! \begin{array}{cc} M^2_Z & 0 \\ 0 & 0  \end{array}  \!\!\right)
    \left(\!\! \begin{array}{c} Z^{\mu} \\ A^{\mu}  \end{array}  \!\!\right)
    \label{noamass}
\end{equation}
\begin{equation}
 \mbox{with} \quad M_Z \; = \; \frac{1}{2} \sqrt{g^2+g'^2} \; v
 \; = \; \frac{M_W}{\cos \theta_W}   \label{mwmzcth}
\end{equation}
  Hence the same weak mixing angle $\theta_W$ that accounts for the electromagnetic 
charge neutrality of the neutrino $\nu$ through the covariant derivative $D_{\mu}$ 
acting on the lepton field $l_L$ in equation~\ref{aonneut},
 deriving from the kinetic term in the Lagrangian for the lepton field in 
equations~\ref{lagklep} and \ref{covdevlep},
 also diagonalises the above mass matrix and leaves the photon field $A_{\mu}$ 
massless through $D_{\mu}$ acting on the Higgs field $\phi$, deriving from the kinetic 
term in the Lagrangian for the  Higgs field in equations~\ref{laghiggs} and 
\ref{covderphi}.

  Considering fluctuations about the vacuum value with $v \to v + H(x)$ in 
equation~\ref{higgsvac} (as neglected in writing down equation~\ref{dmuphi2}) in the 
quantum theory the real field $H(x)$ is associated with a massive scalar particle 
known as the Higgs boson. In terms of the parameters of the theory the Higgs mass is 
determined to be 
 $M_H = \sqrt{2} \, \mu = \sqrt{2\lambda} \, v$. While the vacuum value is empirically 
constrained to the order of the weak scale with, via equation~\ref{masswgv}, $v = 
2\frac{M_W}{g} \sim (\sqrt{2}G_{\! F})^{-\frac{1}{2}} \sim 246$~GeV, where $G_{\! F}$ 
is the Fermi constant, this does not determine the two parameters of the potential in 
equation~\ref{higgspot}. These latter parameters can now be deduced given the 
discovery of the Higgs at the LHC and 
the empirical measurement of $M_H \simeq 125$~GeV \cite{PDG}.

  At tree level the relations in the quantum field theory described in 
equations~\ref{masswgv} and \ref{mwmzcth} lead to the definition of the parameter:
\begin{equation}
   \rho = \frac{M_W^2}{M_Z^2 \cos^2 \theta_W} = 1
   \label{rhopara}
\end{equation}
  The fact that this expression holds approximately for the corresponding empirically 
measured values can be explained in terms of a further symmetry associated with the 
Higgs sector. 
 Expressing the Higgs field components in the form of a bi-doublet, that is the $2 
\times 2$ complex matrix: 
\begin{equation}
  \label{bidoub}
   \Phi \, = \, \frac{1}{\sqrt{2}}\left( \epsilon \phi^{\ast}, \phi \right)
    \, = \, \frac{1}{\sqrt{2}}\left( \begin{array}{cc}
	   {\phi^0}^{\ast}  &  \phi^+  \\
	   -{\phi^+}^{\ast} &  \phi^0  \end{array} \right)
\end{equation}
  with $\epsilon = \binom{\;0 \;\;\, 1}{-1 \;\, 0}$, the Higgs potential term of 
equation~\ref{higgspot} may be rewritten as:
\begin{equation}
  \label{higgspot2}
  V(\Phi) \, = \, -\mu^2 \, \mbox{tr}\, \Phi^{\dag}\Phi\, + \,
                    \lambda \, (\mbox{tr}\,\Phi^{\dag}\Phi)^2
\end{equation}
  This is invariant under the $L\inn \sutw_L$ action $\Phi \to L\Phi$ and $\uo_Y$ 
action $\Phi \to \Phi e^{-\frac{i}{2}\theta\sigma^3}$ with $\theta(x)\inn \rrr$ as 
local gauge transformations.
 While $\phi$ and $\epsilon \phi^{\ast}$ transform in the same way under $\sutw_L$,
 they have opposite hypercharge, with  $\fhy(\phi) = +\fhs$ and  $\fhy(\epsilon 
\phi^{\ast}) = -\fhs$, and hence the 
  generator for $\uo_Y$ transformations here is $\sigma^3$ rather than $\sigma^0$ (see 
for example~\cite{Willen} section 3). 
  The Higgs Lagrangian of equation~\ref{laghiggs}, which is also invariant under these 
gauge transformations, can be written in the form:
\begin{eqnarray}
   \lag_H & = & \mbox{tr} \, (D_{\mu} \Phi^{\dag}\:\! D^{\mu} \Phi) \; - \; V(\Phi)
     \label{laghcust}  \\
   \mbox{where} \quad D_{\mu}\Phi & = &  \pal_{\mu} \Phi
   \, + \, i\frac{g}{2} \, W^{\alpha}_{\mu} \, \sigma^{\alpha}  \Phi
    \,  - \, i\frac{g'}{2} \, B_{\mu} \, \Phi \sigma^3   \label{higgscov2}
\end{eqnarray} 
	is the gauge covariant derivative for the bi-doublet.  In the limit $g' \to 0$ 
this Lagrangian also has an additional, \textit{global},  symmetry denoted $\sutw_R$ 
with action $\Phi \to \Phi R^{\dag}$ for any $R\inn \sutw_R$, as can be seen by cyclic 
permutation of the arguments under the trace, with $\mbox{tr}(R\Phi^\dag\Phi R^{\dag}) 
= \mbox{tr}(R^{\dag}R\Phi^\dag\Phi) = \mbox{tr}(\Phi^{\dag}\Phi)$ for example. This 
symmetry in the Standard Model is considered to be `accidental' in the sense that it 
was not explicitly introduced in constructing the Higgs field to break the electroweak 
symmetry. It enlarges the complete global symmetry of the Higgs field to the action of 
$\sutw_L \times \sutw_R$, as $\Phi \to L\Phi R^{\dag}$ (where $L$ here represents a 
global action of the local $\sutw_L$ symmetry), which is simply the SO(4) symmetry of 
the quantity $\sum_{i=1}^4 \phi_i^2$ described below equation~\ref{higgspot}. The 
vacuum expectation value of equation~\ref{higgsvac}  can be written in the form: 
\begin{equation}
   \label{higgsvac2}   
    \langle \Phi \rangle =  \frac{1}{2}  \left( \begin{array}{cc}
	   v  &  0  \\
	   0  &  v  \end{array} \right)
\end{equation}	
	This vacuum value breaks the global $\sutw_L \times \sutw_R$ down to a single 
$\sutw$ symmetry denoted $\sutw_{L+R}$, with the action $ \langle \Phi \rangle \to L  
\langle \Phi \rangle L^{\dag}$  for $L\inn \sutw_{L+R}$ leaving 
equation~\ref{higgsvac2} invariant.
This is equivalent to the $\soth \subset \mbox{SO}(4)$ symmetry acting on the four 
components $\phi_i$ when taking the values of an arbitrary fixed Euclidean 4-vector, 
such as $(\phi_1,\phi_2,\phi_3,\phi_4) 
  = (0,0,v,0)$ in equation~\ref{higgsvac2}. The global $\sutw_{L+R}$ symmetry itself 
is broken for hypercharge coupling $g' \neq 0$, which involves gauging the $\uo_Y 
\subset \sutw_R$ subgroup via the $\sigma^3$ action of equation~\ref{higgscov2}, which 
is both the hypercharge generator itself and also the third component of the $\sutw_R$ 
action.

  For the Standard Model in the limit $g' \to 0$ the three $W^{\alpha}_{\mu}$ gauge 
fields transform as a triplet under the unbroken $\sutw_{L+R}$ global symmetry, and 
hence the masses gained from electroweak symmetry breaking are identical, with 
$M_{W^{\pm}} = M_{Z^0}$
 (as can be seen from equations~\ref{masswgv} and \ref{mwmzcth} for $g' \to 0$), in 
this limit. For small $g'$ the unbroken $\uo_Q$ symmetry corresponding to the massless 
photon determines a weak mixing angle $\theta_W$
 with $\cos^2 \theta_W = \frac{g^2}{g^2 + {g'}^2}$ which also determines the mass 
ratio of the heavy gauge bosons at tree level according to equation~\ref{rhopara}. 
This relation $\rho = 1$ is protected from radiative corrections by the approximate 
$\sutw_{L+R}$ symmetry, which is hence named `custodial symmetry' \cite{ssvz,Willen}.

  Masses for all three generations of fermions are included in the Standard Model 
Lagrangian by appending gauge invariant terms with Yukawa couplings to the Higgs 
field:
\begin{equation}
 \label{Yukferm}
  \lag_Y \, = \, - \,\Gamma^{ij}_u \, \bar{q}^i_L \, \epsilon \phi^{\ast} \, u^j_R \, 
- \,
           \Gamma^{ij}_d \, \bar{q}^i_L \, \phi \, d^j_R \, - \,
		   \Gamma^{ij}_e \, \bar{l}^i_L \, \phi \, e^j_R \, + \, \mbox{h.c.}
\end{equation} 		   
  (where `h.c.' is the Hermitian conjugate of all the preceding terms). Here the 
Yukawa couplings $\Gamma_u$, $\Gamma_d$ and $\Gamma_e$ are  $3 \times 3$ complex 
matrices in generation space with fermion flavour indices $i,j = \{1,2,3\}$ and hence, 
for example, $u^i_R \equiv \{u_R, c_R, t_R\}$ denotes the three generations of 
$u$-type right-handed quarks. When the Higgs field acquires the vacuum value $\langle 
\phi \rangle$ as expressed with the gauge choice of equation~\ref{higgsvac} the 
fermion states acquire Dirac mass terms via the Yukawa couplings:
\begin{equation}
 \begin{array}{rcl}
 \lag_M & = & - \,M^{ij}_u \, \bar{u}^i_L u^j_R \, - \,
                  M^{ij}_d \, \bar{d}^i_L d^j_R \, - \,
			      M^{ij}_e \, \bar{e}^i_L e^j_R \, + \,  \mbox{h.c.}  \\
 \mbox{where} \quad M^{ij}_{u,d,e} & = & \Gamma^{ij}_{u,d,e} \frac{v}{\sqrt{2}}
 \end{array}
   \label{Masferm}
\end{equation}
   are the three fermion mass matrices. Physical particle states may be identified by 
diagonalising each $M^{ij}$ matrix using independent unitary transformations applied 
to each left and right-handed fermion set via $3 \times 3$ unitary matrices $A^{ij}$, 
such as for:
\begin{eqnarray}
   \bu_L \to \bu'_L & = & A^{\dag}_{u_L} \bu_L \label{ulatran} \\
   \bu_R \to \bu'_R & = & A^{\dag}_{u_R} \bu_R \label{uratran}
\end{eqnarray}
   Hence $\bu'_L \equiv \{u'_L, c'_L, t'_L\}$ and $\bu'_R$ are mass eigenstate fields 
with the masses of the three $u$-type quarks read off from the diagonal elements of:
\begin{equation}
    M'_u = A^{\dag}_{u_L} \, M_u \, A_{u_R} \, = \,
  \left(\!\! \begin{array}{ccc} m_u & 0 & 0 \\
                            0 & m_c & 0 \\
					        0 & 0 & m_t  \end{array} \!\!\right)
\end{equation}
\begin{equation}
 \label{lagdmass}
 \mbox{with} \quad \lag_M = \, - \, m_u \, \bar{u}'_L u'_R
                            \, - \, m_c \, \bar{c}'_L c'_R
							\, - \, m_t \, \bar{t}'_L t'_R \, + \, \mbox{h.c.}
\end{equation}
    as the Lagrangian Dirac mass terms for the $u$-type quarks (with $u', c'$ and $t'$ 
here denoting the individual first, second and third generation $u$-type quarks). The 
$u$-quark itself hence has mass  $m_u = Y_u\frac{v}{\sqrt{2}}$ with the Yukawa 
coupling $Y_u = \Gamma'^{11}_u$ extracted from the diagonalised basis. From 
equation~\ref{masswgv} the $u$-quark mass can be related to the $W^{\pm}$ gauge boson 
mass as:
\begin{equation}
   m_u \, = \, \frac{\sqrt{2}}{g} \, Y_u \, M_W \nonumber  
\end{equation}
\begin{equation}    
   \mbox{with} \qquad \qquad Y_f \, = \, \frac{g \, m_f}{\sqrt{2}\, M_W}
      \qquad \qquad \qquad  
\end{equation}
   where $Y_f$ is the Yukawa coupling for each fermion $f$ to the Higgs field $\phi$, 
including the similar cases for the $d$-type quarks and charged leptons as following 
also from equation~\ref{Masferm}. (The neutrino mass may be treated differently and 
may not involve a Yukawa coupling, \cite{Teub} chapter 7). The couplings $Y_f$ are 
typically small since 
 $m_f \ll M_W$ except for the case of the top quark -- with the mass $m_t$ observed to 
be approximately the sum of $M_W$ and $M_Z$.
  All of the Yukawa couplings are added by hand in order to match the empirically 
determined fermion masses.

  In the physical mass eigenstate basis there is no Yukawa mixing between generations, 
as can be seen in equation~\ref{lagdmass} in comparison to equation~\ref{Masferm} 
where in the latter expression the quark states coupling to the weak $\sutw_L$ gauge 
fields are generally composed of a linear combination of the physical quark states. 
The weak $\sutw_L$ doublets in the quark sector may be written as 
$\binom{u}{\tilde{d}}_{\!\!\;\! L}$,  $\binom{c}{\tilde{s}}_{\!\!\;\! L}$ and 
$\binom{t}{\tilde{b}}_{\!\!\;\! L}$, with the inter-generation mixing expressed purely 
in terms of the $d$-type quark states:
\begin{equation}
 \label{vckmdef}
  \left( \!\! \begin{array}{c} \tilde{d} \\ \tilde{s} \\ \tilde{b} \end{array} \!\! 
\right)
    = \VCKM  \left( \!\! \begin{array}{c} d' \\ s' \\ b' \end{array} \!\! \right)
\end{equation} 
   Here the weak states $\tilde{d}, \tilde{s}, \tilde{b}$ are related to the physical 
states $d',s',b'$ via the $3 \times 3$ unitary Cabibbo-Kobayashi-Maskawa mixing matrix 
$\VCKM = A^{\dag}_{u_L}A_{d_L}$. With five relative global phase transformations 
between the six quarks $(u,d,c,s,t,b)$ only four of the nine parameters of the unitary 
matrix $\VCKM$ are physical. These four parameters describe three real mixing angles 
between the three generations and one complex phase which gives rise to {\it CP} 
violating phenomena. Together with the six quark masses a total of ten physical 
parameters (contributing just over half of the 18 Standard Model parameters listed in 
table~\ref{SMparams}) may hence be deduced from the Lagrangian for the quark sector  
  after the above field redefinitions. (Again, the description of neutrino mixing in 
the leptonic sector is a little different, \cite{Teub} chapter 7).
  
   The weak interaction terms for the quarks with the charged gauge bosons $W^{\pm}$ 
may be described by the Lagrangian:
\begin{equation}
  \label{lagqwint}
 \lag_{qW} \, = \, - \frac{g}{2\sqrt{2}} \, \bar{u}^i \, \gamma^{\mu} ( 1- \gamma^5)
        \, \tilde{d}^i \, W^{-}_{\mu} \, + \, \mbox{h.c}
\end{equation}
 with the implied sum for $i=1,2,3$ over the weak states (and where the Hermitian 
conjugate contains the $W^{+}_{\mu}$ term). Expressing the $d$-type quarks as a linear 
combination of the mass states the above Lagrangian can be written in terms of the six 
physical quarks as:
\begin{equation}
  \label{lagqwckm}
 \lag_{qW} \, = \, - \frac{g}{2\sqrt{2}} \, 
   \left(\!\! \begin{array}{ccc} \bar{u} & \bar{c} & \bar{t} \end{array}  \!\!\right)
      \, \gamma^{\mu} ( 1- \gamma^5)
        \, \VCKM 
	 \left(\!\! \begin{array}{c} d' \\ s' \\ b' \end{array}  \!\!\right)	 
		 \, W^{-}_{\mu} \, + \, \mbox{h.c}
\end{equation}
  In these equations the operator $P_L = \fh(1-\gamma^5)$ of equation~\ref{plop} has 
been put in by hand to project out the left-handed components of the Dirac spinors, 
describing maximal parity violation for the charged weak current. 
This CKM mixing originates from the \textit{mismatch} between the Yukawa and weak 
interactions in the Standard Model Lagrangian, with the corresponding mass and weak 
quark eigenstates for the $u$ and $d$-type quarks related via unitary transformations 
such as equations~\ref{ulatran} and \ref{uratran}. 
 On the other hand the neutral currents are flavour-diagonal and such terms are 
unchanged by the unitary transformations relating the mass and weak states, that is 
with  $A^{\dag}_{u_L}A_{u_L} = \b1_3$ and so on. Hence there are no flavour changing 
neutral currents coupled to the $Z_{\mu}$ or $A_{\mu}$ fields, and only the $W^{\pm}$ 
fields mediate mixing between the generations.


\section{Unification Models and Dynkin Analysis}
\label{dynkin}

  While the action of $\esi$ on $\htho$ studied in chapter~\ref{esihtho}  describes a  
symmetry of time it is also of course desirable that the mathematical structures 
arising in the present theory should bear a close resemblance to the symmetries and 
structures experimentally identified in particle physics. This data is summarised in 
the Standard Model,
as reviewed in the previous two sections,
 which describes the non-gravitational interactions between fundamental particles in 
terms of the gauge symmetry group $\SML$.
  Hence in this section we make a preliminary assessment of the suitability of the Lie 
group $\esi$, both generally and as constructed in chapter~\ref{esihtho}, as a 
unifying symmetry.

 It is well known that the three subgroup components of the Standard Model gauge 
symmetry are related to the series of normed division algebras, as introduced here in 
section~\ref{gfotf} in the context of forms of temporal flow and discussed further in 
section~\ref{oaags}. Indeed, $\uo$ is isomorphic to the complex numbers $\ccc$ of unit 
magnitude under multiplication, while $\sutw$ is similarly isomorphic to the 
quaternions $\hhh$ of unit magnitude, and $\suth$ is the subgroup of $\gt$, the 
automorphism group of the octonions $\ooo$, that leaves invariant a given imaginary 
octonion element. The aesthetic appeal and elegance of such observations have led a 
number of authors to speculate on a direct connection between the existence of these 
unique mathematical objects and the nature of the physical structure of the world (see 
for example \cite{Mori2,Dix1,Gur2,Man3}). However, while identifying a relationship 
between the mathematical properties of the division algebras and features of the 
Standard Model of particle physics much of this work is lacking in any underlying 
\textit{conceptual} motivation for the importance of such mathematical objects in 
nature.

    Since the octonion algebra features significantly in the present paper, in the 
action of the group $\esi \equiv \sltho$ on the space $\htho$, the references cited 
above suggest a reasonable likelihood of identifying some relation between the 
structures of the present theory and those of the Standard Model. Such a 
correspondence will be described in the following chapter. In the present theory we 
have both a clear conceptual understanding of the source of these algebras through the 
symmetry of the flow of time and in turn a well defined constraint on the introduction 
of these algebraic structures into the equations of physics through the relation $\lv$ 
and its symmetries.

  Also in the present theory, as well as aiming to account for the internal gauge 
interactions of the Standard Model through the higher-dimensional structures, 
gravitation is included on the base manifold $M_4$ with a subspace $\htwc \subset 
\htho$ locally identified with the tangent space $\TM_4$ and with the subgroup $\sltc 
\subset \esi$ being the covering group of the external Lorentz group. As described 
towards the end of section~\ref{gcatep} general relativity can be presented in the 
form of a gauge theory with a local Lorentz symmetry constructed in terms of the 
components of both a Lorentz Lie algebra-valued connection $A^a_{\ph{a}b\mu}(x)$ and a 
tetrad field $\teta$. In terms of the covering group it can in turn be considered to 
be an $\sltc$ gauge theory with an $\sltca$-valued  connection, which can accommodate 
a description of both vector and spinor objects in spacetime.

    While the dynamics of such an $\sltc$ `gauge theory' of gravitation 
\cite{Carm1,Carm2} are different to those of a standard Yang-Mills gauge theory, as 
also described in section~\ref{gcatep}, an extension for internal gauge symmetries 
might be more readily achieved with such a theory of gravity. 
 (Considering gravity as a gauge theory contrasts with the Kaluza-Klein approach 
reviewed in chapter~\ref{kktheory} for which an internal gauge theory derives from 
 general relativity with extra spatial dimensions.) Indeed an $\sltc \times\uo$ theory 
of gravitation and electromagnetism can be obtained by introducing an additional 
$e^{i\alpha(x)}$ phase factor element for the group $\uo$. This can be achieved by 
augmenting the set of symmetry actions $S\inn \sltc$ with $\mbox{det}(S)=1$ to include 
also the actions $U=e^{i\alpha(x)} \b1_2 \inn \uo$. The mapping of 
equation~\ref{hshs}, now incorporating also the $\uo$ action  
  $\bh_2 \to U \bh_2 U^{\dag}$, then remains one that preserves the value of 
$\mbox{det}(\bh_2)$ and leaves the metric $g_{\mu\nu}(x)$ on $M_4$ invariant, as for 
the original $\sltc$ action. Further, the $\uo$ action in fact leaves each of the four 
components of $\bh_2$ invariant and hence effectively acts as an `internal symmetry', 
as described also at the end of section~\ref{ltos}.
 Within the set of $\esi$ symmetry actions on $\htho$ the action of $\ssl^1_q$ in 
equation~\ref{sqdiag}, particularly on a type 1 $\htwc \subset \htho$ subspace, is 
most reminiscent of the above $\uo$ symmetry action on $\htwc$ and  this property is 
suggestive for the choice of  the $\uo_Q$ action for the electromagnetic gauge 
symmetry in the present theory.

 By further augmenting the internal degrees of freedom such unification schemes which 
begin with an $\sltc$ theory of gravity can be extended to an $\sltc \times \ul{G}$ 
theory where $\ul{G}$ may be the gauge symmetry group for the internal forces as 
identified experimentally in the Standard Model, that is $\SML$. For such a model 
there remains the task of introducing states which transform as fermions under the 
external $\sltc$ symmetry and under the appropriate representations of the internal 
symmetry as summarised in equation~\ref{multsf}.
 However such an approach, with the appropriate interpretation of the gauge groups and 
their empirically motivated representations, only serves to \textit{describe} gravity 
together with internal field interactions in a more unified framework.

  In the present theory, however, the unification group $\hat{G} = \esi$ 
\textit{includes} the external spacetime symmetry central to general relativity in the 
form of the subgroup $\sltc \subset \esi$. It is then through the distinctive role of 
this subgroup, in the identification of the necessary perceptual background for the 
world, that the larger symmetry is broken down to local gauge groups with 
representations on the broken fragments of the space $\htho$.
 The local gauge groups themselves will be initially identified as the `stability' 
group leaving the space of vectors $\bv_4 \inn \TM_4$, via equation~\ref{vtoh} 
equivalent to $\bh_2 \inn \htwc \subset \htwo$, invariant, generalising from the above 
case of the $e^{i\alpha(x)} \b1_2 \inn\uo$ action on $\htwc$.

 Having at hand the \textit{real} form of $\esi$ acting on $\htho$, as described in 
the previous chapter, a detailed study of this symmetry breaking over $\TM_4$ is 
possible. 
  Initially, however, in this section the symmetry breaking patterns for $\esi$ and 
the question of whether this group is large enough to actually contain both $\sltc$ 
and $\SM$ will be addressed at the level of the \textit{complex} Lie algebras, in 
order to gain an overview, before returning to the specific \textit{real} forms of 
these algebras in the following chapter.

   One of the main motivations for studying the complexified forms of real Lie 
algebras in general is the existence of a concise classification scheme. Indeed, every 
complex simple Lie algebra belongs to one of just four sets of classical algebra 
types, which include the complex forms of the rotation algebras so$(p,q)$, or is 
otherwise identified with one of the five exceptional cases, which include $L(\esi)$. 
A further motivation is that each complex simple Lie algebra has a one-to-one 
correspondence with a `Dynkin diagram', with semi-simple Lie algebras likewise 
corresponding to disconnected Dynkin diagrams. The analysis of such diagrams gives a 
good deal of guidance towards the possible symmetry breaking patterns for a complex 
Lie algebra and its real forms as encountered in the context of a theoretical model 
for physical phenomena.

   Firstly, we briefly review the relationship between Lie algebras and their 
representations. In general, each complex simple Lie algebra, as exemplified by the 
Dynkin diagrams shown later in this section, and taking its place amongst the 
systematic classification of such algebras, may be associated with several real forms, 
with each real algebra in turn associated with one of more Lie group, and finally each 
Lie group possesses an unlimited number of representations. This situation is depicted 
in figure~\ref{LctoR}. 
\vspace{-15pt}
\begin{figure}[htbp]  
\centering
\epsfxsize=14cm
\leavevmode
\epsffile[0 0 1711 355]{\gpath aPfig71e}
\vspace{-15pt}
\caption{\setb Any given complex Lie algebra $L_{\ccc}$, which has a unique Dynkin 
diagram, is in general associated with a multiplicity of real algebra forms 
$L_{\rrr}$, groups $G$ and representations $R$.}
\label{LctoR}
\end{figure}

  While Dynkin analysis at the level of $L_{\ccc}$ is described in this section, in 
this paper we generally deal with the structures of $L_{\rrr}$ and $G$, with notation 
such as so$(p,q)$ used for a real Lie algebra and SO$(p,q)$ for the related Lie group, 
with the distinction being otherwise understood from the context. 
   As an example of the chain of relations in figure~\ref{LctoR} the case for 
$L_{\ccc} = \mbox{so}(10)$ with links through to the $R=\mathbf{16}$ representation, 
of particular interest here and featuring for example in equation~\ref{esidecom} in 
the opening of the following chapter, 
is described in table~\ref{LctoRso}.

\begin{table}[htbp]
\centering
\begin{tabular}{|c|c|c|c|}
 \hline
       $L_{\ccc}$  &   $L_{\rrr}$  &    $G$   &    $R$   \\  
 \hline
                   &                  &     O(1,9)   &      \\			  
                   &     &    SO(1,9)    &    $\mathbf{1}$   \\
                   &    so(10)     &       SO$^+$(1,9)    &  $\mathbf{10}$ \\
$\quad$ so(10) $\marrow$  &  $\;\,$ $\to$ so(1,9) $\marrow$  &    $\to$   
Spin$^+$(1,9) $\marrow$  &	    
    $\;\;\to$  $\mathbf{16}$   $\quad\;\;$        \\
				   &  so(2,8)   &   Spin(1,9)    &   $\mathbf{\overline{16}}$ \\
				   &   $\vdots$   &   Pin(1,9)      &   $\vdots$        \\
   \hline
  \end{tabular}
  \caption{\setb The complex Lie algebra so(10), called $D_5$ in Cartan's notation, 
with corresponding real algebra forms $L_{\rrr}$, groups $G$ and representations $R$ 
(labelled by their dimension), as an example of the general case depicted in 
figure~\ref{LctoR}, with a particular chain of forms discussed in the text highlighted 
by the horizontal arrows.}
\label{LctoRso}
\end{table}

  In developing a theoretical model the initial motivation often begins from the 
left-hand side of figure~\ref{LctoR}, by identifying a complex Lie algebra which 
exhibits an appropriate symmetry breaking pattern to account for the gauge groups of 
the Standard Model as described in previous section; and then the task remains to 
identify the appropriate representations for particle states such as those of 
equation~\ref{multsf}. In this paper such an approach also serves as a useful guide, 
as we describe in this section. However, here our starting point is rather more 
anchored in the right-hand side of figure~\ref{LctoR} since the mathematical form 
$\lv$ strongly motivates the possible representations, with the set of real numbers 
composing the vector $\bv$ already belonging to a representation space transforming 
under the relevant symmetries of $\lv$.  
   
  As a preliminary observation we note that given our use of the $R=\mathbf{27}$ 
representation of the particular group $G = \mbox{E}_{6(-26)}$, this uniquely leads 
back via the real Lie algebra $L_{\rrr} = L(\mbox{E}_{6(-26)})$ to the complex Lie 
algebra $L_{\ccc} = L(\esi)$ as we step from right to left through figure~\ref{LctoR}.
 The structure of symmetry breaking feeding down from the complex Lie algebra is 
largely preserved in terms of semi-simplicity of the algebra and group and in terms of 
the reducibility of the algebra and group representations.  
  Hence we here consider the Dynkin diagrams for the relevant complex Lie algebras and 
the significant Lie subalgebras involved. 

  The `rank' of a Lie algebra is the dimension of the Cartan subalgebra, composed of a 
maximal subset of mutually commuting generators, which is unique up to automorphisms 
of the Lie algebra. For a rank-$n$ Lie algebra there are $n$ `simple roots' in the 
dual root space which is constructed out of the eigenvalues in the adjoint 
representation of the algebra in the Cartan-Weyl basis. The properties of a rank-$n$ 
Lie algebra can be described in terms of geometric relations between these simple 
roots in the Euclidean $\rrr^n$ root-space and 
 encoded in the topology relating the $n$ nodes of the corresponding Dynkin diagram, 
such as those depicted in figure~\ref{dynkine} for the rank-6 $L(\esi)$, rank-5 
so(10), rank-4 su(5) and rank-2 Lorentz Lie algebras. 
\begin{figure}[htbp]  
\centering
\epsfxsize=\maxwidth
\leavevmode
\epsffile[0 0 1791 261]{\gpath aPfig72e}
\vspace{-30pt}
\caption{\setb The four Dynkin diagrams for the (a) $L(\esi)$, (b) so(10), (c) su(5) 
and (d)~Lorentz or $\sltca$ Lie algebras.}
\label{dynkine}
\end{figure}
  For example, the Dynkin diagram for the Lorentz algebra consists of two disconnected 
nodes, meaning that the corresponding two simple roots are at $90^0$ in root space, 
whereas nodes connected by a single line denote an angle of $120^0$. At the level of 
the complexified Lie algebra $L_{\ccc}$ the Lorentz algebra has the semi-simple 
composition $\sutwa \oplus \sutwa$, as described earlier in 
equations~\ref{jpmeja}--\ref{jpmez}, which in this case is \textit{not} respected by 
the corresponding real form $L_{\rrr} = \soota$ of the Lorentz Lie algebra which is 
simple. An explicit basis for the Cartan subalgebra for the real form of $L(\esi)$ of 
importance in this paper was given in equation~\ref{csaset} as represented by vector 
fields on the space $T\htho$.

  Regular subalgebras, that is those respecting the Cartan-Weyl decomposition of the 
complex Lie algebra, may be readily obtained from the Dynkin diagrams. A maximal 
subgroup $G' \subset G$ is one for which there is no intermediate $G^{\prime\prime}$ 
such that $G' \subset G^{\prime\prime} \subset G$ as a series of proper subgroups, 
with a similar definition for the corresponding maximal subalgebra. A regular maximal 
subalgebra can be obtained from a Dynkin diagram by the prescription of removing one 
node and including an extra $\uo$ factor, which also means that the algebra obtained 
is not `semi-simple'. For example figure~\ref{dynkinb} shows a possible symmetry 
breaking pattern for the su(5) algebra for the well-known case~\cite{GeoGla}, as 
alluded to in the previous section, in which the Standard Model local gauge group is 
obtained.
\begin{figure}[htbp]  
\centering
\epsfxsize=12cm
\leavevmode
\epsffile[0 0 1273 115]{\gpath aPfig73e}
\vspace{-13pt}
\caption{\setb Removing a node from the Dynkin diagram for the Lie algebra of the 
group SU(5) reveals a breaking to $\SM$, which motivates the use of SU(5) in unified 
theories.}
\label{dynkinb}
\end{figure}

   Similarly from figure~\ref{dynkine} it can be seen that SO(10) contains SU(5) as a 
subgroup, by removing either of two appropriate end nodes. Hence the full Standard 
Model gauge group can be obtained by first breaking SO(10) to SU(5) and then breaking 
SU(5) as described in figure~\ref{dynkinb}. Hence the 45-dimensional group $\sltwoo 
\equiv \spotn$ constructed in section~\ref{ltos} as the double cover of $\sootn$, 
which is generated by a real form of the complex Lie algebra so(10), is also 
potentially of great interest for internal gauge group unification in particle 
physics.

  In the context of the discussion of section~\ref{esitran} following 
equation~\ref{agsdeps}, the $\gt$ automorphism group of $\ooo$ is reduced to the 
subgroup $\mbox{SU}(3)\subset \gt$ if a complex subspace, for example with the 
imaginary unit $l\inn \ooo$, is fixed, as also alluded to near the opening of this 
section. Similarly the subgroup 
$\suth \subset \gt \subset \sootn$ may be obtained through the selection of a 
preferred subspace $\htwc \subset \htwo$, since this choice also fixes an imaginary 
unit of $\htwo$.  Here the mechanism for such a selection is provided by the nature of 
perception on the base manifold $M_4$ with the vector space $\TM_4 \equiv \htwc$ and 
$\htwc\subset \htwo$ through the identification of an external $\soot \subset \sootn$ 
symmetry. However breaking the rank-2 Lorentz group out of the rank-5 so$(10)$ clearly 
does not leave sufficient symmetry to describe the full rank-4 Standard Model gauge 
group.

  It was also shown in section~\ref{esitran} how 3 copies of $\sltwoo$, described with 
a total of $(3 \times 45) = 135$ generator actions, lock tightly together as an 
independent basis set of 78 generators, summarised in table~\ref{prefbas}, for the 
$\esi$ action on $\mcX \inn \htho$ preserving $\mbox{det}(\mcX)$. This space hence 
describes a highly symmetric form of temporal flow $\lvt$ motivating the study of this 
exceptional Lie group. 

As well as composing a rich symmetry of a multi-dimensional form of $\lv$, additional 
motivation for the use of $\esi$ indeed comes from the fact that this Lie group is 
well known as a good candidate for the unifying symmetry group in models describing a 
unification of the non-gravitational fundamental forces of nature.  Further, unlike 
the two larger exceptional Lie groups, $\ese$ and $\ee$, the group $\esi$ has complex 
representations and these are needed to describe the observed multiplets of states in 
particle physics of equation~\ref{multsf} which are not left-right symmetric.
From figure~\ref{dynkine} it can be seen that $\esi$ contains SO(10) and hence in turn 
SU(5) and finally also the Standard Model gauge symmetry, with the chain of subgroups: 
$\esi \supset \mbox{SO}(10) \supset \mbox{SU}(5)\supset \mbox{SU}(3)\times 
\mbox{SU}(2) \times \mbox{U}(1)$.
 The potential of $\esi$ as a unifying group has been known since the early history of 
the Standard Model of particle physics even as it was still taking shape in the 1970s 
(see for example~\cite{Gur1}) and continues today (see also, for example~\cite{Geor}  
pp.302--308).

 In this case the higher rank of $\esi$ over that of SU(5), with 2 additional Dynkin 
nodes, suggests that in principle the physical phenomena of the rank-2 Lorentz 
transformations might be described alongside the rank-4 Standard Model gauge group 
within the full the rank-6 symmetry group $\esi$. However it is not possible to break 
$\esi$ into the combined Lorentz and Standard Model algebras by the Dynkin analysis 
prescribed above.
 While it can  be shown that $\esi$ contains subgroups such as $\SM  \times 
\mbox{SU}(3)$, for example by removing the central node in figure~\ref{dynkine}(a),
 a similar decomposition but with a rank-2 SU(3) replaced by the rank-2 Lorentz group
 is not possible.
  An alternative prescription for obtaining semi-simple regular maximal subalgebras 
via an intermediate `extended' Dynkin diagram does not help this situation.
 However, to some extent this Dynkin analysis oriented within the Cartan-Weyl basis 
for complex forms of the Lie algebras represents a ball-park picture and is not 
tailored to fit the fine details for a real form of $\esi$ represented within the 
context of a specific theory.

     To study these details not only is the real form of the group action needed but 
also an understanding of how the \textit{dynamics} arises, and the means by which a 
symmetry subgroup of $\lvt$ might be associated with gauge field interactions, in 
order to account for the phenomena observed in the laboratory. In particular the 
structure of the symmetry breaking itself, involving the extended spacetime manifold 
$M_4$, will need to be considered more explicitly. In the meantime, the observation 
that the Lorentz group and Standard Model gauge groups \textit{almost} fit together at 
the level of this \textit{static} Dynkin diagram analysis is an encouraging feature.

   In principle then, the possibility of identifying features of the full gauge 
symmetry group for the strong and electroweak particle interactions for the theory 
presented here based on the $\esi$ symmetry of $\lvt$ is worth pursuing, as we explore 
in the following chapter. It is further noted that the $\mathbf{16}$ representation in 
table~\ref{LctoRso}, that is the Majorana-Weyl spinor introduced in 
section~\ref{secbkkt} and described in the following section, as exemplified by 
 the $\spotn$ spinor $\theta$ of equations~\ref{xoct3} and \ref{xinmcx}, possesses a 
branching pattern under the $\SM \subset \mbox{SO}(10)$ subgroup into representation 
multiplets corresponding to the 15 particle types of a complete generation of Standard 
Model fermions of equation~\ref{multsf} (plus a right-handed neutrino). However a 
\textit{different} approach will be followed here, involving both the incorporation of 
the external Lorentz symmetry within $\spotn$ as well as the extension to the $\esi$ 
symmetry. Indeed we begin in the opening section of the following chapter by 
identifying objects which transform as fermions under the external symmetry.

 We also note here the possible significance of the three possible embeddings of an 
$\htwo$ subspace, as represented by the components $X\inn \htwo$ in 
equations~\ref{type1}, \ref{type2} and \ref{type3}, within the space $\htho$, with 
equivalent symmetry transformation properties, and in particular three copies of the 
$\spotn$ spinor $\theta$ representation. These three embeddings are related by the 
matrix $\mcT$ of equation~\ref{tcycle3}, as  described in section~\ref{esitran}, and 
in terms the octonion triality isomorphism 
 as discussed alongside equations~\ref{mqqbaro} and \ref{trialabc}, relating to the 
rich symmetry of this form of $\lvt$.
 This is suggestive since we shall have to account for three generations of fermion 
families, related through the CKM matrix of equation~\ref{vckmdef} in the case of the 
quarks, which might here be related through the full set of $\esi$ symmetry 
transformations.
 On the other hand only \textit{one}  embedding of $\htwc \subset \htho$ will be 
associated with the local tangent space $\TM_4$ in the symmetry breaking, potentially 
lifting the degeneracy between the three generations of fermions in the present 
theory.

 Again, while the connection between some of these algebraic structures and the 
Standard Model is well known, here there is an underlying motivation for the origin of 
these mathematical forms in a physical theory based on the symmetries of $\lvt$ 
representing a multi-dimensional form of temporal flow.

  Considering then the demands from both ends of figure~\ref{LctoR} at the same time, 
with the choice of $L_{\ccc}$ guided by general features of the Standard Model and the 
space $R$ identified under a highly symmetric form of $\lv$, we naturally converge 
upon the group $\esi$ acting on the representation space $\htho$, such that the matrix 
determinant is invariant, as being of particular interest. Indeed this motivated the 
detailed study in chapter~\ref{esihtho} based on references 
\cite{Man2,Wang,Man4,Man5,Wang2}. Further, the identification of the Lorentz subgroup 
of $\esi$ acting upon the  subspace $\htwc$ representing 4-dimensional spacetime 
explicitly provides the symmetry breaking mechanism through which the broken internal 
subgroups of the larger symmetry may be realised as the local gauge groups. 
 The symmetry breaking was pictured in figure~\ref{mtogmaphr} for the provisional 
model with an $\sootn$ symmetry acting on the form $\lvte$.
    That case for a 10-dimensional spacetime symmetry, now described by $\spotn$ 
acting on $\htwo$, constitutes a significant intermediate stage between the full 
27-dimensional form of temporal flow and the external 4-dimensional spacetime 
structure. 
  
  In order to analyse the physical content of this theory it will be necessary to 
dissect the anatomy of the explicit real form of $\esi$ constructed in 
chapter~\ref{esihtho} in the context of symmetry breaking over the extended $M_4$ 
manifold.
 In the following chapter  we first study the action of the external Lorentz symmetry 
on the full set of $\htho$ components, building on the analysis of 
equation~\ref{vinhinc} presented at the end of section~\ref{lsspin}, and then assess 
how the properties of the 
 internal symmetry, surviving  the symmetry breaking, compare with the Standard Model.


\pagebreak
\chapter{$\esi$ Symmetry Breaking}
\label{chapesb}

\section{External Symmetry on $\htho$}
 \label{extsym} 

  Having at hand a complete mathematical description of the $\esi$ symmetry action 
from chapter~\ref{esihtho}, preserving the determinant on the space $\htho$ as a form 
of $\lv$, the physical significance of various subgroup actions can be considered 
locally with respect to the spacetime manifold $M_4$. In particular a distinguished 
set of symmetry transformations will act on the components of $\bv_{27} \inn \htho$ 
lying in the local spacetime tangent space $\TM_4$. These transformations form the 
subgroup $\sltc$, the double cover of the Lorentz group, which is identified then as 
the \textit{external} symmetry group. This spacetime symmetry is central to general 
relativity, while in the flat spacetime limit these Lorentz transformations form a 
global symmetry on $M_4$ as for the theory of special relativity.
With the flow of time expanded into the 27-dimensional space of $3\times 3$ Hermitian 
octonion matrices $\htho$ there are  23 \textit{extra dimensions} beyond those needed 
to locate events taking place in our 4-dimensional spacetime world.
The explicit action of the external Lorentz symmetry on all components of the space 
$\htho$  will be described this section, based on the real form of $\esi$ as 
constructed in chapter~\ref{esihtho}.

  The form of $\htho$ matrices transforming under the type 1 $\sltc$ and $\sltwoo$ 
subgroups of $\esi$, with the structure described in equation~\ref{mxm3}, is 
compatible with the isomorphism of vector spaces (\cite{Baez1} p.30):
\begin{eqnarray}
    \htho  &   \cong    &   \rrr \oplus  \htwo  \oplus  \ooo^2  \label{horhoo} \\ 
	  \nonumber \\    
	\left( \begin{array}{cc} \!\!\!\!
       \left( \,\,\,\,\,\,\, X \begin{array}{cc} &  \\  &  \end{array} \!\!\!  \right) 
&
    \!\!\!\!\! \left(  \theta \begin{array}{cc} &  \\  &  \end{array} \!\!\!\!\! 
\!\!\!\!\! \right) 
				                         \\ 
	\!\!\!	\left( \,\,\,\,\,\,\, \theta^{\dagger} \begin{array}{cc} &   \end{array} 
\!\!\!  \right)  &			  
					    \!\!\!\!\! n \end{array} \!\!\!\! \right)     
	      &  \to  &   (n , \quad \; X, \quad \; \theta)  \label{xinmcx}    \\   
\nonumber   \\
	  \mathbf{27}_{\mathrm{E}_6}  &  \to  &  (\mathbf{1} + \mathbf{10} + 
\mathbf{16})_{\mathrm{Spin}^+(1,9)}  \label{esidecom}
\end{eqnarray}
  The three parts of this decomposition are respectively the scalar, vector and spinor 
representations of the 10-dimensional spacetime symmetry group $\sootn$, for which the 
covering group is $\spotn \equiv \sltwoo$.
 A spinor representation with both Majorana and Weyl properties is only possible for 
$d= (2,\!\!\!\mod 8)$ spacetime dimensions, as is the case for $\sootn$. The object 
$\theta$ corresponds to the Majorana-Weyl spinor representation, denoted 
$\mathbf{16}$, which can be described by 16 real numbers owing to the reality 
condition for Majorana spinors
 (in general a Majorana spinor $\psi$ is one which is equal to its `charge conjugate' 
$\psi^c$, this reality condition is also possible in 4-dimensional spacetime). 

 As described in~\cite{Baez1} the decomposition of 
equations~\ref{horhoo}--\ref{esidecom} gives a representation of Spin$^+$(1,9)  as 
linear transformations of $\htho$ which do not preserve the Jordan algebra but do, 
importantly for the present considerations, preserve the determinant of $\htho$, as 
presented explicitly in equation~\ref{hthoinv} of section~\ref{esitran}. 
 The relationship between the complex Lie algebra $L_{\ccc} = \mbox{so}(10)$, its real 
forms, the group $\spotn$ and its representations was presented explicitly
 in table~\ref{LctoRso}.
 Similarly as for so(10) in the Dynkin analysis of section~\ref{dynkin} we can 
consider the above decomposition as a mathematically intermediary stage in studying 
the Lorentz subgroup, $\spot\equiv\sltc$, in $\esi$.

  While the 27-dimensional irreducible representation of $\esi$ decomposes as a 
reducible representation of $\spotn$, as shown in equation~\ref{esidecom}, further 
decomposition is to be expected under smaller subgroups such as the external Lorentz 
transformations of 4-dimensional spacetime considered in this section, and also for 
the internal symmetry groups to be identified in the following section.

 We can identify the Lorentz 4-vector $\bv_4=(v^0,v^1,v^2,v^3)\equiv \bh_2$ in the 
upper left-hand $2\times 2$ matrix embedded within the larger $3\times 3$ matrices in 
$\htho$, as was the case for $\htwc \subset \hthc$ in equation~\ref{vinhinc}. The 
relation $\mbox{det}(\mcX) = 1$ with $\mcX \inn \htho$ is preserved under operations 
of $\sltc$ representing the Lorentz group upon this space as: 
\vspace{5pt}
\begin{equation}
   \label{hvvv}
         \mcX \; \to \;
		\left( \begin{array}{c|c} 
        \,\,\,\,\, S   \!      \begin{array}{cc} &  \\  &  \end{array} \!\!\!   &
        \,     0  \begin{array}{cc} &  \\  &  \end{array} \!\!\!\!\!\!\!\!\!\! 
				                         \\  \hline
        \,\,\,\,\,\,  0  \!\! \begin{array}{cc}        &   \end{array}   &	  
		\,  1      \end{array}  \right)  
		\left(\! \begin{array}{cc|c} 		
            h^{00}    & h^{01}\! + \overline{a}(6)   &   c  \\
		  h^{10}\! + a(6)   &    h^{11}         & \overline{b}  \\         	
               \hline										 
         \overline{c}     &   b  &  n      \end{array}  \right)  
		\left( \begin{array}{c|c} 
        \,\,\,\,\, S^{\dag}  \!    \begin{array}{cc} &  \\  &  \end{array} \!\!\!   &
        \,  0  \begin{array}{cc} &  \\  &  \end{array} \!\!\!\!\!\!\!\!\!\! 
				                         \\  \hline
        \,\,\,\,\,\, 0 \!\! \begin{array}{cc}        &   \end{array}   &	  
		\,  1      \end{array}  \right) 
\end{equation}
\vspace{5pt}
\begin{equation}
 \label{hinhtho}
  \mbox{with}    
		 \qquad \qquad
         \begin{array}{cc} h^{00}=v^0+v^3,\quad  &  h^{01}=v^1-v^2l  \\
		                   h^{10}=v^1+v^2l,\quad &  h^{11}=v^0-v^3   \end{array}
					\qquad \qquad \qquad	   
 \end{equation}
   with $S\inn\sltc$, and with `$1$' describing the identity transformation in the 
trivial 1-dimensional representation of this group, acting upon the components of 
$\mcX$ of equation~\ref{xoct3}. This action preserves the value of det$(\bh_2) = h^2$, 
as it is simply the transformation of equation~\ref{hshs}, as well as leaving 
$\mbox{det}(\mcX) = 1$ invariant.  In equation~\ref{hvvv} $a(6)$ denotes the 
6-dimensional imaginary part of $a\inn \htho$ of equation~\ref{xoct3}, that is 
excluding the real $a_1 = v^1$ and imaginary $a_8l = v^2l$ components of $a\inn \ooo$ 
which are associated with the external 4-vector $\bv_4 \inn \TM_4$.

 The four components of the projected $\bv_4(x) \subset \bv_{27}(x)$, forming a 
tangent vector in $\TM_4$ locally on the spacetime manifold $M_4$, transform as the 
components of a Lorentz 4-vector. These components are embedded within the space 
$\htho$ via the $2 \times 2$ matrices $\bh_2 \inn \htwc$.
 While in  section~\ref{lsspin} $\{1,i\}$ denoted the base units for the space $\ccc$,  
for example for $\sigma^2$ in equation~\ref{sigmas} as used in equation~\ref{vtoh} 
(and also in section~\ref{ltos}, for example equation~\ref{xcomp}), here the preferred 
subspace $\ccc\subset \ooo$ basis is taken to be $\{1,l\}$ for $\bv_4$, as indicated 
in equation~\ref{hinhtho}, in conformity with the conventions
 of sections~\ref{esitran} and \ref{laofesi}, and in particular 
equation~\ref{extloract}, and as employed in 
 the following section. Since the $\sltc$ actions, based on this $\{1,l\}$ complex 
subspace are embedded in the `type 1' location of equation~\ref{type1} this group will 
be denoted $\sltc^1$.

  The full set of actions of the real form of $\esi$ on the space $\htho$ was 
constructed in section~\ref{esitran}. 
 With the group action of $\sltc^1$ on $\htho$ in equation~\ref{hvvv}  embedded within 
the type 1 group action of $\sltwoo^1$ on the same space as displayed in 
equation~\ref{mxm3} we can write:
  \begin{equation}
   \label{sltcinesi}
    \soot \equiv \sltc^1 \subset \sltwoo^1 \subset \sltho \equiv \esi
  \end{equation}
   where the first `$\equiv$' strictly applies at the Lie algebra level.
 This shows explicitly how the action of the Lorentz group may be embedded within the 
higher symmetry group $\esi$ acting on the space $\htho$. The direct physical 
interpretation of the former symmetry in the shape of the perceptual background of the 
spacetime manifold $M_4$  provides a direct source for the breakdown of the latter 
symmetry. 

    The six Lorentz group generators as a subset of the 78 $\esi$ generators were 
listed in equation~\ref{extlor6} of section~\ref{laofesi}. They  can be read off from 
the full $\esi$ Lie algebra table~\cite{Wang} and seen to satisfy the $\soot$ algebra 
which is reproduced here in table~\ref{lorwang}.

\begin{table}[htbp]
\centering
\begin{tabular}{|c|cccccc|}
\hline
    $\lbrack \bullet , \bullet \ \rbrack$
    &  $\dot{R}_{z \pqg l}$ & $\dot{R}_{x \pqg z}$ & $\dot{R}_{x \pqg l}$ 
	&  $\dot{B}_{t \pqg x}$ & $\dot{B}_{t \pqg l}$ & $\dot{B}_{t \pqg z}$   \\
	\hline  
  $\dot{R}_{z \pqg l}$ &    $0$    &   $-\dot{R}_{x \pqg l}$   &  $\dot{R}_{x \pqg z}$
                     &    $0$    &    $-\dot{B}_{t \pqg z}$   &  $\dot{B}_{t \pqg l}$  
\\
  $\dot{R}_{x \pqg z}$ &  $\dot{R}_{x \pqg l}$ &    $0$   &  $-\dot{R}_{z \pqg l}$   
                     &  $\dot{B}_{t \pqg z}$ &    $0$   &  $-\dot{B}_{t \pqg x}$   \\ 
  $\dot{R}_{x \pqg l}$ &  $-\dot{R}_{x \pqg z}$ &  $\dot{R}_{z \pqg l}$  &  $0$
                     &  $-\dot{B}_{t \pqg l}$ &  $\dot{B}_{t \pqg x}$  &  $0$    \\  
  $\dot{B}_{t \pqg x}$ &  $0$   &   $-\dot{B}_{t \pqg z}$  &  $\dot{B}_{t \pqg l}$ 
                     &  $0$   &    $\dot{R}_{x \pqg l}$  & $-\dot{R}_{x \pqg z}$  \\
  $\dot{B}_{t \pqg l}$ &  $\dot{B}_{t \pqg z}$ &    $0$    &  $-\dot{B}_{t \pqg x}$
                     & $-\dot{R}_{x \pqg l}$ &    $0$    &   $\dot{R}_{z \pqg l}$  \\
  $\dot{B}_{t \pqg z}$ &  $-\dot{B}_{t \pqg l}$ &  $\dot{B}_{t \pqg x}$  &  $0$
                     &   $\dot{R}_{x \pqg z}$ & $-\dot{R}_{z \pqg l}$  &  $0$      \\
 \hline
  \end{tabular}
  \caption{\setb (Extracted from the $\esi$ Lie algebra table in \protect\cite{Wang}). 
The Lie algebra structure for the set of Lorentz generators of equation~\ref{extlor6}, 
with bracket composition $[\dot{R}_{z \pqg l}^{1}, \dot{R}_{x \pqg z}^{1}] = 
-\dot{R}_{x \pqg l}^{1}$ etc. The type superscripts `1' are omitted in the table 
entries, which are all generators of $\sltc^1$. (Each entry is equivalent to that for 
the corresponding $6 \times 6$ table for the generators 
$\{J^1,-J^2,J^3,-K^1,K^2,-K^3\}$ with the Lie bracket of 
equations~\ref{jjej}--\ref{jkek}, via the correspondence of equation~\ref{msigj}).}
\label{lorwang}
\end{table} 

   The corresponding $2\times 2$ matrix actions for the category 1 boosts and category 
2 rotations can be read off for the case $q=l$ in table~\ref{mtran45} of 
section~\ref{ltos}. Since each of these actions involves the composition of matrix 
elements from a single complex subspace, with base units $\{1,l\}$, and with each of 
$a,b,c \inn \ooo$ (or $p,m,n \inn \rrr$) as elements of $\htho$ appearing in separate 
product terms, the symmetry transformations are equivalent to those based on $\hhh$ 
subalgebras and are hence associative. Consistent with the discussion in the 
paragraphs following equation~\ref{type3} this means that the symmetry group and 
corresponding Lie algebra can be represented in terms of the transformation matrices 
themselves. A matrix representation for the Lorentz Lie algebra is therefore provided 
by defining
  $\dot{M} = \frac{\pal}{\pal \alpha} M \Big|_{\alpha = 0}$ for the corresponding six 
matrix actions in table~\ref{mtran45} (here presented in a different order), that is:
\begin{eqnarray}
  \dot{M}_{z \pqg l} =  \left(\! \begin{array}{cc}
               0  &  -\frac{l}{2}  \\  -\frac{l}{2} & 0
                      \end{array} \!\right)\! , \quad\; 
 & \dot{M}_{x \pqg z} =  \left(\! \begin{array}{cc}
               0  &  +\frac{1}{2}  \\  -\frac{1}{2} & 0
                      \end{array} \!\right)\! ,  & \quad\; 
  \dot{M}_{x \pqg l} =  \left(\! \begin{array}{cc}
               -\frac{l}{2} & 0  \\  0 & +\frac{l}{2}
                      \end{array} \!\right)\! , \qquad  \label{mrotgen}  \\   & & 
\nonumber \\
  \dot{M}_{t \pqg x} =  \left(\! \begin{array}{cc}
               0  &  +\frac{1}{2}  \\  +\frac{1}{2} & 0
                      \end{array} \!\right)\! , \quad\; 
 & \dot{M}_{t \pqg l} =  \left(\! \begin{array}{cc}
               0  &  +\frac{l}{2}  \\  -\frac{l}{2} & 0
                      \end{array} \!\right)\! ,& \quad\; 
  \dot{M}_{t \pqg z} =  \left(\! \begin{array}{cc}
               +\frac{1}{2} & 0  \\  0 & -\frac{1}{2}
                      \end{array} \!\right)\! , \qquad  \label{mboogen}  \\   
\nonumber
\end{eqnarray}
    where the latter three are the boost generators as can be identified by the time 
component `$t$' label in the subscript. Expressing the three Pauli matrices as 
$\sigma^1 = \binom{0 \; 1}{1 \; 0}$,  $\sigma^2 = \binom{0 \, -l}{l \,\; 0}$, 
$\sigma^3 = \binom{1 \,\; 0}{0 \, -1}$, that is equation~\ref{sigmas} with $i$ 
replaced by the imaginary unit $l$, the six elements of this Lorentz Lie algebra can 
be written as:
\begin{equation}
  \left(\! \begin{array}{c} \dot{M}_{z \pqg l} \\
          \dot{M}_{x \pqg z} \\ \dot{M}_{x \pqg l} \end{array} \!\right) =
  \left(\! \begin{array}{c} -\frac{l}{2}\sigma^1 \\ +\frac{l}{2}\sigma^2 
           \\ -\frac{l}{2}\sigma^3  \end{array} \!\right) \sim
  \left(\! \begin{array}{c} +J^1 \\ -J^2 \\ +J^3 \end{array} \!\right), \qquad
  \left(\! \begin{array}{c} \dot{M}_{t \pqg x} \\
          \dot{M}_{t \pqg l} \\ \dot{M}_{t \pqg z} \end{array} \!\right) =
  \left(\! \begin{array}{c} +\frac{1}{2}\sigma^1 \\ -\frac{1}{2}\sigma^2 
           \\ +\frac{1}{2}\sigma^3  \end{array} \!\right) \sim
  \left(\! \begin{array}{c} -K^1 \\ +K^2 \\ -K^3 \end{array} \!\right)
  \label{msigj}
\end{equation}

 The associations with the Lorentz rotation $\{J^a\}$ and boost $\{K^a\}$ generators 
of equation~\ref{jsigksig} are such that with $\{J^2,K^1,K^3\} \to \{-J^2,-K^1,-K^3\}$ 
the Lie algebra of equations~\ref{jjej}--\ref{jkek}
   matches that  
 of the commutators in table~\ref{lorwang}. 
 Hence in the context of the $\sltc^1$ action in equation~\ref{hvvv} these  sign 
conventions, $\{\dot{R}, \dot{B}\}\sim \pm \{J,K\}$, in equation~\ref{msigj} should be 
noted, which at the group level simply corresponds to a sign flip for a subset of the 
six real parameters $\{r_a,b_a\}$ in equation~\ref{stjsk}, and hence in turn will 
relate to the definition of left and right-handed spinors.
 As described in the discussion following table~\ref{werror} in section~\ref{laofesi}
 here the key orientation for such conventions is provided by the $\esi$ Lie algebra 
table of reference \cite{Wang} from which table~\ref{lorwang} is extracted. Ultimately 
a different set of $L(\esi)$ sign conventions may be preferred, in alignment with the 
physical application.

  We next address the action of the external Lorentz symmetry on a general element 
$\mcX\inn\htho$, including the full set of 16 real components of $\theta = 
\binom{c}{\bar{b}} \inn \ooo^2$, that is the 16-dimensional Majorana-Weyl spinor under 
$\sltwoo^1$, composed of the octonion entries $c$ and $\bar{b}$, as introduced  in 
equation~\ref{xoct3} and described after equations~\ref{horhoo}--\ref{esidecom}.

The two-sided $\sltc^1$ action on $\htwo$  in equation~\ref{hvvv} only transforms the 
real diagonal entries $h^{00}$ and $h^{11}$ together with the $h^{10} = a_1 + a_8l$ 
and $h^{01} = a_1 - a_8l$ components of $a\inn \ooo$. The six components of $a(6) \inn 
\mbox{Im}(a)$ remain invariant as may be deduced from the form of the six $\sltc^1$ 
generators in table~\ref{lbrota} or from equation~\ref{sixlorbr} for the case $q=l$.
 (This is equivalent to the invariance of $\ul{\bv}_6$ under $\soot$ for the model of 
figure~\ref{mtogmaphr}).
 Of the additional 17 components in $\htho$ the real diagonal $n$ entry is also 
invariant, as is clear from equation~\ref{hvvv}, while \textit{all} 16 components of 
$b,c \inn \ooo$ transform non-trivially under the one-sided $\sltc^1$ action.

The spinor $\theta_l = \binom{c}{\bar{b}}_{\! l} \inn \ccc^2$ will denote the 
$\{1,l\}$ components of $c$ and $\bar{b}$ in $\theta$, that is $\binom{c}{\bar{b}}\inn 
\ooo^2$ restricted to the $\{1,l\}$ complex subspace. By comparison with 
equation~\ref{vinhinc} this object transforms as a left-handed Weyl spinor  $\psi_L = 
\theta_l$ under the $\sltc^1$ action in equation~\ref{hvvv}. Consistent with the above 
comments on the sign conventions for the Lorentz generators  here we take this
 $S \inn \sltc^1$ action on $\theta_l$ to \textit{define} the left-handed
  spinor representation, guided but not constrained by the standard definitions of 
section~\ref{lsspin}.

 Due to the anticommuting property, for example in equation~\ref{gammalg}, Clifford 
algebras are also related to the division algebras (\cite{Baez1} section 2.3), with 
analogous rotational properties as alluded to following equation~\ref{phiconjr}.
  Indeed it can be shown, for example, that $C(0,2) = \hhh$ 
 for the Clifford algebra associated with the 2-dimensional vector space $\rrr^{0,2}$, 
  while $C(1,3) = \hhh(2)$, that is the Clifford algebra for 4-dimensional spacetime 
is isomorphic to the algebra of $2 \times 2$ quaternionic matrices under 
multiplication. (However, since $C(p,q)$ is in all cases an \textit{associative} 
algebra there are no such isomorphisms involving the octonion algebra).

  As a representation of $C(1,3)$ the algebra $\hhh(2)$ acts, by matrix 
multiplication, on the spinor space $\hhh^2$ rather than the usual Dirac spinor space 
$\ccc^4$. 
 We consider first the quaternionic spinor as a subspace of the octonionic spinor 
 $\theta = \binom{c}{\bar{b}}$ with base units $\{1, l, i, \il\}$:
\begin{equation}
  \label{thccbbh}
   \theta_{\hhh} = \left( \begin{array}{c} c \\ \bar{b} \end{array}  
\right)_{\!\!\!\hhh}  =
        \left( \begin{array}{c} c_1 + c_8l + c_2i + c_7\il \\
		                b_1 - b_8l - b_2i - b_7\il     \end{array}  \right) 
					\inn \hhh^2 \; \subset \ooo^2
\end{equation}
    Upon restriction to the subset of $2 \times 2$ matrix actions of $\sltc^1 \subset 
\sltwoh^1 \subset \hhh(2)$ (which is isomorphic to the group $\spot$ as identified 
within the Clifford algebra $C(1,3) = \hhh (2)$),
 with base units $\{1,l\}$,
 the spinor space $\theta_{\hhh}$ decomposes into two parts:
\begin{equation}
   \theta_{l} = 
        \left( \begin{array}{c} c_1 + c_8l  \\
		                        b_1 - b_8l      \end{array}  \right) \quad \mbox{and} 
\quad
    \theta_i = \left( \begin{array}{c}  c_7\il + c_2i \\
		                     - b_7\il - b_2i     \end{array}  \right) 
		\label{thcth2}
\end{equation}
   which transform independently.
   As was described for equation~\ref{quatrot} the group actions are considered as 
\textit{active} transformations.  Further, under the left action by the imaginary unit 
$l$ the components of $c$ transform as:
\begin{equation}
  \begin{array}{rcl}
     (c_1+c_8l) & \to & l(c_1+c_8l) \;\;\;\!\!\:\!  = (-c_8 + c_1l)   \\
	 (c_7\il+c_2i) & \to & l(c_7\il+c_2i)   = (-c_2\il + c_7i) 
	  \end{array}
	 \label{leftact}
\end{equation}
 and hence the $(c_7, c_2)$ components of $\theta_i$ transform under left 
multiplication by $l$ in an identical manner to the respective components $(c_1, c_8)$ 
of $\theta_{l}$, which is also trivially true for multiplication by the real unit 1. 
This observation applies also to the $\bar{b}$ components of $\theta_{l}$ and 
$\theta_i$, while the structure of the $2 \times 2$ matrix action of $\sltc^1$ applies 
in the same way on both of these objects. In fact the  transformations of $\theta_l$ 
and $\theta_i$ in equation~\ref{thcth2} are identical both for the generators of 
$\sltc^1$ and for the finite group actions, as can be readily seen by explicit 
calculation.  For example applying the Lorentz symmetry rotation matrix $M_{z \pqg 
l}(\alpha)$ from table~\ref{mtran45} to $\theta_l$ and $\theta_i$ results in the 
respective transformations:
\begin{eqnarray} 
   R_{z \pqg l}(\alpha)\, \theta_l & = & 
     \left(\!\!\! \begin{array}{cc} \cos\frac{\alpha}{2} &   
                                   -l\sin\frac{\alpha}{2} \\
        -l\sin\frac{\alpha}{2} & \cos\frac{\alpha}{2}  \end{array}\!\!\! \right) 
		 \left( \begin{array}{c} c_1 + c_8l  \\
		                        b_1 - b_8l      \end{array}  \right)  \nonumber \\
			    & &		\qquad \qquad \qquad   =
 \left( \begin{array}{c}
   (\coha c_1 - \siha b_8) \quad \!   +  \! \quad  (-\siha b_1 + \coha c_8)l  \\
  (\coha b_1 + \siha c_8) \quad\! + \!\quad (-\siha c_1 -\coha b_8)l    \end{array}  
\right)
					  \nonumber    \\
   	R_{z \pqg l}(\alpha)\, \theta_i & = &
	   \left(\!\!\! \begin{array}{cc} \cos\frac{\alpha}{2} &   
                                   -l\sin\frac{\alpha}{2} \\
        -l\sin\frac{\alpha}{2} & \cos\frac{\alpha}{2}  \end{array}\!\!\! \right) 
		\left( \begin{array}{c}  c_7\il + c_2i \\
		                     - b_7\il - b_2i     \end{array}  \right) \nonumber \\
   			 & &		\qquad  \qquad \qquad  = 
  \left( \begin{array}{c}
    \;\; ( \coha c_7 - \siha b_2)\il \;\; + \;\; (\siha b_7 + \coha c_2)i  \\
  (-\coha b_7 + \siha c_2)\il \, + \, (-\siha c_7 -\coha b_2)i \end{array}  \right)
					  \nonumber 			   		  
\end{eqnarray}
   Here it can be seen that the four real coefficients $\{c_1,c_8,b_1,-b_8\}$ of the 
spinor $\theta_l$ map onto the components of $R_{z \pqg l}(\alpha) \theta_l$ in 
precisely the same way that the coefficients $\{c_7,c_2, -b_7, -b_2\}$ of $\theta_i$ 
map onto the components of 
$R_{z \pqg l}(\alpha) \theta_i$. A similar observation applies for  $\theta_l$ and 
$\theta_i$ under the remaining five Lorentz symmetry actions.
Hence as well as the original left-handed Weyl spinor $\theta_{l}$ the components of 
$\theta_i$ \textit{also} transform exactly as a left-handed spinor of $\sltc^1$.
 This representation of $\sltc$ on the two left-handed spinors $\theta_l$ and 
$\theta_i$ of equation~\ref{thcth2} in $\hhh^2$ contrasts with the representation 
constructed in equations~\ref{psitrans2} and \ref{rdsrss} on the left and right-handed 
spinors $\psi_L$ and $\psi_R$ in $\ccc^4$.
 
  Considering the further two quaternionic subspaces with base units $\{1, l, j, 
\jl\}$ and $\{1, l, k, \kl\}$ it can be seen that the original full octonionic spinor 
$\theta = \binom{c}{\bar{b}}$, with 16 real components, reduces to a total of four 
left-handed Weyl spinors under the action of $\sltc^1$, augmenting the set in 
equation~\ref{thcth2} to:
\begin{equation}
   \theta_{l} \! = \! 
        \left( \begin{array}{c} c_1 + c_8l  \\
		                        b_1 - b_8l      \end{array}  \right) \!\! , \;\,
    \theta_i \! = \! \left(\!\! \begin{array}{c}  c_7\il + c_2i \\
		                     - b_7\il - b_2i     \end{array} \!\! \right) \!\! , \;\,
	 \theta_j \! = \! \left(\!\! \begin{array}{c}  c_6\jl + c_3j \\
		                     - b_6\jl - b_3j     \end{array} \!\! \right) \!\! , \;\,
	 \theta_k \! = \! \left(\!\! \begin{array}{c}  c_5\kl + c_4k \\
		                     - b_5\kl - b_4k     \end{array} \!\! \right)
		\label{thcth234} 
\end{equation}
  There is an equivalent decomposition for a corresponding set of conjugate spinors in 
$\theta^{\dag} = (\bar{c} \; b)$ under the 
right action of $S^{\dag}$ on $\htho$ as implied in equation~\ref{hvvv}.
This set of four Weyl spinors in equation~\ref{thcth234} will be important for 
interpreting further symmetries, \textit{internal} to the action of $\sltc^1$ on 
$\htho$, in the following section.

   In this section we have described how the decomposition of the $\mathbf{27}$ 
representation of $\esi$ under the subgroup $\spotn$ of equation~\ref{esidecom} 
further reduces under the subgroup $\sltc^1$ as summarised in table~\ref{sltcreps}.

\begin{table}[htbp]
\centering
\begin{tabular}{|c|c|c|}
 \hline
     $\spotn$       &   $\sltc^1$                &    Components     \\  
 \hline		  
    \: $\mathbf{1}$ \, scalar
	    &  $\qquad$  (0,0) $\;$ scalar            &    $n$            \\
	 $\mathbf{10}$ \, vector
	   &  $ \left\{ \begin{array}{rl}  (\fhs,\fhs) &  \mbox{vector}  \\
                   6\times (0,0) & \mbox{scalars}  \end{array}  \right. $  &         
	$	\begin{array}{c}     \bv_4   \\  a(6)  \end{array} $  \\
	 $\mathbf{16}$ \, spinor
	   & $\quad \! 4\times(\fhs,0)\;\,$ spinors   &    $\theta_{l,i,j,k}$   \\ 
   \hline
  \end{tabular}
  \caption{\setb The further decomposition of the $(\mathbf{1} + \mathbf{10} + 
\mathbf{16})$ representation of $\spotn \subset \esi$ of equation~\ref{esidecom} under 
the subgroup of external 4-dimensional spacetime symmetry $\sltc^1 \subset \spotn$ 
actions of equation~\ref{hvvv}, and the corresponding components of $\htho$ 
transformed.}
\label{sltcreps}
\end{table}  

  This may be compared with the $\sltc^1$ action on the subspace $\hthc \subset \htho$ 
as described in equation~\ref{vinhinc} for which the nine real components of $\hthc$ 
transform as one 4-vector $\bh_2$, one Weyl spinor $\psi_L$ and one scalar $n$ 
(closely relating to $\bv_4$, $\theta_l$ and $n$ respectively in 
table~\ref{sltcreps}). The six extra scalars and three extra spinors in 
table~\ref{sltcreps} result from the additional $27-9 = 18$ real components in 
$\htho$. In both cases each Weyl spinor, as for the space $\ccc^2$, has four real 
parameters.

 The spinor components of $\theta_{l,i,j,k}$ represent `internal' dimensions of the 
space $\htho$ in the sense that, unlike $\bv_4 \inn \TM_4$, they are not tangent to 
the external spacetime $M_4$, but they \textit{do} transform in a non-trivial manner, 
as spinors, under the external $\sltc^1$ symmetry, and in this sense they are not 
purely internal objects. 
 This feature for the cubic form of temporal flow $\lvt$ is hence distinct from that 
seen for a quadratic form with a spacetime symmetry. For example for the 
10-dimensional spacetime form considered in section~\ref{reaic} the external $\soot$ 
symmetry acts on the external $\ol{\bv}_4 \subset \bv_{10}$ components \textit{only}, 
as pictured in figure~\ref{mtogmaphr}(b) and applies also for the corresponding gauge 
field  in equation~\ref{dlvfibte}, as for the external symmetry of any
   higher-dimensional spacetime structure.
  For the present theory based on temporal progression, here taking a cubic form,
 of particular interest in the following section will be the nature of the internal 
symmetry transformations on the four $\sltc^1$ spinors from the final line of 
table~\ref{sltcreps}.


\section{Internal $\suth_c \times \uo_Q$  Symmetry}
  \label{intsym}
  Physically the $\sltc^1$ symmetry studied in section~\ref{extsym} is considered 
`external' as it is the two-to-one cover of the Lorentz group which in the full theory 
acts on the tangent space $\TM_4$ of the extended 4-dimensional spacetime manifold. 
This structure is central to the theory of general relativity and gravitation, as 
described in sections~\ref{riegeo}, \ref{gcatep} and \ref{fdandtd}. On the other hand 
the `internal' symmetry will consist of further subgroups of $\esi$, which will be 
central to the structure of local gauge theories and the Standard Model of particle 
physics as reviewed in the previous chapter.

 At the end of the previous section the branching of the $\mathbf{16}$ representation 
of $\spotn$ into a set of four Weyl spinors under the external Lorentz subgroup 
$\sltc^1$ was described, as listed in table~\ref{sltcreps}. Independently it is also 
known that the same $\spotn$ Majorana-Weyl 16-dimensional representation branches into 
a set of multiplets describing the 15 states of one generation of Standard Model 
quarks and leptons, as listed in equation~\ref{multsf}, together with a right-handed 
neutrino, all expressed uniformly in terms of left-handed fields, under the internal 
subgroup $\SML$, as noted in section~\ref{dynkin}, but it is not the approach we 
follow here. In this section we consider the internal symmetry derived from the 
subgroup $\stab \subset \esi$, defined below, and its relation to the set of four Weyl 
spinors derived from the external symmetry $\sltc^1 \subset \spotn$.

   In contrast to the 6 generators of the external $\sltc^1$ symmetry, of the 
remaining $(78-6)=72$ generators of $\esi$ those which leave all tangent space vectors 
$\bv_4 \inn \TM_4$ untouched may literally be considered to constitute an 
\textit{internal} symmetry, surviving the symmetry breaking,  and are expected to be 
significant for the physics of local gauge theories. While leaving $\TM_4$ invariant 
these internal symmetries will in general have non-trivial actions on the remaining, 
`extra dimensions' within $\bv_{27}$, through which we may seek to identify a relation 
with the  phenomena of physical particle interactions as observed in the laboratory 
and described by the Standard Model.

  Here then, as a preliminary definition, and in contrast to the external symmetry, 
the internal symmetry will be obtained from the set of all $\esi$ actions on $\htho$ 
which leave the four components for any $\bv_4 = (v^0,v^1,v^2,v^3)$ in 
equation~\ref{hinhtho} (that is, $\bh_2 \inn \htwc$ of equation~\ref{hvvv}) invariant. 
These components, including $v^2 \equiv a_8$ associated with the imaginary unit $l$ of 
$a\inn \ooo$, can also be expressed in the combination $(p,m,a_1,a_8)$ with respect to 
the parametrisation of equation~\ref{htho}.  The corresponding symmetry group is  
complementary to the actions of $\sltc^1$ and will be denoted $\stab$ as the stability 
group of all vectors $\bv_4 \inn \TM_4$. By inspection from tables~\ref{lbrota} and 
\ref{ltrota}, for the 78 elements in the preferred basis for the Lie algebra of $\esi$ 
defined on the space $T\htho$, the group $\stab$ is generated by the 31 elements 
listed in table~\ref{stabset}.
 In particular we shall be looking to identify closed subgroups within $\stab$ for 
which each generator is independent of $\sltc^1$ in terms of Lie bracket composition.

\begin{table}[htbp]
\centering
\begin{tabular}{|l|r|}
 \hline  
      Category 1 and 2: Boosts and Rotations   & \#  \\ \cline{2-2}
	  $\qquad (\dot{R}_{x \pqg z}^{2} - \dot{B}_{t \pqg x}^{2}),	
   \qquad   (\dot{R}_{x \pqg z}^{3} + \dot{B}_{t \pqg x}^{3})$	  &   2  \\
      $\qquad (\dot{R}_{z \pqg q}^{2} + \dot{B}_{t \pqg q}^{2}),
   \qquad   (\dot{R}_{z \pqg q}^{3} - \dot{B}_{t \pqg q}^{3})$    &  14  \\
 \hline
     Category 3: Transverse Rotations  &  \\ 
    $\qquad \dot{A}_q$,  \quad   $\dot{G}_l$,  \quad $\dot{S}_{l}^{1}$      &   9  \\
	$\qquad (\dot{G}_q + 2\dot{S}_{q}^{1}) \;\;\; q= \{i,j,k,\kl,\jl,\il \}$   &   6  
\\
 \hline
 \multicolumn{1}{|r|}{Total}    &    31  \\
 \hline
  \end{tabular}
  \caption{\setb The Lie algebra generators of the 31 dimensional group $\stab$. The 
subscript $q$ denotes any of the seven imaginary octonion units $\{i,j,k,\kl,\jl,\il,l 
\}$ unless stated otherwise.}
\label{stabset}
\end{table}

    The 16 vector fields on $T\htho$ generating the Category 1 and 2 elements of 
$\stab$  are written out explicitly in equations~\ref{rbstab1} and \ref{rbstab2} in 
which the invariant action on the 4-dimensional subspace $\htwc \subset \htho$ is 
clear. (In fact they leave all 10 components of $\htwo \subset \htho$ invariant, see 
also \cite{Wang2} equations 4.12(27) and 4.13(28)).
\begin{eqnarray}
 \!\! \dot{R}_{x \pqg z}^{2} - \dot{B}_{t \pqg x}^{2} = 
 \left(\!\! \begin{array}{ccc}
                 \; 0 \; &  \; 0 \;  &    -\bar{a}     \\
	                0    &     0     &    -m          \\
                   -a    &    -m     &    -2b_x
 \end{array} \!\!\right) \!\! ,  &  &
  \dot{R}_{x \pqg z}^{3} + \dot{B}_{t \pqg x}^{3} = 
  \left(\!\! \begin{array}{ccc}
                 \; 0 \; &  \; 0 \;  &    p     \\
	                0    &     0     &    a          \\
                    p    &  \bar{a}  &    2c_x
 \end{array} \!\!\right) \label{rbstab1} \\   
    &   \nonumber \\
 \!\! \dot{R}_{z \pqg q}^{2} + \dot{B}_{t \pqg q}^{2} = 
 \left(\!\! \begin{array}{ccc}
                    0    &     0     &     \bar{a}q     \\
	                0    &     0     &     mq          \\
                  -qa    &   -mq     &   -2b_q
 \end{array} \!\!\right) \!\! ,  &  &
  \dot{R}_{z \pqg q}^{3} - \dot{B}_{t \pqg q}^{3} = 
  \left(\!\! \begin{array}{ccc}
                    0    &     0     &    pq     \\
	                0    &     0     &    aq          \\
                  -pq    & -q\bar{a} &    2c_q
 \end{array} \!\!\right) \label{rbstab2}
\end{eqnarray}

  Of the 15 transverse rotations in table~\ref{stabset} the first 9 are basis vectors 
which explicitly leave the components of $\bh_2$ invariant, while for each of the 
remaining six $(\dot{G}_q + 2\dot{S}_{q}^{1})$ actions the non-zero $\dot{a}_8l$ 
contributions in table~\ref{ltrota} cancel. 

  Although the category 1 and 2 transformations of type 1 as originally composed on 
the 10-dimensional space $\htwo$ each act as a simple rotation or boost in a 
2-dimensional plane the effect on the components of the spinor $\theta$ is less 
straightforward in the full $\htho$ action, as was seen for the case of the external 
symmetry $\sltc^1$ in the previous section. This is also seen for the type 2 and 3 
internal transformations of equations~\ref{rbstab1} and \ref{rbstab2}. Hence these 
actions, together with the 15 internal transverse rotations, stir up the $\theta$ 
components in non-trivial ways.

  Of particular interest is the $\suth$ subgroup introduced below 
equation~\ref{agsdeps} and discussed shortly after figure~\ref{dynkinb} (as described 
in \cite{Wang} pp.115 and 136, following \cite{Gunay}).
 This $\suth$ is defined in terms of the transverse rotations acting on the octonion 
space $\ooo$ alone as the subgroup $\suth \subset \gt$ of the octonion automorphism 
group that leaves one imaginary unit, here $l$, invariant.
The corresponding Lie algebra $\sutha$ is described by the set of 8 generators 
$\{\dot{A}_q, \dot{G}_l\}$  which, as transformations of $\esi$ on the full space 
$\htho$, act on each of the octonion elements $a,b,c \inn \ooo$ in the same way 
leaving invariant the complex $\{1,l\}$ subspaces, and as elements  of 
table~\ref{stabset}
   identified within $\staba$ may be provisionally associated with the colour 
su(3)$_c$ of the Standard Model.
 This algebra is also independent of $\sltc^1$ in terms of the Lie bracket 
composition, that is $\lbrack X,Y \rbrack = 0$ for all $X\inn \sltca^1$ and $Y \inn 
\sutha_c$, and hence we have the semi-simple subgroup:
\begin{equation}
  \label{esisubss}
    \sltc^1 \times \suth_c \subset \esi
\end{equation}
 The Lie algebra composition of the $\{\dot{A}_q, \dot{G}_l\} \inn \sutha_c$ elements  
from the $\esi$ commutation table in \cite{Wang} is reproduced here in 
table~\ref{sutwang}.  
\begin{table}[htbp]
\centering
\begin{tabular}{|c|cccccccc|}
\hline
    $\lbrack \bullet , \bullet \ \rbrack$
  &  $\dot{A}_i$ & $\dot{A}_j$  & $\dot{A}_k$ & $\dot{A}_{\kl}$ 
                             & $\dot{A}_{\jl}$ & $\dot{A}_{\il}$ & $\dot{A}_l$ & 
$\dot{G}_l$ \\ 
  \hline
  $\dot{A}_i$ &    0   & $\dot{A}_k$ & $-\dot{A}_j$ & $-\dot{A}_{\jl}$ &
                 $\dot{A}_{\kl}$  & $\dot{A}_l \!-\! \dot{G}_l$ & $-\dot{A}_{\il}$ & 
$3\dot{A}_{\il}$
  \\
  $\dot{A}_j$ & $-\dot{A}_k$ &    0   & $\dot{A}_i$ & $-\dot{A}_{\il}$ &
                 $-\dot{A}_l \!-\! \dot{G}_l$ & $\dot{A}_{\kl}$ & $\dot{A}_{\jl}$ & 
$3\dot{A}_{\jl}$
  \\
  $\dot{A}_k$ & $\dot{A}_j$ & $-\dot{A}_i$ &    0   & $-2\dot{A}_l$ &
                           $-\dot{A}_{\il}$ & $\dot{A}_{\jl}$ & $2\dot{A}_{\kl}$ &    
0  
  \\
  $\dot{A}_{\kl}$ & $\dot{A}_{\jl}$ & $\dot{A}_{\il}$ & $2\dot{A}_l$ &    0   & 
                          $-\dot{A}_i$  & $-\dot{A}_j$ & $-2\dot{A}_k$ &    0  
  \\
  $\dot{A}_{\jl}$ & $-\dot{A}_{\kl}$ & $\dot{A}_l \!+\! \dot{G}_l$ & $\dot{A}_{\il}$ &  
$\dot{A}_i$ &
                             0    & $-\dot{A}_k$ & $-\dot{A}_j$ & $-3\dot{A}_j$
  \\
  $\dot{A}_{\il}$ & $-\dot{A}_l \!+\! \dot{G}_l$ & $-\dot{A}_{\kl}$ & $-\dot{A}_{\jl}$ 
& $\dot{A}_j$ &
                          $\dot{A}_k$  &    0   & $\dot{A}_i$ & $-3\dot{A}_i$
  \\
  $\dot{A}_l$ & $\dot{A}_{\il}$ & $-\dot{A}_{\jl}$ & $-2\dot{A}_{\kl}$ & 2$\dot{A}_k$ 
&
                          $\dot{A}_j$  & $-\dot{A}_i$ &    0   &    0  
  \\
  $\dot{G}_l$ & $-3\dot{A}_{\il}$ & $-3\dot{A}_{\jl}$ &    0   &    0   &
                          $3\dot{A}_j$  & $3\dot{A}_i$ &    0   &    0  
  \\
 \hline
  \end{tabular}
  \caption{\setb (Extracted from the $\esi$ Lie algebra table in \protect\cite{Wang}). 
The Lie algebra structure for the SU(3)$_c$ generators $\{\dot{A}_q, \dot{G}_l\}$, 
with bracket composition $[\dot{A}_i, \dot{A}_j] = \dot{A}_k$ etc.}
\label{sutwang}
\end{table} 
   The Lie algebra in table~\ref{sutwang} is isomorphic to the su(3) Lie algebra 
represented by the eight $3\times 3$ Gell-Mann matrices listed in table~\ref{gellm}. 

\begin{table}[htbp]
\centering
\begin{tabular}{|ccc|}
 \hline  
    \vspace{-15pt}  & &  \\
$\lambda_1 = \left( \begin{array}{ccc}
                    0    &     1    &   0     \\
	                1    &     0    &   0           \\
                    0    &     0    &   0
 \end{array} \right)$  & 
$\lambda_2 = \left( \begin{array}{ccc}
                    0    &    -i    &   0     \\
	                i    &     0    &   0           \\
                    0    &     0    &   0
 \end{array} \right)$  &  
$\lambda_3 = \left( \begin{array}{ccc}
                    1    &     0    &   0     \\
	                0    &    -1    &   0           \\
                    0    &     0    &   0
 \end{array} \right)$   \\ 
    \vspace{-12pt}  & &  \\
$\lambda_4 = \left( \begin{array}{ccc}
                    0    &     0    &   1     \\
	                0    &     0    &   0           \\
                    1    &     0    &   0
 \end{array} \right)$  &  
$\lambda_5 = \left( \begin{array}{ccc}
                    0    &     0    &  -i     \\
	                0    &     0    &   0           \\
                    i    &     0    &   0
 \end{array} \right)$  &  \\
 \vspace{-12pt}  & &  \\
$\lambda_6 = \left( \begin{array}{ccc}
                    0    &     0    &   0     \\
	                0    &     0    &   1           \\
                    0    &     1    &   0
 \end{array} \right)$  & 
$\lambda_7 = \left( \begin{array}{ccc}
                    0    &     0    &   0     \\
	                0    &     0    &  -i           \\
                    0    &     i    &   0
 \end{array} \right)$  &  
$\lambda_8 = \frac{1}{\sqrt{3}} \left( \begin{array}{ccc}
                    1    &     0    &   0     \\
	                0    &     1    &   0           \\
                    0    &     0    &  -2
 \end{array} \right)$   \\ 
 \vspace{-15pt}  & &  \\
 \hline
 \multicolumn{3}{|c|}{with $\lbrack \lambda_{\alpha}, \lambda_{\beta} \rbrack
    = if_{\alpha\beta\gamma}\lambda_{\gamma}$
       $\qquad$   and   $\qquad$ $f_{123} = 2$, 
		   $f_{458} = f_{678} = \sqrt{3}$, }     \\ 
 \multicolumn{3}{|c|}{  $f_{147}=-f_{156}=f_{246}=f_{257}=f_{345}=-f_{367}=1$}     \\ 		   
		    \hline 
  \end{tabular}
  \caption{\setb The set of eight complex Hermitian Gell-Mann matrices of su(3), with 
representatives of the completely antisymmetric structure constants
  $f_{\alpha\beta\gamma}$ which are non-zero.}
\label{gellm}
\end{table} 

  The two algebras in tables~\ref{sutwang} and \ref{gellm} are identical within the 
choice of sign conventions, numerical coefficients and the fact that the Gell-Mann 
matrices are taken to be Hermitian. The latter property results in an extra factors of 
$i$ accompanying the $\lambda_{\alpha}$ matrices in the algebra isomorphism listed in 
table~\ref{wangel}.
 The factors of $i$ belong to the \textit{same} complex algebra $\ccc$ used in the 
components of the Gell-Mann matrices themselves, but are \textit{independent} of the 
octonion algebra elements on the left-hand side. (That is the isomorphism is between 
the basis $\{\dot{A}_q, \dot{G}_l\}$ and the 
anti-Hermitian matrices $\sim i\lambda_{\alpha}$ rather than directly with the 
Hermitian Gell-Mann matrices.
 This is analogous to the relation between the external $\sltc^1$ generators and the 
conventional Lorentz algebra in equation~\ref{msigj}, where factors of $i$ would also 
appear if the $J^a$ were defined as Hermitian rather than anti-Hermitian in 
equation~\ref{jsigksig}).

\begin{table}[htbp]
\centering
\begin{tabular}{|ccc|}
 \hline  
$\quad \dot{A}_k \sim -i\lambda_1 \quad$ & $\quad \dot{A}_{\kl} \sim -i\lambda_2 
\quad$
      & $\quad\; \dot{A}_l \sim i\lambda_3 \quad\;$ \\  & &  \\
$\dot{A}_i \sim -i\lambda_4$ & $\dot{A}_{\il} \sim i\lambda_5$ &   \\  & & \\
$\dot{A}_{\jl} \sim -i\lambda_6$ & $\dot{A}_{j} \sim -i\lambda_7$ & 
                                            $\dot{G}_l \sim -i\sqrt{3}\lambda_8$ \\
		    \hline 
  \end{tabular}
  \caption{\setb The isomorphism between the su$(3)_c\subset \esi$ Lie algebra basis 
$\{\dot{A}_q, \dot{G}_l\}$ and the eight Gell-Mann matrices $\lambda_{\alpha}$ 
(\protect\cite{Wang} p.137, table 4.5).}
\label{wangel}
\end{table}

  The $3\times 3$ Gell-Mann matrices transform the components of complex vectors 
$\bu\inn \ccc^3$ corresponding, in the context of an SU(3)$_c$ gauge theory, to the 
interactions between `red', `blue' and `green' quark states encountered in quantum 
chromodynamics. Similarly the $\{\dot{A}_q, \dot{G}_l\}$ algebra elements, as 
transformations on the space $\htho$ mix the components of the $\spotn$ spinor $\theta 
= \binom{c}{\bar{b}} \inn \ooo^2$. For example the tangent vector field $\dot{A}_i$ on 
the $\binom{c}{\bar{b}}$ components of $\htho$, obtained from table~\ref{ltrota}, are:
\begin{eqnarray} 
 \!\!\!\!\!\!\!\!\!  
    \dot{A}_i:\left( \begin{array}{c} \dot{c} \\ \dot{\bar{b}} \end{array} \right)
	 \!  & \! = \! & \!
	 \left(  \begin{array}{cccc} 
 \dot{c}_1+\dot{c}_8l,\sb &+\dot{c}_7\il+\dot{c}_2i,\sb &+\dot{c}_6\jl+\dot{c}_3j,\sb 
   &+\dot{c}_5\kl+\dot{c}_4k \\
 \dot{b}_1-\dot{b}_8l,\sb &-\dot{b}_7\il-\dot{b}_2i,\sb &-\dot{b}_6\jl-\dot{b}_3j,\sb  
  &-\dot{b}_5\kl-\dot{b}_4k
  	   \end{array} \right)  \nonumber \\
	 \! & \! = \! & \!
	 \left(  \begin{array}{cccc} 
  0 + 0l, \sb\;\:\, & + 0 \il+0 i, \sb\;\;\, &-{c_5}\jl{-c_4}j, \sb\:\,  
&+{c_6}\kl+{c_3}k \\
  0 - 0l, \sb\;\:\, & - 0 \il-0 i, \sb\;\;\, &+{b_5}\jl{+b_4}j, \sb\:\,  
&-{b_6}\kl-{b_3}k
  	   \end{array} \right)
  \label{aioncb}
\end{eqnarray}

   The components here have been ordered to match those of the four left-handed Weyl 
spinors $(\theta_l, \, \theta_i, \, \theta_j, \theta_k)$ of equation~\ref{thcth234}. 
The fact that each real component of $c$ transforms in the same way as the 
corresponding component of $b$ is expected since $\suth_c$ acts on each of $a,b,c \inn 
\ooo$ in precisely the same way. However, it is also noted that the action $\dot{A}_i$ 
in equation~\ref{aioncb}  respects the 4-way spinor decomposition, with for example  
$\dot{c}_6$ and $\dot{c}_3$ of $\dot{\theta}_j$ taking the respective values of 
$-{c_5}$ and $-c_4$ from the spinor $\theta_k$. This apparently non-trivial 
observation applies to all eight $\suth_c$ generators, which 
hence represent a mixing of the four Weyl spinors, as a structure maintained within 
the
 mixing of the eight real components of the octonion elements.

    The extraction of the components of a spinor $\theta$ into a matrix of real 
numbers will be denoted by $\lbrack \theta \rbrack$. For example, from 
equation~\ref{thcth234} the spinor $\theta_i$ can be mapped to the $2\times 2$ matrix 
of real numbers $\lbrack \theta_i \rbrack = \binom{ c_7 \;\;\; c_2}{-b_7 \, -b_2}$ 
(with components ordered to match those of the spinor $\theta_l$ under $\sltc^1$ 
transformations, as described for equations~\ref{thcth2}--\ref{thcth234}). With this 
notation and the  Lorentz spinor definitions in equation~\ref{thcth234} the above 
equation~\ref{aioncb} can be expressed as:
\begin{eqnarray}
   \dot{A}_i: \lbrack \dot{\theta} \rbrack & = &  (
    \; \lbrack \dot{\theta}_{l} \rbrack, \;\; \lbrack \dot{\theta}_i \rbrack, \;\; 
	                  \lbrack \dot{\theta}_j \rbrack, \;\; \lbrack \dot{\theta}_k 
\rbrack \;
         ) \nonumber  \\
	&  = &    \left(
         \;\;  0_2, \;\;\: 0_2, \:  
	                  -\lbrack {\theta}_k \rbrack, \;\; \lbrack {\theta}_j \rbrack \;
        \right)	
		\label{aionthet}
\end{eqnarray}
   where $0_2$ represents the $2\times 2$ zero matrix. This expression shows 
explicitly how the internal SU(3)$_c$ generator $\dot{A}_i$ mixes the external 
$\sltc^1$ spinors $\theta_j$ and $\theta_k$ identified in the previous section. The 
tangent vectors of all eight generators $\{\dot{A}_q, \dot{G}_l\}$ of 
 SU(3)$_c$ on the spinor space $\theta \inn \ooo^2$ are listed in table~\ref{suttan} 
alongside the actions of the Gell-Mann matrices, using the correspondence in 
table~\ref{wangel}, on the vectors $\bu\inn\ccc^3$. On the left-hand side            
the elements $\{\dot{A}_q, \dot{G}_l\}$ are already expressed as tangent vectors, 
while on the right-hand side the tangents are  
obtained by matrix multiplication of the $\lambda_{\alpha}$ into $\bu \inn \ccc^3$.

\begin{table}[htbp]
\centering
 \hspace*{-12pt}
\begin{tabular}{|c|}
\hline
 $ \begin{array}{c@{(}rrrr@{)}ccc@{(}ccc@{)}}
   &   \lbrack \dot{\theta}_{l} \rbrack, & \lbrack \dot{\theta}_i \rbrack, & 
	   \lbrack \dot{\theta}_j \rbrack,      & \lbrack \dot{\theta}_k \rbrack 
	 & &  & &   \dot{u}_1, & \dot{u}_3, & \dot{u}_3 
\vspace{10pt} \\ 
	 \dot{A}_i = \quad &   0_2, & 0_2, &  
	                  -\lbrack {\theta}_k \rbrack, & \lbrack {\theta}_j \rbrack  
	& \qquad \quad \sim \qquad & \lambda_4 \Rightarrow &   &  u_3, &  0, & u_1 \\
     \dot{A}_{\il} = \quad &   0_2, & 0_2, &  
	                  \lbrack {l\theta}_k \rbrack, & \lbrack {l\theta}_j \rbrack  
	& \qquad \quad \sim \qquad & \lambda_5 \Rightarrow &   &   -iu_3, &  0, & iu_1  
 \vspace{10pt} \\ 
	 \dot{A}_j = \quad &   0_2, &  \lbrack {\theta}_k \rbrack, &
	                       0_2, &   -\lbrack {\theta}_i \rbrack   
	& \qquad \quad \sim \qquad & \lambda_7 \Rightarrow &   &   0, & -iu_3, & iu_2  \\
	\dot{A}_{\jl} = \quad &   0_2, &  -\lbrack {l\theta}_k \rbrack, &
	                       0_2, &   -\lbrack {l\theta}_i \rbrack   
	& \qquad \quad \sim \qquad & \lambda_6 \Rightarrow &   &   0, & u_3, & u_2  
 \vspace{10pt} \\ 
	 \dot{A}_k = \quad &   0_2, &  -\lbrack {\theta}_j \rbrack, &
	                        \lbrack {\theta}_i \rbrack, & 0_2   
	& \qquad \quad \sim \qquad & \lambda_1 \Rightarrow &   &   u_2, & u_1, & 0 \\
	\dot{A}_{\kl} = \quad &   0_2, &  \lbrack {l\theta}_j \rbrack, &
	                        \lbrack {l\theta}_i \rbrack, & 0_2   
	& \qquad \quad \sim \qquad & \lambda_2 \Rightarrow &   &   -iu_2, & iu_1, & 0 
 \vspace{10pt} \\ 
	 \dot{A}_l = \quad &   0_2, &  \lbrack {l\theta}_i \rbrack, &
	                        -\lbrack {l\theta}_j \rbrack, & 0_2   
	& \qquad \quad \sim \qquad & \lambda_3 \Rightarrow &   &   u_1, & -u_2, & 0 \\
	 \dot{G}_l = \quad &   0_2, &  \lbrack {l\theta}_i \rbrack, &
	                    \lbrack {l\theta}_j \rbrack, & -2\lbrack {l\theta}_k \rbrack  
	& \qquad \quad \sim \qquad & \lambda_8 \Rightarrow & \!\! \frac{1}{\sqrt{3}}  &   
u_1, & u_2, & \!\! -2u_3 \\
 \end{array}
 $	 
  \\
 \hline
  \end{tabular}
  \caption{\setb The tangent vector generators for the SU(3) representations on 
$\ooo^2$ and $\ccc^3$.  The column vectors of  $\ccc^3$ are displayed as a row vectors 
for convenience in the table.}
\label{suttan}
\end{table}

   In table~\ref{suttan} a term such as $\lbrack l\theta_i \rbrack$ denotes 
multiplying the spinor $\theta_i$ on the left by $l$ before extracting the 
coefficients of $l\theta_i$ with the $\mbox{Im}(\ooo)$ units ordered as in 
equation~\ref{thcth234}. This notation is used to isolate the mixing effect on the 
real number coefficients, with care for the joint effects of the division algebra 
composition as well as matrix algebra composition. For the case of $\bu \inn \ccc^3$, 
the two real degrees of freedom for each of the $u_1,u_2,u_3$ belong to the 
\textit{same} complex space $\ccc$ (with base units $\{1,i\}$) but occupy 
\textit{different} components of the $1 \times 3$ column matrix vector $\ccc^3$. For 
the case of $\theta \inn \ooo^2$ the four real degrees of freedom for each of the 
$\theta_{l},\theta_i,\theta_j,\theta_k$ belong to a \textit{different} 2-dimensional 
subspace of $\ooo$ (with base units $\{1,l\},\{\il,i\},\{\jl,j\},\{\kl,k\}$ 
respectively) but occupy the \textit{same} components of the $1 \times 2$ column 
matrix vector $\ooo^2$.

  Hence, as seen in table~\ref{suttan}, the six transformations $\dot{A}_q$ ($q\neq 
l$) mix the components of the three Weyl spinors $\theta_i, \theta_j, \theta_k$ in a 
similar manner that the Gell-Mann matrices $\lambda_{\alpha}$ ($\alpha\neq 3,8$) mix 
the three $\ccc^3$ components $u_1,u_2,u_3$, with the correspondence between the 
objects of each representation space depending on the form of the isomorphism in 
table~\ref{wangel}, which is arbitrary up to the automorphism group of su(3).
  In both cases there are two remaining diagonal generators as listed at the bottom of 
table~\ref{suttan}.
 (The physics here is determined by the $\{\dot{A}_q, \dot{G}_l\}$ transformations as 
generators of $\suth_c$ rather than the particular choice of correspondence with the 
$\lambda_{\alpha}$ matrices, as was similarly the case for the external $\sltc^1$ 
action of the previous section as described after equation~\ref{msigj}). 
 In the case of the full set of $\{\dot{A}_q, \dot{G}_l\}$ acting on the components of 
$\htho$ there is a copy of the same set of mixing transformations within the 
components of the Hermitian conjugate spinor $\theta^{\dag} = (\bar{c}\; b)$ of 
equation~\ref{xoct3} which also transforms under the internal SU(3)$_c$ symmetry 
(similarly as described for the $\sltc^1$ spinors below equation~\ref{thcth234}).

 In conclusion the internal SU(3)$_c$ symmetry action in the left-hand column of 
table~\ref{suttan} dovetails neatly with the external $\sltc^1$ spinor structure of 
equation~\ref{thcth234}.
 The mixing action of SU(3)$_c$ in table~\ref{suttan} takes a form summarised as:
\begin{equation}
    \theta  \;  = \; (\, \theta_l, \quad
	  \underbrace{ \theta_i, \quad \theta_j, \quad \theta_k}_{\mbox{\small SU(3)$_c$ 
action}} \, )
	  \label{suthth}
\end{equation}
   which implies that as a gauge theory the $\suth_c$ internal symmetry will mediate 
interactions between the
    Weyl spinors $\theta_i, \theta_j, \theta_k$, transforming under the fundamental 
representation, which in turn will hence be identified with the three colour degrees 
of freedom of the quark states. On the other hand the invariance of $\theta_l$,
 transforming under the trivial representation of SU(3)$_c$, suggests that these 
components should be associated with the leptonic sector of the Standard Model (with 
the subscript $l$ originating from the $\{1,l\}$ base units for $\theta_l$ also then 
serving as a mnemonic for its leptonic character). 
 Further aspects of the Standard Model might then be expected to be uncovered by 
exploring further aspects of the internal symmetry group within $\esi$, which will 
occupy the remainder of this chapter.

   In particular the Standard Model Abelian gauge group $\uo_Q$, underlying Maxwell's 
equations and the phenomena of electromagnetism, might also be sought as an internal 
symmetry within $\stab$.
  Of the 31 generators for the internal symmetry group $\stab$ listed in 
table~\ref{stabset} there is a $(31-8)=23$-dimensional set which as a vector space is 
independent of the internal SU(3)$_c$ generators. Of these 23 there are 3 sets each of 
6 elements:
 \begin{equation}
   (\dot{R}_{z \pqg q}^{2} + \dot{B}_{t \pqg q}^{2}),
   \qquad   (\dot{R}_{z \pqg q}^{3} - \dot{B}_{t \pqg q}^{3}),
   \qquad (\dot{G}_q + 2\dot{S}_{q}^{1})
   \label{wang18} 
 \end{equation}
   with $q\neq l$, totalling 18 elements each of which fails to commute with some of 
the internal SU(3)$_c$ generators in the set $\{\dot{A}_q, \dot{G}_l\}$. As a 
cross-check this observation appears to hold for any linear combination of elements 
selected from the 18 in equation~\ref{wang18}, by further inspection of the $\esi$ Lie 
algebra table \cite{Wang}.
 Hence none of the 18 elements in equation~\ref{wang18} can belong to a group which 
may be appended to the subgroup decomposition $\sltc^1 \times \suth_c$ in 
equation~\ref{esisubss} (in fact the first 12 elements in equation~\ref{wang18} also 
fail to commute with $\sltc^1$).
 This then leaves a set of only $(31-8-18)=5$ internal generators which in terms of 
Lie algebra composition, and not only as a vector space, is independent of su(3)$_c$. 
These are the elements:
 \begin{equation}
   (\dot{R}_{x \pqg z}^{2} - \dot{B}_{t \pqg x}^{2}),
   \quad (\dot{R}_{x \pqg z}^{3} + \dot{B}_{t \pqg x}^{3}),
   \quad (\dot{R}_{z \pqg l}^{2} + \dot{B}_{t \pqg l}^{2}),
   \quad (\dot{R}_{z \pqg l}^{3} - \dot{B}_{t \pqg l}^{3}),
   \quad \dot{S}_{l}^{1}
     \label{wang5}
 \end{equation}  
 Indeed, each of the \textit{nine} individual component parts listed within 
equation~\ref{wang5} commute with all eight elements of the internal su(3)$_c$ basis 
set.
However the first 4 elements in equation~\ref{wang5} each fail to commute with the 
external $\sltc^1$ generators. This leaves $\dot{S}_{l}^{1}$  as the \textit{only} 
$\esi$ Lie algebra generator of $\stab$ which is independent of both $\sltc^1$ and 
$\suth_c$.
  Hence of the many possible $\uo \subset \esi$ subgroups the one generated by  
$\dot{S}_{l}^{1}$ is identified as the most suitable candidate for the internal 
$\uo_Q$ gauge symmetry of electromagnetism.
  
 Moreover the generator $\dot{S}_{l}^{1}$ is also closely associated with the diagonal 
symmetry action $\ssl_l^1$, described by equation~\ref{sqdiag}, and leaves the 
4-dimensional spacetime components in $\htwc$ invariant as a residual of the $\sltwoo$ 
action on $\htwo$ as described at the end of section~\ref{ltos}. A similar internal 
$\uo$ symmetry associated with electromagnetism has been considered for the $\sltc 
\times \uo$ gauge theories as discussed in the opening paragraphs of 
section~\ref{dynkin}.
  While as elements of the vector space $T\htho$  we have $\dot{S}^1_l = \dssl^1_l$, 
as discussed following equation~\ref{slinrr}, the group actions $S_l^{(1)}(\alpha)$ 
and $\ssl_l^{(1)}(\alpha)$ diverge at O$(\alpha^2)$ and in any case, although 
suggestive, this argument alone is insufficient in itself to associate the $\uo_Q$ 
symmetry with  ${S}_{l}^{1}$ out of many possible $\uo \subset \esi$ subgroups. Here 
the main case for this association is the observation that the $\uo$ subgroup   
 ${S}_{l}^{1}$ uniquely both belongs to $\stab$ and at the Lie algebra level is 
independent of the  $\sltc^1 \times \suth_c$  subgroup of equation~\ref{esisubss}.

 Hence here the internal $\uo$ generated by $\dot{S}_{l}^{1}$ is a natural candidate 
to consider for the $\uo_Q$ component of the Standard Model gauge symmetry group.
 From table~\ref{ltrota} it can be seen that the generator $\dot{S}_{l}^{1}$ impacts 
on all 8 real components of both $c$ and $\bar{b}$ of $\theta \inn \ooo^2$. In fact, 
and in comparison with equation~\ref{aioncb}, the tangent vector  $\dot{S}_{l}^{1}$ on 
the spinor components $\theta = \binom{c}{\bar{b}}$ is given explicitly by:
\begin{eqnarray} 
   \!\!\!   \dot{S}_l^1 \! : \!\left( \! \begin{array}{c} \dot{c} \\ \dot{\bar{b}} 
\end{array} \! \right) \!\! & \!\!\! = \!\!\! & \!\!
	 \left( \;\!  \begin{array}{cccc} 
 \dot{c}_1+\dot{c}_8l, \quad\;\, &+\dot{c}_7\il+\dot{c}_2i,  \quad\;\, 
                 &+\dot{c}_6\jl+\dot{c}_3j, \quad\;\,  &+\dot{c}_5\kl+\dot{c}_4k \\
  \dot{b}_1-\dot{b}_8l, \quad\;\, &-\dot{b}_7\il-\dot{b}_2i, \quad\;\,
                 &-\dot{b}_6\jl-\dot{b}_3j, \quad\;\, &-\dot{b}_5\kl-\dot{b}_4k
  	   \end{array} \;\! \right)  \nonumber \\
	\!\!  & \!\!\! = \!\!\! & \!\!
	 \left(  \!\! \begin{array}{cccc} 
  -\frac{3}{2}c_8 + \frac{3}{2}c_1l, & +\fh c_2\il- \fh c_7 i, & +\fh c_3 \jl - \fh 
c_6 j, & +\fh c_4 \kl - \fh c_5 k \\
  \frac{3}{2}b_8 + \frac{3}{2}b_1l, & -\fh b_2\il+ \fh b_7 i, & -\fh b_3 \jl + \fh b_6 
j,  & -\fh b_4 \kl + \fh b_5 k
  	   \end{array} \!\! \right) \quad\;\;   \label{sloncb}    \\ & & \nonumber \\
 \!\!\! \mbox{with}\quad \!\! \lbrack \dot{\theta} \rbrack 
    \;\;\!\! & \!\!\! = \!\!\! & \;\!\!
  \left(\quad +\frac{3}{2}\;\lbrack l\theta_l \rbrack,
        \qquad\quad -\frac{1}{2}\;\lbrack l\theta_i \rbrack,    
        \qquad\quad -\frac{1}{2}\;\lbrack l\theta_j \rbrack,
		\qquad\quad -\frac{1}{2}\;\lbrack l\theta_k \rbrack \quad \right)  \qquad   
  \label{slonthet}
\end{eqnarray}
  which may be compared with the su(3)$_c$ action on $\theta$ in 
equation~\ref{aionthet} and table~\ref{suttan}.
   Here the two components of $c\inn \ooo$ \textit{within} each of the four Weyl 
spinors are mixed, and similarly for the corresponding pair of $\bar{b}\inn \ooo$ 
components, with no mixing of components \textit{between} different spinors. This is 
consistent with the nature of the electromagnetic interaction which does not transform 
between different fermion types.

 A further observation from equation~\ref{slonthet} regards the factor of 
$\frac{3}{2}$ found for the $\theta_l$ spinor in contrast to the factors of $\fh$ 
aligned with the three remaining spinors $\theta_i, \theta_j, \theta_k$. Hence, with 
$\dot{S}_{l}^{1}$ provisionally associated with electromagnetism and by comparison 
with equation~\ref{suthth}, the apparent  `electromagnetic charge' assigned to the 
leptonic sector is \textit{three times} larger than that assigned to the quark sector. 
Associating $\theta_i, \theta_j, \theta_k$ with the three colour states of a $d$-quark 
this observation in principle accounts for the `fractional charge' of magnitude 
$\frac{1}{3}$ as theoretically ascribed and empirically confirmed for $d$-quark states 
relative to the electron charge. Based on this observation we introduce the notation:
\begin{equation}
  \label{sbardot}
      \Sbard^a_l = \frac{2}{3}\dot{S}^a_l
\end{equation}
    (for $a=1,2,3$) such that the above charge values $\frac{3}{2}$ and $\frac{1}{2}$ 
are normalised to $1$ and $\frac{1}{3}$ under $\Sbard^1_l$, representing the generator 
of $\uo_Q$,
for ease of comparison with the Standard Model convention for which the electron 
charge is $-1$. The `bar' through $\Sbard^1_l$ is a mnemonic symbol for this 
normalisation of fractional charges relative to the $e^-$ charge. (The corresponding 
normalisation for components of the \textit{group} action $\Sbar^1_l$, which is not 
needed here, would need to take into account the \textit{nested} composition of 
equation~\ref{srotn}. This group normalisation would hence be different for the single 
action of $\ssl^1_l$ of equation~\ref{sqdiag}).

Hence the subgroup in equation~\ref{esisubss} may be augmented to: 
\begin{equation}
  \label{esisubssu}
    \sltc^1 \times \suth_c \times \uo_Q \subset \esi
\end{equation}
  with the internal group $\suth_c \times \uo_Q$ generated by $\{\dot{A}_q, \dot{G}_l, 
\Sbard^1_l \} \inn \staba$. The action of this larger internal symmetry on the four 
$\sltc^1$ spinors also augments equation~\ref{suthth} as:
\begin{eqnarray}
    \theta  &  = & (\, \theta_l, \quad
	  \underbrace{ \theta_i, \quad \theta_j, \quad \theta_k} \, )  \nonumber \\
  \suth_c & : &  \;\;  \mathbf{1} \qquad \qquad  \mathbf{3} \label{suthuoth} \\
    \uo_Q   & : &   +1 \;\; -\!{\mbox{\small{$\frac{1}{3}$}}}
	     \;\; -\!{\mbox{\small{$\frac{1}{3}$}}}  \;\; 
-\!{\mbox{\small{$\frac{1}{3}$}}}
	  \nonumber
\end{eqnarray}

  With the generator $\Sbard^1_l$ hence associated with electromagnetic charge it is 
instructive to consider this action on the full set of $\htho$ components. From 
table~\ref{ltrota} the diagonal components of $\Sbard^1_l$ are trivial, with $\dot{p} 
= \dot{m} = \dot{n} = 0$, while action on the remaining components $a,b,c \inn \ooo$, 
via equation~\ref{sbardot}, may be summarised as:
\begin{equation}
  \label{sbcharge}
   \Sbard_l^1 \; = \;
   \left( \begin{array}{c} \dot{a} \\ \dot{b} \\ \dot{c} \end{array}
              \right) \; = \; 
			  \left( \!\!\! \begin{array}{rcr}
			   0\,l\,a_{1,l} &\!\! + \!\!& \frac{2}{3}\,l\,a(6)  \\
              -1\,l\,b_{1,l} &\!\! - \!\!& \frac{1}{3}\,l\,b(6)  \\
	          +1\,l\,c_{1,l} &\!\! - \!\!& \frac{1}{3}\,l\,c(6)
			  \end{array} \! \right) 
\end{equation}
	where $a_{1,l} \equiv (a_1 + a_8l)$ and $a(6) \equiv 
	(a_7\il+a_2i+a_6\jl+a_3j+a_5\kl+a_4k)$, with similar expressions for $b$ and $c$, 
following the component order of the spinors in equation~\ref{thcth234}.
 The same definition of $a(6)$ is implied in equation~\ref{hvvv}.
 By comparison with the above discussion leading to equation~\ref{suthuoth} the 
expression for $\Sbard_l^1$ in equation~\ref{sbcharge} incorporates `charges' of 0 and 
$\frac{2}{3}$ for the $\dot{a}$ components, that is we have:
\begin{eqnarray}
    a  &  = & (\, a_{1,l}, \;
	  \underbrace{ a_{\il,i}, \; a_{\jl,j}, \; a_{\kl,k} } \, )  \nonumber \\
  \suth_c & : &  \;\;\;  \mathbf{1} \qquad \qquad \:\! \mathbf{3} \label{suthuoa} \\
    \uo_Q   & : &  \;\;\; 0 \;\;\, +\!{\mbox{\small{$\frac{2}{3}$}}}
	     \;\;\, +\!{\mbox{\small{$\frac{2}{3}$}}}  \;\;\, 
+\!{\mbox{\small{$\frac{2}{3}$}}}
	  \nonumber
\end{eqnarray}
  where the $\suth_c$ action on $a\inn \ooo$ is identical to that on the octonion 
components of $\theta = \binom{c}{\bar{b}}$ in equation~\ref{suthuoth}.
 While physical lepton states are invariant under SU(3)$_c$ and are hence  associated 
with the Weyl spinor $\theta_l$ in equation~\ref{suthuoth}, the neutrino states are 
also invariant under the $\uo_Q$ of electromagnetism, that is with zero charge, and 
are provisionally associated with the $a_{1,l}$ components in equations~\ref{sbcharge} 
and \ref{suthuoa};
 while a set of $u$-quarks with $\frac{2}{3}$ fractional charges is similarly 
associated with the $a(6)$ components.

  However, unlike $\theta=\binom{c}{\bar{b}}$ the $a\inn \htho$ component does 
\textit{not} correspond to a set of $\sltc^1$ Weyl spinors, as can be seen from 
table~\ref{sltcreps}.
 Further, the `neutrino' components $a_{1,l} = a_1 + a_8l = v^1 + v^2l$ have already 
apparently been accounted for as part of the external vector $\bv_4\inn \TM_4$ on the 
base manifold, as described in equations~\ref{hvvv} and \ref{hinhtho}.
 These features clearly require further investigation.

   While in the Standard Model the $e^-$ lepton charge is $-1$ and the $d$-quark 
charge is $-\frac{1}{3}$, with positive charges for their antimatter counterparts, the 
convention and interpretation of the $\pm$-signs  of equations~\ref{slonthet} and 
\ref{suthuoth} will depend upon the conventions used and the identification of 
particle and antiparticle states as relating to the spacetime dynamics of the theory. 
As for GUT theories in which particle and antiparticle states may coexist within the 
same SU(5) multiplet \cite{GeoGla}, see sections~\ref{ewtatsm} and \ref{dynkin}, the 
apparently opposite charges in equation~\ref{slonthet} may  relate, for example, to 
 a combination of `antimatter' electrons and `matter' $d$-quarks in the components of 
$\htho$ (which may in turn ultimately relate to the nature of the asymmetry between 
matter and antimatter in the universe).

  Within the above caveats, aligned with the charges of 1 and $\frac{1}{3}$ for the 
electron and $d$-quark Weyl spinors  of equation~\ref{suthuoth} the respective $\uo_Q$ 
charges of 
  $0$ and $\frac{2}{3}$ in  equation~\ref{suthuoa} correlate  with
 charges of $\binom{\; 0}{-1}$ for the $\binom{\nu}{e}$ lepton doublet and 
$\binom{+2/3}{-1/3}$ for the $\binom{u}{d}$ quark doublet of the Standard Model. 
 In addition the states associated with each \textit{left-handed} doublet of charges 
interact via the exchange of $W^{\pm}$ gauge bosons in the Standard Model.
 Hence it remains to be understood how interactions within each of these doublets may 
be mediated via an $\sutw_L$ symmetry, and how such $\nu$-lepton and $u$-quark 
components of $a\inn \ooo \subset \htho$ gain a Weyl spinor structure under the 
external $\sltc^1$ action.

  While the empirical charge structure of the Standard Model fermions is in principle 
accounted for by a $\uo_Q$ symmetry associated with the generator $\Sbard^1_l$ of 
equation~\ref{sbcharge}, further elaboration of this theory is required in order to 
further reconstruct the pattern of particle multiplets listed in 
equation~\ref{multsf}. 
 Guided by the Standard Model it will be necessary to understand the origin of weak 
interactions in order to address these details. Hence in the following section we 
investigate the possible identification of an SU(2)$_L$ gauge symmetry within the 
structure of the broken $\esi$ action on $\htho$ in the present theory.


\section{Elements of Electroweak Theory}
\label{ewtfesb}

\subsection{$\sutw$ Transformations and $\suth_s$ Symmetry}
\label{strassy}

  Within the set of 31 internal basis elements in table~\ref{stabset} it is possible 
to identify a number of SU(2) subgroups, for example generated by the three elements 
$\dot{G}_q + 2\dot{S}_{q}^{1}$ with $q = \{i,j,k\}$ or a different triplet of 
imaginary units (excluding $l$) belonging to a common line in figure~\ref{octmult} and 
hence generating a quaternion subalgebra. While independent of $\{\dot{A}_q, 
\dot{G}_l\}$  as a vector space none of these su(2) generator sets is independent of 
the su(3)$_c$ algebra in terms of the Lie bracket (that is with $\lbrack X,Y  \rbrack 
= 0$ for all $X\inn \sutha_c$, $Y\inn \sutwa$), as discussed after 
equation~\ref{wang18}.

    It is an open question whether all possible internal symmetry subgroups should 
have physical significance. In the above case the generator $\dot{G}_q + 
2\dot{S}_{q}^{1}$ for $q = i,j$ and $k$ mixes the components of $\theta_l$ with those 
of $\theta_i, \theta_j$ and $\theta_k$ respectively, hence mixing between `leptons' 
and `quarks', and would apparently correspond to `new physics' with respect to the 
Standard Model.
   However this particular $\sutw$ action does not describe a `fundamental 
representation' on the set four Weyl spinors, as was the case for $\suth_c$ on the 
left-hand side of table~\ref{suttan} or for $\uo_Q$ in equation~\ref{slonthet}.

    In any case here we attempt to identify an SU(2) symmetry which, as for the case 
of the $\Sbard_{l}^{1}$ generator identified for equation~\ref{esisubssu} for an 
internal $\uo_Q$ symmetry, is independent of the internal $\suth_c$. The other four 
internal generators in equation~\ref{wang5} form a trivial algebra with zero Lie 
bracket for all products -- although non-zero commutators are obtained if 
$\dot{S}_{l}^{1}$ is included (with $\lbrack (\dot{R}_{x \pqg z}^{2} - \dot{B}_{t \pqg 
x}^{2}), \dot{S}_{l}^{1} \rbrack = \frac{3}{2}(\dot{R}_{z \pqg l}^{2} + \dot{B}_{t 
\pqg l}^{2})$ for example) but this is still insufficient structure to form an su(2) 
algebra.  It is also the case that none of these four elements commute with $\sltc^1$ 
and in fact none of the $31-9 =22$ remaining elements of $\stab$ in 
table~\ref{stabset} commute with the subgroup $\sltc^1 \times \suth_c$  of 
equation~\ref{esisubss}, as implied in the discussion following equation~~\ref{wang5}.
 Further, it is to be expected from the Dynkin analysis described in 
section~\ref{dynkin} that in fact there is no possibility of identifying an  $\sutw$ 
subgroup of $\esi$   which is independent of \textit{both} an external $\sltc$ and an 
internal $\suth \times \uo$ symmetry group.

 However, although the full internal gauge symmetry group of the Standard Model reads 
$\SML$ there are a number of features of  weak interactions associated with $\sutw_L$, 
as observed in high energy physics experiments and written into the Standard Model, 
which qualitatively differ from the strong and electromagnetic interactions associated 
with $\suth_c$ and $\uo_Q$ respectively.  If an internal SU(2) were to be found at 
this stage, at the level of symmetry groups and their representations, in a similar 
manner as for the internal $\suth_c \times \uo_Q$ 
   in the previous section, it seems unlikely that the kind of distinctive properties 
observed for the weak interactions could arise purely in the dynamics of the full 
theory. The differences in empirical properties between the strong and electromagnetic 
interactions themselves originate largely out of the differences between the 
non-Abelian SU(3) and Abelian U(1) symmetries at the group and representation level, 
with many more interactions possible in the quantum theory for the former case. 
However while the non-Abelian group SU(2) is mathematically intermediate in size 
between SU(3) and U(1) the physically observed features associated with the gauge 
group SU(2) are of a quite different nature, as described in section~\ref{ewtatsm} and 
summarised in the following paragraph.

   Firstly the weak interactions violate parity symmetry, prompting the subscript 
`$L$' for the left-handed character of this chiral $\sutw_L$ gauge theory. 
 Secondly, the Standard Model $\sutw_L$ is closely association with a $\uo_Y$ gauge 
symmetry, with $\uo_Q$ surviving the electroweak symmetry breaking, which is in turn 
associated with the Lorentz scalar Higgs field transforming as an $\sutw_L$ doublet 
and  providing the mechanism by which  three gauge bosons, the $W^{\pm}$ and $Z^0$, 
gain a non-zero mass.  Thirdly, the weak interactions mix particle states from the 
three distinct generations of fermions, as described by the CKM matrix.

 Here we initially focus upon the simple fact that weak $\sutw_L$ transformations act 
on fermion doublets of the form $\binom{\nu}{e}$ and $\binom{u}{d}$, which have been 
associated with the $\binom{a}{\theta}$ components of $\htho$ for the present theory. 
This was described at the end of the previous section where it was noted that the 
$\uo_Q$ electromagnetic charges associated with the  $\Sbard^1_l$ action on  $\theta = 
\binom{c}{\bar{b}}$ and the $a$ component are respectively aligned with the charges of 
the ($e$-lepton, $d$-quark) and ($\nu$-lepton, $u$-quarks) particle states.

The type 1 $\sltc^1$ action on 
  the four Weyl spinors of equation~\ref{thcth234}
  is  complemented by $\sltc^2$ and $\sltc^3$ transformations of type 2 and 3, all 
involving quaternion algebra composition with $l\inn \ooo$ being the only imaginary 
octonion unit appearing in the transformation matrices.
 Two SU(2)s are  immediately identifiable in terms of the rotation subgroups of the 
type 2 and type 3 Lorentz groups, as denoted by $\sutw^2$, generated by the set 
$\{\dot{R}_{z \pqg l}^{2},\dot{R}_{x \pqg z}^{2}, \dot{R}_{x \pqg l}^{2}\} $, and 
$\sutw^3$, as generated by $\{\dot{R}_{z \pqg l}^{3},\dot{R}_{x \pqg z}^{3}, 
\dot{R}_{x \pqg l}^{3} \}$.
Neither $\sutw^2 \subset \sltc^2$ nor $\sutw^3 \subset \sltc^3$ is independent of 
$\sltc^1$ within the $\esi$ Lie algebra, with for
example $\lbrack \dot{R}_{x \pqg z}^{2}, \dot{R}_{x \pqg z}^{1} \rbrack = \fhs 
\dot{R}_{x \pqg z}^{3} \neq 0$, and neither of them  forms  a subgroup of   
$\stab$, and hence they do \textit{not} appear to form an \textit{internal} symmetry 
by the original definition which led to table~\ref{stabset}.
 However owing to the properties described below in exploring further the structure of 
these transformations the groups $\sutw^{2,3}$ \textit{are} found to be of some 
interest in relation to the structure of electroweak theory.

  By reference to equations~\ref{type1}, \ref{type2} and \ref{type3} of 
section~\ref{esitran}, and with the spinor components $\theta^a = 
\binom{\theta_1}{\theta_2} \inn \ooo^2$ represented by $\binom{c}{\bar{b}}$, 
$\binom{a}{\bar{c}}$ and $\binom{b}{\bar{a}}$ for the type $a=1,2$ and 3 
transformations respectively, the three types of $M^{(a)} \inn \sltc^a$ action, with 
each set generated by equations~\ref{mrotgen} and \ref{mboogen}, are of the form:
\begin{equation}
 \label{mth3}
     \Bigg( \quad M^{(1)} \quad\! \Bigg)
           \Bigg(  \begin{array}{c}  c \\ \bar{b}  \end{array} 	\Bigg), \qquad
	 \Bigg( \quad M^{(2)} \quad\! \Bigg)
           \Bigg(  \begin{array}{c}  a \\ \bar{c}  \end{array} 	\Bigg), \qquad
	 \Bigg( \quad M^{(3)} \quad\! \Bigg)
           \Bigg(  \begin{array}{c}  b \\ \bar{a}  \end{array} 	\Bigg) \qquad
\end{equation}
    with an equivalent right composition $\theta^{\dag}M^{\dag} = (M\theta)^{\dag}$ 
associated with each action above, as seen in the example of the full type 1  
embedding of equation~\ref{mxm3}. In \textit{all} cases however the group action is by 
\textit{left} translation, that is with group representations $R(g_1)R(g_2) = 
R(g_1g_2)$ as discussed in section~\ref{oaags} after equation~\ref{quatrot}, and 
involves  elements of the non-commutative quaternion algebra.

    The type 1 action of $\sltc^1$  decomposes the space $\theta^1 = 
\binom{c}{\bar{b}} \inn \ooo^2$ into the four Weyl spinors of equation~\ref{thcth234}. 
The transformations ${\sltc^{2,3}}$ of type 2 and 3, with complementary transformation 
matrices also based on the units $\{1,l\}$, similarly respect the octonion 
decomposition aligned to the four base unit sets: 
\begin{equation}
   \{1,l\}, \quad \{\il,i\}, \quad \{\jl,j\}, \quad \{\kl,k\}
   \label{lijksets}
\end{equation}
 based on the same quarternion subalgebras, now for all three of $a,b,c\inn \ooo$. 
  Hence  the subgroups $\sutw^{2,3} \subset \esi$ describe transformations between the  
components of equation~\ref{suthuoth} and those of equation~\ref{suthuoa} respecting 
the alignment of the four component pieces, and hence acting independently on the 
corresponding doublets of leptonic and quark states as appropriate for weak 
interactions. 
 With respect to the embedding of $a,b,c \inn \ooo$ as components of $\htho$ in 
equation~\ref{htho}, the spinor representation mixing actions of $\sltc^{1,2,3}$  can 
also be displayed graphically as: 
\begin{equation}
  \left(
  \setlength{\unitlength}{10pt}
	 \begin{picture}(9.6,5.6)(0,0)
	     \put(4.6,4.5){$\bar{a}$}
		 \put(9.1,4.5){$c$}
		 \put(0.1,0){$a$}
		 \put(9.1,0){$\bar{b}$}
		 \put(0.1,-4.5){$\bar{c}$}
		 \put(4.6,-4.5){$b$} 
	    \thicklines 
		 \put(5.6,4.8){\vector(-1,0){0.1}}
		 \put(8.7,4.8){\vector(1,0){0.1}}
		  \multiput(5.9,4.8)(1,0){3}{\line(1,0){0.5}}
		 \put(1.1,0.3){\vector(-1,0){0.1}}
		 \put(8.7,0.3){\vector(1,0){0.1}}
	      \multiput(1.4,0.25)(0.25,0){28}{$.$}
		 \put(1.1,-4.2){\vector(-1,0){0.1}}
		 \put(4.3,-4.2){\vector(1,0){0.1}}
		  \put(1.1,-4.2){\line(1,0){3.2}}		
		 \put(9.4,4.1){\vector(0,1){0.1}}	 
		 \put(9.4,1.4){\vector(0,-1){0.1}}
		  \put(9.4,1.4){\line(0,1){2.7}}
		 \put(0.4,-0.4){\vector(0,1){0.1}}
		 \put(0.4,-3.5){\vector(0,-1){0.1}}
		  \multiput(0.4,-3.3)(0,1){3}{\line(0,1){0.5}}
		 \put(4.9,4.1){\vector(0,1){0.1}}
		 \put(4.9,-3.4){\vector(0,-1){0.1}}	
		  \multiput(4.75,-3.0)(0,0.25){28}{$.$}
	 \end{picture}
   \right)
    \qquad \mbox{with} \qquad
	 \begin{array}{ll}
	    \setlength{\unitlength}{10pt}
	    \begin{picture}(3,1) 
		 \thicklines    
		 \put(0,0.5){\line(1,0){3}}
		 \put(0,0.5){\vector(-1,0){0.1}}
		 \put(3,0.5){\vector(1,0){0.1}}
	    \end{picture}
	   &  \sltc^1   \\
	    \setlength{\unitlength}{10pt}
	    \begin{picture}(3,1) 
		 \thicklines    
		 \multiput(0.25,0.5)(1,0){3}{\line(1,0){0.5}}
		 \put(0,0.5){\vector(-1,0){0.1}}
		 \put(3,0.5){\vector(1,0){0.1}}
	    \end{picture}
	   &  \sltc^2   \\ 
	   \setlength{\unitlength}{10pt}
	    \begin{picture}(3,1) 
		 \thicklines    
		 \multiput(0.35,0.5)(0.25,0){9}{$.$}
		 \put(0,0.5){\vector(-1,0){0.1}}
		 \put(3,0.5){\vector(1,0){0.1}}
	    \end{picture}
	   &  \sltc^3  
	     \end{array}
		  \label{abcmix}
\end{equation}

 This again shows how the $\binom{c}{\bar{b}}$ spinor components under $\sltc^1$ are 
replaced by $\binom{a}{\bar{c}}$ and $\binom{b}{\bar{a}}$ spinors under $\sltc^2$ and 
$\sltc^3$ respectively, depending on the alignment of the 
$\theta^a=\binom{\theta_1}{\theta_2}$ components in 
equations~\ref{type1}--\ref{type3}. 
 It is the observation that the $\sltc^{2,3}$ actions relate the $\theta^1 = 
\binom{c}{\bar{b}} \inn \ooo^2$ components with the $a\inn \ooo$ component in 
equations~\ref{mth3} and \ref{abcmix}, while respecting the four-way octonion 
decomposition of equation~\ref{lijksets}, that suggests that these transformations 
might be closely related to the weak interactions.

  In section~\ref{extsym} the Weyl spinors $\theta_i, \theta_j, \theta_k$ were 
identified alongside $\theta_l$ in equation~\ref{thcth234}  originating from the 
\textit{one-sided} action of $\sltc^1 \subset \sltwoh^1 \subset \hhh (2)$ on 
$\binom{c}{\bar{b}}_{\hhh} \inn \hhh^2$. The quark spinors $\theta_i, \theta_j, 
\theta_k$ are formed out of subspaces of $\hhh^2$ with quarternion base units 
$\{\il,i\},\{\jl,j\},\{\kl,k\} \inn \ooo$  respectively,  with the actions on these 
objects by  matrices composed  of the base units $\{1,l\}$, completing the 3 sets of 
$\hhh$ subalgebras of the octonions, involving in particular 
 the quaternionic left multiplication by $l$ as demonstrated in 
equation~\ref{leftact}. 
 Although the full set of $\hhh(2)$ matrix actions are not involved this asymmetric 
one-sided action is apparently  incomplete in terms of the set of possible actions of 
the non-commutative quaternion algebra on these components.

  This observation might in principle  relate to a possible mechanism for the origin 
of chirality in $\sutw$ interactions
in the Standard Model. This situation can be contrasted with
 left-right symmetric gauge theories with the internal symmetry group $\sutw_L \times 
\sutw_R \times \uo$  formulated in terms of fields defined over the quaternion algebra 
(see for example \cite{Mori} and the references therein). A mechanism is then required 
through which the symmetry in these parity conserving models is broken to $\sutw_L 
\times \uo$ to match the observed parity violating phenomena of weak interactions.

  Here since the set of  \textit{external} Lorentz transformations of $\sltc^1$ act 
asymmetrically on the \textit{left} on $\theta^1 \inn \ooo^2$ and on the subspaces of  
Weyl spinors $\theta_i, \theta_j, \theta_k$ we may expect to identify a set of  
actions on these spinor components algebraically composed from the \textit{right}, 
which have a complementary effect owing to the non-commuting property of the $\hhh$ 
algebra,   potentially forming a distinct \textit{internal} symmetry, at least with 
regards to the quark states represented by these three Weyl spinors.
  Similarly, in the present context, for actions involving  multiplications by 
elements  belonging to $\sltwoh^a \subset \hhh(2)$ some \textit{chiral} behaviour 
might be expected to arise in this theory as the type $a=2,3$ actions complement the 
symmetry breaking action of the external Lorentz transformations $\sltc^1 \subset 
\sltwoh^1$.
 Further, although only one fermion generation has been considered explicitly, the 
structure of equation~\ref{abcmix} is suggestive in terms of the need to ultimately 
account for the CKM mixing between three generations of fermions.

 However while these possibilities provided some of the initial motivation for 
studying the actions of the groups $\sutw^{2,3}$ the mechanism for the above physical 
phenomena will require further developments.  
  The source of parity violation in the present theory will be described in 
section~\ref{secesef}, having explicitly constructed both left and right-handed Weyl 
spinors by extending the form of temporal flow beyond the action of $\esi$ on $\htho$. 
As will be described in section~\ref{sosmfi} a further expansion to a yet 
higher-dimensional flow of time may be required in order to account for three 
generations of fermions and the phenomena of CKM mixing.

   Here the main motivation for studying the $\sutw^{2,3} \subset \esi$ subgroups is 
the structure of the action on the doublet components of $\htho$ as described for 
equations~\ref{mth3}--\ref{abcmix} above in relation  to the weak interaction 
transformations for doublets of fermions in the Standard Model.
 In this subsection we hence further explore this group structure before focusing on
  a pattern of symmetry breaking that closely parallels the properties of electroweak 
symmetry breaking in the remainder of this section. In particular the subgroup 
$\suth_c \times \sutw^2
 \times \uo^2 \subset \esi$, provisionally considered as an `internal symmetry'
 (where $\uo^2$ is the type 2 equivalent of $\uo^1 = \uo_Q$ identified in the previous 
section),
  is analogous to the Standard Model gauge symmetry $\SML$; with the impingement of 
the action $\sutw^2
 \times \uo^2$ on the external spacetime components of $\htwc \subset \htho$ breaking 
this symmetry down to $\uo_Q$. This will be described in the following subsection and 
constitutes a `mock electroweak theory'. We will then ultimately need to address how 
to combine these structures with the external $\sltc^1$ symmetry, which within the 
$\esi$ structure is \textit{not} independent of the $\sutw^{2,3}$ actions.

  In fact with $\sutw^a \subset \sltc^a$ for $a=1,2,3$ these structures are found 
together with three types of $\uo^a$ action described by $\Sbard^a_l$ for $a=1,2,3$, 
as introduced in equation~\ref{sbardot}, within the full $\esi$ action on the space 
$\htho$. 
The $\suth_c$ action of table~\ref{suttan}, corresponding to the set of eight 
generators $\{\dot{A}_q,\dot{G}_l\}$, not only transforms $a,b,c \inn \ooo \subset 
\htho$ in precisely the same way (table~\ref{ltrota}), acting on the components of 
$a(6)$ as a triplet and $a_{1,l}$ as a singlet (in the notation of 
equation~\ref{sbcharge}), but is also independent of both the $\sltc^{1,2,3}$ and 
$S_l^{1,2,3}$ actions in terms of the $\esi$ algebra Lie bracket. This means that the 
$\sltc^a$ and $S_l^a$ actions may effectively be stripped out and considered  
independently of the SU(3)$_c$ action. This may aid the identification of an internal 
$\sutw_L \times \uo_Y$ symmetry, and its relation with the external Lorentz symmetry 
$\sltc^1$, bearing in mind that the former is expected to be `broken' to the $\uo_Q$ 
symmetry associated with $\Sbard_l^1$.

   The nine generators of the combined type $a=1,2$ and 3 rotations $\sutw^a$ form a 
closed subalgebra of $\esi$, which is  eight dimensional due to the linear dependence 
of the $\dot{R}^a_{x \pqg l}$ generators as displayed in equation~\ref{rsdeprrr}. This 
subalgebra is in fact an $\sutha$, a linearly independent basis for which can be 
described by the eight rotation generators (\cite{Wang} p.128):
\begin{equation}
  \label{rrrrrrrr}
    \sutha_s \, \equiv \,
     \{ \dot{R}^1_{x \pqg l},\,  \dot{R}^2_{x \pqg l},\,  \dot{R}^1_{x \pqg z},\, 
   \dot{R}^2_{x \pqg z},\,  \dot{R}^3_{x \pqg z},\,  \dot{R}^1_{z \pqg l}, \, 
   \dot{R}^2_{z \pqg l},\,  \dot{R}^3_{z \pqg l}  \}  
\end{equation}
  These generate a group denoted $\suth_s$ (where `$s$' denotes the `standard' 
representation   or embedding of this group in $\esi$ \cite{Wang}). 
 As implied above  
within $\esi$ the subgroup $\suth_s$ is independent of the colour subgroup $\suth_c$, 
as generated by the eight elements of  table~\ref{sutwang}, with the Lie bracket 
composition of any element of equation~\ref{rrrrrrrr} with any element of 
$\{\dot{A}_q, \dot{G}_l\}$ being zero.
 The generators of $\suth_c$ are explicitly  `type independent', in that there are no 
type labels on any of the eight generators $\{\dot{A}_q, \dot{G}_l\}$ (\cite{Wang}  
p.128), none of which distinguish between the three types. The subgroup $\suth_s$ is 
also `type independent', in that all three types play an equivalent role, however the 
individual generators do carry type labels as for example in equation~\ref{rrrrrrrr}.

 The group product $\suth_s \times \suth_c \subset \esi$ is a rank-4 subgroup of the 
complete rank-6 symmetry group $\esi$. In fact $\suth_s \subset \slthc_s$ where 
$\slthc_s$ is the 16-dimensional  rank-4 group generated by the type $1,2$ and $3$ 
rotations of equation~\ref{rrrrrrrr} together with the a linearly independent set of 
the   
type $1,2$ and $3$ boosts also based on the $\{1,l\}$ complex subspace. Taking into 
account equation~\ref{bbbdep} we have (\cite{Wang} p.128):
\begin{equation}
     \slthca_s \, \equiv \, \sutha_s \, \cup \,
	  \{ \dot{B}^1_{t \pqg z},\,  \dot{B}^2_{t \pqg z},\,  \dot{B}^1_{t \pqg x},\, 
   \dot{B}^2_{t \pqg x},\,  \dot{B}^3_{t \pqg x},\,  \dot{B}^1_{t \pqg l}, \, 
   \dot{B}^2_{t \pqg l},\,  \dot{B}^3_{t \pqg l}  \}   \label{slthcas}
\end{equation}
  In fact $\slthca_s$ is the closed subalgebra formed collectively out of the three 
types of Lorentz generators $\sltca^a$ for $a=1,2,3$, with group actions as pictured 
in equation~\ref{abcmix}, which also act on the complex subspace $\hthc \subset \htho$ 
formed with base units $\{1,l\}$ with for example the type 1 action of 
equation~\ref{vinhinc}.
   
   At the level of complex Lie algebras $L_{\ccc}$ we have the semi-simple 
decomposition $\slthca \equiv \sutha \times \sutha$, and hence the  rank-6 subgroup 
obtained for this real form of $\esi$:
\begin{equation}
  \slthc_s \times \suth_c \subset \esi
\end{equation}
  is closely related to an  $\suth \times \suth \times \suth \subset \esi$ 
decomposition, which may be readily obtained by the analysis described in 
section~\ref{dynkin} via the extension of the Dynkin diagram for the complex $\esi$ 
Lie algebra
 of figure~\ref{dynkine}(a).
In the present theory it is the $\sltc^1 \subset \slthc_s \subset \esi$ Lorentz 
symmetry of external spacetime that breaks the full $\esi$ symmetry.

 In fact $\esi$ also contains the following rank-6 subgroup 
 (listed as one of a number of possible decompositions from a mathematical point of 
view in \cite{Wang} p.187) which augments equation~\ref{esisubssu}:
\begin{equation}
  \label{uobtzbtz}
  \sltc^1 \times \uo_Q \times \doo_B
 \;  \times  \; \suth_c \subset \esi
\end{equation}
  with $\slthc_s$ broken to $\sltc^1 \times \uo_Q \times \doo_B$, and where $\uo_Q$ is 
generated by 
 $\Sbard^1_l = \frac{2}{3}\dot{S}^1_l$  with
$\dot{S}^1_l = (-\dot{R}^1_{x \pqg l} - 2\dot{R}^2_{x \pqg l})$, via 
equations~\ref{sbardot} and \ref{rsdeprrs}, and $\doo_B$ is generated by 
$(\dot{B}^1_{t \pqg z} + 2\dot{B}^2_{t \pqg z})$. This latter generator is presented 
explicitly in equation~\ref{bbdil} together with a possible physical interpretation of 
the $\doo_B$ subgroup in the context of the present theory as described in 
section~\ref{sectveu}.
 Further contained within this symmetry breaking pattern is the choice of $\sutw^1 
\times \uo_Q \subset \suth_s \subset \slthc_s$ with the identification of $\uo_Q = 
\uo^1$, which, in relation to the three \textit{possible} type $a=1,2,3$  embeddings 
$\sutw^a \times \uo^a \subset \suth_s$ will be seen to be closely related to the 
phenomena of electroweak symmetry breaking in the Standard Model.

  Before describing this connection  we note that within the context of the present 
theory in principle it may be possible to mutually constrain the values of the gauge 
field couplings associated with a range of internal subgroups in terms of the
 normalisation of the underlying simple $\esi$ Lie algebra as expressed by the  
Killing form.
 The Killing metric $K_{\alpha\beta} =  
c^{\rho}_{\ph{\rho}\alpha\sigma}c^{\sigma}_{\ph{\sigma}\beta\rho}$ in terms of the 
algebra structure constants $\cstr$ was introduced in the discussion leading to 
equation~\ref{killmet}. Using this expression
 the components $K_{\alpha\beta}$ of the complete  Killing form for the 78 generators 
of $\esi$ in the preferred basis of table~\ref{prefbas} can in principle be determined 
  directly from the rows of the $\esi$ Lie algebra table in $\cite{Wang}$.
  For example for the $\sutha_c \equiv \{\dot{A_i}, \dot{G_l}\}$ generators of the 
colour symmetry described in the previous section we find:
\begin{equation} 
 K(\dot{A_i},\dot{A_i}) = -48, \qquad \quad K(\dot{A_l},\dot{A_l}) = -48, \qquad  
\quad
 K(\dot{G_l},\dot{G_l}) = -144   \nonumber   
\end{equation}
 The Killing metric elements for $\suth_c$ can be compared with those for the $\sutw^2 
\times \uo^2$ generators identified in $\suth_s$, as adopted in a `mock electroweak 
theory', and in principle used to mutually normalise all coupling constants, including 
$\alpha_s = \frac{g_s^2}{4\pi}$  for the strong interactions, under the unifying 
simple group $\esi$, for comparison with the relative couplings adopted for Standard 
Model gauge group $\SML$. With a view towards studying such a mock electroweak theory 
here we analyse the Killing form for the generators relevant to $\sutha_s$, and 
calculate from the rows of the $\esi$ Lie algebra table in $\cite{Wang}$:
\begin{eqnarray}
      K(\dot{R}^1_{x \pqg l},\dot{R}^1_{x \pqg l}) = -24,
	\quad &  K(\dot{R}^1_{x \pqg l},\dot{S}^1_l)  = 0, & \quad
	      K(\dot{S}^1_l,\dot{S}^1_l)  =-72   \nonumber \\
	K(\dot{R}^1_{z \pqg l},\dot{R}^1_{z \pqg l}) = -24,
\quad  & K(\dot{R}^2_{z \pqg l},\dot{R}^2_{z \pqg l}) = -24, & \quad
	K(\dot{R}^2_{x \pqg z},\dot{R}^2_{x \pqg z}) = -24   \nonumber
\end{eqnarray} 
   The negative values are consistent with the nature of the corresponding group 
actions as `rotations', as described in the opening of section~\ref{laofesi}. The 
bilinear property of the Killing form can be used to deduce further elements as 
appropriate for a change of basis within the linearly dependent set of elements 
$\dot{R}^a_{x \pqg l} $ and 
 $\dot{S}^b_l$ ($a,b \inn \{1,2,3\}$) with:
\begin{equation}
 \begin{array}{rcl}
 \dot{R}^2_{x \pqg l} \, = \, -\fhs \dot{R}^1_{x \pqg l} - \fhs \dot{S}^1_l
  & \quad \Rightarrow \quad &
   K(\dot{R}^2_{x \pqg l},\dot{R}^2_{x \pqg l}) = -24   \\
  \dot{S}^2_l \, = \, +\mbox{\small{$\frac{3}{2}$}} \dot{R}^1_{x \pqg l}
                                   - \fhs \dot{S}^1_l
  & \quad \Rightarrow \quad &
   K(\dot{S}^2_l,\dot{S}^2_l)  =-72 
    \end{array}  \nonumber 
\end{equation}
via equations~\ref{rsdeprrs} and \ref{rsdepsrs},
 while:
\begin{equation}
 \begin{array}{rcl}
 \dot{R}^3_{x \pqg l} \, = \, -\fhs \dot{R}^1_{x \pqg l} + \fhs \dot{S}^1_l
  & \quad \Rightarrow \quad &
   K(\dot{R}^3_{x \pqg l},\dot{R}^3_{x \pqg l}) = -24  \\
  \dot{S}^3_l \, = \, -\mbox{\small{$\frac{3}{2}$}} \dot{R}^1_{x \pqg l}
        - \fhs \dot{S}^1_l
  & \quad \Rightarrow \quad &
   K(\dot{S}^3_l,\dot{S}^3_l)  =-72  
    \end{array}  \nonumber 
\end{equation} 
  Hence the three sets of basis elements $\{\dot{R}^a_{x \pqg l},\, 
\frac{1}{\sqrt{3}}\dot{S}^a_l \}$, for either $a= 1,2$ or 3 have a suitably normalised 
Killing form. Further, from the bilinearity of the Killing form it is also found for 
example that:
\begin{equation}
   K(\dot{R}^2_{x \pqg l},\dot{S}^2_l) = 0 \quad \mbox{while} \quad
   K(\dot{R}^2_{x \pqg l},\dot{S}^1_l) = +36  \nonumber 
\end{equation}
indicating that the Killing form is  not diagonal in the latter basis.

 Alternatively, restricting the computation of $K_{\alpha\beta} =  
c^{\rho}_{\ph{\rho}\alpha\sigma}c^{\sigma}_{\ph{\sigma}\beta\rho}$ to the $\suth_s$ 
subalgebra all elements of the corresponding 
  $8 \times 8$ Killing metric $K_8$ for the subalgebra basis of 
equation~\ref{rrrrrrrr} are determined, with for example:
\begin{equation}
   K_8(\dot{R}^1_{x \pqg l},\dot{R}^1_{x \pqg l}) = -3, \qquad \quad
    K_8(\dot{R}^2_{x \pqg l},\dot{R}^2_{x \pqg l}) = -3,  \qquad \quad
   K_8(\dot{R}^1_{x \pqg l},\dot{R}^2_{x \pqg l}) = +\frac{3}{2}    \nonumber
\end{equation}   
where the latter element is the only non-zero off-diagonal entry of the symmetric 
Killing form. Hence we replace the basis element $\dot{R}^2_{x \pqg l}$ in 
equation~\ref{rrrrrrrr} with $\frac{1}{\sqrt{3}}\dot{S}^1_l$ such that the $\suth_s$ 
basis:
\begin{equation}
  \label{rsrrrrrr}
   \sutha_s \, \equiv \,
    \{ \dot{R}^1_{x \pqg l},\,  \frac{1}{\sqrt{3}}\dot{S}^1_l, \, \dot{R}^1_{x \pqg 
z},\, 
   \dot{R}^2_{x \pqg z},\,  \dot{R}^3_{x \pqg z},\,  \dot{R}^1_{z \pqg l}, \, 
   \dot{R}^2_{z \pqg l}, \, \dot{R}^3_{z \pqg l}  \}
\end{equation}
   has normalised Killing metric $K_8 = -3\,(\b1_8)$, where $\b1_8$ is the $8 \times 
8$ unit matrix.  
As for the generators of $\suth_c$ in table~\ref{wangel} a correspondence may be found 
between the $\suth_s$ generators of equation~\ref{rsrrrrrr} and the representation of 
$\sutha$ in terms of Gell-Mann $\lambda$ matrices, as described here in 
table~\ref{suthsl}. 
   
\begin{table}[htbp]
\centering
\begin{tabular}{|ccc|}
 \hline  
$\quad \dot{R}^1_{z \pqg l} \sim \fh i\lambda_1 \quad$ &
 $\quad \dot{R}^1_{x \pqg z} \sim \fh i\lambda_2 \quad$
      & $\quad\; \dot{R}^1_{x \pqg l} \sim \fh i\lambda_3 \quad\;$ \\  & &  \\
$\dot{R}^2_{z \pqg l} \sim -\fh i\lambda_4$ &
 $\dot{R}^2_{x \pqg z} \sim \fh i\lambda_5$ &   \\  & & \\
$\dot{R}^3_{z \pqg l} \sim \fh i\lambda_6$ & 
$\dot{R}^3_{x \pqg z} \sim \fh i\lambda_7$ & 
    $ \frac{1}{\sqrt{3}}\dot{S}^1_l \sim \fh i \lambda_8$ \\
		    \hline 
  \end{tabular}
  \caption{\setb The isomorphism between the su$(3)_s\subset \esi$ Lie algebra basis 
of equation~\ref{rsrrrrrr} and the eight Gell-Mann matrices of table~\ref{gellm}.}
\label{suthsl}
\end{table}

  The choices of basis elements $\{\dot{R}^a_{x \pqg l}, \, 
\frac{1}{\sqrt{3}}\dot{S}^a_l \}$ for type $a= 2$ or 3 in place of $a=1$ correspond to 
two further possible correlates of the basis matrices $\{\lambda_3,\, \lambda_8 \}$ in 
the Gell-Mann representation of $\sutha$. These three possibilities correspond to 
three closely related ways to embed the subgroup $\sutw \times \uo$ in $\suth$. 
  Here we first study $\sutw^2 \subset \suth_s$ and the corresponding set of 
generators $\{ \dot{R}_{z \pqg l}^2, \dot{R}_{x \pqg z}^2,     \dot{R}_{x \pqg 1}^2 
\}$. These are the type 2 versions of the three actions of equation~\ref{mrotgen} 
which, as described in equation~\ref{msigj}, are respectively associated with  the 
three Pauli matrices $\sigma^1 = \binom{0 \;\; 1}{1 \;\; 0}$,    $\sigma^2 = \binom{0 
\; -l}{l \;\;\; 0}$, $\sigma^3 = \binom{1 \;\;\; 0}{0 \; -1}$,  
 within factors of $\pm \frac{l}{2}$.

 In the Standard Model electroweak theory the $\sutwa_L$ Lie algebra-valued connection 
1-form $\bWW(x) = W^{\alpha}(x)\tau^{\alpha}$,  with  $W^{\alpha}(x) = 
W^{\alpha}_{\mu}(x) \mbox{d}x^{\mu}$, $\tau^{\alpha} = \fh \sigma^{\alpha}$ from 
equation~\ref{tauhsig}  and $\alpha=1,2,3$, is parametrised by the three gauge fields 
$W^{\alpha}_{\mu}(x)$. The charged gauge boson fields $W^{\pm}_{\mu}(x)$ are 
associated with complex linear combinations of the $\sutw_L$ generators $\sigma^{\pm} 
= \fh(\sigma^1 \pm i\sigma^2)$ as was described in equations~\ref{wpmww} and 
\ref{spmss}. Guided by this construction based on $\sutw_L$ generators, here in the 
complex algebra for $\sutw^2 \subset \esi$ we define: 
\begin{equation}
  \label{sigcomb}
   \dot{\Sigma}^{(2)\pm} := \, \dot{R}_{z \pqg l}^2 \; \pm \; i\dot{R}_{x \pqg z}^2
\end{equation} 
  Here the imaginary unit $i\inn \ccc$ in the complexification of the $\esi$ Lie 
algebra  commutes with the elements of  $T\htho$, which are based on an independent 
octonion algebra $\ooo$. This is the standard notion of a complexified Lie algebra 
$L_{\ccc} \equiv L_{\rrr} + iL_{\rrr}$, as for example described for  
figure~\ref{LctoR}, applied here to $L_{\rrr}$ as the real $\esi$ Lie algebra  
represented in the space of vector fields in $T\htho$.  

The generator $\Sbard^1_l$ was associated with the internal symmetry $\uo_Q$ and  
electromagnetic charge in the previous section. As a rotation the corresponding group 
transformation  
  $\Sbar^1_l$ takes the form of unitary $3 \times 3$  matrix actions, as described in 
the opening of section~\ref{laofesi}. Hence considering $i\Sbard^1_l$ to be an 
Hermitian generator in the complexified $\esi$ algebra \textit{real} eigenvalues may 
be obtained under the adjoint representation. In particular, reading off the 
corresponding entries in the Lie algebra table in~\cite{Wang} for the complex element 
of equation~\ref{sigcomb} it is found that:
\begin{eqnarray}
    \lbrack  \dot{S}^1_l \, , \, (\dot{R}_{z \pqg l}^2 \, + \, i\dot{R}_{x \pqg z}^2)
	\rbrack & = &  \mbox{\small $\frac{3}{2}$} \dot{R}_{x \pqg z}^2 \; - \; i 
	         \mbox{\small $\frac{3}{2}$}    \dot{R}_{z \pqg l}^2    \; = \;
	-i\mbox{\small $\frac{3}{2}$} 
	   (\dot{R}_{z \pqg l}^2 \, + \, i\dot{R}_{x \pqg z}^2) \label{srrbrac} \\
 \!\!\!\!\!\!\!\!\!\!\!	\mbox{hence} \qquad 
	 \lbrack  i\Sbard^1_l \, , \, (\dot{R}_{z \pqg l}^2 \, + \, i\dot{R}_{x \pqg z}^2)
	\rbrack & = &   +  (\dot{R}_{z \pqg l}^2 \, + \, i\dot{R}_{x \pqg z}^2) \nonumber 
\\
 \!\!\!\!\!\!\!\!\!\!\! \mbox{and} \qquad \qquad \quad \;\;
	 \lbrack  i\Sbard^1_l \, , \,  \dot{\Sigma}^{(2)\pm} 
	\rbrack     & = &   \pm \dot{\Sigma}^{(2)\pm}    \label{isscomm}
\end{eqnarray} 
  with real charge eigenvalues $\pm 1$.
  Hence the generators $\dot{\Sigma}^{(2)\pm}$ of equation~\ref{sigcomb} are 
associated with the \textit{same} magnitude of $\uo_Q$ charge under $\Sbard^1_l$ as 
was found for the electron in the leptonic components $\theta_l \subset \htho$ as 
described in equations~\ref{sloncb}--\ref{suthuoth}. Since such factors of 
$\frac{3}{2}$ as seen in equation~\ref{srrbrac} are relatively sparse in the $\esi$ 
Lie algebra table~\cite{Wang}, with none appearing for example here in 
table~\ref{ecomm8}, this seems to be a non-trivial correspondence of $\Sbard^1_l$ 
charges.
   In the $\sutha_s$ basis of equation~\ref{rsrrrrrr} the generators 
$\dot{\Sigma}^{(2)\pm}$ are in fact two of the eigenvectors of elements of the Cartan 
subalgebra, which in turn has a  basis $\{\dot{R}^1_{x \pqg l}, \, 
\frac{1}{\sqrt{3}}\dot{S}^1_l\}$, under the adjoint representation in the complex 
$\sutha_s$ algebra. Indeed we find also:
\begin{equation}
 \label{reigen}
   \lbrack  i\dot{R}_{x \pqg l}^1 \, , \,  \dot{\Sigma}^{(2)\pm} 
	\rbrack   \,   =  \,   \pm \, \fhs \dot{\Sigma}^{(2)\pm}  
\end{equation}

More generally for a Lie algebra of rank-$n$ the elements of the Cartan subalgebra 
$\{H_i\}$, 
 $i=1\ldots n$, are mutually commuting and any element, or linear combination of 
elements, in $\{H_i\}$ generates a $\uo$ symmetry. In any representation of the Lie 
algebra the eigenvalues, or `weights', of such a $\uo$ generator can be considered as 
`charges'. In the present case the $\uo_Q$ generator $\Sbard^1_l$, which also belongs 
to the $\esi$ Cartan subalgebra as can be seen from equation~\ref{csaset}, is 
associated with electromagnetic charge.

  In the Cartan-Weyl basis of a complex Lie algebra the eigenvectors $E_{\alpha}$ of 
elements of the Cartan subalgebra  $\{H_i\}$  in the adjoint representation have real 
eigenvalues $\alpha_i$:
\begin{eqnarray}
    \lbrack H_i \:\! ,\:\! E_{\alpha} \rbrack & = & \alpha_i \, E_{\alpha}
	   \label{heacomm} \\
	\mbox{with} \quad \lbrack E_{\alpha} \:\! , \:\! E_{-\alpha} \rbrack & = &
	 (K^{ij} \alpha_j) H_i   \label{eaeacomm}
\end{eqnarray}
 (where $K^{ij}$ are components of the Killing metric restricted to the Cartan 
subalgebra). The eigenvalues, or `weights', $\alpha_i$ of the adjoint representation 
are also called `roots', the full set of which under $\{H_i\}$ is central to the 
classification of complex Lie algebras, as alluded to in section~\ref{dynkin} 
alongside figure~\ref{dynkine}. Since the elements of the Lie algebra form the vector 
space $\mbox{span}(H_i, E_{\alpha})$ upon which the adjoint representation acts, the 
dimension of this representation is equal to the dimension of the Lie algebra itself.
 Generally in a given representation $r$ of a Lie algebra on a vector space $V$ with 
eigenvectors $\vert v \rangle$ and weights $\lambda_i$, that is with:
\begin{eqnarray}
   H_i^{(r)} \vert v \rangle  & \!\! = \!\! & 
      \lambda_i  \vert v \rangle  \nonumber \\
   \!\!\!\!\!\!\!\!\!\!\!\!
   \mbox{then:} \quad\!\!  H_i^{(r)} ( E_{\alpha}^{(r)}\vert v \rangle ) 
       & \!\! = \!\! &
    E_{\alpha}^{(r)}  H_i^{(r)} \vert v \rangle \, + \,
	\lbrack   H_i^{(r)} \! , E_{\alpha}^{(r)}\rbrack \vert v \rangle \, = \,
	(\lambda_i + \alpha_i)  ( E_{\alpha}^{(r)}\vert v \rangle )  \label{raislow}
\end{eqnarray}
   using equation~\ref{heacomm}.
  That is, the $E^{(r)}_{\alpha}$ act as `raising' operators (while the 
$E^{(r)}_{-\alpha}$ act as `lowering' operators) on the eigenstates in the 
representation.

  Hence by comparison of equation~\ref{isscomm} with equation~\ref{heacomm} above the 
complex linear combinations $\dot{\Sigma}^{(2)\pm}$ of equation~\ref{sigcomb} as 
eigenvectors
  $\Sbard^1_l$
 under the $78$-dimensional adjoint representation of $\esi$ indeed have charges of 
$\pm 1$ under the same  generator of  $\uo_Q$ which acts on the $e$-lepton and 
$d$-quark states identified in the $\theta^1=\binom{c}{\bar{b}}$ components of the 
$27$-dimensional representation of the $\esi$ symmetry on the space $\htho$. 
  Further, according to equation~\ref{raislow},  the $\dot{\Sigma}^{(2)\pm}$ actions 
are expected to transform states  in the $\htho$ representation with a change of $\pm 
1$ units of the electron charge. Based on the type 2 subgroup $\sutw^2 \subset 
\sltc^2$ these raising and lowering operations are associated with the 
$\theta^2=\binom{a}{\bar{c}}$ components of $\htho$ as shown explicitly in 
equations~\ref{mth3} and \ref{abcmix}.
 In this subsection we have focussed precisely upon this doublet action of the 
$\sutw^2$ symmetry which appears to be 
  closely related to transformations within the lepton $\binom{\nu}{e}$ and quark 
$\binom{u}{d}$ doublets as mediated by the $W^{\pm}$ gauge bosons in the Standard 
Model.

  In the Cartan-Weyl basis generally the Lie bracket $\lbrack E_{\alpha} \:\! ,\:\! 
E_{-\alpha} \rbrack$ describes an element of the 
Cartan subalgebra, as can be seen from equation~\ref{eaeacomm}. From the $\esi$ Lie 
algebra table in $\cite{Wang}$ we find:
\begin{equation}
   \lbrack \dot{\Sigma}^{(2)+} \:\! ,\:\! \dot{\Sigma}^{(2)-} \rbrack \, = \,
   \lbrack (\dot{R}_{z \pqg l}^2 \, + \, i\dot{R}_{x \pqg z}^2),
           (\dot{R}_{z \pqg l}^2 \, - \, i\dot{R}_{x \pqg z}^2) \rbrack
\,	=  \, - i\dot{S}^1_l \, - \, i \dot{R}_{x \pqg l}^1 
\end{equation} 
  which is indeed in the Cartan subalgebra 
   of equation~\ref{csaset}, for the complexified $\esi$ Lie algebra, and also for the 
complex $\sutha_s$ subalgebra. This is consistent with the identification of the 
$\dot{R}_{x \pqg l}^1$ `charges' for $\dot{\Sigma}^{(2)\pm}$ in equation~\ref{reigen}.
  However it the $\Sbar^1_l$ action that has been associated with the internal 
symmetry $\uo_Q$ in the previous section and in turn the eigenvalues of 
   $\Sbard^1_l$ associated with physical
electromagnetic charges. It is the latter charges of $\pm 1$ for the states 
$\dot{\Sigma}^{(2)\pm}$ which will be provisionally associated with the 
$W^{\pm}_{\mu}(x)$ charged gauge fields in the mock electroweak theory.

In quantum field theory the creation and annihilation operators associated with 
\textit{real} fields do not describe charged particles, rather conserved charges are 
associated with complex fields, or complex linear combinations of real fields,
  as will be described in section~\ref{tranamp}.
  A complex scalar field $\mcY(x)$ has charge $q$ under a $\uo$ symmetry if it 
transforms as
 $\mcY \to  e^{iq\alpha}\mcY$, with $\alpha\inn \rrr$ and $e^{iq\alpha} \inn \uo$, 
with $q$ also labelling the irreducible representation of $\uo$. The derivative of 
this transformation at $\alpha = 0$ can be written as $\pal \mcY / \pal \alpha = 
\dot{\mcY} = +q(i\mcY)$. This has the same form as equation~\ref{slonthet}, which via 
equation~\ref{sbardot} implies the $\uo_Q$ action $\Sbard^1_l$ on the field components 
$\theta_l(x)$ reads 
  $\lbrack \dot{\theta}_l \rbrack = +1 \lbrack l\theta_l \rbrack$, with the complex 
imaginary unit $l$ and charge represented by the real eigenvalue $q=+1$.

As for the electron and $d$-quarks charges identified in the components of $\htho$ it 
remains to be seen how the charges for gauge bosons derived from generators such as  
$\dot{\Sigma}^{(2)\pm}$
   relate to the likelihood of physical processes such as observed in high energy 
physics experiments for the present theory. This will be discussed in 
section~\ref{secdopp} in comparison with standard QFT for which the charges are placed 
by hand into Lagrangian terms, leading to  calculations of transition amplitudes and 
cross-sections. 
 The phenomenon of running coupling, as described in section~\ref{seraps}, will 
ultimately also need to be considered for any comparison between theoretical couplings 
derived from a normalised Killing form for a simple Lie algebra and the couplings 
measured empirically in the laboratory.
  As well as accounting for quantisation a full dynamical theory 
  will also be required, incorporating for example self-interactions for non-Abelian 
gauge fields, as explored in relation to Kaluza-Klein theories here in 
chapters~\ref{kktheory} and \ref{chaputtf}.


\subsection{$\sutw^2 \times \uo^2$ Mixing Angle}
\label{submangle}

  The four type 1 actions $\{\dot{R}_{z \pqg l}^1, \dot{R}_{x \pqg z}^1,
   \dot{R}_{x \pqg l}^1  \},\Sbard^1_l$ generate the group $\sutw^1 \times \uo_Q$. 
Here
   $\sutw^1$, generated by $\{
  \dot{R}_{z \pqg l}^1, \dot{R}_{x \pqg z}^1, \dot{R}_{x \pqg l}^1  \}$,
  is the rotation subgroup of the external Lorentz transformations, as studied in 
section~\ref{extsym}, which commutes with the internal symmetry $\uo_Q$, underlying 
Maxwell's electromagnetic field, generated by $\Sbard^1_l$ as identified in 
section~\ref{intsym}.
  	 For the case of the corresponding set of four type 2 actions  $\{
  \dot{R}_{z \pqg l}^2, \dot{R}_{x \pqg z}^2, \dot{R}_{x \pqg l}^2  \}, \Sbard^2_l,$ a 
similar structure can be identified for $\sutw^2 \times \uo^2$.
In a similar way that $\Sbard^1_l$ commutes with $\sutwa^1$, and indeed with the 
Lorentz group $\sltca^1$, it is also the case that $\Sbard^2_l$ commutes with 
$\sutwa^2$ and hence
 with  $\dot{\Sigma}^{(2)\pm}$ of equation~\ref{sigcomb}:
\begin{equation}
   \lbrack  \Sbard^2_l \, , \,  \dot{\Sigma}^{(2)\pm}	\rbrack = 0
\end{equation}   
   This commutator is consistent with those in equations~\ref{isscomm} and 
\ref{reigen} 
   given the linear dependence obtained from equations~\ref{rsdepsrs} and 
\ref{sbardot}:
\begin{equation}
  \label{srslincom}
   \Sbard^2_l \, =\, \dot{R}_{x \pqg l}^1\, -\, \fhs \, \Sbard^1_l
\end{equation}   
   While the generator $\Sbard^1_l$ is  associated with electric charge $Q$ the 
generator $\dot{R}_{x \pqg l}^2$ is associated with $T^3$, the third component of 
$\sutw^2$.  The linear dependencies in the $\esi$ Lie algebra of 
equations~\ref{rsdeprrs}  and \ref{rsdepsrs} also imply the relation:
\begin{equation}
  \label{srslincom2}
   - \, \Sbard^1_l = \dot{R}_{x \pqg l}^2 + \fhs \Sbard^2_l 
\end{equation}
   which  is closely reminiscent of the relation:
\begin{equation}
  \label{qethy2}
    Q \, = \,  T^3 + \fhs Y   
\end{equation}
	of equation~\ref{qethy}, within the choice of sign conventions.
	This suggests associating $\fh \Sbard^2_l$ with $\fh Y$ as a candidate for the 
generator of the
  \textit{hypercharge}  symmetry $\uo_Y \sim \uo^2$ which commutes with $\sutw^2$, as 
generated by 
 $\{\dot{R}_{z \pqg l}^2, \dot{R}_{x \pqg z}^2, \dot{R}_{x \pqg l}^2  \}$, and as 
provisionally associated with $\sutw_L$ for a mock electroweak theory in the previous 
subsection.
  The generator $\Sbard^2_l$ may also be expressed as the linear combination of type 1 
elements  in equation~\ref{srslincom}, which lies in the Cartan subalgebra  of the 
$\esi$ Lie algebra. Hence the `weights' of $\Sbard^2_l$ may indeed be considered as 
`charges', which are termed hypercharges for the corresponding $\uo_Y$ symmetry.

  More generally opening up consideration of the three $\sltc^a$ actions in the 
previous subsection also motivates an examination of the $\uo$ charge structure 
associated with $\Sbard_l^a$ for all \textit{three} types $a=1,2,3$.
 To understand the relationships between these charges all three generators 
$\Sbard^a_l$, for $a=1,2,3$, from table~\ref{ltrota} with a factor of $\times 
\frac{2}{3}$ from equation~\ref{sbardot},
 are explicitly written out in terms of $T\htho$ components in equation~\ref{sssrdot}.
 Each entry of the form $(x,y)$  represents the factors of $l$ which multiply the 
components of $\htho$ algebraically from the left side -- where $x$ is the `leptonic 
part', that is on the real and $l$ components, while $y$ is the `quark part', that is 
on the remaining six imaginary units of each $a,b,c \inn \ooo$. The components for 
$\Sbard^1_l$ in equation~\ref{sssrdot} contain the same information as 
equation~\ref{sbcharge} rearranged into the $3 \times 3$ matrix of $T\htho$.
 It can be seen here that $\Sbard_l^1+\Sbard_l^2+\Sbard_l^3 = 0$, for each component 
$a,b$ and $c$, consistent with equation~\ref{agsdeps}.
 Also shown are the corresponding components of $\dot{R}_{x \pqg l}^2$ as obtained 
from table~\ref{lbrota}, which can be seen to be consistent with 
equation~\ref{srslincom2}.
\begin{equation}
 \hspace*{-18pt}
  \label{sssrdot}
 \begin{array}{ccc}
        \Sbard_l^1   &   \Sbard_l^2  &   \Sbard_l^3  \\
	  \!\!\!\!\!		  \left( \!\!\!\! \begin{array}{ccc}
			               & \!\!  (0,+\ftts ) \!\! & (1,-\fots )  \\
               (0,+\ftts ) &                        & (1,-\fots ) \\
	          (-1,-\fots ) & \!\! (-1,-\fots ) \!\! &  
			  \end{array} \!\!\!\! \right) \!
			  &  \!\!
			  \left( \!\!\!\! \begin{array}{rcr}
	                       &\!\!  (-1,-\fots ) \!\! & (-1,-\fots ) \\
			   (1,-\fots ) &                        &  (0,+\ftts )  \\
               (1,-\fots ) &  \!\! (0,+\ftts ) \!\! &     
			  \end{array} \!\!\!\! \right) \!
			  & \!\!
			  \left( \!\!\!\! \begin{array}{rcr}
                           &  \!\! (1,-\fots ) \!\! &  (0,+\ftts )  \\
	          (-1,-\fots ) &                        & (-1,-\fots )   \\
			   (0,+\ftts ) & \!\! (1,-\fots )  \!\! &
			  \end{array} \!\!\!\! \right)
              \\
	   	\dot{R}_{x \pqg l}^2  &   &    \\
\!\!\!\!			\left( \!\!\!\! \begin{array}{ccc}
			               & \!\! (+\fhs,-\fhs) \!\! & (-\fhs,+\fhs)  \\
             (-\fhs,-\fhs) &                         & (-1,0 ) \\
	         (+\fhs,+\fhs) & \!\! (+1, 0 ) \!\!      &  
			  \end{array} \!\!\!\! \right) \!\!\!\!\!\!\!\!
			  &   \qquad \qquad  \mbox{all as} \quad 
		  l	 \left( \!\! \begin{array}{rcr}
                           & \;\;\; \bar{a} \;\;\; &  c  \\
	                     a &          & \bar{b}   \\
			       \bar{c} & \;\;\; b  \;\;\;     &
			  \end{array} \!\! \right)  \!\!\!\!\!\!\!\!
			    & \!\!\!\!\!  \inn \; T\htho  \qquad \qquad \qquad
\end{array}
\end{equation}

Hence the $\fhs\Sbard_l^2$ `hypercharge values' of $(-\fhs, 
-\mbox{\small{$\frac{1}{6}$}})$ on the $\bar{a}$ and $c$ components 
 in equation~\ref{sssrdot}
match the Standard Model hypercharge values of $\frac{Y}{2}(l_L) = -\fhs$ and 
$\frac{Y}{2}(q_L) = +\mbox{\small{$\frac{1}{6}$}}$ for the left-handed doublets of 
leptons and quarks  respectively of equation~\ref{multsf} (up to a sign convention, 
which again  will ultimately depend on the definition of particle and antiparticle 
states in spacetime).
 As can be seen in equation~\ref{abcmix} and described in the previous subsection 
these components $(\bar{a}\; c)$ are also linked by the $\sutw^2 \subset \sltc^2$ 
actions and corresponding $\dot{\Sigma}^{(2)\pm}$ operators provisionally associated 
with the $W^{\pm}_{\mu}(x)$ charged gauge fields. Although some of these observations 
are naturally mutually correlated,
  the $(1,2,-\frac{1}{2})_L$ and $(3,2,\frac{1}{6})_L$ pieces of equation~\ref{multsf} 
are hence closely associated respectively with the $\binom{a_{1,l}}{\theta_l}$ and 
$\binom{a(6)}{\theta_{i,j,k}}$
 components of $\htho$  in equations~\ref{suthuoth} and \ref{suthuoa}.

 While right-handed fermion states remain to be  identified, the hypercharges of the 
right-handed fermion singlets in equation~\ref{multsf} are also closely correlated 
with the $\Sbard_l^a$ charges in equation~\ref{sssrdot}. This is expected since 
$Q=\frac{Y}{2}$ for these cases and the electric charge $Q$ is well described by 
$\Sbard_l^1$.  Also in the top row of $\Sbard_l^3$ the values $(1, -\fots)$, $(0, 
+\ftts)$ have the same magnitude as the $\frac{Y}{2}$ values for the right-handed 
singlets $e_R$, $d_R$, $\nu_R$ and $u_R$
 respectively, although these components do \textit{not} correspond to the correct 
electromagnetic charges $Q$ under $\Sbard_l^1$ for those respective fermion states.
  However these observations do suggest opening up consideration of the (correlated) 
charges for all three $\Sbard_l^a$ generators. 
   Indeed as well as $\Sbard_l^2$ the generator $\Sbard_l^3$ should also relate to 
hypercharge as $\sutw^3\times \uo^3$ also forms a possible mock $\sutw_L\times \uo_Y$ 
action with the following linear dependence also  found within the $\esi$ algebra:
\begin{equation}
 \Sbard_l^1  =   \; \dot{R}_{x \pqg l}^3 -\fhs \Sbard_l^3 \label{slin2}  
\end{equation}
   This equation is the type 3 version of equation~\ref{srslincom2}.
  Further linear relations include  $\Sbard_l^1  =  \!\!\! -2\dot{R}_{x \pqg l}^2 \, + 
\, \Sbard_l^3$ and $\Sbard_l^1  =    2\dot{R}_{x \pqg l}^3 \, + \, \Sbard_l^2$ which 
combine non-commuting type 2 and 3 actions on the right-hand side.
Such equations of linear dependence, relating the generators $\Sbard_l^a$ and 
$\dot{R}_{x \pqg l}^b$ for $a,b = \{1,2,3\}$, are fixed by the structure of the $\esi$ 
Lie algebra and closely resemble equation~\ref{qethy} which is constructed to relate 
the electric charge $Q$, third component of weak isospin $T^3$ and hypercharge 
$\frac{Y}{2}$ in the Standard Model. 
 However a fuller understanding of this structure in the present theory will require 
the identification of right-handed fermion states, and in particular such states with 
$T^3=0$.
 The origin of both left and right-handed states, together with their mutual relation 
will  be considered explicitly in  section~\ref{secesef}, while in the meantime we 
further consider the structure of the mock electroweak theory within the $\esi$ 
framework.

	 In particular, moving away from the a \textit{static} analysis of the $\esi$ 
symmetry breaking pattern to a more \textit{dynamic} perspective,  we next
	 study the structure of an $\sutw^2 \times \uo^2$ \textit{gauge theory} based on 
the symmetry generators $\{ \dot{R}_{z \pqg l}^2, \dot{R}_{x \pqg z}^2,
  \dot{R}_{x \pqg l}^2, \Sbard^2_l\}$.
  These act on the doublet components of the type 2 spinor $\theta^{2} = 
\binom{a}{\bar{c}}$ in $\htho$. Restricted to the complex subspace $\ccc\subset \ooo$ 
with $\{1,l\}$ basis units the components   $\theta^{2}_l = \binom{a}{\bar{c}}_{\! l}$  
provisionally represents the lepton doublet $\binom{\nu}{e}$.  
Since these components do not correspond to complete $\sltc^1$ Weyl spinors for either 
the neutrino \textit{or}  the electron part this $\sutw^2 \times \uo^2$ symmetry is 
clearly not directly equivalent to the $\sutw_L \times \uo_Y$ symmetry of electroweak 
theory. However 
the components of $\theta^{2}_l$ do transform under the internal symmetry $\suth_c 
\times \uo_Q$  appropriately to represent such a lepton doublet, as described in 
section~\ref{intsym}, and hence the $\sutw^2 \times \uo^2$ symmetry
  serves as a useful intermediate model, considered as a mock electroweak theory.
   The equations of motion for the corresponding field  $\theta^{2}_l(x)$ in spacetime 
$M_4$ will then involve the gauge covariant derivative (essentially as described in 
section~\ref{fibre}):
\begin{equation}
 \label{covdevth}
   D_{\mu}\theta^{2}_l(x) \, = \, \pal_{\mu} \theta^{2}_l(x) \, + \,
     \tilde{g}\, \tilde{W}^{\alpha}_{\mu}(x) \, \dot{R}^{(2)\alpha}(\theta^{2}_l) 
	 \, + \, \tilde{g}' \, \tilde{B}_{\mu}(x) \, \fh \Sbard^2_l(\theta^{2}_l)
\end{equation}
  where $\alpha=1,2,3$ and $\dot{R}^{(2)\alpha} \equiv  \{ \dot{R}_{z \pqg l}^2, 
\dot{R}_{x \pqg z}^2,
  \dot{R}_{x \pqg l}^2 \}$, and for example $\dot{R}_{z \pqg l}^2(\theta^{2}_l)$ 
denotes the $\theta^{2}_l$ components of $\dot{R}_{z \pqg l}^2$. The couplings 
$\tilde{g},\tilde{g}'$ and the gauge fields $\tilde{W}^{\alpha}_{\mu}(x), 
\tilde{B}_{\mu}(x)$
  associated with the $\sutw^2 \times \uo^2$ gauge symmetry are introduced by analogy 
with the Standard Model case in equation~\ref{covdevlep} and hence  similar notation 
is adopted. However it is important to contrast the corresponding gauge coupling terms 
implied in equations~\ref{covdevlep} and  \ref{covdevth} with for example 
respectively: 
\begin{equation}
  \label{dmucomp}
   D_{\mu}l_L \; \sim \; i \, g' \,  
      B_{\mu}(x) \,\frac{Y}{2}(l_L) \, l_L \qquad \mbox{and} \qquad
   D_{\mu}\theta^{2}_l \; \sim \;   \tilde{g}' \, \tilde{B}_{\mu}(x) \, 
   \frac{\Sbard^2_l}{2} (\theta^{2}_l)
\end{equation}
  In the former case, apart from the conventional factor of `$i$' there are four 
factors: the coupling $g'$, the gauge field $B_{\mu}(x)$, the hypercharge generator 
$\frac{Y}{2} (l_L) = -\frac{1}{2}\binom{1 \; 0}{0 \; 1}$ and the lepton doublet $l_L = 
\binom{\nu}{e}_{\! L}$. In the latter case there are only three factors: with the 
coupling $\tilde{g}'$ and gauge field $\tilde{B}_{\mu}(x)$ having a similar role as 
for the first case, while the third part $\Sbart (\theta^{2}_l)$ corresponds to the 
action of the hypercharge generator represented
  \textit{directly} on the  $\theta^{2}_l = \binom{a}{\bar{c}}_{\! l}$ components of 
$\htho$, which is equivalent to the combination $\fhy(l_L) \, l_L$ for the standard 
case.

  In principle  in the second case the coupling $\tilde{g}'$ may be absorbed into the 
gauge field $\tilde{B}_{\mu}(x)$ since we are here dealing with the pure covariant 
derivatives, as originally expressed in equation~\ref{dlvth2} of 
subsection~\ref{mgjoin} for the gauge field $A_{\mu}(x)$ without any coupling 
constant. Adopting couplings such as $\tilde{g}'=1$ is also compatible with the 
construction of a direct relationship between the curvature for the external linear 
connection and that for the internal gauge connection as described in 
section~\ref{reaic} and equation~\ref{gchift} (with a factor such as $\chi = 1$ in 
principle determined by the geometric structure), 
 in comparison with Kaluza-Klein theory.
 A similar observation applies for the coupling $\tilde{g}$ associated with the gauge 
field $\tilde{W}^{\alpha}_{\mu}(x)$ in equation~\ref{covdevth}. Ultimately both 
$\tilde{g}$ and $\tilde{g}'$ will be absorbed into the relevant gauge fields and 
effectively set equal to one.
  
   In turn the `charges' of individual states will 
  depend upon the representation which is already determined directly by the values of 
$\Sbart(\theta^{2}_l)$ in the second expression of equation~\ref{dmucomp}, which are 
  closely analogous to the case   
for the electromagnetic charges obtained from $\Sbard^1_l(\theta^1)$ in 
equations~\ref{slonthet} and \ref{sbardot} and further discussed towards the end of 
the previous subsection. More generally 
 this will require a suitable mutual normalisation of the generators 
$\dot{R}^{(2)\alpha}$ and $\fh \Sbard^2_l$ based on the Killing form of the full 
$\esi$ Lie algebra, as also described in the previous subsection, to relate the 
charges for the various subgroups of the internal gauge symmetry. With the gauge 
groups represented directly on the space $T\htho$ this structure parallels that 
employed for Kaluza-Klein theory based on homogeneous fibres as described in 
section~\ref{thwhf}.

  For now considering $\tilde{g}$ and $\tilde{g}'$ as free parameters in 
equation~\ref{covdevth} allows a closer comparison with the structure of the 
electroweak theory in the Standard Model for which the couplings $g$ and $g'$ are 
independent. 
  However here neither the $\sutw^2$ generated by $\{ \dot{R}_{z \pqg l}^2, \dot{R}_{x 
\pqg z}^2,
  \dot{R}_{x \pqg l}^2 \}$ nor the  $\uo^2$ generated by $\Sbard^2_l$  are internal 
symmetries in the sense of table~\ref{stabset}, that is within $\stab$, with each of 
these four generators impacting upon the components of the type 1 subspace $\htwc 
\subset \htho$, which represent components of the external spacetime $\TM_4$, as can 
be seen explicitly from the form of these four generators in tables~\ref{lbrota} and 
\ref{ltrota}. The breaking of the full $\esi$ symmetry action on $\htho$ in this 
identification of the type 1 subspace $\htwc$ with the local tangent space of the 
external spacetime hence includes the breaking of the $\sutw^2 \times \uo^2 \subset 
\esi$ subgroup.

The covariant derivative applied to the $\theta^{2}_l$ components  in 
equation~\ref{covdevth} can be 
applied  to the components of $\htho$ more generally  and
 written out explicitly using tables~\ref{lbrota} and \ref{ltrota}. In particular we 
find that applied 
to the type 1 embedding of the $2 \times 2$ matrix of components $X \inn \htwo \subset 
\htho$
 and $\bh_2 \inn \htwc \subset \htho$ this covariant derivative reads respectively:
\begin{eqnarray}
\hspace*{-14pt}
   D_{\mu} X  & = &  \pal_{\mu} X \; + \;  \nonumber \\
\hspace*{-14pt} 
  & &  \!\!\!\!\!\!\!\!\!\!\!\!\!\!\!\!\!\!\!\!\!\!\!\!\!\!\!\!
 \mbox{\small $ 
  \left(\!\!\! \begin{array}{c}  0+0+0+0  \qquad \qquad \qquad \qquad \qquad \;
     \qquad \qquad
  \tilde{g}\tilde{W}^1_{\mu}(\fh cl) +  \tilde{g}\tilde{W}^2_{\mu}(\fh c) + 
\tilde{g}\tilde{W}^3_{\mu}(\fh \bar{a}l)
    + \tilde{g}' \tilde{B}_{\mu} \fh \Sbard^2_l(\bar{a})    \\
 \tilde{g}\tilde{W}^1_{\mu}(-\fh l\bar{c}) +  \tilde{g}\tilde{W}^2_{\mu}(\fh \bar{c}) 
+ \tilde{g}\tilde{W}^3_{\mu}(-\fh la)
    + \tilde{g}' \tilde{B}_{\mu} \fh \Sbard^2_l(a)   \qquad \qquad \qquad \;
 	  \tilde{g}\tilde{W}^1_{\mu}(b_l) +  \tilde{g}\tilde{W}^2_{\mu}(b_x) +0 +0 
\end{array}
	   \!\!\!  \right)
	  $}   \nonumber \\ 
\hspace*{-14pt}
		 & &   \label{dmuhtwo}    \\
\hspace*{-14pt}		 
 D_{\mu} \bh_2  &  = &   \pal_{\mu} \bh_2 \; + \;  \nonumber  \\  
\hspace*{-14pt} 
  & &  \!\!\!\!\!\!\!\!\!\!\!\!\!\!\!\!\!\!\!\!\!\!\!\!\!\!\!\!
  \mbox{\small $ 
   \left(\!\!\! \begin{array}{c}  0  \qquad \qquad \qquad \qquad \qquad \quad
 \frac{\tilde{g}}{2} \Big( \! \tilde{W}^1_{\mu}(c_1l - c_8) + \tilde{W}^2_{\mu}(c_1 + 
c_8l) + \tilde{W}^3_{\mu}(a_1l + a_8) \! \Big)
    + \frac{\tilde{g}'}{2} \tilde{B}_{\mu} (-a_1l - a_8)  \\
  \frac{\tilde{g}}{2}  \Big( \! \tilde{W}^1_{\mu}(-c_1l - c_8) + \tilde{W}^2_{\mu}(c_1 
- c_8l) + \tilde{W}^3_{\mu}(-a_1l + a_8) \!\Big)
    + \frac{\tilde{g}'}{2} \tilde{B}_{\mu} (a_1l - a_8) \qquad \;\;
	  \tilde{g}\tilde{W}^1_{\mu}(b_8) +  \tilde{g}\tilde{W}^2_{\mu}(b_1) \end{array}
	   \!\!\!\right)
	 $}   \nonumber \\
\hspace*{-14pt}	 
	    & &  \label{dmuhtwc}  
\end{eqnarray}
   where the second equation shows that indeed each of the four gauge fields 
$\tilde{W}^{\alpha}_{\mu}(x)$, $\tilde{B}_{\mu}(x)$ has non-zero impact on the 
$\{1,l\}$ components of $X \inn \htwo$, that is on the 4-dimensional vector  $\bh_2 
\inn \htwc$ of equations~\ref{hvvv} and \ref{hinhtho}, unlike the case of 
equations~\ref{rbstab1} and \ref{rbstab2} for example, and hence are not associated 
with a purely internal symmetry in the sense of  $\stab$.
However an orthogonal linear combination of gauge fields may be taken with:
\begin{equation}
 \begin{array}{lcr}
     \tilde{B}_{\mu}  & = &   \cos \thetmt \,
	        \tilde{A}_{\mu} - \sin \theta_{\! M^2} \, \tilde{Z}_{\mu}  \\
	 \tilde{W}^{3}_{\mu}  & = & \sin \theta_{\! M^2} \,
	            \tilde{A}_{\mu} + \cos \theta_{\! M^2} \, \tilde{Z}_{\mu}  \\
 \end{array} 				
     \label{bwthwazt}
\end{equation} 
   by analogy with equations~\ref{azthwbw} and \ref{bwthwaz}, where $\thetmt$ (with 
subscript $M^2$ denoting `mock mixing angle' of type 2) plays a similar role to the 
weak mixing angle $\theta_W$. The corresponding contribution from the 
$\tilde{B}_{\mu}(x)$ and $\tilde{W}^3_{\mu}(x)$ fields to the $\bar{a}_{1,l}$ 
components in the top-right element of equation~\ref{dmuhtwc} is then:
\begin{equation}
 \hspace*{-15.5pt}
  D_{\mu} \bar{a}_{1,l}\;\; =\;\ldots\; +  
  \qquad \qquad \qquad \qquad \qquad   \qquad \qquad \qquad \qquad \qquad 
  \qquad \qquad \qquad \qquad \qquad   \qquad   \nonumber 
\end{equation}
\vspace{-20pt}
\begin{equation}   
 \hspace*{-15.5pt}    
      \frac{\tilde{g}}{2} \sin \thetmt  \tilde{A}_{\mu} (a_1l + a_8)
    + \frac{\tilde{g}'}{2} \cos \thetmt  \tilde{A}_{\mu}  (-a_1l - a_8)
	+ \frac{\tilde{g}}{2} \cos \thetmt  \tilde{Z}_{\mu} (a_1l + a_8)
    - \frac{\tilde{g}'}{2} \sin \thetmt  \tilde{Z}_{\mu}  (-a_1l - a_8)
	\label{aazza}
\end{equation}

  Hence the gauge field $\tilde{A}_{\mu}(x)$ represents a purely internal field, with 
no action on the external $\bh_2 \inn \htwc$ components, provided:   
\begin{eqnarray}
        \tilde{g} \sin \thetmt & = & \tilde{g}' \cos \thetmt  
		   \nonumber \\
    \mbox{that is:} \qquad \qquad \tan \thetmt & = &  \frac{\tilde{g}'}{\tilde{g}}  
	               \qquad \qquad \qquad \quad  \label{tantwt}
\end{eqnarray}
  This relation is closely analogous to equation~\ref{tantw} for electroweak theory
 in the Standard Model. However here in the case of equation~\ref{tantwt} neither a 
Lagrangian formalism, using for example equation~\ref{lagwwwb}, nor a Higgs field is 
required to break the $\sutw^2 \times \uo^2$ symmetry down to a $\uo$ symmetry 
associated with the gauge field $\tilde{A}_{\mu}(x)$.    
 Considering more generally the $\tilde{A}_{\mu}(x)$ field part of the covariant 
derivative $D_{\mu}$ of equation~\ref{covdevth}, via equation~\ref{bwthwazt}, on all 
of the components of $\mcX \inn \htho$, with for example
 $\dot{R}^{2}_{x \pqg l} \equiv \dot{R}^{2}_{x \pqg l}(\mcX)$,  we have:
\begin{eqnarray}
   D_{\mu}\mcX(x) &  = & \pal_{\mu} \mcX(x) \, + \,
     \tilde{g}\, \sin \thetmt \, \tilde{A}_{\mu}(x) \, \dot{R}^{2}_{x \pqg l} \, + \, 
\tilde{g}' \, \cos\thetmt \, \tilde{A}_{\mu}(x) \, \fh \Sbard^2_l \nonumber \\
  &  = & \pal_{\mu} \mcX(x) \, + \,
     \tilde{g}\, \sin \thetmt \, \tilde{A}_{\mu}(x) \, \dot{R}^{2}_{x \pqg l} \, + \, 
\tilde{g}\, \tan \thetmt \, \cos\thetmt \, \tilde{A}_{\mu}(x) \, \fh \Sbard^2_l 
\nonumber \\
   &  = & \pal_{\mu} \mcX(x) \, + \,
     \tilde{g}\, \sin \thetmt \, \tilde{A}_{\mu}(x) \, ( \dot{R}^{2}_{x \pqg l} \, + 
\,
	 \fh \Sbard^2_l    ) \nonumber \\
	 &  = & \pal_{\mu} \mcX(x) \, + \,
     \tilde{g}\, \sin \thetmt \, \tilde{A}_{\mu}(x) \, ( - \Sbard^1_l    )
	   \label{achargee} 
\end{eqnarray}
 where the final line is fixed by the \textit{linear dependence} of 
equation~\ref{srslincom2} for the generators of the $\esi$ Lie algebra. The gauge 
field $\tilde{A}_{\mu}(x)$ is hence  associated with  $\Sbard^1_l$ which as an element 
of $\staba$ has been identified as the generator of the internal gauge symmetry 
$\uo_Q$ of electromagnetism  in the previous section
 (see the discussion following equation~\ref{wang5}).
 The zero charge of the $\nu$-lepton,
   associated with the $a_{1,l}$ components in equation~\ref{suthuoa},
    is here taken to be entirely equivalent to the fact that the action 
$\dot{\Sbar}^1_l$ does not impinge on the $\{1,l\}$ components of $a\inn \ooo \subset 
\htho$.
 The apparently ambiguous nature of these $a_{1,l}$ components, which have been 
associated \textit{both} with the neutrino state \textit{and} with part of the vector 
space $\htwc \equiv \TM_4$ on the external spacetime, will be resolved in 
section~\ref{secesef}.

 The lines of equation~\ref{achargee} are closely analogous to those of 
equation~\ref{amucoup} from electroweak theory, with $\tilde{A}_{\mu}(x) \sim 
A_{\mu}(x)$ and $- \Sbard^1_l \sim Q$. The apparent electromagnetic coupling 
$\tilde{e}$ may  be identified directly in equation~\ref{achargee} as:
\begin{equation}
  \label{sintwt}
   \tilde{e} = \tilde{g} \sin \thetmt
\end{equation} 
   which is also analogous to equation~\ref{egsint} in the Standard Model.

   We next employ a basis for the $\esi$ algebra with a normalised Killing form with  
components proportional to the unit $78 \times 78$ matrix.
In this case it will be possible to determine the value of the mixing angle $\thetmt$ 
in the breaking of the $\sutw^2 \times \uo^2$ symmetry to $\uo_Q$. 
 For the $\sutha_s$ subalgebra such a normalised basis is provided by  
equation~\ref{rsrrrrrr} with three possible choices of $\{\dot{R}^a_{x \pqg l}, \, 
\frac{1}{\sqrt{3}}\dot{S}^a_l \}$, for type $a= 1,2,3$, for the first two elements. 
   The covariant derivative of equation~\ref{covdevth} may be rewritten  in the 
normalised Killing form basis, with $\frac{1}{\sqrt{3}}\dot{S}^a_l = 
\frac{\sqrt{3}}{2} \Sbard^a_l$  (via equation~\ref{sbardot})
 and with the couplings $\tilde{g}$ and $\tilde{g}'$ absorbed into the gauge fields 
as:
\begin{equation}
 \label{covdevth2}
   D_{\mu}\theta^{2}_l(x) \, = \, \pal_{\mu} \theta^{2}_l(x) \, + \,
     \tilde{W}^{\alpha}_{\mu}(x) \, \dot{R}^{(2)\alpha}(\theta^{2}_l) \, + \, 
\tilde{B}_{\mu}(x) \, \frac{\sqrt{3}}{2} \Sbard^2_l(\theta^{2}_l)
\end{equation}
   where again $\alpha=1,2,3$ and $\dot{R}^{(2)\alpha} \equiv  \{ \dot{R}_{z \pqg 
l}^2, \dot{R}_{x \pqg z}^2,
  \dot{R}_{x \pqg l}^2 \}$.
  The gauge fields $\{\tilde{W}^3_{\mu}(x), \tilde{B}_{\mu}(x) \}$ aligned with the 
generators  $\{\dot{R}^2_{x \pqg l}, \, \frac{\sqrt{3}}{2} \Sbard^2_l \}$ may be 
expressed in a new basis with gauge  fields $\{\tilde{Z}_{\mu}(x), \tilde{A}_{\mu}(x) 
\}$ aligned with the generators  $\{\dot{R}^1_{x \pqg l}, \, \frac{\sqrt{3}}{2} 
\Sbard^1_l \}$. In this basis the `internal' gauge field $\tilde{Z}_{\mu}(x)$ is 
associated with $\dot{R}^1_{x \pqg l}$ which as a generator of $\sltc^1$, as 
originally listed in equation~\ref{extlor6}, is in fact a purely external action! 
However here we are dealing with a mock electroweak theory for which some 
inappropriate features may be observed, as was the case for the ambiguity of the 
$a_{1,l}$ components noted above. In any case the electromagnetic gauge field 
$\tilde{A}_{\mu}(x)$ associated with $\frac{\sqrt{3}}{2} \Sbard^1_l$ in the new basis 
does represent a purely internal action. Transferring to the new basis we have:
\begin{eqnarray}
   \tilde{W}^{3}_{\mu}\, \dot{R}^{2}_{x \pqg l}  \; + \; 
   \tilde{B}_{\mu} \, \frac{\sqrt{3}}{2} \Sbard^2_l  & \Rightarrow & 
    \tilde{Z}_{\mu} \, \dot{R}^{1}_{x \pqg l} 
  \; + \; \tilde{A}_{\mu} \, \frac{\sqrt{3}}{2} \Sbard^1_l   \nonumber   \\
 \!\!\!\!\! \mbox{hence:} \qquad \qquad \qquad \quad
  \sin \thetmt \, \tilde{A}_{\mu} \, \dot{R}^{2}_{x \pqg l} + 
  \cos \thetmt \, \tilde{A}_{\mu} \, \frac{\sqrt{3}}{2} \Sbard^2_l & = & 
           \tilde{A}_{\mu} \, \frac{\sqrt{3}}{2} \Sbard^1_l  \nonumber \\
  \sin \thetmt \, \tilde{A}_{\mu} \, (-\fh \dot{R}^{1}_{x \pqg l} - 
                                    \frac{3}{4} \Sbard^1_l ) + 
  \cos \thetmt \, \tilde{A}_{\mu} \, \frac{\sqrt{3}}{2} 
                          (\dot{R}^{1}_{x \pqg l} - \fh \Sbard^1_l ) & = & 
           \tilde{A}_{\mu} \, \frac{\sqrt{3}}{2} \Sbard^1_l  \label{wsbsas}
\end{eqnarray}
  where in the second line the orthogonal transformation of equation~\ref{bwthwazt} 
has been applied to the left-hand side and
   only the $\tilde{A}_{\mu}$ field part has been retained on both sides.  
Equations~\ref{rsdeprrs} and \ref{rsdepsrs}, together with equation~\ref{sbardot}, 
have been used for the bottom line. By equating the basis vector $\dot{R}^{1}_{x \pqg 
l}$ and $\Sbard^1_l$ parts separately in this final line above it can be deduced that:

\begin{equation}
 \label{sincosm}
      \sin \thetmt = - \frac{\sqrt{3}}{2}  \quad \mbox{and} \quad \cos\thetmt = - \fh
\end{equation}
and hence:
\begin{equation}
 \label{sinmock}
      \sin^2 \thetmt = \frac{3}{4}   \qquad \mbox{with} \qquad \thetmt = 240^0
\end{equation}
   as the mixing angle. Performing a similar analysis for the type 3 case of  $\sutw^3 
\times \uo^3$ breaking to $\uo_Q$ leads to a similar result, except with $\sin 
\thetmth = + \frac{\sqrt{3}}{2}$ and $\thetmth = 120^0$. Setting $\tilde{g} = 1$ the 
magnitudes of the coupling constants to substitute into equations~\ref{covdevth} and 
the bottom line of equation~\ref{achargee} in order to match the normalised 
expressions of equations~\ref{covdevth2} and the right-hand side of 
equation~\ref{wsbsas} are, relative to $\tilde{g}$:
\begin{equation}
\label{ggesett}
   \tilde{g}  \qquad : \qquad \tilde{g}' = \sqrt{3}\,\tilde{g}
              \qquad : \qquad \tilde{e} = \frac{\sqrt{3}}{2}\,\tilde{g} 
\end{equation}
  These values are consistently obtained from equations ~\ref{tantwt} and \ref{sintwt} 
by substituting in the value of $\thetmt$ from equations~\ref{sincosm} and 
\ref{sinmock}, with a similar observation applying for the type 3 case.

 This analysis is useful for comparison with the Standard Model for which the gauge 
groups $\sutw_L$ and $\uo_Y$ are not obtained from a single unifying group and hence 
the respective gauge couplings $g$ and $g'$ of equation~\ref{covdevlep} are 
independent.
 Indeed equation~\ref{covdevth2} above may be compared with the form of the covariant 
derivative of a left-handed doublet of leptons in the Standard Model, which from 
equations~\ref{tauhsig} and \ref{covdevlep} can be written as:
\begin{equation}
      \quad D_{\mu}  =  \pal_{\mu}\, +\, ig  W^{\alpha}_{\mu}(x)  \fh \sigma^{\alpha}
                   \, - \,i g' B_{\mu}(x) \fh \sigma^0  
\end{equation}
  In this case the third component of weak isospin $T^3 = \fh \sigma^3$ and 
hypercharge $\frac{Y}{2} = -\fh \sigma^0$ combine to form the charge operator $Q = 
\binom{0 \;\;\; 0}{0 \; -1}$ via equation~\ref{qethy} for the lepton doublet.
 In this particular case for equation~\ref{amucoup} we have
 the weak mixing combination:
\begin{equation}
     ig \, W^{3}_{\mu} \, \fh \sigma^3
                   \quad\, - \quad\,i g' \, B_{\mu}\, \fh \sigma^0
				      \quad \Rightarrow \quad 
				   ie \, A_{\mu} \, Q     \nonumber 
\end{equation}
   which may be directly compared with:
\begin{equation}
  \tilde{W}^{3}_{\mu}\, \dot{R}^{(2)}_{x \pqg l}  \quad\;\; + \quad\;\; 
  \tilde{B}_{\mu} \, \frac{\sqrt{3}}{2} \Sbard^2_l
    \quad \Rightarrow \quad  \tilde{A}_{\mu} \, \frac{\sqrt{3}}{2} \Sbard^1_l 
   \nonumber
\end{equation}
   from the top line of equation~\ref{wsbsas}.   
   In the former case
   the set of $2 \times 2$ matrix actions $ 
\{\fhs\sigma^1,\fhs\sigma^2,\fhs\sigma^3,\fhs\sigma^0\}$, as well as forming a basis 
for elements of the vector space $\htwc \subset \ccc(2)$,  forms a basis for the Lie 
algebra $\sutw_L \times \uo_Y$ with the normalisation convention 
$\mbox{tr}(\tau^{\alpha}\tau^{\beta}) = \fh\delta^{\alpha\beta}$, here  including 
$\tau^0 = \fh \sigma^0$ with $\alpha,\beta = 0\ldots 3$. The couplings $g$ and $g'$ 
are introduced in this basis.
 For the empirically measured case the electroweak mixing angle is determined to be 
$\sin^2 \theta_{W} \simeq 0.23$ at the energy scale of $M_Z$ \cite{PDG}, with 
corresponding electroweak couplings
 from equations~\ref{tantw} and \ref{egsint}  approximately in the proportions: 
\begin{equation}
\label{ggesets}
        g    \qquad : \qquad g' \simeq 0.55 \,g
              \qquad : \qquad e  \simeq 0.48 \,g 
\end{equation}

   Given the unit electron charge for the leptonic component $\theta^1_l$ in 
equation~\ref{suthuoth} the action of the generator $\Sbard^1_l$ is analogous to that 
of the unit $2 \times 2$ matrix $\sigma^0$ of equation~\ref{sigmas}. Similarly the 
normalisation of the type 1 actions $\{
  \dot{R}_{z \pqg l}^1, \dot{R}_{x \pqg z}^1, \dot{R}_{x \pqg l}^1  \}$, as seen in 
equations~\ref{mrotgen}  and \ref{msigj}, parallels the set of $2 \times 2$ matrices 
$\{\tau^1, \tau^2, \tau^3\} = \{\fhs\sigma^1,\fhs\sigma^2,\fhs\sigma^3\}$.
   Transferring this analysis to the type 2 case, the set of generators $\{ 
  \dot{R}_{z \pqg l}^2, \dot{R}_{x \pqg z}^2, \dot{R}_{x \pqg 
l}^2,\fhs\dot{\Sbar}^2_l\}$ also parallels 
 the set of matrix actions $ \{\fhs\sigma^1,\fhs\sigma^2,\fhs\sigma^3,\fhs\sigma^0\}$.

 Since this is the generator normalisation used initially in equation~\ref{covdevth} 
it may naively be expected that the couplings obtained for the $\sutw^2 \times \uo^2$ 
symmetry breaking via the constraint of the $\esi$ algebra Killing form in 
equation~\ref{ggesett} may be directly compared with the corresponding values for 
$\sutw_L \times \uo_Y$ electroweak theory obtained empirically as displayed in 
equation~\ref{ggesets}. The significant differences in these values hinges on the 
differing values for the calculated  $\sin^2 \thetmt =\frac{3}{4}$
 of equation~\ref{sinmock} and the empirical 
 $\sin^2 \theta_W \simeq 0.23$. However, as emphasised earlier in this subsection the 
$\sutw^2$ symmetry does \textit{not} act on $\sltc^1$ Weyl spinors in the appropriate 
way to describe weak interactions, and here we are dealing with a provisional `mock 
electroweak theory', which nevertheless exhibits some of the features associated with 
corresponding structures of the Standard Model such as the
 identification of a mixing angle itself.
  
  It is also noted that in the mock theory the calculated value of $\sin^2 \thetmt 
=\frac{3}{4}$ effectively corresponds to a `unification scale' whereas the empirical 
value of 
 $\sin^2 \theta_W \simeq 0.23$ is determined at the practical energy scale of $M_Z 
\sim 10^{2}\,$GeV. In standard quantum field theory the phenomena of `running 
coupling' for an Abelian compared with a non-Abelian gauge theory implies that the 
ratio $g'$:$\;\!g$ increases with the energy scale as will be described in 
section~\ref{seraps} and depicted in figure~\ref{runcup}. Hence the need of a 
quantisation scheme for the present theory, as alluded to at the end of the previous 
subsection and as proposed in chapter~\ref{newapp}, with the consequence of running 
coupling, may be one factor leading to the large calculated mixing angle for the 
present theory. This observation would apply even if the gauge group $\sutw_L \times 
\uo_Y$ were to be correctly identified in the theory.

  In any case in this subsection it has been demonstrated how the relative couplings 
of the internal gauge groups may in principle be related through unification within 
the simple Lie group $\esi$. Finally here we consider how a more realistic electroweak 
theory might be constructed within this framework.
 In the above we have assumed a symmetry breaking pattern of $\sutw^2 \times \uo^2 \to 
\uo_Q$, whereas these subgroups are actually embedded in a larger symmetry breaking 
structure with $\suth_s \to \uo_Q$. That is, instead of equation~\ref{covdevth2} we 
might rather begin with the gauge covariant derivative:  
\begin{equation}
   D_{\mu}\theta^{2}_l(x) \, = \, \pal_{\mu} \theta^{2}_l(x) \, + \,
     W^{\alpha}_{\mu}(x) \, \dot{R}^{\alpha}(\theta^{2}_l) 
   \nonumber
\end{equation}
  where now $\alpha=1\ldots 8$ summing over the full basis of eight $\suth_s$ 
generators in equation~\ref{rsrrrrrr}. Here all three embeddings of $\sutw^a \times 
\uo^a \subset \suth_s$, for type $a=1,2,3$, must come into play with the choice of 
$\sutw^1 \subset \sltc^1$ as the rotation subgroup of the Lorentz group acting on 
external spacetime breaking the symmetry.

  While this symmetry breaking structure requires further study the fact that the mock 
electroweak symmetry $\sutw^a \times \uo^a$ may be embedded in $\suth_s$  in 
\textit{two} ways, of type $a=2$ or $a=3$, while the symmetry $\sutw^1 \times \uo^1$ 
has only \textit{one} embedding, of type $a=1$, may be of some significance. Within 
$\sutha_s$ the three $\uo^a$ generators are linearly dependent by 
equation~\ref{agsdeps}, while by equation~\ref{rsdeprrr} only the $\dot{R}^a_{x \pqg 
l}$ part of the three $\sutw^a$ generators are linearly dependent.
 These observations 
 offer a hint that for an internal $\sutw^a$ combining types $a=2$ and 3 the ratio of 
the effective coupling $\tilde{g}$ to that for the effective $\uo_Q$ coupling 
$\tilde{e}$ maybe somewhat larger than that for the type $a=2$ case alone which led 
the final expression of equation~\ref{ggesett}, once the linear dependencies of the 
generators are taken into account, which may result in a  closer correspondence with 
the Standard Model case in equation~\ref{ggesets}.

 Type 2 gauge fields:
\begin{equation}
 \label{wpmcomb} 
 \tilde{W}^{(2)\pm}_{\mu}(x) =
    \tilde{W}^{(2)1}_{\mu}(x) \mp i\tilde{W}^{(2)2}_{\mu}(x)
\end{equation}
    may be associated with the type 2 generators $\dot{\Sigma}^{(2)\pm}$ of
equation~\ref{sigcomb} in the complex $\sutw^2$ subalgebra (by comparison with 
equations~\ref{wpmww} and \ref{spmss} for the Standard Model, although neglecting 
possible factors of $\frac{1}{\sqrt{2}}$ or $\fh$ here).
  As described above and as can be seen from table~\ref{lbrota} the relevant 
generators $\dot{R}^2_{z \pqg l}$ and $\dot{R}^2_{x \pqg z}$  mix the $a$ component of 
$\htho$ with the $c$ component only.
 The fact that in the mock theory the $\sutw^2 \times \uo^2$ symmetry acts on the 
$\theta^{2} = \binom{a}{\bar{c}} \subset \htho$ components, and \textit{not} physical 
fermion doublets, is one reason not to expect the calculated mixing angle to match the 
empirical case.
That is, while the type 2 symmetry $\sutw^2 \times \uo^2$ has some of the properties 
associated with the electroweak symmetry $\sutw_L \times \uo_Y$, the $\sutw^2$ 
transformations do not relate the $a\inn \htho$ component to \textit{both} components 
of $\theta^{1}=\binom{c}{\bar{b}}$.

 On the other hand type 3 gauge fields $\tilde{W}^{(3)\pm}_{\mu}(x) = 
   \tilde{W}^{(3)1}_{\mu}(x) \pm i\tilde{W}^{(3)2}_{\mu}(x)$  may be  associated with 
similar generators  in the complex $\sutw^3$ subalgebra:
\begin{equation}
 \label{sig3pm}
   \dot{\Sigma}^{(3)\pm}  =  \dot{R}^3_{z \pqg l} \mp i \dot{R}^3_{x \pqg z}
\end{equation}
  The $\pm$ signs are chosen such that the generators  $\dot{\Sigma}^{(3)\pm}$, as for  
$\dot{\Sigma}^{(2)\pm}$ in equation~\ref{isscomm}, carry charges of $\pm$1, that is:
\begin{equation}
  \lbrack i \Sbard^1_l \, , \,  \dot{\Sigma}^{(3)\pm} \rbrack \, = \,
     \pm  \dot{\Sigma}^{(3)\pm}
\end{equation}
 Adding to the discussion towards the end of the previous subsection,
  together  $\dot{\Sigma}^{(2)\pm}$ and $\dot{\Sigma}^{(3)\pm}$ describe four of the 
six eigenvectors of the Cartan subalgebra in the Cartan-Weyl basis for the adjoint 
representation of the complexified $\sutha_s$ algebra.
 The full set of six eigenvectors are sometimes denoted
  $U^{\pm}$, $V^{\pm}$ and $T^{\pm}$ in the  $\sutha$ root space diagram
  (as for example in the context of the $\suth$ flavour symmetry between $u$, $d$ and 
$s$-type quarks).

   For the case of equation~\ref{sig3pm}, as can also be seen from table~\ref{lbrota}, 
the generators $\dot{R}^3_{z \pqg l}$ and $\dot{R}^3_{x \pqg z}$  mix the $a$ 
component of $\htho$ with the $b$ component only, that is within the $\theta^3 = 
\binom{b}{\bar{a}} \subset \htho$ components as shown in equation~\ref{abcmix}.  Hence 
it appears that physical
 charged gauge boson fields $W^{\pm}_{\mu}(x)$ must indeed be related to both type 2 
$\dot{\Sigma}^{(2)\pm}$ and type 3 $\dot{\Sigma}^{(3)\pm}$ operators to act on fermion 
doublets.
 Ultimately 
 the interactions of the physical $W^{\pm}$ particle states will need to be 
appropriately oriented with respect to physical fermion states. The latter will in 
turn require a possible $\sltc^1$ Weyl spinor interpretation of the $a\inn \htho$ 
components, which have been provisionally associated with neutrino and $u$-quark 
states according the internal $\suth_c \times \uo_Q$ transformations of 
equation~\ref{suthuoa}.

 The possible means of identifying Weyl spinor states for the
 $\nu$-lepton and $u$-quarks  within the $a\inn \ooo \inn \htho$ components will be 
addressed in section~\ref{subehafws}. The identification of both left and right-handed 
Weyl spinors together with the Dirac representation of the external $\sltc^1$ symmetry 
will then be described in section~\ref{secesef}. Finally the possibility of 
identifying three generations of fermions and the phenomena of CKM mixing will be 
outlined in section~\ref{sosmfi}.
All of the above features may need to come together in order to fully identify the
 physical $\sutw_L$ symmetry together with standard phenomena of electroweak theory 
within the context of the present theory.
 In the meantime in the following subsection we study further suggestive features of 
the $\sutw^2 \times \uo^2$ mock electroweak theory based within the $\esi$ framework, 
and in particular concerning the source of  finite mass  for the both the gauge bosons 
and the fermion states. Then we shall briefly
 consider further possible $\sutw \subset \esi$ subgroups as candidate components of 
an electroweak symmetry, before extending beyond $\esi$ in the following chapter.


\subsection{Origin of Mass and Higgs Phenomena}
\label{suboomahp}

 The empirical weakness of the weak interaction relative to electromagnetic phenomena 
owes not to the value of the coupling $g$, in equation~\ref{covdevlep} for example, 
which is around twice the value of $e$, equation~\ref{ggesets}, but to the large 
values for the masses of the $W^{\pm}$ and $Z^0$ gauge bosons. Although in this 
chapter we are dealing primarily at the level of the Lie algebra structure, together 
with the simple dynamic expressions introduced in the previous subsection,  it will be 
considered here how mass terms for particle states may originate in the symmetry 
breaking structure, not only for the massive gauge bosons but also for leptons and 
quarks in the full theory.
 Here $\tilde{W}^{\pm}$ and $\tilde{Z}^0$ gauge bosons will be provisionally 
associated with the appropriate fields of the $\sutw^2 \times \uo^2$ mock electroweak 
theory, and hence we first look in more detail at the field $\tilde{Z}_{\mu}(x)$.
     
   The gauge field $\tilde{Z}_{\mu}(x)$ appearing in the top line of 
equation~\ref{wsbsas} was identified along with $\tilde{A}_{\mu}(x)$ as aligned to the 
choice of basis elements $\{\dot{R}^1_{x \pqg l}, \, \frac{\sqrt{3}}{2} \Sbard^1_l 
\}$.
 As described earlier the apparent association of the `internal' field 
$\tilde{Z}_{\mu}(x)$ with the `external' generator $\dot{R}^1_{x \pqg l}$ is one of a 
number of significant caveats associated with the mock electroweak theory.
 Through the orthogonal transformation of equation~\ref{bwthwazt}; that is with 
$\tilde{Z}_{\mu} = \cos \thetmt \, \tilde{W}^3_{\mu} - \sin \thetmt \, 
\tilde{B}_{\mu}$,  in analogy with electroweak theory and equation~\ref{azthwbw}, as 
for the `photon' field $\tilde{A}_{\mu}(x)$, the field $\tilde{Z}_{\mu}(x)$  is 
associated with a linear combination of the generators 
$\dot{R}^2_{x \pqg l}$ and $\Sbard^2_l$.
  Since in the $\esi$ Lie algebra $\lbrack 
\Sbard^1_l, \, \dot{R}^2_{x \pqg l} \rbrack = 0$ and  $\lbrack 
\Sbard^1_l, \, \Sbard^2_l \rbrack = 0$ any such linear combination of
  $\dot{R}^2_{x \pqg l}$ and $\Sbard^2_l$
 has zero electromagnetic charge. 
 Hence the $\tilde{Z}^0$ gauge boson and $\tilde{\gamma}$ photon, associated with the 
fields $\tilde{Z}_{\mu}(x)$ and $\tilde{A}_{\mu}(x)$ respectively, are neutral, unlike 
the case of the charged $\tilde{W}^{\pm}$ gauge bosons associated with fields 
$\tilde{W}^{(2)\pm}_{\mu}(x)$ of equation~\ref{wpmcomb} corresponding to the 
generators $\dot{\Sigma}^{(2)\pm}$ of equations~\ref{sigcomb} and \ref{isscomm}.
 This is also the case when such linear combinations are extended to include the type 
3 form of these generators since also $\lbrack 
\Sbard^1_l, \, \dot{R}^3_{x \pqg l} \rbrack = 0$ and  $\lbrack 
\Sbard^1_l, \, \Sbard^3_l \rbrack = 0$.

  From the type 2 $\tilde{W}^3_{\mu}$ and $\tilde{B}_{\mu}$ terms in 
equation~\ref{dmuhtwo}
it can be seen that the transformations associated with the fields 
$\tilde{Z}_{\mu}(x)$ and $\tilde{A}_{\mu}(x)$  mix the components of $a\inn \ooo$ 
within $\htwo$. This is unlike the more involved transformations associated with the 
fields $\tilde{W}^{(2)1}_{\mu}(x)$ and $\tilde{W}^{(2)2}_{\mu}(x)$, as can also can be 
seen  in table~\ref{lbrota}  for the corresponding generators    $\dot{R}^2_{z \pqg 
l}$ and $\dot{R}^2_{x \pqg z}$   which mix components of $\htwo \subset \htho$ with 
those not in $\htwo$.
  Isolating the interaction of the $\tilde{Z}_{\mu}$ field with the $a$ and 
$\theta^{1} = \binom{c}{\bar{b}}$ components separately may allow a determination of 
the coupling of the $\tilde{Z}^0$ to the lepton pairs as well as  quark pairs, which 
might be directly compared with the electromagnetic coupling of the photon to the same 
components as summarised in equations~\ref{slonthet}, \ref{suthuoth} and 
\ref{suthuoa}.

  This may be more straightforward than for interactions involving the 
$\tilde{W}^{\pm}$ gauge bosons as here not only is the interaction restricted to 
single components  but also one generation of fermion states may suffice since there 
are no flavour changing neutral currents in the Standard Model, as described at the 
end of section~\ref{ewtatsm}.
Hence a more detailed study of interactions for  the field $\tilde{Z}_{\mu}(x)$ in 
comparison with the field  $\tilde{A}_{\mu}(x)$ may prove enlightening in comparison 
with the relevant properties of the Standard Model described in section~\ref{ewtatsm} 
and in particular with respect to the determination of the relative couplings.

  This may involve  linear combinations of type 2 and type 3 actions on $\theta^{1} = 
\binom{c}{\bar{b}}$ with generators of the weak neutral field $\tilde{Z}_{\mu}(x)$ 
being complementary to the generator $\Sbard^1_l$ of the electromagnetic field 
$\tilde{A}_{\mu}(x)$ with respect to the full $\suth_s \subset \esi$ symmetry, as 
considered towards the end of the previous subsection, with the $\tilde{Z}_{\mu}(x)$
field  associated with a different linear combination of charge neutral $\suth_s$ 
generators.
 Ultimately however for comparison with weak neutral interactions described in the 
Standard Model via equation~\ref{zmucoup} both
 left-handed and right-handed fermions will need to be identified, such that $T^3 = 0$ 
for right-handed states, and this itself will require an extension beyond the study of 
$\esi$ on the space $\htho$. Such an extension will also be required to identify
  the physical $\sutw_L \times \uo_Y$ symmetry, independent of the external $\sltc^1$ 
generators, and fully account for both $W^{\pm}$ and $Z^0$ interactions.

    In the meantime  here we consider broader features of the mock electroweak theory 
as described in the previous two subsections, and in particular  how masses may arise 
for gauge bosons through  the impingement of the $\sutw^2 \times \uo^2 \subset \esi$ 
symmetry on the 4-dimensional subspace $\htwc \subset \htho$ associated with the 
tangent space $\TM_4$ of the external spacetime. Having  in mind comparisons with the 
Standard Model we return to the convention of equation~\ref{covdevth} with coupling 
parameters $\tilde{g}$ and $\tilde{g}'$ in place of employing generators normalised 
according to the $\esi$ Killing form.

In equation~\ref{aazza} the coupling of the $\tilde{A}_{\mu}(x)$ and 
$\tilde{Z}_{\mu}(x)$ fields to the $\bar{a}_{1,l} = a_1 - a_8l$ components of $\htwc 
\subset \htho$ was extracted. The constraint $\tan \thetmt = \tilde{g}'/\tilde{g}$ was 
derived in equation~\ref{tantwt} in order for the $\tilde{A}_{\mu}(x)$ contribution to 
vanish. With this constraint the impingement of the field $\tilde{Z}_{\mu}(x)$ on the 
$\bar{a}_{1,l}$ subcomponent part of $\htwc$  from equation~\ref{aazza} can be 
written:
\begin{eqnarray}
 D_{\mu} \bar{a}_{1,l}\;\; =
  & \ldots & + \frac{\tilde{g}}{2} \cos \thetmt  \tilde{Z}_{\mu} (a_1l + a_8)
    - \frac{\tilde{g}'}{2} \sin \thetmt  \tilde{Z}_{\mu}  (-a_1l - a_8)  \nonumber \\
  & = & \fh \, (\tilde{g} \cos \thetmt + \tilde{g}' \sin \thetmt) \, \tilde{Z}_{\mu} 
\,
     (a_1l + a_8)  \nonumber \\
  & = & (\tilde{g} \cos \thetmt + \tilde{g} \tan \thetmt \sin \thetmt)
   \, \tilde{Z}_{\mu} \,
    \fh (a_1l + a_8)  \nonumber \\
  & = & \frac{\tilde{g}}{\cos \thetmt} \: \tilde{Z}_{\mu} \: \fh (a_1l + a_8)
  \label{zmixaa}
\end{eqnarray}
   This compares with the impingement of the 
   type 2 fields $\tilde{W}^{(2)\pm}_{\mu}(x)$ 
   of  equation~\ref{wpmcomb}, as
composed of $\tilde{W}^{(2)1}_{\mu}(x)$ and $\tilde{W}^{(2)2}_{\mu}(x)$,  on the same 
off-diagonal elements of $\htwc$ in equation~\ref{dmuhtwc}, which is proportional to 
$\tilde{g}/2$. Hence the coupling of the corresponding gauge fields to the 
subcomponent $\bar{a}_{1,l}$ of
$\htwc \subset \htho$  is in the following ratio:
\begin{equation}
  \begin{array}{ccccc}
     \tilde{Z}_{\mu} & \quad : \quad & \tilde{W}^{\pm}_{\mu} & \quad : \quad &
	                        \tilde{A}_{\mu} \\
	 \frac{\tilde{g}}{\cos \thetmt}  &
	  \quad : \quad &  \tilde{g} & \quad : \quad & 0    
  \end{array}  
   \label{zwamass}
\end{equation}

This suggests, given the Standard Model expression for $M_W$ in equation~\ref{masswgv} 
and its relation to $M_Z$ in equation~\ref{mwmzcth}, that the interaction with 
components of $\htwc$  
 originating here in equation~\ref{dmuhtwc} is closely related to the masses of the 
gauge bosons in the present theory (within the caveats that the impingement on all 
four components of $\htwc$ may need to be addressed and factors of 2 or $\sqrt{2}$  
may appear for some terms in a more thorough analysis, but here we are merely noting 
certain general features of the mock electroweak theory).
 This structure arises here without the need to introduce a Lagrangian or a 
custom-built  scalar Higgs field $\phi$.

Mass terms such as for equations~\ref{masswgv} and \ref{mwmzcth}, arising in the 
Standard Model Lagrangian, are quadratic in the gauge boson fields due to the 
quadratic composition $ (D_{\mu}\phi)^{\dag}D^{\mu} \phi$  constructed for the Lorentz 
invariant initial Lagrangian $\lag_H$ in equation~\ref{laghiggs}. In the present 
theory the composition of the gauge fields with the components of $\htho$ in 
expressions such as $\dmo$ has a different structure, linear in the gauge fields. Here 
the concept and nature of particle  `mass' is yet to be identified, and will require 
an understanding of quantisation and physical particle states as will be described in 
chapter~\ref{newapp}. 
 However the impingement of the `internal' $\sutw^{2,3}$ symmetry upon the components 
of the external spacetime tangent space $\TM_4$ is expected to correlate closely with 
the phenomenology of the massive ${W^{\pm}}$ and ${Z^0}$, involving the kinematic 
properties of these gauge boson states in spacetime, and hence accounting for the 
short-range nature of the weak interaction.
  If gauge boson masses may be obtained through these interactions this raises the 
question of how further elements of the present theory might correspond to the Higgs 
sector of the Standard Model.

   In the present theory the Lorentz $\soot$ symmetry acts on the form $L(\bv_4) = 
(v^1)^2 - (v^2)^2 - (v^3)^2 - (v^4)^2 =h^2$ of equation~\ref{lorform2} with the 
components of $\bv_4 \inn \htwc$ embedded in $\htho$ under the $\sltc^1 \subset \esi$ 
subgroup action.  
In a suitable choice of frame a Lorentz 4-vector can be expressed as $\bv_4 = 
(v^0,0,0,0)$ which in turn can be written as
 \begin{equation} 
  \label{v4vac}
    \bh_2 =   \left( \begin{array}{cc}
	   v^0  &  0  \\
	   0  &  v^0  \end{array} \right)
\end{equation}	
	that is with the three components $v^1=v^2=v^3 = 0$ in equations~\ref{hvvv} and 
\ref{hinhtho}. This 4-vector is invariant under the $\sutw^1 \subset \sltc^1$ 
transformations $\bh_2 \to S \bh_2 S^{\dag}$, which preserve the form of 
equation~\ref{v4vac}, with
 the type 1 rotations $S \inn \sutw^1 \subset \sltc^1$  generated by
		$\{\dot{R}_{z \pqg l}^{1},\dot{R}_{x \pqg z}^{1}, \dot{R}_{x \pqg l}^{1}\}$; 
that is the subset of Lorentz generators in equation~\ref{extlor6} leaving the $t 
\equiv v^0$ component in equation~\ref{extloract} fixed.

 As described in section~\ref{ewtatsm} for the Higgs sector an $\sutw$ custodial 
symmetry originates as a subgroup of the SO(4) symmetry of the form of the potential 
$V(\phi) = f(\phi_1^2 + \phi_2^2 + \phi_3^2 + \phi_4^2)$ implicit in 
equation~\ref{higgspot}. 
 The vacuum value of 
 the Standard Model Higgs field can be expressed in terms of the bi-doublet $\Phi$
  of equation~\ref{bidoub} in the form of equation~\ref{higgsvac2}, that is with
   $\langle \Phi \rangle = \fh \binom{v\; 0}{0\; v}$, which is   
 invariant under the transformations $ \langle \Phi \rangle \to L \langle \Phi \rangle 
L^{\dag}$ with $L\inn \sutw_{L+R}$, where $\sutw_{L+R}$ is  the custodial symmetry, 
highlighting the close similarity to the symmetry of a given Lorentz 4-vector $\bh_2 
\inn \htwc$ such as that in equation~\ref{v4vac} in the present theory.

 Rather than a Higgs complex doublet field $\phi$ and Lagrangian $\lag_H$ with 
`accidental' global SO(4) symmetry (for $g' \to 0$), here we have the Lorentz 4-vector 
$\bv_4$ with an external $\soot$ symmetry for the form $L(\bv_4) = h^2$ which is 
`spontaneously broken' by the non-zero particular projected value of
   $\bv_4 \inn \TM_4$. 
Here the value $v^0 \neq 0$ in equation~\ref{v4vac} is simply the magnitude of the 
Lorentz 4-vector $\bv_4$, projected out of the components of $\bv_{27} \inn \htho$, 
onto the tangent space $\TM_4$; without the need for a `Mexican hat potential' such as 
equation~\ref{higgspot} to provide the mechanism for `spontaneous symmetry breaking' 
and induce a non-zero `vacuum value' for the field.
   This choice of vacuum value for $\bv_4 \inn \TM_4$ is in addition to the $\esi$ 
symmetry breaking through the necessary choice of a Lorentz subgroup $\sltc^1 \subset 
\esi$ associated with the external spacetime and as explored in the earlier sections 
of this chapter.

 Here the Lorentz symmetry itself, expressed with $\bh_2 = \bv_4 \! \cdot \! \bsig_4$  
as $\bh_2 \to S\bh_2 S^{\dag}$ in equation~\ref{vtoh} and \ref{hshs}, is reduced to 
the choice of $S\inn \sutw^1 \subset \sltc^1$ for a particular 4-vector $\bv_4 \inn 
\TM_4$, as a close analogy to the custodial symmetry $\sutw_{L+R} \subset \sutw_L 
\times \sutw_R$ for the Higgs case as described following equation~\ref{higgsvac2}.
 However the present theory may need to be developed beyond the model based on the 
$\sutw^2 \times \uo^2$ symmetry towards a more standard $\sutw_L \times \uo_Y$ 
electroweak theory before a more precise correlate of the `custodial symmetry' might 
be identified.

In the Standard Model electroweak theory three of the four  $\sutw_L \times \uo_Y$ 
generators  are spontaneously broken since they change the vacuum expectation value of 
the Higgs field ($\langle \phi \rangle$ in equation~\ref{higgsvac} or $\langle \Phi 
\rangle$ in equation~\ref{higgsvac2}), while maintaining the minimum of the Higgs 
potential ($V(\phi)$ in equation~\ref{higgspot} or $V(\Phi)$ in 
equation~\ref{higgspot2}).
The three degrees of freedom of the Higgs field associated with the broken generators 
give rise to the mass terms for the $W^{\pm}$ and $Z^0$ gauge bosons in the 
Lagrangian.  
  Fluctuations around the vacuum $v+H(x)$ in the fourth degree of freedom are 
associated with a mass term for the Higgs scalar particle, which is also proportional 
to the vacuum value of $v \simeq 246$~GeV, as described shortly after 
equation~\ref{mwmzcth}.
	 The unbroken $\uo_Q$ generator $Q$ leaves both $\langle \phi \rangle$ and 
$V(\phi)$ invariant, with the vacuum carrying zero electric charge, as described just 
after equation~\ref{higgsvac}.

	In the present theory, while the  $\bv_4 \inn \htwc$ vacuum value of 
equation~\ref{v4vac} is also invariant under the $\uo_Q$ symmetry, in fact with the 
$\Sbar^1_l$ action leaving all components of $\htwc \subset \htho$ unchanged,  the 
$\sutw^2 \times \uo^2$ generators associated with the $\tilde{W}^{\pm}_{\mu}$ and 
$\tilde{Z}_{\mu}$ fields    mix  the $\bv_4$ components in $\htho$ such that  $\vert 
\bv_4 \vert$ is not invariant, unlike the case for $V(\phi)$ in the standard 
electroweak theory as described above.
 While here we are dealing with a mock theory 
  the possibility of associating three of the four degrees of freedom for 
$\delta\bv_4(x)$ (in particular for the spatial components $\{v^1(x), 
v^2(x),v^3(x)\}$) with longitudinal components for the gauge bosons and hence masses 
for the  $\tilde{W}^{\pm}$ and $\tilde{Z}^0$ particles may assist in the 
identification of the physical $\sutw_L \times \uo_Y$ electroweak theory within the 
present framework.
  In any case here fluctuations in the Lorentz scalar magnitude $\vert \bv_4 \vert$ 
(closely related to   variation in the remaining temporal component $v^0 + 
\tilde{H}(x)$)   will be associated with the  Higgs field and corresponding massive 
boson particle state.  
That is, $\tilde{H}(x) \sim \delta \vert \bv_4(x) \vert$ in the present theory is 
provisionally correlated with the scalar field $H(x)$ in the Standard Model.

 In the full dynamical quantum theory it will of course be necessary to explain how 
the phenomenology of the Standard Model Lorentz scalar Higgs field and particle state, 
as observed in the laboratory, may be derived in detail from the components of the 
fundamental 4-vector field $\bv_4$, which will be referred to in this context as a 
`vector-Higgs'.
 This structure brings to mind
 other models for which there is no fundamental Higgs scalar field, with the latter 
for example composed out of fermion states. Hence here 
 we briefly review some of the properties of  technicolor models (\cite{Suss1,Suss2}, 
see also \cite{ssvz}) for comparison and contrast with the present theory.

  For QCD with two flavours $Q_{L,R} = \binom{u}{d}_{\! L,R}$ in the massless fermion 
limit the manifold of vacuum states, that is the $2 \times 2$ matrix of scalars 
$\langle Q_L \ol{Q}_R \rangle \neq 0$, breaks the global symmetry of the Lagrangian 
resulting in three Goldstone bosons corresponding to the three pion states $\pi^{\pm}$ 
and $\pi^0$. Coupling the quarks to $\sutw_L \times \uo_Y$ this gauge symmetry is 
broken by the vacuum since the $\sutw$ only couples to the left-handed fermions, while 
a $\uo_Q$ gauge symmetry is preserved. The symmetry breaking generates masses for the 
corresponding $W^{\pm}$ and $Z^0$ gauge bosons which are, however, too small compared 
to the empirical values since the pion decay constant $f_{\pi}$ is only around 93~MeV.

  Motivated by these observations and difficulties associated with a fundamental 
scalar Higgs in the Standard Model, a \textit{new} strongly interacting sector of 
fermions called `techniquarks' is postulated which couple to a new `technicolor' gauge 
symmetry SU$(N)_{tc}$. The techniquarks $T_{L,R} = \binom{U}{D}_{\! L,R}$ also 
transform under $\sutw_L \times \uo_Y$ but are singlets under the standard colour 
symmetry $\suth_c$. Scalar combinations of $T$ and $\overline{T}$ condense in the 
vacuum owing to the new strong technicolor interaction. As for the QCD case above the 
vacuum state is termed a `condensate' by analogy with phenomena in condensed matter 
physics, and in particular the formation of BCS pairs of electrons in 
superconductivity.

  For such a  model the technipion decay constant may be taken to be $F_{\Pi} \simeq 
246$~GeV  to replicate the masses of the $W^{\pm}$ and $Z^0$ as previously obtained 
with a scalar Higgs sector. The Standard Model relation $M_{W}/M_{Z} = \cos \theta_W$
   of equation~\ref{mwmzcth}
 is also reproduced. The techniquark Lagrangian includes kinetic terms of the form:
\begin{equation}
\label{lagtech}
  \lag_{tc} \sim \overline{T}_{L,R} \, \gamma^{\mu} (\pal_{\mu} \: + \:
   ig_{\! N} G_{tc\, \mu} \: + \: ig W_{\mu} \: + \: ig' B_{\mu}) \, T_{L,R}
\end{equation}
  with technicolor gauge field $G_{tc\, \mu}$ and coupling $g_{\! N}$ as well as the 
$\sutw_L \times \uo_Y$ gauge fields and couplings. In the quantum field theory masses 
for the $W^{\pm}$ and $Z^0$ gauge bosons are generated by the corrections introduced 
into the corresponding gauge boson propagators through the interaction terms in 
equation~\ref{lagtech}, with massless technipions effectively appearing as the 
longitudinal components of massive $W^{\pm}$ and $Z^0$  bosons. The low energy 
behaviour can be described by an effective phenomenological Lagrangian for the vacuum 
expectation value $\Psi(x) =  \langle T_L \ol{T}_R \rangle   \neq 0$   with:
\begin{equation}
\label{lagtechv}
  \lag_{\Psi} \sim F^2_{\Pi} \, \mbox{tr}(D_{\mu}\Psi^{\dag} \:\! D^{\mu}\Psi)
\end{equation}
  For the two techniquark model the scalar $\Psi(x)$ is a $2 \times 2$ matrix which 
plays the role of the scalar doublet of Higgs fields. The form of 
equation~\ref{lagtechv} is analogous to the kinetic term in the Higgs Lagrangian in 
the form of equation~\ref{laghcust} for the bi-doublet field $\Phi(x)$. 

  Masses for ordinary quarks and leptons are introduced by replacing the scalar Higgs 
in the Standard Model Yukawa terms of equation~\ref{Yukferm} by techniquark bilinears 
resulting in 4-fermion interactions with quartic terms such as the scalar:
\begin{equation}
\label{fmtech}
  \lag \sim \overline{Q}_L \, (\b1_2 T_L\ol{T}_R) \, Q_R
\end{equation}
  Here $Q_{L,R}$ are ordinary quarks which gain mass when the techniquarks form a 
condensate 
  $\langle T_L \ol{T}_R \rangle   \neq 0$. A suitable variety of quartic interactions 
and coupling parameters are needed to reproduce the empirical values for the standard 
quarks and leptons. Higher-order Lagrangian terms such as 6-fermion interactions may 
also be considered.
  
   As a theory of electroweak symmetry breaking without a fundamental scalar Higgs the 
above technicolor model has some resemblance with the present theory. The structure of 
the $2 \times 2$ scalar condensate  $\langle T\overline{T} \rangle$ as composed out of 
fermions indeed bares some resemblence to the spinor decomposition of the vector 
$\bh_2 = \chi\chi^{\dag} + \phi \phi^{\dag}$ of equation~\ref{htwocp}. However the 
latter expression merely represents the \textit{algebraic substructure} within the 
components of $\bh_2$ without the need of a new \textit{technicolor interaction} with 
gauge group SU$(N)_{tc}$ to condense fermions into a single object. 
 Hence for the present theory  the association of the scalar Higgs with the scalar 
magnitude $\vert \bv_4 \vert$ of the `vector-Higgs' $\bv_4 \equiv \bh_2$ is analogous 
to technicolor models with a scalar condensate composed of a new set of fermions, in 
that in both cases the need to postulate a fundamental scalar Higgs field is avoided.  
The absence of any observation of states belonging to a technihadron spectrum rules 
out a number of technicolor models.

  Unlike the case of the Standard Model
  for the present theory the mass for the $W^{\pm}$ and $Z^0$ states is expected to 
arise from terms in $\dmo$ which are linear in the gauge fields, as suggested in part 
by the relative couplings listed in equation~\ref{zwamass} and
 described earlier in this subsection.   These gauge field interaction terms are in 
fact similar in structure to those of the Lagrangian of equation~\ref{lagtech} and 
hence the origin of the gauge boson masses in the present theory also resembles the 
corresponding structure of the technicolor model. At a suitably low energy scale the 
present theory might also be compatible with an effective Lagrangian term quadratic in 
the gauge fields similar to equation~\ref{lagtechv} for the technicolor case.

 An origin for the masses of the fermion states  in the present theory is 
  also required, 
  as a  correlate of the `Yukawa interactions' introduced in the Standard Model 
Lagrangian. As described above for the $W^{\pm}$ and $Z^0$ gauge bosons, mass terms 
for the fermions might also be expected to involve a form of coupling with the 
external components $\bv_4 \inn \htwc \subset \htho$, which have been shown to exhibit 
properties analogous to those of the Standard Model Higgs field.
	 The expression of the full form of $\lvt$ as the determinant of $\mcX \inn \htho$ 
matrices, as written out in equation~\ref{detpmn}, includes the terms $p\vert b 
\vert^2$,  $m\vert c \vert^2$ and $n\vert a \vert^2$.
 With the projected components on $\TM_4$ related under $\lvfh$ in 
equation~\ref{lorform2}
 and adopting the $\bv_4 \equiv \bh_2$ components of
  equation~\ref{v4vac} with $v^0 = h$ 
  embedded within $\htho$ as for equations~\ref{hvvv} and \ref{hinhtho} the 
determinant can be written as:   
\begin{equation}
	\det(\mcX) \; = \;  h^2n - h \vert b \vert^2 
	        - h \vert c \vert^2  - n \vert a \vert^2
			  + 2\mbox{Re}(\bar{a}\bar{b}\bar{c})  \; =\; 1   \label{hexpan2}
\end{equation}  
  The $b,c$ internal components of $\htho$ hence have a multiplicative `coupling'  
with the
  vacuum value $h$ from the
 components of $\bv_4$ in the form $h(b\bar{b} + c\bar{c})$. These   are analogous to 
the Yukawa coupling terms between the Higgs field and fermion fields in Standard Model 
Lagrangian of equation~\ref{Yukferm}. This suggests that the fermion masses may be 
proportional to $h$, in a similar way that they are proportional to the Higgs field 
vacuum value $v$, equation~\ref{Masferm}, in the Standard Model.

  Since $\lvt$ is invariant under the transformations of $\esi$, and hence also under 
the external and internal subgroups, in the dynamics of the theory the actual mass 
terms may correspond to gauge invariant expressions as for the case of the Lagrangian 
approach.
Possible quartic or higher-order terms within a higher-dimensional  form of $\lv$ as a 
source of mass for the standard fermions, considered towards the end of 
section~\ref{secesef}, are also analogous  to the technicolor Lagrangian terms of the 
form of equation~\ref{fmtech}, at least with regard to their non-standard quartic 
nature.

 As for the case of the massive gauge bosons, for which quadratic mass terms do not 
arise in the basic elements of the present theory as discussed above, ultimately 
comparison between this theory and the Standard Model should be made at the level of 
empirical phenomena rather than a Lagrangian, which in any case is absent in the 
present theory. In addition to the field dynamics the role of mass in the calculations 
of quantum field theory  and its relation to `renormalisation' and physical particle 
states as studied in high energy physics experiments will need to be understood, as 
will be discussed in chapter~\ref{newapp}.

  Although only one generation of fermions has so far been considered in relation to 
the components of $\htho$, in the discussion following equation~\ref{abcmix} in 
subsection~\ref{strassy} it was hinted that for the full theory the existence of three 
generations of physical fermion states might ultimately be correlated with the 
existence of three types of $\sltwoo \subset \esi$ subgroup action, as introduced in  
equations~\ref{type1}--\ref{type3}.
From this perspective, given
 the asymmetric structure of the three terms $h b\bar{b}$, $h c\bar{c}$ and 
$na\bar{a}$ with respect to $h$ in equation~\ref{hexpan2}, and the need for 
renormalisation in the full theory, it is possible that the physical mass eigenstates 
of empirically studied particles will not be aligned neatly with the type $a=1,2$ and 
$3$ spinor $\theta^a$ components of $\htho$. Instead the choice of the external $\bv_4 
\inn \htwc\subset \htho$ may be skewed relative to the three generations of physical 
fermions, which may each then be related to $\bv_4$ via a \textit{continuous} (defined 
in \cite{Wang} p.127, as alluded to here after equation~\ref{type3}) rather than 
\textit{discrete} type transformation, leading to the spectrum of masses observed for 
the leptons and quarks.

	In the Standard Model the phenomena of CKM mixing in the quark sector relates to a 
mismatch between weak interaction and mass eigenstates as was reviewed in 
section~\ref{ewtatsm}.
 In this section we have established a correlation between the weak interaction and 
the  subgroups $\sutw^2$ and $\sutw^3$ of $\esi$ in the context of the present theory. 
	 If the mass states of three generations of quarks are skewed into the components 
of $\htho$ via a continuous type transformations as described above this contrasts 
with the charged weak interaction of the $\tilde{W}^{\pm}$ gauge bosons associated 
with the $\sutw^{2,3}$ actions which constitute a discrete type complement to the 
external $\sltc^1$ symmetry.
 This structure hence provides a possible basis for the mismatch between weak and mass 
eigenstates responsible for the CKM mixing between three generations of quarks, with a 
similar structure accounting for neutrino oscillations in the leptonic sector.

   As described earlier the  transformations for the symmetry group $\suth_c \subset 
\esi$, generated by  $\{\dot{A}_q, \dot{G}_l\}$, act on each of the three $a,b,c \inn 
\ooo$ components of $\htho$ in exactly the same way, in  manner that is independent of 
both discrete and continuous type transformations  (as contrasted with the $\suth_s$ 
actions after equation~\ref{rrrrrrrr}). In the present context this symmetry of the 
SU(3)$_c$ action on the three octonion components in $\htho$, 
together with its independence from the $\sltc^{1,2,3}$ and $S_l^{1,2,3}$  
transformations,
  is likely to be physically relevant for the observation of three generations of 
fermions, at least for the quark content and the corresponding phenomena of CKM mixing 
between generations with each of the three generations of quarks subject to an 
identical coupling to the $\suth_c$ strong interaction gauge bosons.

 However while the existence of three generations of fermions may ultimately be 
correlated with the three types of embedding of the $\theta^{1,2,3}$ components in 
$\htho$, as presumed for the discussion above, a somewhat larger space will be 
required to explicitly house all of the degrees of freedom, as we shall explore in the 
following chapter.
 The expansion of the form of temporal flow $\lv$ will be accompanied by a 
corresponding expansion of the group of symmetry transformations, opening up the 
possibility of identifying an internal $\sutw_L \times \uo_Y$ symmetry matching all 
the properties of the Standard Model.

   Finally in this section
   we consider further possible candidates for  the Standard Model $\sutw_L$ gauge 
symmetry  in terms of generators confined to the $\esi$ Lie algebra in the present 
theory. We return to $ \{\dot{R}_{z \pqg l}^{a},\dot{R}_{x \pqg z}^{a}, \dot{R}_{x 
\pqg l}^{a}, \dot{B}_{t \pqg x}^{a},
     \dot{B}_{t \pqg l}^{a}, \dot{B}_{t \pqg z}^{a}  \}$ as the three sets of six  
generators for $\sltc^{a}$ for each of $a=1,2,3$ 
 (the $a=1$ set was listed in equation~\ref{extlor6}) with the Lorentz Lie algebra of 
table~\ref{lorwang} satisfied in all three cases, as considered in 
subsection~\ref{strassy}. 
 In particular we look more generally to construct explicit SU(2) subgroups out of the 
collection of 12 generators of this form with $a=2$ or 3.
	 These form a subset of the 16 generators for the $\slthca_s$ subalgebra described 
in equation~\ref{slthcas} and presented explicitly within table~\ref{lbrota}, taking 
$q=l$, including the elements  $\dot{R}_{x \pqg l}^{2,3}$
  and $\dot{B}_{t \pqg z}^{3}$
 which do not belong to the preferred 78-dimensional basis for $\esi$.

   As for the case of the six generators of the Lorentz algebra, listed 
equation~\ref{extlor6} and table~\ref{lorwang}, in the \textit{complexified} Lie 
algebra the  $\sltc^2$ subalgebra of type 2  is also isomorphic to $\sutw \times 
\sutw$. The generators of these two $\sutw$s may be denoted $A^a$ and $B^b$, in 
correspondence with equations~\ref{jpmeja}--\ref{jpmez} and \ref{mrotgen}--\ref{msigj} 
(within the choice of sign conventions as noted for the latter equations), with:
\begin{eqnarray}
  \{A^1,A^2,A^3\} & = &
  \{\fh(\dot{R}_{z \pqg l}^{2} + i\dot{B}_{t \pqg x}^{2}),
    \fh(\dot{R}_{x \pqg z}^{2} + i\dot{B}_{t \pqg l}^{2}), 
	\fh(\dot{R}_{x \pqg l}^{2} + i\dot{B}_{t \pqg z}^{2}) \}  \nonumber  \\
 \mbox{and} \quad  \{B^1,B^2,B^3\} & = &
  \{\fh(\dot{R}_{z \pqg l}^{2} - i\dot{B}_{t \pqg x}^{2}),
    \fh(\dot{R}_{x \pqg z}^{2} - i\dot{B}_{t \pqg l}^{2}), 
	\fh(\dot{R}_{x \pqg l}^{2} - i\dot{B}_{t \pqg z}^{2}) \}  \nonumber  \\
 \mbox{such that:} \quad \lbrack i\dot{\Sbar}^1_l \, ,\, (A^1  \pm iA^2) \rbrack & = &
  \pm  (A^1 \pm iA^2)  \nonumber \\
 \mbox{and} \quad \lbrack i\dot{\Sbar}^1_l \, ,\, (B^1  \pm iB^2) \rbrack & = &
  \pm  (B^1 \pm iB^2)    \nonumber
\end{eqnarray}
 with the latter two expressions hence describing charge eigenstates. Such eigenstates 
might  in principle be correlated with charged gauge bosons $\tilde{W}^{\pm}$ as 
described in the previous two subsections.
   A similar analysis follows for the  $\sltc^3$ subalgebra of type 3. In addition to 
this by using the full set of 12 generators for both $\sltc^2$ and $\sltc^3$ two 
further $\sutw$s can be identified in the complexified algebra in this case with  
$A^a$ and $B^b$ composed as:
\begin{eqnarray}
   A^1 \,  = \, 
    \frac{1}{\sqrt{2}}(\dot{R}_{z \pqg l}^{2} + \dot{R}_{z \pqg l}^{3}
	    + i\dot{B}_{t \pqg x}^{2}  + i\dot{B}_{t \pqg x}^{3} )
		& \qquad &
   B^1 \,  = \, 
    \frac{1}{\sqrt{2}}(\dot{R}_{z \pqg l}^{2} + \dot{R}_{z \pqg l}^{3}
	    - i\dot{B}_{t \pqg x}^{2}  - i\dot{B}_{t \pqg x}^{3} )	
		 \nonumber \\
   A^2 \, = \, \frac{1}{\sqrt{2}}(\dot{R}_{x \pqg z}^{2} + \dot{R}_{x \pqg z}^{3}
	    + i\dot{B}_{t \pqg l}^{2}  + i\dot{B}_{t \pqg l}^{3} )
	    & \qquad &
   B^2 \, = \, \frac{1}{\sqrt{2}}(\dot{R}_{x \pqg z}^{2} + \dot{R}_{x \pqg z}^{3}
	    - i\dot{B}_{t \pqg l}^{2}  - i\dot{B}_{t \pqg l}^{3} )
		 \nonumber \\
   A^3 \, = \, \quad\;\: (\dot{R}_{x \pqg l}^{2} + \dot{R}_{x \pqg l}^{3}
	         + i\dot{B}_{t \pqg z}^{2}  + i\dot{B}_{t \pqg z}^{3} ) 
	    & \qquad &
   B^3 \, = \, \quad\;\: (\dot{R}_{x \pqg l}^{2} + \dot{R}_{x \pqg l}^{3}
	         - i\dot{B}_{t \pqg z}^{2}  - i\dot{B}_{t \pqg z}^{3} ) 			 
			\nonumber \\ & &   \label{jjj12}
\end{eqnarray} 
 However in this case none of the linear combinations $(A^1 \pm i A^2)$ or $(B^1 \pm i 
B^2)$ is a charge eigenstate of $i\dot{\Sbar}^1_l$ under the adjoint representation in 
the complexified $\esi$ algebra. In any case in order to identify a candidate for the 
$\sutw_L$ gauge symmetry of the Standard Model a \textit{real} $\sutw$ subalgebra of 
the \textit{real} form of $\esi$ is required. Such a compact real form of $\sutw$ can 
be obtained from a combination of the type 2 and 3 rotation generators with:  
\begin{equation}
  \{J^1,J^2,J^3\} = \{
    {\sqrt{2}}(\dot{R}_{z \pqg l}^{2} + \dot{R}_{z \pqg l}^{3} ),
	{\sqrt{2}}(\dot{R}_{x \pqg z}^{2} + \dot{R}_{x \pqg z}^{3} ),
		2	(\dot{R}_{x \pqg l}^{2} + \dot{R}_{x \pqg l}^{3} ) \} \nonumber
\end{equation} 
  However again here a complex linear combination of $J^1$ and $J^2$ fails to form a 
charge eigenstate under $i\dot{\Sbar}^1_l$. It can also be noted that the third 
generator $J^3$ is in fact equal to $-2\dot{R}_{x \pqg l}^{1} $, by 
equation~\ref{rsdeprrr}, which is a generator of the type 1 $\sutw^1$ rotation 
subgroup and hence not even independent of the external Lorentz symmetry $\sltc^1$ in 
terms of the vector space of generators. A similar observation applies to $A^3$ and 
$B^3$ in equation~\ref{jjj12}, and indeed was also noted for the gauge field
$\tilde{Z}_{\mu}(x)$ associated with $\dot{R}^1_{x \pqg l}$ for the mock $\sutw^2 
\times \uo^2$ theory before equation~\ref{wsbsas}. 
 These observations are not surprising since the Dynkin analysis for the Lie algebra 
of $\esi$ in section~\ref{dynkin} suggests that it is not possible to append any 
$\sutw$ subgroup alongside an $\sltc \times \suth \times \uo \subset \esi$ 
decomposition, as recalled near the opening of subsection~\ref{strassy}.

  However, of the possible $\sutw$ structures examined within the $\esi$ algebra, 
which in some sense are  complementary to the type 1 Lorentz subgroup $\sltc^1$, the 
subgroups $\sutw^2$ and $\sutw^3$ are the most promising in terms of properties 
\textit{resembling} the $\sutw_L$ gauge symmetry of the Standard Model, as has been 
described in this section.
 These observations supplement the identification of the subgroup 
    $\sltc^1 \times \suth_c \times \uo_Q \subset \esi$ in equation~\ref{esisubssu} 
which exhibits properties correlating closely with features of the Standard Model as 
described in sections~\ref{extsym} and \ref{intsym}.
 These observations also helped motivate the detailed study of the subgroup $\sutw^2 
\times \uo^2$  in this section in an attempt to account for aspects of electroweak 
theory within the scope of the $\esi$ action on the form of $\lvt$ in the present 
theory.

While a number of features of this mock electroweak theory resemble those of the 
Standard Model the lack of a complete match, together with the knowledge that the full 
Standard Model external and internal symmetry cannot be accommodated within $\esi$, 
now motivates the consideration of a higher-dimensional form of temporal flow,  with a 
higher degree of symmetry, with the goal of incorporating the physical $\sutw_L$ gauge 
symmetry. The aim will be to retain the significant traits of electroweak theory as 
identified in this section, within the breaking of the $\esi$ symmetry of $\lvt$ over 
the external spacetime $M_4$, in developing  a higher-dimensional 
  expression. As a further feature in reconstructing the full details of the Standard 
Model it will be necessary to explain how a set of Weyl spinors might be obtained from 
the $a\inn \ooo \subset \htho$ components listed in equation~\ref{suthuoa} for the 
$\nu$-lepton and $u$-quark states. This will be the topic of section~\ref{subehafws}. 
In section~\ref{secesef} an explicit higher-dimensional form of $\lv$ will be 
presented resulting in the identification of both left and right-handed Weyl spinors. 
Finally, bearing in mind the need to incorporate three generations of fermions, the 
possibility of a further expansion will be described in section~\ref{sosmfi}, with the 
features of the Standard Model so far identified in the context of the present theory 
then summarised.


\pagebreak
\chapter{Further Dimensions}
\label{secfd}

\section{Expanding $\htho$ and Further Weyl Spinors}
\label{subehafws}

  In aiming towards the identification of a physical $\sutw_L$ symmetry acting on 
doublets of $\sltc^1$ Weyl spinors in the present theory we first recall how the 
symmetry $\esi$ acting on the space $\htho$ relates to lower-dimensional forms of 
temporal flow expressed as $\lv$.
 In particular the $\esi$ symmetry of 
 the cubic form $\det(\mcX) = \lvt$, with $\mcX \inn \htho$, may be contrasted with 
the case of taking the full symmetry of $\lv$ to be the group $\spotn$ acting on the 
space $\htwo$, intermediate in size between $\htwc$ and $\htho$, such that the 
quadratic form $\det(X)$, with $X=\binom{p\;\;\bar{a}\;\!}{a\; m}\inn \htwo$, is 
preserved as the full form of temporal flow.
  With $\spotn$ being the double cover of $\sootn$ acting on the form $\lvte$ this is 
essentially the model described in section~\ref{reaic} as depicted in 
figure~\ref{mtogmaphr}.
 In this case there is an $\sltc\subset\spotn$ subgroup,
  based on the choice of an imaginary octonion unit such as $q=i$ for 
equation~\ref{xcomp},
  which acts as the external symmetry of 4-dimensional spacetime upon the subspace 
$\htwc\subset\htwo$ with the two-sided action of equation~\ref{mxmass} similarly as 
for the 10-dimensional case. This breaks the set of 45 generators of 
$\sltwoo\equiv\spotn$ acting on the space $\htwo$ to an internal 
 $\mbox{Stab}_2(T \! M_4)$ set of symmetry operations which here consist purely of 
transverse rotations amongst the remaining six imaginary units of the $a\inn \ooo$ 
component of $\htwo$.
 This is again sufficient to contain $\suth_c \times \uo_Q$ as an internal symmetry 
group.
  Indeed it can be seen from the Dynkin diagram of figure~\ref{dynkine}(b), by 
removing the central node with the most connections, that the Lie algebra so(10) has a 
breaking pattern to $\sltca \times \sutha \times \uoa$.

  However in the present theory we are not restricted to the consideration of extra 
\textit{spatial} dimensions, which might lead to the study of such a $\spotn$ symmetry 
of 10-dimensional spacetime. Here we are dealing with a higher-dimensional form of 
\textit{temporal} flow, allowing the structure of the above paragraph to be augmented 
to the group $\esi$ acting as the symmetry of a cubic form on the 27-dimensional space 
$\htho$. This larger structure incorporates three interlocking $\spotn$ actions, with 
associated representations on three spinor spaces $\theta^a\inn\ooo^2$, for $a=1,2$ or 
3, identified in the additional components as described in section~\ref{esitran}. 
 Within this structure the components of $\theta^1 = \binom{c}{\bar{b}}$ and $a$ 
within $\htho$ transform under the internal $\suth_c \times \uo_Q \subset \stab$ 
symmetry as a generation of leptons and quarks, with the appropriate fractional 
charges, as summarised in equations~\ref{suthuoth} and \ref{suthuoa}  of 
section~\ref{intsym}.
 The three-way embedding of $\sltwoo \subset \sltho$ is analogous to the empirical 
observation of three generations of leptons and quarks, although it remains to be seen 
whether these features do actually correlate.

  Further, in augmenting the $2 \times 2$ matrices in $\htwo$, upon which the symmetry 
of 10-dimensional spacetime may be represented,  to the $3 \times 3$ matrices of 
$\htho$, with the structure of a temporal symmetry, Weyl spinor states are identified 
in the $\theta^1$ components of the additional column of this matrix, as listed in 
equation~\ref{thcth234}, under the external $\sltc^1$ symmetry. 
 This is analogous to the motivation of the original Kaluza-Klein theories 
\cite{Kaluza,Klein} in which the $4\times 4$ metric $g_{\mu\nu}$ describing the 
gravitational field is augmented to the case of a 5-dimensional spacetime such that 
the four components $A_{\mu} = g_{\mu 5}$ in the extra column of the $5 \times 5$ 
metric describe the electromagnetic field.
(This structure was later generalised to incorporate the non-Abelian case in a larger 
spacetime with gauge field components $\omega^{\alpha}_{\ph{\alpha}a}$
 included in the metric, as described for equation~\ref{gmetug} in 
section~\ref{lccop}).
   In the present theory we identify fermion states in the extra temporal components, 
rather than gauge bosons in the additional metric components.

 For the case of $\esi$ acting on $\lvt$ not only is the above set of four Weyl 
spinors under the external $\sltc^1$ symmetry identified in the $\theta^1$ components, 
as described in section~\ref{extsym}, but they are also seen to be aligned with the 
internal $\suth_c \times \uo_Q$ transformation properties of the electron and a 
triplet of $d$-quark states, as described in section~\ref{intsym}.  In the present 
theory rather than generalising to a higher-dimensional spacetime here  the 
augmentation is applied to a multi-dimensional form of temporal flow, the concept of 
which is further compared and contrasted with the Kaluza-Klein approach in 
chapter~\ref{chaputtf}. These temporal structures are not restricted to a quadratic, 
or even cubic, form and in principle may be extended to a homogeneous polynomial form 
of arbitrary order.

  In the previous section we assessed the possibility of identifying the structure of 
electroweak theory in the breaking of the $\esi$ symmetry on the components of 
$\htho$. 
  For example the gauge field components corresponding to $\tilde{W}^{\pm}$ and 
$\tilde{Z}^0$ gauge bosons, identified in a mock electroweak theory based on the 
$\sutw^2 \times \uo^2 \subset \esi$ symmetry,
  impinge on the external $\htwc \equiv \TM_4$ components as described in 
equation~\ref{dmuhtwc}, which led to equation~\ref{zwamass}, providing a possible 
mechanism for identifying gauge boson mass terms
 analogous to the standard Higgs sector.
 In addition  
   the physical $W^{\pm}$ states of the Standard Model  act as charge raising and 
lowering transformations in interactions with left-handed doublets of leptons and 
quarks, that is $\binom{\nu}{e}_{\! L}$ and $\binom{u}{d}_{\! L}$ respectively for the 
first generation of fermions as listed in equation~\ref{multsf}, where \textit{each} 
component $\{\nu,e,u,d\}$ is a left-handed Weyl spinor under the external $\sltc$ 
symmetry.

One aim of the present theory has been to derive the spectrum of particle states of 
the Standard Model, and in particular the above doublets of left-handed fermions, from 
the components of $\htho$ under the broken $\esi$ action. Towards this end the action 
of the $\tilde{W}^{\pm}_{\mu}(x)$ fields associated with both the $\sutw^2$ and 
$\sutw^3$ transformations on the $e$-lepton and $d$-quark states, which have already 
been associated with the components of $\theta^1 = \binom{c}{\bar{b}}$ through 
equation~\ref{suthuoth}, should serve as a useful guide.
 As can be seen from equations~\ref{mth3} and \ref{abcmix} the $\sutw^2$ and $\sutw^3$ 
actions mix the components of $\theta^1 = \binom{c}{\bar{b}} \inn \ooo^2$ with the 
$a\inn \ooo$ component. Further,  the subcomponents of $a$ transform in the same way 
as those of $b$ and $c$ under the internal $\suth_c$ symmetry, containing a colour 
singlet and a colour triplet, and the corresponding elements of $a$ have the correct 
electromagnetic charges of $0$ and $\frac{2}{3}$  under the $\uo_Q$ generator 
$\Sbard^1_l$ to described the $\nu$-lepton and $u$-quark respectively as can be seen 
in equations~\ref{sbcharge} and \ref{suthuoa} and reviewed above. However, as 
described in table~\ref{sltcreps} under the external Lorentz transformations of 
$\sltc^1$ the `leptonic' components of $a$ transform as part of the vector $\bv_4$ 
while the `quark' components $a(6)$ are scalars; and hence these components appear to 
be unsuitable to describe fermion states. In this section we focus on the possible 
means of constructing these further required spinors.

  To see how such spinors may potentially arise and account for the $\nu$-lepton and 
$u$-quark states the 10 real components of $X\inn \htwo$, embedded in the type 1 
location of $\htho$ as depicted in equations~\ref{xoct3}, \ref{mxm3} and \ref{type1}, 
may be provisionally  composed in terms of the 16 real components of a new object 
$\thX = \binom{\bar{r}}{s} \inn \ooo^2$, with $r,s\inn \ooo$ and
\begin{equation}
  \label{xththx}
 X= \thX \thX^{\dag}
\end{equation}
  Hence the vector $X$ is considered to be the square of the spinor $\thX$, in the 
form as originally presented in equation~\ref{ththvec}.
  The compatibility relationship between the vector and spinor actions for the 
octonion case in equation~\ref{compati} also applies here since the $2\times 2$ 
transformation matrices $M \inn \sltwoo^1$ are required to have this property:
\begin{equation}
  \label{mxmththx}
  M \, X \, M^{\dag} = M(\thX \thX^{\dagger})M^{\dagger} 
   =  (M\thX)(M\thX)^{\dag}
\end{equation}
  In particular this shows that under the Lorentz transformations via $M=S \in 
\sltc^1$ the components of $\thX = \binom{\bar{r}}{s}$ decompose into a set of four 
Weyl spinors, as is the case for $\theta^1 = \binom{c}{\bar{b}}$ under the 
\textit{same} transformations as described in equations~\ref{thccbbh}--\ref{thcth234}. 
This hence shows how in principle further left-handed Weyl spinors may be indeed be 
identified within the $\esi$ action on $\lvt$ by opening up further dimensions through 
the decomposition of equation~\ref{xththx}.

  In terms of the real $p,m$ and octonion $a,r,s$ components equation~\ref{xththx} can 
be written in more detail as:
\begin{equation}
  \label{xththfull}
   X \, = \, \left( \begin{array}{cc} p & \bar{a} \\ a & m \end{array} \right) 
     \, = \, \thX\thX^{\dag}
	 \, = \, \left( \begin{array}{c} \bar{r} \\ s \end{array}  \right)
             \left( \begin{array}{cc} r & \bar{s} \end{array}  \right)
	 \, = \, \left( \begin{array}{cc} \bar{r}r & \bar{r}\bar{s} \\
	               sr & s\bar{s} \end{array} \right)
\end{equation}
  that is with $p = \vert r \vert^2$, $m=\vert s \vert^2$ and $a=sr$ (and as may be 
compared with equation~\ref{ththvec}).
 The fact that there are 16 real components of $\thX$ given the original 10 real 
components of $X$ is compatible with the underlying conceptual motivation of the 
present theory for which an $n$-dimensional form of temporal flow, such as $\lvt$, is 
derived given the original 1-dimensional progression of time, and represents what is 
essentially a \textit{further} generalisation and extension of this idea to a still 
\textit{higher-dimensional} structure. This structure provides a means to identify a 
set of Weyl spinors from the external $\sltc^1$ Lorentz action on $X$ which might in 
principle be associated with physical particle states. 

  We shall consider how these
     new spinors identified within the components of $\thX$  may correlate with the 
first generation $\nu$-lepton and $u$-quark states,   in a similar way that the 
$e$-lepton and $d$-quark states were identified within the components of $\theta^1$ 
according to equation~\ref{suthuoth}. These fermion states corresponding to $\sltc^1$ 
Weyl spinors will be required to be mutually oriented within the components of $\thX$ 
and $\theta^1$ with respect to  $\tilde{W}^{\pm}$ gauge bosons which mix the 
corresponding leptonic or quark states, raising or lowering the electromagnetic charge 
of the fermion state by one unit. This $\sutw$ mixing of Weyl spinors \textit{between} 
the components of $\thX$ and $\theta^1$ should be analogous to the $\suth_c$ mixing of 
the $\theta_{i,j,k}$ Weyl spinors \textit{within} the $\theta^1$ components as 
described on the left-hand side of table~\ref{suttan}.

   The close relationship between the 10-dimensional vector $X\inn\htwo$ and 
10-dimensional vectors of the form $\theta \theta^{\dag} \inn \htwo$ is also exhibited 
by the 10-dimensional Lorentz inner product in the final term of equation~\ref{detx3} 
for $\det (\mcX )$ with $\mcX \inn \htho$. 
 The 10-dimensional type 1 Lorentz transformations $\sltwoo^1$ leave both terms of 
$\det (\mcX)$, that is both $\det (X)n$ and $2X\cdot(\theta^1 {\theta^1}^{\dag})$, 
invariant. However the subgroups $\sutw^{2,3} \subset \esi$ mix the components of 
these two terms, as well as mixing components between $X$ and $\theta^1$, and it is 
these properties which might be studied in order to describe for example a $u 
\leftrightarrow d$-quark interaction in terms of  $\thX \leftrightarrow \theta^1$ 
components.

 Substituting the $\thX = \binom{\bar{r}}{s}$ components $r$ and $s$ in place of $p,m$ 
and $a$ from equation~\ref{xththfull} into the expression for $\det(\mcX )$ in 
equation~\ref{detpmn} leads directly to:
\begin{equation}
  \det({\mathcal X}) = \vert r \vert^2 \vert s \vert^2 n
   - \vert r \vert^2  \vert b \vert^2 - \vert s \vert^2 \vert c \vert^2 
       - n\vert sr \vert^2  
                     + 2 \mbox{Re}(\bar{r}\bar{s}\bar{b}\bar{c})    
			\label{detrsn}
\end{equation}
   Since for any $r,s$ in the division algebra $\ooo$ we have
     $\vert s \vert \vert r \vert  =  \vert sr \vert$ the first and fourth terms above 
cancel, leaving a quartic expression in $r,s,b,c \inn \ooo$.
 Hence in principle equation~\ref{detrsn} describes a homogeneous form $\lv$ with 32 
dimensions, namely the real components of $\{r,s,b,c\} \inn \ooo$, with a symmetry 
group deriving from the action of $\esi$ on $\lvt$.

 However, one significant difference between any elements $X\inn \htwo$ and $\theta 
\in \ooo^2$ with $\theta \theta^{\dag} \inn \htwo$ is that in the former case $\det(X) 
\inn \rrr$ may take arbitrary real values while in the latter case we necessarily have 
$\det(\theta\theta^{\dag}) = 0$, as was described in equation~\ref{ththdet0}. This can 
be seen here from the right-hand side of equation~\ref{xththfull} for which 
$\det(\thX\thX^{\dag}) = \vert r \vert^2 \vert s \vert^2 - \vert r \vert^2 \vert s 
\vert^2 = 0$, for any $\thX = \binom{\bar{r}}{s}$, and accounts for the cancellation 
of the two quintic terms in equation~\ref{detrsn}. It also clearly implies that the 
decomposition of $X$ as suggested in equations~\ref{xththx} and \ref{xththfull} is not 
possible for the general case if $\det(X) \neq 0$.

 This apparent incompatibility may be remedied by further generalising 
equations~\ref{xththx} and \ref{xththfull} by introducing an additional spinor $\phX = 
\binom{\bar{r}'}{s'} \inn \ooo^2$, with identical transformation properties as the 
original $\thX$ in equation~\ref{mxmththx}, such that:
\begin{eqnarray}
 X & = & \thX\thX^{\dag} \, + \, \phX\phX^{\dag}  \label{xththpx} \\
 \mbox{with} \qquad M \, X \, M^{\dag}  & = & M(\thX \thX^{\dagger})M^{\dagger} \, + 
\,
                                        M(\phX \phX^{\dagger})M^{\dagger}  \nonumber  
\\
										 & = &
                                   (M\thX)(M\thX)^{\dag} \, + \,
								   (M\phX)(M\phX)^{\dag}   \label{mxmththp}
\end{eqnarray} 
 This introduces a further 16 real parameters in $\phX$ which transform as a further 
set of four Weyl spinors under the Lorentz actions with $M = S \inn \sltc^1$. The 
value of $\det(\thX\thX^{\dag}  + \phX\phX^{\dag}) \inn \rrr$ may now be compatible 
with the determinant of $X$ in the general case.
  This is analogous to the case of the 4-dimensional Lorentz vector decomposition in 
equation~\ref{htwocp}, with the spinor substructure of the vector $X$ potentially 
providing a source of microscopic physical structure, as suggested after 
equation~\ref{htwocp} for the 4-vector field $\bv_4(x)$. The form of $\mcX$ in 
equation~\ref{htho} and $\det(\mcX)$ in equation~\ref{detx3} when substituting in 
equation~\ref{xththpx} become:
\begin{eqnarray} 
 \label{hthoext}
   \mcX & = & 
   \left( \begin{array}{ccc}
    \vert r \vert^2 + \vert r' \vert^2 & \bar{r}\bar{s} + \bar{r'}\bar{s'}   & c  \\
                             sr + s'r' & \vert s \vert^2 + \vert s' \vert^2  & \bar{b}        
\\
                               \bar{c} &        b                            & n 
          \end{array}  \right)   \inn \htho    \\
	\det(\mcX) & = &  \det(\thX\thX^{\dag}  + \phX\phX^{\dag})n \, + \,
	       2(\thX\thX^{\dag}  + \phX\phX^{\dag})\cdot (\theta^1\theta^{1\dag}) 
		   \label{xththfull2}
\end{eqnarray} 
    Here the first part of the expression for $\det(\mcX)$ contains quintic terms, 
which now do not cancel in general as they did in equation~\ref{detrsn}, while the 
second part contains further quartic terms. This expression hence represents an 
\textit{inhomogeneous} polynomial form, and hence deviates from  the  form of $\lv$ in 
equation~\ref{lv} of section~\ref{gfotf} on incorporating a further higher-dimensional 
dissolving of specific components, such as those of $X$ above. The potential physical 
consequences of such a mathematical possibility in relation to the  original 
homogeneous form of $\lvt$ requires further clarification. While such inhomogeneous 
expressions may be explored upon examination they appear indeed inconsistent with the  
underlying  conceptual basis employed in deriving equation~\ref{lv}, as relating to  
infinitesimal intervals of temporal flow $\delta s$. On the other hand expressions 
such as equation~\ref{xththfull2} may represent an intermediate step towards the 
derivation of a higher-dimensional homogeneous form, such as a purely quintic 
expression for $\lv$, as will be proposed hypothetically in section~\ref{sosmfi} in 
the light also of the physical motivation described below.

 The need to generalise from equation~\ref{xththx} motivated the introduction of a 
combination of  spinors, $\thX$ and $\phX$, in equation~\ref{xththpx}.
 This latter expression  $X= \thX\thX^{\dag}
  +\phX\phX^{\dag}$, with $\thX$ and $\phX$ having identical transformation properties 
under the external and internal symmetry actions,
   suggests that for example the first generation $u$-quark \textit{and} the second 
generation $c$-quark
  states might be accommodated in the $\thX$ and $\phX$ components respectively. 
 
 As described in the previous section the $\sutw^{2,3} \subset \sltc^{2,3}$ actions 
indicated in equation~\ref{abcmix} transform  the octonion components   $(a) 
\leftrightarrow (b,c)$ in a seemingly asymmetric way.
 Hence, with $a=sr$ from equation~\ref{xththfull}, a combination of 
$\dot{\Sigma}^{(2)\pm}$ and $\dot{\Sigma}^{(3)\pm}$, of equations~\ref{sigcomb} and 
\ref{sig3pm}, appear to be needed in order to transform any of the four Weyl spinors 
located in the components of $\theta^1= \binom{c}{\bar{b}}$ into the doublet partner 
located within the corresponding components of 
$\thX= \binom{\bar{r}}{s}$.
  
 Together the  observations of the above two paragraphs
 suggest the possibility of a Cabibbo-like mixing between the first two generations of 
quarks. 
 Empirically the gauge action $\sutw_L$ mixes the quark states $(u) \leftrightarrow 
(d\cos\theta_c + s\sin\theta_c)$, where $s$ denotes the strange quark and $\theta_c$ 
is the Cabibbo angle which may be generalised to the full CKM matrix for three 
generations, as described towards the end of section~\ref{ewtatsm}.
 In the Standard Model the coupling between the first and third quark generations is 
very small. In the present theory quark exchanges via the $\tilde{W}^{\pm}$ bosons 
associated with the $\sutw^{2,3}$ symmetry may open up a full set of possible states 
associated for example with $X = \thX\thX^{\dag}
  +\phX\phX^{\dag} +\psX\psX^{\dag} $, further augmenting equation~\ref{xththpx} and 
lifting the degeneracy to a complete set of three generations interacting via a 
CKM-like mixing, although a similar degeneracy will also need to be identified 
relating to the $\theta^1$ components. In principle
 this offers a possible  means of accommodating  three generations of fermions into 
the theory which may \textit{not} 
  relate directly to the existence of the three types of $\sltwoo$ embedded within 
$\sltho$ as described in equations~\ref{type1}--\ref{type3}.

  The full study of these phenomena, as discussed in the previous subsection, will 
require the identification of each of the physical mass states, as observed in the 
laboratory, 
in relation to the components of 
  $\theta^1 = \binom{c}{\bar{b}}$ and  $\thX = \binom{\bar{r}}{s}$ for example. Around 
equation~\ref{hexpan2} it was proposed that fermion masses will relate to the
   degree of coupling with the scalar magnitude $\vert \bv_4 \vert = h$, and in 
particular the
    `vacuum value' of $h$ in the projection of $\bv_4\inn \TM_4$, by analogy with 
Higgs phenomena  
   in the Standard Model. Since terms containing both the $a\inn \ooo$ and the $v^0=h$ 
components of $\mcX \inn \htho$ do not appear in $\mbox{det}(\mcX)$ in 
equation~\ref{hexpan2} an explicit higher-dimensional form of $\lv$, such as 
introduced in the following section, may be needed for further study of possible mass 
terms.
   On the other hand composition with the scalar field $n(x)$, such as for the $\vert 
a \vert$ term in equation~\ref{hexpan2},  might also provide a source of fermion mass 
terms.

   With the possible generalisation of equation~\ref{xththpx} and the above weak 
interactions in mind it is also necessary to determine the internal $\suth_c \times 
\uo_Q$ symmetry transformations of the $\thX$ components, with similar transformations 
implied for $\phX$. 
 The $\mbox{SO}(8) \subset \sltho$ subgroup can be generated by the composition of $3 
\times 3$ matrices $\mcM^{(1)}$ of type 1 in the form of equation~\ref{mqqbaro} based 
on the $2 \times 2$ matrices $M = \binom{q \;\; 0}{0 \;\; \bar{q}}$.
 Acting via the conjugation $X \to M X M^{\dag}$ on $X$ in equation~\ref{xththfull} 
the component $a\inn \ooo$ transforms under the vector representation of SO(8) while 
the $\thX$ components $s\inn \ooo$ and $r\inn \ooo$ transform individually via the 
spinor and dual spinor representations of SO(8) as can be seen via 
equation~\ref{mxmththx}.

  These three 8-dimensional representations are mutually related through triality maps 
described in the opening of section~\ref{earlyo} and  around equation~\ref{trialabc}  
-- the triality structure in the present mathematical context is also described in 
more detail in (\cite{Wang} pp.77--80 and 120--126).
  Further, 
elements of the $\gt \subset \mbox{SO}(8)$ octonion automorphism subgroup transform 
the vector, spinor and dual spinor, here represented by
the octonions $a, s$ and $r$ respectively, in precisely the same way
 via symmetric, left and right multiplication by the same sequence of octonions.
  This property of termed `strong triality' in (\cite{Wang} p.123).
  
 Hence under the colour gauge symmetry $\suth_c \subset \gt \subset \mbox{SO}(8)$ each 
of the octonion components of $\thX = \binom{\bar{r}}{s}$ transform in the same way as 
the component $a$, and hence also in the same way as the octonion components of  
$\theta^1 = \binom{c}{\bar{b}}$. This means that the four Weyl spinors, obtained from 
the reduction of $\thX$ under the external $\sltc^1$ action, transform as a leptonic 
singlet and quark triplet under the internal $\suth_c$, just as is the case for 
$\theta^1$ as summarised in equation~\ref{suthth}. There then remains the question of 
how the $\uo_Q$ charges for the $\thX$ Weyl spinors compare to those for the 
$\theta^1$ Weyl spinors deduced for equation~\ref{suthuoth}. 

   The electromagnetic $\uo_Q$ generator $\Sbard^1_l$ is contained in the group 
$\mbox{SO}(7) \subset \mbox{SO}(8)\subset \sltwoo^1$, but unlike the $\suth_c$ 
generators $\{\dot{A}_q,\dot{G}_l\}$ it is not contained within the subgroup $\gt 
\subset \mbox{SO}(7)$.
 Although $\dssl^1_l = \dot{S}^1_l$ the simpler unnested single group action 
$\ssl^1_l$ of equation~\ref{sqdiag} is not used here since it is not constructed as a 
`compatible' action in the sense of equation~\ref{compati} or \ref{mxmththx}. We hence 
employ $S^1_l$ as a member of the preferred basis incorporated into the $\esi$ Lie 
algebra composition as discussed following equation~\ref{sqdiag}.

 The group action of ${S}^1_l(\alpha)$ on the components of 
$X = \binom{p\;\;\bar{a}}{a\;m}$ and $\thX = \binom{\bar{r}}{s}$ may be determined 
from  table~\ref{opairs} and equation~\ref{srotn}  together with table~\ref{mtran45} 
and equation~\ref{rrsmat} 
  as the type 1 nested compositions:
\begin{eqnarray}
  X & \to & R^1_{\il , i}(\alpha) \,\scirc\,
      R^1_{\jl , j}(\alpha)  \,\scirc\, R^1_{\kl , k} (\alpha)\, X
                            \nonumber \\
 \binom{p\;\;\: \bar{a}}{a \;\; m} & \to &  \nonumber \\
   & &  \!\!\!\!\!\!\!\!\!\!\!\!\!\!\!  \!\!\!\!\!\!\!\!\!\!\!\!\!\!\!\!\!\!\!\!
   M_{\il , i2} (  M_{\il , i1} ( M_{\jl , j2} (  M_{\jl , j1} (M_{\kl , k2} (  M_{\kl 
, k1} 
        \binom{p\;\;\: \bar{a}}{a \;\; m}
	M^{\dag}_{\kl , k1}) M^{\dag}_{\kl , k2})   M^{\dag}_{\jl , j1}) M^{\dag}_{\jl , 
j2}) 
	 M^{\dag}_{\il , i1}) M^{\dag}_{\il , i2}  \nonumber \\
	   & & \nonumber \\
  \thX & \to & R^1_{\il , i}(\alpha) \,\scirc\, 
     R^1_{\jl , j}(\alpha)  \,\scirc\, R^1_{\kl , k} (\alpha)\,\thX
                          \nonumber \\
   \binom{\bar{r}}{s} & \to &
     M_{\il , i2} (  M_{\il , i1} ( M_{\jl , j2} (  M_{\jl , j1} (M_{\kl , k2} (  
M_{\kl , k1} 
	    \binom{\bar{r}}{s}  ))))) \label{phxtran}
\end{eqnarray}  
  where the expression for the $\thX$ transformation is a consequence of 
equation~\ref{xththx} or \ref{xththpx} together with the compatibility of the 
${S}^1_l(\alpha)$ action as defined in equation~\ref{compati} or \ref{mxmththx}.
    The following notation for the 6 nested actions, with factors of $\pm 1$ 
accumulated into the initial negative signs, is introduced to simplify subsequent 
equations:   
\begin{eqnarray}
  N_6(  & \equiv & -(\il \cos \frac{\alpha}{2} + i \sin\frac{\alpha}{2})\big( \il 
\big(
                     (\jl \cos \frac{\alpha}{2} + j \sin\frac{\alpha}{2})\big( \jl 
\big(
					 (\kl \cos \frac{\alpha}{2} + k \sin\frac{\alpha}{2})\big( \kl  
\nonumber \\
  )N^{\dag}_6 & \equiv & -\kl \big) (\kl \cos \frac{\alpha}{2} + k 
\sin\frac{\alpha}{2}) \big)
                           \jl \big) (\jl \cos \frac{\alpha}{2} + j 
\sin\frac{\alpha}{2}) \big) 
                           \il \big) (\il \cos \frac{\alpha}{2} + i 
\sin\frac{\alpha}{2}) \nonumber
\end{eqnarray}
  From equations~\ref{phxtran}
     the three octonion components $a,s$ and $r$ then transform  under  
${S}^1_l(\alpha)$ in the manner (as may be compared with equation~\ref{trialabc}):
\begin{equation}
 \begin{array}{rcl}
   a & \to &  N_6( a )N^{\dag}_6     \\
   s & \to &  N_6( s )))))       \\
   \bar{r} & \to &  N_6( \bar{r} )))))    \\
   \mbox{hence} \quad r & \to & ((((( r )N^{\dag}_6
  \end{array} 
    \label{asrtri}
\end{equation}

  While here we have symmetric, left and right multiplication on $a$, $s$ and $r$ 
respectively by the same sequence of octonions, as described by `$N_6($' and 
`$)N^{\dag}_6$', these three actions are \textit{not} mutually related by triality 
since they do not describe the same transformation on $\ooo$. Indeed since the action  
${S}^1_l(\alpha)$ is \textit{not} part of the $\gt \subset \mbox{SO}(8)$ subgroup it 
does not exhibit the property of `strong triality', but rather participates in the 
SO(8) triality structure collectively when further generators are considered.

 Taken at face value equations~\ref{asrtri} imply that the electric charge identified 
with $\Sbard^1_l$  for the $\bar{r},s$ components here hence \textit{differs} from 
that for the $a$ component. This is an undesirable feature which means that the 
$\binom{0}{2/3}$ charge structure observed for the  component parts of $a$ under 
$\Sbard^1_l$ in equations~\ref{sbcharge}
 and \ref{suthuoa}, as sought for $\nu$-lepton and $u$-quark fermion states, has 
apparently been \textit{lost} for the set of $\sltc^1$ Weyl spinors in $\thX = 
\binom{\bar{r}}{s}$. In fact, as expected from the compatibility of the $S^1_l$ group 
action, the $\uo_Q$ transformations  of the spinor  $\thX = \binom{\bar{r}}{s}$ are 
identical to those of the spinor $\theta^1=\binom{c}{\bar{b}}$ 
  and hence both spinors possess  the same $\Sbard^1_l$ charge
 values of 1 and $\frac{1}{3}$, as described for equation~\ref{suthuoth}, which have 
been associated with the $e$-lepton and $d$-quark states.
 Indeed, as implied in the discussion around equation~\ref{mxmththx}, the components 
of $\thX$
  transform in precisely the same way as those of $\theta^1$ under the action of the 
subgroup 
  $\sltc^1\times \suth_c \times \uo_Q \subset \sltwoo^1$  owing to the compatibility 
requirement of all $M\inn \sltwoo$ group transformations.

 A possible solution would be to maintain the same $S^1_l (\alpha)$ action on $X$ 
while redefining the transformation properties of $\thX$ under the $\uo_Q$ subgroup.
 That is, with $X \to \thX\thX^{\dag}$  in equation~\ref{xththx} or \ref{xththpx}, on 
 expanding the 10-dimensional space for $X$ to the 16-dimensional space for $\thX$ 
there is a degree of redundancy  in the transformation properties of $\thX$ under 
$\uo_Q$ provided $X$ transforms in the same way. While equation~\ref{phxtran} 
represents the simplest assumption for the action of $S^1_l (\alpha)$ on the 
components of $\thX$, based on the notion of compatibility in equation~\ref{mxmththx}, 
in principle there may be further choices such as:
\begin{eqnarray}
   X & \to & N_6( \, \b1_2 \; X \; \b1_2 \, )N^{\dag}_6 \nonumber \\
  & = &  N_6( \, \b1_2 \; \thX \thX^{\dag} \; \b1_2 \, )N^{\dag}_6 \nonumber \\
 \mbox{with} \quad \thX & \to &  N_6( \, \b1_2 \; \thX  )N^{\dag}_6
   \label{thtranew}
\end{eqnarray}
  rather than $\thX  \to N_6( \b1_2 \, \thX )$,
  although with care needed to take into account the non-associative
   properties of  octonion composition. On employing equation~\ref{thtranew} in place 
of equation~\ref{asrtri} the action of the $\uo_Q$ symmetry $S^1_l(\alpha)$ on $a,s,r 
\inn \ooo$ would be expressed uniformly as:
\begin{eqnarray}
   a & \to &  N_6( a )N^{\dag}_6   \nonumber  \\
   s & \to &  N_6( s )N^{\dag}_6   \label{chcont} \\
   r & \to &  N_6( r )N^{\dag}_6   \nonumber 
\end{eqnarray}

    Hence in this case the $\binom{0}{2/3}$ charge structure of $a$ under $\Sbard^1_l$ 
would also apply to the  $r,s \inn \ooo$  components of $\thX$. These $\uo_Q$ charges 
for $\thX$ are here \textit{contrived} by inserting further nested $N_6^{(\dag)}$ 
actions for  $\thX^{(\dag)}$ in the appropriate places for equation~\ref{thtranew}. 
However the introduction of the 16 real component object $\thX = \binom{\bar{r}}{s}$ 
itself, obtained from the 10 real component object $X$ in equation~\ref{xththx}, 
\textit{is}  
contrived and defined in order to construct a possible set of doublet partners for the 
$\theta^1 = \binom{c}{\bar{b}}$ components under the action of  $\tilde{W}^{\pm}$ 
charge raising and lowering operators.
  With physical states associated with definite representations under the Lorentz 
symmetry of 4-dimensional spacetime $M_4$, such as the above Weyl spinors within 
$\theta^1$ and $\thX$, the purpose here is to demonstrate the mathematical 
\textit{possibility} of recovering weak interactions between fermions, such as those 
mediated via $\tilde{W}^{\pm}$ gauge bosons, from within the present theory. A 
mathematical \textit{justification} for  transformations such as those of
 equation~\ref{chcont} might ultimately be sought within a natural higher-dimensional 
homogeneous form  $\lv$.

   Hence $\thX = \binom{\bar{r}}{s}$ provisionally describes a possible mathematical 
construction which possesses the appropriate transformation properties under the 
external symmetry $\sltc^1$ and internal symmetry $\suth_c \times \uo_Q$ to represent 
the neutral $\nu$-leptons and charge-$\frac{2}{3}$ $u$-quarks. These latter states are 
related to the charge-1 $e$-leptons and charge-$\frac{1}{3}$ $d$-quarks of $\theta^1 = 
\binom{c}{\bar{b}}$ via $\tilde{W}^{\pm}$ interactions. The relation between the 
fermion states and the unit charge raising and lowering action of the 
$\tilde{W}^{\pm}$ oriented with respect to the $\thX$ components serves to mutually 
  motivate and aid the determination of both the Weyl spinor states and the internal 
gauge symmetry. From this point of view, with $X\inn \htwo$ composed in the form of 
equation~\ref{xththpx}, the $\tilde{W}^{\pm}$ and $\tilde{Z}^0$ may derive from a weak 
$\sutw \times \uo$ action which is less directly related to the $\sutw^{2,3} \times 
\uo^{2,3}\subset \esi$ subgroups than suggested in the previous section.

  To conclude the above discussion, while equation~\ref{xththpx} describes a possible 
way to include the required further Weyl spinor states there are several questions 
which remain to be resolved -- these include the means by which 
equation~\ref{xththfull2} might be incorporated into a higher-dimensional  
\textit{homogeneous} form of $\lv$ with a larger symmetry group incorporating an 
appropriate electroweak $\sutw_L \times \uo_Y$ subgroup action and the means by which 
the electromagnetic charges for the $\nu$-lepton and $u$-quark states may be retained 
from the $\Sbard^1_l$ action on the components of the original $a \inn \ooo$ component 
of $\htho$. Further, while here we are working at the level of the basic group and 
representation structure, a leading question in the full theory will be to understand 
the nature of physical particle states in general, and in particular for the gauge 
bosons, three generations of fermions and also a Higgs state as empirically observed.

  The above identification of fermions by opening up the  10-dimensional vector $X 
\inn \htwo$ according to equation~\ref{xththpx} can similarly  be applied to the 
4-dimensional Lorentz vector $\bh_2 \inn \htwc$, for the subspace $\htwc \subset 
\htwo$, according to the decomposition of equation~\ref{htwocp}. This latter Weyl 
spinor substructure of the \textit{vector} $\bh_2 = \chi\chi^{\dag} + \phi\phi^{\dag}$ 
in terms of the \textit{spinors} $\chi,\phi$ has some analogy with
  composite Higgs and technicolor models in which fermion states are combined in 
\textit{scalar} condensates in the vacuum, hence replacing the fundamental scalar 
Higgs of the Standard Model, as reviewed in subsection~\ref{suboomahp}.
 Here opening up the $\bh_2 \inn \htwc$ components to form spinors in this way 
incorporates the $a_{1,l}$ components of equations~\ref{sbcharge} and \ref{suthuoa}, 
leaving the set of $\sltc^1$ Lorentz scalars in $a(6)$ which transform under the 
internal $\suth_c \times \uo_Q$ symmetry as a colour triplet of $u$-quarks. In 
principle each of the three  scalars 
   $a_{\il,i}$, $a_{\jl,j}$ and $a_{\kl,k}$ of equation~\ref{suthuoa} might be 
composed in terms of a suitable scalar product of Weyl spinors of the form 
$\chi^{\dag}\chi$ with the aim of describing the fermion nature of $u$-quarks under 
the external symmetry.

 These spinor decompositions involve $\ccc$ or $\hhh$ subalgebras of $a \inn \ooo$. 
Hence, 
 in comparison with equation~\ref{thtranew}, 
   an extra intermediate factor of the form $e^{-l\beta}e^{l\beta}$
  might be inserted for the decomposed vector $\bh_2 \to \chi\chi^{\dag}$ in 
 augmenting the $\Sbar^1_l(\alpha)$   action, 
 to ensure the charge neutrality of the candidate neutrino states, avoiding any  
complication due to the non-associative nature of the octonions. With the possibility 
of a similar insertion for the $u$-quark states  in principle the $\uo_Q$ charges 
under $\Sbar^1_l$ of $\binom{0}{2/3}$, as originally found for the 
$\binom{a_{1,l}}{a(6)}$ components in equation~\ref{suthuoa}, might be maintained 
under the spinor decomposition of these components which might hence indeed be 
associated with $\binom{\nu}{u}$ fermion states. Again, while such a structure might 
be mathematically contrived as a proof of principle, ultimately the aim will be to 
account for the external and internal symmetry properties of all Standard Model 
fermion states in a natural manner in the components of a higher-dimensional 
homogeneous form of $\lv$.

  In keeping the $\binom{\nu}{u}$ particle type interpretations aligned with the 
$\binom{a_{1,l}}{a(6)}$ components in this way, based on the spinor decomposition of 
$\bh_2 \inn \htwc$ (unlike the case for the collective decomposition $X \to \thX 
\thX^{\dag} \inn \htwo$ as originally considered in equation~\ref{xththx}), also 
suggests that exchanges with the corresponding  doublet partners $\binom{e}{d}$ in the 
components of $\theta^1$ might be mediated by $\tilde{W}^{\pm}$ states closely 
associated with the $\sutw^{2,3} \subset \esi$ actions as described in 
section~\ref{ewtfesb} for the mock electroweak theory. 
  Based on the $\{1,l\}$ base units these $\sutw^{2,3}$ actions preserve the 4-way 
decomposition of octonion components as listed in equation~\ref{lijksets} for the 
transformations between the $a,b,c \inn \ooo$ components of $\htho$ of the kind 
described in equations~\ref{mth3} and \ref{abcmix}. This is consistent with an 
electroweak $\sutw$ gauge symmetry action on independent lepton and quark doublets as 
accommodated respectively within the 
  $\{1,l\}$ and $(\{\il,i\}, \{\jl,j\}, \{\kl,k\})$ components of both $a$ and 
$\theta^1 = \binom{c}{\bar{b}}$.

  In particular the neutrino state is associated with the $a_{1,l}$ components, which 
also form part of the 4-vector $\bh_2 \inn \htwc$ as projected onto the external 
spacetime $\TM_4$.
 In the present theory the scalar degree of freedom 
 $\vert \bh_2 \vert = \sqrt{\det(\bh_2)}$, or an alternative scalar combination of the 
spinor components $\chi,\phi$ in the implicit substructure of $\bh_2$ described above, 
will provide a candidate for the origin of the observed Higgs particle as described in  
subsection~\ref{suboomahp}.  This apparent inconsistency with the degrees of freedom 
of   $\bh_2 \inn \htwc$ seemingly required to play a double role as the correlate of 
both the neutrino and the Higgs will be resolved in the following section.

  The approach of the present theory is to gently coax the known properties of the 
Standard Model out of the symmetry breaking structures of forms of $\lv$ over the base 
manifold $M_4$,
 with an awareness of the known empirical features
 while being conscious of not contriving them ultimately for the complete theory.
  However the possibility of contriving an augmented structure based on the components 
of $a \inn \htho$ under the $\esi$ action transforming as a $\nu$-lepton and $u$-quark 
under $\sltc^1 \times \suth_c \times \uo_Q \subset \esi$, as described in this 
section, and with the further possibility of incorporating a second and third 
generation through additional spinors such as described for equation~\ref{xththpx}, is 
at least consistent with the possibility that a higher-dimensional form of $\lv$, for 
example with an $\ee$ symmetry as will be considered in section~\ref{sosmfi}, might 
naturally contain these structures. 
 Similarly the empirical properties of left and right-handed spinors, as we recap 
below, will contribute to the motivation for the study of an $\ese$ symmetry of a 
higher-dimensional form of time in the following section.

  In the Standard Model Lagrangian each fermion kinetic term, such as 
equation~\ref{lagklep}, or interaction term, such as equation~\ref{lagqwint}, contains 
either left or right-handed fermion states, while the Yukawa or Dirac mass terms 
combine opposite chiralities, as for example in equations~\ref{Yukferm} or 
\ref{lagdmass}. In all cases the operators $P_L = \fh(1-\gamma^5)$ or $P_R = 
\fh(1+\gamma^5)$, of equations~\ref{plop} and \ref{prop}, may be used to project out 
the respective left or right-handed chirality states from a 4-component Dirac spinor.

    The factors of $P_L = \fh(1-\gamma^5)$ which appear in all fermion terms involving 
the $\sutw_L$ gauge symmetry,  such as in equations~\ref{lagqwint} and \ref{lagqwckm}, 
are placed in the Lagrangian by hand in order to replicate the parity violating 
phenomena observed empirically for the weak interaction. This parity violation is 
maximal for the case of interactions via $W^{\pm}$ gauge bosons but non-maximal for 
$Z^0$ interactions, which is associated with a linear combination of $\sutw_L$ and 
$\uo_Y$ generators,
 equation~\ref{zmucoup}, as described in section~\ref{ewtatsm}. It is this asymmetry 
in the chiral structure, with different weak isospin transformations for left- and 
right-handed fields, implying that no fermion state transforms under the complex 
conjugate representation of that of any other fermion, which necessitates the 
introduction of Yukawa couplings to the Higgs field, as for equation~\ref{Yukferm}, in 
order to include fermion mass terms in the Lagrangian.

  In the present theory we have described how the components of 
$\theta^1=\binom{c}{\bar{b}}$ form the set of four left-handed Weyl spinors of 
equation~\ref{thcth234} under the external Lorentz symmetry $\sltc^1$. In this section 
a similar decomposition has also been identified for the components of 
$X=\binom{p\;\;\bar{a}\;\!}{a\; m}$, for example via  the spinor  
$\thX=\binom{\bar{r}}{s} \inn \ooo^2$ as introduced in equation~\ref{xththx}.
  Hence the projection operator $P_L = \fh(1-\gamma^5)$ has not been introduced since 
only \textit{left}-handed Weyl spinors under $\sltc^1$ have so far been considered.
 There then remains the question of how \textit{right}-handed Weyl spinor counterparts 
may identified within this framework, and related to the above left-handed components 
in a single 4-component Dirac spinor $\psi(x)$ to describe, for example, a physical 
electron state.

   Here we began with the cubic form $\det (\mcX) = 1$, with $\mcX \inn \htho$, as a 
27-dimensional expression of temporal flow $\lvt$. Determinant preserving $\esi$ 
transformations were then considered on this space, with for example $\mcX \to \mcM 
\mcX \mcM^{\dag}$ for the $3 \times 3$ matrices of equations~\ref{mxm3} and 
\ref{type1} incorporating any of the $2 \times 2$ matrices of category 1 or 2 in 
table~\ref{mtran45}, for the unnested case. With $\overline{\mcM}$ representing the 
matrix $\mcM$ with each entry replaced by its octonion conjugate, as described in 
equation~\ref{octabar}, transformations of the form:
\begin{equation}
  \mcX \to \overline{\mcM} \, \mcX \, \overline{\mcM}^{\dag}
\end{equation}
   clearly also leave the value of $\det(\mcX)$ invariant. With $\esi$ a symmetry of 
the form of time $\lvt$ the \textit{two} possible representations $\mathbf{27}$ and  
$\overline{\mathbf{27}}$ are equally valid while only one of them has been used so 
far.

    Correspondingly the set of six actions with $\overline{\mcM}$, for $\mcM \inn 
\sltc^{1}$, provides an alternative choice for the type 1 Lorentz transformations 
acting on the vector components $\bv^4 \inn \htwc \subset \htho$, with $\bh_2 \to 
\ol{S} \, \bh_2 \, \ol{S}^{\dag}$ in place of equation~\ref{hshs}. In turn these 
transformations are represented on $\theta^1$ as a set of four \textit{right}-handed 
Weyl spinors. The group $\sltc$, as for the full group $\esi$, has complex 
representations, and the  actions of   
 $S\inn\sltc^1$ and $\ol{S}$ describe distinct sets of transformations of $\theta^1$, 
as explained in section~\ref{lsspin}. This means that the left and right-handed 
transformations are not equivalent to each other but are instead mutually related as 
described in equations~\ref{rlssrrss} and \ref{rlsrrs}. In the context of the present 
theory both 
 $S\inn\sltc^1$ and $\ol{S}$  act as symmetry transformations leaving the form $\lvt$ 
invariant, and hence both the left \textit{and} right Weyl spinor compositions of 
$\theta^1$ should in principle play a role.  

  The apparent asymmetry in the choice of the $\esi$ $\mathbf{27}$ or  
$\overline{\mathbf{27}}$ representation to express  $\lvt$,  with a corresponding 
choice of left or right-handed representations of $\sltc^1 \subset \esi$,  the need to 
clearly identify both left- and right-handed fermions $\psi_L$ and $\psi_R$, in 
particular with reference to an $\sutw_L$ gauge symmetry, and the existence of a 
homogeneous \textit{quartic}  form as a candidate for a higher-dimensional temporal 
flow 
in the form $\lvfs$  all point to consideration of  the group $\ese$ as a symmetry of 
time,
as will be described in the following section.



\section{$\ese$ Symmetry and the Freudenthal Triple System}
\label{secesef}

  The introduction of further dimensions in the previous section and the observation 
of the  quartic expression of equation~\ref{detrsn}, with the extension to 
equation~\ref{xththfull2} including quintic terms also, suggests the possibility of a 
higher-dimensional expression for the flow of time generalising beyond the cubic form 
$\lvt$ described in chapter~\ref{esihtho}. A higher-dimensional  \textit{homogeneous} 
polynomial form  is desired, in conformity with the derivation of equation~\ref{lv} in 
chapter~\ref{sym}. While the determinant preserving symmetry of the space $\mcX \inn 
\htho$ describes the lowest-dimensional non-trivial representation of $\esi$ the 
smallest non-trivial representation of the exceptional Lie group E$_7$ is 
56-dimensional and may be constructed in terms the elements $x$ of the Freudenthal 
triple system $F(\htho)$ (\cite{Krute,Borst,Rios}, \cite{Baez1} p.48). 

  In the above references and related publications these mathematical structures are 
applied  in two very different contexts -- namely the classification of black hole 
solutions in string theory and the entanglement of qubits in quantum information 
theory -- with a correspondence between these applications identified through the 
mathematical forms they share. Neither application is relevant for the present 
discussion. While much of the literature describes a more general algebraic framework 
or particular cases involving for example the `split octonions' $\ooo_s$ or takes an 
underlying field of integers $\zzz$, here we are interested in the octonion $\ooo$ 
case over an underlying field of real numbers $\rrr$ as we summarise in the following.

  In order to describe the Freudenthal triple system $F(\htho)$  it is useful to first 
introduce further definitions regarding the exceptional Jordan algebra $\htho$ itself. 
The \textit{structure group} Str$(\htho)$ leaves the cubic norm $\det(\mcX)$ of 
equations~\ref{detx3} and \ref{detpmn} invariant up to a real scalar factor, that is:
\begin{equation}
 \label{strdef}
   \mbox{Str}(\htho) \; = \;
    \{ g\inn \mbox{GL}(\htho)\, \vert\, \det (\sigma_g(\mcX))
	 = \lambda(g) \det (\mcX), \; \forall \mcX \inn \htho  \}
\end{equation}
  with $\lambda \inn \rrr$ depending only on $g$. The norm preserving subgroup with 
$\lambda=1$ is identified as the reduced structure group Str$_0(\htho) \equiv \sltho$. 
This latter symmetry corresponds to the 27-dimensional representation of $\esig$ as 
described in detail in chapter~\ref{esihtho}. 

  A \textit{trace bilinear map} may be defined for any elements $\mcX,\mcY \inn \htho$ 
of the Jordan algebra, mapping $\htho \times \htho \to \rrr$ with:
\begin{equation}
 \label{trabimap}
 (\mcX, \mcY) \; = \; \mbox{tr}(\mcX \circ \mcY)
\end{equation}
  where on the right-hand side the $\circ$ denotes the Jordan algebra product of 
equation~\ref{joralg}. An \textit{adjoint} $s^{\ast}$ for any transformation $s(\mcX)$ 
with $s\inn \esi$ may be defined with respect to the above trace bilinear form such 
that:
 \begin{equation}
  \label{sadjoint}
   (s(\mcX), \mcY ) = (\mcX, s^{\ast}(\mcY))  \qquad \forall \mcX, \mcY \inn \htho
 \end{equation}
 Along with the Jordan product there is a second natural composition for the elements 
of $\htho$ which is called the \textit{Freudenthal product} and may be defined by:
\begin{equation}
  \mcX \wedge \mcY  =  \mcX \circ \mcY - \fh 
    \big( \mbox{tr}(\mcX)\mcY + \mbox{tr}(\mcY)\mcX  \big) + \fh
	\big( \mbox{tr}(\mcX) \mbox{tr}(\mcY) - \mbox{tr}(\mcX \circ \mcY) \big)\b1_3
	 \, \inn \htho  \;
\end{equation}
 For any $\mcX \inn \htho$ a \textit{quadratic adjoint map} $\htho \to \htho$ can be 
defined in terms of the Freudenthal product  as:
\begin{eqnarray}
  \mcX^{\sharp} & = & \mcX \wedge \mcX \\
\mbox{or explicitly:} \quad \mcX^{\sharp}  & = & 
       \mcX^2 - \mbox{tr}(\mcX)\mcX +\fh \lbrack\mbox{tr}(\mcX)^2 -
	   \mbox{tr}(\mcX^2)\rbrack \, \b1_3		\label{quadmap}									  			 
\end{eqnarray}
 This `sharp' operation satisfies the relations $(\mcX^{\sharp})^{\sharp} = \det(\mcX) 
\mcX$
 and $\mcX \circ \mcX^{\sharp} = \det(\mcX) \b1_3$. The linearisation of the quadratic 
adjoint is written as:
\begin{eqnarray}
  \mcX \times \mcY & = & (\mcX + \mcY)^{\sharp} \, - \,
      \mcX^{\sharp} \, - \, \mcY^{\sharp}  \\
     & \equiv &  2\mcX \wedge \mcY									  			 
\end{eqnarray}
  For the elements of $\htho$ the quadratic adjoint is in fact the classical adjoint, 
that is the transposed cofactors of $\mcX \inn \htho$ which, for the components of 
$\mcX$ presented in equation~\ref{htho} or \ref{hthoxy} below, can be written 
explicitly as the $3 \times 3$ matrix:
\begin{equation}
 \label{hthoshp}
\mcX^{\sharp} = 
   \left( \begin{array}{ccc}
       mn - \vert b \vert^2 \, & \, cb - n\bar{a} \, & \, \bar{a}\bar{b} - mc  \\
       \bar{b}\bar{c} - na \, & \, pn - \vert c \vert^2 \, & \, ac - p\bar{b}        
\\
       ba - m\bar{c} \, & \, \bar{c}\bar{a} - pb \, & \, pm - \vert a \vert^2 
          \end{array}  \right) \; \inn \htho
\end{equation}

  The vector space $F(\htho)$ has 56 real components and may be introduced according 
to Freudenthal's construction with the vector space composition (which may be compared 
with the further decomposition of equation~\ref{horhoo}):
\begin{equation}
  \label{fvecspa}
  F(\htho) \; \cong \; \htho \, \oplus \, \htho  \, \oplus \, \rrr \, \oplus \, \rrr
\end{equation}
  Correspondingly elements $x \inn F$, with $F=F(\htho),$ are generally written in the 
form of a `$2 \times 2$ matrix' as:
\begin{equation}
  \label{ftscomp}
   x =  \left(\begin{array}{cc} \alpha & \mcX \\
                          \mcY & \beta \end{array} \right), \qquad
		\mbox{with} \;\;   \mcX,\mcY \inn \htho,
						   \quad \alpha,\beta \inn \rrr
\end{equation}
\begin{equation}
 \label{hthoxy}
 \mbox{and} \qquad \quad  \mcX = 
   \left( \begin{array}{ccc}
       p & \bar{a} & c  \\
       a &   m     & \bar{b}        \\
 \bar{c} &   b     & n 
          \end{array}  \right), \qquad 
   \mcY = 
   \left( \begin{array}{ccc}
       P & \bar{A} & C  \\
       A &   M     & \bar{B}        \\
 \bar{C} &   B     & N 
          \end{array}  \right)
\end{equation}
  here with the real $P,M,N$ and octonion $A,B,C$ components of $\mcY$ distinguished 
from the lower case counterpart components of $\mcX$. 
 A non-degenerate bilinear antisymmetric quadratic form mapping $F\times F \to \rrr$ 
may be defined on this space which acts on $x=\binom{\alpha \; \mcX}{\mcY \:\, \beta}, 
y =
 \binom{\gamma \; \mcW}{\!\mcZ \;\, \delta} \inn F$ as:
\begin{equation}
 \{x,y\} \; = \; \alpha\delta \, - \, \beta\gamma \, + \, 
    (\mcX, \mcZ) \, - \, (\mcY, \mcW)  \label{biasqf}
\end{equation}
  Of more significance for the present theory  
  there is also a homogeneous quartic norm $q:F \to \rrr$ defined on the components of 
$x\inn F$ as follows:
\begin{equation}
   q(x) \; = \; -2\lbrack \alpha\beta - (\mcX,\mcY)\rbrack^2 \, - \,
       8\lbrack\alpha \det(\mcX) + \beta\det(\mcY) - (\mcX^{\sharp},
	                                          \mcY^{\sharp})\rbrack \label{fquartic}									  			 
\end{equation}
   where all the necessary definitions contained within this expression are inherited 
from those for the Jordan algebra $\htho$ as described above. The quadratic and 
quartic forms of equations~\ref{biasqf} and \ref{fquartic} may be used in turn to 
define a trilinear mapping of the space $F\times F\times F \to F$, which is the triple 
product by which the `Freudenthal triple system' gains its name.
     When written out explicitly in terms of the real and octonion components of 
equations~\ref{ftscomp} and \ref{hthoxy} there are a large number of quartic terms in 
$q(x)$. In fact, via equations~\ref{trabimap} and \ref{hthoshp}, and cross-checking 
with~(\cite{Borst} equation~9.51), we have for equation~\ref{fquartic}:
\begin{eqnarray}
 q(x) \; = \; -2 \!\! & \!\!\!\! \Big[ \!\!\!\! & \!\!  \alpha\beta - pP - mM - nN -
    2\big(\,\langle a,A \rangle + \langle b,B
	   \rangle + \langle c,C \rangle \, \big) \, \Big]^2 \nonumber \\
	-8 \!\! & \!\!\! \Big[ \!\!\! & \!\! \beta PMN +\alpha pmn 
	   -pPmM - pPnN -nNmM     \nonumber \\
 & & + \; (pm - \beta N)\vert A \vert^2 \, + \, (PM - \alpha n)\vert a \vert^2 
\nonumber \\
 & & + \; (mn - \beta P)\vert B \vert^2 \, + \, (MN - \alpha p)\vert b \vert^2 
\nonumber \\ 
 & & + \; (np - \beta M)\vert C \vert^2 \, + \, (NP - \alpha m)\vert c \vert^2 
\nonumber \\
 & & + \; 2\beta \, \mbox{Re}(\bar{A}\bar{B}\bar{C})
               \, + \,  2\alpha \, \mbox{Re}(\bar{a}\bar{b}\bar{c})      \nonumber \\
 & & - \; \vert a \vert^2 \vert A \vert^2 \, - \,
          \vert b \vert^2 \vert B \vert^2 \, - \,
	      \vert c \vert^2 \vert C \vert^2   \nonumber \\
 & & - \; (cb - n\bar{a})(\bar{B}\bar{C} - NA) \,\; - \,
                 (CB - N\bar{A})(\bar{b}\bar{c} - na)  \nonumber \\
 & & - \; (ac - p\bar{b})(\bar{C}\bar{A} - PB) \,\;\; - \,
                 (AC - P\bar{B})(\bar{c}\bar{a} - pb)  \nonumber \\
 & & - \; (ba - m\bar{c})(\bar{A}\bar{B} - MC) \, - \,
                 (BA - M\bar{C})(\bar{a}\bar{b} - mc)\, \Big]  \label{fquartall}                        
\end{eqnarray} 
 where the inner product $\langle a,A \rangle  = \frac{1}{2}(a\bar{A} + A\bar{a})$, 
which has the property $\langle a,A \rangle = \langle \bar{a},\bar{A} \rangle$, was 
defined in equation~\ref{octinner}. Equations~\ref{fquartic} and \ref{fquartall} for 
the quartic form $q(x)$ are the analogue of equations~\ref{detx3} and \ref{detpmn} 
respectively for the cubic form $\det(\mcX)$. Clearly there are many more terms for 
the above quartic from in equation~\ref{fquartall} as an extension from the cubic form 
of equation~\ref{detpmn}. 

   The group Inv($F$) of all invertible transformations $\sigma$ in $F$ preserving the 
above quartic norm with $q(\sigma(x)) = q(x)$, as well as the bilinear form of 
equation~\ref{biasqf} with $\{\sigma(x), \sigma(y)\} = \{x,y\}$, is also denoted 
Aut$(F)$ since it in turn forms the automorphism group of the trilinear product 
defined for the  Freudenthal triple system. This group is found to be the non-compact 
real form E$_{7(-25)}$ of the exceptional Lie group E$_7$. Hence, in particular, under 
this symmetry group the invariance of the quartic form $q(x)$, as a homogeneous 
polynomial, describes a possible 56-dimensional form of temporal flow which may be 
denoted $L(\bv_{56})=1$. The possible physical implications of this form and the 
accompanying E$_7$ symmetry will be assessed in the remainder of this section and 
summarised in the following one.

 The symmetry of the  cubic form $\lvt$, in the form of $\det(\mcX)$ or $\det(\mcY)$, 
is contained within this structure as can be seen from equation~\ref{fquartic}. In 
fact the elements $\mcX$ and $\mcY$, with 54 real components in total, may be 
considered to represent a `complexification' of the space $\htho$, with both the 
27-dimensional representation of $\esi$ and its complex conjugate contained within the 
E$_7$ action on $q(x)$.  
 Including the actions of the subgroup $\esi \subset \mbox{E}_7$  on the elements  
$x \to s(x) \inn F$ the transformations of full symmetry $\ese \equiv \mbox{Inv}(F)$ 
may be categorised in terms of  four sets. With $s \inn \esi$, $\lambda \inn \rrr$ and 
$C,D \inn \htho$ these are~\cite{Krute,Borst,Rios}: 
\begin{eqnarray}
 \!\!\! T(s): \,  \left(\begin{array}{cc} \alpha & \mcX \\
                          \mcY & \beta \end{array} \right) & \!\! \to \!\! &
		  \left(\begin{array}{cc} \alpha & s(\mcX) \\
                          {s^{\ast}}^{-1}(\mcY) & \beta \end{array} \right)		
						   \label{ftsstran}		   \\ 
 \!\!\! \lambda: \,   \left(\begin{array}{cc}\alpha & \mcX \\
                          \mcY & \beta \end{array} \right) & \!\! \to \!\! &
  \left(\begin{array}{cc}\;\! \lambda^{-1}\alpha\; &\; \lambda^{\frac{1}{3}}\mcX \;\! 
\\
                \;\! \lambda^{-\frac{1}{3}}\mcY \;& \; \lambda \,\beta \end{array}\;\!
				   \right)	\label{lamtran} \\
\!\!\! \phi(C): \,   \left(\begin{array}{cc} \alpha & \mcX \\
                          \mcY & \beta \end{array} \right) & \!\! \to \!\! &
		  \left(\begin{array}{cc}
 \!  \alpha+(\mcY,C) + (\mcX,C^{\sharp}) + \beta \det(C) \, & \, \mcX + \beta C \! \\
        \!  \mcY + \mcX\times C + \beta C^{\sharp} \, & \, \beta \end{array} \! 
\right)		
						\label{phicact}     \\ 
\!\!\!  \psi(D): \, \left(\begin{array}{cc} \alpha & \mcX \\
                          \mcY & \beta \end{array} \right) & \!\! \to \!\! &
		  \left(\begin{array}{cc}
  \! \alpha \, & \, \mcX + \mcY\times D + \alpha D^{\sharp} \! \\
  \! \mcY + \alpha D \, & \,
      \beta + (\mcX,D) + (\mcY,D^{\sharp}) + \alpha \det(D) \end{array} \! \right)
	  \quad
	    \label{psidact}	
\end{eqnarray}   
  where $s^{\ast}$ is the adjoint of $s\inn \esi$ as defined in 
equation~\ref{sadjoint}. The set of actions ${s^{\ast}}^{-1}$ in 
equation~\ref{ftsstran} is equivalent to the complex conjugate of the representation 
defined by the set of actions $s\inn \esi$ on $\htho$. Under the subgroup $\esig 
\subset {\mbox{E}_{7(-25)}}$ the space $F$ decomposes into the reducible 
representation (\cite{Borst} equations~9.45 and 9.46):
\begin{equation}
  \label{esetoesi}
    \mathbf{56}_{\mathrm{E}_7} \, \to \, 
	(\mathbf{27} + \mathbf{\overline{27}} + \mathbf{1} + \mathbf{1})_{\mathrm{E}_6}
\end{equation}	
  compatible with the structure of equation~\ref{fvecspa}
  (and can be compared with the further reduction under $\spotn$ in 
equation~\ref{esidecom}). The 78 actions of $\esi$ combined with the single dilation 
action $\lambda$ of equation~\ref{lamtran} applied to an 
$\htho \subset F$ subspace together form the 79-dimensional group Str$(\htho)$ as 
defined in equation~\ref{strdef}.  The 27 independent actions of $\phi(C)$ together 
with the further 27 for $\psi(D)$ in equations~\ref{phicact} and \ref{psidact} further 
augment the $\esi$ symmetry to complete the full $(78+1+27+27)=133$-dimensional 
exceptional Lie group $\ese$. (Building up the symmetry structure this way is 
analogous to augmenting the $\ff$ algebra by the $D^B$ maps in equation~\ref{drbdecom} 
to complete the full $\esi$ symmetry.)
In addition to the continuous actions of equations~\ref{ftsstran}--\ref{psidact} a 
discrete symmetry $\tau := \phi(-\b1_3)\psi(\b1_3)\phi(-\b1_3)$ such that:
\begin{equation}
  \tau: \,  \left(\begin{array}{cc} \alpha & \mcX \\
                          \mcY & \beta \end{array} \right) \quad \to \quad
		  \left(\begin{array}{cc} -\beta & -\mcY \\
                            \mcX & \alpha \end{array} \right)		
						   \label{tautran}		   
\end{equation}
  with $\tau^2(x) = -x$, may also be defined. Since $\psi(C) = \tau\, \phi(-C) \, 
\tau^{-1}$ the set of actions $\phi$ and $\psi$ are conjugate with respect to $\tau$. 
Between equations~\ref{ftsstran} and \ref{tautran} the further relationship $\tau \, 
T(s) = T({s^{\ast}}^{-1})\, \tau$ is also found.
  
  At the Lie algebra level the actions $\dot{s} \inn L(\esi)$ may be divided into the 
14 elements of $L(\gt)$, denoted $D^G$, and the action of the 64 tracefree octonion 
matrices $x_0$, denoted $D^S$ in equation~\ref{dsgdecom}. The latter set further 
divides into the 26 boosts with Hermitian $x_0$ and 38 rotations with anti-Hermitian 
$x_0$. Such decompositions were also discussed in the opening three paragraphs of 
section~\ref{laofesi}. The subgroup $\gt$ itself may also be obtained through 
sequences of nested rotations as described in section~\ref{esitran}. The \textit{dual} 
representation of $L(\esi)$ may be obtained by defining the action $\dot{s}'(\mcX)$ 
for each $\dot{s} \inn L(\esi)$ such that
  (\cite{DrayMW} equation~4):
\begin{equation}
  \label{sadjointa}
   (\dot{s}(\mcX), \mcY ) = -(\mcX, \dot{s}'(\mcY))  \qquad \forall \mcX, \mcY \inn 
\htho
\end{equation}
  which may be contrasted with equation~\ref{sadjoint} at the group level. For the 
$L(\esi)$  maps $\mcX \to \mcx_0 \mcX + \mcX \mcx_0^{\dagger}$ the dual 
transformations correspond to:
\begin{eqnarray}
    x'_0 & = & x_0 \qquad \mbox{for rotations} \label{dualrot} \\
    x'_0 & = & -x_0 \qquad \mbox{for boosts}  \label{dualboo}
\end{eqnarray}
  that is with $x'_0  =  -x_0^{\dag}$ in general (\cite{DrayMW} equation~5). 
  It also follows that $\dot{s}' = \dot{s}$ for the $L(\gt)$ actions derived from 
transverse rotations. Applied to the subalgebra $L(\esi) \subset L(\ese)$ acting on 
the elements of the Freudenthal triple system these 78 generators form the first of 
the four sets of $\ese$ actions at the Lie algebra level (corresponding to 
equations~\ref{ftsstran}--\ref{psidact} at the group level) which may be listed as the 
infinitesimal transformations of $x=\binom{\alpha \; \mcX}{\mcY \:\, \beta}$ 
(\cite{DrayMW} section 2):
\begin{eqnarray}
  T(\dot{s}): & \quad & 
		  \left(\begin{array}{cc} 0 & \dot{s}(\mcX) \\
                           \dot{s}'(\mcY) & 0 \end{array} \right)		
						   \label{ftsstrana}		   \\ 
  \dot{\lambda}: & \quad &   
  \left(\begin{array}{cc}  \!
              -\dot{\lambda}\alpha  &  \frac{1}{3}\dot{\lambda}\mcX  \\
                \!  -\frac{1}{3}\dot{\lambda}\mcY  &  \dot{\lambda} \beta
				 \end{array}
				   \right)	\label{lamtrana} \\
 \phi(\dot{C}): & \quad &   
		  \left(\begin{array}{cc}
   (\mcY,\dot{C})  \, & \, \beta \dot{C}  \\
          \mcX\times \dot{C}  \, & \, 0 \end{array}  \right)		
						\label{phicacta}     \\ 
  \psi(\dot{D}): & \quad &
		  \left(\begin{array}{cc}
   0 \, & \, \mcY\times \dot{D} \\
  \alpha \dot{D} \, & \,
         (\mcX,\dot{D})  \end{array} \right)
	    \label{psidacta}	
\end{eqnarray}   
  with $\dot{s} \inn L(\esi)$, $\dot{\lambda} \inn \rrr$ and $\dot{C}, \dot{D} \inn 
\htho$. Higher-order terms such as $C^{\sharp} = C \wedge C$ and the cubic norm 
$\det(C)$ appear for the finite group actions of equations~\ref{phicact} and 
\ref{psidact}.

   Having extended beyond the $L(\esi)$ subalgebra to the full $L(\ese)$ we next focus 
on the generators of the 4-dimensional spacetime Lorentz subgroup $\sltc^1 \subset 
\esi \subset \ese$ of type 1 as studied in section~\ref{extsym}. As for all $\esi$ 
transformations for the actions of the Lorentz subalgebra $\sltca^1 \subset L(\esi)$  
the dual transformations $\dot{s}'$ in equation~\ref{ftsstrana} have identical 
rotation generators to $\dot{s}$ while the boosts are reversed, by 
equations~\ref{dualrot} and \ref{dualboo}. As was described for 
equations~\ref{leftrep} and \ref{rightrep} of section~\ref{lsspin} reversing the sign 
of the boosts, there parametrised by $b_a$, is precisely the operation which 
interchanges between the $L$ and $R$ representations of $\sltc$.
  
  Hence while the components of $\theta_l$ in $\theta^1$ within  $\mcX  \inn \htho$, 
defined in equation~\ref{thcth2}, transform as a \textit{left}-handed Weyl spinor 
under $\sltc^1$ the corresponding components of $\theta_{\!\lag} = 
\binom{C_1 + C_8l}{B_1 - B_8l}$ within the $\theta^1$ component of $\mcY \inn \htho$, 
extracted from equation~\ref{hthoxy}, transform as a \textit{right}-handed  Weyl 
spinor under the same $\sltc^1 \subset \esi \subset \ese$ action.
 (The subscript `$\lag$' on $\theta_{\!\lag}$ denotes both the use of the imaginary 
unit $l$ and the identification of the `leptonic' components of $\theta^1 = 
\binom{C}{\bar{B}}$ in $\mcY$, as will be seen below. In general the superscript `1' 
is not appended to components such as $\theta_l$ and $\theta_{\!\lag}$ since they are 
unambiguously extracted from  `type 1' $\theta^1$ components,
  while a superscript is included for the `type 2' or `type 3' case as for
  $\theta^2_l$ in equation~\ref{covdevth} for example).
 Considered as an action of  $2 \times 2$ matrices $S\inn \sltc^1$  on the 2-component 
Weyl spinors $\theta_l$ and $\theta_{\!\lag}$, extracted from the corresponding 
$\theta^1$ components of $\mcX$ and $\mcY$
respectively, and  using equation~\ref{rlsrrs}, the action of equation~\ref{ftsstran} 
may be summarised as:
\begin{equation}
   \left(\begin{array}{c}  \theta_l \\ \theta_{\!\lag} \end{array} \right)
			 \; \to \;			  
		 \left(\begin{array}{cc} S  &  0  \\
                          0 & {S^{\dag}}^{-1} \end{array} \right)
         \left(\begin{array}{c} \theta_l \\ \theta_{\!\lag}  \end{array} \right)
\end{equation}
  This is precisely the Lorentz transformation of a 4-component Dirac spinor $\psi$ as 
described in equations~\ref{psitrans2} and \ref{rdsrss}.  Alternatively the above 
expression could be obtained directly at the group level from equation~\ref{ftsstran} 
using the definition of the adjoint $s^{\ast}$ in equation~\ref{sadjoint} applied 
directly to the $\sltc^1 \subset \esi$ group transformations. 

 As explained in section~\ref{extsym} the components of $\theta^1$ within $\mcX \inn 
\htho$ under the action of $\sltc^1$ actually decompose into a set of four left-handed 
Weyl spinors
 $\{\theta_l, \theta_i, \theta_j, \theta_k\}$ as listed in equation~\ref{thcth234}. 
 Hence equation~\ref{ftsstran} contains both the original representation of $\sltc^1$ 
on $\mcX$, which contains the set of four left-handed Weyl spinors in the $\theta^1$ 
components, simultaneously with an equivalent of the complex conjugate representation 
on $\mcY$, which hence contains a corresponding set of four right-handed Weyl spinors, 
which may be denoted $\{\theta_{\! \lag}, \theta_I, \theta_J, \theta_K\} \subset 
\mcY$. Correspondingly a set of four 4-component Dirac spinors have hence been 
identified with:
\begin{equation}
 \label{diraclr}
   \psi = \left(\begin{array}{c}  \psi_L \\ \psi_R \end{array} \right) \; = \;
     \left(\begin{array}{c}  \theta_l \\ \theta_{\!\lag} \end{array} \right), \quad
	 \left(\begin{array}{c}  \theta_i \\ \theta_I \end{array} \right), \quad
	 \left(\begin{array}{c}  \theta_j \\ \theta_J \end{array} \right)
	 \quad \mbox{or} \quad
	 \left(\begin{array}{c}  \theta_k \\ \theta_K \end{array} \right) 
\end{equation}
\begin{equation}
 \label{diraclrt}
 \mbox{with} \qquad \psi = \left(\begin{array}{c}  \psi_L \\ \psi_R \end{array} 
\right)
			 \; \to \;			  
		 \left(\begin{array}{cc} S  &  0  \\
                          0 & {S^{\dag}}^{-1} \end{array} \right)
         \left(\begin{array}{c}  \psi_L \\ \psi_R \end{array} \right)
\end{equation}
  under $S\inn \sltc^1 \subset \esi \subset \ese$ transformations in each case.
  
  The above analysis applied to the $\theta^1$ components of $\mcX$ and $\mcY$ 
similarly applies for the left-handed $\sltc^1$ Weyl spinors contained within $\thX$ 
under the decomposition of equation~\ref{xththx} or \ref{xththpx}. In this case a 
corresponding set of four right-handed spinors are found in the components of $\thY$ 
obtained in turn under a decomposition which may be denoted  $Y= \thY \thY^{\dag}$ for 
the $\htwo \subset \htho$ components of $\mcY$. A similar observation applies for the 
alternative spinor decomposition of $Y$ beginning with the $\htwc \subset \htwo$ 
subspace as described towards the latter part of the previous section. 

  The internal $\suth_c \times \uo_Q$ symmetry, described in section~\ref{intsym}, is 
composed as a subgroup of $\esi$ purely out of the subset of rotations. Hence, by the 
discussion around equation~\ref{dualrot} above, these actions are identical on the 
components of $\mcX$ and $\mcY$ in equation~\ref{ftsstran}. Hence in turn the 
$\suth_c$ action on the components of $\mcX$, including upon the $\theta^1$ components 
as detailed in table~\ref{suttan} and summarised together with the $\uo_Q$ action in 
equation~\ref{suthuoth}, is identical for the corresponding components of $\mcY$, and 
the corresponding $\uo_Q$  charges for the respective subcomponents of 
equation~\ref{hthoxy} are also the same. Hence the $\psi_L$ and $\psi_R$ components 
carry matching $\suth_c \times \uo_Q$ transformation properties for the set of four 
Dirac spinors in equation~\ref{diraclr}
(justifying the identification of both $\theta_l$ and $\theta_{\!\lag}$ as leptonic
  components). 
  Similarly the $\sutw^{2,3} \times \uo^{2,3} \subset \esi$ rotations, for the  mock 
electroweak theory described in section~\ref{ewtfesb}, also act on the $\mcX$ and 
$\mcY$ components of $x \inn F(\htho)$ in the same way.

  While the total number of dimensions has been increased from 27 to 56 it remains the 
case that only a single set of 4 dimensions will describe the external spacetime. This 
can be chosen as an $\htwc \subset \htho$ subset of components  $\bv_4 \subset \mcX$, 
under an $\sltc \subset \esi$ action, \textit{or} as an $\htwc \subset \htho$ subset 
of components  $\bv_4 \subset \mcY$, transforming under the complex conjugate 
representation, but not \textit{both}. Here we choose 
  $\bv_4 \equiv \bh_2 \inn \htwc$ as embedded within the $Y=\binom{P \; \bar{A}}{A \; 
M} \inn \htwo$ components of $\mcY$ in equation~\ref{hthoxy} to represent external 
spacetime,
 with Lorentz transformations hence described by:
\begin{equation}
\label{hshsdual}
  \bh_2 \to  \bh_2^{\prime} =  {S^{\dag}}^{-1} \, \bh_2 \, S^{-1}
\end{equation}
  rather than equation~\ref{hshs}, under the action of $S\inn \sltc^1 \subset \esi$.
 The complex subspace with base units $\{1,l\}$ still underlies 
 both the $\sltc^1$ subgroup and the subspace for the vectors $\bh_2 \inn \htwc$.    
 These $\bh_2$ components of $\mcY$ will also now be taken to
  form the `vector-Higgs' 
   correlated with the phenomena of the Standard Model Higgs sector and  Yukawa 
couplings, as was described for the original case of $\lvt$ in 
subsection~\ref{suboomahp}. Here for the case of $\lvfs$ this now implies that 
\textit{none} of the 27 components of $\mcX \inn \htho \subset F(\htho)$ are 
identified with components of the external spacetime vectors $\bv_4 \inn \TM_4$.

In particular this means that in addition to the $d$-quark and charged lepton 
components of left-handed Weyl spinors in $\theta^1 \subset \mcX$, potentially both 
$u$-quark \textit{and} neutral lepton left-handed Weyl spinors might be identified in 
the $X$ components of $\mcX$ as described in the previous section. The $a\inn \ooo$ 
component of $\mcX$ has the correct $(0,\frac{2}{3})$ charge structure to describe 
($\nu$-lepton, $u$-quark) particle states, as seen in equations~\ref{sbcharge} and 
\ref{suthuoa}, and is now free to accommodate both states.
 However while the corresponding imaginary $A(6)$ components of $\mcY$ also have an 
$\Sbard_l^1$ charge of $\frac{2}{3}$, the $A_{1,l} = (A_1 + A_8l)$ part of $A\inn 
\ooo$ in $\mcY$ is \textit{occupied} by the above components $\bh_2 \inn \htwc$, 
representing the vector-Higgs and external spacetime, 
 as depicted in equation~\ref{fhthopart}.
\begin{equation}
 \hspace*{-20pt}
  \label{fhthopart}
  \left(\!\!\!\!\! \begin{array}{cc}   \alpha &   \!\!\!
     \left(\!\! \begin{array}{c|c} 
	      \, X \!\!\sim\thX\thX^{\dag} 
		  		   \!\!\!\!\! \begin{array}{cc} &  \\  &  \end{array} \!\!\!   &
        \,  \theta^1  \begin{array}{cc} &  \\  &  \end{array} \!\!\!\!\!\!\!\!\!\! 
				                         \\  \hline
        \,\,\,\,\,\,\,\, {\theta^1}^{\dagger} \!\!\!\! \begin{array}{cc}  
		      &   \end{array}   &	  
		\,  n      \end{array}  \!\! \right)_{\mbox{$\!\!\!\!\!\;\mcX$}}   \\
		\left(\!\!  \begin{array}{c|c} 	
	\, Y \!\!\sim\thY\thY^{\dag}
	  \!\!\!\!\! \begin{array}{cc} &  \\  &  \end{array} \!\!\!   &		  
        \,  \theta^1  \begin{array}{cc} &  \\  &  \end{array} \!\!\!\!\!\!\!\!\!\! 
				                         \\  \hline
        \,\,\,\,\,\,\,\, {\theta^1}^{\dagger} \!\!\!\! \begin{array}{cc} 
		         &   \end{array}   &	  
		\,  N      \end{array} \!\! \right)_{\mbox{$\!\!\!\mcY$}}
             \!\!\!          & \beta  \end{array} \!\!\!\!\!\right)
\quad \sim \quad
  	 \left(\!\!\!\!\! \begin{array}{cc}    &   \!\!\!
     \left(\!\! \begin{array}{c|c} 
         \begin{array}{c} \;\;\, \mbox{`}\nu_L\mbox{'} \;\;\,  \\
		   \mbox{`}u_L\mbox{'}  \end{array}    &
          \!\!  \begin{array}{c}    e_L   \\ d_L    \end{array} \!\! 
				                         \\  \hline
        \begin{array}{cc}  
		           &   \end{array}   &	  
		          \end{array}  \!\! \right)_{\mbox{$\!\!\!\!\!\;\mcX$}}   \\
		\left(\!\! \begin{array}{c|c} 
         \begin{array}{c} \! \bv_4 \equiv \bh_2  \!  \\ 
		     \;\;    \mbox{`}u_R\mbox{'} \;\;  \end{array}    &
          \!\!  \begin{array}{c}    e_R   \\ d_R    \end{array} \!\! 
				                         \\  \hline
        \begin{array}{cc}  
		           &   \end{array}   &	  
		          \end{array}  \!\! \right)_{\mbox{$\!\!\!\mcY$}}
             \!\!\!          &  \end{array} \!\!\!\!\!\right) 
\end{equation}

This provisionally provides an explanation for the existence of the left-handed 
neutrino $\nu_L$ while the corresponding right-handed state $\nu_R$ is prohibited, at 
least at the level of the basic symmetry structures, as a feature of the breakdown of 
left-right symmetry through the identification of external spacetime in the breaking 
of the full symmetry of  $\lvfs$.
 This observation is accompanied by the caveat concerning the Weyl spinor composition 
of the components of $X\subset \mcX$ and $Y \subset \mcY$.
 With this in mind, and hence with quote marks  placed on the $\nu_L$, $u_L$ and $u_R$ 
states, the relation between the component 
 structure for elements of $x \inn F(\htho)$, in the form of equations~\ref{ftscomp} 
and \ref{hthoxy}, and the first generation of Standard Model fermions is summarised in 
  equation~\ref{fhthopart}.

  As described in the previous section, in order to obtain left-handed Weyl spinors in 
the components of $X= \binom{\:\!\!p \;\! \; \bar{a}}{a \; m}$ a further decomposition 
is required, as for example in equation~\ref{xththx} or \ref{xththpx}; with a similar 
decomposition of $Y=\binom{P \; \bar{A}}{A \; M}$, as for example $Y = 
\thY\thY^{\dag}$ with 
$\thY = \binom{\bar{R}}{S} \inn \ooo^2$, also required to obtain the corresponding 
right-handed spinors within the components of $Y\subset \mcY$. 
 With $a=sr$ in equation~\ref{xththfull} or $a=sr+s'r'$ in equation~\ref{hthoext} for 
the $a \inn \ooo$ component of $X$, and similarly with $A=SR$ for example for the $A 
\inn \ooo$
component of $Y$, this decomposition is related to the octonion property of triality 
for SO(8) transformations, as described near the opening of section~\ref{earlyo} and 
around equation~\ref{trialabc}.  In fact, with the $\esi$ `rotations' acting in the 
same way on the $\mcX$ and $\mcY$ components by equation~\ref{dualrot} and following 
the discussion before equation~\ref{phxtran} in the previous section, the triality 
symmetry implies that each of $a,s,r,A,S,R \inn \ooo$ transform in precisely the same 
way under the action of any $\suth_c \subset \mbox{SO}(8)$ transformation.

 The Lorentz spinor structure under the external $\sltc^1$ symmetry may also be 
obtained under an alternative decomposition of $X$ and $Y$  based on the $\htwc 
\subset \htwo$ subspaces, as for example in equation~\ref{htwocp}, as also described 
in the previous section. This possibility may also be relate to the structure of the 
technicolor models reviewed subsection~\ref{suboomahp}. 
 With the external 4-vector $\bh_2 \inn \htwc$ accommodated within the $Y$ components 
and left-handed neutrino $\nu_L$ to be accommodated in the $X$ components ultimately a 
\textit{different} decomposition of the $X$ and $Y$ components may be involved in  
consistently accounting for the corresponding empirically observed phenomena.
 These phenomena require the correct matching of the internal $\suth_c \times \uo_Q$ 
action to the observed fermion multiplets of equation~\ref{multsf}.
  Indeed, as also described in the previous section, some care is needed in order to 
maintain the $\Sbard_l^1$ charge structure correlating with $\nu$-lepton and $u$-quark 
states in the spinor decomposition.  
 Ideally a yet higher-dimensional form of $\lv$ may prove the best guide for 
uncovering this structure in a mathematically natural manner.

  While further components are needed to unfold the full spinor structure, under the 
enlargement of the symmetry group from $\esi$ to $\ese$ on the temporal form $\lvfs$  
we next consider the possible identification of an internal $\sutw_L$ action within 
the $\ese$ symmetry structure.
 The Dynkin diagram for the rank-7 Lie algebra $\ese$ is compared with that for the 
rank-6 Lie algebra E$_6$ in figure~\ref{dynkinee}. 
\begin{figure}[htbp]  
\centering
\epsfxsize=\maxwidth
\leavevmode
\epsffile[0 0 2114 291]{\gpath aPfig91e}
\caption{\setb The  Dynkin diagrams for the (a) $L(\esi)$, (b) $L(\ese)$ and (c) 
$L(\ee)$ Lie algebras, which may be contrasted with those for the subalgebras listed 
in figure~\ref{dynkine}.}
\label{dynkinee}
\end{figure}

  Unlike the case for $\esi$, the Lie algebra $\ese$ does contain a rank-6 subgroup 
corresponding to the combined external Lorentz symmetry and internal gauge symmetry of 
the Standard Model, that is:
\begin{equation}
 \label{esedecom}
   \sltc \; \times \; \suth \times \sutw \times \uo \;\;\, \subset \;\;\, \ese
\end{equation} 

  The description of the internal symmetry, defined in section~\ref{intsym} as the 
stability group of the external $\htwc \equiv \TM_4$ spacetime  components and adapted 
here with respect to the external  components of $\bh_2 \subset \mcY $, will be 
augmented beyond the 31 $\esi$ generators of table~\ref{stabset} to a complete set for 
$\stabse \subset \ese$. These will include for example the actions $\psi(\dot{D})$ of 
equation~\ref{psidacta} for which $\dot{D}(\bh_2) = 0$ as well as any linear 
combination of the four sets of $\ese$ generators in 
equations~\ref{ftsstrana}--\ref{psidacta} which sum to zero on the four
 projected components of $\bh_2 \subset \mcY$ in equation~\ref{fhthopart}. An $\sutw 
\subset \stabse \subset \ese$ subgroup, independent of the $\sltc^1 \times \suth_c$ 
symmetry, acting upon the left-handed spinors of $\mcX$ and, together with 
  the identification of a further $\uo$
  action, completing the $\ese$ decomposition of equation~\ref{esedecom} may be 
considered as a candidate for the $\sutw_L \times \uo_Y$ gauge symmetry of the 
Standard Model. Indeed such an $\sutw \subset \stabse \subset \ese$, having not been 
identified within the $\esi$ generators, being internal to the components $\bh_2 
\subset \mcY$ whilst acting freely on $\mcX$ and hence constructed asymmetrically in 
terms of $\phi(\dot{C})$ and $\psi(\dot{D})$, would be expected to have an asymmetric 
action on the left and right-handed spinors identified in equation~\ref{fhthopart}.

Empirically it is the gauge bosons of an $\sutw_L$ gauge symmetry which mediate 
interactions within doublets of quarks $\binom{u}{d}_{\! L}$ and leptons 
$\binom{\nu}{e}_{\! L}$. Hence the identification of such an internal symmetry within 
the present theory may  be a valuable guide to the full identification of left-handed 
$u$-quark and $\nu$-lepton states
 in equation~\ref{fhthopart}
 given that we have already identified left-handed $d$-quark and $e$-lepton states 
within the $\theta^1$ components of $\mcX$. 
  With a different action on the $\mcX$ and $\mcY$ components of $x = 
  \binom{\alpha \;\; \mcX}{\mcY \;\; \beta}$
   in principle  the identification of such an $\sutw_L \subset \ese$ gauge symmetry 
  is free to act on the left-handed doublets derived for example from the components 
of $\binom{\thX}{\theta^1}_{\! L}$ identified within $\mcX$, without impinging upon 
the external spacetime components of $\mcY$. This hence provides a free channel for 
charged weak transitions within the leptonic $\binom{\nu}{e}_{\! L}$ and quark 
$\binom{u}{d}_{\! L}$ doublets which may be extracted from equation~\ref{fhthopart}. 
  The analysis of such an $\sutw_L$ action
   relating to $W^{\pm}$ gauge boson interactions,
   consistent with the appropriate $\suth_c \times \uo_Q$ transformations and charges 
for the left-handed states,  may also clarify the structure of left-handed spinors 
themselves within the $X$ components.
More generally, guided by standard electroweak theory, the identification of the 
$\nu$-lepton and $u$-quark left-handed spinors in the components of $X$ will be 
mutually related to a determination of the composition of an internal
 $\sutw_L \times \uo_Y \subset \ese$ symmetry action itself.
 
   Towards this end, and in contrast with the opening of section~\ref{intsym}, an 
internal symmetry might be defined as any group $\ul{G}$ consistent with the subgroup
  decomposition $\sltc^1 \times \ul{G} \subset \ese$ for which the set of $\sltc^1$ 
spinors transform under the trivial or fundamental representations of $\ul{G}$. That 
is, while the external $\sltc^1$ symmetry partitions the components of $\lvh$ into 
irreducible pieces, including the spinors $\theta_{l,i,j,k}$ of 
equation~\ref{thcth234} and table~\ref{sltcreps} each composed of four real 
components, the internal symmetry $\ul{G}$ respects this partitioning  in treating the 
Weyl spinors as individual components of a representation of $\ul{G}$. This definition 
excludes for example the $\sutw$ generated by $\dot{G}_q + 2\dot{S}_{q}^{1}$ for $q = 
i,j$ and $k$ which, as described in the opening of subsection~\ref{strassy}, does not 
transform the spinors  $\theta_{l,i,j,k}$ as a fundamental representation, but does 
still include the internal $\suth_c \times \uo_Q$ symmetry as identified in 
section~\ref{intsym}, with the actions on the spinors 
 described in table~\ref{suttan} and equation~\ref{slonthet} as
 summarised, via equation~\ref{sbardot}, in equation~\ref{suthuoth}. The question then 
regards the uniqueness of this $\suth_c \times \uo_Q$ action or the existence of 
further internal symmetry groups which possess a similarly tidy action on the spinors.

  At the same time the action of $\sutw_L \subset \ese$  might still be expected to be 
closely related to the subgroups $\sutw^{2,3}\times \uo^{2,3} \subset \esi$ acting on 
the components of $\mcX$, since the latter have desirable properties in relation to 
electroweak theory as described in the `mock electroweak theory' of 
section~\ref{ewtfesb}.
 These include the  $\Sbard^1_l$ charges for the $\tilde{W}^{\pm}$ and $\tilde{Z}^0$ 
gauge bosons, for example for $\dot{\Sigma}^{(2)\pm}$  in equation~\ref{isscomm}, and 
the similarity of the linear dependencies for the corresponding $\esi$ generators, as 
seen for example in equation~\ref{srslincom2}, to the structure of 
equation~\ref{qethy2} for the Standard Model. In attempting to fit an $\sutw_L \times 
\uo_Y$ symmetry into the $\esi$ analysis the generator $\Sbard^2_l$ was also found to 
provide the correct hypercharges for the left-handed fermion states in $\mcX$ as 
described following equation~\ref{sssrdot}. An $\sutw_L \times \uo_Y \subset \ese$ 
symmetry action will differ for the $\mcX$ and $\mcY$ components of $x\inn F(\htho)$, 
with for example presumably $Q = \frac{Y}{2}$ required for the right-handed spinors in 
$\mcY$ as singlets of $\sutw_L$.

  A quantitative test of the $\ese$ symmetry breaking structure might be found in a 
calculation of the electroweak mixing angle $\theta_W$, following a similar derivation 
that led to $\sin^2 \thetmt = \frac{3}{4}$ in equation~\ref{sinmock} for the mock 
electroweak theory within the $\esi$ structure. 
 As described in section~\ref{intsym} the relative coupling of the $\uo_Q$ gauge 
symmetry to the fermions, in terms of the fractional charges of the quarks, already 
matches the observed values.
 The relative value of the internal $\suth_c$ coupling  to the spinor components of 
equation~\ref{fhthopart}, in comparison with the electroweak couplings, with respect 
to a normalised $L(\ese)$ Killing form could also in principle be calculated.

 The explicit structure of the $\esi$ symmetry actions on the cubic form of $\mcX \inn 
\htho$, obtained by generalisation of Lorentz transformations on quadratic forms 
\cite{Wang,Man4,Man5,Wang2}, could ideally be further generalised to obtain the 
structure of the $\ese$ symmetry actions on the quartic form of $x \inn F(\htho)$.
 This would involve an additional $133-78=1+27+27 = 55$ generators from 
equations~\ref{lamtrana}--\ref{psidacta} now expressed as tangent vectors to the 
56-dimensional space of $F(\htho)$. In principle the application of 
equation~\ref{rrbrac} for
 the full set of $\ese$ actions on $x \inn F(\htho)$
 could in turn  be used to determine the full $133 \times 133$ $L(\ese)$ table, 
building upon the $78 \times 78$ $L(\esi)$ table in \cite{Wang}.
 Another approach to such a construction might be based on the identification of 
$L(\mbox{E}_{7(-25)})$ with the Lie algebra of the symplectic group Sp$(6,\ooo)$ as 
described in \cite{DrayMW}.

  An $\sutw_L$ action might then be sought using 
  the new generators, either solely or in combination with the original 78 $\esi$ 
generators, acting on a set of left-handed Weyl spinors, via a spinor decomposition of 
$X$, identified within the components of $\mcX$ in equation~\ref{fhthopart}, as guided 
by the nature of electroweak interactions for the fermions.
 Since the $\sltc^1$ Lorentz symmetry and $\suth_c \times \uo_Q$ internal symmetry 
have already been identified in sections~\ref{extsym} and \ref{intsym} within the 
$\esi$ actions of equation~\ref{ftsstrana} it may be possible to use this as a 
starting point to more directly construct an $\sutw_L$ symmetry with appropriate 
properties out of the further generators listed in 
equations~\ref{ftsstrana}--\ref{psidacta}. That is in seeking a particular $\sutw_L 
\subset \ese$ action as represented on the components of $\mcX$ in 
equation~\ref{fhthopart} the full $133 \times 133$ Lie algebra table for $\ese$ may 
not be required.

  On the other hand the study of the complete algebra, and the subalgebras it 
contains, may be necessary to both identify the actions $\sutw_L \times \uo_Y$ 
corresponding to electroweak theory and to determine the weak mixing angle 
$\sin^2\theta_W$ for the present theory.
   Even in this case a `quantisation' of the theory to describe the phenomena of 
`running coupling' may be necessary in order to make comparison with the  value of 
$\sin^2 \theta_W \simeq 0.23$ as empirically determined at the energy scale of $M_Z$, 
as alluded to shortly after equation~\ref{ggesets} 
in subsection~\ref{submangle}.
  This full picture may also be needed to include the $\suth_c$ interactions in this 
comparison, given the differing behaviour of the running coupling associated with each 
of the three components of $\SML$ in the Standard Model as sketched in 
figure~\ref{runcup}.

  In constructing an $\sutw_L \subset \ese$ action with an appropriate action on the 
components of $\mcX$ in equation~\ref{fhthopart}, including upon the four 
$\theta_{l,i,j,k}$ left-handed Weyl spinors, as part of an 
  $\sltc^1 \times  \suth_c \times \sutw_L \times \uo_Y$ subgroup decomposition, as an 
exemplification of equation~\ref{esedecom}, the $\sutw_L$ action might also be found 
to act non-trivially on the $\mcY$ components of equation~\ref{fhthopart} and in 
particular impact upon the external $\bh_2 \inn \htwc \equiv \TM_4$ components. 
  This is analogous to the $\doo_B \subset \esi$ action in the decomposition of 
equation~\ref{uobtzbtz} which, although independent of $\sltc^1$ in the Lie algebra, 
with the generator of equation~\ref{bbdil}, clearly impacts upon the external 
spacetime components.

 In the present theory it is proposed that some of the differing properties of the 
internal gauge interactions associated with $\sutw_L$ compared with  $\suth_c \times 
\uo_Q$ arise  since the latter forms a subgroup of $\stab \subset \esi$, and even of 
$\stabto \subset \sltwoo$ considered as a subgroup of a 10-dimensional spacetime 
symmetry
 as described in the opening of section~\ref{subehafws}, while  the former is only to 
be identified as a subgroup of $\ese$, acting on a quartic form of temporal flow, such 
that
 $\sutw_L$ is \textit{not} a subgroup of $\stabse$. This structure is further proposed 
to be closely related to the phenomena of electroweak symmetry breaking in the 
Standard Model, based on the study of the mock electroweak theory described for the
 $\sutw^{2,3} \times \uo^{2,3}$ subgroups
 of $\esi$, acting on a cubic form, as described in section~\ref{ewtfesb}. 
 
   These features of a higher-dimensional temporal form $\lvh$ of cubic or higher 
polynomial order are distinct from those of a quadratic spacetime form. For the model 
considered in section~\ref{reaic} with the quadratic form $\lvte$, representing a 
10-dimensional form of time which can also be interpreted as a higher-dimensional 
spacetime structure, the external $\soot$ and internal $\sox$ components of the broken 
full $\sootn$ symmetry act \textit{independently} on the external $\ol{\bv}_4$ and 
internal $\ul{\bv}_6$ components of temporal flow $\bv_{10}$, respectively, as 
depicted in figure~\ref{mtogmaphr}(b). 
 On extension to the cubic form of time $\lvt$ the external $\soot$ symmetry was found 
to also act on the extra `internal' dimensions of $\theta = \binom{c}{\bar{b}}$, 
identifying a set of four Weyl spinors, as described in section~\ref{extsym} and 
contrasted with the 10-dimensional spacetime case at the end of that section. Here we 
make the complementary observation that a component of the \textit{internal} symmetry 
$\ul{G}$, in the subgroup decomposition $\sltc \times \ul{G} \subset \hat{G}$ with 
$\hat{G} = \esi$
 or $\ese$, can itself act on the projected \textit{external} $\bv_4 \inn \TM_4$ 
spacetime components. This possibility, for a cubic or higher form of temporal flow, 
  is proposed to underlie the origin of mass for the corresponding gauge bosons.
  
  With significant physical properties deriving from the combined action of the 
external and internal symmetry on both the external and internal temporal components 
the present theory deviates significantly from models based on a higher-dimensional 
spacetime. In particular these observations mark a departure from the resemblance with 
Kaluza-Klein theories,  as reviewed in chapter 4 and incorporated into the geometric 
structures of the present theory in section 5.1, which may assist in the aim of 
deriving a relation between the external and internal geometry, in the form of 
equation~\ref{gchift}, in the context of the present theory alone.

   Similarly as for the proposed $\sutw_L \times \uo_Y \subset \ese$ subgroup
  the  $\sutw^{2,3}\times \uo^{2,3} \subset \esi$  actions are \textit{not} contained 
within $\stab \subset \esi$. However in section~\ref{ewtfesb} the impingement of these 
actions on the $\bh_2 \inn \htwc$ components of
  $\mcX$ were seen to be analogous in structure to the Higgs mechanism of electroweak 
symmetry breaking and the origin of the masses for the $W^{\pm}$ and $Z^0$ gauge 
bosons. 
The  object $\bh_2$, now accommodated in the $\mcY$ components in 
 equation~\ref{fhthopart}, together with the properties of its components, is now 
considered as the `vector-Higgs', providing the source for the empirically observed 
Higgs phenomena.
 
 The corresponding components `$\bh_2$'$\subset X \subset \mcX$, in the complementary 
$\htwc \subset \htwo \subset \htho$ subspace of $x \inn F(\htho)$ can be opened up by 
a spinor decomposition, as described in the previous section, to account for the 
$\nu_L$ fermion state. 
  On the other hand while the vector-Higgs
   $\bh_2 \subset Y \subset \mcY$ \textit{could} be interpreted to be composed of 
spinors, in the form of equation~\ref{htwocp} and by analogy with technicolor models 
for example, the physical expression of these components is directly in terms of a 
tangent vector $\bv_4 \inn \TM_4$ in the external 4-dimensional spacetime. Hence in 
particular a right-handed neutrino $\nu_R$ cannot be accommodated in the $\mcY$ 
components.

 The two kinds of interaction for the $\sutw_L\times \uo_Y$ gauge fields on the $\mcX$ 
and $\mcY$ components of equation~\ref{fhthopart}  contain  analogous structures to 
the Standard Model Lagrangian terms respectively for the weak interactions of 
left-handed fermions, such as  equations~\ref{lagklep} and \ref{covdevlep}, and weak 
coupling to the Higgs field, such as equations~\ref{laghiggs} and \ref{covderphi} -- 
with the same mixing angle $\theta_W$ applying in both cases as described after 
equation~\ref{mwmzcth}. For the present theory 
  the $\sutw_L\times \uo_Y$ symmetry breaks to $\uo_Q$ as generated by $\Sbard^1_l 
\inn L(\esi) \subset L(\ese)$ which acts upon $\mcX$ and $\mcY$ in the same way. In 
the case of the $\mcX$ components the $\Sbar^1_l \equiv \uo_Q$ action misses the 
$\nu_L$ components of equation~\ref{fhthopart} accounting for the charge neutrality of 
the neutrino, which is described in the Standard Model in terms of 
equations~\ref{lagklep}--\ref{tantw}. In the case of the $\mcY$ components the 
$\Sbar^1_l \equiv \uo_Q$ action misses the $\bh_2$ components of 
equation~\ref{fhthopart} here potentially accounting for the 
massless nature of the photon,  as suggested for
 the gauge field $\tilde{A}_{\mu}(x)$ for the mock electroweak theory in 
equation~\ref{zwamass}, and as constructed for the Standard Model in 
equation~\ref{noamass}.

   These two different aspects of electroweak theory may hence here be described 
together in terms of the broken $\ese$ action on the $\mcX$ and $\mcY$ components of 
$L(\bv_{56})=q(x)=1$ in equation~\ref{fhthopart}. While the $\sutw_L \times \uo_Y 
\subset \ese$ action may differ on the $\mcX$ and $\mcY$ components, involving the 
asymmetric actions of equations~\ref{phicact}  and \ref{psidact}, a unique mixing 
angle $\theta_W$ and surviving $\uo_Q$ symmetry with the same action on $\mcX$ and 
$\mcY$ should result from the symmetry breaking  over the external $\bh_2 \subset 
\mcY$ components.
 In principle a complementary $\sutw_R \times \uo'_Y \subset \ese$ subgroup might also 
be identified, by reversing the contributions from equations~\ref{phicact}  and 
\ref{psidact}, however such a symmetry, acting on right-handed doublets of quarks 
$\binom{u}{d}_{\! R}$, may be heavily suppressed due to a larger impact on the 
external $\bh_2 \subset \mcY$ components. 

 In the quantum theory the propagators for the gauge fields will 
 attain a finite mass through interaction with the external $\bh_2$ components. 
Naturally a `quantisation' scheme and particle concept will be needed in order to 
assess the properties of the particle content of the theory
  (as will be developed in chapter~\ref{newapp}), with all physical particle states 
transforming under well-defined representations of both the external and internal 
symmetry.

  In identifying an $\sutw_L \times \uo_Y \subset \ese$ symmetry
 it will be desirable to maintain the features of the electroweak theory studied in 
section~\ref{ewtfesb}, in particular with a degree of 
  impingement on the $\bh_2 \subset \mcY$ components accounting for the corresponding 
gauge boson masses. 
 This is counter to the provisional assumption in the opening of section~\ref{intsym} 
that an internal symmetry should belong to the stability group $\stab$ of the external 
spacetime components 
 $\bh_2 \inn \htwc \equiv \TM_4$. Rather here  in a decomposition such as 
equation~\ref{esedecom} the emphasis is upon 
  defining internal subgroups  through the structure of
   their well-defined representations on the external
  $\sltc^1$ spinors.

  The internal $\suth_c \times \uo_Q$ symmetry may also be motivated in this way, as a 
component of the $\ese$ decomposition with well-defined representations on the spinor 
components, as seen in equation~\ref{suthuoth} for example. In this case the fact that 
it also \textit{happens} that $\suth_c \times \uo_Q \subset \stabse$ is responsible 
for the fact that the gauge bosons associated with QCD and QED \textit{happen} to be 
massless. 
  It may also be the case that an internal  $\sutw_L \times \uo_Y \subset \ese$ 
symmetry might also impinge on any of the scalars $\alpha, \beta, n$ and $N$ of 
equation~\ref{fhthopart}, all of which are invariant under the $\suth_c \times \uo_Q$ 
action. The possible physical consequences of these scalar components remains to be 
seen, whether in terms of masses for the gauge bosons and fermions or other effects 
(as will be considered in chapter~\ref{anpocs}).
 It also remains to be seen whether the action of an  $\sutw_L \times \uo_Y \subset 
\ese$
  on the $\bh_2\subset \mcY$ components may be more closely analogous to the action of 
the Standard Model group $\sutw_L \times \uo_Y$ on the Higgs complex doublet $\phi$ 
than was the case for the mock electroweak theory.

  In the Standard Model fermion masses are introduced through Yukawa couplings to the 
Higgs field, as described in the Lagrangian of equation~\ref{Yukferm}. In the present 
theory there is neither a fundamental scalar Higgs field nor an explicit Lagrangian, 
however amongst the long list of quartic terms in the expression for $q(x) \equiv 
\lvfs$ in equation~\ref{fquartall} the top line includes the terms:
\begin{equation}
 q(x) \sim (\alpha\beta - ph - mh - nN)\;(\langle b,B
	   \rangle \, + \, \langle c,C \rangle ) 
  \label{qxmass}
\end{equation}
  with the `vacuum value' $P=M = v^0 = h$ substituted in using equation~\ref{v4vac} 
applied to the
  `vector-Higgs'
   $\bh_2 \subset Y \subset \mcY \inn \htho$ components of $x\inn F(\htho)$. Terms of 
this form contain both left-handed $\binom{c}{\bar{b}} \subset \mcX$ and right-handed 
$\binom{C}{\bar{B}} \subset \mcY$ components and in this sense are 
 reminiscent of the Standard Model Lagrangian mass terms deriving from 
equation~\ref{Yukferm}.
 The terms of equation~\ref{qxmass}, 
 in potentially contributing to the fermion masses in the full theory, supersede   the 
cubic terms such as $h(b\bar{b} + c\bar{c})$ obtained from the form $\det (\mcX) 
\equiv \lvt$ as described for equation~\ref{hexpan2}. In both cases the full set of 
terms transform under the broken symmetry of $\lvh$ rather than under the symmetry of 
a Lagrangian. The introduction of non-standard mass terms, as extracted from 
equation~\ref{fquartall}, is not unprecedented  as can be seen by comparison with the 
quartic term of equation~\ref{fmtech} for the technicolor model described in 
subsection~\ref{suboomahp}.

A key part of developing the present theory will be the identification of the 
empirically observed properties of the neutrino sector. 
Of particular interest will be to identify a description of neutrino oscillations, and 
contrast that structure with the CKM mixing in the quark sector. These structures are 
also expected to relate closely to the identification of fermion masses.
  The low value of the left-handed neutrino mass may correlate in this theory with the 
lack of a  right-handed  counterpart  in the components of $\mcY$.
 Again with reference to equations~\ref{ftscomp}, \ref{hthoxy} and \ref{fhthopart}, 
and based on the structure of equation~\ref{suthuoa}, the neutrino is associated with 
the 
$a_{1,l}$ components of $\mcX$. As possible contributions  to the particle masses, in 
addition to equation~\ref{qxmass} above, equation~\ref{fquartall} contains the quartic 
terms:
\begin{eqnarray}
   q(x) & \sim & PM \vert a \vert^2  \, + \, MN \vert b \vert^2 \, + \,
               NP \vert c \vert^2  \nonumber \\
	    & = &  h^2 	\vert a \vert^2  \; + \; 
		h N ( \vert b \vert^2 \, + \,  \vert c \vert^2)	\label{qxmassnu}   
\end{eqnarray}
 where the second line again follows on substituting the vacuum value $P=M = v^0 = h$. 
Hence, 
 as well as relating to the lack of right-handed $A_{1,l}$ neutrino components in 
$\mcY$, 
  the low mass of the neutrino in comparison with the electron might depend upon the 
magnitude of 
  $h$ in comparison with that of the scalar field $N(x)$.
   In any case the quartic `mass terms' for the neutrino state do differ from those 
for the charged lepton, and in a somewhat more complicated manner than suggested by 
the terms of equation~\ref{qxmassnu} alone.
  Although the $u$-quarks have both left and right-handed
 components there are also differences between such quartic terms for the $u$ and $d$ 
quark states identified in equation~\ref{fhthopart}; in this case required to account 
for the smaller empirical mass difference with $\frac{m_u}{m_d} \sim 0.5$ \cite{PDG}. 
The proximity of the $u$ and $d$ quark masses may in fact be related to the attainment 
of a stable vacuum value for $\lvfh$, as will be discussed in section~\ref{sectveu}, 
and which may also correlate  with the low mass for the neutrino.

  With the ambiguity over the possible \textit{mathematical} ways in which to 
decompose  the components of $X,Y \inn \htwo$ into a set of spinors, as described in 
the previous section,
  ultimately an understanding of the nature of the empirically observed particle 
states as originating  out of the present theory may be required in order to motivate 
a natural \textit{physical} choice for such a decomposition. 
 The full identification of  scalar, spinor and gauge boson particle states will 
require consideration of a means of `quantisation' for the present theory, and we 
pursue that direction in the following two chapters. 

  The scheme in equation~\ref{fhthopart} accounts for one family of quarks and leptons 
with the appropriate transformations under the  internal $\suth_c\times \uo_Q$ 
symmetry and external $\sltc^1$ symmetry, within the above  caveat for the $u$-quark 
and $\nu$-lepton fermion states.   
In addition  the particle states yet to be identified include the second and third 
generation of fermions, as related through CKM mixing in the case of the quark states, 
and their relation to the massive gauge bosons associated with electroweak theory in 
the Standard Model. In the following section we speculate on the possible nature of a 
yet higher-dimensional form of temporal flow in principle capable of accommodating 
these phenomena.

 
\section{$\ee$ Symmetry and the Standard Model}
\label{sosmfi}

  The extension from $\esi$ acting on $\lvt$ to $\ese$ acting on $\lvfs$ can be 
considered as a continuation of the progression to higher-dimensional forms of 
temporal flow which began with the $\sltc$ Lorentz symmetry of the quadratic form 
$\lvf$ on 4-dimensional spacetime.
   This progression, the first stages of which were also described in the opening of 
section~\ref{subehafws}, is summarised here in table~\ref{lvftolvfs}.
\begin{table}[htbp]
\centering
\begin{tabular}{|l|ccccr|}
 \hline
        &  form & dimensions       & space  & symmetry & \# generators   \\
 \hline
 $\lvf$ & quadratic &  4 spacetime & $\bv_4 \inn \htwc$  & $\sltc$ 
                                                            &  $6 \qquad \;$ \\
 $L(X)\:\!=\:\! 1$ & quadratic &  10 spacetime & $X \inn \htwo$  & $\sltwoo$ 
                                                            &  $45 \qquad \;$ \\
 $L(\mcX)\:\!=\:\! 1$ & cubic & 27 temporal & $\mcX \inn \htho$ & $\mbox{E}_{6(-26)}$ 
                                                            &  $78 \qquad \;$ \\ 
 $L(x)\:= \, 1$ & quartic & 56 temporal & $x \inn F(\htho)$ & $\mbox{E}_{7(-25)}$ 
                                                            &  $133 \qquad \;$ \\ 
   \hline
  \end{tabular}
  \caption{\setb Four-dimensional spacetime, as a form of temporal flow itself, may be 
embedded in a progression of higher-dimensional temporal forms.}
\label{lvftolvfs}
\end{table} 
 
 The highest dimensional form of temporal flow $\lvfs$ has a symmetry breaking pattern 
to $\mbox{E}_{6(-26)} \subset \mbox{E}_{7(-25)}$  with
 the representations of equation~\ref{esetoesi}
	as exhibited by the structure of equation~\ref{ftsstran}. This is analogous to the 
further breaking pattern of $\esi$ to $\sltwoo \equiv \spotn$, as described by the 
representations of equations~\ref{horhoo}--\ref{esidecom}, which is also implied in 
the structure of left-hand side of equation~\ref{fhthopart}. The $\sltwoo$ symmetry of 
10-dimensional spacetime is an intermediate stage on the way down to the Lorentz 
$\sltc$ symmetry which further decomposes the representation space into a Lorentz 
4-vector, Weyl spinors and Lorentz scalars, as described in table~\ref{sltcreps} and 
now applied to both $\mcX$ and $\mcY\inn \htho$, with the external Lorentz 4-vector 
$\bv_4 \inn \TM_4$ accommodated within the $\mcY$ components in 
equation~\ref{fhthopart}, where two further Lorentz scalar  components $\alpha$ and 
$\beta$ are also identified. 

  Apart from the three additional scalars, $N$, $\alpha$ and $\beta$ in 
equation~\ref{fhthopart},
   the increase in dimension from 27 to 56 does not contain any redundancy in terms of 
comparison with the structures of the Standard Model. Most of the additional 29 
dimensions are interpreted as an augmentation from 2-component  Weyl spinors to 
4-component Dirac spinors, together with a separation in the identification of the 
left-handed neutrino state and
 the external spacetime $\htwc \equiv \TM_4$ components.

 At the level of the Dynkin diagrams of figures~\ref{dynkinee}(a) and (b) the $\ese$ 
algebra marks a minimal extension from $\esi$, but one which together with the 
$\mathbf{56}$ representation as described in the previous section may be sufficient to 
account for much of the structure of Standard Model symmetries and particle spectrum 
with little further augmentation.  This additional augmentation is needed to account 
for the identification of the $u$-quark  and $\nu$-lepton spinor components which will 
require a further decomposition of the $X \subset \mcX$ and $Y \subset \mcY$ 
components, with for example  $X =  \thX\thX^{\dag} +   \phX\phX^{\dag} + \ldots$ as 
described in equations~\ref{xththx} and \ref{xththpx} of section~\ref{subehafws}. The 
features of these two means of augmenting the form of temporal flow to a higher 
dimension, as described in the two previous sections, will need to be combined in the 
complete theory. One possible means of achieving this will be described in the present 
section.

   In terms of the dimension of the underlying space, as listed for the sequence of 
forms $\lv$ in table~\ref{lvftolvfs}, we first note that a further expansion from $56$ 
to $\sim\,$80
real components would be sufficient incorporate Weyl spinors for the $\nu_L$, $u_L$ 
and $u_R$ states of equations~\ref{fhthopart}. This is deduced by observing that 
$a\inn \ooo$ of equation~\ref{suthuoa} has 8 real components while a set of four Weyl 
spinors requires a total of 16 real components, or alternatively by noting that
the decomposition of the form $X =  \thX\thX^{\dag}$ involves an augmentation from 10 
to 16 real components.
 With a complete generation of Standard Model fermions then accounted for the second 
and third generations might also be  directly incorporated under a further 
augmentation from 80 to $\sim\,$240 real components.

  Given the progression to larger symmetry groups summarised in table~\ref{lvftolvfs} 
from a mathematical point of view it is also natural to consider whether the Lie group 
$\ee$, as the largest exceptional Lie group, represented on a quintic homogeneous form 
$\lv$, may  mark one further and final possible step in this sequence.
 (While we refer to such a hypothetic `quintic' form, essentially an order greater 
than quartic is implied).
 With the smallest non-trivial representation of $\ee$ being 248-dimensional, this 
possibility is particularly worth consideration in light of the observations of the 
previous paragraph. 
  In a similar way that extending the symmetry from $\esi$ to $\ese$ led to the 
incorporation of right-handed as well as left-handed fermion states,  
     ideally a further extension to $\ee$ would subsume both the $\ese$ symmetry of 
the structure in equation~\ref{fhthopart} and explicitly  incorporate also the 
$u$-quark  and $\nu-$lepton spinor states and a full three generations of fermions
 all under a higher-dimensional form of $\lv$ with an $\ee$ symmetry.

  The smallest non-trivial representation  of $\esi$ is the \textbf{27} which can be 
expressed as the symmetry of the cubic form $\lvt$, while for $\ese$ the  \textbf{56} 
representation,
 again the lowest-dimensional non-trivial representation, can be expressed as the 
symmetry of the quartic form $\lvfs$. However the \textbf{248} representation for 
  $\ee$ is expressed in terms of the adjoint representation on the 248-dimensional 
$\ee$ Lie algebra itself, with no clear interpretation in terms of a symmetry of a 
form of temporal flow $\lv$.
 Indeed the Lie algebra $\ee$ can be essentially introduced in terms of its action on 
itself,  and constructed in purely \textit{algebraic} terms which may involve the 
octonions \cite{Baez1},
  with the absence of any \textit{geometric} motivation or application which might be 
related to a form $\lv$.

 The group $\ee$ \textit{can} be defined as the symmetry group of a 57-dimensional 
manifold based on $F(\htho)+\rrr$, known as the `extended Freudenthal triple system' 
equipped with extra geometric structure \cite{Rios}, with E$_7 \subset \mbox{E}_8$ now 
identified as a subgroup.
 These mathematical objects have some  connection with the structures of M-theory 
\cite{Borst,Rios}, and indeed $\ee$ features heavily in some branches of theoretical 
physics as for example in $\ee \times \ee$ heterotic string theory.  
  There is also some debate in the literature concerning whether or not the structure 
of the $\ee$ Lie algebra alone is large enough to fully describe the Standard Model 
together with gravity (see for example~\cite{Eenoyes}).

 For the present theory with an $\ee$ symmetry acting on a \textit{hypothetical} form 
of temporal flow, which may be denoted $\lvtfe$, three generations of fermions 
together with a vector-Higgs could be accommodated within the 248 temporal components, 
as outlined above, while the external Lorentz group and internal
 gauge symmetries could all in principle be identified within the $\ee$ group actions. 
Without the need to employ a supersymmetry there are no SUSY states or set of `mirror' 
particles of any kind, although it is of course possible that new physics might be 
predicted with consequences that might be tested.
 However in the present theory the primary guiding principles are driven by 
\textit{conceptual} ideas, rather than taking a
 fundamental motivation from a notion of \textit{mathematical} elegance.
  Hence here it is the possible forms of temporal flow $\lv$, together with their  
symmetries, which lead the development of the theory, and this may or may not involve 
the Lie group $\ee$.
 The progression towards higher dimensions in table~\ref{lvftolvfs} does however 
strongly hint towards consideration of $\ee$, hence allowing this one lead  from the 
perspective of  mathematical beauty,
 we consider the possible marriage of this `aesthetically pleasing mathematics' with 
the underlying conceptual form of the present theory.
 
  The fact that the smallest non-trivial representation of the 248-dimensional $\ee$ 
Lie algebra is expressed as the adjoint representation does not itself preclude the 
possibility that a \textbf{248} representation may \textit{also} be identified in 
terms of the symmetries of a quintic form $\lvtfe$. By comparison for example the 
smallest non-trivial faithful representation of SO(3) is the adjoint representation on 
the 3-dimensional so(3) Lie algebra, but in this case there is \textit{also} a 
fundamental \textbf{3} representation preserving the magnitude of vectors $\bv_3 \inn 
\rrr^3$ in a 3-dimensional Euclidean space.  
 Indeed the SO(3) symmetry of the 3-dimensional form $\lvth$ of equation~\ref{flow3d} 
was the example of a symmetry of a multi-dimensional form of temporal flow with which 
we began in section~\ref{perc}. In this case the two representations of SO(3) are 
closely related
 since the bilinear Killing form on the elements of the so(3) Lie algebra has the same 
structure as the quadratic scalar product in the space $\rrr^3$ as used in forming the 
magnitude $\vert \bv_3 \vert$.

  A quintic form underlying $\lvtfe$, invariant under an $\ee$ symmetry action, may 
not be as closely related to the adjoint representation and the $L(\ee)$ algebra 
structure, unless such a quintic form (or more generally a homogeneous polynomial form 
of order greater than four) might be related to the bilinear Killing form on $L(\ee)$ 
in some way.
 Further, given the progression from the cubic polynomial form $\mbox{det}(\mcX)$ 
  of equation~\ref{detpmn} as an expression of $\lvt$ with an $\esi$ symmetry to the 
terms of the quartic form $q(x)$ of equation~\ref{fquartall} underlying the form 
$\lvfs$ with an $\ese$ symmetry, a possible quintic form for $\lvtfe$ with an $\ee$ 
symmetry may be a considerably more complicated mathematical object still. Hence it is 
perhaps conceivable that such a structure has not been identified through purely 
algebraic means, even over fifty years after the corresponding $\esi$ and $\ese$ 
structures were first realised.
 On the other hand if such a mathematical structure does exist, namely a quintic form 
$\lvtfe$ with an $\ee$ symmetry, then as for the other forms of table~\ref{lvftolvfs} 
it \textit{would} naturally apply for the present theory, based on multi-dimensional 
forms of temporal flow, and further physical consequences would be \textit{expected} 
to be uncovered in this further progression.

In reference \cite{Asch}, as an example of a more geometrical approach, all of the 
classical Lie groups are accounted for as isometry groups of  bilinear or sesquilinear 
forms and the first four exceptional Lie groups, $\gt$, $\ff$, $\esi$ and $\ese$, are 
described as isometry groups constructed for cubic or quartic forms, but with $\ee$ 
essentially absent from the discussion. More generally little reference has been 
identified in the literature in which a 248-dimensional representation of $\ee$ is 
described in terms of an action on a quintic, or any other homogeneous polynomial, 
form. However in \cite{CedP, GarG}
  a polynomial of degree \textit{eight} which is invariant as a 248-dimensional 
representation of the compact real form of $\ee$ is described, and is closely related 
to an invariant polynomial for the real form $\mbox{E}_{8(8)}$. For the present theory 
it is then an open question whether an octic form with an $\ee$ symmetry might  
contain the quartic form with $\ese$ symmetry. Such a natural extension consistent 
with the form of temporal flow $\lv$ may also be required to have a symmetry described 
by a non-compact real form of $\ee$ in order that temporal causality may be respected 
 for physical structures identified on the base manifold $M_4$, 
 with a local $\soot \subset \ee$ symmetry, as will be discussed in 
  section~\ref{secuni}.

   Considering the possible real forms of $\ee$ more generally, a suitable candidate 
would be 
  $\eeg$ since the following maximal subgroups involving the exceptional Lie groups 
are well known (see for example~\cite{Ramo}):
\begin{equation}
  \begin{array}{rcl}
   \eseg \times \mbox{SU}(1,1) & \subset & \eeg   \\
   \esig \times \mbox{SO}(1,1) & \subset & \eseg  
   \end{array}  \label{esixtoee}
\end{equation}
   This suggests the employment of the chain of non-compact real forms
   $\esig \to \eseg \to \eeg$ as symmetry groups for the corresponding forms of the 
sequence
   $\lvt \to \lvfs \to \lvtfe$, where the first two stages have been described here in 
chapter~\ref{esihtho} and section~\ref{secesef} respectively, while the third form 
remains hypothetical. As for the structure of the first two stages it seems likely 
that a construction of the final form in this progression will involve the algebraic 
structure of the octonions in a significant way.

 At the level of the complex Lie algebras the corresponding Dynkin diagrams for 
$\esi$, $\ese$
  and $\ee$ are displayed alongside each other in figure~\ref{dynkinee}. The Lie group 
generated by the rank-8 $L(\ee)$ algebra is large enough to contain a rank-8 
decomposition of the form:
\begin{equation}
 \label{eedecomp}
   \sltc \times \suth \times \sutw \times \sutw \times \uo \times \uo \subset \ee
\end{equation}
  as can be shown by straightforward  analysis of the Dynkin diagrams involved.
 Hence while the degrees of freedom of the components of $\bv_{248}$, as an extension 
from $\bv_{56} \equiv x \inn 
  F(\htho)$ of equation~\ref{fhthopart},  are sufficient to contain a full three 
generations of Standard Model fermions and a vector-Higgs, the $\ee$ symmetry group is 
comfortably large enough to describe the external Lorentz symmetry together with the 
internal $\SML$
 gauge group.
 
  While the higher-dimensional extensions of section~\ref{subehafws} were 
\textit{contrived}, for example via equation~\ref{xththpx} leading to the 
inhomogeneous expression of   equation~\ref{xththfull2}, in order to describe the 
further necessary spinors and generations for the Standard Model, ideally these 
structures will be found to arise \textit{naturally}  within a homogeneous form 
$\lvtfe$ under an $\ee$ symmetry broken over an external 4-dimensional spacetime 
$M_4$.  This natural structure should include a full set of
   $\suth_c \times \uo_Q$ transformations and charges aligned with the $\sltc^1$ 
spinors,  completing the structure identified within the $\esi$ action in 
equations~\ref{suthuoth} and \ref{suthuoa}, and in particular supplying a mathematical 
justification for the electromagnetic charges of the $\nu$-lepton and $u$-quark spinor 
states through a $\uo_Q$ action which might be related to the form of 
equation~\ref{chcont} for example.

 Towards the end of the previous section an $\sutw_R \times \uo'_Y$ subgroup was 
considered as a possible complementary alternative to $\sutw_L \times \uo_Y$ within a 
decomposition of $\ese$ in the form of equation~\ref{esedecom}, however the $\ee$ 
breaking structure of equation~\ref{eedecomp} can in principle accommodate both 
subgroups together. 
   In general a decomposition of $\ee$ in the form of equation~\ref{eedecomp}, arising 
from the symmetry breaking through a choice of $\sltc^1$ on the external spacetime 
$\TM_4$, will contain internal symmetry groups acting on the set of spinors which do 
not belong to the stability group $\stabee \subset \ee$. This may include for example 
an `$\sutw_R$', acting asymmetrically on the $\mcX$ and $\mcY$ components of the 
$\bv_{56} \subset \bv_{248}$ subspace in equation~\ref{fhthopart} or other gauge 
groups with a significant impingement on the vector-Higgs 
 $\bv_4 \equiv \bh_2 \inn \htwc \equiv \TM_4$ components and  hence associated with 
very massive gauge bosons, which are hence as yet unobserved as are the corresponding 
gauge interactions. 
 An internal $\sutw_L \times \uo_Y$ is anticipated which is also broken through a 
degree of impingement on the vector-Higgs components, resulting in the empirically 
observed massive
 $W^{\pm}$ and $Z^0$ gauge bosons and associated  electroweak phenomena, as a 
progression from the `mock electroweak theory' described in section~\ref{ewtfesb}.

  Given the projection of $\bv_4 \inn \TM_4$ with an external $\sltc^1$ symmetry and 
the set of fermions identified in the residual $\bv_{248}$ components transforming 
under 
 the internal $\suth_c \times \uo_Q$ symmetry,  the further internal $\sutw_L$ 
symmetry action will be sought as acting on doublets of  quark $\binom{u}{d}_{\! L}$ 
and lepton $\binom{\nu}{e}_{\! L}$  left-handed Weyl spinors, and not only for the 
first but also the second and third generation of Standard Model fermions. Masses for 
the fermions are anticipated to arise through interaction with the vector-Higgs as 
expressed in the quintic terms of $\lvtfe$ as an extension from the quartic terms such 
as those of equations~\ref{qxmass} and \ref{qxmassnu} for the $\lvfs$ case. A 
misalignment between the $\sutw_L$ weak doublet states and the mass eigenstates for 
the quark sector is expected to give rise to the phenomena of CKM mixing, as described 
for the Standard Model towards the end of section~\ref{ewtatsm}, with a related 
consideration leading to the phenomena of neutrino oscillations in the lepton sector.

 It would be possible to attempt to embed the structures of the Standard Model, as 
alluded to above, into the components of a quintic form $\lvtfe$ with an $\ee$ 
symmetry \textit{if} the latter structure was already known and described in the 
literature. This would continue the approach adopted for  the $\esi$ symmetry of 
$\lvt$ and $\ese$ on $\lvfs$, as based on the corresponding mathematical structures 
originally discovered in the 1950s \cite{Chev} and 1960s \cite{Freud} respectively, 
for which the consequences of symmetry breaking over $M_4$ have been studied here in 
chapter~\ref{chapesb} and section~\ref{secesef}.

  Alternatively the mathematical structure of $\ee$ acting on a quintic form 
underlying $\lvtfe$, if it exists, might itself be \textit{constructed}  
 through its application in the present theory as a form of temporal flow based on a 
knowledge of the empirical properties of the Standard Model.
 That is, continuing the progression of table~\ref{lvftolvfs} through the Standard 
Model structure identified in the components of $F(\htho)$ under the broken $\ese$ 
symmetry in equation~\ref{fhthopart}, and \textit{using} the need to identify spinor 
components for the $\nu$-lepton  and $u$-quarks, together with three generations of 
fermions oriented under an
 $\sutw_L$ action and relating to CKM mixing, all in a structural correspondence with 
the Standard Model, might lead to the identification of a suitable underlying 
248-dimensional space. The study of this mathematical structure, incorporating the 
subspaces of $\htho$ and $F(\htho)$ under the subgroups $\esi$ and $\ese$ 
respectively, may lead to the identification of an $\ee$ symmetry represented on the 
form $\lvtfe$, which might then be rigorously studied as an objective mathematical 
entity in its own right.  Such an interplay between the development of physical 
theories and mathematical structures has a long history of stimulating mutually 
beneficial progress, as for the parallel development of gauge theories and the 
structure fibre bundles reviewed in chapter~\ref{rogaeom}.  
 
 Essentially this is the approach we have set out to follow in section~\ref{subehafws} 
in attempting to open up further components to account for further spinors and further 
generations through augmentation such as that in equation~\ref{xththpx}. The aim is 
then to
  \textit{combine} that form of extension with the augmentation to the action of 
$\ese$ on $F(\htho)$ described in section~\ref{secesef} in seeking an $\ee$ action on 
a 248-dimensional space such that a 
 homogeneous 
quintic, or higher order, norm $\lvtfe$ is invariant. As well as aiming to incorporate 
the essential structure of the Standard Model, in progressing in this way it is also 
possible that new features will appear for $\ee$ acting on $\lvtfe$ as the full form 
of temporal flow.

  For the subgroup action of $\sltwoo^1 \subset \esi \subset \ese$ on the components 
of $F(\htho)$ in equation~\ref{fhthopart} the subspace elements $X \inn \htwo$ and $Y 
\inn \htwo$
 transform as 10-dimensional spacetime vectors, and need to be opened up to identify a 
   \textit{spinor} substructure as discussed in section~\ref{subehafws}. However under 
the corresponding 
  $\sltwoo^1 \subset \ee$ subgroup on the hypothetical form $\lvtfe$ the object 
$\theta^1$ (denoted $\theta$ in equation~\ref{xinmcx}) might be directly identified 
along with further Majorana-Weyl spinors, including $\thX$ and $\phX$ of 
equation~\ref{xththpx}, all as naturally occurring representations within the enlarged 
structure and without any direct \textit{vector} representations in the components of 
$\bv_{248}$. Further, as for the four-way decomposition of $\theta^1$ in 
equation~\ref{thcth234} the Majorana-Weyl spinor $\thX$, for example, will decompose 
into a set of four left-handed Weyl spinors 
 $\{\theta_{Xl}, \theta_{Xi},\theta_{Xj},\theta_{Xk} \}$ under the Lorentz subgroup
  $\sltc^1 \subset \sltwoo^1$; with an internal $\sutw_L \subset \ee$ symmetry action 
sought on doublets of the $\sltc^1$ Weyl spinors identified within 
$\binom{\thX}{\theta^1}$.

  In this case in place of \textit{decomposing} a vector into spinor representations, 
as for equation~\ref{xththpx}, for the $\sltc^1 \subset \ee$ action the need is rather 
to 
 \textit{construct} a 4-component vector $\bv_4$ to be locally associated with the 
external tangent space $\TM_4$. This may be achieved by going the other way and 
composing together right-handed spinors for example (such as effectively associated 
with a subset of the $Y$ components
  in equation~\ref{fhthopart}), under $\sltc^1 \subset \ee$ to form the 4-vector:
\begin{equation}
 \label{htwofuse}
  \bh_2 \;=\; \theta_{Y\!\lag}(\theta_{Y\!\lag})^{\dag} \;+\; 
\phi_{Y\!\lag}(\phi_{Y\!\lag})^{\dag}
\end{equation}
  This is essentially equation~\ref{htwocp}, interpreted as composing the right-hand 
side to form the left-hand side rather than as a decomposition of the latter.
 Here the Weyl spinors are fused together through the projection of the full temporal 
flow onto the external spacetime $M_4$ as an arena for perception in the world, with 
the local Lorentz symmetry acting on the 4-vectors $\bh_2 \equiv \bv_4 \inn \TM_4$ 
which also forms the vector-Higgs in the present theory. While this structure is 
analogous to the formation of a scalar Higgs in technicolor models, as a condensate of 
a set of proposed techniquarks interacting under an SU$(N)_{tc}$ gauge symmetry as 
reviewed in subsection~\ref{suboomahp},
 in the present theory it is the identification of the geometric form of an external 
spacetime
 as an innate feature of perception, as described in section~\ref{perc}, which 
necessarily draws together spinor components into a 4-vector composition. This 
4-vector $\bh_2$  under $\sltc^1 \subset \sltwoo^1 \subset \esi \subset \ese$ can be 
seen directly 
 in the components of the various forms of $\lv$ 
for the progression in table~\ref{lvftolvfs}, as shown explicitly for example in the 
components of $F(\htho)$ on the right-hand side of equation~\ref{fhthopart}, now 
considered as intermediate stages on the way to the full form $\lvtfe$.

  The fusing of $\uo_Q$ charge neutral Weyl spinors $\theta_{Y\!\lag}$ and 
$\phi_{Y\!\lag}$ (perhaps also with a third spinor $\psi_{Y\!\lag}$) to form the 
external spacetime vector $\bh_2\equiv \bv_4 \inn \TM_4$ in equation~\ref{htwofuse}, 
through the requirement of perception on the extended manifold $M_4$, is consistent 
with absence of physical particle states corresponding to the right-handed neutrino 
(for all three generations). On the other hand the complementary   $\theta_{X\! l}$, 
$\phi_{X\! l}$ and $\psi_{X\! l}$ spinors under $\sltc^1 \subset \ee$ remain free as a 
full set of three generations of left-handed neutrinos. This analysis is similar to 
that described for equation~\ref{fhthopart} under the $\ese$ symmetry, with  $\bh_2 
\inn \TM_4$ accommodated in the $\mcY$ components and the $\nu_L$-neutrino derived 
from the $\mcX$ components, for the first generation only.  

   Although provisional, this discussion for the hypothetical action of $\ee$ on 
$\lvtfe$ describes a possible marriage of a full form of temporal flow, deduced on the 
basis of  mathematical elegance as a further progression from the sequence in 
table~\ref{lvftolvfs},
 together with the basic conceptual framework of the present theory, with a knowledge 
of the Standard Model structure presiding over the union.

 Whether or not $\ee$ will ultimately feature in a significant way for the present 
theory remains to be seen. Here the primary focus is upon homogeneous forms of 
temporal flow expressed as $\lv$, as derived in section~\ref{gfotf}, and if it happens 
that $\ee$ does not form a symmetry group of such an object then it seems unlikely 
that this largest exceptional group will play an important role in this theory. 
However such homogeneous forms are known for Lie groups as large as $\ese$, as we have 
described in this paper.  
 We end this chapter with a summary of the Standard Model features, based on the gauge 
symmetry $\SML$, which have  been identified up to this stage within the breaking of 
known symmetries of $\lv$ forms through the extraction of an external Lorentz 
symmetry. These are described in relation to the progression of higher-dimensional 
forms of temporal flow listed in table~\ref{lvftolvfs}.

\begin{itemize}
\item  $\lvf$:   External spacetime Lorentz symmetry $\sltc$ acting on $\bv_4 \inn 
\TM_4$.
 The Lorentz transformations on 4-dimensional spacetime are subsequently  
 identified with the subgroup $\sltc^1 \subset \esi$ within the larger symmetry, as 
generated by  the basis elements    $\{\dot{B}_{t \pqg z}^{1}, 
  \dot{R}_{x \pqg l}^{1}, \dot{B}_{t \pqg x}^{1},
     \dot{B}_{t \pqg l}^{1}, \dot{R}_{x \pqg z}^{1}, \dot{R}_{z \pqg l}^{1}\}$
	 for $\sltca^1$  of equation~\ref{extlor6}.

\item  $L(X)=1$:  Internal symmetry $\suth_c \times \uo_Q$ actions may be identified 
in $\stabto \subset \sltwoo$. In the context of the subsequent $\esi$ action this 
symmetry is generated by the basis elements $\{\dot{A}_q, \dot{G}_l\} + \Sbard_l^1$ 
acting on the components
 $X = \binom{p \;\; \bar{a}}{a \;\; m}\inn \htwo \subset \htho$, with the 
transformations of the $a\inn \ooo$ components as described in equation~\ref{suthuoa}.
 (This form is closely related to the $\lvte$ model of figure~\ref{mtogmaphr}).

\item $L(\mcX)=1$:  The additional $\theta^1 = \binom{c}{\bar{b}} \inn \ooo^2 \subset 
\htho$ components ($\theta$ in equation~\ref{xoct3}) transform under the external 
$\sltc^1$ as 4 left-handed Weyl spinors $\theta_l,\theta_i,\theta_j,\theta_k$ as 
subspaces of $\ooo^2$ (equation~\ref{thcth234}). These spinors neatly dovetail with 
the corresponding internal
 $\suth_c \times \uo_Q \subset \stab \subset \esi$ actions, as deduced from 
table~\ref{suttan} and
  equation~\ref{slonthet} and summarised in equation~\ref{suthuoth}, hence identifying 
a charged lepton singlet and $d$-quark triplet.

  Although the group $\esi$, acting on $\mcX \inn \htho$, is not large enough to 
contain an additional internal $\sutw$ symmetry a number of the more esoteric 
properties of $\sutw_L \times \uo_Y$ electroweak theory  are reflected in the action 
of the  type 2 subgroup  $\sutw^2 \times \uo^2 \subset \esi$ generated by 
$\{\dot{R}_{z \pqg l}^2,\dot{R}_{x \pqg z}^2, \dot{R}_{x \pqg l}^2\} + \Sbard_l^2$, 
and similarly for the corresponding type 3 case, which complement the type 1 external 
$\sltc^1$ actions, as described in section~\ref{ewtfesb}.
These properties include the doublet actions of equations~\ref{mth3} and \ref{abcmix},
 a `mock electroweak' symmetry breaking pattern
 leading to the mixing angle deduced for equation~\ref{sinmock}
  and the potential origin of  gauge boson masses
   for  the broken $\sutw^2 \times \uo^2$ generators deriving from an impingement on 
the external spacetime components as described for equation~\ref{zwamass}.
 Fermion mass terms are similarly considered to arise
 through interactions with the projected external $\bv_4 \inn \TM_4$ components
  under $\lvt$
as described for equation~\ref{hexpan2}.
   The vector $\bv_4 \inn \TM_4$ itself is
  considered to constitute a `vector-Higgs', with the degree of freedom of the 
magnitude $\vert \bv_4 \vert$ provides a candidate for the empirically observed scalar 
Higgs.

 In addition to the $\sltc^1$ spinors identified from the components of $\theta^1$ 
   a corresponding set of 4 left-handed Weyl spinors may be identified within the 
components of $\thX$ for example, upon introducing the decomposition $X= 
\thX\thX^{\dag}$ of equation~\ref{xththx}. These further Weyl spinors can be 
interpreted as the components of a neutrino and triplet of $u$-quarks, although care 
is needed to maintain the necessary electromagnetic charges of $0$ and $\frac{2}{3}$ 
as explained around equation~\ref{chcont}, and further some of these components 
coincide with the external $\bv_4 \inn \TM_4$, which has provisionally been associated 
with the above vector-Higgs.

\item $L(x)=1$:  Containing now the $\esi$ $\mathbf{27}$ and $\overline{\mathbf{27}}$ 
representations, equation~\ref{esetoesi}, the external Lorentz symmetry $\sltc^1 
\subset \ese$ can be taken to act on the $\bv_4 \equiv \bh_2 \subset \mcY \inn \htho 
\subset F(\htho)$ components, which continue to both represent external spacetime and 
also account for the Higgs sector, as depicted in equation~\ref{fhthopart}. The 
left-handed electron and $d$-quark Weyl spinors of the $L(\mcX)=1$ case above now have 
right-handed counterparts, combining in 4-component Dirac spinors, as described in 
equations~\ref{diraclr} and \ref{diraclrt}.
A left-handed neutrino (along with a set of $u_L$-quark spinors) might now be 
identified by expanding the $X\inn \htwo \subset \htho \subset F(\htho)$ components, 
while a right-handed counterpart may be excluded by the external $\bh_2 \inn \TM_4$ 
components of $\mcY$ (while a set of $u_R$-quark spinors remain), as also indicated in 
equation~\ref{fhthopart}.

  The internal $\suth_c \times \uo_Q \subset \ese$ symmetry acts on the $\mcX$ and 
$\mcY$ components of equation~\ref{fhthopart} in the same way. Further internal 
symmetries may be sought which also act on the set of spinors
 in the shape of trivial or fundamental representations. In particular an internal 
 $\sutw_L \subset \ese$, with an asymmetric action on the $\mcX$ and $\mcY$ 
components, might now be accommodated within the larger group.
 An explicitly left-right asymmetric coupling to fermion doublets for an internal 
symmetry $\sutw_L \times \uo_Y \subset  \ese$ may  be possible for this structure, 
with further analysis of the $\ese$ algebra required. The $\uo_Q$ action, surviving 
the mock electroweak symmetry breaking over $\TM_4$, is identical on the $\mcY$ and 
corresponding $\mcX$ components, accounting for the massless nature of the photon in 
the first case and the charge neutrality of the left-handed neutrino in the second 
case.  
\end{itemize}
  
  In augmenting the full symmetry from $\esi$ to $\ese$, and hence embedding the 
Lorentz symmetry in the latter, there is a  \textit{two-way} choice regarding the 
embedding of the external spacetime  $\htwc \subset F(\htho)$ in either the $\mcX$ or 
$\mcY$ components, with the latter option taken in equation~\ref{fhthopart} as 
described in the text and alluded to above. The necessary asymmetry in this choice is 
then ultimately responsible for the left-right asymmetry observed for physical 
phenomena, and in particular leads to the parity violating properties of the weak 
interaction.

 The Lie group $\ese$ does not have complex representations and is hence unsuitable as 
a unification group for the \textit{purely internal} symmetry structure of the 
Standard Model, as mentioned in section~\ref{dynkin}.
However the external Lorentz symmetry does have complex representations and including 
the $\sltc$ action \textit{within} the structure of the $\ese$  action on $F(\htho)$ 
in this asymmetric way in turn implies a left-right asymmetry for the action of the 
residual internal symmetry.  
   In the present theory this mechanism provides the source of parity violating 
phenomena
    (rather than such phenomena arising from the complementary actions of 
$\sutw^{2,3}$ with respect to $\sutw^1$ in terms of non-commutative  quaternion 
subalgebras, as had been briefly considered in subsection~\ref{strassy} shortly after 
equation~\ref{abcmix} as guided by \cite{Mori} for example).
 In addition to the $\sutw_L$ action the hypercharge symmetry $\uo_Y$ also remains to 
be specifically identified, although the latter derives from the $\Sbard^{2,3}_l$ 
generators for the mock electroweak theory as described for equation~\ref{srslincom2}
  for example.
 
 Further, the \textit{three} possible embeddings of $\sltwoo$ acting on $\htwo$ and 
$\ooo^2$ according to equations~\ref{type1}--\ref{type3} within the structure of the 
$\esi$ action on $\htho$ may relate to the empirical observation of \textit{three} 
generations of fermions.  
    The embedding of a choice of  Lorentz symmetry $\sltc^1 \subset \esi$ acting on 
the type 1 subset $\htwc \subset \htho$ breaks the discrete three-way symmetry between 
the type $1,2$ and 3 actions described in equations~\ref{type1}--\ref{type3}, hence 
also breaking the continuous type transformation symmetry. This in turn will lift the 
degeneracy of the three generations of fermions and may be related to the phenomena of 
CKM mixing between the quark states. 
  However in order to explicitly accommodate three generations of Standard Model 
fermions the extension to $\ese$ symmetry on $\lvfs$ may need to be further augmented 
to an  $\ee$ action on $\lvtfe$, incorporating a spinor expansion of the original 
components in a form such as equation~\ref{xththpx} with also a third term 
$\psX\psX^{\dag}$. The possibility of this further extension to $\ee$, which is 
currently hypothetical, has been the main topic of this section.

  The above observations, through to the $\ese$ action on $F(\htho)$, currently mark 
the point of closest approach between the present theory and the empirical world of 
elementary particle phenomena recorded in high energy physics experiments. A possible 
extension to an $\ee$ action of $\lvtfe$ is suggested partly on aesthetic mathematical 
grounds and partly through the known structure of the Standard Model itself considered 
in the context of the present theory.

  Here we have largely only considered a somewhat `static' picture based on the 
structures of the forms $\lv$ and the corresponding symmetry groups, with emphasis on 
the
 explicit structure of the
 Lie group  $\esi$ acting on the space $\htho$. For this case in addition to the terms 
arising from the expansion of equation~\ref{hexpan2} the constant value of $\lvt$ will 
be expressed `dynamically' on an extended spacetime manifold $M_4$ as the zero 
covariant derivative $\dmo$. The terms of the latter expression resulting from the 
symmetry breaking  contain the internal gauge fields $Y_{\mu}(x)$, as was described 
for the $\lvte$ model in equation~\ref{dlvfibte} -- which includes an interaction 
between the gauge field $Y_{\mu}$ and the internal $\ul{\bv}_6$ components. The cubic 
temporal form $\lvt$ does not have an interpretation as a higher-dimensional spacetime 
form and in this case, through the terms of $\dmo$, an internal gauge field can also 
impinge upon the \textit{external} 4-dimensional spacetime components of $\bv_4 \inn 
\TM_4$.
 It is through this impingement that massive gauge bosons are anticipated to arise
  as described for the mock electroweak theory in subsection~\ref{suboomahp}, with the 
field $\bv_4(x)$ termed the vector-Higgs through association with Higgs phenomena.
  In the full theory the masses for the $W^{\pm}$ and $Z^0$ gauge bosons of the 
Standard Model might be identified in this manner, while the fermion masses may arise 
through the composition of fermion components with the vector-Higgs under the full 
form $\lvh$.

    Within the expansion of $\dmo$ there are also 
terms of the form $h(b Y_{\mu} \bar{b} + c Y_{\mu} \bar{c})$, by comparison with 
equation~\ref{hexpan2}, with similar terms deriving from the quartic norm in the 
$\ese$ case, describing a coupling between the gauge field $Y_{\mu}(x)$ and the 
fermion components within $\htho$.
 In this way the internal gauge field $Y_{\mu}(x)$, taking values for example in the 
SU(3)$_c$ Lie algebra, will mix the components of the Weyl spinors, such as those of 
 the set
 $\{\theta_i, \theta_j, \theta_k\}$ in equation~\ref{suthth}, creating the possibility 
of field interactions. Ultimately the consequences of the mutual couplings of all 
fields in the terms of $\lvh$ and $\dmoh$ will need to be assessed for the full form 
of temporal flow.

   The initial dynamical equations for this theory
  derived from the relation between  the  geometry of the external spacetime and the 
curvature of the internal gauge fields, as
    deduced in section~\ref{reaic} and
  culminating in equation~\ref{gchift},    
     as guided by  the structure of Kaluza-Klein theories. Hence the gauge fields, 
such as $A_{\mu}(x), W^{\pm}_{\mu}(x)\ldots$, in being closely related to the 
spacetime geometry $G^{\mu\nu} = f(A,W,\ldots)$ in the form of equation~\ref{gety}, 
seem to take some priority over possible fermion states which may be identified in 
turn through the field  interactions, as will be described in  chapter~\ref{newapp}. 
 That is \textit{given} for example an initial $W^{\pm}_{\mu}(x)$ field in turn 
fermion fields $\psi(x)$ within doublets such as $\binom{\nu}{e}_{\! L}$ or 
$\binom{u}{d}_{\! L}$ will be drawn into relation with
 the external spacetime geometry
  $G^{\mu\nu} = f(A, W, \psi, \ldots)$ 
 from the components of $x \inn F(\htho)$ via interactions with a gauge fields as an 
example of the generalisation described for equation~\ref{getypsi} in 
section~\ref{subwal}.
 In section~\ref{secpotnt}  a direct relation between the spacetime geometry and the 
magnitude of the vector-Higgs field with $G^{\mu\nu} = f(\bv_4)$ will also be derived, 
leading to a further and more direct link with fermions through the terms of $\lvh$.

  An understanding of the empirical consequences of all of the  possible field 
interactions, and the phenomena of high energy physics in general, will require a full  
dynamical and \textit{quantum} expression of the theory.
 This will include an understanding of how macroscopic `mass' as central to general 
relativity through the field equation $G^{\mu\nu} = - \kappa T^{\mu\nu}$ is related to 
particle `mass' as observed in the laboratory, and an exposition of a unified 
conceptual basis for describing both gravitational and quantum phenomena more 
generally. Within this unified framework 
the concept and the nature of physical elementary particles themselves 
 might be addressed. A  quantised theory dynamically expressed on the spacetime 
manifold $M_4$   will also be required in order to
  deduce the empirical particle spectrum  as well as to
  express kinematic quantities such as the masses of the particle states, for fermions 
as well as gauge and the Higgs bosons.

  In quantum field theory (QFT) the particle masses feature in `propagators' while 
charges and coupling constants appear in interaction `vertex' terms. Both of these 
objects are intrinsic to calculations of cross-sections via the transition amplitude 
${\mathcal M}_{fi}$, as will be described in the following chapter. 
 The propagator factors in calculations of process probabilities contain various 
kinematic quantities with the dimension of mass. For example the Feynman propagator 
for the scalar Higgs field has a particularly simple form, $i/(p^2 - M^2_H)$ where $p$ 
is the 4-momentum, which may provide a guide for the role of mass terms for the 
present theory. Here the effective incorporation of finite mass into the propagators 
for massive gauge bosons is expected to be related to that for technicolor models as 
described  between equations~\ref{lagtech} and \ref{hexpan2} in 
subsection~\ref{suboomahp}.

  Similarly while equations~\ref{suthuoth} and $\ref{suthuoa}$ describe the correct 
$\uo_Q$ charge structure for a generation of leptons and quarks, it will need to be 
understood how these `charges' enter into cross-section calculations and hence 
   actually account for the
  electromagnetic charge structure as observed for particle states in high energy 
physics (HEP) experiments.
   The interpretation of such QFT calculations within the context of the present 
theory will need to be addressed before the concepts of charge and mass can be fully 
comprehended here.
 The nature of field interactions and the concept of particles themselves will also 
need to be addressed in the course of this study, as we explore in the following two 
chapters.

  Rather than beginning with fields or particles which are then postulated to 
\textit{have} various properties and forms of interaction, in the present theory we 
begin essentially with a composition of, or coupling between, components of the full 
form $\lvh$ together with the generators of the symmetry transformations. Here 
`masses' and `charges' originate in the terms of the expressions for $\lvh$ 
 and $\dmoh$.
 Only once these mathematical relations are expressed in terms of dynamical equations 
over the manifold $M_4$, with spacetime geometry $G^{\mu\nu}(x) = f(Y,\hat{v})$ in the 
notation of equation~\ref{getypsi}, might
particle states themselves be identified as a phenomenon arising out of the field 
interactions. In turn the observable characteristics of such particle phenomena  might 
be determined.

   The particle properties, including masses and mixing parameters, although arising 
from the underlying interactions of the fields, are not necessarily expected to be 
literally read off directly from the $\esi$ or $\ese$ symmetry breaking level. Indeed 
some particle characteristics, such as their behaviour under \textit{CPT} 
transformations and the identification of antiparticles necessarily requires a theory 
expressed in an extended spacetime.
  In dealing with the bare $F(\htho)$ components together with the algebraic form of 
the $\ese$ symmetry actions it can only be expected to uncover a shadow of the full 
variety of Standard Model phenomena at this level. However it is also desirable that 
this shadow should possess identifiable features, such as the correct fractional 
charges and a left-right asymmetry, that may plausibly underlie the empirical data.  
The dynamic aspects of the  theory  and a quantisation scheme will need to be 
developed in order to make more rigorous comparisons with the full variety of 
laboratory phenomena.

  In the meantime,  a collection of general properties of the Standard Model  have 
already been identified in the study of the breaking of the $\esi$ symmetry of $\lvt$ 
in chapter~\ref{chapesb} and $\ese$ symmetry of $\lvfs$ as presented in 
section~\ref{secesef}. 
 In particular the structure of the $\ese$ symmetry on the components of $F(\htho)$ 
when broken over $\TM_4$ makes significant contact with the Standard Model, as also 
summarised in this section, and further inroads may be possible by further exploring 
this structure. However the aim here is to avoid the possibility of contriving the 
appearance of Standard Model properties, but rather to be primarily guided by the 
development of the theory itself, albeit very much in the light of known empirical 
phenomena.
A number of features, including the identification of $u$-quark and $\nu$-lepton 
spinors and their $\sutw_L$ interactions with the $d$-quarks and $e$-leptons 
respectively, the `Yukawa couplings' and origin of mass, the structure of three 
generations of fermions  and the mixing between them,
remain to be better understood.

   The progression towards higher-dimensional forms of time listed in 
table~\ref{lvftolvfs}, together with the need to fill out the empirical picture, hints 
at the possibility of uncovering an $\ee$ symmetry action on a quintic form $\lvtfe$ 
as the final `Russian doll' in the sequence of enveloping symmetries of time, as we 
have described in this section.
  However, as well as extensions to higher dimensions a quantised theory and an 
understanding of physical particle states, as considered in the following two 
chapters, will be needed to identify further details of the Standard Model from within 
the present theory for 
 a thorough comparison with and testing against the empirical data. 
  Until then the extent to which the $\ese$ stage is sufficient or otherwise to 
account for the Standard Model will not be completely clear.

In this regard the main question concerns the identification of the structure of 
particle-like states within the theory before returning to further assess the 
correspondence between the present theory and empirical data, and then progress 
towards making predictions which may be tested. Before comprehending the particle 
concept it will be necessary to understand how in the present theory quantum phenomena 
arise together with the mathematical structures of quantum field theory which are 
intrinsic to calculations of high energy physics processes.  Hence in the following 
chapter we begin by reviewing the standard machinery of QFT as applied for HEP 
experiments.

\pagebreak

\chapter{Particle Physics}
  \label{pp}
\section{High Energy Physics Experiments}
  \label{sechepe}

  The concept of particle phenomena as observed in HEP experiments in the context of 
the theory presented in this paper will be examined here and in the following chapter. 
In this theory the world appears in our experience necessarily within the geometrical 
confines imposed in order for it to actually be perceived through the flow of time,
  with the geometrical conditions for the perceived 4-dimensional spacetime world  
projected out of a  general higher-dimensional progression in time. 
 The arbitrary nature inherent in a degenerate set of possible geometric solutions 
manifests itself as quantum and particle phenomena -- such as observed in the detector 
apparatus of high energy physics (HEP) experiments, and through which we interact with 
and experience the world in general. This perspective, introduced in this section, 
will be described more thoroughly in the next chapter.

 The phenomena of particles are observed in the laboratory in the limit of near 
`vacuum' conditions  as elementary transitions of the world as recorded in detector 
components. 
  Similar phenomena will be manifest more generally in a curved spacetime associated 
with an arbitrary distribution of matter, however in the flat spacetime limit of the 
near vacuum, approximating the laboratory environment as considered here, these 
phenomena may be simpler to categorise.   The `particles' observed in HEP experiments 
are states of matter that arise in this simplifying limit, rather than the fundamental 
`building blocks' of matter itself. 

  It is the aim of experimental high energy physics -- employing huge and 
technologically complex  macroscopic physical structures in the form of `particle' 
accelerators, colliders and detectors coupled with sophisticated computer software and 
data analysis (see for example~\cite{sld}) --  to detect and analyse the most delicate 
and minimal transitions of the state of the perceived physical world. In this way the 
nature and properties of the elementary particles ascribed to such transitions are 
empirically determined -- for example, the relative degree of interaction between 
particular gauge boson and quark fields in the case of~\cite{sld}. In the present 
theory internal symmetries and fermion states have been 
identified at the level of the broken $\ese$ symmetry action on the multi-dimensional 
temporal form $\lvfs$, as described in section~\ref{secesef}, and will relate to the 
components of the corresponding gauge fields and quark or lepton fields respectively, 
subject to dynamical constraints in extended spacetime. While significant contact has 
been made with structures of the Standard Model, as summarised in 
equation~\ref{fhthopart} and section~\ref{sosmfi},
  it will be necessary to identify in detail the mathematical correlate of HEP 
phenomena within the present framework in order to establish a closer relation between 
the theoretical and experimental environment and hence further assess the  validity of 
the theory.   

  It should be kept in mind that the events recorded in a high energy physics 
experiments are not \textit{actively made to happen} by physicists, rather the 
complete experimental apparatus is designed and built to \textit{passively} make 
highly refined observations of the course of nature. The most elementary and minute  
transitions of the physical world, expressed for example in terms of  gauge or fermion 
fields, are  isolated and amplified through such experiments as exemplified in 
figure~\ref{figsld}. Such a process, or `event',
 may involve `jets' of many final state particles as displayed in figure~\ref{figsld} 
or
 could be as simple as that sketched in figure~\ref{eemm} in the following section.

\begin{figure}[htbp]  
\centering
 \hspace*{-48pt}
\epsfxsize=7.2cm
\leavevmode
\epsffile[0 0 550 504]{\gpath SLDdetbw} 
\hspace{0.1cm}
\epsfxsize=8.5cm
\leavevmode
\epsffile[0 0 1049 788]{\gpath sldevente} 
\hspace*{-48pt}
\caption{\setb The most delicate changes of the macroscopic state of physical 
structures such as HEP detectors are interpreted in terms of `particle tracks' 
composed out of a series of such minimal detectable transitions, here exemplified in 
an event recorded by the SLD collaboration~\protect\cite{SLDweb}.}
\label{figsld}
\end{figure}

 All `material' objects, such as particle detectors, are apparently infused with and 
seemingly `composed of'  field transitions. The environment of a HEP experiment is 
such that a particular series, or chain, of macroscopic transitions of the apparatus 
can be reconstructed, via amplified signals and computer algorithms, as a particle 
track. At the elementary microscopic level the particular components of 
equation~\ref{fhthopart} involved,  as developed so far up to the action of $\ese$ on 
$\lvfs$ for the present theory, will determine the particle type, for example an 
electron or $d$-quark, with properties
  such as the observed bending of an electron track in a magnetic field or the 
manifestation of quarks in hadronic states
 determined by the coupling to the gauge field components. Similarly, the appearance 
of a \textit{set} of such particle tracks, as seen in figure~\ref{figsld}, is 
correlated through the higher-order interactions with other fields, such as that of 
the $Z^0$ gauge boson field, hence making connection with mathematical calculations in 
the corresponding theoretical framework.

  Although the higher-order interactions may be complicated  empirically the unique 
properties of elementary particles, such as the masses of the electron and muon for 
example, are independent of the external material environment (for example with the 
particle production and detection apparatus made of copper, silicon or other elements) 
as far as we can observe  (excepting cases such as an `effective mass' in a solid 
state device for example). These properties are measured to be the same in all the 
variety of experiments that have been set up to induce them, and also as they have 
been observed for a range of particle states in natural events such as cosmic ray 
showers. This robustness arises presumably since there is a universal `vacuum' limit. 
Hence, although ordinary matter is complex, we expect to be able to isolate the robust 
and invariant quantities that describe the observed particle properties in the 
appropriate limit for theoretical calculations.

  The eventual aim will be to calculate the effects seen in particle physics 
experiments in terms of  transitions between the fields to determine the properties of 
the observed elementary particles. These include their masses and spins which  
categorise the particle transformations under the  Poincar\'e symmetry of 
4-dimensional Minkowski spacetime, assuming an approximately flat base manifold $M_4$.
   An `electron', for example, will be associated with particular field 
transformations under both a spinor representation of the global external Lorentz 
symmetry over $M_4$ and a particular representation of the internal symmetry of the 
local gauge group, with for example unit charge relative to other particle states 
under the $\uo_Q$ action of electromagnetism.

 Part of the defining notion of a particle is its local nature. A particle is an 
entity, whether in experiment or in theory, which causally connects and relates two 
spacetime events or interactions. In HEP experiments the chain of interactions can be 
traced from the production of the initial particle beams, through interactions with 
guiding magnets and accelerating components, into the interaction region of the 
collider and out into a spray of detector hits and signals to be recorded and 
analysed. Knowledge of the spacetime location of the directly detected interactions 
allows the reconstruction of kinematic quantities, such as the invariant mass or 
electric charge, of the particles ascribed to these observations.

  The ultimate ambition here will be to describe what the `in' and `out' particle 
states in HEP experiments actually \textit{are}, physically understood and 
mathematically expressed, as well as to account for the process taking place in the 
spacetime volume of the interaction region. In the spirit of this theory these 
phenomena, as for all physical processes in spacetime, will be `enveloped' by the 
structure of the spacetime geometry as related to the other fields through $G^{\mu\nu} 
= f(Y,\hat{\bv})$ of equation~\ref{getypsi}, as described in section~\ref{subwal}. It 
will be necessary to understand the precise general form of the right-hand side of 
this expression to address the question of what an elementary particle, such as an 
electron, \textit{is}. For completeness this question will also include the physical 
nature of particle states such as quarks which are not observed to propagate  
macroscopically as independent objects in spacetime.
 
 In standard field theory an independent flat spacetime background is given as an 
arena upon which fields may be arbitrarily added. Gauge invariance of a Lagrangian 
function composed of the fields is then postulated as a means to introduce 
interactions between fields, as described in sections~\ref{subfal} and \ref{ewtatsm}. 
This construction is transferred to the corresponding quantum field theory (QFT) in 
which the gauge transformations mix internal components of the field operators such as 
$\hat{\phi}(x)$. The field itself may be quantised by applying canonical commutation 
relations, by analogy with non-relativistic quantum mechanics, to the infinite degrees 
of freedom of the field, and particle creation and annihilation operators  
$a^{\dag}(\bp)$  and $a(\bp)$ identified, as will be reviewed later in this chapter.

  In contrast, in the present theory all elementary structures arise out of the 
interplay of multi-dimensional forms of the flow and symmetry of time expressed in 
$\lv$. The higher-dimensional mathematical form of temporal flow $\lvfs$ gives rise to 
the components of fields locally in interaction when  perceived in  physical 
4-dimensional extended spacetime $M_4$ in a manner consistent with the underlying 
fundamental ordered one-dimensional flow of time. With the action of $\ese$ on 
$\bv_{56} \inn F(\htho)$ broken over $M_4$ and the derivative $\dmofs$ in turn 
fragmented through this 4-dimensional projection the physical manifestation of a 
degeneracy of causally linked exchanges between fields describing multiple solutions 
under $G^{\mu\nu}(x)$ will be identified as the origin of indeterministic 
\textit{interactions}. It is these interactions which give rise to apparent 
\textit{particle} phenomena, such as quarks and leptons, as objects of study in high 
energy physics experiments.
 
  Hence the aim is then to understand how such \textit{discrete} particle phenomena 
arise out of the fundamental elements of the theory, \textit{without} needing to 
impose creation and annihilation operators, or using similar ad hoc quantisation 
techniques, to describe this particle-like behaviour. Rather the mathematical 
structures of the present theory are intended to match the physical structure of the 
world down to the most elementary level. Here particles should be \textit{derived} as 
a phenomenon arising out of the possibility of multiple field solutions under 
$G^{\mu\nu}(x)$ on $M_4$. 

  The principle goal of the following chapter will be to consider how the new theory 
describes the phenomena observed in high energy physics experiments, yet without the 
conceptual problems -- for example regarding the particle interpretation -- of quantum 
field theory. In particular this essentially means to be able to match the 
cross-section calculations for particle interactions in QFT except with both an 
underlying motivation for the nature of probabilities in these processes and a clearer 
understanding of the particle concept itself.
 
  Quantum field theory, although incomplete, provides a set of pragmatic tools and 
strategies which have achieved great empirical success, and hence much of the 
mathematical machinery is expected to remain of importance. The preliminary and 
general nature of QFT allows for the successful elements to be extracted for 
comparison with the present theory. It is the agreement between calculations based on 
scattering matrix amplitudes in QFT and cross-sections measured in the laboratory that 
needs to be accounted for in the context of the present theory, and hence in the 
remainder of this chapter we review some of the standard textbook material on the 
structure of such calculations for reference in the following chapter.


\section{Cross-section Calculation}
\label{crosss}

  In this chapter we consider how quantum field theory (see for example~\cite{Kaku, 
Pesk, ManSh}) is employed in practice to calculate cross-sections for processes 
observed in high energy physics experiments, for example in proton machines such as 
the LHC, but in particular for the kind of  events detected in electron-positron 
colliders as depicted in figure~\ref{figsld}. 
 The cross-section $\sigma (e^+e^- \to X)$ for a particular process quantitatively 
represents the likelihood for the production of the final state $X$. The description 
of this final state in general combines a particular collection of outgoing particles, 
or of `jets' containing a spray of particles as for the event in figure~\ref{figsld}, 
together with a particular range of kinematic or geometric characteristics. 

    The aim here will be to present the cross-section for such processes and then 
strip down this expression to identify how the basic structure of QFT is used to 
calculate the probability of such events. In the following chapter we describe how 
such calculations might be reconstructed  in the context of the present theory. 
Given the cross-section $\sigma (e^+e^- \to X)$ the predicted event rate $R$ (for $N$ 
events per $t$ seconds) is simply:
\begin{equation}
   R \, \equiv \, \frac{dN}{dt} \, = \, L\sigma
    \label{ratels}
\end{equation}
   which also \textit{defines} the luminosity value $L$ at which the machine is 
operating while producing the events. In practice `bunches' of incoming particles are 
directed through the interaction region of the experiment, with bunches of $n_-$ 
electrons facing oncoming bunches of $n_+$ positrons (where the apparent number 
$n_{\pm}$ of particles per bunch can be closely estimated from the total charge or 
energy carried by the bunch). With the effective two-dimensional overlap, normal to 
the beam direction, of the opposing bunches given by the area $A$ and the rate of 
bunch crossings given by the frequency $f$, in the laboratory centre-of-mass frame, 
the luminosity is simply:
 \begin{equation}
    L = \frac {f n_+ n_-}{A}
	\label{lumif}
 \end{equation}     
  If this luminosity $L$, in units of $\mbox{cm}^{-2}\mbox{s}^{-1}$, is known in 
addition to the cross-section $\sigma$, in units of $\mbox{cm}^2$, then the rate of 
\textit{detection} of the corresponding events will be $R$ in equation~\ref{ratels} 
multiplied by the total efficiency $\varepsilon$ for the experimental apparatus to 
observe such events. In practice $L$ itself is \textit{measured} using the detection 
rate $\varepsilon R$ for a process for which $\sigma$ in equation~\ref{ratels} is both 
well-known and sufficiently high to achieve a small statistical uncertainty for $L$.
     In quantum electrodynamics (QED) the cross-section $\sigma (e^+e^- \to 
\mu^+\mu^-)$, for the process depicted below in figure~\ref{eemm} and described 
subsequently, is one of the simplest to calculate and is well known. It was used as a 
reference point for $e^+ e^-$ colliders in the 1970s in order to measure the 
cross-section for hadronic final state production relative to muon pairs as a function 
of centre-of-mass energy.

   The approach taken in this chapter is to begin with \textit{observable} quantities 
in HEP experiments, writing down the \textit{general} expression for the cross-section 
as below, and then show how this is related to calculations in QFT through computation 
of the $S$-matrix. This in turn will lead to consideration of the elementary 
interaction terms in the Lagrangian and a description of the procedure of calculation 
aided by Feynman diagrams and rules. We begin then with the differential cross-section 
$d\sigma (e^+e^- \to X)$ for a general process  at an $e^+e^-$ collider experiment 
(see for example \cite{Pesk} p.106):
\begin{equation}
  d\sigma = \frac{1}{4E_1 E_2 \vert \bv_1 - \bv_2 \vert}
  \;\;  \vert {\mathcal M}_{fi} \vert^2 \;\; (2\pi)^4 \, \delta^4 \Big( \sum_f p_f - 
\sum_i p_i \Big)
    \;\;  \prod_f \frac{d^3 \bp_f}{(2\pi)^3 \, 2E_f}
	\label{diffcrs}
\end{equation}
  where $E_{1,2}$ and $\bv_{1,2}$ are the energy and 3-velocity of the particles in 
the two opposing incoming beams, $E_f$ and $\bp_f$ are the energy and 3-momentum for 
each final state particle and $p_i$ and $p_f$ are the 4-momenta of each initial and 
final state particle ($i=1,2$ and $f=1,\ldots, N_f$), all in the centre-of-mass frame. 
A  further combinatoric factor may be needed, for example to account for initial or 
final state particle spins for an unpolarised cross-section, as for 
equation~\ref{dcseemm} below, or a factor of $1/n!$ for a total cross-section with $n$ 
identical particles in the final state.  

    The only non-kinematic quantity in equation~\ref{diffcrs} is the transition 
amplitude ${\mathcal M}_{fi}$ (where here the subscript $fi$ labels the overall 
process) which contains the dynamics of the transformation between the initial and 
final particle states.
 The relativistic state normalisation of
  equations~\ref{pcreate} and \ref{pqbrac} below 
   will be employed and is consistent with the Lorentz invariance of ${\mathcal 
M}_{fi}$ as constructed in the following section.
 Everything to the right of $\vert {\mathcal M}_{fi} \vert^2$ in 
equation~\ref{diffcrs} is the `Lorentz invariant phase space' term $d\Phi$ for the 
final state. 
 The only factor on the right-hand side of equation~\ref{diffcrs} which is not Lorentz 
invariant is the initial state flux factor $(4E_1 E_2 \vert \bv_1 - \bv_2 
\vert)^{-1}$, however this term is invariant under Lorentz boosts along the beam 
direction. Indeed $d\sigma$, on the left-hand side of this equation, transforms as a 
two-dimensional cross-sectional area under Lorentz transformations. When composed with 
the luminosity $L$ of equation~\ref{lumif} in equation~\ref{ratels} the event rate $R$ 
exhibits a simple special relativistic time-dilation effect under a change of Lorentz 
frame, as for any physical `clock'.

   The cross-section $\sigma$ can be considered as the effective cross-sectional area 
within scattering range of each particle in the beam, or as the number of scattering 
events per unit time, per unit volume, per unit flux density of the incoming beams. 
Indeed the above cross-section formula can be calculated by considering the 
interaction to take place over a finite time period $T$ in a finite spatial volume 
$V$, which contains purely free fields in the limits $t \to \pm \infty$ relative to 
the interaction time around $t=0$. Factors of $T$ and $V$ cancel in the final result 
of equation~\ref{diffcrs}. Alternatively, a more detailed approach may be followed in 
which the incoming states are modelled as wave packets localised in space  
(\cite{Pesk} pp.102--106). In this case the final result for $d\sigma$ is independent 
of the shape of the wave packets.

  For either way of deriving this formula the transition amplitude ${\mathcal M}_{fi}$ 
itself in equation~\ref{diffcrs} is calculated for the idealised case of `in' and 
`out' plane wave states of definite momentum extending throughout spacetime. The 
resemblance of these states to the concept of a  particle is somewhat limited due to 
the absence of localisation, however their use in the determination of  ${\mathcal 
M}_{fi}$, and in turn the cross-section for particle interactions, may be followed 
pragmatically.

  The transition amplitude is determined by the matrix element between the initial 
$e^+e^-$ free field state represented by $\vert \bp_{1}, \bp_{2} \rangle_{\mathrm in}$ 
for $t\to -\infty$ and a particular final state $\vert \bq_{1}, \bq_{2} \ldots 
\bq_{N_f} \rangle_{\mathrm out}$ for $t \to +\infty$, in the respective `in' and `out' 
Fock space bases for the incoming and outgoing particles states. While neither of 
these two bases are simply related to a further Fock basis for interacting fields, 
since they both represent the free-field case they are isomorphic to each other. This 
isomorphism is described by the unitary operator $S$, connecting the `in' and `out' 
bases such that $\vert \bsl{P} \rangle_{\mathrm in} = S\vert \bsl{P} \rangle_{\mathrm 
out}$ with $\bsl{P}$ denoting any state. Unitarity is required here to conserve 
probabilities, with the transition probability being determined by the squared modulus 
of the amplitude, that is $\vert {\mathcal M}_{fi} \vert^2$, by a basic 
\textit{postulate} of quantum theory, as discussed further below.

 This situation can be expressed in a single `interaction picture' basis $I$ with an 
initial state $\vert i \rangle_I$ evolving in time from $t=-\infty$, through 
interactions as described by the $S$-matrix, to be measured in the final state  $\vert 
f \rangle_I$ at  $t=+\infty$ with a probability determined by the matrix element:
\begin{equation}
   S_{fi} \; = \;
  {}_{\mathrm out}\langle f \vert i \rangle_{\mathrm in}
  \;  = \; {}_{\mathrm out}\langle f \vert S \vert i \rangle_{\mathrm out}
   \;  = \; {}_{\mathrm in}\langle f \vert S \vert i \rangle_{\mathrm in}
  \;	\equiv \; {}_I\langle f \vert S \vert i \rangle_I
   \label{sfioutin}
\end{equation}
   where we subsequently drop the subscripts $I$ since the interaction picture, 
described further in the following section, will be used throughout. The $S$-matrix 
can be written:
\begin{equation}
   S = \b1 + iT
    \label{soneit}
\end{equation}
   where $\b1$ represents the trivial identity operation and $iT$, with the 
conventional $i=\sqrt{-1}$ factor, represents the non-trivial interaction part of the 
$S$-matrix. It is this latter part $iT = S - \b1$ which is of most interest and its 
matrix element between the initial and final states can be written, with $p_I = \sum_i 
p_i$ and $p_F = \sum_f p_f$, as:
\begin{equation}
   \langle f \vert iT \vert i \rangle \; = \; (2\pi)^4\, \delta^4 (p_F - p_I) \; i 
{\mathcal M}_{fi}
   \label{sfimfi}
\end{equation}
  which isolates the transition amplitude ${\mathcal M}_{fi}$. Expressions for 
$i{\mathcal M}_{fi}$  will later be associated with Feynman diagrams which in turn may 
be obtained directly from the Lagrangian for the field theory. Hence the transition 
amplitude is identified from the matrix element in equation~\ref{sfimfi} by factoring 
out an ever-present total 4-momentum conserving delta function. Such delta functions 
arise as a consequence of treating the external particles as idealised states of 
definite momentum.   
 
   In deriving the expression for the
  cross-section a factor of $\vert \langle f \vert S \vert i
    \rangle \vert^2$ is incorporated which hence contributes two factors of 
$(2\pi)^4\, \delta^4 (p_F - p_I)$; one of which may be interpreted as the spacetime 
interaction volume $VT$ and cancels with other factors of $V$ and $T$ in the final 
result of equation~\ref{diffcrs}.
   In this expression for the differential cross-section  the surviving delta function 
is included in the Lorentz invariant phase space $d\Phi$ when composed with the final 
factor of $\prod_f  d^3 \bp_f / ((2\pi)^3 \, 2E_f)$.

    This latter object is a statistical factor representing the density of final 
states in `small' regions of phase space between $\bp_f$ and $\bp_f + d^3\bp_f$ for 
each outgoing particle. These regions are constrained by the delta function for the 
total 4-momentum when integrating over the final state degrees of freedom of the 
differential cross-section. The factors of $1/E_f$ arise in the phase space from the 
relativistic state normalisation of equation~\ref{pcreate}.  The first factor in 
equation~\ref{diffcrs} arises in a related way and represents the flux density for the 
incoming colliding particle beams.
 
  The overall expression is such that the cross-section $\sigma$ essentially 
represents the probability of individual particle on particle interactions and is 
hence correctly normalised for equations~\ref{ratels} and \ref{lumif}. Bearing in mind 
these latter equations together with equation~\ref{diffcrs} the total differential 
event rate can be written:
\begin{eqnarray}
  dR & \! = \! & \bigg(\frac {f \, n_+ n_-}{A} \cdot \frac{1}{4E_1 E_2 \vert \bv_1 - 
\bv_2 \vert}\bigg)
     \, \cdot \,
      \vert {\mathcal M}_{fi} \vert^2 \; \cdot \;
	  \bigg( (2\pi)^4 \, \delta^4 (p_F - p_I)
          \prod_f \frac{d^3 \bp_f}{(2\pi)^3 \, 2E_f}\bigg)   \nonumber   \\
		&   &  \label{diffevr}   
\end{eqnarray}
  as a composition of three parts. The factor in the first brackets contributes to the 
likelihood of events occurring given the properties of the incoming beams from a 
purely statistical point of view. In a similar way the Lorentz invariant phase space 
$d\Phi$ in the final set of large brackets represents the range of possible outgoing 
state configurations as a further natural statistical factor. These two factors hence 
arise out of consideration of the basic classical laws of probability, essentially 
with the probability simply being proportional to the sum of the `number of ways' that 
something can happen. A further combinatoric factor is possible, such as a sum over 
outgoing particle spin states, as alluded to after equation~\ref{diffcrs}. 
Observations made in the experiment depend also on the efficiency $\varepsilon$ of the 
detector, as alluded to after equation~\ref{lumif}. 
 Further, classical statistical methods are used to analyse the data to complete the 
measurements of physical quantities with the results presented along with their 
statistical and systematic  uncertainties.

 The point of this discussion is to highlight the contrast between this list of 
classical probabilistic factors and the middle term $\vert {\mathcal M}_{fi} \vert^2$ 
of equation~\ref{diffevr} with which they are composed and which has rather different 
characteristics. Historically this final factor originated from non-relativistic 
quantum mechanics for which the transition probability for a state described by the 
normalised wavefunction $\Psi(\bx,t)$ to be measured in the normalised eigenstate 
$\Phi_i(\bx,t)$ is represented by the squared modulus $\vert A \vert^2$ of the 
amplitude $A = \langle \Phi_i(\bx,t) \vert \Psi(\bx,t) \rangle$, that is  the overlap 
integral
\begin{equation}
 A = \int \Phi_i^{\ast}(\bx,t) \Psi(\bx,t) d^3\bx
  \label{ampqmpp}
\end{equation}
  This construction of a $\textit{probability}$ is a postulate of quantum theory, 
apparently quite different to the notion of probability as being a measure of the 
`number of ways' that something can happen, as encountered in all non-quantum walks of 
life. This form of quantum probability was itself originally introduced to represent 
the likelihood for locating a particle at the spatial position $\bx$ by the value of 
$\vert \Psi(\bx,t) \vert^2$, and dates from the formative years of quantum theory in 
the mid 1920s.

   As an example the production of muon pairs in the process $e^+e^- \to \mu^+\mu^-$, 
as depicted in figure~\ref{eemm}, will be considered.
\begin{figure}[htbp]  
\centering
\epsfxsize=11cm
\leavevmode
\epsffile[0 0 982 469]{\gpath aPfig102e}
\caption{\setb A schematic diagram for the transition from an $e^+ e^-$ incoming state 
to a $\mu^+ \mu^-$ outgoing state. In the text the purely QED process is considered.}
\label{eemm}
\end{figure}
  The cross-section formula of equation~\ref{diffcrs} simplifies for this case of 
scattering to a two-particle final state. The $\delta^4$ function constrains $\vert 
\bp_f \vert$  in the centre-of-mass frame to the same fixed value for each outgoing 
particle and, taking the approximation that all particle masses are sufficiently below 
the centre-of-mass energy $\sqrt{s}$ and hence can be neglected, the differential 
cross-section reduces to:
\begin{equation}
  \frac{d\sigma}{d\Omega} \; = \; \frac{\vert {\mathcal M}_{fi} \vert^2}{64\,\pi^2\,s}
   \label{dsdotwo}
\end{equation}
  where $\Omega$ is the solid angle within which the $\mu^-$ is produced. 
   For the unpolarised process $e^+e^- \to \mu^+\mu^-$ there is a further combinatoric 
factor corresponding to an average over the initial electron spin states and sum over 
the final muon spin states, with $\vert {\mathcal M}_{fi} \vert^2$  above then 
replaced by:
\begin{equation}
   \frac{1}{4}\sum_{\mbox{\scriptsize spins}}\vert {\mathcal M}_{fi} \vert^2 \; = \;
      e^4 \, (1+\cos^2\theta)
	  \label{mfimu}
\end{equation}
   This is for the lowest non-trivial order of perturbation in the QFT, for which  the 
unpolarised differential cross-section is hence
 given by (\cite{Pesk} pp.8 and 137):
\begin{equation}
  \frac{d\sigma}{d\Omega} \, = \, \frac{\alpha^2}{4s}(1+\cos^2\theta)
   \label{dcseemm}
\end{equation}   
  with fine structure constant $\alpha = e^2/4\pi \simeq 1/137$, where $e$ is the 
charge of the electron, conventionally taken to be negative. In equation~\ref{dcseemm} 
$s$ is the square of the centre-of-mass energy and $\theta$ is the polar angle of the 
final state $\mu^-$, as depicted in figure~\ref{eemm}.
 In deriving equation~\ref{dcseemm} it is assumed not only that $s \gg m^2_{\mu^-}$, 
and hence the lepton masses are neglected, but also that $s$ is sufficiently below 
$M^2_{Z}$, so that a contribution from the weak interaction can also be neglected.
  In particular this means that the centre-of-mass energy is assumed to be
   somewhat lower than that for the experiment in figure~\ref{figsld}, which operated 
on the $Z^0$ resonance.
 In this case, for a purely QED process, the lowest-order calculation can be 
associated with the Feynman diagram of figure~\ref{fdeemm} featuring an intermediate 
`virtual photon'.

\begin{figure}[htbp]  
\centering
\epsfxsize=11cm
\leavevmode
\epsffile[0 0 1260 433]{\gpath aPfig103e}
\caption{\setb Feynman diagram for the process $e^+e^- \to \mu^+\mu^-$ to lowest order 
in QED perturbation theory. In such diagrams the external lines on the left-hand side 
represent incoming particle states, while those on the right-hand side represent 
outgoing particles.
 (The direction of the arrows on the external lines is explained under `item 3' in the 
discussion of Feynman diagrams in section~\ref{fraot}, while the causal structure of 
the two vertices will also be discussed later, for example alongside 
figures~\ref{feyndel}(b) and \ref{delf2way} in section~\ref{subpac}.)}
\label{fdeemm}
\end{figure}

 On integrating over the solid angle the total cross-section is found to be: 
\begin{equation}
   \sigma (e^+e^- \to \mu^+\mu^-)    \, = \, \frac{4 \pi \alpha^2}{3s}
    \label{dcseemmt}
\end{equation} 

 This cross-section, based on the leading order process depicted by the Feynman 
diagram in figure~\ref{fdeemm} agrees with observations in HEP experiments to within 
about 10\%. Most of this discrepancy is accounted for by the next order in 
perturbation theory (\cite{Pesk} p.8), with excellent agreement between the data and 
theory for a more thorough calculation. 

  The $\cos^2 \theta$ angular dependence in equation~\ref{dcseemm} arises in $\vert 
{\mathcal M}_{fi} \vert^2$  from the spin-$\fh$ property of the initial and final 
state particles. The actual calculations involving interaction processes in QED are 
made significantly more complicated by the presence of Lorentz spinor and vector 
fields, with the derivation of the right-hand side of equation~\ref{mfimu} for example 
being non-trivial. Since we are here interested in the probability interpretation of 
the transition amplitude ${\mathcal M}_{fi}$ in the following section we consider in 
detail a simpler, but closely analogous, model based on interacting scalar fields in 
order to extract the essential mathematical structure that is used in the calculation 
of such probabilities in a more transparent manner.

\section{Transition Amplitudes}
\label{tranamp}

  For the remainder of this chapter we consider a scalar model for an interacting 
field theory with three scalar fields, including one real field $\hat{\phi}(x)$ and 
two complex fields $\hat{\mcX}(x)$ and $\hat{\mcY}(x)$, that is with a total of five 
real field  components, with the quanta of the complex fields being interpreted as 
charged particles. (The analogy with the HEP process described in the previous section 
being constructed here may be briefly summarised by comparing the Feynman diagrams in 
figures~\ref{fdeemm} and \ref{fdxxyy} for the respective lowest-order calculations). 
Both real and complex \textit{free} fields can be expressed in terms of a 
corresponding annihilation and creation operator expansion which for the fields 
$\hat{\phi}(x)$ and $\hat{\mcX}(x)$ can be written as:
\begin{eqnarray}
   \hat{\phi}(x) & = & \int \frac{d^3 \bsl{p}}{(2\pi)^3} \frac{1}{\sqrt{2\omp}} \,
          \Big( a(\bp) \, e^{-ip\cdot x} \, + a^{\dag}(\bp) \, e^{+ip\cdot x} \Big)
		     \label{kgosol2}  \\
   \hat{\mcX}(x) & = & \int \frac{d^3 \bsl{p}}{(2\pi)^3} \frac{1}{\sqrt{2\omp}} \,
     \Big( b_{\mcX}(\bp) \, e^{-ip\cdot x} \, + d_{\mcX}^{\dag}(\bp) \, e^{+ip\cdot x} 
\Big)
		     \label{kgosolx}  \\
   \hat{\mcX}^{\dag}(x) & = & \int \frac{d^3 \bsl{p}}{(2\pi)^3} \frac{1}{\sqrt{2\omp}} 
\,
     \Big( d_{\mcX}(\bp) \, e^{-ip\cdot x} \, + b_{\mcX}^{\dag}(\bp) \, e^{+ip\cdot x} 
\Big)
		     \label{kgosolxd}  
\end{eqnarray}   
    with  ${p^{0}=\omp=+\sqrt{\bsl{p}^2+m^2}}$ in all three expressions. 
  The mass $m$ for each field will be associated with the corresponding particle 
states which  are identified in the following.	
	In QFT the Fourier field coefficients such as $a(\bp)$ and  $a^{\dag}(\bp)$ in 
equation~\ref{kgosol2} are taken to be linear operators acting on the Fock space of 
particle states.  The `quantisation' of the free field is completed by 
\textit{imposing} commutation relations on these operators:
\begin{equation}
   \begin{array}{c}
    \lbrack a(\bsl{p}),  a^{\dag}(\bp') \rbrack \, = \, (2\pi)^3 \, \delta^3 (\bp - 
\bp') 
	   \\
    \lbrack a(\bp),  a(\bp') \rbrack =0, \qquad
	\lbrack a^{\dag}(\bp),  a^{\dag}(\bp') \rbrack =0 
	  \end{array} \label{aacomr}
\end{equation}
  By imposing these relations, largely by analogy with the quantum mechanical simple 
harmonic oscillator, the spectrum of states possesses a ladder structure with 
$a^{\dag}(\bp)$ interpreted as \textit{creating} a particle of momentum $\bp$ and 
$a(\bp)$ \textit{annihilating} such a state. Hence in turn the $e^{\pm ip\cdot x }$ 
Fourier modes of equation~\ref{kgosol2} are associated with particle quanta of mass 
$m$ the creation or annihilation of which are attributed to the free scalar field 
$\hat{\phi}(x)$.
This structure marks an attempt to achieve direct contact with the concept of 
particles by modelling their discrete nature, although the associated Fourier modes 
are clearly not localised in space. With the vacuum represented by the state $\vert 0 
\rangle$ in the Fock space the annihilation operator acts as $a(\bp)\vert 0 \rangle = 
0$ while a single particle state $\vert \bp \rangle$ is created as:
\begin{equation}
    \vert \bp \rangle \, = \, \sqrt{2\omp} \; a^{\dag}(\bp)\vert 0 \rangle  
\label{pcreate}
\end{equation}  
  such that, given the vacuum normalisation $\langle 0 \vert 0 \rangle = 1$, we have:
\begin{equation}
   \langle \bp \vert \bq \rangle = 2\omp (2\pi)^3 \delta^3(\bp-\bq)
    \label{pqbrac}
\end{equation}
  which is Lorentz invariant, justifying the choice of normalisation factor employed 
in equation~\ref{pcreate}.

	  Analogous relations to equations~\ref{aacomr} hold for each pair of operators, 
namely $b_{\mcX}(\bp), b_{\mcX}^{\dag}(\bp)$ and $d_{\mcX}(\bp), 
d_{\mcX}^{\dag}(\bp)$, for the $\hat{\mcX}(x)$ field of equations~\ref{kgosolx} and 
\ref{kgosolxd}. These two pairs of operators, with the corresponding two sets of 
commutators, are interpreted as generating two types of particle states, with 
$b_{\mcX}^{\dag}(\bp)$ and $b_{\mcX}(\bp)$ respectively creating and annihilating 
$\mcX^-$ particles, and similarly with $d_{\mcX}^{\dag}(\bp)$ and $d_{\mcX}(\bp)$ for 
$\mcX^+$ antiparticles. (The $\uo$ charges associated with the particles and 
antiparticles are actually $+1$ and $-1$ respectively, however the charge unit is 
chosen to be negative. This is  by analogy with the convention adopted for the 
electron field, with $e < 0$ as described after equation~\ref{dcseemm}, with 
negatively charged particles and positively charged antiparticles, that is positrons).

 The field $\hat{\mcY}(x)$ and its conjugate $\hat{\mcY}^{\dag}(x)$ can be similarly 
expanded in terms of corresponding creation and annihilation operators by direct 
analogy with equations~\ref{kgosolx} and \ref{kgosolxd} and $\hat{\mcY}^{\pm}$ 
particle states similarly described. The normalisation of these single particle states 
follows the convention of equation~\ref{pcreate} and hence we define the creation 
operators:
\begin{eqnarray}
  \hat{B}^{\dag}_{\mcX}(\bp)  & = &  \sqrt{2\omp} \: b_{\mcX}^{\dag}(\bp) \qquad 
\mbox{with} \qquad  
  \hat{B}^{\dag}_{\mcX}(\bp)\vert 0 \rangle    =  \vert \bp_{\mcX^-} \rangle  
                          \label{bdagcr} \\
  \hat{D}^{\dag}_{\mcX}(\bp)  & = &  \sqrt{2\omp} \: d_{\mcX}^{\dag}(\bp) \qquad 
\mbox{with} \qquad  
  \hat{D}^{\dag}_{\mcX}(\bp)\vert 0 \rangle    =  \vert \bp_{\mcX^+} \rangle   \\
  \hat{B}^{\dag}_{\mcY}(\bp)  & = &  \sqrt{2\omp} \: b_{\mcY}^{\dag}(\bp) \qquad 
\mbox{with} \qquad  
  \hat{B}^{\dag}_{\mcY}(\bp)\vert 0 \rangle    =  \vert \bp_{\mcY^-} \rangle   \\
  \hat{D}^{\dag}_{\mcY}(\bp)  & = &  \sqrt{2\omp} \: d_{\mcY}^{\dag}(\bp) \qquad 
\mbox{with} \qquad  
  \hat{D}^{\dag}_{\mcY}(\bp)\vert 0 \rangle    =  \vert \bp_{\mcY^+} \rangle   
                          \label{bdagcr4}
\end{eqnarray}
    with corresponding conjugate annihilation operators. These may be considered as 
subcomponents   of the operator fields $\hat{\mcX}(x)$ and $\hat{\mcY}(x)$, as for the 
operator in equation~\ref{pcreate} with respect to the field $\hat{\phi}(x)$. We 
stress that here in sections~\ref{crosss}--\ref{fraot} we are describing the standard 
constructions of a quantum field theory
 (as described in more much detail in~\cite{Kaku, Pesk, ManSh} for example)
 and in the following chapter we shall need to describe how the corresponding elements 
arise in the context of the new theory presented in this paper.

   The Lagrangian for the model under consideration here consists of three free field 
parts, each of which is essentially a Klein-Gordon Lagrangian, for the fields 
$\hat{\phi}(x)$, $\hat{\mcX}(x)$ and $\hat{\mcY}(x)$, with mass parameters $m_{\phi}$, 
$m_{\mcX}$ and $m_\mcY$ respectively, together with an interaction part 
$\lag_{\mathrm{int}}$ consisting of polynomial functions of the fields:
\begin{eqnarray}
  \lag & = & \lag_{\phi}\; +\; \lag_{\mcX}\; +\; \lag_{\mcY}
    \; +\; \lag_{\mathrm{int}}  \nonumber  \\ 
  \mbox{with} \quad  \lag_{\phi} & = & \fhs \pal_{\mu}\hat{\phi} \, 
\pal^{\mu}\hat{\phi}
       \; - \; \fhs m_{\phi}^2 \, \hat{\phi}^2   \nonumber \\
	  \lag_{\mcX} & = & \pal_{\mu}\hat{\mcX}^{\dag} \, \pal^{\mu}\hat{\mcX}
       \; - \;    m_{\mcX}^2\, \hat{\mcX}^{\dag}\hat{\mcX}   \nonumber \\
	  \lag_{\mcY} & = & \pal_{\mu}\hat{\mcY}^{\dag} \, \pal^{\mu}\hat{\mcY}
       \; - \;    m_{\mcY}^2\, \hat{\mcY}^{\dag}\hat{\mcY}   \nonumber \\
  \mbox{and} \quad
    \lag_{\mathrm{int}} & = & -g\hat{\phi}\hat{\mcX}^{\dag}\hat{\mcX}
	                    \, - \,  g\hat{\phi}\hat{\mcY}^{\dag}\hat{\mcY}  
	 \label{lagfint}
\end{eqnarray}
   where $g$ is the interaction coupling constant.
   Since the Lagrangian must be a real function a complex field appears in each term 
symmetrically with its conjugate field; for example $\lag_{\mcX}$ contains the mass 
term $ -m^2_{\mcX}\hat{\mcX}^{\dag}\hat{\mcX}$. It is the invariance of this total 
Lagrangian under the \textit{global} $\uo$ symmetry with $\hat{\mcX} \to 
e^{i\alpha}\hat{\mcX}$ and $\hat{\mcX}^{\dag} \to e^{-i\alpha}\hat{\mcX}^{\dag}$ (and 
similarly for the complex $\hat{\mcY}(x)$ field) that implies a conserved $\uo$ charge 
as described above, consistent with Noether's theorem as briefly reviewed alongside 
equation~\ref{noether} in  section~\ref{subfal}.

  The simple QFT model described here is not a gauge theory and
   in equation~\ref{lagfint} the interaction terms are added by hand. By contrast in 
QED or scalar electrodynamics the coupling of the charged fields to the 
electromagnetic field $A_{\mu}(x)$ is induced by the requirement of a \textit{local} 
$\uo$ gauge invariance of the Lagrangian, as also described in  section~\ref{subfal} 
and exemplified in the final term of equation~\ref{lagdym}, although an arbitrary 
coupling constant can still be 
employed.
 In the Standard Model non-Abelian gauge theories are also incorporated through such 
expressions as for example in equations~\ref{lagklep}
 and \ref{covdevlep} of section~\ref{ewtatsm}.   In all cases such Lagrangian terms 
imply interactions since the fields mutually influence one another in equations 
derived from the principle of extremal  action. 
     Here with the additional interaction terms of equation~\ref{lagfint} the 
Euler-Lagrange equations of motion from equation~\ref{eula}, derived by varying 
$\hat{\phi}(x)$, $\hat{\mcX}^{\dag}(x)$, $\hat{\mcX}(x)$, $\hat{\mcY}^{\dag}(x)$ and 
$\hat{\mcY}(x)$ respectively as five independent fields subject to the constraint 
$\delta \!\int \! {\mathcal L}\;\!d^4 x = 0$ (in a flat spacetime) are non-linear in 
the fields:
\begin{eqnarray}
     (\square \, + \, m_{\phi}^2)\hat{\phi}(x)  & = &
	     - \, g\hat{\mcX}^{\dag}\hat{\mcX} 
	       - \, g\hat{\mcY}^{\dag}\hat{\mcY}  	  \label{kgphxx} 
	 \\
	 (\square \, + \, m_{\mcX}^2)\hat{\mcX}(x)  & = &
	     -g\hat{\phi}\hat{\mcX}  
		 \qquad\quad \mbox{and} \;\; \mbox{with}
		 \quad  \hat{\mcX} \to \hat{\mcX}^{\dag}   \label{kgxphx}
	 \\ 
	 (\square \, + \, m_{\mcY}^2)\hat{\mcY}(x)  & = &
	     -g\hat{\phi}\hat{\mcY} 
		 \qquad\quad \mbox{and} \;\; \mbox{with}
		   \quad  \hat{\mcY} \to \hat{\mcY}^{\dag}  \label{kgyphy}
\end{eqnarray} 
      and impossible to solve exactly. Neglecting the $\lag_{\mathrm{int}}$ terms in 
equation~\ref{lagfint}, that is in the limit for the coupling $g \to 0$, each of 
equations~\ref{kgphxx}--\ref{kgyphy} reduces to the free Klein-Gordon equation for 
which fields of the form in equations~\ref{kgosol2}--\ref{kgosolxd} provide exact 
general solutions.

    Equations~\ref{kgphxx}--\ref{kgyphy} correspond to the `Heisenberg picture' in 
which all of the time dependence is ascribed to the operator fields, while for the 
`Schr\"{o}dinger picture' the time dependence would apply purely to the states. In all 
cases in quantum theory the time evolution is determined by the Hamiltonian operator 
$H$ which may be expressed as the sum of a free field part $H_0$ and in interaction 
part $H_{\mathrm{int}}$. In the `interaction picture' the time dependence of all 
operators is determined by $H_0$ only, with the corresponding evolution of free 
operator fields such as $\hat{\phi}(x)$ then readily handled (as for 
equation~\ref{kgosol2} as a solution of equation~\ref{kgphxx} with $g=0$) while 
$H_{\mathrm{int}}$ governs the evolution of the states.
   In the interaction picture the aim is to express the transition amplitude, and 
hence the scattering probability, purely in terms of free fields. (This structure will 
be significant for making a link with the conceptual picture of the present theory, as 
will be discussed in `item 3)' of section~\ref{secdopp} for example.)

  For the model QFT under consideration here
  the evolution of the states is closely related to the interaction terms of 
equation~\ref{lagfint}.  Indeed if there are no time derivatives in the Lagrangian 
density $\lag_{\mathrm{int}}$ the interaction Hamiltonian $H_{\mathrm{int}}$ can be 
written simply as:
\begin{equation}
   H_{\mathrm{int}} \: = \: \int d^3 \bx \, 
    {\mathcal H}_{\mathrm{int}} \: = \: -\int d^3 \bx \, \lag_{\mathrm{int}}
	\label{htlagx}
\end{equation}
     In the interaction picture the initial state 
 $\vert i \rangle$ evolves according to the unitary operator $U$ into the state  
$\vert \Psi(t) \rangle  \equiv U(t,-\infty)\vert i \rangle$ at time $t$, with the 
equation of motion:
\begin{equation}
  \label{stateev}
   i \frac{d}{dt} \vert \Psi(t) \rangle  =   H_{\mathrm{int}}(t) \vert \Psi(t) \rangle 
\end{equation}
  with the Hamiltonian $H_{\mathrm{int}}(t)$ defining the time evolution.
 The scattering amplitude is obtained from the overlap of the state $\vert \Psi(t) 
\rangle$ evolved to  $t=+\infty$ with the given final state  $\vert f \rangle$, that 
is the matrix element:
  \begin{equation}
    S_{fi} = \langle f \vert U(+\infty, -\infty) \vert i \rangle 
	  \label{sfifi}
  \end{equation}
  In the interaction picture the `initial value problem' for $U(t,-\infty)$ is posed 
by the initial condition $U(-\infty,-\infty) = \b1$ together with the equation of 
motion obtained directly from equation~\ref{stateev}: 
\begin{equation}
     i \frac{d}{dt} U(t,-\infty) = H_{\mathrm{int}}(t)U(t,-\infty)    \label{uevolve}
\end{equation}

As an Hermitian operator the  Hamiltonian $H$ acts as the infinitesimal generator of a 
one-parameter unitary group. This unitary symmetry is employed in QFT to model the 
conservation of probability in scattering processes.
    A solution to equation~\ref{uevolve},  which might naively be expected to take the 
form $U(t,-\infty) \sim e^{-itH_{\mathrm{int}}}$, when taking into account the time 
dependence and operator action 
  can be obtained by iteration (and checked by direct substitution into 
equation~\ref{uevolve}) and then restructured using the time-ordered product $T$ of 
operators. Considering the evolution for any time interval from $t_0$ to $t$ it is 
found that:
\begin{eqnarray}
  & \!\!\!\!\!\!\!\!\!\!\!\!\! U(t,t_0) \!\!\!\!\!\!\!\!\! &
           \,\,\,\,\,\, =  \nonumber \\
 & \!\!\!\!\!\! & \b1 \quad + \quad (-i) \int_{t_0}^t dt_1 \:  H_{\mathrm{int}}(t_1) 
  \quad + \quad (-i)^2  \int_{t_0}^t dt_1 \int_{t_0}^{t_1} dt_2 \, 
H_{\mathrm{int}}(t_1) 
	     \, H_{\mathrm{int}}(t_2)   \nonumber  \\
   & \!\!\!\!\!\!  & \qquad  \, + 
	 \quad (-i)^3 \int_{t_0}^t dt_1 \int_{t_0}^{t_1} dt_2 \int_{t_0}^{t_2} dt_3  \,
  H_{\mathrm{int}}(t_1) \, H_{\mathrm{int}}(t_2) \, H_{\mathrm{int}}(t_3) 
       \!\! \quad + \ldots 
	 				\qquad	\;\; \;\;\;    \label{uiter0} \\
	   & & \nonumber \\											 
  & \!\!\!\!\!\!  = & \b1 \quad - \quad i \int_{t_0}^t dt_1 \: T \lbrack 
H_{\mathrm{int}}(t_1) \rbrack
     \quad - \quad \fh \int_{t_0}^t dt_1 \int_{t_0}^t dt_2 
		    \: T \lbrack H_{\mathrm{int}}(t_1)  H_{\mathrm{int}}(t_2) \rbrack  
											\nonumber \\ 
  & \!\!\!\!\!\!  & \qquad  \, + \quad
    \frac{i}{3!} \int_{t_0}^t dt_1 \int_{t_0}^{t} dt_2 \int_{t_0}^{t} dt_3  \, 
  T \lbrack H_{\mathrm{int}}(t_1)     H_{\mathrm{int}}(t_2)
         H_{\mathrm{int}}(t_3)   \rbrack
   \quad + \, \ldots 
   									     \label{uiter1} \\
		  & & \nonumber \\										 
  & \!\!\!\!\!\!  = & \sum_{n=0}^{\infty} \frac{(-i)^n}{n!} 
      \int_{t_0}^t dt_1 \, \int_{t_0}^t dt_2 \ldots \int_{t_0}^t dt_n \;
	   T \lbrack H_{\mathrm{int}}(t_1) \, H_{\mathrm{int}}(t_2) \ldots 
H_{\mathrm{int}}(t_n) 
	    \rbrack  \qquad \label{ufulltt}  \\
		 & & \nonumber \\		
  & \!\!\!\!\!\!  = &  T \lbrack \exp(-i\int_{t_0}^t dt' \: H_{\mathrm{int}}(t') ) 
			  \rbrack   \label{utexp}  
\end{eqnarray}

  The factor of $\fh$ appears on the right-hand side in equation~\ref{uiter1} since 
the extended integral does everything that is needed for the corresponding term in 
equation~\ref{uiter0} twice. This generalises to the factor of $1/n!$ in 
equation~\ref{ufulltt} for the corresponding combinatorial over-counting for the 
higher-order terms. 
  The final expression above is a useful shorthand notation for 
equation~\ref{ufulltt}. The $S$-matrix, as introduced in equation~\ref{sfioutin}, is 
then defined, on taking $t_0 = -\infty$ and $t=+\infty$, as the unitary operator:
\begin{equation}
  \label{smatrix}
   S = U(+\infty,-\infty) = T e^{-i\int_{-\infty}^{+\infty} dt \, H_{\mathrm{int}}(t) 
}
\end{equation}
   which appeared in equation~\ref{sfifi} for the transition amplitude for a 
particular process. Hence the $S$-matrix contains the information needed to calculate 
the probability of scattering from one plane wave state to another. In the interaction 
picture the basis for the external plane wave states is expressed in terms of the same 
sets of annihilation and creation operators which provide the coefficients of the 
Fourier expansions of the fields $\hat{\phi}(x)$, $\hat{\mcX}(x)$ and $\hat{\mcY}(x)$, 
of equations~\ref{kgosol2}--\ref{kgosolxd} for example, through which in turn 
$H_{\mathrm{int}}$ and hence the $S$-matrix is expressed in equation~\ref{smatrix}, 
containing all the information about the interaction.

    For the case of $\lag_{\mathrm{int}}=0$ the interaction Hamiltonian is zero and 
trivially $S = \b1$. In the general case with $\lag_{\mathrm{int}} \neq 0$  
equations~\ref{sfifi} and \ref{smatrix} together describe a time-ordered chain of 
field operations between the initial and final states. This time ordering is explicit 
in equation~\ref{uiter0} owing to the temporal limits for each integral and the order 
of the interaction Hamiltonian operators in the integrand. 
Essentially the $S$-matrix represents everything that can happen at all intermediate 
times between the initial and final free states according to the $\lag_{\mathrm{int}}$ 
terms. 
 As will be described below, when the calculation is restructured with a more 
symmetric set of temporal limits for each integral in equation~\ref{uiter1} the time 
ordering with $T$ ensures \textit{causal} relations are maintained through this chain, 
with Hamiltonian field operators acting in the correct sequence with intermediate 
field states first being created before being annihilated. 

   Given this iterative solution for $U(t,t_0)$ described in 
equations~\ref{uiter0}--\ref{utexp}, for $(t,t_0)=(+\infty,-\infty)$, the assumption 
of \textit{perturbation theory} is that the first few terms provide a good 
approximation to the exact full expression. This may be possible if the magnitude of 
the first few terms in equation~\ref{sfifi} decreases (or if there are cancellations 
between large terms) with increasing order $n$, as defined in equation~\ref{ufulltt}, 
which will generally be the case if the coupling constant, such as $g$ in 
equation~\ref{lagfint} or $\alpha$ in equation~\ref{dcseemm} for the case of QED, is 
sufficiently small. Even in this case many terms will lead to divergent integrals in 
QFT which will need to be accounted for by renormalisation.  However, even this does 
\textit{not} imply that the expression for $U(+\infty,-\infty)$, and in turn $S_{fi}$, 
will converge for large $n$. Nevertheless  the first few terms of perturbation theory 
do lead to calculations that have a well-defined meaning in that they generate 
quantities that can be compared with experiment, as is the case for muon pair 
production in the Standard Model as described towards the end of the previous section.

   By analogy with the real process $e^+e^- \to \mu^+\mu^-$  here for the scalar field 
model we consider the scattering process $\mcX^+\mcX^- \to \mcY^+\mcY^-$, that is, 
using equations~\ref{bdagcr}--\ref{bdagcr4}, between:
\begin{eqnarray}
  \mbox{the initial state $\mcX^+\mcX^-$:} \qquad  \vert i \rangle 
     & = & \hat{D}^{\dag}_{\mcX}(\bp_2) 
                       \hat{B}^{\dag}_{\mcX}(\bp_1) \vert 0 \rangle
            \qquad \mbox{at} \quad t=-\infty  \nonumber \\
   \mbox{and final state $\mcY^+\mcY^-$: } \qquad \vert f \rangle
     & = & \hat{D}^{\dag}_{\mcY}(\bq_2)
             \hat{B}^{\dag}_{\mcY}(\bq_1) \vert 0 \rangle
            \qquad \mbox{at} \quad t=+\infty  \nonumber 
\end{eqnarray}

 Hence for the process $\mcX^+\mcX^- \to \mcY^+\mcY^-$ under consideration in the 
scalar field model the transition amplitude of equation~\ref{sfifi},
 via equation~\ref{smatrix}, can be written:
\begin{equation}
   S_{fi} =  \langle 0 \vert \,  \hat{B}_{\mcY}(\bq_1)  \hat{D}_{\mcY}(\bq_2) \;
	      T \lbrack \exp \big(-i\int_{-\infty}^{+\infty} dt \, H_{\mathrm{int}}(t)
		       \big) \rbrack  \;
		     \hat{D}^{\dag}_{\mcX}(\bp_2)
             \hat{B}^{\dag}_{\mcX}(\bp_1) \vert 0 \rangle
			 \label{sfifull}
\end{equation}
  where $H_{\mathrm{int}}$ is expressed in terms of a polynomial in the interaction 
picture operator fields $\hat{\phi}(x)$, $\hat{\mcX}(x)$ and $\hat{\mcY}(x)$. These 
are free-fields evolving simply under $H_0$ and can be expanded in terms of creation 
and annihilation operators, that is by substituting the free fields of 
equations~\ref{kgosol2}--\ref{kgosolxd} (as well as for $\hat{\mcY}(x)$ and 
$\hat{\mcY}^{\dag}(x)$) into equations~\ref{lagfint} and \ref{htlagx} in turn, hence 
linking the initial and final states in equation~\ref{sfifull}. The general problem 
then in the interaction picture is to evaluate terms of the form:
\begin{equation}
     \int   dt_1, dt_2 \ldots dt_n  T \lbrack H_{\mathrm{int}}(t_1) \, 
H_{\mathrm{int}}(t_2) \ldots H_{\mathrm{int}}(t_n) 
	    \rbrack
		 \label{thhh}
\end{equation}
  between the external particle Fock states.
  This calculation can be somewhat simplified by noting that these terms, together 
with the initial and final state creation operators in equation~\ref{sfifull}, are 
sandwiched between vacuum states which have the property $a(\bp)\vert 0 \rangle = 0$ 
and $\langle 0 \vert a^{\dag}(\bp) = 0$ for an arbitrary annihilation operator 
$a(\bp)$ and its conjugate. Hence the goal is to use the commutation relations for 
such operators, for example equation~\ref{aacomr}, to extract the residual non-zero 
terms from equation~\ref{sfifull}.
 This is achieved by decomposing the time-ordered product into a combination of 
\textit{normal-ordered} terms and \textit{contractions}, which takes a simple form for 
the product of two field values:
  \begin{equation}
      T(\hat{\phi}(x)\hat{\phi}(y))
	        \quad = \quad : \! \hat{\phi}(x)\hat{\phi}(y) \! :  \; + \; 
\stackrel{\cont}{\hat{\phi}(x)\hat{\phi}}\!\!\!\;(y) 
			\label{phicont}
  \end{equation}
    Here the final term is the  contraction which can be \textit{defined} as the 
difference between the time-ordered product and the normal-ordered product of the 
field values. The normal-ordered product, denoted by the colon braces $: \! \hat{F} \! 
:$, is defined such that all annihilation operators are placed to the right of all 
creation operators in each term, and hence $\langle 0 \vert \! : \! \hat{F} \! : \! 
\vert 0 \rangle = 0$, that is the vacuum expectation value (v.e.v.) for the 
normal-ordered product of any collection $\hat{F}$ of fields is zero. The contracted 
product in  equation~\ref{phicont} is a scalar multiple of the identity operator 
$\b1$, as can be shown by considering the case for $x^0 > y^0$ and for $x^0 < y^0$. 
For example:
\begin{eqnarray}
 \!\!\! \!\!\!\! & \stackrel{\cont}{\hat{\phi}(x)\hat{\phi}}\!\!\!\;(y) \!\!\!\!\!\!  
     \!\!\!\!   & \,\,\,\,\,\,\,\,\, = \;\,       
                 T(\hat{\phi}(x)\hat{\phi}(y))
         \;\, - \; : \! \hat{\phi}(x)\hat{\phi}(y) \! : 
		 \qquad \quad \mbox{which for the $x^0 > y^0$ case:} 
		  \nonumber \\	    
 \!\!\!\! & = \!\!\!\!\!\!  &  \!\!  \!\!\!\!	
      \int \!\! \frac{d^3 \bsl{p}}{(2\pi)^3} \frac{d^3 \bsl{q}}{(2\pi)^3} 	  
	   \frac{1}{\sqrt{2\omp}}\frac{1}{\sqrt{2\omega_{\bsl{q}}}}  \,
          \Big\{ \! \big( a(\bp)  e^{-ip\cdot x} \! + \! a^{\dag}(\bp)  e^{+ip\cdot x} 
\big)
		\!  \big( a(\bq)  e^{-iq\cdot y} \! + \! a^{\dag}(\bq)  e^{+iq\cdot y} \big)
		 \;    \nonumber \\
 \!\!\!\! & \!\!\!\!\!\!   &   \!\!\!\!
  \qquad \qquad \qquad \qquad \qquad \quad
    - \, \big( a(\bp)a(\bq)  e^{-ip\cdot x} e^{-iq\cdot y} 
	     +  a^{\dag}(\bq)a(\bp)  e^{-ip\cdot x} e^{+iq\cdot y} 
		 \nonumber \\
 \!\!\!\! & \!\!\!\!\!\!   &    \!\!\!\!
 \qquad \qquad	\qquad \qquad	\qquad \quad \;
		 + \, a^{\dag}(\bp)a(\bq)  e^{+ip\cdot x} e^{-iq\cdot y} 
		 +  a^{\dag}(\bp)a^{\dag}(\bq)  e^{+ip\cdot x} e^{+iq\cdot y} \big) \! \Big\}
		    \nonumber \\
 \!\!\!\! & = \!\!\!\!\!\!   &  \!\!\!\!
     \int \frac{d^3 \bsl{p}}{(2\pi)^3} \frac{d^3 \bsl{q}}{(2\pi)^3} 	  
	   \frac{1}{\sqrt{2\omp}}\frac{1}{\sqrt{2\omega_{\bsl{q}}}}  \,
	    \Big( a(\bp)a^{\dag}(\bq) \, e^{-ip\cdot x} e^{+iq\cdot y} 
		  -   a^{\dag}(\bq)a(\bp) \, e^{-ip\cdot x} e^{+iq\cdot y} \Big)
	      \nonumber \\
 \!\!\!\! & = \!\!\!\!\!\!   &  \!\!\!\!
     \int \frac{d^3 \bsl{p}}{(2\pi)^3} \frac{d^3 \bsl{q}}{(2\pi)^3} 	  
	   \frac{1}{\sqrt{2\omp}}\frac{1}{\sqrt{2\omega_{\bsl{q}}}}  \,
	   \lbrack  a(\bp), a^{\dag}(\bq) \rbrack \, e^{-ip\cdot x} e^{+iq\cdot y}
	      \nonumber \\
 \!\!\!\! & = \!\!\!\!\!\!   &  \!\!\!\!  
     \int \frac{d^3 \bsl{p}}{(2\pi)^3} \frac{d^3 \bsl{q}}{(2\pi)^3} 	  
	   \frac{1}{\sqrt{2\omp}}\frac{1}{\sqrt{2\omega_{\bsl{q}}}}  \,
	   (2\pi)^3 \delta^3(\bp - \bq)\, e^{-ip\cdot x} e^{+iq\cdot y}
	      \nonumber \\ 
 \!\!\!\! & = \!\!\!\!\!\!   &  \!\!\!\! 
     \int \frac{d^3 \bsl{p}}{(2\pi)^3}   \frac{1}{2\omp}
	    \, e^{-ip\cdot (x-y)}   
	\quad \qquad	\qquad \qquad \mbox{(for $x^0 > y^0$)}   \label{ppcontx}
\end{eqnarray} 
  which is a scalar quantity, and with $e^{-ip\cdot (x-y)}$ replaced by $e^{+ip\cdot 
(x-y)}$  in the concluding line found for the case $x^0 < y^0$.
 Hence taking the v.e.v. of  equation~\ref{phicont}, with the normalisation $\langle 0 
\vert 0 \rangle = 1$, shows that:
\begin{equation}
    \langle 0 \vert T(\hat{\phi}(x)\hat{\phi}(y)) \vert 0 \rangle \; = \;  
\stackrel{\cont}{\hat{\phi}(x)\hat{\phi}}\!\!\!\;(y)
	\label{tphiphi}
\end{equation}   
   which is an object also known as the `Feynman propagator' for the field 
$\hat{\phi}(x)$.
 The complete contractions for the fields of equations~\ref{kgosol2}--\ref{kgosolxd} 
can be written as:
\begin{eqnarray}
   \stackrel{\cont}{\hat{\phi}(x)\hat{\phi}}\!\!\!\;(y) & = &
    i\int \frac{d^4 k}{(2\pi)^4} \; \frac{e^{-ik\cdot (x-y)}}{k^2-m_{\phi}^2 
+i\varepsilon}  
       \label{ppcont} \\
   \stackrel{\contd}{\hat{\mcX}(x)\hat{\mcX}^{\dag}}\!\!\!\;(y) & = & 
     i\int \frac{d^4 k}{(2\pi)^4} \; \frac{e^{-ik\cdot (x-y)}}{k^2-m_{\mcX}^2 
+i\varepsilon}  
       \label{xxcont}
\end{eqnarray}  
  as will be explained in the following section, see for example equation~\ref{feynp}, 
where the role of $\varepsilon$ will also be described.
  While the above functions are identical the latter case can be interpreted as 
representing 
 $\mcX^-$ \textit{particle} propagation for $x^0 > y^0$ and $\mcX^+$ 
\textit{antiparticle} propagation for  $x^0 < y^0$, since in the latter case the 
antiparticle creation operator $d^{\dag}_{\mcX}(\bp)$ of equation~\ref{kgosolx} acts 
first at the earlier time $x^0$.

    The generalisation of equation~\ref{phicont} for higher-order compositions of 
fields, in particular for those occurring in equation~\ref{thhh}, is given by Wick's 
theorem. This expresses the $T$-product as a sum of terms involving permutations of 
normal-ordered products composed with contracted field pairs. Many of these terms 
vanish when taking the v.e.v. due to their normal-ordered part, leaving residual terms 
expressible as a product of pair-wise contractions, that is Feynman propagators.

  However, the terms in the Wick expansion of the $T$-ordered product in 
equation~\ref{sfifull} do not act on the vacuum directly due to the operators for the 
initial and final states and hence it is necessary to consider also the more trivial 
contractions such as (by substituting in for example equations~\ref{kgosolx} and 
\ref{bdagcr}):
\begin{eqnarray}
      \stackrel{\contd}{\hat{\mcX}(x) \hat{B}^{\dag}_{\mcX}}\!\!\!\;(\bp)
	  & = & \langle 0 \vert  \hat{\mcX}(x) \hat{B}^{\dag}_{\mcX}(\bp) \vert 0 \rangle
	    \nonumber \\
	  & = &  
	     \langle 0 \vert 
	  \int \frac{d^3 \bsl{q}}{(2\pi)^3} \frac{1}{\sqrt{2\omega_{\bsl{q}}}} \,
     \big( b_{\mcX}(\bq) \, e^{-iq\cdot x} \, + d_{\mcX}^{\dag}(\bq) \, e^{+iq\cdot x} 
\big)
	  \,  \sqrt{2\omp}b^{\dag}_{\mcX}(\bp) \vert 0 \rangle  
		\nonumber \\
	  & = &  \langle 0 \vert 
	  \int \frac{d^3 \bsl{q}}{(2\pi)^3} {\sqrt{\frac{\omp}{\omega_{\bsl{q}}}}} \;
	  e^{-iq\cdot x} \, (2\pi)^3 \delta^3(\bq - \bp)  \vert 0 \rangle  
		\nonumber \\
	  & = &  \langle 0 \vert \, e^{-ip\cdot x} \, \vert 0 \rangle  
	    \; = \; e^{-ip\cdot x}  \label{xbcont}  \\
	  \stackrel{\conte}{\hat{D}_{\mcY}(\bq) \hat{\mcY}}\!\!\!\;(x)
	  & = & \langle 0 \vert  \hat{D}_{\mcY}(\bq) \hat{\mcY}(x) \vert 0 \rangle
	 \;  = \; e^{+iq\cdot x}   \label{dycont}
\end{eqnarray}
     which can be interpreted as the position space representation of the one-particle 
wavefunctions for the respective initial and final single particle states.  These have 
a simple form since there is no time dependence for the operators  
$\hat{B}^{\dag}_{\mcX}(\bp)$ and $\hat{D}_{\mcY}(\bq)$ and the order of products in 
these two expressions is given  explicitly with creation operators for initial state 
particles acting first and those for final state particles acting last in temporal 
order.

   For example, substituting $\int dt \, H_{\mathrm{int}}(t)  = -\int d^4 x 
\lag_{\mathrm{int}}(x)$   from equation~\ref{htlagx},  with the interaction Lagrangian 
$\lag_{\mathrm{int}}$ of equation~\ref{lagfint},  into equation~\ref{sfifull}  the 
lowest-order non-trivial term in the perturbative expansion,  corresponding to $n=2$ 
in equation~\ref{ufulltt}, for $S_{fi}$ will include a contribution from the 
expression: 
\begin{eqnarray}
 \hspace*{-14pt}
 & &  \!\!\!\!\!\!\!\!\!\!  S_{fi} \vert_{n=2} \, =   \nonumber  \\		
 \hspace*{-14pt}  
 & &  \!\!\!\!\!\!\!\!
   -\frac{g^2}{2} \, \langle 0 \vert  \,  \hat{B}_{\mcY}(\bq_1) \hat{D}_{\mcY}(\bq_2)  
\;
	      T  \Big(\int d^4x \, \hat{\phi}(x) \hat{\mcX}^{\dag}(x) \hat{\mcX}(x)
		          \int d^4y \, \hat{\phi}(y) \hat{\mcY}^{\dag}(y) \hat{\mcY}(y)
		       \Big)   \;
		     D^{\dag}_{\mcX}(\bp_2)
             B^{\dag}_{\mcX}(\bp_1) \vert 0 \rangle   \nonumber  \\			 
 \hspace*{-14pt}
 & &	\!\!\!\!\!\!\!\!\!\! =  
     -\frac{g^2}{2} \, \int d^4x \, d^4y \;
		   \stackrel{\conte}{\hat{B}_{\mcY}(\bq_1) \hat{\mcY}^{\dag}}\!\!\!\;(y) \;
		  \stackrel{\conte}{\hat{D}_{\mcY}(\bq_2) \hat{\mcY}}\!\!\!\;(y) \;
		  \stackrel{\cont}{\hat{\phi}(x)\hat{\phi}}\!\!\!\;(y) \;
		  \stackrel{\conte}{\mcX^{\dag}(x) D^{\dag}_{\mcX}}\!\!\!\;(\bp_2) \;
		  \stackrel{\contd}{\mcX(x) B^{\dag}_{\mcX}}\!\!\!\;(\bp_1)    \nonumber  \\
 \hspace*{-14pt} & &		   \label{xxyylow}
\end{eqnarray}

   As an alternative to expressions such as equation~\ref{xxyylow} the operators 
creating the initial and final states, such as $B^{\dag}_{\mcX}(\bp_1)$ and the 
Hermitian conjugate of $\hat{D}^{\dag}_{\mcY}(\bq_2)$, can also be expressed in terms 
of functions of  free fields, such as $\hat{\mcX}(x)$ or $\hat{\mcY}(x)$. In this case 
an additional Feynman propagator is introduced for each external particle state as 
expressed in the LSZ reduction formula (\cite{Pesk} p.227). In this form each 
contribution in the expansion of the scattering amplitude is expressed as the Fourier 
transform of the v.e.v. of a $T$-product of free fields, that is of a Green's function 
(or correlation function). This full LSZ expression may be needed for example for a 
consistent treatment of ultraviolet divergences in higher perturbative orders. Here we 
deal essentially  with the `truncated' Green's function, describing the internal 
interactions, in order to abstract out the general structure needed to calculate the 
transition amplitude, as required to make connection with the present theory in the 
following chapter.

   Each non-zero term in the transition amplitude can be represented by a Feynman 
diagram. In practice QFT calculations of such terms \textit{begin} with the 
corresponding Feynman diagrams as constructed from a small set of rules. For example 
the lowest-order non-trivial term described in equation~\ref{xxyylow} corresponds to 
the diagram in figure~\ref{fdxxyy}.
\begin{figure}[htbp]  
\centering
\epsfxsize=11cm
\leavevmode
\epsffile[0 0 1282 442]{\gpath aPfig104e}	 
\caption{\setb Feynman diagram for the process $\mcX^+\mcX^- \to \mcY^+\mcY^-$ to 
lowest order in perturbation theory in the scalar model; closely analogous to the 
diagram for the QED process $e^+e^- \to \mu^+\mu^-$ shown in figure~\ref{fdeemm} for 
which the general comments in the caption apply also here.}
\label{fdxxyy}
\end{figure}

  More generally
   the essence of the transition amplitude calculation can be distilled out into a 
collection of Feynman rules and diagrams as will be described in section~\ref{fraot} 
and  table~\ref{frulesp} for the scalar model. These may be obtained either from the 
canonical quantisation route, as described above (taking care to handle fermion state 
operator  anticommutators correctly in the case of the Standard Model) or the path 
integral approach  to QFT. Here we are interested in the origin of the Feynman rules, 
which may be written down from the Lagrangian density for a particular model, for 
comparison with the present theory. From this point of view the approach of canonical 
quantisation will prove to be more illuminating, in particular through the 
intermediate stage of equation~\ref{uiter0} as will be described in the following 
chapter. On the other hand the formalism of the path integral, while pragmatically 
serving as a valuable calculational tool for QFT, seems to provide less in the way of 
relevant conceptual insight for the present theory.

   By substituting the contractions in the form of 
equations~\ref{ppcont}--\ref{dycont} the leading-order term of the transition 
amplitude expressed in equation~\ref{xxyylow} can be written out explicitly as (with 
the integrals  covering all terms to the right of the integral signs):
 \begin{eqnarray}			 
	S_{fi} \vert_{n=2} & = &   -\frac{g^2}{2} \, \int d^4x \, d^4y \,\,
	  e^{+iq_1\cdot y} \,  e^{+iq_2\cdot y} \;
	  i \!\! \int \frac{d^4 k}{(2\pi)^4} \; \frac{e^{-ik\cdot (x-y)}}{k^2-m_{\phi}^2 
     +i\varepsilon} \;
	  e^{-ip_2\cdot x} \,  e^{-ip_1\cdot x}
		      \nonumber  \\
	  	& = &   -\frac{g^2}{2} \, i\!\int d^4x \, d^4y \, \frac{d^4 k}{(2\pi)^4} \;
		 	  \frac{1}{k^2-m_{\phi}^2 +i\varepsilon} \;
			e^{i(k+q_1+q_2)\cdot y} \,  e^{-i(k+p_1+p_2)\cdot x}
			  \nonumber  \\ 
	    & = &   -\frac{g^2}{2} \, i\!\int  d^4y \, \frac{d^4 k}{(2\pi)^4} \;
		 	  \frac{1}{k^2-m_{\phi}^2 +i\varepsilon} \;
			e^{i(k+q_1+q_2)\cdot y} \,  (2\pi)^4\,   \delta^4(k+p_1+p_2)
			  \nonumber  \\ 
	    & = &   -\frac{g^2}{2} \, i\!\int  d^4y \;
		 	  \frac{1}{(-p_1-p_2)^2-m_{\phi}^2 +i\varepsilon} \;
			e^{i(q_1+q_2-p_1-p_2)\cdot y} 
			  \nonumber  \\ 
		 & = &   -\frac{g^2}{2} \;
		     \frac{i}{(p_1+p_2)^2-m_{\phi}^2 +i\varepsilon} \;
		 	  (2\pi)^4\, \delta^4(q_1+q_2-p_1-p_2)  \label{sfixmom}
\end{eqnarray}
    where the three integrals over $d^4x$, $d^4k$ and $d^4y$ have been carried out in 
the third, fourth and fifth lines above respectively. The final expression is 
relatively simple and explicitly shows how such terms of the matrix element $S_{fi}$ 
are functions of the coupling $g$, the particle masses and the momentum variables. 
Indeed since HEP experiments generally prepare initial particles in momentum states 
and measure the final particles also in particular momentum states such calculations 
are simplified by beginning with momentum space Feynman rules, as will be described in 
section~\ref{fraot}. In this case the scattering matrix is calculated in terms of 
momentum space Green's functions which are related to  the corresponding position 
space functions, such as equation~\ref{ppcont}, by a Fourier transform  (see also 
equation~\ref{delfmom} in the following section).

  In explicit calculations  the final integral over position space always leads to an 
overall 4-momentum conserving delta function, as for the bottom line in 
equation~\ref{sfixmom}. This is factored out and not included in the definition of the 
transition amplitude ${\mathcal M}_{fi}$  as was described for equation~\ref{sfimfi}, 
and hence this delta function is also not included in the Feynman rules for 
$i{\mathcal M}_{fi}$.
 Further, in equation~\ref{xxyylow} only complementary halves of $\lag_{\mathrm{int}}$ 
from equation~\ref{lagfint} have been employed under each integral. The reverse choice 
corresponds to swapping the coordinate labels $x$ and $y$ on the two vertices of the 
Feynman diagram in figure~\ref{fdxxyy}. Hence the complete expression for $S_{fi} 
\vert_{n=2}$ based on equations~\ref{xxyylow} and \ref{sfixmom} will contain a further 
equivalent contribution with the dummy variables $x$ and $y$ interchanged.
  More generally an amplitude $i{\mathcal M}_{fi}$ will be associated with each 
topologically distinct Feynman diagram, with the permutation of $n!$ ways of 
associating the $n$ interactions with $n$ vertices for an $n^{\mathrm th}$ order 
diagram cancelling the $\frac{1}{n!}$ factor in the expansion of 
equation~\ref{ufulltt}. 
 This cancellation is generally incorporated into the Feynman rules for a quantum 
field theory, including the case of the model QFT considered here as will be described 
in the opening of section~\ref{fraot} (see the discussion of `rule 6' following 
table~\ref{frulesp}).

 With the above observations on mind, and
 by reference to equations~\ref{sfioutin}--\ref{sfimfi}, the transition amplitude for 
this leading-order term can be extracted from equation~\ref{sfixmom} (now including 
also the $x\leftrightarrow y$ case) for the Feynman diagram of figure~\ref{fdxxyy} 
(drawn \textit{without} the explicit $x,y$ labels) as:
\begin{eqnarray}
 {\mathcal M}_{fi}  & = &  -g^2 \frac{1}{(p_1+p_2)^2-m_{\phi}^2 +i\varepsilon}
                          \label{mfiexplt} \\
 \mbox{and hence} \quad
    \vert {\mathcal M}_{fi} \vert^2  & = & \frac{g^4}{s^2}
    \label{mfisca}
\end{eqnarray}
 where for the second equation it has been 
  assumed that $s = (p_1+p_2)^2 \gg m_{\phi}^2$, and also $\varepsilon$ has been set 
to zero as will be explained in the following section.
  The differential cross-section for $\mcX^+\mcX^- \to \mcY^+\mcY^-$ scattering to 
lowest non-trivial order is then obtained by substituting this transition amplitude 
into equation~\ref{dsdotwo} for this two-particle final state to find
$\frac{d\sigma}{d\Omega} = \frac{g^4}{64\pi^2s^3}$.

  The purpose of this section has been to show explicitly how such transition 
amplitudes, featuring in the general cross-section and hence event rate formulae of 
equations~\ref{diffcrs} and \ref{diffevr}, are calculated.  In the case of muon 
production the contribution from the lowest-order transition amplitude in 
equation~\ref{mfimu} is rather different to the analogous case for the scalar model in 
equation~\ref{mfisca}. In the case of $e^+e^- \to \mu^+\mu^-$ the coupling $e = 
\sqrt{4\pi\alpha}$ is dimensionless, unlike the case for $g$ in the scalar model, and 
(combined with the kinematic normalisation factors for the Dirac spinor and 
electromagnetic fields) this leads to an absence of $s$ in equation~\ref{mfimu}, while 
for equation~\ref{mfisca} there is no $\theta$ dependence since the model deals with 
scalar fields only. 
  However, underlying these differences the essential elements of quantum field theory 
going into these calculations are very similar. In the following section we explore 
further the basic ingredients and structure of the transition amplitude in the context 
of the scalar field model.

\section{Propagators and Causality}
  \label{subpac}

   Central to the calculation of the amplitude in equation~\ref{sfifull}, via Wick's 
theorem for the general $T$-ordered product of several fields, is the Feynman 
propagator. This was introduced for the scalar field $\hat{\phi}(x)$ in 
equations~\ref{phicont}--\ref{ppcont} and is generally denoted by the symbol 
$\Delta_F$ (`delta F') with a conventional factor of $i$ (or by $D_F  \equiv  
i\Delta_F$ as for \cite{Pesk}) in the expression:
\begin{eqnarray}
   i\Delta_F(x-y) & = & \langle 0 \vert \, T(\hat{\phi}(x)\hat{\phi}(y))\, \vert 0 
\rangle \label{delf1} \\
                  & = & \langle 0 \vert \, \theta(x^0-y^0)\,\hat{\phi}(x)\hat{\phi}(y) 
  \;  + \; \theta(y^0-x^0)\,\hat{\phi}(y)\hat{\phi}(x) \,  \vert 0  \label{delf2} 
\rangle  \quad
\end{eqnarray}
    The $\theta$-function takes the value $\theta(t) = 1$ for $t>0$ and $\theta(t) = 
0$ for $t<0$ (with the value $\theta(0) = \fhs$ less significant since $\theta(t)$ is 
generally used under a time integral; see also the discussion of 
equation~\ref{thetrep} below) and explicitly expresses the time ordering of the field 
product. The Hamiltonian 
 $H_{\mathrm{int}}$ is composed of a product of free fields in the interaction picture 
with the scalar field $\hat{\phi}(x)$ having the Fourier expansion of 
equation~\ref{kgosol2}.
The field $\hat{\phi}(x)$ can be constructed as a sum of positive and negative 
frequency parts, $\hat{\phi}(x) = \hat{\phi}^+(x) \, + \, \hat{\phi}^-(x)$, with 
$a(\bp)$ and $a^{\dag}(\bp)$ operator coefficients respectively:
\begin{eqnarray}
    \hat{\phi}^+(x) & = & \int \frac{d^3 \bsl{p}}{(2\pi)^3} \frac{1}{\sqrt{2\omp}} \,
           a(\bsl{p}) \, e^{-ip\cdot x}   \label{phipos}    \\
    \hat{\phi}^-(x) & = & \int \frac{d^3 \bsl{p}}{(2\pi)^3} \frac{1}{\sqrt{2\omp}} \,
           a^{\dag}(\bsl{p}) \, e^{+ip\cdot x} \label{phineg}
\end{eqnarray}  
   The $e^{-ip\cdot x}$ components are termed `positive frequency' since as 
wavefunctions they would represent states of positive energy under the  quantum 
mechanical operator $H \equiv \hat{E} = i\hbar\pal / \pal t$ (as implied for the same 
operator in equation~\ref{hamphi} of section~\ref{qpagig} we generally employ natural 
units with $\hbar=1$ and $c=1$ in this paper). Similarly the $e^{+ip\cdot x}$ modes 
are termed `negative frequency'.  
 Hence decomposing $\hat{\phi}(x)$ into a sum of the  positive and negative frequency 
parts, with $\hat{\phi}^+(x)\vert 0 \rangle = 0$ and $\langle 0 \vert \hat{\phi}^-(x) 
= 0$,  equation~\ref{delf2} for the scalar Feynman propagator can be written:
\begin{eqnarray}
 & \hspace{-30pt} i\Delta_F(x-y) \hspace{-50pt} & \hspace{30pt} =  \nonumber \\  
   & & \theta(x^0-y^0)\;
            \langle 0 \vert \,\hat{\phi}^+(x)\hat{\phi}^-(y)\,\vert 0 \rangle 
	\; + \; \theta(y^0-x^0)\;
	 \langle 0 \vert \,\hat{\phi}^+(y)\hat{\phi}^-(x)\,\vert 0 \rangle \label{delf2w} 
\\
				& = & \theta(x^0-y^0)\;\langle 0 \vert \,
			\lbrack \hat{\phi}^+(x),\hat{\phi}^-(y)  \rbrack \, \vert 0 \rangle 
		    \; + \; \theta(y^0-x^0)\;\langle 0 \vert \,
			\lbrack \hat{\phi}^+(y),\hat{\phi}^-(x)  \rbrack \, \vert 0 \rangle 
			\qquad\quad\; \label{vaclr} \\
				& = & \theta(x^0-y^0)\; i\Delta^+(x-y) 
			\; + \; \theta(y^0-x^0)\; i\Delta^+(y-x)  \label{delfpp}
\end{eqnarray}
    In the final line above the function $\Delta^+(x-y)$ can be defined in terms of 
the commutator of the positive and negative frequency parts of the field and then 
written out explicitly using equations~\ref{phipos} and \ref{phineg}:
\begin{eqnarray}
 \!\!\!\!\!\!\!\!\!
 i\Delta^+(x-y) \!\! & = & \!\! \lbrack \hat{\phi}^+(x),\hat{\phi}^-(y) \rbrack   \\
                & = & \!\! \int \frac{d^3 \bsl{p}}{(2\pi)^3} \frac{1}{\sqrt{2\omp}} \,
		  	      \int \frac{d^3 \bsl{q}}{(2\pi)^3} \frac{1}{\sqrt{2\omega_{\bsl{q}}}} 
\,
          \lbrack a(\bp), a^{\dag}(\bq) \rbrack \, e^{-ip\cdot x} \, e^{+iq\cdot y}
		         \;\;\;            \label{useaacom}  \\
		        & = & \!\! \int \frac{d^3 \bsl{p}}{(2\pi)^3} \frac{1}{2\omp} \,
				     e^{-ip\cdot x} \, e^{+ip\cdot y}   \label{delplus}     
\end{eqnarray}
  where the constraint on the energy components, such as $p^0 = +\omp = 
+\sqrt{\bsl{p}^2 + m^2}$, is understood in these expressions, and 
equation~\ref{aacomr} has been used in the final line -- which agrees with 
equation~\ref{ppcontx} for the $x^0>y^0$ case as expected. Again here, since 
$\Delta^+(x-y)$ is simply a function rather than an operator, the vacuum normalisation 
$\langle 0 \vert 0 \rangle = 1$ has been used to factor out the vacuum states in 
equation~\ref{vaclr} above to obtain equation~\ref{delfpp}. Integrals of the form 
{\large $\int$}$\!\frac{d^3 \bsl{p}}{(2\pi)^3} \frac{f(p)}{2\omp}$ are Lorentz 
invariant provided $f(p)$ is a general Lorentz invariant function (\cite{Pesk} p.23, 
equation~2.40), and hence from equation~\ref{delplus} it can be seen that the function 
$\Delta^+(x-y)$ is Lorentz invariant. Together with the function:
\begin{equation}
   i\Delta^-(x-y)  =  \lbrack \hat{\phi}^-(x),\hat{\phi}^+(y) \rbrack
                   = - \lbrack \hat{\phi}^+(y),\hat{\phi}^-(x) \rbrack =  
-i\Delta^+(y-x)
					   \label{delmin}
\end{equation}
  these can be written in the manifestly Lorentz invariant form:
\begin{equation}
   i\Delta^{\pm}(x-y) = \pm \int \frac{d^4 p}{(2\pi)^4} \,  e^{-ip\cdot (x-y)} \;
                       \theta(\pm p^0) \, 2\pi \delta (p^2 - m^2)
					   \label{dpmd4}
\end{equation}

  The objects $\theta$ and $\delta$ are `generalised functions', or `distributions', 
which typically only make full mathematical sense when composed with regular functions 
in an integrand. A representation of the $\theta$-function will be given below. In one 
dimension the Dirac $\delta$-function can be defined by the property:
 \begin{equation}
   \label{dirdel}
   \int dx \, f(x) \, \delta(x-x')  = f(x')
 \end{equation}
   which is essentially to substitute the value $x=x'$ into any function $f(x)$. The 
one-dimensional $\delta$-function can be represented by the following expression, 
which has the subsequent properties (while generally in the text denoting 
four-parameter objects, $x$ and $k$ each represent a single real variable in 
equations~\ref{dirdel}--\ref{deldel}):
\begin{eqnarray}
   \delta(x-x') & = & \frac{1}{2\pi} 
         \int_{-\infty}^{+\infty} dk \, e^{\pm i k(x-x')} 
     \label{delfrep} \\
   \mbox{with} \qquad\!\!\! \int_{-\infty}^{+\infty} dx  \, \delta(x-x')   & = &  1, 
\nonumber \\
   \mbox{and} \qquad\;\!\!\! \int dx \, f(x) \, \delta(g(x)) & = &  \sum_i \, 
\frac{f(a_i)}{\vert g'(a_i)\vert}  
      \qquad\! \mbox{with} \;\; g(x)=0 \;\; \mbox{for} \;\; x=\{a_1, a_2\ldots\}  
\nonumber \\
	 \mbox{i.e.} \qquad \qquad \qquad \quad\!\!\!
	  \delta(g(x)) & \equiv & \sum_i \,\frac{\delta(x-a_i)}{\vert g'(a_i)\vert}
	     \nonumber  \\
	 \mbox{e.g.} \qquad \qquad \quad \;\;\!\!\! \delta(x^2 - a^2) & \equiv & 
	  \frac{1}{2a}\big(\,\delta(x-a)\,+\,\delta(x+a)\,\big)\big\vert_{a\ge 0} 
	     \label{deldel} 
\end{eqnarray}
    The final expression above can be substituted into equation~\ref{dpmd4} and the 
$p^0$ integral performed to show that it is equivalent to the expression for 
$\Delta^+(x-y)$ in equation~\ref{delplus} and to that for $\Delta^-(x-y)$ via 
equation~\ref{delmin} for the $p^0<0$ case.

 The expression for $\Delta^+(x-y)$ in equation~\ref{dpmd4} describes the positive 
energy and `on-mass-shell' momentum space overlap integral of the plane waves, or 
wavefunctions, $e^{-ip\cdot x}$ and $(e^{-ip\cdot y})^{\ast}$. In quantum theory the 
probability for a particle originating at the spacetime location $y$ to be found at 
the location $x$ is represented precisely by this amplitude (which via a Fourier 
transform is analogous to the wavefunction transition amplitude of 
equation~\ref{ampqmpp}). In quantum field theory the form of this amplitude 
$i\Delta^+(x-y) = \langle 0 \vert 
 \phi^+(x),\phi^-(y) \vert 0 \rangle$, from equations~\ref{delf2w} and \ref{delfpp}, 
indeed suggests the propagation of a particle \textit{created} at $y$ and 
\textit{annihilated} at $x$. Since the spacetime locations $x$ and $y$ are arbitrary 
$x$ may be either later \textit{or} earlier than $y$.

  The Feynman propagator can be expressed either in terms of operators acting on the 
vacuum state, equations~\ref{delf1} and \ref{delf2}, or in terms of plane waves as 
described in equations~\ref{delfpp} and \ref{delplus}, with the bridge between these 
forms of $\Delta_F(x-y)$ provided by the intermediate equations. In either case a 
temporal ordering is introduced via the $\theta$-functions.

   For $x^0 > y^0$ the Feynman propagator is simply $\Delta_F(x-y) = \Delta^+(x-y)$, 
from  equation~\ref{delfpp}, and hence represents the amplitude for a positive energy 
particle to propagate forward in time from $y$ to $x$. On the other hand the `negative 
energy'  part in equation~\ref{dpmd4}, with $p^0<0$ and $\theta(-p^0) = 1$, represents 
a propagation from $x$ to $y$ in the $x^0 < y^0$ part of $\Delta_F(x-y)$ and in QFT is 
interpreted as an \textit{antiparticle} of \textit{positive} energy carried forward in 
time from $x$ to $y$.
 As described following equation~\ref{xxcont} for the complex scalar field case and 
for 
  $x^0<y^0$ the operator $\hat{\mcX}$ acts before $\hat{\mcX}^{\dag}$ with 
$d^{\dag}_{\mcX}$ creating an antiparticle; while for the real scalar field 
$\hat{\phi}$ there is no distinction between  particle  and  antiparticle states.
  Hence $\Delta_F(x-y)$ can be consistently interpreted as only representing 
propagation forwards in time. Further, from  equation~\ref{delfpp} $\Delta_F(x-y)$ is 
clearly symmetric in $x$ and $y$, as is the above interpretation.

  In actual calculations all spacetime location variables,
   such as $\{x,y\}$ for the propagator $\Delta_F(x-y)$, 
   will appear under an integral, such as the $\int d^4x \, d^4y$ in the first line of 
equation~\ref{sfixmom}, over all spacetime (including regions outside the light cone
  with $(x-y)^2 < 0$) hence showing explicitly how all possible time orderings are 
included equally. These integrals  
 essentially represent a Fourier transform to momentum space, allowing for a 
simplification of the calculations in terms of the momentum space Feynman rules as 
will be presented in the following section. 

  Hence the Feynman propagator $\Delta_F(x-y)$ combines wave-like functions $e^{\pm i 
p\cdot x}$ and particle-like operators $a^{(\dag)}(\bp)$ of the field $\hat{\phi}(x)$ 
with structures of \textit{causality} through the $\theta$-functions, for example in 
equation~\ref{delf2w}; --  apparently elements required to describe the dynamics of 
exchanges between fields in an interacting theory. It is represented pictorially by an 
internal line in a Feynman diagram such as figure~\ref{feyndel}(b).
\begin{figure}[htbp]  
\centering
\epsfxsize=13.1cm
\leavevmode
\epsffile[0 0 1522 655]{\gpath aPfig105e}
\vspace{-10pt}
\caption{\setb (a) The function $\Delta^+(x-y)$ represented as the creation, 
propagation and annihilation of a particle state from $y$ to $x$ in spacetime. (b) The 
internal line Feynman propagator between two spacetime points, representing
equation~\ref{delfpp}. No time ordering is implied in either diagram.}
\label{feyndel}
\end{figure}
  Such diagrams do not represent literal particle trajectories but should merely be 
interpreted as mnemonic symbols for mathematical terms such as $\Delta_F(x-y)$ which 
form the basis of perturbative calculations for an interacting QFT. Indeed the form of 
$\Delta_F(x-y)$ results from the \textit{restructuring} of the $S$-matrix calculation 
of equation~\ref{uiter0}, which describes an explicitly causal chain of operator 
actions, to the form of equation~\ref{uiter1} with $\theta$-functions implicitly 
introduced to impose the apparent time ordering required for mathematical 
\textit{consistency with} the first equation.
  
  Hence with the Feynman propagator $\Delta_F(x-y)$ employed to aid 
\textit{calculation} in this way there need not be any direct \textit{physical} 
interpretation of this object.  
However, due to the time ordering, the Feynman propagator can be considered to 
represent the internal part of \textit{both} `processes' depicted in 
figure~\ref{delf2way} below, in which a specific time direction is indicated. While 
the latter diagram in particular represents a purely mathematical element of the 
calculation both of these `processes' are implied in a single Feynman diagram, such as 
figure~\ref{feyndel}(b), for which there is no explicit temporal direction relating 
the two vertices. 

\begin{figure}[htbp]  
\centering
\epsfxsize=\maxwidth
\leavevmode
\epsffile[0 0 1974 565]{\gpath aPfig106e}
\caption{\setb The two terms in equation~\ref{delf2w} for the Feynman propagator 
$\Delta_F(x-y)$  describe respectively the two  internal $\hat{\phi}$ field 
`processes' depicted here.
 In (a) an internal particle state propagates from $y$ to $x$ while in (b) an internal 
antiparticle propagates from $x$ to $y$, however with no distinction between particle 
and antiparticle states for a real scalar field such as $\hat{\phi}$.
 In (b) $\phi$, $\mcY^+$ and $\mcY^-$ particle states are created out of the vacuum at 
$x$.}
\label{delf2way}
\end{figure} 

  The propagator $\Delta_F(x-y)$ depends only on the 4-vector difference $(x-y)$. The 
functions $\Delta^{\pm}(x-y)$, and hence also $\Delta_F(x-y)$, are non-zero outside 
the light cone region, $(x-y)^2 < 0$, where they decay exponentially. While the 
$\Delta^{\pm}(x-y)$ are Lorentz invariant the function $\theta(x^0-y^0)$ is 
\textit{not} Lorentz invariant for spacelike separations outside the light cone. 
However the combination of both terms in equation~\ref{delfpp} \textit{is} Lorentz 
invariant.

  The generalised function $\theta(t)$ itself can be expressed in the Fourier, or 
integral, representation as:
\begin{equation}
  \theta(t) = \lim_{\eta \to 0^{\splus}} \; \frac{i}{2\pi} \, \int_{-\infty}^{+\infty}
        \frac{e^{-ist}}{s + i\eta} \; ds
		\label{thetrep}
\end{equation}
  which as a \textit{distribution} is differentiable everywhere (unlike the closely 
related Heaviside \textit{function} $H(t)$ defined with $H(t)=1$ for $t\ge 0$ and 
$H(t)=0$ for $t<0$). In fact:
\begin{equation}
  \frac{d\theta(t)}{dt} = \lim_{\eta \to 0^{\splus}} \; \frac{i}{2\pi} \, \int
        \frac{-is \,e^{-ist}}{s + i\eta} \; ds  
		\; = \; \frac{1}{2\pi} \, \int e^{-ist} ds   \; = \; \delta(t) 
\end{equation}
  from the representation of the $\delta$-function in equation~\ref{delfrep}. The 
substitution of the $\theta$-function into equation~\ref{delfpp} for $\Delta_F(x-y)$ 
is aided by first making the change of integration variable $s \to k^0 - \omega$, with 
finite real constant $\omega$, in equation~\ref{thetrep} so that:
\begin{eqnarray}
     \theta(t) & = & \lim_{\eta \to 0^{\splus}} \; \frac{i}{2\pi} \, 
\int_{-\infty}^{+\infty}
        \frac{e^{-i(k^0-\omega)t}}{k^0-\omega + i\eta} \; dk^0   \nonumber \\
	 & = & \lim_{\eta \to 0^{\splus}} \; \frac{i}{2\pi} \, e^{+i\omega t} \, 
\int_{-\infty}^{+\infty}
        \frac{e^{-ik^0t}}{k^0-\omega + i\eta} \; dk^0   \nonumber \\
 \hspace{-20pt} \mbox{and hence:} \qquad 
      \theta(t) \, e^{-i\omega t} & = & \lim_{\eta \to 0^{\splus}} \; \frac{i}{2\pi} 
\, \int_{-\infty}^{+\infty}
        \frac{e^{-ik^0t}}{k^0-\omega + i\eta} \; dk^0   \label{thetexp}
\end{eqnarray}
 
  This expression for the $\theta$-function, along with equation~\ref{delplus} for the 
function $\Delta^+(x-y)$, can be substituted into equation~\ref{delfpp} for the 
Feynman propagator as follows:  
\begin{eqnarray}
\Delta_F(x-y) & \! = \! & \theta(x^0-y^0)\; \Delta^+(x-y) 
			\; + \; \theta(y^0-x^0)\; \Delta^+(y-x)  \label{delfpmp} 
			  \\
			  & \! = \! &  \theta(x^0-y^0)\; 
			  (-i)\int \frac{d^3 \bsl{p}}{(2\pi)^3} \frac{1}{2\omp} \,
				       e^{+i\bsl{p}\cdot (\bsl{x}-\bsl{y})} \, e^{-ip^0\cdot 
(x^0-y^0)} \nonumber \\
			 & \!\!  &	\!\!\!\!\!   +  \,
			  \theta(y^0-x^0)\; 
			  (-i)\int \frac{d^3 \bsl{p}}{(2\pi)^3} \frac{1}{2\omp} \,
				       e^{+i\bsl{p}\cdot (\bsl{y}-\bsl{x})} \, e^{-ip^0\cdot 
(y^0-x^0)}
   \, \Big\vert_{p^0 = +\omp = +\sqrt{\bsl{p}^2+m^2}}  \nonumber
\end{eqnarray}
  Since $\{x,y\}$ are fixed for each value of $\Delta_F(x-y)$ the $\theta$-function 
can be moved inside the $d^3 \bp$ integral and with $p^0 = +\omp$, which is constant 
for each value of the 3-vector $\bp$,
equation~\ref{thetexp} above may be substituted into the square brackets below:
\begin{eqnarray}
 \!\!\!\!\!\!\!\!\!\!\!\!\!\!\!\!\!\!
 \Delta_F(x-y) \!\! & = & \!\! 
      (-i)\int \! \frac{d^3 \bsl{p}}{(2\pi)^3} \frac{1}{2\omp} \,
	 e^{+i\bsl{p}\cdot (\bsl{x}-\bsl{y})}  \, \Big\lbrack \, \theta(x^0-y^0) \,
	          e^{-i\omp\cdot (x^0-y^0)}\,\Big\rbrack   \nonumber \\
    \!\!  & & \!\!\!\!\!		  + 
					  (-i)\int \! \frac{d^3 \bsl{p}}{(2\pi)^3} \frac{1}{2\omp} \,
	 e^{+i\bsl{p}\cdot (\bsl{y}-\bsl{x})}  \, \Big\lbrack\, \theta(y^0-x^0) \, 
	       e^{-i\omp\cdot (y^0-x^0)}\,\Big\rbrack 
					  \label{deltwothet}    \\
 \!\! & = &  \!\!     (-i)\int \! \frac{d^3 \bsl{p}}{(2\pi)^3} \frac{1}{2\omp} \,
	 e^{+i\bsl{p}\cdot (\bsl{x}-\bsl{y})}  \, \Big\lbrack \, \lim_{\eta \to 
0^{\splus}} \; \frac{i}{2\pi} \, \int \!
        \frac{e^{-ik^0(x^0-y^0)}}{k^0-\omp + i\eta} \; dk^0 \, \Big\rbrack   \nonumber 
\\
 \!\! & & \!\!\!\!\!  +
		              (-i)\int \! \frac{d^3 \bsl{p}}{(2\pi)^3} \frac{1}{2\omp} \,
	 e^{+i\bsl{p}\cdot (\bsl{y}-\bsl{x})}  \, \Big\lbrack \, \lim_{\eta \to 
0^{\splus}}
	        \; \frac{i}{2\pi} \, \int \!
        \frac{e^{-ik^0(y^0-x^0)}}{k^0-\omp + i\eta} \; dk^0 \, \Big\rbrack 
		      \label{deltwokint}
\end{eqnarray}

  Hence the 3-momentum integral has been enlarged to a 4-parameter integral by 
including the full unrestricted range of the $k^0$ variable associated with the 
$\theta$-function integral.
 That is while the $p^0$ component of the 4-vector $p$ is constrained to the value
  $\omp = +\sqrt{\bp^2 + m^2}$, the free $k^0$ integration variable is introduced from 
equation~\ref{thetexp}. In relabelling the 3-momentum $\bp$ by the 3-vector $\bk$ the 
above final expression is seen to take the form of an \textit{apparent} 4-momentum 
integral:
\begin{eqnarray}
    \Delta_F(x-y)
 & = &       \lim_{\eta \to 0^{\splus}} \int
		  \frac{d^4k}{(2\pi)^4 2\omega_{\bsl{k}}} 
		\Big\lbrack \, \frac{e^{-ik \cdot (x-y)}}{k^0-\omega_{\bsl{k}} + i\eta} +
		        \frac{e^{+ik \cdot (x-y)}}{k^0-\omega_{\bsl{k}} + i\eta} \, 
\Big\rbrack
			\label{delxybkkb}	\\
  & = &    \lim_{\eta \to 0^{\splus}} \int
		  \frac{d^4k}{(2\pi)^4} \, e^{-ik \cdot (x-y)} \,
 \Big\lbrack \, \frac{1}{2\omega_{\bsl{k}}} \Big( \frac{1}{k^0-\omega_{\bsl{k}} + 
i\eta}
    +	 \frac{1}{-k^0-\omega_{\bsl{k}} + i\eta} \Big) \, \Big\rbrack  \nonumber \\
    & = &   \lim_{\eta \to 0^{\splus}} \int
		  \frac{d^4k}{(2\pi)^4} \, e^{-ik \cdot (x-y)} \,
 \Big\lbrack \, \frac{\omega_{\bsl{k}} - i\eta}
       {\omega_{\bsl{k}}((k^0)^2 - \omega_{\bsl{k}}^2 + 2i\omega_{\bsl{k}}\eta - 
\eta^2)}
	    \, \Big\rbrack 
	 \nonumber \\				
	& = &   \lim_{\varepsilon \to 0^{\splus}} \int
		  \frac{d^4k}{(2\pi)^4} \; \frac{e^{-ik \cdot (x-y)}}{k^2 - m^2 + 
i\varepsilon}
		   \label{feynp}
\end{eqnarray}
   Here the second line is obtained by reversing the sign of all 4 integration 
variables in the second term in square brackets in equation~\ref{delxybkkb}.
   The third and final lines follow after some straightforward algebra,  with the new 
limiting parameter $\varepsilon \simeq + 2\omega_{\bsl{k}} \eta$ introduced, and with 
the limit $\varepsilon \to 0^{\splus}$ for the integral understood even if not 
explicitly stated.
 Through substituting  $\omega_{\bsl{k}}^2 = \bk^2 + m^2$ (see equation~\ref{delfpmp})
 into the third line, and  
 with $k^2 = (k^0)^2 - \bk^2$  in the final line, $k$ is treated as a Lorentz 
4-vector.
 This is the expression for the Feynman propagator scalar function quoted in 
equation~\ref{ppcont} (with a factor of $i$ from equation~\ref{delf1}). This function 
of the spacetime difference $(x-y)$ may also be written:
\begin{equation}
  \begin{array}{rcl}
  \Delta_F(x-y) & = & {\displaystyle \int \frac{d^4k}{(2\pi)^4} \, e^{-ik \cdot (x-y)} 
\, \widetilde{\Delta}_F(k) }
  \vspace{6pt}  \\  
  \mbox{with} \qquad  \widetilde{\Delta}_F(k)  & = & 
                    {\displaystyle   \frac{1}{k^2 - m^2 + i\varepsilon} }
   \end{array}   \label{delfmom}
\end{equation}
  being the momentum space representation of the Feynman propagator, obtained as the 
coefficients in the Fourier decomposition of the position space function.

   Unlike the 4-momentum integral expression for $\Delta^{\pm}(x-y)$ in 
equation~\ref{dpmd4}, for the Feynman propagator in equation~\ref{feynp} there is no 
`mass-shell' condition with a $\delta(k^2 - m^2)$ function, and with 4 independent 
`momentum' variables  the Feynman propagator represents `states' which are generally 
`off-shell'. This situation motivates the term `virtual particle' in referring to the 
`propagating entity'. 
  On the other hand $\Delta_F(x-y)$ is constructed in equation~\ref{delfpmp} out of 
elements which \textit{are} on-shell with energy $\omp = +\sqrt{\bp^2 + m^2}$, with 
the off-shell interpretation for the full expression arising through the incorporation 
of the $\theta$-functions.

   Equation~\ref{feynp} follows from the structure of $\Delta^{+}(x-y)$, which is 
found  through  $\lbrack a(\bp), a^{\dag}(\bp')\rbrack$ commutators  appearing for 
example in the expansion of terms in equation~\ref{uiter0} between vacuum states to 
determine a scattering amplitude, together with the $\theta$-functions, which are 
deployed when the calculation is reorganised with the time ordering $T$ of 
equation~\ref{uiter1}. Hence the notion of `virtual particle states' may be considered 
to be a purely mathematical construction arising from this reworking of the 
calculation.

In equation~\ref{thetrep} the $\theta$-function is defined by a contour integration in 
the complex plane. This involves a combination of Cauchy's theorem and the residue 
theorem -- respectively for integration contours surrounding a region of the integrand 
function which is regular or containing singularities, together with Jordan's lemma 
for the vanishing of particular $e^{-ist}$ contour integrals depending on the sign of 
the real parameter $t$ in the complex $s$-plane. The result is that $\theta(t)$ can be 
expressed in equation~\ref{thetrep} with the horizontal integration contour $C$ of 
figure~\ref{contours}(a), in which the pole in the integrand at $s = -i\eta$ is also 
shown.
\begin{figure}[htbp]  
\centering
\epsfxsize=13.3cm
\leavevmode
\epsffile[0 0 1869 758]{\gpath aPfig107e}
\vspace*{-10pt}
\caption{\setb Integration contours (a) in the complex $s$-plane for $\theta(t)$ 
defined in equation~\ref{thetrep} and (b) in the complex $k^0$-plane for the Feynman 
propagator $\Delta_F(x-y)$ in equation~\ref{feynp}. The single pole in the first case 
and pair of poles in the second case are also indicated.}
\label{contours}
\end{figure}

   The single pole in the integrand function for $\theta(t)$ carries over into two 
poles in the complex plane (since there are two $\theta$-functions in 
equation~\ref{deltwothet} leading to equation~\ref{delxybkkb}) for the integrand in 
equation~\ref{feynp} for $\Delta_F(x-y)$. In this latter equation (which was derived 
from equation~\ref{deltwokint}) it is understood that the $k^0$ integration should be 
carried out first following the straight contour $C$ along the real axis in 
figure~\ref{contours}(b). Using Cauchy's theorem this contour integral can be 
`analytically continued' by a $90^0$ counterclockwise rotation to the imaginary $k^0$ 
axis without encountering any poles. Under this `Wick rotation' to Euclidean 4-space 
(with $k^0$ replaced by $k^4 = ik^0$ to form a Euclidean 4-vector with $\bk$) the 
parameter $\eta$ (and hence $\varepsilon$ in equation~\ref{feynp}) may be discarded. 

    Alternatively equation~\ref{feynp} and the real $k^0$ integration in 
figure~\ref{contours}(b) is equivalent setting $\varepsilon = 0$ and performing the 
resulting integral:
\begin{equation}
 \Delta_F(x-y)  =      \int_{C_F}
		  \frac{d^4k}{(2\pi)^4} \; \frac{e^{-ik \cdot (x-y)}}{k^2 - m^2}
		   \label{feyncf}
\end{equation}
   following the contour $C_F$ with an implied limit of infinitesimal detours below 
the first then above the second pole on the real axis as displayed by the thick line 
in figure~\ref{delcont}. Although these expressions are equivalent 
equation~\ref{feynp} is generally quoted in preference to equation~\ref{feyncf} since 
the $i\varepsilon$  term in the former case serves to explicitly indicate the side on 
which the contour avoids the poles. 
\begin{figure}[htbp]  
\centering
\epsfxsize=13.3cm
\leavevmode
\epsffile[0 0 1819 913]{\gpath aPfig108e}
\vspace*{-10pt}
\caption{\setb The six functions $\Delta^{\pm}(x-y)$, $\Delta(x-y)$
  and $\Delta_{F,R,A}(x-y)$ described in the text can be defined by the integration 
along six different contours ($C^{\pm}$, $C$ and $C_{F,R,A}$ respectively) in the 
complex $k^{0}$-plane for the same integrand function presented in 
equation~\ref{feyncf}.}
\label{delcont}
\end{figure}

   Maintaining the same integrand while adapting the contour $C_F$ employed in 
equation~\ref{feyncf} in a total of six different ways leads to the expression of a 
total of six different functions, all related to $\Delta_{F}(x-y)$, and each then 
defined here in a related \textit{mathematical} form. However, the primary importance 
is given to the $C_F$ contour and the Feynman propagator in QFT since this object 
arises prominently in the calculation of scattering amplitudes. The three contours 
$C_F$, $C_R$ and $C_A$ hug the real axis in figure~\ref{delcont} with the integral 
determined in the limit of vanishingly small detours around the poles. However these 
integrals do \textit{include} these infinitesimal detours are \textit{not} the Cauchy 
principle values of the integrals which `hop over' the poles in this limit and would 
then be identical for `$C_F$', `$C_R$' and `$C_A$'.
  
  The three remaining contours $C$, $C^+$ and $C^-$ can be taken anywhere in the 
complex plane, so long as they navigate around the poles with the topology indicated 
in figure~\ref{delcont}. These contour integrals in the complex $k^0$-plane simply 
have the values of $-2\pi i$ times the residues enclosed, with a negative sign 
relative to the residue theorem which is based on anticlockwise circulating contours. 
It is again understood that this complex $k^0$ integral is performed first in 
equation~\ref{feyncf} for the respective contours, before the remaining real $\int d^3 
\bk$, in defining the $\Delta$, $\Delta^+$ and $\Delta^-$ functions.

 Here the outer contour $C$, encompassing both poles in figure~\ref{delcont}, 
represents the Lorentz invariant singular function $\Delta(x-y)$. This function can be 
introduced in the discussion of causality relating to field interactions and defined 
directly in terms of the field commutator:
\begin{eqnarray}
    i\Delta(x-y) & = & \lbrack \hat{\phi}(x), \hat{\phi}(y) \rbrack \label{delxypp} \\
	 & = & \lbrack \hat{\phi}^+(x), \hat{\phi}^-(y) \rbrack \; 
	                + \; \lbrack \hat{\phi}^-(x), \hat{\phi}^+(y) \rbrack \nonumber \\
     & = & i\Delta^+(x-y) \; + \; i\Delta^-(x-y)  \label{deleqpm}  \\
	    & = &  \int \frac{d^4k}{(2\pi)^4} \, \varepsilon(k^0)
     \, 2\pi \, \delta(k^2 - m^2) e^{-ik\cdot (x-y)}  \label{deld4} 
\end{eqnarray}
   using equations~\ref{delmin} and \ref{dpmd4} and with $\varepsilon(k^0) = +1,0,-1$ 
for 
   $k^0 \! > \! 0, k^0 \! = \! 0, k^0 \! < \!0$ respectively.
  Equation~\ref{deleqpm} is consistent with the residue theorem with the integral 
contour $C$ in figure~\ref{delcont} enclosing both poles, which are separately 
enclosed by $C^+$ and $C^-$.  With $\Delta(x-y)=0$ for $(x-y)^2 < 0$, unlike the case 
for the individual $\Delta^{\pm}(x-y)$ components, this function represents causality 
in field interactions through equation~\ref{delxypp}, in the sense that it implies 
$\hat{\phi}(x)$ and $\hat{\phi}(y)$ operate independently of each other outside the 
light cone. Each of these three functions satisfies the Klein-Gordon equation: 
\begin{equation}
    (\square_x \, + \, m^2)\,\Delta^{(\pm)}(x-y) = 0
\end{equation}
   where the differential operator $\square_x$ acts on the spacetime variables 
corresponding to $x$, and $m$ in the above is understood to be the mass $m_{\phi}$ 
associated with the scalar field $\hat{\phi}(x)$. In the spatial plane $x^0 - y^0 = 0$ 
the function $\Delta(x-y)$ also satisfies the time derivative equation
  $\partial_0 \, \Delta(\bx-\by,0) = -i\delta^3(\bx-\by)$ which, via 
equation~\ref{delxypp}, and the conjugate field $\hat{\pi}(x) = \pal_0 \hat{\phi}(x)$, 
is consistent with the equal-time field commutation relation:
\begin{equation}
  \label{ppcomr}
    \lbrack \hat{\phi} (\bx,t), \hat{\pi} (\by,t) \rbrack \, = \, i \, \delta^3 (\bx - 
\by) 
\end{equation} 

  Here we have arrived at this expression by employing the commutation relation 
 $[a(\bp),a^{\dag}(\bq)] = (2\pi)^3\delta^3 (\bp - \bq)$ in order obtain 
equation~\ref{deld4} from equation~\ref{delxypp} via equation~\ref{useaacom}.
 However the `canonical' commutation relation of equation~\ref{ppcomr} may be 
postulated ahead of equation~\ref{aacomr} as the field quantisation rule, as a 
generalisation from the non-relativistic quantum mechanical
 relation $\lbrack \hat{x}^a, \hat{p}^b  \rbrack  =  i \hbar \delta^{ab}$ for $a,b = 
\{1,2,3\}$ in the three spatial dimensions.

   In contrast to the three $C^{(\pm)}$ contours for the three $\Delta^{(\pm)}$ 
functions in figure~\ref{delcont} 
 the three remaining contour integrals essentially follow the real $k^0$ axis, 
differing only in their means of bypassing the two poles as described above. Although 
figure~\ref{delcont} provides a neat mathematical way of summarising these six 
functions it is important to understand their conceptual meaning and the relationships 
between them.

    In particular the two functions $\Delta_R(x-y)$ and $\Delta_A(x-y)$ are the 
`retarded' and `advanced' parts of the Lorentz invariant singular function 
$\Delta(x-y)$, that is:
\begin{eqnarray}
   \Delta_R(x-y) & = & \quad \theta(x^0-y^0) \, \Delta(x-y)  \qquad \quad 
      (  = 0 \quad \mbox{for} \quad x^0 < y^0 ) \;\;\;  \label{delrbc} \\
   \Delta_A(x-y) & = &  -\,\theta(y^0-x^0) \, \Delta(x-y)  \qquad \quad 
      (  = 0 \quad \mbox{for} \quad x^0 > y^0 ) \;\;\; \label{delabc}
\end{eqnarray} 
 Both of these functions of course vanish outside the light cone since $\Delta(x-y)$ 
does. The function $\Delta_R(x-y)$ also vanishes for $x^0 < y^0$ into the past while 
$\Delta_A(x-y)$ vanishes into the future.
  In solutions for a classical theory both retarded and advanced waves can be 
identified, with the latter then being eliminated on the grounds of causality. Bearing 
in mind the antiparticle interpretation described earlier in this section, the 
retarded and advanced functions are of comparable significance in quantum field 
theory.
  These two functions, along with the Feynman propagator $\Delta_F(x-y)$, are Green's 
functions which satisfy the inhomogeneous Klein-Gordon equation:    
\begin{equation}
    (\square_x \, + \, m^2)\,\Delta_{F,R,A}(x-y) = -\delta^4(x-y)
	\label{kginhom}
\end{equation}

   The conventional factor of $i$ introduced in equation~\ref{delf1} is chosen so that 
such a factor is absent in the above equation.  The choice of detours around the poles 
for the contour integration in figure~\ref{delcont} reflects different choices of 
boundary conditions for solutions to the differential equation~\ref{kginhom}, such as 
the vanishing of the functions into the past or the future described in 
equations~\ref{delrbc} and \ref{delabc}. The relation of the Feynman propagator to the 
retarded and advanced Green's functions can be seen from figure~\ref{delcont} to be:
\begin{equation}
    \Delta_F(x-y) \; = \;
   \Delta_R(x-y, \theta{(k^0)}) \; + \; \Delta_A(x-y,\theta{(-k^0)}) \label{delfrpa}
\end{equation}
  That is, with the $\theta(\pm k^0)$-functions understood to be attached to the 
integrand in the right-hand side of equation~\ref{feyncf}, the contour integral for 
the Feynman propagator $\Delta_F$ follows the advanced contour $C_A$ in the negative 
frequency $k^0<0$ half-plane  and the retarded contour $C_R$ for positive frequency 
$k^0>0$. 
 Alternatively, on attaching the $\theta(\pm k^0)$ to the integrand of 
equation~\ref{deld4}
which is then substituted into equations~\ref{delrbc} and \ref{delabc} and in turn 
into equation~\ref{delfrpa} the resulting expression is found to be identical to 
equation~\ref{delfpp}, with the latter expressing $\Delta_F(x-y)$ in terms of the 
$\Delta^{\pm}(x-y)$ functions.

  Retarded and advanced propagators are employed in quantum field theory to study 
solutions to the equations of motion.
  For example,  with regard to the scalar model of section~\ref{tranamp}, expressions 
such as:
\begin{equation}
   \hat{\phi}(x) =   \int d^4y \, \Delta_R(x-y) \, g
   \hat{\mcX}^{\dag}(y)\hat{\mcX}(y)
   \label{phirmcx}
\end{equation} 
 may be considered.
   The retarded propagator $\Delta_R(x-y)$  satisfies equation~\ref{kginhom}, which 
applied to equation~\ref{phirmcx} yields:
\begin{equation}
   \label{phxxonly}
   (\square_x + m^2) \, \hat{\phi}(x) = -g\hat{\mcX}^{\dag}(x)\hat{\mcX}(x)
\end{equation}
  as an  equation of motion for the quantum field $\hat{\phi}(x)$ with source term 
  $-g\hat{\mcX}^{\dag}(x)\hat{\mcX}(x)$. This is equation~\ref{kgphxx}, for the two 
fields $\hat{\phi}(x)$ and $\hat{\mcX}(x)$ of the scalar model, which in the previous 
section was   derived from the Lagrangian of equation~\ref{lagfint}. 
  This method of obtaining solutions to equations of motion via Green's functions was 
originally employed for classical field theories. For the classical case the 
right-hand side of equation~\ref{phxxonly} may act as a source of disturbance 
generating a wave motion for the corresponding classical field $\phi(x)$ on the 
left-hand side, while in the quantum case the right-hand side may act as a source for 
the production of particles of the quantum field $\hat{\phi}(x)$.

\section{Feynman Rules and Optical Theorem}
\label{fraot}

   The various systematic procedures involved in calculating a given transition 
amplitude for a given interacting quantum field theory can be conveniently summarised 
in a small set of rules, which are most simply expressed in the momentum space 
representation, obtained in turn for the Feynman propagators in their Fourier 
expansions. The Feynman rules associate mathematical elements of the calculation with 
graphical elements in a diagram representing a particular contribution to the 
transition amplitude. These rules are written down here for the scalar model with the 
interaction Lagrangian of equation~\ref{lagfint} in table~\ref{frulesp}. These rules 
resemble those for the simpler interacting field theory based on a single scalar field 
$\hat{\phi}(x)$ with the interaction Lagrangian $\lag_{\mathrm{int}} = 
-\frac{\lambda}{4!}\hat{\phi}^4$ (\cite{Pesk} p.115), which is often presented as a 
model QFT. 

\begin{table}[htb]
\centering
\begin{tabular}{|lcc|}
 \hline
    1. For each propagator: &
	 \setlength{\unitlength}{10pt}
	 \begin{picture}(8,3)
	     \multiput(1,0)(1,0){6}{\line(1,0){0.5}}
		 \put(3.5,0.7){$\hat{\phi}$}
	 \end{picture} or
	 \begin{picture}(8,3)     
		 \put(1,0){\line(1,0){5.5}}
		 \put(2.6,0.7){\small $\hat{\mcX},\hat{\mcY}$}
		 \thicklines
		 \put(4,0){\vector(1,0){0.1}}
	 \end{picture}
	  & {\Large $\frac{i}{k^{\mbox{\scriptsize 2}}
	                     -m^{\mbox{\scriptsize 2}}+i\varepsilon}$}  \\
	  2. For each vertex: & 
	  \setlength{\unitlength}{10pt}
	    \begin{picture}(8,4)
	     \multiput(0,0)(1,0){4}{\line(1,0){0.5}}
		 \put(4,0){\circle*{0.4}}
	     \put(4,0){\line(2,1){3}}		 
		 \put(4,0){\line(2,-1){3}}
		 \thicklines
		 \put(5.86,0.93){\vector(2,1){0.1}}
		 \put(5.34,-0.67){\vector(-2,1){0.1}}
	 \end{picture}
	   & $-ig$  \\
	  3. For each external line: & 
	  \setlength{\unitlength}{10pt}
	 \begin{picture}(8,4)
	     \multiput(1,0)(1,0){6}{\line(1,0){0.5}}
		 \put(3.5,0.7){$\hat{\phi}$}
		 \put(1,0){\circle*{0.4}}
	 \end{picture} or
	 \begin{picture}(8,4)
		 \put(1,0){\line(1,0){5.5}}
		 \put(2.6,0.7){\small $\hat{\mcX},\hat{\mcY}$}
		 \put(1,0){\circle*{0.4}}
		 \thicklines
		 \put(4,0){\vector(1,0){0.1}}
	 \end{picture}
	 & $1$ \\
	 & &   \\
	 \multicolumn{2}{|l}
	 {4. Impose 4-momentum conservation at each vertex:}
	              &    $\sum_a k_a = 0$ \\
	 & &   \\
	 \multicolumn{2}{|l}
	 {5. Integrate over each unconstrained loop momentum $k$:} 
	       & {\Large $\int
		    \frac{d^{\mbox{\scriptsize 4}}k}{(2\pi)^{\mbox{\scriptsize 4}}}$}          
\\
	 & &   \\
	 \multicolumn{2}{|l}
	 {6. Multiply by the symmetry factor:}
	       &   1   \\ 
     & &   \\
  \hline
  \end{tabular}
  \caption{\setb The Feynman rules in momentum space for the scalar model, relating 
mathematical terms and instructions to the elements of a Feynman diagram, each of 
which contributes to a transition amplitude $i{\mathcal M}_{fi}$.}
\label{frulesp}
\end{table}

  Representing possible terms in the transition amplitude by the possible topologies 
of graphical diagrams greatly assists the bookkeeping involved in the calculation. 
While terms in the expansion of $S = T e^{-i\int dt \, H_{\mathrm{int}}(t) }$ of 
equation~\ref{smatrix} can be pictured this way the internal lines should not be 
literally interpreted as representing trajectories of `virtual particles', indeed 
there is no reference to location at all in the momentum space Feynman rules. Rather 
the topology of the diagrams describes the structure of possible mathematical terms.
 Here we make some further comments on these rules 1--6 as listed in 
table~\ref{frulesp}:
\begin{itemize}
   \item[1.]  Each line, whether internal or external, is associated with a particular 
field type. The direction of an arrow on a line can be used to distinguish a particle
 from an antiparticle when relevant, as described in `item 3' below. 
 The propagator term is $i\widetilde{\Delta}_F(k)$ from equation~\ref{delfmom}, where 
the factor of $i$ follows from the convention of equation~\ref{delf1}.

   \item[2.]  The coupling $g$ is added by hand in equation~\ref{lagfint} and hence 
for the interaction Hamiltonian in equation~\ref{htlagx}. The factor of $-i$ 
originates from equation~\ref{utexp} and in turn from the evolution 
equation~\ref{uevolve}.

   \item[3.] The factors in these first three items are multiplied together. The 
external lines can be labelled with the on-mass-shell 4-momentum $k$, with an arrow on 
the line following the momentum transfer (into or out of the terminating vertex) for a 
particle and in the opposite direction for an antiparticle (with a similar convention 
for internal lines), as depicted in figures~\ref{fdeemm} and \ref{fdxxyy}.

   \item[4.] The momentum conservation for each vertex arises from the $\int d^4x$ 
over spacetime associated with each of $n$ factors of $\lag_{\mathrm{int}}(x)$ in the 
$n^{\mathrm th}$ order of perturbation, with the $x$-dependence in the integrand 
purely in terms of the form $e^{i(\sum_a k_a)\cdot x}$, with the $\sum_a k_a$ summing 
over all lines connected to the vertex. This is seen for example for $n=2$ in 
equation~\ref{sfixmom}, where all the various factors for the $S_{fi}\vert_{n=2}$ term 
of equation~\ref{xxyylow} are composed. 

   \item[5.] The loop integrals over $\int d^4 k$ tend to diverge leading to the need 
for renormalisation, as will be discussed below for figure~\ref{fdxxyy4} and also in 
section~\ref{seraps}. The loop integral includes the full independent range $-\infty < 
k^0 < +\infty$, arising originally from equations~\ref{thetexp}--\ref{feynp} as 
described in the previous section. 
In other quantum field theories there may also be a discrete sum over field indices 
such as spin.

   \item[6.]  This factor is simply the exponential expansion coefficient of 
equations~\ref{ufulltt} and \ref{utexp} multiplied by  $n!$ from the number of ways  
the dummy integration variables $\{x,y\ldots \}$ can
 label the $n$ vertices of the Feynman diagram. In other theories there may also be 
symmetry factors for permutations of identical particles, as for example in the 
$\hat{\phi}^4$ theory (\cite{Pesk} p.93).  
\end{itemize}

  Bearing in mind equations~\ref{soneit} and \ref{sfimfi} each Feynman diagram 
corresponds to a contribution to the $S$-matrix without the overall factor of 
$(2\pi)^4\, \delta^4 (p_F - p_I)$, that is the transition amplitude $i{\mathcal 
M}_{fi}$. Since the amplitude appears as $\vert {\mathcal M}_{fi} \vert^2$ in the 
cross-section calculation of equation~\ref{diffcrs} the overall factor of $i$ is 
sometimes neglected. 

   The above rules can be applied to the Feynman diagram of figure~\ref{fdxxyy}, 
representing the lowest-order term for the process $\mcX^+\mcX^- \to \mcY^+\mcY^-$. 
Reading off the Feynman rules in table~\ref{frulesp} for this diagram we find 
directly: 
\begin{equation}
   i{\mathcal M}_{fi} = -g^2\, \frac{i}{(p_1+p_2)^2-m_{\phi}^2 +i\varepsilon}  
     \label{xxyylowo}
\end{equation}
    This is the same expression for the transition amplitude as obtained in 
equation~\ref{mfiexplt} by explicit calculation, as it should be. The Feynman rules, 
as applied above, strip out the essence of such calculations. 

   We recall here that the transition probability is obtained from the square of the 
absolute value of the transition amplitude, by a basic postulate of quantum theory, as 
discussed around equation~\ref{ampqmpp}. The transition amplitude itself is strictly 
composed of all of the terms in the expansion of equation~\ref{smatrix}, of which only 
the lowest-order non-trivial term for $n=2$ has been accounted for in 
equation~\ref{xxyylowo}. It is an assumption of perturbation theory that the 
subsequent inclusion of terms of higher order into the sum gives a rapidly improving 
approximation to physical quantities such that very few orders are needed in practice. 
One aim of the following chapter is to understand how this procedure works in the 
context of the theory presented in this paper, but here we first explore a 
next-to-leading order term in the standard QFT approach for the scalar model. One of 
several contributions to the transition amplitude for $n=4$ is described by the 
Feynman diagram in figure~\ref{fdxxyy4}.

\vspace{2pt}
\begin{figure}[htbp]  
\centering
\epsfxsize=12.5cm
\leavevmode
\epsffile[0 0 1543 396]{\gpath aPfig109e}
\vspace{-2pt}
\caption{\setb Feynman diagram for the process $\mcX^+\mcX^- \to \mcY^+\mcY^-$ for a 
possible higher-order perturbation. At this `next-to-leading order' level an 
unconstrained internal loop momentum $r$ appears, depicted here for the $\hat{\mcX}$ 
field.}
\label{fdxxyy4}
\end{figure}

  In this case reading off the instructions from table~\ref{frulesp} `rule 5' is 
invoked for the freedom in the internal loop momentum $r$ which is not constrained by 
the application of `rule 4', leading to the amplitude contribution:
\begin{equation}
   i{\mathcal M}_{fi} = g^4  \,
       \Big( \frac{i}{(p_1\!+\!p_2)^2-m_{\phi}^2 +i\varepsilon} \Big)^2
	   \int   \frac{d^4 r}{(2\pi)^4} \,
	   \frac{i}{r^2-m_{\mcX}^2+i\varepsilon} \;
	   \frac{i}{(p_1\!+\!p_2\!-\!r)^2-m_{\mcX}^2+i\varepsilon}
     \label{xxyynexo}
\end{equation}  
   In QFT such loop momentum integrals are frequently divergent, as is the case here 
and for similar terms in the $\hat{\phi}^4$ scalar model, giving infinite and hence 
meaningless answers if taken at face value. This leads to the need for a program of 
`renormalisation' in order to extract useful results out of these calculations.

   In practice the divergent internal loop integrals  are first made finite by 
introducing a parameter to smooth the integrand or act as a cut-off to the integration 
range, a process known as `regularisation'. The theory is then renormalised, 
essentially by calibration against  an empirical input, before the regularising 
parameters are eliminated. The aim is to achieve finite predictive quantities in this 
way for comparison with further physical measurements, such as the observation of 
`running coupling' which is a consequence of renormalisation as will be described in 
section~\ref{seraps}.

  In the natural units we are adopting, with $\hbar = 1$ and $c=1$, any physical 
quantity can be expressed in units of mass, that is with dimension $M^D$, where the 
mass dimension $M$ is reciprocal to that of length and time, that is $M^1 \equiv 
L^{-1} \equiv T^{-1}$.
  The success of the renormalisation procedure generally depends upon the power of the
   mass dimension $D$ for the coupling parameter itself.  Since $\int \! \lag \, d^4 x 
$ represents the `action' which is a dimensionless quantity with $D=0$, the Lagrangian 
density $\lag$ has dimension $D=4$, which is also consistent with 
equation~\ref{htlagx} since the Hamiltonian $H$ has dimension $D=1$. If the coupling 
parameter in the interaction Lagrangian has $D\ge 0$ such a theory is probably 
renormalisable, whereas theories with $D<0$, such as gravitation for which Newton's 
constant $G_{\! N}$ has $D=-2$, are non-renormalisable.

Hence the renormalisation procedure works for quantum field theories with 
dimensionless coupling constants, such as QED and the Standard Model in general and 
also the scalar model with $\lag_{\mathrm int} = -\frac{\lambda}{4!} \hat{\phi}^4$. 
For the scalar model considered here with  
 $\lag_{\mathrm{int}} = -g\hat{\phi}\hat{\mcX}^{\dag}\hat{\mcX} - 
     g\hat{\phi}\hat{\mcY}^{\dag}\hat{\mcY}$ the full Lagrangian of 
equation~\ref{lagfint}
 implies that the coupling $g$ has dimension $D=+1$, and hence the theory can be 
renormalised. Such a theory with $D>0$ may even be `super-renormalisable' and contain 
no infinities at all after some order of perturbation.

   Since a cross-section $\sigma$ has the dimension $L^2$ the right-hand side of 
equation~\ref{diffcrs} must also have the overall dimension $D=-2$, which is also the 
dimension of the initial state flux factor in this equation. For a two-particle final 
state the   
 Lorentz invariant phase space $d\Phi$ is dimensionless, implying that the amplitude 
${\mathcal M}_{fi}$ itself should also have $D=0$ in this case. 
 This is consistent with the dimensionless coupling $e$ of QED in equation~\ref{mfimu} 
and with the coupling $g$ having $D=1$ for the scalar model in equations~\ref{mfisca}, 
\ref{xxyylowo} and \ref{xxyynexo}.
 More generally the dimension of the transition amplitude ${\mathcal M}_{fi}$ will 
depend upon the multiplicity of the final state and the conventions employed for 
initial and final state normalisation, consistent with the composition of factors 
forming the cross-section having the appropriate net dimension, as is the case for 
equation~\ref{diffcrs}.

   Higher-order corrections, as appearing for the internal propagator for the field 
$\hat{\phi}$ of figure~\ref{fdxxyy} when dressed as in figure~\ref{fdxxyy4}, will also 
be important for the external particle states. This applies also for calculations in 
QED and Standard Model QFT calculations in general. Although the theory begins by 
describing \textit{free} field states it is not possible in the physical world to 
decouple the electron field from the electromagnetic field (or the $\hat{\mcX}$ field 
from the $\hat{\phi}$ field in the scalar model) since they are intrinsic elements of 
a single \textit{interacting} system.

Any parameters, such as masses $m_{\phi}$ and $m_{\mcX}$ in the model here, ascribed 
to a free field will be unphysical and unmeasurable. Instead a finite set of 
fundamental physical parameters can be operationally defined as those quantities which 
are directly measurable in the laboratory. The self-interaction effects for the 
observed particle states are absorbed into these measured parameters, with the Fock 
space of initial and final states (in the interaction picture basis) assumed to 
represent precisely the observed masses and charges of physically produced or detected 
particles in matrix element $\langle f \vert S \vert i \rangle$ calculations.  These 
renormalised parameters obey the fundamental conservation laws of external and 
internal symmetries in collision processes.
 The physical renormalised mass is not the same object as the `bare' mass parameter 
appearing in the Lagrangian of the theory.

  As well as the obvious necessity to `tame the infinities' for calculations of 
physical quantities, the finite results obtained must also respect the basic 
requirement of probability conservation, namely that the total probability for 
\textit{something} to happen must always be equal to 1.  This fundamental principle 
translates in quantum theory into the  unitarity
  of the $S$-matrix, with the restrictions of this condition having implications for 
the relationship between physical quantities such as the cross-section $\sigma$ and 
the structure of the transition amplitude ${\mathcal M}_{fi}$ as will be described 
here.

  The unitarity  of the $S$-matrix of equation~\ref{smatrix}, that is the property 
$SS^{\dag} = S^{\dag}S = \b1$, together with the definition of the operator 
$T=i(\b1-S)$ in equation~\ref{soneit}, hence with $T^{\dag} = -i(\b1 - S^{\dag})$, 
implies that:
\begin{eqnarray}
  TT^{\dag}\, = \, T^{\dag}T \, & = & \, i(T^{\dag} - T)  \\
  \mbox{and therefore:} \qquad  \langle f \vert TT^{\dag} \vert i \rangle
         \, & = & \, i \langle f \vert T^{\dag} \vert i \rangle  - 
		             i \langle f \vert T \vert i \rangle    \label{tttt}     
\end{eqnarray}

    Inserting a sum over a complete set of intermediate states $\vert m \rangle$ the 
left-hand side of this expression can be written as:
\begin{equation}
   \langle f \vert TT^{\dag} \vert i \rangle \, = \,
      \sum_m \, \left( \prod_{j=l}^{r_m} \, 
	                   \int \! \frac{d^3\bsl{k}_j}{(2\pi)^3 2E_j} \right) \;
	  \langle f \vert T \vert m \rangle   \langle m \vert T^{\dag} \vert i \rangle    
	  \label{ttsumtt} 
\end{equation}   
  where $r_m$ is the number of particles in each state $\vert m \rangle$ and 
$\frac{d^3\bsl{k}_j}{(2\pi)^3 2E_j}$ is the invariant phase space element for the 
particle state normalisation adopted, as described in section~\ref{crosss} and 
required here for the insertion of the unit operator $\b1$ between $T$ and $T^{\dag}$.
    The two terms on the right-hand side of equation~\ref{tttt} can be written as:   
\begin{eqnarray}
  i\langle f \vert T \vert i \rangle \, & = & \, 
                        i{\mathcal M}_{fi} \, (2\pi)^4\,   \delta^4 (p_F - p_I)
                     \label{tmfi1} \\
  i\langle f \vert T^{\dag} \vert i \rangle \, & = & \, 
                         i{\mathcal M}^{\ast}_{if} \, (2\pi)^4\, \delta^4 (p_F - p_I)
                    \label{tmfi2}
\end{eqnarray}
  These are obtained directly from equation~\ref{sfimfi}, which can also be applied to 
the right-hand side of equation~\ref{ttsumtt} and hence substituted into 
equation~\ref{tttt} along with equations~\ref{tmfi1} and \ref{tmfi2} to find:
\begin{eqnarray}
      \sum_m &  \!\! {\mathcal M}_{fm} \, (2\pi)^4\,   \delta^4 (p_F - p_M) \;
	            {\mathcal M}^{\ast}_{im}  & \!\!\! \Big\lbrack (2\pi)^4\, \delta^4 
(p_M - p_I) \;
		   \prod_{j=l}^{r_m} \, \int \! \frac{d^3\bsl{k}_j}{(2\pi)^3 2E_j} \Big\rbrack      
   \qquad\qquad \nonumber \\
	 && {} \!  = \,  (i{\mathcal M}^{\ast}_{if} - i{\mathcal M}_{fi})\, 
		      (2\pi)^4\,   \delta^4 (p_F - p_I) 
			  \label{mmmmfull}
\end{eqnarray}   
    This is a non-linear relationship between transition amplitudes, with a product on 
the left and a sum on the right-hand side, resulting from the unitarity of the 
$S$-matrix. Given the second $\delta$-function on the left-hand side the first one 
$\delta^4 (p_F - p_M)$ may be replaced by  $ \delta^4 (p_F - p_I) $,  which hence 
cancels with the $\delta$-function on the right-hand side. The term in square brackets 
is simply the Lorentz invariant phase space $d\Phi$, as described for 
equations~\ref{diffcrs} and \ref{diffevr}, here for the intermediate states, and hence 
equation~\ref{mmmmfull} can be written simply as:
\begin{equation}
   \sum_m  \left( {\mathcal M}_{fm}    {\mathcal M}^{\ast}_{im}  \int \! d\Phi \right)
    \, = \,  i({\mathcal M}^{\ast}_{if} - {\mathcal M}_{fi})
	 \label{mmmm}
\end{equation}

  Considering a two-particle initial state and setting $\vert f \rangle = \vert i 
\rangle$, corresponding to elastic forward scattering at a HEP collider with the final 
state being identical to the initial state, and by comparison with 
equation~\ref{diffcrs}, the left-hand side above is then identical to the 
\textit{total} cross-section for the transition from an initial state $\vert i 
\rangle$ to any state $\vert m \rangle$, up to an initial state flux factor, which 
again relates to the state normalisation. That is, with  $\vert f \rangle = \vert i 
\rangle$ and since $\vert {\mathcal M}_{im} \vert = \vert {\mathcal M}_{mi} \vert$, 
equation~\ref{mmmm} becomes:
\begin{eqnarray}
 \sum_m  \left( \vert {\mathcal M}_{mi} \vert^2  \int \! d\Phi \right)
  & = & 2\,\mbox{Im}({\mathcal M}_{ii})  \label{msqmii} \\
 \equiv \; 4E_1E_2 \vert \bv_1 - \bv_2 \vert  \,  \sigma_{\mathrm tot}
                            & = & 2\,\mbox{Im}({\mathcal M}_{ii})    
\end{eqnarray}
   where equation~\ref{diffcrs}, with an implied integration over the phase space for 
each final state to obtain the total cross-section 
$\sigma_{\mathrm tot}$, has been substituted in for the left-hand side in the second 
line.
   (Here $\mbox{Im}({\mathcal M}_{ii})$ is of course a \textit{real} number, as for 
the standard definition of the imaginary part of a complex number, in contrast to the 
definition of the imaginary part of an octonion as described immediately before 
equation~\ref{octinner}).
   The flux factor can be expressed in terms of the total centre-of-mass energy $E_T$ 
$(=\sqrt{s})$ and the momentum of either initial particle in the centre-of-mass frame 
$\vert \bp_i \vert$ (noting however that this factor is not fully Lorentz invariant, 
as described
after equation~\ref{diffcrs}), such that the total cross-section can finally be 
written as:
\begin{equation}
   \sigma_{\mathrm tot}(i \to \mbox{anything})   =   \frac{\mbox{Im}({\mathcal 
M}_{ii})} {2 E_T \vert \bp_i \vert}  
    \label{optith}   
\end{equation}     
     
  This relationship, along with its derivation, is a form of the `optical theorem' 
(\cite{Pesk} p.231, equation~7.50). It is a consequence of the $S$-matrix unitarity 
condition in scattering experiments, which in turn expresses basic properties of the 
laws of probability, and has further implications for observable quantities. Here it 
shows how the \textit{total} cross-section for the production of any final state is 
directly related to the \textit{imaginary} part of the forward scattering amplitude, 
up to the normalisation factor in equation~\ref{optith}. By equations~\ref{soneit} and 
\ref{sfimfi} the imaginary part of ${\mathcal M}_{ii}$ corresponds to the non-trivial 
real part of $\langle i \vert S \vert i \rangle$, with many intermediate processes 
contributing.
    The significance of this result in the context of the present paper is that it 
demonstrates a \textit{linear} relationship between a  cross-section, that is the 
likelihood of an event occurring, and an amplitude.

  The generalised optical theorem as expressed in equation~\ref{mmmm} can also be 
applied to the case of a single particle initial state. On again setting $\vert f 
\rangle = \vert i \rangle$ in this case an expression for the total decay rate 
$\Gamma$ can be identified as:
\begin{equation}
 \label{optigam}
   \Gamma (i \to \mbox{anything}) = \frac{\mbox{Im}({\mathcal M}_{ii})}{m_i}
\end{equation}
 where $m_i$ is the mass of the initial state particle. For a single particle the tree 
level contribution to ${\mathcal M}_{ii}$ is just the propagator 
$\widetilde{\Delta}_F(k)$ of equation~\ref{delfmom}. 
 For $\varepsilon \to 0$ this function is real except when the particle is on-shell, 
with the consequence that $\mbox{Im}(1/(k^2 - m^2 + i\varepsilon)) \sim \delta (k^2 - 
m^2)$.

  This observation can be generalised for higher-order perturbations.
  In fact the application of the optical theorem in a quantum field theory can also be 
demonstrated in terms of Feynman diagrams, where it can also be proved to all orders 
of perturbation theory by applying `cutting rules' (\cite{Pesk} pp.232--236, 
\cite{Velt} pp.183--196). An example obtained by relabelling the Feynman diagram in 
figure~\ref{fdxxyy4} to represent an amplitude for the forward scattering process 
$\mcX^+\mcX^- \to \mcX^+\mcX^-$, with identical incoming and outgoing particles and 
momenta, via two $\hat{\phi}$ field propagators and a $\hat{\mcY}$ field internal loop 
is shown here in figure~\ref{fdxxxx4}. 
\begin{figure}[htbp]  
\centering
\epsfxsize=12.9cm
\leavevmode
\epsffile[0 0 1555 560]{\gpath aPfig1010e}
\caption{\setb A Feynman diagram for the forward scattering process $\mcX^+\mcX^- \to 
\mcX^+\mcX^-$, with a `cut line' drawn through the intermediate loop propagators of 
the $\hat{\mcY}$ field.}
\label{fdxxxx4}
\end{figure}

  By careful analysis of the singularities  that occur when internal propagators go 
on-mass-shell under internal loop momenta integrals, twice the \textit{imaginary} part 
of the amplitude can be obtained by summing over the `cutting' possibilities (only one 
for the diagram  in figure~\ref{fdxxxx4}, shown by the vertical dashed line) and 
replacing the term in the Feynman rule for each propagator that may be simultaneously 
put on-shell by the cut as:
\begin{equation}
    \frac{i}{k^2-m^2+i\varepsilon} \, \to \, 2\pi i \, \delta(k^2-m^2)
	 \label{cutrule}
\end{equation}
  (with the sign and factors of $2$ and $i$ depending on the conventions adopted)
  before performing the $\int d^4 r$ over the loop 4-momentum.
  Hence the imaginary part of a loop amplitude is obtained by placing the intermediate 
states on-shell together, as may have been expected from the optical theorem itself 
since the final states for cross-sections and decay rates, equations~\ref{optith} and 
\ref{optigam} respectively, consist of on-shell particles.
 Each way of placing intermediate states on-shell together, as for 
figure~\ref{fdxxxx4}, is called a `cut' after Cutkosky, with the above cutting rules 
providing a method to compute the  imaginary part of a transition amplitude in 
general.

   The cutting rules for obtaining the imaginary part of the transition amplitude for 
a given Feynman diagram can be derived by summing over sets of replacements of each 
Feynman propagator $\Delta_F$ by either $\Delta_F, \Delta_F^{\ast},\Delta^+$ or 
$\Delta^-$  in the Feynman rules. This calculational tool involves a sum over 
permutations of selected vertices which determine the kind of replacement for each 
$\Delta_F$ (see for example~\cite{Velt} p.186). Indeed it can be seen that replacing 
$\Delta_F$ of equation~\ref{feynp} with $\Delta^{\pm}$ from equation~\ref{dpmd4} 
incorporates the substitution of equation~\ref{cutrule} together with the introduction 
of a factor of $\theta (\pm k^0)$. This latter factor relates to the time ordering of 
the corresponding vertices and the resulting interpretation as an apparent particle or 
antiparticle propagating forwards in time between the two vertices. While in 
equation~\ref{delfpp} or \ref{delfpmp} the Feynman propagator was constructed out of 
two $\Delta^{\pm}$ components, here it is  taken apart again and a single on-shell 
part retained.

  This on-mass-shell condition is expressed by the $\delta$-function in 
equation~\ref{dpmd4}.
 On performing the $k^0$ integral this constraint leads to the form of 
equation~\ref{delplus}  
  which in the present context represents the phase space factor for real final state 
external particles, on-mass-shell and with positive energy, in cross-section or decay 
rate calculations.
The remaining propagators $\Delta_F$, for example for the two internal $\hat{\phi}$ 
field lines in figure~\ref{fdxxxx4} are unchanged, representing their usual 
(non-physical) aid to calculation as described in the previous two sections.

  For example for the Feynman diagram in figure~\ref{fdxxxx4}, by adapting 
equation~\ref{xxyynexo} and applying the substitutions from equation~\ref{cutrule} to 
the 
 basic Feynman rules of table~\ref{frulesp} we obtain:
\begin{eqnarray}
  & 2 & \!\!\!\!\mbox{Im}
 \left( {\mathcal M}(\mcX^+\mcX^- \to \mcX^+\mcX^-) \right) \nonumber  \\
  & & {}
   = g^4  \,
       \left( \frac{i}{(p_1+p_2)^2-m_{\phi}^2 +i\varepsilon} \right)^{\! 2}
	   \int \!  \frac{d^4 r}{(2\pi)^4} \,
	     2\pi i \;\! \delta(r^2 - m_{\mcY}^2)   \,
		 2\pi i \;\! \delta((k_1-r)^2 - m_{\mcY}^2) \quad	\nonumber   \\
		 & & \label{cutmm}
\end{eqnarray}
  The latter integral can be more easily performed under the substitution of the 
original 4-momenta $q_1$ and $q_2$, as indicated in figure~\ref{fdxxxx4}, in place of 
the integral over $r$, by including a 4-momentum constraint in:
\begin{equation}
   \int   \frac{d^4 r}{(2\pi)^4} \, \equiv \, 
    \int   \frac{d^4 q_1}{(2\pi)^4} \,  \int   \frac{d^4 q_2}{(2\pi)^4}  \,
	   (2\pi)^4 \, \delta^4(q_1 + q_2 -k_1)   \nonumber
\end{equation}
 This leads to:
\begin{eqnarray}
\hspace{-20pt}  & \!\! &
     2\, \mbox{Im}
	\left( {\mathcal M}(\mcX^+\mcX^- \to \mcX^+\mcX^-) \right)  \nonumber  \\
\hspace{-20pt}  & \!\! & 
   = -{g^4}  \,
       \left( \frac{i}{(p_1+p_2)^2-m_{\phi}^2 +i\varepsilon} \right)^{\!\! 2}
	 \int \frac{d^4 q_1}{(2\pi)^4} \,  \int   \frac{d^4 q_2}{(2\pi)^4}   \nonumber \\
\hspace{-20pt} & &  \qquad \qquad \qquad 
       \qquad \qquad \qquad \qquad \; 
	   2\pi  \delta(q_1^2 - m_{\mcY}^2)   \, 2\pi  \delta(q_2^2 - m_{\mcY}^2) \,
		(2\pi)^4 \,   \delta^4(q_1 + q_2 -k_1)	\nonumber  \\
\hspace{-20pt}   & \!\! &  
  =  -{g^4}  \,
       \left( \frac{i}{(p_1+p_2)^2-m_{\phi}^2 +i\varepsilon} \right)^{\!\! 2}
	\int  \frac{d^3 \bsl{q}_1}{(2\pi)^3} \frac{1}{2\omega_{\bsl{q}_1}} \,
	\int  \frac{d^3 \bsl{q}_2}{(2\pi)^3} \frac{1}{2\omega_{\bsl{q}_2}} \,
	    (2\pi)^4 \, \delta^4(q_1 + q_2 -k_1)  \nonumber \\
\hspace{-20pt}	& & \label{optifey}
\end{eqnarray}
  by applying equation~\ref{deldel} to obtain the bottom line with  $q_{1,2}^0 > 0$. 
Here the initial $q^0_{1,2}$ part of the $\int \frac{d^4 q_{1,2}}{(2\pi)^4}$ integrals 
over $\delta(q_{1,2}^2 - m_{\mcY}^2)$ place the momenta $q_{1,2}$ on-shell resulting 
in integrals of 
the form  $\int \frac{d^3 \bsl{q}_{1,2}}{(2\pi)^3} \frac{1}{2\omega_{\bsl{q}_{1,2}}}$, 
that is over the relativistic phase space. Together with the overall 
$(2\pi)^4\delta^4(p_F - p_I)$ for 4-momentum
 conservation implied in the final delta function this identifies the Lorentz 
invariant phase space factor $d\Phi$ for a two-body $\mcY^+\mcY^-$ final state in 
equation~\ref{optifey}.
 The remaining factor, before the first $\int$ sign, can be identified with $\vert 
{\mathcal M}_{fi} \vert^2$ for the scattering amplitude ${\mathcal M}_{fi}$ of 
equation~\ref{xxyylowo}
 for the process $\mcX^+ \mcX^- \to \mcY^+\mcY^-$ at the level of the Feynman diagram 
depicted in figure~\ref{fdxxyy}, and hence
 (swapping the two sides of equation~\ref{optifey}):
\begin{equation}
  \vert {\mathcal M}(\mcX^+\mcX^- \to \mcY^+\mcY^-) \vert^2 \int d\Phi \, = \,
   2\,\mbox{Im}\left({\mathcal M}(\mcX^+\mcX^- \to \mcX^+\mcX^-)\right)
  \label{optifey1}
\end{equation}

  This equation verifies the optical theorem relation of equation~\ref{msqmii} for the 
$\mcY^+\mcY^-$ final state contribution to the total cross-section in $\mcX^+\mcX^-$ 
collisions for the Feynman diagram analysis at this level of perturbation theory.  A 
further contribution for a $\mcX^+\mcX^-$ final state can be obtained in a very 
similar manner based on an intermediate $\mcX^+\mcX^-$ loop, in place of the 
$\mcY^+\mcY^-$ loop, 
  in figure~\ref{fdxxxx4}.
 The above argument applies to arbitrary loop diagrams and this Feynman diagram 
approach based on the cutting rules can be used to prove the optical theorem to all 
orders of perturbation theory (\cite{Pesk} pp.235--236), with care for combinatoric 
factors and the consistency of the conventions used in general.
  
  Our main point here has been to review the relation between a physical cross-section 
and an expression \textit{linear} in a component of an amplitude, namely the imaginary 
part of the forward scattering amplitude, both in terms of the total cross-section in 
equation~\ref{optith} and at the level of individual processes as implied in 
equation~\ref{optifey1}. These expressions relate to the optical theorem and the 
unitarity constraint which in turn represents the basic property that the total 
probability must always equal one.
 This structure will provide a means to connect calculations of the likelihood of 
scattering processes for the present theory with the techniques of quantum field 
theory, as we shall describe in section~\ref{secdopp}. In the meantime, in the 
following section, we assess the nature of basic field interactions in the context of 
the present theory.


\pagebreak
\chapter{A Novel Conception of HEP Processes}
\label{newapp}

\section{Degeneracy of Spacetime Solutions}
\label{secdos}

  In this chapter we consider how the probabilistic nature of quantum phenomena arises 
in the context of the present theory, and in particular in the environment of 
laboratory experiments.
The main goals will be to relate the calculation of cross-sections, for example, for 
the present theory with the corresponding formalism of QFT and to address the related 
question concerning the nature of particle phenomena generally. Here the probability 
for a particular process will be a measure of the degeneracy of field states 
describing the mathematical form  of a particular 4-dimensional geometry, that is 
through the symmetry of possible local reinterpretations of fields such as the gauge 
field $Y_{\mu}(x)$ or fermion field $\psi(x)$ (denoted without `hats', since these are 
\textit{not} quantum field operators here) under the same Einstein tensor 
$G^{\mu\nu}(x)$.
 The spacetime geometry is locally completely insensitive to
 reinterpretations of the underlying fields, that is exchanges between components of 
the fields  
$\delta Y(x) \leftrightarrow \delta\psi(x)$,
 which leave $G^{\mu\nu}(x)$ locally unchanged,
 while the geometric contracted Bianchi identity $G^{\mu\nu}_{\ph{\mu\nu};\mu} = 0$ 
remains globally valid.

    The field interactions proceed by a kind of `Chinese whispers' of field 
indistinguishability, as a degenerate mathematical possibility underlying the 
spacetime geometry. This leads directly to the indeterminate nature characteristic of 
quantum phenomena.
 Probabilities, in the form of cross-sections and decay rates, will then arise in 
proportion to the sum of the `number of ways' in which such underlying field 
descriptions are possible.

  We begin however by considering a particular case for which $G^{\mu\nu}(x)$ is a 
function of a single internal gauge field.
   Based on the breaking of the full symmetry of $\lvh$ into external and internal 
parts over an extended base manifold $M_4$ a relationship between the external 
geometry described by $G^{\mu\nu}(x)$ and internal gauge fields $Y_{\mu}(x)$ over 
$x\in M_4$ was developed in chapters~\ref{sym}--\ref{chaputtf}. Structures are 
identified analogous to those of Kaluza-Klein theory in leading to 
equation~\ref{einfiek}-\ref{einym}, which is conjectured to arise out of the geometric 
constraints of the present theory culminating in equation~\ref{gchift},  for which the 
practical normalisation convention $\chi=\frac{\kappa}{2}$ may be adopted.
For the present theory with an external linear connection 
$\Gamma^{\rho}_{\ph{\rho}\mu\nu}(x)$ and an internal gauge field $A_{\mu}(x)$, 
deriving from an internal Abelian $\uo$ gauge symmetry, the relation between the 
external Riemannian curvature and internal gauge curvature is expressed  in 
equation~\ref{geinm}, 
 via the above connection with classical Kaluza-Klein theory,
 and as reproduced here:
\begin{equation}
    -\frac{1}{\kappa}G^{\mu\nu}   =  +F^{\mu}_{\ph{\mu}\rho}F^{\rho\nu}
	        + \frac{1}{4} g^{\mu\nu} \, F_{\rho\sigma}F^{\rho\sigma}   \label{geinm2}    
\end{equation}	

  Under the Bianchi identity $\gmo$ this relation \textit{implies} the source-free 
homogeneous Maxwell equation~\ref{maxfein}, as explained in section~\ref{subwal}.
 Here, within the context of the present theory the generator of the internal $\uo_Q$ 
gauge symmetry of electromagnetism
  is identified with the element $\Sbard^1_l$ within the set of $\esi$ Lie algebra 
actions as described in section~\ref{intsym}. 
 The electromagnetic gauge field $A_{\mu}(x)$, associated with the $\uo_Q$ generator,
 is in turn identified with the field  $\tilde{A}_{\mu}(x)$ as described in and 
following equation~\ref{achargee}.
 We next consider
  the form of the free field $A_{\mu}(x)$ as a solution of Maxwell's equation, which 
can be written in terms of the gauge field itself as:
\begin{equation}
 \label{maxafree}
   \square A^{\mu}(x)=0 
\end{equation}

   The energy-momentum tensor for the electromagnetic field can be obtained directly 
through the definition $T^{\mu\nu}:= -\frac{1}{\kappa}G^{\mu\nu}$ with the geometry 
$G^{\mu\nu}(x)$ determined in terms of the electromagnetic field tensor $F^{\mu\nu}$ 
according to equation~\ref{geinm2} above.
 (A normalisation convention setting $\kappa = 8\pi G_{\! N} = -1$ might also be 
adopted in order to emphasise the equivalence of both sides in this definition 
$T^{\mu\nu} := G^{\mu\nu}$).
Here the electromagnetic gauge field itself is analysed under the assumption of an 
approximately flat spacetime.
 A real field $A^{\mu}(x)$  can be expressed in terms of Fourier components, which in 
terms of trigonometric functions and a single 4-vector $k$ takes the form:
\begin{eqnarray}
 A^{\mu}(x) & = &  A^{\mu}_c(\bk) \, \cos k\!\cdot\! x \, + \,A^{\mu}_s(\bk) \,
   \sin k\!\cdot\! x \label{acsin} \\ 
  & &  \qquad        \bigg[ \, \equiv \,   
\frac{A^{\mu}_c}{\cos\left(\tan^{-1}\frac{A^{\mu}_s}{A^{\mu}_c}\right)}
	   \cos \left(k\!\cdot\! x - \tan^{-1}\frac{A^{\mu}_s}{A^{\mu}_c}\right) \; \bigg]   
\label{aonecos}  \\
      & = & A^{\mu}(\bk) \, e^{-ik\cdot x} \quad + \quad  {A^{\mu}}^{\ast}(\bk) \, 
e^{+ik\cdot x}
	    \label{aeepm} \\
     & \equiv & C^{\fh}\,\varepsilon^{\mu}_r(\bk)\,A_r(\bk)\,e^{-ik\cdot x} 	 +
	            C^{\fh}\,\varepsilon^{\mu}_r(\bk)\,A_r^{\ast}\,(\bk)e^{+ik\cdot x}   
\label{kcoeff}
\end{eqnarray}

   In equation~\ref{aonecos} the gauge field is expressed in terms of a single cosine 
function, that is in the form 
$A^{\mu}(x) = A^{\mu}\cos (k \! \cdot \!  x + \lambda)$, with no sum implied over the 
index $\mu=0,1,2,3$, using the trigonometric identity $\cos\alpha\cos\beta \pm 
\sin\alpha\sin\beta = \cos(\alpha \mp \beta)$.  Equation~\ref{aeepm} follows from 
equation~\ref{acsin} with $A^{\mu}(\bk) = \fh(A^{\mu}_c(\bk)+ iA^{\mu}_s(\bk))$, and 
with $A^{\mu}(\bk)$ expressed as $C^{\fh}\,\varepsilon^{\mu}_r(\bk)\,A_r(\bk)$ in the 
final line.

  Hence for each 4-vector $k$ each of the four vector components of $A^{\mu}(x)$ can 
be associated with the real coefficients  $A^{\mu}_c(\bk)$ and $A^{\mu}_s(\bk)$ or the 
complex coefficients $A^{\mu}(\bk)$  and ${A^{\mu}}^{\ast}(\bk)$, either pair of which 
can be considered to be independent in terms of possible field interactions as 
described in the following section. With all physical phenomena being invariant under 
spacetime translations, and hence with no preferred set of coordinates $\{x\}$, and 
hence with $A^{\mu}_s(\bk) \neq 0$
 in general,  equation~\ref{aonecos} forms a relatively cumbersome expression for the 
free field and will not be employed further.

 The constant coefficient $C^{\fh}$ is introduced with the square root in 
equation~\ref{kcoeff} since each factor of $A^{\mu}(x)$ appears quadratically in the 
expression for $G^{\mu\nu}$ in equation~\ref{geinm2}. In line with textbook solutions 
to Maxwell's equations, and anticipating the quantum field analysis, four polarisation 
vectors $\varepsilon^{\mu}_r(\bk)$ are introduced, representing a 4-vector object for 
each of $r=0,1,2,3$ (see for example~\cite{ManSh} section~5.1). These provide a 
\textit{constant} basis for the 4-vector $A^{\mu}(\bk)$, analogous to the 
\textit{variable} tetrad components $\tetu$
 describing four vector fields for $a=0,1,2,3$
 as a basis for tangent 4-vectors to the manifold $M_4$, as employed in 
section~\ref{fdandtd} for example. 
 A standard choice of basis is such that the polarisation vectors are real and 
orthogonal, with respect to the Minkowski metric $\eta_{\mu\nu} = 
\mbox{diag}(1,-1,-1,-1)$, that is:
\begin{equation}
   \varepsilon_{r}(\bk) \cdot \varepsilon_s(\bk) =   \varepsilon_{r \, 
\mu}(\bk)\varepsilon^{\mu}_s(\bk) =
	      \left\{ \begin{array}{ll}
			    \;\;\:  \delta_{rs}  &  \; r=0  \\
				 -\delta_{rs}  & \; r=1,2,3  \end{array} \right. \label{poldel}
\end{equation} 
   More specifically a standard basis in a given reference frame can be taken with:
\begin{eqnarray}
     \varepsilon^{\mu}_0(\bk) & = & (1,0,0,0)   \label{varep0} \\
	 \varepsilon^{\mu}_r(\bk) & = & (0, \bvare_r (\bk))  \qquad r=1,2,3  \\
	 \mbox{with} \quad \bk\cdot\bvare_r(\bk) & = & 0 
	   \qquad \qquad \quad \;\; r = 1,2 \label{varep2}  \\
     \mbox{and} \quad  \bvare_3(\bk) & = & \bk/\vert \bk \vert     \label{varep3}
\end{eqnarray}  
  The $r=0$ case is the scalar, or timelike, polarisation vector, while $r=1,2$ 
represent  transverse polarisation vectors and the case 
 for $r=3$ is called the longitudinal polarisation vector.
  (With respect to a 3D scalar product with metric $\delta_{ij} = 
\mbox{diag}(+1,+1,+1)$ the 3-vector parts of the transverse polarisation vectors 
satisfy $\bvare_r(\bk) \cdot \bvare_s(\bk) = +\delta_{rs}$ for $r=1,2,3$ according to 
equation~\ref{poldel}; while equation~\ref{varep2} for $r=1,2$ is also valid for the 
4D scalar product with
 $k \cdot \varepsilon_{r}(\bk) = k_{\mu}\varepsilon^{\mu}_r(\bk) =  0$).

 The full set of four polarisation states $\varepsilon^{\mu}_r(\bk)$ for $r=0,1,2,3$ 
provides a Lorentz covariant description  for the 4-component vector field 
$A^{\mu}(x)$ and suggests the possibility of four kinds of photon states corresponding 
to these four degrees of freedom.
However, in the standard theory, the requirement of gauge invariance, which allows 
some field excitations to be transformed to zero, together with the massless condition 
$k^2=0$ for the free electromagnetic field, result in there being only \textit{two} 
physical photon states corresponding to the transverse polarisation states 
$\varepsilon^{\mu}_1(\bk)$ and $\varepsilon^{\mu}_2(\bk)$.

  Maxwell's equation in the form of equation~\ref{maxafree} is obtained from 
equation~\ref{maxfein} under the Lorenz gauge condition $\pal_{\mu} A^{\mu} = 0$, as 
described for the inhomogeneous case of equations~\ref{maxj} and \ref{maxaj} in 
section~\ref{subfal}. In turn  equation~\ref{kcoeff} forms a solution of the free 
field case of equation~\ref{maxafree} provided that the massless condition $k^2=0$ 
holds. Further the Lorenz gauge condition itself, applied to equation~\ref{kcoeff}, 
requires that $k_{\mu}\varepsilon^{\mu} =0$ and hence, from 
equations~\ref{varep0}--\ref{varep3}, the transverse polarisation states are clearly 
permitted.

   When substituted into equation~\ref{geinm2}  the gauge field $A^{\mu}(x)$ of 
equation~\ref{kcoeff}, with $A_r(\bk)\neq 0$ for either $r=1$ or $r=2$ only,  yields a 
large number of terms, most of which are zero due to the conditions $k \cdot 
\varepsilon_{1,2}(\bk)  = 0$, for these transverse states, and $k^2=0$ (for example no 
$F_{\rho\sigma}F^{\rho\sigma}$ terms remain) leading to:
\begin{eqnarray}
  T^{\mu\nu} := -\frac{1}{\kappa}G^{\mu\nu} 
     \!  & \!\! = \!\! & \!  + C \, k^{\mu}k^{\nu} \, 
	\Big(2\vert A_r \vert^2 \, 
	   - \, A_r^2 e^{-2ik\cdot x} \,
	    -\, A_r^{\ast 2} e^{+2ik\cdot x}\Big) \nonumber \\
     \!  & \!\! = \!\! & \!   +2 C \, k^{\mu}k^{\nu}  \Big(\vert A_r \vert^2   - 
	  \big(\mbox{Re}(A_r^2) \cos 2 k \!\cdot\! x + 
	   \mbox{Im}(A_r^2) \sin 2 k \!\cdot\! x \big) \Big)  \qquad \quad  \nonumber \\
	 \!  & \!\! = \!\! & \!   +2 C \, k^{\mu}k^{\nu} \,
	 \vert A_r \vert^2  \Big(1+ \cos (2k\!\cdot \!x +\alpha)\Big)
	   \label{geinawave}
\end{eqnarray}
 The form of the final line, with $\alpha \inn \rrr$, follows by a similar argument 
that led to equation~\ref{aonecos}, in terms of a single real cosine function. 
 Taking the 4-vector $k=(k^0, 0, 0, k^0)$, representing the propagation of the 
electromagnetic wave in the direction of the $x^3$ coordinate (with transverse 
polarisation vectors
$\varepsilon^{\mu}_1(\bk) = (0,1,0,0)$ and $\varepsilon^{\mu}_2(\bk) = (0,0,1,0)$ for 
example),
the variable part of $T^{\mu\nu}(x)$ is sketched alongside that for the gauge field 
$A^{\mu}(x)$ in figure~\ref{gacos} as projected onto the spatial coordinate $x^3$ on 
$M_4$.
   	 
\begin{figure}[htbp]  
\centering
\epsfxsize=13.5cm
\leavevmode
\epsffile[0 0 1700 455]{\gpath aPfig111e}
\caption{\setb The energy-momentum tensor $T^{\mu\nu}(x) := 
-\frac{1}{\kappa}G^{\mu\nu}(x)$ is modulated by a non-negative cosine function as 
depicted above, corresponding to an electromagnetic vector field $A^{\mu}(x)$ in the 
form of a plane wave.}
\label{gacos}
\end{figure}

  Although the field function $A^{\mu}(x)$ is presented in the Lorenz gauge, the form 
of the Einstein tensor $G^{\mu\nu(x)}$ is gauge invariant. Indeed this was one of the 
conditions used to derive the relation between the external and internal geometry, as 
described for example in the discussion following equation~\ref{gamsetkk} in 
section~\ref{reaic}, in leading to equation~\ref{geinm2} itself. Hence in turn the 
energy-momentum tensor $T^{\mu\nu}(x)$, of equation~\ref{geinawave} and 
figure~\ref{gacos}, is
  naturally gauge invariant. 

  As for the electric $E_i$ and magnetic $B_i$ field components of the gauge invariant 
electromagnetic field tensor $F$ in equation~\ref{febmat}, the geometry of the 
Einstein tensor $G^{\mu\nu}(x)$ in equation~\ref{geinawave} represents an unambiguous 
physical feature associated with the electromagnetic wave. 
  Further, as discussed shortly after equation~\ref{tffgff} the  scalar curvature $R$ 
associated with any electromagnetic field vanishes, with the Ricci tensor $R^{\mu\nu}$ 
hence identified with the Einstein tensor $G^{\mu\nu}$. Hence equation~\ref{geinawave} 
describes a `wave of Ricci curvature', which is complementary to the usual notion of a 
gravitational wave, with the latter composed of purely Weyl curvature in the Ricci 
vacuum as described after equation~\ref{cbian}, also in section~\ref{subwal}.

  Indeed under the assumption of an approximately flat spacetime, as employed for 
equation~\ref{gbiana} below, and with $R^{\mu\nu}=G^{\mu\nu}$ given by 
equation~\ref{geinawave} it can be seen from equation~\ref{cbian} that 
$K^{\rho\sigma\mu}=0$, that is the source of Weyl curvature vanishes for this 
geometry.
 As implied in the discussion before equation~\ref{rdecom} the Weyl curvature vanishes 
for any conformally flat geometry, and hence a metric of the form: 
\begin{equation}
 \label{gmetwave}
g_{\mu\nu}(x) = 
  (1+\beta \cos 2 k \!\cdot\! x) \eta_{\mu\nu}
\end{equation}  
   for a small value of $\beta \inn \rrr$, provides a candidate solution underlying an 
Einstein tensor in the form of equation~\ref{geinawave}.
 Indeed, assuming the Levi-Civita connection of equation~\ref{gtoGam}, via equations 
\ref{ritencon} and \ref{riccicon}  it can be seen that the scalar curvature $R$ 
vanishes for such a metric if $k^2 = 0$, and that the resulting  $G^{\mu\nu}(x) = 4 
\beta k^{\mu}k^{\nu} 
  \cos 2 k \!\cdot\! x $, to first order in $\beta$, exhibits a corresponding 
oscillatory behaviour, although more work is needed to obtain the precise form for a 
metric underlying the Einstein tensor of equation~\ref{geinawave}.

  The physical spacetime curvature described by $G^{\mu\nu}(x)$ in 
equation~\ref{geinawave} is assumed to be very small. As described in 
section~\ref{subwal}, alongside
 equation~\ref{geinm}, while the Einstein tensor is theoretically directly related to 
the internal gauge field, in the form of equation~\ref{geinawave} for example, in 
units appropriate for laboratory measurements the Einstein equation can be written 
$G^{\mu\nu} = - \kappa T^{\mu\nu}$, where the normalisation constant $\kappa$ is a 
very small number. Hence while the energy-momentum carried by the electromagnetic wave 
may be readily detected the distortion of the spacetime geometry away from Minkowski 
flatness is extremely small and utterly unobservable via any direct means. In turn the 
plane wave description of equations~\ref{acsin}--\ref{kcoeff}, modelled on the flat 
spacetime case, can be used to a very good approximation.
 The divergence of the Einstein tensor 
  in equation~\ref{geinawave} can be expressed in terms of the energy-momentum tensor 
in this approximately flat spacetime limit with $T^{\mu\nu}_{\ph{\mu\nu}\! ;\mu} \to
 T^{\mu\nu}_{\ph{\mu\nu}\! ,\mu}$ in Cartesian coordinates, as described in the 
opening of section~\ref{subwal}.  Consistent with the Bianchi identity $\gmo$ this 
object can be seen to vanish, due to the condition $k^2=0$, as would be expected:
\begin{equation}
    T^{\mu\nu}_{\ph{\mu\nu}\! ,\mu} =  2C \, k^{\mu}k^{\nu} \, 
	 \vert A_r \vert^2  \Big(\!\! - \! 2k_{\mu}\sin (2k\!\cdot \!x +\alpha)\Big)
	    = 0 \label{gbiana}
\end{equation}

  In the context of the present theory while the polarisation requirement 
$k_{\mu}\varepsilon^{\mu} = 0$ can again be seen to be a consequence of imposing the 
Lorenz gauge condition on such plane wave solutions for the field $A^{\mu}(x)$, the 
`momentum' requirement $k^2=0$ is a consequence of the necessity for the free field 
solution to satisfy   
   the geometric Bianchi identity as for equation~\ref{gbiana}. 
   That is as a plane wave the free field is necessarily `massless'
    in order to identify a consistent solution for $G^{\mu\nu}(x)$ in the form of 
equation~\ref{geinm2}, that hence might occur in nature.
  Equivalently the requirement $k^2 = 0$ could be considered to be a consequence of 
the homogeneous Maxwell equation~\ref{maxafree}, which \textit{itself} is a direct 
consequence of the Bianchi identity $\gmo$ given equation~\ref{geinm2}, as described 
for equation~\ref{maxfein}.
 Equation~\ref{gbiana} also of course directly implies energy-momentum conservation, 
$T^{\mu\nu}_{\ph{\mu\nu},\mu} = 0$, for the gauge field described in 
equation~\ref{kcoeff} as employed in equation~\ref{geinawave}.

 Recalling that in general relativity the components $T^{\mu\nu}(x)$ represent the 
energy-momentum \textit{density},
   the field $A^{\mu}(x)$ in a spatial volume $V$ carries  4-momentum $P^{\mu}$  which 
may  be expressed as:
\begin{equation}
   P^{\mu} \, = \, \int_V \, d^3 \bx \, T^{\mu 0}  \label{pmuvtmu}
\end{equation} 
   In the present theory the energy-momentum is always fundamentally determined by the 
Einstein tensor $G^{\mu\nu}(x)$ through the Einstein equation $T^{\mu\nu} := 
G^{\mu\nu}$. As also described in the opening paragraphs of section~\ref{subwal} this 
is in contrast to the Lagrangian approach for which an energy-momentum tensor 
$t^{\mu\nu}$ can be defined giving rise to a conserved 4-momentum, in the form of 
equation~\ref{pmuvtmu}, as described in equation~\ref{tmnneo} and the subsequent 
discussion of  section~\ref{subfal}.

   The components $P^{\mu}$ are locally four conserved  quantities which transform 
amongst each other  covariantly as a 4-vector under Lorentz transformations.  Setting 
$\vert A_r \vert^2 = 1$, with the real coefficient $C^{\fh}$  taking care of the field 
normalisation in equation~\ref{kcoeff}, and substituting the top line of 
equation~\ref{geinawave} into the above expression, taking into account the vanishing 
of the integral of the $e^{\pm i k\cdot x}$ terms for suitably defined 
boundary conditions for the volume $V$, yields:
\begin{equation}
   P^{\mu} \, = \, \int_V \, d^3 \bx \, 2C \, k^0 k^{\mu} \, = \, 2\,V\,C\,k^0k^{\mu}
    \label{kextract}
\end{equation} 
   Hence by setting the coefficient $C= \frac{1}{2Vk^0}$ the  4-vector $k^{\mu}$ in 
the Fourier component can be identified with the 4-momentum $P^{\mu}$ of the field in 
the  volume $V$. Such an object with field values localised to within the volume $V$  
might naively be considered to represent a `particle',  
 although a less simplistic particle concept that emerges in the present theory will 
be described in section~\ref{seraps}.

Hence in turn $C^{\fh} = \frac{1}{\sqrt{2Vk^0}}$ is the normalisation required in 
equation~\ref{kcoeff}, given the transverse polarisation vectors described in 
equations~\ref{poldel}--\ref{varep3}, taking $\vert A_r \vert = 1$ and considering the 
field in the volume $V$ to possess 4-momentum $P^{\mu}\equiv k^{\mu}$.
   The origin of this field normalisation factor $C^{\fh}$ here therefore is in the 
interpretation of the Einstein tensor as the energy-momentum, and in particular with 
$G^{00}(x)$ as the energy density 
 in the local reference frame
 $T^{00} := -\frac{1}{\kappa}G^{00} \sim Ck^0k^0 \, \propto \, \frac{k^0}{V}$ relating 
the $k^0$ component of the Fourier expansion of the $A^{\mu}(x)$ gauge field directly 
to the physical energy $P^0$ carried by the field (that is, the classical Hamiltonian 
$H$).
  This construction is independent of the spatial volume $V$ which, being arbitrary
   within the choice of the boundary conditions, should cancel in all calculations of 
physical quantities when interactions are considered, as it does for cross-section 
calculations in QFT as described in the discussion following equation~\ref{diffcrs}.

  In the present theory the energy of a real field, such as $A^{\mu}(x)$,  is obtained 
directly by substitution of the field into the appropriate expression for the 
right-hand side of $T^{\mu\nu} := -\frac{1}{\kappa}G^{\mu\nu}$. Taking a 
complex-valued expression for the field $A^{\mu}(x)$ in the form of the first term on 
the right-hand side of equation~\ref{kcoeff}, for example, leads to the subsequent 
expression for the energy-momentum tensor:
 \begin{eqnarray}
    A^{\mu}(x) &  = & C^{\fh}\,\varepsilon^{\mu}_r(\bk)\,A_r(\bk)\,e^{-ik\cdot x} 
\label{acomp1}  \\
	  \Rightarrow T^{\mu\nu} & = & - C \, k^{\mu}k^{\nu}\,A_r^2 \,e^{-2ik\cdot x}
	\label{tcomp}  
 \end{eqnarray}
   which, while consistent with $T^{\mu\nu}_{\ph{\mu\nu}\! ,\mu} = 0$, is a 
\textit{complex} tensor and hence does \textit{not} represent a \textit{real} 
energy-momentum tensor $T^{\mu\nu}$, or a \textit{real} geometric tensor $G^{\mu\nu}$.  
In addition here $A^{\mu}(x)$ is required in any case to be real in order to represent 
the real components of a $\uo_Q$ Lie algebra-valued vector field, that is a classical 
macroscopic gauge field.

 Alternatively, the first term on the right-hand side of  equation~\ref{acsin}, for 
example, is real and does, alone, produce a real energy-momentum tensor:
\begin{eqnarray}
   A^{\mu}(x) & = &  C^{\fh}\,\varepsilon^{\mu}_r(\bk)\,A_{c\,r}(\bk) \,
    \cos k\!\cdot\! x 
            \label{areal1} \\
   \Rightarrow T^{\mu\nu} & = & + C \, k^{\mu}k^{\nu}
      \, A_{c\,r}^2 \, \sin^2 k \!\cdot\! x         \label{treala}
\end{eqnarray} 
 as a special case of equation~\ref{geinawave} (with $A_r = \fh A_{c\,r} \inn \rrr$).

     All three expressions for $A^{\mu}(x)$ in equations~\ref{acsin} (or 
\ref{kcoeff}), \ref{acomp1} and \ref{areal1} also necessarily satisfy Maxwell's 
equation $\square A^{\mu} = 0$  since this is implicit in the identity $\gmo$ when
 applied to equation~\ref{geinm2} as was described in equation~\ref{trmaxw} and the 
subsequent discussion as reviewed above.
 In all cases a solution with $A^{\mu}$ dependent upon the Fourier mode 4-vector $k$
 expressed in the Lorenz gauge requires 
a polarisation vector with $k_{\mu}\varepsilon^{\mu} = 0$, and with  $G^{\mu\nu}$ in 
the form of equation~\ref{geinm2} the geometric Bianchi identity  $\gmo$ implies 
$k^2=0$.

  For the standard treatment of a massive vector field with  $k^2= m^2 \neq 0$, as for 
the case of a massive gauge vector boson, the plane wave expansion in the form of any 
of equations~\ref{acsin}--\ref{kcoeff} can again be employed, and the Lorenz gauge 
condition again implies  $k_{\mu}\varepsilon^{\mu} = 0$ for the polarisation vector. 
In this case however the remaining gauge freedom, subject to $\pal_{\mu} A^{\mu}=0$, 
cannot be used to uncover a cancellation between the scalar and longitudinal 
components of polarisation. Hence for massive gauge bosons there are three possible 
states, with the longitudinal degree of freedom appended to the two transverse 
polarisation states.
   In this case Maxwell's equation is replaced by an expression incorporating a mass 
term:  
 \begin{equation}
    (\square + m^2)A^{\mu} = 0   \label{sqmassa}
 \end{equation} 	
 
  In the context of the present theory on substituting the free field in the form of 
equation~\ref{kcoeff} into the expression $G^{\mu\nu} = f(A)$ of equation~\ref{geinm2}
 the Bianchi identity $\gmo$ in the form of equation~\ref{gbiana} is no longer 
satisfied for this new case with $k^2 \neq 0$. This suggests that the direct 
relationship between the Einstein tensor and a gauge field of the form of 
equation~\ref{gchift} and \ref{gety}, as employed for equation~\ref{geinm2},  no 
longer holds, but rather a more general expression is to be sought, as suggested by 
the form of equation~\ref{getypsi} in section~\ref{subwal}. In the present context 
this latter expression $G^{\mu\nu} =  f(A, \hat{\bv})$  implicitly incorporates the 
consequences of \textit{interactions} between the gauge field $A^{\mu}(x)$ and 
components of the temporal flow under the full form $\lvh$. 

  Indeed in subsection~\ref{suboomahp} it has been suggested that in the present 
theory gauge boson masses arise through an impingement of the corresponding internal 
symmetry on the external vector $\bh_2 \equiv \bv_4 \inn \TM_4$ of 
equation~\ref{v4vac}, which forms the components of a `vector-Higgs'. This argument 
was constructed in part by analogy with technicolor models, with longitudinal 
components for massive gauge bosons obtained when the propagators are corrected for 
the field interactions, as described following equation~\ref{lagtech}.
  In the context of the present theory while equation~\ref{geinm2} together with the 
identity $\gmo$ implies equation~\ref{maxafree}, the form of $G^{\mu\nu} =  f(A, 
\hat{\bv})$ under the same identity is expected to be consistent with 
equation~\ref{sqmassa}.

  The $A^{\mu}(x)$ gauge field associated with electromagnetism is in fact massless. 
In the present theory in terms of the corresponding $\uo_Q \subset \esi$ internal 
symmetry this property is attributed to the fact that the $\uo_Q$ 
   generator $\Sbard^1_l$ does not impact upon the external $\bv_4 \inn \TM_4$ 
components of $\bv_{27}$, as presented for example in equation~\ref{zwamass}.
 However there are interactions between $A^{\mu}(x)$ and other temporal components 
which suggest that the free field expansion and equation~\ref{geinm2} will not 
represent the full picture.

  Indeed, in the present theory the gauge field $A^{\mu}(x)$ and the associated 
internal $\uo_Q$ symmetry are not considered as basic entities in themselves, rather 
they are introduced since they act on components within the  form  $\lvh$. These 
latter components include the Dirac spinors $\psi$ of equations~\ref{diraclr} and 
\ref{fhthopart}, as identified in the components of $\bv_{56}$ in the extension to the 
$\ese$ symmetry of the full form $\lvfs$ in section~\ref{secesef}, which unlike the 
$\bv_4$ components do transform non-trivially under the $\uo_Q$ action.
 In principle these temporal components provide a greater freedom for building the 
spacetime geometry, now with an underlying degeneracy of possible $\delta A(x) 
\leftrightarrow \delta \psi(x)$ field `redescriptions', always consistent with the 
Bianchi identity $\gmo$ for the external spacetime.

  Objects transforming as a 4-vector can be constructed out of Dirac spinors in the 
form of 
   $\ol{\psi} \gamma^{\mu}  \psi$ via the conjugate field $\ol{\psi} = 
\psi^{\dag}\gamma^0$
 as introduced for equation~\ref{lagdym}  in section~\ref{subfal}. As for the standard 
theory the relationship between $\psi$ and $\ol{\psi}$ is expected to relate to the 
dynamics of fermions and antifermions for physical particle states propagating in 
spacetime. 
   For the present theory `interactions' between the vector field $A^{\mu}(x)$ and a 
fermion field $\psi(x)$ take the form of
    vector field `redescriptions' as provisionally sketched in figure~\ref{apsirelab}.
\begin{figure}[htbp]  
\centering
\epsfxsize=\maxwidth
\leavevmode
\epsffile[0 0 2161 683]{\gpath aPfig112e}
\caption{\setb Field redescriptions: (a) The \textit{same function} of spacetime is 
associated with the field $\ol{\psi}(x)\gamma^{\mu}\psi(x)$ before time $t_2\in T$ and 
with the field $A^{\mu}(x)$ at later times, here in the spatial volume $V$. (b) The 
field $A^\mu(x)$ in the spacetime volume $VT$ is redescribed as the field 
$\ol{\psi}(x)\gamma^{\mu}\psi(x)$ from time $t_1\in T$.}
\label{apsirelab}
\end{figure}

	While the field function is relabelled $A^{\mu}(x) \to 
\ol{\psi}(x)\gamma^{\mu}\psi(x)$ at time $x^0=t_1$ in figure~\ref{apsirelab}(b), the 
\textit{function} form itself is independent of the choice of $t_1$, as is the local 
geometric structure $G^{\mu\nu}= f(A,\psi)$ in spacetime.  Figure~\ref{apsirelab}(b) 
does not represent the field $A^{\mu}(x)$ `turning into' the field 
$\ol{\psi}(x)\gamma^{\mu}\psi(x)$ at time $t_1$, rather this possible redescription is 
everywhere \textit{implicit} in $A^{\mu}(x)$ as a function on $M_4$.
 For example the  plane wave described by $A^{\mu}(x)$ in figure~\ref{gacos} might be 
redescribed in terms of the field $\ol{\psi}(x)\gamma^{\mu}\psi(x)$ at any time $t_1$ 
with the
 external form of $T^{\mu\nu} := G^{\mu\nu}$ remaining unchanged 
 throughout the spacetime volume $VT$.

    This notion of field indistinguishability is closely analogous in spirit to the 
`arithmetic indistinguishability' of  the multi-dimensional form $\lv$ from  the 
original one-dimensional  temporal flow within which  the form $\lv$ is ever implicit, 
as described in section~\ref{gfotf}. That is one-dimensional time $s$ innately 
contains possible `redescriptions', such as $s^2 = (x^1)^2+(x^2)^2+(x^3)^2$, which may 
potentially be interpreted as geometric or spatial structures. Further, when projected 
onto the base manifold $M_4$ in the full theory the form $\lvh$ will provide 
constraints on possible field redescriptions, represented by $A^{\mu}(x) 
\leftrightarrow \ol{\psi}(x)\gamma^{\mu}\psi(x)$ here, as will be described for 
equation~\ref{dlvpbbz} below for example.

  Although \textit{locally} indistinguishable, the possibility of local field 
redescriptions such as depicted in figures~\ref{apsirelab}(a) and (b) will lead to 
\textit{globally} distinguished and observable phenomena on $M_4$. This includes the 
possible outcomes of a `Schr\"odinger's cat' type  experiment, as will be described in 
section~\ref{qpagig}. This is possible since different field descriptions point 
towards a different set of subsequent possible field redescriptions propagating in the 
broader spacetime environment, always under the constraint $\gmo$. The relative 
\textit{probability} for a specific observable effect will depend directly upon the 
degeneracy of the local underlying possible field descriptions, as we shall explore in 
the following section.

In the full theory the spacetime geometry with metric $g_{\mu\nu}(x)$ and Einstein 
tensor $G^{\mu\nu}(x)$ are continuous and smooth over $M_4$ and have `surveillance' 
over the other fields, as described in section~\ref{subwal} in the discussion shortly 
before equation~\ref{cbian} for example.
 As well as shaping the equations of motion for macroscopic fields and entities this 
surveillance will also constrain the form of microscopic field interactions and 
exchanges.
 While the original form of $T^{\mu\nu} := G^{\mu\nu} = f(A)$ is expected to be 
associated with photon states in some way,
  further redescriptions of a form suggested by the sketches of figure~\ref{apsirelab} 
will ultimately introduce matter terms $T^{\mu\nu}$ primarily associated with the 
spinor $\psi$ field components, to be associated with electron states for example.

	 The precise mathematical form of the field redescriptions 
    remains to be fully understood. However this  structure, as pictured in 
figure~\ref{apsirelab}, brings to mind Huygen's principle for the description of a 
field at a later time as propagated from earlier times and the form of the retarded 
propagator, closely relating to $\Delta_R (x-y)$ of equation~\ref{delrbc} for the 
scalar case. Here however, rather than fields propagating through a 
\textit{pre-existing} spacetime background, the 4-dimensional spacetime $M_4$ with 
geometry $G^{\mu\nu} = f(A,\psi)$ is  \textit{constructed} in terms of the fields. In 
any case we provisionally represent the structures of figure~\ref{apsirelab}(a) and 
(b) respectively by the mathematical relations:
\begin{eqnarray}
   A^{\mu}(x) & = & \int d^4y \, D^{\mu\nu}_r(x-y) \, \ol{\psi}(y)\gamma_{\nu} \psi(y)
     \label{vecredes}   \\
   A^{\mu}(x) & = & \int d^4y \, D^{\mu\nu}_a(x-y) \, \ol{\psi}(y)\gamma_{\nu} \psi(y)
     \label{vecredesa} 
\end{eqnarray}

 The role of the functions $D^{\mu\nu}_r(x-y)$ and  $D^{\mu\nu}_a(x-y)$ is
 hence to provide a more rigorous account of the field exchanges  
  depicted graphically and somewhat naively in figure~\ref{apsirelab}.
 Here, by analogy with the case of standard electrodynamics, 
  the `redescription function' $D^{\mu\nu}_r(x-y)$ is analogous to the retarded 
propagator $D^{\mu\nu}_R(x-y)$ for a vector field. Such a redescription may also take 
place `into the past', that is by analogy with the advanced propagator 
$D^{\mu\nu}_A(x-y)$, equivalent to the field exchange $A^{\mu}(x) \to 
\ol{\psi}(x)\gamma^{\mu} \psi(x)$ forward in time, as described in terms of 
$D^{\mu\nu}_a(x-y)$ in equation~\ref{vecredesa} and depicted in 
figure~\ref{apsirelab}(b).

   Hence we provisionally identify the functions $D^{\mu\nu}_{r,a}(x-y)$ in 
equations~\ref{vecredes} and \ref{vecredesa} with the propagators 
$D^{\mu\nu}_{R,A}(x-y)$.
  The vector retarded propagator for the massless gauge boson case can be defined in 
terms of the corresponding scalar propagator $\Delta_R(x-y)$  as:
\begin{eqnarray}
   D^{\mu\nu}_R(x-y) & = & \lim_{m\to 0} \lbrack -g^{\mu\nu} \Delta_R(x-y) \rbrack 
         \label{drfromdr} \\
   \mbox{with} \qquad \square_x  D^{\mu\nu}_R(x-y) & = &  g^{\mu\nu}\delta^4(x-y)
\end{eqnarray}
 hence following from equation~\ref{kginhom}, and with a similar construction for the 
advanced propagator. In turn
equation~\ref{vecredes} implies:
\begin{equation}
 \square A^{\mu} =  \ol{\psi}\gamma^{\mu} \psi
  \label{maxhere}
\end{equation}
  that is Maxwell's equation $F^{\mu\nu}_{\ph{\mu\nu};\mu} = j^{\nu}$ of 
equation~\ref{maxj}  for the inhomogeneous case with source current $j^{\mu} = 
\ol{\psi}\gamma^{\mu}\psi$.

  This relation, deriving from  equation~\ref{vecredes}, 
 is incompatible with the classical expression of equation~\ref{geinm2}, which led to 
$\square A^{\mu} = 0$. This generalisation from equation~\ref{maxafree} with the 
addition of the source term $j^{\mu}$ on the right-hand side is analogous to the 
extension with a mass term $m^2$ on the left-hand side in equation~\ref{sqmassa}, in 
both cases arising out of interactions between the gauge field $A^{\mu}(x)$ and 
components of the temporal flow under $\lvh$.
  In both cases this involves 
 opening up a more general relation between the spacetime geometry and the internal 
fields, and correspondingly more general properties of matter described by 
the energy-momentum tensor $T^{\mu\nu} := G^{\mu\nu} = f(Y,\bvh)$, as outlined in the 
discussion around equation~\ref{getypsi} in section~\ref{subwal}.
 For the  generalisation to $G^{\mu\nu} = f(A,\psi)$ considered here the evolution of 
the \textit{both} fields, $A^{\mu}(x)$ and $\psi(x)$, will in turn be shaped in 
conformity with the Bianchi identity  $G^{\mu\nu}_{\ph{\mu\nu};\mu} = 0$ (which led to 
the source-free Maxwell equation for the electromagnetic field $A^{\mu}(x)$ alone in 
equation~\ref{maxafree}).

   From equations~\ref{deld4} and \ref{delrbc} the scalar retarded propagator as 
appearing in equation~\ref{drfromdr} can be written as:
\begin{equation}
   \Delta_R(x-y) =  -i \theta(x^0 - y^0)
   \int \frac{d^4k}{(2\pi)^4} \, \varepsilon(k^0)
     \, 2\pi \, \delta(k^2 - m^2) e^{-ik\cdot (x-y)}  
   \label{delrlong}
\end{equation}
  This function contains similar features to those required for $D^{\mu\nu}_r(x-y)$ in 
equation~\ref{vecredes}, including a $\theta$-function for the temporal ordering of 
the field redescription, which takes place at time $t_{2}$ in  
figure~\ref{apsirelab}(a).
  Further, this propagator is employed to obtain field solutions in the form of 
equation~\ref{phirmcx}, which also applies for classical fields as described at the 
end of     section~\ref{subpac}, and which is closely analogous to 
equation~\ref{vecredes} for the field redescription above.

  As used above in deriving equation~\ref{maxhere}
the retarded propagator $\Delta_R(x-y)$ of equation~\ref{delrlong} satisfies 
equation~\ref{kginhom}.
Indeed in QFT the propagators $\Delta_{F,R,A}(x-y)$ may be introduced as inverse 
functions for the operator $(\square_x + m^2)$ in equation~\ref{kginhom}, with 
appropriate boundary conditions,  motivated by the search for solutions to 
differential equations of motion for the fields, such as equation~\ref{phxxonly}. 
These equations of motion are themselves derived via the Euler-Lagrange 
equation~\ref{eula} given an original postulated Lagrangian as the starting point, 
which led for example to equation~\ref{kgphxx} (incorporating equation~\ref{phxxonly}) 
for the scalar model. A very similar situation applies for the QFT employed for the 
Standard Model in particle physics.

  In the standard theory the complete Lagrangian, including the interaction terms, is 
subject to the Euler-Lagrange equation collectively. For example the combined  Maxwell 
and Dirac Lagrangian, given by equation~\ref{lagdym}  for the internal $\uo_Q$ case, 
under variation of the gauge field $A^{\mu}(x)$ and its spacetime derivatives 
$\pal_{\nu}A^{\mu}(x)$ leads, in the Lorenz gauge, directly to:   
\begin{equation}
   \square A^{\mu} = \ol{\psi} \gamma^{\mu} \psi =: j^{\mu}  \label{apsij}
\end{equation}
  as implied in equations~\ref{jglocur} and \ref{ymols}.
  In order to arrive at this expression the variation of \textit{both} the 
$F_{\mu\nu}F^{\mu\nu}$ and $j^{\mu}A_{\mu}$ parts implied in equation~\ref{lagdym} are 
mutually related by appearing in the \textit{same} Lagrangian object under a single 
Euler-Lagrange equation.

   By contrast the form of the redescription propagator $D^{\mu\nu}_r(x-y)$ of 
equation~\ref{vecredes} is \textit{not} motivated on the grounds of finding solutions 
for equations of motion such as equation~\ref{apsij}, but rather in the present theory 
it is conceptually motivated on the grounds of a degeneracy of field solutions under 
the construction of the spacetime geometry $G^{\mu\nu} = f(A,\psi)$ over $M_4$. 
 In fact here there is no similar direct expression 
   with an explicit source term for the microscopic case, as there is in the standard 
theory with equation~\ref{apsij} above.
 In the present theory  simple differential equations such as equation~\ref{maxhere} 
arise as a \textit{consequence} of the possibility of mutual field redescriptions at 
the microscopic level. 
 However, apparent source terms in these expressions might be identified which are 
reminiscent of those seen in the field equations of motion for the Standard Model. A 
generalisation of the gauge-fermion field interactions described in
equations~\ref{vecredes}--\ref{maxhere} for non-Abelian gauge symmetries  for 
comparison with the general case of equations~\ref{jglocur} and \ref{ymols} in 
section~\ref{subfal} could also be considered.

 Here, rather than an interaction Lagrangian or Hamiltonian relating the different 
fields as for QFT, the form of temporal flow $\lvh$ places mutual constraints on field 
values and provides \textit{selection rules} for possible `transitions' linking 
possible initial, intermediate and final  states.
 Here the field interaction terms appear not in a \textit{single} Lagrangian function 
but rather
 through a \textit{range} of constraint equations, which may be provisionally listed 
as: 
\begin{equation}
 \label{conequa}
  \lvh; \qquad \dmoh; \qquad G^{\mu\nu} =f(Y); \qquad \gmo
\end{equation}
 Of these $\lvh$, as a  scalar invariant, is perhaps most closely related to a 
standard Lagrangian, however in being constrained to the fixed scalar value 1 
 further field interactions are implied in the terms of $\dmoh$.
 The third of these constraints is the relation between the external geometry and 
internal degrees of freedom consisting purely of gauge fields, that is 
equation~\ref{gchift}, and relates closely
 to Kaluza-Klein theories as described in section~\ref{reaic}.
 Together with $\gmo$  further geometric structures such as the Bianchi identity 
$\mbox{D}F=0$ for the internal gauge fields constrain the equations of motion.

 Underlying the more general spacetime geometry $G^{\mu\nu} = f(Y, \bvh)$, it is the 
possibility of gauge-fermion 
  field redescriptions such as expressed in equations~\ref{vecredes} and 
\ref{vecredesa} as considered here for the Abelian case, consistent with the selection 
rules of equations~\ref{conequa},
 which leads to the identification of the current $j^{\mu} := \ol{\psi} \gamma^{\mu} 
\psi$ in equation~\ref{maxhere}, which is identical in form to equation~\ref{apsij}.
 In addition to the vector field transitions described above, spinor field 
redescriptions may also be considered with for example:
\begin{equation}
   \psi(x)  =  \int d^4y \, S_r(x-y) \, \Asl(y) \psi(y)
   \label{spinredes} 
\end{equation}
where $\Asl = \gamma^{\mu}A_{\mu}$, by analogy with equation~\ref{vecredes}. As for 
the vector case in figure
\ref{apsirelab} this field redescription is also possible for the reverse temporal 
ordering. The spinor redescription function $S_r(x-y)$ is here closely related to the 
spinor retarded propagator which may be expressed as  $S_R(x-y)  =  (i 
    \pasl_{\! x} + m)\Delta_R(x-y)$ in terms of the scalar propagator of 
equation~\ref{delrlong}, which satisfies the relation
 $(i \pasl_{\! x} - m)S_R(x-y) = \delta^4 (x-y)$ (\cite{Pesk} p.63).

 In a similar way that equation~\ref{vecredes}  led to equation~\ref{maxhere}, that is 
Maxwell's equation with a source term,  here equation~\ref{spinredes} leads to the 
Dirac equation, also with a source term, assuming that the properties of the spinor 
redescription function $S_r(x-y)$ are similar to the propagator $S_R(x-y)$. In this 
case  the action of 
 $(i \pasl_{\! x} - m)$ on both sides of equation~\ref{spinredes} results in:
 \begin{equation}
   \label{diracli}
    (i \pasl - m)\psi = /\!\!\!\! A \psi
 \end{equation}
  This is the Dirac equation that was obtained in section~\ref{subfal} via the Dirac 
Lagrangian in leading to equation~\ref{diracl}, here with the convention for the gauge 
covariant derivative $D_{\mu} = \pal_{\mu} + iA_{\mu}$. 
 
  In the context of the present theory the mass $m$ terms in these equations will also 
be introduced through field interactions. In the case of equation~\ref{sqmassa} for a 
massive gauge field the mass arises from the impact of the gauge symmetry upon the 
components external vector-Higgs $\bv_4 \inn \TM_4$, as recalled in the discussion 
after equation~\ref{sqmassa}, as introduced in the terms of $\dmoh$ of 
equation~\ref{conequa}. Mass terms for fermions on the other hand will be incorporated 
through the constraint of $\lvh$ itself in equation~\ref{conequa}, in the form of 
Yukawa-like couplings between the fermion components and the same vector-Higgs, as 
described for equation~\ref{hexpan2} in subsection~\ref{suboomahp} in the case of the 
form $\lvt$ and for equation~\ref{qxmass} in section~\ref{secesef} in the case of the  
form $\lvfs$.
 Both for gauge bosons and fermions the interaction mass terms will correct the form 
of the
  corresponding Feynman propagators in the quantum theory.
 However, in focussing on the gauge-fermion interactions in the following  we neglect 
the mass terms  and hence equation~\ref{diracli} reduces to simply (within a 
conventional factor of $i$):
 \begin{equation}
   \label{Dirish}
    /\!\!\!\pal \psi = \;\! \Asl \psi 
 \end{equation}

   In the present theory  field redescriptions occur if permitted by the constraint 
equations~\ref{conequa}, which effectively provide interaction selection rules. For 
the case of an electromagnetic gauge field $A^{\mu}(x)$ associated with the internal 
$\uo_Q$ symmetry generated by $\Sbard^1_l \inn L(\esi)$, as described for example in 
equations~\ref{sloncb}--\ref{sbardot} of section~\ref{intsym},
interactions between the gauge and fermion fields can be identified in the expression  
$D_{\mu} L(\bv_{27}) = 0$, here taking the conserved quantity $\lvt$ as the full form 
of temporal flow. This is analogous to the expression for $D_{\mu} L(\bv_{10}) = 0$ in 
equation~\ref{dlvfibte}, as described towards the end of section~\ref{secbkkt}, for 
the $\sootn$ model, while here for the $\esi$ symmetry of the form $\lvt$ of 
equation~\ref{detpmn} the expression $D_{\mu} L(\bv_{27}) = 0$ includes terms of the 
form:
  \begin{eqnarray}
  \label{dlvpbbz}
  D_{\mu} L(\bv_{27}) & \!\! = \!\! & \ldots \; + \;
   p b (\pal_{\mu}\bar{b}\, + \, \dot{s}_f A_{\mu} \bar{b}) \; + \; 
   m \bar{c}(\pal_{\mu}c\, + \, \dot{s}_f A_{\mu} c) \; + \;\ldots = 0 
    \qquad \quad \\
     & \!\! = \!\! & \ldots \; + \;   h\, {\theta^1}^{\dag} D_{\mu}\theta^1   \; + \; 
\ldots = 0
	   \label{dlvpbbzth}
\end{eqnarray}
 In the first line $\dot{s}_f$ carries the $\Sbard^1_l$ charges of the corresponding 
fermion components and
   in the second line the values $p=m=v^0=h$ via equation~\ref{v4vac} and the spinor 
$\theta^1 = \binom{c}{\bar{b}}$ of equation~\ref{xoct3} have been substituted in.
  The spinor $\theta^1$ decomposes into the four Weyl spinors $\theta_{l,i,j,k}$ of 
equation~\ref{thcth234} under the external $\sltc^1$ symmetry, each of which is 
augmented to a Dirac spinor $\psi$ of equation~\ref{diraclr} upon extension to the 
$\ese$ symmetry of $\lvfs$. The Dirac spinor for the `electron' field for example will 
consist of the 4-component object:
\begin{equation}
 \label{psicol}
  \psi = \left( \!\! \begin{array}{c}
                    c_1 + c_8l  \\
					b_1 - b_8l  \\
					C_1 + C_8l  \\
					B_1 - B_8l   \end{array}
   \!\! \right)
\end{equation}
  in the notation of equation~\ref{hthoxy}.
  Having identified $\psi(x)$ in the components of $F(\htho)$ its conjugate 
$\ol{\psi}$ can also be constructed, and both fields expanded in terms of plane waves 
with complex coefficients, as was the case for the electromagnetic wave in 
equation~\ref{kcoeff}.  
 The nature of particle and antiparticle states will ultimately need to be addressed 
in relation to such field expansions, although here we deal directly with the fields 
and their mutual exchanges.

 Hence in generalising from equation~\ref{dlvpbbzth} for the $\ese$ symmetry case the 
expression $D_{\mu} L(\bv_{56}) = 0$ will contain terms incorporating factors of the 
form $\psi^{\dag} D_{\mu} \psi$ involving a juxtaposition of gauge  and fermion 
fields, with the latter identified in the components of $F(\htho)$. In the present 
theory field exchanges in the form of equation~\ref{spinredes}, with the ensuing 
equations of motion such as equation~\ref{Dirish}, are required to be compatible with 
the constraints such as $D_{\mu} L(\bv_{56}) = 0$.

 The precise means of implementing these constraints remains to be well understood, 
although the terms identified are analogous to the form of those found in the Standard 
Model Lagrangian. Further, the mutual redescriptions of the field functions are 
considered to be  discrete, as suggested by the  provisional picture of  
figure~\ref{apsirelab}, which is   
  reminiscent of the actions of the creation and annihilation operators in the 
expansion of quantum fields which appear through an interaction Lagrangian in 
expressions such as 
 equation~\ref{xxyylow} in a quantum field theory.

  Here equations of motion such as equation~\ref{Dirish}, derived from the field 
redescription of equation~\ref{spinredes}, must be filtered through the selection 
rules such as equation~\ref{dlvpbbzth}, deriving from equations~\ref{conequa}, with a 
corresponding range of charges.  This is one factor leading to differences in the 
likelihood of a particular process to occur. Specifically
 the relative factors of $\dot{s}_f$ for different fermion components in 
equation~\ref{dlvpbbz} will  relate to the relative number of ways in which such a 
process may be channelled via equation~\ref{spinredes}, which takes the same form for 
all such processes, as will be described further after figure~\ref{mumudd} in the 
following section. Hence the factors of $\dot{s}_f$, obtained from the components of 
$\Sbard^1_l$ in equations~\ref{sloncb}--\ref{sbardot}, with $\vert \dot{s}_f \vert = 
1$ and $\vert \dot{s}_f \vert = \frac{1}{3}$ provisionally associated with charged 
leptons and $d$-type quarks respectively in section~\ref{intsym} provide a factor of 
three in the relative interaction strength between these fermion states and the 
electromagnetic field, that is with an apparent `fractional charge' of $\frac{1}{3}$ 
for the $d$-quark relative to the unit electron charge.

  For both equation~\ref{dlvpbbz} in the present theory and equation~\ref{lagfint} in 
the model quantum field theory  an \textit{interaction} is mediated since changes is 
one field influence another field through their mutual composition in these 
expressions, with the constraint of $D_{\mu} L(\bvh) = 0$ in the former case and 
through the Euler-Lagrange equation of motion derived form the total Lagrangian in the 
latter case.
In the present theory equations of motion with field interactions are induced through 
consistency with equations~\ref{conequa} rather than directly as Euler-Lagrange 
equations of motion from a Lagrangian with interaction terms.

 Interactions between gauge and fermion fields arise for the Standard Model through 
the Lagrangian approach by requiring the invariance of the total Lagrangian $\lag$ 
under local internal symmetry transformations, such as with the gauge group $\uo_Q$ in 
the case of electromagnetism. This implies an equivalence or indistinguishability 
between for example a photon and an $e^+e^-$ pair, with $A^{\mu} \leftrightarrow 
\ol{\psi} \gamma^{\mu}\psi$, or between an electron and an electron-photon pair, with 
$\psi \leftrightarrow  /\!\!\!\!A \psi$; which implies the possibility of physical 
interactions between the fields. Similarly in the present theory it is the property of 
invariance of the form $\lvh$ with respect to the internal symmetry, dynamically 
expressed over $M_4$ through terms such as those of equation~\ref{dlvpbbz}, that 
allows  interchanges between gauge and fermion field components corresponding to a  
multitude of possible solutions  for the geometric form $G^{\mu\nu} = f(A, \psi)$ in 
4-dimensional spacetime.

In the present theory it is the possibility of such multiple solutions with coupling 
between the $A^{\mu}(x)$ and $\psi(x)$ fields implied in $\dmoh$ terms that  leads to 
the identification of the current $j^{\mu} := \ol{\psi} \gamma^{\mu} \psi$ in 
equation~\ref{maxhere}. The fields $A^{\mu}$ and  $\ol{\psi} \gamma^{\mu} \psi$ 
mutually appear in the field redescriptions of 
equations~\ref{vecredes} and \ref{vecredesa}
 which are also subject to  the selection rules  implied in $\dmoh$  and incorporated 
into a world geometry, with the form of $G^{\mu\nu}(x)$ generalised from 
equation~\ref{geinm2} but always with $\gmo$ as a further constraining identity.

   The further constraint $G^{\mu\nu} = f(Y)$ listed in equations~\ref{conequa}, 
referring to the direct relation between the external and internal geometry as 
expressed in equation~\ref{gchift}, itself will generalise to incorporate gauge-gauge 
field exchanges for the case of a  non-Abelian internal symmetry. 
 That is,  
   for a gauge field $Y^{\mu}(x)$ associated with a non-Abelian internal gauge 
symmetry 
   with:
\begin{eqnarray}  
    -\frac{1}{\kappa} G^{\mu\nu} &  =  & 
   F^{\alpha\mu }_{\ph{\alpha\mu}\rho}F_{\alpha}^{\ph{\alpha}\rho\nu}
	                +\frac{1}{4} g^{\mu\nu}  F^{\alpha}_{\ph{\alpha}\rho 
\sigma}F_{\alpha}^{\ph{\alpha}\rho\sigma}   \label{gfagain} \\
 \mbox{and}  \quad
  F^{\alpha}_{\ph{\alpha}\mu\nu} & = & \partial_{\mu}Y^{\alpha}_{\ph{a}\nu} - 
	         \partial_{\nu}Y^{\alpha}_{\ph{a}\mu} + \cstr Y^{\beta}_{\ph{Y}\mu} 
A^{\gamma}_{\ph{a}\nu}  
\end{eqnarray}  
   with the latter from equation~\ref{falfbcp},
      there will be  possible gauge field redescriptions consistent with the cubic and 
quartic terms of $G^{\mu\nu} = f(Y)$, namely:
\begin{equation}
  \begin{array}{lcr}
      \pal Y \, YY \quad  \mbox{terms} \quad & \Rightarrow & \quad
	    Y \leftrightarrow YY \quad \mbox{exchanges}  \\
	     YYYY \quad  \mbox{terms} \quad & \Rightarrow & \quad
	    YY \leftrightarrow YY  \quad \mbox{exchanges}  \end{array}
		 \label{nonaredes} 	  
\end{equation}
  Mutual gauge field redescriptions channelled through these constraints will  
    augment the form of equation~\ref{gfagain}, similarly as for 
equations~\ref{vecredes} and \ref{vecredesa} and again under the identity $\gmo$, 
allowing for further possible solutions for the extended spacetime geometry.

 In the Standard Model such cubic and quartic gauge field interaction terms for a 
non-Abelian gauge field similarly appear  through terms quadratic 
 in the curvature tensor $F$, in this case via a Lagrangian in the form of 
equation~\ref{lagym}.
 In a quantum field theory for describing particle phenomena, such as for the Standard 
Model, there are certain constraints placed on the form of the Lagrangian. In general 
all possible terms which are allowed by gauge invariance and other symmetries of the 
theory should be included, but there should be no terms involving coupling constants 
with negative dimension $D$, in order to construct a renormalisable theory, as 
described after equation~\ref{xxyynexo} in section~\ref{fraot}. For QCD (quantum 
chromodynamics) in addition to equation~\ref{lagym} the Lagrangian term:
\begin{equation}
  \lag = \frac{\alpha_s}{4\pi} \, \theta \, F_{\alpha}^{\ph{\alpha}\mu\nu}
    \past{F}^{\alpha}_{\ph{\alpha}\mu\nu}
     \qquad \mbox{with} \qquad \past{F}^{\alpha}_{\ph{\alpha}\mu\nu}
	     = \fh\varepsilon_{\mu\nu\rho\sigma}
	   F^{\alpha\,\rho\sigma}
	   \label{stocp}
\end{equation}
 is \textit{also} admitted. Here $\alpha_s = \frac{g_s^2}{4\pi}$ is the strong 
coupling while the index $\alpha$ corresponds to the Lie algebra values and 
$\past{F}^{\alpha}_{\ph{\alpha}\mu\nu}$ is the dual field strength tensor, as 
originally introduced for the electromagnetic field in equation~\ref{hodged}.
  The $\theta$-parameter is sometimes considered as the $19^{\mathrm {th}}$ parameter 
of the Standard Model along with the 18 others (as summarised later in 
table~\ref{SMparams} of section~\ref{secrhp}). However this Lagrangian term implies 
\textit{CP} violation for strong interactions, contradicting empirical observations, 
unless the $\theta$-parameter is unnaturally very small. This is the `strong 
\textit{CP} problem' in the Standard Model, which indicates that the Lagrangian 
approach may contain too many terms, leading to effects not seen in nature. 

   In the present theory gauge field interactions have a different origin. The 
expression for $G^{\mu\nu}$ in terms of the gauge field strength 
$F^{\alpha}_{\ph{\alpha}\mu\nu}$ as implied in equation~\ref{conequa}  in the form of 
\ref{gfagain} can be rewritten in a form similar to equation~\ref{geinms}, with a term 
quadratic in the dual field strength. However, as noted after equation~\ref{tffgff}, 
there is no term of the form in equation~\ref{stocp} and hence the strong \textit{CP} 
problem is potentially sidestepped in this Lagrangian-free theory.

 Regardless of the nature of the underlying gauge or fermion field content,  
 the object $G^{\mu\nu}(x)$, describing the spacetime geometry of $M_4$, is a 
real-valued tensor, while the identity $\gmo$ is a real-valued vector. Similarly the 
constraints $\lvh$ and $\dmlvh$ are a real-valued scalar and real-valued vector 
respectively. The collection of these objects, as listed in equations~\ref{conequa}
 (with $G^{\mu\nu} = f(Y)$ interpreted as a local constraint),
 is analogous to the collection of terms in a single real-valued scalar Lagrangian, 
and in the present theory they will also be interrelated through the full dynamics. 
However,  as for a real-valued Lagrangian, the components of fields underlying these 
objects may be mathematically analysed into complex-valued parts, such as the Fourier 
modes for the electromagnetic field in equations~\ref{aeepm} and \ref{kcoeff}.

  More generally, as described in subsection~\ref{mgjoin} and equation~\ref{aegasth}, 
a real-valued gauge field $Y(x)$ on $M_4$ was originally obtained as the pull-back of 
the Maurer-Cartan 1-form defined on the manifold of an  unbroken symmetry group 
$\hat{G}$.
 Subsequently internal gauge fields deriving from the symmetry breaking over the base 
manifold were considered, as appearing in the final term of equation~\ref{dlvfib}  for
 example. 
 Although only the complete real field $Y(x)$ represents a
 \textit{macroscopic} gauge field (as discussed after equation~\ref{tcomp}),
 the functional form of the gauge field $Y^{\alpha}_{\ph{\alpha}\mu}(x)$ may be 
analysed into complex Fourier components. Similarly, the 56 real components of a 
vector $\bv_{56} \inn F(\htho)$ under $\lvfs$,
 including the various fermion subcomponents $\psi(x) \subset
  \bv_{56}(x)$,
 when expressed as functions over $M_4$ may be 
 analysed into complex Fourier modes.
  Further, in principle such  \textit{complex} $e^{\pm i k\cdot x}$ Fourier mode 
components of the fields  $Y(x)$ and $\psi(x)$, or a hybrid combination, might be 
composed at the \textit{microscopic} level to form \textit{real} expressions for 
objects such as $G^{\mu\nu}(x)$ and $D_{\mu}L(\bv_{56}(x))=0$ over $M_4$.

   While a crucial observation for the present theory is that the spacetime associated 
with any field propagation is \textit{not} flat, as pictured in figure~\ref{gacos} 
with $G^{\mu\nu} = -\kappa T^{\mu\nu}$ for example, here the geometry is assumed to be 
sufficiently close to flat in order to employ such a plane wave expansion in 
essentially Cartesian coordinates, as described before equation~\ref{gbiana}.
 The field redescription functions, featuring in equations~\ref{vecredes}, 
\ref{vecredesa} and \ref{spinredes} for example, are closely related to the retarded 
propagator  $\Delta_R(x-y)$ of equation~\ref{delrlong}. This latter function itself is 
expressed as an integral over 
$e^{-ik\cdot (x-y)}$  Fourier modes, suggesting that in turn the exchanges and 
interactions between the components of fields such as $Y(x)$ and $\psi(x)$ might also 
be most conveniently analysed in terms of $e^{\pm i k\cdot x}$ Fourier modes, as is 
the case for the field expansions in quantum field theory. That is the association of 
the functions 
$D^{\mu\nu}_{r,a}(x-y)$  with the propagators $D^{\mu\nu}_{R,A}(x-y)$
as provisionally suggested after equations~\ref{vecredes} and \ref{vecredesa}, as for 
the association of the function $S_r(x-y)$ in equation~\ref{spinredes} with the 
propagator $S_R(x-y)$, may involve analysis of the corresponding field structures in 
terms of complex-valued components. 
 These structures in the present theory will be linked with the cross-section 
calculations of QFT in the following section.

 In all cases the mutual field exchanges are required to be consistent with the full 
set of
  constraints of equations~\ref{conequa}, with the  geometric condition $\gmo$ in 
4-dimensional spacetime implying 4-momentum conservation through the definition of 
energy-momentum as $T^{\mu\nu} := G^{\mu\nu}$. The underlying one-dimensional form of 
temporal progression  is reflected in the structure of a  \textit{causal} sequence of 
field redescriptions, as expressed by the $\theta$-function in $\Delta_R(x-y)$ of 
equation~\ref{delrlong}, while the $\delta$-function in that expression relates to the 
appropriate matching of  Fourier modes for the general case, for which a finite mass 
$m$ may result from further field interactions.
 Each possible field redescription itself, for individual Fourier modes such as $e^{- 
i k\cdot x}$, may \textit{provisionally} be associated by analogy with QFT with an 
element of a Feynman diagram, namely a vertex diagram of the kind listed
 in `rule 2' of table~\ref{frulesp}, as depicted in the examples of 
figure~\ref{fvhere}.

\begin{figure}[htbp]  
\centering
\epsfxsize=12.5cm
\leavevmode
\epsffile[0 0 1405 355]{\gpath aPfig113e}
\caption{\setb Three Feynman vertex diagrams correlated with the possible field 
exchanges (a) $A \leftrightarrow  \ol{\psi} \psi$, (b) $\psi \leftrightarrow A \psi$ 
and (c)  $Y \leftrightarrow  YY $,
  as associated with equations~\ref{vecredesa}, \ref{spinredes} (strictly with $S_r$ 
replaced by $S_a$ here, since the implied time ordering is from left to right in these 
diagrams) and the cubic terms of equation~\ref{nonaredes} respectively; with a 4-way 
gauge field vertex also possible for the quartic terms of the latter equation.}
\label{fvhere}
\end{figure}

  The field redescription of equation~\ref{spinredes}, associated with 
figure~\ref{fvhere}(b),  
is directly suggested by the form of the terms of $D_{\mu} L(\bv_{27}) = 0$ in 
equations~\ref{dlvpbbz} and \ref{dlvpbbzth} via equation~\ref{Dirish}, although a 
higher-dimensional full form such as $\lvfs$ will be needed for more explicit details.
 More generally the juxtaposition of a gauge field and quadratic fermion field factor 
in  the terms of $D_{\mu} L(\bv_{56}) = 0$ may lead to interactions between this 
combination of fields with various spacetime orientations, while sharing the same 
vertex topology, resulting in the exchange of figure~\ref{fvhere}(a) for example.

 Similarly,
 as well as identifying particular particle states in the components of $F(\htho)$ the 
distinction between particles and antiparticles,  together with their different 
dynamic behaviour, will require a full consideration of the fields under the 
symmetries of extended 4-dimensional spacetime. The provisional correlation of 
 the combination of the fermion field $\psi$ and its conjugate $\ol{\psi}$  with the 
combination of a fermion and antifermion pair, as  discussed after 
equation~\ref{psicol}, will be dependent upon the temporal orientation of the field 
components on the extended  manifold $M_4$.

  The association between terms of the constraints in equation~\ref{conequa} and the 
form on an interaction Lagrangian, as emphasised by the Feynman vertices of 
figure~\ref{fvhere},  raises the question of how calculations for quantities such as 
cross-sections, as measured in HEP experiments, might be determined in the present 
theory and how such calculations might be related to the Feynman rules more generally. 
This will form the topic of the following section.



\section{Determination of Process Probability}
\label{secdopp}

  Here we make a provisional connection between the calculation of physical quantities 
such as cross-sections, as described in the previous chapter, and the notion of a 
degeneracy of field redescriptions underlying the corresponding processes, as 
introduced in the previous section. Since such calculations in quantum field theory 
have achieved great success in comparison with empirical HEP observations a relation 
between the present theory and the mathematical structures and tools of QFT will be 
desirable. 

  First we consider as an example a field sequence $\ol{\psi}\gamma^{\mu}\psi \to 
  \ol{\varphi}\gamma^{\mu} \varphi$, where $\psi,\varphi \subset \bv_{56} \inn 
F(\htho)$ denote fermion components, with the interaction taking place in the spatial 
volume $V$ over a time period $T$ via an intermediate $A^{\mu}(x)$ field state. This 
situation is depicted in figure~\ref{xxphyy} which essentially consists of a 
juxtaposition of figures~\ref{apsirelab}(a) and (b) where the initial and final 
fermion types may differ in general.

\begin{figure}[htbp]  
\centering
\epsfxsize=10cm
\leavevmode
\epsffile[0 0 1054 514]{\gpath aPfig114e}
\caption{\setb The transition from an initial $\ol{\psi}\gamma^{\mu}\psi$ field state 
to a final $\ol{\varphi}\gamma^{\mu} \varphi$ state via an intermediate description of 
the field function in terms of a mathematically equivalent $A^{\mu}(x)$ field state.}
\label{xxphyy}
\end{figure}

   In this section we consider interactions at the level of such field exchanges.
    As alluded to at the end of the previous section the structure of physical 
particle states in spacetime, including both particle and antiparticle states, is yet 
to be identified in this theory. Further, the inclusion of the second and third 
generation fermions
 may require a further extension of the full form $\lvh$, as suggested for example in 
section~\ref{sosmfi}.
 However a field state such as $\ol{\psi}\gamma^{\mu}\psi$  is provisionally 
considered to represent fermion pairs such as $e^+e^-$ or $\mu^+ \mu^-$ leptons
or $d\bar{d}$ or $t\bar{t}$ quarks for example. 
 Hence the  field sequence in figure~\ref{xxphyy} mimics a HEP collision process such 
as $e^+e^- \to \mu^+\mu^-$. In the following section the physical nature of the actual 
incoming and outgoing particle states observed in HEP phenomena will be considered.

    In the analogous QFT calculation the initial and final `particle' states are 
represented by complex plane waves, that is Fourier modes of the form $e^{\pm ik\cdot 
x}$, 
 as discussed for equations~\ref{xbcont} and \ref{dycont} for example.
 This is similar to the picture initially considered here in figure~\ref{xxphyy}
  with the field functions in spacetime expanded in terms of Fourier modes such as 
those of equation~\ref{kcoeff}. However,
   in the present theory the incoming, interacting and outgoing field states conform 
everywhere to an expression of the spacetime geometry described by the real tensor 
$G^{\mu\nu}(x)= f(A, \psi, \varphi)$.

    For the case of $d$ discrete intervals of time $\Delta t_i$ during which the field 
exchanges between $t=0$ and $t=T$ in figure~\ref{xxphyy} may occur the total number of 
ways $N$ in which the overall transition may proceed is simply:
\begin{equation}
  N \; = \;  \sum_{i=1}^{d} \left( R(t_1 \inn \Delta t_i: \, A
   \to \ol{\varphi}\varphi)  \sum_{j=1}^{i-1} R(t_2 \inn \Delta t_j: \,
	     \ol{\psi}\psi \to A) \right)
		  \label{desnow}
\end{equation}
  with $R$ denoting `redescription' such that
     $R(\Delta t_i: \, A  \to \ol{\varphi}\varphi) \equiv 1$ simply
    expresses the fact that the corresponding field exchange
	 $A^{\mu} \leftrightarrow \ol{\varphi} \gamma^{\mu}\varphi$
	 is allowed during the time interval $\Delta t_i$. More generally $R(t)$ will take 
the value $1$ if the corresponding field exchange is allowed, according to the 
constraint equations~\ref{conequa} as described in the previous section, and $0$ if it 
is not.

   For the process with incoming field state $\ol{\psi}\gamma^{\mu}\psi$ the total 
field function is already distributed everywhere in $V$ from time $t=0$ in 
figure~\ref{xxphyy}, and as a function in spacetime it is indistinguishable from that 
of the outgoing $\ol{\varphi}\gamma^{\mu}\varphi$ field state at $t=T$. The field 
redescription applies \textit{everywhere} in $V$ simultaneously at any time such as 
$t_2$ or $t_1$ since this simply involves a reinterpretation of the \textit{same} 
field function, with nothing physically changing in $VT$. Hence from the point of view 
of the spacetime geometry and $G^{\mu\nu}(x)$ alone it would be possible to link the 
states $\ol{\psi}\gamma^{\mu}\psi$ and $\ol{\varphi}\gamma^{\mu}\varphi$ directly, 
without an intermediate $A^{\mu}(x)$ field description. This is prevented in the 
present theory by the absence of selection rule being provided by constraints such as 
$\dmlvh$ which determine whether $R(t) = 1$ or $R(t) = 0$ 
for a particular field redescription.

  This is closely analogous to the Lagrangian approach in QFT as described for example 
for the scalar model where the absence of a  coupling term of the form 
 $\hat{\mcX}^{\dag}\hat{\mcX} \hat{\mcY}^{\dag}\hat{\mcY}$  in the interaction 
Lagrangian of equation~\ref{lagfint}, and hence
 via equation~\ref{htlagx} in $H_{\mathrm{int}}(t)$,
  implies that the collision process
$\mcX^+\mcX^- \to \mcY^+\mcY^-$ requires an intermediate $\phi$ state as depicted in 
the Feynman diagram of figure~\ref{fdxxyy}.
  Similarly the lack of a direct electron-muon coupling in the Standard Model 
Lagrangian leads to consideration of scattering processes via an intermediate photon, 
such as depicted in figure~\ref{fdeemm}, which will be seen to be analogous to 
figure~\ref{xxphyy} for the present theory.

 Taking equation~\ref{desnow} to the continuum limit, as implied in  
figure~\ref{xxphyy}, a measure of the total degeneracy $D$ can then be expressed as:
\begin{equation}
  D(T,0) \; =  \; \int_0^T dt_1 \int_0^{t_1} dt_2 \; R(t_1) R(t_2)
  \label{degnow}
\end{equation}
   The structure of this equation has some similarity to the second-order term in the 
expansion of the time evolution operator $U(t,t_0)$ in quantum field theory. In the 
interaction picture, with interaction Hamiltonian $H_{\mathrm{int}}$, the operator
 $U$ satisfies the differential equation~\ref{uevolve}, as described in 
section~\ref{tranamp},  with the iterative solution for $U(t,t_0)$ displayed in 
equation~\ref{uiter0}.

In equation~\ref{uiter0} the factors of the Hamiltonian operator $H_{\mathrm{int}}$ in 
each term naturally stand in time order, with the earliest to the right and latest to 
the left, in virtue of the time integration limits. As explained in 
section~\ref{tranamp} this expansion of the time evolution operator $U(t,t_0)$ can be 
written in the familiar more compact form of equations~\ref{utexp}, via 
equations~\ref{uiter1} and \ref{ufulltt}, using the $T$-product of operators which 
\textit{imposes} time ordering over a broadened, and more symmetric, range of time 
integrals.  In particular the second-order term in equation~\ref{uiter0} can be 
replaced by that in equation~\ref{uiter1} since:
 \begin{equation}
    \int_{t_0}^t dt_1 \int_{t_0}^{t_1} dt_2 \, 
	        H_{\mathrm{int}}(t_1) \, H_{\mathrm{int}}(t_2)
 \; \equiv \; \frac{1}{2}  \int_{t_0}^t dt_1 \int_{t_0}^{t} dt_2 \, 
    T\big(H_{\mathrm{int}}(t_1)  H_{\mathrm{int}}(t_2)\big)
   \label{useco}
 \end{equation}

   It is the similarity between equation~\ref{degnow}, as a measure of the degeneracy 
or number of ways in which to describe the field transition sequence 
$\ol{\psi}\gamma^{\mu}\psi \to  A^{\mu} \to
  \ol{\varphi}\gamma^{\mu} \varphi$, and the left-hand side of equation~\ref{useco} 
that provides a further preliminary entry point for the present theory into the 
workings of QFT. In a similar way that the field exchange of figure~\ref{apsirelab}(b) 
has been provisionally associated with the Feynman vertex diagram of 
figure~\ref{fvhere}(a), the $A^{\mu}(x)$  internal field stage of figure~\ref{xxphyy} 
might be associated with the Feynman propagator corresponding to the internal line of 
the Feynman diagram in figure~\ref{fdeemm} for example, via the relation between 
equations~\ref{degnow} and \ref{useco} described above.   

 Equations~\ref{uiter0} and \ref{uiter1} are matched on a term by term basis and hence 
the terms of the perturbative expansion of equation~\ref{utexp} match those of 
equation~\ref{uiter0}. In turn the higher-order terms of equation~\ref{uiter0} can be 
associated with higher-order sequences of field redescriptions, such as depicted in 
figure~\ref{xxpxhyy} below. In the Feynman rules for the mathematical elements 
associated with a Feynman diagram at order $n$ in perturbation theory the factor of 
$1/n!$ in equation~\ref{ufulltt} cancels against a factor of $n!$ from the possible 
vertex permutations, as described shortly after equation~\ref{sfixmom} and summarised 
for `rule 6' in the opening of section~\ref{fraot}. Hence in the Feynman rules for the 
second-order term correlated with the right-hand side of equation~\ref{useco} the 
factor of $\fh$ does not appear.

  Via the above associations
   the field exchange sequence described in figure~\ref{xxphyy} is analogous to the 
Feynman diagram in figure~\ref{fdxxyy} for the corresponding scalar model QFT 
calculation. 
While a possible physical interpretation of the Feynman propagator $\Delta_F(x-y)$ in 
terms of `virtual particles' is conceptually dubious, as discussed in 
section~\ref{subpac} (for example after equation~\ref{delfmom}),  
 this object is a key part of calculations in QFT and we return to this propagator -- 
which in the scalar field case may be expressed for the internal field operator 
$\hat{\phi}(x)$ in canonical QFT by the equation: 
 \begin{equation}
  i\Delta_F(x-y) \, = \, \langle 0 \vert \, T(\hat{\phi}(x)\hat{\phi}(y))\, \vert 0 
\rangle
  \label{delftpp}
 \end{equation}
 as we began with equation~\ref{delf1} in section~\ref{subpac}. This object arose when 
the transition amplitude calculation was restructured with the time evolution operator 
$U(\infty,-\infty)$ in the form of equation~\ref{uiter1} placed between vacuum states,
 in particular for the second-order term.
 This object hence consists of terms implicitly containing time-ordered field 
products, such as  $T(\hat{\phi}(x)\hat{\phi}(y))$ in the right-hand side of 
equation~\ref{useco}.

   The time ordering implies that $\Delta_F(x-y)$ consists of two parts, associated 
with $\theta(x^0- y^0)$ and $\theta(y^0- x^0)$, as described in 
equations~\ref{delf2}--\ref{delfpp} and as represented by the two diagrams in 
figure~\ref{delf2way}. From the point of view of the concept of field redescriptions 
in the present theory the first diagram, figure~\ref{delf2way}(a), can be physically 
motivated as representing the field redescription causal sequence such as 
$\ol{\psi}\gamma^{\mu}\psi \to A^{\mu} \to
 \ol{\varphi}\gamma^{\mu}\varphi$  as depicted in figure~\ref{xxphyy} while the second 
diagram, figure~\ref{delf2way}(b), represents a figment of the \textit{mathematical 
restructuring} of the calculation, leading in turn to the notion of intermediate 
`virtual particle' states.

 Nevertheless, via the above chain of argument each case of an intermediate 
$A^{\mu}(x)$ field state, as depicted in figure~\ref{xxphyy}, may be 
\textit{provisionally} associated with a corresponding Feynman propagator 
$D^{\mu\nu}_F(x-y)$. That is, intermediate field redescriptions such as that in 
figure~\ref{xxphyy}
 may be associated with the `virtual particle' states as represented by the internal 
line in figure~\ref{fdxxyy}, and in Feynman diagrams in general, considered as a 
restructuring of a calculation which is here fundamentally based on an underlying 
conceptual notion of a degeneracy of field descriptions. 

   Associating a Feynman propagator  with each intermediate causal redescription, such 
as that with the field $A^{\mu}(x)$ in figure~\ref{xxphyy} as described above, 
supplements the set of interaction vertices associated with the constraints of 
equations~\ref{conequa}, as exemplified in figure~\ref{fvhere}.
 Hence with propagators identified in addition to the vertices these objects may be 
combined to form Feynman diagrams more generally.
  Beginning from the idea that the probability of an observable process is a measure 
of the number of ways in which it can occur, summing over all possible intermediate 
field redescriptions as for example in equation~\ref{degnow}, the aim is to 
effectively reproduce a full set of Feynman rules, for comparison with the Standard 
Model version of table~\ref{frulesp}, and further to use this relation in order to 
make calculations of empirical quantities such as cross-sections.

  Regarding the Feynman vertices  the key to understanding how   $\dmo$ terms, for 
example, are to be used in place of a Lagrangian here may be found in the coupling 
strength, which is put in by hand in the Lagrangian case. In equation~\ref{dlvpbbz} 
the value of $\dot{s}_f$  for the leptonic states is 3 times larger than for the quark 
states, as determined in section~\ref{intsym} and noted in the previous section.
 The question then is how this mathematical factor of 3 corresponds to an empirical 
factor of 3 in `electric charge' with an underlying explanation in terms of the 
degeneracy for the number of ways a process can occur. Consider the processes 
described by the Feynman diagrams in figures~\ref{mumudd}(a) and (b), either of which 
may be correlated with, while not literally representing,
 the field sequence depicted in figure~\ref{xxphyy} as described above.

\begin{figure}[htbp]  
\centering
\epsfxsize=13.5cm
\leavevmode
\epsffile[0 0 1495 565]{\gpath aPfig115e}
\caption{\setb Feynman diagrams for the electromagnetic processes (a) $e^+e^- \to 
\mu^+\mu^-$ and (b) $e^+e^- \to d\bar{d}$, together with a `higher-order correction' 
via (c) a radiated photon and (d) a gluon exchange between the final state quarks 
respectively.}
\label{mumudd}
\end{figure}

  In the calculation of the degeneracy for a process, as initially described for 
figure~\ref{xxphyy}, the number of possibilities depends upon the total time $T$ 
available for the process, as can be seen in equation~\ref{degnow}.
  For a quantum field theory, the spacetime volume factor $VT$ for an interaction 
cancels in cross-section and decay rate calculations, as described in 
section~\ref{crosss} following equation~\ref{diffcrs}, essentially since the effective 
values of $V$ and $T$ in external spacetime are the same for all possible processes. A 
similar cancellation might be expected for calculations based on field degeneracies in 
the present theory.  
 On the other hand, unlike the case for the common \textit{external} dimensions of the 
interaction, here for the present theory, the effective `charge volume' $C$ in the 
\textit{internal} space dimension varies from process to process, as indicated by the 
differing values of $\dot{s}_f$ in equation~\ref{dlvpbbz}, and does not cancel  in 
such calculations. 
 In QFT these three spaces are closely related, as seen for example in the {\it CPT} 
theorem, while in the present theory they are mutually related through the structure 
and symmetries of the underlying temporal flow in the form $\lvh$. A more precise 
expression for the way in which the relative charges channel the relative likelihood 
for different field exchanges,  and indeed a fuller understanding of the relation of 
the present theory to the Lagrangian approach in general, requires further study, as 
was also discussed after equation~\ref{psicol}.

   Given the `virtual photon' mediating both processes in figures~\ref{mumudd}(a) and 
(b) further \textit{internal} degeneracy, as for example in figure~\ref{xxpxhyy} 
below,  will be essentially the same for both cases and not effect the relative rates. 
That is the branching fractions or relative cross-sections for competing processes 
will depend on the \textit{differences} in the number of ways, and this may be 
dominated by the factors of $\vert\dot{s}_f\vert = 1$ or $\vert\dot{s}_f\vert = 
\frac{1}{3}$ associated with the final state vertex in figures~\ref{mumudd}(a) and (b) 
respectively. Differences may also arise due to the mass of the final state particles
(upon which the final state phase space depends), relating to further possible field 
interactions with the components of the  vector-Higgs $\bh_2 \equiv \bv_4 \inn \TM_4$,  
and more generally due to higher-order field exchange possibilities, such as those 
represented in figures~\ref{mumudd}(c) and (d); as will be further discussed in the 
following section.

  In the full theory the possible Feynman diagrams will generalise corresponding to 
the range of gauge fields and further interactions identified for a full set of  
internal symmetries
 as studied in chapters~\ref{chapesb} and \ref{secfd}, and which show a significant 
resemblance to the structures of the Standard Model of particle physics.
  For example,
  in figure~\ref{mumudd}(d) an SU(3)$_c$ gauge field exchange is included. There are 
eight internal SU(3)$_c$ generators, as described in section~\ref{intsym} and listed 
down the left-hand side of table~\ref{suttan}. Unlike the $\uo_Q$ action in 
equation~\ref{dlvpbbz} these mix the components of $\theta^1 = \binom{c}{\bar{b}}\inn 
\ooo^2$ between different Weyl spinors hence introducing interactions between the 
corresponding quark states. The identification of an $\sutw_L \subset \ese$ (or within 
a larger symmetry of time such as $\ee$), also mediating between the external 
$\sltc^1$ Weyl spinors in $F(\htho)$ (or within a  higher-dimensional form of time 
such as $\lvtfe$) will provide a further internal gauge symmetry action central to an 
understanding of electroweak theory within the present theory.

 The measure of degeneracy in equation~\ref{degnow} can be generalised to higher-order 
sequences of $A^{\mu},\psi,\varphi$ field exchanges which mirror the general expansion 
to higher-order perturbations for QFT in equation~\ref{uiter0}; with the Hamiltonian 
operator $H_{\mathrm{int}}(t)$ in the latter case replaced by the `redescription 
parameter' $R(t)$, as determined by the constraints of equations~\ref{conequa}, in the 
former case.
The temporal sequence of figure~\ref{xxpxhyy}  provides an example of the ways in 
which the causal sequence of figure~\ref{xxphyy} may be generalised for nested 
sequences of field indistinguishability to arbitrary high order.

\begin{figure}[htbp]  
\centering
\epsfxsize=13.1cm
\leavevmode
\epsffile[0 0 1627 505]{\gpath aPfig116e}
\caption{\setb The transition from an initial $\ol{\psi}\gamma^{\mu}\psi$ state to a 
final $\ol{\varphi}\gamma^{\mu}\varphi$ state via an intermediate description of the 
field function in terms of a sequence mathematically equivalent $A^{\mu}(x) \to  
\ol{\psi}(x)\gamma^{\mu}\psi(x) \to A^{\mu}(x)$ field states.}
\label{xxpxhyy}
\end{figure}
 
  The corresponding degeneracy for the chain of field interpretations in 
figure~\ref{xxpxhyy}, as an augmentation of equation~\ref{degnow}, is expressed as :
\begin{eqnarray}
  D(T,0) & \! = \! & \int_0^T dt_1 \int_0^{t_1} dt_2 \int_0^{t_2} dt_3 \int_0^{t_3} 
dt_4
           \; R(t_1: \, A \to \ol{\varphi}\varphi)
		      R(t_2:\,\ol{\psi}\psi \to A) \qquad  \nonumber \\
  & & \qquad \qquad \qquad  \qquad \qquad \quad \;\;
         R(t_3:\,A\to \ol{\psi}\psi) R(t_4:\,\ol{\psi}\psi\to A) 
  \label{degnow4}
\end{eqnarray}
 While the sequence of field descriptions pictured in figure~\ref{xxphyy} can be 
correlated with the Feynman diagram of figure~\ref{fdxxyy}, via equations~\ref{degnow} 
and \ref{useco},
 the higher-order process of figure~\ref{xxpxhyy} is similarly analogous to the form 
of figure~\ref{fdxxyy4}, representing the $T$-ordered expression for this fourth-order 
term for the scalar QFT model. A similar correspondence may be identified between 
field sequences for the present theory and Feynman diagrams in QED, as  depicted in 
figure~\ref{eenestmm} in the following section for example. These figures represent 
steps in the direction of connecting the structures of the present theory with Feynman 
diagrams and rules more generally.

   Intuitively the extra sums over the two additional intermediate times, 
   labelled $t_3$ and $t_2$ in figure~\ref{xxpxhyy}
    and equation~\ref{degnow4}, will lead to a relative `infinity' of new ways in 
which the overall event may proceed from the initial to the final state. However the 
degeneracy \textit{measure} $D$ for both equations~\ref{degnow} and \ref{degnow4} is 
actually finite. On the other hand the intermediate state composed of $\psi$ and 
$\ol{\psi}$ between $t_3$ and $t_2$
 involves two field contributions   \textit{simultaneously}, each of which may be 
expanded into Fourier modes independently with a
 combined product of the form $ \sim e^{-i(p_1+p_2)\cdot x}$ which, although the total 
$p_1+p_2$ is constrained, leads to a further infinity in the degeneracy of the 
internal 4-momentum.
 In this case the integral sum over $p_1$ is unlimited, unlike the situation for the 
time integrals, and  is expected to be reflected in the divergent momentum loop 
integrals, as for example in equation~\ref{xxyynexo} for the scalar model, in the 
correspondence with QFT calculations. For the present theory, as for QFT, such 
divergences might be expected to cancel when observable quantities such as branching 
\textit{ratios} are appropriately normalised, as will be described in the following 
section, with such observables ultimately dominated by the charges involved in the 
final interaction of the sequence as discussed above.

   In the present theory the world geometry is necessarily described by the 
\textit{real} tensor $G^{\mu\nu}(x)$ which itself in principle may be composed out of 
the \textit{real} or \textit{complex} components of fields, such as $A^{\mu}(x)$ and 
$\psi(x)$.
  Regarding the degeneracy count itself it is an open question concerning whether 
there is a unique or optimal way in which possible field redescriptions should be 
counted, consistent with the constraint equations~\ref{conequa}. This question 
concerns both the domain of the field functions, as a patchwork of regions in 
spacetime or in momentum space for example, and also the form of the field functions.
 Here we are analysing the degeneracy count in terms of complex Fourier modes on the 
base manifold $M_4$. In this sense each $e^{\pm ik\cdot x}$ component is not 
considered as an independent \textit{physical} field, rather this decomposition 
provides a \textit{mathematical} means of identifying a set of mutually independent 
field solutions which may be summed over.

   In describing the transitions between fields such as $\psi(x)$, $A^{\mu}(x)$ and 
$\varphi(x)$ it is possible that linear combinations of \textit{real} sine and cosine 
expansion terms, rather than \textit{complex} $e^{\pm ik\cdot x}$ parts, might be 
employed to preserve the identity of \textit{real}, and hence physical, fields   under 
the spacetime geometry $G^{\mu\nu}(x)$ subject to the constraint 
equations~\ref{conequa} everywhere. For example
 considering the \textit{real} Fourier components $A^{\mu}_c(\bk)$ and 
$A^{\mu}_s(\bk)$  of the field in equation~\ref{acsin} to be exchanged independently 
maintains a real condition for the field $A^{\mu}(x)$ which at every stage may compose 
an intermediate, but physical, gauge field coupled to the fermion fields consistent 
with $\dmoh$.

  However here we have described field interactions such as $A^{\mu} \leftrightarrow  
\ol{\psi}\gamma^{\mu}\psi $ in terms of the indistinguishability of complex Fourier 
modes of the fields, as expanded for example in equations~\ref{aeepm} and \ref{kcoeff} 
for the electromagnetic field, in part since this provides a closer link with the 
framework of QFT. 
 Indeed, as alluded to towards the end of the previous section, many of the tools 
involved in QFT, such as the various propagators and the $\delta$-function of 
equation~\ref{delfrep} and the $\theta$-function of equation~\ref{thetrep}, are 
conveniently expressed in terms of complex Fourier modes.
 Further, complex components of gauge fields have already been considered with regard 
to the charged gauge bosons $\tilde{W}^{(2)\pm}_{\mu}(x)$ of equation~\ref{wpmcomb} in 
section~\ref{ewtfesb}, by analogy with the standard electroweak gauge fields 
$W^{\pm}_{\mu}(x)$ of equation~\ref{wpmww} in section~\ref{ewtatsm}, which relate to 
the corresponding physical interactions with Lorentz spinors. Hence here the field 
redescriptions will be analysed  in terms \textit{complex} Fourier modes  in the 
determination of a \textit{real} measure or count of the degeneracy of field 
solutions.

  As described in the previous section both parts of equation~\ref{kcoeff} are 
required to identify a field state carrying \textit{real} energy-momentum, which in 
the present theory is determined by the form of the field under $T^{\mu\nu} := 
-\frac{1}{\kappa}G^{\mu\nu}$.
   Hence transitions between fields must necessarily link \textit{both} of the $e^{\pm 
ik\cdot x}$ parts
with the \textit{external} 4-momentum $k$, as identified through 
equation~\ref{kextract}, matched under an everywhere real $T^{\mu\nu} := 
-\frac{1}{\kappa}G^{\mu\nu}$ energy-momentum tensor, subject to the identity $\gmo$, 
and also with the \textit{internal} representations of the field components matching 
under the 
 constraints of equation~\ref{conequa} in general, with the 
form $\lvh$   broken over the base manifold. 

   With $A^{\mu}(\bk) = \fh(A^{\mu}_c(\bk)+ iA^{\mu}_s(\bk)) \inn \ccc$ a general 
complex number in equation~\ref{aeepm} transitions in the field $A^{\mu}(x)$ can be 
considered to take place treating $A^{\mu}(\bk)$ and ${A^{\mu}}^{\ast}(\bk)$ as 
independent 
degrees of freedom in terms of possible exchanges with complex Fourier modes of the 
fermion fields.
 This implies the possibility of  intermediate \textit{complex} fields such as 
$A^{\mu}(x)$ and   $\psi(x)$
 while  hybrid combinations of these gauge and fermion fields mutually form under 
\textit{real} objects such as $G^{\mu\nu}(x)$ and $\dmoh$.

Hence the temporal sequence of redescriptions should be considered independently for 
the complex $e^{-ik\cdot x}$ and $e^{+ik\cdot x}$ parts such that, for example, the 
processes represented in figures~\ref{xxphyy} and \ref{xxpxhyy} may be generalised for 
this independence, as depicted for example in figure~\ref{xxphyy2}.

\begin{figure}[htbp]  
\centering
\epsfxsize=12cm
\leavevmode
\epsffile[0 0 1436 647]{\gpath aPfig117e}
\caption{\setb The transition from an initial $\ol{\psi}\gamma^{\mu}\psi$ state to a 
final $\ol{\varphi}\gamma^{\mu}\varphi$ state generalised for an intermediate 
description of the field function in terms of the complex Fourier modes $e^{-ik\cdot 
x}$ and $e^{+ik\cdot x}$ independently in time.}
\label{xxphyy2}
\end{figure}

 With the need to account for both sets of possible sequences as exemplified in 
figure~\ref{xxphyy2} the probability $P$ for the overall process $\ol{\psi}\psi \to 
\ol{\varphi}\varphi$ is proportional to $D_+ \times D_-$, where $D_+$ represents the 
degeneracy of ways via $e^{+ik\cdot x}$ exchanges and $D_-$ that for the 
$e^{-ik\cdot x}$ mode exchanges, each of which has a structure similar to that in 
equation~\ref{degnow} or \ref{degnow4}.
  Alternatively the process probability could be expressed in terms of the 
degeneracies $D_{c}$ and $D_{s}$  representing the number of field exchanges relating 
to the cosine and sine Fourier modes
 as alluded to above,
 with for example $A^{\mu}_c(\bk)$ and $A^{\mu}_s(\bk)$ of equation~\ref{acsin} 
independent, and    with $P \propto D_+D_- \equiv D_cD_s$. In this case all fields are 
real-valued and hence can be interpreted as physical entities at all times, however 
here we pursue the equivalent calculation based on the complex decomposition.

  Earlier in this section we have described a correlation between the form of a 
degeneracy count $D(T,0)$ and the anatomy of a Feynman diagram, with for example 
figure~\ref{xxphyy} compared with figure~\ref{mumudd}(a) or (b), via the structure of 
the expansion of the QFT operator $U(t,t_0)$ of equations~\ref{uiter0}--\ref{utexp}. 
Here the \textit{underlying physical} basis of  probability calculations is found in 
the field degeneracies, with the use of $T$-ordered products in QFT, via the 
$\theta$-functions, simply implementing a restructuring of the calculations.
 Hence in turn  the representation of the Feynman propagator in figure~\ref{delf2way} 
should \textit{not} be interpreted as two possible physical processes. 
  On the other hand the fact that the underlying field redescriptions are free to take 
place independently for \textit{both} the $e^{-ik\cdot x}$ and $e^{+ik\cdot x}$ field 
components, as depicted for example for the process $\ol{\psi}\gamma^{\mu}\psi \to 
\ol{\varphi}\gamma^{\mu}\varphi$ in figure~\ref{xxphyy2}, does extend the range of 
possible field redescriptions and hence will have physically observable consequences.
 With \textit{both} sets of field redescriptions for the $e^{\pm i k \cdot x}$ Fourier 
modes  required to link the initial and final states the process probability takes the 
form $P \propto D_+D_-$, and we hence now wish to determine a correlate for this 
product in QFT.

   In figure~\ref{xxphyy2} the field states at $t=0$ and $t=T$ (and hence also for 
$t\to \pm \infty$) represent real external particle states, that is on-mass-shell 
particles.
 This suggests that the diagram in figure~\ref{xxphyy2} can be `unfolded' to represent 
an extension of a \textit{linear} degeneracy count, having the same basic structure as 
figure~\ref{xxphyy} or \ref{xxpxhyy}, but with a `fold line' denoting on-shell states. 
The corresponding unfolded diagram is depicted in figure~\ref{eemmeeee}(a). This field 
sequence 
correlates with the structure of the Feynman diagram of figure~\ref{eemmeeee}(b), with
 the fold line mapped to the cut line -- for which the propagators are simultaneously 
placed on-mass-shell, as originally described for figure~\ref{fdxxxx4}. 

\begin{figure}[htbp]  
\centering
\epsfxsize=13.5cm
\leavevmode
\epsffile[0 0 1945 962]{\gpath aPfig118e}
\caption{\setb (a) The unfolding of figure~\ref{xxphyy2}, with a corresponding `fold 
line' and reparametrised time intervals. (b) 
 A correlated Feynman diagram for the forward scattering process $e^+e^- \to e^+e^-$, 
with a `cut line' drawn through the intermediate loop propagators of the muon field.}
\label{eemmeeee}
\end{figure}

  According to the `cutting rules', as also described in section~\ref{fraot}, the 
imaginary part of the transition amplitude associated with a Feynman diagram is 
obtained by summing over the cutting possibilities. These involve adapting the Feynman 
rules
 for each possibility
 by placing the `cut' virtual states on-mass-shell -- and hence open to interpretation 
as external particle states --  via equation~\ref{cutrule}, which essentially replaces 
each corresponding Feynman propagator $\Delta_F$ by
 one of the $\Delta^{\pm}$ function components described in
  equations~\ref{delfpp}--\ref{dpmd4}.
  From the unfolding of figure~\ref{xxphyy2} the initial and final field states in 
figure~\ref{eemmeeee}(a) are equivalent, and hence the cutting rules applied to the 
corresponding figure~\ref{eemmeeee}(b) yields the imaginary part of the forward 
scattering amplitude for the $e^+ e^-$ initial state $\vert i \rangle$, namely in fact 
$2\:\!\mbox{Im}({\mathcal M}_{ii})$, as contributed by placing the cut line on the 
intermediate $\mu^+\mu^-$ state for this Feynman diagram.

  The important observation of the present theory is that $\mbox{Im}({\mathcal 
M}_{ii})$ is a \textit{real} number, and hence might be directly compared with event 
probabilities with contributions of the form  $P \propto D_+D_-$ based on a count of 
the `number of ways' in which an observed process might arise. Adding all possible 
contributions for all possible final states, exemplified by the process in 
figure~\ref{xxphyy2}, then correlates, via the generalisation of 
figure~\ref{eemmeeee}, with the imaginary part of the forward scattering amplitude for 
all possible Feynman diagrams for the full perturbative expansion. The resulting real 
number    
$\mbox{Im}({\mathcal M}_{ii})$ is in turn directly related to the total cross-section 
$\sigma$,  equation~\ref{optith}, via the optical theorem as described in 
section~\ref{fraot}. Hence we arrive at a provisional relationship between a 
degeneracy count and a physical observable.

  As described towards the end of section~\ref{fraot} the optical theorem can be 
proven to all orders of perturbation through the analysis of Feynman diagrams. The cut 
pictured in figure~\ref{eemmeeee}(b) represents one contribution to 
$\mbox{Im}({\mathcal M}_{ii})$ for this diagram, with a second contribution provided 
by placing the cut instead through the $d\bar{d}$ fermion loop. Hence by the above 
discussion the determination of $\mbox{Im}({\mathcal M}_{ii})$ via the cutting rules 
for this diagram correlates with a sum of a $D_+D_-$ field sequence for both a $\mu^+ 
\mu^-$ final state and a $d\bar{d}$ final state.
  Similarly for equation~\ref{optifey1} the imaginary part of the Feynman diagram of 
figure~\ref{fdxxxx4} was determined corresponding to opening up a  $\mcY^+ \mcY^-$ 
final state, with a further contribution to $\mbox{Im}({\mathcal M}_{ii})$ at this 
order of perturbation obtained by replacing the loop in figure~\ref{fdxxxx4} with a 
$\mcX^+ \mcX^-$ state, as described after equation~\ref{optifey1}.

 The Feynman diagram with the cut of figure~\ref{eemmeeee}(b), in placing the 
  $\mu^+\mu^-$ pair on-mass-shell and via the optical theorem, contributes to the 
cross-section $\sigma (e^+e^- \to \mu^+\mu^-)$. However the structure of 
  $\mbox{Im}({\mathcal M}_{ii})$, in summing over the cuts, generally incorporates a 
collection of final states from which individual cross-sections for particular 
processes need to be untangled, as they are for the sum on the left-hand side of 
equation~\ref{msqmii} for example. Also, as alluded to in the caption comments, the 
fold-line in figure~\ref{eemmeeee}(a) should in principle be constrained to the 
`half-time' point to accurately represent the degeneracy count of 
figure~\ref{xxphyy2}.
 Further, we have considered the degeneracy count, based on particular sequences of 
fields leading to a particular final state, to represent a measure of the probability  
$P \propto D_+D_-$ for a particular process. However,  in order to actually represent 
a \textit{probability} this count needs to be determined relative to the total 
degeneracy for \textit{all possible} final states, which will provide the overall 
normalisation and which so far we have not taken into account.
 In looking to address these points we recap  how a particular final state is 
extracted and an event probability determined in the context of all possible outcomes 
in the framework of a QFT, with the aim of establishing a more precise link with 
similar calculations for the present theory. 

 As described in section~\ref{tranamp} in QFT the initial state $\vert i \rangle$ 
evolves through a period of field interactions into the state $\vert \Psi (\infty) 
\rangle = S \vert i \rangle$ according to the $S$-matrix of equation~\ref{smatrix}. 
This evolution is governed at each moment by the equation of motion expressed in 
equation~\ref{stateev} in which the  interaction Hamiltonian $H_{\mathrm{int}}(t)$ 
contains all possible field interactions. Hence $\vert \Psi (\infty) \rangle$ in turn 
contains \textit{all possible} final states. Since $H_{\mathrm{int}}(t)$ is Hermitian 
the evolution of the state in equation~\ref{stateev} is a unitary transformation, and 
hence if the initial state normalisation is chosen with $\langle i \vert i \rangle = 
1$  this is preserved such that 
 $\langle \Psi(t) \vert \Psi(t) \rangle = 1$ at any time $t$.
  On inserting a sum over a complete orthonormal set of
   similarly normalised
   final states $\vert f \rangle$
   we have $\sum_f  \langle \Psi(t) \vert f \rangle \langle f \vert \Psi(t) \rangle = 
1$,
  and in particular in the aftermath of the interaction, we have:
\begin{equation}
 \label{fsumuni}
   \sum_f \vert \langle f \vert \Psi(\infty) \rangle \vert^2 = 1
\end{equation} 
 as a mathematical identity.
  Hence the objects $\vert \langle f \vert \Psi(\infty) \rangle \vert^2$, in the sense 
of consisting of a set of positive real numbers that sum to unity, do have the 
\textit{property} of representing probabilities, \textit{and} in a structure which 
implicitly contains information about all possible final states.
 
  A relationship between the degeneracy $D(T,0)$ of equation~\ref{degnow} and the 
second order term of $U(t,t_0)$ of equation~\ref{useco} was described for a particular 
field sequence leading to a particular final state, as pictured in 
figure~\ref{xxphyy}. However, in general a degeneracy count associated with all terms 
of the entire $S$-matrix is desired in order to express everything that can happen, 
according to the field redescriptions permitted by the constraints of 
equations~\ref{conequa} in place of an interaction Hamiltonian, and hence incorporate 
all possible outcomes. This suggests a `complexification' of the probability 
calculation based on the degeneracy count such that the unitarity constraint, that is  
$SS^{\dag} = \b1$ in QFT, might effectively be employed to normalise the total 
probability for any outcome to unity.

 The subcomponent degeneracy counts $D_+$ and $D_-$, originally considered to provide 
a measure of the probability $P \propto D_+D_-$, are each real numbers.  
 The probability for any process is a positive real number $P\in \rrr$ from 0 to 1,
  as for any probability, and as for the square root of this quantity $p = \sqrt{P}$. 
However in principle it may be possible to consider a complexification of the 
underlying calculation, represented by $p \to \tilde{p} \inn \ccc$, such that $P = 
\tilde{p}^{\ast}\tilde{p}$. This is considered to be essentially the case in quantum 
theory where unitary symmetry is used to model the properties of probabilities, and in 
the case of QFT the role of the above complex quantity $\tilde{p}$ is played by the 
transition amplitude ${\mathcal M}_{fi}$.

   Specifically, the likelihood of an event in QFT is proportional to the squared 
modulus of the transition amplitude, as extracted from the terms of 
equation~\ref{fsumuni} via equation~\ref{sfimfi}, and as introduced in 
equation~\ref{diffcrs}. With the cross-section for a HEP process, for example, linked 
to the imaginary part of the forward scattering amplitude via the optical theorem 
expression of equation~\ref{optith} and this latter object, as the real number 
$\mbox{Im}({\mathcal M}_{ii})$,  correlated with a degeneracy count $D_+ D_-$, as 
described for figure~\ref{eemmeeee}, we have the following chain of associations:
\begin{equation}
      P  \propto  D_+ \, D_-  \sim 
	     \mbox{Im}({\mathcal M}_{ii}) \sim  \vert {\mathcal M}_{fi} \vert^2        
    \label{pddddm}
\end{equation}
 Here, in order for calculations in the present theory to converge with the formalism 
of QFT, the process probability on the left-hand side is linked with the QFT 
calculation 
  on the right-hand side
 via the mediation of $\mbox{Im}({\mathcal M}_{ii})$. The provisional connection on 
the side of the present theory with  $D_+D_-$ has been  described above and the 
connection through the optical theorem with $\vert {\mathcal M}_{fi} \vert^2 $ on the 
side of QFT was described in section~\ref{fraot}.

  While the structure of QFT on right-hand side of equation~\ref{pddddm} exhibits the 
basic property of probability conservation, via equation~\ref{fsumuni}, the input from 
the present theory on the left-hand side provides an explanation of the underlying 
\textit{physical nature} of the probabilities in terms of the relative degeneracy of 
the field redescriptions involved -- that is the `number of ways' in which the event 
may happen. Essentially the progression from left to right in equation~\ref{pddddm} 
represents a complexification of the calculation in order to employ unitarity to 
gather a normalised expression of the degeneracy count from which particular final 
states might be extracted with a combined probability of unity.

 The fact that the degeneracy count for field redescription sequences may be 
correlated with Feynman diagrams, as described for figure~\ref{eemmeeee}, together 
with the fact that the optical theorem can be demonstrated order by order in 
perturbation theory via the analysis of Feynman diagrams,
 as described in section~\ref{fraot},
 suggests that the structure of equation~\ref{pddddm} might be explored further for 
low orders of perturbation. Indeed the assumption of perturbation theory, provided the 
coupling constant is sufficiently small, is that only the first few terms of the 
expansion of the $S$-matrix of equation~\ref{smatrix} are required for precise 
calculations.

   In principle here it might be possible to work backwards from  QFT Feynman rules, 
such as those in table~\ref{frulesp} based on the Fourier expansions of quantum fields 
such as $\hat{\phi}(x)$ in the interaction picture, via the construction of the 
Feynman propagator $\Delta_F(x-y)$ as implied in the right-hand side of 
equation~\ref{useco}, and use the analogy between the left-hand side of that 
expression and equation~\ref{degnow} to make a detailed connection with the present 
theory. This connection, employing also the optical theorem, should also provide a 
guide for deducing a more rigorous mathematical expression for the underlying 
conceptual picture of the present theory, with the spacetime geometry $G^{\mu\nu}(x)$ 
constructed in terms of  fields such as $A^{\mu}(x)$ and $\psi(x)$ as one of many 
possible solutions.

  On understanding the parallels between QFT and the present theory and making the 
connection from the right-hand side of equation~\ref{pddddm} the aim would be to 
extract from the constraints of equations~\ref{conequa}  effective Lagrangian terms 
within the framework of the QFT formalism, expressed in the flat spacetime of special 
relativity. On importing aspects of the present theory into QFT in this way, with 
field redescriptions expressed in terms of the algebra of creation and annihilation 
operators, the aim would be to follow through calculations such as cross-sections 
using the familiar machinery of QFT.

  In this section we have largely considered the alternative route beginning with the 
provisional picture
  described in figures~\ref{xxphyy}, \ref{xxpxhyy} and \ref{xxphyy2}  for the present 
theory leading to the simple relation $P\propto D_+D_-$ for a process probability, 
with $D_+,D_- \inn \rrr$. Through comparing  the structure of figures~\ref{xxphyy2} 
and \ref{eemmeeee}(b), via figure~\ref{eemmeeee}(a),
 and making the association $D_+D_- \sim \mbox{Im}{\mathcal M}_{ii}$ this calculation 
might be `complexified' as guided by the optical theorem of QFT. In particular  a 
unitarity constraint could be employed to effectively normalise the  process 
probability calculation for all possible outcomes, as expressed in terms of an 
amplitude ${\mathcal M}_{fi} \inn \ccc$. This complex transition amplitude may then in 
turn be determined as described in the previous chapter, and in particular in terms of 
the Feynman propagator $\Delta_F(x-y)$ and Feynman rules, such as those of 
table~\ref{frulesp}.

  This approach is anchored in left-hand side of equation~\ref{pddddm}, with the aim 
of first motivating all development from the perspective of the present theory in 
itself. On establishing a link with the framework of QFT various techniques, such as 
the employment of unitarity in probability calculations, might be extracted  from QFT 
and adapted for use in the framework of the present theory. It may also be possible 
learn from the relation of QFT to phenomena in condensed matter physics, as we allude 
to in the following section. Here the aim is to understand the nature of physical 
particle states and determine cross-sections and other observable quantities within 
the environment of the present theory, for which the spacetime geometry accompanying 
empirical phenomena is not flat. However in a suitable limit the present theory may 
approximate to the form of a QFT in flat spacetime.

 The plausibility of either approach, from the left or right side of 
equation~\ref{pddddm}, rests on the identification of connections between the present 
theory and QFT which straddle the parallel development of the theories. Such a 
correspondence will be summarised in points 1) to 7) below. 
    The ultimate aim here would be to comprehend and follow through a complete 
calculation in the present theory, without any arbitrary reference to standard QFT, 
and to establish a direct connection with HEP empirical phenomena. However,
using the canonical approach to QFT as a close guide is a reasonable strategy since  
it has been used widely and successfully  in practice to obtain results for comparison 
with experiment.

 In the present theory there have been two distinct considerations:

\begin{itemize}
  \item[(a)]  The nature of field redescriptions and an understanding of the permitted 
elementary exchanges, such as depicted in figure~\ref{apsirelab}, according to the 
various equations of constraint in the theory. This was the topic of the previous 
section. 

   The fields such as $A^{\mu}(x)$ and $\psi(x)$ are not introduced onto a 
pre-existing 4-dimensional manifold $M_4$, rather spacetime itself, with the spacetime 
geometry $G^{\mu\nu}(x) = f(A,\psi)$, is shaped by the possibilities of the field 
descriptions. Hence figure~\ref{apsirelab} should not be interpreted too literally but 
rather a more dynamical mathematical expression of field redescriptions is desired. 
This might take the form of equations~\ref{vecredes} or \ref{vecredesa} (or 
\ref{spinredes} for the spinor case) in terms of retarded or advanced Green's 
functions, provided these expressions are compatible with constraints deriving from 
equations~\ref{conequa}.

  \item[(b)]  The calculation of the probability of observable processes, for example 
in HEP experiments, based on a count of the possible internal field degeneracies 
underlying the process, as depicted for example in figure~\ref{xxphyy2}. This has been 
the topic of the present section.
 
  Again here the sequence of $A^{\mu}(x)$ and $\psi(x)$ fields in 
figure~\ref{xxphyy2},  superposed as if upon a pre-existing spacetime, presents a 
somewhat naive and mechanical picture for the degeneracy count. A more conceptually 
and mathematically rigorous expression of this count may be required to describe the 
multiplicity of ways in which the geometry of spacetime $G^{\mu\nu} = f(A,\psi)$ may 
be fabricated out of these fields.

\end{itemize}

    One of the initial aims has been to establish a correspondence between the basic 
elements of the present theory and those of calculations in QFT.
   In QFT the construction of the transition amplitude ${\mathcal M}_{fi}$ generally 
breaks down into very simple elements as described by the Feynman rules, as listed in 
table~\ref{frulesp} of section~\ref{fraot} for the scalar model.   Hence the goal is  
  to explain how the `number of ways' approach of figures~\ref{xxphyy}, \ref{xxpxhyy} 
and \ref{xxphyy2} leads to the Feynman rules which determine the quantity $\Mfi$, and 
understand why $\vert \Mfi \vert^2$ should determine the probability for various 
processes as expressed in cross-section or decay rate calculations.
   
    The parallels identified between the present theory and QFT are  listed here. 
 The first six items below loosely correlate with the respective Feynman rules of 
table~\ref{frulesp}  and the subsequent discussion in section~\ref{fraot}.

\begin{itemize}
  \item[1)] The number of ways a series of field redescriptions may unfold through a 
one-dimensional temporal progression with degeneracy $D$,  with terms such as 
equations~\ref{degnow} and \ref{degnow4}, is analogous to the perturbative expansion 
of the time evolution operator $U(t,t_0)$ of equation~\ref{uiter0} in QFT.
 The `number of ways' integral sum is naturally normalised by the linear uniform flow 
of time, with `one way' for each equal discrete temporal interval $\Delta t_i$ in 
equation~\ref{desnow} taken to the continuum limit $\Delta t_i \to 0$ for 
equation~\ref{degnow}. This symmetry between equal time intervals implies a flat prior 
probability distribution as a basis for a Bayesian statistical approach. It then needs 
to be understood how the propagator $\Delta_F$ of equation~\ref{delftpp}, taking the 
form of equation~\ref{delfmom}, arises as an effective momentum space prior 
probability distribution when the calculation is restructured as for QFT.  

   As simply a set of real parameters in the expansion of a field into Fourier modes 
the variables $k \in \rrr^4$, which may be interpreted as 4-momentum under $T^{\mu\nu} 
:= -\frac{1}{\kappa}G^{\mu\nu}$, as described in the previous section for $G^{\mu\nu} 
= f(A)$ in leading from equation~\ref{kcoeff} to equation~\ref{kextract}, may also 
appear in factors relating to process probabilities as a result of calculations based 
on underlying field degeneracy. 
  This is the case for cross-section calculations in QFT with factors of the Feynman 
propagator $\tilde{\Delta}_F(k) = 1/(k^2-m^2 +i\varepsilon)$ effectively appearing as 
a \textit{weight} factor, as for example in equation~\ref{mfiexplt}.  
 Hence in the restructuring of process calculations for the present theory, via the 
introduction of $T$-ordering in equation~\ref{uiter1} and the resulting Feynman 
propagators,  
such  prior probability distributions should also appear through this connection with 
QFT.

  \item[2)] The redescription expansion is moderated by the need for consistency with 
the constraint equations. These include the higher-dimensional form of temporal flow 
$\lvh$ with $D_{\mu}L(\bvh(x))=0$ and the original form of the external geometry  
$G^{\mu\nu}(x) = f(Y)$ with $\gmo$ throughout; as listed in equations~\ref{conequa} 
and all effectively acting as selection rules for field interactions. Collectively 
these constraints are analogous to  a Lagrangian, including in particular the 
$\lag_{\mathrm{int}}$ terms in QFT as associated with the vertices in Feynman 
diagrams.
 For the present theory the `number of ways' a process may occur is taken to be 
proportional to the couplings implicit in the constraints, such as the factors of 
$\dot{s}_f$ in the terms of $\dmo$ in equation~\ref{dlvpbbz}, as also discussed after 
figure~\ref{mumudd}.

  In QFT the structures correlating with (a) and (b), listed above for the present 
theory,   are seemingly inextricably linked. The interaction Lagrangian, which is 
closely associated with the selection rules
 provided by $\dmlvh$ for example in  (a),  appears explicitly in the $S$-matrix, 
through equations~\ref{htlagx} and \ref{smatrix}, which is used in the determination 
of event probabilities for item (b) above. That is in QFT the mathematical structure 
of possible field interactions is embedded in the structure of event probability 
calculations. 
  Effectively this is achieved through the mechanism of `quantisation' itself, with 
the expansion of the fields in terms of creation and annihilation operators, which 
essentially converts a classical composition of fields in an interaction term into a 
selection rule for contributions to the $S$-matrix.

  In calculations of the transition amplitude $\Mfi$ the commutation relations, such 
as equations~\ref{aacomr}, ensure the correct matching and avoid unwanted cross-terms 
in compositions of the interaction Lagrangian or Hamiltonian $H_{\mathrm{int}}(t)$ in 
the terms of equation~\ref{uiter0} and its time-ordered form in 
equations~\ref{uiter1}--\ref{utexp}. 
 The sequences of creation and annihilation operators placed between vacuum states 
also ensures causality in QFT calculations in the sense that any intermediate state 
must always be created before it is annihilated to yield a non-zero matrix element 
$S_{fi}$.  
 Sequences of creation and annihilation operators from the interaction Lagrangian 
embedded in $S_{fi}$ ultimately determine relative probabilities in the context of all 
possible processes.

  A similar method of `quantisation' might be employed in the present theory in order 
to incorporate the constraints of equations~\ref{conequa} as selection rules for 
chains of field redescriptions between initial and final states in a degeneracy count,
 through the structure of $R(t)$ in equations~\ref{degnow} and \ref{degnow4} for 
example.

  \item[3)] A free field solution for $A^{\mu}(x)$ under $G^{\mu\nu} = f(A)$
  in the form of equation~\ref{geinm2}
   may be expanded in terms of $e^{\pm ik\cdot x}$ Fourier modes as described in 
equation~\ref{kcoeff}, as consistent with Maxwell's equations under $\gmo$. Exchanges 
between fields such as $A^{\mu} \leftrightarrow \ol{\psi}\gamma^{\mu}\psi$ are 
considered in terms of the complex Fourier modes of the fields. Similarly for QFT 
calculations as presented in chapter~\ref{pp} using the interaction picture, as 
discussed after equation~\ref{kgyphy}, between the initial and final plane waves of 
the form $e^{\pm ik\cdot x}$ the state evolution  is mediated by an expansion of free 
fields of the form in equations~\ref{kgosol2}--\ref{kgosolxd}, which are solutions of 
the Klein-Gordon equation for the scalar model.

 In the canonical quantisation approach to QFT, as described in chapter~\ref{pp},  
annihilation and creation operators, such as $a(\bp)$ and $a^{\dag}(\bp)$, are 
associated with the Fourier modes $e^{- i p\cdot x}$ and $e^{+ i p\cdot x}$ of the 
field respectively, as seen in equation~\ref{phipos} for $\hat{\phi}^+(x)$, 
equation~\ref{phineg} for $\hat{\phi}^-(x)$ and equation~\ref{kgosol2} for the 
complete free scalar field $\hat{\phi}(x)$.
In  a QFT calculation the complex plane waves of the form $e^{\pm ik\cdot x}$ 
representing the incoming and outgoing particle states  are linked by a chain of 
creation and annihilation operators
 for a variety of fields to determine the transition amplitude as described for 
example in equation~\ref{sfifull}.  
 This structure, employed  throughout the calculations in the interaction picture, 
provides a close analogy with the present theory.  

  The quantum field $\hat{\phi}(x)$ of equation~\ref{kgosol2} does not represent a 
solution of the equations of motion given an interaction, nor does it represent a 
physical entity in any context. Rather this expansion $\hat{\phi}(x)$ carries the 
potential for all possible transitions for the corresponding classical field in terms 
of Fourier components. This is the interpretation in the present theory, for which 
such quantum field expansions might be employed in the construction of chains of field 
redescriptions, expressed in terms of complex Fourier modes and employed in a 
degeneracy count for any process.

  \item[4)] The geometric constraint $\gmo$ over the external 4-dimensional spacetime, 
with energy-momentum $T^{\mu\nu} := -\frac{1}{\kappa}G^{\mu\nu}$, implies the 
conservation of 4-momentum for all possible field redescriptions (in the flat 
spacetime limit considered here, as discussed before equation~\ref{gbiana}).
 In QFT calculations the time integral $\int dt$ over the interaction Hamiltonian 
$H_{\mathrm{int}}$ is replaced by a manifestly Lorentz invariant spacetime integral 
$\int d^4 x$ over the interaction Lagrangian density $\lag_{\mathrm{int}}$ via 
equation~\ref{htlagx} which, as seen for example in the lines of 
equations~\ref{sfixmom}, leads to the constraint of 4-momentum conservation for each 
interaction vertex as expressed by the $\delta^4$-functions.

  Whether spacetime integrals, as a generalisation of purely temporal integrals, might 
feature in a generalisation of the field redescription degeneracy count for solutions 
underlying a particular geometry $G^{\mu\nu}(x)$ is open to consideration.
 However here the
 field exchanges have been considered to take place purely through a temporal 
progression, consistent with the notion of a fundamental one-dimensional progression 
in time that underpins the conceptual basis of the whole theory.
 In any case,
 in the present theory 4-momentum conservation is ensured through the prevailing 
relation $T^{\mu\nu} := -\frac{1}{\kappa}G^{\mu\nu}$ and the identity $\gmo$ which 
hold throughout spacetime and in particular for local exchanges of the underlying 
fields. For such exchanges applied to the Fourier modes such as $A^{\mu}(x) \sim  
e^{-ik\cdot x}$ 
 and $ \ol{\psi}(x)\gamma^{\mu}\psi(x) \sim  e^{-ip_1\cdot x}e^{-ip_2\cdot x}$ for 
example
the 4-momentum conservation in a $A^{\mu} \leftrightarrow \ol{\psi}\gamma^{\mu} \psi$  
field redescription takes the form of the mutual condition  $k = p_1 + p_2$. This is 
essentially implied in the requirement that \textit{locally} the spacetime geometry  
$G^{\mu\nu}(x)$ itself is unchanged for such an underlying field redescription.

  \item[5)]   In the present theory an infinity in the degeneracy count occurs when 
for example the intermediate $A^{\mu}(x)$ field state in figure~\ref{xxphyy} is 
augmented for a further intermediate redescription in terms of a pair of fields,  such 
as $A^{\mu} \to \ol{\psi}\gamma^{\mu}\psi \to A^{\mu}$ as shown in 
figure~\ref{xxpxhyy}. Here the degeneracy count of equation~\ref{degnow4} will be 
further augmented as the field $A^{\mu}(x) \sim e^{-ik\cdot x}$ is replaced by the 
field
$\ol{\psi}(x)\gamma^{\mu}\psi(x) \sim  e^{-ip_1\cdot x}e^{-ip_2\cdot x}$ up to a 
mutual freedom in the share of the total 4-momentum  between $p_1$ and $p_2$, 
accounting for an infinite degeneracy of solutions, as described after 
equation~\ref{degnow4}.

 This is closely analogous  to the ambiguity in the 4-momentum carried by an internal 
loop in a Feynman diagram, such as that in figure~\ref{fdxxyy4} leading to the 
divergent momentum integral $\int d^4 r$ in equation~\ref{xxyynexo}, and as frequently 
encountered in QFT.
   In both cases a means of `renormalisation' is required in order to obtain a finite 
calculation.
By matching such infinities in the present theory with the analogous quantities in QFT 
a similar program of renormalisation might be obtained for the present theory, 
although with a different interpretation as will be described in the following 
section. Indeed, the degeneracy count for any given process in any case stands in need 
of a `normalisation' with respect to the count of the number of ways in which 
\textit{anything} can happen.

 \item[6)]  Various combinatoric factors due to permutations of interactions for 
higher-order field redescriptions,  or symmetries between identical particle states, 
will need to be assessed for the present theory and related to the corresponding 
factors based on the analysis of Feynman diagrams  in QFT. Discrete sums over field 
degrees of freedom such as spin in QFT
  also reflect the number of ways a process may occur.

 \item[7)]   The need to match both the $e^{-ik\cdot x}$ \textit{and} $e^{+ik\cdot x}$ 
complex Fourier modes of the fields, through independent chains of degeneracies $D_+$ 
and $D_-$, underlying a real expression of $\lvh$ and $G^{\mu\nu} = f(Y,\bvh)$, means 
that an overall event probability is of the form $P \propto D_+ \times D_-$ as 
described for figure~\ref{xxphyy2} (rather than $P \propto D$ alone from `item 1)' 
above).  
  For practical calculations it is the relative ratios of the  degeneracies  for the 
range of possible processes that is needed to obtain actual probabilities with $\sum_F 
P_F = 1$, for a sum over all possible final states $F$ arising from an
 initial state interaction, including the case for which the final state is identical 
to the initial state.

 The calculation of $D_+D_-$ is correlated with the determination of  
$\mbox{Im}({\mathcal M}_{ii})$ in QFT, as described for figure~\ref{eemmeeee}, which 
via a complexification of the calculation and the optical theorem is then closely 
related to the 
amplitude squared $\vert {\mathcal M}_{fi} \vert^2$ in QFT as described for 
equation~\ref{pddddm}.  
 Expressed this way the unitary symmetry applying to the complex amplitudes ${\mathcal 
M}_{fi}$ models the conservation of the total probability, implicitly normalising the 
degeneracy count for all possible processes. The fact that renormalisation is required 
in QFT shows that this application of unitarity is only partially successful, and does  
not necessarily automatically normalise the degeneracy count completely. 
 Indeed even for a renormalisable QFT finite calculations might not be achievable at a 
very high order of perturbation, and in general a more watertight method of 
normalisation might be sought for the present theory.

\end{itemize}

   For a complete calculation in this theory, putting all of the pieces together, 
    the actual value of the probability $P_F$ for a process yielding the final state 
$F$ is determined by the \textit{relative},
	rather than \textit{absolute},   number of ways in which it can occur, essentially 
as is the case for the probabilities of classical physics. 
 For example degeneracy counts over the infinite possibilities in the timing of field 
redescriptions, such as those in equation~\ref{degnow}, may be independent of the 
choice of the external fields, as for example in figure~\ref{xxphyy} which may 
describe the leptonic or quark final states for figure~\ref{mumudd}(a) or (b).
 More generally  the infinities in the count of the number of ways will be in common 
for a range of competing processes and will cancel in the calculation of physical 
quantities such that the total probability for any outcome will \textit{necessarily} 
satisfy the requirement $\sum_{F} P_{F} = 1$. Some care will then be needed in this 
theory to deal with infinities that arise in the stages of such calculations. 
  However, since all probabilities are normalised by the total degeneracy for any 
process the 
 bound $0 \le P_{F} \le 1$ will apply trivially. A relative infinity of ways to 
produce one particular final state $F$ will result in a probability $P_{F} = 1$, which 
may be problematic in terms of comparison with the corresponding empirical value, but 
it is not possible for the theory to yield a nonsensical infinity for the calculated 
value.

 The calculation of probabilities via a complexification  may prove an effective 
technique to apply for the present theory, once the relation between the underlying 
real number measure of degeneracy  and the QFT calculation through 
equation~\ref{pddddm} has been fully understood. In this translation of the 
calculation a 
 `unitarity' condition will model probability conservation, consistent with kinematic 
factors appearing through the propagators, as described for `item 1)' above, provided 
the ultimate expression for the probability is a dimensionless quantity.

	 While the seven points listed above express a close parallel between structures 
in the present theory and perturbative calculations in QFT, as well as obtaining the 
Feynman rules for ${\mathcal M}_{fi}$ the full cross-section expression is needed for 
comparison with empirical data.	
   For QFT the structure of the cross-section $\sigma$ was introduced in 
equation~\ref{diffcrs} as a product of three factors, namely the amplitude squared 
$\vert {\mathcal M}_{fi} \vert^2$ together with the initial state flux factor and the 
final state Lorentz invariant phase space $d\Phi$.  Various normalisation factors such 
as the volume $V$ and time interval $T$ of the interaction cancel in forming this 
expression.

 The  probability for a process, whether expressed in terms of a degeneracy count or 
not, should be a dimensionless quantity, as is the transition amplitude ${\mathcal 
M}_{fi}$ for the two-body final states considered in chapter~\ref{pp}.
 In general  ${\mathcal M}_{fi}$ need not be a dimensionless quantity provided the 
cross-section has the dimension of a length squared, as for example in 
equation~\ref{dcseemmt}, and as described in the discussion following 
equation~\ref{xxyynexo}.

   The present theory  may involve a different breakdown across the three factors 
composing the expression for the cross-section, compared with that displayed for 
example in equation~\ref{diffevr}, with the form of the appropriate normalisation for 
all three factors, including those for the initial  state flux factor and final state 
phase space, possibly differing also from the QFT case. For the present theory, as for 
QFT, it is ultimately the calculated cross-section that is required to be of the 
appropriate form in the context of equation~\ref{ratels}. 
 The normalisation of factors required for consistency with the cross-section having 
the mass dimension $D=-2$ will be closely correlated with the normalisation employed 
to obtain dimensionless probabilities that sum to unity.  

 As described for equation~\ref{diffevr} in section~\ref{crosss}  the event rate is 
proportional to the initial state luminosity and flux factors and final state phase 
space, as would be expected based on a \textit{classical} notion of probability. In 
this section we have argued for the replacement or interpretation of the central term 
$ \vert {\mathcal M}_{fi} \vert^2$ in this expression in the form of a quantity 
representing an underlying measure of the degeneracy of ways in which the process may 
occur, in terms of sequences of field redescriptions, and hence constituting a further 
purely \textit{statistical} factor having  essentially the same character as a 
classical probability.

   Explicitly, in this section we have considered field interactions in terms of 
possible field redescriptions,
	 involving the $e^{\pm ik \cdot x}$ Fourier modes of the fields, as expressed for 
example in equations~\ref{vecredes} and \ref{spinredes} of the previous section, 
causally linked together to mediate observable processes, conceived as a field 
sequence such as depicted in figure~\ref{xxphyy} or \ref{xxpxhyy} and combined as for 
figure~\ref{xxphyy2}, and as allowed by the form of constraint
 equations~\ref{conequa} such as $\dmlvh$ and $\gmo$.
This however leads to a picture of the extended spacetime geometry $G^{\mu\nu}(x)$ 
itself constructed as one solution out of a myriad of possible ways based on local 
field description degeneracy, again subject to the constraint equations, not only for 
HEP processes but everywhere throughout the 4-dimensional spacetime world.

   This implies  a conception of HEP phenomena, such as an $e^+e^- \to \mu^+\mu^-$ 
event, supported by the underlying field exchanges which seamlessly also support the 
macroscopic physical world including the detector apparatus itself. In turn the 
physics of quantum mechanics is seamlessly connected to the world of classical 
physics. In the  section~\ref{qpagig} we further explore this conceptual picture 
within which quantum and particle phenomena are found alongside macroscopic objects 
and gravitation in a unified framework.

While the above seven points provide a useful guide into the workings of such 
calculations ultimately a stand-alone approach within the present theory may be 
desired.
  In this way the aim is to achieve  explicit calculations  for comparison with HEP 
processes such as $e^+e^-$ collisions for the full theory.
 To make a detailed comparison between the present theory and HEP data ultimately the 
particle concept, and in particular the nature of the `in' and `out' states at a 
collider experiment, will need to be understood within the context of the present 
theory.
 This will require an understanding of the nature of physical particle states 
propagating in spacetime in general, relating to a fully `renormalised' expression of 
field exchanges, rather than representing particle states in the form of simple 
$e^{-ik \cdot x}$ plane waves as for QFT. This direction will be explored in the 
following section.


\section{Renormalisation and Particle States}
\label{seraps}

   The relation $G^{\mu\nu} = f(A)$  derives from the internal $\uo_Q \subset \ese$  
action within the isochronal symmetry of $\lvfs$, and is  expressed explicitly in 
equation~\ref{geinm2} as determined through the analogy with  Kaluza-Klein theory as 
described in section~\ref{reaic}. In deriving directly from the basic structure of the 
theory, through equation~\ref{aegasth} applied to the full symmetry group, 
 the internal gauge field component $A^{\mu}(x)$ itself, which  appears in expressions 
such as $\dmofs$ for the broken full symmetry,  can be considered as  a `bare' or
elementary field at the  
 `microscopic' level from the point of view of QFT.
  This same field, implicit in equation~\ref{geinm2} 
   and hence satisfying the relation $\square A^{\mu} = 0$ of equation~\ref{maxafree}, 
in the Lorenz gauge,
  is also essentially the classical gauge potential of Maxwell's electrodynamics of 
1864, associated with directly observable laboratory effects. In this sense, again 
from the point of view of QFT, 
 the gauge field $A^{\mu}(x)$ can be considered as a `dressed' or renormalised field 
at the `macroscopic' level. The question then arises as to how these two views of the 
same field $A^{\mu}(x)$ are consistent. 

  For a non-Abelian gauge field $Y^{\mu}(x)$ the relation $G^{\mu\nu} = f(Y)$ of 
equations~\ref{conequa}, that is the classical field expression of 
equation~\ref{gchift}, contains self-interaction terms as described in 
equations~\ref{gfagain}--\ref{nonaredes}. Hence, even from the perspective of gauge 
fields alone, the macroscopic form for $G^{\mu\nu} = f(Y)$ will be necessarily shaped 
and corrected as a consequence of the multiple solutions for the spacetime geometry, 
as built upon a degeneracy of underlying gauge field redescriptions, with the 
constraint $\gmo$ holding throughout the base manifold.
 However, in the full theory the gauge field $A^{\mu}(x)$, associated with the 
internal Abelian $\uo_Q$ gauge group, is also not free since it couples to fermions 
through the constraints of equations~\ref{conequa}, as seen in the terms of 
equation~\ref{dlvpbbz} for example. Through field exchanges as considered for 
equation~\ref{vecredes} 
 Maxwell's equation is modified  to the form of equation~\ref{maxhere}, with a source 
term deriving from the fermion components. Hence it is necessary to consider the 
macroscopic form of the spacetime geometry $G^{\mu\nu}(x) = f(A,\psi)$, and understand 
how this relates to the original classical expression for $G^{\mu\nu}(x) = f(A)$ and 
also to empirical phenomena.

  Empirically electromagnetic waves are observed to propagate `at the speed of light' 
effectively according to Maxwell's equation $\square A^{\mu} = 0$, with solutions such 
as that in equation~\ref{kcoeff} for the transverse polarisation states $r=1$ or $2$ 
and with $k^2 = 0$. Hence the overall form of the function $G^{\mu\nu}(x)$, on the 
left-hand side of equation~\ref{geinawave} with $\gmo$ implied in 
equation~\ref{gbiana}, appears to be completely \textit{transparent} to underlying 
exchanges of indistinguishable fields, with possible intermediate stages similar to 
those of figure~\ref{xxpxhyy} or \ref{xxphyy2}, which percolate down through higher 
orders with the spacetime geometry $G^{\mu\nu}(x)$ always preserved over the possible 
field redescriptions. That is, unlike the general case, the underlying gauge-fermion 
field redescriptions appear to make little or no impression on the spacetime geometry 
associated with an electromagnetic wave -- with  $G^{\mu\nu}(x) = f(A,\psi) \simeq 
f(A)$ which takes the shape of $T^{\mu\nu} := -\frac{1}{\kappa}G^{\mu\nu}$ as depicted 
in figure~\ref{gacos} for example.

 In QED  these higher-order solutions are described in  terms of  photon self-energy 
contributions, as shown for example in the Feynman diagrams of figure~\ref{photsel}.

\begin{figure}[htbp]  
\centering
\epsfxsize=14cm
\leavevmode
\epsffile[0 0 1987 182]{\gpath aPfig119e}
\caption{\setb A series of possible Feynman diagrams which `dress' the original `bare' 
photon propagator, which itself corresponds to the first diagram alone.}
\label{photsel}
\end{figure}  

  The particles observed in experiments correspond to renormalised states of the 
fields. The quanta of the electromagnetic field are massless, even 
 though the higher-order corrections to the photon propagator in figure~\ref{photsel} 
contain virtual particles such as $e^+e^-$ and $d\bar{d}$ pairs. In QED the 
preservation of the bare photon mass $m_{\gamma} = 0$, and hence the equation of 
motion $\square A^{\mu}(x) = 0$, for the renormalised field is explained in terms of 
Ward identities (see for example \cite{Pesk}). This observation in QED is analogous  
to the transparency of the geometry $G^{\mu\nu} = f(A)$ to higher-order microscopic 
field redescriptions in the present theory, maintaining the macroscopic field 
condition $k^2 = m^2 =0$, and a correlated mathematical explanation might be sought 
here.

    In the standard theory of QED the behaviour of the field $A^{\mu}(x)$ deviates 
from that in classical electrodynamics due to the properties of low energy $e^+e^-$ 
pairs.
 In HEP experiments  
  an effective internal structure of the photon is manifested in `two-photon 
collisions', such as the process $\gamma\gamma \to c\bar{c}$ induced and observed at 
$e^+e^-$ colliders. In such experiments  the photon expresses itself in revealing the 
internal structure of its dressed state.
 Equivalent  empirical effects are expected to arise from  the principles of the 
present theory, with the internal structure of matter composed of endless possible 
 internal `bare' field redescriptions. Here for example 
   solutions for $G^{\mu\nu} = f(A,\psi)$ may take the effective macroscopic form of 
an electromagnetic field alone, such as the wave solution in equation~\ref{kcoeff}, 
while implicitly containing a myriad of possible field components and hence carrying 
the potential for the associated interactions as seen for example in two-photon 
collisions.

  The mathematical divergences associated with higher-order loop diagrams in QED are 
tamed by accepting the non-physical nature of quantities such as `mass' and `charge' 
in the bare Lagrangian and instead aligning the physical parameters of the 
renormalised theory with empirical values of mass and charge, as described briefly 
following equation~\ref{xxyynexo}. 
 The effect of combining an empirically measured generic coupling parameter $g$ with 
quantum corrections determined in theory, through the machinery of renormalisation in 
QFT, leads to the observable phenomenon of `running coupling' in which the parameter 
$g$ is found to depend on the energy scale $E$ as described by the `renormalisation 
group equation':
\begin{equation}
     \frac{d}{d \ln \! E} \: g(E) \, = \, \beta (g(E))  \label{regreq}
\end{equation}
   The function $\beta$ depends upon the particular theory. In the Standard Model 
$\beta$ is positive for the $\uo_Y$ gauge group and negative for the non-Abelian 
internal symmetries resulting in the running coupling shown qualitatively in 
figure~\ref{runcup}.

\begin{figure}[htb]  
\centering
\epsfxsize=12.5cm
\leavevmode
\epsffile[0 0 1333 855]{\gpath aPfig1110e}
\caption{\setb The running coupling $g'$, $g$ and $g_s=\sqrt{4\pi \alpha_s}$ 
respectively for the $\uo_Y$, $\sutw_L$ and strong $\suth_c$ gauge interactions in the 
Standard Model. Extrapolated from their laboratory values over a number of orders of 
magnitude in energy scale, via equation~\ref{regreq} with conventional $\beta$ 
functions, the three parameters mutually intersect, although not simultaneously, at 
around $10^{14}$--$10^{16}\,$GeV (\protect\cite{Pesk} p.787).}
\label{runcup}
\end{figure}

   The energy dependence of the coupling $g$, representing the general case in 
equation~\ref{regreq}, is independent of the bare Lagrangian parameters, and also 
independent of the regularisation method and parameters used to temporarily suppress 
the divergences in the process of renormalisation.

  As described in section~\ref{fraot} generally a quantum field theory is 
renormalisable, and finite results may be obtained for comparison with experiment, if 
the coupling parameter $g$ is of mass dimension $M^{D}$ with $D \ge 0$. 
 All of the couplings for the Standard Model, such as $g$ and $g'$ in 
equation~\ref{covdevlep}, have $D=0$ and the corresponding QFT is \textit{just} 
renormalisable. Even here though for the renormalised Standard Model divergences 
remain in the sense that the expansion series for equation~\ref{smatrix} in 
equation~\ref{sfifi} does not generally converge at higher orders, although the 
problem does not become apparent until terms of approximately order 137 in the case of 
QED for example (\cite{Pen} p.681, this is the point alluded to at the end of `item 
7)' in the previous section), far beyond the first few orders needed for calculations 
in practice.

   The structure and tools of QFT have a broad scope of applications and do not 
necessarily describe the fields or particle states of a `fundamental' theory. An 
\textit{effective} quantum field theory is one which is only valid as a physical 
theory below a certain energy threshold and describes particle states appropriate 
within that energy range. Such an effective QFT, for example a theory for nucleon-pion 
scattering, is necessarily an approximation to nature, with different physics and new 
particle states observed at higher energy. The interpretation of particles associated 
with an effective field theory, such as nucleon and pion states, as `fundamental' 
particles is hence unsatisfactory.

  Renormalisable QFTs such as the Standard Model are also considered to be low energy 
effective field theories. The form of the renormalisation group equation, and contact 
with empirical observations, is insensitive to high-energy, short-distance phenomena, 
which are also unknown. Hence  QFT provides a \textit{phenomenological} framework for 
particle physics with fields in the Standard Model Lagrangian transforming under the 
$\SML$ gauge group describing the \textit{types} of particles that are observed in 
high energy physics experiments. The theory applies over a wide energy range and 
provides a unifying framework incorporating weak and strong, in addition to 
electromagnetic, interactions. The corresponding quanta of the Standard Model quantum 
fields describe the particle states of leptons, quarks, gauge bosons and the Higgs, 
all of which from an empirical point of view appear to be elementary. However this is 
\textit{not} a conclusion that can be drawn from the QFT for the Standard Model 
itself.

 A more fundamental theory is needed to ascertain the true elementary structures of 
nature.
  The renormalisation for the QFT of the Standard Model has had great pragmatic 
success in particle physics but, as well as being insensitive to the method by which 
divergences are `cut-off', in general has very little to say regarding the structure 
of an ultimate high energy theory. Hence the results of the Standard Model 
renormalised QFT are plausibly consistent with an underlying theory for which 
interaction probabilities are fundamentally expressed in terms of a degeneracy count 
of possible redescriptions of the underlying field function as proposed in this paper.
 The present theory aims to describe  the \textit{actual} nature and behaviour of 
physical entities 
 down to arbitrarily short distances and up to any energy scale.

   Indeed the present theory is intended to be a fundamental, rather than an 
effective,
   theory, in contrast with the Lagrangian approach, as has already been emphasised in 
section~\ref{subwal} and as will be discussed further in section~\ref{secrhp}.    
	The present theory is
	 also completely `renormalisable' in an essentially trivial way since 
probabilities are constructed simply in terms of the relative `number of ways' field 
solutions may be obtained. 
 These involve nested sequences going down through higher orders of field exchanges, 
as depicted for example in figure~\ref{eenestmm}(a), which itself 
 represents a higher-order extension from the form of figure~\ref{xxpxhyy} for a 
single $e^{-ik\cdot x}$ field component sequence, here depicted alongside the 
associated Feynman diagram.
\vspace{-3pt}
  \begin{figure}[htbp]  
\centering
\epsfxsize=13.8cm
\leavevmode
\epsffile[0 0 1977 868]{\gpath aPfig1111e}
\caption{\setb (a) A higher-order sequence of field exchanges for the process $e^+e^- 
\to \mu^+\mu^-$ together with (b) the correlated Feynman diagram with internal loops.
 This figure is similar to figure~\ref{eemmeeee}, except here with differing initial 
and final states and without a fold or cut line.}
\label{eenestmm}
\end{figure}

   Even considering the degrees of freedom of the field redescription timings $t_i$ 
the sum of possible ways is infinite, as described  following equation~\ref{degnow4}. 
Further, the  internal 4-momentum freedom for the $d\bar{d}$ field state, for example, 
in figure~\ref{eenestmm}(a)  translates into the divergent momentum integral for the 
corresponding $d\bar{d}$ virtual particle loop in the Feynman diagram of 
figure~\ref{eenestmm}(b), as described in `item 5)' of the previous section.
  For yet higher orders this structure 
 implies a nested product of infinite sums and integrals which would appear to more 
and more dominate  calculations for more and more `dressed' diagrams. However it is 
conceivable that such infinite degeneracy counts largely cancel, resulting in a 
non-trivial finite calculation of cross-sections or branching ratios.

For example, by  relabelling  the final state, figure~\ref{eenestmm}(a) can be 
considered to represent a field sequence underlying either an $e^+e^- \to \mu^+\mu^-$ 
or $e^+e^- \to d\bar{d}$ event, amongst other possibilities.
  Since the intermediate redescriptions in figure~\ref{eenestmm}(a) are applicable for 
both  processes $e^+e^- \to \mu^+\mu^-$ and $e^+e^- \to d\bar{d}$  the relative 
`branching ratio' to obtain the final state $ d\bar{d} $ is simply:
\begin{equation}
 \label{brddmm}
   \mbox{BR}(e^+e^- \to d\bar{d}) = \frac{\mbox{sum of ways for }\, d\bar{d}}
          {\mbox{sum of ways for } \, \{d\bar{d} \mbox{ or } \mu^+\mu^- \}}  
		 \; \to \; \mbox{`}\left( \frac{\infty}{\infty} \right)\mbox{'}
\end{equation}
  For either final state there is an infinite degeneracy of intermediate states owing 
to the implied unconstrained 4-momenta for example. 
 These infinities clearly cancel in calculations such as equation~\ref{brddmm} 
 since there is a similar, in fact here identical, `degree of divergence' in each 
case.
 Indeed generally in forming measurable branching fractions  cancellation between 
common factors will provide the main source of normalisation.  Further normalisation 
factors will be involved in deriving event rates and cross-sections such as 
$\sigma(e^+e^- \to \mu^+\mu^-)$, as described towards the end of the previous section.

 A similar situation arises in QFT with for example  
  the $4^{\mathrm{th}}$ order Feynman diagram  in figure~\ref{fdxxyy4} together with 
the same diagram with the final state relabelled by $\mcX^+\mcX^-$, for the processes 
$\mcX^+\mcX^- \to \mcY^+\mcY^-$ and $\mcX^+\mcX^- \to \mcX^+\mcX^-$ respectively, both 
of which contain loops with infinite degrees of freedom in terms of the corresponding 
momentum integrals. In QFT the methods of renormalisation lead to finite 
cross-sections and branching fractions for comparison with the empirical data.
In fact in QFT tree level diagrams, such as figure~\ref{fdxxyy}, already give a good 
approximation
for the  rates of such processes, provided the coupling constant is sufficiently 
small. This is the case for QED in which the cross-section calculation based on the 
tree-level diagram in figure~\ref{fdeemm} gives a good approximation for the process 
$e^+e^- \to \mu^+\mu^-$ as described towards the end of section~\ref{crosss}. 

  For the present theory based fundamentally on degeneracy counts the interpretation 
of equation~\ref{brddmm} 
    may be contrasted with the case of Newtonian calculus in which the ratio 
$\frac{\delta y \to 0}{\delta x \to 0}$ has a well defined meaning and value since 
$\delta{y}$ and $\delta{x}$ tend to zero in a related manner through a continuous 
function $y=f(x)$. Here, in a similar and yet complementary situation, the limit of 
the ratio $\frac{\sum\! y \to \infty}{\sum \! x \to \infty}$ in equation~\ref{brddmm} 
gives a finite and well defined result due to the close relationship between the 
divergence in the numerator and that in the denominator.

   This a very literal notion of (re)normalisation in calculating probabilities. It is 
analogous to everyday cases such as the probability of hitting the `20' on a 
dartboard. There are an infinite number of ways in which the point of the dart can 
land on the surface of the 20 segment. However this infinity is normalised by the 
infinite number of ways of landing in any other region such that the total probability 
is finite and approximately $\frac{1}{20}$ (for a suitably random dart thrower). 
Alternatively the sum over points may be quantified as a finite integral over surface 
area, rather like the finite integral over possible field redescription times $t_i$ in 
equation~\ref{degnow} for figure~\ref{xxphyy} as a measure of the sum of ways to 
describe the underlying field function.

  The above analogy demonstrates the close association of classical and quantum 
probabilities in the present theory as will be discussed further in the following 
section. The cancellation  in equation~\ref{brddmm} not only applies for the infinite 
degeneracy of field redescription times $t_i$ but also  for the unrestricted internal 
momentum freedom, implicit in the $d\bar{d}$ internal state of 
figure~\ref{eenestmm}(a) for example, which is also infinite in terms of a real-valued 
$\int d^4 k$ measure.  
  
   In practice calculations of branching fractions and cross-sections may be much more 
readily performed by noting the symmetry of the system (analogous for example to the 
equal sizes of the twenty segments on a circular dartboard in the metaphor described 
above). In the case of QFT unitary symmetry, in calculations based on complex 
amplitudes, is applied to model the conservation of probability; and yields successful 
results when supplemented by the techniques of renormalisation. However these 
calculations, founded on postulated complex-valued entities,
  miss the physical meaning of the infinities as a real-valued degeneracy in the 
number of ways a process can occur. Hence in the present theory renormalisation based 
on a real degeneracy count is expected to be closely related to QFT renormalisation 
based on complex objects, such as amplitudes and propagators, with similar conclusions 
except with finite results necessarily to all orders in the present theory.

    While generating finite results when normalised for specific processes it is 
plausible that
 the sums and integrals over the myriad of continuous possibilities, such as for 
figure~\ref{eenestmm}(a), and for an endless range of higher-order field sequences, 
 may have residual effects such as the dependency on the energy scale of physically 
measurable
  interaction strengths as described by the running coupling in figure~\ref{runcup}.
   Underlying differences in branching ratios such as equation~\ref{brddmm} will then 
depend directly upon differences in the `bare' couplings associated with the field 
redescriptions such as $A^{\mu} \leftrightarrow \ol{\psi}
 \gamma^{\mu} \psi$ for example. These include the $\dot{s}_f$ real coefficient 
factors of magnitude $1$ or $\frac{1}{3}$ for the $\uo_Q$ coupling in 
equation~\ref{dlvpbbz},  applied for the outer layer of field exchanges, that is in 
the external vertices of the Feynman diagram as for example in figure~\ref{mumudd}(a) 
and (b) and as described there in the subsequent text, and will apply here for the 
final field redescription at time $t=t_1$ in figure~\ref{eenestmm}(a). (In QED there 
are Ward identities which both preserve the bare value of the photon mass 
$m_{\gamma}=0$, as alluded to earlier in this section with reference to 
figure~\ref{photsel}, and also which preserve the ratios of charges through 
renormalisation, and again a correspondence might be sought with the structures of the 
present theory.)

   The field redescriptions underlying the many solution possibilities 
    are profusely diffused throughout spacetime, from the
   temporal origin of the universe in the Big Bang, shaping the initial conditions for 
the evolution of the cosmos as considered in the following two chapters, to the 
quantum effects observed in laboratory experiments such as that represented in 
figure~\ref{figsld} and described further in the following section. As well as the 
photon `self-energy' contributions of figure~\ref{photsel} the field redescriptions 
`dress' the initial and final state particles for an event observed at a collider 
experiment. These higher-order solutions include the final state processes suggested 
by the Feynman diagrams in figures~\ref{mumudd}(c) and (d). Since the $d\bar{d}$ 
fields undergo strong $\suth_c$ interactions, producing an observed final state $\pi^+ 
\pi^-$ pair for example, \textit{objectively} it might be expected that many more  
spacetime world solutions with a $d\bar{d}$ compared with a $\mu^+\mu^-$ final state 
might be identified, in the context of a grand ensemble of all possible $G^{\mu\nu}(x) 
= f(Y,\bvh)$ solutions on $M_4$.
 This consideration would suggest that the branching ratio of equation~\ref{brddmm} 
should effectively be unity, owing to the apparent relative infinity of ways to 
produce a $d\bar{d}$ rather than a $\mu^+\mu^-$ final state.
 As pointed out in the  discussion after `item 7)' in the previous section such a 
conclusion for the present theory, although being internally  consistent, would appear 
to be drastically incompatible with  empirical phenomena. 

  However laboratory phenomena, as for all observations, \textit{subjectively} evolve 
progressively in time. At the time $t_1$ of the final field redescription in 
figure~\ref{eenestmm}(a) the likelihood of a field exchange will depend upon the 
$\uo_Q$ coupling $\dot{s}_f$ regardless of what has happened before or  what can 
happen after. Hence the charge value of 1 or $\frac{1}{3}$ will dominate the 
cross-section.
 Subsequent field redescriptions and interactions for the  final state produced, as 
represented in terms of the Feynman diagrams in figures~\ref{mumudd}(c) and (d) for 
example, will not affect the branching fraction calculation for equation~\ref{brddmm}
 other than through their implications for a final state phase space factor, which in 
the present theory correlates with the range of spacetime geometries associated with a 
particular set of final state particles.

 More generally out of the grand ensemble of all possible $G^{\mu\nu} = f(Y,\bvh)$ 
solutions it might be expected that a typical world would be dominated by strong 
$\suth_c$ interactions and corresponding forms of matter, since a relatively much 
larger range of field redescriptions are possible, via the set of eight 
self-interacting  gluons, compared with other  kinds of Standard Model interactions. 
However we do not apprehend a full 4-dimensional universe all together in its full 
temporal extent as a given object, rather we subjectively sample a possible world 
progressively through time. The corresponding progressive accumulation of 
probabilities selects a type of possible  $G^{\mu\nu} = f(Y,\bvh)$ solution which is 
extremely  rare in the context of the full ensemble, with a sparser more open form of 
matter shaped by a more democratic contribution from the components of $\SML$ gauge 
interactions.

 That is the  $G^{\mu\nu} = f(Y,\bvh)$ solution that we observe is selected with all 
probabilities oriented with respect to an underlying one-dimensional temporal flow 
from the past to the future, moulding the matter content and laws of physics for such 
a universe, with the structure of causality built into the world we perceive. As for 
perception of the world in space and time itself, this subjective causal aspect of 
observations is a further necessary \textit{a priori} structure through which we 
experience the world, as will be discussed further in chapter~\ref{chaptoot}.

  While the accumulation of probabilities along a causal path through a choice of 
world solutions  shapes the macroscopic properties of matter on the large scale, the 
probabilities locally determine the relative likelihood to achieve different outcomes 
such as for example the event $e^+e^- \to d\bar{d}$ or $e^+e^- \to \mu^+\mu^-$ in a 
HEP experiment as described above.
 Once the final state particles, such as $\pi^+\pi^-$ or $\mu^+\mu^-$, are formed and 
propagate through spacetime to the extent that the macroscopic shape of $G^{\mu\nu} = 
f(Y,\bvh)$ diverges the relative degeneracy count of field redescriptions for 
different processes under the same $G^{\mu\nu}(x)$ geometry no longer applies. That is 
a branching ratio such as equation~\ref{brddmm} is determined by the relative number 
of world solutions effectively within a local finite spacetime volume (similarly as 
represented by $VT$ for the QFT calculations described in section~\ref{crosss}) with a 
common local geometry described by $G^{\mu\nu}(x)$, regardless of what can happen 
after the final states form.

 In quantum mechanics the
 causal sequence of probabilities is reflected in the evolution of the wavefunction 
$\Psi$ in equation~\ref{hamphi} below as punctuated by apparent `collapses'  of the 
wavefunction, as will be discussed in the following section. As will also be described 
further in the next section the underlying statistical origin of quantum phenomena in 
the present theory is very similar in nature to that for a classical statistical 
system, with outcomes essentially determined by the `number of ways' in which 
something can happen. 
 The causal accumulation of probability, that is the temporal ordering property as 
described above for quantum phenomena in the present theory, naturally also applies 
for systems of classical physics. In the classical world the temporal ordering of 
probabilities underlies the  second law of thermodynamics for example, which will be 
considered in relation to the very early universe in section~\ref{sectveu}.

  The statistical approach underlying quantum phenomena in the present theory, 
fundamentally based on a real-valued degeneracy of field possibilities, has a 
microscopic structure analogous to that studied in the classical physics of critical 
phenomena. There the forces and behaviour of basic elements of condensed matter 
systems, such as magnets or fluids, are sufficiently well known to be modelled and 
parametrised.
 There is also a close relationship between such systems and quantum field theory at 
the phenomenological level -- in fact a 
  correspondence can be identified between renormalisation in QFT  and the theory of 
critical phenomena which leads to a principle of \textit{universality} for statistical 
fluctuations, which is equivalent to the cut-off independence in QFT (\cite{Pesk}  
p.268). However, although the empirical tests in HEP have been very successful, in the 
case of the QFT for the Standard Model the short-distance physics, only provisionally 
represented by field parameters in the bare Lagrangian, is essentially unknown, as 
alluded to earlier in this section.

  Potentially the present theory extends the analogy between HEP phenomena and 
critical phenomena conceptually as well as mathematically, with the microscopic world 
being `modelled' on the idea of underlying field redescriptions. This makes a closer 
relation to the theory of condensed matter systems than for standard QFT, with the 
latter founded pragmatically on calculations based on complex transition amplitudes.
 
   In principle the present theory reaches down without limit into the microworld 
revealing an internal structure in terms of nested multiple field solutions continuing 
indefinitely in almost fractal-like manner, analogous to the perturbative expansion of 
the QFT time-evolution operator expressed as an infinite series in 
equation~\ref{uiter0}. On the other hand the scope of the theory in principle also 
feeds upwards and seamlessly into the phenomena of condensed matter physics itself, 
with magnetic and fluid properties emerging at the macroscopic level, and into the 
realm of classical physics and classical probabilities, as alluded to above and 
described further in the following section.

 Out of the construction of the spacetime geometry over sequences of field 
degeneracies, of arbitrary high order, it is suggested that the phenomena of particle 
states themselves  arise, apparently propagating through field configurations in 
spacetime in the fully `renormalised' theory and mutually interacting, accounting for 
the phenomena observed in HEP experiments. This picture of particle states brings to 
mind the excitations of `phonons' in the medium of a solid state device, with here the 
colourful variety of Standard Model particle types arising out of the variety of 
underlying internal field interactions allowed by the broken form of $\lvfs$ and 
$\dmofs$ and the constraint equations~\ref{conequa} on $M_4$ in general.

As for the Standard Model, in the present theory particle masses arise through the 
interactions of the corresponding field with a `Higgs' field. Here a vector-Higgs 
field is associated with the components of $\bh_2 \equiv \bv_4 \subset \bv_{56}$
 of equation~\ref{fhthopart}
 projected onto the local tangent space $\TM_4$, with the effective Higgs 
phenomenology provisionally identified as described in subsection~\ref{suboomahp}.
As well as the selection of the external $\TM_4$ subspace component of $F(\htho)$ here 
`spontaneous symmetry breaking' is also realised in terms of a particular choice of 
vector field $\bv_4(x)$ which may `point' in an arbitrary direction at any given 
location $x\inn M_4$.
 This structure may be closely relate to the statistical methods employed in 
spontaneous symmetry breaking for critical phenomena, as for example associated with 
the properties of ferromagnetism.
  Recalling that the Higgs mechanism was developed from the early 1960s through 
analogy with spontaneous symmetry breaking phenomena as originally conceived in 
condensed matter physics this observation sees the Higgs concept returning to familiar 
territory.

 Low energy effective phenomena might also arise and be related to the Standard Model, 
which itself may considered to be an effective field theory as discussed after 
figure~\ref{runcup}.
 In this case while some components of $\bv_{56} \inn F(\htho)$ such as the Dirac 
spinors $\psi$ might correlate directly with elementary fermion states, the 
vector-Higgs $\bv_4$ components may correlate less directly with the  empirically 
observed scalar Higgs particle. 
  In the Standard Model this latter state is itself treated as a `fundamental 
particle'
  in the effective theory
 with symmetry breaking modelled by a scalar Higgs field $\phi$ in the contrived 
potential of equation~\ref{higgspot} as described in section~\ref{ewtatsm}.
 In the present theory the degree of freedom $\vert \bv_4 \vert$ is considered as a 
candidate for a field underlying the observed Standard Model scalar Higgs  particle, 
which hence does 
  not correspond to a fundamental scalar field in the components of $\lvfs$ projected 
over $M_4$.  While the scalar condensates of technicolor models, described shortly 
before equation~\ref{lagtech} in subsection~\ref{suboomahp}, are analogous to BCS 
pairs of electrons bound through  interactions with phonons in solid state devices, a 
different relation to condensed matter systems might be sought for the present theory 
since here technicolor gluons are not required to bind the scalar Higgs  together.

  While the microworld is infused with field function redescriptions, such as $A^{\mu} 
\leftrightarrow \ol{\psi}\gamma^{\mu}\psi$, in the multiple solutions under 
$G^{\mu\nu}(x)$ physically transmitted real particle states, such as electrons and 
photons, as detected in HEP experiments propagate over macroscopic distances with 
measurable and regular properties such as mass $m$, charge $e$, spin $s$ and average 
lifetime $\tau$.
 These  features, which \textit{define} the particle types, are regular and 
reproducible and hence must to some degree arise as the properties of self-sufficient 
discrete entities, in the sense of being generally independent of the conditions under 
which they are produced and the environment within which they are observed.   Such 
real propagating particle states
 are associated with
 a distinct impression in the spacetime geometry, that is the form of $G^{\mu\nu} = 
f(Y,\bvh)$, itself.
 In propagating over macroscopic distances particle states, such as photons and 
electrons, are revealed through their observable apparent interactions between each 
other and with the elements of macroscopic apparatus.

   An electron state in the $e^-$ beam of a particle accelerator for example is in 
constant interaction with the electromagnetic fields produced by the accelerating, 
bending and focussing components of the machine, via the elementary field exchanges 
depicted in figure~\ref{elecsel}(a).
 Even for a freely propagating electron  interactions with an electromagnetic
  field $A^{\mu}(x)$ are present in terms of internal `self-energy' possibilities, 
similar to those for the photon in figure~\ref{photsel}, as shown here for a free 
electron state in figure~\ref{elecsel}(b). 
\begin{figure}[htbp]  
\centering
\epsfxsize=12.9cm
\leavevmode
\epsffile[0 0 1728 428]{\gpath aPfig1112e}
\vspace{-17pt}
\caption{\setb Exchanges between the electron field, described in terms of the 
components of the fermion field $\psi(x)$, and the electromagnetic gauge field 
$A^{\mu}(x)$, for (a) an
interaction with experimental equipment via an external photon and (b)
 a self-energy contribution for a free electron in terms of an internal photon.}
\label{elecsel}
\end{figure}

   Both situations depicted in figure~\ref{elecsel} are submerged within a saturation 
of multiple solution possibilities for $G^{\mu\nu} = f(Y,\bvh)$ such that the 
empirically observed electron state emerges out of this myriad of interactions as an 
apparently robust and reproducible discrete entity. Such a particle entity may be 
guided and to some degree localised, propagating in a 4-dimensional spacetime 
expression of the underlying 1-dimensional temporal flow with properties shaped out of 
the full higher-dimensional form $\lvh$.
 The particle states exhibit probabilistic behaviour, of the form modelled by quantum 
theory,
 as inherited from the probabilistic nature of the underlying degenerate set of 
possible field configurations, as described in the previous section.

  For the case of the plane wave electromagnetic field $A^{\mu}(x)$ of 
equation~\ref{kcoeff} an explicit form of the spacetime geometry was derived in 
equation~\ref{geinawave}, with $G^{\mu\nu} \sim \kappa
         \frac{k^{\mu}k^{\nu}}{Vk^0}$ via the coefficient $C$ extracted from 
equation~\ref{kextract}. This geometry for the field in a spatial volume $V$ was 
provisionally  associated with a `photon' of 4-momentum $P^{\mu} = k^{\mu}$ and 
$k^2=0$. 
 In this naive picture the photon propagates as a kind of `microscopic gravity wave', 
as suggested by equation~\ref{geinawave}, consisting of purely Ricci curvature as 
described after figure~\ref{gacos} and suggesting a metric $g_{\mu\nu}(x)$ of a form 
similar to equation~\ref{gmetwave}. 
  The  4-momentum carried by such a `particle state' is naturally `quantised' in the 
sense that the parameter $k\in \rrr^4$ in the $e^{\pm i k\cdot x}$ Fourier modes 
appears in the expression $T^{\mu\nu} := -\frac{1}{\kappa} G^{\mu\nu} = f(A)$,
 that is equation~\ref{geinawave},
  since $G^{\mu\nu}(x)$ is a function of the spacetime derivatives of the gauge field 
$A^{\mu}(x)$.

 The actual nature of physical particle structure is expected to be rather more 
elaborate than initially suggested by this  picture of plane waves in a given volume, 
which was initially motivated in part by analogy with QFT as recapped at the end of 
the previous section. For the present theory, unlike the external states
  the  intermediate states of field redescriptions, over which the form of the local 
spacetime geometry $G^{\mu\nu}(x)$ is unchanged, may however indeed involve
 independent complex wave components. As described in section~\ref{secdopp} a hybrid 
set of $e^{-ik\cdot x}$ and $e^{+ik\cdot x}$ mode  field exchanges  in such 
interactions, as depicted in figure~\ref{xxphyy2}, correlates with the apparent 
`amplitude squared' rule for the associated interaction probabilities. 
 On the other hand it is in the nature of a `particle' to possess properties quite 
distinct from plane waves.

   Whether or not considered in terms of wave packets or within a volume $V$ a 
`particle' here is also not considered to be a `point-like' entity, but rather a  
state of fields as a function on $M_4$ dynamically prescribed through the conditions 
of $G^{\mu\nu}= f(Y,\bvh)$ and $\lvh$. Higher energy particle transitions may be 
possible in ever smaller effective volumes $V$, correlated with higher 4-momentum $k$, 
without limit, with an apparent `size' or structure never observable for the initial 
and final state `entities' in processes such as $e^+e^- \to \mu^+\mu^-$. Indeed such 
`particle interactions' are manifestations of field redescriptions which effectively 
apply throughout a finite volume $V$ simultaneously, as indicated in and described for 
figure~\ref{xxphyy} for example, without reference to any point-like particle 
structure at all.

  The fact that particle phenomena become apparent for interactions on very short 
distance scales, relative to macroscopic laboratory equipment, may be due to there 
being a greater likelihood for field functions to be
  indistinguishable  for small spacetime volumes. On the other hand
 the idealised case of transition amplitude $\Mfi$ calculations in QFT effectively 
considers plane waves defined in the limit $V \to \infty$, with factors of $V$ 
cancelling for observable quantities such as cross-sections. 
  In the present theory
  the role of an apparent volume $V$ with regards to  particle interactions,
  and the discrete `quantised' nature of particle states and interactions more 
generally, requires further understanding. 

 The factor of $\frac{1}{A}$ in the event rate formula of equation~\ref{diffevr}, from 
the expression for the  luminosity in equation~\ref{lumif},
  makes intuitive sense when picturing the incoming beam components as `bunches of 
particles'. However, here the question is how a greater intensity of field 
interactions, apparently corresponding a smaller area $A$, increases the production 
probability for final state particles, with the particle concept \textit{itself} 
deriving from the underlying field interactions. 
 The relation of the macroscopic to the microscopic world through a program resembling 
renormalisation will be key to addressing these questions.

   As described in section~\ref{subwal}, and reviewed in the opening of the following 
section, the generalisation from the classical solution $G^{\mu\nu} = f(A)$, closely 
relating to Kaluza-Klein theory,  will modify the macroscopic form of $G^{\mu\nu}(x)$ 
in way that incorporates the charge density $\sigma(x)$ in the current $J^{\mu} = 
\sigma u^{\mu}$ of equation~\ref{jeqsigu}, the material density $\rho(x)$ in 
$-\frac{1}{\kappa}G^{\mu\nu} = \rho u^{\mu}u^{\nu}$ from equation~\ref{gtruu2} and the 
structure of matter $T^{\mu\nu} := -\frac{1}{\kappa} G^{\mu\nu}$ more generally. 
 
  From this point of view elementary particle states, such as the electrons and muons 
observed in HEP experiments, can be considered \textit{as}
 quantum transitions within the macroscopic world, which is geometrically described by 
$G^{\mu\nu} = f(Y,\bvh)$.
 With gravitation encompassing quantum phenomena this describes an environment one 
layer outside the traditional approach to QFT for which the particle states are simply 
\textit{given} as the initial and final states of particle interactions.
We hence return to the conception of the physical world as described 
section~\ref{sechepe} for the experiment depicted in figure~\ref{figsld} for example. 
In the meantime, in chapter~\ref{pp}, we have dismantled the QFT cross-section 
calculation in order to identify a correspondence with the present theory, as 
summarised in points `1)--7)' of the previous section; with the ultimate aim of 
reassembling such calculations in light of the present theory and fully accounting for 
the observed particle phenomena.

 Together with the observations of chapters~\ref{chapesb} and \ref{secfd} for the 
breaking of higher-dimensional forms of $\lvh$
 we may  hope to gain some insight into the reasons for the observed properties of the 
various particle types without having to merely write them in by hand based on 
empirical findings. The abstract Fock space of QFT is not required, with creation and 
annihilation of particles through mutual exchanges now being firmly grounded in the 
field state of the macroscopic world. Such a state may consist in the physical 
components of experimental apparatus themselves, which exhibit essentially classical 
behaviour, providing a framework to make firm calculations and predictions for the 
properties of the apparent particle transitions recorded.

   An electron state is then consistent both with the idea that nested multiple field 
solutions, generalising from  figure~\ref{elecsel}(b), continue indefinitely down on 
the microscopic level together with the spacetime geometry satisfying 
$G^{\mu\nu}=f(Y,\bvh)$  and $\gmo$ at the macroscopic level, with a form of 
`renormalisation' relating the two levels. 
In the case of the electromagnetic field the massless `renormalised' field has a close 
resemblance to the bare field $A^{\mu}(x)$ of equation~\ref{kcoeff}, as described in 
the opening of this section. Further, from the perspective of Kaluza-Klein theory, the 
external geometry is directly related to the internal gauge fields through $G^{\mu\nu} 
= f(Y)$ in the form of equation~\ref{gchift} for example. On the other hand, in the 
absence of an expression of the form $G^{\mu\nu} = f(\psi)$,
 the physical fermion particle states of an `electron field' for example appear to 
have a somewhat more distant resemblance to the bare $\psi \subset \bv_{56}$ 
subcomponents with which they were originally identified through the action of $\ese$ 
on the components of  $F(\htho)$ broken over $\TM_4$, as summarised in 
equation~\ref{fhthopart}.

  That is,
rather than being described directly by the $\psi(x) \subset \bv_{56} \inn F(\htho)$ 
field components projected onto $M_4$ the form of $G^{\mu\nu}=f(Y,\bvh)$ for an $e^-$
 particle state observed in HEP experiments will be shaped through interactions with 
other fields, such as $A^{\mu}(x)$,  
   resulting in a  `renormalised' or `dressed' state.
 This was suggested towards the end of section~\ref{sosmfi} where it was also hinted 
however that fermion particle states might be identified more directly through 
interactions of a spinor $\psi(x)$ field and the vector-Higgs $\bv_4(x)$, initially 
shaping a geometry more simply of the form $G^{\mu\nu} = f(\bvh)$, as will be 
described in section~\ref{secpotnt}.

  The $\psi \leftrightarrow \bv_4$ field exchanges between the spinor and vector-Higgs 
fields, consistent with the constraint $\lvh$ of equations~\ref{conequa}, are also 
proposed to give rise to the generation of `mass' for the fermion states. The 
corresponding interaction terms, as described for equation~\ref{qxmass}, are 
reminiscent of Yukawa couplings of a fermion field to the Higgs field in 
equation~\ref{Yukferm} for the Standard Model.
 However while such a `bare mass' might be identified at the level of $\lvh$, the 
physical mass, and indeed the  \textit{concept} of mass itself, 
 as an observable quantity is only defined for the macroscopic dressed state as 
described in terms of the energy-momentum  $T^{\mu\nu} := G^{\mu\nu}$.
   For a free electron state  in the complete theory the aim will be to identify the 
corresponding macroscopic form of $G^{\mu\nu}=f(Y,\bvh)$, and to understand how $k^2 = 
m^2_e$ arises as a robust observable quantity for such a state, as deriving from the 
underlying interactions of the `bare' fields. Further light will be shed on the nature 
and origin of mass in the presentation of cosmology in the context of the present 
theory, in particular towards the end of section~\ref{secpotnt} and opening of 
section~\ref{sectveu}.

 As well as carrying energy-momentum density in  $T^{\mu\nu} := G^{\mu\nu}$ particle 
phenomena are observed through the transfer of discrete values of 4-momenta $k$, ever 
enveloped within a spacetime geometry and consistent with $\gmo$, such that the total 
initial and total final momenta match in processes such as   $e^+e^- \to \mu^+\mu^-$. 
A full understanding of the nature of such interactions, as provisionally described in 
sections~\ref{secdos} and \ref{secdopp}, is of course intimately  related to an 
understanding of the nature of the particle concept itself. This may require a full 
exploration of the relation between the present theory, quantum field theory and 
condensed matter physics as alluded to in this section.

   In general terms
 to understand what \textit{is} an electron state or what \textit{is} a muon state, as 
observed in HEP processes such as displayed in figure~\ref{figsld} or \ref{eemm}, it 
is necessary to think of the full 4-dimensional spacetime picture in relation to the 
underlying field component redescriptions. This will be described further for 
figure~\ref{eemmeds} in the following section and connects to the broader question 
concerning the incorporation of a theory accounting for the quantum properties of 
fields and particles alongside general relativity in a consistent framework, in the 
form of the theory presented in this paper.
 The conception of particle phenomena for the present theory will also be discussed 
further alongside figure~\ref{gtovac} in section~\ref{secrhp} of the concluding 
chapter, with particle states correlated with the emergence of discrete topologies for 
geometric solutions for $G^{\mu\nu} = f(Y,\bvh)$ in the near vacuum limit.


\section{Quantum Phenomena and Gravitation Unified}
  \label{qpagig}

  In the present theory we begin with a 1-dimensional temporal progression and hence 
need to \textit{build} a 4-dimensional spacetime $M_4$ with geometry $G^{\mu\nu}(x)$ 
out of the structure and symmetries of the underlying multi-dimensional form of 
temporal flow \mbox{$\lvh$}. The degeneracy of possible solutions for the ways in 
which this may be achieved results in the indeterminacy of empirical observations in 
our world and other apparent `quantum' phenomena, as studied for example in HEP 
experiments.

  Beginning with the electromagnetic field $A^{\mu}(x)$  in section~\ref{secdos} the 
possibility of alternative solutions involving the fermion field $\psi(x)$ underlying 
the spacetime geometry were expressed in terms of the field redescriptions of 
equations~\ref{vecredes} and \ref{vecredesa}.
These equations satisfy equation~\ref{maxhere} in which  the current $j^{\mu} = 
\ol{\psi}\gamma^{\mu} \psi$ may be considered as a source term. 
  Such  `microscopic' field redescriptions via the mutual exchanges $A^{\mu} 
\leftrightarrow   j^{\mu}$ are incorporated into the spacetime geometry, generalising 
from the   classical  relation $G^{\mu\nu} = f(A)$ of equation~\ref{geinm2} as 
originally derived through association with Kaluza-Klein theory in 
section~\ref{reaic}.

  The observable world is awash with the interchanges between the $A^{\mu}(x)$ and  
$\psi(x)$  fields, together with higher-order redescriptions through which the fields 
may interact,
  saturating the world, as described alongside figure~\ref{elecsel} in the previous 
section. This gives rise to a rather fluid mathematical creation of \textit{matter} as 
perceived through these exchanges and hence the properties and forms of the 
`macroscopic' material world are conditioned by them. 
 This describes the general relativistic limit pertaining to tangible physical objects 
that take shape on  $M_4$ \textit{over} the collective contribution of the internal 
fields, such that the apparent composition of the Einstein tensor may be written 
simply as:
\begin{equation}
 \label{cet}
    G^{\mu\nu} = -\kappa T^{\mu\nu}(Y, \bvh) 
\end{equation}
 This is equation~\ref{getypsi} of section~\ref{subwal} expressed in a form which 
emphasises the implicit field composition of the material world.
   The effective energy-momentum tensor $T^{\mu\nu}$ on $M_4$ may take different forms 
in terms of the  apparent macroscopic matter distribution on the manifold, but it must 
always be fundamentally composed out of the interplay of the underlying fields, 
mutually subject to the constraint equations~\ref{conequa}.  For example the 
energy-momentum tensor might describe a perfect fluid and the Einstein tensor will be 
macroscopically composed as described in equation~\ref{gtruup}, as we shall consider 
for the cosmological scales of the universe, alongside equations~\ref{cosgtruup} and 
\ref{CosLem}, in the following chapter. 
 The general form of that equation for the structure of $G^{\mu\nu}(x)$ incorporates  
macroscopic `pressure' $p(x)$ terms and  defines the scalar field $\rho(x)$ which in 
general relativity is identified with the familiar notion of `matter density'.
 For the case of a pressureless perfect fluid we  have:
\begin{equation}
\label{einf}
  -\frac{1}{\kappa}G^{\mu\nu} =: T^{\mu\nu}_{\epsilon} = \rho u^{\mu}u^{\nu}
\end{equation}
 that is equation~\ref{gtruu2},
 with $T^{\mu\nu}_{\epsilon}$ interpreted as the effective energy-momentum tensor for 
a pressureless fluid.
   In the original formulation of general relativity such an energy-momentum tensor, 
through the above field equation, would be interpreted as the  `source of curvature' 
on the manifold.
     This correspondence with general relativity was explored in more detail in 
section~\ref{subwal} where it was described how
  equation~\ref{einf} leads to the geodesic equation of motion for this form of 
matter, that is equation~\ref{trgeod}, owing to the Bianchi identity 
$G^{\mu\nu}_{\ph{ab};\mu} = 0$,   without the need to introduce the geodesic 
constraint as an additional postulate of the theory.

  The extension of the classical field relation of equation~\ref{geinm2}, which 
implies the homogeneous Maxwell equation $\square A^{\mu} = 0$ of 
equation~\ref{maxafree}, as shown for equation~\ref{maxfein}, with the inclusion of 
the charged matter term $\rho u^{\mu} u^{\nu}$ on the right-hand side of 
equation~\ref{gtruum} is an example of a break away from the pure Kaluza-Klein 
relation of $G^{\mu\nu} = f(A)$.
 This deviation from a free electromagnetic field alone is here considered at 
 the level of macroscopic phenomena, which overlays the microscopic field interactions 
which led to equations~\ref{sqmassa} and \ref{maxhere} in section~\ref{secdos} and as 
described in the opening of the previous section.
   Applying $\gmo$ to the full expression in equation~\ref{gtruum} led to the 
incorporation of  a charged current $J^{\mu}$, with $\square A^{\mu} =: J^{\mu}  = 
\sigma u^{\mu}$ as defined in equation~\ref{jeqsigu}, and to the identification of the 
Lorentz force law of equation~\ref{rellorg} as a deviation from the purely 
gravitational geodesic flow.
Here the charged current $J^{\mu}$  corresponds to that
  observed in macroscopic classical experiments, typically for the non-relativistic 
limit such as  performed by Faraday in the $19^{\mathrm th}$ century.
  Hence in addition to the apparent matter density $\rho$ in equation~\ref{einf}  
effective  macroscopic phenomena 
  also involve the  charge density $\sigma$ in $J^{\mu} = \sigma u^{\mu}$.
  Both the macroscopic and microscopic currents are conserved, with 
$J^{\mu}_{\ph{\mu};\mu}= 0$ as described following equation~\ref{rellorg} in 
section~\ref{subwal} also applying for
 $j^{\mu} = \ol{\psi}\gamma^{\mu} \psi$ of equation~\ref{maxhere} (as originally 
expressed for equation~\ref{jmupgp} in section~\ref{subfal} for the Lagrangian 
approach).

 Within  this limiting case of general relativity described  above, that is neglecting 
explicit quantum phenomena,  if the 
approximation of a flat spacetime for which $G^{\mu\nu}(x) \simeq 0$ may be assumed 
then the theory of special relativity will apply to the laws of physics. Further 
beyond that limit the motion of bodies for physical systems in which the relative 
velocities are small compared with the speed of light may be analysed using classical 
Newtonian mechanics.
    Local energy-momentum conservation in all physical processes is ensured under the 
Bianchi identity $G^{\mu\nu}_{\ph{\mu\nu};\mu} = 0$, since 
 the energy-momentum tensor $T^{\mu\nu}$  is identified with  the spacetime geometry 
$G^{\mu\nu}$, regardless of the magnitude of the spacetime curvature. As described in 
the opening of section~\ref{subwal} this observation applies in particular in 
approaching the flat spacetime limit with $T^{\mu\nu}_{\ph{\mu\nu},\mu} = 0$, and will 
also apply for the further limit of the non-relativistic case; with the corresponding 
energy-momentum conservation also encompassing all underlying quantum phenomena in all 
cases.

   Microscopic transitions of internal fields such as $A^{\mu} \leftrightarrow 
    \ol{\psi}\gamma^{\mu}\psi$, and quantum processes in general, may be recorded in 
macroscopic devices, generally in the form of amplified electronic signals. 
 All such macroscopic equipment is also itself 
  composed over field interactions in the form of equation~\ref{cet} and
 effectively described by an appropriate classical energy-momentum tensor 
$T^{\mu\nu}$, for the solid state devices typically employed, and at a basic level a 
tiny `detector recoil' will accompany any production or detection of particle states 
as a consequence of 4-momentum conservation.
 The equations governing the evolution and interactions of the microscopic world hence 
merge into the equations of motion for `classical' objects, such as described by 
geodesic trajectories or the Lorentz force law. This framework will then shed some 
light on a key question concerning the relation of quantum mechanics to the world of 
classical physics.

	In the previous section it has been outlined how empirically observed particle 
states, such as electrons and muons, might be identified in parallel with a program of 
`renormalisation' for the present theory, and merge seamlessly into the state of the 
macroscopic environment. 
In figure~\ref{eemmeds} a typical high energy physics process, as described in 
section~\ref{crosss} and already depicted in figure~\ref{eemm}, is contrasted with a 
typical experiment involving non-relativistic quantum theory.
\vspace{-3pt}
\begin{figure}[htbp]  
\centering
\epsfxsize=\maxwidth
\leavevmode
\epsffile[0 0 1912 724]{\gpath aPfig1113e}
\vspace{-32pt}
\caption{\setb (a) The process  $e^+e^- \to \mu^+\mu^-$ as observed in HEP experiments 
for which the cross-section can be calculated in QFT. (b) The double-slit experiment 
in which a single electron is detected on the screen according to a probability 
distribution determined in non-relativistic quantum mechanics.}
\label{eemmeds}
\end{figure}

  In figure~\ref{eemmeds}(a) a particular event is detected, a final state 
$\mu^+\mu^-$ pair at an angle $\theta$, presumably mediated by one of many possible 
intermediate sequences of field states such as represented in figure~\ref{xxphyy2}. In 
figure~\ref{eemmeds}(b) an electron is detected at a particular location $A$ out of a 
continuum of possibilities including $B,C \ldots$.
   In standard quantum theory both of these processes are assumed to take place
    against a flat background of space and time, which for figure~\ref{eemmeds}(a) is 
Minkowskian and for figure~\ref{eemmeds}(b) is Newtonian. In the present theory 
however the base manifold curvature, although smooth, is finite and non-flat 
essentially everywhere in 4-dimensional spacetime, with both processes depicted in 
figure~\ref{eemmeds} representing particular features of a global $G^{\mu\nu} = 
f(Y,\bvh) \neq 0$ solution. 

  Events of the kind sketched in figure~\ref{eemmeds}(a) are readily observed by 
experiments of the kind depicted in figure~\ref{figsld} for example. In this case both 
the macroscopic SLD detector and microscopic $e^+e^- \to \mu^+\mu^-$ interaction are 
uniformly enveloped within a particular solution for $G^{\mu\nu} = f(Y,\bvh)$. The 
interaction region of such an experiment for such an event will locally have a 
spacetime geometry $G^{\mu\nu}$ of a similar  
form to that for $T^{\mu\nu} := G^{\mu\nu}$  pictured in figure~\ref{gacos} and as 
represented by the wavy lines in figure~\ref{eemmeds}(a). As described in 
section~\ref{secdos} the associated metric solution $g_{\mu\nu}(x)$ will have 
properties closely relating to the metric of equation~\ref{gmetwave}; and the 
underlying field redescriptions, as represented for example by figure~\ref{xxphyy2}  
in section~\ref{secdopp}, will necessarily respect this external physical geometric 
form.

  Similarly the system described in figure~\ref{eemmeds}(b) will be enveloped within a 
particular $G^{\mu\nu} = f(Y,\bvh)$ solution, with a non-flat metric $g_{\mu\nu}(x)$ 
description. For a sufficiently high intensity source with a stable interference 
pattern observed on the screen a wave-like solution for $G^{\mu\nu}(x)$ will permeate 
the spaces between the elements of apparatus. The lower intensity case, with a single   
 electron detected on the screen as indicated in the figure, will correspond to a  
different 4-dimensional world solution for $G^{\mu\nu}(x)$. While both wave-like and 
particle-like solutions are shaped by an enveloping geometry with $\gmo$ uniformly 
throughout space and time, the underlying indeterministic character of the field 
redescriptions become evident as discrete particle phenomena emerge at low intensity. 
In all cases the direct identification of $-\kappa T^{\mu\nu} := G^{\mu\nu}
 = f(Y,\bvh)$ implies that the field equation of general relativity is faithfully 
reproduced, even for the case of a single particle state exhibiting the underlying 
quantum behaviour. 

 While a solution $G^{\mu\nu} = f(Y,\bvh)$ envelopes the full 4-dimensional system 
depicted in 
figure~\ref{eemmeds}(b), including the macroscopic apparatus,
 the indeterminacy of the single particle process lies in the perfect symmetry of 
possible field solutions underlying the smooth function $G^{\mu\nu}(x)$  locally at 
the source $S$, which is the same for any possible outcome.
 This situation is then very similar to that in figure~\ref{eemmeds}(a), with the 
source $S$ corresponding to the interaction region, as represented by the rectangular 
box, and with the range of outcomes $A,B,C\ldots$ corresponding to the angular range 
$0 < \theta < \pi$. The comparison is even more direct if the intermediate double-slit 
screen is removed from the apparatus in   
figure~\ref{eemmeds}(b).

 In all cases the full 4-dimensional solution $G^{\mu\nu} = f(Y,\bvh)$ encompassing 
the entire system is intrinsically shaped through a 1-dimensional causal accumulation 
of probabilistic outcomes wherever the geometry $G^{\mu\nu}(x)$ is locally expressible 
in terms of a degeneracy of underlying field functions. The inclusion of the 
double-slit screen in figure~\ref{eemmeds}(b) is accompanied by a more complicated 
spectrum of single particle solutions, as might be expected since the full system 
\textit{is} more complicated. In this case the relative probabilities, while depending 
crucially on the underlying field degeneracy at $S$, turns out to be    
 weighted by the interference pattern as shown.

 The spacetime curvature itself is far too small to be directly detectable, for 
example by geodesic deviation, although the fact that $G^{\mu\nu}(x)$ \textit{is} 
non-zero in these laboratory experiments is crucial in the present theory. As 
described above the local spacetime curvature associated with the interaction region 
in figure~\ref{eemmeds}(a) will be analogous to that for the free electromagnetic wave 
as derived in equation~\ref{geinawave} and pictured in figure~\ref{gacos}. This 
curvature will naturally be higher in cases of higher energy density such as at the 
interaction region of the LHC, where it remains also far too small to be observable.

With $T^{\mu\nu} := G^{\mu\nu}$ the spacetime curvature is also indirectly made 
apparent through the presence of energy-momentum. For example, since energy-momentum 
is everywhere conserved in line with the identity $G^{\mu\nu}_{\ph{\mu\nu};\mu}=0$,  a 
small recoil of the electron source $S$ in figure~\ref{eemmeds}(b) will causally 
precede the detection of an electron at $A$. 
  Indeed, in principle  `elementary' particles might be observed with detectors in a 
way analogous to `Brownian motion' with macroscopic matter `recoiling' against the 
elementary transitions of the fields within which it is immersed, bringing out the 
properties of both the particles and material objects themselves.

Considering a thought experiment with a very lightweight source $S$ situated at a very 
long distance from the detector screen in figure~\ref{eemmeds}(b) in principle an 
observation of the momentum recoil of $S$ could precede the detection of the signal at 
$A$ (and for apparatus consisting of the source and screen alone the prediction of the 
hit location on the screen would be very direct). The total momentum of the system, 
including the source, double-slit screen and detection screen, will be conserved.  The 
same quantum interference pattern would still appear on the screen given a large 
number of such events. 

 With the  momentum recoil of the macroscopic source $S$ too small to be measurable in 
practice for the process depicted in figure~\ref{eemmeds}(b)  the first and only sign 
of the event will be through the amplification of an electronic signal at $A$. 
Pragmatically the possible observable outcomes
 can be represented in terms of an electron wavefunction $\Psi(\bx)$  evolving 
according to Schr\"{o}dinger's equation until collapsing to 
   zero at $B,C \ldots$ at the moment when the electron is detected at $A$.
  In the present theory such a description in terms of an apparently non-local action 
of wavefunction collapse represents our knowledge of the state of the system rather 
than its underlying physical evolution. 

  In standard non-relativistic quantum mechanics the basic principle of the 
conservation of energy and momentum is considered to hold together with the constraint 
that no signals may be transmitted faster than light. For the case of the experiment 
depicted in figure~\ref{eemmeds}(b) this leads to the question concerning the location 
of the `conserved energy' during the intermediate period between the emission of a 
particle of a given energy at $S$ and the later detection of a particle of the same 
energy at $A$. The corresponding energy-momentum cannot be carried by the wavefunction 
for example, due to the discontinuous nature of the wavefunction collapse.

  In the present theory the `energy-momentum' is distributed throughout in terms of 
the 4-dimensional geometry $T^{\mu\nu}:=G^{\mu\nu}$. Energy-momentum conservation is 
everywhere implied in the identity $\gmo$, with nothing being transmitted faster than 
the speed of light -- as defined by the
 light cone structure which arises through the projection of the full form $\lvh$ onto 
the manifold $M_4$ as described in section~\ref{fdandtd}.   
 Since a solution $G^{\mu\nu}(x) = f(Y,\bvh)$ primarily describes the shape of a 
particular \textit{spacetime geometry} it may have a highly counter-intuitive 
distribution when \textit{interpreted} through $T^{\mu\nu}:= G^{\mu\nu}$ as an 
apparent flow of `\textit{matter}' \textit{through space}. Some forms of geometry do 
possess a form with a natural interpretation in terms of energy-momentum, as expressed 
for the macroscopic example in equation~\ref{einf} for a pressureless fluid.
 However, more generally rather more arbitrary geometries are permitted, provided 
$\gmo$, and some form of continuous geometry $G^{\mu\nu}(x)$ will be associated with 
the single particle process depicted in figure~\ref{eemmeds}(b).

  For the present theory the question concerns the manner in which everywhere 
continuous solutions for a geometry $G^{\mu\nu}(x)$ can be apparently channelled in 
certain \textit{discrete} and \textit{localised} ways which give the impression of 
`particle' transitions. That is, locally the energy-momentum $T^{\mu\nu}:=G^{\mu\nu}$ 
can be interpreted as the emission or detection of a discrete particle, for example at 
$S$ or $A$ respectively in  figure~\ref{eemmeds}(b). The answer presumably lies in the 
nature of the \textit{fixed} constraint equations~\ref{conequa}, such as $\dmoh$, 
which channel the underlying field redescriptions in a limited number of ways and in 
turn determine the properties of the particle transitions which emerge, as alluded to 
in the previous section.

   All empirical phenomena, whether naturally occurring or constructed in physical 
experiments such as that in figure~\ref{figsld}, will be enveloped under a geometry 
$G^{\mu\nu}(x) = f(Y,\bvh)$. Since in the laboratory setting this non-trivial 
geometry, that is any deviation from Minkowski spacetime,  is  completely unobservable 
the consequences of this perspective may be pursued by considering more extreme cases, 
such as the thought experiment described earlier for figure~\ref{eemmeds}(b) with a 
very lightweight source $S$ far removed from the detector screen, as well as
 by analysing phenomena physically realised in practice. 
   Any beam of electromagnetic radiation carries energy and is hence associated with 
spacetime curvature as for the case of standard general relativity
 and as depicted in figure~\ref{gacos} for example.
 In a further thought experiment intense beams of light, for example produced by 
lasers, could in principle be configured such that the tiny geodesic deviation of a
 suitable test body projected through the curved spacetime associated with the laser 
beam and over a large distance might be observed, without any  photons of the beam 
being detected.

 This situation may be contrasted with the empirical observation of the deflection of 
light itself in the gravitational field of the sun, as first reported just a few years 
after the formulation of general relativity. In these cases, for both the above 
thought and practical experiments, there is an `interaction' between light and gravity 
\textit{without} the detection of any photons or the need to appeal to any properties 
associated with quantum theory.
 For the above thought experiment similar observations would hold for an intense beam 
of particles such as electrons in place of the lasers, and leads to the conclusion 
that the electron field associated with the event of detecting  even single electron 
in figure~\ref{eemmeds}(b) will indeed be accompanied by a small, although utterly 
undetectable, spacetime curvature.

 As described in section~\ref{secdopp} the spacetime curvature $G^{\mu\nu}(x)$ is 
always a real valued tensor field but may be constructed out of a hybrid of complex 
components, such as the $e^{\pm i k \cdot x}$ Fourier modes, of the underlying fields 
such as $A^{\mu}(x)$ and $\psi(x)$ as depicted in figure~\ref{xxphyy2} for example. 
The matching of both the $e^{-i k \cdot x}$ and $e^{+i k \cdot x}$ parts coming 
together into a real-valued function 
for a single field such as $\psi(x)$ may correlate with detection events, such as at 
$A$ in figure~\ref{eemmeds}(b), through which apparently propagating particle states, 
such as electrons, are revealed. More generally the concept and nature of elementary 
particles  needs to be fully addressed, as was discussed in the previous section and 
will be further elaborated in section~\ref{secrhp}.

   While the understanding of the nature of particles as observed in the laboratory 
requires further work, it is clear in the present theory that there are no `graviton' 
states since the gravitational field itself is \textit{not} quantised in any sense. In 
fact general relativity provides a classical description of the geometry of the 
external perceptual framework of the world which fully accounts for the phenomena of 
gravitation. There is no given \textit{flat} 4-dimensional spacetime  manifold and 
hence no `force' of gravity as an apparent empirical addition on top of such a 
Minkowski spacetime.
 In turn there is no place for gravitons as `carriers' of such a gravitational force 
and no compelling motivation to consider any form of quantisation of gravity.

  In the present theory `quantisation' is a phenomenon that applies to the fields 
underlying spacetime solutions of the form $G^{\mu\nu} = -\kappa T^{\mu\nu}(Y, \bvh)$, 
that is
equation~\ref{cet} above. Only the right-hand `matter' side of this expression is 
effectively quantised, as a consequence of the degeneracy of field redescriptions, 
involving interchanges of  gauge and fermion fields for example, which underlie the 
solution. To attempt to impose `quantisation rules' on the left-hand `geometry' side 
of this expression would be to quantise the same object $G^{\mu\nu} \equiv T^{\mu\nu}$ 
\textit{twice} in two different ways. The external geometry $G^{\mu\nu}(x)$ itself 
implicitly incorporates a choice of $T^{\mu\nu}(Y, \bvh)$ and effectively  the 
identification of this 4-dimensional spacetime geometry
  is itself the \textit{mechanism of quantisation} for all non-gravitational fields.
That is, the possibility of multiple solutions of the form $G^{\mu\nu}=f(Y, \bvh)$ 
underlying the external geometric framework for perception of objects in the world is 
the \textit{reason} why the fields implicit in the energy-momentum tensor 
$T^{\mu\nu}:=G^{\mu\nu}$ are quantised, with no similar argument applying to the 
degrees of freedom of the spacetime geometry itself.

  Quantum field theory is however formulated against a flat spacetime background and  
we may also consider  the corresponding limit for the present theory. 
 For the respective theories of general relativity and quantum fields the geometry of 
the spacetime manifold and that of the internal gauge fields are \textit{independent} 
constructions. In the present unifying theory the relation between them is identified 
through a larger, all encompassing, symmetry group $\hat{G}$ for the full general form 
of the flow of time $L(\hat{\bv})=1$, linking the external and internal forces of 
nature, as for example seen in equations~\ref{gchift} and \ref{geinm2}. Their 
distinctive, complementary, features arise in the breaking of the full symmetry group 
over the base manifold $M_4$, as depicted in figure~\ref{mtogmaphr} for the
 $\lvte$ model. Considering the full forms $\lvt$ and $\lvfs$ in turn the surviving 
local gauge symmetry and resulting field interactions in this theory have been 
compared with corresponding features  of the Standard Model in chapters~\ref{chapesb} 
and \ref{secfd}.
 The properties of these interactions will be drawn out and made apparent through 
discrete particle phenomena, which themselves can only be fully explored in the 
present theory when the associated minute deviations from a flat geometry are fully 
embraced,  
 as described above and in the previous section.

  In a curved spacetime there are generally no preferred choices of Lorentz frames and 
through the local freedom in $l(x)\inn \soot$ general relativity can be interpreted as 
having some relation to gauge theory, as alluded to towards the end of 
section~\ref{gcatep}.  
  However, in the limit of a flat linear connection, with $\Gamma(x)\to 0$ in a 
Minkowski coordinate system,  it is also meaningful to define a global external gauge, 
that is a global basis for components of a tangent vector field in $\TM_4$, with a 
single choice of $l\inn \soot$. While the equations of physics are gauge covariant we 
expect them to take a particularly simple form when such a natural global gauge is 
possible. This is the case in special relativity and also in Newtonian physics for 
which a Galilean reference frame is typically preferred. In the limit of a flat 
spacetime  the freedom of local symmetry in $l(x)\inn \soot$ has effectively been 
\textit{broken} to the much more restricted freedom of a global symmetry on $M_4$.

  In fact, in the spirit of the present theory as introduced in section~\ref{perc}, 
the requirement of perception implies that the local Lorentz symmetry freedom of the 
local reference frames as a function of $x \inn M_4$ acts globally 
 over macroscopic scales
to a good approximation and hence is essentially broken down from a local to a merely 
global symmetry, and hence with far fewer degrees of freedom. This is the reverse of 
the usual case seen in gauge field theories in which a global symmetry is generalised 
to become a local symmetry leading to the interactions described in the Standard Model 
of particle physics for example.

  It is this  assumption of what is essentially a   hole in the full symmetry 
$\hat{G}$ 
 of $\lvh$ 
  carved out by the \textit{global} $\soot$ symmetry on $M_4$  that allows the 
deployment of a global Minkowski coordinate frame on the base manifold that in turn 
allows the expansion of each field as a sum over the linearly independent functions of 
a Fourier series on the base manifold, as described for the electromagnetic field for 
example in equations~\ref{acsin}--\ref{kcoeff}. The question concerning the relation 
between the interactions of such fields in the present theory in this limit and 
calculations performed in quantum theory has been considered in the previous sections 
of this chapter and is further elaborated in the following.

 Generally in physics there are numerous examples in which observable quantities 
parametrised by real numbers are analysed through expressions involving the algebra of 
complex numbers. To take a simple example an oscillating quantity such as the electric 
current in a wire of the form $I = I_0 \cos \omega t \inn \rrr$ can be expressed as  
$I = I_0 \mbox{Re}(e^{i\omega t}) \inn \rrr$. The straightforward mathematical 
properties of objects such as $e^{i\omega t} \inn \ccc$, under multiplication and 
differentiation for example, may then be exploited in calculations before the 
underlying physically  real (in the sense of `existing') part is extracted in terms of 
the mathematically real (in the sense of $\rrr$) part at the end of the calculation.  

   In quantum theory complex analysis is used directly from the foundations. Via 
either canonical quantisation or the path integral approach as the starting point for 
QFT, Feynman rules and the complex transition amplitude ${\mathcal M}_{fi}$  may be 
constructed on the way to extracting real-valued cross-sections or decay rates at the 
end of a calculation. Similarly the postulates of non-relativistic quantum mechanics 
are couched in terms of complex mathematical objects from the beginning -- with a 
complex wavefunction $\Psi(\bx)$ or state vector in a Hilbert space completely 
defining the dynamical state of a quantum system and empirical predictions obtained in 
terms of the real eigenvalues of Hermitian operators.

   These structures for quantum theory appear quite distinct from other applications 
of complex analysis in physics which, as for the example of the electric current $I = 
I_0 \mbox{Re}(e^{i\omega t})$ above, \textit{begin} with real-valued quantities. In 
this sense, by comparison, quantum theory appears to hang in the air, apparently 
lacking a more tangible conceptual foundation. The present theory aims to supply such 
an underlying physical basis for quantum theory in terms of the relative frequency of 
possible solutions for fabricating the 4-dimensional spacetime $M_4$ itself, with the 
geometry $G^{\mu\nu}(x)= f(Y,\bvh)$, as provisionally expressed in terms of the 
probability $P \propto D_+D_-$  of equation~\ref{pddddm}. Linked to the QFT 
probability $\vert {\mathcal M}_{fi} \vert^2$ via the structure of the real-valued 
quantity $\mbox{Im}({\mathcal M}_{ii})$ and the optical theorem, as discussed for 
figure~\ref{eemmeeee} and summarised in points `1)--7)' of section~\ref{secdopp}, the 
aim is to build the theory up from \textit{beneath} QFT through a complexification of 
the underlying probability computation, which is based on the degeneracy of solutions, 
on the way adopting some the mathematical machinery of QFT itself. 

   Historically  QFT was developed in the late 1920s on the coat-tails of the original  
quantum mechanics by promoting the wavefunction to an operator field (which was 
sometimes called `second quantisation', although there is still only \textit{one} 
quantisation). 
 By Fourier analysing the vector potential $A^{\mu}(x)$, as a free-field solution of 
Maxwell's equations, into normal modes and applying a quantum mechanical harmonic 
oscillator treatment to each mode independently photons, as quanta of the 
electromagnetic field, were the first `particle states' to be studied in a QFT. 
 Since in the present theory we began by making contact with QFT in the environment of 
HEP experiments the connection in the other direction, with quantum mechanics arising 
as a limit of QFT, should also be considered.
For example, the retarded propagator or Green's function of QFT, which enters the 
present theory as described for equation~\ref{vecredes}, can be taken to the 
non-relativistic limit in which it is found to be identical to the transition 
amplitude for single particle transitions in quantum mechanics. (The possibility of 
such a connection can be inferred from the relation of the function $\Delta_R(x-y)$ to 
$\Delta^+(x-y)$ through equations~\ref{deleqpm} and \ref{delrbc} and  the structure of  
$\Delta^+(x-y)$ in equations~\ref{delplus} and \ref{dpmd4}  in comparison with 
equation~\ref{ampqmpp}
 as discussed after equation~\ref{deldel}).

  While the path integral approach has not been found useful for establishing the link 
between the present theory and QFT, as alluded to before equation~\ref{sfixmom} in 
section~\ref{tranamp}, 
  the relationship between QFT and QM is perhaps most readily seen in terms of the 
path integral approach for which the same basic postulates apply in both cases. The 
transition amplitude $K$ is treated as a fundamental object and  identified with the 
sum over `all possible paths' of a phase factor $e^{iA/\hbar}$, where $A$ is the 
classical action associated with the path (see for example~\cite{Kaku} chapter~8). The 
mathematical properties of this phase factor are exploited in the structure of the 
theory, with the transition probability postulated to take the form $P = \vert K 
\vert^2$ such that the basic law of probability conservation is upheld. Both QFT, with 
a mathematical formalism for generating expressions associated with Feynman diagrams 
through higher-order functional derivatives, and the single particle theory of QM for 
the non-relativistic case, which can be generalised with higher-order Green's 
functions to describe multi-particle systems, may be derived from the path integral 
approach.
 The two cases of spontaneous symmetry breaking, in condensed matter physics and the 
Higgs sector of the Standard Model, alluded to in the previous section may also be 
described in very similar ways mathematically using the path integral approach, which 
is otherwise here seen as a useful formal method of performing calculations rather 
than relating to the conceptual basis for the present theory.

    In the present theory objects such as the Schr\"odinger wavefunction $\Psi(\bx)$ 
in QM and transition amplitudes, such as ${\mathcal M}_{fi}$ for QFT, are also 
considered as  mathematical constructions for pragmatic use in the relevant 
calculations of real observable quantities such as event  probabilities. 
 As complex representations none of these mathematical objects directly represent 
physical entities such as fields or particle states, although complex Fourier modes of 
the fields have been employed in the degeneracy count as represented for example in 
figure~\ref{xxphyy2}.

 For the present theory the fundamental objects in spacetime are the real-valued 
fields directly drawn out from the components of $\lvh$, and its symmetry actions, 
over the base manifold $M_4$, as described in the opening of this section. The 
energy-momentum possessed by such fields is strictly defined through $T^{\mu\nu} := 
G^{\mu\nu} = f(Y,\bvh)$, as described throughout this chapter, in terms of the 
4-dimensional spacetime geometry description via the Einstein tensor effectively 
composed of the underlying fields. Field interactions and transitions follow from the 
degeneracy of possible solutions. This definition of energy-momentum in $T^{\mu\nu}$ 
is independent of the field content, applying to the quantum as well as classical 
physics case, with 4-momentum conservation corresponding in all cases to the identity 
$G^{\mu\nu}_{\ph{\mu\nu};\mu}=0$.

 While in the present theory we begin with the form $\lvh$ and then identify the 
spacetime geometry over an extended manifold $M_4$, in QFT the starting point is a 
flat spacetime manifold itself.
  From this point of view in standard quantum theory the presence of energy-momentum 
$T^{\mu\nu}\neq 0$ alongside the flat spacetime assumption $G^{\mu\nu}=0$ not only 
directly contradicts the central field equation of general relativity but also evades 
any possibility of a unifying theory of quantum mechanics with gravitation.
  That is, since in QFT a flat 4-dimensional spacetime is a \textit{given} background 
arena the conceptual origin of indeterminate quantum phenomena as a degeneracy of
solutions for the underlying spacetime structure itself is entirely missed.

 Hence in quantum theory wavefunctions and amplitudes are introduced at the outset and 
unitary symmetry imposed in order to \textit{model} the probabilities of such 
phenomena. This approach dates back to matrix mechanics, presented by Heisenberg in 
1925, in resorting to a mathematical framework aimed at coherently linking observable 
phenomena without an underlying conceptual and physical motivation as the basis. On 
the one hand the present theory provides such an underlying basis for quantum 
phenomena in terms of a degeneracy of field solutions for the spacetime geometry, and 
on the other hand it should also be able to account for the original quantum mechanics 
of Heisenberg and Schr\"{o}dinger in the non-relativistic limit.

  Although the curvature of the spacetime geometry $G^{\mu\nu} = - \kappa 
T^{\mu\nu}\neq 0$ is unobservably small by many orders of magnitude on the scale of 
atomic or HEP phenomena the standard equations of quantum mechanics and QFT, in 
assuming a flat spacetime background, \textit{do} depend on the existence of a smooth 
continuum of spacetime points $x\inn M_4$ with the structure of a global Minkowski 
metric $\eta_{ab}$ on the manifold, since this is \textit{required} to give meaning to 
the location of wavefunction or operator field values in these theories. In the 
present theory this continuum takes the full metric form $g_{\mu\nu}(x)$ of general 
relativity, which is determined by the fields themselves, describing a spacetime which 
is only approximately flat.

  In the mathematical formalism of QFT the points $x$ of  an independent flat 
spacetime background are mapped into operators $x \to \hat{\phi} (x)$ which are 
defined by their action on the states of the system. Quantisation rules are imposed on 
the dynamical degrees of freedom of the field $\hat{\phi}(x)$ itself, as described for 
equations~\ref{kgosol2} and \ref{aacomr} in section~\ref{tranamp}, while the spacetime 
location $x$ is simply a parametrisation for the field in terms of a set of real 
number coordinates $\{x\}$. In non-relativistic quantum mechanics the operator 
$\hat{x}^a$, appearing in the discussion below equation~\ref{ppcomr}, represents the 
spatial location of a particle state, while there is no operator corresponding to 
time. However in QFT, which is invariant under the transformations of special 
relativity, there are no operators corresponding to either time or space.
Since these quantities are clearly `observables' quantum theory as it stands is not a 
\textit{universal} theory, rather extended spacetime provides an `external' background 
arena \textit{for} QFT, as it does for classical mechanics in the non-relativistic 
limit.

 General relativity \textit{is} the theory of external space and time and is itself 
\textit{not} a theory standing in need of quantisation, either on empirical or 
necessary theoretical grounds. While some approaches to `quantum gravity' seek to 
include gravitation and spacetime geometry \textit{within} the framework of an 
extended quantum theory, here in the present theory the phenomena of quantisation 
arise \textit{beneath} the surveillance of gravitation, with the geometric degrees of 
freedom associated with general relativity  hence outside the domain of quantum 
theory. As a consequence, for example, there are no gravitons for this theory, as 
discussed earlier in this section.

 The concept of time plays a central role both in relativity theory and in quantum 
mechanics. In general relativity the proper time, with the interval $d\tau$ of 
equation~\ref{dtgxx} for infinitesimal displacements, can be used to parametrise a 
series of events on the manifold, such as those that map out the spacetime trajectory 
of a physical object with 4-velocity flow $u^{\mu} = dx^{\mu}/d\tau$ as described in 
section~\ref{fdandtd}. In quantum mechanics the temporal evolution of a state is 
determined by the Hamiltonian operator $H$ (as introduced before 
equation~\ref{htlagx}),
which also describes the energy of the state, with the wavefunction $\Psi(t)$ for 
example
 in the time-dependent Schr\"{o}dinger equation satisfying:
\begin{equation}
\label{hamphi}
     i  \frac{\partial}{\partial t} \Psi =  H\Psi
\end{equation}
 (as exemplified in equation~\ref{stateev} for the evolution of the state vector in 
QFT,
 and alluded to after equation~\ref{phineg} with $H \equiv \hat{E}$ in quantum 
mechanics).
   One of the main difficulties with background dependent approaches to quantum 
gravity that apply the superposition principle of quantum theory to spacetime 
geometries, or make quantum transitions from one to another, is that, owing to the 
principle of general covariance in general relativity, there is no well defined way to 
map points in one spacetime to those of another. Labelling the points with coordinates 
does not help since under general covariance coordinates are of no physical 
significance, as described in section~\ref{gcatep}. This, in particular, means that 
there is no unique way to specify a map from a temporal derivative on one spacetime 
manifold to a temporal derivative on another. Hence the temporal evolution of a 
quantum state in equation~\ref{hamphi} cannot be transferred in any meaningful way 
between different spacetimes. This absence of a well defined independent temporal 
parameter is known as the `problem of time' in quantum gravity.

   It is a problem which does not arise within the present theory since here gravity 
itself is \textit{not} quantised and there is no `superposition of spacetimes'. 
  As for classical general relativity, here the emphasis is on complete 
\textit{four-dimensional} solutions for the spacetime geometry satisfying $G^{\mu\nu} 
= f(Y,\bvh)$, with indeterminacy and the probabilistic nature of quantum phenomena 
inherent fundamentally  in the degeneracy of many possible field solutions which 
underlie  the world geometry.
 Hence there is no difficulty in identifying a universal one-dimensional time 
parameter (such as the proper time in the local frame of any given observer) and the 
`problem of time' does not arise here, as it does for theories which place the 
temporal evolution as conceived  in a quantum theory at the forefront.
Rather here quantum effects arise \textit{underneath} gravity, and can be consistently 
parametrised in terms of coordinates on the single spacetime background of perception, 
with the ready availability of unambiguous temporal derivatives. 
  The structure of quantum theory is hence fused within the structure of general 
relativity, with 4-dimensional spacetime infused throughout with a 1-dimensional 
causal progression in time as employed for example in equation~\ref{hamphi}.

   Geometric structures, including the causal structure of spacetime, are described by 
degrees of freedom expressed in the metric $g_{\mu\nu}(x)$ and tetrad $\tetu$ fields, 
consistent with the Riemann tensor $R^{\rho}_{\ph{\rho}\sigma\mu\nu}(x)$, on a single 
spacetime manifold $M_4$.
 Indeed, this manifold is \textit{itself}  constructed out of the more fundamental 
underlying notion of progression in  temporal flow $s$ as expressed through $\lvh$. 
The projection onto the manifold results in relative time dilation phenomena as 
described in section~\ref{fdandtd}.
   In the 4-dimensional continuum \textit{each} observer carries a clock which 
provides a time parameter which may be applied in quantum experiments in the 
laboratory or for observations in general within the spacetime arena; with temporal 
parameters for mutual observers simply related by relativistic transformations. 

 The probabilistic nature in terms of what can happen as the \textit{outcome} of a 
laboratory experiment involving quantum phenomena
 motivates the construction of the quantum state or wavefunction
 locally parametrised through a 1-dimensional progression in time according to  
equation~\ref{hamphi} for each local observer
  and the employment of the associated quantum theoretical tools.

   In the present theory we begin with \textit{real} fields such as $A^{\mu}(x)$ of 
equations~\ref{acsin}--\ref{kcoeff}, with a complex decomposition into parts such as 
$e^{-ik\cdot x}$ which seem to resemble  a complex wavefunction $\Psi(\bx,t)$. 
 Indeed, similarly as applied to the positive frequency modes  of the quantum field  
component
  $\hat{\phi}^+(x)$ as described after equations~\ref{phipos} and \ref{phineg},  the 
Hamiltonian operator $H$ of equation~\ref{hamphi} can be applied to the complex 
component of the classical electromagnetic field in equation~\ref{acomp1} resulting 
in:
\begin{equation}
 \label{haka}
  H\:\! A^{\mu}(x) \, = \, k^0 \:\! A^{\mu}(x)
\end{equation}
  with eigenvalue $k^0$.  This is identical to the energy of the normalised 
electromagnetic field within the volume $V$ as described following 
equation~\ref{kextract}, which was extracted through the relations $-\kappa T^{\mu\nu} 
:= G^{\mu\nu} = f(A)$.
 Equation~\ref{haka} exemplifies how the `operator plus wavefunction' description can 
offer a concise way to extract properties such as the 4-momentum from the field as a 
useful tool for calculations. 
 Although the real-valued field $A^{\mu}(x)$ of equation~\ref{kcoeff} precisely 
describes  the actual field, if it may be reconstructed uniquely as the 
`realification' of a complex component such as equation~\ref{acomp1} (that is, by 
adding that equation to its complex conjugate) then the latter in principle carries 
all the information concerning the real physical  field, and also satisfies the same 
equation of motion as the real field as described after equation~\ref{treala}.

   However, in quantum mechanics the wavefunction $\Psi(\bx)$ represents a single 
observable particle,
  applying to a physical electron state in figure~\ref{eemmeds}(b) for example,  
    and it remains to be seen explicitly how such particle states may be described in 
terms of underlying fields and their interactions in the present theory. Hence an 
understanding of field renormalisation and the nature of observable particles, as 
discussed in the previous section,  will need to be further developed  in order to 
establish the full connection between the mathematical objects of the present theory 
and the pragmatic devices of quantum theory. 

   While representing a single particle a wavefunction $\Psi(\bx,t)$  is in general a 
continuous function of the spatial coordinates with intrinsically non-local properties 
in terms of the temporal evolution as a measurement is made and therefore exhibits a 
non-particle-like structure itself.
 In a measurement of position the quantum particle is observed to be in a particular 
spatial location which is determined by the wavefunction, to the extent that the 
squared modulus of this complex function $\vert \Psi(\bx,t) \vert^2$ determines the 
probability to detect the particle at that location, as alluded to after 
equation~\ref{ampqmpp}, with the wavefunction immediately `collapsing' to that 
measured point. 
 While carrying information concerning various physical quantities, when combined with 
the appropriate quantum mechanical operator, the wavefunction $\Psi(\bx)$, unlike the 
field $A^{\mu}(x)$ does not itself \textit{carry} energy  or any other physical 
attribute and hence there is no \textit{physical} discontinuity for the experiment of 
figure~\ref{eemmeds}(b) when the electron is observed at a particular location. 
Rather, in the interpretation of the present theory, the energy is contained in 
components of $T^{\mu\nu} := G^{\mu\nu}$ which is continuous everywhere in these 
experiments, as described  earlier in this section.

   From this point of view the quantum mechanical wavefunction reflects our best 
\textit{knowledge} of the \textit{range} of world solutions our current empirical 
situation is consistent with; it evolves in a determined way $U$ through passage of 
laboratory time, as governed by equation~\ref{hamphi} for a given Hamiltonian 
operator, in such a way as to yield probabilistic predictions for which particular 
solution state we shall find ourselves observing at the time of the next measurement 
(see for example \cite{Pen} p.592). Since the wavefunction is a non-physical entity, 
the so-called `collapse' or `reduction' $R$ of the wavefunction merely represents the 
change in our knowledge when such an observation is made, and is not itself a 
constituent property of the physical world.

  Hence the apparent conceptual difficulties concerning  `wavefunction collapse' are a 
somewhat artificial feature of quantum mechanics since the change in evolution law 
from the unitary $U$ for the wavefunction $\Psi$, describing a superposition of 
states, to reduction $R$ selecting an eigenstate $\Psi_i$ in the measurement, is just 
a  pragmatic  device for calculation (similarly as for the employment of the 
transition amplitude $\Mfi$ in QFT ) and does not directly describe the behaviour of a 
physical entity, such as represented by the gauge field $A^{\mu}(x)$ for example.

  From the perspective of the subjective laboratory view with a sequence of events 
seemingly evolving in time upon a given 4-dimensional background manifold some quantum 
phenomena appear mysterious, such as the `spooky action at a distance' as predicted 
and observed for Einstein-Podolski-Rosen (EPR) experiments.
  Such experiments demonstrate that quantum phenomena \textit{cannot} be accounted for 
by an underlying theory which is both local and deterministic, as constructed in terms 
of 
 `hidden variables' for example. In the present theory however the phenomena of EPR 
correlations and quantum entanglement in general are all sown into the fabric of the 
full 4-dimensional spacetime solutions under the geometry $G^{\mu\nu}(x)$. These 
observations are hence in principle naturally accounted for without any non-local 
interactions or behaviour and without hidden variables but with  indeterminacy fully 
embraced as a manifestation of the local degeneracy of possible fields, including the 
gauge field $A^{\mu}(x)$ and fermion field $\psi(x)$, necessarily featuring in 
solutions for $G^{\mu\nu} = f(Y,\bvh)$. Here the underlying fields such as 
$A^{\mu}(x)$ are intimately involved in the \textit{construction} of the spacetime 
geometry, rather than introduced separately as classical waves spreading out over a 
\textit{pre-existing} $M_4$ background.

 The local causality in the present theory incorporates the restriction that signals 
cannot  propagate faster than the speed of light, with special relativity holding 
locally as for general relativity.
  (In principle a form of the `equivalence principle', as described in 
section~\ref{gcatep}, might be adopted, but the employment of a `torsion-free' 
external geometry is a simplifying and provisional \textit{assumption} both for 
general relativity and for the present theory, as discussed in section~\ref{fdandtd} 
and also section~\ref{secuni}). Here `causality' means of course that the range of 
probabilities for possible future states, and not the \textit{actual} future state 
itself, is determined locally by the present state.

 Although the local redescriptions of the fields such as depicted in 
figures~\ref{xxphyy}, \ref{xxpxhyy} and \ref{xxphyy2} are arbitrary within the 
constraints of equations~\ref{conequa}  the overall theory is `deterministic' in the 
sense that \textit{all} possible worlds, all solutions, potentially exist. On the 
other hand events in the single solution of our world do necessarily appear 
indeterministic -- `God does play dice' from the point of view of observations in our 
universe.

     In the case of Schr\"{o}dinger's famous thought experiment the outcome can only 
be to perceive an alive \textit{or} a dead cat (\cite{Pen} p.808), while
 an entity described by the quantum
 state `$\mid$alive$\rangle \;+ \,\mid$dead$\rangle$' cannot be observed.  
The present framework incorporates a theory of perception through which 
  each of the two possible macroscopic states corresponds to a separate 
$G^{\mu\nu}(x)$ world, each necessarily observed subject to 
$G^{\mu\nu}_{\ph{ab};\mu}(x)=0$ and constructed out of the flow of time with $\lvh$, 
such that we cannot perceive \textit{both} large scale states simultaneously since 
they describe \textit{different} worlds. 
 The more practical experiments with an electron being detected at $A,B,C\ldots$ in 
figure~\ref{eemmeds}(b), or the muon detected at an angle $\theta$ 
figure~\ref{eemmeds}(a), 
 are associated with a spectrum of different worlds.

  Whatever the relative probability of the two alternative outcomes as determined by 
the 
 apparatus of a `Schr\"odinger's cat' type experiment,
  it is possible to consider two \textit{sets} of worlds each of which consists of a 
`coarse-grained' ensemble characterised by one of the two possible outcomes. More 
generally  we inhabit one of a much larger ensemble of possible worlds, each 
distinguished by the resolution of a vast number of locally indeterministic processes 
intrinsic to the 4-dimensional world solutions. With the range of worlds resulting 
from the many ways to construct $G^{\mu\nu}=f(Y,\bvh)$ over a 4-dimensional base 
manifold each solution, each universe,  is as real as ours. (This statement carries 
the caveat that each universe should support observers, in the sense described in 
chapter~\ref{chaptoot}).

 The availability of `many solutions' for $G^{\mu\nu}(x)$ in spacetime responsible for 
the indeterminacy in such experiments is reminiscent of the `many worlds' 
interpretation of quantum mechanics. However, here the theory \textit{has}  many 
solutions by nature, this feature is \textit{not} an interpretation of the theory. In 
the many worlds interpretation of quantum mechanics the wavefunction is taken 
literally as a real entity with the above observations of both `an  alive and  a dead 
cat' effectively interpreted as  a bifurcation of \textit{our} world as one of many 
such divisions in a `branching universe'. In the present theory the other worlds  
might be thought of existing  `out there' in a realm of possible mathematical 
solutions, unlike the more intimate picture of the many worlds interpretation.

   Here there is also no essential observer participation in `wavefunction collapse' 
in the sense of the `many minds' interpretation of quantum mechanics, rather the 
wavefunction, as a non-physical entity, is our own pragmatic construction employed to 
predict the likelihood of future events.
 On the other hand in the present theory the observer does have an innate role  in 
shaping the overall theory through the subjective nature of perception on the base 
manifold, which implies the breaking the full $\lvh$ symmetry and the ensuing physical 
structures. This perspective is influenced by the Kantian philosophy concerning the 
\textit{a priori} nature of perception in the form of space, time and causality, as 
will be further elaborated in  chapter~\ref{chaptoot} and in particular in the opening 
paragraphs of section~\ref{secauf}.

   During the early history of quantum mechanics  the meaning of the formalism in 
terms of the `Copenhagen interpretation', was a natural, pragmatic and provisional way 
of addressing the conceptual difficulties raised. This also marked a relatively 
conservative break away from the world of classical mechanics, combining the quantum 
with the classical aspects of the world  in a way that upheld the classical behaviour 
of experimental apparatus and the classical notion that physics exclusively  studies 
the properties of a \textit{single} universe, although now, however, one with an 
intrinsic element of uncertainty. While the postulates and mathematical structure of 
quantum theory has remained essentially intact and unchanged since the 1920s, the 
debate over the \textit{interpretation} of the theory continues into the  
 $21^{\mathrm{st}}$ century.  
 
  The main difficulty with the Copenhagen interpretation is the `measurement problem' 
concerning the grey area of interface between classical apparatus and the quantum 
system under investigation and the nature of the apparent `wavefunction collapse'. 
This issue is highlighted by the  `Schr\"odinger's cat' thought experiment and helped 
motivate the later many worlds interpretation alluded to above. In the present theory 
the measurement problem is resolved through the seamless employment of a classical 
notion of probability, defined in terms of the number of ways an event can happen, all 
the way down from the macroscopic apparatus to the underlying microscopic field 
redescriptions. This theory hence unifies the notion of probability for the classical 
and quantum domains, as applies for example to the experiments depicted in 
figure~\ref{eemmeds}.

   As well as having a common underlying origin  the \textit{meaning}  of the 
probability of an outcome for a quantum process (involving for example an experiment 
in figure~\ref{eemmeds} or the fate of Schr\"odinger's cat) and for a classical 
process (such as the roll of a dice or the toss of a coin) is subjectively the same, 
in terms of for example how we might make choices dependent upon such outcomes.
   In both the quantum and classical cases the outcome probability is calculated based 
on our knowledge of the set-up of the system before the experiment is performed.
 However there is also a significant objective difference in the nature of quantum and 
classical chance even in the context of the present theory.
 The difference is that in quantum theory the outcome is fundamentally unknowable in 
advance, whereas for the classical case the probability merely represents the 
practical limitations of our knowledge and our ignorance of the precise details of the 
initial conditions. The actual outcome of such classical experiments would in 
principle be calculable and fully determined if we could gather sufficient data and 
muster the necessary computational power (the improving accuracy of weather 
forecasting with improving technology provides an example). On the other hand, 
although in the many solutions there are many worlds and essentially everything that 
\textit{can} happen \textit{does} happen in some universe, quantum phenomena from our 
perspective in our world are objectively and inherently indeterministic.

For a given observed event, for a process such as $e^+e^- \to \mu^+\mu^-$ pictured in 
figure~\ref{eemmeds}(a), the question can be asked whether a \textit{particular} 
sequence of field exchanges \textit{actually} mediates the process between the initial 
and final states. In terms of the field sequence  $\ol{\psi}\gamma^{\mu}\psi \to 
\ol{\varphi}\gamma^{\mu}\varphi$  in figure~\ref{xxpxhyy} for example this corresponds 
to the question of whether there are particular values for $t_4, t_3, t_2$ and $t_1$, 
whether the intermediate $\ol{\psi}\gamma^{\mu}\psi$ field state represents a 
$\mu^+\mu^-$, $d\bar{d}$ or other fermion pair between $t_3$ and $t_2$, and  the value 
of the corresponding unconstrained internal 4-momentum degrees of freedom. In turn 
there is an endless list of possible field sequences, with field exchanges separated 
by intervals of time down to $\delta t \to 0$.

   These possibilities are not observable, but it is precisely the fact that they 
signify \textit{distinct} descriptions of the overall process that 
\textit{contributes} to the total probability to observe the event which \textit{is} 
statistically measurable. In a similar way that \textit{one} particular outcome of 
many possibilities is observed, such as a $\mu^+\mu^-$ or $\tau^+\tau^-$ final state 
at an angle $\theta$ to the incoming $e^-$ beam in figure~\ref{eemmeds}(a), from a 
philosophical point of view it is \textit{consistent} to think of the internal process 
as following \textit{one} particular sequence, such as via a $\mu^+\mu^-$ or 
$d\bar{d}$ internal fermion state in figure~\ref{xxpxhyy} for example, with particular 
values for the continuous degrees of freedom described above.
 (Although since there is an endless number of infinitely nested possible field 
redescriptions, as alluded to in the previous section, the idea of singling out `one' 
such sequence may be poorly defined).
 This is again analogous to the classical case in which the outcome of the roll of a 
dice, for example, is the result of \textit{one} particular dynamical path taken by 
the dice out of an infinite range of possibilities --  a path which although not 
predictable \textit{is}, however, observable to within practical limits of precision 
for the classical system. 

  This interpretation is of course required to also be consistent with all 
observations of quantum phenomena. These include interference effects, such as 
described in figure~\ref{eemmeds}(b), apparently well accounted for in terms of a
 superposition of  wavefunctions, which in turn feature in the course of the 
calculations involving complex number algebra, but
 which don't individually generally represent a particular `way' in which a process 
occurs.
   It will be necessary to trace a path from the many solutions picture of degeneracy 
in the present theory to the QFT Feynman rules for cross-section calculations based on 
the amplitude $\Mfi$, through equation~\ref{pddddm} as described in 
section~\ref{secdopp}, and further to the postulates of quantum mechanics and the 
construction of the wavefunction $\Psi(\bx)$ for the non-relativistic limit, in order 
to see how such phenomena (and their quantum mechanical description) are compatible 
with the present theory.

  The QFT calculation for the event rate at an $e^+e^-$ collider, for processes such 
as depicted in figure~\ref{eemmeds}(a), was presented in equation~\ref{diffevr} and 
described in section~\ref{crosss}. A doubling of the incoming luminosity, for example 
by doubling the bunch crossing frequency $f$ in equation~\ref{crosss}, or a doubling 
of the available final state phase space, in the final term of equation~\ref{diffevr}, 
leads to a direct doubling of the observed event rate. On the other hand on 
\textit{adding} new intermediate processes interference between the complex amplitudes 
${\mathcal M}_{fi}$ may lead to a \textit{reduction} of the event rate. Indeed, in 
practice the phenomenology predicted as a result of adding new hypothetical processes 
in such a calculation is sometimes studied in order to explain the observation of a 
lower than expected cross-section. However, according to the basic principles of the 
present theory the addition of new processes will only add to the `number of ways' 
through which to bridge an initial to a final state and always serve to increase 
cross-sections and decay rates.

 The question then may be asked how apparent interference phenomena arise in the 
present theory with probabilities based on degeneracy counts which always accumulate 
in a positive sense. However, it should be noted that there is no one-to-one 
correspondence between  components of the degeneracy count $D$ and contributions to 
the transition amplitude  ${\mathcal M}_{fi}$. Rather these two means of calculating 
the \textit{total} probability are 
 \textit{collectively} related by a correspondence of the form of 
equation~\ref{pddddm}, which in particular implies a mechanism for normalising the 
degeneracy count through a complexification of the calculation.

 Interference phenomena in quantum theory are more explicitly presented in the 
experiment of figure~\ref{eemmeds}(b). As alluded to above this system can be analysed 
in terms of two wavefunctions, each emanating from one of the two intermediate slits, 
and added together to form the pattern of constructive and destructive interference 
generating the probability distribution for events observed on the final screen.  
Again there is no direct analogue of the `superposition of wavefunctions'   in the 
present theory, and again there is no one-to-one correspondence between wavefunctions 
and elements of a degeneracy count.

  In the present theory such a degeneracy count is also \textit{not} based on the 
`number of ways' in which an electron, as a particle state, could pass through the 
slits,  but rather on the number of underlying field solutions for 
$G^{\mu\nu}=f(Y,\bvh)$ given the degeneracy of field redescriptions underlying the 
common geometry $G^{\mu\nu}(x)$ for the source $S$. Particle phenomena themselves 
arise as an apparent feature of these solutions. In fact, strictly speaking it is the 
phenomena of particle emission or detection, for example from the source $S$ or at the 
point $A$ on the screen in figure~\ref{eemmeds}(b), that emerge in these solutions, 
with no continuous trajectory of a particle-like entity  ever observed.
Only the particle-like interactions are ever actually directly recorded. 

 Even for the events of sophisticated experiments such as depicted in 
figure~\ref{figsld} 
 the apparent `tracks' of particles are reconstructed from a series individual 
detector hits, in particular in a tracking chamber. 
 `Joining the dots' in this way creates an illusion of continuous particle 
trajectories, as was presumed for the incoming and outgoing particle states sketched 
in figure~\ref{eemm} for example. 
  The theory is hence required to explain how field solutions for 
$G^{\mu\nu}=f(Y,\bvh)$
 incorporate apparent particle emission and detection phenomena, which in many cases 
create the impression of intermediate particle trajectories -- as an interpretation in 
part based on a close analogy with the properties of classical particles.   
Since the effective local field  interaction  volume can be  arbitrarily small  the 
associated elementary  particle states have no apparent size, consistent with a 
point-like interpretation.
 
  In conclusion, for the present theory particle effects and the probabilistic nature 
of quantum phenomena generally arise out of the merging of two \textit{necessary} 
features of the world. On the one hand the world we inhabit 
 must be \textit{perceivable}, as expressed mathematically in terms of geometric 
structures on an extended manifold such as $M_4$. On the other hand all such 
mathematical structures derive from a fundamentally one-dimensional \textit{temporal} 
progression which may be expressed in terms of  a general multi-dimensional form  
$\lvh$ together with its  symmetries.
 Resolving these two  requirements in a compatible manner leads to the equations of 
motion and physical properties of the tangible material world as perceived in 
spacetime and incorporating the phenomena of `quantum mechanical' transitions deriving 
from the degeneracy of  underlying field solutions.

 While the underlying field components of  $L(\hat{\bv}) = 1$ on $M_4$ are in 
principle subject to the \textit{full} symmetry degrees of freedom of the 
multi-dimensional  form of time  the geometrical interpretation needed to support the 
perceptual frame of the world requires the identification of a Riemannian geometry on 
the base manifold of the appropriate mathematical form with a \textit{lower} symmetry.
 Here, as for quantum theory in general, symmetry rather than scale is the key to 
quantum processes; although (as discussed shortly after figure~\ref{elecsel} in the 
previous section) with a higher degree of field symmetry more likely to be encountered 
on a `microscopic' scale  quantum phenomena are most frequently associated with such 
dimensions. The spirit of the principles of quantum mechanics is hence preserved in 
this new theory in unification with gravitation, with the identification $-\kappa 
T^{\mu\nu} := G^{\mu\nu}$ expressing  the field equation of general relativity. 

   The similar nature of 
  the interplay between the larger symmetry and the broken symmetry in the present 
theory to
  the situation in quantum mechanics can be exemplified by the Zeeman effect. The 
energy levels of the hydrogen atom are split by the presence of a uniform magnetic 
field, as a preferred direction in 3-dimensional space \textit{reducing} or breaking 
the rotational symmetry of the system from SO(3) to SO(2). Passing a beam of electrons 
through a magnetic field configured to select a certain spin state provides a further 
example. 
 Generally, in all cases of a measurement of a quantum mechanical system a structure 
of \textit{lower} symmetry, such as the configuration of laboratory equipment, is 
imposed upon the intrinsically \textit{higher} symmetry of the unobserved state.

 In the present theory quantum phenomena arise through the unavoidable \textit{a 
priori} imposition of the lower symmetry of 4-dimensional spacetime upon the general 
flow of time as a prerequisite for perception and observation in the world itself.
  Through our innate faculty to organise and interpret our experiences in the world 
through a coherent global geometrical manifold $M_4$ (playing the part of the 
directional magnetic field in the analogy with the Zeeman effect)
the full $\ese$ symmetry of $\lvfs$ is broken
 down to the local external symmetry $\soot$ together with the internal gauge group  
$\SML$ (which, as the surviving symmetries, collectively play the part of SO(2) in the 
Zeeman analogy).
  However, while in the Zeeman effect the magnetic field direction is  a 
\textit{particular choice} of experimental setup, in perception the Lorentz frame, 
within an approximately global $\soot$ symmetry, is a \textit{necessary form} for all 
physical experience of the world and hence applies to \textit{all} experiments and 
observations.

  Further, while the $\sotw$ symmetry of the uni-directional magnetic field imposed on 
a hydrogen atom with $\soth$ symmetry results in a discrete splitting of the atomic 
energy levels, the surviving $\soot \times \SML$ external and gauge symmetry of the 
4-dimensional perceptual field imposed over the full flow of time $\lvfs$ with an 
$\ese$ symmetry will be correlated with a discrete set of possible transitions of the 
microscopic world which determines the spectrum of elementary particles.
 (Strictly speaking the `surviving symmetry' is $\soot \times \suth_c \times \uo_Q$ 
since the electroweak symmetry $\sutw_L \times \uo_Y$ is itself broken down to $\uo_Q$ 
through its action on the external spacetime components of the `vector-Higgs' $\bh_2 
\equiv \bv_4 \inn \TM_4$. As described in section~\ref{secesef} the electroweak 
symmetry is also yet to be explicitly identified in terms of $\ese$ generators).
 In general the resulting phenomena will be exhibited in the observed properties of 
particles in HEP experiments as well as in the non-relativistic limit of quantum 
mechanics itself, as exemplified in figures~\ref{eemmeds}(a) and (b) respectively.

 While the physical structures of both gravitational and quantum theory are ever 
present in nature it is possible to consider  
 the limiting cases of the present theory as applicable to the corresponding empirical  
observations. The limit in which classical general relativity emerges on the one hand 
and a complementary limit through which an apparent quantum field theory emerges on 
the other hand can be described  in terms of two significant symmetries for our world 
with the external Lorentz group $\ol{H}=\soot$ (in the notation of 
section~\ref{hdasb}) as a subgroup of $\hat{G}=\ese$, with the latter being the 
symmetry of the full form of temporal flow $\lvfs$ as described in 
section~\ref{secesef}. These alternative limits can be characterised by the role of 
the linear connection $\Gamma(x)$ on the spacetime manifold $M_4$, as described in 
table~\ref{limits}.

 \begin{table}[htbp]
\centering
\begin{tabular}{|c|c|c|}
 \hline
    Symmetry     &         GR limit      &      QFT limit  \\  
 \hline					
  $\ol{H} = \soot$  &   local symmetry on $M_4$    &   global symmetry on $M_4$   \\
                  &   generally  $\Gamma(x) \neq 0$   &  can take  $\Gamma(x) = 0$ 
exactly \\
 \hline
  $\hat{G} = \ese$      &  effective macroscopic matter  
                  &  local $\ese /\soot$ symmetry   \\			  
 & $\;\; T^{\mu\nu}(x) := G^{\mu\nu}(x) = f(Y,\bvh) \;\; $ &  $\Rightarrow$ $Y(x)$ 
gauge fields  \\
  \hline
  \end{tabular}
  \caption{\setb Limits in which general relativity and quantum field theory arise.
  The employment of a local or global freedom for Lorentz frames with 
  $l(x) \inn \soot$ was also discussed earlier in this section in relation to gauge 
theory.}
\label{limits}
\end{table}

 The fact that GR and QFT emerge as almost exclusive complementary limits is not 
surprising given the notorious incompatibility of the respective mathematical theories 
and difficulties in uniting them under a single framework. However there is 
necessarily a trace of overlap even in the limiting cases. In the GR limit quantum 
effects are always locally present underneath the effective energy-momentum tensor 
which describes the apparent matter distribution, with macroscopic material properties 
shaped by the underlying quantum world. Similarly in the QFT limit particle 
interactions are clearly associated with regions of matter density and hence a minute 
but finite spacetime curvature is involved, which is a critical observation from the 
perspective of the present theory.

  As well as shedding light on the respective limits, the present theory may also 
address conceptual problems for physical systems where both gravitational and quantum 
effects are significant. For example the difficulties seen in some approaches to 
quantum gravity such as the `problem of time', as described earlier in this section, 
and the non-renormalisable nature of quantised gravity, as implied in the discussion 
following equation~\ref{xxyynexo} in section~\ref{fraot}, are avoided here since 
gravity itself is not quantised. 
 While one 
 aim of the present theory is to explore particle physics phenomena in the flat 
spacetime limit with Riemann curvature tensor components 
$R^{\rho}_{\ph{\rho}\sigma\mu\nu}(x) \to 0$, as an approximation to laboratory 
conditions to test the theory, the case for `quantum transitions' and `particle 
effects' for  $R^{\rho}_{\ph{\rho}\sigma\mu\nu}(x) \neq 0$, and in general for a 
\textit{highly} curved spacetime, is intended to be fully accounted for in this 
inclusive theory.

  The general form of the relation $T^{\mu\nu}(x) := G^{\mu\nu}(x) = f(Y,\bvh)$  in 
table~\ref{limits} 
 will be applicable even in locations of the universe with extreme spacetime 
curvature, such as in the vicinity  black holes and during the `Big Bang' epoch. For 
example, an environment in which \textit{both} gravitational and quantum effects are 
expected to be significant arises for the phenomenon of the emission of Hawking 
radiation (1974) in the highly curved spacetime in the proximity of a black hole, and 
similarly for the Unruh effect (1976) in which an observer undergoing a uniform high 
acceleration in the `vacuum' of a flat spacetime can detect thermal radiation. Quantum 
and particle effects should be calculable with the present theory in such 
environments, and also for Big Bang cosmology -- which will be discussed in the 
following two chapters.

  In QFT the Fock space representation is generally only valid for free fields in flat 
spacetime. The Fourier expansion of the field $\hat{\phi}(x)$ in 
equation~\ref{kgosol2} relies on the Poincar\'e symmetry of flat spacetime for the 
preferred basis of normal modes $e^{\pm ip\cdot x}$ and a corresponding preferred 
vacuum state $\vert 0 \rangle$. Particle excitations are built upon this ground state 
via the operators $a^{\dag}(\bp)$ and $a(\bp)$.
In flat Minkowski spacetime only global inertial frames of reference are used for 
which the particle content of a state, implied in the Fourier components, agrees for 
all observers.
 
 This construction is not possible in curved spacetime for which the reference frames 
of global coordinate systems are necessarily non-inertial.
For QFT in curved spacetime there is generally no  unique set of normal modes, which 
results in different inequivalent expressions of a particular QFT without a unique 
vacuum state, and the particle interpretation in turn becomes ambiguous. Hence in 
general there is no objective possibility of identifying either a vacuum or specific 
particle state for QFT in general relativity. However, interference between normal 
modes expressed in different general coordinate systems has the physical consequence 
that real particles may be created by gravitational fields. 

  Indeed physical particle states produced by gravitational fields or, equivalently, 
by accelerated observers are in principle detectable and hence \textit{do} represent 
real objective phenomena which in principle should be consistently accounted for in a 
complete theory. Similarly the particle states observed in high energy physics 
experiments are empirical objective entities. In all cases the detection of particle 
effects hinges on the nature of particle or field \textit{interactions}, without which 
the particles could not be observed. In the present theory it remains then to fully 
understand  the nature of particle phenomena, and their apparent physical interactions 
in general, as emerging out of the underlying  interactions of fields, as represented 
by a degeneracy of redescriptions, 
as we began to address in the previous section and will further consider in 
section~\ref{secrhp} in the discussion of figure~\ref{gtovac}. 

  In the present theory  the use of the Fourier transform expansion in 
equation~\ref{kcoeff} is merely an effective approximation that arises in the 
\textit{limit} of a flat Minkowski spacetime, and in which the apparent particle 
effects might most simply be analysed. Elementary particles are not fundamental 
entities out of which the world is built, they are a robust phenomenon that arises in 
the flat spacetime (and near vacuum) limit, as alluded to in the opening of 
section~\ref{sechepe} and as studied in experiments such as depicted in 
figure~\ref{figsld}. The properties of  `particles' may be less robust in highly 
curved spacetime, and more difficult to calculate than in the fixed limit of flat 
background manifold,  but there is no fundamental conceptual difficulty.

 The field redescription $\ol{\psi}\gamma^{\mu} \psi \to A^{\mu} \to 
\ol{\varphi}\gamma^{\mu} \varphi$ of figure~\ref{xxphyy} is presumed to be locally 
enveloped in a 
 spacetime geometry $G^{\mu\nu}(x)$ which takes a form resembling that of 
$T^{\mu\nu}:=G^{\mu\nu}$  in figure~\ref{gacos}. If such an interaction takes place in 
the
  prevailing environment of a highly curved spacetime, for example in the proximity of 
a black hole, then to a certain extent all of the fields, $\psi(x)$, $\varphi(x)$ and 
$A^{\mu}(x)$, will be `bent the same way' and hence  processes  such as depicted in 
figure~\ref{xxphyy} might be largely unaffected. Similar underlying field 
redescriptions in combination with immense  gravitational tidal forces might then 
provide a description of black hole evaporation in the context of the present theory.

  The question can also be asked concerning the nature of phenomena for yet more 
extreme spacetime curvature, such as in the region of a black hole `singularity' or
 generally
 corresponding to a yet higher scale of energy.  
 In the context of figure~\ref{runcup}  the GUT scale, at around $10^{15}\,$GeV, in 
marking a point of gauge coupling unification ought to be of significance for the 
present theory in terms of the phenomena of the internal forces, while the external 
gravitational field will be treated in the same manner as for the low energy 
phenomena.
 Further, in the present theory
 gravity itself is not `quantised', there are no `graviton' particles, and the Planck 
scale  at around $10^{19}\,$GeV may just be a dimensional quirk  with no particular 
significance.  
 Hence arbitrarily high energy densities and arbitrarily high spacetime curvature 
might be considered in the present theory in a continuous manner without limit.

  In summary, from the point of view of the present theory the postulate in quantum 
theory that an event probability is determined by the square of the absolute value of 
an `amplitude', with unitary symmetry imposed to ensure the structure is consistent 
with the basic laws of probability, should be considered as a \textit{provisional} 
construction standing in need of an underlying conceptual basis and physical 
explanation. Such an explanation would be preferred in place of any theoretical 
`postulate', and here it lies in the idea of the natural degeneracy inherent in the 
number of ways local field solutions may be found for $G^{\mu\nu} = f(Y,\bvh)$ for 
processes such as those observed in figure~\ref{eemmeds} and more generally. 

 This is the key to combining general relativity and quantum phenomena in a single 
complete and unified theory. Indeed, given the prohibitive conceptual and mathematical 
difficulties encountered in attempting to unify these two pillars of $20^{\mathrm th}$ 
theoretical physics it seems likely that a significant concept or postulate on at 
least one side must yield some ground. Here the \textit{definition} of probability in 
terms of amplitudes in quantum theory seems a reasonable place for this, with the 
amplitudes and wavefunctions of quantum theory then representing calculational tools 
employed in an \textit{intermediate} complexification of a computation. This approach 
has been exemplified by unravelling the QFT event rate calculation of 
equation~\ref{diffevr} and making the case for replacing the contribution from the 
amplitude ${\mathcal M}_{fi}$ by a quantity based on a degeneracy count $D$ via the 
associations of equation~\ref{pddddm}.

 This foundation also unifies the notion of probability with the classical concept in 
the sense of essentially referring to the `number of ways' that a process can occur 
given a particular initial state or situation. However, while classical probabilities 
concern the number of ways that things can happen \textit{in} spacetime $M_4$, 
  quantum probabilities concern the more fundamental question of the number of ways in 
which the spacetime manifold $M_4$ \textit{itself} can be constructed with a world 
geometry described by $G^{\mu\nu}(x) = f(Y,\bvh)$.  
 Further, in principle this approach to quantum phenomena also leads to a 
clarification of the meaning of `renormalisation' as discussed for 
equation~\ref{brddmm} in the previous section.

  For theories which postulate extra spatial dimensions, such as the Kaluza-Klein 
theories described in chapter~\ref{kktheory}, our 4-dimensional spacetime world is 
contained within the larger space, for example as a 4-dimensional \textit{brane} 
embedded within the higher-dimensional \textit{bulk} manifold or with the extra 
dimensions being `compactified', as discussed in section~\ref{secbkkt}. 
 For the present theory founded on one-dimensional temporal flow the extended
 physical world is perceived through the structure and symmetries of the 
 multi-dimensional form $\lvh$, with the degeneracy of solutions for constructing such 
a 4-dimensional world underlying the phenomena of quantum theory while the external 
spacetime geometry itself conforms with the structure of general relativity.

 As well as combining general relativity and quantum theory in a consistent framework 
within which the two theories are separately preserved in essence, the complete 
conceptual theory is based on sound intuitive principles, founded upon the ever 
pervading multi-dimensional  form of temporal flow $\lvh$ rather than upon seemingly 
arbitrary, mysterious or purely pragmatic assumptions. The theory should of course 
also be able to make predictions and be found to be in full agreement with all 
observations.
 Such a correspondence has been initiated in chapters~\ref{chapesb} and \ref{secfd}
 with regards to comparison with the Standard Model of particle physics.
 Further, on incorporating all physical scales, including that of HEP phenomena,  in 
principle the present theory is expected to be profusely testable.
 
  All the underlying fields in nature, which underlie for example electron and photon 
particle states, are in continual interaction through mutual  indistinguishability  
under the external geometry $G^{\mu\nu}(x)$ --  from the interaction region of an HEP 
experiment such as that in figure~\ref{figsld}, to atoms and molecules, through to 
biological organisms, planets, stars and galaxies, with the underlying processes 
moulding a smooth and continuous geometry $G^{\mu\nu} = f(Y,\bvh)$ with all the 
quantum phenomena embedded within and
 in turn, through $T^{\mu\nu} := G^{\mu\nu}$, shaping the structure and  apparent 
phenomena of the material world.

  As well as the extreme environment of a highly curved spacetime alluded to above, 
the complementary question concerning  the nature of the `vacuum state' can also be 
considered.
  Even in the apparent vacuum, away from tangible physical matter, in general a form 
of $G^{\mu\nu} = f(Y, \bvh)$ must be present throughout $M_4$ in order to describe the 
spacetime geometry. This structure might in principle implicitly include a form of 
effective `vacuum energy',  incorporated into the spacetime geometry  and describing 
the effects of a cosmological constant $\Lambda$, at least to a good approximation, 
and hence in turn accounting for observations of the large scale structure of the 
universe.
Indeed, more generally,
 as well as terrestrial laboratory phenomena the present theory has also been 
developed with the cosmological scale in mind, and hence in the following two chapters 
we review aspects of cosmology in the context of the new theory.


\pagebreak
\chapter{Cosmology}
\label{chapcos}

\section{The Large Scale Structure of the Universe}
 \label{tlssotu}

  While the previous chapter focussed on the application of the present theory to the 
smallest observable scales, regarding in particular the  quantum field and particle 
phenomena studied in high energy physics experiments, here we return to consider 
general relativity and gravitation, continuing the thread from sections~\ref{subwal} 
and \ref{fdandtd} in the light of the intermediate chapters, as applied up to the 
largest empirically accessible scale of the observable universe and beyond. 
 In the context of the large scale structure of 4-dimensional spacetime the right-hand 
side of equation~\ref{getypsi} can generally be considered to describe the effective 
macroscopic form of apparent matter terms, with $-\kappa T^{\mu\nu} :=  G^{\mu\nu} = 
f(Y,\bvh)$  for this equation, that is in the GR limit as summarised in 
table~\ref{limits}, with the practical normalisation factor of $-\kappa$ inserted. 
However an understanding of the impact upon  the spacetime geometry of the underlying 
microscopic fields and their interactions will also be directly relevant both for the 
universe at the present epoch as well as in its much earlier history. Indeed since the 
energy density in the early universe reaches and surpasses that attainable in high 
energy physics experiments, the environment of the immediate aftermath of the Big Bang 
may itself provide a possible test arena for theoretical particle physics through  any 
imprint which the corresponding phenomena may leave in the structure of the cosmos 
which is still observable today.

  In the following two sections we review some of the main features of  standard 
textbook cosmology, as deduced from and motivated by  empirical observations. In the 
following chapter we then
   collect and describe a series of observations concerning the present theory which, 
at a qualitative level at least, correlate with a number of aspects of modern 
cosmology.
 Without making a quantitative argument in terms of cosmological parameters these 
aspects include the dark sector of implied matter and energy in the universe and the 
origin and nature of the Big Bang and the very early universe itself.

  The rather direct application of the conceptual scheme described in the previous 
chapters to the cosmological scale will first be outlined briefly in this section. 
This application is possible since the general picture of the standard cosmological 
model of the evolution of the universe according to Einstein's field equation of 
general relativity, given broad underlying assumptions concerning the large scale 
structure of spacetime, is naturally compatible with the present framework.

 Based on the translation symmetry represented in figure~\ref{spillout} we described
   in sections~\ref{gfotf} and \ref{perc}  how the perceptual background of a flat 
$\soth$ symmetric spatial manifold $M_3$ could be effectively derived through the 
structure and symmetries of the flow of time expressed in the form $\lvth$ of 
equation~\ref{flow3d}.
  On extending this model to the case of a full $\hat{G}=\sofi$ symmetry of $\lvfi$ 
projected over $M_3$, as described for figure~\ref{mtogmaph} in section~\ref{hdasb}, a 
finite external (and also internal) curvature was obtained.
 Subsequently a $\hat{G}=\sootn$ model  for the case of a 4-dimensional spacetime 
$M_4$ as pictured in figure~\ref{mtogmaphr} was described in section~\ref{reaic}, 
again introducing
    minor distortions from a flat geometry corresponding to the effects of general 
relativity.
	These geometric distortions are presumed to be undetectable in everyday experience 
--
that is out of the degrees of freedom of the full symmetry group  of $\lvh$ projected 
onto the base manifold $M_4$ we require the local $\soot \subset \hat{G}$ subgroup to 
be broken down to an approximately global symmetry of the 4-dimensional spacetime 
manifold, incorporating an approximately Euclidean 3-dimensional space, forming the 
backdrop for our perception of physical objects in the world. 

This requirement is borne out by our observations of the world around us on the scale 
of the solar system for which the non-Euclidean effects of general, as well as 
special, relativity are indeed \textit{imperceptible}. The non-Euclidean effects such 
as the deflection of starlight passing close to the sun are well beyond the reach of  
 casual observation. On the other hand
local observations such as the accelerating fall of an apple from a tree might at 
first sight be ascribed to a `force of gravity' active within a flat arena of space 
and time, rather than to an effect of a curved spacetime arena itself.
 The apparent flatness of the local geometry both from the point of view
 of our everyday experience of the world and also for most scientific experiments 
accounts for the fact that the existence of a non-Euclidean element of 3-dimensional 
space combined with 1-dimensional time was not recognised, through centuries of 
scientific developments, until the early 1900s. 

  Carrying the same principle of our innate requirement of perception in the world 
  to the largest scale in which we encompass everything in our observable universe it 
seems natural to ask how it could be possible for \textit{our} existence and 
experiences to influence in any way the shape or form of the universe over regions 
measuring billions of light-years across. However, a central point of the work 
presented in this paper is that here we consider the whole universe to be the physical 
manifestation that is created through and \textit{within} the possibility of our 
experiencing it and is therefore shaped by the necessary form of that possibility, as 
we shall describe further in chapter~\ref{chaptoot}. 
 The initial naive picture that hence comes to mind is then based upon the assumption 
of an
 approximately Euclidean background extending to the largest observable scale, 
neglecting the (generally imperceptible) local variations from flatness, with the flow 
of time propagating through a 4-dimensional manifold as depicted in 
figure~\ref{cosmos}. This picture represents the largest scale realisation, for our 
own 4-dimensional universe, of the general idea introduced in figure~\ref{twodworld} 
of section~\ref{perc} for the model 3-dimensional world.  

\begin{figure}[htb]  
\centering
\epsfxsize=10cm
\leavevmode
\epsffile[0 0 1209 1123]{\gpath aPfig121e}
\caption{\setb Propagation of galaxies, clusters of galaxies and large scale physical 
structures through the $M_4$ spacetime manifold, with the temporal dimension directed 
from left to right and one spatial dimension suppressed.}
\label{cosmos}
\end{figure}

  We further recall that in the full theory the components of the 4-dimensional vector 
field $\bv_4(x)$ on $M_4$ are considered to be locally embedded in a
 higher-dimensional form of temporal flow $\lvt$ via the space of
  $\bv_{27} \equiv \mcX \inn \htho$ matrices through equations~\ref{hvvv} and 
\ref{hinhtho} of section~\ref{extsym}, and in turn within the form $\lvfs$ via the 
elements $x\inn F(\htho)$ in the form of equation~\ref{fhthopart} as described in 
section~\ref{secesef}. 
    From a purely \textit{mathematical} point of view the intermediate 4-dimensional 
case is readily
bypassed in generalising from a 1-dimensional temporal progression to
 higher-dimensional forms, here represented by a 27-dimensional and on to a 
56-dimensional form of time with a full $\hat{G}=\esi$ and $\hat{G}=\ese$ symmetry 
respectively. 
However, in order to \textit{physically} experience or perceive any structures 
implicit within the general form of time a lower-dimensional part, with mathematical 
properties isomorphic to the geometrical forms required to define the perception of 
objects in the world, is
 projected out, or syphoned off, from the full general form of temporal flow. 
 
  In our world this has been taken to be achieved through extracting
   $\v_4 \equiv \bh_2 \inn \htwc \subset \htho \subset F(\htho)$,  with the quartic 
form $\lvfs$ for $\bv_{56} \inn F(\htho)$ having an $\ese$ symmetry, and projecting 
the vector component $\bv_4 \subset \bv_{56}$ onto $\TM_4$. The form  $L(\v_4) = 
\mbox{det}(\bh_2) = h^2$ has the symmetry group $\sltc \subset \ese$ which, as 
described in section~\ref{lsspin},  is the double cover of the external Lorentz group 
acting on Lorentz 4-vectors. While the representations of the complementary  internal 
symmetry upon the components of $\lvt$ and $\lvfs$ are reminiscent of Standard Model 
properties, as described in  
chapter~\ref{chapesb} and section~\ref{secesef} respectively, an extension for example 
to an $\ee$ symmetry of a form $\lvtfe$, as outlined hypothetically in 
section~\ref{sosmfi}, may be needed to fully incorporate the structure of the Standard 
Model.

In principle this projection, on employing the associated 4-dimensional translation 
symmetry of the form $\lvfs$, opens out the local Lorentz subgroup into an 
approximately global symmetry on the $M_4$ manifold, breaking the full local $\ese$ 
symmetry while at the same time actually generating the spacetime manifold itself. 
That is the $M_4$ manifold is created in the act of the symmetry breaking projection 
from $\bv_{56} \inn F(\htho) \to \bv_4 \inn \TM_4 \equiv \htwc$, with $v^a(x)$ for 
$a=0,1,2,3$ being the components of the Lorentz tangent vector field in a 
\textit{local} Minkowski coordinate frame on the manifold as described in 
section~\ref{fdandtd}. In terms of the initial picture, deriving from the translation
 symmetry as described for figure~\ref{spillout}, the metric 
$g_{\mu\nu}(x)=\mbox{diag}(1,-1,-1,-1)$ may in fact be adopted \textit{globally}. This 
continues to be the case to a good approximation in practice even when planets, stars 
and galaxies are incorporated as depicted in figure~\ref{cosmos}.  However, while 
maintaining the approximation of neglecting  the local distortions correlated with 
matter in these forms, it will \textit{not} be possible to adopt a global
Minkowski frame when considering the overall cosmological point of view.

 That is, while compatible with an approximately flat $\soot$ frame locally, on the 
scale of the solar system for example, on larger spacetime scales through to the vast 
arena of the universe studied in cosmology there is no longer a necessity for the 
geometric form of spacetime  to describe a perceptual frame even in approximation. 
Further, extrapolating beyond our possible experience or observation of the world the 
4-dimensional geometrical interpretation may itself at some point break down 
altogether. This may apply to extreme regions such as black holes at any epoch and the 
structure of 
 the very early history of the universe and the Big Bang.
 The effects of particle physics studied in an Earth-bound laboratory, and in the 
previous chapters, are also likely to play a significant role when extrapolated to the 
extreme conditions of a highly non-Euclidean spacetime geometry, as alluded to in the 
previous section.

  In the following section the standard cosmological model and a range of possible 
large scale metric solutions will be reviewed, before turning to the very early 
universe in section~\ref{secinf}. 
 This  will provide a basis for the perspective of the present theory to be presented 
in the following chapter.


\section{The Standard Model of Cosmology}
\label{sectsmoc}

  While the Standard Model for particle physics has been constructed in recent decades 
in parallel with the findings of high energy physics experiments, the underlying tools 
of quantum field theory were originally developed in the 1920s through to the 1940s. 
The framework for cosmological models was originally developed over a similar period 
following soon after the publication of general relativity in 1915 and through to the 
1930s, although again here the `standard model of cosmology' has only become 
established in recent decades in the light of the empirical data revealed with modern 
observational technology. In this section we examine  the picture of the cosmos and 
the standard cosmological model that has emerged out of this work  (see for 
example~\cite{Pea,Narl}).

  The standard approach incorporates general relativity, as reviewed in 
sections~\ref{riegeo} and \ref{gcatep}, for which the empirical observation that 
spacetime curvature is strongly correlated with the presence of matter is expressed 
through the field equation $G^{\mu\nu} = -\kappa T^{\mu\nu}$. This equation postulates 
the equality of the Einstein tensor $G^{\mu\nu} = R^{\mu\nu} - \frac{1}{2} R 
g^{\mu\nu}$ with the energy-momentum tensor $T^{\mu\nu}$, to within a constant of 
proportionality.
 This approach is here summarised in terms of three quotes from \cite{Rob}:
 
\begin{itemize}
 \item 
 `We wish to relate the curvature of spacetime to the presence of matter, since 
gravity appears in the neighbourhood of matter' (\cite{Rob} p.232).
 The proportionality constant is determined for weak fields by comparison with 
Newton's theory of gravity and found to be $\kappa = 8\pi G_{\! N}$, where $G_{\! N}$ 
is Newton's constant, as described for equation~\ref{Eins}.

\item 
  `It will be assumed that the metric in a nearly empty universe is nearly Minkowski'
    (\cite{Rob} p.229). Essentially this implies that a flat spacetime arena $M_4$ is 
presupposed before the introduction of matter. Within relatively local portions of the 
universe a  flat Minkowski spacetime can act as a boundary condition in regions 
sufficiently far from matter, as for the example of the Schwarzschild solution in 
equation~\ref{ttrtp}.

\item
  `...the vanishing of the divergence of $G^{\mu\nu}$ as a mathematical identity 
implies the vanishing of the divergence of $T^{\mu\nu}$' (\cite{Rob} p.232). That is 
in light of the contracted Bianchi identity $\gmo$ this conclusion follows immediately 
given that the Einstein field equation is assumed to hold.
\end{itemize}

 The divergence-free nature of $T^{\mu\nu}$ can be interpreted as the conservation of 
energy and momentum in the limit of an approximately flat spacetime, 
as described in the opening of section~\ref{subwal}, since for a suitable choice of 
coordinates with linear connection $\Gamma \to 0$ we have 
$T^{\mu\nu}_{\ph{\mu\nu},\mu}=0$.
 On the other hand,  this equation for the conservation of energy and momentum is 
often cited as a starting point and \textit{then} expressed in a general curved 
spacetime as $\tmo$, and it is this observation that then justifies the introduction 
of $T^{\mu\nu}$ 
   on the right-hand side of the field equation itself, with the Einstein tensor on 
the left-hand side, since it \textit{happens to be} also the case that $\gmo$ as the 
contracted Bianchi identity. 
 Consistent with this requirement an additional divergence-free term may be 
postulated, associated with a `cosmological constant' $\Lambda$, as may be necessary 
to account for the empirically observed evolution of the universe, yielding the full 
standard field equation as quoted in equation~\ref{einlamt} and reproduced here (where 
we turn here to a convention of generally using lower indices in such expressions):
\begin{equation}
\label{Einfield}
    G_{\mu\nu} + \Lambda g_{\mu\nu} = - \kappa T_{\mu\nu}.
\end{equation}
 
  In 1922 Aleksandr Friedmann made two classes of assumptions in order to obtain 
solutions for the spacetime structure of the universe as a whole. The first class 
required that the gravitational field should satisfy the equation~\ref{Einfield}, that 
is the Einstein field equation including the cosmological constant term (Friedmann 
considered the case for both arbitrary $\Lambda$ as well as $\Lambda=0$),  with matter 
represented as a pressureless fluid with energy-momentum tensor $T_{\mu\nu} = \rho 
u_{\mu}u_{\nu}$ where $\rho$ is the proper  density of matter. In 1927 Georges 
Lema\^{i}tre, working independently of Friedmann, considered the more general case by 
including a spatially isotropic pressure term and hence treating matter as a perfect 
fluid with an energy-momentum tensor in the form of equation~\ref{gtruup}, that is:
\begin{equation}
 \label{cosgtruup}
   T_{\mu\nu} = (\rho + p) u_{\mu}u_{\nu} - p g_{\mu\nu}
\end{equation} 
   where $p$ is the pressure and here $u_{\mu}$ represents the 4-velocity of the flow 
of galaxies, as depicted for example in figure~\ref{cosmos}.
  This energy-momentum is then substituted into Einstein's equation~\ref{Einfield} to 
give:
\begin{equation}
\label{CosLem}
    G_{\mu\nu} + \Lambda g_{\mu\nu}= -\kappa (\rho + p) u_{\mu}u_{\nu} + \kappa p 
g_{\mu\nu}.
\end{equation}
 We now know that the contribution of radiation pressure to the evolutionary dynamics 
of the universe is most significant for around  the first 10,000 years of its history,
 with the contribution of the matter density becoming comparable around 50,000 years 
after the Big Bang and subsequently
 increasingly dominating over the radiation term. Hence the idealisation of Friedmann, 
treating the flow of galaxies as a dust or pressureless fluid with $p=0$, makes a very 
good approximation for modelling the cosmic evolution, particularly since the epoch of 
the `decoupling' of matter from radiation around 372,000 years \cite{PDG}  after the 
Big Bang, still relatively early in the 13.8 billion year history of the universe.

  The second class of assumptions made by Friedmann in order to obtain a solution 
concern the nature of preferred coordinate systems and more direct restrictions on the 
form of the metric deriving from symmetries imposed on the spacetime.
 Based on the picture of galaxies pursuing non-intersecting world lines, for which 
figure~\ref{cosmos} represents only a particular special case, 3-dimensional spacelike 
hypersurfaces, orthogonal to and parametrised by a global timelike coordinate $t$,  
are assumed to have a uniform $t$-dependent 3-dimensional scalar curvature $R_3(t)$ 
independent of the location on a given 3-dimensional spatial surface. The unambiguous 
cosmic time $t$ is taken to be the proper time as measured for any given galaxy.
 The `Copernican view', that here on Earth we do not inhabit a central or preferred 
location of the universe, is subsumed into the `cosmological principle' which asserts 
that at any given cosmic time $t$ the universe on large scales is spatially 
homogeneous and isotropic about any location.

   From an observational point of view at the present epoch the assumption of 
homogeneity may be justified by the smallness of fluctuations in the distribution of 
galactic clusters on scales larger than a few 100 Mpc (megaparsecs, where 1 parsec is 
around 3.26 light-years) in an observable universe with distance scales of up to the 
order of the Hubble radius: 
\begin{equation}
  \label{hubrad}
R_H := c/H_0 \simeq 3000h^{-1}\;\mbox{Mpc}
\end{equation}
 with $h\simeq 0.7$ and the Hubble constant $H_0$ defined below for 
equation~\ref{hubcon}. Similarly, the assumption of isotropy may be justified by the 
evenness of the cosmic microwave background (CMB) radiation to within of order 1 part 
in $10^5$ as observed over the full coverage of the sky from the Earth. 

   The mathematical basis for the assumptions of the cosmological principle was 
studied thoroughly by H.P. Robertson and independently by A.G. Walker in the 1930s. 
The 3-dimensional hypersurfaces for constant $t$ are everywhere orthogonal to a 
congruence of geodesics given by the integral curves of the vector field $\partial/ 
\partial t$. For each solution the hypersurface curvature $R_3(t)$, while it can vary 
with time, remains always either positive (3-sphere), negative (hyperboloid) or zero 
(for a spatially flat universe). The Robertson-Walker line element is the most general 
spacetime metric compatible with homogeneity and isotropy and can be expressed in 
terms  of intervals of  proper time $\tau$ as: 
\begin{equation}
\label{CosRob}
   d\tau^2 = dt^2 - a^2(t)\left[ \frac{dr^2}{(1-kr^2)} + r^2(d\theta^2 + 
\mbox{sin}^2\theta d\phi^2) \right] 
\end{equation}
 where the parameters $a(t)$ and $k$ will be described below.
   With the world line of any given idealised galaxy expressed in terms of constant 
3-dimensional spatial spherical coordinates $\{r, \theta, \phi\}$  the full 
4-dimensional set $\{t, r, \theta, \phi\}$ describes a comoving coordinate system with 
the cosmic time parameter $t$ equivalent to the proper time $\tau$ elapsed for the 
galaxy.

  As for any metric for 4-dimensional spacetime here the convention is to take the 
components of $g_{\mu\nu}(x)$ to have the dimension of length squared, that is $[ 
g_{\mu\nu} ] = [ d\tau^2 ] = L^2$ (with the dimension of length $L$ equivalent to that 
of time $T$ since $c=1$, while in the notation of the discussion following 
equation~\ref{xxyynexo} the mass $M$ dimension of $g_{\mu\nu}$ is $D=-2$). In turn  
the components of the metric inverse have the dimension $[g^{\mu\nu}] = L^{-2}$. (We 
note that in the present theory the internal Killing metric components, such as 
$g_{\alpha\beta}$ in equation~\ref{ggghlb} for the case of a Kaluza-Klein metric,  are 
\textit{not} interpreted as representing a physical length in a higher-dimensional 
space in the present theory, however here we are dealing with purely external metric 
components in 4-dimensional spacetime).
  This convention is consistent with the principle of general covariance in general 
relativity, as described in section~\ref{gcatep}, which implies that in general no 
physical significance can be attached to a set of coordinates, which consists of 
numerical parameters of dimension $L^0$.
As implied in the name, only when the manifold is endowed with a `metric' are lengths 
defined.
 In fact all parameters on the right-hand side of equation~\ref{CosRob}, including the 
scale factor $a(t)$, can be considered to be dimensionless quantities.
 Since an implicit factor of $g_{00} = 1$ carrying the dimension $L^2$ accompanies  
the $dt^2$ term in equation~\ref{CosRob}, the cosmic time coordinate $t$ may be 
\textit{interpreted} as having the dimension of $T\equiv L$. A similar interpretation 
might be applied to  spatial coordinates in certain cases, in particular for Euclidean 
coordinates $\{x,y,z\}$ in the limit of a flat spacetime with Minkowski metric.

  The sign of the dimensionless real number $k$ in equation~\ref{CosRob} indicates the 
sign of the 3-space curvature. For $k=0$ the spatial hypersurfaces are flat, although 
even in this case the 4-dimensional curvature will generally be finite.
  For non-zero values of $k$ the coordinate $r$ may be redefined as $r\to r/\vert k 
\vert^\frac{1}{2}$, and the scale factor as $a \to a\vert k \vert^\frac{1}{2}$, such 
that the thus normalised values of $k=+1,-1$ and $0$ represent the positive, negative 
and zero spatial curvature solutions respectively.

 The simplifying assumptions of the cosmological principle have hence reduced the 10 
parameters of the unknown metric $g_{\mu\nu}(x)$ down to a single real parameter 
$a(t)$ along  with a discrete set of three possible values for $k$ in 
equation~\ref{CosRob}. Together $a(t)$ and $k$ characterise the Robertson-Walker line 
element which itself represents a trial solution for cosmological models. The specific 
form of the line element will be determined by the dynamics provided by 
equation~\ref{CosLem}, which depends in turn on the choice of cosmological constant 
$\Lambda$ and the `equation of state' relating $\rho$ and $p$ in
   the energy-momentum tensor on the right-hand side of Einstein's field equation.
   Equation~\ref{cosgtruup}, with $\rho$ and $p$ functions of $t$ only, is in fact the 
most general form of energy-momentum tensor consistent with the requirements of 
homogeneity and isotropy as expressed in the cosmological principle, which is also 
respected by the $\Lambda$ term in equation~\ref{CosLem}, with $g_{\mu\nu}(x)$ in the 
form of equation~\ref{CosRob}.
 The resulting differential equations in the single independent variable $t$ may be 
solved for $a$, $\rho$ and $p$, each of which is a function of $t$ only owing to the 
homogeneity assumption.

  The Einstein tensor is constructed from the Riemann curvature tensor in terms of the 
components of the Ricci tensor as $G_{\mu\nu} = R_{\mu\nu} - \frac{1}{2}R g_{\mu\nu}$, 
as introduced after equation~\ref{rrbianc} and via equation~\ref{riccicon}; the  
Riemann tensor is a function of the linear connection $\Gamma$ as expressed in 
equation~\ref{ritencon}, and the torsion-free Levi-Civita connection of 
equation~\ref{gtoGam} is employed as also described in section~\ref{riegeo}. The 
components of the metric tensor $g_{\mu\nu}$ implied in equation~\ref{CosRob} are:
\begin{equation}
\label{gcoscomp}
  g_{00} = 1, \qquad g_{11} = \frac{-a^2}{(1-kr^2)},
    \qquad g_{22} = -a^2r^2, \qquad  g_{33} = -a^2r^2 \sin^2 \theta
\end{equation}
 These can be substituted into the above chain of relations, via the linear 
connection, to determine the  components of the Ricci tensor $R_{\mu\nu}$ and scalar 
curvature $R = g^{\mu\nu}R_{\mu\nu}$ as (see for example \cite{FosN} p.151): 
\begin{eqnarray}
  R_{00} & = &  3 \frac{\ddot{a}}{a}  \label{rzzaa}  \\
  R_{11} & = &  -(a\ddot{a}+2\dot{a}^2+2k)/(1-kr^2)   \nonumber  \\
  R_{22} & = &  -(a\ddot{a}+2\dot{a}^2+2k)r^2   \nonumber  \\
  R_{33} & = &  -(a\ddot{a}+2\dot{a}^2+2k)r^2\sin^2 \theta   \nonumber  \\
  R & = &  6 \left(
         \frac{\ddot{a}}{a} + \frac{\dot{a}^2}{a^2} + \frac{k}{a^2} \right)  
		  \label{rsixaak}
\end{eqnarray}
   with both $g_{\mu\nu}=0$ and $R_{\mu\nu}=0$ for $\mu\neq \nu$, and the notation
  $\dot{a}=da/dt$ and $\ddot{a}=d^2a/dt^2$ has been employed. Further following the 
standard procedure and completing the chain of relations from the metric $g_{\mu\nu}$ 
to the Einstein tensor $G_{\mu\nu}$ the above expressions for $R_{\mu\nu}$ and $R$ are 
substituted into the field equation~\ref{CosLem}, which includes the cosmological term 
and energy-momentum in the form of a perfect fluid, with components of the galactic 
flow 4-velocity $u_{\mu} =
 g_{\mu\nu}\frac{dx^{\nu}}{d\tau} = (1,0,0,0)$ in the comoving coordinates, to find 
for the $G_{00}$ and $G_{11}$ components respectively:
\begin{eqnarray}
    \frac{\dot{a}^2}{a^2} \, + \, \frac{k}{a^2} \, - \, \frac{1}{3}\Lambda 
	    &  = & \frac{\kappa}{3} \rho   \label{cosg00} \\
  2\frac{\ddot{a}}{a} \, + \,  \frac{\dot{a}^2}{a^2} \, + \, \frac{k}{a^2}
     \, - \, \Lambda  &  = & -\kappa p   \label{cosg11}
\end{eqnarray}
  Only the above two independent non-trivial equations result since the equations for 
the $G_{22}$ and $G_{33}$ components are each identical to that for $G_{11}$ in 
equation~\ref{cosg11}, due to the symmetries of the cosmological principle, while the 
set of six equations for $G_{\mu\nu}$ with $\mu \neq \nu$ are identically zero on both 
sides. In equations~\ref{cosg00} and \ref{cosg11} the parameters
   $a$, $\rho$ and $p$ are functions of the cosmic time $t$ while $\Lambda$, $\kappa$ 
and $k$ are constants.

  Multiplying equation~\ref{cosg00} by $a^3$, differentiating the full resulting 
expression with respect to $t$ and replacing the left-hand side by 
equation~\ref{cosg11} multiplied by $\dot{a}a^2$ leads to the relations:
\begin{eqnarray}
   \frac{d}{dt}( \rho a^3) & = & -3p\dot{a}a^2  \, = \,  -p\frac{d}{dt}a^3 
      \label{rhodt}  \\
   \mbox{that is:} \qquad \frac{d}{da}( \rho a^3) & = & -3pa^2  \nonumber
\end{eqnarray}
  These equations may also be derived directly from the identity $\tmo$ for the 
perfect fluid  energy-momentum tensor of equation~\ref{cosgtruup} given the metric 
components of equation~\ref{gcoscomp} (see for example \cite{FosN} pp.152--153, with 
the same result holding if a $\frac{\Lambda}{\kappa}g_{\mu\nu}$ term is included in 
$T_{\mu\nu}$ since $\left(
 \frac{\Lambda}{\kappa}g^{\mu\nu} \right)_{\! ;\mu} = 0$). Alternatively the 
constraint $\tmo$ can be combined with equation~\ref{cosg00} in order to 
\textit{derive} equation~\ref{cosg11}.

   This apparent redundancy between the Einstein field equation and the expression 
$\tmo$ is expected since, as alluded to above, the form of the field equation  
$G_{\mu\nu} = -\kappa T_{\mu\nu}$ can itself be motivated by the divergence-free 
identity which applies to both sides and contains equivalent information.
 In fact in \textit{defining} $-\kappa T_{\mu\nu} := G_{\mu\nu}$,
 which is the interpretation implied in the third bullet point near the opening of 
this section,
 the identity $\tmo$ is simply a \textit{copy} of the contracted Bianchi identity 
$\gmo$ which is an intrinsic property of the Einstein tensor
 (see also for example \cite{MTW} p.729).

 The apparent `conservation law' $\tmo$ can not be directly interpreted as the 
`conservation of total energy-momentum' unless the 4-dimensional geometry involved is 
asymptotically a flat Minkowski spacetime, which is \textit{not} generally the case 
for the Robertson-Walker line element of equation~\ref{CosRob}. In general a suitably 
flat spacetime can be identified for local regions of the universe, as suggested in 
the second bullet point near the opening of this section, and more specifically leads 
to energy-momentum conservation when applied in the laboratory setting, such as for 
the QFT limit summarised in table~\ref{limits} towards the end of 
section~\ref{qpagig}.

  As described above in comparison with the contribution from radiation pressure the 
universe   
 has been matter dominated since a relatively short time after the Big Bang. Hence 
considering the pressure-free case of dust with $p=0$ equation~\ref{rhodt} implies 
that $\rho a^3$ is constant in time. In this case the matter density at any epoch can 
be written as
 $\rho = \rho_0 \frac{a^3_0}{a^3}$, where a subscript `0' on a quantity such as the 
density $\rho_0$ or scale factor $a_0$ denotes the present day value at cosmic time 
$t=t_0$. Generally the boundary condition $a(0) = 0$ will be employed, with the cosmic 
time $t=0$ designating the origin of the universe at the `Big Bang'. For such a 
cosmology the present cosmic time $t_0$ hence denotes the current  ago of the 
universe.

  The physical spatial distance $d(t)$ between any two galaxies at a given cosmic time 
$t$ is simply $d(t) = a(t)\Delta\Sigma$ where $\Delta\Sigma$ is the comoving 
`coordinate distance' between the galaxies (when interpreted with care for the length 
$L$ dimension as introduced through the metric as described above, similarly as 
discussed for the interpretation of the Schwarzschild solution around 
equation~\ref{ttonly}). The physical speed of one of the galaxies relative to the 
other is $v = \frac{d}{dt}d(t)$ which leads directly to the expression:
\begin{equation}
  \label{hublaw}
   v = \frac{\dot{a}(t)}{a(t)}d(t) = H(t)d(t)
\end{equation}  
   where $H(t) := \frac{\dot{a}(t)}{a(t)}$ is the Hubble parameter, which generally 
varies with time. Equation~\ref{hublaw} expresses Hubble's law which states that at 
any given epoch $t$ the relative speed between any two galaxies on the corresponding 
spatial hypersurface is directly proportional to the distance $d(t)$ between them, 
with the constant of proportionality given by the Hubble parameter $H(t)$ at that 
cosmic time. Hubble's law is a direct consequence of the form of the  Robertson-Walker 
line element of equation~\ref{CosRob} with variable $a(t)$ and says nothing about the 
actual dynamics, that is the function $a(t)$ itself, for the universe. At the present 
epoch the Hubble parameter is called the Hubble constant (since it is the same 
everywhere in space) $H_0 = H(t_0) = \frac{\dot{a}_0}{a_0}$ and is empirically found 
to take the value:
\begin{equation}
 \label{hubcon}
  H_0 = 100h \; \mbox{km} \, \mbox{s}^{-1} \mbox{Mpc}^{-1}
\end{equation}
  with $h = 0.673 \pm 0.012$  \cite{PDG}, as employed above in equation~\ref{hubrad}.

 The functional form of $a(t)$ itself may be determined from equation~\ref{cosg00}, 
which is also called the Friedmann equation. 
 While considering the case with $p=0$ if the cosmological constant is also neglected 
by setting   $\Lambda = 0$  the Friedmann equation for any cosmic time $t$ may be 
written as:
\begin{equation}
 \label{friedeq}
  H^2 \, + \, \frac{k}{a^2} \, = \, \frac{\kappa}{3} \rho
\end{equation}
  The particular value of $\rho = \rho_c = \frac{3H^2}{\kappa}$ is called the 
`critical density' and corresponds to a solution with $k=0$, that is a universe which 
is spatially flat at any epoch. This solution is known as the Einstein-de Sitter model 
and describes an ever-expanding universe with scale factor  $a(t) \propto 
t^{\frac{2}{3}}$, as listed in
 table~\ref{cosevo} (in contrast the radiation dominated case with $p\neq 0$ and 
equation of state $p = \frac{1}{3}\rho$ results in a dynamics with $a(t) \propto 
t^{\frac{1}{2}}$ for $k=0$, as also listed in the table). 
\begin{table}[htbp]
\centering
\begin{tabular}{|r|c|c|c|c|}
\hline
  FLRW model: & $ \;\! G_{\mu\nu} = - \Lambda g_{\mu\nu} \;\! $ 
              & $\;$Matter $\rho\neq 0  \;$    
              & $\!\!$Radiation $p\neq 0  \!\!$ 
		      & $\:\!\!\! R_{\mu\nu} = \lambda(t)v_{\mu}v_{\nu} \:\!\!\!$ 
      \\	\hline  
  $p= \epsilon \rho: \;\; \epsilon =$ &  $-1$ & $0$ & $+\frac{1}{3}$ & $+1$   \\
  $a(t) \propto$  & $e^{\sqrt{\frac{\Lambda}{3}}(t-t_0)}a_0$  &  $t^{\frac{2}{3}}$
            &  $t^{\frac{1}{2}}$  &  $t^{\frac{1}{3}}$  \\
  $\rho(t) \propto$ & constant & $t^{-2} \sim a^{-3}$	
      & $t^{-2} \sim a^{-4}$	& $t^{-2} \sim a^{-6}$	    
     \\
 \hline
  \end{tabular}
  \caption{\setb Four FLRW (Friedmann-Lema\^{i}tre-Robertson-Walker) 
   cosmological models for $k\! =\!0$ with an energy-momentum  \mbox{$T_{\mu\nu} = 
-\frac{1}{\kappa}G_{\mu\nu}$}
 in the form of equation~\ref{cosgtruup}, with $p$ and $\rho$ related via the equation 
of state 
 $p= \epsilon \rho$,  corresponding to a universe dominated by a cosmological term
  (see below), matter, radiation and through $R_{\mu\nu} = \lambda(t)v_{\mu}v_{\nu}$
   (for section~\ref{secpotnt}) respectively.
 The evolution of the scale factor $a(t)$ and effective matter density $\rho(t)$ are 
obtained as solutions for equations~\ref{cosg00} and~\ref{cosg11}. }
\label{cosevo}
\end{table}

 Equation~\ref{friedeq} can be rearranged  in the form:
\begin{equation}
  \label{ommatter}
     \frac{k}{a^2} \, = \, H^2 (\Omega_M \, - \, 1)
\end{equation} 
   on introducing the matter density parameter $\Omega_M = \frac{\rho}{\rho_c} =
    \frac{\kappa \rho}{3H^2}$. For $\Omega_M > 1$ the spatial curvature is positive, 
$k=+1$, and the evolution equation for $a(t)$, that is the Friedmann 
equation~\ref{cosg00} with $\Lambda = 0$,
 shows that the universe will inevitably collapse back down to the condition $a=0$, 
while for $\Omega_M < 1$ the spatial curvature is negative, $k=-1$, and the evolution 
equation for $a(t)$ shows that the universe will expand forever as for the $\Omega_M = 
1$ case, with the latter then representing the critical value upon which the ultimate 
destiny of the universe depends. While the Einstein-de Sitter universe with $\Omega_M 
= 1$ describes the unique spatially flat case with $\rho = \rho_c = 
\frac{3H^2}{\kappa}$ for a given $H(t)$ at any cosmic time $t$, for the spatially 
non-flat cases with $k=+1$ and $k=-1$ there is a continuous range of solutions with 
$\rho > \rho_c$ and $\rho < \rho_c$ respectively, for any given values of $\rho_c$ and 
$t$.

 At the present epoch the density parameter for ordinary baryonic matter alone, which 
is largely readily visible in the form of galaxies of stars and clouds of dust and 
gas, is 
  observed to have a value of $\Omega_{B_0} = 0.050 \pm 0.002$ \cite{PDG} which, being 
much less than unity, would imply that we inhabit a universe with spatial curvature 
$k=-1$ if such matter were the sole source of gravitation. The contribution of `dark 
matter', which is needed to explain the rotation dynamics of stars within galaxies as 
well as the dynamics of individual galaxies within clusters, is found to be larger 
with density parameter $\Omega_{D_0} = 0.265 \pm 0.011$ \cite{PDG}, implying a total 
matter density parameter at the present epoch of $\Omega_{M_0} \simeq 0.315$. However 
this total is still inconsistent with independent observations, namely of the angular 
anisotropy of the CMB radiation, which suggest that the universe is very close to 
being spatially flat with $k=0$.

  Since $\Omega_{M_0}$ falls well short of the total value needed to account for the 
observed spatial flatness, and since this quantity is only sensitive to gravitating 
matter associated with clustering up to the largest scales probed, a significant 
contribution from relativistic particles or a vacuum energy term is implied. With a 
negligible contribution from the CMB radiation itself of $\Omega_{R_0} \simeq 5.5 
\times 10^{-5}$ \cite{PDG} (and with an even smaller contribution predicted for relic 
neutrinos from the Big Bang) we continue to assume $p=0$ but allow the cosmological 
constant $\Lambda$ to take a finite value in equation~\ref{cosg00}, which can be 
divided by $H^2= \frac{\dot{a}^2}{a^2}$ and rearranged in the form:
\begin{equation}
    \Omega_M \, + \, \Omega_{\Lambda} \, = \, 1 \, + \, \frac{k}{\dot{a}^2} 
	  \label{ommpoml} 
\end{equation}
\begin{equation}
	\mbox{with} \quad 
	 \Omega_M  = \frac{\kappa\rho}{3H^2} \quad \mbox{and} \quad
	 \Omega_{\Lambda}  :=  \frac{\Lambda}{3H^2}  \label{ommldef}
\end{equation}
  Analysis of the Hubble diagram for distant supernovae of type SN Ia independently 
determines a value of $\Omega_{\Lambda_0} = 0.685 \pm 0.017$ \cite{PDG}. Hence, as can 
be seen from equation~\ref{ommpoml}, the empirical observations of $\Omega_{M_0} +  
\Omega_{\Lambda_0} \simeq 1.000$ 
 and of spatial flatness consistent with $k=0$ from the CMB anisotropy are in 
excellent agreement.

  However while these observations are mutually consistent it appears coincidental 
that the contributions from  $\Omega_{M_0}$ and $\Omega_{\Lambda_0}$ are of the same 
order of magnitude at the present epoch. In particular from equation~\ref{ommldef} and 
the empirical values of the density parameters the present overall matter density in 
the universe is approximately half that of the vacuum energy, with $\rho_0 \simeq \fhs 
\Lambda / \kappa$. Since $\rho$ was much larger in the earlier universe and is ever 
decreasing into the future, and since the matter density $\rho$ describes matter which 
is heavily clumped into clusters of galaxies and the stars within, while $\Lambda / 
\kappa$ (which may be generically termed `dark energy') is of an apparently very 
different nature, both constant in time and uniformly distributed in space, the 
approximate coincidence of their present average values, within a factor of two, is 
notable. It is also observed that within the  $\Omega_{M_0}$ contribution itself  the 
dark matter content is around five times that of the baryonic matter, which is assumed 
to be a feature largely independent of cosmic time.
 An understanding of the origin of the above empirical observations will require a 
theoretical understanding of the nature of the dark sector itself.

  A solution for the large scale cosmic geometry must also be consistent with 
equation~\ref{cosg11}, which can be employed to further analyse the dynamics. 
Substituting $\frac{\dot{a}^2}{a^2} + \frac{k}{a^2}$ from equation~\ref{cosg00} into 
this second dynamic equation leads directly to the relation:
\begin{equation}
  \label{cosacc}
  \frac{\ddot{a}}{a} \, = \, -\kappa \left( \frac{\rho + 3p}{6}  \right)
      \, + \, \frac{\Lambda}{3} 
\end{equation} 
  An era of accelerating expansion of the universe, that is with $\ddot{a}(t) > 0$, 
provides a formal definition of cosmic `inflation'. 
 From the above equation it can be seen that this is the case for $(\rho + 3p) < 0$ 
for $\Lambda = 0$, or for $\Lambda > 0$ if $\rho$ and $p$ are relatively small, or 
some combination of these factors.
 The dynamics can be described in terms of the `deceleration parameter', defined as $q 
:= -\frac{a\ddot{a}}{\dot{a}^2}$. 
 Taking the case $p=0$, using equation~\ref{ommldef} and dividing 
equation~\ref{cosacc} by $H^2$ the  deceleration parameter
  is found to be related to the density parameters as:
\begin{equation}
 \label{qdefom}
   q \, = \, \frac{\Omega_M}{2} \, - \, \Omega_{\Lambda}
\end{equation}     
   At the present epoch,  as for the previous several billion years, with the value of 
$\frac{\Omega_M}{2} < \Omega_{\Lambda}$ and $q<0$ the expansion of the universe is 
accelerating, and at an increasing rate. In contrast during the first few billion 
years of cosmic evolution the values were such that $\frac{\Omega_M}{2} > 
\Omega_{\Lambda}$ with
 $q > 0$ and the rate of expansion of the universe was, temporarily, slowing down -- 
as sketched in figure~\ref{aevolve}.
\begin{figure}[htbp]  
\centering
\epsfxsize=11cm
\leavevmode
\epsffile[0 0 1141 629]{\gpath aPfig122e}
\vspace{-10pt}
\caption{\setb A qualitative depiction of the evolution of the cosmological scale 
factor $a(t)$
 as a function of cosmic time $t$, up to and beyond the present epoch $t_0$.}
\label{aevolve}
\end{figure}

   In the future as the matter density $\rho(t)$ and the value of $\Omega_M$ decrease 
with the expanding universe the cosmological constant $\Lambda$ will increasingly 
dominate the large scale evolution of the cosmos. With $p=0$, $k=0$, $\Lambda > 0$ and 
taking the limit $\rho \to 0$ equation~\ref{cosg00} becomes simply $\dot{a}^2 = 
\frac{1}{3}\Lambda a^2$. Hence if such a  cosmic epoch begins at time $t=t_{\Lambda}$ 
the scale factor increases as $a(t) = 
\exp\left(\sqrt{\frac{1}{3}\Lambda}\:(t-t_{\Lambda})\right)a(t_{\Lambda})$, which is 
also consistent with equation~\ref{cosacc}. For a cosmology entirely determined by a 
cosmological constant then  $t_{\Lambda} = 0$ and, setting $a(0) = 1$ for this 
scenario, this describes the de Sitter model with line element:
\begin{equation}
\label{CosdeS}
   d\tau^2 = dt^2 \, - \,  e^{At} d\Sigma^2_{k=0}
\end{equation}
  where $A = 2\sqrt{\frac{1}{3}\Lambda}$ is a constant and $d\Sigma^2_{k=0}$ 
represents the
 3-dimensional spatial part of the line element in equation~\ref{CosRob} for the 
spatially flat case with $a^2(t) = e^{At}$. Since if $\Lambda = 0$ ordinary matter on 
the right-hand side of equation~\ref{Einfield} does not yield a solution in the form 
of equation~\ref{CosdeS} this special case for FLRW cosmology with an exponential 
expansion factor was originally considered to represent a matterless vacuum with 
Einstein equation $G_{\mu\nu} + \Lambda g_{\mu\nu} = 0$, which is equivalent to the 
Ricci tensor being constrained to the form   $R_{\mu\nu} = \Lambda g_{\mu\nu}$ with 
constant $\Lambda$.

However,
  since the `vacuum' Einstein equation can be written with the cosmological term on 
the right-hand side as 
 $G_{\mu\nu} = - \Lambda g_{\mu\nu} \equiv -\kappa T_{\mu\nu}(\Lambda)$
   the exponential expansion observed for our universe at the present epoch is 
generally attributed to `vacuum energy' or `dark energy', in contrast to `dark matter' 
and as alluded to in the discussion following equation~\ref{ommldef} above. By direct 
comparison with equation~\ref{cosgtruup} the object $T_{\mu\nu}(\Lambda)$ may be 
interpreted as a non-standard form of energy-momentum for a `fluid' with an energy 
density $\rho_{\Lambda} = \Lambda / \kappa$ and pressure $p_{\Lambda} = 
-\rho_{\Lambda} = -\Lambda / \kappa $ which are constant in time as well as space even 
as the universe evolves. This substitution, replacing $\Lambda$ with effective values 
of $\rho_{\Lambda}$ and $p_{\Lambda}$, can also be applied directly in the evolution 
equations~\ref{cosg00} and \ref{cosg11}, from which the accelerating expansion may be 
deduced via equation~\ref{cosacc} since $(\rho_{\Lambda} + 3p_{\Lambda})<0$ for 
$\Lambda > 0$ (consistent of course with employing $\Lambda$ itself directly in
 equation~\ref{cosacc}).
 The quantities $\rho_{\Lambda}$ and $p_{\Lambda}$ remain constant in time even if an 
energy-momentum tensor with $T_{\mu\nu}\neq 0$ for ordinary matter is included in the 
field equation~\ref{Einfield}, as is the case in equation~\ref{CosLem}.

   The above de Sitter model of equation~\ref{CosdeS} was introduced in 1917 and 
originally thought to represent a static solution until it was shown how test 
particles would fly apart from each other in such a universe. In the same year
    a truly static universe model was proposed by Einstein, also with  $p=0$ and 
$\Lambda > 0$ as for the de Sitter model but in this case with a finite matter density 
$\rho$ tuned to solve equations~\ref{cosg00} and \ref{cosg11} with the constraint 
$\dot{a} = \ddot{a} =0$. The solution for the Einstein model 
 requires a positive curvature $k=+1$ and
 a constant density $\rho = \frac{2}{\kappa}\Lambda$ for ordinary matter fixed for all 
time as the universe neither expands nor contracts.  

  From an observational point of view an initial data set of measurements of 
significant redshifts for a number of nebulae was observed by V.M. Slipher as early 
1917, that is  the same year the above models were proposed. In the early 1920s the 
brightest nebulae were resolved into stars, including those of the Cepheid type 
allowing Edwin Hubble to estimate their distances out to several million light-years. 
At this time it was established that the nebulae are in fact further distant galaxies 
comparable in size to our own and the visible scale of the cosmos was vastly 
augmented.
 That galaxies are receding away from our own Milky Way with velocities proportional 
to their distance from us, consistent with equation~\ref{hublaw}, was first discovered 
by  Hubble in 1929. 

   Following the empirical conclusion from the 1920s that the universe is expanding 
and Eddington's theoretical observation in 1930 that the static Einstein model is 
unstable a policy of dropping the cosmological constant term $\Lambda$ was generally 
adopted. This led in particular to the Einstein-de Sitter model of 1932, with 
$\Lambda=0$, $k=0$, $p=0$ and $\rho = \rho_c = \frac{3H^2}{\kappa}$  evolving in time, 
as described above following equation~\ref{friedeq}.
As described alongside equations~\ref{ommpoml} and \ref{ommldef} observations in 
cosmology dating from the 1990s have  resulted in the reintroduction of a  $\Lambda>0$ 
term, which is now incorporated into the standard model of cosmology.
 (The present domination of this term over the matter density, composed of both 
baryonic and dark matter, with $\rho_0 \simeq \frac{1}{2\kappa}\Lambda$ contrasts with 
above finely balanced Einstein universe for which
 $\rho = \frac{2}{\kappa}\Lambda$). 
 While the impact of the cosmological constant on the more recent evolution of the 
universe is clearly visible in figure~\ref{aevolve} the much earlier radiation 
dominated period, while also forming a key part of the standard model, in spanning a 
period of less than 50,000 years after the Big Bang is far too brief to feature on the 
linear scale adopted in this figure. In the following section we motivate and review 
some of the theoretical ideas applied to the yet far earlier universe.


\section{Inflationary Theory} 
\label{secinf}
 
  The redshift $z$ observed for distant galaxies by Hubble, and through to present day 
observations now extending out across several billion light-years,  is defined by the 
relation:
\begin{equation}
  \label{zshdef}
   1+z \equiv \, \frac{\lambda_0}{\lambda_e} \, = \, \frac{a(t_0)}{a(t_e)} 
\end{equation}
 where $t_e$ is the cosmic time of the emission of radiation from a distant galaxy 
with  
wavelength $\lambda_e$ (which can be deduced from well-known patterns of spectral 
lines) and $t_0$ is the present cosmic time at which we detect the radiation and 
measure the wavelength to be $\lambda_0$ in our galaxy. With the value of $z_0 = 0$ 
for the present epoch and adopting the convention $a(t_0) = 1$
 there is a simple relationship between the redshift $z$ at an earlier epoch  and the 
corresponding scale factor $a(t_e)$ at that time.
 Since for our universe $a(t)$ is an ever increasing function of time, as can be seen 
in figure~\ref{aevolve}, the value of the redshift $z$ can be used to label the 
earlier epochs of our expanding universe.
 Hubble's observations of a positive redshift are explained via equation~\ref{zshdef} 
by the simple fact that $a(t_e)$ was smaller in the past, while an increasing value of 
$a(t)$ at any given time $t$ implies a positive value for the Hubble parameter defined 
in equation~\ref{hublaw}.

   This cosmological redshift does \textit{not} arise from the Doppler effect, which 
only strictly applies in a local or extended \textit{flat} region of spacetime, but 
from the passage of light through a \textit{curved} 4-dimensional spacetime. As a 
further example the non-Euclidean geometry of spacetime also accounts for the 
gravitational redshift or blueshift resulting from the propagation of light away from 
or towards, respectively, a massive object, which is also an effect of general 
relativity.

  Although the 4-dimensional geometry of our universe is  far from (pseudo-) 
Euclidean, the observations described in the previous section indicate that the 
3-dimensional spatial hypersurfaces at any given cosmic epoch $t$ appear to be very 
flat. While the hypothetical Einstein universe was found by Eddington to be balanced 
precariously, as alluded to at the end of  the previous section, there is also an 
apparent instability concerning the state of the universe we actually observe. If the 
value of the total density parameter $\Omega$ is not exactly equal to one in a  matter 
or radiation dominated universe, such as we have described for the first few billion 
years of our own cosmos, this value will diverge away from unity as the universe 
evolves. Taking equation~\ref{ommatter}, generically replacing $\Omega_M$ by a density 
parameter $\Omega$ and with $H^2 = \frac{\kappa \rho}{3\Omega}$, as for the form of 
equations~\ref{ommldef} for example, leads directly to the relation:
\begin{equation}
  \label{flatprob}
     \left( \Omega^{-1} - 1 \right) \rho a^2 = -\frac{3k}{\kappa}
\end{equation} 
  from which different conclusions may be reached depending on the equation of state 
for the apparent matter density $\rho$, given that the right-hand side of this 
expression is a constant. In particular for an expanding FLRW universe that is matter 
dominated or radiation dominated the quantity $\rho a^2$ decreases with cosmic time in 
proportion to $a^{-1}$ or $a^{-2}$ respectively, as can be seen from the columns of 
table~\ref{cosevo}. Hence it can be seen from equation~\ref{flatprob} that a value of 
$\Omega \neq 1$ will diverge further from unity as such a universe evolves.

  That is any small deviation of the density parameter  $\Omega$ from the value of 
unity at an earlier epoch with a large redshift $z$  will have been  greatly amplified 
by the present day, such that in going back to the extreme case of the Planck epoch of 
$t \simeq 10^{-43}$ seconds after the Big Bang an apparent fine tuning of the density 
parameter  to about 1 part in $10^{60}$
 is required  in order to be consistent with the present day observation of spatial 
flatness for the universe  (\cite{Pea} p.323). The need for an explanation of this 
precise tuning of the initial spatial flatness condition arising out of  the Big Bang 
is known as the `flatness problem'.

  That the 4-dimensional geometry can be highly curved even for a spatially flat 
cosmology with $k=0$ is particularly evident in the early universe.  For the matter 
dominated case it can be seen by substituting terms from equations~\ref{cosg00} and 
\ref{cosg11} into equation~\ref{rsixaak} that the scalar curvature of the spacetime is 
simply $R= \kappa \rho$ 
 (as would be expected from the paragraph following equation~\ref{gtruu}).
 Hence as $t \to 0$, in principle to an epoch even earlier than the Planck time, with 
the scale factor $a(t) \to 0$ while $\rho \to \infty$ the scalar curvature $R$ 
diverges to infinity at what is referred to as the `initial singularity' at $t=0$.  
  It is sometimes noted that the standard cosmological model hence predicts the 
paradoxical origins of the universe in such an initial singularity, studied by S.W. 
Hawking, G.F.R. Ellis and R. Penrose around 1965--70, a point at which general 
relativity, which governs the model itself, breaks down. However any conclusions drawn 
from the structure of gravitation at the Planck scale are inevitably uncertain given 
the as yet unknown role of quantum phenomena in such an extreme environment. As 
described in section~\ref{qpagig} for the present theory gravitation itself is not 
quantised and hence in principle the Planck scale will be of less significance and not 
represent a barrier to further extrapolation to arbitrarily early times, as will be 
considered in section~\ref{sectveu}.

  It is also informative to write the Robertson-Walker line element 
 of equation~\ref{CosRob} with the cosmic time coordinate $t$ transformed to a 
conformal time parameter $\eta = \int_0^t \frac{dt'}{a(t')}$ as:
\begin{equation}
\label{Coseta}
   d\tau^2 =  a^2(\eta)\left[d\eta^2 \: - \: d\Sigma^2_k \right] 
\end{equation}
    where $d\Sigma^2_k$  represents the spatial part of the line element inside the 
square brackets of equation~\ref{CosRob} with $k=+1,0$ or $-1$. Hence by adopting the 
conformal time coordinate $\eta$ in equation~\ref{Coseta} the scale factor $a(\eta)$ 
can be seen as a special case of a conformal transformation, which more generally 
takes the form $g_{\mu\nu}(x) \to f(x) g_{\mu\nu}(x)$ where $f(x)$ is an arbitrary 
real function of spacetime (a very different example of which was considered in 
equation~\ref{gmetwave}).
 For the case $k=0$ the Robertson-Walker line element is hence related to a flat 
4-dimensional spacetime via a conformal transformation. It can also be shown, using a 
further suitable coordinate transformation, that the geometry for each of the $k = \pm 
1$ cases is also conformally flat (\cite{Pea} p.71). Hence for all FLRW models the 
4-dimensional geometry of the universe, with the metric of equation~\ref{CosRob}, is 
conformally flat, which implies the vanishing of the Weyl curvature tensor, introduced 
before equation~\ref{rdecom}, that is $C_{\rho\sigma\mu\nu}(x) = 0$, even though 
components of the Ricci curvature $R_{\mu\nu}(x)$ part of the Riemann tensor may 
attain arbitrarily large values in the very early universe.

  The initial singularity of the Big Bang is a spacelike boundary of spacetime in our 
distant past, represented by the horizontal wiggly line in the conformal diagram of 
figure~\ref{confbb}. In such a diagram all null-rays, that is with proper time line 
element $d \tau^2 =0$, are drawn at $45^{\circ}$ and hence the causal properties of 
the spacetime are made apparent. The vertical axis of such a diagram is linear in the 
conformal time $\eta$ with the horizontal axis representing  comoving coordinate 
distances $\Delta\Sigma$, consistent with equation~\ref{Coseta}. In 
figure~\ref{confbb} epochs on the vertical axis are labelled by the cosmic time $t$, 
although of course not to scale, and intervals of the horizontal axis at any given 
epoch can be converted to  physical proper distances $a(t)\Delta \Sigma$,
 as described  before equation~\ref{hublaw}.
 A ray of light emitted at time $t=t_e$ and reaching us now at $t \equiv t_0$ will 
have  travelled the comoving coordinate distance 
\begin{equation}
 \label{parthor}
   \Sigma_p(t_0,t_e) \, = \, \Delta \eta \, = \, \int_{t_e}^{t_0} \frac{dt'}{a(t')}  
\end{equation}
  where $\delta t'/a(t')$ is the coordinate distance traversed in a small interval of 
cosmic time $\delta t'$.
Hence any signal emitted beyond this distance at time $t_e$  will not have been able 
to reach us yet and hence in turn $\Sigma_p(t_0,t_e)$ is termed  the `particle 
horizon'. For any given $t_e$ the particle horizon grows with time $t=t_0$ from the 
perspective of the observer $b$ in figure~\ref{confbb}. The particle horizon can also 
be defined as the proper distance $R_p(t_0,t_e) = a(t_0) \Sigma_p(t_0,t_e)$ on the 
3-dimensional spatial hypersurface at the time $t_0$, that is $R_p(t_0,t_e) = 
a(t_0)\Delta \eta$, where $\Delta \eta$ is the conformal time elapsed between $t=t_e$ 
and $t=t_0$.

\begin{figure}[htbp]  
\centering
\epsfxsize=12cm
\leavevmode
\epsffile[0 0 1455 691]{\gpath aPfig123e}
\caption{\setb Conformal diagram depicting the past light cone from our present 
location $b$ at cosmic time $t=t_0$ extending back to the Big Bang singularity at 
$t=0$. The recombination era $t=\trec \; (\simeq t_0/37,000)$ is also indicated. As a 
plot of conformal time $\eta$ versus comoving coordinate distance $\Sigma$ the diagram 
is a 2-dimensional representation of a 4-dimensional spacetime.}
\label{confbb}
\end{figure}

  At the present epoch $t_0$ the largest particle horizon $R_p(t_0)$ corresponds to 
signals emitted at the time of the Big Bang. Setting $t_e =0$ the integral in 
equation~\ref{parthor} converges provided the equation of state is such that $\rho(t)$ 
decreases at least as fast as $a^{-2}(t)$, as it does for a matter or radiation 
dominated universe as seen in table~\ref{cosevo}.
 For various cosmological models the particle horizon, obtained from 
equation~\ref{parthor}, is generally greater than $t_0$ itself since $a(t)$ tends to 
be smaller for earlier times $t<t_0$. For a radiation dominated universe the particle 
horizon has a value of $R_p(t_0) = 2t_0$, while for the matter dominated case the 
value is  $R_p(t_0) = 3t_0$.
  (For the Einstein-de Sitter model with $k=0$ the age of the universe can be 
expressed as $t_0 = \frac{2}{3H_0}$ and the particle horizon is twice the Hubble 
radius,
   $R_p = 3t_0 = \frac{2}{H_0} = 2R_H$, with the latter defined in 
equation~\ref{hubrad}.)

   For our own universe the present particle horizon is determined to be 
 $R_p(t_0) \simeq 46$ billion light-years, which is greater than  $3t_0$, where $t_0 
\simeq 13.8$ billion years is the age of the universe, owing to increasing impact of 
the cosmological $\Lambda$ term at the present epoch. The proper distance  $R_p(t_0)$ 
represents the edge of the observable universe in terms the present distance to 
objects on the 3-dimensional spatial hypersurface at cosmic time $t=t_0$, \textit{not} 
of course as we actually might see them via light emitted in the distant past. While 
$R_p(t_0,t_e) = a(t_0) \Sigma_p(t_0,t_e)$ is the present particle horizon for 
observing events from time $t=t_e$, the proper distance to such an event on the 
horizon at the time of signal emission was $a(t_e) \Sigma_p(t_0,t_e)$. 
 For comparing particle horizons at different epochs  comoving coordinate distances 
$\Delta\Sigma$, that is intervals of the horizontal axis in conformal diagrams,  will 
be move convenient, as we describe in the following.

  As well as the Big Bang at $t=0$ and the present era $t=t_0$ the time of 
`recombination' $t=\trec$ is also labelled in figure~\ref{confbb}. This is the epoch 
around 372,000 years after the Big Bang, with a redshift of $z\simeq 1090$ and as the 
temperature dropped below around 4,000$\,$K, during  which the residual electrons, 
which had not annihilated with positrons, combined with protons and other light nuclei 
to form neutral atoms, mainly hydrogen and helium (the name `recombination' is 
somewhat inaccurate as this process is an initial combination of such objects, unless 
thought of as a return to charge neutral states in a different form to that at $t=0$). 
Since there is only an extremely small interaction between an external electromagnetic 
field and neutral atoms this also marks the era of decoupling between radiation and 
matter alluded to after equation~\ref{CosLem}. Photons from this decoupling epoch have 
effectively been propagating freely since $t=\trec$, relatively early in the 13.8 
billion year history of the universe, until detected in the present
  as the observed  CMB radiation now redshifted to a temperature below $3\,$K.

   From our perspective  photons composing the CMB radiation were emitted from 
anywhere on the 2-sphere of our past light cone in 4-dimensional spacetime at the time 
$t=\trec$.
Two points   $u$ and $v$  on the continuous surface of a  2-sphere can be arbitrarily  
close together,
 unlike the points $u$ and $v$ in figure~\ref{confbb} on the  past light cone of this 
2-dimensional representation of spacetime.
 If the comoving coordinate distance between $u$ and $v$ at $t=\trec$ is greater than 
twice the particle horizon $\Sigma_p(\trec)$ then the two spacetime points have never 
been in causal contact. Hence from our perspective $b$, with both $u$ and $v$ observed 
on our particle horizon $\Sigma_p(t_0,\trec)$, there is no reason to expect a 
homogeneity of physical quantities such as the CMB temperature as measured and 
compared for such  regions $u$ and $v$
 which have not been in causal contact with each other. In fact the particle horizon 
at the recombination era  $\Sigma_p(\trec)$  only subtends of order $1^{\circ}$ in the 
sky from our present perspective $b$ on Earth. The difficulty in contriving an 
assumption of homogeneity as an initial condition of the hot Big Bang to account for 
the observed uniformity of the CMB temperature to within 1 part in $10^5$ over all 
angles of the sky is known as the `horizon problem'.

   In place of postulating homogeneous initial conditions across causally separated 
spatial regions of the very early universe the only means by which the temperatures at 
$u$ and $v$ might be related through a process of thermalisation is to arrange for the 
possibility of causal contact in their past. This requires a mechanism through which 
the Big Bang epoch effectively retreats back further below the recombination era  in 
the conformal diagram of figure~\ref{confbb}, as demonstrated in figure~\ref{confbba}. 
This in turn can be achieved by a sufficient rescaling of proper spatial distances 
with $a(\tinf-\epsilon) \lll a(\tinf+\epsilon)$, where $\epsilon$ may be a very short 
time interval, effectively   `miniaturising' 3-dimensional space during the epoch 
$t<\tinf$. In this case the horizontal displacements in figure~\ref{confbba} labelled 
by comoving coordinate intervals $\Delta \Sigma$ now represent much shorter physical 
proper distances $a(t)\Delta \Sigma$ for $t<\tinf$ and a given null-ray propagating 
for a given cosmic time interval $\Delta t$ occupies a somewhat larger portion of the 
vertical axis which is linear in conformal time intervals $\Delta \eta \sim 
\frac{\Delta t}{a(t)}$. Hence the Big Bang epoch at $t=0$ is pushed back in the 
conformal diagram to accommodate this rescaling.  Hence in turn 
the comoving coordinate distance traversed by null-rays in a fixed cosmic time 
interval during this early epoch before $t=\trec$ can in principle comfortably 
encompass the present particle horizon
 $\Sigma_p(t_0,\trec)$  at $t=t_0$ for signals emitted at $t=\trec$ (see for 
example~\cite{Pen} pp.744--747).  

\begin{figure}[htbp]  
\centering
\epsfxsize=\maxwidth
\leavevmode
\epsffile[0 0 2028 951]{\gpath aPfig124e}
\caption{\setb Conformal diagram depicting a similar cosmic history as 
figure~\ref{confbb} with the same three values of $t=0$, $t=\trec$ and $t=t_0$ but 
with the addition of a further epoch $t=\tinf$ during which the scale factor $a(t)$ is 
`inflated' by an enormous degree in a  short period of cosmic time.}
\label{confbba}
\end{figure}

  Evolving forwards in time from the Big Bang the rapid expansion of the universe 
scale factor $a(t)$ at the epoch $\tinf$, which in principle solves the horizon 
problem,  is termed `inflation', as a particular case of an accelerating expansion 
described generically after equation~\ref{cosacc}.  The question then still remains 
regarding the physical mechanism behind such a radical transformation of the spacetime 
geometry at that very early epoch. Guided by the de Sitter model with the line element 
of equation~\ref{CosdeS} describing an exponential expansion with scale factor $a(t) 
\propto \exp\left(\sqrt{\frac{1}{3}\Lambda}\;t\right)$ one way to achieve inflation is 
with a very large, but only temporarily active, cosmological term of the form $\Lambda 
g_{\mu\nu}$ in Einstein's field equation~\ref{Einfield}.

  On introducing a new scalar field $\varphi(x)$ (which is unrelated to the scalar 
Higgs field $\phi(x)$ of the Standard Model of particle physics described in 
section~\ref{ewtatsm}) a false vacuum state obtained for a suitable potential 
$V(\varphi)$ can model the effect of a cosmological term via
 an energy-momentum tensor $T_{\mu\nu}$ with a term of the form $V(\varphi)g_{\mu\nu}$ 
(such as in equation~\ref{tmninf} below). That is, a potential $V(\varphi,T)$, as a 
function of the field $\varphi(x)$ and temperature $T(x)$, may be contrived such that 
the high temperature vacuum state $\varphi = 0$ becomes a `false vacuum' as the 
universe achieves a `supercooled' condition below a certain critical temperature $T_c$ 
in the very early universe. The phase transition to the new true vacuum state with 
$\varphi \neq 0$ for $T<T_c$ may involve either quantum mechanical tunnelling through 
an intermediate potential barrier (`old inflation') or a gradual roll down a potential 
slope (`new inflation'). In either case the potential function $V(\varphi,T)$ may be 
suitably contrived in order that the true vacuum is not immediately attained and the 
energy of the false vacuum state dominates the cosmological evolution equations for a 
brief period of time. This cosmic time period of  
 $\tinf \sim 10^{-35}$---$10^{-32}$~seconds can be long enough for the scale factor 
$a(t)$ to increase by a factor of $\sim 10^{30}$ or more, effectively solving the 
horizon problem by the rapid inflation of a small homogeneous region of the very early 
universe (see for example \cite{Pea} chapter~11).

  While the de Sitter model of equation~\ref{CosdeS} assumes a spatially flat universe 
with $k=0$, the evolution of the scale factor $a(t)$ resulting from a cosmological 
term $\Lambda g_{\mu\nu}$ in the field equation can also be determined for the cases 
of $k= \pm 1$ with spatial curvature. It is found that for $k=+1,0$ and $-1$ the scale 
factor  evolves as $a(t) \propto \cosh\left( \sqrt{\frac{\Lambda}{3}}\,t \right),\: 
\exp \left(\sqrt{\frac{\Lambda}{3}}\,t\right)$ and $ \sinh \left( 
\sqrt{\frac{\Lambda}{3}}\,t\right)$ respectively, and hence the $k=\pm 1$ solutions in 
time converge towards the de Sitter solution with $k=0$ and constant Hubble parameter 
$H(t) = \sqrt{\frac{\Lambda}{3}}$ (\cite{Pea} p.326).
 This convergence towards a density parameter $\Omega$ of unity can also be seen from 
equation~\ref{flatprob} since the equation of state for a cosmological term implies 
that $\rho a^2 \propto a^2$, as can be seen from table~\ref{cosevo}, which hence 
rapidly increases during inflation, driving $\Omega \to 1$. 
 Hence during inflation solutions for $a(t)$ with finite spatial curvature rapidly 
approach the purely exponential expansion solution with $k=0$, that is the de Sitter 
model for a flat universe with the $\Lambda g_{\mu\nu}$ term simulated by the energy 
of the false vacuum during the inflationary period. The brief inflationary era $\tinf$ 
described in the previous paragraph, and depicted in figure~\ref{confbba}, is 
sufficient to suppress any non-zero spatial curvature by a factor of around $10^{60}$ 
or more, hence in principle solving the flatness problem described after 
equation~\ref{flatprob}, in addition to solving the horizon problem.

   Inflationary theory was initially proposed by Alan Guth in 1980, precisely to 
address the horizon problem while also accounting for the flatness problem.
 In fact the strong bias towards spatial flatness is sometimes considered to have been 
a successful \textit{prediction} of the theory.
 The hypothetical period of inflation at $\tinf$ drives the total density parameter 
$\Omega$ extremely close to unity in the early universe such that the subsequent 
radiation dominated era of thousands of years and matter dominated era of billions of 
years have been insufficient to prise the value of $\Omega$ away from the value of 
one, as described following equation~\ref{flatprob}, to any measurable degree. The 
more recent and increasingly dominant effect of the apparently presently active 
cosmological term $\Lambda g_{\mu\nu}$ is again tending to bind the density parameter 
yet closer to unity, although this effect has thus far been too weak to account for 
the observation of spatial flatness without the much earlier and much more dramatic 
inflationary epoch.

 However unlike the cosmological term $\Lambda g_{\mu\nu}$ which accounts for the 
present day relatively pedestrian accelerating expansion of the universe the much 
earlier period of rapid inflation is required to terminate, and such a change in 
conditions is generally ascribed to a phase transition as introduced above. The 
original `old inflation' model employed a first order phase transition via quantum 
tunnelling from the false to the true vacuum once the temperature had dropped 
sufficiently to allow penetration through the potential barrier. However the quantum 
nature of the transition results in bubble formation and corresponding large 
inhomogeneities that are not observed. This `graceful exit problem' can be solved by 
`new inflation' which ends via a transition from the false vacuum at a local maximum 
in the potential at $\varphi=0$, that is through a second order phase transition, 
which proceeds more nearly simultaneously throughout the universe. An almost flat 
potential around $\varphi = 0$ can result in a `slow roll' down to the true vacuum at 
the potential minimum, still allowing sufficient time for a dramatic inflationary 
expansion.

  Amongst a range of inflationary models proposed  `chaotic inflation' in principle 
also solves the graceful exit problem. In this model
    the potential of the scalar field can take a much simpler form such as
	$V(\varphi) = m^2 \varphi^2$ or 
	 $V(\varphi) = \lambda \varphi^4$ with a single minimum at $\varphi=0$.
 Under a large range of possible initial conditions in the primordial chaos in some 
regions the value of $\varphi(x)$
  may be far from the minimum. Such a value, required to be essentially uniform over a 
region of space of order the present day Hubble radius, can stimulate an inflationary 
period.
 A large inflation factor is possible provided that the constant $\lambda$ for example 
is chosen such that the potential function is sufficiently shallow to allow a 
sufficiently delayed roll down to the true vacuum value at $\varphi = 0$.
  As the true vacuum is attained and inflation ends our observable universe is 
contained within a single bubble, one of many resulting from the initial chaotic 
conditions. Even if the scalar field $\varphi$ begins with a value close to the 
minimum at zero quantum fluctuations can drive this value further from the minimum 
resulting in a self-sustaining `stochastic inflation', or even motivating 
consideration of an `eternal inflation' model.

   For any of the above inflationary models an energy-momentum tensor can be derived 
from a standard Lagrangian for a scalar field, namely $\lag = \fh 
\pal_{\mu}\varphi\pal^{\mu}\varphi - V(\varphi)$ including a kinetic as well as the 
potential term, via Noether's theorem as described for equation~\ref{tmnneo} (some 
care is needed with the interpretation of translation invariance since here we are 
clearly \textit{not} dealing with a globally flat Minkowski spacetime, however 
equation~\ref{tmnneo} may be applied for sufficiently small spacetime regions by the 
strong equivalence principle described in section~\ref{gcatep} and then generalised 
for the result below on replacing $\eta_{\mu\nu}$ by $g_{\mu\nu}$) leading directly to 
(\cite{Pea} p.329):
\begin{equation}
 \label{tmninf}
   T_{\mu\nu} \, = \, \pal_{\mu}\varphi\,\pal_{\nu}\varphi \, - \, \fhs 
      \pal_{\rho}\varphi\pal^{\rho}\varphi\, g_{\mu\nu} \, + \,  V(\varphi)g_{\mu\nu}
\end{equation}
  In addition to the cosmological term for a temporarily finite (and uniform at least 
over the spatial extent of the observable universe)
   value $V(\varphi) \equiv \frac{\Lambda}{\kappa}$,  with an effective equation of 
state $p_{\Lambda} = -\rho_{\Lambda}$ ($=-V(\varphi)$), driving the exponential 
expansion, there are also kinetic terms in the derivatives of the scalar field 
$\varphi(x)$.
  An equation of motion for $\varphi(x)$ can be derived as the Euler-Lagrange equation 
for the stationarity of the action $S = \int \lag \sqrt{\vert g \vert} d^4 x$ which, 
since the metric $g_{\mu\nu}(x)$ incorporates the scale factor $a(t)$, is found to 
include a Hubble drag term of the form 
 $H \dot{\varphi}$ (\cite{Pea} p.331). 
 
  If after the Planck time the universe is initially radiation dominated then as the 
temperature drops below the critical temperature $T_c$ inflation begins to dominate 
and the radiation is rapidly redshifted. During the vacuum driven expansion the 
universe is essentially devoid of matter and radiation, with the scalar field 
$\varphi$ completely dominating towards the end of inflation, however any coupling 
between $\varphi$ and matter fields leads to a further drag term in the equation of 
motion for $\varphi$.
 As the minimum of $V(\varphi)$ is approached
 the dynamic equations drive rapid  oscillations, which are dampened by the drag 
terms. This in turn fuels a
   reheating in the post-inflation era as the vacuum energy is converted into 
interacting particles, including the familiar states of the Standard Model. This 
period of transition to essentially zero vacuum energy,
 in which the energy is transferred from the scalar field $\varphi$ to ordinary matter 
and radiation via their mutual interactions, may also be the time during which any 
mechanism that generates an asymmetry between matter and antimatter, as still 
manifestly observed today, may act. The origin of dark matter might also turn out to 
be associated with the termination of inflation. 
 This epoch then merges into the beginning of the radiation and then matter dominated 
FLRW periods of the standard cosmological model as described in the previous section, 
with the initial conditions set by the inflationary expansion.

 In de Sitter spacetime, as for that of inflation, the event horizon (which is 
distinct from the particle horizon) is of finite size, as for the case of back holes. 
This means that the conditions for producing Hawking radiation, as alluded to towards 
the end of section~\ref{qpagig}, are also present during inflation. In turn the 
possibility arises that quantum fluctuations can become frozen into residual classical 
deformations in the latter stages of inflation. In turn these classical fluctuations 
will modulate the density of the radiation and matter produced at the end of 
inflation, seeding the evolution of large scale structure as eventually manifested in 
galactic formations.
 Similar fluctuations during the inflationary epoch are also predicted to generate a 
background of gravitational waves which still propagate today and which, although 
being much more difficult detect, are in principle observable through the large scale 
CMB anisotropies
 which may provide a signature for the metric distortions of the gravity waves.

 A significant degree of fine tuning is required for any model of inflation based on 
the properties of a postulated scalar field $\varphi(x)$ in order to achieve a match 
with a range of empirical observations, which is somewhat unsatisfactory since 
inflationary theory was designed to avoid the necessary fine tuning as initially 
implied by the horizon problem and flatness problem. There is also no understanding of 
the origin of the vast difference between the magnitude of the effective  cosmological 
term associated with inflation due to $V(\varphi)$, which is of $O(10^{-10})$ in 
natural units, and  apparent cosmological constant $\Lambda$ of the present epoch, 
which is of $O(10^{-120})$ in natural units. Indeed the 
 unaccounted for
 magnitude of the latter number itself, the `cosmological constant problem'  is one of 
the biggest puzzles in physics, as already alluded to briefly at the end of 
section~\ref{lccop}.

 A further significant issue regarding the standard model of cosmology, which is not 
addressed by inflation, relates to the origin of the very special conditions of the 
Big Bang in that the entropy of the early universe must have apparently been extremely 
low, despite the high degree of thermalisation achieved for the degrees of freedom of 
the electromagnetic field. The degrees of freedom of the gravitational field may be 
described by the Weyl tensor
   (\cite{Pen} section~28.8), although both the Ricci curvature and Weyl curvature 
parts of the Riemann tensor exhibit the effects of gravity.
   The Weyl curvature and its distorting tidal effect tend to increase as matter 
gravitationally clumps into dense regions, diverging to infinity in the neighbourhood 
of a black hole. The entropy associated with a black hole is correspondingly extremely 
high, attaining values much higher than that associated with ordinary thermal entropy.
 On the other hand, as described following equation~\ref{Coseta}, in the idealisation 
of the FLRW cosmological models the spacetime is conformally flat with zero Weyl 
curvature. This suggests that if the universe originates in a state very close to an 
FLRW model the initially low entropy may correlate with the very low Weyl curvature, 
both of which then tend to increase as matter progressively clumps together as the 
universe evolves.

  More generally the `Weyl Curvature Hypothesis', proposed by Roger Penrose in 1979  
(\cite{Pen} section~28.8), asserts that  $C_{\rho\sigma\mu\nu}(x) = 0$, or is at least 
very close to zero, as a constraint on the initial singularity. Hence the universe 
shares at least this property, of conformal flatness,  with the FLRW models in the 
early stages. (In principle this constraint might be further augmented by the 
condition $k=0$  as the universe evolves into a spatially flat model due to a 
subsequent period of inflation). This hypothesis of zero Weyl curvature for the 
initial singularity of the Big Bang is then in stark contrast to the situation for the 
terminal singularities of black holes as alluded to above.

  This very special condition of the Big Bang represents an enormous constraint of low 
entropy on the initial conditions which in turn provides a suitable point of departure 
for the second law of thermodynamics. Gravitation, in comparison to all other fields, 
hence appears to have had a very special status, aloof from  thermalisation  in the 
Big Bang, with the
second law of thermodynamics
 only later exercised through the gravitational degrees of freedom.
  While inflation, as described for figure~\ref{confbba}, provides the breathing space 
for ordinary matter, including the electromagnetic field, to reach thermal equilibrium 
in the aftermath of the hot Big Bang, the question remains to explain why gravitation 
should apparently be treated in such a radically different manner to the other forces 
of nature.
 The theory presented in this paper may shed some light on these questions since, as 
discussed in the previous chapter, here the gravitational field itself is not 
quantised and is hence different from all other fields in this respect.

  Further, while for a range of given initial conditions  inflationary theory can 
solve the horizon problem, which was introduced in figure~\ref{confbb},
 by opening up a suitable  spacetime volume to allow
  points such as $u$ and $v$ to exhibit the same temperature through
  thermalisation, as described for figure~\ref{confbba}, 
 the structure of these diagrams indicates that there may be a more fundamental 
difficulty with this picture. Namely, since \textit{any} two different points such as 
$x$ and $y$ on the spacelike surface of the initial singularity at $t=0$, as indicated 
in figure~\ref{confbb} for example, have clearly never been in causal contact with 
each other it is difficult to conceive how the Big Bang could be effectively 
`triggered'  simultaneously across this potentially infinite 3-dimensional 
hypersurface.
 This observation applies even if initial properties, such as the temperature, are 
very different at $x$ and $y$. It also applies in exactly the same way for 
figure~\ref{confbba} and inflation is of no help in addressing this `start-up 
problem'.

   On the other hand if the Big Bang can be considered as a `spacelike event', 
encompassing the points of a large region of the initial spatial hypersurface, then 
there seems no reason to suppose that the simultaneous `cause' of the Big Bang at 
points such as $x$ and $y$ in figure~\ref{confbb} might not also `cause' them to have 
the same properties such as temperature. Indeed, the notion of a simultaneous start-up 
along the $t=0$ spacelike surface which endows different points
  with \textit{different} properties, implying the application of a range of possible 
start-up conditions and resulting in an uneven temperature distribution, seems 
somewhat more contrived.
That is, it seems any two points like $x$ and $y$ on the initial singularity  
\textit{must} be related in order for the universe to start-up at both, and any 
solution to this problem may well itself entail a high degree of homogeneity in the 
very early universe and solve the horizon problem without the need for inflation. A 
source of later fluctuations and inhomogeneity will then still be needed to account 
for the origin and formation of the galactic structures seen today.

  However, even without the issue of the uncertain role of quantum phenomena under the 
extreme gravitational conditions of the very early universe, care is needed in the 
extrapolation  to the earliest epoch. For most FLRW models as the cosmic time 
approaches the moment of the Big Bang $t \to 0$ the scale factor also approaches zero 
$a(t) \to 0$, indeed the boundary condition $a(0) = 0$ is adopted for various dynamic 
solutions, as described before equation~\ref{hublaw}.
 In this limit any finite comoving coordinate distance $\Delta \Sigma$ corresponds to 
a vanishing physical proper distance $a(0) \Delta \Sigma = 0$. With the horizontal 
axes in figures~\ref{confbb} and \ref{confbba} representing coordinate distances this
 naive analysis implies that the observable universe at present came from a physically 
vanishingly small region of the initial singularity. In turn  the initial singularity, 
represented by the wiggly line in these figures, might perhaps be interpreted as a 
fully causally connected entity in the limit $t=0$, amending the strict causal 
structure of the conformal diagrams in this extreme case. 

   However, since the spacelike coordinate distances are unlimited in magnitude even 
as $a(t) \to 0$ any proper distance in the limit $t \to 0$ is obtained as the product 
of one number in principle approaching infinity with another approaching zero, a 
situation which approaches the meaningless. Hence rather than speculating upon `how 
many angels can dance on the head of a pin', what is really needed is a more complete 
understanding of what \textit{happens} in the Big Bang, what \textit{causes} it to 
happen and even \textit{why} there should be a universe at all.


\pagebreak
\chapter{A Novel Perspective on Cosmological Structure}
 \label{anpocs}

\section{The Dark Sector}
 \label{secpotnt}

 Within the context of the present theory the external geometric structure of the 
world is intimately associated with a subjective perceptual requirement, forged out of 
a multi-dimensional form of temporal flow expressed as $\lvh$, rather than being an 
apparently arbitrary feature of an objective universe independent of the need for 
perception. Indeed the specific identification of 3-dimensional spatial expanses with 
an approximately global $\soth$ symmetry would seem to be a somewhat redundant and 
unnecessary feature of such an inanimate mathematical entity. On the other hand the 
extent of spatial flatness for the observable universe, as described in the previous 
chapter, goes far beyond that utilised for perception by sentient beings on the planet 
Earth.

 Further, given the observed Hubble constant of equation~\ref{hubcon} at the present 
epoch, in a period of 100 years the fractional change in the scale factor is $\Delta 
a_0 / a_0 \simeq 0.7 \times 10^{-8}$. Hence on the scale  of a human lifetime the 
 Robertson-Walker line element of equation~\ref{CosRob}, for the $k=0$ case, describes 
a flat Minkowskian spacetime to within 1 part in $10^8$, with the expanding universe 
seemingly hanging suspended as a vast spatial expanse through a given human interval 
of cosmic time.
 For the horizontal time axis representing a duration of 100 years figure~\ref{cosmos} 
would then represent an accurate snapshot of our universe at the present epoch.

  However the breakdown of global Lorentz symmetry beyond our 100 year thick slice of 
the universe is readily observed in the cosmological redshift. This redshift, defined 
in equation~\ref{zshdef} and as first observed by Hubble and others and now probing 
distant galaxies reaching back over billions of years in cosmic time, uncovers the 
non-Euclidean geometry of the cosmos as summarised by the evolution of the scale 
factor $a(t)$ depicted in figure~\ref{aevolve}.

 The question then is the extent to which the present theory might account for the 
observations of such large scale structure in cosmology, and the phenomena of the dark 
sector more generally as summarised in section~\ref{sectsmoc}, as we shall explore in 
this section. In section~\ref{secinf} it was described how the origin of spatial 
flatness and the cosmological principle of homogeneity and isotropy, beyond the 
pragmatism of assumptions employed for FLRW models, can in principle be accounted for 
by the theory of inflation in the very early universe. In the following section the 
evolution of the very early universe and the nature of the Big Bang itself will be 
considered here within the context of the  projection of spacetime out of the general 
form of temporal flow for the
present theory. In section~\ref{secuni} the extent to which cosmological and other 
physical parameters might be explicitly constrained by the theory will also be 
considered.

  The pure flow of time $s$, underlying the multi-dimensional form of temporal flow 
through $\lvh$, is directly related to the proper time $\tau$ elapsed from the point 
of view of any timelike trajectory through spacetime, as described in 
section~\ref{fdandtd}.
 Time dilation effects for $\tau$, as implied in the metric $g_{\mu\nu}(x)$ such as 
that for the Schwarzschild solution of equation~\ref{ttrtp}, are directly equivalent 
to those for $s$.
 A similar observation applies for the Robertson-Walker metric of 
equations~\ref{CosRob} and \ref{gcoscomp} and hence for an idealised galaxy based 
observer, with constant comoving coordinates $\{r, \theta, \phi\}$, the fundamental 
time parameter $s$ in being to proportional to $\tau$ is in turn equivalent to the 
cosmic time parameter $t$.
Only in this special case under the assumptions of an FLRW model might $s$ be 
associated with a preferred universal temporal parameter, namely the cosmic time $t$, 
for observers attached to idealised galaxies in the context of such a model.

  However, the fundamental temporal flow $s$ itself does not represent a unique 
universal parameter. In the context of large scale structure a local parameter $s$, 
subject to each observer, depends upon the relative motion of the observer with 
respect to a galaxy or the relative finite peculiar velocity of the galaxy itself, in 
precisely the same way as the proper time $\tau$ in special relativity. Similarly the 
parameter $s$ will depend upon the location of the observer with respect to a local 
source of gravity, such as any massive body or even a black hole, again exactly as for 
the proper time $\tau$, in this case as for general relativity.

  The relative time dilations for a community of $N$ observers, each of whom is 
associated with a  personal flow of pure time $s_I \equiv \tau_I$ (for $I=1\ldots N$, 
generalising from the case of `twin $A$' and `twin $B$' described at the end of 
section~\ref{fdandtd}), distributed anywhere in the universe dovetail together in a 
mutually consistent manner. The particular temporal parameter $s_I$ for a given 
observer describes the `fundamental' flow of time underlying the mathematical 
structure of the multi-dimensional form $\lvh$  through which the physical processes 
of the universe unfold from the perspective of \textit{that} observer.
 In this sense \textit{each} $s_I$ is a universal temporal parameter, as noted in the 
discussion of the `problem of time' in section~\ref{qpagig} following 
equation~\ref{hamphi}.

 Locally the flow of time $s\equiv \tau$ parametrises the evolution of fields, such as 
a gauge field $Y(x)$ or fermion field $\psi(x)$ and microscopic quantum phenomena 
generally, as well as the dynamics of macroscopic entities, such as a dust cloud 
described by the energy-momentum tensor $T_{\mu\nu} = \rho u_{\mu}u_{\nu}$ or 
classical matter generally. Either quantum or classical processes  may be utilised in 
the construction of a physical clock which may in turn be employed to \textit{measure} 
the proper time $\tau$ itself and hence observe time dilation effects. With the 
microscopic quantum properties of matter underlying, and in harmony with, the 
macroscopic geometry of gravitational phenomena there is no `problem of time'  in this 
picture, as described in section~\ref{qpagig}, with gravity itself \textit{not} 
quantised.

  In general relativity local coordinates can always be found such that for any 
4-dimensional metric, such as that in equation~\ref{CosRob}, the line element can be 
expressed through a local Minkowski metric with $d\tau^2 = \eta_{ab}dx^{a}dx^b$. In 
the present theory such a local structure \textit{derives} from a 4-dimensional form 
of temporal flow $ds^2 = \frac{\eta_{ab}}{h^2}dx^{a}dx^b$, that is 
equation~\ref{vvvhids}
 which is equivalent to equation~\ref{lorform2},
  that is the expression:
\begin{equation}
 \label{lorfourh}
  L(\bv_4) \, = \, h^2
\end{equation}  
  This latter structure is embedded within a higher-dimensional form such as $\lvt$ or 
$\lvfs$ as described in chapter~\ref{chapesb} and section~\ref{secesef} respectively. 
It is the higher-dimensional form which both sets the normalisation for the temporal 
flow $s$ and gives rise to a range of many possible solutions for an extended 
4-dimensional spacetime, with geometry
  $G_{\mu\nu} = f(Y,\bvh)$ as described for equation~\ref{getypsi}, incorporating 
quantum phenomena in the degeneracy of solutions as described in chapter~\ref{newapp}.

  Hence with the geometry $G_{\mu\nu}(x)$ and the spacetime manifold $M_4$ itself 
together drawn out of the structures implicit in $\lvh$, with solutions such that 
$G_{\mu\nu} = f(Y,\bvh) \neq 0$ in general, there is no presumption of taking a flat 
background manifold as a starting point or expectation of obtaining such a Minkowskian 
spacetime geometry.  With the external curvature related to the internal curvature as 
the symmetries of $\lvh$ are projected over $M_4$, as conjectured in 
section~\ref{reaic} in comparison with Kaluza-Klein theory, there \textit{is} a 
solution with both zero external and zero internal curvature, as implied in 
equation~\ref{gchift} for example with $G_{\mu\nu} = f(Y) = 0$. Even in this case the 
assumption, as applied in section~\ref{fdandtd},  that the value of $\lvfh(x)$ of 
equation~\ref{lorfourh}, as projected out of $\lvh$, is constant throughout spacetime  
is required to obtain a flat spacetime manifold.  
  The consequences of a variation in the value of $h(x)$, as alluded to at the end of 
section~\ref{fdandtd}, will be considered shortly and will contribute, along with the 
freedom of the gauge fields and quantum transitions, to a solution for  $G_{\mu\nu} = 
f(Y,\bvh)$ which is non-zero in the general case.

   Our \textit{a priori} predisposition to mentally project a flat background of space 
and time onto the world in order to perceive objects in it will be consistent with the 
above mathematical structure provided an effective \textit{assumption} of 
$G_{\mu\nu}(x)=0$ is a sufficiently good approximation at least for the region of the 
world we locally inhabit. 
  As discussed in section~\ref{tlssotu} this means for example that
  the local observation of a falling apple can be accounted for in terms of a  `force 
of gravity'
   superposed upon an apparently flat arena of space and time, which in practice is 
both as precise as and much simpler than a full explanation in terms of spacetime 
curvature.
 On the global cosmological scale the observed accelerating expansion of the universe 
not only contradicts the assumption of a flat `vacuum'  geometry, but is also 
counter-intuitive given the terrestrial bias of associating gravity  with a universal 
force of attraction.

   In the present theory the question does not concern what needs to be added to a 
flat background manifold to produce the effects of terrestrial gravity or the 
introduction of  an apparent vacuum energy to account for the accelerating expansion 
of the universe, but rather, in all cases involving gravitation, to ask what is the 
form of $G_{\mu\nu} = f(Y,\bvh)$ in general. This observation applies to both everyday 
material objects such as apples and trees and also in the apparent absence of tangible 
matter in the case of the dark sector for cosmology. This approach in the present 
theory can be summarised in the following three points  (which may be contrasted 
respectively with the  three points listed near the opening of section~\ref{sectsmoc}  
for the standard theory):

\begin{itemize}
  \item 
 Rather than beginning with a flat spacetime $G_{\mu\nu} =0$ and then introducing 
terms such as $T_{\mu\nu}$ or $\Lambda g_{\mu\nu}$ through Einstein's field 
equation~\ref{Einfield} as an apparent \textit{source} of curvature, with matter in 
some sense actively perturbing the otherwise flat geometry, here the energy-momentum 
tensor is \textit{defined} through the Einstein equation, that is $-\kappa T_{\mu\nu} 
:= G_{\mu\nu}$, with the external geometry itself determined through the relation  
$G_{\mu\nu} = f(Y,\bvh)$  out of the underlying flow of time in the form $\lvh$
 (as for the example of equation~\ref{geinawave} and figure~\ref{gacos}).

 \item
 Hence there is no  flat spacetime background, acting as a boundary condition, as an 
apparent \textit{consequence} of the absence of matter. Originating from our 
apparently innate bias to conceive of such a flat spacetime as a given entity, this 
assumption  in part underlies the apparent mystery of the cosmological constant, 
requiring the term $\Lambda g_{\mu\nu}$ to be added to the field equation in a 
seemingly ad hoc manner to account for the empirical observation.

\item
  On the third point quoted from \cite{Rob} in section~\ref{sectsmoc}, a similar 
interpretation applies here.  The identity $\tmo$ follows trivially from the 
definition of $T_{\mu\nu} := G_{\mu\nu}$ given the geometric Bianchi identity $\gmo$. 
Indeed, the \mbox{reverse} interpretation of the Einstein equation with $G_{\mu\nu} := 
T_{\mu\nu}$ implying that matter somehow \textit{causes} spacetime curvature is more 
problematic since an independent justification is then required for the relation 
$\tmo$ in a general curved spacetime, while the identity $\gmo$ does not require any 
such external support. 
\end{itemize}

  Regarding the accelerating expansion of the universe the question then boils down to 
what in the structure of  $G_{\mu\nu} = f(Y,\bvh)$  might account for this 
observation. Ultimately a full understanding will be required for the general 
macroscopic form for $G_{\mu\nu}(x)$ constructed over a degeneracy of underlying local 
field exchanges $\delta Y \leftrightarrow \delta \bvh$, in principle incorporating 
some of the machinery of a quantum field theory as described in chapters~\ref{pp} and 
\ref{newapp}.
 Both the matter density $\rho$ and radiation pressure $p$, for 
equation~\ref{cosgtruup} substituted into equation~\ref{Einfield} to obtain 
equation~\ref{CosLem}, represent possible macroscopic forms of  $G_{\mu\nu}(x)$ which, 
while also entailing classical thermodynamic phenomena, are dependent upon the 
statistical range of possible exchanges for the microscopic fields. 
 Arising out of the degeneracy of possible field solutions the conceptual origin of 
quantum and particle phenomena in the present theory differs to that in standard QFT 
as described in chapter~\ref{newapp}. Correspondingly the notion of a `vacuum state' 
is also different. Indeed the failure of calculations of the value for the vacuum 
energy  in QFT to match the empirical value for $\Lambda$ (typically by 120 orders of 
magnitude, as discussed towards the end of the previous section, see also for 
example~\cite{Pesk} pp.790--791) provides a further argument for the need to reassess 
the underlying structure of QFT itself, in particular in relation to the theory of 
gravitation. The possibility of addressing the cosmological constant problem within 
the context of the present theory was raised at the end of section~\ref{qpagig}.

 While a number of features of the broken $\ese$ action on the components of 
$F(\htho)$ projected over $M_4$ explicitly match features of the Standard Model of 
particle physics, as described for equation~\ref{fhthopart} and summarised in the 
bullet points in section~\ref{sosmfi}, in this chapter we shall describe more 
qualitatively potential connections between features of the present theory and those 
of the standard cosmological model and theories of the very early universe.
  
  As alluded to above a correlation between the external curvature and internal gauge 
fields $Y(x)$, expressed generically as $G_{\mu\nu} = f(Y)$, via the action integral 
of equation~\ref{teinhil},  was described in section~\ref{reaic} through a comparison 
with the framework of Kaluza-Klein theory.
 Further, towards the end of section~\ref{seraps} it was implied that an external 
geometry of a form which might ideally be expressed as $G_{\mu\nu} = f(\psi)$, 
corresponding for example to the electron field $\psi(x)$ for figure~\ref{eemmeds}(a) 
and (b) in section~\ref{qpagig}, may arise from the fermion components within the 
space $F(\htho)$ for the 56-dimensional vectors under $\lvfs$ through interactions 
with the gauge fields or more directly via an expression of the form $G_{\mu\nu} = 
f(\bvh)$. Similarly, without yet having a fully developed quantised theory,  the 
possible physical manifestation of further components in the space $F(\htho)$ may be 
considered.

  In addition to the Lorentz vector $\bv_4$ and Lorentz spinor components  of an 
element of $F(\htho)$, transforming under the external subgroup $\sltc^1 \subset \esi 
\subset \ese$, identified in equation~\ref{fhthopart} there are four Lorentz scalar 
components $\alpha,\beta,n$ and $N$ which may also contribute to shaping the external 
geometry through  $G_{\mu\nu} = f(Y,\bvh)$. In principle any of these four scalars, or 
even the scalar magnitude $\vert \bv_4 \vert$ projected onto $M_4$, could contribute 
to the macroscopic geometry.
 Each of the Lorentz scalars $\alpha,\beta,n,N$ and $\vert \bv_4 \vert$ also transform 
trivially under the internal $\suth_c \times \uo_Q$ gauge groups identified in 
section~\ref{intsym} and incorporated into the $\ese$ symmetry in 
section~\ref{secesef}. Hence, while the specific nature of $\sutw_L \times \uo_Y$ 
actions on these, or any other, components of $F(\htho)$ is not yet known, in lacking 
both strong and electromagnetic interactions any of these scalar fields  might 
contribute to the dark sector in cosmology.

  For example a constant value for a scalar field such as $\alpha,\beta,n, N$ or  
$\vert \bv_4 \vert$ projected over spacetime might be associated with the constancy of 
the scalar $\Lambda$ 
 for an effective  cosmological constant term $\Lambda g_{\mu\nu}$ in the field 
equation~\ref{Einfield} deriving from at least one of these fields.
Interactions between $\alpha,\beta,n, N$ and $\vert \bv_4 \vert$ implied under the 
terms of the constraint $\lvfs$ may underlie empirically observed gravitational 
effects, in particular with the first four of these scalar fields coupled to the 
vector-Higgs $\bv_4$ in this way. Similar interactions under $\lvfs$ also relate to 
the fermion masses as described for equation~\ref{qxmass}, and in particular for the 
low neutrino mass alongside equation~\ref{qxmassnu}.

  The identification of these scalars in the components of the full form  $\lvfs$ 
projected over $M_4$ is analogous to the appearance of a multiplet of scalar fields 
deriving from the components of a higher-dimensional metric in some forms of 
Kaluza-Klein theory,
 via a non-Killing metric $\Phi$ on the gauge group $G$ as alluded to towards the end 
of section~\ref{thwhf}, as the geometry is `reduced' over a 4-dimensional spacetime. 
In the present theory there is no higher-dimensional physical metric but, as for the 
scalars of Kaluza-Klein theories, here also scalar fields deriving from the breaking 
of the full form of temporal flow $\lvfs$ may have implications for cosmology.

  While ordinary matter, subject to the Standard Model internal gauge symmetry $\SML$, 
clumps together with an energy density $\rho(x)$ an essential requirement for a 
cosmological term is that, while locally having a much lower energy density than 
ordinary matter, it should have a largely even effect over cosmological scales in the 
apparent `vacuum' of spacetime. While here not making a quantitative or specific 
argument for the $\Lambda g_{\mu\nu}$ term in the field equation the presence of a 
number a scalar fields in the theory, any of which may impact upon the external 
geometry, provides a source for investigation.

  If a scalar field deriving from a component such as $N$ in $F(\htho)$  does give 
rise to a geometry of the form $G_{\mu\nu} = -\Lambda g_{\mu\nu}$ the effective energy 
density $T_{\mu\nu}:= -\frac{1}{\kappa}G_{\mu\nu}$ in the form of a perfect fluid with 
constant energy density $\rho_{\Lambda} = \frac{\Lambda}{\kappa}$, and equation of 
state
  $p_{\Lambda}=-\rho_{\Lambda}$, might appear as a form of `dark energy'  arising as 
an apparent vacuum state,
 as described after equation~\ref{CosdeS} in section~\ref{sectsmoc}. The dynamical 
implications of such a term, as implied in equation~\ref{CosdeS} and summarised for 
the first FLRW model listed in table~\ref{cosevo} in section~\ref{sectsmoc}, are well 
known to qualitatively match the empirical observation of the accelerating expansion 
of the universe at the present epoch.

  Regarding the projection of the components of $\bv_{56} \inn F(\htho)$ onto the base 
manifold, and again without here making a rigorous argument, a symmetric rank-2 
energy-momentum tensor \textit{could} be constructed as $T_{\mu\nu} = 
\frac{1}{\kappa}\lambda v_{\mu}v_{\nu}$, where $\lambda$ is a real constant and 
$v_{\mu}(x) = g_{\mu\nu}v^{\nu}$ are the components of the Lorentz 4-vector $\bv_4 
\subset \bv_{56} \inn F(\htho)$ projected onto $\TM_4$
 with magnitude $\vert \bv_4 \vert =h$. This proposal is also motivated by analogy 
with the energy-momentum for dust $T_{\mu\nu} = \rho u_{\mu}u_{\nu}$, as contained in 
equation~\ref{cosgtruup} for $p=0$, with the 4-velocity $\bu$ (with $\vert \bu \vert 
=1$) representing the motion of idealised galaxies in the FLRW models. The field 
equation for $T_{\mu\nu} = \frac{1}{\kappa}\lambda v_{\mu}v_{\nu}$ can be written as 
$G_{\mu\nu} + \lambda v_{\mu}v_{\nu} =0$, which has a similar appearance to the field 
equation  $G_{\mu\nu} + \Lambda g_{\mu\nu} =0$ for the de Sitter model. On assuming 
the timelike flow of $\bv_4$ to be aligned with the galactic flow, as is the case for 
the 4-velocity $\bu$, the components of $\bv_4$ are simply 
$v_{\mu} = hu_{\mu}$, which are numerically the same as $v^{\mu} = (h,0,0,0)$ on 
employing the metric of equations~\ref{CosRob} and \ref{gcoscomp} and the comoving 
coordinates $\{t,r,\theta,\phi \}$.
 The substitution of  $T_{\mu\nu} = \frac{1}{\kappa}\lambda v_{\mu}v_{\nu}$ into the 
field equations~\ref{cosg00} and \ref{cosg11} then leads to the identical situation as 
the matter dominated case except here with an apparent matter density $\rho = 
\frac{1}{\kappa} \lambda h^2$. For constant $h(x)$ these equations do \textit{not} 
lead to a solution unless $\lambda(t)$ is allowed to vary with cosmic time as for the 
parameter $\rho(t)$, in which case this model is identical to the matter dominated 
case as listed in the middle column of table~\ref{cosevo}.

  Alternatively, since the full geometry is described by the Riemann tensor (which for 
example is also directly correlated with the internal curvature through relations on a 
bundle space such as equations~\ref{rijkl} and \ref{gamsetkk} in the manner of a 
Kaluza-Klein theory)
 the Ricci tensor, defined with components $R_{\mu\nu} = 
R^{\sigma}_{\ph{\sigma}\mu\nu\sigma}$, might be considered to be geometrically more 
fundamental than the Einstein tensor in terms of having a direct link with the 
underlying fields such as $N(x)$ or $\bv_4(x)$ deriving from the components of 
$F(\htho)$ in equation~\ref{fhthopart}.
 For the case of  a constant scalar $N$ giving rise to a cosmological constant 
$\Lambda$  postulating the relation  $R_{\mu\nu} = \Lambda g_{\mu\nu}$ implies 
directly that $G_{\mu\nu} := R_{\mu\nu} - \frac{1}{2}Rg_{\mu\nu} = -\Lambda 
g_{\mu\nu}$, which is identical to the case of the first model in table~\ref{cosevo} 
already considered above.
 On the other hand postulating the relation $R_{\mu\nu} = \lambda v_{\mu}v_{\nu}$ as a 
possible vacuum limit does lead to a new scenario.  Substituting this expression, with 
$\bv_4 = (h,0,0,0)$ again aligned with the comoving coordinates, into the $R_{00}$ 
component obtained from the Robertson-Walker line element in equation~\ref{rzzaa} 
leads immediately to the relation $3\frac{\ddot{a}}{a} = \lambda h^2$. This expression 
describes an exponentially expanding universe for constant $\lambda > 0$
 and provides some of the 
 motivation for originally considering a $\lambda v_{\mu}v_{\nu}$ term in the field 
equations.

  However, a solution is of course required to be consistent with all components of 
the field equation. The fundamental role of the Einstein tensor is essentially due to 
the contracted Bianchi identity $\gmo$. The relation $R_{\mu\nu} = \lambda 
v_{\mu}v_{\nu}$ implies in turn 
 $G_{\mu\nu} := R_{\mu\nu} - \fh Rg_{\mu\nu} = \lambda v_{\mu}v_{\nu}
    - \fh \lambda h^2 g_{\mu\nu}$ which via the definition 
	$T_{\mu\nu} := -\frac{1}{\kappa}G_{\mu\nu}$ leads to an
	 effective energy-momentum tensor in the form of equation~\ref{cosgtruup}
 for this model with the equation of state $p_{\lambda} =
  \rho_{\lambda} = -\frac{\lambda h^2}{2\kappa}$. This contrasts with the de Sitter 
model with a $\Lambda$ term for which
  $p_{\Lambda} = -\rho_{\Lambda} = -\frac{\Lambda}{\kappa}$, as reviewed above. 
However the differing signs means that for the case of $R_{\mu\nu} = \lambda(t) 
v_{\mu}v_{\nu}$ a solution 
 for equations~\ref{cosg00} and \ref{cosg11}
is only possible if $\lambda$ is negative
 (that is $p_{\lambda} = \rho_{\lambda} > 0$) and allowed to vary in time, with the 
result listed in the final column of table~\ref{cosevo} in section~\ref{sectsmoc}.
 Hence rather than being able to account for an accelerating expansion this hypothesis 
describes a more extreme
  deceleration than either the matter or radiation dominated models.
   Only the first case listed in table~\ref{cosevo} describes an accelerating 
expansion, with $(\rho + 3p) < 0$ as discussed following equation~\ref{cosacc}.

  The redundancy between equations~\ref{cosg00}, \ref{cosg11} and the expression 
$\tmo$ was highlighted by equation~\ref{rhodt}, and similarly  here the identity 
$\tmo$ itself for $T_{\mu\nu} = \frac{1}{\kappa}\lambda v_{\mu}v_{\nu}$  or 
$T_{\mu\nu} = - \frac{1}{\kappa}(\lambda v_{\mu}v_{\nu}
    - \fh \lambda h^2 g_{\mu\nu})$ 
 prohibits a constant value for~$\lambda$. However in principle a full solution for 
$G_{\mu\nu} = f(Y,\bvh)$ may involve a range of contributions individually in the form 
of those  in table~\ref{cosevo} as well as others besides. In this case there will be 
a string of terms effectively composing the energy-momentum tensor which 
\textit{collectively} are required to satisfy $\tmo$, a relation which in general may 
no longer hold  for a particular contribution. This is very similar to the situation 
as described for equation~\ref{gtruum} in section~\ref{subwal} for which a synthesis 
of charged matter and the electromagnetic field  led to the Lorentz force law under 
the constraint $\tmo$. Here effectively a synthesis of several terms may arise within 
$G_{\mu\nu} = f(Y,\bvh)$
on the cosmological scale.

  The 4-vector $\bv_4$ in a $\lambda v_{\mu}v_{\nu}$ term could also be considered to  
have non-zero spatial components which might in principle relate to the formation of 
large scale structure in the universe and open up possibilities not available for a 
purely scalar degree of freedom in a $\Lambda g_{\mu\nu}$ term. However this in turn 
would imply the complication of loosening the FLRW assumptions of homogeneity and 
isotropy in the definition of the metric in equation~\ref{CosRob}. Even within those 
assumptions the possibilities with finite spatial curvature $k=\pm 1$, as well the 
purely $k=0$ solutions of table~\ref{cosevo}, might be further considered. More 
generally, if the general structure of $-\kappa T_{\mu\nu} := G_{\mu\nu} = f(Y,\bvh)$ 
on the largest scales of the universe can be established it will be a case of 
refitting the cosmological data with the parameters of the new model.

   However, unlike the need to provisionally postulate explicit terms such as $\Lambda 
g_{\mu\nu}$ or $\lambda v_{\mu}v_{\nu}$ in the Einstein equation, as potentially 
effectively arising from a Lorentz scalar such as $N$ or the Lorentz vector $\bv_4$ in 
the components of $\bv_{56} \inn F(\htho)$ projected over $M_4$, there is a much more 
direct and intrinsic way in which this projection can shape the 4-dimensional 
spacetime geometry. We describe this observation, and its possible implications for 
the large scale structure of the universe, for the remainder of this section.

  Earlier in this section, as for the discussion in section~\ref{fdandtd}, the 
gravitational time dilation effects for $s\equiv \tau$ have been considered to result 
entirely from the metric $g_{\mu\nu}(x)$ as might be obtained through the Einstein 
equation~\ref{Einfield}, such as the case of the Schwarzschild solution of 
equation~\ref{ttrtp}, that is essentially for cases with a known form of 
energy-momentum tensor. So far a \textit{constant} magnitude has been assumed for  
$\vert \bv_4 \vert^2 = L(\bv_4) = h^2$ in equation~\ref{lorfourh} in the projection of 
$\bv_4 \subset \bv_{56}$ onto $\TM_4$.  However in principle all fields on $M_4$ may 
vary, within the necessary constraints such as $\lvfs$, similarly as the internal 
gauge field $Y(x)$ can vary under the constraint that the action integral of  
equation~\ref{teinhil} should remain stationary, that is $\delta \tilde{I}=0$, for 
example.
 Since the components of $\bv_4 \inn \TM_4$ represent the injection of the pure 
temporal flow $s$ into the base manifold $M_4$ any \textit{variation} in $\vert \bv_4 
\vert$ will itself have some impact on the spacetime geometry. Here we begin by 
considering this impact upon an otherwise flat manifold.

  Hence we first return to the translation symmetry of the form $\lvfh$ under the four 
degrees of freedom $\{x^0, x^1, x^2, x^3\} \inn \rrr^4$, as originally depicted for 
the 3-dimensional case in figure~\ref{spillout}. Here the constant vector field $\bv_4 
= (h,0,0,0)$, in conformity with a constant value for $h$ and with $v^0 = dx^0/ds = 
h$, is aligned with the global Lorentz frame as represented in 
figure~\ref{vtovary}(a). This first figure depicts the uniform translation symmetry 
implicit in the form $L(\bv_4)$ as described in equation~\ref{rspill}, which contrasts 
with the case in figure~\ref{vtovary}(b) in which the magnitude $h(x)$ of the 4-vector 
$\bv_4 (x)$ is free to vary.
  
\begin{figure}[htbp]  
\centering
\epsfxsize=12.5cm
\leavevmode
\epsffile[0 0 1759 687]{\gpath aPfig131e}
\caption{\setb The vector field $\bv_4$ subject to $\lvfh(x)$  for (a) the original 
translation symmetry over $\rrr^4 \equiv M_4$ with constant $h(x)$ and global Lorentz 
symmetry and (b) with $h(x)$ variable and only local Lorentz symmetry. In both cases 
the flow $\bv_4$ is aligned to the timelike coordinate $x^0$ while $x^i$ with 
$i=1,2,3$ represents the three spacelike coordinates.}
\label{vtovary}
\end{figure}

  For the present theory the local metric $g_{\mu\nu}(x)$ is
 projected from the form $L(\bv_4) = \eta_{ab}v^av^b = h^2$,
 framing the local injection of temporal flow into the base manifold. However with the 
local coordinate $x^0$ of figure~\ref{vtovary} representing the fundamental flow of 
time according to the relation $\delta s = \delta x^0/h$, the expression $\lvfh$, 
subject to the constraint $\lvh$, also sets the scale for temporal flow in the local 
frame.
 That is,
  the $x^0$ coordinate representation of time will vary with the value of $h$. With 
 $\delta s = \delta x^0/h$ the pure time $s$ will effectively flow more slowly in 
regions of large $h$, corresponding to the vectors $\bv_4$ with a larger magnitude in 
figure~\ref{vtovary}(b), and more quickly in spacetime regions with a lower value of 
$h$.
 More generally the relation $\lvfh$ can be rearranged in the form $ds^2 = 
\frac{\eta_{ab}}{h^2}dx^a dx^b$, that is the final relation in equation~\ref{vvvhids}, 
with the spacetime metric extracted as:
\begin{equation}
  \label{gwarph}
 g_{\mu\nu}(x) = \frac{1}{h^2(x)}\eta_{\mu\nu}
\end{equation}
  when expressed in the global coordinates on the extended manifold $M_4$
  (see also the discussion of equation~\ref{taufroms} below).
 This physical metric $g_{\mu\nu}$, related to flat spacetime through the conformal 
transformation $\eta_{\mu\nu} \to \eta_{\mu\nu}/h^2(x)$, describes a non-Euclidean 
manifold incorporating time dilation effects.
 As for  general relativity, while general coordinate systems are arbitrary and 
unphysical, local inertial frames with $g_{\mu\nu}(x) = \eta_{\mu\nu}$ and 
$\pal_{\sigma}g_{\mu\nu} = 0$ do have physical significance. Such an inertial frame 
may be identified globally for figure~\ref{vtovary}(a) but only locally for  
figure~\ref{vtovary}(b).
 By the strong equivalence principle the laws of physics according to special 
relativity apply in a local inertial reference frame. As described in 
section~\ref{gcatep} the  weak equivalence principle is sufficient to incorporate the 
notion that all gravitational effects can be transformed away in a sufficiently small 
spacetime volume, and can be interpreted as implying that the torsion vanishes.

  While the unphysical nature of coordinate systems in general relativity is 
encapsulated  under general covariance, as also described in section~\ref{gcatep}, any 
coordinate system  can be represented  by the parameter space grid of 
figure~\ref{onecoord}(a). In the special case of Minkowski spacetime such a coordinate 
grid can be mapped onto the 4-dimensional manifold such that the metric has constant 
components $g_{\mu\nu}(x) = \eta_{\mu\nu}$, as is the case for the spacetime 
underlying the constant vector flow depicted in figure~\ref{vtovary}(a). In this case 
a unique family of coordinate charts are identified through the parameter space of 
translation symmetry of the form $L(\bv_4) = h^2$, as described in 
equation~\ref{rspill}, and related to each other via global Lorentz transformations. 
On the other hand in projecting the coordinate grid of figure~\ref{onecoord}(a) onto 
the spacetime underlying figure~\ref{vtovary}(b) the simplest expression for the 
metric takes the form of equation~\ref{gwarph}.

  For the metric of either figure~\ref{vtovary}(a) or (b) obtained in this way through 
the underlying injection of temporal flow $s$ into the spacetime manifold, as for the  
case of a metric determined as a solution to Einstein's equation as considered in 
section~\ref{fdandtd},  the proper time $\tau$ recorded by physical clocks is again 
tied to the fundamental flow of time $s$. This is the case since the laws of physics, 
including those utilised by the structure of clocks, unfold through the underlying 
temporal flow $s$ and hence the proper time $\tau \equiv s$ exhibits the equivalent 
time dilation effects due to variation in $\lvfh$, as was the case for other sources 
of temporal dilation. The question then concerns the more specific nature of this 
relation between $\tau$ and $s$, as originally discussed at the end of 
section~\ref{fdandtd} and earlier in this section.

   The fundamental temporal flow $s$ is modelled by the real line and hence can be 
represented by the values of a pure real number $s \inn \rrr$, intervals of which can 
be expressed in terms of a set of real parameters of arbitrarily high dimension, as 
described for equation~\ref{sbits}, which is an essential observation for the present 
paper. On the other hand the proper time $\tau$ represents intervals of 4-dimensional 
spacetime on the manifold $M_4$ and is expressed by a real number associated with the 
dimension of length $L$ (which is equivalent to the dimension of time $T$ since 
natural units are employed, and in a sense it would be more appropriate to use $T$ as 
we are ultimately dealing with multi-dimensional forms of temporal flow). Hence the 
constant factor $\gamma$ relating the pure 1-dimensional temporal flow $s$ to a 
corresponding empirically observable progression in proper time $\tau = \gamma s$ is 
one which carries the dimension of length $L$.
 Hence in turn proper time intervals for the spacetime geometry underlying 
figure~\ref{vtovary}(a) or (b) can be expressed through:
\begin{equation}
 \label{taufroms}
   d\tau^2 \, = \, \gamma^2 ds^2  \, = \, \frac{\gamma^2}{h^2}
                                     \eta_{\mu\nu} dx^{\mu} dx^{\nu}
\end{equation}
      Here then the metric $g_{\mu\nu} = \frac{\gamma^2}{h^2}   \eta_{\mu\nu}$ 
explicitly carries the dimension of $L^2$, as was described for the general case in 
the discussion following equation~\ref{CosRob} in section~\ref{sectsmoc}. Since the 
underlying temporal flow $s$ is not directly observed, and the scale of the real line 
parametrising $s$ is in any case arbitrary, once empirical units, such as metres, are 
chosen for $\tau$ the coordinate parameters can in turn be chosen such that the metric 
takes a convenient form. For example in the case of  figure~\ref{vtovary}(a) 
pseudo-Euclidean coordinates can be chosen such that
 $g_{\mu\nu}(x) = \eta_{\mu\nu}$ everywhere.
 
  While the value of the factor $\gamma$ is of no meaning, its significance lies in 
representing a constant relation $\tau = \gamma s$.
 In practice setting $\gamma=1$ can be interpreted as choosing the arbitrary scale of 
$s \inn \rrr$ to match the practical parametrisation of the proper time $\tau$. In 
this case the basic metric $g_{\mu\nu}$ from equation~\ref{taufroms} is that of 
equation~\ref{gwarph}.
 With $h(x)$ varying the constancy of $\gamma$ in equation~\ref{taufroms} implies that 
in general it is not possible to find any coordinates such that $g_{\mu\nu}(x) = 
\eta_{\mu\nu}$ globally for the scenario in figure~\ref{vtovary}(b), although this 
relation is always possible locally, as also suggested by the equivalence principle.

  Hence variation in the value of $h(x)$ on $M_4$ directly modifies the effective 
metric $g_{\mu\nu}(x)$, warping the spacetime geometry  that  underlies the vector 
field in figure~\ref{vtovary}(b) for example. Assuming the geometry to be described in 
terms of a torsion-free linear connection the corresponding  Levi-Civita connection 
$\Gamma$ can be constructed as a function of the metric of equation~\ref{gwarph} via 
equation~\ref{gtoGam} and in turn the components of the  full Riemannian curvature 
tensor $R^{\rho}_{\ph{\rho}\sigma\mu\nu}$ of equation~\ref{ritencon} computed.
 In turn the Einstein tensor, for the conformal geometry $g_{\mu\nu} = 
\theta(x)\eta_{\mu\nu}$ with a real scalar field $\theta(x) = h^{-2}(x)$,  is found 
explicitly (and cross-checked with a related calculation in
  \cite{HawkEl} pp.42 and 76) to take the form:
\begin{equation}
 \label{gmnconf}
  G_{\mu\nu} =
  -\frac{3}{2}\theta^{-2} \pal_{\mu}\theta \pal_{\nu}\theta +
  \frac{3}{4} \theta^{-2} \pal_{\rho}\theta \pal^{\rho}\theta \, g_{\mu\nu} +
      \theta^{-1} \pal_{\mu} \pal_{\nu}\theta  -
	  \theta^{-1} \square  \theta \, g_{\mu\nu} 
\end{equation} 
   A similar expression, with a different set of coefficients, is obtained as a 
function of $h$ under the substitution $\theta \to h^{-2}$, as for any other scalar 
field related to $\theta$ by a simple power expression. The derivation of this 
expression for $G_{\mu\nu}$ follows the same chain of relations that led to the form 
of $G_{00}$ and $G_{11}$, appearing alongside the corresponding $\Lambda g_{\mu\nu}$ 
terms on the left-hand side of equations~\ref{cosg00} and \ref{cosg11} respectively, 
given the  metric form of equation~\ref{gcoscomp} and via the Ricci tensor 
$R_{\mu\nu}$ and scalar curvature $R$. However here equation~\ref{gmnconf} represents 
a direct warping of the spacetime geometry due to  the variation in $\lvfh$ which 
implies equation~\ref{gwarph}, without the need to employ further assumptions 
regarding the form of an energy-momentum tensor in Einstein's equation in order to 
extract a solution.

   Hence this construction can be contrasted with the usual determination of a metric 
$g_{\mu\nu}$ in general relativity. There the metric is extracted as a solution to the 
set of second order differential equations contained in the Einstein field 
equation~\ref{Einfield} under assumptions of symmetry regarding both the matter 
distribution and the form of the metric itself. This was the approach taken for the 
Schwarzschild solution of equation~\ref{ttrtp} and also for the cosmological models 
based on the Robertson-Walker line element of equation~\ref{CosRob}. Here in contrast 
the form of the metric $g_{\mu\nu} = \frac{1}{h^2}\eta_{\mu\nu}$ implies a linear 
connection $\Gamma$ and Riemannian curvature $\bR$ and hence \textit{leads to} the 
Einstein tensor $G_{\mu\nu} = f(\bv_4)$ as a \textit{consequence} of the variation in 
$\lvfh(x)$ under the constraint $\lvh$, rather than as an equation to solve for the 
metric.

  In practice the distribution $h(x)$ might be constrained by observations of the 
corresponding gravitational effects, in a similar way that the constant $k$ and scale 
factor $a(t)$ of the Robertson-Walker line element of equations~\ref{CosRob} and 
\ref{gcoscomp} are determined through empirical observations, found to be consistent 
with $k=0$ and deducing the structure depicted in figure~\ref{aevolve} for example. In 
this sense the procedure to constrain the actual function $h(x)$ is very similar to 
the standard approach for general relativity, that is by matching 
equation~\ref{gmnconf} with empirical observations. On the other hand in this case it 
may also prove possible to calculate both the typical value of $h(x)$, and the typical 
range of variation in this magnitude, as constrained for example under the relation 
$\lvh$, within the theory itself.

 In addition to the warped spacetime $G_{\mu\nu} = f(\bvh)$ of equation~\ref{gmnconf} 
geometries of the form $G_{\mu\nu} = f(Y)$, relating the external curvature to the 
internal gauge fields as described in section~\ref{reaic}, are also possible. The 
combined general expression $G_{\mu\nu} = f(Y,\bvh)$ can be interpreted to incorporate 
a contribution from the gauge fields $Y(x)$ which determine the metric via the 
differential field equations $G_{\mu\nu} = f(Y)$, while the geometry $G_{\mu\nu} = 
f(\bvh)$ concerns the direct impact of $L(\bv_4) = h^2(x)$ on the metric in the form 
of equation~\ref{gwarph}.
 As described in chapter~\ref{newapp} ordinary matter exhibiting quantum phenomena 
will arise out of an underlying degeneracy of solutions for $G_{\mu\nu} = f(Y,\bvh)$ 
given the field exchanges such as $\delta Y \leftrightarrow \delta \bvh$ allowed 
according to the selection rules summarised in equations~\ref{conequa}.

 Since the fermion components $\psi(x)$ in $F(\htho)$ are correlated with the 
components of $\bv_4 \equiv \bh_2$ in $F(\htho)$ under the constraint $\lvfs$ as 
described for equation~\ref{qxmass}, fermion terms may explicitly appear through field 
exchanges of the form $\delta \bv_4 \leftrightarrow \delta \psi$. That is, these 
interactions may directly give rise to the effective geometry $G_{\mu\nu} = f(\psi)$ 
alluded to earlier in this section and towards the end of section~\ref{seraps}, and as 
applied for the external geometric structure associated with the fermion fields 
underlying the processes depicted in figures~\ref{eemmeds}(a) and (b) for example. In 
general the full set of microscopic field redescription possibilities, consistent with 
the constraint equations, will need to be taken into account to determine the form of 
macroscopic geometry $G_{\mu\nu} = f(Y,\bvh)$ as shaped through a local degeneracy of 
field solutions as described in chapter~\ref{newapp}.

 The gravitational time dilation effect, that is the relative slowing of time in the 
vicinity of a massive object, described by the       
Schwarzschild solution for the metric in equations~\ref{ttrtp} and \ref{ttonly}, can 
be ascribed to the presence of the massive object itself in general relativity.
 Accordingly the situation for regions in figure~\ref{vtovary}(b) with relatively 
large values of $h=\vert \bv_4 \vert$, and hence a relative slowing of the flow of 
time, might be `reverse engineered' to identify an apparent presence of `matter' in 
such a region. 
That is, for the geometry $G_{\mu\nu} = f(\bv_4)$ in equation~\ref{gmnconf} it is 
possible to define an associated energy-momentum tensor through $-\kappa 
T_{\mu\nu}^{D} := G_{\mu\nu} = f(\bv_4)$.
 Here $T_{\mu\nu}^{D}$ does not then represent `ordinary  matter' which is  built upon 
a degeneracy of gauge and fermion field solutions for the geometry $G_{\mu\nu}(x)$ 
over $M_4$, and in particular made `visible' through the $\uo_Q$ electromagnetic 
interactions,
 but rather an underlying warping of spacetime geometry itself. Hence, while not 
describing baryonic matter, the implicit energy-momentum $T_{\mu\nu}^{D}$ is a 
candidate for the `dark matter' of the universe.

With  $T_{\mu\nu}^{D}$ defined in this way for the Einstein tensor $G_{\mu\nu}$  of 
equation~\ref{gmnconf} deriving directly from the metric $g_{\mu\nu}$ of 
equation~\ref{gwarph},  this in fact follows the procedure
  for obtaining solutions for Einstein's equation by cataloguing $(g_{\mu\nu}(x), 
T_{\mu\nu}(x))$ pairs  as  outlined towards the end of section~\ref{subwal}.
  Here however it is the form of the metric that is physically motivated and not 
arbitrary while the resulting energy-momentum tensor need not necessarily correspond 
to any known form of matter.

  The term `dark matter' implies a kind of `phantom source' of gravitation, which is 
only detectable through its manifestation as a structure of spacetime geometry, and 
indeed the above definition of $T_{\mu\nu}^{D}$ essentially describes a purely 
gravitational phenomenon. In general a structure described in the relation $-\kappa 
T_{\mu\nu} := G_{\mu\nu}$ may or may not be detectable as `matter' and may or may not 
be detectable as `gravity'. For example ordinary baryonic matter in the form of stars 
or planets is both visible as matter $T_{\mu\nu}$ and evident as gravity $G_{\mu\nu}$. 
On the other hand baryonic matter in the form of tables and chairs, while clearly 
exhibiting a number of properties of matter, does not give rise to any detectable 
gravitational effects. Contrary to that situation `dark matter' in the form of 
$G_{\mu\nu} = f(\bv_4)$ might produce very significant gravitational phenomena without 
being associated with any apparent material effects at all.
 For example, as alluded to above, the variation in $h(x)$ might be determined through 
observations of galactic rotation curves and gravitational lensing effects rather than 
an explicit empirical detection of a `dark matter' distribution
 (as would be possible for example for a cloud of dust on a galactic scale).
 A fourth case is conceivable in which a definite mathematical form of $-\kappa 
T_{\mu\nu} := 
G_{\mu\nu} = f(Y,\bvh)$ has evaded detection both as a material and a gravitational 
entity. 

  Since the material effects of the local ordinary matter distribution  present 
themselves more immediately than the corresponding gravitational phenomena, 
historically the sense that universal gravity is a property to be associated with 
matter was a natural point of view to adopt. In turn the Einstein equation $G_{\mu\nu} 
= -\kappa T_{\mu\nu}$, influenced by the Newtonian gravity which arises in the 
appropriate limit, was initially interpreted to imply that in some sense matter 
`causes' the curvature of spacetime . \textit{That} interpretation is considered to be 
a `reverse engineering'  from the perspective adopted in this paper in which the 
energy-momentum tensor is simply defined through $-\kappa T_{\mu\nu} := G_{\mu\nu}$, 
with the spacetime geometry determined primarily as a solution for  $G_{\mu\nu} = 
f(Y,\bvh)$ subject to the constraint equations (see also the discussion in the opening 
paragraphs of section~\ref{subwal}).

  In terms of the spacetime solution in the particular region of the early universe, 
through
  the mutual gravitation of dark matter and ordinary baryonic matter the effects of  
 $G_{\mu\nu} = f(\bv_4)$, as a network of creases in the underlying fabric of 
spacetime, might have guided the formation of galaxies and galactic clusters.
 The properties of these structures are then visible today through the motions of 
galaxies within clusters and the rotation curves of stars within galaxies, all still 
in mutual gravitational interaction with the dark matter. This interplay between 
baryonic and dark matter is depicted in figure~\ref{darkevo}, where the final stage 
labelled (e) corresponds to the kind of structures observed through to the present 
epoch as also represented in figure~\ref{cosmos}.
\begin{figure}[htb]  
\centering
\epsfxsize=14cm
\leavevmode
\epsffile[0 0 1910 1315]{\gpath aPfig132e}
\caption{\setb (c) Fluctuations in the magnitude $h(x)$ of the vector field $\bv_4$, 
represented by the vertical arrows (as for figure~\ref{vtovary}), in the early 
universe (d) gravitationally merge along with baryonic matter, represented as points 
of dust, as the universe evolves leading to (e) the formation of large scale galactic 
structures as observed through to the present epoch. (Earlier epochs will be 
represented in figure~\ref{vphaset}(a),(b) in the following section).}
\label{darkevo}
\end{figure}

  In the standard cosmological model it is known that the dark matter cannot be 
baryonic due to the abundances of the light elements resulting from nucleosynthesis in 
the early universe.
Dark matter composed of relic particles from the Big Bang must also be weakly 
interacting in order to have evaded direct detection.
 In the case of `cold dark matter' (CDM) the relic particles have a low thermal 
velocity leading to a hierarchical formation of structure through the merger of 
smaller initial units beginning in the early universe. This description is consistent 
with the picture in figure~\ref{darkevo}, except that for `dark matter' in the form of 
variations in $\vert \bv_4 \vert$ there are seemingly no associated `particle' 
phenomena at all.

  Since $\bv_4(x)$ is a 4-vector field, as well as fluctuations in $\vert \bv_4 
\vert$, ascribed to the temporal component $v^0$ in figures~\ref{vtovary}(b)  and 
\ref{darkevo}, in principle there may be variations in the spatial components $v^i$ 
also (as suggested earlier in this section for the case in the final column of 
table~\ref{cosevo} for a spacetime geometry incorporating a $\lambda v_{\mu}v_{\nu}$ 
term in the Einstein field equation) which could be pictured as a horizontal component 
for the vectors in these figures. Such spatial fluctuations are counter to the 
assumption of strict homogeneity and isotropy, as indeed are variations in the 
magnitude $h(x)$, but they could potentially be a factor in the observed peculiar 
motions of galaxies and clusters of galaxies and might even be associated with a `dark 
flow' if observations of such phenomena were to be established.

   Further, while fluctuations towards higher values of $\vert \bv_4 \vert$, that is a 
larger value for $\lvfh$, correspond to regions of spacetime with an apparent slowing 
of the flow of time $\tau \equiv s$, and hence associated with  `dark matter', regions 
with a smaller value for $\lvfh$ will have a complementary spacetime geometry with a 
faster rate of temporal flow and in principle the opposite gravitational effect. Such 
regions may hence tend to open up cosmic `voids' between the galactic clusters and 
play an important role in the  structural evolution process represented in 
figure~\ref{darkevo}. On the largest observable scales such a gravitational repulsion 
might also be a factor in the composition of the apparent `dark energy'.

   It is a very familiar idea that a 2-dimensional surface embedded within a 
3-dimensional space will generally have an intrinsic curvature, such as the surface of 
a ball for example. 
  Here we are considering the embedding of 4-dimensional spacetime within the 
structures of a higher-dimensional form of temporal flow $\lvh$, and it is again 
natural to expect that in general a finite intrinsic curvature for the 4-dimensional 
manifold might result. It is hence proposed that such intrinsic curvature for the 
spacetime geometry, closely correlated with variation in the component values of the 
projected 4-vector  $\bv_4 \subset \bvh$ onto $\TM_4$, constitutes at least a 
significant factor in
 accounting for the observed phenomena of the `dark sector' in cosmology.

  In general relativity spacetime curvature might be considered to account for the 
\textit{origin} of mass in general by interpreting Einstein's field equation 
essentially as a definition of energy-momentum  $-\kappa T_{\mu\nu} := G_{\mu\nu}$, as 
we have in this paper and as reviewed above. This is in contrast with the Standard 
Model of particle physics in which the Higgs field $\phi(x)$ and the Higgs mechanism 
of spontaneous symmetry breaking, as described in section~\ref{ewtatsm}, is 
responsible for the origin of mass
 through field interactions for a theory framed in a flat spacetime background.

 In the present theory with  a continuous variation in the magnitude of the underlying 
field $\bv_4(x)$ on $M_4$, implying time dilation effects and shaping the spacetime 
geometry, an apparent `mass' might be associated with this field through $-\kappa 
T^{D}_{\mu\nu} := G_{\mu\nu} = f(\bv_4)$  as described above, 
 and hence the field $\bv_{4}(x)$ can be considered as the \textit{source} of this 
apparent mass. In subsection~\ref{suboomahp}, and in particular in the discussion 
around equation~\ref{hexpan2}, and similarly around equation~\ref{qxmass} in 
section~\ref{secesef}, the \textit{same} field $\bv_4(x)$ has been associated with  
Higgs phenomena in conveying masses to the fermion and gauge boson fields via possible 
$\delta \bv_4 \leftrightarrow \delta \psi$ and  $\delta \bv_4 \leftrightarrow \delta 
Y$ interactions respectively, compatible with the constraints summarised in 
equations~\ref{conequa}.

  Combining these observations suggests that the physical mechanism through which a 
$\psi(x)$ or $Y(x)$ field interaction with the `vector-Higgs' field $\bv_4(x)$  
results in a `mass' for the fermions or gauge bosons respectively is through the 
effect on the local geometry due to the projection of the field $\bv_4(x)\inn \TM_4$ 
out of $\bvh(x)$ under the full form $\lvh$. Such a $\delta \bv_4$ interaction may 
locally correspond to a further  geometric  effect on top of the continuous $\bv_4(x)$ 
variation. In the present theory such external gravitational effects  will be 
compatible with underlying quantum effects through the above field redescriptions of 
the form $\delta \bv_4 \leftrightarrow \delta \psi$ and  $\delta \bv_4 \leftrightarrow 
\delta Y$, which add to the list of possible interaction vertices of 
figure~\ref{fvhere} in the correspondence with Feynman rules for a quantum field 
theory.
  In the context of QFT the relation between the `bare mass' associated with these 
field interactions and the measured mass of physical particle states, as considered 
near the end of section~\ref{seraps}, will depend upon this impact on the external 
spacetime geometry.

 In this theory while both dark matter and  Higgs phenomena are directly associated 
with the field $\bv_4(x)$ the dark matter is of course not \textit{composed} of Higgs 
particles.  Rather the Higgs interactions via \textit{discrete} $\delta \vert \bv_4 
\vert$ exchanges are closely associated with visible baryonic matter in the universe, 
and as observed in high energy physics experiments, while dark matter relates  to a 
\textit{continuous} variation in the underlying field $\bv_4(x)$.
 Even if $\lvfh(x)$ was constant on large scales, hence with no dark matter phenomena, 
`strongly coupled' field interactions with the vector-Higgs field $\bv_4$ and Higgs 
particles would still be observable in the laboratory. This observation is compatible 
with the apparently `weakly interacting'
 effects of dark matter  as a gravitational phenomenon that arises through variations 
in $h(x)$ on the galactic scale.  
  That is, while the spacetime geometry resulting from the continuous $\bv_4(x)$ 
variation need not itself be associated directly with any quantum or particle 
phenomena, the scalar $\delta \vert \bv_4 \vert$ interactions of the same everywhere 
pervading field may give rise to the detected Higgs particle states.

 More generally the question remains to understand  whether specific quantum or 
particle effects might be associated with the dark sector, and how the geometric 
phenomena arising from the injection of the temporal flow into the spacetime manifold 
relate to the properties of the familiar Standard Model particle states. 
 Together these phenomena shape the world geometry as described collectively under  
 $G_{\mu\nu} = f(Y,\bvh)$. The external geometry will involve the  conformal 
transformation of equation~\ref{gwarph}, which generates Ricci curvature and leads to 
equation~\ref{gmnconf}, together with more general solutions over a degeneracy of 
underlying internal field exchanges, resulting in a combination and interplay of both 
Ricci and Weyl curvature in general.

  Dark matter is empirically observed to be associated with galactic clusters, and 
hence the value of $h(x)$ is expected to be larger in such regions and lower in 
inter-galactic space, as sketched in figure~\ref{darkevo}(e). Given that copious 
photons of light and other Standard Model particles can be detected on Earth after 
being  transmitted through such regions, in travelling  from distant galaxies, it 
appears that the properties of such particles must be physically robust for small 
variations of $h(x)$ to some degree.

 The development of the full physical form and consequences of the expression 
$G_{\mu\nu} = f(Y,\bvh)$ will require a greater understanding of the incorporation of 
quantum phenomena as introduced in chapter~\ref{newapp}. The full implications of the 
theory, derived either via direct calculation or simulation, for particle physics as 
well as cosmology will depend both on the degree of variation of $\lvfh(x)$ and the 
typical value of $h(x)$ itself at the present epoch. With material properties and the 
laws of physics likely to have some dependence on the value of $h(x)$ it may be that 
the Standard Model of particle physics requires a certain apparent `tuning' of this 
parameter in order to allow the formation of ordinary baryonic matter itself. This 
raises the question more generally of the possible uniqueness, or otherwise, of the 
`physical constants' as observed in the world, both for Standard Model of
 particle physics and in terms of the cosmological parameters, as we shall discuss in 
section~\ref{secuni}.

 In the following section we first  consider the possibility that both the average 
value and the fluctuations in $h(x)$ may have been very different in the very early 
universe, leading to a `phase transition' to an average value of $h(x)$ compatible 
with the nature and properties of Standard Model interactions in particle physics and 
which has remained stable to the present day. In this scenario the phase transition 
may mark a point of convergence upon the familiar laws of physics in 4-dimensional 
spacetime more generally. These may include the second law of thermodynamics expressed 
in terms of the degrees of freedom of familiar interacting particles which are 
themselves produced in the phase transition.

  In summary, the `novel perspective' in the title of this chapter refers to the
   manner in which the intrinsic geometry of the 4-dimensional spacetime backdrop for 
cosmology is shaped through the projection of the extended $M_4$ manifold out of the 
full multi-dimensional form of progression in time $\lvh$.
 While the present theory based on general forms of time is very simple there are a 
number of features such as the projection of the vector-Higgs field $\bv_4$ onto 
$\TM_4$, generating  the conformal transformation of equation~\ref{gwarph} 
parametrised for example by the dilation symmetries described in the opening of the 
following section, and a set of elementary scalar fields  $\alpha,\beta,n$ and $N$, as 
described earlier in this section, which potentially correlate with large scale 
cosmological phenomena in particular associated with the dark sector. These features 
identified within the components of $\bv_{56} \inn F(\htho)$ for the 56-dimensional 
form $\lvfs$ are complementary to the features identified for the Standard Model of 
particle physics as summarised in equation~\ref{fhthopart} and in 
section~\ref{sosmfi}. Since the known phenomena of HEP cannot account for the dark 
sector in cosmology  new features, such as identified in this section, are indeed 
required to account for the cosmological parameters. 
 There then remains the question concerning the degree to which the mathematical 
structures described in this section might compare quantitatively with empirical 
observations of the large scale physical structure of the universe. 

  A first step will be to seek a guide through a comparison between the approach of 
the present theory and geometric models aimed at accounting for the dark sector of 
cosmology in the existing literature. While papers involving conformal gravity (see 
for example \cite{Mann,Nesb}) may account for elements of the dark sector in geometric 
terms, such models appear somewhat different to the approach described in this 
section.
  In replacing the Einstein-Hilbert action by a conformally invariant action based on 
the Weyl tensor these papers do however implicitly incorporate geometric 
transformations of the kind in equation~\ref{gwarph} and hence may relate to the 
structures of the present theory. 
 The present theory both aims to avoid the employment a Lagrangian formalism and does 
not propose a `modified gravity' of any kind. In fact here the Einstein equation is 
identified as a fundamental feature embedded within the definition of energy-momentum 
through the expression $-\kappa T_{\mu\nu} := G_{\mu\nu} = f(Y,\bvh)$, which also 
provides the interpretation of the Einstein equation in the context of the present 
theory. While one aim of this paper has been to avoid `postulating' a Lagrangian of 
any form, the Einstein-Hilbert action of equation~\ref{einhil} for general relativity, 
for the vacuum case with $\lag=0$ and $\Lambda =0$, has been  adapted in 
section~\ref{reaic} to  facilitate a provisional  connection between the external and  
internal geometry arising from the symmetries of $\lvh$ broken over the base manifold 
$M_4$, as described for equations~\ref{teinhil} and \ref{einhilnol} and guided by 
Kaluza-Klein theory.

  Further, rather than devising a scheme tailored to match empirical observations, 
here we begin with an underlying conceptual motivation and foundation for the theory.  
  Once this theory has been sufficiently developed a full cosmological model might be 
established leading for example to a calculation for the density parameters 
$\Omega_B$,
 $\Omega_D$ and $\Omega_{\Lambda}$, as introduced following equation~\ref{ommatter} in 
the previous chapter. In principle the cosmological data itself might
 then be refit within the context of the theory in order to test these ideas 
quantitatively.



\section{The Very Early Universe}
\label{sectveu}

  The highest-dimensional form of temporal flow considered in any detail in this paper 
is the form $\lvfs$ with $\ese$ symmetry, as introduced in section~\ref{secesef}. If 
any of the four scalar components $\{\alpha,\beta,n,N \}$ of $\bv_{56} \inn F(\htho)$ 
in equation~\ref{fhthopart} is found to be associated with an effective cosmological 
term $\Lambda g_{\mu\nu}$ in equation~\ref{Einfield} then the magnitude of this 
component will be correlated with the magnitude of the acceleration of the cosmic 
expansion, which may be arbitrarily small. 
Similarly the magnitude and variation of the projected $\bv_4 \inn \TM_4$ components 
may directly correlate with the properties of dark matter or dark energy, while as a 
`vector-Higgs' the field $\bv_4$ also generates mass terms for the fermions and gauge 
bosons and underlies Higgs phenomena in general, as also reviewed in the previous 
section.
  Interactions between the scalars $\{\alpha,\beta,n,N \}$ and  $\bv_4$ might also 
generate massive weakly-interacting scalar states as a possible contribution to the 
dark sector.

  One means of varying $\vert \bv_4 \vert$ can be described via a simple dilation 
symmetry as a subgroup of $\sootn$ for the
  model described in section~\ref{reaic} for a 10-dimensional form $\lvte$ projected 
over $M_4$ as pictured in figure~\ref{mtogmaphr}. This dilation symmetry  acts on the 
components of $\bv_{10}$ such that the magnitudes of the external $\ol{\bv}_4$ and 
internal $\ul{\bv}_6$ vectors are  traded subject to the constraint $L(\bv_{10}) = 
\vert \ol{\bv}_4 \vert^2 +  \vert \ul{\bv}_6 \vert^2 = 1$. This variation in  $h = 
\vert \bv_4 \vert $ can also be described in terms of a one-parameter subgroup denoted 
$\mbox{D}(1)_X \subset \sltwoo$ acting as a dilation symmetry on the components of
 $\bv_{10} \equiv X \inn \htwo$ in equation~\ref{xoct} preserving $L(\bv_{10})=\det 
(X) = 1$.

  In terms of the largest form of temporal identified
 another possibility for `tuning' the magnitude of the $\bv_4$ components under the 
constraint
  of the  56-dimensional form $\lvfs$ lies in the dilation symmetry, which will be 
denoted $\mbox{D}(1)_{\lambda} \subset \ese$, as parametrised by $\lambda\inn \rrr$ 
and introduced in equation~\ref{lamtran}. This symmetry acts upon all 27 components of 
$\mcY \inn \htho$ of the 56-dimensional space $F(\htho)$ in a uniform way, and not 
only on the $\bv_4 \subset \mcY$ subset of components in equation~\ref{fhthopart}.

 As an intermediate case
 a further dilation of components can be identified within the $\esi$ symmetry on the 
$\htho$ subspaces of $F(\htho)$ and will be denoted $\mbox{D}(1)_B$, as generated by 
the linear combination of boosts $\btzt$ and as introduced in equation~\ref{uobtzbtz}. 
From table~\ref{lbrota} this $\esi$ generator as a vector field in the tangent space 
$T\htho$ has the form:
\begin{equation}
 \label{bbdil}
  \btzt \; = \; 
  \left(\!\! \begin{array}{ccc}
                    +p    &    +\bar{a}      &   -\fh c     \\
	                +a    &   +m       &   -\fh \bar{b}      \\
                 -\fh \bar{c} &   -\fh b   &   -2n 
 \end{array} \!\;\!\!\right)
\end{equation} 
  Hence under the action of $\mbox{D}(1)_B$ on the components of $\mcX\inn \htho$ in 
equation~\ref{xoct3} the 10 components of $X$ are inflated while the remaining 17 
components of $\theta$ and $n$ are deflated, or vice versa. The consistency of this 
$\mbox{D}(1)_B$ action with the preservation of $L(\bv_{27}) = \det(\mcX)$ can be seen 
directly from the form of $\det (\mcX)$ in equation~\ref{detx3} together with the 
generator coefficients 
 in equation~\ref{bbdil}. For this particular linear combination of boosts the rank-6 
$\esi$ Lie algebra contains a rank-6 subalgebra decomposition, which in terms of the 
corresponding Lie groups can be written as:  
\begin{equation}
 \label{bbinesi}
   \sltc^1 \times \suth_c \times \uo_Q \times \mbox{D}(1)_B \subset \esi
\end{equation}
 As mentioned for the same decomposition in equation~\ref{uobtzbtz} the mathematical 
structure of this Lie subalgebra is described in \cite{Wang} (the first example in 
Appendix D, p.187). 
 In the present paper the subgroup $\sltc^1$ has been identified with the external 
symmetry of spacetime $M_4$ in section~\ref{extsym}, and $\suth_c \times \uo_Q$ as the 
internal symmetry subgroup within $\stab$ in section~\ref{intsym}, in each case with a 
corresponding physical interpretation. Clearly  $\mbox{D}(1)_B$ is not  a subgroup of 
$\stab$ due to the action on the $\bv_4 \subset X$ components in equation~\ref{bbdil}, 
however since $\mbox{D}(1)_B$ is independent of $\sltc^1 \times \suth_c \times \uo_Q$, 
as described in equation~\ref{bbinesi}, it may also be of physical significance.

  Regardless of the means of varying the 4-vector magnitude $\vert \bv_4 \vert = h$, 
whether via $\mbox{D}(1)_X$,  $\mbox{D}(1)_B$, $\mbox{D}(1)_{\lambda}$ or otherwise, 
the geometric impact of relatively high values of $\lvfh(x)$ projected out of the full 
form $\lvh$ over extended regions of the cosmos 
 is considered to form a candidate for the effects of dark matter, as described in the 
previous section. As well as these small variations in $h(x)$, the implications of a 
much larger time-dependent evolution in the scalar value $h(t)$, as averaged over the 
3-dimensional spatial hypersurfaces  as a function of cosmic time $t$, can be 
considered for the very early universe, as we describe in this section. For example we 
shall consider the progression from a  value approaching zero $h(t) \to 0$ for $t \to 
0$ towards a stable average value $h(t_v) = h_0$ in a period of time associated with 
the epoch of the Big Bang, with $h_0 = \vert \bv_4 \vert$ hence also denoting the 
present day average value. On adopting a normalisation factor of $\gamma=1$ (a further 
natural option would be to set $\gamma = h_0$) relating the fundamental temporal flow 
$s$ with the proper time $\tau$, as described for equation~\ref{taufroms},  
 the basic metric  deformation can be described by equation~\ref{gwarph}. An interval 
of proper time $d\tau$ can then be expressed as:
\begin{equation}
\label{hconfac}
     d\tau^2 =  ds^2  \, = \, \frac{1}{h^2(t)}
	  \left[  dt^2 \, - \, d\Sigma^2 \right]
\end{equation}
 where $d\Sigma$ represents a Euclidean 3-dimensional spatial element.
  While  this line element has the form of a conformal transformation dependent upon 
$h(t)$,
  similarly as for the case of equation~\ref{Coseta} for the FLRW models with 
conformal time parameter $\eta$, we continue to think of $t$ as the `cosmic time' 
parameter. Given that the value of $h(t)$ is only considered to differ from the 
present value $h_0$  for $t < t_v$ in the very early universe, and that elementary 
physical structures will be unfamiliar during that epoch as the nature of the 
projection of $\lvh$ over that region of $M_4$  is correspondingly also   different, 
the form of any `physical clock' and the measure of time itself will need further 
consideration for this earliest era. Hence in the above mathematical expression we 
keep track of the coordinate time $t$ in place of defining a new temporal parameter.

   The identification of $\mbox{D}(1)_B \subset \esi$ alongside other subgroups in 
equation~\ref{bbinesi} is analogous to the proposed subgroup $\sutw_L \times \uo_Y 
\subset \ese$, as a candidate for the gauge symmetry underlying the left-handed weak 
interactions, as described in equation~\ref{esedecom} and the subsequent discussion.
 In the case of  $\sutw_L \times \uo_Y$ the fact that this symmetry is not independent 
of $\stab$ (while it is independent of $\sltc^1 \times \suth_c $) leads to the 
phenomena of `electroweak symmetry breaking' through the interaction of the $\sutw_L  
\times \uo_Y $ gauge fields and the $\bv_4 \inn \TM_4$ vector-Higgs field, breaking 
the symmetry down to $\uo_Q$.
 For the future development of the present theory it will be important to gain an 
understanding of the interplay between electroweak symmetry breaking (as well as the 
role of the unification scale described for figure~\ref{runcup}) and the breaking of 
the $\mbox{D}_{B}(1)$ symmetry action (or other dilation symmetry) in the very early 
universe.

   A `gauge field' associated with the $\mbox{D}(1)_B$ dilation symmetry, through 
interaction with the components of $\bv_{56}$, might itself drive an inflationary 
effect in the very early universe. The very different physical environment associated 
with
a very different magnitude of $\bv_4 \subset \bv_{56}$ projection onto $\TM_4$ in the 
very early universe might also  in principle incorporate some of the effects of a 
cosmological inflation. In either case, following the inflationary epoch the 
$\mbox{D}(1)_B$ symmetry action would effectively  be broken in a `phase transition' 
as the value of $\vert \bv_4 \vert$ is stabilised and the parameters of the Standard 
Model of particle physics are established.
 That is, with field interactions leading to a mutual stabilisation of both the value 
of 
 $\vert \bv_4 \vert = h_0$ and the Standard Model parameters.

 In this theory, with the field $\bv_4$ also closely associated with  Higgs phenomena, 
these parameters include the masses of the fermions through couplings implied in the 
constraint $\lvfs$, as recalled towards the end of the previous section.  The 
$\mbox{D}(1)_B$ symmetry, as generated by the $\esi$ Lie algebra element of 
equation~\ref{bbdil}, applies to both $\mcX$ and $\mcY \inn \htho$ in 
equation~\ref{fhthopart} and with a uniform action on all components of $Y \subset 
\htwo$, not just the $\bv_4 \inn \htwc$ subspace.  The $X$ and $Y$ components in 
equation~\ref{fhthopart} carry the $u$-quark and $\nu$-lepton states  while the 
$d$-quark and $e$-lepton states reside in the $\theta^1_{\mcX}$ and  $\theta^1_{\mcY}$ 
components. 
 Hence the stabilisation  of the magnitude of $\bv_4 \inn \TM_4$ in the projection out 
of $\lvfs$, as the $\mbox{D}(1)_B$ symmetry is broken in the early universe, will 
establish the observed masses for the  $u$-quark and $\nu$-lepton states  in 
comparison with those for the 
 $d$-quark and $e$-lepton states. The `tuning' to these values may be automatic if 
there is some mechanism underlying the stability for the corresponding value of 
$\lvfh$
 (see also the discussion at the end of section~\ref{secesef}).

   As described in section~\ref{sosmfi} a yet higher-dimensional form of temporal flow 
may be required to fully identify the  $u$-quark and $\nu$-lepton states 
 as $\sltc^1$ fermions as well as to identify the second and third generation of 
Standard Model fermions. The structure of such a higher-dimensional symmetry of time, 
possibly involving an $\ee$ symmetry on a form of temporal flow $\lvtfe$ of greater 
than quartic order, may hence be needed to address the question of the stability of 
$\lvfh(t)$.

  As well as  the fermions the masses for the $W^{\pm}$ and $Z^0$ gauge bosons from 
the $\sutw_L \times \uo_Y$ sector of the theory and the Higgs mass itself will also be 
established at the epoch of this phase transition. 
 Fermion pairs such as $e^+e^-$ might be produced  via the decay of a heavy gauge 
boson, as associated with the Feynman vertex of figure~\ref{fvhere}(a), or via other  
underlying field exchanges of the form $\delta Y \leftrightarrow \delta\psi$ 
generalising from a geometric solution $G_{\mu\nu} = f(Y)$ for the spacetime geometry. 
 However the energy density of the early universe in the form of $-\kappa T_{\mu\nu} 
:= G_{\mu\nu} = f(\bv_4)$ might also be converted into fermion states through 
underlying   $\delta \bv_4 \leftrightarrow \delta\psi$ field exchanges, under $\lvh$ 
as implied in the  previous section.

  As well as the production of such fermion pairs
 new interactions, for example under the full form $\lvtfe$, may be significant in the 
high energy density environment of the very early universe. Such a form, involving 
quintic or higher order field composition terms, might involve the  production  of 
leptons and quarks in the same interactions, with potentially a mechanism for creating 
an asymmetry between matter and antimatter acting during this very early  epoch 
through to the phase transition.
 A further possibility might involve gauge bosons, for example from a `beyond the 
Standard Model' $\sutw \subset \ee$ subgroup of the full symmetry of time (in 
principle identified through an explicit symmetry breaking decomposition in the form 
of equation~\ref{eedecomp} acting on the components of $\lvtfe$), which might mediate 
interactions between leptons and quarks in an analogous manner to the `$X$ and $Y$' 
gauge bosons of an SU(5) GUT model.
The origin of this  imbalance of matter over antimatter remains to be understood, but 
the underlying  asymmetry in the directed flow of time, the parity asymmetry arising 
from in the choice of the $\bv_4 \inn \TM_4$ projection out of the components of 
$F(\htho)$ in equation~\ref{fhthopart}, 
 a mechanism for combined lepton plus quark production or even {\it CP} violation in 
the quark sector, all in the context of the present theory, may play a part here.

 In the standard theory by the time the temperature of the universe has cooled to 
$10^{12}\,$K at around $10^{-4}$ seconds after the Big Bang quarks and gluons no 
longer form a component of a relatively weakly interacting plasma, along with leptons 
and photons, but become confined in hadronic states. A proton to photon ratio of 
around $10^{-9}$ to one is established at this epoch, with a negligible contribution 
from antiprotons. Similarly most of the initial electrons and positrons mutually 
annihilate leaving a residual $e^-$ contribution, balancing the residual $p^+$ states, 
leading to the much later recombination era as electrons combine with nuclei forming 
neutral atoms around 372,000 years after the Big Bang, marking the origin of the CMB 
radiation as observed today and as described in section~\ref{secinf}. 
  An understanding of the origin of the
 imbalance between matter and antimatter states in the very early universe, accounting 
for the predominance of `matter' states as still observed today, within the context of 
the present theory may also aid in the identification of the mathematical structure of 
the currently hypothetical $\ee$ action on the form $\lvtfe$, augmenting the input 
from the required Standard Model properties as discussed in section~\ref{sosmfi}.

  For the case of $\mcX \inn \htho$ with constant $L(\bv_{27}) = \det(\mcX) = 1$ and 
given a very small initial value of $\vert \bv_4 \vert$ for a projected $\bv_4 \subset 
\mcX$ a very large value for the scalar $n$ is permitted, as can be seen from 
equation~\ref{detx3}. 
 Similarly, in the context of the form $\lvfs$ and equation~\ref{fhthopart},
 with $\bv_4 \inn \TM_4$ projected from the $\mcY$ components,
  a very large value of  the scalar field $N(x)$ may be achieved if the dilation 
symmetry $\mbox{D}(1)_B$, with the generator of equation~\ref{bbdil}, is involved in 
obtaining a very small value of $\vert \bv_4 \vert$. Further, if this latter value is 
obtained via the dilation symmetry $\mbox{D}(1)_{\lambda}$
of equation~\ref{lamtran} then a very large value for either the scalar field $n(x)$ 
or $\beta(x)$ from equation~\ref{fhthopart} can result. Hence if a cosmological term 
$\Lambda g_{\mu\nu}$ is derived for equation~\ref{Einfield} in the early universe, 
with the scalar $\Lambda$ closely related to any of the scalars $N$, $n$ or $\beta$ 
then a temporary but rapid inflationary dynamics might be obtained.

  The action of a dilation symmetry  $\mbox{D}(1)_B$ or $\mbox{D}(1)_{\lambda}$, or 
some combination, may  slide the value of $\vert \bv_4 \vert = h \simeq 0$, 
corresponding to an initial unstable `inflationary state' in the very early universe, 
towards a preferred solution under $\gmyv$ with a stable value for $\vert \bv_4 \vert 
= h_0$ via interactions with other fields on $M_4$.
 In the very early universe  $\bv_4(x)\simeq 0$, correlated with a very large value 
for a scalar field such as $N(x)$ or $\beta(x)$, might account for inflationary 
phenomena, with the largest inflation driven for example by $N \to \infty$ as $\vert 
\bv_4 \vert \to 0$ for $t \to 0$. After the magnitude of $\bv_4$ has grown in time the 
same field with small variations around $\vert \bv_4 \vert = h_0$ might account for 
dark matter effects, as described in the previous section, while a residual, now small 
value for $N(x)$  or $\beta(x)$ might account for the dark energy term $\Lambda 
g_{\mu\nu}$ at the present epoch, as a greatly suppressed remnant of the early 
inflationary era. A  field such as $N(x)$  driving inflation in the very early 
universe must be weakly coupled at the present epoch in order to have evaded detection 
in the laboratory. 
 On the other hand the stable and complementary scalar field $h_0(x) = \vert \bv_4 
\vert$ is associated with the vector-Higgs field $\bv_4$,  giving rise to  phenomena  
which are evident in experiments.

   At the time of the phase transition $t=t_v$ the energy of the vector-Higgs field 
$\bv_4$ is transferred to fermion and gauge particle states under the external 
geometry $G_{\mu\nu} = 
f(Y,\bvh)$ solution. In addition to the stable value of $\vert \bv_4 \vert = h_0$ the 
masses of the fermions will be established under terms of $\lvh$, as described for the 
case of $\lvfs$ in equations~\ref{qxmass} and \ref{qxmassnu} at the end of 
section~\ref{secesef}.
 Hence the  low value of the cosmological constant  $\Lambda$ may be correlated with 
the low value of the neutrino mass, and the pattern of fermion masses more generally, 
according to the balance between the stable scalar values for $\vert \bv_4 \vert = 
h_0$, $n$, $N$, $\alpha$ and $\beta$ in equation~\ref{fhthopart}, although again a 
full form such as $\lvtfe$ may be required for the full picture. 
  
  Under the assumptions applied for FLRW models, as described in 
section~\ref{sectsmoc}, while for $t>t_v$ a radiation dominated solution for the line 
element of equation~\ref{CosRob}
 initially emerges, for $t<t_v$ both the scale factor $a(t)$ of that equation and the 
conformal factor $h(t) = \vert \bv_4 \vert$ of equation~\ref{hconfac} combine together 
to form the line element:
\begin{equation}
\label{hconfaca}
     d\tau^2 \, = \,  ds^2  \, = \, \frac{1}{h^2(t)}
	  \left[  dt^2 \, - \, a^2(t)\, d\Sigma^2 \right]
\end{equation}
 In general it will be necessary to solve the 4-dimensional geometry $G_{\mu\nu} = 
f(Y,\bvh)$ to determine the dynamical form of $g_{\mu\nu}(x)$ both for $t>t_v$  for 
the evolution from a radiation to a matter dominated universe and on to the era of 
dark energy dominance and potentially also for an `inflationary' epoch for $t<t_v$, 
both in principle involving an evolution of the scale factor $a(t)$ driven by an 
effective $\Lambda g_{\mu\nu}$ term in  equation~\ref{Einfield} induced by a scalar 
component such as $n, N, \alpha$ or $\beta$.
 As described above such a scalar field might take a very large value in the early 
universe as balanced against $h(t) \to 0$ as $t \to 0$ under the constraint $\lvh$. As 
was described for equation~\ref{hconfac} the parameter $t$ is considered to represent 
`cosmic time' rather than `conformal time', even for $t<t_v$. This convention is 
further justified here with the scale factor $a(t)$  incorporated into the more 
complete expression in equation~\ref{hconfaca}.

  However, for $t<t_v$ in addition to a possible $\Lambda g_{\mu\nu}$ term there are  
new features that arise in the present theory.
 The initial low value for $\vert \bv_4 \vert = h(t)$ in equation~\ref{hconfaca} 
implies a relatively rapid flow of the fundamental time $s$ through the spacetime 
manifold as parametrised by the
 cosmic time $t$. This property  is complementary to the relative slowing of time in 
the later universe  associated with small fluctuations to relatively high values of 
$\vert \bv_4 \vert$ distributed in space, with the resulting gravitational effects   
ascribed to apparent regions of `dark matter' as described for figures~\ref{vtovary}
 and \ref{darkevo}. The complementary case with much lower values of $\vert \bv_4 
\vert$ in the very early universe may
  imply an effective expansion of spacetime  (which might also apply to a smaller 
degree in regions of the later universe correlating with  `voids' between galactic 
clusters, as discussed after figure~\ref{darkevo} also in the previous section).

  In the case of the very early universe the conformal scaling via the factor 
$h^{-2}(t)$ in equation~\ref{hconfaca} as $h(t)$ becomes smaller for $t \to 0$ means 
that intervals of the `comoving coordinates' $\{t, r, \theta, \phi\}$ represent 
greater physical spacetime volumes as $t \to 0$. For this conformal geometry an 
infinite spacetime volume may be inscribed within a finite coordinate boundary (as 
might be represented for example by the `Circle Limit' woodcuts of M.C. Escher 
described in \cite{Pen} pp.33--34).

  Allowing an infinite passage of time in the past in this way with $\tau \equiv s \to 
-\infty$ as $t \to 0$ may itself not help solve the `horizon problem' since spatial 
volumes are dilated by the same factor. That is, the conformal diagram of 
figure~\ref{confbb}, representing the causal structure for the evolution of the 
universe since the `initial singularity' at $t=0$, is unchanged by variation in $h(t)$ 
alone.
 As for the standard approach to solving the horizon problem it appears necessary to 
`miniaturise' physical spatial displacements relative to temporal intervals at the 
earliest epoch via for example an inflationary dynamics of the scale factor $a(t)$, as 
has been applied for figure~\ref{confbba}.  Hence although in the present theory much 
more time may be available in the very early universe both $h(t)$ and the scale factor 
$a(t)$, which will be mutually correlated in the dynamics, play significant roles in 
equation~\ref{hconfaca}.

 Nevertheless, in the present theory the magnitude of any inflationary effect, in 
terms of the increase in the scale factor $a(t)$, and its period of duration may be 
somewhat different than in the original theory of inflation. Here the non-uniformity 
in the way that the underlying flow of time $s$ is injected into the spacetime 
manifold as $h(t) = \vert \bv_4 \vert$ evolves may have consequences which partially, 
or even totally, remove the need for a rapid `inflation'. A much smaller value for 
$h(t)$ in the very early universe relative to the present day value will also mean 
that the properties of physical structures are likely to be very different, compared 
with those of the Standard Model for example. These differing structures may also in 
principle imply uniform characteristics, such as  `temperature', across the initial 
singularity, with little if any time required to attain the high degree of  `thermal 
equilibrium' as observed today for the CMB radiation across the full spatial extent of 
the observable universe.

 In assuming the vector field $\bv_4(x)$ to take the particular unstable value in the 
very early universe with $\bv_4(x)\to 0$ as $t\to 0$ the dynamics of the cosmic 
evolution, described globally through $\gmyv$, will change with the phase transition 
at the time at which the stable value $\vert \bv_4 \vert = h_0$ emerges at $t=t_v$. 
This potentially abrupt change in the nature of the dynamics may be accompanied by a 
reduction in symmetry, in particular regarding the effective breaking of the dilation 
symmetry, composed of a combination of the groups $\mbox{D}(1)_{X}$, $\mbox{D}(1)_{B}$ 
and $\mbox{D}(1)_{\lambda}$ described in the opening of this section for example.
 In the original inflation theory there is no relation between the postulated scalar 
inflaton field $\varphi(x)$ and the Standard Model scalar Higgs field $\phi(x)$ of 
particle physics. In the present theory  `inflation' in the early universe is 
correlated with a very small value for $\bv_4(x)$, while Higgs phenomena derive from 
the present stable vector field with  $\vert \bv_4 \vert = h_0$, with the two values 
of the same field $\bv_4$ related through the action of the dilation symmetries in the 
very early universe. 

  There are however some models in the literature for which inflation in the very 
early universe \textit{is} correlated with the Standard Model Higgs field, and the 
properties of conformal transformations, in some way, as for example in 
\cite{Garc,Bezr,Brax}. These references typically incorporate a coupling between the 
Higgs and gravitational fields by postulating a new interaction term in a Lagrangian 
of the form 
  $\lag \sim \xi \phi^{\dag}\phi R$, where $\phi$ is the scalar Higgs field, $R$ is 
the scalar curvature and $\xi$ is a new  coupling parameter. In the theory presented 
in this paper however the Higgs sector is more intimately associated with gravity 
since variations in the magnitude
 $h(x) = \vert \bv_4 \vert$ of the vector-Higgs field $\bv_4(x)$ directly impact upon 
the external spacetime geometry via a change in the metric of the form described by 
equation~\ref{gwarph}.

  The picture of the very early universe in the present theory does also have close
  parallels with the original inflationary models described in section~\ref{secinf}. 
The scalar magnitude $h = \vert \bv_4 \vert$ for the initial projection of $\bv_4 \inn 
\TM_4$ out of the components of $\bvh$ in the very early universe is analogous to the 
initial value of the scalar field $\varphi$ in inflationary theory. For example in  
`old' or `new' inflation the initial value $\varphi = 0$ becomes a `false vacuum' as 
the potential $V(\varphi, T)$ is modified with the dropping cosmic temperature $T$ 
until subsequently a stable condition with $\varphi = \varphi_0 \neq 0$ is achieved. 
In the present theory the consequences can be considered for a state with $h=\vert 
\bv_4 \vert \simeq 0$ in the very early universe followed by a continual range  of 
projections of $\lvh$ over $M_4$ until the stable value $h = \vert \bv_4 \vert =h_0 
\neq 0$ is achieved, particularly in terms of the form of the geometric solution 
$\gmyv$.

 In the present theory interactions between the components of the vector field 
$\bv_4(x)$ and for example the fermion field
  $\psi (x)$  under the $\lvfs$ terms, together with gauge fields $Y(x)$ via terms in 
the expansion of $\dmofs$, may compose a thermal system incorporating an effective 
temperature dependent potential $V(h,T)$. Such interactions will also generate `drag 
terms' in the dynamics of the evolution of $h(t)$ leading to damping effects 
 accompanying a possible
  period of  oscillations about the potential minimum  as energy is transferred to 
Standard Model particle states created and `reheated'  as the point of stabilisation 
with $h(t) = h_0$ at time $t=t_v$ is approached. This describes the `phase transition' 
at the end of an inflationary period, leaving a residual dark energy contribution, 
arising for example from a much reduced and stable value for the scalar field $N(x)$, 
in addition to the Standard Model particle spectrum as observed today. A radiation 
dominated FLRW cosmology emerges at this time $t=t_v$ out of the `Big Bang' with the 
initial conditions of the  standard cosmological model having been set.

  In beginning with $L(\bv_4) = h^2 \simeq 0$ and converging towards $L(\bv_4) = 
h^2_0$  at time $t=t_v$ via interactions under the terms of the full form of $\lvh$ 
this picture is closely analogous to the second order phase transition of the `new 
inflation' model described in section~\ref{secinf}, involving a `slow roll' down an 
effective potential slope $V(h,T)$ on the way to achieving the stable value. For the 
present theory the `potential energy' associated with the original unstable value of 
$h \simeq 0$, following the analogy of the `false vacuum' state in the new inflation 
model, might itself effectively provide a direct source of an inflationary expansion.

 The variation of $\vert \bv_4 \vert = h(x)$ alone modifies an otherwise flat 
Minkowski spacetime via the conformal transformation $g_{\mu\nu}(x) = \theta(x) 
\eta_{\mu\nu}$, with $\theta(x) = h^{-2}(x)$ from equation~\ref{gwarph}. Via the 
Levi-Civita connection $\Gamma(x)$ this results in the Einstein tensor $G_{\mu\nu}(x)$ 
explicitly
 presented in equation~\ref{gmnconf}. The shaping of the geometry $G_{\mu\nu} = 
f(\theta)$ from beneath in this way contrasts with the geometry $G_{\mu\nu} = -\kappa 
T_{\mu\nu}(\varphi)$ deriving from the energy-momentum \textit{source} 
$T_{\mu\nu}(\varphi)$ of equation~\ref{tmninf}, which in turn was derived from a 
Lagrangian for the postulated scalar field $\varphi(x)$ in the original  inflation 
theory. Despite this difference in origin there is a close similarity between the 
kinetic terms in the field $\varphi(x)$ in equation~\ref{tmninf} and the first two 
terms in the field $\theta(x)=h^{-2}(x)$ in equation~\ref{gmnconf}, suggesting the 
possibility of a similar field dynamics for the two models. 

  For inflationary theory, in addition to the kinetic term drag terms are also 
introduced into the Lagrangian, as described following
   equation~\ref{tmninf}, and relate to the physical phenomenon of post-inflation 
reheating during which the energy of the false vacuum is converted into interacting 
particles.
 A similar effect may arise in the present theory, with an effective potential  
$V(h,T)$ and `drag terms'  for the new theory deriving from interaction terms implicit 
in the form $\lvh$ as described above. The favoured minimum in $V(h,T)$ (which may be 
largely independent of the effective temperature $T$) will correspond to the stable 
value $\vert \bv_4 \vert = h_0$, without  the need to \textit{contrive} an appropriate 
form for the potential $V(\varphi,T)$ as is the case for inflationary theory, since 
all the couplings of the present theory are effectively implied within the constraint 
equations~\ref{conequa}.

  An alternative proposal for the present theory features initial conditions with  
$\vert \bv_4 \vert \gg h_0$ with potentially large fluctuations in the components of 
the field $\bv_4(x)$,  perhaps  accompanied by a large value for the scalar field 
$\alpha(x)$ from equation~\ref{lamtran} providing the source of an inflationary 
$\Lambda g_{\mu\nu}$ term in the very early universe.
  Amongst the many possible solutions for $\gmyv$ in principle arbitrarily extreme 
spacetime geometries may occur, but without necessarily being supported throughout the 
full expanse of the manifold $M_4$. Such an extreme structure may describe the initial 
geometry in the Big Bang, where the conditions may even be somewhat `chaotic' as an 
extended spacetime solution is first shaken out of the mathematical possibilities 
implied in the form $\lvh$ and its symmetries, assuming the present universe to have 
evolved from such a state. With potentially a large range of possibilities for $\vert 
\bv_4(x) \vert \gg h_0$
 the initial value for $\bv_4(t)$ for $t\to 0$ may be required to be fairly uniform 
over a spatial extent of order the Hubble radius, as for the initial value of the 
scalar field $\vert \varphi \vert > \vert \varphi_0 \vert $ in models of `chaotic 
inflation', as also described in section~\ref{secinf}.

  However, while there may be a range of possible `false vacuum' initial conditions 
for the projection of $\bv_4 \subset \bvh$ onto $\TM_4$ in the present theory, the 
`post-inflationary' stable value of  $\vert \bv_4 \vert = h_0$, as coordinated with 
the parameters of the Standard Model of particle physics, may still be uniquely 
determined. The possible range of initial values for $\vert \bv_4 \vert \gg h_0$
 in principle implies a range of inflationary effects and a corresponding range of 
properties for the later evolution of the universe, some of which may be compatible 
with the present day universe as actually observed, and in particular with both the 
horizon and flatness problems resolved as for standard inflationary theory.

 This raises the question of the uniqueness of the present theory, which will be 
discussed more generally in the following section. For the case of $h(t) = \vert \bv_4 
\vert \to 0$
  as $t \to 0$ the scalar function $\theta = h^{-2}$ diverges. Hence even in this 
case,  if $\theta$ is interpreted as the inflationary field, the present theory may be 
interpreted in a manner analogous to chaotic inflation, with $\theta$ effectively 
taking a broad range of  large values in the `primordial chaos' of the very early 
universe. Although these options, in relation to chaotic inflation, might be 
considered further, here we explore in more detail the implications of taking $h(t) = 
\vert \bv_4 \vert \to 0$
  as $t \to 0$ as a potentially unique starting point.

  Hence this general structure involving a transition from the initial condition with 
$\bv_4(x) \to 0$, as depicted in figure~\ref{vphaset}(a), towards the stable state 
with $\vert \bv_4 \vert = h_0$, with a value which may also be determined uniquely, 
and as represented in figure~\ref{vphaset} at stage (c),
 may be essentially unambiguous. 
 The state immediately emerging from the phase transition in figure~\ref{vphaset}(c), 
along with ordinary matter represented by the sprinkling of points of dust, was also 
represented in figure~\ref{darkevo}(c), although with the spatial fluctuations of the 
field $\bv_4(x)$ neglected in the earlier figure.

\begin{figure}[htb]  
\centering
\epsfxsize=14cm
\leavevmode
\epsffile[0 0 1923 1196]{\gpath aPfig133e}
\caption{\setb (a) Beginning with $\bv_4(x)\simeq 0$ at the temporal origin of 
4-dimensional spacetime,
 (b) the value of $\vert \bv_4 \vert = h$ grows, with potentially large fluctuations 
in both magnitude and direction, until the phase transition with (c) a stable value 
attained for $\bv_4(x)$ with small fluctuations about the components $(h_0,0,0,0)$ for 
$\bv_4$ in the comoving cosmological frame $\{t,r,\theta,\phi\}$. (Later epochs are 
depicted in figure~\ref{darkevo}(d),(e) in the previous section).} 
\label{vphaset}
\end{figure}

 The picture of the phase transition between (b) and (c) in figure~\ref{vphaset} is 
analogous to the that associated with the property of ferromagnetism in a piece of 
iron. The atoms in the iron can be considered as forming a lattice of a very large 
number of randomly oriented magnets for temperatures $T>T_c$ above the critical value. 
This is similar to the situation in figure~\ref{vphaset}(b), except that the atomic 
magnets would be represented by 3-dimensional spatial vectors of a uniform constant 
magnitude. Upon cooling to a temperature $T<T_c$ it is energetically favourable for 
neighbouring magnetic vectors to align, with an analogous phenomena applying for the 
vector field $\bv_4(x)$ as the stable value $\vert \bv_4 \vert = h_0$ is attained as 
depicted in  figure~\ref{vphaset}(c), with small fluctuations about the average 
4-vector value of $\bv_4(x)$ greatly exaggerated in the diagram.

  For the present theory the symmetry breaking in the phase transition, both in terms 
of the actions of the dilation symmetry for the magnitude of $\bv_4(x)$ and 
fluctuations in the orientation of this 4-vector, further suggests a close 
relationship between $\bv_4(x)$ and the Standard Model Higgs field. Indeed the stable 
vacuum value $\bv_4 = (h_0, 0,0,0)$  is precisely the same 4-vector as that in 
equation~\ref{v4vac} (where a pseudo-Euclidean basis $\{t,x,y,z\}$
 for $\bv_4$ was employed),
  with $v^0 = h_0$, as described in subsection~\ref{suboomahp}. 
  The `vacuum symmetry' is broken as the vector-Higgs field $\bv_4(x)$ takes a 
magnitude and particular direction in spacetime which, on average, is presumed to be 
essentially aligned with the preferred cosmological frame parametrised by comoving 
coordinates   $\{t,r,\theta,\phi\}$. 

That is, the comoving cosmological frame is aligned with the average distribution of 
visible matter, which in turn is presumed to have been formed and evolved in line with 
the underlying flow $\bv_4(x)$ through the spacetime manifold $M_4$. The degree of 
correlation between local fluctuations in the flow $\bv_4$ and  peculiar motions on a 
galactic scale at the present epoch is an open question. 
 While the laws of physics are locally Lorentz invariant actual physical structures 
clearly are not, and this also applies to the large scale structure of the universe. 
For example a directional relative blueshift and redshift  for the detected CMB 
radiation depends upon the local choice of Lorentz frame for the observer. In our case 
these shifts are due to our local motion within our galaxy, and can be readily 
corrected for in the CMB maps.

   As for the inflationary theories described in section~\ref{secinf} quantum 
fluctuations and potentially Hawking radiation in the inflationary epoch  generate 
inhomogeneities in the very early universe which may become frozen as classical 
fluctuations in energy density at the end of inflation, ultimately seeding the 
formation of galactic structures. In the present theory these quantum effects include 
interactions between $\bv_4(x)$ and other fields, such as those for the fermions $\psi 
(x)$ and gauge bosons $Y(x)$ as well as scalar fields such as $N(x)$, all subject to 
the constraint equations~\ref{conequa} in forming the overall external geometric 
solution $G_{\mu\nu} = f(Y, \bvh)$ in spacetime. Fluctuations in the value of
 $h(x) = \vert \bv_4(x) \vert$ directly impact upon the spacetime geometry, as 
described for equation~\ref{gwarph}, and hence in particular may generate large scale 
structure when amplified as the scale factor $a(t)$ rapidly grows.

   Fluctuations in the spatial components of $\bv_4(x)$ could also in principle have a 
large effect during the evolution of the very early universe as represented by the 
stage of figure~\ref{vphaset}(b), particularly for the case of a pre-inflation 
spatially `miniaturised' world, with a relatively very small value of $a(t)$, as 
described for figure~\ref{confbba}.
 A calculation of how such fluctuations might stir up the primordial geometry would 
involve taking into account all components of the field $\bv_4(x)$ to determine 
$G_{\mu\nu} = f(Y,\bvh)$, rather than just the magnitude  $\vert \bv_4 \vert = h = 
\theta^{-\fh}$ as was the case for equation~\ref{gmnconf}.
 Such a cosmological model, with fluctuations in both the magnitude and direction of 
$\bv_4(x)$ impacting upon the large scale structure, would differ from the forms 
derived in equations~\ref{cosg00} and \ref{cosg11}, for $G_{00}$ and $G_{11}$ 
respectively, for which homogeneity and isotropy were assumed, unless a statistical 
average is taken for the large scale structure conforming to those assumptions.

   As the vector field $\bv_4(x)$ stabilises through the phase transition to the stage 
depicted in figure~\ref{vphaset}(c) small residual variations in the components of 
$\bv_4(x)$ might still remain, and be found to be finely grained on the scale of the 
observable universe. This residual fingerprint of the earlier fluctuations is the 
network of creases in the fabric of spacetime as has been  described in the previous 
section for the same epoch as also depicted in figure~\ref{darkevo}(c).
At this point, and throughout the remaining evolution of the cosmos, this small 
residual variation in the components of $\bv_4(x)$
 might account for the phenomena of dark matter, and even a dark flow, as also 
suggested in the previous section. That is,
 with sufficient deviation from the assumptions of uniformity of the cosmological 
principle these residual variations might seed the early formation of galaxies and 
clusters of galaxies through gravitational merging into the courser structures 
observed at the present epoch. This cosmic imprint
 in the underlying spacetime geometry, arising from fluctuations in the very early 
universe, is interpreted as a manifestation of `cold dark matter' in particular, as 
was described in the previous section for structures observed through to the present 
epoch as depicted in figure~\ref{darkevo}(e).

   As noted above,
 on top of these geometric effects of a continuous variation in the field $\bv_4(x)$ 
in the present theory the same field is responsible for the Higgs sector in particle 
physics through interactions or exchanges with other fields. More generally, 
throughout the history of the universe quantum transitions, in the form of $\delta 
\bv_4 \leftrightarrow \delta \psi$
 or $\delta \bvh \leftrightarrow \delta Y$ field exchanges underlying the multiple 
possible solutions for the spacetime geometry $\gmyv$, with the constraints of 
equations~\ref{conequa} applying everywhere on $M_4$, will shape the evolution of the 
cosmos, including the epoch of the very early universe.
  This shaping includes both the impact of observable fluctuations as described above 
as well as the physical implications arising from the statistical average of the 
microscopic interactions.

As described in chapter~\ref{newapp} the direct association of the likelihood of an 
observable quantum event with  the `number of ways' 
 in which the same empirical effect can be achieved, quantified in terms of the 
degeneracy of underlying field solutions for the same external local geometry $\gmyv$, 
unifies the quantum process notion of probability itself with the classical concept. 
While a time-ordered accumulation of probabilities in the quantum case is relevant for 
cross-section calculations, in the case of classical phenomena it gives rise to the 
second law of thermodynamics as quantified by an ever increasing value of entropy for 
any evolving thermodynamic state. In the present theory all such thermodynamic 
phenomena are played out \textit{in} time and do not themselves \textit{drive} an 
`arrow of time', as will be clear in the following chapter.

  As alluded to above the structure of the very early universe may allow sufficient 
breathing space for the thermalisation of the particle degrees of freedom in the epoch 
before the phase transition in figure~\ref{vphaset}, as is the case for the 
pre-inflation environment in figure~\ref{confbba} as described in 
section~\ref{secinf}. However some care is needed in applying the principles of 
thermodynamics and statistical mechanics, familiar from their application in the flat 
spacetime environment of the laboratory for example, in the potentially highly curved 
and dynamic spacetime of the very early universe. Even basic notions such as 
`temperature' or a `black body spectrum' may be hard to define in such an extreme 
environment.
 The approach may be justified to some extent by applying  thermodynamics within small 
spacetime regions which  approximate to local inertial frames, and adopting the strong 
equivalence principle, given a sufficient number of `particles' and `particle 
interactions', or underlying field exchanges, within such a region to apply 
statistical methods.  Further, the properties of the `particles' and fields themselves 
in the era before the phase transition may be very different to the familiar Standard 
Model particles and fields that emerge out of the Big Bang.

 It is also noted that the universe, and in particular the structure of the very early 
universe, is a  \textit{single} system for empirical study. Hence thermodynamic 
arguments, which consider an \textit{ensemble} of systems each of which might form a 
small component \textit{within} the universe, as for example employed for laboratory 
experiments, may not apply for the potentially unique system composing the precursor 
to and immediate aftermath of the Big Bang.
 That is the observable universe today may have evolved from state in the very early 
universe which is too small or simple to incorporate a statistical average, and which 
might in fact  be dominated by the effect of a single `fluctuation'.

 Further, as was described towards the end of section~\ref{secinf}, beyond the 
`horizon problem' there is apparently a `start-up problem' in the need to choreograph 
a vast number of spacelike separated `bangs' along the initial singularity, either in 
figure~\ref{confbb} or \ref{confbba}, in order to effectively simultaneously trigger 
the `Big Bang' itself at cosmic time $t=0$. Analogous to a synchronised display of 
fireworks there might be range of `temperatures' across the range of `bangs' creating 
inhomogeneous initial conditions. However the Big Bang is not such a terrestrial event 
and there seems no reason why it should not be in the nature of the start-up to 
generate essentially homogeneous thermodynamic conditions over the entire spacelike 
hypersurface at $t=0$, upon which local fluctuations may be identified in terms field 
exchanges underlying the multiple solutions for $\gmyv$. In particular the different 
nature of gravity, associated with the smooth geometry $G_{\mu\nu}(x)$, compared with 
the quantum phenomena associated with the internal field interactions, may play an 
important role in this structure.

  In the immediate aftermath of the phase transition of figure~\ref{vphaset}(c) with 
the `vacuum energy' being converted into  Standard Model particles through transitions 
of the form  $\delta \bv_4 \leftrightarrow \delta \psi$ under $\lvfs$, with familiar 
microscopic quantum properties, \textit{many} more degrees of freedom may open up. The 
entropy content of the observable universe emerging from this epoch will depend on the 
reheating effects of the drag terms implicit in $\lvh$ combined with the kinetic terms 
implied in equation~\ref{gmnconf}, which were discussed earlier in this section and 
similarly as described following equation~\ref{tmninf} for inflationary theory. 
However, as also noted in section~\ref{secinf}  the gravitational field appears to 
have had a very special role in the Big Bang and very early universe in being aloof 
from the thermalisation process. 

The strong equivalence principle (as reviewed in section~\ref{gcatep} and adopted 
above) in part demonstrates how the characteristics of gravity fundamentally differ 
from the other forces of nature. The properties of local inertial frames are key to 
the structure of general relativity, with all physical phenomena other than gravity 
behaving in such a frame as if gravity were completely absent, while gravity itself is 
described by the geometry of the extended spacetime. The differences between gravity 
and other physical phenomena will be significant for addressing issues for the early 
universe, including also the `flatness problem' as well as the `horizon problem' and 
an understanding of  the role of entropy.

   Within the present theory the special status of the external gravitational field 
further derives from the fact that it is of a quite different, `unquantised' nature in 
comparison with the internal gauge $Y(x)$ and fermion $\psi(x)$ fields. The external 
geometry, described for example in terms of the metric components $g_{\mu\nu}(x)$ or 
linear connection $\Gamma(x)$, does not partake in the statistical physics of the 
internal fields which lies beneath continuous geometric solutions of the form $\gmyv$.
  In the expression $-\kappa T_{\mu\nu} := G_{\mu\nu}$ the right-hand side describes 
the smooth external geometry, with all quantum mechanical properties of matter 
implicitly underlying the energy-momentum tensor on the left-hand side. Such quantum 
phenomena, based on an degeneracy of field solutions, can generally be described to a 
good approximation within local inertial frames, as was the case in 
sections~\ref{secdos} and \ref{secdopp}. 
 While the electromagnetic field, for example, exhibits thermal properties through the 
underlying field interactions the external gravitational field, being aloof from such 
interactions, has a very different relation with thermodynamic phenomena, and also, 
being unquantised,  does not directly partake in quantum fluctuations.

  The phenomenon of Hawking radiation, as discussed towards the end of 
section~\ref{qpagig}, arises for quantised fields in the classical curved spacetime of 
a black hole exterior, with the consequence for example that a black hole with mass 
$O(10^6)\,$kg will evaporate in approximately one second. Such phenomena involve the 
quantum mechanical description of the vacuum but the gravitational field itself is not 
quantised, and lead to 
 a study of the thermodynamic and entropy properties of black holes.
 Similar properties of the vacuum may arise for the present theory, since gravity is 
not quantised here,
  and also be important in the study of the thermodynamic and entropy properties of 
the very early universe, in particular during the inflationary period.

For the case in which the initial geometry is dominated by variation in the value of 
$\vert \bv_4 \vert = h(x)$, with a metric of the form $g_{\mu\nu}(x) =  
h^{-2}(x)\eta_{\mu\nu}$
 in equation~\ref{gwarph}, the spacetime geometry of equation~\ref{gmnconf} is 
conformally flat, even for arbitrarily large variations in the scalar field $h(x)$. 
For such a geometry the Weyl curvature tensor vanishes, $C_{\rho\sigma\mu\nu}(x)= 0$, 
consistent with the proposal of the Weyl curvature hypothesis as motivated and 
described towards the end of section~\ref{secinf}. 
 Hence this observation may account for the `cosmological problem'  concerning the 
extraordinarily special state of the Big Bang to 1 part in $10^{10^{123}}$
 (according to \cite{Pen} p.777) as required for the low entropy initial conditions 
which underlie the subsequent evolution of the cosmos consistent with second law of 
thermodynamics.
 
 Through interactions and fluctuations of the form $\delta \bv_4 \leftrightarrow 
\delta \psi$, in particular with the transfer of energy from the vacuum to Standard 
Model particle states towards the end of the inflationary period corresponding to 
figure~\ref{vphaset}(c), a non-conformally flat geometry will emerge incorporating 
Weyl curvature, and hence the propagation of gravitational waves for example, as well 
as Ricci curvature. As described in section~\ref{secinf} (with reference to \cite{Pen} 
section~28.8) the entropy of the gravitational field might be expressed in terms of 
the degrees of freedom of the Weyl curvature and hence contribute to the increase in 
entropy from this time.
  As also described in section~\ref{secinf}, following equation~\ref{Coseta}, 
 all FLRW models are consistent with $C_{\rho\sigma\mu\nu}(x)= 0$ but  require 
something like an initial period of inflation to explain why observations are 
consistent with $k=0$, that is with spatial flatness. Similarly for the present theory 
an inflationary evolution for $a(t)$ in equation~\ref{hconfaca} in the very early 
universe may relate to this observation.

  As described in section~\ref{seraps}, and depicted in
figure~\ref{runcup}, the three coupling parameters of the Standard Model gauge group 
$\SML$ approximately converge at an energy scale of $O(10^{15})\,$GeV. This 
unification scale will mark a significant threshold in the early universe, and it will 
be important to understand how it relates to the epoch of the phase transition in 
figure~\ref{vphaset}(c) for the present theory.
  The interplay between the dilation symmetry, such as $\mbox{D}(1)_B$, and 
electroweak symmetry, together with the nature of their breaking, will also be key, as 
described shortly after equation~\ref{hconfac} with reference to 
equation~\ref{bbinesi}. The electroweak symmetry $\sutw_L \times \uo_Y$
 is broken by its action on $\bv_4 \inn \TM_4$, with the stable value for the 
magnitude
  $\vert \bv_4 \vert = h_0$  arising out of the Big Bang at $t=t_v$. For $t< t_v$, and 
in particular for $t\to 0$ with  $\vert \bv_4 \vert \ll h_0$ via the action of the 
dilation symmetry, the properties of the electroweak symmetry and the Higgs sector 
more generally will be somewhat different, for example with regards to the pattern of 
particle masses. To address the complete symmetry breaking picture it will be required 
to explicitly identify the electroweak symmetry  $\sutw_L \times \uo_Y$ within the 
full $\ese$, or $\ee$, symmetry of the full form $\lvh$, in relation to the dilation 
symmetries, such as $\mbox{D}(1)_B$, and the $\sltc^1 \times \suth_c \times \uo_Q$ 
symmetry already identified, completing the development of these structures described 
in chapters~\ref{chapesb}
 and \ref{secfd}.

  As also alluded to towards the end of section~\ref{qpagig}  for a theory of `quantum 
gravity', with the degrees of freedom of the gravitational field quantised, 
significant effects are expected at the Planck energy scale $E_P = \left( \frac{c^3 
\hbar}{G_{\! N}} \right)^{\!\fh} \simeq 
  1.2 \times 10^{19}\,$GeV. For any description of the very early universe in the 
context of such a theory all classical field concepts in turn fail at epochs earlier 
than the Planck time
 $t_P = \left( \frac{G_{\! N}\hbar}{c^5} \right)^{\!\fh} \simeq 
  5  \times 10^{-44}$ seconds. 
 However the energy scale $E_P$ is considered to be of no special significance for the 
present theory and the time scale $t_P$, representing for example the extremely early 
universe, in principle presents no barrier for this theory. Hence the nature of the 
universe down through epochs at arbitrary cosmic times $t<t_P$ might be studied within 
the context of the present theory. This leads essentially to two broad possibilities 
as depicted in figures~\ref{earlyu}(a) and (b). In these diagrams  $t=t_v$ (presumably 
with $t_v \gg t_P$) denotes the epoch of the phase transition at which there is a 
convergence to the average value $L(\bv_4) = \vert \bv_4 \vert^2 = h_0^2$, as 
represented in figure~\ref{vphaset}(c). 
  For either figure~\ref{earlyu}(a) or (b) the epoch of the `Big Bang' can be 
identified with the time $t=t_v$ or more generally with the period from $t=0$ to 
$t=t_v$ and the state emerging at that latter time.

\begin{figure}[htbp]  
\centering
\epsfxsize=13.5cm
\leavevmode
\epsffile[0 0 1855 959]{\gpath aPfig134e}
\caption{\setb Two scenarios for the relation between the temporal origin of the 
universe and the fundamental flow of time with (a) $s\to -\infty$ for $t<0$ and (b) 
$s\to -\infty$ for positive values of $t \to 0$. The width of each figure for $t > 0$ 
represents the spatial scale factor $a(t)$ as a function cosmic time $t$, neither of 
which are drawn to scale.} 
\label{earlyu}
\end{figure}

 In the first case for figure~\ref{earlyu}(a) the time $t=0$ can be considered to be 
the moment at which an extended 4-dimensional spacetime world first emerges out of the 
forms of the pure temporal flow $s$ as identified through the geometric relation 
$\gmyv$. This is the point in time at  which  extended and potentially infinite 
3-dimensional spatial hypersurfaces may be identified as an offshoot out of the 
multi-dimensional form of temporal flow $\lvh$
and a spacetime geometry with metric $g_{\mu\nu}(x)$ established, although with a 
significant deviation from flatness possible both for the 4-dimensional curvature and 
for the 3-dimensional hypersurfaces. 
 Considering a time $t>0$ in figure~\ref{earlyu}(a) 
and retracing the temporal flow  backwards the time $t=0$ marks the point at which the 
geometrical interpretation in terms of a 4-dimensional extended manifold,  supported 
by the mathematical structure and symmetries of the form $\lvh$, completely breaks 
down.

   Before $t=0$ in figure~\ref{earlyu}(a) the parameter $t$ no longer represents a 
coordinate on the manifold $M_4$, while the fundamental flow of time $s \to -\infty$ 
continuous without any limit as expressible through a general form $\lvh$, as always, 
but without any projection of  $\bv_4 \inn \TM_4$ components onto an extended 
manifold. Here, as depicted for example in figure~\ref{vphaset}(a), we have considered 
the case with $\bv_4(x) \simeq 0$ in the very early universe. In the context of 
figure~\ref{earlyu}(a) beginning with $\bv_4(x) = 0$ at $t=0$ with $\vert \bv_4(x) 
\vert =h(x)$ generally growing with $t>0$ in the very early universe, as depicted in 
figure~\ref{vphaset}(b), the time $t=0$ could be considered as the epoch at which a 
fragment of temporal flow under $\lvh$ is `syphoned off' into the thereby created 
spacetime manifold $M_4$. However it is also conceivable that this point of spacetime 
creation at $t=0$ can be accompanied by arbitrary values for $\bv_4(x) > 0$, in 
principle even with
 $\vert \bv_4 \vert \gg h_0$.
 
   The width in both figures~\ref{earlyu}(a) and (b) represents the spatial scale 
factor $a(t)$ of equation~\ref{hconfaca}, under the presumption of a solution with 
$a(t) \to 0$ as $t\to 0$ and some form of inflationary expansion leading up to 
$t=t_v$, not drawn to scale. The behaviour of the ratio $\frac{a(t)}{h(t)}$, and in 
particular whether this fraction tends towards zero, infinity or is finite as $t\to 
0$, will be significant for understanding the nature of the geometry of the manifold 
$M_4$ in this limit, according to the spacetime structure described by 
equation~\ref{hconfaca}. The geometry in this limit will also be important in relation 
to the horizon problem and the `start-up problem'  as discussed for 
figures~\ref{confbb} and \ref{confbba} in section~\ref{secinf}. This might be best 
approached via a redefined cosmic time parameter such as $\bar{t}$ with $\delta 
\bar{t} = \frac{\delta t}{h(t)}$
 and with the  line element of equation~\ref{hconfaca} correspondingly replaced by:
\begin{equation}
\label{hconfacha}
     d\tau^2 \, = \, 
	     d\bar{t}^{\, 2} \, - \,\frac{a^2(t)}{h^2(t)} \, d\Sigma^2 
\end{equation}
  As has been discussed earlier, care is needed for the meaning of `cosmic time' for 
the epoch $t<t_v$, whether parametrised by $t$ or $\bar{t}$, since physical clocks 
will be of a somewhat different nature for the very early universe, and indeed do not 
exist in any form for $t<0$.
 In any case a more complete theory is required to avoid the dangers of speculating on 
the number of angels that might be accommodated upon the head of a pin, as noted at 
the end of section~\ref{secinf}.

   The above comments also apply for  the  scenario  depicted in 
figure~\ref{earlyu}(b), for which necessarily $\bv_4(x) \to 0$ at the spacelike edge 
of the manifold $M_4$ in the past at $t=0$. For this second picture the relation 
between the flow of the fundamental time parameter $s$ and the `cosmic time' 
coordinate $t$ is sketched in figure~\ref{svrst}.

\begin{figure}[htb]  
\centering
\epsfxsize=11cm
\leavevmode
\epsffile[0 0 1327 860]{\gpath aPfig135e}
\vspace{-10pt}
\caption{\setb For the scenario depicted in figure~\ref{earlyu}(b) the projection 
$\bv_4(x) \in \TM_4$ converges to zero
 for $t\to 0$ in the very early universe. While $t=0$ marks a coordinate boundary to 
the 4-dimensional spacetime $M_4$ the range of the fundamental temporal flow $-\infty 
< s < +\infty $
 is tucked away and entirely contained within this manifold. Adopting the approximate 
components $(h,0,0,0)$ for $\bv_4$ in the comoving frame the phase transition $t=t_v$ 
marks the point at which $v^0=\vert \bv_4(x) \vert = h(t) = h_0$ stabilises.} 
\label{svrst}
\end{figure}

 For this scenario if  $\bv_4(x) \in \TM_4$ converges to zero  in an appropriate 
manner as $t\to 0$ then as $s \to -\infty$ the 
 spacelike hypersurface at $t=0$, potentially an `initial singularity' as $a(t) \to 
0$, is never attained and all of the fundamental flow of time $-\infty < s < +\infty $  
is absorbed into the extended spacetime $M_4$ of the universe.
 With $s \to -\infty$ without limit at the temporal coordinate origin on $M_4$
 the structure for $t < t_v$ in figure~\ref{earlyu}(b) and \ref{svrst} might be 
pictured poetically as the bottomless waterfall at  the end of time. With the familiar 
structures of the Standard Model of particle physics emerging in the phase transition, 
from this epoch and for all times $t > t_v$ the fundamental time flow $s$ is 
equivalent to both the proper time $\tau$  and also the cosmic time $t$ for idealised 
observers in the context of an FLRW cosmological model, as described near the opening 
of section~\ref{secpotnt}.

  For the case of the scenario depicted in figure~\ref{earlyu}(b) the present day 
universe is, in a sense, infinitely old in terms of the fundamental time parameter 
$s$. However for the picture in \textit{both}   figure~\ref{earlyu}(a) and (b) the 
physical and mathematical structures can be traced back to arbitrarily early times for 
$s \to -\infty$, with the difference being that for (a) physical structures are no 
longer defined for $t<0$ while for (b) parameter values $t \le 0$ are outside the 
domain of the underlying temporal flow $s$. In both cases physical structures relating 
to the Standard Model of particle physics arise out of the Big Bang at $t=t_v$. This 
is the point in time at which we can effectively `start the clock' with $s\equiv \tau 
\equiv t$, as might be measured through familiar physical processes, now determined to 
stretch back through around 13.8 billion years of cosmic evolution.
 Such an apparent temporal origin for the laws of physics in our 4-dimensional world 
may be necessary for  consistency with an environment supporting biological life at 
the present epoch. Here we refer in particular to the second law of thermodynamics 
which implies the universe is still evolving away from the particularly low entropy 
state conceivably corresponding to the nature or uniformity of the gravitational field 
in the very early universe, as described above.

   For either scenario depicted in figure~\ref{earlyu} the cosmic evolution itself is 
a feature of the full macroscopic 4-dimensional spacetime  $\gmyv$, as shaped by  
microscopic field interactions in the form of the local degeneracies of fields 
underlying the possible solutions, consistent with the constraint 
equations~\ref{conequa}, as described in chapter~\ref{newapp}. 
  As discussed in section~\ref{qpagig} in combining gravitation with quantum theory 
the notion of a 4-dimensional spacetime solution of general relativity takes 
precedence over the 1-dimensional propagation of an apparent quantum state, with the 
latter described in terms of a local time coordinate, hence also circumventing the 
`problem of time' encountered by some approaches to quantum gravity. As also concluded 
in section~\ref{qpagig} the nature of probability in quantum processes is essentially 
the same as that for classical systems, at heart formulated in terms of the `number of 
ways' that an  empirical effect may be produced. 

  On the large scale, with many underlying degrees of freedom, the interplay of both 
quantum and classical statistical phenomena will contribute to the shaping of the 
cosmological solution for $\gmyv$.
  This solution will also incorporate macroscopic contributions to the geometry in the 
form of $G_{\mu\nu} = f(Y)$ of equation~\ref{gchift}, by comparison with Kaluza-Klein 
theory as described in section~\ref{reaic}, and of the form $G_{\mu\nu} = f(\bvh)$ of 
equation~\ref{gmnconf} from variations of $\vert \bv_4 \vert =h(x) = 
\theta^{-\frac{1}{2}}(x)$ in the projection of $\lvh$ onto $M_4$, as described in this 
chapter.   
  A correspondence with the techniques of `renormalisation' in quantum field theory 
might in principle be developed in order to study the relation between the macroscopic 
external geometry and the underlying `bare' fields, as has been described in 
section~\ref{seraps}.

  The question then concerns how the combination of all of the above geometrical and 
statistical factors in determining a solution for $\gmyv$, with $T_{\mu\nu} := 
G_{\mu\nu}$ providing the interpretation of
  equation~\ref{Einfield},    might collectively account for the observed cosmic 
evolution, compatible in approximation with the assumptions of the FLRW models  
  and the metric form of equations~\ref{CosRob} and \ref{gcoscomp}, together with the 
large scale galactic structures. While observations of the latter structures require 
an apparent `dark matter' component, on the largest scale the solution $\gmyv$ is 
required to account for the apparent effects of `dark energy', for example in the form 
of an effective cosmological term $\Lambda g_{\mu\nu}$ in the Einstein field equation. 
 As for the earlier inflationary epoch, the modern era parameter $\Lambda$ may not be 
entirely constant, but with any variation such that $(\Lambda g^{\mu\nu})_{;\mu} \neq 
0$ exactly compensated by an apparent effective energy-momentum tensor with 
$T^{\mu\nu}_{\epsilon\ph{\nu};\mu} \neq 0$ consistent with $\gmo$ and 
equation~\ref{Einfield}.
 This possibility was alluded to in the previous section in the discussion regarding 
table~\ref{cosevo}, and with reference to a similar observation for 
equation~\ref{gtruum}.
 In the present theory the total energy-momentum tensor $T_{\mu\nu}:=G_{\mu\nu}$ is 
defined to incorporate any possible `dark energy'  cosmological term, and indeed the 
full solution $\gmyv$.

  In describing the overall cosmological evolution in the spirit of the FLRW models 
the metric of the line element in equation~\ref{CosRob} or \ref{hconfaca} underlying 
the full 4-dimensional solution $\gmyv$ will incorporate the expansion of the 
universe, including that of the present day, in terms of the scale factor $a(t)$. The 
perspective adopted here is \textit{not} that the universe is expanding now 
\textit{because} it was expanding in the past, analogous to the kinematic propagation 
of the flight of a cannonball from one moment to the next along its trajectory, in 
either case raising the question of \textit{how} it was set in motion in the first 
place. Rather here the very early universe is conceived of as one particular 
\textit{region} of the full \textit{four}-dimensional spacetime manifold $M_4$, which 
happens to exhibit properties such as $a(t) \to 0$ and $h(t) \to 0$ as the coordinate 
parameter $t \to 0$,   consistent with the overall $\gmyv$ external geometry solution.

 This is analogous to thinking of the Earth as being in orbit around the sun at the 
present day \textit{not} as a kinematic consequence of the fact that it was in orbit 
one year ago or a billion years ago but since the 4-dimensional spacetime trajectory, 
featuring an approximately elliptical orbit, exists as a geodesic solution for a 
4-dimensional Schwarzschild spacetime.
 In fact since the Bianchi identity $\gmo$  implies geodesic motion, as described for 
equation~\ref{trgeod} in section~\ref{subwal}, the full spacetime geometry of an 
entire planetary system can be conceived of as a particular 4-dimensional solution for 
$G_{\mu\nu}(x)$.
 The idealised Schwarzschild solution itself describes an infinite and eternal 
4-dimensional spacetime with $G_{\mu\nu}(x) = 0$ everywhere, except for the point at 
the centre of spherical spatial symmetry, with the metric of equation~\ref{ttrtp}. 
While the components of this Schwarzschild metric are constant in time but vary as a 
function of the radial coordinate $r$ via the factors of $(1-2\:\!G_{\! N}M/r)$, the 
geometry of the Robertson-Walker metric for an FLRW  cosmological solution is 
independent of the spatial coordinates but varies with the time coordinate through the 
scale factor $a(t)$. Both cases represent full 4-dimensional spacetime geometries.

  In the present theory both $a(t)$ and $h(t)$ in the line element of 
equation~\ref{hconfaca} 
  shape the geometry for the very early universe with $t < t_v$, with a correlated 
evolution of these parameters associated with a period of inflation. The comparison, 
earlier in this section, with the `new inflation' model represents an analogy for the 
present theory, however the `slow roll' down from $h(t) \simeq 0$ for $t\to 0$ to the 
stable average value $h(t_v) = h_0$ may or may not end with a series of `oscillations' 
as the minimum of the effective potential $V(h,T)$ is achieved. In any case, given the 
correlation between $a(t)$ and $h(t)$, it is conceivable that spatial regions  with 
residual small positive fluctuations $h(x) > h_0$ may have `inflated' a little longer 
leaving a value of the scale factor $a(x)$ also slightly larger than the average value 
at the end of inflation.

  For the large scale evolution of the observable universe for any time $t>t_v$, with 
quantities averaged over each
 3-dimensional spatial hypersurface, the value $h(t) = h_0$ remains constant and 
stable while $a(t)$ continues to increase, parametrising the expansion of the universe 
as sketched in figure~\ref{aevolve}. However on the local scale of galaxies and 
galactic clusters it is the correlated distribution in space of $a(x)$ and $h(x)$, 
initially established at $t\simeq t_v$, that might be associated with dark matter.  
  That is, evolving forward to the present day, the effects of dark matter might be 
attributed to regions with small fluctuations of $h(x) > h_0$ 
 together with a correlated spatial profile in $a(x)$, rather than simply the 
conformal scaling alone of equation~\ref{gwarph} as suggested following 
figure~\ref{vtovary} in the previous section.  In this way, generalising from 
equation~\ref{hconfaca} for spacetime variation of $h$ and $a$, the line element takes 
the form:
\begin{equation}
\label{hconfacas}
	    d\tau^2 \, = \, 
	     \frac{1}{h^2(x)}dt^2 \, - \,\frac{a^2(x)}{h^2(x)} \, d\Sigma^2 
\end{equation}
    This structure opens up a greater degree of independence between the temporal and 
spatial components of the metric, with for example $g_{00}(x)$ relatively low and 
$g_{ii}(x)$ for $i=1,2,3$ relatively high in spatial regions where both $h(x) > h_0$ 
and $\frac{a(x)}{h(x)}$ are relatively high, which in this sense is more reminiscent 
of the Schwarzschild solution  of equation~\ref{ttrtp}, and which also may have 
geometric properties more characteristic of a distribution of an apparent form of 
`matter' than variation of $h(x)$ alone.

   While, given an initially flat spacetime, the purely conformal action of $h(x)$ 
only generates Ricci curvature, the metric of equation~\ref{hconfacas} will generate 
both Ricci and Weyl curvature contributions extended throughout the spacetime manifold 
$M_4$, both in regions of galactic clusters and the voids between.  
 Having the variation of both $h(x)$ and $a(x)$ in equation~\ref{hconfacas}  increases 
the potential to match the observations of galactic motions and rotation curves, 
together with gravitational lensing effects, as a candidate for dark matter in 
interaction with the distribution of ordinary baryonic matter. On the yet larger  
scale of cosmological evolution these contributions to the dynamics of the universe 
might also be compared with the measured density parameters $\Omega_{D}$ and 
$\Omega_B$, in addition to $\Omega_{\Lambda}$, as introduced in 
section~\ref{sectsmoc}, as part of a global fit to the cosmological data.

  In summary, the large scale structure and cosmological evolution of the universe are 
to be identified generally as aspects of a full 4-dimensional solution for the 
spacetime geometry
 $\gmyv$. There is no presupposition of a flat spacetime manifold. In projecting an 
extended 4-dimensional spacetime $M_4$ out of the full multi-dimensional form $\lvh$ 
of the fundamental temporal flow $s$ large scale geometric distortions might be 
expected, which in turn may correlate with the observations ascribed to inflation, 
dark energy and dark matter, as reviewed in the previous chapter. There remains, of 
course, the need for a more complete theory and a much more thorough analysis, but in 
the meantime the possible variation of the magnitude of the projected 4-vector $\bv_4 
\inn \TM_4$ and the identification of several scalar fields $\alpha,\beta,n$ and $N$ 
from the components of $\lvfs$ indicates the potential for the application of the 
present theory to these cosmological questions.  

  The above discussion applies for the geometry of the 4-dimensional spacetime 
manifold $M_4$ whether in the context of the scenario depicted in 
figure~\ref{earlyu}(a) or (b). 
 However, compared with the first scenario of figure~\ref{earlyu}(a) that in  
figure~\ref{earlyu}(b) is more symmetric in time in the sense that both the limit for 
$s \to -\infty$ as well as for $s \to +\infty$ is incorporated within the 
4-dimensional spacetime solution $\gmyv$, as depicted in figure~\ref{spminf}.

\begin{figure}[htb]  
\centering
\epsfxsize=14cm
\leavevmode
\epsffile[0 0 1177 1033]{\gpath aPfig136e}
\vspace{-5pt}
\caption{\setb As parametrised by the fundamental temporal flow $s$ the  spacetime 
manifold underlying the physical universe can be of infinite extent without boundary 
in time as well as in space for the scenario of figures~\ref{earlyu}(b) and 
\ref{svrst}.} 
\label{spminf}
\end{figure}

 We inhabit a region of this eternal and infinite spacetime located within the period 
of several tens of billions of years following the phase transition at $t=t_v$ during 
which complex physical structures supporting biological life can be found, as 
represented in \ref{spminf}(e) and corresponding to the epoch of 
figure~\ref{darkevo}(e).
 It may be that both the far future through to $s \to +\infty$ as well as the far past 
with $s \to -\infty$  may become progressively less structurally  varied and eventful 
compared with the present epoch.  
For $s \to +\infty$ the universe may evolve into a relatively uneventful interplay 
between slowly evaporating massive black holes and thermal radiation, as depicted in 
figure~\ref{spminf}(f), while for $s \to -\infty$ there may be an equally uneventful 
asymptotic progression with $\vert \bv_4 \vert \to 0$, as depicted in 
figures~\ref{vphaset}(a) and \ref{spminf}(a). 
 In this picture  a physical understanding of the structure of the universe for both 
$s \to +\infty$ and $s \to -\infty$ may be equally open to study.

  On the other hand there then  remains the question concerning the \textit{reason 
why} the universe should exist at all. In the context of the scenario in 
figure~\ref{earlyu}(a), as for the standard cosmological models discussed in 
section~\ref{sectsmoc},
 in tracing the cosmological history back through the epoch of the very early universe
 this question can be phrased in terms of the \textit{cause} of the Big Bang and the 
nature of the temporal origin of the universe itself. However, with everything, 
including the Big Bang \textit{happening in time}, and with all physical structures in 
the universe for the present theory built entirely upon the notion of the 
one-dimensional flow of time $s$, there will still remain the question of the 
foundation of this apparently fundamental temporal entity itself,
 a question which applies equally for the scenario in figure~\ref{earlyu}(b).
 This will form the topic for the following chapter. In the meantime, in the following 
section, we consider the extent to which the properties and laws of physics of the 
universe, as depicted for example in figure~\ref{spminf}, might or might not be unique 
within the conceptual notions and mathematical constraints of the present theory.



\section{Uniqueness}
\label{secuni}

  In this section we consider several topics concerning the extent to which the 
particular properties as empirically observed  for the universe might be either  
necessarily determined or down to chance, within the context of the present theory, 
beginning with the values of the large scale cosmological parameters.
   Without a full understanding of their underlying origin, the fact that  the density 
parameters are observed to take the values   $\Omega_{B_0} = 0.050 \pm 0.002$,
 $\Omega_{D_0}= 0.265 \pm 0.011$ and $\Omega_{\Lambda_0}= 0.685 \pm 0.017$~\cite{PDG}, 
as reviewed in section~\ref{sectsmoc}, mutually within an order of magnitude or so of 
each other at the present epoch, given the apparent possibility for each to range over 
many orders of magnitude, is striking. On the other hand given a universe dominated by 
either a cosmological constant $\Lambda$ or matter density $\rho$ term the Friedmann 
equation~\ref{cosg00}, particularly for the $k=0$ case with $H^2 = \frac{1}{3}(\Lambda 
+ \kappa \rho)$, shows that the Hubble parameter is essentially determined by 
$\Lambda$ or $\rho$ respectively, and is clearly not an independent observable.

 At the present epoch for our universe, which is consistent with $k=0$ and with the 
cosmological term beginning to dominate, it is then to be expected  that $\Lambda \sim 
R_H^{-2}$ are of the same order of magnitude, where $R_H$ is the Hubble radius 
introduced in equation~\ref{hubrad}. This observation is a direct consequence of the 
field equation~\ref{Einfield} which leads to the dynamical solution for  the metric 
structure of equation~\ref{CosRob}, including the case of a $\Lambda$ dominated 
universe.
 If the history of the scale factor $a(t)$ is such that $H_0^{-1}$ approximates the 
current age of the universe, which is the case for our universe with the cosmic 
evolution sketched in figure~\ref{aevolve}, then $R_H$ will be of the same order as 
the scale of the observable universe hence in turn relating $\Lambda^{-\frac{1}{2}}$ 
to this scale given the dominance of the $\Lambda$ term at the present epoch.

 While the constant $\Lambda$ in equation~\ref{Einfield}, considered as a geometrical 
effect, has the length dimension of $L^{-2}$ the equivalent `vacuum energy density' 
   $\rho_{\Lambda} = \Lambda / \kappa$   
   has the dimension $ML^{-3} \equiv L^{-4}$ and may be directly compared with the 
mass density $\rho$ for both ordinary and dark matter. It should be noted though that 
on substituting $\frac{\dot{a}^2}{a^2} + \frac{k}{a^2}$ from equation~\ref{cosg00} 
into equation~\ref{cosg11}, for the $\Lambda$ dominated case, it is the extra factor 
of $-\Lambda$ in the second equation which leads to a positive value for $\ddot{a}$ in 
the case of positive vacuum energy density $\rho_{\Lambda} > 0$. This difference  can 
be interpreted as a consequence of the effective `equation of state' for dark energy, 
with $p_{\Lambda} = -\rho_{\Lambda}$, as also implied in equations~\ref{cosacc} and 
\ref{qdefom}.

  While the $\Lambda$ term is beginning to dominate, 
  the present day values of $\rho_{\Lambda}$ and $\rho_M$ (with the latter composed of 
both baryonic and dark matter together) still have a comparable impact on the large 
scale cosmological dynamic equations. The value of $\rho_{\Lambda} \simeq 5.8 \times 
10^{-27}$~kg$\,$m$^{-3}$ is apparently uniform in space and time, and hence the same 
locally as well as globally, and can be compared with the global value of
 $\rho_{M} \simeq 2.6 \times 10^{-27}$~kg$\,$m$^{-3}$, which includes a contribution 
from  $\rho_{B} \simeq 0.4 \times 10^{-27}$~kg$\,$m$^{-3}$,  at the present epoch. 
However  the
 value of $\rho_B$ changes significantly with the cosmic epoch while  local values for 
density of ordinary baryonic matter, such as for the planet Earth with $\rho_{B_E} 
\simeq 5,500$~kg$\,$m$^{-3}$, are much more stable in time. The magnitude of the 
stable terrestrial ratio of
 $\rho_{B_E} / \rho_{\Lambda} \simeq 10^{30}$ then provides a measure of the 
apparently very different nature of ordinary matter and dark energy.

  Another well known apparently natural `large number' in physics concerns the order 
of magnitude of the Standard Model couplings of particle physics in comparison to the 
strength of the gravitational interaction. For example the ratio of the classical 
electrostatic force between an electron and a proton to the classical gravitational 
force between them has a value of $O(10^{39})$ to one. This empirical observation was 
also alluded to near the opening of section~\ref{subwal} in motivating the need to 
introduce practical normalisation factors in studying the implications of 
equation~\ref{gchift} in the laboratory environment.  In the present theory, with 
general relativity and the Standard Model relating to the external and internal 
structures of $\lvfs$ respectively, the relative strengths of the corresponding 
interactions in general will be related to the identification and interpretation of 
equation~\ref{gchift}, which in turn is related to the geometric structures of 
Kaluza-Klein theory. The fact that the gravitational field is not `quantised', and 
hence does not exhibit the running coupling of figure~\ref{runcup} for example, 
further distinguishes gravity from the Standard model forces in the present theory.

 The differing strengths of gravitational and internal gauge forces
should also be connected in some way with the relative magnitudes  of the components, 
such as those of the vector $\bv_4$ or spinors $\psi$, within $\bv_{56} \inn F(\htho)$ 
of equation~\ref{fhthopart} in the symmetry breaking projection over $M_4$.
 With the forms of matter and dark energy also relating to structures within $\lvfs$ 
and its symmetries these relative magnitudes for the components of $\bv_{56}$ may also 
determine the widely differing local values of $\rho_{B_E}$ and $\rho_{\Lambda}$, with 
the value of $\Lambda$ possibly relating to the value of a scalar field such as 
$N(x)$, $n(x)$ or $\beta(x)$ projected out of the components of $F(\htho)$.
 Hence the  symmetry breaking pattern of $\ese$ on $\lvfs$ down to an external 
$\sltc^1$ acting on $\bv_4 \inn \TM_4$ together with the internal structures and the 
details of the projection of the components of $\bv_{56}(x)$ over $M_4$ may underlie 
the empirical observation of both of  the above large numbers.

   It is the relative weakness of gravity that allows structures to form on  large 
scales, from the formation of stable planetary bodies through to clusters of galaxies. 
On the other hand the relative strength of the internal forces shapes the smaller 
scale structures from terrestrial geology down through biological and chemical systems 
to the elements of particle physics. Immersed in the relatively small scale biological 
structures our perspective is one of a spacetime which is flat to a very good 
approximation upon which an apparent `force of gravity' is observed to determine the 
motion of material objects such as apples and cannonballs, as described in 
section~\ref{tlssotu} and before the bullet points in section~\ref{secpotnt}.

For all of the reasons of the above paragraph a world in which the elementary 
interactions of the Standard Model of particle physics are of a much greater strength 
than that of gravitation is `anthropically' favoured. Such a preference may correlate 
with a certain value, or range of values, for the magnitude $\lvfh(x)$ in the 
projection of the $\bv_4 \subset \bv_{56}$ components onto $\TM_4$, and in turn 
underlie the empirical observation of $\rho_{B_E} \ggg \rho_{\Lambda}$ locally on 
Earth and for concentrations of baryonic matter generally.
With the global density $\rho_M$ of the combination of ordinary and dark  matter
 (assuming `dark matter' to behave in a similar manner to baryonic matter in this 
respect)
  declining from a potentially divergent value in the initial singularity and 
seemingly asymptotically approaching zero in the future, the observation that 
$\rho_{M_0} \sim \rho_{\Lambda}$ are of the same order at the present epoch, an 
apparently arbitrary point in cosmic time, appears to be essentially coincidental.

  This determination of $\rho_{M_0} \sim \rho_{\Lambda}$ has some analogy with the 
observation that $r_m \sim r_s$ at the present epoch, where $r_m$ and $r_s$ are the 
apparent sizes of the moon and the sun respectively as viewed from the Earth. The 
value of $r_m$ has been declining since the formation of the Earth-moon system as the 
average distance between these two bodies increases by $O(1\,$cm) every year 
due to the nature of the gravitational interaction between the two bodies. Hence the 
present situation in which the moon is apparently \textit{just} large enough to create 
a total solar eclipse is largely coincidental. However there are  anthropic arguments, 
with the distance of the Earth from the sun being in the `habitable zone' (not too 
near and too hot while also not too far and too cold) and similarly for the distance 
of moon from the Earth resulting in a magnitude of tides which may have aided the 
early development of biological life, which make such an apparent coincidence much 
more likely. Similarly there  may be underlying  anthropic reasons  involving the 
nature of cosmological evolution which make the observation of   
$\rho_{M_0} \sim \rho_{\Lambda}$ more probable during a cosmic epoch supporting 
biological life.

  In summary, in the present theory the observation of $\rho_{B_E} \ggg 
\rho_{\Lambda}$ is expected to be correlated with the observation that Standard Model 
forces are far greater in strength than the gravitational force. Indeed the 
cosmological term $\Lambda g_{\mu\nu}$ might be considered effectively as a geometric 
perturbation within general relativity as the large scale external spacetime structure 
$M_4$ is identified through the projection of $\bv_4 \inn \TM_4$ out of $\lvh$, rather 
than an internal effect underlying the solution $\gmyv$.
  The relation between $\Lambda$ and the Hubble radius, described near the opening of 
this section, may also hint at a geometric origin for the cosmological term.
 As well as the great difference in strength, the rather different nature of 
gravitational compared with internal gauge forces is further emphasised  in the 
present theory by the fact that the degrees of the freedom of the gravitational field, 
describing the external spacetime geometry, are not quantised here.

 With dark matter associated with the external geometric consequences of a variation 
in the magnitude $h(x) = \vert \bv_4 \vert$, as described in the previous two 
sections, here the dark sector in general is associated with locally `weakly 
interacting' general relativistic effects. In the context of a solution for the full 
4-dimensional cosmological geometry `density parameters' such as $\Omega_D$ and 
$\Omega_{\Lambda}$ may not have the same meaning as for the standard theory, since for 
example the above candidate for `dark matter' may not evolve in time in the same way 
as the baryonic matter density and the above  origin for `dark energy' may not imply a 
constant value for $\Lambda$. In any case the present observation of    $\Omega_{D_0} 
\!\sim \Omega_{\Lambda_0}$ may be a consequence of a correlated geometric origin for 
the associated empirical effects, collectively arising from the warping of the  
manifold $M_4$ in the projection out of $\lvh$,  while the proximity of $\Omega_{B_0}$ 
to these values may in part be due to an element of coincidence as described above in 
the analogy with the apparent size of the moon and the sun.

   In developing the present theory further gravitational or material effects may be 
derived in studying the general structure of $\gmyv$ beyond those of the empirically 
observed baryonic matter and dark sector. It \textit{would} seem to require a 
significant coincidence if \textit{all} such effects are of a measurable magnitude and 
hence observable at the present epoch. If there are physical consequences of the 
relation $\gmyv$ which have not yet been detected, and which may be beyond the reach 
of any practical observation, this itself would \textit{partly} account for the 
apparent coincidence of $\rho_{M_0} \sim \rho_{\Lambda}$.
 That is,  these two latter quantities may  form a subset of effects which 
collectively comprise a list of mutual contributions to $G_{\mu\nu}(x)$ at present, 
with a range of other potential terms having much lower density parameters and hence 
remaining undetected. For example if the empirically deduced cosmological term itself 
had been just one order of magnitude smaller it would have been far harder to detect. 
On the other hand while a contribution to the cosmic evolution  
  of the form $R_{\mu\nu} = \lambda(t)v_{\mu}v_{\nu}$ (as described in 
section~\ref{secpotnt} and listed in the 
 final column in table~\ref{cosevo}) has not been observed such a term, with a 
sufficiently low value of $\lambda (t)$, might in principle form part of the large 
scale spacetime solution. With a larger range of such contributions it is more likely 
for any two of them, such as $\rho_{M_0}$ and $\rho_{\Lambda}$, to take similar values 
and be mutually observable.

  In chapter~\ref{newapp} the degeneracy of multiple possible local field solutions 
underlying the spacetime geometry $\gmyv$ was described as the origin of 
indeterministic  quantum phenomena in general. However in terms of constructing a 
solution there may also be a degeneracy in terms of the average projected values of 
the components of for example $\bv_4(x)$ and $\psi(x)$ out of $F(\htho)$ globally over 
$M_4$. In this case there may be only a small certain range of values which lead to 
physical properties of matter capable of supporting life as we know it. Even with this 
degree of anthropic selection to `dial in' certain ratios of the components of 
$\bv_{56} \inn F(\htho)$, via the dilation symmetries described in the opening of 
section~\ref{sectveu} for example, since only a small number of `free' parameters are 
involved in the projection of $\bv_4 \inn \TM_4$ under the fixed structures of $\lvfs$ 
the theory would still be highly constrained, and hence in principle still capable of 
making predictions which might be tested. 
 In section~\ref{sectveu} the point of view was adopted that the interactions under 
the constraints of the theory are such that a unique stable value of $\vert \bv_4 
\vert = h_0$ is achieved, resulting in a phase transition in the very early universe, 
implying an even greater degree of predictability for the theory.
  
   For the scenario described in figures~\ref{earlyu}(b), \ref{svrst} and \ref{spminf} 
at the end of the previous section a unique asymptotic condition with $h(t) = \vert 
\bv_4 \vert \to 0$ as $t \to 0$ has also been presumed. The ensuing progression from 
$h(t) \to 0$ to the stable value $h(t_v) = h_0$ was compared with  models of `new 
inflation'. It is also possible to consider a range of starting conditions for  $h(t) 
< h_0$ as $t \to 0$ and even a broad range of values $h(t) > h_0$ for $t \to 0$, as 
might be associated with the scenario depicted in figure~\ref{earlyu}(a), and evoking 
a comparison with models of `chaotic inflation'. In turn a range of long term 
cosmological conditions might emerge out of the subsequent phase transition,
 even given the same stable value for $h(t_v) = h_0$,
  and hence in principle with a degree of anthropic selection implied for our own 
habitable universe.

  This raises the question of the degree of uniqueness regarding other aspects of the 
theory. 
 With the general form of the function $L(\v)$ determined, as described in 
section~\ref{gfotf},  it is a well defined mathematical problem to identify particular 
forms and then consider the reasons why certain of these may be significant for the 
physical world. Two such significant forms that we have identified are $L(\v_{56})$ 
with an $\ese$ symmetry acting on elements of $F(\htho)$ and $L(\v_4)$ with the 
Lorentz symmetry acting on the 4-dimensional tangent space $\TM_4$ on the base 
manifold.
  With respect to the larger form symmetry breaking over $M_4$ identifies the smaller 
form via the chain $L(\bv_4) \to L(\bv_{10}) \to L(\bv_{27}) \to L(\bv_{56})$, rather 
like a sequence of Russian dolls, with a corresponding chain of subgroups $\soot 
\subset \sltwoo \subset \esi \subset \ese$ as summarised in table~\ref{lvftolvfs} in 
the opening of section~\ref{sosmfi}.

  Alternatively a progression of forms $L(\bv_4) \to L(\bv_{9}) \to L(\bv_{27}) \to 
L(\bv_{56})$ aligned with the subgroup chain $\soot \subset \slthc \subset \esi 
\subset \ese$
 might be considered by expanding the Lorentz symmetry action of $\sltc$ on 
 $\bv_9 \equiv \mcX \inn \hthc$ in equation~\ref{vinhinc} to an $\slthc$ symmetry of 
the 9-dimensional form $L(\bv_9)=\det (\mcX) =1$. The $\slthca$ Lie algebra basis of 
equation~\ref{slthcas} explicitly demonstrates how this structure is naturally 
embedded within the $\sltho \equiv \esi$ action on the 27-dimensional space $\htho$ at 
the next stage of the sequence.

 At either end of this chain it may  be asked why these two particular forms are 
selected out of a large array of possibilities -- why the projection should be onto a 
4-dimensional spacetime manifold and why the highest-dimensional form $\lvh$ should be 
represented by a 
  quartic expression in 56 dimensions with an $\ese$ symmetry.
   It could be considered whether further worlds, different to our own and of course 
not observable by us, could be created out of other possible mathematical forms of 
$L(\bv)$.
 That is, whether the forms $\lvfh$ and $\lvfs$ are largely identified as choices that 
agree with our world, or whether either or both of these are determined by physical 
stability or mathematical symmetry arguments for example.

  By extension from the 3-dimensional  model world of section~\ref{perc} and 
figure~\ref{twodworld} with an $\soth$ symmetry one way to construct a 4-dimensional 
world would be to embed 3-dimensional spatial hypersurfaces within a 4-dimensional 
base manifold with local tangent vectors $\bv'_4(x)$ satisfying the form 
$L(\bv'_4)=(v^1)^2+(v^2)^2+(v^3)^2+(v^4)^2 = 1$
 (appending one dimension to the model case of equation~\ref{flow3d}) with an 
$\mbox{SO}(4)$ symmetry.
However while  geometric curvature and even particle trajectories might be defined
 in such a world, given the local SO(4) symmetry on $M_4$
 there is no consistent definition and distinction of a `temporal' direction compared 
with `spatial' displacements. 
 In principle one of the four dimensions could be arbitrarily declared to represent an  
apparent \textit{temporal} component, however due to the nature of the symmetry 
between the four components the causal structure on $M_4$ would  not be well defined. 

  However given that the 4-dimensional manifold arises as a multi-dimensional 
manifestation of the ordered 1-dimensional flow of time itself and  the necessity for 
the temporal causal structure to be retained on the manifold, the form 
$L(\bv_4)=(v^0)^2-(v^1)^2-(v^2)^2-(v^3)^2 = 1$ of equation~\ref{flow4d} is naturally 
preferred. As described in section~\ref{fdandtd}
 the metric of Lorentz signature implied in this form locally defines a `light cone' 
structure on the extended manifold, which hence distinguishes timelike from spacelike 
directions on $M_4$. The symmetry preserving this form $\lvf$ is the non-compact 
Lorentz group $\soot$, which is also the group which preserves the causality structure 
on a Minkowski spacetime~\cite{Zeeman}. 
 The local symmetry of the 3-dimensional spatial hypersurfaces is identified as the 
$\soth$ subgroup of the Lorentz group. These observations concerning causality appear 
decisive in favour of the `pseudo-Euclidean' form $\lvf$ over the above `Euclidean' 
alternative $L(\bv'_4)=1$.

   A more accurate model for chapter~\ref{sym}  would have involved the 3-dimensional 
Lorentz group group SO$^+(1,2)$ acting on the form $L(\bv_3) = 
(v^0)^2-(v^1)^2-(v^2)^2$ projected onto the tangent space of $M_3$, as a subgroup of 
the full symmetry \mbox{$\hat{G}=\mbox{SO}^+(1,4)$} acting on a 5-dimensional  
Lorentzian form for example. This would be necessary to identify \textit{timelike} and 
\textit{spacelike} vectors and temporal causality on the base manifold $M_3$. However  
dealing with the simplified Euclidean model in chapter~\ref{sym} was sufficient to 
demonstrate the relation between the external and internal symmetry in the present 
theory, with  the same conceptual ideas applying  for the case of the real world with 
the 4-dimensional Lorentz group on the base space $M_4$ as described in 
section~\ref{reaic},
  leading to the connection with Kaluza-Klein theory as also discussed in that 
section.

  As well as the local Lorentz symmetry, which also holds to a good approximation on 
for example the scale of the solar system in the case of our world,
we may also consider whether the base manifold $M_n$ is required to have 
$n=\mbox{\textit{four}}$ spacetime dimensions. If we attempt to construct another 
possible world using similar reasoning to that presented in this paper then we would 
expect something similar to general relativity, that is gravitation, to arise out of 
the geometrical properties on the base manifold of the world, independent of its 
dimension. One important factor may be that while for a 4-dimensional spacetime base 
manifold robust, stable planetary orbits around a massive object, such as a star, are 
to be found in the solutions to the equations of gravitation, this does not arise for 
other dimensions of base space.

 This was shown to be the case for the motion of a body near a massive gravitating 
object, as the source of the Schwarzschild solution for general relativity, in an 
$n$-dimensional spacetime by F.R. Tangherlini in 1963. While for $n=4$ the metric 
solution takes the form of equation~\ref{ttrtp} the functional  form of 
$g_{\mu\nu}(x)$ depends on the value of $n$.
  Although theoretically a circular orbit may be permitted in some cases for $n>4$, 
the slightest perturbation, for example from the impact of a `meteor' or the 
gravitational influence of a third body, would cause the `planet' to wander out of 
orbit and into a path forever receding to larger distances or spiralling inwardly 
until colliding with the central `star'. The same conclusion, that a stable orbit is 
only possible for $m=3$ spatial dimensions, was also found by Paul Ehrenfest in 1917 
for Newton's theory of gravity in which the gravitational potential is determined as a 
solution of the $m$-dimensional Poisson equation (which was introduced for the $m=3$ 
case above equation~\ref{Eins} in section~\ref{gcatep}). 

   Clearly the stability of the elliptical orbit of the Earth around the sun is 
necessary for life on our planet in our world, although this does not imply that the 
equivalent stability is absolutely necessary for life in another world with $n\neq 4$ 
spacetime dimensions. For example the `chemistry' in such a world would be vastly 
different from our own and the relative time scale for the development of life 
structures to the time scale of planetary motions may also be vastly different -- 
potentially allowing a civilisation to evolve out of the primordial chemical soup 
stirring on the planet in the time it takes to glance past a star, even assuming such 
an encounter with a low entropy source in the form of stellar `nuclear' energy is 
necessary.

    Having then decided upon the 4-dimensional Lorentz group on the $M_4$ manifold to 
break the  full symmetry there are still issues concerning the degree to which 
assumptions made about the form of the linear connection $\Gamma(x)$ on $M_4$ are 
necessary. In the present theory the base manifold $M_4$ derives from the projected 
form $L(\bv_4)$ and hence
 regarding the local geometry there are a range of local coordinate systems  at any 
point on $M_4$ for any of which the metric has a Minkowski form $g_{\mu\nu}(x) = 
\delta^a_{\ph{a}\mu}\delta^b_{\ph{b}\nu} \eta_{ab}$. In turn, in deriving from an 
$\soot$-valued connection form the linear connection $\Gamma(x)$ will be metric 
compatible, as also discussed in section~\ref{fdandtd}.  

  As for general relativity, in a 4-dimensional spacetime there is enough freedom in 
general coordinate transformations  to set $\partial_{\lambda} g_{\mu\nu}(x) = 0$ at 
any given point $x \inn M_4$.
 However, as reviewed in section~\ref{gcatep} (and also following 
equation~\ref{gwarph}) general relativity goes further by asserting the
 `equivalence principle' -- according to which the gravitational field can be 
transformed away at any given point, that is, a local inertial frame can be 
constructed such that not only
  $g_{\mu\nu}(x) = \delta^a_{\ph{a}\mu}\delta^b_{\ph{b}\nu} \eta_{ab}$ and
 $\partial_{\lambda} g_{\mu\nu}(x) = 0$ but also $\Gamma(x) = 0$ for any given $x\inn 
M_4$. This means for example that there exists everywhere a local coordinate system in 
which a geodesic trajectory as described in equation~\ref{geotra} for the 4-vector 
$\bu$, with components $u^{\mu} = dx^{\mu}/d\tau$, takes the simple form $d\bu/d\tau = 
0$. Since the torsion tensor $\bT$, with the components of equation~\ref{trmngg}, must 
be zero in all coordinate systems if it is zero in any of them, such a linear 
connection $\Gamma(x)$ is  necessarily torsion-free.

  In the present theory an extended frame of reference throughout which both 
$\partial_{\lambda} g_{\mu\nu}(x) \simeq 0$ and $\Gamma(x) \simeq 0$ is preferred for 
the anthropic purpose of framing an environment for perception. This is certainly 
consistent with the existence of local coordinates such that the equivalence principle 
holds with both $\partial_{\lambda} g_{\mu\nu}(x) = 0$ and $\Gamma(x) = 0$ at any 
given $x\inn M_4$, and taking the torsion to be zero may be a very good approximation. 
However given the mathematical basis for what we are taking as the act of perception
 it seems perhaps artificial to impose the extra restriction on the connection that it 
should necessarily be torsion-free or, further, require that the strong equivalence 
principle in general should hold. It may be that there is a non-vanishing contribution 
to physical phenomena from torsion which has so far been beyond the reach of 
observation --  for example any contribution to the connection coefficients   
$\Gamma^{\lambda}_{\ph{\lambda}\mu\nu}(x)$ asymmetric in the  $\{\mu,\nu\}$ indices 
would have no effect on the simple geodesic motion of equation~\ref{geotra} --  and 
neglecting it has therefore been of no consequence.

    This is also the case in general relativity where setting the torsion equal to 
zero acts as a simplifying assumption. Both in general relativity and the present 
theory the linear connection $\Gamma$ is a metric connection with $\nabla g =0$, but 
this does not imply that the torsion should vanish. In the Einstein-Cartan version of 
general relativity the more general geometry with finite torsion is considered (with 
extra dimensions such a generalisation is also significant for the Kaluza-Klein 
theories reviewed in section~\ref{olcop}). In this case while the spacetime curvature 
is still related to the energy-momentum of matter through the Einstein equation the 
torsion is a function of the spin current of matter. Unlike curvature the torsion does 
not propagate in the matter-free vacuum and the two theories are identical in such an 
 environment. Further, given that the spin density is small for ordinary matter in the 
universe the two theories have been experimentally indistinguishable, and this itself  
justifies adopting $\bT = 0$  as a simplifying assumption.

  In the present theory it is an open question whether the linear connection $\Gamma$ 
is necessarily symmetric and torsion-free, and if so to explain why this is the case. 
More generally the question regards whether the spacetime geometry and forms of matter 
consistent with $\gmyv$ contain the structures of torsion and a spin current. In the 
meantime as for general relativity the assumption $\bT = 0$ may be adopted to simplify 
some of the mathematical expressions, with in particular the Levi-Civita connection of 
equation~\ref{gtoGam} hence being employed.
 This is analogous to adopting the simplifying conditions which underlie the 
Robertson-Walker line element of equation~\ref{CosRob} in order to study models for 
the evolution of the universe as a whole, even though the assumptions of the 
cosmological principle clearly do not hold exactly. 
 The degree to which the large scale structure deviates from the conditions of 
homogeneity and isotropy may itself not be a uniquely restricted property of the 
universe.

  Even for the $\bT=0$ case, in constructing the external geometry in terms of the 
internal fields, as well as the 10 components of the Einstein tensor in the form of 
$\gmyv$ itself the full 20 independent components of the Riemann curvature tensor may 
also explicitly depend on those fields with
 $R^{\rho}_{\ph{\rho}\sigma\mu\nu} = f(Y,\bvh)$. This will include the Weyl tensor 
components
  $C^{\rho}_{\ph{\rho}\sigma\mu\nu} = f(Y,\bvh)$ and hence the identity
   $C_{\sigma\mu} = C^{\rho}_{\pho \sigma\mu\rho} = 0$, as described towards the end 
of section~\ref{riegeo}, as well as the Bianchi identity $\gmo$, will also apply 
implicitly for the internal fields. The 10 degrees of freedom of 
$C^{\rho}_{\ph{\rho}\sigma\mu\nu}(x)$ are still considered to represent the `vacuum' 
in the sense that they complement the 10 degrees of freedom of $-\kappa T_{\mu\nu} := 
G_{\mu\nu} = R_{\mu\nu} - \frac{1}{2}Rg_{\mu\nu}$ in the decomposition of 
equation~\ref{rdecom} for the full Riemann tensor $R^{\rho}_{\ph{\rho}\sigma\mu\nu} = 
f(Y,\bvh)$.

  In classical general relativity while matter is identified with the content of the 
energy-momentum tensor $T_{\mu\nu}$ the vacuum geometry with $G_{\mu\nu} = 0$ and 
$C^{\rho}_{\ph{\rho}\sigma\mu\nu} \neq 0$ still carries energy, in the form of gravity 
waves for example, as also discussed after equation~\ref{cbian} in 
section~\ref{subwal}. Hence energy can propagate through the `vacuum' of spacetime 
even when not expressed in terms of any underlying internal fields. In contrast
 it is also possible in the present theory that there may be fields on $M_4$, for 
example from some of the components of $F(\htho)$ underlying the form $\lvfs$, at 
least in some regions of spacetime, that may not directly contribute to the spacetime 
structure of  $G_{\mu\nu} = f(Y,\bvh)$ at all, and hence which do not carry 
energy-momentum.

  While for general relativity the Einstein equation~\ref{einlamt} can be derived from 
the Einstein-Hilbert action of equation~\ref{einhil}, it can be shown, as demonstrated 
by
   Cartan, Weyl and others, that the most general divergence-free symmetric 2-index 
tensor constructed from the metric and its derivatives up to second order is a linear 
combination of $G_{\mu\nu}$ and $g_{\mu\nu}$ (see for example~\cite{Weyl2} appendix 
II). This consideration itself leads to  Einstein's equation~\ref{einlamt} and 
\ref{Einfield}, with $\Lambda$  a free parameter, as essentially the only admissible 
field equation for a geometric theory of gravity consistent with a divergence-free 
energy-momentum tensor on the right-hand side.
   Regardless of the method of derivation the significance of the Einstein equation 
derives largely from the contracted Bianchi identity $\gmo$, which then necessarily 
applies to  
 the energy-momentum tensor. On the other hand symmetries in the apparent distribution 
of matter can be employed to assist the search for solutions, with for example 
equation~\ref{cosgtruup}, with $\rho$ and $p$ being functions of cosmic time only, 
being the most general energy-momentum tensor consistent with the assumptions of the 
cosmological principle, as described in section~\ref{sectsmoc}.

 In the present theory  energy-momentum is \textit{defined}  through $T_{\mu\nu} := 
G_{\mu\nu}$. While $\gmyv$ incorporates  ordinary matter, a possible $\Lambda 
g_{\mu\nu}$ term, dark matter phenomena and the structure of the very early universe 
collectively into an apparent $T_{\mu\nu}(x)$ there may be further geometric or 
material phenomena, arising out of the internal fields or their interactions, which 
have not yet been detected. This possibility was alluded to earlier in this section in 
the discussion of the observation of $\rho_{M_0} \sim \rho_{\Lambda}$, 
 as exemplified by a potential contribution originating from a term of the form 
$R_{\mu\nu} = \lambda(t)v_{\mu}v_{\nu}$,
 and now incorporates also the possibility of finite torsion.
  The potential for new phenomena will be of particular interest if yet
   higher-dimensional forms of $\lvh$ are considered, with a corresponding larger 
symmetry,  which may also be needed to fully account for known Standard Model particle 
phenomena.

    The Lie group $\esi$ was originally selected as a candidate symmetry for the full 
form of $\lvh$, in the context of the present theory, in part since it is already of 
well known interest in relation to the observed gauge groups of elementary particle 
theory,
 as reviewed in section~\ref{dynkin}. However, it was primarily chosen for detailed 
study as it acts on a relatively high dimensional vector space, with 27 dimensions 
compared with the four on the base manifold $M_4$, and stands out as exhibiting 
particularly rich mathematical structures,  involving for example the triality 
symmetry of the octonions and three interlocking actions of $\sltwoo$ as described in 
chapter~\ref{esihtho}, through which to channel the temporal flow via the components 
of $\v_{27}$ under
 the constraint  of the 27-dimensional cubic form 
$\lvt$. Expressed in terms of the octonions, which themselves form the largest of the 
normed division algebras, this form of $\lvh$ provides a unique structure.
The existence of elaborate mathematical properties within the substructures of the 
$\esi$ symmetry acting on $\htho$ matrices is perhaps the reason why $\esi$ stands out 
as a kind of significant mathematical resonance amongst other possible symmetries of 
temporal forms in yet higher dimensions.
  By comparison for example higher-dimensional spacetime symmetries 
$\mbox{SO}^+(1,m)$, acting on quadratic Lorentzian forms with an arbitrarily large 
number $m$ of spatial dimensions, arguably exhibit a somewhat less elaborate  
structure.

 In section~\ref{secesef} the analysis was extended to the smallest non-trivial 
representation of E$_7$  realised as an action on the 56-dimensional space $F(\htho)$ 
preserving a certain quartic form $\lvfs$ and incorporating the octonions in two 
independent  $\htho$  subspaces.
 Building upon the properties identified for the symmetry breaking of the $\esi$ 
action on $\htho$ described in chapter~\ref{chapesb}, the structure of the broken 
$\ese$ action on $F(\htho)$ when projected over $M_4$ has a number of properties 
reminiscent of the Standard Model, as summarised in equation~\ref{fhthopart}. 
 However it is  still very much an open question as to which other symmetry groups 
should perhaps be considered  and what observable effects they may have on our own 
world. These effects might be manifested in particle physics phenomena through the 
prediction of additional states,
 or the determination of the properties of known states, which might be observed in 
high energy physics experiments.

  The hypothetical extension to an $\ee$ symmetry on a 248-dimensional form $\lvtfe$, 
as described in section~\ref{sosmfi}, would in principle be large enough to 
incorporate the full set of known Standard Model states, including all three 
generations of the fermions.
 Given that $\ee$ is the largest exceptional Lie algebra, terminating the chain of 
Dynkin diagrams depicted in figure~\ref{dynkinee} of section~\ref{secesef}, such a 
form of temporal flow might uniquely complete the sequence of extensions listed in 
table~\ref{lvftolvfs} of section~\ref{sosmfi}. As a continuation of that sequence, and 
also in particular to contain the non-compact Lorentz group $\soot$ as the local 
symmetry of the causal structure on $M_4$ as discussed earlier in this section, this 
may involve the non-compact real form $\eeg$ as described for equation~\ref{esixtoee}.

 The present theory is based on the observation that the one-dimensional progression 
in time, via the elementary arithmetic properties of the real line $\rrr$, can be 
expressed in terms of variables in an arbitrary number of dimensions. In principle the 
same observation might be applied to each of the $n$ real components underlying 
an $n$-dimensional form of temporal flow $\lvn$.   
   For the case of the orthogonal group O$(n)$ in the limit $n\to \infty$ certain 
properties related to the octonions  make various calculations  more tractable.
 In his study of the homotopy groups of the topological group O$(\infty)$ in 1957 
Raoul Bott discovered the isomorphism $\pi_{i+8}(\mbox{O}(\infty)) \cong 
\pi_i(\mbox{O}(\infty))$. Such period 8 structures, which are also seen for Clifford 
algebras and known generally as `Bott periodicity', are all closely related to the 
8-dimensional octonions. Similar periodicity structures may become relevant in the 
exploration of higher-dimensional forms of $\lvh$,
 for which octonion elements explicitly feature, 
 and might even be important for calculations relating to the degeneracy of solutions 
underlying $\gmyv$, involving a higher-order nesting of field redescriptions, which 
underlie quantum and particle phenomena. 

   The progression towards higher-dimensional forms of $\lvh$ described above may 
uncover a uniquely determined mathematical structure. Given also that the 
4-dimensional Lorentzian form $L(\bv^4) = h^2$ projected into $M_4$ may  necessarily 
provide the means of breaking the higher symmetry the laws of physics observed in our 
universe might in turn be uniquely determined. 
 Even in this case our universe does \textit{not} represent the unique manifestation 
of such a world, but rather  one of a vast number of possible solutions for the 
external 
 geometry $\gmyv$,
 built upon an underlying degeneracy of local internal field descriptions 
 as expounded in chapter~\ref{newapp}. While events at a HEP experiment, such as 
depicted in figure~\ref{figsld}, exhibit the intrinsic structure of quantum 
uncertainty, the spectrum and properties of the particles identified in the laboratory 
may be unique. On the other hand in principle there might still be solutions for 
multiple universes with a range of large scale cosmological structures depending on 
the nature of the overall $\gmyv$ solution, in particular with regard to the apparent 
conditions in the very early universe. 

 It nevertheless will be required to carve out of the full form $\lvh$ a universe like 
ours, such as depicted in figure~\ref{spminf} and described at the end of the previous 
section, incorporating all of the observed large scale structure and the phenomena of 
the Big Bang.
  Regardless of the degree of uniqueness of such a world, 
 in being constructed out of the multi-dimensional forms of temporal flow, it derives 
in turn from the priority of one-dimensional temporal flow as the underlying basis of 
the universe. 
 Hence the conceptual question remains regarding the origin of this one-dimensional 
structure itself, as we consider in the following chapter. 


\pagebreak
\chapter{The Origin of Time}
\label{chaptoot}

\section{Two Loose Ends in the Theoretical Sciences}
\label{sectleitp}

  The aim of theoretical physics at a fundamental level could be described as a 
program to uncover the basic scientific principles of the world, the consequences of 
which
    encompass \textit{all} empirical phenomena.
  From the objective point of view  the existence of the universe, and its matter 
content, began with the Big Bang and evolved according to equations of motion, as 
governed by the fundamental principles, for billions of years as the matter condensed 
into galaxies, stars and planets, some of which are conducive for biological life, 
until eventually conscious observers such as ourselves in turn evolved, with the 
ability to contemplate the world and the cosmos around us. Two of the most pressing 
kinds of questions raised by this picture concern the nature of (1) the Big Bang and 
(2) conscious life: 

\begin{itemize}
\item[(1)] What can we say about the universe \textit{before} the Big Bang? How and 
why does the Big Bang occur? How is spacetime itself created? Can the `initial 
singularity' be avoided? 
What determines the particular initial conditions? How is matter created and what 
determines its properties? Why are the laws of physics the way they are? 
\item[(2)] Given that a material universe is created and set in motion subject to the 
physical laws, how is it possible to mould the conscious \textit{experiences} of 
observers, aware of themselves and the world around them, out of inert, lifeless, 
material substance of a seemingly qualitatively entirely different nature? 
\end{itemize}
  
   It seems inevitable that any physical theory must be founded on a `loose end' 
concerning the basic elements of the theory. This is the case whether these basic 
entities consist of particles, fields, strings, spacetime, extra dimensions, or some 
combination of these or further concepts, and is generally justified on the grounds 
that `one has to start somewhere'. A similar argument could be made for the present 
theory founded on the concept of time. This paper has presented the mathematical 
development of this theory, beginning with the general form of temporal flow $\lv$ as 
deduced for equation~\ref{lv}, through the construction of a physical world in space 
and time for comparison with observations, leading to a discussion of the possible 
uniqueness of this structure in the previous section -- which addresses some of the 
points of item (1) above. However, no matter how far progress might be made with the 
elucidation of  empirical phenomena the theory is incomplete so long as there remains  
the question regarding the origin of temporal flow itself, as represented by the loose 
end on the left-hand side of figure~\ref{loosends}.

\begin{figure}[htbp]  
\centering
 \hspace*{-15pt}
\epsfxsize=15.4cm
\leavevmode
\epsffile[0 0 2178 296]{\gpath aPfig141e}
\caption{\setb Beginning with the notion of one-dimensional progression in time, via 
the general mathematical form $\lv$, both an extended spacetime manifold and the 
physical bodies perceived within it are in turn derived. The two loose ends concern 
respectively the origin of time itself and the subjective experience of the observer.}
\label{loosends}
\end{figure}
  
    With the basic entity having such a simple structure, namely a one-dimensional 
ordered flow of time modelled by the real line $\rrr$, this first loose end is 
particularly striking for the present theory.  By comparison a theory founded for 
example upon the basic entities of a set of fields in spacetime begins with a great 
deal of structure, and can to a large  extent  be considered as a study of the 
phenomenology of fields in spacetime. However here since the  fundamental temporal 
flow, represented by the real line, cannot be readily decomposed into simpler elements 
it is very natural to raise the question of its origin, and in turn  there is a 
greater sense of obligation to address the issue of a foundation for the present 
theory.

   Given a description of the physical world, whether founded on the notion of time or 
other basic concepts, containing bodies which can be observed, the second loose end,  
as depicted
  on the right-hand side of figure~\ref{loosends},
 regards the question of how it is possible for an entity to be \textit{aware} of an 
observation. This question concerns the issue of how `we', as beings conscious of  
observations and thoughts, are embedded within the structures of the world. The 
physical structure of the organic brain is closely associated with this latter loose 
end as an apparent vehicle for self-reference  capable of encoding subjective 
experiences within the physical world. In this section we consider how such a 
structure might be modelled or explained in terms of mathematical or physical 
elements, before returning to the first loose end of figure~\ref{loosends}.

   The idea that conscious experience can arise out of physical structures on the 
spacetime manifold $M_4$ should not be too controversial since it is essentially 
implied in most approaches to fundamental physics. If based on a quantum field theory, 
as applied in the Standard Model of particle physics for example,
 all properties of matter ultimately arise from the properties of the basic fields and 
their mutual interactions.
  Hence the microscopic properties of matter   underlie the structure of macroscopic 
objects in the world including both inanimate objects such as  rocks and pencils as 
well as biological structures such as flowers and brains.
The self-reflective, self-conscious activity of the human brain must therefore be 
supported by the underlying elements of the theory and the structures which they 
generate in spacetime. This is essentially the case for \textit{any} physical theory, 
  since it is evident  that beings conscious of experience arise in the same world as 
described by the theory. Both aspects of this world, that is the subjective mental 
phenomena as well as the objective material phenomena, are then in principle amenable 
to theoretical analysis.

  On the practical side,
  since the early history of computing, with devices designed or constructed first of 
mechanical and later electronic components, comparisons have been drawn between 
`artificial intelligence' and the workings of the naturally occurring physical 
structure of the brain. Indeed, the design of a computer as envisaged by Alan Turing 
in the 1930s and 1940s was based on modelling the action of the human mind with the 
ambition to `build a brain'  out of electronic  components. This came with the 
significant advance in the design whereby programs as well as data could be stored in 
symbolic form, allowing both to be modified and manipulated by the `universal 
machine'. 
  On the more philosophical side Turing demonstrated that there are questions 
involving the performance of a universal machine which are intrinsically 
`non-computable' for the device.
Turing also came to the conclusion that the actions of a human brain are `computable'; 
with such thought processes then being amplified through the actions of the human 
body. 

  The notion of computability for  physical devices has a close parallel in the field 
of pure mathematics, regarding in particular the demonstration by 
  Kurt G\"{o}del  a few years earlier that propositions can be constructed in an 
arithmetical calculus which are intrinsically unprovable within the calculus. 
   It is this latter analysis we consider here in order to then describe a model for a 
self-referencing   subjective state.

 Proposition VI of G\"{o}del's 1931 paper, \textit{On Formally Undecidable 
Propositions of Principia Mathematica and Related Systems I}~\cite{Godel} can be 
paraphrased: `All consistent axiomatic formulations of number theory include 
undecidable propositions'; that is, there are true statements of number theory which 
its methods of proof are too weak to demonstrate. The argument can be applied to any 
calculus (that is a formal system consisting of a set of axioms and rules of 
inference) powerful enough to express the basic arithmetic (with addition and 
multiplication) of the natural numbers ($0,1,2,\ldots$). Hence any such formal system 
is `incomplete'. The essential idea employed by G\"{o}del was to find a way to use 
mathematical reasoning to explore mathematical reasoning itself (see for example 
\cite{NN,geb}).

  Following a chain of deductions which begins with a construction known as `G\"{o}del 
numbering'
  a formula called $G$ (after G\"{o}del)  is derived which is the mirror image within 
the arithmetical calculus of the meta-mathematical statement that: `The formula $G$ is 
not demonstrable'. G\"{o}del was able to show that if in fact $G$ is demonstrable then 
its formal contradictory $\sim \! G$ (i.e. `not $G$') would also be demonstrable, 
leading to an obvious inconsistency. He proved that if the formal system \textit{is} 
consistent then $G$ is  formally \textit{undecidable}; that is, neither $G$ nor $\sim 
\! G$ can be deduced from the axioms and rules of the calculus. 

     It can however be seen by meta-mathematical reasoning that $G$ is in fact a true 
proposition of the calculus.
 Hence $G$ is a true arithmetical formula and in fact expresses a certain property of 
all natural numbers. Hence  an arithmetical truth has been discovered which can not be 
deduced formally from the axioms and rules of inference of the calculus. Any calculus 
incorporating arithmetic is incomplete in this way (in the original historical context 
this signalled the demise of David Hilbert's challenge to prove the contrary). 
Although we are free to simply add $G$ as an extra axiom for the formal system, in 
this case a different true undecidable arithmetical formula $G'$ could be constructed 
from the augmented calculus. Again, adding $G'$ as an axiom we would still be able to 
construct a true undecidable $G''$, and so on; that is, for any augmented set of 
axioms and rules it will always be possible to construct further undecidable 
propositions -- the calculus is `essentially incomplete'.

 The essential points of G\"{o}del's theorem for our purposes are summarised here:
\begin{itemize}
 \item The symbols, axioms, rules, theorems and general expressions of a calculus or 
formal system capable of expressing arithmetic can be mapped onto a subset of the 
integers by G\"{o}del numbering.
 \item Meta-mathematical statements about expressions of the calculus are associated 
with a mirror image within the arithmetic itself.
 \item Assuming that the calculus is consistent, formulas such as $G$ can be 
constructed   which can be shown to be true while being formally undecidable -- it is 
not possible to prove either $G$ or $\sim \! G$ within the calculus. 
 \item Augmenting the calculus with new axioms such as $G$ leads to a new calculus for 
which new undecidable formulas such as $G'$ can be found; completeness of arithmetic 
can not be achieved, it is `essentially incomplete'.
 \item The consistency of the calculus can not be proved from within the system, but 
it can be demonstrated by meta-mathematical reasoning outside the system. 
\end{itemize}     
  
 We next ask how the above considerations may be of relevance in the theoretical 
sciences and in particular in relation to the theory investigated in this paper.
 The  general mathematical form $\lv$ was derived in equations~\ref{vonedo}--\ref{lv}
 on considering the  notion of progression in time  to have a structure isomorphic to 
the algebra of the real numbers $\rrr$, including the basic arithmetic operations of 
$+$ and $\times$.
 Since the natural numbers $\mathbb{N}$ are embedded as a subset of the real numbers 
the mathematical calculus concerned with $\lv$ is certainly sufficient to express the 
usual rules of arithmetic for the non-negative integers. Further,
in developing this physical theory  certain mathematical structures arising from the 
forms and symmetries of $\lv$ have been taken to be isomorphic to the structures that 
we perceive in the physical world.

  It is a world in which we find both natural and manufactured machines and devices 
which are in some cases capable of expressing statements about mathematics, and in 
particular about the kind of mathematical calculus that underlies the world. 
 Since the physical world can be expressed in mathematical terms capable of describing 
the behaviour of objects and devices in the world  exhibiting for example structures 
(such as the human brain) powerful enough to perform arithmetic operations and support 
states of self-reference, then it seems that `formally undecidable propositions' must 
inevitably arise in the application of these mathematical structures. We may then 
consider the possibility that the manifestation of such mathematical phenomena  in the 
world  is in the form of our  own conscious experience of being in an `undecided 
state', with the above list of five points correlated with the corresponding list 
below: 

\begin{itemize}
\item  There is a necessary isomorphism between the physical structure of everything 
in the material world, including brains, and mathematical structures expressible in 
the calculus  underlying the expression $\lv$.
\item  The human brain is capable of making meta-mathematical statements about 
structures deriving from the mathematics of $\lv$, which therefore necessarily have a 
mirror image in structures deriving from the $\lv$ calculus itself.
\item  We \textit{experience} questions we can ask of ourselves in making a choice, 
such as ``Shall I pick up the pen or the pencil in front of me?'' as being 
\textit{undecidable} (that is, we cannot predict our own \textit{future} actions).
\item  In making a choice, for example in picking up the pencil, we find ourselves in 
a new state for which a further horizon of similarly undecidable questions perpetually 
arise.
\item Our experiences are organised and synthesised into a self-consistent and 
coherent awareness of the world. 
\end{itemize}

 This is indeed, of course, very far from being a definitive analysis of the 
phenomenon of our conscious experience in the world. The intention here is rather 
merely to consider the close analogy with the elements that go into the construction 
of G\"{o}del's theorems. That there may be a more significant relation between these 
two cases is suggested by their close structural similarity, the fact that they are 
both grounded in mathematical considerations involving self-reference and the fact 
that potentially highly complicated mathematical expressions arise in both cases. We 
observe further that in considering a choice it is precisely our `undecided' state 
that we are aware of.

  For this  preliminary discussion of this phenomenon in the context of the present 
theory we proceed with the following simple experiment.
 For clarity of exposition the discussion is presented in terms of \textit{my} 
experiences in the world, where \textit{my} and \textit{I} refer to any individual, 
such as the person currently reading this text. 
 I can place, for example, a pen and a pencil on the table in front of me and allow 
myself to deliberate for several seconds over the question ``shall I pick up the pen 
or the pencil?'', while filtering out other thoughts as far as possible. In performing 
such an experiment the experience is one of initially having an awareness of being in 
an `undecided' state, in which I may `change my mind' several times almost as if 
compelled along on a wave of reasoning guided by practical or aesthetic judgements 
concerning, for example, the utility of the pencil or the colour of the pen, and then, 
quite suddenly, as if I have to let go, I find myself in the `decided' state of having 
chosen the \textit{pencil} and hold it in my hand (in fact, the more casually or 
lazily I make the choice the more it feels \textit{determined} by the rational course 
of the world, including subconscious processes, with my conscious deliberation being a 
kind of internally reflecting resistance to that flow). That we can readily do this 
kind of `thought experiment' and attempt to observe what \textit{happens} when the 
choice is made serves to emphasise just how central the phenomenon of conscious 
decision making is in the world. A general physical theory of the world should then 
ideally have something to say about this phenomenon or be able to offer a good reason 
why it does not. 

	 Here we comment on the fundamental difference between questions we can ask of the 
kind ``will the apple fall off the tree?'' and of the kind ``shall I pick up the 
apple?''. The former question about the external world, not involving self-reference, 
is `undecided' to the extent that we lack the relevant knowledge about the physical 
state of the objects concerned -- we simply await the resolution of the question as 
carried externally in the inertia of the world
 (and with a similar interpretation applying for the outcome of indeterministic 
quantum processes, as depicted in figure~\ref{eemmeds}(b) for example).
 For the latter question regarding whether or not to pick up the apple, in attempting 
to predict our own future action based on our internal thoughts we are conscious of 
falling over ourselves in search of the answer until we \textit{experience} the 
resolution.

 To proceed further we consider a self-referential mathematical system  $R$  which is 
assumed to be correlated with a physical brain state. For such a given formal system  
$R$ in principle a large number $n$ of undecidable propositions $G_i$, with $i = 
1\ldots n$, might be formulated; as represented in figure~\ref{rtordash}(a). If any 
one of these is taken to be absorbed into the mathematical structure as a new axiom 
then a new formal system $R'$ with a new horizon of internally undecidable 
propositions $G'_i$ arises, as depicted in figures~\ref{rtordash}(b) and (c).

\begin{figure}[htbp]  
\centering
\epsfxsize=\maxwidth
\leavevmode
\epsffile[0 0 1982 511]{\gpath aPfig142e}
\caption{\setb Expansion of a formal system $R$ as the `undecidable' proposition $G_i$ 
is incorporated as a new axiom.}
\label{rtordash}
\end{figure}

 From the subjective point of view  the system $R$ represents a self-reflective 
conscious state of mind, which is in constant interaction with the subconscious mind 
and the world beyond, which are also represented by mathematical structures and 
provide a reservoir of information and data in the environment $E$ which might enter 
conscious thought as represented by figure~\ref{rtordash}(b), effectively 
corresponding to a realisation of the truth of $G_i$. The subjective correlate of a 
single step $R \to R'$ is considered to be the experience of making a \textit{choice}.
 The overall mathematical structure with the progression $R \to R' \to R''$ drawn out 
through the interaction between the conscious state $R$ and the broader environment 
$E$ is outlined in figure~\ref{rindata}. 

\begin{figure}[htbp]  
\centering
\epsfxsize=14cm
\leavevmode
\epsffile[0 0 2000 588]{\gpath aPfig143e}
\caption{\setb In terms of mathematical objects the formal system $R$ corresponds to a 
subset of a larger environment $E$  which provides the source of information and data 
for the progression depicted in figure~\ref{rtordash}.}
\label{rindata}
\end{figure}

  The essential feature of figures~\ref{rtordash} and \ref{rindata} is that 
\textit{any} change in the system $R$, due to the interaction between $R$ and the 
mathematical forms of $E$, results in a \textit{progression}. The all-encompassing 
mathematical environment $E$ in figure~\ref{rindata} can be thought of as a static 
sculpture of mathematical objects. Within this structure the self-referential systems 
with $\ldots R \subset R' \subset R''\ldots$ carve out a one-dimensional ordered 
progression.
 A given self-referential state $R'$ within this sequence absorbs the state $R$ 
accompanied by one of its undecidable propositions $G_i$,  now included as an axiom 
within $R'$, with respect to which $R$ represents the `past'. Similarly $R'$ is itself 
in turn contained within $R''$ with the latter state incorporating a resolution of an 
undecided proposition of the state $R'$, that is one of the $G'_i$, and hence 
representing the `future'. Along with this terminology, with for example the 
self-reflective state $R'$ corresponding to the `present' experience, incorporating 
$R$ in the past and drawn towards $R''$ in the future, the structure represented in 
figures~\ref{rtordash} and \ref{rindata} is
postulated as a model for subjective \textit{temporalisation}.

 To follow the above analogy with G\"{o}del's theorem closely then would be to say 
that our experienced state of being undecided finds resolution by absorption into a 
new state in which a particular choice, or corresponding new `axiom', is included. The 
possibilities to incorporate further new axioms in the attempt to resolve a perpetual 
state of undecidability leads to an ordered progression (incorporating $\ldots G$, 
$G'$, $G''\ldots$ into $\ldots R$, $R'$, $R''\ldots$ respectively) which is therefore 
structurally identical to, and proposed as the origin of, our \textit{experience} of 
temporality.

  It is important to emphasise here that it is the mutual association of the $R$ 
states in this \textit{ordered} series that \textit{has itself} a temporal structure. 
It is not a question of being situated at $R'$ for example and asking how it is 
possible to move on to $R''$, since for something to \textit{move} presupposes an 
already existing flow of time with respect to which the motion takes place. Rather it 
is the unambiguously ascending logical order of this series itself which, having a 
structure that can be mapped onto and modelled by the one-dimensional ordered real 
line $\rrr$, reveals the form of time itself. We have `time' \textit{already} in its 
pure and simplest essence as an ordered progression in this abstract series.

From the objective point of view a state $R$
  corresponds to  a limited physical system in the world, correlated in particular 
with features of a physical brain. Via physical processes new data can be introduced 
through the interaction between, effectively, the conscious brain and the subconscious 
brain, as well as with the rest of the physical world, as will be described further 
below and in the following section.
Future actions are not fully determined by or contained within the self-reflective 
state $R$ itself. The physical process of the subconscious intervening in the 
conscious deliberation, as 
 modelled by the progression from $R$ to $R+G_i$ and attainment of the new state $R'$ 
depicted in figure~\ref{rtordash},
 correlates with the  subjective experience of `letting go' after a period of `falling 
over oneself' in debating whether to pick up the pen or the pencil.
 From an internal subjective point of view the conscious mind is ignorant of the 
choice \textit{until} it is made and the individual finds himself holding the pencil 
rather than the pen for example, contributing to his sense of temporalisation. 
   The passage of time and conscious experience more generally may feel somewhat 
mysterious since we do not generally perceive the objective structures and 
interactions represented in figure~\ref{rindata}, only their internal subjective 
correlate.

  Evidently our thoughts are not really as clear cut or `binary' as suggested in the 
example above when confronted with a simple choice such as ``Shall I pick up the pen 
or the pencil in front of me?''.
It is not that we are really considering an isolated possible future state 
corresponding to each alternative $G_i$ in figure~\ref{rtordash}(a). Rather, there is 
an enormous ensemble of possible future states which may be divided into two sets, 
each with a vast range of members: 
\begin{equation}
  \begin{array}{ll}
     \mbox{A)}  & \{\mbox{I shall pick up the pen $+$ $X$}\}   \\
     \mbox{B)}  & \{\mbox{I shall pick up the pencil $+$ $Y$}\}  \end{array}
	\label{aorbpen}
\end{equation}   
  where $X$ and $Y$ each refer to possible features of a state of mind in addition to 
whether or not I hold the pen or pencil respectively. 
The idea here is not so much that individual undecidable  propositions $G$ uniquely 
correspond to simple thoughts or actions such as `I pick up the pen'. Rather it is to 
be considered that there is a vat of an \textit{enormously} large number of correlated 
$G$-like statements $G_i$ (with $i=1\ldots n$ and $n$ an extremely large number) 
relevant to a particular brain state. A subset  of the $G_i$ will incorporate the 
statement `I pick up the pen' amongst other actions, others will incorporate the 
statement `I pick up the pencil' amongst other actions, while still further subsets of 
the $G_i$ will represent the cases `I pick up both'
 or `I do not pick up anything'. Each of these ghostly undecidable propositions $G_i$ 
points towards a possible extension of my self-reflective state. Such extended systems 
draw us in and as we progress from one state to another, augmented, state our sense of 
temporality is created.

In the course of this dynamical stream of temporalisation I shall find myself coming 
into a state of picking up the pen or pencil, depending on the choice of the possible 
$G_i$. This set of potential $G_i$ is itself of course very dynamic, as represented 
for example by the set $G'_i$ in figure~\ref{rtordash}(c), and evolves in turn with 
the incorporation of new axioms, or choices, and new information into my system, 
corresponding to the ever evolving set of my possible future actions.

That the nature of subjective awareness may be correlated with the mathematical notion 
of the undecidable in self-referencing systems opens the door to a more thorough 
investigation.
 However, technically, in the context of the physical world, it may be that 
`computability', rather than the closely related notion of `decidability', is a more 
directly relevant concept to employ, since we know that the laws of physics in our 
world are such that `computing machines' (both artificial and organic) are supported. 
That is, we are directly dealing with the states of such `devices' in the physical 
world rather than with abstract mathematical symbols in a formal system, although 
there is a close structural parallel between the two cases. The discussion has been 
framed in terms of `decidability' partly due to the similarity of the language used to 
express the \textit{experience} of making a choice; that is, in making a choice we are 
primarily conscious of being in an undecided state.
  On the other hand given this coincidence of language terms  some caution is needed 
in order to avoid being misled into taking the connection too literally.

  It is indeed very much open to question how far to take the analogy between the 
structures pertaining to G\"{o}del's theorem and the subjective process of decision 
making, although  there is some degree of correspondence as indicated by the two sets 
of bullet points listed earlier in this section. With contradictory `undecided' 
propositions from sets A and B being simultaneously entertained in 
equation~\ref{aorbpen}, corresponding for example to $G_i$ and $G_j$ with $i \neq j$, 
this structure does seem to have some important differences also with the above 
mathematical correlate, since for G\"{o}del's theorem the `undecidable' describes the 
relation between  $G_i$ and $\sim \! G_i$, with the  
proposition $G_i$ representing an unprovable   but \textit{true} statement. Although 
this implies that to some degree  G\"{o}del's construction should be taken 
metaphorically here,
 the employment of a  mathematical framework with self-referential  structures is 
still very relevant.

  As well as the subjective interpretation the structure in figure~\ref{rindata} must 
also correlate with a physical manifestation.
 From this objective perspective the laws of physics must support  a kind of inertia 
in the substructure of the physical brain, corresponding to the subconscious mind, 
that carries the subject into just \textit{one} of the array of `true' states either 
in set A or in set B
  of equation~\ref{aorbpen};
  that is into a new structure of self-reference such that the other options (in 
particular those in set B or set A respectively) become manifestly false propositions. 
 The wiring of the subconscious mind in this sense will govern to a large degree the 
patterns of behaviour of an individual.

Naturally, we are taking this to be a phenomenon that our thoughts are thoroughly and 
continuously saturated with, rather than a discrete set of deliberations such as 
``hmmm, shall I pick up the pen or pencil?''. That is, many of our `choices' in this 
sense are simply the train of thoughts at the forefront of our mind that continually 
bubble up even when we are not \textit{trying} to think. Most of these thoughts are 
not directly accompanied by an external bodily action such as picking up an object or 
not. For example each process of `changing my mind', as described for the thought 
experiment shortly after the second set of bullet points above, is also a choice, even 
when not accompanied by a decisive external action.

 An analogy between our thought processes and the mathematical structures underlying 
G\"{o}del's theorem has been elucidated by other authors.  
 In the preface to reference~\cite{geb} (p.7) Hofstadter refers to elementary 
expositions involving a self-referencing loop leading to undecidable propositions, 
such as  $G$ considered above, as containing: 

\begin{quotation}
 
   \ldots only the most bare-bones strange loop, and it resides in a system whose 
complexity is pathetic, relative to that of an organic brain. Moreover, a formal 
system is static; it doesn't change or grow over time. A formal system does not live 
in a society of other formal systems, mirroring them inside itself, and being mirrored 
in turn inside its ``friends''. \ldots there is no counterpart to time, no counterpart 
to development, let alone to birth and death.  

\end{quotation}

    For Hofstadter, it is the self-referential and mirroring properties of the brain, 
giving rise to abstract structures similar to the `strange loops' encountered in 
demonstrating G\"{o}del's theorem, that is central to the emergence of an animated 
conscious `I' from the inanimate particles of matter of the brain. As suggested by 
Hofstadter, for the present theory also, a mathematical structure somewhat more 
complicated than that required to demonstrate G\"{o}del's theorem might be needed to 
account for these phenomena.

    In this paper, we consider that the possibility for such systems to change and 
grow is not only something that objectively takes place in time; but moreover it is 
the ordered nature implicit in such a series of potentially related states that 
describes 
  temporalisation itself.  
 It is the possible existence of this \textit{ordered} progression of systems which, 
through its simple structural isomorphism to an \textit{ordered} one-dimensional 
mathematical series (that can be mapped onto the real line $\rrr$),  itself 
corresponds to our immediate experience of time. The resolution of undecidable 
propositions from one system to another corresponds to the progression of choices we 
find ourselves making, with varying degrees of awareness. From the subjective point of 
view these choices are \textit{not} deterministic, in the usual sense of the word, 
since they are not something that \textit{happens in time}; rather they are the 
\textit{generators of temporality} itself.

   The progression depicted in figure~\ref{rindata} only has one direction. This 
underlies our experience of an apparent `arrow of time' which   
  corresponds simply to the one-way nature of this process (always with the 
possibility of losing knowledge of the world as our memory becomes frayed at the 
edges, it being supported by an imperfect physical device and following behind in the 
wake of our new experiences). 
 The phrase `arrow of time' is somewhat misleading since it implies the possibility of 
time having the opposite sense, that is flowing in the `other' direction, effectively 
  as if an empirical time parameter $t$ could be seen to be reversed with
 $t \to -t$. However this is not the case as the progression in figure~\ref{rindata} 
possesses only a single direction, which may be associated with $+t \inn \rrr$. The 
sequence of self-reflective states $R$, subjectively experienced as the flow of time, 
creates the inertia of the external world carrying physical objects. These objects 
include, for example, the components of a clock which can be used to measure the 
`time' $t$. The fact that we can imagine, or even construct, a physical clock to `run 
backwards', or for example watch scenes of a movie played backwards, creates the 
illusion of an alternative possible sense for time.
 However, we can only detect that a physical clock is running backwards since it 
operates \textit{relative} to the fundamental underlying ordered progression of time.

  The purely  mathematical structure of figure~\ref{rindata}, encapsulating 
   the  experience of a 1-dimensional temporal progression, can itself be encoded 
within the physical structures of a 4-dimensional spacetime world as depicted in 
figure~\ref{rinworld}.
 Here the structures in $M_4$ can be considered to represent a \textit{static} 
4-dimensional physical sculpture, as a manifestation of the static mathematical 
sculpture described for figure~\ref{rindata},
 within which a chain of states, having a one-to-one isomorphic correspondence with a 
self-reflective experience of a one-dimensional temporal flow, is embedded.  

\begin{figure}[htbp]  
\centering
\epsfxsize=8cm
\leavevmode
\epsffile[0 0 1141 1087]{\gpath aPfig144e}
\vspace{-10pt}
\caption{\setb A representation of the progression of the self-reflecting state  of 
figure~\ref{rindata} as translated into an extended physical environment $M_4$.}
\label{rinworld}
\end{figure}

  This origin of our experience of 1-dimensional time is analogous to the origin of 
our perception of 3-dimensional space. In general an abstract mathematical structure 
might be interpreted in several possible ways, whether geometrical or not.
  The arena for spatial perception arises out of a possible interpretation of the 
mathematical structure and symmetries of the multi-dimensional form $\lv$, and in 
particular the properties of the components $v^i$, with $i=1,2,3$, of the form 
  $L(\bv_4) = (v^0)^2-(v^1)^2-(v^2)^2-(v^3)^2$ with an $\mbox{SO}(3) \subset \soot$ 
symmetry, in terms of extended geometrical forms, as was described more generally in 
chapter~\ref{sym}. This incorporates  the perception of physical objects in an 
extended 3-dimensional space, as represented for example on the hypersurface planes in 
figure~\ref{twodworld}. 
 The structures of that figure, interpreted as a 4-dimensional spacetime, can be
superposed  on the manifold $M_4$ of figure~\ref{rinworld}, within which 
  more complicated mathematical structures also arise out of the full mathematical 
form $\lvh$ when projected onto the 4-dimensional base manifold. These further 
mathematical objects, described for example in terms of fields on $M_4$, 
incorporate series of self-referring elements which have a structural isomorphism  
with an experience of a directed 1-dimensional flow in time.

  For the world sketched in figure~\ref{rinworld} the mathematical structures in 
spacetime hence have both the necessary mathematical properties to give rise to 
perception of objects \textit{in space}, that is in a 3-dimensional geometrical volume 
(represented by 2-dimensional planes in figure~\ref{twodworld}), \textit{as well as} 
the experience of events \textit{in time}, in a direction geometrically `orthogonal' 
to the 3-dimensional spatial hypersurfaces on the manifold $M_4$, which possesses the 
local $\soot$ symmetry of the form $\lvfh$ of equation~\ref{lorform2}. Hence both 
spatial \textit{and} temporal forms of perception are encoded in the mathematical 
structures of the world. 
 That is, we consider not only that the structures obtained from $\hat{G}/\soot$ 
symmetry breaking for $\lvh$ projected over $M_4$ can be equivalent to the geometrical 
shapes we experience in space, but that they also incorporate self-referential 
mathematical structures which may be isomorphic to the self-reflecting progression 
that we experience subjectively as the flow of time.

  In turn this one-dimensional temporalisation itself provides the source of dynamical 
laws through the breaking of the multi-dimensional form of temporal flow $\lvh$ over 
the 4-dimensional spacetime, generating the physical laws on $M_4$ with which physical 
structures in general, and those depicted  in figure~\ref{rinworld} in particular, 
must be compatible. Hence the progression $R\to R' \to R''$ identified within the 
mathematical environment $E$ in figure~\ref{rindata} must be consistent with the 
seemingly \textit{inevitable} progression  $R\to R' \to R''$ of states in the physical 
spacetime environment $M_4$ in figure~\ref{rinworld} described as an apparent 
consequence of the laws of physics.

   The inertia of the physical world conforming to these laws of nature carries with 
it both the subconscious and conscious components of the brain and with them a 
`decision' already shaped in the former is swept into a new self-reflecting state of 
the latter, for which an `undecided state of mind' is now experienced as being 
resolved. More generally, information and data in the broader environment of $M_4$ in 
figure~\ref{rinworld}, as labelled by $E$ in figure~\ref{rindata}, which implicitly 
includes both the subconscious element and anything else distinguished from the 
self-reflecting $R$ state considered, can in principle contribute to the progression.

  An analogy can be made between the self-reflective system $R$ in the extended 
environment $M_4$ and a thermodynamic system $B$ embedded within the same larger 
environment on the manifold $M_4$.
While interactions between $R$ and the further structures in $M_4$ lead inevitably to 
the progression of the self-reflective state $R\to R' \to R''$, for example in terms 
an increase in subjective `information', the interaction between $B$ and the broader 
environment leads  inexorably 
 to an increase in the total entropy $S$. In fact
 \textit{both} the incorporation of a G\"{o}del statement $G_i$ into $R$,
  as depicted in 
 figure~\ref{rtordash}, as well as the case of increasing entropy might be considered 
as analogies for the phenomenon of temporalisation. The first example may also carry 
some elements above a mere metaphor, while for the second example an increase in 
entropy will accompany the physical process underlying the subjective experience of 
time. 
 
  For a sufficiently complex system such as a human brain the complete immersion of 
the self-reflective state within the wider environment might effectively generate a 
continuous temporalisation. Indeed, while for figures~\ref{rtordash}--\ref{rinworld} a 
series of discrete steps has been described, subjectively we generally 
\textit{experience} a continuous flow of time without any gaps or jumps. For example, 
while watching a ball roll along a table, essentially obeying Newton's first law of 
motion, we observe a smooth progression relative to our internal sense of temporality. 
It is this continuous subjective experience of the flow of time, as modelled by the 
one-dimensional real line $\rrr$, that forms the basic entity of the present theory.
 
  Indeed, although subjective experience in general exhibits a correlation with 
objective phenomena it is not explicitly described by the latter phenomena. For 
example, the sensation of `green' is associated with radiation from an interval of the 
electromagnetic spectrum with a wavelength of around 500~nm in physical interaction 
with the cells of the human eye and the resulting neural activity in the brain. 
However the subjective experience of `green' is \textit{not} explicitly contained in 
the description at any level of detail of these objective physical processes. 
Similarly here, the subjective experience of a \textit{continuous flow in time} is 
associated with the physical structures implied in figure~\ref{rinworld}. However it 
is not necessarily the case that a \textit{continuous} sequence needs to be identified 
in a physical system based on the progression $R\to R' \to R''$ in order for it to 
underlie a subjective experience of time which can be accurately modelled by the 
continuous real line $\rrr$.

  The irreversibility of conscious choices, the origin of the `arrow of time', is 
echoed in the irreversibility of many physical systems which are all governed by 
equations derived from the general mathematical form of progression in time. For 
example  
the second law of thermodynamics itself arising as a statistical \textit{consequence} 
of a progression of states, as alluded to in section~\ref{sectveu}.
 An essential difference is that while entropy increase is solely something which 
\textit{happens in time}, the physical progression $R\to R' \to R''$ of 
figure~\ref{rinworld} is directly correlated with a subjective experience of time 
which \textit{drives} the temporal flow itself. In addition to systems of classical 
physics, quantum phenomena are also subject to the underlying ordered flow of time 
which is infused into the base manifold $M_4$. 
 Calculations of probabilities and cross-sections for quantum processes depend on the 
accumulation of the possible field degeneracies conforming to a causal sequence in 
time,
 building upon time-ordered expressions as described for equations~\ref{degnow} and 
\ref{degnow4} and more generally in sections~\ref{secdopp} and \ref{seraps}.

  In addition to the fundamental temporal progression itself there are a large number 
of apparently one-dimensional quantities which may be constructed out of the physical 
structures on $M_4$. As well as the example of entropy $S$ these include the 
temperature $T$ of a body or even the `time' $t$ recorded by a mechanical clock. 
However each of these quantities correlates solely with the objective collective 
actions of molecular motions in spacetime and each defines a measurable property of 
the \textit{four-dimensional} world. Even the time $t$ recorded by the clock is 
\textit{not} a 1-dimensional geometric entity, but rather signifies a certain 
coincidence between the hands of the clock and the numerals on its dial in 
3-dimensional space.

  In fact no purely 1-dimensional phenomenon can be objectively inscribed within a 
4-dimensional spacetime without reference to the extended $M_4$ manifold or  physical 
processes within it. While the self-reflective physical structure in 
figure~\ref{rinworld} is similarly diffused in spacetime, objectively in terms of the 
firing of brain neurons for example, the subjective \textit{experience} of time is of 
a different character. Unlike a physical quantity such as $S$, the mental process of 
experiencing time is a \textit{purely} 1-dimensional phenomenon and in this subjective 
sense it is \textit{not} located within spacetime.
 The progression  $R\to R' \to R''$ is subjectively fused in mind into a purely 
1-dimensional experience of a qualitatively different nature to, and hence 
distinguished from, the 4-dimensional spacetime arena. 

  This one-dimensional structure \textit{is} the origin of time in the world, in the 
form of subjective temporalisation, and provides the foundation which underlies the 
general mathematical form of temporal flow $\lv$ and the physical laws in spacetime 
itself.
 It is through attempting to address the second loose end on the right-hand side of 
figure~\ref{loosends} that a source has been identified for the first loose end  on 
the
left-hand side.



\section{A Universal Foundation} 
\label{secauf}
 
   For a description of the universe in terms of a purely objective theory a 
4-dimensional background arena for events in spacetime, as for the case of general 
relativity, might be postulated as a fundamental entity or perhaps derived from a 
higher-dimensional spacetime.
 This is consistent with the observation that \textit{all} physical events in the 
world have \textit{both} a spatial \textit{and} a temporal  location in the universe.  
 For the present theory it is noted, however, that while we observe such events in 
spacetime our subjective \textit{experience} in the world is more fundamentally 
temporal than spatial.
  While many experiences are accompanied by a sense of both time and space, 
\textit{all} appear to exhibit a temporal aspect while some, such as the experience of 
listening to a piece of music or of simply thinking itself, lack any accompanying 
sense of an extended spatial arena. This observation, along with the simplification of 
founding a theory on one dimension rather than four, provided a source of motivation 
for the present theory.

 The sensation of time that accompanies all subjective experiences may be modelled 
mathematically by a continuous interval of the real line $\rrr$, which is precisely 
the same one-dimensional  structure of temporal flow considered objectively as a 
presence which underlies the structure of the entire universe. That is, innate within 
the expression for this temporal flow in the multi-dimensional form $\lv$ of 
equation~\ref{lv}, along with its symmetries, the form of the physical universe 
throughout an expanse of both time and space is supported.
 As discussed towards the end of the previous section, both temporal causality and a 
spatial geometry, deriving from the form $\lv$, are infused throughout the manifold 
$M_4$.

  At the mathematical level the unfolding of this structure is analogous to some 
degree to the properties of the Mandelbrot  set, in that a highly complex pattern is 
identified through a very simple mathematical expression. 
 A further analogy we consider here is  the simple differential equation 
 $\frac{\pal^2 y}{\pal x^2} + y = 0$ with the possible solution $y= \sin x$. This sine 
wave is typically represented as a graph in the 2D plane incorporating for example a 
horizontal axis for values of $-\pi \le x \le +\pi$. However innate in the expression 
$\frac{\pal^2 y}{\pal x^2} + y = 0$ the actual mathematical solution is of course 
present \textit{throughout} the infinite real line for  $-\infty < x < +\infty$, even 
though we typically only picture a small portion of this solution. Similarly while the 
extended `spatial' arena corresponding to the translation symmetry of $\lv$ is 
pictured for a finite region in figure~\ref{spillout} this purely mathematical 
structure is of infinite extent in all $n$ dimensions. This observation still applies 
when the construction of the spacetime arena is generalised for the geometry $\gmyv$, 
as one of many possible solutions involving differential equations in, and a 
degeneracy of,  the underlying fields. As for the sine wave above, this purely 
mathematical solution has no limit for the coordinate parameters on the base manifold 
(including in fact the  particular solution for $G_{\mu\nu}(x)$ of 
equation~\ref{geinawave} which is itself  described by a simple sine wave function as 
represented over an interval of $x^3$ in figure~\ref{gacos}), while as for the 
Mandelbrot set the structure which emerges in general may be highly complex.

  These mathematical patterns and structures  on $M_4$ arise through the projection of 
the full form $\lvh$ onto the base manifold and the associated symmetry breaking.
 Through an innate subjective interpretation of certain entities  on $M_4$
 there arises for us
a vivid impression of material phenomena which appear to be detached and hovering 
outside us in  an apparently spatial expanse. This entire \textit{perceived} world is 
however mathematically enfolded within the one-dimensional subjective temporal 
progression through which everything in the world is observed. (This description is 
very much influenced by the notion of the \textit{a priori} necessity for both 
temporal and spatial forms of experience, and their mutual relation, as elucidated by 
Immanuel Kant).   
 The observation that spatial structures through the form $\lv$ can be implicitly 
enfolded within the experienced one-dimensional flow of time that accompanies all of 
our perceptions in the world completes the initial motivation for the present theory 
described in the opening paragraph of this section.

   The form $\lv$ itself is derived from within the notion of a `moment of time', 
divided into infinitesimal intervals, as described in section~\ref{gfotf}.
  From the mathematical point of view the solution $\gmyv$ over the manifold $M_4$ is 
a structure implicit \textit{within} the full form $\lvh$ and its symmetries,  
   describing the geometry of an infinite expanse of 4-dimensional spacetime (as for 
example implied for figure~\ref{spminf}) which does \textit{not} `take time' to unfold 
across the cosmos, rather it underlies all cosmic structure, similarly as the solution 
$y= \sin x$ is implicit within the expression
  $\frac{\pal^2 y}{\pal x^2} + y = 0$ across the full extent of the $x$-axis.
This mathematical structure 
 logically precedes the laws of physics and the   
 properties of physical objects perceived in the world.
  As a manifestation of the underlying mathematical structures 
  these physical properties \textit{include} causal relations in general, 
    incorporating for example 
   the dynamic evolution of the fields, on $M_4$.  The causal and spatial 
relations between physical events unfolding in the world, which \textit{do} `take 
time', create the sense of a world \textit{outside} accommodating all of the apparent 
material phenomena. 
 All physical structures are subject to the laws of physics, which derive from the 
underlying mathematical forms, which apply for example to the phenomena depicted in 
figures~\ref{rindata} and \ref{rinworld} which in turn have both an objective and a 
subjective interpretation.

  The logical precedence of the elements of the theory described above is unpacked in 
the following sequence:

\begin{itemize}
\item[(1)] The objective starting point of the theory is one-dimensional progression 
in time with a mathematical structure isomorphic to an interval of the real line 
$\rrr$.
\item[(2)] From the basic arithmetic properties of $\rrr$ a general multi-dimensional 
flow in time subject to the constraint $\lv$ can be derived.
\item[(3)] The identification of extended geometrical structures from the form and 
symmetries of $\lv$ provides a basis for the necessary arena for perception, that is a 
subjective experience of a spatial expanse.
\item[(4)] In breaking the symmetry of the full form $\lvh$ over the $M_4$ base 
manifold
  the properties of  material phenomena are sculptured and made visible in conformity 
with 
   the resulting laws of physics.
\item[(5)]  The material objects in the world include the complex structures of 
physical devices, such as brains,  capable of performing mathematical operations and 
encoding a progression of states of self-reference governed by the  physical laws.  
\item[(6)]   The sequence of self-referential states, linked through a contiguous 
resolution of their associated `undecidable propositions', correlate with subjective 
thoughts and experiences, ever accompanied by the sense of an ordered flow in time.
\item[(7)]   The subjective temporalisation may be modelled by an interval of the real 
line $\rrr$ having a one-dimensional mathematical structure identical to that in item 
(1). 
\end{itemize}

   The first four points listed above form the main thrust of this paper
    from the opening chapters through to and including chapter~\ref{anpocs}, while the 
remainder of the above chain has been the topic of the present chapter.
 In this paper the self-reflective structures depicted in figure~\ref{rinworld} and  
discussed in the previous section are proposed as the means through which subjective 
experiences arise,
 although  this may be a vast  simplification, or even largely a metaphor, for the 
actual mechanism. In any case,
 the existence of a sequence such as that described in the latter four points above, 
beginning  with an empirically observed physical world and leading to self-reflective 
conscious experience in the world, is incontrovertible to the extent that it is  
evident that the presence of conscious beings is amongst the known phenomena of the 
physical world.
 This observation applies for \textit{any} physical theory, as discussed shortly after 
figure~\ref{loosends}, although the details of the theoretical mechanism that 
underlies the subjective thought process remains open to investigation.

  For any physical theory built upon essentially any postulated entities, such as 
fields or particles and a background of spacetime, the universe can be described in 
mathematical terms as a `static' 4-dimensional object, for example in the form of a 
spacetime diagram for the entire cosmos, which includes within it the full history of 
each human brain and all other material entities. However this is clearly not the way 
we \textit{see} the universe, rather the 4-dimensional spacetime structure of the 
brain must prescribe  our subjective perception of the universe as dynamically 
evolving through a progression in time.
  Hence for any such physical theory  the above segment of argument in points (5) and 
(6) can still be applied, however 
 there then remains dangling  the prominent loose end that there is no apparent 
justification for the origin and properties of the initially postulated physical 
entities themselves, other than that they may be contrived pragmatically, for example 
in terms of a Lagrangian function in spacetime, to match the empirical data from 
observations and experiments. 

  On the other hand the key observation for the present theory is that the final link, 
item (7) in the above chain, representing the fact that temporalisation is contained 
as an ever present feature of subjective experience in the world, reconnects the chain 
to the initial link of item (1) at the top. Hence not only is a mechanism for the 
origin of time conceivable, providing a foundation for the left-hand loose end of 
figure~\ref{loosends}, but this temporalisation itself arises through  self-reflecting 
structures, identified in the physical world itself, which account for our subjective 
experiences in general and the right-hand loose end  of figure~\ref{loosends}.  
The chain then naturally closes into the cycle depicted in figure~\ref{tcycle}.

\begin{figure}[htb]  
\centering
\epsfysize=10cm
\leavevmode
\epsffile[0 0 1591 1393]{\gpath aPfig145e}
\setlength{\unitlength}{25pt}
	    \begin{picture}(7.0,0.0)(0.0,0.0)
	\put(1.0,12.1){\Large $2$}
  {\large
	 \put(0.2,11.35){$\lv$}
	 \put(-0.19,10.65){mathematical}
	 \put(0.7,9.95){form}
	\put(5.8,12.1){\Large $3$}
	 \put(5.6,11.35){$M_4$}
	 \put(5.1,10.65){extended}
	 \put(5.0,9.95){spacetime}
	\put(-1.4,7.9){\Large $1$}
	 \put(-2.66,7.15){1-dimensional}
	 \put(-2.2,6.45){temporal}
	 \put(-1.65,5.75){flow}
	\put(8.2,7.9){\Large $4$}
	 \put(7.23,7.15){matter and}
	 \put(7.62,6.45){laws of}
	 \put(7.59,5.75){physics}
	\put(0.9,3.7){\Large $6$}
	 \put(-0.22,3.0){thoughts and}
	 \put(0.05,2.3){experience}
	 \put(0.4,1.6){of time}
	\put(5.8,3.7){\Large $5$}
	  \put(5.5,3){self-}
	 \put(4.8,2.3){referencing}
	 \put(4.63,1.6){structures $R$}
   }
	 \end{picture}
\vspace{-30pt}	
\caption{\setb A self-contained `time cycle' leading from the notion of progression in 
time, through the general mathematical form $\lv$ and perception of physical 
structures in spacetime, to self-reflective entities incorporating experience of a 
progression in time and hence completing the cycle.}  
\label{tcycle}
\end{figure} 
	
	It is a feature of the present theory that the two loose ends of 
figure~\ref{loosends} can be mutually tied up in this way. From a mathematical point 
of view each of the six stages in figure~\ref{tcycle} is contained within the previous 
stage, supplying a foundation for all of the structures of the theory. This system can 
then be considered to establish a `universal foundation' for the present theory.
 The entire system is self-supporting in the sense  that whenever we ask ``where does 
$X$ come from?'', where $X$ can be time, space, matter, conscious beings, or anything 
at all, the question can be answered in terms of something else \textit{within} the 
system.
	Without the need for any external foundation or assumptions and entire structure 
in figure~\ref{tcycle} hence detaches itself and floats free.

   This figure does not, of course, express an impossible cyclic chain of cause and 
effect relating the six stages in a temporal sense. Indeed time itself
  is contained as one link \textit{within} this cycle hence incorporating also the
  notion of 
 temporal causality \textit{within} this structure, and in particular for the physical 
laws in node (4).  Rather each  connection between neighbouring stages has the logical 
nature of  a structural isomorphism, more precisely in the sense that the properties 
of node $(i+1,\!\!\!\mod 6)$ are contained within the structure of node (i),
 with the net effect  of  expressing a self-contained and consistent mathematical and 
physical system. 
 While providing a chain of concepts for the benefit of deliberation the  six nodes of 
figure~\ref{tcycle} can be considered to collapse down to a single entity. This entity 
contains an entire universe created through the temporalisation represented by the 
structure in figure~\ref{rindata} which itself is entirely enveloped within the same 
physical world.

   In the opening of this section it was noted that while empirically everything 
happens in \textit{spacetime} from the subjective point of view  \textit{time} is a 
more fundamental mode of experience than space. This observation provides part of the 
original motivation for basing the present investigations on a general form of 
temporal flow $\lv$, together with its symmetries, rather than upon a 4-dimensional, 
or even higher-dimensional, spacetime structure. The further observation here that it 
is in the nature of time itself to provide the link connecting the two loose ends in 
the theoretical sciences described in the previous section and in 
figure~\ref{loosends} adds further circumstantial support for this approach.

 While the means of supporting spatial perception arises from a very direct 
interpretation of the geometric forms implicit in the mathematical properties of 
$\lv$, as described in chapter~\ref{sym} and corresponding to nodes $(2)\to (3)$ of 
figure~\ref{tcycle}, the means of generating temporal experience arises from the far 
more complex mathematical structures represented in figure~\ref{rindata}, as described 
in the previous section and corresponding to nodes $(5)\to (6)$ of 
figure~\ref{tcycle}.
 The figure as a whole can be seen as an interplay between 1-dimensional and 
multi-dimensional forms of time. The underlying mechanism for  obtaining an extended 
4-dimensional world out of 1-dimensional temporal flow, summarised in the upper half 
of figure~\ref{tcycle}, differs from the far more complex structures required to 
identify a purely 1-dimensional entity out of the 4-dimensional physical world, as 
summarised in the lower half of the figure. (As discussed at the end of the previous 
section a simple physical clock, for example, does \textit{not} possess any 
intrinsically 1-dimensional geometric structure). It is the very different nature of 
the mechanism for obtaining multi-dimensional forms and extended spacetime from 
1-dimensional time on the one hand and for identifying temporal flow itself out of the 
higher-dimensional and spacetime structures on the other hand, as required for 
subjective perception and experience, that opens up the non-trivial system of     
figure~\ref{tcycle}.

 The contrast between the \textit{objective} features of temporal progression,  
identified  as the \textit{simplest} element of figure~\ref{tcycle},  and the 
\textit{subjective} experience of   progression in time  arising out of the 
\textit{most complex} structures in this system, 
 while both aspects  of time, in nodes (1) and node (6) respectively, share    
 the identical structure of an interval of the real line, underlies the enigmatic 
quality of the concept of time itself. References to the seemingly more philosophical 
nature of time in the physics literature are rare but not entirely absent.
 Near the beginning of the introduction to his \textit{Space--Time--Matter} Hermann 
Weyl writes~(\cite{Weyl2} p.1):
   
\begin{quotation} 

       Since the human mind first wakened from slumber, and was allowed to give itself 
free rein, it has never ceased to feel the profoundly mysterious nature of 
time-consciousness, of the progression of the world in time, -- of Becoming. It is one 
of those ultimate metaphysical problems which philosophy has striven to elucidate and 
unravel at every stage of its history.
\end{quotation}

  While the upper half of figure~\ref{tcycle}, that is the chain of nodes (1)--(4), 
represents the development of the theory within the traditional scope of physics, the 
entire scheme, including the lower half of the figure, is fully incorporated within 
the sphere of scientific study more generally. Indeed experiments are performed, 
dating for example from those conducted by the neurologist Benjamin Libet in the early 
1980s, concerned with the relation between the physical brain and conscious actions, 
that is essentially nodes (5) and (6) respectively in figure~\ref{tcycle}. In such 
experiments physical cerebral activity is found to \textit{precede} a conscious 
awareness of intention typically by around 300 milliseconds or more.     
  
  For the present theory a conscious intention, or choice, is associated with the 
origin of temporalisation, as described for figures~\ref{rtordash} and \ref{rindata}. 
This leads to the   multi-dimensional form of temporal flow $\lvh$ through which 
derives the mathematical structure underlying the entire physical universe on $M_4$, 
incorporating its full eternal temporal extent both into the past and into the future. 
This physical universe includes in particular the brain state 300 milliseconds before 
the conscious choice was experienced, and indeed at any other time, as embedded within 
$M_4$ as represented in figure~\ref{rinworld}. Hence the overall scheme presented here 
is fully consistent with the experimental findings of Libet and others. More generally 
the full cycle of figure~\ref{tcycle}, including all of the nodes and links, is fully 
amenable to theoretical and scientific investigation.

 As a preliminary discussion
  the remarks  made here on the origin of our temporal experience
 and the phenomenon of consciousness, together with their  mutual association,
  are necessarily somewhat speculative. However, it is meaningful to formulate such 
questions, the worldview presented in this paper provides a new arena through which 
the construction of figure~\ref{tcycle} seems inevitable, and this provides a firm 
mathematical basis for a possible scientific enquiry into the nature of subjective 
phenomena compatible with the basic structure of the present theory.

 Most fields of scientific study are rooted in node (4) of figure~\ref{tcycle}, in 
that the natural starting point for any scientific investigation is observation of the 
physical world around us. For the physical sciences the general aim is to deduce the 
basis of the underlying structure of the world, extrapolating inwards as for example 
in the direction of nodes $(3) \to (2)$ for the present theory,  while on the other 
hand the biological sciences, for example, also study the world at face value and 
extrapolate outwards, which might include the properties of nodes $(5) \to (6)$ in 
figure~\ref{tcycle}. However, unlike the case for other scientific theories in 
general,  in the present theory it is natural to extrapolate one step further, both 
inwards and outwards,  to establish the final link in node (1) and hence complete the 
circuit.  Here the overall structure of figure~\ref{tcycle} then has the shape of 
providing an answer to the general question ``why is there something rather than 
nothing?'', rather than merely displacing it.

  From the mathematical perspective
 while  an exposition of the structures in figure~\ref{tcycle} could begin with any 
given node the simple mathematical structure of time, as a 1-dimensional progression 
modelled by the real line $\rrr$, provides a convenient entry point into the study of 
the whole system.
 In particular the unique properties of a real interval  provide an unambiguous 
structure upon which to develop the theory, 
as will be emphasised  later in this section. 
 If the properties of the real line $\rrr$ are considered to define the axioms, which 
in general underlie all the expressions which may be derived in a formal system, then 
the self-contained structure of figure~\ref{tcycle} might be thought of as a 
mathematical system which `contains its own axioms'. 

 From this point of view
 as a single entity of self-creation the time cycle in figure~\ref{tcycle} can be 
considered firstly as a purely mathematical structure which can be described in terms 
of the six nodes displayed with each one mathematically identical to, or contained 
within, the previous node of the chain. This picture can then be `coloured in' with 
both the objective material features of a physical world and the subjective 
experienced aspects of self-reflective thoughts and perceptions.

   The subjective experiences, as much as the objective material phenomena observed, 
are an irreducible component of this system.
Indeed it is the \textit{experience} of time, as well as of space, that generates 
necessary links in the time cycle of  figure~\ref{tcycle}. 
 Such a world cannot \textit{exist} unless it is \textit{experienced}. The two loose 
ends,  left exposed in many conceptual worldviews, relating to the origin of conscious 
experience and the origin of the material world are interwoven into one coherent 
system.
   Here the emphasis does not weigh heavily upon a pre-existing material content of 
any kind, but rather takes an overall more balanced view within which `matter' is 
identified with a form of experience shaped in `mind'.

The apparent distinction between mind and body arises in part since the spatially 
distributed matter we experience appears to exist \textit{out there}, however here the 
concepts of
`mind' and `matter' are intimately intertwined within one system.
 We have no need to postulate two wholly different kinds of substance and ponder how 
they interact, such as through the pineal gland in the brain in the worldview of 
Ren\'{e} Descartes. Rather mind and matter are different aspects of the same 
self-contained system: the conscious mind being bound with the structures of mental 
activity and temporality embedded in the physical world, while spatially extended 
matter itself is carved out of the multi-dimensional algebraic properties of time.
   Hence both sides of the philosophical dichotomy between mind and matter are
 accounted for and the points of view of both the idealist and the materialist 
democratically 
 amalgamated into this structure.
  We are not spirits haunting Earthly bodies, and neither are we machines in search of 
a soul.

	While forming components of one overarching framework both the objective structure 
of the physical world and the subjective forms of experience in the world can be 
described in terms of theoretical elements, and each is sufficiently distinctive and 
well defined to seemingly take on a `life of its own'. From the point of view of the 
present theory the materialist is grounded in node (4) of figure~\ref{tcycle} and can 
construct a relatively short, physically motivated, argument to account, via node (5), 
for the realm of the idealist in node (6). On the other hand the idealist, based upon 
the subjective experiences of node (6), is required to make a more lengthy detour, via 
the conceptual and mathematical structures of nodes (1), (2) and (3), in order to 
arrive at the materialist's realm in node (4). This asymmetry in 
 the apparent directness of 
 mutual explanatory power perhaps in part accounts for the predominance of the 
materialist, `a spade is a spade', philosophy that has underpinned most progress in 
the history of science, in addition to its practical utility.

	Outside the present chapter of this paper, as for the vast majority of work in 
theoretical physics in general, the focus has been with the study of a mathematical 
description or model of the physical \textit{material} world, here through equations 
such as $\lvh$ and $\gmyv$. 
	However \textit{mental} phenomena, such as our awareness of the physical world and 
decision making actions, are also very much a feature of the universe and in principle 
equally amenable to theoretical analysis, as discussed above.

   To recap, in the present theory the mathematical structure described in 
figure~\ref{rindata} models our conscious self-reflective state and ever present 
feeling of not knowing for sure quite what we shall do in the next moment. This 
perpetual uncertainty as to  our own thoughts or actions resolves momentarily in a 
choice `$G_i$' opening up a new horizon of uncertainty, as represented in the 
progression of figure~\ref{rtordash}. The self-reflective state is inexorably drawn 
through the series $\ldots R \to R' \to R''\ldots$ of figure~\ref{rindata} correlating 
with an internal experience of a sequence of thoughts, aspects of which have a complex  
mathematical representation, but in all cases associated with a subjective experience 
of a simple one-dimensional temporal flow.

  Within this structure the term \textit{freewill}, as used without hesitation in 
everyday language, is identified as this `experience of choice' as one feature of the 
overall system of figure~\ref{tcycle}. Everything that happens objectively in the 
physical world follows in the wake of this subjective temporalisation phenomenon. The 
historical philosophical debate concerning `freewill versus determinism'  becomes more 
strictly a question of `freewill versus the laws of physics' in the context of modern 
day science. The laws of physics include `indeterministic' quantum phenomena as a 
feature of the objective world which in the present theory  are \textit{not} 
correlated with the subjective act of making a conscious choice. Indeed the 
intrinsically random transitions of quantum effects are of a wholly      different 
nature to  rational decision making or freewill.
 On the other hand quantum properties are a major component of the laws of physics, 
and it is this full package of physical laws which determine all physical structures. 
These include the physical state of the brain which evolves in time according to the 
laws of physics, exhibiting properties which \textit{do} correlate with the 
interaction between the conscious and subconscious mind as implied in 
figure~\ref{rinworld} and hence providing the vehicle to carry self-reflective 
experiences.  

  The traditional philosophical difficulty in reconciling freewill and the laws of 
physics derives from the observation that an apparently independent objective world 
evolving according to a set of deterministic laws (together with random quantum 
phenomena) seems to leave no room for the notion of freewill.  However, here in the 
present theory, since the physical world is brought into being \textit{through} a 
subjective temporalisation sufficient breathing space opens up for the concept of 
freewill -- not as a secondary phenomenon on top of a given physical world, but as an 
irreducible feature in dynamic interplay with it, as summarised in the time cycle of 
figure~\ref{tcycle}.
   An element of the philosophical confusion concerning
 these issues arises as there is considerable ambiguity in the meaning of the  term 
`freewill' in itself. The present theory provides a context within which the notion of 
freewill might be more precisely defined. Within the system of figure~\ref{tcycle} the 
means by which the world is \textit{experienced} in mind is as important as the 
 \textit{empirical} forms of matter, with freewill being a property of the former 
while the laws of physics are a property of the latter. 

  It seems of course counter-intuitive to suggest that the great expanse and `weight' 
of the entire physical universe might be created through and carried in a single 
waking moment of thought. However, as described near the opening of this section, the 
mathematical structures underlying a solution for $\gmyv$ innate in the form of 
temporal flow $\lvh$ are perfectly `weightless' and infinitely delicate, effortlessly 
supporting an entire cosmic history throughout the full expanse of the physical
manifestation of the universe. If the laws of physics in this spacetime are compatible 
with the local evolution of a physical brain as depicted in figure~\ref{rinworld} 
which encodes the self-reflective sequence of figure~\ref{rindata} which in turn 
represents a subjective experience of a one-dimensional temporal progression 
isomorphic to the ordered real line in node~(1) of figure~\ref{tcycle} then the 
circuit closes and the experiencing being locates himself at a particular place in a 
particular world (in this chapter in this context pronouns such as `himself' or `his' 
refer  to a non-gender-specific being in any world). This spacetime location will be 
within the habitable epoch of the cosmological evolution as depicted in 
figure~\ref{spminf}(e),
 and  most likely upon a planet orbiting within the habitable zone of a suitable solar 
system as considered in section~\ref{secuni},  for the case of our own universe. The 
poets have more readily conceived of such a world, as for
example in the often quoted opening lines from William Blake's \textit{Auguries of 
Innocence} of 1803:

\begin{quotation}
\noindent
  \hspace*{3.3cm}   To see a world in a grain of sand \\
  \hspace*{3.3cm}   \mbox{  } And a heaven in a wild flower, \\
  \hspace*{3.3cm}   Hold infinity in the palm of your hand \\
  \hspace*{3.3cm}   \mbox{  } And eternity in an hour.
\end{quotation}

 Here, not limited by poetic licence, we require only a moment rather than an hour 
through which the entire universe may be perceived. The contention of the present 
theory sees the world and the heavens, together with an infinite expanse of space and 
an eternal temporal duration all held within a moment of time. The completion of this 
picture is depicted in figure~\ref{tcycle} with the experience of time itself 
contained within the structures of the physical universe.

While the entire physical universe is created through the experience of a single 
temporal moment, the moment itself is not unique. The circuit of figure~\ref{tcycle} 
can be closed by any one of a large number of possible local structures representing 
the progression of figure~\ref{rindata}, each  embedded within the physical world and 
each associated with a moment of experienced time.
 Indeed if the physical world is capable of supporting such a structure at all then in 
principle there may be many examples.
 This generalisation is depicted in figure~\ref{tcycles}.
\vspace{8pt}
\begin{figure}[htb]  
\centering
\epsfxsize=11.5cm
\leavevmode
\epsffile[0 0 1601 1387]{\gpath aPfig146e}
\setlength{\unitlength}{25pt}
	    \begin{picture}(7.0,0.0)(0.0,0.0)
	\put(1.0,12.1){\Large $2$}
	 \put(0.29,10.65){\Large \lv}
	\put(5.8,12.1){\Large $3$}
	 \put(5.3,10.65){\Large space}
	\put(-1.4,7.9){\Large $1$}
	 \put(-1.85,6.48){\Large time}
	\put(8.2,7.9){\Large $4$}
	 \put(7.6,6.48){\Large matter}
	 \put(-2.15,1.52){\Large $A_1$}
	 \put(-1.09,2.73){\Large $A_2$}
	 \put(0.18,3.77){\Large $A_3$}
	 \put(1.55,4.63){\Large $A_4$}
	 \put(3.05,5.3){\Large $A_i$}
	 \put(0.58,2.0){\Large $B$}
	\put(5.8,3.65){\Large $5$}
	 \put(5.2,2.25){\Large brains}
	 \end{picture}
\vspace{-27pt}
\caption{\setb Rather than the single experience of time represented by node (6) in 
figure~\ref{tcycle} in general a large number of physical structures on $M_4$ may 
generate moments of temporalisation, each represented by one of the small boxes 
labelled by $A$ or $B$ here and each of which completes the circuit of a time cycle 
for the same physical world of node (4).}
\label{tcycles}
\end{figure}

  In particular the set of temporal moments labelled by the series $A_1 \ldots A_i$ in 
figure~\ref{tcycles} might form a contiguous structure in the sense of the embedding 
of the progression of figure~\ref{rinworld} within the physical world on $M_4$. The 
corresponding temporal experience correlated with this structure is the sensation of a 
`sliding now' for a particular individual. Similarly while the `spark' that creates 
the universe from within can be any moment $A_i$ associated with such an individual, 
it could also belong to any other being, such as the temporal experience represented 
by $B$ in figure~\ref{tcycles}.
 In general the experiences of a community of beings $A,B,C\ldots$ may be inscribed 
within the same manifestation of a physical world, as depicted in figure~\ref{alltog}. 

\begin{figure}[htb]  
\centering
\epsfxsize=10cm
\leavevmode
\epsffile[0 0 1147 1083]{\gpath aPfig147e}
\caption{\setb A depiction of the trajectories of several self-reflecting `beings' $A, 
B$ and $C$ experiencing life in the $M_4$ spacetime manifold.}
\label{alltog}
\end{figure}

  Any structure of temporalisation, such as those represented by each `${\scriptstyle 
X}$' in figure~\ref{alltog} can take the place of node (6) in figure~\ref{tcycle} and 
complete the circuit which also incorporates the physical world itself  which a 
community of beings such as $A,B$ and $C$ cohabit. Each of these individuals observes 
a time-ordered progression of states of material entities, from stars and planets to 
tables and chairs and other individuals, distributed in a 4-dimensional spacetime 
as originally envisaged in figure~\ref{twodworld}.
 While a given observer $A$ experiences the subjective freewill of his own 
self-reflective state, and an internal temporalisation as represented in 
figure~\ref{rindata}, from his perspective 
both the subconscious as well as the conscious elements of the brains of the other 
beings $B$ and $C$ are unambiguously seen to partake seamlessly in the physical flow 
of events in the world. 
That is, the behaviour of the other, progressing in parallel and
   as represented for example in figure~\ref{rinworld}, conforms to the  basic laws of 
physics \textit{exactly} as any other physical entity such as the tables and chairs 
carried along in the inertia of the world. With a perfectly reciprocal account given 
from the internal subjective point of view of $B$ or $C$ the mutually consistent 
perspectives of all individuals dovetail together within the common physical world.

  In conformity with this symmetry between $A,B$ and $C$ in terms of a perspective on 
freewill and the laws of physics each observer carries a personal experienced 
fundamental time parameter $s$. This temporal flow $s$ is equivalent to the proper 
time $\tau$ recorded by a physical clock in the frame of the individual, as related by 
the constant factor $\gamma$ described for equation~\ref{taufroms}.  
 The progression in time $s \equiv \tau$ for any given individual is related to that 
associated with each of the other observers through the dilation effects of both 
special and general relativity in spacetime, as described for the `twins' $A$ and $B$ 
towards the end of section~\ref{fdandtd} and generalised near the opening of 
section~\ref{secpotnt}, again in a completely reciprocal manner.

  For any individual the seemingly vast potential arena for the flow of time in the 
universe at large contrasts sharply with the observation that we experience time at an 
apparently brief moment. The concept of `now' can be identified subjectively with the 
`present moment', which consists of a small duration rather than a point in time. 
 This leads us to pose the question -- given such a vast expanse of time -- ``why is 
it \textit{now}!?''; as opposed to, say, some time last week.
 This question is particularly challenging for theories of the world which posit an 
initial extensive and objective spacetime arena upon which the laws of physics are 
mathematically constructed from an independent perspective \textit{outside} spacetime, 
such that the physical laws governing  \textit{all} phenomena have perfect symmetry 
with respect to translation of location in either space or time. Within such a 
framework `now' is generally conceived objectively as a point in time, as a 
mathematical point of the real line. 
 While we have a wide choice over \textit{where} to make an observation
the fact that we necessarily observe the world as it \textit{is now}, at this 
\textit{particular} point in time,  appears to explicitly break the time translation 
symmetry.

 The problem disappears when we consider the meaning of `now' within the theory 
presented in this paper. 
 Our self-referencing awareness involves the physical structure of a small region of 
the world which is sufficiently complex to support `undecided states', but further 
complex structure carried in the physical world, in particular that of the 
subconscious brain, holds the resolution to such states and draws conscious awareness 
into the wider world in the process of temporalisation. Beyond the brain we find also 
the human body, the habitable environment and the entire physical world unfolding 
through the physical realisation of temporality creating a \textit{situation} in which 
the conscious being exists. Since every situation is an experience and every 
experience is an experience \textit{now} the logical meaning of the word `now' in this 
system is entirely redundant (although, of course, it has a practical purpose in 
everyday language). The fact that it is \textit{now}, rather than some time last week, 
is simply that I \textit{am this} experience, whereas the situation for a particular 
individual at a particular time last week \textit{is that} experience. The apparent 
problem is then largely an issue of the assumptions made in the use of language 
regarding the identity of an individual (the `I am') as something more attached to a 
bodily form than to an experience.

   The fact that it \textit{feels} like `now' comes from the fact that the world 
exists `all at once' -- that we can conceive of a past and future progression within 
which we place ourselves in the present, \textit{now}. However, \textit{past} and 
\textit{future} are \textit{not} periods of a pre-existing external and independent 
world-time; rather the past and future refer to locations within the universe with 
respect to the perceiving being whom experiences the situation -- it is a description 
of the experience which partitions a self-reflecting conscious state into a concrete 
\textit{past} and an uncertain \textit{future} as a necessary structural form of a 
thinking being.  
(This aspect of the worldview being described here is philosophically close to the 
standpoint of existentialism, and is influenced in part by the philosophy of Jean-Paul 
Sartre).

  I have to experience the world \textit{now} in a similar way that I also find myself 
\textit{here} at a particular spatial location in the world. While the `body' of the 
whole world is created through the structure of our being, \textit{here} is where my 
eyes, and other sense organs of the human body, locate me spatially relative to other 
physical objects in the world. To necessarily exist \textit{here} and \textit{now} is 
simply the statement of having to be the centre of reference for an experience in a 
world. This central vantage-point is 
 essentially 
the location of the physical manifestation of the associated thought processes, as 
represented in figure~\ref{rinworld}, within the extended spacetime manifold $M_4$.

  From the perspective of any individual such as $A$ the universe created through any 
given  experienced moment, such as $A_3$ in figure~\ref{tcycles}, not only mutually 
supports the contiguous moments of the `sliding now' and into the span of the current 
day, but also the moments of yesterday, and the past in general, and those of 
tomorrow, and the future in general. As well as the spacetime separation between 
moments experienced by $A$ and $B$ each individual is also separated from himself in 
time, corresponding to the moments marked `${\scriptstyle X}$' on the trajectory of 
$A$ in figure~\ref{alltog} for example. The identical universes generated from $A$'s 
experienced moments on Monday, Tuesday, Wednesday and so on resonate together into a 
single life history. Any moment of temporalisation not only brings the corresponding 
present self into being but also the physical structure for all the past and future 
`selves' in the life history of the same individual.
   This system is hence comfortably compatible with the experimental findings of Libet 
and others as noted in the discussion following figure~\ref{tcycle}. 
 The mutual  relations between any of $\ldots A_1, A_2 \ldots A_i \ldots$ from a 
single life history dovetail together, as with the moments of any other beings, such 
as $B$ and $C$, in the same physical world.

 The exhaustive spacetime coverage of the universe created through each temporalised 
moment $A_i$ together with the inertia of the derived material processes in the world 
maintains the physical manifestation of any individual during non-waking hours or 
through different shades of consciousness. In this way historically separated waking 
moments are seamlessly stitched together over periods of years alongside those of 
other beings immersed in the same world.

  The existence of different shades of consciousness, such as the experience of 
dreams, suggests that a rigid geometric framework in space may not be essential for 
some forms of perception, although dream sequences are closely associated with waking 
experiences. The question regards whether spatial perception is required in some form 
in order to complete the circuit of the time cycle in figure~\ref{tcycle}. As 
discussed in section~\ref{tlssotu} and section~\ref{secpotnt} (before the bullet 
points) our \textit{a piori} imposition of an extended 3-dimensional frame for our 
perceptions in the world does \textit{not} perfectly match the non-Euclidean geometry 
of the world -- which we however effectively interpret as being flat while certain 
phenomena are ascribed to an apparent force of gravity. We very rarely perceive solely 
events within a local inertial reference frame, such as within an orbiting spacecraft, 
however such an idealised limiting geometry is not required in order for us to be able 
to interpret and organise our perceptions of the world in a manner compatible with the 
presumption of a flat Euclidean frame of reference.

  In addition to providing a spatial orientation for vivid conscious experiences of 
the world, with material objects obeying physical laws of motion within the perceptual 
framework, the general laws of physics themselves, which shape all material 
properties, arise from the projection of the full form of temporal flow $\lvh$ onto 
the base manifold $M_4$ . Complex mathematical and physical structures which arise in 
this breaking of 
$\lvh$ and its full symmetry over $M_4$ accommodate the mechanism for self-reflective 
conscious experience itself, as described for figures~\ref{rindata} and 
\ref{rinworld}.
The physical laws deriving from the symmetry breaking  hence not only maintain 
objective material objects in the world but also \textit{images} of the same objects 
which can be maintained in our subjective thoughts even while the object is not being 
directly perceived (as for example in a dream or in a waking moment in which we simply 
look away from the physical object while still thinking about it).

 Hence the laws of physics derived from the symmetry breaking of $\lvh$ over
  the base manifold $M_4$  give rise to both the structure of conscious 
self-reflective states and the material phenomena, perceived against the $M_4$ 
background, which constitute objects \textit{of} consciousness.
 This then describes the primary requirement of the symmetry breaking of $\lvh$ in 
order to complete the time cycle in figure~\ref{tcycle}, that is to open up structures 
that may be presented as objects of conscious experience together with the 
self-reflective elements capable of contemplating such objects. In our world these 
structures are obtained through the
projection of $\lvh$ over a locally approximately flat 4-dimensional spacetime $M_4$ 
which incorporates a 3-dimensional spatial arena for the perception of objects.

   In principle we can enquire what it might be like to be immersed in a highly curved 
spacetime environment of a different world. Without the support of an effectively 
Euclidean spatial orientation it would be harder to organise our incoming sensory data 
and difficult to predict the physical consequences of our own actions and to engage in 
such a world generally. The likelihood of errors of judgement in this respect is much 
lower in the local environment of an apparently flat spacetime combined with the very 
regular patterns of motion deriving from  Newtonian gravity, as we encounter in our 
own world.

  It seems very natural to us that space \textit{ought} to have Euclidean properties, 
as witnessed by the historical perseverance of the geometrical laws of Euclid 
formulated in ancient Greece, which until the early $20^{\mathrm th}$ century were 
assumed to describe the real world. While applying to an excellent approximation in 
the local environment of the Earth and solar system, the assumption of a flat 
spacetime geometry breaks down for large scale cosmological structures. In general any 
manifold with two or more dimensions can exhibit arbitrarily large curvature at any 
point, as is the case  for our 4-dimensional universe for which the curvature diverges 
in the proximity of the initial singularity or a black hole.
 However the curvature of any 1-dimensional manifold is trivially zero and the 
geometry necessarily `Euclidean'. Hence an interval of the \mbox{1-dimensional} real 
line $\rrr$, as a unique and robust structure, and as a parametrisation of the 
subjective experience of temporal flow, provides an unambiguous starting point from 
which the present theory has been developed in this paper.

 This discussion raises the questions considered in section~\ref{secuni} regarding 
whether or not 
  the  symmetry of $\lvh$  is uniquely required to be broken over a  4-dimensional 
spacetime  $M_4$ and whether structures identified in the symmetry breaking are 
required to be  compatible with the notion of perception as conceived in our world. 
  Whether a complete time cycle of the kind in figure~\ref{tcycle} incorporating 
self-reflecting beings \textit{without} an \textit{a priori} spatial perception of any 
form could exist, or even whether there are conscious organisms within our own world 
that completely lack any spatial awareness, may be difficult 
 questions to address. Such self-reflecting creatures may still necessarily require an 
$M_4$ base space to break the full $\lvh$ form in order to physically exist (as do all 
non-sentient
 biological life forms in our world), yet without employing a subjective spatial 
interpretation of the
  3-dimensional structures on $M_4$. In a similar way \textit{we} require the extra 
dimensions of the form $\lvh$ in order to physically exist ourselves, yet without our 
being directly aware of them.
 
  Here we recall that the term `perception' is being employed \textit{not} just in the 
narrow sense of that which we are visually aware of in the moment. It refers more 
generally to an organising faculty for \textit{all} the data about the world that 
enters  and our thoughts through all of our senses. This data is accumulated both 
directly, for example through the experience of vision or touch, as well as 
indirectly, for example via intermediate objects,  tools of experimentation or the 
accounts of other people. This data concerns aspects of the world  in spacetime 
ranging from our immediate locality, down to the minute microscopic scales explored in 
HEP experiments, out to regions very remote from us and through to the limit of 
observations relating to the structure and evolution of the cosmos. Perception is a 
form of knowledge that encompasses everything we can understand about the world in 
space and time, in principle anything associated with nodes (3), (4) and (5) in 
figure~\ref{tcycle}.

  Moulded by this form of perception 
 physical structures as we experience them exhibit the effortless complexity inherent 
in the breaking of $\lvh$ over the infinite expanse of $M_4$ as depicted for example 
in figure~\ref{spminf}. As alluded to in the opening of this section such a structure 
is analogous to the endless delicacy of the Mandelbrot set, which arises from the 
iteration of a simple mathematical expression in the complex plane. In both cases an 
inexhaustible variety of fine detail can be observed wherever we choose to `zoom in' 
and examine for example biological forms in the physical world or the spiralling 
patterns of the Mandelbrot set. 
 At the shortest physical distances probed the properties of elementary particles 
emerge over an underlying fractal-like structure of field solutions, as described in 
section~\ref{seraps}, while
at the other end of the scale, throughout the expanse of the observable universe, 
 we perceive the manifestation of the laws of physics in the swirling patterns of 
galaxies and galactic clusters.  The observations of cosmology, on this largest scale, 
are contained within the physical world of node (4) in figure~\ref{tcycle}, which in 
turn provides a context for understanding the `cause' of the Big Bang and the origin 
of the universe more generally within the `system of the world' presented in this 
paper.

\section{A Context for Cosmology}
\label{secacfc}

  The big picture for the present theory, as represented by the time cycle of 
figure~\ref{tcycle} which  sees the conscious observer in a dynamic interplay with the 
entire physical universe as an irreducible, integral component of the world, offers a 
very different perspective to the `Copernican view', which sees humanity playing a far 
less significant role in the cosmos.
 The fact that the physical manifestation of humanity represents a tiny contribution 
to the total matter content of the Earth, which itself is in orbit around a far more 
massive sun, which in turn is one of countless stars distributed through the galactic 
structures of the universe all serves to cement the Copernican worldview concerning  
our apparent insignificance in the grand scheme of things. This is a misconception of 
the nature of the cosmos from the point of view of the present theory.
 On the other hand here there are potentially a vast number of subjective experiences 
which may complete the circuit for any physical universe, as described for 
figure~\ref{tcycles}, and each objective physical universe is one of a potentially 
vast number solutions of the form $\gmyv$ capable of supporting self-reflective 
temporalising beings.

  In any case the `cosmological principle', as described in section~\ref{sectsmoc}, is 
valid for our universe in being sufficiently consistent with empirical observations to 
provide a valuable aid in finding solutions for the spacetime geometry on the largest 
scales observable.
 Such an entire solution for a physical universe, represented in a spacetime of  
unlimited 4-dimensional extent as depicted in figure~\ref{spminf} and with a geometry 
expressed as $\gmyv$ in the present theory, is created as  
 a mathematical possibility \textit{within} the system of figure~\ref{tcycle}.
 The nature of this geometric solution is very much in the spirit of general 
relativity for which a spacetime geometry such as the Schwarzschild solution of 
equation~\ref{ttrtp}, although typically employed to determine a planetary orbit about 
a star, represents an infinite 4-dimensional spacetime.

  Here the possibility of the overall mathematical solution represented in 
figure~\ref{tcycle} is the \textit{reason why} the universe exists. Our local 
perspective of observing the flow of cause and effect in the everyday physical world 
leads by analogy to the presumption that the universe itself must have been created 
either by an event in time or by an event coinciding with the beginning of time. For 
any creation event in time the question then ever remains regarding the cause of 
\textit{that} event while for a creation event at $t=0$ the nature of an event without 
an apparent cause is certainly no less problematic from a conceptual point of view. In 
either case
 there are an array of further conceptual difficulties regarding the origin our own 
universe, such as the `start-up problem', as described towards the end of 
section~\ref{secinf}.

  In the present theory the creation of the universe is not something that `happens' 
in the Big Bang, or temporally before it, rather the very early universe and the Big 
Bang correspond to a  certain region of the  spacetime geometry  at a particular epoch 
of the full 4-dimensional solution. This early epoch is beyond the horizon of our 
direct experience but its existence depends upon the self-reflective temporalising 
experience that arises in the history of the universe, as does everything in the 
cosmos.
 All the  physical structure and conditions of the universe,  including that for all 
future as well as past epochs and throughout   the vast spatial expanse both within 
and beyond  our observational reach at any epoch, are brought into being through the 
nature of a temporalising entity, which in turn is supported within the physical 
world, as depicted here in figure~\ref{cosevolve}.

\begin{figure}[htbp]  
\centering
\epsfxsize=\maxwidth
\leavevmode
\epsffile[0 0 1935 2566]{\gpath aPfig148e}
\caption{\setb The physical universe contains its own means of creation as perceived 
through the window of an interval of pure time $\Delta s$ subjectively experienced by 
the observer within the world. This picture is in contrast to standard cosmology for 
which the observable universe evolves from a vanishingly small spatial extent 
$a(t)\Delta \Sigma$  at $t=0$. Here the scenario described for figure~\ref{earlyu}(b) 
has been depicted.}
\label{cosevolve}
\end{figure}

  All experience in general is played out through a moment in time, including our 
perception of the physical world, with the structure of the universe being 
mathematically described by a solution for $\gmyv$ as conforming to the full form of 
temporal flow $\lvh$. Hence here the structure of the  entire universe is derived 
mathematically through a moment of time, 
 typically conceived as a duration of order one second as
represented by a small one-dimensional interval $\Delta s$, as the window through 
which it is perceived for example by the observer in the centre of 
figure~\ref{cosevolve}. This contrasts with the standard cosmological models for which 
the entire observable universe evolves physically from a vanishingly small 
3-dimensional spacelike hypersurface of size $a(t) \Delta \Sigma \to 0$ for $t\to 0$, 
as described for figures~\ref{confbb} and \ref{confbba} in section~\ref{secinf}, 
corresponding to the point at the base of figure~\ref{cosevolve}.

  For the standard approach \textit{all} of the field content, particle properties and 
physical laws in general need to be \textit{added} onto the spacetime in order to 
determine the evolution of the universe from the initial spacelike state, which is 
presumed to exhibit suitable initial conditions.
 However for the present theory all of the fields and physical laws \textit{derive} 
from the structure and symmetries of $\lvh$ through the necessary projection onto 
$M_4$ in framing our perception of the world, including the Standard Model particle 
properties as identified in chapters~\ref{chapesb} and \ref{secfd}. Here the apparent 
`initial conditions' for $t\to 0$ 
 represent a particular region of the full spacetime solution as required in order 
that the overall solution contains a habitable epoch such that the circuit in 
figure~\ref{tcycle} closes, and hence in principle the initial conditions might appear 
improbable from the standard perspective.

  Indeed, as described in the previous two sections, developing the theory from an 
interval of one-dimensional temporal flow $\Delta s$  brings with it the possibility 
of constructing a universal foundation as represented in figure~\ref{tcycle}, while 
beginning with a spacelike 
 hypersurface $\Delta \Sigma$ at $t=0$ leaves questions open concerning not only the 
source of physical laws and the nature of the initial conditions but also the origin 
of spacetime itself. Further the existence of the temporal moment is evident, in fact 
we are perhaps more intimately familiar with our experience of it than of anything 
else in the world, while the hypothetical initial spacelike state of the universe is 
an extremely remote entity.  
 Hence overall, the notion of the present theory that everything is `perceived through 
a moment of time' is perhaps not less reasonable than the standard picture for which 
everything `evolves from a point of space'.

 The system constructed in figure~\ref{tcycle}, for which figure~\ref{cosevolve} 
represents a particular manifestation such as our own world,  can be considered as 
being  centred fundamentally upon addressing the question of how it is possible to 
have subjective experiences of a world. As described in the previous section such 
experiences always take place \textit{here} and \textit{now} in the world, with 
everything else we can say about the universe, whether at some distance in space or 
extrapolated through time into the future \textit{or} the past, necessarily consistent 
with the fact that we experience the world here in the present moment. The environment 
we experience in the present incorporates, amongst other things, observations based on 
a geometrical spacetime manifold; in particular we are \textit{able} to perceive a 
world \textit{since} it is cast against an approximately flat spatial background. 
However, there is no reason to expect the mathematical preservation of such an 
approximately flat pseudo-Euclidean spacetime indefinitely into the past as we 
extrapolate beyond the horizon of our direct physical experience of the world. 
  The geometry of the very early universe for example, being beyond our immediate 
perception, with a potentially extreme spacetime curvature, is not required to be 
compatible with our $\textit{a priori}$ imposition of a flat framework of space and 
time within which to organise our impressions of the world and plot our actions within 
it.

 Hence neither an approximation to spatial flatness nor any other constraint on the 3 
or 4-dimensional geometry is required for the early universe regions of 
figures~\ref{earlyu} or \ref{cosevolve}, in particular in approaching $t\to 0$. In 
fact at  earlier times there remains no requirement for the identification of a 
3-dimensional spatial or 4-dimensional spacetime manifold structure of any kind, as is 
the case for the scenario depicted in figure~\ref{earlyu}(a). However while the 
identification of the manifold $M_4$ itself, together with the projection $\bv_4 \inn 
\TM_4$, may break down at an epoch before the Big Bang the general form of temporal 
flow $\lvh$ remains ever valid for any value of the fundamental temporal parameter, 
even for $s \to -\infty$ as described for figure~\ref{earlyu}(a) in 
section~\ref{sectveu}.

  From this point of view while the universe can be considered to be infinitely old, 
in terms of the value $s \to -\infty$, the cosmic time $t=0$ can be considered to be 
the point in time at which a 4-dimensional spacetime manifold $M_4$ unfolds from the 
form $\lvh$, as depicted in figure~\ref{earlyu}(a). The familiar laws of physics in 
4-dimensional spacetime, including  the second law of thermodynamics expressed in 
terms of the degrees of freedom of Standard Model particle states and the 
gravitational field, may first be collectively applied as they emerge from the Big 
Bang at $t=t_v$; for either the
 scenario of figure~\ref{earlyu}(a) or (b) as also described in section~\ref{sectveu}.
  While we do not directly interact with the very early universe we are intimately 
connected with it not only through observations in cosmology but also through the need 
for the conditions of both stellar and biological evolution to arise and be consistent 
with the support of self-reflective beings at the present epoch.

 In order to achieve this in addition to the microscopic field and particle 
interactions underlying the macroscopic gravitational structure $\gmyv$, as 
empirically observed in the high energy physics laboratory and the cosmos 
respectively, at an intermediate scale the laws of physics implicit in this solution 
must necessarily be compatible with the development of the structures of molecular 
biology, such as DNA, which underpin the evolution of life.
 The complex biological structures implicit in figure~\ref{rinworld}, correlated with 
the subjective experience of temporalisation, must themselves arise in the material 
dynamics of the universe in a manner consistent with the laws of physics in 
4-dimensional spacetime. That is, the physical universe we observe must support not 
only the formation and history of the solar system but also the evolution of 
biological life on Earth and the birth and development of specific self-reflective 
beings as manifested  in human form and as represented in the centre of 
figure~\ref{cosevolve} for our world.

  It could be asked: if the whole universe is brought into being through an experience 
of it \textit{here} and \textit{now}, why does it appear that biological evolution, 
leading \textit{up to} the human race was necessary? Why not have readily formed 
humans along with the Earth and our local environment \textit{suddenly} appearing, 
along with the
 identification of the $M_4$ manifold itself dating from the `cosmic time' $t=0$
 just a few centuries or even a few minutes ago? However, the full extent of our 
spacetime world, including everything causally related to us from the past, must 
conform to the
 form of our  perception in spacetime through the breaking of $\lvh$ over $M_4$ and 
the 
 consequential  laws of physics as implicit in the solution $\gmyv$. The flow of the 
world in our past and into the future must obey these laws \textit{and} also be 
consistent with our biological form as observers in the present.

 Such an overall solution might be much more likely achieved through a very simple 
initial state followed by a prolonged cosmic and biological evolution as shaped by the 
laws of physics, rather than the apparently more direct route via a highly improbable 
`initial state', in the form for example of a  `snapshot' of the universe taken a few 
minutes ago, which may in any case be prohibited through contradiction 
 with the necessary laws of physics.
  This would still be the case even if the `snapshot' only met the minimal requirement 
of preserving the complex form of the local environment, in which case the large scale 
cosmos would also most probably look very different to our universe. 
  At the other extreme  the universe, as an extended spacetime manifold, may not have 
a temporal origin at all in the sense that arbitrarily early times with fundamental 
time parameter $s \to -\infty$ might be contained within $M_4$. This is the seemingly 
more natural
    scenario depicted in figure~\ref{earlyu}(b) for which the `initial state' 
corresponds to the asymptotic conditions as $s \to -\infty$  and $t \to 0$, as also 
described for figure~\ref{svrst}.

  On the other hand the conditions in the universe observable today, even neglecting 
the consequences of the cosmic expansion, could not have prevailed indefinitely into 
the past.
 The laws of physics, in particular the second law of thermodynamics, demonstrate that 
it is not possible to sustain an everlasting immortal species on the Earth, and itself 
implies a necessarily finite lapse of time into the past to an apparent origin for our 
physical universe, which is also consistent with the observed expansion of the 
universe. Hence human life forms must have been moulded out of the state of the 
physical world at an apparent temporal origin of the 4-dimensional universe, that is 
the time at which the familiar laws of physics were established,  culminating in a 
physical evolutionary process which in our case involves the processes of genetic 
mutations and natural selection. This apparent temporal origin
 must itself have an explanation in terms of the overall theory, and is here 
associated with the phase transition at the end of the Big Bang, that is at $t=t_v$ in 
the scenario of figure~\ref{earlyu}(a) or (b) and as also indicated in 
figure~\ref{cosevolve}.

   That `there was evolution' is a statement from our perspective within the universe, 
which itself can be considered from an \textit{outside} perspective as an `atemporal' 
static 4-dimensional  entity, about the world as a whole and the structure it must 
have for us to exist \textit{here} and \textit{now} in 4-dimensional spacetime.  To 
`visualise' the whole universe it is convenient to return to the three-dimensional 
spacetime analogy and combine the content of figures~\ref{rinworld}, \ref{alltog} and 
\ref{cosevolve}. Through the circuit of figures~\ref{tcycle} and \ref{tcycles}
 life draws itself into being out of the `mathematical vacuum'. While the laws of 
nature on our spacetime manifold are carved out of the general form of progression in 
time, the actual physical forms we encounter in the universe, whether in our present 
or uncovered from our past, are moulded to conform with the possibility of our own 
bodily existence and conscious experience within it.

  Many features of the world that we observe, such as our existence within a community 
of beings (the experiences of whom mutually dovetail together as described for 
figure~\ref{alltog}) rather than finding ourselves in isolation, are the way they are 
since the world in which we find ourselves situated must accommodate a physical 
sequence of events, including for example an evolutionary and social history, leading 
up to the form of each individual experience.

 All matter of the universe is brought into existence through our experience and 
perception of it as being mathematically, and hence physically, connected to the 
necessity that the experience itself exists. Hence all of our body organs, blood 
vessels and so on, as well as the human brain, necessarily come into being through the 
mechanisms and processes that give rise to life, in terms of its physical parts, along 
with the entire biological world, through the logical and rational requirement that we 
must be physically sustained within the world which we experience.  
  The seemingly great improbability of life in terms of the complexity of biological 
structures such as sensory organs and the nervous system is essentially irrelevant. If 
such a biological system is physically \textit{possible} at all and represents a 
self-reflective temporalising structure within the mathematical system of 
figure~\ref{tcycle} then it will draw itself into being and \textit{exist}  as the 
realisation of an underlying mathematical necessity.

 The constraints on the form of such a mathematical solution will be all the more 
stringent if there are essentially no free parameters in the breaking of the full form 
of temporal flow $\lvh$ over a base manifold $M_n$ (for an $n$-dimensional world 
solution). However, as for any mathematical problem, whether or not a solution 
actually exists does \textit{not} depend upon the apparent difficulty of the problem. 
Whether or not a degree of tuning is possible for the symmetry breaking parameters 
(and whether or not $n=4$), as considered in section~\ref{secuni}, and regardless of 
the extent to which the apparent `initial conditions' of the universe might be 
constrained, life will find a way if any solution for the structure in 
figure~\ref{tcycle} exists, no matter how difficult or how remote the possibility of 
such a solution might seem to us.
 
  Conscious life draws itself into being, through a self-supporting system, within the 
constraints of the mathematical form of the physical world it engages with. 
 This is not necessarily a straightforward feat to achieve, in the sense of the 
non-trivial mathematical and physical structures required.
 Indeed, the fact that our ability to physically experience the world relies on the 
support of a human body which is enormously complex on the scale of the fundamental 
laws of physics (gravitational and quantum particle) is itself evidence of the 
difficulties of embedding the physical manifestation of a conscious life within a 
physical world constructed within the constraints imposed by the underlying 
mathematical progression in time experienced by the conscious beings themselves.

  As a solution to the cycle of figure~\ref{tcycle} the physical world of node (4) can 
be described in the mathematical terms of the full 4-dimensional spacetime structure 
$\gmyv$ as moulded in conformity with the simple state of a moment of time in node 
(1). Alternatively the structure of the universe can be described effectively in the 
physical terms of a dynamic evolution from an apparently initial state at $t=0$, as 
discussed for figure~\ref{cosevolve}.  
 Expressed in this latter way the physical development of the organic form of 
conscious beings out of a comparatively far simpler physical state at the apparent 
temporal origin of the world, 
  according to precisely determined laws of physics,
  not surprisingly requires a relatively long period of biological evolution on a 
planet such as the Earth in a stable orbit around a star such as the sun. Hence the 
fact that we find ourselves in a universe at a spacetime location such that the sun is 
149 million kilometres  away in space and the Big Bang is 13.8 billion years away from 
us in our temporal past, as depicted in figure~\ref{cosevolve}, have similar 
explanations: both are required of our physical environment in order that we, 
conscious beings, can consistently exist \textit{here} and \textit{now}. The observed 
vastness of the cosmos that surrounds us beyond the solar system is in some sense a 
byproduct arising from the non-triviality of realising a solution for  the overall 
structure in figure~\ref{tcycle}, albeit a byproduct
 which is entirely `weightless' from a  mathematical point view as described towards 
the opening of section~\ref{secauf}.

  Analogous observations would apply to worlds other than our own, drawn into 
existence as a solution for the general form of figure~\ref{tcycle}, insomuch as it 
would seem surprising for a `simple' solution to exist.
 The question concerning the uniqueness of our world, as considered in 
section~\ref{secuni}, requires consideration of other worlds that could be created by 
and through other self-reflective beings.
 Since the 1950s philosophers in this world have sometimes enquired ``what is it like 
to be a bat?'', which is very difficult to answer since, amongst other things, bats 
and humans have different forms of sense perception. This kind of question becomes yet 
much harder if we attempt to enquire ``what is it like to experience a different 
possible world to our own?'' Here we refer to a different world with different laws of 
physics and perhaps even a base manifold  with an intrinsically highly non-Euclidean 
geometry or a different dimensionality to ours.

   All the varieties of other possible worlds with different laws of physics still 
have significant features in common, assuming they fall within the general framework 
described in this paper, involving  a multi-dimensional form of temporal flow.
 The full form $\lvh$ may in a strict sense represent the greatest possible dissolving 
of the temporal flow via an infinite dimensional channelling through $\bvh \inn 
\rrr^{\infty}$,
 as alluded to towards the end of section~\ref{secuni},
 and hence be unique for \textit{all} worlds.
 Symmetries such as $\esi$ acting on the form $\lvt$ with $\bv_{27} \equiv \mcX \inn 
\htho$ or
   $\ese$ acting on the form $\lvfs$ with $ \bv_{56} \equiv x \inn F(\htho)$   may 
also represent  significant mathematical resonances which dominate the actual physics 
observed in any  universe. 
The physical laws themselves are then effectively determined in breaking the full 
symmetry, for example through the identification of a subgroup acting on the subspace 
of vectors $\bv_n \subset \bvh$  projected onto the tangent space of an 
$n$-dimensional base manifold $M_n$. 
This smaller space, together with the symmetry group for $L(\v_n)$, is broken out of 
the larger space and symmetry group of $\lvh$ in the formation of a global background 
manifold which acts as a geometrical reference frame for events perceived in the 
world.
 This background provides the relief against which apparent material objects are 
brought to the attention of the self-reflective beings through the laws of physics 
resulting
 from the breaking of the full symmetry group $\hat{G}$.

 Whether there is only one such kind of world, of which our own would then be a 
particular manifestation, or several, which might even be catalogued, is likely to be 
difficult to determine
(perhaps even much more so than categorising all possible biological life forms given 
the laws of physics \textit{within} our own universe, whether on the Earth or 
elsewhere). Certainly for any world to be \textit{possible}  in this framework is 
equivalent to the statement that it must actually \textit{exist}, and in this case our 
variety of universe would not be entirely unique. However, we would not be able to 
communicate with other worlds, or the creatures living within them, and there is no 
question of interference with the internal consistency of our own world.

    While the existence of other worlds with different laws of physics is an open 
question, there will be, according to this theory, many possible solutions for a 
geometry $\gmyv$ on a 4-dimensional spacetime manifold $M_4$, apart from our own 
world, which share the same laws of physics and will also internally support conscious 
life under circumstances similar to those in which we find ourselves on Earth. The 
notion of `many worlds' as an  \textit{interpretation} of quantum mechanics is 
distinct from, although implicit within, the overall framework presented in this 
paper, where here we are referring to the `many \textit{solutions}' embedded within 
the theory, as discussed in section~\ref{qpagig}. Different solutions for $\gmyv$ 
involve $\delta Y \leftrightarrow \delta \bvh$ field exchanges in principle anywhere 
on the spacetime manifold $M_4$, even back to the Big Bang epoch for cosmic time $0 < 
t <t_v$, and when considered from a dynamic point of view our universe in some sense 
might be considered to have `branched' from another possible solution at each quantum 
event.

    Such quantum transitions, which are indeterministic from the perspective of a 
single universe, taking place in the very  early universe might serve to seed the 
eventual formation of stars, galaxies and large scale structure generally, as alluded 
to in sections~\ref{secinf} and \ref{sectveu}. That is, in part due to the causal 
temporal accumulation of such probabilistic events, the impact of a quantum 
fluctuation for $t<t_v$ on the overall structure of the universe might generally be 
far more dramatic than a similar `branching' resulting from a `Schr\"{o}dinger's cat' 
type experiment performed at the present epoch. While many solutions for $\gmyv$ might 
be considered to be mutually related by such `branching' events, in the present theory 
each solution is primarily interpreted as an independent full 4-dimensional spacetime 
solution in its own right.

  Much of modern science adopts an essentially  materialist worldview
   in line with our Newtonian heritage. 
  From this objective point of view
   with the universe  seen as a fundamentally material phenomenon, created in the Big 
Bang as an inanimate physical entity with the various seemingly arbitrary parameters 
of cosmology and particle physics, it appears extremely fortunate for us that such a 
world can both support
  biological life and  
lead to the development of our own society, culminating in our own personal human 
form, through a series of chance events. In particular life itself, as we know it, 
would be impossible given a small change in any of a range of the empirically measured 
physical parameters. 
  
As usually presented this means that, for example, the laws of physics are required to 
be such that the chemical elements necessary for life on Earth could be manufactured 
in the hot Big Bang -- which successfully accounts for the relative abundances of the 
light nuclei, D, ${}^3$He, ${}^4$He and ${}^7$Li, cooked up from a hot soup of protons 
and neutrons in the first few minutes -- together with the much later generation of 
the heavier elements through stellar nucleosynthesis. The latter stage is possible 
thanks to a seemingly fortuitous energy level of the carbon nucleus that allows the 
three-body reaction $3\,{}^{4}\textrm{He} \to {}^{12}\textrm{C}$ to proceed at a 
reasonable rate. In 1953, in a famous case of anthropic principle reasoning, the 
necessity of this carbon resonance was \textit{predicted} by Fred Hoyle, to account 
for our own presence in the world as a carbon-based life form dependent on the heavy 
elements. Given this motivation the resonance was then
  experimentally observed shortly afterwards.   
  
   However, for the present theory the universe, through the structure of 
figure~\ref{tcycle}, is born out the intimate interplay between conscious beings and 
the physical world. The complexity of the resulting physical structures within such a 
solution creates the illusion of the fortuity of our own existence. Due to the 
non-trivial nature of solutions achieving a completion of the time cycle in 
figure~\ref{tcycle} any possible physical world is likely to appear highly complex, as 
discussed above. Hence beings in any such world are likely to require a number of 
parameters to describe empirical findings in their world, as for the Standard Model of 
particle physics in our world for example. Hence in turn, with the physical support 
for known biological life forms apparently collapsing under a hypothetical change in 
the empirical parameters, beings in such a world might consider themselves lucky. 
Given the familiarity of our own world as a starting point we can readily conceive of 
many ways in which a physical world could not support life, through small 
perturbations to the properties of our own universe, but it is much harder to conceive 
of very different worlds with very different solutions for supporting the structures 
of conscious life.

    Hence here there is a major \textit{contrast} between the present theory and 
various forms of the anthropic principle, which are generally subject to criticism due 
to their \textit{lack} of predictive power. For example based on the anthropic 
principle a theory may postulate the existence of a very large ensemble of different 
universes with different initial conditions, physical constants or laws of nature -- 
then the fact that our universe is necessarily a member of the ensemble in which the 
structures for life can form necessarily greatly restricts the possible structure of 
the physical laws and conditions that we can observe. However the potentially vast 
range of physical properties for the worlds of the whole ensemble, most of which are 
presumably not observed by any being, is in no way limited by this principle. 

Here the present theory is `anthropic' to a more extreme
 extent in the sense that the only worlds that \textit{exist at all} are those that 
can be brought into being through a conscious, temporalising observer, and in this 
case we may hope to discover the \textit{opposite} conclusion that the laws of physics 
are \textit{necessarily} determined,
 or at least highly constrained, 
 by this requirement (although naturally there will always be the more trivial 
anthropic matter of the local selection of a habitable environment, such as the Earth, 
within such a world).  That is, rather than postulating a large ensemble of typically 
inanimate physical universes with a range of parameters, one of which happens to 
provide a suitable environment for ourselves, we draw our own world into existence, 
sculpting the physical contents of the world out of the possibilities inherent in 
perceiving a world through the forms of $\lv$ as a solution for figure~\ref{tcycle}.

   The number of parameters needed to describe the projection of $\bv_4 \inn \TM_4$ 
out of the components of $\mcX \inn \htho$ or $x \inn F(\htho)$, involving for example 
the dilation symmetries described in the opening of
section~\ref{sectveu}, is much less than the number of parameters needed to describe 
the empirical data as observed in particle physics and cosmology. Hence the present 
theory in principle will be highly predictive 
 even if there are some possibilities for the variation of certain parameters within 
the mathematical constraints of the theory, that is with an anthropic degree of tuning 
for the  parameters involved in the projection of  $\bv_4 \inn \TM_4$ out of $\lvh$.
 However, in chapter~\ref{anpocs} we have generally presumed that 
interactions between the fields as determined  by the  constraint 
equations~\ref{conequa}
 result in a fixed and stable value around 
 $\vert \bv_4 \vert =  h_0$ emerging from the phase transition at $t=t_v$ in the very 
early universe, as described for figure~\ref{vphaset}(c). The 
 potential uniqueness of the laws of physics and particle properties arising at this 
time suggests the present theory should be profusely testable.

  Here the  perspective  is to consider the `early' universe  to be an object of study 
as a limiting extrapolation \textit{from} our present experience in the world rather 
than as an objective self-sufficient  physical state that happens to be the causal 
origin leading up \textit{to} the present conditions in our world, as has been 
described for figure~\ref{cosevolve}.
  With a solution for the time cycle structure of figure~\ref{tcycle} taking priority 
and founding the theory, not only is nothing needed as a temporal antecedent of the 
Big Bang to \textit{cause} the universe to exist, but the particular conditions of the 
Big Bang and  early universe are \textit{shaped} by the overall consistency of the 
solution within the structural constraints implied in figure~\ref{tcycle}. These 
constraints on the apparent `initial conditions' of the early universe are ultimately 
manifested in the  physical and biological processes required to support 
self-reflective life forms at the present epoch, as described above.

   Here we take, possibly rather indirect, measurements of cosmological structure 
including that for the earliest epochs of the universe, as  for laboratory experiments 
in particle physics, as being extensions of our world experience -- quantitatively 
differing from the nature of everyday experience in the world more generally, but in 
all cases subject to the same laws of physics and all within the same system. From the 
basic experiences of thinking, listening to music, walking down the street and 
watching an apple fall from a tree to performing experiments and studying the 
structures of biology, chemistry and physics on all scales, there is a continuity from 
the notion of experience through to, and incorporating, the practice of experimental 
and empirical observations. In the present theory both the notions of scientific 
observations and subjective experiences more generally are drawn together and unified 
as particular manifestations of experience in time.

  If the present theory were to be founded on a purely \textit{objective} notion of 
one-dimensional temporal flow, as modelled by the real line $\rrr$, the sequence for 
figure~\ref{tcycle} could still be constructed linking the nodes $(1),(2)\ldots (6)$ 
but \textit{without} the final link between nodes (6) and (1). For such a theory the 
\textit{subjectively} experienced time of node (6) would hence be derived from a long 
chain of non-trivial steps $(1),(2)\ldots (6)$ beneath which the fundamental objective 
temporal entity of node (1) would be very much hidden from our immediate view of the 
physical world, and would not be an entity we might directly perceive. 
 However, this unnatural duplication of the concept of time in nodes (1) and (6) is 
avoided through the actual perspective of the present theory which has been developed 
\textit{beginning with} the notion of subjective experienced time.
 Indeed temporal flow is not something that we `see' in the world, as is the case even 
for the elementary forms of space,
 rather it is an innate characteristic of our engagement in the world.
That is temporal flow
is not a property of the world which we need to set out to discover, as for the 
`hidden' structures of material phenomena or particle interactions of node (4) for 
example.  
 Rather we \textit{do} directly 
  perceive the underlying temporal flow of node  (1) since it is identified with our 
immediate experience of time in node (6), hence in turn completing cycle of 
figure~\ref{tcycle}.

    The overall system of figure~\ref{tcycle} is perhaps best understood by thinking 
through the cycle of six nodes and links in turn, beginning from any point, but the 
structure can be contracted down in a number of ways including a more minimal scheme 
describing  an interplay between experience and the empirical, or essentially between 
subjective temporal flow and the objective laws of physics as associated with nodes 
(1) and (4) respectively.  
 Ultimately the full set of six nodes coincide as six facets of the internal structure 
of the possibility of conscious experience, conceived as a unified whole, essentially 
adopting the philosophical outlook of existentialism as alluded to in the previous 
section.
As discussed in the previous section, from this point of view the possibility of an 
experience is a more fundamental concept than the individuals who believe they have 
them, and with the laws of physics, which shape both the physical individual and his 
environment, also determined through the constraints on the possible forms an 
experience can take within the system of figure~\ref{tcycle}.

   From the philosophical perspective of materialism, which is grounded largely in 
node (4) of figure~\ref{tcycle}, the `problem of consciousness' arises since the 
concept of subjective experiences seems to be of a  qualitatively different nature to 
anything studied in the realm of the physical world. While from this point of view 
consciousness appears mysterious and beyond the reach of the physical sciences, it 
nevertheless remains the case that conscious experience is a very real phenomenon of 
the world, and indeed it is the feature of the world with which we are most intimately 
familiar. Hence an inclusive scientific theory should either have something to say 
regarding the nature of consciousness or provide a good explanation as to why it 
should not, as suggested shortly before figure~\ref{rtordash}. On this basis the 
speculative structure of figures~\ref{rtordash}--\ref{rinworld} 
 has been studied here in section~\ref{sectleitp}. One possible justification for not 
addressing this question regards the complexity of the human brain, being beyond the 
current scope of an exhaustive scientific understanding.

 On the other hand the nature of subjective experience can be very simple, as 
exemplified by the `thought experiment' involving picking up a pen or pencil as also 
described in section~\ref{sectleitp}. This suggests that the broad objective physical 
correlate of such experiences might also be described in terms far simpler than those 
required to give an account of the detailed structure of the brain.
 Together with the practical experiments of Libet and others discussed in 
section~\ref{secauf} it is clear that the phenomena of conscious thought are in any 
case open to study. 
 Indeed research into consciousness is a scientific field of study in its own right, 
although one which is not traditionally closely linked with physics. It's relevance 
for the present theory lies in the close relationship between the nature of 
consciousness and the structures proposed to complete the cycle of 
figure~\ref{tcycle}. In return the perspective of the present theory, in which 
consciousness is closely associated with temporalisation and related to the physical 
world through figure~\ref{tcycle}, might in principle be of  value for the 
corresponding area of study in neurology, for which a firmly materialist standpoint is 
commonly adopted.

  It is suggested here that
 consciousness is not something that can be fully explained as a phenomenon arising 
solely  \textit{within} a pre-existing physical world, as would be required from a 
purely materialist
 perspective. Subjective experiences cannot be directly described in terms of 
objective matter, but rather correlate with certain mathematical structures which 
underlie the physical world within the context of the system depicted in 
figure~\ref{tcycle}.
 On the other hand
 the content of the physical world is not fully contained \textit{within} the horizon 
of our conscious observations, as might be the case for the pure idealist. We can 
conceive of an infinite expanse of the physical world in space and time  
\textit{beyond} the horizon of our direct experience as supported by the full 
mathematical solution for $\gmyv$ implied in the structure of figure~\ref{tcycle}, 
which itself provides the context for all the structures of cosmology.
  Conscious experience is an irreducible feature of the world and the means by which a 
mathematically possible universe is realised through the intimate interplay between 
the subjective and objective aspects of figure~\ref{tcycle}.

 Much of the apparent mystery of `consciousness' owes to the fact that nothing exists 
without its support and hence it is impossible to step back and isolate the phenomenon 
`in itself'. Everything that exists or happens does so \textit{within} the context of 
consciousness, even our awareness of a discussion of consciousness itself, with the 
phenomena of the physical world ultimately inseparable from the experiences of 
temporalising beings. A theory which, on the contrary, attempts to construct a notion 
of consciousness entirely within the limits of a given independent physical world, 
implying that such a world can `exist' even in the absence
 of such sentient beings,  is necessarily dealing with an incomplete system. Rather, 
while also supporting the physical correlate of conscious mental phenomena, the 
physical world is itself engulfed within the sphere of conscious experience, as 
implied in the relations depicted in figure~\ref{tcycle}.

 For the above materialist worldview in addition to the difficulty in constructing an 
explanation of consciousness upon a given physical world, as alluded to also for the 
right-hand end of figure~\ref{loosends} and discussed more generally in 
section~\ref{sectleitp}, on the other hand there remains the second major loose end 
regarding a foundation for the physical theory itself. 

 An appeal to `beautiful mathematics' is often made either explicitly or implicitly as 
a significant motivating force in theoretical physics, promoting a sense that nature 
`ought' to make use of aesthetically pleasing mathematical structures. While some 
successes may be cited, notably for example regarding the Dirac equation for a fermion 
field (quoted here in equations~\ref{diracl} and \ref{diracli} with a gauge field 
interaction included), the achievements of this approach, in terms of discovering 
empirical phenomena that match a beautiful mathematical theory (applied in particle 
physics or cosmology), have been particularly limited in recent decades. This approach 
also has serious philosophical difficulties, regarding not least the highly subjective 
notion of `beautiful mathematics' itself and the means through which physical entities 
in the world should relate to the mathematical components of the theory. 

  Alternatively an objective physical theory might be founded upon a conceptual idea 
regarding the nature of an inanimate physical world, which will subsequently be 
formulated and developed in mathematical terms in order to derive testable 
consequences for the theory. Examples of this approach include the description of 
gravitation in terms of a curvature of 4-dimensional spacetime in general relativity, 
or the properties of discrete particle-like entities interacting in a flat spacetime. 
However it is difficult to conceive of any physical concept which does not itself 
stand in apparent need of a further underlying explanation. Progress may be proposed, 
for example with gravitation and the geometry of our world in 4-dimensions arising out 
of a more fundamental higher-dimensional spacetime or with particle phenomena deriving 
from a field theory, but at some point the basic physical entities, together with 
perhaps a Lagrangian formalism  or a quantisation procedure, is essentially 
`postulated' as an apparently necessary starting point.

 The foundations of such a theory can be justified provisionally on the grounds that  
`one has to start somewhere', as alluded to in the opening of section~\ref{sectleitp}, 
provided the theory satisfies a criterion of empirical success. Based upon that 
success we learn what a more fundamental theory should effectively look like in a 
certain environment or under certain limiting conditions, such as those for general 
relativity or quantum field theory as described for table~\ref{limits} in 
section~\ref{qpagig}.
 Whether an objective physical theory is founded chiefly upon mathematical, conceptual 
or empirical grounds (and in practice in some combination) the foundational loose end 
is generally accompanied by questions concerning the nature of the origin of the 
universe in the Big Bang, which is needed in order `to get the ball rolling' in the 
first place, as summarised in point (1)  in the opening of section~\ref{sectleitp}.

 The approach of the present theory, with respect to the two loose ends of 
figure~\ref{loosends}, is to fully embrace the subjective element of our engagement in 
the world. With all experience in the world having a temporal aspect the theory is 
founded purely on the notion of a one-dimensional flow of `time' as a  necessary 
component of both the subjective and objective world. Since time \textit{is} a feature 
of the world, which we experience \textit{directly} without any intermediate 
interpretation, this offers an extremely conservative starting point for a theory. To 
be aware of anything at all is to experience an irreducible moment in time, as a basic 
aspect of thought and experience generally. With all thinking having a necessarily  
temporal dimension
 we have essentially retreated to the minimal observation that,
   with a twist on the famous words of Descartes, 
   `I think therefore I temporalise'.
 This  provides the  mathematical basis for a full physical theory which supports the 
entire structure of the universe as perceived \textit{in an experience} itself. 
   In its simplicity this starting point is largely devoid of any arbitrary aspects, 
unlike the case for most theories which are motivated on  mathematical or conceptual 
grounds which are purely objective.

 Through the dual subjective and objective nature of time, both modelled on the same 
mathematical real line, this theory can ultimately also supply its own foundation, 
tying up the two loose ends of figure~\ref{loosends} in the shape of 
figure~\ref{tcycle}.
 Although here the theory is motivated from the direction of a conceptual argument, 
based upon temporal flow, rather than from the direction of `beautiful mathematics', 
the mathematical structure represented in  figure~\ref{tcycle} has itself a degree of 
elegance in its simplicity and self-contained nature. However instead of beginning 
with mathematical beauty together with the presumption of its necessary application to 
the physical world, here the realisation of the mathematically elegant structure 
described in  figure~\ref{tcycle} \textit{contains its own
 inevitability}, in that it incorporates both self-reflective intelligent entities and 
its own foundation.

 Further, this structure provides a context within which an entire universe, as 
depicted for example in figures~\ref{spminf} and \ref{cosevolve} and
   supported by a spacetime manifold $M_4$ of infinite extent, forms part of the 
overall solution, hence incorporating all features of the physical world, from the 
microscopic to cosmological scales, including the Big Bang and  events in the 
arbitrarily distant past. Although the subjective aspects are necessary to conceive of 
the whole system and help motivate the initial foundation of the theory in terms of 
the flow of time, the structures contained within nodes (1)--(4) of 
figure~\ref{tcycle} can be essentially treated as an objective physical theory, as has 
been the case for the large majority of the work presented in this paper, which may be 
measured against observation in the empirical world as for any other theory.

  While the simplicity and elegance of the mathematical structure of nodes 
  \mbox{(1)--(4)} might itself be  considered,
  in order to fully justify the present theory not only on conceptual grounds but also 
from the perspective of the mathematical elegance of figure~\ref{tcycle} as a whole a 
more rigorous mathematical account of the lower half of the chain through nodes 
$(4)\to (5) \to (6) \to (1)$ might be desirable. However, all elements of the cycle 
are open to such an exploration, and
 in section~\ref{sectleitp} we described a possible approach to uncovering a 
mathematical correlate of self-reflective subjective thoughts and decision making.

  There we also noted a close analogy between the mathematical structures relating to 
G\"{o}del's notion of decidability and the properties of physical devices relating to 
Turing's notion of computability. Following Turing and the ambition to develop 
artificial intelligence it is conceivable to attempt to build a machine exhibiting the 
properties self-reflective conscious experiences and creative thought. The design of 
such a machine might include a complicated arrangement of malleable and adaptable 
electronic, and even biological, components capable of internal development, as well 
as an array of sensory input devices and means of interacting with the environment. 
Given the design on paper, for the machine to actually `exist' it would
then need to be built, requiring the physical assembly of the necessary technological 
components.
 Only when manufactured in this way could we declare, in the words of Dr. 
Frankenstein, that ``it's alive!''.

   If the machine could think and have  experiences in a  similar way that we do, it 
might also ask itself how the physical universe and its place in the world came into 
being, and might also be drawn to a conclusion in the form of the system described in 
figure~\ref{tcycle}. For the case of this artificial intelligence the full physical 
environment must  include not only \textit{our} biological evolution but also the 
particular human inventors and technicians with the ability to design and construct 
the machine.

  On the other hand if we consider directly the purely mathematical construction of 
self-reflecting elements relating to G\"{o}del's theorem or a similar theoretical 
structure, rather than taking the computing route of Turing, the conclusion is 
somewhat different. In this case we might design a particular mathematical system 
capable of describing self-reflective states and which also contains its own 
foundation as sketched in figure~\ref{tcycle}. 
 This mathematical structure, as for \textit{any} logically possible mathematical 
construction, is in principle a free creation for our mind to think about abstractly 
and objectively from an independent point of view. While we can \textit{discover} such 
a logically coherent structure in this case any ambition to \textit{build} such an 
entity would be meaningless (unless it could be mapped onto the design of a practical 
machine as described above). 
  However, since the kind of structure depicted in figure~\ref{tcycle} has the 
characteristic that it \textit{contains} thoughts and experiences of internal elements 
all within the same structure together with its own foundation it is in the nature of 
this mathematical system to spontaneously realise its own existence,  detached from 
any external support. The contention here then is that our own experiences in our own 
universe are a particular manifestation of precisely such a self-illuminating world.


\pagebreak

\chapter{Towards a Complete Theory}
\label{chaptow}

\section{Summary and Future Directions}
\label{secsafd}

 The underlying unifying principle for the theory is simply the observation that 
everything takes place through progression in time. Based upon this principle in this 
paper 
 we have explored the extent to which the empirical phenomena of the physical world 
might be accounted for. In the previous chapter we have described how physical 
structures in the world might themselves inscribe subjective experience of progression 
in time
  and hence act as the source of \textit{temporalisation} itself.
 Regarding the general structure of the theory, we first  summarise here the main 
novel ideas presented as the foundation for the physical world as described in detail 
in the preceding chapters.

 The mathematical possibility of a multi-dimensional flow in time is expressed through 
the general mathematical form of progression in time $\lv$ as derived for 
equation~\ref{lv}.
   The creation of an extended spacetime manifold out of the flow of time is possible 
through an innate subjective interpretation of a subset of the algebraic structures 
incorporated within $\lv$ in terms of a  geometrical representation.
 This \textit{spatialisation} of the world is considered a subjective phenomenon 
insofar as it is through it that experience of a physical world by sentient beings is 
possible.
 The description of the geometry of the resulting extended external spacetime is 
identified with that for general relativity, as applying for all physical scales.
 
  Since the extended frame for perception is constructed out of a substructure of the 
full form of temporal flow described by $\lvh$ a natural mechanism for breaking the 
higher, unifying, symmetry of time arises.  Non-gravitational fields and interactions 
are induced on the spacetime manifold through the residual components of the full form 
and symmetry of $\lvh$. The possibility of a degeneracy of solutions for the external 
spacetime geometry underlies the phenomena of quantum theory and particle physics. The 
breaking of  explicit full symmetry groups for candidate forms for $\lvh$ over the 
4-dimensional spacetime base space is found to yield 
 structures closely correlating with features of the Standard Model of particle 
physics.

  A significant novel feature of this theory is that the spacetime manifold is not 
postulated as a starting point, rather it is grounded as a possible structure within 
the multi-dimensional flow of time, arising out of the translation symmetry inherent 
in the form $\lvh$. In other background-free theories one main difficulty is to 
explain the origin of such an extended spacetime structure. Hence most theories employ 
a pre-existing 4-dimensional manifold, or a higher-dimensional spacetime arena in 
which to embed the former, and then introduce fields or other mathematical entities 
upon the manifold. Since here we extended the symmetry group of $\lvf$ to act on a 
higher-dimensional form of time, absorbing the 4-dimensional one, these ideas could 
also be considered as a theory with \textit{extra dimensions}. However, here they are 
not extra dimensions of a spacetime, although algebraic forms or symmetries  which 
also have such a geometrical interpretation (including the temporal form $\lvte$ with 
the symmetry $\sootn$  considered in figure~\ref{mtogmaphr} for the model of 
section~\ref{reaic}) may happen to arise in the mathematics.
  On the other hand this theory can also be conceived as a rather more economical 
approach with \textit{fewer dimensions}, in that the world emerges from a 
one-dimensional progression of time.

 The physical theory presented in this paper, based on the notion of a fundamental 
underlying progression in time taking the general form $\lv$, has progressed along 
four main fronts, as depicted in figure~\ref{fronts4}. In this concluding chapter 
these theoretical developments are summarised along with a discussion of how they are 
mutually related and might be combined together in progressing towards a complete 
theory.
\begin{figure}[htb]  
\centering
\hspace*{2pt}
\epsfysize=10cm
\leavevmode
\setlength{\unitlength}{25pt}
   \begin{picture}(7.0,13.0)(0.0,0.1)	 
 {\large		
     \put(2.35,6.75){\LARGE $\lv$} 
   \thicklines	 
	 \put(5.35,6.9){\vector(1,0){1}}
	 \put(1.95,6.9){\vector(-1,0){1}}
	 \put(3.65,7.9){\vector(0,1){1}}
	 \put(3.65,5.9){\vector(0,-1){1}}
	\put(3.25,12.1){\LARGE $(2)$}
     \put(1.6,11.35){$\ese$ Action on $F(\htho)$}
	 \put(1.05,10.65){$\lvfs$, $\dmofs$}
	 \put(1.9,9.95){(Standard Model)}
	\put(-2.75,7.9){\LARGE $(1)$}
     \put(-4.5,7.15){Isochronal Symmetry}
	 \put(-5.6,6.45){$G_{\mu\nu} = f(Y)$, $\gmo$, on $M_4$}
	 \put(-4.7,5.75){(Kaluza-Klein Theory)}
	\put(9.25,7.9){\LARGE $(3)$}
	 \put(8.1,7.15){Many Solutions}
	 \put(7,6.45){$\delta Y \leftrightarrow \delta \bv_{56}$ Redescriptions}
	 \put(7.1,5.75){(Quantum Field Theory)}
	\put(3.25,3.7){\LARGE $(4)$}
	 \put(1.4,3.0){Large Scale Structure}
	 \put(0.4,2.3){$G_{\mu\nu}=f(\bv_{56})$ and Generalised}
	 \put(1.3,1.6){(Standard Cosmology)}
  }
	 \end{picture}
\vspace{-30pt}	
\caption{\setb Developed from the original underlying notion of the primary role of 
temporal flow these four areas  of progress  (1), (2), (3) and (4) have been described 
in detail in this paper in chapters 2--5, 6--9, 10--11 and 12--13 respectively. (In 
each case the main guide from established physical theory is appended 
parenthetically).}
\label{fronts4}
\end{figure}

  The four fronts of the theory described in figure~\ref{fronts4} contain  aspects of 
the interplay between the various forms of the flow of time considered, from 
one-dimensional temporal causality itself  up to the largest form $\lvfs$,  the full 
symmetry of which is broken over the  base manifold $M_4$.
  Individually these four fronts exhibit the following principal features:

\begin{itemize}
 \item[(1)] Motivated by the notion of perception over a 4-dimensional base manifold 
$M_4$
 four extended external dimensions are initially identified through translation 
symmetries
  of the full form $\lvh$. Subgroups of `rotational' symmetries of $\lvh$ imply the 
identification of gauge fields on $M_4$ relating to both the external and internal 
geometry 
 and the unifying framework of a principle fibre bundle for general relativity and 
classical gauge theory can be constructed. 
 With the external and internal geometry correlated as the full symmetry of $\lvh$ is 
broken in the projection over $M_4$ this structure, with the four external dimensions 
identified as above rather than with the `extra' dimensions being `compactified', is 
reminiscent of non-Abelian Kaluza-Klein theories.

\item[(2)]  Motivated by its mathematically rich structure out of the infinite 
possible forms of $\lv$, a 56-dimensional form of temporal progression $\lvfs$ with a 
high degree of symmetry is identified through the action of the group $\ese$ in 
preserving a quartic form defined on the space $F(\htho)$, containing the determinant 
preserving action of $\esi$ on the space $\htho$.
 When broken over the external $M_4$ base manifold the residual internal gauge group 
contains features of the symmetry $\SML$ acting upon components of $F(\htho)$, 
including subspaces identified as spinors under the local external Lorentz symmetry 
$\sltc$ with charges under an internal $\uo_Q$ symmetry,
 which are reminiscent of the Standard Model of particle physics.

 \item[(3)] Conforming with the underlying one-dimensional causal flow of time the 
degeneracy of field solutions for the world geometry $G_{\mu\nu}(x)$, consistent with 
the broken form of temporal flow expressed dynamically on the base manifold via 
expressions such as $\dmofs$, selection rules for exchanges between gauge $Y(x)$ and 
spinor $\psi(x)$ fields may be obtained. This leads to interaction phenomena with a 
mathematical structure reminiscent of calculations employing the time evolution 
operator $U(t,t_0)$ in a quantum field theory based upon a given Lagrangian.

\item[(4)] In constructing the base manifold $M_4$ out of the full form $\lvfs$ and 
its symmetries variation in the magnitude of the projected subspace vectors
 $\bv_4(x) \inn \TM_4$, with $\vert \bv_4 \vert^2 = L(\bv_4) = h^2(x)$, itself 
generates a non-flat external geometry.
 The general solution for the 4-dimensional geometry  $\gmyv$ might also incorporate a 
cosmological term in principle deriving from the scalar components of $F(\htho)$. 
Collectively the resulting large scale structure of the cosmos may correlate with the 
observed phenomena of the dark sector and properties of the very early universe, that 
is in a manner reminiscent of the standard cosmological model and inflationary theory. 
\end{itemize}

  Hence the theory represents new directions of research in fundamental physics 
branching into several areas. At the same time the main part of this work sits 
comfortably within the existing infrastructure of theoretical and experimental 
physics. The mathematical framework has been adopted entirely from that used in much 
of contemporary theoretical physics, with the novel input more in the nature of the 
overall conceptual picture.

   The essential theoretical ingredients to account for the Standard Model of particle 
physics and large scale cosmological structure, while sidestepping the Lagrangian 
formalism   and also providing a conceptual basis for the `quantisation' of the 
fields, are
in principle all found in the structures of the present theory.
  All four of the above fronts are directly related to consideration of the basic idea 
expressed in the general form of temporal flow $\lv$, and are mutually related to each 
other.
 The immediate future direction and main aim for further study on each front is first 
\mbox{summarised here:}
 
\begin{itemize}
 \item[(1)]  
  Use the mutual relationship between the external and internal curvature in 
originating from  symmetries of the same full form $\lvh$ projected over $M_4$, 
described in terms of the differential geometry of the structure of a  fibre bundle, 
to derive the  relation $G_{\mu\nu} = f(Y)$ in the form of equation~\ref{gchift} 
without any explicit application of an action integral such as equation~\ref{teinhil} 
as adapted from  Kaluza-Klein theory.

\item[(2)] 
 Determine a higher-dimensional form of temporal flow and corresponding symmetry to 
build upon the features of the Standard Model identified in the action of $\ese$ on 
$\lvfs$ when broken over $M_4$ as summarised in equation~\ref{fhthopart}. For example 
a presently hypothetical $\ee$ action on a full form $\lvtfe$ might be sought, the 
structure of which will be guided by fields and interactions of the Standard Model 
Lagrangian yet to be accounted for. 

\item[(3)] 
 Use a statistical approach to HEP phenomena with probabilities based upon field 
degeneracy, building upon the relationship with quantum field theory described for 
equation~\ref{pddddm} and possibly employing the analogy between the properties of 
condensed matter systems and QFT, to develop the theory through to the calculation of 
cross-sections and the identification and conceptual understanding of particle states 
without imposing quantisation rules. 

\item[(4)] 
  Build upon the geometry $G_{\mu\nu}(x)$ of equation~\ref{gmnconf}, deriving from a 
variation of the magnitude $L(\bv_4) = h^2(x)$, to a full general form $\gmyv$ 
incorporating also scalar fields and applied for the large scale structure of the 
universe, in order to make a more quantitative comparison between the present theory 
and empirical observations in cosmology; with one aim being to deduce which scenario, 
such as that in figure~\ref{earlyu}(a) or (b), applies for the very early universe.
\end{itemize}

  The main prediction of the theory at present is a mathematical one concerning the 
existence of an $\ee$ symmetry acting upon a quintic or higher order form $\lvtfe$ as 
alluded to in front (2) above. This structure, as an extension from the $\ese$ action 
on $\lvfs$, when broken over $M_4$ should incorporate further Standard Model 
properties such as three generations of fermions, as motivated in detail in 
section~\ref{sosmfi}.
 More generally the
 overall aim is to fuse the above four areas together in a full unified theory, and 
assess the consequences and possible predictions of the theory that can be further 
compared with and tested against
 empirical data from HEP experiments, cosmology and other observations.
 We begin here by observing the following relations between the four theoretical 
branches
 summarised in figure~\ref{fronts4}.

\begin{itemize}
 \item[(1+3)] 
  The key motivation for front (1) is the identification of a smooth external geometry 
$G_{\mu\nu}(x)$ on $M_4$ as an arena for perception in the world. Since there is no 
similar requirement regarding the need for a `smooth' internal geometry of gauge 
fields it would be more natural to \textit{begin} with the structure of fronts (1+3) 
combined, as implied in the relation $G_{\mu\nu} = f(Y,\bvh)$ as a possible solution 
for the world geometry on $M_4$.
 A finely fragmented and fractal-like structure of field exchanges $\delta Y 
\leftrightarrow \delta \bvh$  underlies the smooth external spacetime arena, with  
$G^{\mu\nu}_{\ph{\mu\nu};\mu}=0$ maintained  as a geometric identity. 
  In this way the degeneracy of many possible solutions brings the phenomena of 
general relativity and quantum theory together at the same time in the process of 
identifying the base manifold itself, rather than beginning with a `classical theory' 
of the form $G_{\mu\nu} = f(Y)$ which is then `quantised'.

  The relation between the initial theoretical  `bare' fields and empirically observed 
`dressed' fields was also described in the opening of section~\ref{seraps}.
  Indeed, a geometrical relation of the form $G_{\mu\nu} = f(Y)$ might still be 
identifiable for macroscopic fields, such as the empirically observed electromagnetic 
field.
 Out of the complete framework the standard theories alluded to parenthetically for 
fronts (1) and (3) in figure~\ref{fronts4} may be shown to emerge in the appropriate 
limits: namely  Kaluza-Klein theory in a curved spacetime as an example of the 
macroscopic field limit of general relativity 
 and QFT in the limit of a flat spacetime for microscopic fields, as described for 
table~\ref{limits} in section~\ref{qpagig}.

\item[(2+4)]
   In the present theory the phenomena of electroweak symmetry breaking and in 
particular the masses of particle states observed in the laboratory arise out of 
interactions between the components of the vector-Higgs field $\bv_4(x)$ and other 
fields such as the fermions $\psi(x)$ identified in the components of $F(\htho)$ 
through the terms of the quartic form $\lvfs$. On the other hand cosmological 
structure depends on variation in the magnitude $\vert \bv_4 \vert = h(x)$ as $\bv_4 
\inn \TM_4$ is projected out of the full form $\lvfs$ over $M_4$, which itself 
provides a geometric explanation of the origin of mass in terms of an effective 
energy-momentum tensor defined in $-\kappa T_{\mu\nu}:=G_{\mu\nu} = f(\bv_{56}) \neq 
0$.
 Hence these two notions  of mass are intimately related via the field $\bv_4(x)$. 
 
  The dilation symmetries, acting on the components of $F(\htho)$ as discussed in the 
opening of section~\ref{sectveu}, change the value of $\vert \bv_4 \vert$ and may be 
significant in relation to the mechanism of electroweak symmetry breaking in the very 
early universe. The physics of the very early universe may also guide the 
identification of a higher-dimensional form of time, such as the hypothetical  
$\lvtfe$ with $\ee$ symmetry. In particular the mechanism for generating a 
matter-antimatter asymmetry might be determined by interaction terms implicit in the 
form $\lvtfe$ or involve a further internal gauge field deriving from the $\ee$ 
action, as also discussed in section~\ref{sectveu}. Hence the structure of the full 
form $\lvh$ is closely linked with an understanding of significant 
 questions in cosmology.

 \item[(1+2)]  In equation~\ref{Daaa} of chapter~\ref{esihtho} the generators of the 
symmetry of a 27-dimensional form of $\lvt$ were introduced as operators that 
annihilate the cubic norm $\det(\mcX)$ with $\bv_{27} \equiv \mcX\inn \htho$. A 
complete basis for this  78-dimensional Lie algebra of $\esi$, as represented by 
vectors of the tangent space  $\dot{R} \inn T\htho$, is listed in tables~\ref{lbrota} 
and \ref{ltrota} at the end of section~\ref{laofesi}.
 Such a `static' generator can be pulled back to a Lie algebra valued 1-form 
$Y_{\mu}(x)$ on  $M_4$, as initially described in subsection~\ref{mgjoin}, and appears 
in `dynamic' expressions on the base manifold.
Kaluza-Klein models based on fibres identified with homogeneous spaces were reviewed 
in section~\ref{thwhf}, and might provide additional insight in comparison with the 
closely related theories constructed on principle fibre bundles described in 
sections~\ref{lccop}
 and \ref{olcop}.

  With regards to the model described for figure~\ref{mtogmaphr} in 
section~\ref{reaic}, with the full symmetry group $\sootn$ acting on the form $\lvte$ 
over $M_4$,  the structure of the Lie algebra for $\sootn$ can itself be expressed in 
terms of vector fields on the space of 10-dimensional vectors $\bv_{10} \equiv X \inn 
\htwo$ with $\det(X) = 1$, based on the opening of section~\ref{ltos}.  With $\htwo 
\subset \htho$ embedded as a subspace  a close connection is made with the above case 
for $\esi$ acting upon the homogeneous space composed of vectors $\bv_{27} \equiv \mcX 
\inn \htho$ of unit determinant.
 The $\ese$ action on $F(\htho)$, broken over the 4-dimensional base space $M_4$, 
represents a higher-dimensional extension of this structure, while the full form of 
$\lvh$ that provides the actual setting for a description of the real world is open to 
further investigation. Hence branch (1) relates to branch (2) of figure~\ref{fronts4} 
essentially in the choice of $\lvh$ and the corresponding full symmetry group over the 
base manifold $M_4$.

 \item[(2+3)]  Taking the example of the $\ese$ case, 
 the generators of the internal symmetry action $\dot{R} \inn TF(\htho)$ give rise to 
the gauge fields $Y_{\mu}(x)$ on the base space while the components of $\bv_{56}  
\inn F(\htho)$ are also intimately related to the base manifold through the 
translation symmetry over $x\inn M_4$ as originally described for 
figure~\ref{spillout}.
 Hence since  $F(\htho)$ forms the representation space of $\ese$ 
 the gauge fields $Y_{\mu}(x)$ naturally couple with components of $\bv_{56}(x)$, 
including the spinor fields $\psi(x)$.
 The dynamics of the interaction between the components of $\bv_{56} \inn F(\htho)$ 
and the gauge  fields, under the constant form $\lvfs$, is subject to the constraint  
$D_{\mu}L(\bv_{56})=0$, expressed  through the  covariant derivative $D_{\mu} \sim 
\pal_{\mu} + Y_{\mu}$ (as for the $\esi$ example in equation~\ref{dlvpbbz}).
  In this way  interaction terms similar in form to those introduced for 
$\lag_{\mathrm {int}}$ in the Lagrangian approach for the Standard Model  are 
identified.
 Arising from symmetry breaking over the base manifold $M_4$ the possible $\delta Y 
\leftrightarrow \delta \psi$ exchanges of field components are also constrained by the 
set of  degenerate solutions under the same local external geometry $\gmyv$.

 All observed fermion states interact with at least one gauge boson
 via terms of $\dmofs$, as applied for the electron self-energy interaction in 
figure~\ref{elecsel}(b) for example. Hence the external geometric structures relating 
to the $\psi(x)$  components will be shaped by the bare gauge fields such as 
$A_{\mu}(x)$ with which they interact. With the bare gauge fields subject to 
$G_{\mu\nu} = f(Y)$ from the isochronal Kaluza-Klein relation the physical fermion 
particle states will emerge through modifications to the geometry $G_{\mu\nu}(x)$
 due to $\delta Y \leftrightarrow \delta \psi$ interactions. In turn 
  the question of the form of $G_{\mu\nu} = f(\psi)$ for electron, muon and further 
particle  states might be considered. This form of solution should also extrapolate to 
the non-relativistic limit, such as for the implied electron state linking $S$ and $A$  
in figure~\ref{eemmeds}(b) for example.

 \item[(3+4)]  Given also the non-trivial geometry $G_{\mu\nu}=f(\bv_{56})$ from 
$L(\bv_4)=h^2(x)$ variation the implications of further field interactions of the form 
 $\delta \bv_4 \leftrightarrow \delta \psi$ under the constraint $\lvfs$  will also 
contribute to the form of $G_{\mu\nu} = f(\psi)$.
 These interactions with the vector-Higgs field $\bv_4(x)$ are expected to relate to 
the origin of fermion masses, with the details giving rise to the mass difference 
between the electron and $d$-quark states for example. In order to investigate the 
mass differences 
 between the three generations of fermions, such as between the electron and muon, a 
higher-dimensional form such as $\lvtfe$ may be required. The equality of the 
empirically observed electric charge across the generations may relate to the role of 
`Ward identities' in the QFT limit.

  With the relation $G_{\mu\nu} = f(\bv_{56})$ generalised for multiple solution field 
exchanges  under the form $G_{\mu\nu} = f(Y, \bv_{56})$ essentially all matter 
$T_{\mu\nu} := G_{\mu\nu}$ is expected to be associated with quantum phenomena, 
 with the variety material forms observed in the universe shaped according to the 
probabilistic nature of the underlying field composition.  
  The relative probabilities of local solutions for  $G_{\mu\nu} = f(Y, \bv_{56})$ are 
determined through a `number of ways' statistical count of the underlying field 
redescriptions, essentially as for the determination of probabilities for classical 
systems. This leads to a unified approach to quantum and classical thermodynamic 
properties, which in particular will be significant for studying the evolution from 
$t=0$ to the phase transition at $t=t_v$, as the stable value $L(\bv_4)=h^2_0$ is 
attained in the very early universe, as described for figure~\ref{vphaset}. 
 This may also mark an epoch of fermion production via
$\delta \bv_4 \leftrightarrow \delta \psi$ exchanges as the properties of the Standard 
Model of particle physics emerge in the phase transition.

 \item[(4+1)]
 While we have considered beginning with the classical geometric relations  
$G_{\mu\nu}=f(\bv_{56})$ or $G_{\mu\nu}=f(Y)$ more generally these two means of 
obtaining finite external curvature will be combined in a general solution for 
$\gmyv$. In the full theory field interactions of the form  $\delta Y \leftrightarrow 
\delta \bv_4$, resulting from the action of the corresponding gauge symmetry on the 
external components $\bv_4 \inn \TM_4$, will relate closely to the identification of 
gauge boson masses and the phenomena of electroweak symmetry breaking generally. 

  In principle the theory might rather \textit{begin} with the full general form of 
$\gmyv$, fully incorporating quantum phenomena and completing the program described 
for fronts (1+3) combined above, as will be required to fully account for both the 
large scale structure in cosmology and the phenomena observed in the HEP laboratory. 
While the pure `bare' forms of the relations
 $G_{\mu\nu}=f(\bv_{56})$ or $G_{\mu\nu}=f(Y)$ may not be found in nature, due to the 
possibility of underlying field interactions, each of these relations may play a role 
in an appropriate classical field limit.
\end{itemize}

 Hence the aim from the developments in figure~\ref{fronts4} is to generalise from (1) 
the  geometric structure of gravitational and gauge fields deriving from the 
isochronal symmetry of $\lvfs$ to incorporate interactions with the field components 
of (2) $\bv_{56}$ itself subject to the dynamic relation $\dmofs$ derived from the 
action of the full symmetry of $\ese$  on $F(\htho)$ broken over $M_4$, taking into 
account the impact of (4) variation in the projected value of $\vert \bv_4 \vert = 
h(x)$, to arrive at a general form of solution for $\gmyv$ over (3) a degeneracy of  
`quantum' field redescriptions  underlying an external geometry with 
$G^{\mu\nu}_{\ph{\mu\nu};\mu}=0$  everywhere, which itself provides one of the 
constraint equations~\ref{conequa}.

 Collectively progress on fronts (1), (3) and (4) of figure~\ref{fronts4} can be 
considered together under the ambition of accounting for the empirical properties of a 
quantum field theory \textit{without} applying standard quantisation rules for the 
present theory.
 These three fronts all relate to the identification of a smooth geometry 
$G_{\mu\nu}(x)$ constructed in terms of fields extended on the spacetime manifold 
$M_4$, the identification of which, as the background for perception in the world, 
itself motivates this construction.
 This area of research, guided by the analogy between QFT and condensed matter 
systems, might proceed based on a provisional assumption for the full symmetry of the 
full form $\lvh$ such as the $\esi$ case.

  In fact for this purpose a yet simpler, but non-trivial, model could be considered 
based on $\hat{G} = \slthc$ as the full symmetry  of time acting upon elements 
 $\bvh = \bv_9 \inn \hthc$ such that $L(\bv_9) = \det (\bv_9) = 1$ is invariant. 
 This structure incorporates a subgroup action $\sltc \subset \slthc$ on the 
subcomponents
 of $\bv_4 \equiv \bh_2 \inn \htwc$, identified with the external tangent space 
$\TM_4$, as described for
 equation~\ref{vinhinc} at the  end of section~\ref{lsspin}. The structure of the 
resulting symmetry breaking to $\sltc \times \uo \subset \slthc$ over the base 
manifold $M_4$ may be sufficient to study a model accommodating both general 
relativity together with a form of quantum electrodynamics deriving from the internal 
$\uo$ symmetry. 
  On generalising from the complex space $\ccc$ to the octonions $\ooo$ the symmetry 
action $\slthc$ is itself contained as a subgroup of $\sltho \equiv \esi$ as 
explicitly demonstrated by the generator composition of equations~\ref{rrrrrrrr} and 
\ref{slthcas} 
in subsection~\ref{strassy}. In this way the form $L(\bv_9) =  1$ naturally takes its 
place in the progression 	$L(\bv_4) \to L(\bv_{9}) \to L(\bv_{27}) \to L(\bv_{56})$ 
discussed in   section~\ref{secuni}.

 Independently of combining the above three fronts, that is (1), (3) and (4),  further 
progress may be made on the structure of front (2) itself which, although the subspace 
of vectors $\bv_4\inn \htwc$ is associated with the external spacetime, considers the 
symmetry structure of $\lvh$  without 
 explicitly projecting the components into fields over $M_4$.
  This further study concerns, for example, the explicit identification of an internal
   $\sutw_L \times \uo_Y \subset \ese$ subgroup together with a determination of 
$\sin^2 \theta_W$ and the study of electroweak properties within the theory based on 
the form \mbox{$\lvfs$}. However the larger ambition for front (2) will be the 
identification of the full general form of temporal flow, involving for example an 
$\ee$ symmetry of the currently hypothetical form $\lvtfe$.
  The progression of table table~\ref{lvftolvfs} and the known structure of
  equation~\ref{esixtoee} together with the general discussion of 
   section~\ref{sosmfi} strongly hints towards the real form $\eeg$ as a candidate to 
be sought for such a full symmetry.

  A more thorough understanding of quantum phenomena in spacetime and a
 determination of the  full form of $\lvh$ are hence the two main branches  to be 
pursued  
 en route to the formation of a complete theory incorporating all four fronts of 
figure~\ref{fronts4}, with the aim to account both for cosmological observations and 
the properties of the Standard Model of particle physics
 through the structure of $\lvh$, and without  introducing a Lagrangian
 or any other arbitrary postulates for any point of the theory.

\section{Reconstructing HEP Phenomena}  
\label{secrhp}

  For contrast with the present theory the general recipe for constructing a standard 
field theory is summarised in the following three stages. This involves in particular 
employing  a Lagrangian, such as equation~\ref{lagdym} or as described in 
section~\ref{ewtatsm} for the Standard Model, to introduce interactions into the 
theory in order to describe the phenomena observed in HEP experiments.

\begin{itemize}
\item[(a)]  Together with the Lorentz group for the external spacetime symmetry, a 
gauge group is selected, generally motivated on empirical grounds, to describe the 
internal symmetry of the model.  The field content of the theory, in terms of the 
field transformation properties as a choice of the representations of the symmetry 
groups, is also determined in order to comply with the findings of experiments. 

\item[(b)] A scalar Lagrangian as a function of the fields is written down, invariant 
under the symmetries of the theory, with various caveats on the general form of the 
terms -- for example to ensure the renormalisability of the quantum version of the 
theory. The Lagrangian function is used in conjunction with the principle of extremal 
action to determine the equations of motion for the fields. 

\item[(c)]  The classical theory can be quantised for example by introducing 
  field operators $\hat{\phi}(x)$, commutation relations and a Fock space of particle 
states such as $\vert \bp \rangle$ as reviewed in the opening of 
section~\ref{tranamp}.
The framework of QFT is built upon a flat spacetime background as a given entity.
\end{itemize}

   From the point of view taken here the introduction of a scalar Lagrangian function 
in item (b) above is conceptually a particularly poorly motivated aspect of the 
standard theory.
The roots of the Lagrangian approach originate historically in the study of classical 
mechanics for non-relativistic material bodies, reproducing Newton's Laws of Motion in 
a more general framework.
 Later, further pragmatic progress and empirical success was achieved in generalising 
this framework to
  incorporate field theories and also to derive relativistic field equations in the 
Minkowski spacetime of special relativity. The Lagrangian approach is also employed 
for the quantised fields of QFT in a flat spacetime on the one hand, and in general 
relativity, with the geometric $R\sqrt{\vert g \vert}$ Lagrangian term based on the 
Ricci scalar $R$ for example in equation~\ref{einhil}, in a curved spacetime  on the 
other hand.

    However there is no underlying conceptual justification for the invention of such 
a scalar field, the integral of which over a set of spacetime coordinates should be 
stationary under field variations, either for a classical or quantum theory. In the 
QFT for the Standard Model it is the empirical observation of the effects of local 
gauge groups through their representations on apparent particle multiplets that guides 
the construction the Lagrangian, \textit{taylored} to generate the desired equations 
of motion. That the Lagrangian framework should remain valid for a unified theory of 
quantum phenomena and gravitation is a further assumption built upon an uncertain 
foundation.

  By contrast with the Lagrangian approach, in the present theory a fundamental scalar 
function  which is not only stationary but constrained to a particular scalar value is 
readily identified, that is $\lvfs$. Although general empirical features, such as the 
required rank of a unification group as described in section~\ref{dynkin}, serve as a 
useful guide for the study of $\esi$ and $\ese$ as a symmetry of time, here empirical 
details of the Standard Model are \textit{uncovered} in the structure of the external 
and internal broken symmetry action on the components of the spaces $\htho$ and 
$F(\htho)$, as described in chapters~\ref{chapesb} and \ref{secfd}.
 Further, the equations of motion for the fields on $M_4$ can be derived purely as a 
consequence of the constraints of the theory, which are  summarised in 
equations~\ref{conequa}.
 For example Maxwell's equation~\ref{maxhere} and the Dirac equation~\ref{diracli} 
result from the degeneracy of field solutions subject to the constraints, as described 
in section~\ref{secdos}.
  Hence in contrast to the recipe for a standard field theory listed above in 
(a)--(c), the necessary ingredients arise naturally in the present framework as listed 
below:

\begin{itemize}
\item[(A)]   All the main symmetries considered must form a group or subgroup of a 
symmetry of time, that is of the equation $\lv$. The Lorentz group is motivated by its 
pseudo-Euclidean structure as required for external perception, while the internal 
gauge groups are identified  in the breaking of the higher, richer, symmetry such as 
$\ese$ over the base manifold $M_4$.   
 The representations are already essentially determined since the Lorentz and $\ese$ 
groups are selected \textit{by} their actions upon the vector spaces $\htwc$ and 
$F(\htho)$ respectively, with the broken internal gauge groups acting upon multiplets 
of $\sltc^1 \subset \ese$ Weyl spinors.

\item[(B)]  Equations of motion are constrained by the fundamental requirement  
\mbox{$L(\bv_{56}) = 1$}
  which further implies $\dmofs$, as listed in equations~\ref{conequa}. 
  Further constraints on the equations of motion for the fields are governed by the 
relation  $\gmyv$, consistent with the Bianchi identities for the external and 
internal symmetries. This structure over $M_4$ naturally arises as required to frame a 
world of physical perception, in a geometrical space and time, out of the general form 
of temporal flow.
 Field `interactions' are implied at the outset in the form of the above expressions 
over the base manifold, in terms of gauge $Y(x)$ and spinor $\psi(x)$ fields for 
example, leading to expressions such as equation~\ref{dlvpbbz}.

\item[(C)] In the present theory the phenomena of quantisation correspond to the 
degeneracy of the multiple solutions implied in the expression $\gmyv$, consistent 
with $\lvfs$, as has been summarised in the previous section.  That is, the fields are 
intrinsically involved in creating the non-trivial geometry $G_{\mu\nu}(x)$ of the 
base manifold itself. It then remains to be described how the particle phenomena seen 
in HEP experiments, in particular the nature of the initial and final particle states,  
arise out of these field exchanges in spacetime. 
\end{itemize}

  The non-gravitational fields on $M_4$ derive from the symmetries and components of 
the `extra dimensions' of temporal flow, in a manner analogous to the employment of 
the additional degrees of freedom in theories based on extra spacetime dimensions such 
as Kaluza-Klein theories.
  Here the equations of motion are simply equations for the variation of the 
mathematical structures which arise as projected onto the 4-dimensional base manifold 
and parametrised by the underlying 1-dimensional temporal flow. They are \textit{not} 
equations of motion for some other body or entity introduced independently of time 
itself. 

  The field and particle content of the theory will be determined by the choice of the 
full and external forms of temporal flow, here taken to be $\lvfs$ and 
 \mbox{$L(\bv_{4}) = h^2$} on $M_4$ with their respective symmetries of $\ese$ and 
$\sltc^1$ (with the latter originally identified as a subgroup of $\esi$ as described 
for equation~\ref{sltcinesi}). 
 The mathematical and conceptual limitations on the choice of these significant forms 
and the component normalisation such as $h^2$, and hence the observed field and 
particle properties induced through the symmetry breaking, were considered in the 
section~\ref{secuni}. There questions were raised concerning the  uniqueness of the 
present theory  and the extent to which it is constrained given, for example, the 
possibility of further higher-dimensional forms of temporal flow.

 Here, with $\ese$ taken to describe the symmetry of the full 56-dimensional form of 
temporal flow, for the complete theory the full set of broken $\lvfs$ and $\dmofs$ 
terms may be written out. All empirical effects must then be consistent with these 
equations together with the local geometrical forms $G_{\mu\nu} = f(Y)$ and 
$G_{\mu\nu} = f(\bv_{56})$, the latter of which augments the set in 
equation~\ref{conequa}, as combined globally under the solution $\gmyv$ together with 
the identity $\gmo$ framing the spacetime manifold.
 Hence the set of possible field couplings, as expressed through causal sequences of 
degenerate field redescriptions, must conform to this set of equations. These 
equations, essentially acting as selection rules, are listed in the left-hand column 
of table~\ref{coneqco} alongside
examples of possible terms and the associated field interactions or empirical effects 
in the remaining columns.

\begin{table}[htbp]
\centering
\begin{tabular}{|l|l|l|}
 \hline
     Equations & Terms  &  Field Interactions and Phenomena  \\  
 \hline		
    $\lvfs$ & $ \sim vv\ol{\psi}\psi$ 
	        &  Yukawa-type couplings for fermion masses \\
   &		&  involving vector-Higgs $\bv_4$ components \\  \hline
  $\dmofs$	&  $ \sim vv\ol{\psi}Y\psi$
            & gauge-fermion interactions for internal forces   \\  
   &        & also gauge-$\bv_4$ coupling for $Z^0$,$W^{\pm}$ masses
     \\   \hline
	$G_{\mu\nu} = f(\bv_{56})$ &  equation~\ref{gmnconf}
	                    &  significant for geometry of dark sector  \\
   &        &       and evolution of the very early universe \\  \hline
   	$G_{\mu\nu} = f(Y)$ & $\sim FF$
	                    & with $F = \mbox{d}Y + \fhs \lbrack Y,Y \rbrack$,
						  equation~\ref{fdapaa},  have  \\
   &        &  gauge field cubic and quartic self-coupling \\ \hline 			
	$\gmo$  & $T^{\mu\nu}_{\ph{\mu\nu};\mu}(Y,\bvh) = 0$
	        &  conservation of energy-momentum and   \\
   &        &   constraint on field equations of motion      \\ 
  \hline
  \end{tabular}
  \caption{\setb The set of constraints in the first column determine the field 
interactions and associated field equations of motion, in place of an imposed 
Lagrangian.}
\label{coneqco}
\end{table} 

  The interactions described in the right-hand column bare a close resemblance to 
those placed by hand in the Standard Model Lagrangian, however the corresponding field 
terms in table~\ref{coneqco} arise naturally in the present theory. Collectively  the 
constraints in table~\ref{coneqco} expressed over the spacetime manifold $M_4$ replace 
the need to introduce a scalar Lagrangian function.
  With respect to local internal symmetry transformations all of the equations in 
table~\ref{coneqco} are gauge invariant  while they transform covariantly under 
external Lorentz transformations as scalar, vector or tensor representations. This 
latter feature, as well as the fact that there are several equations, distinguishes 
this theory from the scalar Lagrangian approach, and indeed the present theory will 
need to be fully worked out independently of the standard framework.

 Given a sufficient understanding of how field degeneracy in the present theory 
relates to quantum phenomena it may be possible to deduce \textit{effective} 
Lagrangian terms from the constraints of the equations listed in table~\ref{coneqco} 
and import these structures into the framework of a QFT employing a Lagrangian 
approach.  This substitution of fields and interactions derived from the present 
theory into the standard procedure summarised in items (a), (b) and (c) above  might 
be provisionally followed all the way through to standard QFT calculations such as 
cross-sections.
However, the alternative approach, with the emphasis on a complete understanding of 
the present theory, would be much preferred in the long term, with the formalism of a 
QFT Lagrangian later identified in a suitable limit of the complete theory.

 For the present theory the meaning of \textit{quantisation} itself is to be found in 
the  degeneracy of field solutions, without following a standard QFT approach such as 
attaching creation $a^{\dag}(\bp)$ and annihilation $a(\bq)$ operators to the field 
components and applying canonical commutation rules. However in the process of 
calculation the field couplings arising from the equations in table~\ref{coneqco} may 
be associated with vertex diagrams, as was described for a few cases in 
figure~\ref{fvhere},
 as one part of the correspondence with Feynman rules described more completely in 
section~\ref{secdopp}. That is, while the present theory is constructed on a firm 
conceptual foundation,  the empirical successes of QFT suggests that a 
complexification of a calculation and the employment of the mathematical tools of QFT, 
such as amplitudes and unitary evolution, might also be applied pragmatically here. 
Hence the optimal approach may be to straddle both perspectives -- pursuing the 
development of the present theory while incorporating calculational tools from QFT.

Between the macroscopic structure of the external geometry $\gmyv$ and the internal 
microscopic field interaction exchanges, consistent with the equation $\dmofs$ for 
example, nested layers of multiple solutions will shape the physical manifestation of 
the theory in a way reminiscent of  `renormalisation' techniques in QFT.
 While the particle concept and HEP calculations may be motivated from within the 
present theory mathematical tools extracted and adapted from QFT will play an 
important role in the development of the complete theory and the   
 establishment of a detailed comparison with empirical measurements.

  Since the physical couplings and masses measured for 
   HEP phenomena correspond to renormalised states it isn't expected that the full 
features of the Standard Model should be seen directly in the bare broken terms of 
$\ese$ on $F(\htho)$
 for example.  In QFT the bare Standard Model Lagrangian, with the Higgs field added 
in a relatively unnatural way, does mimic the processes of HEP to some extent.
 For the present theory, intended as an underlying fundamental theory,  the fact that 
a number of features placed by hand into the Standard Model Lagrangian have already 
been reproduced, as summarised in section~\ref{sosmfi},  suggests that further 
specific details of empirical phenomena might be uncovered for the complete theory.
 These empirical details include in particular the 18 free parameters of the Standard 
Model as summarised in table~\ref{SMparams}.

\begin{table}[htbp]
\centering
\begin{tabular}{|l|c|l|}
 \hline
   SM Parameters    &  \#   &   Origin in present theory  \\  
 \hline					
    Fermion Masses  & 9 &   $\psi \leftrightarrow \bv_4$ coupling  in $\lvh$ terms  \\
	                &   &   equations~\ref{hexpan2} and \ref{qxmass}    \\
   \hline
  Gauge Couplings   & 3 &   $\psi \leftrightarrow Y$ coupling  in $\dmlvh$ terms  \\
                    &   &   equations~\ref{dlvpbbz} and \ref{dlvpbbzth}   \\
  \hline
  Higgs Potential   & 2 &   $\bv_4 \inn \TM_4$ projected from full $\bvh$   \\
                    &   &   equation~\ref{v4vac}, with $\vert \bv_4 \vert = h_0$ 
stable  \\
  \hline
  Quark Mixing CKM  & 4 &   mass and gauge couplings for 3 generations   \\
                    &   &   may require `$\ee$ on $\lvtfe$'   \\
  \hline
  \end{tabular}
  \caption{\setb The 18 parameters of the Standard Model and their correspondence in 
the present theory. All essentially originate as couplings implied in $\lvh$ as 
exemplified in the above equation references, including further parameters for the 
neutrino sector.}
\label{SMparams}
\end{table}

  The QCD $\theta$-parameter, introduced in equation~\ref{stocp} and which is 
consistent with zero empirically,  is not included in the table since the 
corresponding field interaction terms do not arise in the present theory, as described 
in section~\ref{secdos}. On the other hand the new structures
 presented in this paper may imply new kinds of interaction terms  which do have 
empirical consequences. 
 As well as identifying new processes the present theory may be tested through its 
ability to reproduce the details of known phenomena through the interactions listed in 
table~\ref{coneqco}.

   As noted in that table,
  these include observations of the large scale structure in cosmology, which may 
relate to variation in the magnitude $\vert \bv_4 \vert$ under 
  \mbox{$G_{\mu\nu} = f(\bv_{56})$}.
 In addition to accounting for the Standard Model particle properties the complete 
theory would aim to provide a match for the cosmological data, including the density 
parameters $\Omega_B$, $\Omega_D$ and $\Omega_{\Lambda}$ introduced in 
section~\ref{sectsmoc}, and the structure of the cosmic evolution generally.
  In particular the Lorentz scalar components $\alpha, \beta, n$ and $N$ of $F(\htho)$ 
in equation~\ref{fhthopart}, which also transform trivially under the 
 $\suth_c \times \uo_Q \subset \ese$ gauge group while effectively acquiring mass 
through interactions with the vector-Higgs $\bv_4$ under the terms of $\lvfs$, may 
contribute to the dark sector in cosmology, as discussed in section~\ref{secpotnt}.

  Other known phenomena are not explicitly expressed in table~\ref{coneqco}. 
 An example is provided by the CKM quark mixing parameters alluded to in 
table~\ref{SMparams}, which can be expressed explicitly in the Standard Model 
Lagrangian as described for equation~\ref{lagqwckm}. 
In the Standard Model the phenomena of CKM mixing arise for the three generations of 
quarks  due to the mismatch between the Yukawa and gauge couplings, as described 
towards the end of section~\ref{ewtatsm}. While  fermion masses and gauge couplings 
arise in the present theory as indicated in the upper half of table~\ref{SMparams}, 
the further necessary ingredient  of three generations required for CKM mixing may 
require a further extension to for example an $\ee$ symmetry acting upon the 
hypothetical  form $\lvtfe$ as discussed in section~\ref{sosmfi}.

  Further parameters for three generations of neutrino masses and corresponding mixing 
phenomena are also needed as a known extension to the  Standard Model, and are 
presumed to have a similar origin as described above for the quark sector in the 
present theory.
As also suggested in section~\ref{sosmfi}  the $\sutw_L$ internal symmetry may play an 
essential role in distinguishing three generations of fermion states. 
 It will also be required to identify neutrino and $u$-type quark states that 
transform as $\sltc^1$ Weyl spinors and hence form $\sutw_L$ doublet partners with 
charged lepton and $d$-type quark Weyl spinors respectively, which may also involve 
the identification of a full $\ee$ symmetry action on $\lvtfe$.

  The phenomena of electroweak symmetry breaking arise since the $\sutw_L \times 
\uo_Y$
 symmetry action itself also impinges on the components of the external vector-Higgs 
field $\bv_4 \inn \TM_4$. These
 interactions of the $\sutw_L \times \uo_Y$ gauge fields 
   account for the massive nature of the $Z^0$ and $W^{\pm}$ gauge bosons as described 
in subsection~\ref{suboomahp}.
  That is the masses of all particles, fermions and gauge bosons,  are here postulated 
to originate through field interactions with the components of $\bv_4(x) \inn \TM_4$ 
rather than with a 
 fundamental scalar Higgs field. The large mass of the $Z^0$ and $W^{\pm}$ bosons, of 
the same order as that of the empirically observed Higgs boson, will need to be 
understood
 in the context of the present theory. Indeed,
  the Higgs particle state itself will also need to be identified within this theory, 
echoing the empirical search for the Higgs which concluded successfully in 2012 at the 
Large Hadron Collider.

In the Standard Model the masses for the $Z^0$, $W^{\pm}$ and Higgs boson can be 
expressed in terms of gauge coupling and Higgs parameters of the left-hand side 
table~\ref{SMparams} as described in section~\ref{ewtatsm}. 
 The scalar Higgs field $\phi$ exhibits self-coupling, with terms such as 
$\phi^{\dag}\phi$ and $(\phi^{\dag}\phi)^2$ in the Lagrangian potential  of 
equation~\ref{higgspot}, as contrived to break the symmetry of the vacuum.
 Within the new approach the scalar Higgs is provisionally identified with the 
magnitude $h(x)$ of the vector-Higgs $\bv_4(x)$ as projected onto $\TM_4$ such that  
the relation  $L(\bv_4) = \vert \bv_4 \vert^2 =  h^2$ is directly identified within 
the full form $\lvh$.
  Cubic and quartic field couplings, within the terms of $\lvt$ and $\lvfs$ 
respectively,
  involving the  components of $\bv_4$ (coupled with combinations of the four scalar 
fields from the $\alpha$, $\beta$, $n$ and $N$ components of $F(\htho)$ for example, 
as can be seen in equation~\ref{fquartall}), generate an effective potential $V(h,T)$, 
which may be dependent upon an apparent temperature $T$, as described in 
section~\ref{sectveu}. 
 For the new approach yet further possible interactions will arise for higher-order 
field exchanges or a higher-dimensional full form of time.
  An initial unstable value of $h(t)$ has been considered for the extreme spacetime 
environment of the very early universe as discussed in section~\ref{sectveu} in 
relation to inflationary theory, with the stable value $h(t) = h_0$ achieved at cosmic 
time $t=t_v$ marking a phase transition. 

  In chapters 6--9 of this paper the emphasis has been on the identification of known 
Standard Model properties from within the structure of the present theory, as 
summarised in the four bullet points and further discussion in section~\ref{sosmfi}.
 The further ambition is to develop the theory to the point of making new
empirical predictions that  might be tested in existing and future laboratory 
experiments in particle physics as well as through observations in cosmology. Such 
theoretical predictions could be worked out concurrently with the running of the LHC 
in time to anticipate new effects that may appear in the data analysis. The 
predictions might also influence the design specifications for the future 
International Linear Collider.

  For the present theory in addition to breaking the full symmetry of $\lvh$ through 
the choice of the projected vector $\bv_4 \inn \TM_4$, with the stable value of $\vert 
\bv_4 \vert = h_0$,
 symmetry breaking is also exhibited through the choice of particular components for 
the vector-Higgs $\bv_4$  in the local tangent space on the 4-dimensional manifold. 
This choice, represented in figure~\ref{vphaset}(c) with exaggerated fluctuations 
about the mean value, is analogous to the choice of component contributions for the 
Standard Model scalar Higgs vacuum value in equation~\ref{higgsvac}. However, due to 
the difference in underlying structure, differences between the Standard Model Higgs
 phenomena and predictions of the present theory might be observable in the laboratory 
environment.

  In considering the hypothetical structure of an $\ee$ action on a form
   \mbox{$\lvtfe$} the possibility of identifying the external spacetime vector $\bh_2 
\equiv \bv_4 \inn \TM_4$ by fusing together a set of two or three right-handed spinors
 $\{\theta_{Y\!\lag}, \phi_{Y\!\lag}, \psi_{Y\!\lag}\} \inn \ccc^2$ under $\sltc^1 
\subset \ee$  was described alongside equation~\ref{htwofuse} in section~\ref{sosmfi}.
 In turn there are a number of ways of identifying scalars from the components of the 
above three spinors, including the scalar magnitude $\vert \bv_4 \vert = h$.
 This in principle opens up the possibility of identifying additional Higgs-like 
states, beyond the earlier possible scalar states that might be associated with the 
$\alpha$, $\beta$, $n$ and $N$ components of $F(\htho)$ for the $\ese$ case.
 In addition to a direct search for such scalar states at the LHC
 an $e^+e^-$ collider tuned to operate as a `Higgs factory' might be sensitive to  
some of the observable consequences.     
 Since the employment of the three spinors in this way corresponds to the empirical 
absence of a set of three generations of right-handed neutrinos, these structures may 
also impact upon the neutrino sector in a manner beyond the Standard Model. 
 
 Considered in general terms the extension to an $\ee$ symmetry itself also suggests 
the possibility of new gauge bosons beyond the Standard Model deriving from the extra 
$\sutw \times \uo$ that is appended to the familiar Standard Model symmetry in the 
rank-8 decomposition of equation~\ref{eedecomp}  in section~\ref{sosmfi}. However, the 
first objective is a mathematical one in identifying the predicted $\ee$ action on a 
quintic or higher order form $\lvtfe$ itself, as highlighted in the previous section, 
and to assess the further extent to which known Standard Model properties might be 
recovered before considering additional empirical consequences in great detail. In the 
meantime the general manner in which particle states might be described from a 
conceptual point of view can be further elaborated as we now consider.

  Under the assumption of a global flat spacetime in the laboratory the Lorentz 
symmetry may be augmented to the 10-parameter Poincar\'{e} group and particle states 
classified by their mass $m$  and spin $s$ (or helicity $h$ for $m=0$)   according to 
the values of $(m^2)$ and  $(m^2)s(s+1)$ (for $m\neq 0$) they take respectively  for 
the two Casimir operators $P_{\mu}P^{\mu}$ and $W_{\mu}W^{\mu}$, where $W^{\mu}$ is 
the Pauli-Lubanski vector.
 This applies to all particle states, including hadrons composed of quarks and the 
Higgs scalar which is presumed to be composed out of the collection of non-scalar 
field components of the vector-Higgs field $\bv_4$ in the present theory, as recalled 
above (with an analogous construction for technicolor models reviewed in 
subsection~\ref{suboomahp}).

The four Weyl spinors of equation~\ref{thcth234} identified in the components of 
$\theta^1$ in  section~\ref{extsym}
 relate to projected components of the larger Dirac spinors, which in turn can be 
 identified within the components of $F(\htho)$ in equation~\ref{fhthopart} under the 
action of $\sltc^1 \subset \ese$.
 The fermions of the Standard Model are Dirac spinors, with differing properties for 
the projected left and right-handed Weyl spinor parts as reviewed in 
chapter~\ref{rotsm}.
 These different properties arise here through the
 necessarily asymmetric embedding of the vector-Higgs $\bv_4 \inn \TM_4$ with respect 
to the $\mcX, \mcY \inn \htho$ subspaces of $F(\htho)$ and the resulting 
   asymmetric action of an internal $\sutw_L \subset \ese$ symmetry on these 
components in equation~\ref{fhthopart}.

 As described in section~\ref{intsym} alongside the $\uo_Q$ symmetry of 
electromagnetism 
  the broken $\esi$ symmetry on the space $\htho$ also includes $\suth_c$ as a pure 
internal symmetry, to be associated with massless gauge bosons, the gluons of QCD, in 
the Standard Model. In subsection~\ref{submangle} it was described how this $\uo_Q$ 
symmetry  survives the breaking of an $\sutw^2 \times \uo^2 \subset \esi$ symmetry in 
a `mock electroweak theory', as a provisional guide towards the identification of an  
$\sutw_L \times \uo_Y$ symmetry within $\ese$ or $\ee$ acting on the full temporal 
form $\lvh$ in the complete theory.

Combining the above external properties under the Poincar\'e symmetry with  full set 
of
internal quantum numbers according to the transformation properties under $\SML$ will 
lead to a classification of  particle states for a more thorough comparison with  the 
Standard Model framework.  That the enormous wealth of experimental data in high 
energy physics all points to a concise and simple table of a relatively small number 
of elementary particles, the fermions and bosons, as summarised in the Standard Model 
of particle physics with the 18 parameters of table~\ref{SMparams} above, further 
motivates the aim to determine such particle properties in the present theory by 
taking a mathematical limit or approximation that mirrors the physical conditions to 
be found in such laboratory experiments.

 In order to make contact with terrestrial laboratory experiments in HEP it will be 
necessary to proceed from the ideas presented in this paper through practical 
calculations for processes such as those in figures~\ref{figsld} and \ref{eemmeds} and 
beyond to more general, and even novel, applications. In the particular case of 
figure~\ref{figsld}  out of the general solutions $G_{\mu\nu} = f(Y, \bvh)$ over $M_4$ 
the emergence of the initial $e^+$ and $e^-$  states, an intermediate
$Z^0$ boson and the final state particles will need to be described.
  Out of the annihilation of the particle and antiparticle in
 the centre-of-mass system in figure~\ref{eemmeds}(a)  a large number of field 
transmutations are possible, whether through a  photon or a $Z^0$ boson state,  
allowing a large number of possible $\delta \psi \leftrightarrow \delta Y$ field 
exchanges and further states to be produced. These include the leptonic final state 
depicted in figure~\ref{eemmeds}(a) as well as the hadronic jets seen in 
figure~\ref{figsld}, resulting from quark pair production,  
 together with all the particle states within the jets. These and further particle 
phenomena need to be accounted for within the structure and constraints of the present 
theory, as has been described in chapter~\ref{newapp}.

 One way of approaching the nature of particle  states might be to consider the simple 
decay process $Z^0 \to e^+e^-$ via $\delta Y \leftrightarrow \delta \psi$ field 
exchanges resulting in the propagation of two independent fermions. This would also 
require an understanding of the $Z^0$ gauge boson mass in terms of $\delta Y 
\leftrightarrow \delta \bv_4$ interactions, incorporated into a solution $G_{\mu\nu} = 
f(Y)$ for a massive gauge field with $k^2 = m^2 \neq 0$, possessing a third 
polarisation state $\varepsilon^{\mu}_3$, and which satisfies equation~\ref{sqmassa}.
 Similarly a Higgs decay process such as $H \to e^+e^-$ could be studied directly in 
terms of $\delta \bv_4 \leftrightarrow \delta \psi$ field exchanges, closely relating 
to the mechanism  for fermion production during the phase transition at $t=t_v$ in the 
very early universe described in section~\ref{sectveu}.

 On the other hand a purely QED process might be considered with the electromagnetic 
field $A^{\mu}(x)$ interacting with fermions. Since the photon is massless a possible 
approach would be to take 
   a superposition, or sum, of electromagnetic fields, each in the form of 
equation~\ref{kcoeff}, mimicking the situation of a two-photon collision and hence 
able to produce fermion pairs, as alluded to near the opening of section~\ref{seraps}.
  The nature of a single intermediate photon state, effectively with $k^2 \neq 0$, in 
the centre-of-mass frame of an $e^+e^-$ collider might also be considered. The 
production of fermions would be required to proceed through
  field redescriptions of the form $A^{\mu} \leftrightarrow   
\ol{\psi}\gamma^{\mu}\psi$, as initially discussed for figure~\ref{apsirelab}, 
consistent with the constraint equations~\ref{conequa} under a  geometric solution for 
$\gmyv$.

  A consistent normalisation of the fields will be required in field exchanges of the 
form $A^{\mu} \leftrightarrow  \ol{\psi}\gamma^{\mu}\psi$, under the local geometry 
$\gmyv$ with $\gmo$, linking external and intermediate field states. 
 This will relate the $C^{\frac{1}{2}}$ coefficient and polarisation vectors 
$\varepsilon^{\mu}_{r}(\bk)$ for the electromagnetic field, as introduced in 
equation~\ref{kcoeff},   
 to the spinor coefficients for a Dirac field $\psi(x)$. In the standard theory there 
are four independent solutions to the free Dirac equation  labelled by the 4-component 
coefficients $u^{1,2}(p)$ and $v^{1,2}(p)$, with for example $\psi(x) = u^1(p)e^{- i 
p\cdot x} $
 which may be normalised by kinematic factors of energy and mass
 (see for example \cite{Pesk} sections 3.3 and 5.2).
  Similarly in the present theory the coefficients of the electron field $\psi(x)$ for 
example will contain energy $p^0$ and mass $m$ factors which will need to match those 
for the normalisation coefficients of the electromagnetic field $A^{\mu}$ in Fourier 
mode expansion exchanges between the fields under $G_{\mu\nu} = f(A,\psi)$. In all 
cases such `kinematic factors' arise from `numerical parameters' such as $p \inn 
\rrr^4$ in the Fourier modes $e^{\pm i p\cdot x}$ themselves. As well as being 
mutually compatible these normalisation factors will ultimately translate into the 
appropriate dimensions for cross-section calculations, as described towards the end of 
section~\ref{secdopp}

 In the environment of HEP experiments it is generally assumed that the spacetime is 
flat and a Minkowski coordinate system employed such that
 the external Lorentz
  connection has components $A^a_{\ph{a}b\mu}(x) = 0$, corresponding to a  linear 
connection $\Gamma (x) = 0$ by equation~\ref{AtoG}. Transforming 
under the global Lorentz symmetry  the components of the 
 4-component Dirac spinors $\psi(x)$ are
 normalised as alluded to above.
  The Lorentz connection $A^a_{\ph{a}b\mu}(x)$  acts on a Dirac spinor $\psi(x)$  
through the associated spinor connection as a representation of the Lorentz symmetry. 
This structure can also be applied to the
 more general case of a curved spacetime, 
 employing a spinor bundle over $M_4$ to express the dynamics of the Lorentz 
connection, with $A^a_{\ph{a}b\mu}(x) \neq 0$ in general, on the base manifold in 
relation to spinor fields.
 As described towards the end of the previous section, a starting point might be to 
develop a minimal model based on the full symmetry $\hat{G}=\slthc$ acting on $\bv_9 
\inn \hthc$ leaving the form $L(\bv_9) = 1$ invariant. For this model fermion states 
derive from the Weyl spinor $\psi_L$ in equation~\ref{vinhinc} in interaction with an 
internal $\uo$-valued gauge field, in principle describing a model for QED.

  As well as classifying particle states such as gauge bosons and fermions in a 
representation space according to their transformation properties under the external 
and internal symmetry groups and their possible interactions, the structure of 
tangible physical particles in spacetime as detected in experiments can also be 
investigated. Physical particles evidently transfer energy and momentum, which can be 
described by the tensor $T_{\mu\nu}(x)$ and is presumed to be conserved in 
4-dimensional spacetime. In the present theory energy-momentum is \textit{defined} by 
the relation $T_{\mu\nu} := G_{\mu\nu}$ (within a practical normalisation factor of 
$-\kappa$), and hence the transfer of a finite amount of energy must necessarily be 
associated with $G_{\mu\nu} \neq 0$ and hence a non-flat spacetime, while the identity 
$\gmo$ also ensures energy-momentum conservation throughout. In turn this tangible 
spacetime form of a particle is expressed in terms of the underlying fields as a 
solution for $\gmyv$.
  This smooth external geometry represents a macroscopic `dressed' or `renormalised' 
object  constructed out of the underlying microscopic `bare' field exchanges.
 Representing the electron beam in a HEP accelerator for example, 
 observable properties  associated with the energy-momentum for the electron field are 
carried by the tensor: 
\begin{equation}
 \label{tfromgyv} 
T_{\mu\nu} := \gmyv
\end{equation}

    This is equation~\ref{getypsi} of section~\ref{subwal}, where a particular vector 
space representing the full temporal flow $\bvh$ may be substituted in.
  The expression $G_{\mu\nu} = f(Y, \bv_{56})$ implies an underlying  innumerably  
nested sequence of indistinguishable field descriptions under $G_{\mu\nu}(x)$. This 
geometry is entirely constructed out of field components derived from $\lvfs$ and the 
corresponding $\ese$ symmetry actions. However, in this theory, it seems quite 
possible that some components of the fundamental form $\lvfs$ and the gauge fields may 
exist on $M_4$ \textit{without} contributing to the geometry field $G_{\mu\nu}$. With 
$T_{\mu\nu} := G_{\mu\nu}$ this would imply that not all fields in spacetime have 
energy-momentum in the sense of $T_{\mu\nu} \neq 0$. This possibility was discussed in 
section~\ref{secuni} and compared to the case of gravity waves
 which, while associated with a geometry with $G_{\mu\nu}=0$,
carry energy via a finite Weyl curvature as described after equation~\ref{cbian} in 
section~\ref{subwal}.
 Here we consider the measurable phenomena of HEP particle types and properties to be 
determined by the mutual constraints of equations~\ref{conequa} applied to the 
underlying fields and  conveyed via energy-momentum in the form of the generalised 
expression 
 of equation~\ref{tfromgyv},
 as originally employed for the special case of the free electromagnetic field leading 
to figure~\ref{gacos} in section~\ref{secdos}.

 While a significant correlation between the structures of the present theory and 
calculations in QFT has been identified as described in sections~\ref{secdos} and 
\ref{secdopp}, a key question remains regarding the precise conceptual form and 
mathematical expression of the nature of field quantisation. One major aspect concerns 
whether the projected field components themselves are effectively fragmented into 
discrete elements distributed over spacetime and related via $\delta Y(x)$ and $\delta 
\bvh(x)$ differences, as has typically been conceived as the theory has developed, 
with the components of the external gravitational field composing the only smooth and 
continuous functions on $M_4$. An alternative view might see \textit{all} fields 
smooth and continuous on $M_4$, with discrete exchanges only in the local 
contributions to $G_{\mu\nu}(x)$ in equation~\ref{tfromgyv} consistent with 
equations~\ref{conequa}, considered as `excitations' of the fields and giving rise to 
observations of apparent quantum phenomena.
   A full understanding of this description of such quantum phenomena in the context 
of the present theory is one of the two main branches to be pursued as summarised at 
the end of the previous section.

  With \textit{all} physical entities described by equation~\ref{tfromgyv}, subject to 
constraints such as $\lvh$, this includes solutions that incorporate the phenomena of 
apparent particle effects, as discussed in section~\ref{seraps}. These solutions must 
describe the discrete emission and detection of the \textit{same} conserved 4-momentum 
$p$ with $p^2 = m^2$ and conserved charges, arising from the internal field 
constraints, giving the rather mechanical \textit{impression} of an intermediate 
`classical particle' or projectile of some form. As discussed in section~\ref{qpagig} 
the `particle tracks' that we construct by joining up detector hits, as depicted in 
figure~\ref{figsld} for example, reinforces this illusion of an independent 
particle-like entity pursuing a continuous trajectory.

  One way to approach the nature of the actual physical structure underlying such 
particle-like
   phenomena is to begin by considering a general state of macroscopic matter 
described by  $\gmyv$, as represented by the `bulky' geometry of 
figure~\ref{gtovac}(a), which might represent for example the matter content 
$T_{\mu\nu} := G_{\mu\nu}$ of ordinary `table and chairs'. Subsequently a progression 
down to a more minimal field content underlying a solution of $\gmyv$ can be 
considered, down to a stage that does not simply gradually fade away towards 
$G_{\mu\nu}(x) = 0$, but rather solutions for geometric structure emerge that take on 
the shape of a discrete set of topologies due to the discrete constraints on the 
underlying fields. In this case a somewhat `tubular' structure might arise as the 
vacuum limit is approached, as represented in figure~\ref{gtovac}(b).  These near 
vacuum conditions  correspond for example to the environment created in HEP 
experiments as described near the opening of section~\ref{sechepe}.

\vspace{10pt}  
\begin{figure}[htbp]  
\centering
\epsfxsize=\maxwidth
\leavevmode
\epsffile[0 0 1918 550]{\gpath aPfig152e}
\caption{\setb Representations of 4-dimensional solutions for $\gmyv$ for (a) the 
general case of ordinary extended matter (b) the discrete structure emerging as 
permitted by the underlying field constraints as the vacuum state is approached.}
\label{gtovac}
\end{figure}

  The pattern of inner lines in figure~\ref{gtovac}(b) are analogous to the contours 
on a map representing the altitude of a continuous physical terrain, with the geometry 
$G_{\mu\nu}(x)$ being perfectly smooth and continuous, as also for 
figure~\ref{gtovac}(a)  and all other cases. Hence this geometry might more accurately 
be represented by a continuous shading. Considered as a full 4-dimensional spacetime 
solution the contour tubes in the near vacuum region in  figure~\ref{gtovac}(b) 
connect and are \textit{continuous with} macroscopic entities such as HEP accelerators 
and detectors, as represented by the outer structure in the same figure. 
 The inner structure in figure~\ref{gtovac}(b), with time directed from left to right, 
might represent for example the overall particle interaction process $e^+e^- \to 
\mu^+\mu^-$, via an intermediate $\gamma$ or $Z^0$ state, which is typically pictured 
in terms of particle trajectories in 3-dimensional space as depicted in 
figure~\ref{eemmeds}(a) and described in section~\ref{qpagig}.

  While shaped by the discrete enveloping topology the spacetime geometry for such a 
process will also be modulated by a wave-like structure of a form similar to 
equation~\ref{geinawave} and figure~\ref{gacos},   corresponding to a particular 
4-momentum transfer.
 As also described in section~\ref{secdos} the spacetime metric $g_{\mu\nu}(x)$ itself
  associated with this modulation is presumed to take a form similar to 
equation~\ref{gmetwave}. In the overall solution of equation~\ref{tfromgyv} for such a 
process the left-hand side  `$T_{\mu\nu} := G_{\mu\nu}$' of the equation describes 
both the kinematic properties of the interaction via the energy-momentum tensor 
$T_{\mu\nu}(x)$ and the smooth external geometry $G_{\mu\nu}(x)$ as for general 
relativity. Through the right-hand side `$f(Y,\bvh)$' of the same equation  all 
quantum properties are sown into this structure in the form of an underlying set of 
discrete field redescriptions of the form $\delta Y \leftrightarrow \delta \bvh$, 
subject to the constraints such as $\lvh$, which determine in turn the possible set of 
discrete particle types and interactions that can be observed in HEP experiments.

   That is, while $G_{\mu\nu}(x)$ is perfectly smooth and continuous there is both a 
discrete set of apparent particle types and a discrete set of possible topologies, 
corresponding for example to  $n$-particle final states, that may be obtained for the 
near-vacuum solutions.
 This structure hence provides a coherent conception of the nature and properties of 
particle states observed in the laboratory. For example a continuous range of 
conserved momenta is available for the apparent emission and detection of a fermion 
state  within the discrete 
 constraint $p^2= m^2$, corresponding to an apparent particle mass $m$ which arises 
from the underlying interactions between the particular fermion field $\psi(x)$ and 
the vector-Higgs field $\bv_4(x)$.

 The metric $g_{\mu\nu}(x)$ for the external geometry depicted in 
figure~\ref{gtovac}(b)
  represents a particular solution for $\gmyv$ on the macroscopic scale of HEP 
experiments, similarly as the 
  Schwarzschild metric of equation~\ref{ttrtp} represents a particular  macroscopic 
solution on a much larger scale.
   Unlike the large scale case, for which the precise trajectory of planetary orbits 
and the deflection of light passing near the sun is observable, it is clearly not 
possible in practice to send `test particles' through the laboratory environment of 
figure~\ref{gtovac}(b) in order to map out the spacetime curvature (although such a 
project can be readily conceived in terms of a  thought experiment, as for that 
involving geodesic deviation due to the geometry of intense beams of light as 
described in  section~\ref{qpagig}).

 However, crucially for the present theory, this non-trivial external geometry with 
metric $g_{\mu\nu}(x)$ is a physical characteristic of a possible solution for $\gmyv$ 
and the test of this proposal, which will require all elements of the full theory, 
will rest on the ability to identify HEP processes which are actually observed and to 
predict new phenomena.  This will involve both the determination of the internal 
quantum numbers of the apparent particle types, as implied in the underlying field 
structure $f(Y,\bvh)$ for such a process, and in particular the apparent kinematic 
constraints on the 4-momentum $p$ transferred, where with $p^2 = m^2$ and 
$T_{\mu\nu}:=G_{\mu\nu}$ the invariant mass $m$ provides a direct characterisation of 
the external geometry itself.

  Within the field constraints more generally a range of topologies which are rather 
more complicated than that depicted in figure~\ref{gtovac}(b) will arise. 
 For example the  process recorded in   
figure~\ref{figsld}  is identified as an $e^+e^- \to Z^0 \to b\bar{b}$ event in the 
analysis of~\cite{sld}. Such a process typically involves `particle tracks', as shown 
in the event picture, each of which apparently emanates from one a sequence of 
vertices, each of which in turn is associated with the $Z^0$ boson itself or a $B$ or 
$D$ hadron in a subsequent decay chain. With generally five such decay vertices for 
each such event mutually separated by typically a few millimetres, within the volume 
of the detector for which the closest devices are a few centimetres from the 
interaction point, the topology of the apparent particle-like structure described by 
the solution $\gmyv$ will be relatively intricate for these processes.

  Yet other forms of solutions for $\gmyv$ may  appear  less `particle-like' as for 
the case of an  $e^-$ state apparently simultaneously `passing through both slits'  in 
the experiment depicted in figure~\ref{eemmeds}(b). The overall geometry 
$G_{\mu\nu}(x)$ for the set-up of figure~\ref{eemmeds}(b) for the case of a high 
intensity electron beam, with the full interference pattern clearly observed on the 
final screen, will be of a macroscopic form as described for figure~\ref{gtovac}(a) 
above. As the intensity is turned down, corresponding to a transition towards a near 
vacuum solution as exemplified in figure~\ref{gtovac}(b), an overall geometry will 
emerge incorporating the transfer of an apparent single $e^-$ particle from the source 
$S$ to the detector hit $A$ in figure~\ref{eemmeds}(b) in continuity with the 
structure of the apparatus of the double-slit experiment.
 The geometry of such a solution serves to emphasise the fact that a `particle' should 
not be  considered as a kind of localised entity in the form of an `energy-knot' 
propagating in 3-dimensional space (see for example \cite{Weyl2} pp.202--204), but 
rather as an apparent phenomenon associated with a particular kind of smooth extended 
4-dimensional solution for $\gmyv$ constructed over the underlying field 
possibilities.
 Similar 4-dimensional spacetime solutions will also incorporate the phenomena of 
quantum entanglement and EPR experiments as discussed in section~\ref{qpagig}.
 In many cases however a solution for \mbox{$\gmyv$} will take a form consistent  
   with the notion of a localised propagating particle-like entity.

   Although in the present theory there are also no fundamental `string-like' objects, 
there may be some relation to string theory (for which there are also no fundamental 
particle entities) in that diagrams with a similar topology to  that of the inner 
structure in figure~\ref{gtovac}(b) also appear in relation to string theory 
calculations. Here however rather than describing the trajectory  and interactions of 
a set of closed strings the tubular contours in figure~\ref{gtovac}(b) purely 
represent the structure of an extended 4-dimensional geometry. 
 In string theory such a diagram correlates with the `tree level' process as 
represented by the Feynman diagram of figure~\ref{fdeemm} for example, while for the 
present theory 
 figure~\ref{gtovac}(b) represents the full physical process with arbitrarily nested 
field exchanges implied under the solution $\gmyv$.
 However, although the inner structure of figure~\ref{gtovac}(b) in relating to a 
process such as $e^+e^- \to \mu^+\mu^-$ has a very different physical and conceptual 
meaning to analogous diagrams featuring in string theory, 
  some of the mathematical properties of topological structures in 4-dimensional 
spacetime   might be jointly applicable.

  Rather than the phenomena of a discrete spectrum of particles being determined by 
the vibrations and tension of hypothetical strings, here such phenomena are generated 
by the possibility of actual underlying field redescriptions subject to the 
constraints of equations~\ref{conequa}.
As considered above a practical  starting point may be to identify QED processes 
involving electron-photon interactions, such as with Bhabha or Compton scattering 
events, in this unified framework alongside general relativity. This study might begin 
with a model based on the full symmetry $\slthc$ for the form $L(\bv_9) = 1$ before 
generalising to the octonion case with a full $\sltho \equiv \esi$ symmetry acting on 
the form $\lvt$.
 The action of the internal $\uo_Q \subset \esi$ symmetry generated by  $\Sbard^1_l$ 
on the spinor components of $T\htho$, as seen for example in equations~\ref{dlvpbbz} 
and
 \ref{dlvpbbzth} in the terms of the field constraint equation
 $\dmo$,  gives rise to the phenomena of electrodynamics. 
 The precise manner in which the factors of $\vert \dot{s}_f \vert = 1$ or $\vert 
\dot{s}_f \vert = \frac{1}{3}$ in this expression translate into the corresponding 
factor of three in charge ratio for physical renormalised  particle states,
 as discussed for figure~\ref{mumudd} in section~\ref{secdopp} in the context of 
cross-section calculations, will need to be determined alongside the full 
understanding of the 
  structure of quantum phenomena and particle states themselves.

The above QED phenomena will generalise for the complete internal symmetry identified  
in the breaking of the $\esi$ symmetry over the extended external $M_4$ manifold, and 
then further with the full symmetry of time identified as $\ese$ or even $\ee$ on the 
full form of temporal flow $\lvh$.
 The insight gained from the $\uo_Q$ case might then be extended for the remaining 
internal generators to identify further features of the Standard Model and beyond as 
they arise naturally out of the complete theory.
 It is likely that the full package will be required with all the features of 
figure~\ref{fronts4} combined together, and the full set of possible fields and field 
interactions incorporated, in order to determine specific quantities such as the 
electron mass and the full set of Standard Model parameters as summarised in 
table~\ref{SMparams}, including the neutrino sector, generally.

    In conclusion, the field and particle content of the present theory, in terms of 
figure~\ref{fronts4}, includes the external gravitational and internal gauge fields 
which arise from the symmetries of $\lvh$ and are  mutually related as described for 
`front~(1)', together with
 the fermion and `vector-Higgs' fields identified from the $F(\htho)$ components 
studied for `front~(2)'. Consistent with the gauge invariance of the constraint 
equations the non-gravitational fields mutually interact to form combinations under 
possible solutions $G_{\mu\nu} = f(Y,\bvh)$ for the world geometry on $M_4$ as 
described for `front~(3)',
 taking into account the intrinsic warping of the spacetime geometry due to variation 
in
  $\vert \bv_4(x) \vert$ and the role of the scalar field components as studied for 
`front~(4)'.
 In order to develop this theory further and establish full contact with the results 
of HEP experiments, as well as with empirical observations in cosmology and physical 
phenomena more generally, the four fronts of figure~\ref{fronts4} will need to be 
further developed and combined as provisionally outlined in the previous section.

\section{Concluding Remarks}

   While emphasising the possibilities for progressing outwards from the structure of 
figure~\ref{fronts4} the present theory is based upon the multi-dimensional form of 
temporal flow $\lvh$, at the centre of the figure, which in turn derives from the 
simple structure of  one-dimensional  progression in time as described in 
section~\ref{gfotf}. With both the familiar four dimensions  of the extended spacetime 
manifold $M_4$ and the `extra dimensions', which are associated with the properties of 
physical objects in spacetime, deriving from a single temporal dimension the question 
concerning the origin of time itself is inevitable. A naive further reduction down to 
`zero dimensions' together with a contrived argument to generate one dimension is not 
considered here to be of any great value. 
 On the other hand the observation that the arithmetic properties of multiple 
dimensions 
 are implicit within the arithmetic structure of the real line $\rrr$, as described in 
section~\ref{gfotf},  provides a natural and major motivation for the present theory. 
A second founding motivation for the entire theory is the apparent necessity for any 
and every subjective \textit{experience}, including our observations of the physical 
world, to take place in time. This conception of the theory itself implies a 
subjective nature for the origin of time and  leads to the conclusions described in 
chapter~\ref{chaptoot}, and in particular to the `universal foundation' for the theory 
depicted in figure~\ref{tcycle}.

  This overall structure can be considered as a  \textit{system} rather than just a 
\textit{theory} (in the usual sense of the word) -- it is intended not merely to 
represent the world by a \textit{model}, but rather it aims to describe the way the 
world actually \textit{is}, and how it is possible for it to \textit{be}. This is in a 
similar spirit that a biologist, for example, might describe the system of a living 
organism -- although finding such a metaphor for the whole system is particularly 
problematic due to its unique and all-embracing nature.

   It is a system founded upon general experience of \textit{living in the world} as 
well as upon knowledge gained from the high energy physics laboratory together with 
cosmology and from scientific observations in general. Indeed all such experiments and 
observations are just a refined and specialised form of our experience in the world. 
While the primary aim has been to demonstrate a unified theory that can account for a 
wealth of scientific data, and thereby also provide a means of verification of the 
ideas, it has also been considered desirable to incorporate the nature of experience 
itself in the world. This leads to a unification not only of experimental findings but 
also of science as a whole with our experiences of the world in general. Hence 
although much of the presentation has involved scientific knowledge, from particle 
physics to cosmology, the overall conceptual scheme arrived at is that of a world 
which one can feel oneself to be immersed or engaged within while walking down the 
street.

  While the physical laws and structures of the 4-dimensional world are carved out of 
the general flow of time, as filtered by the spacetime form of perception, the actual 
physical objects we encounter, such as complex organic life forms, are moulded to 
conform with the possibility of our actual existence in the world. The apparent 
stability of the perceived physical forms -- from inter-galactic structures to the 
insect world on Earth -- gives the illusion of a robust universe, independent of 
conscious life, constructed upon an independently existing material substratum, a 
notion upon which the early development of science also built its foundations. It is 
an illusion which continues to yield enormous practical advances in navigating our way 
around the physical world.

   Both time and space are direct forms of subjective experience of mathematical 
structures in the world, through which the physical world itself is created and 
sustained as 
 incorporated in node (4) of figure~\ref{tcycle}.
 Although a more rigorous mathematical description of all aspects of this structure is 
to be sought this does \textit{not} imply that the system of the world is itself 
fundamentally a `mathematical object'. Rather, as is the case in general, mathematics 
provides a precise and concise means of describing and elaborating both physical and 
abstract structures.    
 It is conceivable though that there may be essential properties of complex entities 
in the physical world such as the structure of the human brain which cannot be 
transcribed into a mathematical language which is both precise and concise enough for 
an exhaustive and practical description.
 Such a physical entity is of course `still there' even if it cannot be succinctly 
expressed in mathematical terms, in which case
  a mathematical approximation to nature might still be employed for practical 
purposes.

 For the present theory mathematics offers a precise, quantitative language for the 
scientific study of the conceptual, organic interplay between the physical world and 
conscious observer as represented in figure~\ref{tcycle}. However, while there is 
considerable scope for further mathematical development of the theory the time cycle 
structure can be conceptually and logically  coherent even if it may be humanly 
difficult to comprehend or develop a precise mathematical description of certain 
elements, such as for nodes (5) and (6) of figure~\ref{tcycle}, or if such an element 
does not directly correlate with a mathematical expression in a sense that we might 
recognise from familiar textbook maths. These elements of the theory may be 
correspondingly harder to both investigate in full detail as well as model in 
mathematical terms. Regardless of these practical difficulties the fact remains 
ultimately that we do `see' the world through a one-dimensional progression in time 
(in a similar sense that we see some objects as `green' as described towards the end 
of section~\ref{sectleitp}).
 This continuous temporal progression is inseparably  fused together with all 
subjective experience as a fundamental characteristic of all experiences.

  Taking the 1-dimensional flow of time in node (1) of figure~\ref{tcycle} to be 
modelled accurately by an interval of the mathematical real line $\rrr$ can itself be 
considered as a provisional assumption. This can be justified since 
\textit{experience} of a moment of time has the very simple structure of a continuous 
one-dimensional progression which may be uniquely and unambiguously represented by the 
properties of the real line. This assumption may be further justified by empirical 
tests of the consequences of the theory derived, via nodes (2) and (3), for node (4) 
of figure~\ref{tcycle}. 

 Again, further stepping around the cycle in this figure, the employment of  tractable 
mathematical language may fall short of providing an accurate and unambiguous account 
of the full nature of the self-reflective structures $R$   
 which are central to nodes (5) and (6) of figure~\ref{tcycle}.
 The use of the mathematical structures relating to G\"{o}del's theorem and 
undecidable propositions $G$ in  section~\ref{sectleitp} marked a provisional attempt 
to model such a structure, although in a manner that seems far too simplistic.

  However, further mathematical development of this aspect of the theory is both 
desirable and possible, with the aim of identifying a more precise description of the 
progression of self-reflective physical states, as crudely represented in 
figure~\ref{rinworld}, in mathematical terms. This may involve a degree of 
approximation based on a statistical approach to the phenomena of systems composed of 
many parts, by analogy for example with the thermodynamic properties of entropy. Even 
if such a mathematical structure remains somewhat elusive the conceptual ideas 
regarding the notion of subjective temporalisation might in principle be tested to 
some extent against  empirical findings in the field of neuroscience. Some of the 
ideas presented might also be of relevance in the field of artificial intelligence (as 
initially discussed at the end of section~\ref{secacfc}) featuring for example the 
design of a  device as a 4-dimensional entity in spacetime incorporating a structure 
of internal temporalisation -- that is a machine not just programmed to do things in 
time but also capable of internally representing a potentially subjective temporal 
structure itself.

  In contrast to these more speculative elements of the theory the full mathematical 
expression of the upper half of figure~\ref{tcycle}, beginning with the objective flow 
of time modelled by an interval of the one-dimensional real line $\rrr$ and leading 
via the multi-dimensional form of temporal flow $\lvh$ and its symmetries to the 
extended spacetime arena of the physical world as constructed through one of a myriad 
of solutions for the expression $\gmyv$, is in principle highly testable and has also 
been by far the main focus of the present theory.
  Amidst the resulting quantum phenomena the full theory can be applied to the 
observations of HEP experiments as modelled by the techniques of QFT and expressed in 
the form of the Standard Model of particle physics, and here arising from the $\ese$ 
symmetry of a 56-dimensional form of time.
The external theory of general relativity, describing gravitational phenomena, is here 
unified with the internal theories of gauge fields and particle physics through the 
projection  of the form $\lvfs$ and breaking of its symmetry in the identification the 
spacetime manifold $M_4$ as an arena for perception in the world.

    Through these ideas the present theory also incorporates the subjective way in 
which \textit{we experience} an apparently classical world of Newtonian material 
objects.
  Although having its origins in the fundamental notion of progression in time and 
perception in space the theory has developed with large scale cosmology and the 
Standard Model of laboratory particle phenomena in mind,
 resting heavily upon knowledge accumulated by the experimental and theoretical 
communities over recent decades to draw out the system of the world presented in this 
paper. 
 The theory is expected to be profusely testable   in terms of determining
 the extent to which the known form of the physical world can be ascertained  from the 
basic conceptual ideas of the theory  in addition to making new predictions for as yet 
unobserved phenomena which might be discovered.
 Indeed the properties already deduced from the theory, in matching a number of 
features of the Standard Model mark a first success for the theory. This success is 
summarised in section~\ref{sosmfi} where further progress is proposed in seeking an 
$\ee$ symmetry of an appropriate form $\lvtfe$  as a mathematical prediction of the 
theory. 

 The other principle area for study in the next stage of developing the theory is 
towards
 a more detailed understanding of the application of statistical methods and 
renormalisation techniques  for the present theory in relation to QFT.
 The phenomena of `running coupling' will be of relevance here and the extrapolation 
of the three gauge couplings from the laboratory energy scale may encounter `new 
physics' in terms of new interactions or states identified in the theory on the way up 
to the GUT scale. Consistency with the unification of the gauge couplings hence  will 
also provide a test of this theory.
 The Planck scale seems to be of no great significance for the present theory since 
gravity is not quantised.

 Returning again to figure~\ref{fronts4}, with the theory developed from the notion of 
a multi-dimensional form of time $\lv$, front~(1) has shown how a Kaluza-Klein related 
unification between gravitational and internal gauge fields can arise naturally out of 
an underlying isochronal symmetry, rather than an isometry, for a world perceived over 
a 4-dimensional spacetime manifold. 
The results presented for front~(2) already establish a substantial connection with 
empirical data in the form of several basic features of the Standard Model.
Within the same framework, generalised for multiple solutions, front~(3) has described 
how the calculational tools of quantum field theory might be incorporated, again 
originating out of the basic principles of this new theory.
 In addition to accounting for small scale laboratory phenomena, culminating in the 
particle concept described  for figure~\ref{gtovac}(b), the large scale structure of 
cosmology 
 is also addressed in front~(4), including the remote reaches of the very early 
universe, leading to the conception of the cosmos summarised in 
figure~\ref{cosevolve}.
  While well defined areas of further development have been identified the 
 progress made and properties uncovered  in all directions, together with the 
simplicity inherent in the founding notion of the flow of time,  add to the overall 
plausibility of the theory.


\pagebreak


\par}

\end{document}